%% file: main.tex
\documentclass[a4paper,12pt,diffusion]{book}

\input{structure.tex} 

\usepackage[T1]{fontenc}
\usepackage{textcomp}

\usepackage{caption}
\usepackage{moresize}
\usepackage{fix-cm}    
\newcommand\HUGEE{\@setfontsize\Huge{30}{30}}
\usepackage{dsfont}
\usepackage{cancel}
\usepackage{dirtytalk}
\usepackage{tikz-cd}
\usepackage{adjustbox}
\usepackage{tabularx}

\usepackage{amsmath}
\usepackage{accents}
\newcommand*{\dt}[1]{%
   \accentset{\mbox{\large .}}{#1}}
\newcommand*{\ddt}[1]{%
   \accentset{\mbox{\large .\hspace{-0.0ex}.}}{#1}}
   \newcommand*{\dddt}[1]{%
   \accentset{\mbox{\large .\hspace{-0.0ex}.\hspace{-0.0ex}.}}{#1}}

\renewcommand{\over}[2]{\left(\frac{#1}{#2}\right)}

\newcommand\blfootnote[1]{%
  \begingroup
  \renewcommand\thefootnote{}\footnote{#1}%
  \addtocounter{footnote}{-1}%
  \endgroup
}

\def\beq{\begin{equation}}
\def\eeq{\end{equation}}
\def\bea{\begin{eqnarray}}
\def\eea{\end{eqnarray}}

\def\IR			{{\mathsmaller{\rm IR}}}
\def\FO			{{\mathsmaller{\rm FO}}}
\def\FD			{F_{\mathsmaller{D}}}
\def\FY			{F_{\mathsmaller{Y}}}
\def\fV			{f_{\mathsmaller{V}}}
\def\GammaV		{\Gamma_{\mathsmaller{V}}}
\def\tauV		{\tau_{\mathsmaller{V}}}
\def\mV			{m_{\mathsmaller{V}}}
\def\MDM		{M_{\mathsmaller{\rm DM}}}

\def\Muni		{M_{\rm uni}}
\def\MuniSM		{M_{\rm uni}^{\mathsmaller{\rm SM}}}
\def\MuniSMs	{M_{\rm uni,0}^{\mathsmaller{\rm SM}}}
\def\MPl		{M_{\mathsmaller{\rm Pl}}}
\def\mZ			{m_{\mathsmaller{Z}}}

\def\gD			{g_{\mathsmaller{D}}}
\def\gSM		{g_{\mathsmaller{\rm SM}}}

\def\gSMdec		{g_{\mathsmaller{\rm SM}}^{\rm dec}}

\def\xfo		{x^\FO}
\def\xfoSM		{x^\FO_{\mathsmaller{\rm SM}} }
\def\sSM		{s_{\mathsmaller{\rm SM}}}
\def\SSM		{S_{\mathsmaller{\rm SM}}}
\def\DSM		{D_{\mathsmaller{\rm SM}}}
\def\DSMbar		{\bar{D}_{\mathsmaller{\rm SM}}}
\def\TD			{T_{\mathsmaller{D}}}
\def\TSM		{T_{\mathsmaller{\rm SM}}}
\def\TtildeSM	{\widetilde{T}_{\mathsmaller{\rm SM}}}
\def\TSMdecay		{T_{\mathsmaller{\rm SM}}^{\mathsmaller{\rm decay}}}
\def\TtildeSM	{\widetilde{T}_{\mathsmaller{\rm SM}}}

\def\rtilde		{\tilde{r}}
\def\gtildeD	{\tilde{g}_{\mathsmaller{D}}}
\def\gtildeSM	{\tilde{g}_{\mathsmaller{\rm SM}}}

\def\aD			{\alpha_{\mathsmaller{D}}}

\def\aSM        {\alpha_{\mathsmaller{\rm SM}}}
\def\ann		{{\rm ann}}
\def\BSF		{\mathsmaller{\rm BSF}}

\def\vrel       {v_{\rm rel}}
\def\DIS			{{\mathsmaller{\text{DIS}}}}

\def\FO			{{\mathsmaller{\text{FO}}}}
\def\LO			{{\mathsmaller{\text{LO}}}}
\def\NLO			{{\mathsmaller{\text{NLO}}}}

\def\SM			{{\mathsmaller{\text{SM}}}}
\def\SC			{{\mathsmaller{\text{SC}}}}
\def\TC			{{\mathsmaller{\text{TC}}}}
\def\RH			{{\mathsmaller{\text{RH}}}}

\def\mDM		{m_{\mathsmaller{\text{DM}}}}

\def\mV		{m_{\mathsmaller{\text{V}}}}

\def\Tstart		{T_{\mathsmaller{\text{start}}}}

\def\TRH		{T_{\mathsmaller{\text{RH}}}}

\newcommand{\MeV}{\,\mathrm{MeV}}

\newcommand{\dilaton}{\chi}
\newcommand{\dilatonvev}{f}

\newcommand{\Tnuc}{T_{\rm nuc}}

\newcommand{\updownarrows}{\uparrow\mathrel{\mspace{-1mu}}\downarrow}

\renewcommand{\upuparrows}{\uparrow\uparrow}

\newcommand{\DP}{\gamma_{\mathsmaller{\rm D}}}

\def\gwp		{\gamma_{\mathsmaller{\text{wp}}}}
\def\gwc		{\gamma_{\mathsmaller{\text{wc}}}}
\def\gcp		{\gamma_{\mathsmaller{\text{cp}}}}
\def\ECM		{E_{\mathsmaller{\text{CM}}}}

\usepackage{chngcntr}
\usepackage{etoolbox}
\usepackage{appendix}

\definecolor{AppendixColor}{RGB}{0, 128, 128}
\AtBeginEnvironment{subappendices}{%
\chapter*{Appendix}
\addcontentsline{toc}{chapter}{{\normalsize Appendices}}

\counterwithin{figure}{section}
\counterwithin{table}{section}
}

\usepackage[sort&compress,numbers,merge]{natbib}

\usepackage{chapterbib}
\usepackage[final]{pdfpages}

\definecolor{Sienna}{RGB}{136, 45, 23}
\definecolor{Cornsilk}{RGB}{255, 248, 220}
\definecolor{LightGoldenrod}{RGB}{255, 238, 139}
\definecolor{Gold}{RGB}{255,215,0}
\definecolor{SaddleBrown}{RGB}{139, 69, 19}
\definecolor{coverTextColor}{RGB}{255, 248, 220}

\usepackage{xintexpr }



\begin{document}

\renewcommand{\hbar}{\mathchar'26\mkern-7mu h}
\def\x{0}

%

%
%
%
%
%
%

\thispagestyle{empty}
\begin{tikzpicture}[remember picture,overlay]

\node[fill=black,minimum width=\paperwidth,
minimum height=\paperheight,anchor=south]%
at (current page.south) {Malaysian \LaTeX\ User Group};

\node[text width=18cm,fill=Sienna,text=white,font=\LARGE\bfseries,
text=Cornsilk,minimum width=\paperwidth,
minimum height=4em,anchor=north,align=center]%
at (current page.north){\LARGE For graduate students and researchers};

\node[draw=Sienna,line width=1mm,rounded corners=.55cm,inner ysep=20pt, inner xsep=-5pt, text width=20cm, text=coverTextColor,font=\HUGE\bfseries, align=center] at (7,-0.7) {\fontsize{32pt}{0pt}\selectfont Beyond the Standard Model Cocktail };

\node[text width=20cm, text=coverTextColor,font=\LARGE\bfseries,align=center] at (7,-4) { \Large A modern and comprehensive review of the major open \\ puzzles in theoretical Particle Physics and Cosmology with a focus on Heavy Dark Matter
};

\node[anchor=north west, inner sep=0pt] at (-3.,-6.5) {\includegraphics[width=1\paperwidth]{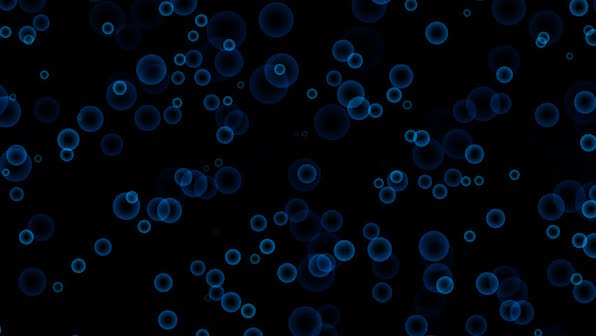}};

\node[text width=18cm, text=coverTextColor,font=\LARGE\bfseries,] at (7.5,-20.5) { \Huge Yann Gouttenoire};
\node[text width=18cm, text=coverTextColor,font=\LARGE\bfseries,] at (7.5,-22.) { \Large G\'{e}raldine Servant};
\node[text width=18cm, text=coverTextColor,font=\LARGE\bfseries,] at (7.5,-23.5) { \Large Filippo Sala};


\vspace*{2\baselineskip}

{\bfseries\itshape\color{LightGoldenrod!50!Gold}
\fontsize{36pt}{46pt}\selectfont
The Wonderful Calmness\par
of Still Life Photos\par}

\vspace*{2\baselineskip}

\node[fill=Sienna,font=\LARGE\bfseries,
text=Cornsilk,minimum width=\paperwidth,
minimum height=3em,anchor=south]%
at (current page.south) {\Large Deutsches Elektronen-Synchrotron DESY and Hamburg University};

\xintifboolexpr { \x = 2}
  {
\node[text width=18cm,fill=Sienna,text=white,font=\LARGE\bfseries,
text=Cornsilk,minimum width=\paperwidth,
minimum height=10em,anchor=north]%
at (current page.north){\Huge PhD thesis \LARGE written after a 3-year research study in DESY with the aim of receiving a doctoral degree from \Huge Hamburg University};
\node[draw=Sienna,line width=1mm,rounded corners=.55cm,inner ysep=15pt, inner xsep=-10pt, text width=20cm, text=coverTextColor,font=\HUGE\bfseries, align=center] at (7,-3.5) {\fontsize{32pt}{0pt}\selectfont Beyond the Standard Model Cocktail \\   \huge Focus on Heavy Dark Matter};
}
{
}

\LARGE\bfseries\color{SaddleBrown!30!black}

\end{tikzpicture}

\thispagestyle{empty}
\newpage

\xintifboolexpr { \x = 0}
{
%
%
%
%
%
%
%
}
{
\begin{center}
\bfseries \huge Supervisors’s Foreword
\end{center}
\vspace{2cm}

This PhD thesis manuscript  provides a remarkable and complete survey of the important questions at the interface between theoretical particle physics and cosmology.
It spans a very large range of physics in early universe cosmology. Out of the eight main chapters, four of them are reviews and the other four discuss new research results.  It goes  through all major open puzzles in theoretical particle physics beyond the Standard Model and presents the status of the art in the proposed solutions.  
The clear logic in the organization of the material, the fact that both standard and more advanced computations are carried out in detail, and the extensive bibliography, make this volume a highly valuable reference for both researchers and students entering this field. 

The first topic reflects many aspects in Dark Matter physics, from the production mechanism with non-trivial effects (Sommerfeld, bound state formation, dilution by entropy injection), to precise phenomenology and indirect detection signatures. 
Over the past decades, Weakly-Interacting-Massive-Particles around the TeV scale have been among the most studied dark matter candidates. Motivated by the absence of experimental evidence for such particles, this thesis explores the possibility that dark matter is much heavier than what is conventionally assumed. The author shows that it is possible to have thermal dark matter with mass beyond the 
unitarity upper limit of about 100 TeV, and motivates data analysis in this mass range by a plethora  of current telescopes, like ANTARES, and future ones, like CTA, LHAASO and KM3NeT. 

The second topic is the analysis of supercooled confinement phase transitions during which the dark matter particle acquires its mass. It is massless outside the bubble and in the form of free quarks while it is composite and massive inside the bubble. This leads to very interesting dynamics that had never been addressed in this context before (string fragmentation,  cosmological deep inelastic scatterings) and that have implications for the final relic abundance of dark matter as a function of the amount of supercooling.  These effects  also affect the calculation of the bubble wall velocity, which is in turn relevant for the determination of the associated gravitational-wave signal and for the production of other early-universe relics. 
Supercooled cosmological phase transitions have become a hot topic in the last few years. Yann Gouttenoire has contributed to bring new insights in this field.

The other line of investigation concerns gravitational waves from cosmic strings.  
While the broad literature on this topic over the last two decades mainly focuses on the predictions within the standard cosmological paradigm, this work shows that the shape of the spectrum can be drastically different from usual predictions due to a variety of physical effects, such as non-standard equations of state due to intermediate matter eras or intermediate inflationary eras as motivated by either heavy unstable particles or supercooled cosmological phase transitions respectively. 
It is shown how to use the potential measurement of a gravitational-wave spectrum produced by cosmic strings to extract information on very high energy physics and in particular derive bounds on the mass and lifetime of heavy unstable particles. These limits would extend well beyond what can presently be inferred from Big Bang Nucleosynthesis constraints.

Research in particle physics beyond the Standard Model has become highly interdisciplinary and much closer to cosmology.
In this thesis, Dr Yann Gouttenoire presents original work that illustrates the non-trivial connections between many topics, and provides a very comprehensive review of this thriving field. We are, in fact, already using this text as a precious reference for both our research work, and for the new students that we are introducing to particle cosmology.\\
\\
\begin{flushright}
G\'{e}raldine Servant and Filippo Sala.
\end{flushright}

}

~\vfill
\thispagestyle{empty}
\newpage

\begin{center}
\bfseries \huge Abstract
\end{center}
\vspace{1.5cm}

This book provides a thorough survey of the important questions at the interface between theoretical particle physics and cosmology.
After discussing the theoretical and experimental physics revolution that led to the rise of the Standard Model in the past century, this volume reviews all major open puzzles, like the hierarchy problem, the small value of the cosmological constant, the matter-antimatter asymmetry, or the dark matter problem, and presents the state-of-the-art in the proposed solutions, with an extensive bibliography.
This manuscript emphasises the fields of thermal dark matter, cosmological first-order phase transitions and gravitational-wave signatures. Comprehensive and encyclopedic, this book could be a rich resource for both researchers and students entering the field.

Written after a 3-year doctoral program in DESY, in addition to the reviews composing two third of the material, one third presents the original PhD research work of the author.
Weakly Interacting Massive Particles (WIMPs) around the TeV scale have since long been among the best-motivated and most studied Dark Matter (DM) candidates. However, the absence of experimental evidence for such particles either in colliders, at telescopes or in underground laboratories, has stimulated model-building and  studies of new methods of detection beyond the WIMP paradigm. This thesis performs several  steps in this direction.

We are in particular interested in the case where DM is much heavier than a TeV. A well-known obstacle for such a realization is the unitarity bound on the annihilation cross-section which constrains the mass of thermal DM to be smaller than $\sim$100 TeV. However, the unitarity bound can be evaded in presence of entropy injection which dilutes the DM abundance. In this thesis, we investigate two possible sources of entropy injection.

First, we study the entropy injection following reheating after an early matter era, when a heavy spectator field, which we choose to be the DM mediator, dominating the energy density of the universe, decays into radiation. We study in detail the corresponding constraints from indirect detection, Cosmic Microwave Background (CMB) and 21-cm, and we show that experimentalists have interests to extend the thermal DM constraints to DM masses beyond the 10/100 TeV range.

Second, we study the entropy injection following reheating after an early stage of vacuum domination generated by a supercooled confining first-order phase transition. Considering the well-motivated scenario where DM is a composite state of a new  confining force, we found that a variety of new effects, e.g. string fragmentation and deep-inelastic-scattering in the early universe, play an important role for setting the final DM abundance. In both cases, we show that we can increase the DM mass up to the EeV scale, 4 orders of magnitude larger than the unitarity bound.

Such scenarios involve non standard cosmologies (either matter era or inflationary era inside the radiation era) and we show that these can be probed using the would-be imprints on the Gravitational-Waves (GW) spectrum from Cosmic Strings (CS) if observed with future GW detectors. In this thesis, we study in detail the computation of the GW spectrum from CS in the presence of non-standard cosmology and the associated constraints on DM models responsible for such a change of cosmology. 


\xintifboolexpr { \x = 0}
{
}
{
}

\includepdf[pages=1]{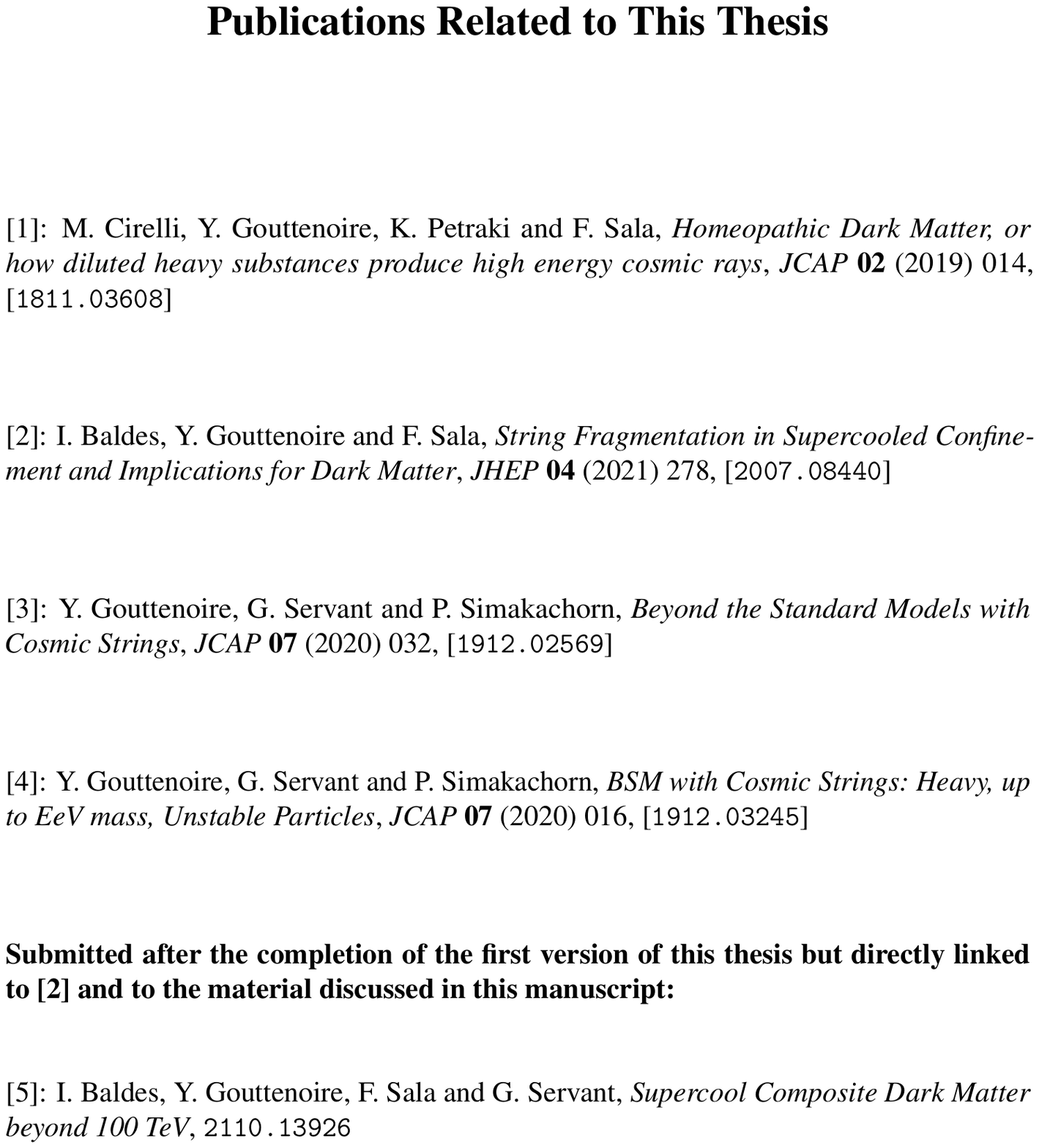}

\newpage

\thispagestyle{empty}

\begin{center}
\bfseries \huge Acknowledgements
\end{center}
\vspace{2cm}

This is with a sincere feeling of gratitude, that I am acknowledging G\'{e}raldine Servant. She first played an important role in my academic life when I was completing my master of theoretical physics in Paris and that the number of PhD subjects seemed infinite. Among the numerous researchers that I visited throughout that year, she was the one, whom enthusiasm for physics, as well as sympathy for the disoriented student I was, convinced me that I could confidently go into her supervision for a PhD, which I consider now as one of the best choices I have ever done in my life. I am indebted for the careful consideration she has manifested during the whole completion of the PhD,  for instance by planning a numerous amount of conferences and summer schools per year or by always being a precious help with administration.  I am indebted for the time she invested in crucial times, like during postdoc applications, and for dissecting every single word of this thesis manuscript with valuable comments and corrections added. 
On the scientific side, I have been impressed by her ambition and determination throughout the completion of projects and by the resilience that she has manifested in presence of difficulties. Her formidable experience and her remarkable mastery in model-building were rewarding numerous times. She has been an extraordinary project designer, with my three main independent projects all being connected to each other in an elegant and coherent manner. I am very grateful to her for having successfully transmitted to me her enthusiasm for fantastic fields of research like cosmological first-order phase transitions and gravitational waves of primordial origin.

I am also indebted to G\'{e}raldine for having introduced me to Filippo Sala, who supervised my master thesis and who co-supervised my PhD thesis. The impact of Filippo on the young physicist I have become since I have met him is absolutely consequent. I have learnt so much from his critical thinking, constantly pushing me to be more skeptical about my claims and to think harder. Filippo is a careful thinker, a perfectionist, who gives a great attention to every single detail in his work. He has constantly been an inspiration, by showing me how to become more meticulous, more rigorous and more accurate. Besides that, I have always had a lot of fun thinking about physics with him, like when spending long days of confinement wondering about what happens when a quark approaches a wall boundary of confined phase. I also thank Filippo for having carefully read my thesis and for the numerous valuable comments added. Finally, I thank Filippo for having been helpful and diligent on the administrative side at multiple times.

I am deeply grateful to G\'{e}raldine, Filippo and Marco Cirelli for having invested so much time and energy - about 200 emails - into trying to build a cotutelle between Hamburg university and Sorbonne university. I would like to especially thank Marco for having moved heaven and earth in Sorbonne university and EDPIF doctoral school. I thank him for having addressed them a 11-page report to state that in 2020 it is impossible to have a PhD affiliated with two universities because the two doctoral certificate templates do not look the same, and because in France the doctoral certificate needs to be printed on a special parchment paper. I thank Brando Bellazzini and Benoit Estienne for having accepted to play the role of scientific tutor and mentor at EDPIF doctoral school.

I would like to thank Peera Simakachorn for the absolutely admirable dedication into our projects, which ended up with very thorough and detailed studies, with two of our common papers being 96 and 151 pages.
I thank Iason Baldes for many valuable physics discussions throughout our projects. I thank Kallia Petraki for teaching me Sommerfeld effects and bound state formation while hosting me in her office in LPTHE, and for many insightful physics discussions.

I thank Thomas Konstandin for having accepted to be a reviewer of the thesis, as well as Marco Cirelli, Dieter Horns and G\"unter Sigl for having accepted to be part of the jury. I also thank the anonymous third reviewer for having agreed to award the Summa Cum Laude. I thank Felix Giese and Marco Hufnagel for having translated the abstract in German.

To complete my PhD in the extremely rich and dynamic environment of DESY in Hamburg was a golden opportunity. With about 100 members in the theory group, we are exposed to a tremendous amount of seminars, conferences and interesting discussions. I am grateful to every colleagues and friends who have made the atmosphere so friendly and my time in Hamburg so unique. 
The list would be long and I am sure that I don't need to cite their name for them to know what I think about them. I would simply thank my three partners in crime Adrien, Akshansh and Henrique for unforgettable moments, and Jorinde for her company during the difficult confinement phase. I also thank CrossFit WestGym and Erika for great times.

\thispagestyle{empty}
Je remercie mes amis Claude et Mathieu pour des moments incroyables durant notre master, merci en particulier à Claude pour les nuits passées \`{a} discuter de physique dans les escaliers du b\^{a}timent G du campus de l'ENS Cachan.
Je remercie mon ami Sylvain pour avoir \'{e}t\'{e} un excellent bin\^{o}me de travail pendant nos 2 ans à Orsay et pendant notre ann\'{e}e d'agr\'{e}gation. Une interaction scientifique de qualit\'{e} et une discipline exceptionnelle nous ont permis de tenir le rythme effr\'{e}n\'{e} des 4 mois de pr\'{e}paration des épreuves orales, et je pense que l'on peut en \^{e}tre vraiment fier. Une pens\'{e}e pour mes amis agr\'{e}gatifs pour avoir rendu la pr\'{e}paration du concours si formidable. Merci \`{a} Arnaud Le Diffon pour des corrections de le\c{c}ons de qualit\'{e}.

Je remercie l'équipe enseignante du Magist\`{e}re de Physique Fondamentale d'Orsay pour m'avoir autant subjugu\'{e}. Ces deux ann\'{e}es étaient incroyablement passionnantes, bien au-del\`{a} de mes attentes. La liste de personnes qui m'ont marqu\'{e}es durant ces deux ans est longue. Je voudrais simplement remercier P\'{e}lini, J\'{e}r\'{e}my, Bacchus et Leila pour tous nos moments de rire et de complicité. 

Je remercie mes professeurs Bruno Dernoncourt et Jean-Paul Roux pour m'avoir captiv\'{e} pour la physique ainsi que mon professeur de math\'{e}matiques Pierre-Jean Hormi\`{e}re pour avoir cru en mon potentiel et m'avoir recommand\'{e} le Magist\`{e}re d'Orsay.
Je remercie ma bande de sup 1 cr\'{e}tins et l'internat Claude Fauriel pour avoir fait de nos deux ans de pr\'{e}pa une aventure humaine exceptionnelle, qui restera \`{a} jamais grav\'{e}e dans nos m\'{e}moires. Merci Jonathan pour avoir transform\'{e} l'internat en \'{e}cole de spectacle.
Je remercie mon professeur de physique-chimie de premi\`{e}re et terminale au lyc\'{e}e Albert Triboulet, Mr. Thibault,  pour ses qualit\'{e}s de p\'{e}dagogie exceptionnelle.

Je remercie ma bande de riders pour des moments inoubliables dans les bois du Forez ou au cours des nombreux \'{e}t\'{e}s pass\'{e}s \`{a} Ch\^{a}tel. Je remercie Lois pour notre amiti\'{e} sinc\`{e}re. Je remercie la bande des Rozierois, qui est comme une seconde famille. Je remercie Justine pour son soutien et son admiration sans faille alors que j'\'{e}tais si focalis\'{e} sur les \'{e}tudes. Je remercie mon fr\`{e}re, Houda, \'{E}douard, Henrique, Sarah et Luis, mais aussi Eva, \'{E}lisa, Florian, Mathilde, Cyrielle, Cl\'{e}ment, Simon et Antoine, pour leur compagnie durant le confinement \`{a} Rozier. Je remercie Bertand Ph\'{e}lut pour les cours de tennis. Je remercie Dounia pour sa grandeur d'\^{a}me.

Je d\'{e}die mon travail \`{a} ma famille. D'abord \`{a} ma maman, disparue en 2008, pour son amour inconditionnel et sa g\'{e}n\'{e}rosit\'{e} in\'{e}galable. \`{A} mon papa pour son soutien inconditionnel, pour sa curiosit\'{e} hors-du-commun, et  pour m'avoir transmis son int\^{e}ret pour les grandes questions. Ensuite \`{a} mon fr\`{e}re pour notre complicit\'{e} sans \'{e}gal et notre admiration inconditionnelle l'un pour l'autre.
Je d\'{e}die \'{e}galement mon travail \`{a} Corinne, mes grand-parents, mes tantes, mes oncles, mes cousines et mes cousins pour leur gentillesse et leur soutien, et en particulier \`{a} mes grand-m\`{e}res pour leur g\'{e}n\'{e}rosit\'{e} infinie.
~\vfill
\thispagestyle{empty}
\newpage

%
%
%
%

\chapterimage{blackboard} 

\pagestyle{empty} 

\tableofcontents 

\cleardoublepage 

\pagestyle{fancy} 

\include{chap1}
\setcounter{NAT@ctr}{0}
\include{chap2}
\setcounter{NAT@ctr}{0}

\include{chap3}
\setcounter{NAT@ctr}{0}

\include{chap4}
\setcounter{NAT@ctr}{0}
\include{chap5}
\setcounter{NAT@ctr}{0}
\include{chap6}
\setcounter{NAT@ctr}{0}

\include{chap7}
\setcounter{NAT@ctr}{0}

\include{chap8}
\setcounter{NAT@ctr}{0}

\include{chap9}
\setcounter{NAT@ctr}{0}

\include{chap10}
\setcounter{NAT@ctr}{0}


%
\newpage
\thispagestyle{empty}

\begin{center}
\bfseries \huge About the Author
\end{center}
\vspace{2cm}

\begin{figure}[h!]
\begin{center}
\includegraphics[width=0.43\textwidth]{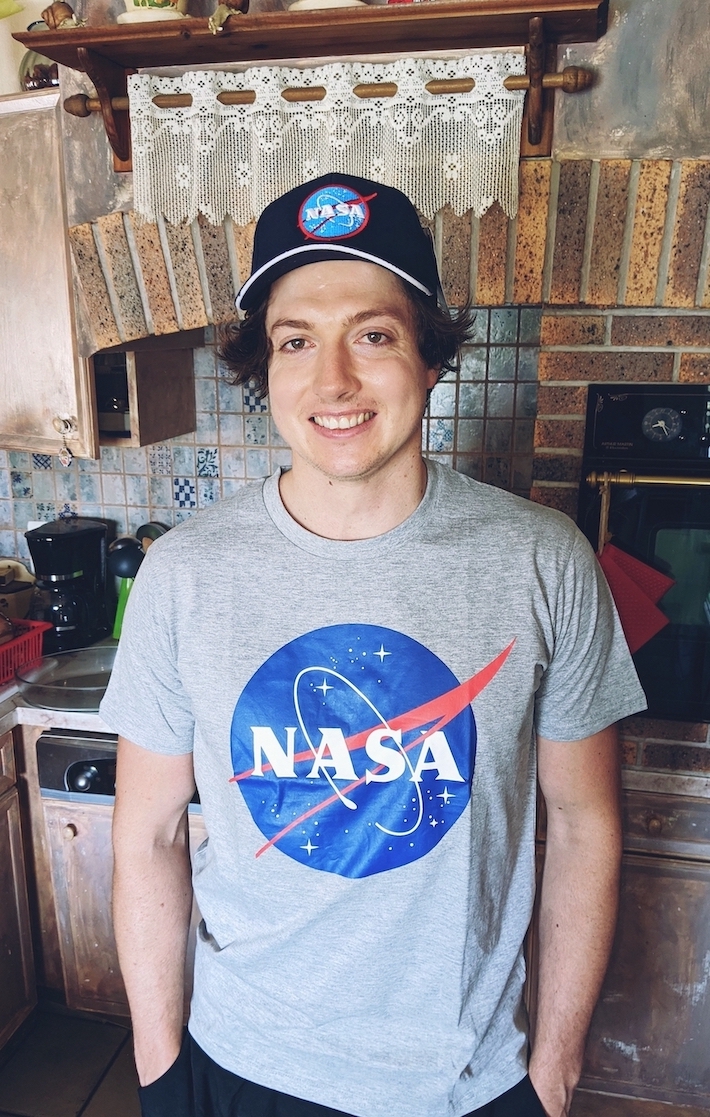} 
\end{center}
\captionsetup{labelformat=empty,listformat=empty}
\caption{}
\end{figure}

\vspace{1cm}

During his Bachelor, Yann Gouttenoire had the opportunity to benefit from the excellent quality of courses at the Magistere of Fundamental Physics in Orsay, France.  The next year in 2015, he received a 2-year scholarship at \'{E}cole Normale Sup\'{e}rieure Paris-Saclay. The first year, he got awarded the ``Agr\'{e}gation'' while in the second year he earned his MS in Theoretical Physics from \'{E}cole Normale Sup\'{e}rieure rue d'Ulm. In 2017, he joined DESY in Hamburg, one of the world's leading center for high-energy particle-physics research, where he studied the topics presented in this book. He defended his PhD on the 29th of September 2020 after which he got awarded the highest distinction from the University of Hamburg (``Summa Cum Laude''). In 2021, he earned an international Postdoctoral Fellowship in Tel Aviv University founded by the Azrieli Foundation.




\end{document}

%% file: structure.tex
\usepackage{tikz-feynman}
\usepackage{mathtools,relsize}
\usepackage{slashed} 
\usepackage[normalem]{ulem}
\usepackage{placeins}
\usepackage{lscape}
\usepackage{multirow}
\usepackage{breakcites}

\usepackage{graphicx} 
\graphicspath{{Pictures/}} 

\usepackage{lipsum} 

\usepackage{tikz} 

\usepackage[english]{babel} 

\usepackage{enumitem} 
\setlist{nolistsep} 

\usepackage{booktabs} 

\usepackage{xcolor,color} 
\definecolor{ocre}{RGB}{0,162,255}

\usepackage{geometry} 

\usepackage{avant} 
\usepackage{mathptmx} 

\usepackage{microtype} 
\usepackage[utf8]{inputenc} 
\usepackage[T1]{fontenc} 

\usepackage{calc} 
\usepackage{makeidx} 
\makeindex 

\usepackage{titletoc} 

\contentsmargin{0cm} 

\titlecontents{part}
	[0cm] 
	{\addvspace{20pt}\bfseries} 
	{}
	{}
	{}

\titlecontents{chapter}
	[1.25cm] 
	{\addvspace{12pt}\large\sffamily\bfseries} 
	{\color{ocre!60}\contentslabel[\Large\thecontentslabel]{1.25cm}\color{ocre}} 
	{\color{ocre}} 
	{\color{ocre!60}\normalsize\;\titlerule*[.5pc]{.}\;\thecontentspage} 

\titlecontents{section}
	[1.25cm] 
	{\addvspace{3pt}\sffamily\bfseries} 
	{\contentslabel[\thecontentslabel]{1.25cm}} 
	{} 
	{\hfill\color{black}\thecontentspage} 

\titlecontents{subsection}
	[1.25cm] 
	{\addvspace{1pt}\sffamily\small} 
	{\contentslabel[\thecontentslabel]{1.25cm}} 
	{} 
	{\ \titlerule*[.5pc]{.}\;\thecontentspage} 

\titlecontents{figure}
	[1.25cm] 
	{\addvspace{1pt}\sffamily\small} 
	{\thecontentslabel\hspace*{1em}} 
	{} 
	{\ \titlerule*[.5pc]{.}\;\thecontentspage} 

\titlecontents{table}
	[1.25cm] 
	{\addvspace{1pt}\sffamily\small} 
	{\thecontentslabel\hspace*{1em}} 
	{} 
	{\ \titlerule*[.5pc]{.}\;\thecontentspage} 

\titlecontents{lchapter}
	[0em] 
	{\addvspace{15pt}\large\sffamily\bfseries} 
	{\color{ocre}\contentslabel[\Large\thecontentslabel]{1.25cm}\color{ocre}} 
	{}  
	{\color{ocre}\normalsize\sffamily\bfseries\;\titlerule*[.5pc]{.}\;\thecontentspage} 

\titlecontents{lsection}
	[0em] 
	{\sffamily\small} 
	{\contentslabel[\thecontentslabel]{1.25cm}} 
	{}
	{}

\titlecontents{lsubsection}
	[.5em] 
	{\sffamily\footnotesize} 
	{\contentslabel[\thecontentslabel]{1.25cm}}
	{}
	{}

\usepackage{fancyhdr} 

\fancypagestyle{plain}{
	\fancyhead{}%
}

\makeatletter
\renewcommand{\cleardoublepage}{
\clearpage\ifodd\c@page\else
\hbox{}
\vspace*{\fill}
\thispagestyle{empty}
\newpage
\fi}

\usepackage{amsmath,amsfonts,amssymb,amsthm} 

\newtheoremstyle{ocrenumbox}
{0pt}
{0pt}
{\normalfont}
{}
{\small\bf\sffamily\color{ocre}}
{\;}
{0.25em}
{\small\sffamily\color{ocre}\thmname{#1}\nobreakspace\thmnumber{\@ifnotempty{#1}{}\@upn{#2}}
\thmnote{\nobreakspace\the\thm@notefont\sffamily\bfseries\color{black}---\nobreakspace#3.}} 

\newtheoremstyle{blacknumex}
{5pt}
{5pt}
{\normalfont}
{} 
{\small\bf\sffamily}
{\;}
{0.25em}
{\small\sffamily{\tiny\ensuremath{\blacksquare}}\nobreakspace\thmname{#1}\nobreakspace\thmnumber{\@ifnotempty{#1}{}\@upn{#2}}
\thmnote{\nobreakspace\the\thm@notefont\sffamily\bfseries---\nobreakspace#3.}}

\newtheoremstyle{blacknumbox} 
{0pt}
{0pt}
{\normalfont}
{}
{\small\bf\sffamily}
{\;}
{0.25em}
{\small\sffamily\thmname{#1}\nobreakspace\thmnumber{\@ifnotempty{#1}{}\@upn{#2}}
\thmnote{\nobreakspace\the\thm@notefont\sffamily\bfseries---\nobreakspace#3.}}

\newtheoremstyle{ocrenum}
{5pt}
{5pt}
{\normalfont}
{}
{\small\bf\sffamily\color{ocre}}
{\;}
{0.25em}
{\small\sffamily\color{ocre}\thmname{#1}\nobreakspace\thmnumber{\@ifnotempty{#1}{}\@upn{#2}}
\thmnote{\nobreakspace\the\thm@notefont\sffamily\bfseries\color{black}---\nobreakspace#3.}} 
\makeatother

\newcounter{dummy} 
\numberwithin{dummy}{section}
\theoremstyle{ocrenumbox}
\newtheorem{theoremeT}[dummy]{Theorem}

\newtheorem{exerciseT}{Exercise}[chapter]
\theoremstyle{blacknumex}
\newtheorem{exampleT}{Example}[chapter]
\theoremstyle{blacknumbox}

\newtheorem{definitionT}{Definition}[section]
\newtheorem{corollaryT}[dummy]{Corollary}
\theoremstyle{ocrenum}

\RequirePackage[framemethod=default]{mdframed} 

\newmdenv[skipabove=7pt,
skipbelow=7pt,
backgroundcolor=black!5,
linecolor=ocre,
innerleftmargin=5pt,
innerrightmargin=5pt,
innertopmargin=5pt,
leftmargin=0cm,
rightmargin=0cm,
innerbottommargin=5pt]{tBox}

\newmdenv[skipabove=7pt,
skipbelow=7pt,
rightline=false,
leftline=true,
topline=false,
bottomline=false,
backgroundcolor=ocre!10,
linecolor=ocre,
innerleftmargin=5pt,
innerrightmargin=5pt,
innertopmargin=5pt,
innerbottommargin=5pt,
leftmargin=0cm,
rightmargin=0cm,
linewidth=4pt]{eBox}	

\newmdenv[skipabove=7pt,
skipbelow=7pt,
rightline=false,
leftline=true,
topline=false,
bottomline=false,
linecolor=ocre,
innerleftmargin=5pt,
innerrightmargin=5pt,
innertopmargin=0pt,
leftmargin=0cm,
rightmargin=0cm,
linewidth=4pt,
innerbottommargin=0pt]{dBox}	

\newmdenv[skipabove=7pt,
skipbelow=7pt,
rightline=false,
leftline=true,
topline=false,
bottomline=false,
linecolor=gray,
backgroundcolor=black!5,
innerleftmargin=5pt,
innerrightmargin=5pt,
innertopmargin=5pt,
leftmargin=0cm,
rightmargin=0cm,
linewidth=4pt,
innerbottommargin=5pt]{cBox}

\makeatletter
\renewcommand{\@seccntformat}[1]{\llap{\textcolor{ocre}{\csname the#1\endcsname}\hspace{1em}}}                    
\renewcommand{\section}{\@startsection{section}{1}{\z@}
{-4ex \@plus -1ex \@minus -.4ex}
{1ex \@plus.2ex }
{\normalfont\large\sffamily\bfseries}}
\renewcommand{\subsection}{\@startsection {subsection}{2}{\z@}
{-3ex \@plus -0.1ex \@minus -.4ex}
{0.5ex \@plus.2ex }
{\normalfont\sffamily\bfseries}}
\renewcommand{\subsubsection}{\@startsection {subsubsection}{3}{\z@}
{-2ex \@plus -0.1ex \@minus -.2ex}
{.2ex \@plus.2ex }
{\normalfont\small\sffamily\bfseries}}                        
\renewcommand\paragraph{\@startsection{paragraph}{4}{\z@}
{-2ex \@plus-.2ex \@minus .2ex}
{.1ex}
{\normalfont\small\sffamily\bfseries}}

\newcommand{\@mypartnumtocformat}[2]{%
	\setlength\fboxsep{0pt}%
	\noindent\colorbox{ocre!20}{\strut\parbox[c][.7cm]{\ecart}{\color{ocre!70}\Large\sffamily\bfseries\centering#1}}\hskip\esp\colorbox{ocre!40}{\strut\parbox[c][.7cm]{\linewidth-\ecart-\esp}{\Large\sffamily\centering#2}}%
}

\newcommand{\@myparttocformat}[1]{%
	\setlength\fboxsep{0pt}%
	\noindent\colorbox{ocre!40}{\strut\parbox[c][.7cm]{\linewidth}{\Large\sffamily\centering#1}}%
}

\renewcommand{\part}[1]{
\ifnum \c@secnumdepth >-2\relax%
\refstepcounter{part}%
\addcontentsline{toc}{part}{\texorpdfstring{\protect\@mypartnumtocformat{\thepart}{#1}}{\partname~\thepart\ ---\ #1}}
\else%
\addcontentsline{toc}{part}{\texorpdfstring{\protect\@myparttocformat{#1}}{#1}}%
\fi%
}

\newif\ifusechapterimage
\usechapterimagetrue
\newcommand{\thechapterimage}{}%
\newcommand{\chapterimage}[1]{\ifusechapterimage\renewcommand{\thechapterimage}{#1}\fi}%
\newcommand{\autodot}{.}
\def\@makechapterhead#1{%
{\parindent \z@ \raggedright \normalfont
\ifnum \c@secnumdepth >\m@ne
\if@mainmatter
\begin{tikzpicture}[remember picture,overlay]
\node at (current page.north west)
{\begin{tikzpicture}[remember picture,overlay]
\node[anchor=south west, inner sep=0pt] at (0,-10) {\ifusechapterimage\includegraphics[width=\paperwidth]{\thechapterimage}\fi};
\draw[anchor=west,text width=16cm] (\Gm@lmargin-1cm,-8.5cm) node[line width=2pt,rounded corners=15pt,draw=ocre,fill=white,fill opacity=0.5,text=black,text opacity = 1,font=\LARGE\bfseries,inner sep=15pt] {\huge\sffamily\bfseries\color{black}\thechapter\autodot~#1\strut};
\end{tikzpicture}};
\end{tikzpicture}
\else
\begin{tikzpicture}[remember picture,overlay]
\node at (current page.north west)
{\begin{tikzpicture}[remember picture,overlay]
\node[anchor=south west, inner sep=0pt] at (0,-10){\ifusechapterimage\includegraphics[width=\paperwidth]{\thechapterimage}\fi};
\draw[anchor=west] (\Gm@lmargin-1cm,-8.5cm) node [line width=2pt,rounded corners=15pt,draw=ocre,fill=white,fill opacity=0.5,inner sep=15pt]{\strut\makebox[22cm]{}};
\draw[anchor=west] (\Gm@lmargin-1cm,-8.5cm) node {\huge\sffamily\bfseries\color{black}#1\strut};
\end{tikzpicture}};
\end{tikzpicture}
\fi\fi\par\vspace*{270\p@}}}

\def\@makeschapterhead#1{%
\begin{tikzpicture}[remember picture,overlay]
\node at (current page.north west)
{\begin{tikzpicture}[remember picture,overlay]
\node[anchor=south west, inner sep=0pt] at (0,-10) {\ifusechapterimage\includegraphics[width=\paperwidth]{\thechapterimage}\fi};
\draw[anchor=west] (\Gm@lmargin-2cm,-9cm) node [line width=2pt,rounded corners=15pt,draw=ocre,fill=white,fill opacity=0.5,inner sep=15pt]{\strut\makebox[22cm]{}};
\draw[anchor=west] (\Gm@lmargin+.1cm,-9cm) node {\huge\sffamily\bfseries\color{black}#1\strut};
\end{tikzpicture}};
\end{tikzpicture}
\par\vspace*{270\p@}}
\makeatother

\usepackage{hyperref}
\hypersetup{hidelinks,backref=true,pagebackref=true,hyperindex=true,colorlinks=false,breaklinks=true,urlcolor=ocre,bookmarks=true,bookmarksopen=false}

\usepackage{bookmark}
\bookmarksetup{
open,
numbered,
addtohook={%
\ifnum\bookmarkget{level}=0 
\bookmarksetup{bold}%
\fi
\ifnum\bookmarkget{level}=-1 
\bookmarksetup{color=ocre,bold}%
\fi
}
}

%% file: chap1.tex
\chapterimage{spaceman_readytogo} 
\chapter{Introduction}


The current knowledge of particle physics has been theoretically achieved in the early 70s and summarized in a concise theoretical model called the Standard Model of Elementary Particles (SM).
The discovery at the Large Hadron Collider (LHC) in 2012 \cite{Aad:2012tfa, Chatrchyan:2012xdj} of the last missing piece of the SM, the Higgs boson, quantum excitation of the field giving the particles their mass, has finally validated the entire SM. In spite of its elegant simplicity, the SM suffers from important theoretical issues. Why is gravity so weak with respect to the other forces while we naively expect quantum corrections to equilibrate their strengths? Why do neutrinos have non-vanishing masses while the SM predicts they should be massless ? Why is the electron $300$ $000$ times lighter than the top quark ? Why does the strong force seem to be invariant under Charge-conjugation and Parity ?

The production of elementary particles as heavy as the Higgs boson at colliders requires a huge energy, around the TeV scale. This corresponds to the plasma temperature in the very first moments of the universe, $\sim 10^{-13}~$seconds after the Big-Bang. Hence, the study of elementary particles at very small length scales is closely connected to the study of the early universe a fraction of a second after the Big-Bang. A good understanding of the phenomena taking place at the earliest epochs is necessary to explain the state of the universe today, as for instance the abundance of light elements.

In spite of its remarkable agreement with the data of Planck satellite, the Standard Model of Cosmology suffers from important theoretical issues. Only $5$\% of the energy content of our universe has been theoretically identified. We observe that $26$\% is made of Dark Matter (DM), a fluid responsible for making the galaxies spinning faster and whose nature is unknown. The remaining $69$\% constitutes the energy of the vacuum, responsible for the accelerated expansion of the universe, and for which the theoretical prediction from the SM of elementary particles is off by $120$ orders of magnitude. Another important challenge is how to explain why our universe is only filled with matter and does not also contain anti-matter. 

The discovery by LIGO in 2015 \cite{TheLIGOScientific:2016qqj} of Gravitational Waves (GW) of astrophysical origin has opened a new avenue of investigation of our universe. With future GW experiments like SKA, LISA, CE or ET, there is hope for discovering GW of cosmological origin, produced during the early universe, at the time when the universe was still opaque to light.

The main goal of the research carried out during this thesis is to investigate the possibility that the DM of the universe is composed of particles much heavier than the TeV scale, to motivate such a scenario and to present distinct ways to probe it. As this is a very interdisciplinary topic, we will also discuss important topics in early universe dynamics such as cosmological phase transitions and gravitational-wave probes of new physics beyond the Standard Model. My main new contributions can be summarized as follows:
\begin{itemize}
\item
Effect of entropy injection from the decay of the DM mediator on the DM abundance and implications for its mass prediction and searches at indirect detection experiments.
\item
Investigation of  new dark matter production mechanisms during a confining supercooled first-order phase transition: string fragmentation and     deep-inelastic scattering.
\item
Probing a non-standard matter era induced by DM mediators, or a second period of inflation induced by the supercooled first-order phase transition generating the DM mass, using gravitational waves from cosmic strings.
\end{itemize}

Out of the eight main chapters in this thesis, four of them (Chapters \ref{chap:secluded_DM}, \ref{chap:SC_conf_PT}, \ref{chap:cosmic_strings} and \ref{chap:DM_GW_CS}) are reporting my new contributions that have led to four research articles. The other chapters (Chapters \ref{chap:SM_particle}, \ref{chap:SM_cosmology}, \ref{chap:DM} and \ref{chap:1stOPT}) are detailed reviews which may a priori appear disconnected from my main work but they were included as they reflect work that I have been carrying, and on which future upcoming articles will be based. Most of the figures are mine, using existing results in the literature, although a large number of these figures have never appeared in the literature.

\paragraph{Plan of the thesis.}
The plan of the thesis is described below.

\paragraph{Chap.~\ref{chap:SM_particle}: Standard Model of Elementary Particles.}
Chap.~\ref{chap:SM_particle} introduces the SM and  tries to highlight the main historical details, either theoretical breakthroughs or experimental discoveries, leading to its completion. We introduce the different open problems and discuss  the main proposed solutions. 

\paragraph{Chap.~\ref{chap:SM_cosmology}: Standard Model of Cosmology.}
Chap.~\ref{chap:SM_cosmology} starts by introducing the Standard Model of cosmology and computing the abundance of baryons, neutrinos and the time of photon decoupling. We show the necessity for a period of inflation in order to explain the striking flatness and homogeneity of our universe. We motivate the use of Gravitational-Waves (GW) as a new search strategy for probing the early universe.
Open problems are introduced together with the main proposed solutions.

\paragraph{Chap.~\ref{chap:DM}: Thermal Dark Matter.}
Chap.~\ref{chap:DM} is a review on thermal Dark Matter. After having discussed the two production mechanisms of thermal Dark Matter, freeze-in and freeze-out, we focus on the second on, which is motivated by the WIMP miracle. We compute the DM abundance from the eV range to the $100$~TeV range and we present the constraints from Lyman-$\alpha$, CMB, Big-Bang Nucleosynthesis (BBN), indirect detection, direct detection. We then give more attention to the case where DM is heavy, a limit where non-perturbative Sommerfeld effects are important and where unitarity is saturated, and which is the case of interest for Chap.~\ref{chap:secluded_DM} and follow-up studies of Chap.~\ref{chap:SC_conf_PT} (e.g. \cite{Baldes:2021aph}).  All the figures in the chapter are mine (Fig.~\ref{fig:feyn_XX_ZZ}, \ref{FI_VS_FO_Boltzmann}, \ref{FI_VS_FO}, \ref{fig:DMabundance_WIMP}, \ref{fig:sommerfeld-ladder-feynman} and \ref{fig:sommerfeld_factor}).

\paragraph{Chap.~\ref{chap:secluded_DM}: Homeopathic Dark Matter.}
\textit{Based on studies performed between May 2017 and January 2019 with Marco Cirelli, Kalliopi Petraki and Filippo Sala.} \\
\textbf{Publication \cite{Cirelli:2018iax}: } \\
\vspace{0.5cm}

Due to the \textbf{unitarity bound}, Dark Matter (DM) can not be heavier than $\sim 100$~TeV if its abundance is set by the standard freeze-out scenario, otherwise it overcloses the universe. Nevertheless, we show that we can extend the unitarity bound to the EeV range by introducing a long-lived heavy DM mediator which decays into radiation before BBN, hence \textbf{injecting entropy} and \textbf{diluting DM}.
Large DM mass corresponds to large coupling constant, implying the presence of non-perturbative Sommerfeld effects which enhance the indirect-detection constraints. We consider the possibility of having heavy thermal DM in a benchmark toy model where a $U(1)_{D}$ is kinetically coupled to $U(1)_Y$. We also study the associated phenomenology. We find that entropy injection opens a large parameter space with DM masses up to \textbf{EeV scale}, far beyond the standard unitarity bound. This motivates future indirect-detection experiments aiming to explore cosmic rays beyond the TeV range, e.g. {\sc Cta, Lhaaso, Km3net, Herd, Iss-Cream}. However, the region far beyond $100$ TeV  (e.g. $10$ PeV) is far beyond the reach of standard methods: colliders, telescopes and direct detection, which motivates new methods of detection, see Chap.~\ref{chap:DM_GW_CS}.
The Chap.~\ref{chap:secluded_DM} is a minimal adaptation of the publication \cite{Cirelli:2018iax}.

\paragraph{Chap.~\ref{chap:1stOPT}: First-order Cosmological Phase Transition.}
The Chap.~\ref{chap:1stOPT} is devoted to 1st-Order cosmological Phase Transitions (1stOPT), and motivates the research work exposed in Chap.~\ref{chap:SC_conf_PT} and published in \cite{Baldes:2020kam} as well as other ongoing works, e.g. \cite{Baldes:2021aph}. We start by considering the physics involved during \textbf{bubble nucleation}: we present the standard techniques for computing an \textbf{effective potential} at finite-temperature, and we remind how to use it in order to compute the \textbf{bounce action}, numerically or analytically. Second, we consider the physics involved during \textbf{bubble propagation}: we present various methods, valid in different regimes, for computing the \textbf{bubble wall velocity}. A special attention is given to the computation of the friction pressure in the ballistic regime, at leading order and next-to-leading-order in gauge coupling constant. Finally, we discuss the physics involved during \textbf{bubble collision}. We remind how to compute the resulting GW emission, using the most recent results in the literature. Particularly, we discuss how to compute the energy transfer to sound-waves with arbitrary speed of sound. we conclude the chapter by introducing the two classes of \textbf{nearly-conformal potential }present in the litterature leading to supercool first-order phase transitions. First, we introduce the \textbf{Coleman-Weinberg} potential, valid in the weakly-coupled regime. Second, we introduce the \textbf{light-dilaton} potential, which assumes a strongly-coupled regime, or its 5D-holographic dual, the Goldberger-Wise potential, which assumes the existence of a warped extra-dimension. For each of these models, we compute the nucleation temperature and GW parameters $\alpha$ and $\beta$ , both numerically and analytically. All the figures in the chapter are mine (Fig.~\ref{fig:Veff_dim6}, \ref{fig:O3_different_methods_comparison}, \ref{fig:thick_vs_thin_wall}, \ref{fig:Dim6_Tn_Lambda}, \ref{fig:DeltaP_VS_gamma_T_vs_R}, \ref{eq:DeltaV_alpha_contours_v_LO}, \ref{fig:OmegaGW_scalarField_Litterature}, \ref{fig:SC_PT_constraints}, \ref{fig:detonation_deflagration_profile}, \ref{fig:efficiency_sw_RMS_fluid_velocity_vs_vw}, \ref{fig:SW_PT_constraints}, \ref{fig:CW_gX_Tnuc_alpha_beta} and \ref{fig:light_dilaton_msigma_Tnuc_alpha_beta}).

\paragraph{Chap.~\ref{chap:SC_conf_PT}: Supercooled Composite Dark Matter.}
\textit{Based on studies performed between October 2018 and July 2020 with Iason Baldes, Filippo Sala and Geraldine Servant.}\\ 
\textbf{Publications \cite{Baldes:2020kam,Baldes:2021aph}: } \\
\vspace{0.5cm}

First-order phase transitions driven by \textbf{nearly-conformal potential} are known to predict a potentially large hierarchy between the critical temperature, when the two minima coincide, and the nucleation temperature, when the phase transition (PT) completes. For intermediate temperatures, the universe is dominated by the vacuum energy of the PT, which leads to a short period of inflation, also known as \textbf{supercooling}. An interesting cosmological consequence is the dilution of any relic produced beforehand. The dilution of the DM abundance had previously only been studied in the case of a supercooled PT driven by a Coleman-Weinberg potential, generated by quantum corrections in the weakly-coupled regime. The aim of our study is to consider the case where the supercooled PT is driven by a nearly-conformal potential arising from strongly-coupled dynamics. A possible realization is the light-dilaton potential, motivated by holography. In contrast to the weakly-coupled scenario where DM already exists before the PT, in the strongly-coupled regime, the DM is \textbf{composite} and is only formed when the techni-quanta enter inside the expanding bubble of confined phase. In the limit of large supercooling, the distance between techni-quarks, set by the inverse temperature, is large with respect to the confining scale. Hence, it is energetically favorable for the techni-quarks to form a \textbf{flux tube} with the bubble wall instead of with their neighbors. 
This has three important consequences.
First, due to the large kinetic energy of the incoming quark in the wall frame, the flux tube breaks into many hadrons, similarly to the \textbf{string fragmentation} which follows the decay of a $Z$-boson into a $q\bar{q}$ pair in the SM.
Second, due to charge conservation, a quark is \textbf{ejected} from the bubble.
Third, the rest frame of the gluon string between the incoming quark and the wall is boosted with respect to the plasma frame. Hence, the string fragments have large momenta and can undergo \textbf{deep-inelastic-scattering} (DIS) with the different mediums which they encounter, which enhances further the abundance of composite states. Such mediums can be the diluted bath, the particles from the preheated bath or the string fragments from the neighboring bubbles. We show that the second possibility is the most relevant.
As a consequence of these new effects which we point out, the resulting DM abundance, assuming that DM is a composite state of the confining sector, is very different from the trivial case where it only receives a dilution factor. Due to the enhanced DM production, we find that in order to get the correct DM abundance, the required supercooling must be larger than in the weakly-coupled scenario.
Chap.~\ref{chap:SC_conf_PT} is entirely devoted to this study and is almost entirely identical to the publication \cite{Baldes:2020kam}.

\paragraph{Chap.~\ref{chap:cosmic_strings}: Beyond the Standard Models with Gravitational Waves from Cosmic Strings.}
\textit{Based on studies performed between October 2018 and April 2020 with Geraldine Servant and Peera Simakachorn.}\\
\textbf{Publication \cite{Gouttenoire:2019kij}: }\\ 
\vspace{0.5cm}

Cosmic Strings (CS) are topological defects arising from the breaking of a $U(1)$ symmetry. The reason why they have been the subject of so many studies for the last 40 years is probably due to their universal behavior called \textbf{scaling regime}. The fraction of energy density stored in long cosmic strings, $\Omega_{\infty} \propto \rho_{\rm \infty}/\rho_c$ with $\rho_\infty$ the energy density of infinitely-long strings and $\rho_c$ the critical density, is conserved through the evolution of the universe, from formation of the network until today. This results from a conspiracy between Hubble expansion and energy loss into loop formation. Hence, CS do not overclose the universe, as the other topological defects, domain walls and monopoles, would do, neither they redshift away: CS are always there.

An important consequence is that CS constitute a \textbf{long-standing} source of GW, which leads to a \textbf{scale-invariant} GW spectrum (a power law). Note that the GW are not emitted by infinitely-long strings which have a conserved topological charge, but they are instead emitted by \textbf{loops}.
A second intriguing property is that in a radiation-dominated universe, the GW spectrum is \textbf{flat} in frequency. A property which we can impute to the presence of a fortuitous cancellation between the loop formation rate and the redshift factor. However, this is not the case anymore if the loop formation occurs in a different background, e.g. kination and matter domination where the slope turns to $f^1$ and $f^{-1}$, respectively. Also, if loop formation occurs during a period of inflation, the scaling regime is violated and loop formation freezes. In our study, we propose to use the would-be detection of GW spectrum from CS by future GW experiments to scan for the presence of a \textbf{non-standard cosmology} in the early universe. 

In Chap.~\ref{chap:cosmic_strings}, we first review in great details the computation of the GW spectrum from CS including deviation from the scaling regime due to change of cosmology and deviation from the Nambu-Goto approximation due to quantum effects in the presence of small-scale structures.
Then, we provide the energy scales of an early matter era, as well as the energy scale and duration of a second period of inflation, which we could detect with future GW interferometers.
Chap.~\ref{chap:cosmic_strings} is a minimal adaptation of the publication \cite{Gouttenoire:2019kij}. The sections dealing with kination and metastable cosmic strings have been removed. The figures Fig.~\ref{fig:loopTraj} and Fig.~\ref{fig:Impact_loop_size_distrib_Ringeval_vs_BlancoPillado} are mine and have been added during the redaction of the manuscript.

\paragraph{Chap.~\ref{chap:DM_GW_CS}: Probe Heavy Dark Matter with Gravitational Waves from Cosmic Strings.}
\textit{Based on studies performed between October 2018 and April 2020 with Geraldine Servant and Peera Simakachorn.}\\
\textbf{Publication \cite{Gouttenoire:2019rtn}: }\\ 
\vspace{0.5cm}

In Chap.~\ref{chap:DM_GW_CS}, we suppose that the early period of non-standard cosmology introduced in the previous chapter is induced by particle dynamics. First, we consider that the \textbf{early matter-domination era} is induced by a \textbf{heavy unstable relic}. We give model-independent constraints on the lifetime $\tau_X$ and the product mass times abundance $m_X\,Y_X$ of the relic, which extends the current BBN contraints up to $\tau_X \lesssim 10^{-17}~\rm s$. Then, we focus on the model of \textbf{Homeopathic DM} presented in Chap.~\ref{chap:secluded_DM} and we show that the large DM mass region which was opened by the entropy dilution is within the reach of GW experiments if a string network of tension $G\mu \gtrsim 10^{-15}$ is detected. 
Finally, we apply our model-independent constraints on \textbf{second period of inflation} to the model of \textbf{Supercooled Composite DM} presented in Chap.~\ref{chap:SC_conf_PT}. We show that the regions of the parameter space leading to the correct DM abundance are within the reach of future GW experiments.
Being sensitive to large DM mass, this new method of detection offers an excellent complementary with standard techniques based on colliders, indirect-detection and direct-detection whose sensitivities generally fade away beyond the TeV scale.
Chap.~\ref{chap:DM_GW_CS} is a minimal adaptation of the publication \cite{Gouttenoire:2019rtn}. The sections dealing with scalar oscillating moduli, scalar particles produced gravitationally or though the Higgs portal have been removed. Sec.~\ref{sec:SC_DM_CS} together with its Fig.~\ref{fig:SCDM_GWCS}, were not included in the publication \cite{Gouttenoire:2019rtn} but have instead been realized for the purpose of those notes.

We conclude the thesis manuscript in Chapter 10.
Technical details are reported in a number of Appendices.


%

\xintifboolexpr { \x = 2}
  {
  }
{
\medskip
\small
\bibliographystyle{JHEP}
\bibliography{thesis.bib}
}

%% file: chap2.tex
\chapterimage{subatomic-universe} 
\renewcommand{\arraystretch}{1.5}
\chapter{Standard Model of Elementary Particles}
\label{chap:SM_particle}

The Standard Model (SM) is the theory which describes at the quantum level three of the four fundamental forces - electromagnetism, weak interaction and strong interaction - and all the known elementary particles. It is currently the most rigorous theory of particle physics, with an  unprecedented level of precision and accuracy in its predictions. It is the result of the collaborative work of the greatest physicists in the last century, whom at least seventy of them earned the Nobel prize, since the introduction of the photon by Einstein in 1905 \cite{einstein1905einstein} or the unification of Quantum Mechanics and Special Relativity by Dirac in 1928 \cite{Dirac:1928hu, Dirac:1928ej}. 
The SM has been theoretically achieved after three major breakthroughs in the end 60s / early 70s. 
\begin{enumerate}
\item
The unification of electromagnetism and weak interactions under the gauge group $SU(2)\times U(1)$ by Glashow in 1961 \cite{Glashow:1961tr}, followed by the incorporation of spontaneous symmetry breaking $SU(2)_L\times U(1)_Y \rightarrow U(1)_{\rm e.m.}$ by Weinberg and Salam in 1967 \cite{Weinberg:1967tq,Salam:1968rm} in order to provide a mass for the vector bosons without violating gauge invariance. Weinberg and Salam used the Higgs mechanism developed by many authors in 1964 \cite{Englert:1964et, Higgs:1964ia,Guralnik:1964eu,Higgs:1964pj,Higgs:1966ev}, which was itself inspired from the Nambu-Goldstone theorem \cite{Nambu:1960tm,Goldstone:1961eq} formalized in Quantum Field Theory in 1962 by Goldstone, Weinberg and Salam \cite{Goldstone:1962es}. The renormalizability of such a massive gauge theory is proved in 1972 by  't Hooft and his adviser Veltman who developed the technique of dimensional regularization \cite{tHooft:1971qjg, tHooft:1971akt, tHooft:1972tcz}.
\item
Second, the theoretical justification by Gross, Wilczek and Politzer in 1973 \cite{Gross:1973id, Politzer:1973fx} that Quantum Chromodynamics, a Yang-Mills gauge theory with color triplet quarks and color octet gluons, proposed by Gell-Mann et al. the previous year \cite{Fritzsch:1972jv, Fritzsch:1973pi}, manifests asymptotic freedom, in excellent agreement with proton-electron scattering, and hence is the correct gauge theory for describing strong interactions.
\item
Third, the prediction of a third generation of quarks in order to explain CP violation by Kobayashi and Maskawa in 1973 \cite{Kobayashi:1973fv}.
\end{enumerate}
The Standard Model of elementary particles is reviewed in many excellent books, e.g. \cite{Itzykson:1980rh, Huang:1982ik,Halzen:1984mc,Cheng:1984vwu,Cheng:2000ct,Zinn-Justin:1989rgp,Brown:1992db,Kaku:1993ym, Ryder:1985wq, Donoghue:1992dd,Peskin:1995ev,Weinberg:1995mt, Weinberg:1996kr, Huang:1998qk,Kiselev:2000ts,Atkinson:2002tc,Zee:2003mt,Morii:2004tp, Maggiore:2005qv,Nair:2005iw,  Burgess:2007zi,Cottingham:2007zz,Dine:2007zp,Srednicki:2007qs, Griffiths:2008zz, Bettini:2008zz,McMahon:2009zz,Barnes:2010zzb,Langacker:2010zza,Mann:2010zz,Nagashima:2010zz,Nagashima:2010jma,Nagashima:2014tva,Boyarkin:2011zza,Boyarkin:2011dfa,Robinson:2011lia,Tully:2011zz,Schwartz:2013pla,AlvarezGaume:2012zz,Shifman:2012zz,Shifman:2019fvj,Banks:2014twn,Lancaster:2014pza,Manoukian:2016sxi,Manoukian:2016jpj, bookKane2017,Fritzsch:2017zku,Hamilton:2017gbn,Vergados:2017xtv,Vergados:2017rfe,Wells:2017uxd,Coleman:2018mew,Ecker:2019vda,Gelis:2019yfm,Schmitz:2019hxk,Strickland:2019uej,Strickland:2019pdt,Strickland:2019tnd,Nair:2018cuy,Peskin:2019iig,Larkoski:2019jnv,Iliopoulos:2021mze,Donoghue:2022azh}. 
For historical complements about the period giving rise to the SM, we call attention to \cite{Hoddeson:1997hk,pickering1999constructing,baggott2011quantum,Farmelo:2019pgv,Frampton:2020kki,schweber2020qed}.

\begin{figure}[h!]
\centering
\raisebox{0cm}{\makebox{\includegraphics[width=0.7\textwidth, scale=1]{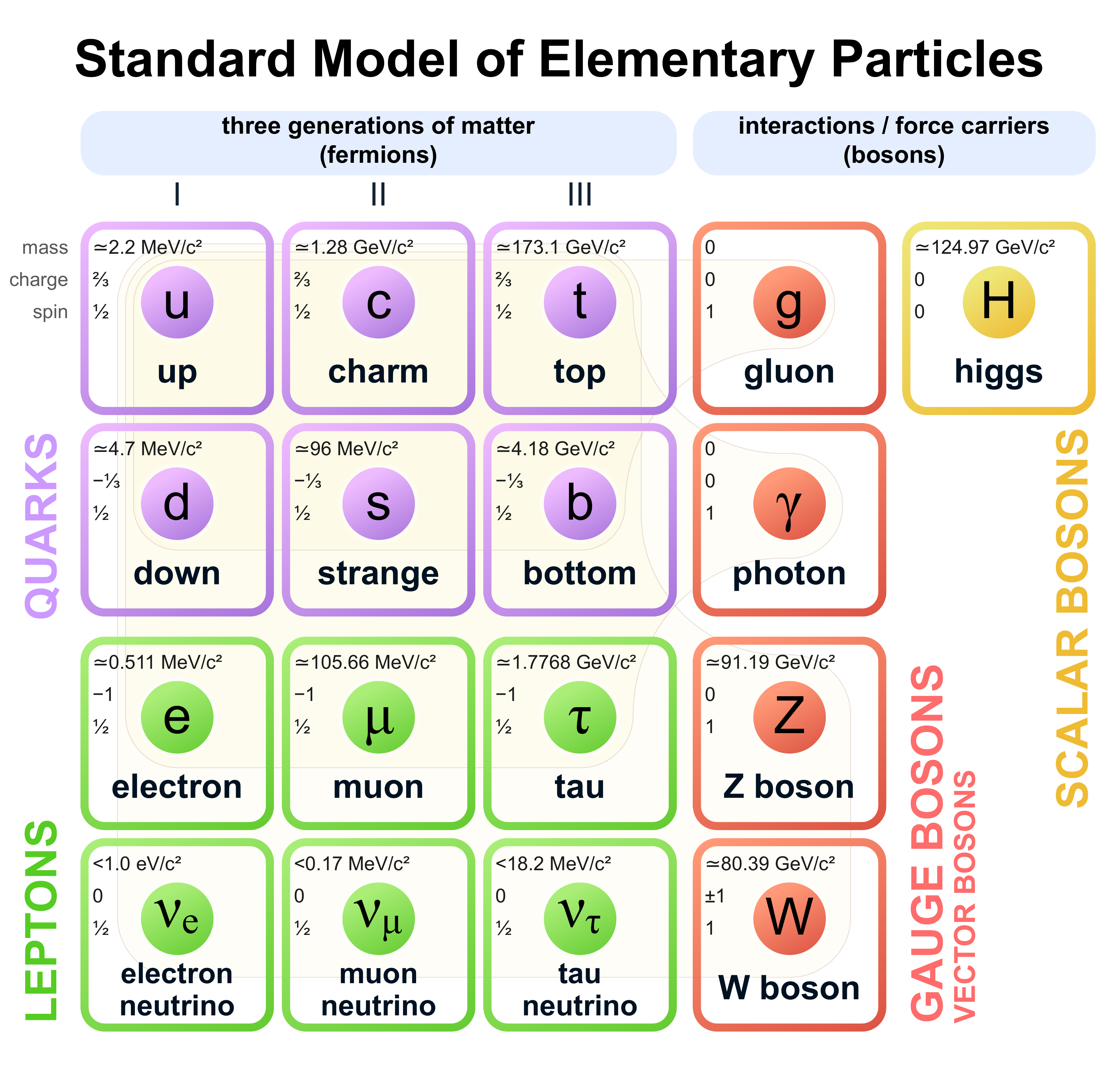}}}
\caption{\it \small Elementary Particles of the Standard Model.}
\label{fig:Standard_Model_of_Elementary_Particles}
\end{figure}

\section{Fields and symmetries}
\subsection{The Lorentz representations}
\label{sec:lorentz_representations}
When Dirac unifies Quantum Mechanics and Special Relativity, he adds the \textbf{spinor} $\psi$, with spin $j=1/2$, to the list of fields compatible with Special Relativity
\begin{equation}
\label{eq:field_lorentz}
\textbf{Spin 0}:~ \rm scalar ~\phi(x), \qquad \textbf{Spin 1/2}: \rm spinor ~\psi_{\sigma}(x), \qquad \textbf{Spin 1}:~ \rm vector ~A_{\mu}(x),
\end{equation}
with $\sigma = 1,2$ and $\mu = 1..4$. The spin $j=0,\, 1/2,\, 1,..$ labels the different representations under the group $SU(2)$ of spatial rotation. Thanks to the spin-statistics connection\footnote{Fierz and Pauli formulate the connection between spin and statistics in 1939/1940 \cite{Fierz:1939zz, Pauli:1940zz}.}, $j$ also tells how the corresponding states occupy the available phase space: even spins obey to the Bose-Einstein statistics whereas odd spins obey to the Fermi-Dirac statistics. Hence, we call them \textbf{bosons} and \textbf{fermions}, respectively.

Actually, there are two nonequivalent spinorial representations of the Lorentz group: the \textbf{right-handed Weyl spinors} $\psi_{R}$ and the \textbf{left-handed Weyl spinors} $\psi_{L}$.\footnote{The Lorentz algebra $\rm so(1,\,3)$ can be decomposed as the tensor product of two commuting $\rm su(2)$ subalgebras
\begin{equation}
\rm so(1,\,3) = su(2) \oplus su(2).
\end{equation} 
Therefore, irreducible representations of the Lorentz group can easily be constructed from the irreducible representations of $\rm su(2)$, which are labeled by $j = 0,\, 1/2,\, 1,$ and so on. Hence, the scalar representation of the Lorentz group is $(0,\,0)$, the spinorial representations are $(0,\, \frac{1}{2})$ and $(\frac{1}{2},\,0)$, the 4-vector representation is $(\frac{1}{2},\, \frac{1}{2})$, ...}
$\psi_R$ and $\psi_L$ are related by a \textbf{parity transformation} $\vec{x} \to - \vec{x}$, so they transform the same way under spatial rotation but in the opposite way under a Lorenz boost.  It is often convenient to merge the two two-component Weyl spinors into a four-component \textbf{Dirac spinor}
\begin{equation}
\psi =  \begin{pmatrix}  \psi_L \\ \psi_R \end{pmatrix},
\label{eq:dirac_spinor_def}
\end{equation} 
and to introduce the \textbf{Dirac conjugate}
\begin{equation}
\bar{\chi} = \chi^{\dagger} \gamma^0, \qquad \text{with} \quad \chi^{\dagger} = {}^{t}\chi^*,
\end{equation}
such that $\bar{\chi}\psi$ is a scalar and $\bar{\chi}\gamma^\mu\psi$ is a 4-vector, while $\bar{\chi}\gamma^5\psi$ is a pseudo-scalar (changes sign under parity) and $\bar{\chi}\gamma^5\gamma^\mu\psi$ is a pseudo-4-vector. We have introduced the Dirac matrices $\gamma^\mu$ which obey to the Clifford algebra
\begin{equation}
\{\gamma^\mu,\, \gamma^\nu \} = 2 \eta^{\mu\nu}, \qquad \text{and} \qquad \gamma^5 = i\gamma^0 \gamma^1\gamma^2 \gamma^3.
\end{equation}
A convenient representation is the Weyl basis, already assumed when writing Eq.~\eqref{eq:dirac_spinor_def}
\begin{equation}
\gamma^0 =  \begin{pmatrix}  0 & 1 \\ 1 & 0 \end{pmatrix}, \qquad \gamma^i =  \begin{pmatrix}  0 & \sigma^i \\ -\sigma^i & 0 \end{pmatrix}, \qquad \gamma^5= \begin{pmatrix}  -1 & 0 \\ 0 & 1 \end{pmatrix},
\end{equation}
where $\sigma^i$ are the Pauli matrices.

\subsection{The gauge interactions}
Simply speaking, the Standard Model is a set of fields among the list given in Eq.~\eqref{eq:field_lorentz}, interacting with each other. The list of quantum fields constituting the SM is shown in Fig.~\ref{fig:Standard_Model_of_Elementary_Particles}. A powerful aspect is that the field content as well as their interactions are entirely dictated by symmetries.
The three forces are described by the gauge structure
\begin{equation}
\label{eq:gauge_structure}
SU(3)_c\times SU(2)_L\times  U(1)_Y  \quad \rightarrow \quad SU(3)_c\times U(1)_{\rm e.m.}
\end{equation}
where each generator of each Lie Group defines a force mediator, each of them being a vector boson of spin $1$. More precisely, the 8 generators of $SU(3)_c$ define the 8 \textbf{gluons}
\begin{equation}
SU(3)_c: \quad G_{\mu}^a, \qquad a= 1...8,
\end{equation}
mediators of the strong force responsible for the binding of the nucleus, with gauge coupling $g_3$.
The three generators of $SU(2)_L$ as well as the unique generator of $U(1)_Y$, which we denote by
\begin{equation}
SU(2)_L: \quad W_{\mu}^a, \qquad a =1..3, \quad \rm and \quad U(1)_Y: \quad B_{\mu},
\end{equation}
mediate the electroweak interactions.
The \textbf{vector bosons} $\mathbf{W^{\pm}}$, responsible for the beta decay are given by linear combinations of $W^1_\mu$ and $W^2_\mu$.
The \textbf{photon}, mediator of the electromagnetism force, and well as the \textbf{vector boson} $\mathbf{Z^0_\mu}$, responsible for neutrino elastic scattering, are given by linear combinations of $W_\mu^3$ and $B_\mu$. We provide more details in Sec.~\ref{sec:EWSB}.
\\

\begin{table}[h!]
\centering
\hspace{-1cm}
\begin{tabular}{ |c c  c c c c|}
 \hline
 & & $\rm SU(3)_c$ &$\rm  SU(2)_L $&  $\rm U(1)_Y$ & Spin \\
 \hline
  &   $\rm B_\mu $ & $\bold{1}$ & $\bold{1}$  & 0 &1   \\
Gauge fields & $W_\mu^a$  &  $\bold{1}$ & $\bold{3}$  & 0 &1   \\
 &   $\rm G_\mu^a $ & $\bold{8}$ & $\bold{1}$  & 0 &1   \\
 \hline 
  &   $\rm Q_L $ & $\bold{3}$ & $\bold{2}$  & $+\frac{1}{6} $&$\frac{1}{2}$  \\
Quarks & $\rm u_R$  &  $\bold{3}$ & $\bold{1}$  & $+\frac{2}{3} $& $\frac{1}{2}$   \\   
 &   $\rm d_R $ & $\bold{3}$ & $\bold{1}$  &$ -\frac{1}{3} $&$ \frac{1}{2}$  \\
 \hline 
  &   $\rm L_L $ & $\bold{3}$ & $\bold{2}$  &$ -\frac{1}{2}$ & $\,\, \frac{1}{2} $  \vspace{-0.4cm}  \\ 
  Leptons \vspace{-0.4cm} & & & & & \\
 & $\rm e_R$  &  $\bold{3}$ & $\bold{1}$  & $ -1$& $\frac{1}{2}$   \\   
  \hline 
  Higgs &$\rm H$  &  $\bold{1}$ & $\bold{2}$  & $+\frac{1}{2}$ & $0$  \\
\hline
\end{tabular}
\caption{\it \small The fields of the Standard Model and their symmetry transformation.}
\label{table:SM_content}
\end{table}

\subsection{The matter content}
In addition to the gauge fields which mediate the interactions and are adjoint representations of the gauge groups, there are the matter fermions, which are fundamental representations of the gauge groups through which they interact. On the one hand, there are \textbf{quarks}, which transform as triplet under $SU(3)_c$ and come as three $SU(2)_L$ doublet $Q_L^n$ and six $SU(2)_L$ singlets $u_R^n$ and $d_R^n$ where $n=1,\,2,\,3$ is the generation index
\begin{equation}
Q_L^{1,2,3} = \begin{pmatrix}  u_L \\ d_L \end{pmatrix}, \, \begin{pmatrix}  c_L \\ s_L \end{pmatrix}, \begin{pmatrix}  t_L \\ b_L \end{pmatrix}, \quad u_R^{1,2,3} = u_R,\, c_R,\, t_R \quad {\rm and} \quad d_R^{1,2,3} = d_R,\,s_R,\,b_R,
\end{equation}
Their \textbf{hypercharges} $\mathbf{Y}$, eigenvalues under $U(1)_Y$, are $+1/6$, $+2/3$ and $-1/3$. The corresponding \textbf{electric charges} $\mathbf{Q}$, after electroweak symmetry breaking, are given by the Gell-Mann–Nishijima formula $Q = \tau_3 + Y,$ rederived in Eq.~\eqref{eq:GellMannNishijim} of Sec.~\ref{sec:EWSB}. The quantity $\tau_3$ is the eigenvalue under the third generator of $SU(2)_L$, also known as the \textbf{weak isospin}. 
On the second hand, there are \textbf{leptons} which transform as singlets under $SU(3)_c$, and come as three $SU(2)_L$ doublet $L_L^n$ and three $SU(2)_L$ singlets $e_R^n$ 
\begin{equation}
L_L^{1,2,3} = \begin{pmatrix}  \nu_e \\ e_L \end{pmatrix}, \, \begin{pmatrix}  \nu_\mu \\ \mu_L \end{pmatrix}, \begin{pmatrix}  \nu_\tau \\ \tau_L \end{pmatrix} \quad {\rm and} \quad e_R^{1,2,3} = e_R,\, \mu_R,\, \tau_R .
\end{equation}
Their \textbf{hypercharges} $\mathbf{Y}$ are $-1/2$ and $-1$.
A first curiosity is the absence of \textbf{right-handed neutrino} $\mathbf{\nu_R}$ in the SM. A second curiosity is the emergence of three \textbf{generations} $n=1,\,2,\,3$ of quarks $(u,\,d)$, $(c,\,s)$ and $(t,\,b)$ as well as three generations of leptons $(\nu_e, \, e)$, $(\nu_\mu,\, \mu)$, $(\nu_\tau,\, \tau)$. As shown in Fig.~\ref{fig:Standard_Model_of_Elementary_Particles}, the first generation contains the lightest fermions while the third generation contains the heaviest ones.

\subsection{The Higgs field}
Finally, the last particle in the SM is the \textbf{Higgs boson}. It comes as a complex scalar, doublet under $SU(2)_L$ with hypercharge $+1/2$ and spin $0$ called the \textbf{Higgs multiplet}
\begin{equation}
H = \begin{pmatrix}  \phi^+ \\ \phi^0 \end{pmatrix},
\end{equation}
where $\phi^+$ and $\phi^0$ are two complex scalar fields. 

The arrow in Eq.~\eqref{eq:gauge_structure} indicates that the vacuum of the universe in which we live, lies in a particular gauge configuration of $SU(2)_L\times U(1)_Y$, chosen arbitrarily when in the early universe, the \textbf{Higgs field} $H$ acquires a non-zero \textbf{vacuum expectation value} $\left< H \right>\neq 0$. We talk about \textbf{spontaneous} or \textbf{dynamical} breaking of the electroweak symmetry. This is the Higgs mechanism developed in 1964, which gives the masses to the gauge bosons $W^{\pm},\,Z^0$, the quarks and the leptons\footnote{Except for the neutrinos which are considered massless and only left-handed in the Standard Model.} , while preserving the gauge symmetry at the quantum level. We provide more details in Sec.~\ref{sec:EWSB}.

A summary of the field content of the SM is given in Table.~\ref{table:SM_content} and in Fig.~\ref{fig:Standard_Model_of_Elementary_Particles}.

\section{The Standard Model in a nutshell}
\subsection{The Lagrangian}
\label{sec:lagrangian}
\paragraph{The golden rules:} 
The dynamical evolution of the fields, including both their free evolution and their interactions, is encoded in the Lagrangian density $\mathcal{L}$. It is a function of the quantum fields introduced in the previous section, whose construction in the SM follows two golden rules
\begin{enumerate}
\item
$\mathcal{L}$ must be \textbf{invariant} under the gauge structure $SU(3)_c\times SU(2)_L\times  U(1)_Y$.
\item
$\mathcal{L}$ must be \textbf{renormalizable}, so it must only include operators up to dimension four.\footnote{Given an operator $\mathcal{O}$ with dimension $4+\delta>4$, since the lagrangian density $c\, \mathcal{O}$ must have dimension $4$, the coupling constant $c$ must have the appropriate scaling $c \propto \Lambda^{-\delta}$, where $\Lambda$ is some energy scale. Hence the transition amplitude of any process occuring at energy $E$, at order $n$ in perturbation, depends on the quantity $\left(\frac{E}{\Lambda} \right)^{\delta\, n}$, which is divergent for $E \gg \Lambda$, and even more divergent than the perturbation order $n$ is large. Hence, the cancellation of the UV divergences would need an infinity of counterterms and the theory is not renormalizable.  }
\end{enumerate}
\paragraph{The Lagrangian: } 
\label{paragraph:SM_lagrangian}
We obtain:
\begin{align}
\mathcal{L}_{\rm SM} = &- \frac{1}{4}B_{\mu \nu} B^{\mu \nu}- \frac{1}{4}W_{\mu \nu}^a W^{a\mu \nu} - \frac{1}{4}G_{\mu \nu}^a G^{a\mu \nu} \label{eq:line1}\\
 & + i \bar{L}_{\rm L}^n \slashed{D} L_{\rm L}^n + i \bar{e}_{\rm R}^n \slashed{D} e_{\rm R}^n+ i \bar{Q}_{\rm L}^n \slashed{D} Q_{\rm L}^n+ i \bar{u}_{\rm R}^n \slashed{D} u_{\rm R}^n+ i \bar{d}_{\rm R}^n \slashed{D} d_{\rm R}^n \label{eq:line2}\\[0.3 cm]
  &+ \mu^2 \left|H\right|^2 + \lambda \left|H\right|^4. \label{eq:line3} \\[0.3cm]
 &+\left(D_{\mu} H  \right)^\dagger\left(D^{\mu} H  \right) - \left( Y_{mn}^e \bar{L}_{\rm L}^m H e_{\rm R}^n+ Y_{mn}^d \bar{Q}_{\rm L}^m H d_{\rm R}^n+  Y_{mn}^u \bar{Q}_{\rm L}^m \tilde{H} u_{\rm R}^n  + h.c. \right) \label{eq:line4}\\
&- \theta_1\frac{g_1^2}{32\pi^2} B^{ \mu \nu} \tilde{B}_{ \mu\nu} 
- \theta_2\frac{g_2^2}{32\pi^2} W^{a \mu \nu} \tilde{W}_{ \mu\nu}^a - \theta_3\frac{g_3^2}{32\pi^2}
 G^{b \mu \nu} \tilde{G}_{\mu \nu}^b \label{eq:line5}\\
 \end{align}
 
\paragraph{Gauge fields kinetic terms:} 
The first line in Eq.~\eqref{eq:line1} contains the kinetic terms for the gauge fields $X_\mu$ with the strength tensors $F_{\mu\nu}$ being 
\begin{equation}
F_{\rm \mu \nu} = \partial_\mu X_\nu - \partial_\nu X_\mu+ g f^{abc} X_\mu^b X_\nu^c,
\end{equation}
where $f^{abc}$ are the \textbf{structure constant} of the corresponding Lie group, defined from the algebra of the generators
\begin{equation}
\left[\tau^a,\,\tau^b\right] = i\,f^{abc}\,\tau^c.
\end{equation}
Particularly, for $U(1)_Y$ and $SU(2)_L$, we have $f_1^{abc}=0$ and $f_2^{abc}=\epsilon^{abc}$, where $\epsilon^{abc}$ is the Levi-Civita tensor.

\paragraph{Fermion kinetic terms:} 
The second line in Eq.~\eqref{eq:line2} contains the kinetic terms for the quarks and leptons, assuming the Feynman notation $\slashed{D} = \gamma^{\mu} D_\mu$. Invariance under the gauge symmetries is satisfied thanks to the \textbf{gauge-covariant derivatives} which contain the interactions between the fermions and the gauge fields
\begin{equation}
D_\mu = \partial_\mu - i g_1\,B_\mu\,Y - i g_2\,W_\mu^a\,\tau_2^a - i g_3\,G_\mu^a\,\tau^a_3,
\end{equation}
where the $X_\mu^a$ are the gauge bosons, $\tau^a$ are the generators of the corresponding gauge symmetries and $g$ are the corresponding gauge couplings.

\paragraph{Global symmetries:} \label{par:global_sym_SM} In the absence of the Yukawa matrices $Y^d$, $Y^u$ and $Y^e$, cf. Eq.~\eqref{eq:line4}, the Lagrangian of the SM is invariant under a large $U(3)^5$ global symmetry, where $3$ is the number of flavors and $5$ is the number of kinetic terms in Eq.~\eqref{eq:line2}, which can be decomposed as \cite{DAmbrosio:2002vsn}
\begin{align}
\label{eq:global_sym_SM}
&\qquad U(3)^5=SU(3)_q^3 \times SU(3)_l^2 \times U(1)^5 \\
\text{where}\qquad& \qquad \qquad \qquad   \notag\\
&\qquad SU(3)_q^3 = SU(3)_{Q_L} \times SU(3)_{u_R} \times SU(3)_{d_R},\notag \\
&\qquad  SU(3)_l^2 = SU(3)_{L_L}^3 \times SU(3)_{e_R},\\
&\qquad  U(1)^5=U(1)_{Q_L}\times U(1)_{u_R} \times U(1)_{d_R} \times U(1)_{L_L} \times U(1)_{e_R} \notag.
\end{align}
These global symmetries have not been imposed directly on the lagrangian, hence they are called \textbf{accidental symmetries}.
The five $U(1)$ symmetries, which rotate the fermion fields in a flavor-blind way, can be reshuffled as $U(1)^5=U(1)_B \times U(1)_L \times U(1)_Y \times U(1)_{d_R} \times U(1)_{e_R}$.
In the presence of the Yukawa interactions, we can check that the global symmetry group gets broken to
\begin{align}
U(3)^5 \longrightarrow U(1)_{B}\times U(1)_{L_e} \times U(1)_{L_\mu} \times U(1)_{L_\tau},
\end{align}
where we dropped the $U(1)_Y$ since it is gauged.
We have introduced the baryonic and leptonic symmetries defined by
\begin{equation}
l_{\rm e,\, \mu,\, \tau}\xrightarrow{U(1)_{\rm L_e,\, L_\mu, \, L_\tau}} e^{i\alpha_{\rm e,\, \mu,\, \tau}}l_{\rm e,\, \mu,\, \tau} \quad {\rm and} \quad q_{n}\xrightarrow{U(1)_{\rm B} }e^{i\frac{\alpha}{3}}q_{\rm n}, \quad {\rm with} ~n=1,\,2,\,3.
\end{equation}
From the \textbf{Noether theorem}, there exists corresponding baryonic and leptonic charges $B$, $L_e$, $L_\mu$ and $L_\tau$, which must be conserved\footnote{The are however broken by non-perturbative processes, see Sec.~\ref{sec:Matter-anti-matter-asymmetry}} in all processes. They are the \textbf{baryon number} and \textbf{lepton numbers}.

\subsection{Quantum Chromodynamics}
\label{sec:QCD}
\paragraph{The birth of QCD: }
A lot of progress as been made since Yukawa predicted in 1935 the existence of the $\pi$ meson as the mediator responsible for the force binding protons and neutrons together in the nucleus \cite{Yukawa:1935xg}. 
In 1961, Gell-Mann and Ne'eman interpreted the \textbf{hadrons} as \textbf{bound-states} of three `fictitious leptons' \cite{GellMann:1961ky, Neeman:1961jhl} which Gell-mann later called \textbf{quarks} $u,\,d,\,s$ \cite{GellMann:1964nj}. It was then possible to catalog all the possible hadrons as being different representations of the group $SU(3)_{\rm f}$, where $f$ denotes the number of \textbf{flavors}.\footnote{In the SM, there are six flavors of quarks $(d,\,u,\,s,\,c,\,b,\,t)$.} Gell-Mann received the Nobel prize in 1969. Seminal articles are collected in the book \cite{gell2018eightfold}.    \\
In 1965, the non-observation of hadrons based on the tensor product $3\otimes 3$ but instead based on $3\otimes \bar{3}$ led Han and Nambu to postulate that quarks appear under three \textbf{colors} and that hadrons are singlet under a second $SU(3)_{\rm c}$ symmetry \cite{Han:1965pf}. \\
In 1968, Bjorken proposed that the structure functions for the electron-proton deep-inelastic scattering should scale at high-energy \cite{Bjorken:1968dy,Bjorken:1969ja}. Simply speaking, the energy dependence of the cross-section should mimick the one for electron-muon scattering. The scaling properties where observed at SLAC the same year \cite{Bloom:1969kc}. This pushed Feynman to develop the \textbf{parton model} in which the proton is described by a collection of point-like constituents from which the electron scatters incoherently \cite{Feynman:1969ej}. \\
In 1973, using \textbf{renormalization group equations}, Gross, Wilczek and Politzer proved that the non-abelian $SU(3)_c$ Yang-Mills gauge theory \cite{Yang:1954ek} with color triplet quarks and color octet gluons, proposed by Fritzsch, Gell-Man and Leutwyler the previous year \cite{Fritzsch:1972jv, Fritzsch:1973pi}, becomes weaker at high energy, a property called \textbf{asymptotic freedom} \cite{Gross:1973id, Politzer:1973fx}. Then, the theory of Quantum Chromodynamics (QCD) was universally recognized as the correct fundamental theory of hadrons and they shared the Nobel prize in 2004. The gluons were discovered in 1979 at the experiment PETRA taking place at DESY \cite{Barber:1979yr}. Their existence was inferred from the presence of a third jet in $e^+e^-$ collision.  \\

\paragraph{Running of the strong coupling : }
Upon neglecting the quark masses, \textbf{renormalization group equations} at one loop read \cite{Caswell:1974gg}
\begin{equation}
\beta_3(\mu) \equiv \mu \frac{d}{d\mu} g_{3}(\mu) = -\frac{g_3^3(\mu)}{16\pi^2}\left(\frac{11}{3}N_c - \frac{2}{3} n_f \right),
\end{equation}
where $\mu$ is the renormalization scale, $N_c=3$ is the number of colors and $n_f$ is the number of quarks lighter than  $\mu$ (up to $6$ in the SM). 
The solution is
\begin{equation}
\label{eq:running_alpha_s}
\alpha_3(\mu) \equiv \frac{g_3^2(\mu)}{4\pi} = \frac{12\pi}{(11N_c-2n_f)\ln \left( \mu^2/\Lambda^2\right),}
\end{equation}
where $\Lambda$ is an integration constant. 

\paragraph{Asymptotic freedom: }

At high energy $\mu \gg \Lambda$, the strong coupling constant $\alpha_s$, solution of the renormalization group equation in Eq.~\eqref{eq:running_alpha_s}, tends to zero. This is the seminal result of Gross, Wilczek and Politzer showing that QCD is \textbf{asymptotically free}. Hence, perturbation theory can be applied at energies $\mu \gg \Lambda$

\paragraph{Color confinement: }
\label{par:color_confinement}

The solution of the renormalization group equation in Eq.~\eqref{eq:running_alpha_s} exhibits a second important feature of QCD. The dimensionless coupling $g_3$ is related to a free dimensionful parameter, the integration constant $\Lambda$, which is a free mass scale, fixed by measurements of $\alpha_s$. This property is known as \textbf{dimensional transmutation}. At that scale, the coupling constant diverges and the theory becomes strongly-coupled. Below the scale $\Lambda$, we expect the quarks and gluons to confine, giving rise to a mass gap in the particle spectrum.  The later is given by the vacuum expectation value of the correlator $\bar{q}q$, or \textbf{meson condensate}
\begin{equation}
\left< \bar{q} q \right> \simeq \Lambda^3.
\end{equation}
This is the flip side of asymptotic freedom, called \textbf{infrared slavery} or \textbf{color confinement}. However, even though it has been experimentally verified as a failure of free-quark searches, it has only been verified in lattice numerical simulations \cite{Greensite:2011zz} and still remain to be theoretically proven. Proving the existence of mass gap in Yang-Mills theories is defined as one of the Millenium problem.\\
From injecting the measurement of the strong coupling at the Z pole \cite{Tanabashi:2018oca}
\begin{equation}
\alpha_3 (m_Z)=0.1179 \pm 0.0010,
\end{equation}
and from setting $n_f=5$, we obtain $\Lambda \simeq 90~$MeV. Renormalization group equations at 3 loops \cite{Tanabashi:2018oca} give $\Lambda \simeq 1.4~\rm GeV$. Instead of relying on perturbation theory, the QCD scale $\Lambda$ is better computed in lattice simulations \cite{Aoki:2019cca}
\begin{equation}
\Lambda = 211(10)~{\rm MeV}.
\end{equation}
During the late 70s and early 80s,  it is shown that the phenomenology of color confinement can be reproduced by modeling gluon field lines as Nambu-Goto strings \cite{nambu1969quark,  Nambu:1978bd, Goto:1971ce, Casher:1974vf, Artru:1974hr}. The Lund group \cite{Andersson:1978vj, Andersson:1983jt, Andersson:1983ia} gives a successful description of jet fragmentation in terms of string fragmentation, which is now at the core of the very widespread high-energy event generator code PYTHIA \cite{Sjostrand:2006za}.\\
In 1999, Yang-Mills gauge theory in the large N super-conformal limit are shown to be dual to anti de Sitter space time \cite{Maldacena:1997re}. A new method of investigation of strongly-coupled theories, as being dual of gravitational theories, is opened.

\paragraph{The theory of mesons.}
Below the Landau pole $\mu \lesssim \Lambda_{\rm QCD}$, the strong coupling constant diverges and perturbation theory can not be used to explain certain phenomena observed in Nature, as for instance nuclear physics. Instead, an Effective Field Theory (EFT) of nuclear interactions can be constructed based on symmetries of the fundamental QCD Lagrangian. For introductory textbooks on effective field theory, the reader is referred to  \cite{Petrov:2016azi,Burgess:2020tbq}. Let us consider a simple theory with only up and down quarks, which in good approximation can be assumed to be massless,
\begin{align}
\label{eq:fonda_lagrang}
\mathcal{L} = - \frac{1}{4}\left( F_{\mu\nu}^a\right)^2 + i\bar{u} \slashed{D} u + i\bar{d} \slashed{D} d.
\end{align}
Upon introducing left-handed and right-handed component $\psi^{\rm R/L}=\frac{1}{2}\left( 1 \pm \gamma_5 \right) \psi$, we can see that the Lagrangian
\begin{equation}
\label{eq:fonda_lagrang_2}
\mathcal{L} = - \frac{1}{4}\left( F_{\mu\nu}^a\right)^2 + i\bar{u}_{\rm L} \slashed{D} u_{\rm L} + i\bar{d}_{\rm R} \slashed{D} d_{\rm R},
\end{equation}
is manifestly invariant under independent $SU(2)$ rotations of the left and right components
\begin{equation}
\begin{pmatrix}  u_{\rm L} \\  d_{\rm L}\end{pmatrix} \rightarrow U_{\rm L}\begin{pmatrix}  u_{\rm L} \\  d_{\rm L}\end{pmatrix} , \qquad \begin{pmatrix}  u_{\rm R} \\  d_{\rm R}\end{pmatrix} \rightarrow U_{\rm R}\begin{pmatrix}  u_{\rm R} \\  d_{\rm R}\end{pmatrix} ,
\end{equation}
where $U_{\rm R} \in S(U)_{\rm L}$  and $U_{\rm R} \in SU(2)_{\rm R}$.
However, in the ground state of QCD observed in our universe, the quark billinears have a non zero expectation value 
\begin{equation}
\left<\bar{u}u \right> =\left< \bar{d} d\right> = V^3 \simeq \Lambda_{\rm QCD}^3.
\end{equation}
Therefore the \textbf{chiral symmetry} $SU(2)_{\rm R}\times SU(2)_{\rm R} $ is spontaneously broken to the \textbf{vectorial group}, in which left and right components rotate the same way
\begin{equation}
SU(2)_{\rm R}\times SU(2)_{\rm R}  = SU(2)_{V}.
\end{equation}
The remaining symmetry under $SU(2)_{V}$ is known as \textbf{isospin}.
The dynamics of  spontaneous chiral symmetry breaking can be described in the linear sigma model
\begin{equation}
\label{eq:linear_sigma_model}
\mathcal{L}_{\rm eff} \subset  \left| \partial_\mu \Sigma \right|^2 + m^2 \left| \Sigma \right|^2 - \frac{\lambda}{4} \left| \Sigma \right|^4, 
\end{equation}
where $\Sigma_{\rm ij}$ is a set of fields, known as the \textbf{Goldstone matrix}, transforming as a \textbf{bi-fundamental} of $SU(2)_{\rm R}\times SU(2)_{\rm R} $
\begin{equation}
\Sigma \rightarrow U_{\rm L}\Sigma U_{\rm R}^\dagger.
\end{equation}
We can check that Eq.~\eqref{eq:linear_sigma_model} is minimized for
\begin{equation}
\left< \Sigma \right> = \frac{v}{\sqrt{2}}\begin{pmatrix}  1 & 0 \\  0 & 1 \end{pmatrix}, \qquad \textrm{with}\quad v= \frac{2m}{\sqrt{\lambda}}.
\end{equation}
We can choose to parameterize the Goldstone matrix as
\begin{equation}
\Sigma(x) = \frac{v + \sigma(x)}{\sqrt{2}} \exp \left(2i \frac{\tau^a \pi^a(x)}{F_\pi}   \right),
\end{equation}
where the \textbf{pion decay constant} $F_{\pi}= v$ is chosen such that $\pi^a$ are canonically normalized. The later can be measured from the pion decay and one finds $F_{\pi}\simeq 130$~MeV \cite{Tanabashi:2018oca}. Under a chiral transformation $U_{\rm L} = \exp(i\theta_{\rm L}^a \tau^a)$ and $U_{\rm R} =  \exp(i\theta_{\rm R}^a \tau^a)$, we can check that $\sigma$ is invariant while
\begin{equation}
\label{eq:pi_tranf}
\pi^a \rightarrow \pi^a + \frac{F_{\pi}}{2}(\theta_{\rm L}^a - \theta_{\rm R}^a)- \frac{1}{2} f^{\rm abc}(\theta_{\rm L}^b + \theta_{\rm R}^b) \pi^c + \cdots,
\end{equation}
where $f^{\rm abc} = \epsilon^{\rm abc}$ are the structure constants of $SU(2)_{\rm V}$. We can see that $\pi^a $ transforms as an adjoint of $SU(2)_{\rm V}$ but non-linearly - they \textbf{shift} at leading order - under the broken symmetry transformation $\theta_{\rm L}^a \neq \theta_{\rm R}^a$. Upon transforming the fundamental Lagrangian in Eq.~\eqref{eq:fonda_lagrang_2} under $\theta_{\rm L}^a = - \theta_{\rm R}^a = \theta(x)$ where $\theta(x)$ depends on $x$, we obtain
\begin{equation}
\mathcal{L} \rightarrow \mathcal{L} + j^{\mu5a} \partial_\mu \theta,
\end{equation}
where $ j^{\mu5a}$ is the \textbf{conserved Noether current}
\begin{equation}
\label{eq:chiral_current}
 j^{\mu5a}= \bar{Q} \gamma^\mu \gamma^5 \tau^a  Q, \qquad \textrm{with}\quad Q = \begin{pmatrix}  u \\  d \end{pmatrix} .
\end{equation}
Now upon applying the exact same transformation on the effective lagrangian $\mathcal{L}_{\rm eff} = \frac{1}{2}(\partial \pi)^2 + \cdots$, and using Eq.~\eqref{eq:pi_tranf}, we obtain that the \textbf{chiral current} in Eq.~\eqref{eq:chiral_current}, in the effective description matches to
\begin{equation}
 j^{\mu5a}= F_{\pi}\partial_\mu \pi.
\end{equation}
By acting it on the meson state $\left| \pi \right>$, we obtain
\begin{equation}
\left<0\right|  j^{\mu5a} \left| \pi \right> = ip^\mu F_{\pi}e^{-ipx}.
\end{equation}
The conservation of the chiral current $ \partial_\mu j^{\mu5a} = 0$, implies that the mesons $\pi$ are massless, so we are able to identify them as \textbf{Nambu-Goldstone bosons} of the chiral symmetry breaking.
Next we want to use symmetries to write all the possible terms in the effective Lagrangian, in addition to those in Eq.~\eqref{eq:linear_sigma_model}. Since $\sigma$ does not transform under any symmetry, we get rid of it by taking $m\to \infty$ and $\lambda \to \infty$,  while holding $F_\pi$ fixed
\begin{equation}
\frac{\sqrt{2}}{v}\Sigma(x) \rightarrow U(x) \equiv \exp \left[2i \frac{\tau^a \pi^a}{F_{\pi}}  \right] = \exp \left[ \frac{i}{F_\pi}  \begin{pmatrix}  \pi^0 & \sqrt{2} \pi^-\\  \sqrt{2} \pi^+ & - \pi^0\end{pmatrix} \right],
\end{equation}
where $\pi^0 = \pi^3$ and $\pi^\pm = \frac{1}{\sqrt{2}} \left(\pi^1 \pm i\pi^2   \right)$. Such a theory with decoupled $\sigma$ is known as \textbf{non-linear sigma model}. From this, we can write the most general effective Lagrangian invariant under $SU(2)_{\rm R}\times SU(2)_{\rm R} $,  known as the \textbf{chiral lagrangian} 
\begin{align}
\label{eq:chiral_lagrangian}
\mathcal{L}_{\rm eff} \subset \frac{F_\pi^2}{4} \textrm{tr} \left[ (D_\mu U)(D_\mu U)^\dagger \right] + L_1 \textrm{tr} \left[ (D_\mu U)(D_\mu U)^\dagger \right]^2 \\+ L_2 \textrm{tr} \left[ (D_\mu U)(D_\nu U)^\dagger \right]\textrm{tr} \left[ (D_\mu U)^\dagger(D_\nu U) \right]\\+ L_3 \textrm{tr} \left[ (D_\mu U)(D_\mu U)^\dagger (D_\nu U)(D_\nu U)^\dagger \right]+\cdots
\end{align}
where the covariant derivative only contains electroweak gauge fields but not the gluons. The derivation of the theory of hadrons below the confining scale using symmetry of the fundamental lagrangian was pioneered by Callan, Coleman, Wess and Zumino (CCWZ) in 1969 \cite{Coleman:1969sm,Callan:1969sn}. See \cite{Leutwyler:1993iq,Ecker:1994gg,Pich:1995bw, Scherer:2002tk, Scherer:2012xha} for reviews.
In fact, the chiral symmetry is explicitly broken by the presence of the tiny up and down masses, 
\begin{equation}
\label{eq:fund_lagrang_mass}
\mathcal{L} \subset \bar{q} M_q q, \qquad M_q=  \begin{pmatrix}  m_u &0 \\  0 & m_d \end{pmatrix} 
\end{equation}
and the mesons $\pi$ are said to be \textbf{pseudo}-Nambu-Goldstone bosons. In order to compute the mass of the mesons, we can use the \textbf{spurion} trick. We trade the Lagrangian couplings responsible for the explicit breaking - here the mass matrix $M_q$ - by fake fields which transform in such a way that the Lagrangian looks symmetric. We assume that $M_q$ transforms under a bi-fundamental of the chiral transformation $M_q \to U_{\rm L} M_q U_{\rm R}$. The leading $SU(2)_{\rm R}\times SU(2)_{\rm R} $ invariant term which we can add is 
\begin{align}
\mathcal{L}_{\rm eff}&\supset \frac{V^3}{2}\textrm{tr}\left[M_qU + M_q^\dagger U^\dagger \right] \notag\\ \label{eq:eff_lag_mass_pion}
&\supset V^3(m_u + m_d) - \frac{V^3}{2 F_{\pi}^2}(m_u + m_d)(\pi_0^2 + \pi_1^2 + \pi_2^2) + \mathcal{O}(\pi^3).
\end{align}
The prefactor $V^3/2$ is fixed such that the vacuum energy in the effective lagrangian in Eq.~\eqref{eq:eff_lag_mass_pion} matches the vacuum energy in the fundamental lagrangian in Eq.~\eqref{eq:fund_lagrang_mass}. We conclude that the pion mass follows the relation
\begin{equation}
\label{eq:pion_mass}
m_\pi^2 = \frac{V^3}{F_{\pi}^2} (m_u + m_d),
\end{equation}
known as the \textbf{Gell-Mann-Oakes-Renner} relation \cite{Gell-Mann:1968hlm}. Using $V \simeq 250~\rm MeV$ \cite{Ioffe:2005ym}, $F_{\pi}\simeq 92~$MeV and $m_\pi = 135$~MeV, the sum of the up and down quark masses is predicted to be $m_{u}+m_d \simeq 10 ~$MeV. This is however only the leading order contribution to the pion mass and it overestimates the measured value for up and down quark masse, $m_{u}+m_d \simeq 6.83_{-0.43}^{+0.97}~$MeV,  by more than $20\%$ \cite{Tanabashi:2018oca}.

\subsection{Electroweak Symmetry Breaking}
\label{sec:EWSB}

\paragraph{The birth of the Glashow-Weinberg-Salam model:}
Fermi, who died in 1954 did not get the chance to experience the major revolutions that his theory of weak interactions which he introduced in 1934 to explain beta decay, went through during the 60s. Before this, weak interactions were described using a \textbf{non-renormalizable} four-fermion interaction between left-handed currents, see e.g. in 1958 the formulation of the `V-A' theory by Feynman and Gell-mann \cite{Feynman:1958ty} or Sudarshan and Marshak \cite{Sudarshan:1958vf}.  \\
A model unifying electromagnetism and weak interactions under the gauge group $SU(2)_{\rm L}\times U(1)_{\rm Y}$  was first proposed by Glashow in 1961  \cite{Glashow:1961tr}. But the theory was not yet renormalizable.\\
In 1960, Nambu and Goldstone discovered, in the context of superconductivity, that whenever a symmetry is broken spontaneously, \textbf{massless modes} are generated \cite{Nambu:1960tm,Goldstone:1961eq}. Shortly after, Goldstone, Salam and Weinberg proved the statement in the framework of Quantum Field Theory \cite{Goldstone:1962es}. In 1964, Brout, Englert, Higgs, Guralnik, Hagen and Kibble provided a mechanism which gives a mass to a vector boson without violating gauge invariance \cite{Englert:1964et, Higgs:1964ia,Guralnik:1964eu,Higgs:1964pj,Higgs:1966ev}. Nowadays, this is known as the \textbf{Higgs mechanism} and it can be formulated as follows: if the local symmetry is spontaneously broken, the massless Nambu-Goldstone bosons become the longitudinal modes of the gauge fields which therefore acquire a mass. \\
Finally, this is in 1967 that the model of Glashow, unifying electromagnetism and weak interactions under the gauge group $\mathbf{SU(2)_{\rm L}\times U(1)_{\rm Y}}$ was improved by Weinberg and Salam who incorporated the spontaneous symmetry breaking into $\mathbf{U(1)_{\rm em}}$ \cite{Weinberg:1967tq,Salam:1968rm}. The modern theory of Electroweak interactions was born. All three shared the Nobel prize in 1979. The $W^{\pm},\,Z^0$ were discovered in 1983 with two corresponding Nobel prizes in 1984. \\
The renormalizability of such a massive gauge theory was proved in 1972 by  't Hooft and his adviser Veltman who developed the technique of dimensional regularization \cite{tHooft:1971qjg, tHooft:1971akt, tHooft:1972tcz}. They received the Nobel prize in 1999.

 \paragraph{Spontaneous symmetry breaking:} 
The third line of the SM Lagrangian in Eq.~\eqref{eq:line3}  contains the Higgs potential, which after minimization, is responsible, when $-\mu^2<0$, for a \textbf{vacuum expectation value} for the Higgs multiplet, which we can write, without loss of generality
\begin{equation}
\label{eq:VEV_Higgs}
\left<H\right> = \begin{pmatrix}  0 \\  \frac{v}{\sqrt{2}} \end{pmatrix},
\end{equation}
with
\begin{equation}
v = \sqrt{\frac{-\mu^2}{\lambda}}\simeq 246~\rm GeV.
\end{equation}
The expansion of the Higgs field around its vacuum background, gives
\begin{equation}
\label{eq:higgs_vac_background}
\left<H\right> = \begin{pmatrix}  \pi_1 + i\pi_2 \\  \frac{v+h}{\sqrt{2}} + i \pi_3 \end{pmatrix},
\end{equation}
where $\pi_1$, $\pi_2$ and $\pi_3$ are the $4-1=3$ \textbf{massless} modes predicted by the Nambu-Goldsone theorem and $h$ is the \textbf{massive} excitation of the Higgs field, commonly known as the Higgs boson. First theorized in 1964 \cite{Englert:1964et, Higgs:1964ia,Guralnik:1964eu,Higgs:1964pj,Higgs:1966ev}, the Higgs boson has been discovered at the Large Hadron Collider in 2012 at a mass of $m_{\rm h}\simeq 125~$GeV \cite{Aad:2012tfa, Chatrchyan:2012xdj}. Englert and Higgs get the Nobel prize in 2013.\\
From injecting Eq.~\eqref{eq:higgs_vac_background} into Eq.~\eqref{eq:line3}, we read the Higgs mass
\begin{equation}
m_{\rm h} = \sqrt{-2\mu^2}.
\end{equation}

\paragraph{Generation of the vector boson masses:} 
The fourth line of the SM Lagrangian in Eq.~\eqref{eq:line4} contains the interactions with the Higgs boson responsible for the mass of the different fields. First, there is the kinetic term for the Higgs boson which contains the interactions with the vector bosons $W^{\pm}$ and $Z^0$ responsible for their mass. Second, there are the Yukawa terms responsible for the mass of the quarks and leptons.  We have introduced
\begin{equation}
\label{eq:Htilde}
\tilde{H} = i\sigma_2 H^*,
\end{equation}
where $\sigma_2$ is the second Pauli matrix. $\tilde{H}$ has the opposite hypercharge as $H$ but the same transformation rules under $SU(2)_L$.\footnote{The field $\tilde{H} = i\sigma_2 H^*$ has opposite hypercharge as $H$, and so transforms the opposit way as $H$ under $U(1)_Y$. However, since the fundamental representation of $SU(2)$ is pseudo-real (e.g. \cite{Zee:2016fuk}),
\begin{equation}
\sigma_2 \left(e^{i\vec{\phi}.\vec{\sigma}}\right)^*\sigma_2 = e^{i\vec{\phi}.\vec{\sigma}},
\end{equation}
then $\tilde{H}$ transforms exactly the same way as $H$ under $SU(2)_L$,
\begin{align}
&H \xrightarrow{SU(2)_L} e^{i\vec{\phi}.\vec{\sigma}}H, \\
\Leftrightarrow \quad & H^*   \xrightarrow{SU(2)_L}  \left(e^{i\vec{\phi}.\vec{\sigma}}\right)^*H^* , \\
\Leftrightarrow \quad &\tilde{H}   \xrightarrow{SU(2)_L}  -i\sigma_2\left(e^{i\vec{\phi}.\vec{\sigma}}\right)^* \overbrace{i\sigma_2  i\sigma_2}^{=-1} H^* =  e^{i\vec{\phi}.\vec{\sigma}}\tilde{H}.
\end{align}}
Upon assigning to the Higgs field its vacuum expectation value in Eq.~\eqref{eq:VEV_Higgs}, we obtain 
\begin{align}
\left|D_\mu H \right|^2 &=  g_2^2\frac{v^2}{8} \begin{pmatrix}  0&1 \end{pmatrix} \begin{pmatrix}  \frac{g_1}{g_2} B_\mu + W_\mu^3 & W_\mu^1-iW_\mu^2 \\ W_\mu^1+iW_\mu^2 &  \frac{g_1}{g_2} B_\mu + W_\mu^3 \end{pmatrix} \begin{pmatrix}  \frac{g_1}{g_2} B_\mu + W_\mu^3 & W_\mu^1-iW_\mu^2 \\ W_\mu^1+iW_\mu^2 &  \frac{g_1}{g_2} B_\mu + W_\mu^3 \end{pmatrix} \begin{pmatrix} 0 \\ 1\end{pmatrix} \notag \\
\label{eq:mass_matrix_EW}
&= g_2^2\frac{v^2}{8} \left[ (W_\mu^1)^1 +(W_\mu^2)^2 + \left(\frac{g_1}{g_2}B_\mu - W_\mu^3\right)^2 \right].
\end{align}
The diagonalization of the mass matrix after electroweak symmetry breaking defines the boson $Z^0$ and the photon $A_\mu$
\begin{align}
\label{eq:bosonZdef}
&Z_\mu = W_\mu^3 \cos{\theta_{\rm w}} -B_\mu \sin{\theta_{\rm w}}, \\
\label{eq:photondef}
&A_\mu = W_\mu^3 \sin{\theta_{\rm w}} + B_\mu \cos{\theta_{\rm w}},
\end{align} 
where $\theta_{\rm w}$ is the Weinberg angle
\begin{equation}
\sin{\theta_{\rm w}}= \frac{g_1}{\sqrt{g_1^2+g_2^2}}.
\end{equation}
As it can be inferred from the next section, $W^1$ and $W^2$ do not have proper electric charge, instead we define the generators
\begin{equation}
\label{eq:taupm_W}
\tau^{\pm} = \frac{1}{\sqrt{2}} ( \tau^1 \pm i \tau^2 ),
\end{equation}
with $[\tau^3,\,\tau^\pm] = \pm \tau^\pm$ such that the vector bosons 
\begin{equation}
\label{eq:Wpm}
W_\mu^\pm = \frac{1}{\sqrt{2}} (W_\mu^1 \mp i W_\mu^2),
\end{equation}
have electromagnetic charge $\pm 1$.  
The breaking of the electroweak gauge interactions to electromagnetism, $SU(2)_{\rm L}\times U(1)_{\rm Y} \rightarrow U(1)_{\rm e.m.}$, also known as \textbf{Electroweak Symmetry Breaking} (EWSB) or Glashow-Weinberg-Salam model, is explicit from Eq.~\eqref{eq:mass_matrix_EW}, where we can see that the photon is massless whereas the boson $W^\pm$ and $Z^0$ obtain the masses
\begin{equation}
m_{\rm W}^2 = \frac{g_2^2}{2} v^2 \qquad {\rm and}\qquad m_Z^2 = \frac{g_1^2+g_2^2}{2}v^2.
\end{equation}

\paragraph{Generation of the Fermion masses:}
Repeating the same steps for the Yukawa terms, the quarks and leptons get the masses
\begin{equation}
m_{i} = \frac{Y_{i}}{\sqrt{2}} v,
\end{equation}

\paragraph{Fermion interactions:}
Before EWSB, the interactions between the fermions and the electroweak bosons, contained in the fermion kinetic terms, in the second line of the SM Lagrangian, in Eq.~\eqref{eq:line2}, are
\begin{equation}
\mathcal{L} \supset  g_2 W^1_\mu J_\mu^{\rm 1} + g_2 W^3_\mu J_\mu^{\rm 2} + g_2 W^3_\mu J_\mu^{\rm 3} +  g_1 B_\mu J_\mu^{\rm Y}
\end{equation}
with
\begin{align}
J_\mu^{\rm i} = \sum_f \bar{\psi}_f^L \gamma^\mu \tau^i \psi_f^L, \quad {\rm and} \quad J_\mu^{\rm Y} = \sum_f Y_f^L \bar{\psi}_f^L \gamma^\mu \psi_f^L+\sum_f Y_f^R \bar{\psi}_f^R \gamma^\mu \psi_f^R.
\end{align}
After EWSB, they become
\begin{equation}
\mathcal{L} \supset g_2 W^+_\mu J_\mu^{\rm +} + g_2 W^-_\mu J_\mu^{\rm -} + \frac{e}{\sin \theta_{\rm w}} Z_\mu J_\mu^{\rm Z} + e A_\mu J_\mu^{\rm EM}
\end{equation}
with 
\begin{align}
&J_\mu^{\rm Z} = \frac{1}{\cos \theta_{\rm w}} \left( J_\mu^3 - \sin^2 \theta_{\rm w} J_\mu^{\rm EW}\right), \\
&J_\mu^{\rm EM} = \sum_f Q_f \left( \bar{\psi}_f^L \gamma^\mu \psi^L_f + \bar{\psi}_f^R \gamma^\mu \psi_f^R \right), \\
&J_\mu^{\rm +} = \sum_f \bar{\psi}_f^L \gamma^\mu \tau ^+ \psi^L_f , \\
&J_\mu^{\rm +} = \sum_f \bar{\psi}_f^L \gamma^\mu \tau ^- \psi^L_f , 
\end{align}
where we have defined the electric charge\footnote{The first relation of Eq.~\eqref{eq:GellMannNishijim} was already derived in the mid-50s, by Gell-Mann, Nakano and Nishijima \cite{Nakano:1953zz,Gell-Mann:1956iqa}.} $Q_f$ and the electromagnetic coupling $e$
\begin{equation}
\label{eq:GellMannNishijim}
Q_f = T^3 +Y, \quad {\rm and} \quad 
e = g_2 \sin \theta_{\rm w} = g_1 \cos {\theta_{\rm w}}.
\end{equation}
The current $J_\mu^{\rm Z}$ which has charge $0$, is called the \textbf{neutral current}.

\subsection{Weak CP violation}

\paragraph{A flavor mixing mass matrix:} 
The mass terms for the quarks can be written in a $3\times 3$ matrix form
\begin{align}
\mathcal{L} \supset & -\frac{v}{\sqrt{2}}Y_{\rm mn}^d \, \bar{d}_L d_{R} -\frac{v}{\sqrt{2}}Y_{\rm mn}^u \, \bar{u}_L u_{R} +h.c.\\
\supset &  -\frac{v}{\sqrt{2}} \left[ \bar{d}_L Y_d d_R +\bar{u}_L Y_u u_R \right] + h.c.
\end{align}
Hence the fermions from different flavors interact with each other in a complicated way, which calls for a diagonalization of the mass matrix. The Yukawa matrices $Y$ are not hermitian but $YY^\dagger$ are. So it can be diagonalized with unitary matrices
\begin{equation}
\label{eq:diagonalization_Yuk_bi-unitary_transf}
Y_dY_d^\dagger = U_d M_d^2 U_d^\dagger,\quad {\rm and} \quad  Y_uY_u^\dagger = U_u M_u^2 U_u^\dagger,
\end{equation}
where $M_d = \frac{\sqrt{2}}{v}{\rm diag}(m_d,\,m_s,\,m_b)$ and $M_u=\frac{\sqrt{2}}{v}{\rm diag}(m_u,\,m_c,\,m_t)$.
We can freely introduce other unitary matrices $U_R$ and write
\begin{equation}
 Y_d  = U_L^d M_d \left( U_R^d \right)^{\dagger}, \quad {\rm and} \quad Y_u  = U_L^u M_u \left( U_R^u \right)^{\dagger},
\end{equation}
such that the Yukawa couplings read
\begin{equation}
 \mathcal{L} \supset -\frac{v}{\sqrt{2}} \left[ \bar{d}_L U_L^d M_d \left( U_R^d \right)^{\dagger} d_R +\bar{u}_L  U_L^u M_u \left( U_R^u \right)^{\dagger} u_R \right].
\end{equation}

\paragraph{Mass eigenstates and the CKM matrix:} 
\label{par:CKM}
Now, we go to the mass basis, where the mass matrix is diagonal, by performing the unitary transformations
\begin{align}
 &d_L \rightarrow U^d_L \, d_L,  \qquad  \, u_L \rightarrow U^u_L \, u_L, \\
&d_R \rightarrow U^d_R \, d_R,  \qquad u_R \rightarrow U^u_R \, u_R.
\end{align}
The matrices being unitary, all the terms of the Lagrangian which are flavor diagonal are preserved by the transformations above. Only the terms of interaction with the bosons $W^\pm$ which mix the flavors are not preserved and become
\begin{equation}
\label{eq:CKM_lagrangian}
 \mathcal{L} \supset \frac{g_2}{\sqrt{2}} \left[  W_\mu^+ \bar{u}_L^i \gamma^\mu \left(V\right)^{ij} d_L^j + W_\mu^- \bar{d}_L^i \gamma^\mu \left( V^{\dagger} \right)^{ij} u_L^{j} \right].
\end{equation}
where $V=\left(U_L^u\right)^\dagger U_L^d$ is the \textbf{Cabiddo-Kobayashi-Maskawa (CKM) matrix}. 

Leptons do not have such flavor mixing interactions in the SM since there are no neutrino mass terms. Hence, we are free to re-arrange $\nu_L$ in order to cancel the would-be $\left(U_L^\nu \right)^\dagger U_L^e$, analog of the CKM matrix in the lepton sector. However, such a matrix is needed to explain the neutrino oscillations, see Sec.~\ref{sec:neutrino_oscillations}.

Note that the CKM matrix is unitary. In 1970, Glashow, Iliopoulos and Maiani used the unitarity property of the CKM matrix to predict the existence of a fourth quark `c' in order to account for the suppression of flavor-changing neutral currents \cite{Glashow:1970gm}. The first charmed hadron, the $J/\psi:c\bar{c}$, was detected in 1974 at SLAC and BNL.

\paragraph{The origin of the $CP$ violation:} 
\label{par:CP_violation_CKM}
The CKM matrix is a complex unitary $3\times 3$ matrix and so has $9$ real degrees of freedom. If it was a real $O(3)$ matrix it would have $3$ real parameters. Hence a $U(3)$ matrix has $3$ rotation angles and $6$ complex phases. Actually, we can freely reparameterize the phases of the 6 quarks to set 5 phases to zero (nothing changes if all the rotations are the same, indeed this is due to baryon number conservation).  So the physical parameters are 3 angles and 1 phase.   Hence, we can parameterize the CKM matrix as follows \cite{Chau:1984fp}
\begin{align}
V_{\rm CKM} &= \begin{pmatrix} 1 & 0 & 0 \\ 0 & c_{23} & s_{23} \\ 0 & -s_{23} & c_{23}  \end{pmatrix}  \begin{pmatrix} c_{13} & 0 & s_{13} \, e^{i\delta} \\ 0 & 1 & 0 \\ -s_{13}\,e^{i\delta} & 0 & c_{13}  \end{pmatrix}  \begin{pmatrix} c_{12} & s_{12} & 0 \\ -s_{12} & c_{12} & 0 \\ 0 & 0 & 1  \end{pmatrix}  \notag\\
\label{eq:CKM_matrix}
&=\begin{pmatrix} c_{12} \,c_{13} & s_{12}\, c_{13} & s_{13} \,e^{-i\delta}\\ -s_{12} \,c_{23} - c_{12} \,s_{23} \,s_{13} \,e^{i\delta} & c_{12} \,c_{23} - s_{12}\, s_{23} \,s_{13}\,e^{i\delta} & s_{23}\, c_{13} \\ s_{12}\,s_{23} - c_{12}\,c_{23} \,s_{13} \,e^{i\delta} & -c_{12}\,s_{23}-s_{12}\,c_{23}\,s_{12}\,e^{i\delta} & c_{23}\,c_{12}  \end{pmatrix} 
\end{align}
with $s_{ij} = \sin {\theta_{ij}}$ and $c_{ij} = \cos{\theta_{ij}}$. 
The complex phase $\delta$ is responsible for the \textbf{CP violation}.\footnote{The Lagrangian must be invariant under $CPT$ \cite{Streater:1989vi, Kostelecky:1998ic}, where $C$, $P$ and $T$ stand for charge conjugation, mirror transformation and time reversal. So a $CP$-transformation is equivalent to a $T$-transformation. But $T$ is anti-unitary (it sends $i$ to $-i$) so any imaginary coefficient implies CP violation (assuming the field operators preserve $T$).} However, this is just an arbitrary choice of normalization and it is useful to define basis-invariant quantity which encodes the CP violation in the electroweak quark sector. From Eq.~\eqref{eq:diagonalization_Yuk_bi-unitary_transf}, we can see that if $\left[Y_uY_u^\dagger,\,Y_dY_d^\dagger\right]=0$, then we can find a unitary transformation $U=U^d_L=U^u_L$, acting on both down-type and up-type quarks, $d_L\rightarrow U d_L$ and $u_L\rightarrow U u_L$, which diagonalizes $Y_uY_u^\dagger$ and $\,Y_dY_d^\dagger$ simultaneously, and then the CKM matrix is trivial $V=\mathds{1}$. Hence, the CP violation must be encoded in the commutator
\begin{equation}
iC = \left[ Y_uY_u^\dagger,\,Y_dY_d^\dagger\right] = \left[  U_u M_u^2 U_u^\dagger,\, U_d M_d^2 U_d^\dagger\right] = U_u\left[  M_u^2 ,\, V M_d^2 V^\dagger\right] U_u^\dagger.
\end{equation}
The determinant of $C$ is invariant under any unitary transformation of the quarks and can be computed with the Vandermond formula \cite{Jarlskog:1985ht, Jarlskog:1985cw}
\begin{align}
\label{eq:CP_violation_invariant}
 {\rm det}~ C= -16 \,T\,B\,J, &\\
\text{where}\quad T&= (m_t^2-m_u^2)(m_t^2-m_c^2)(m_c^2-m_u^2)/v^6, \\
B&= (m_b^2 - m_d^2)(m_b^2 - m_s^2)(m_b^2 - m_s^2)(m_s^2-m_d^2)/v^6, \\
J&= s_{12} s_{23} s_{31} c_{12}c_{23}c_{31}^2\sin{\delta},
\end{align}
where $J$ is known as the \textbf{Jarlskog invariant}.\footnote{The basis-invariant formulation of $CP$ violation, ${\rm det}~C$ in Eq.~\eqref{eq:CP_violation_invariant}, has been introduced by Cecilia Jarlskog in 1985. She first introduced a formula containing linear masses \cite{Jarlskog:1985ht} before correcting with squared mass two months later \cite{Jarlskog:1985cw}.} $CP$ is violated if and only if ${\rm det}~C \neq 0$, otherwise the CKM matrix can be made real with any rephasing of the quark fields.

The entries of the CKM matrix have been measured from the decay of many hadrons \cite{Tanabashi:2018oca}
\begin{equation}
\begin{pmatrix} V_{ud} & V_{us} & V_{ub }\\ V_{cd} & V_{cs }& V_{cb} \\ V_{td}&V_{ts}&V_{tb} \end{pmatrix} = \begin{pmatrix} 0.97\pm 0.0001 & 0.22\pm 0.0004 & 0.0037 \pm 0.0001 \\ 0.22 \pm 0.0004 & 0.97 \pm 0.0001 & 0.042 \pm 0.0009 \\ 0.0090 \pm 0.0002 & 0.041 \pm 0.0007 & 0.999 \pm 0.00003 \end{pmatrix} .
\end{equation}
The corresponding magnitude for the mixing angled as well as the $CP$ phase are \cite{Beringer:1900zz}
\begin{equation}
\theta_{12} \simeq 13.02^\circ \pm 0.04^\circ, \quad \theta_{23} \simeq 2.36^\circ, \pm 0.08^\circ \quad \theta_{13} \simeq 0.20^\circ \pm 0.02^\circ, \quad \delta_{CP} \simeq 69^\circ \pm 5^\circ,
\end{equation} 
while the Jarlskog invariant is
\begin{equation}
J=(2.96 \pm 0.20)\times 10^{-5}.
\end{equation}
Hence, even though the $CP$ phase in the parameterization of Eq.~\eqref{eq:CKM_matrix} is close to $90^\circ$, the actual basis-invariant $J$ is small. Also for some applications, e.g. baryogenesis discussed in Sec.~\ref{par:EWbaryo}, the physical quantity is ${\rm det}~C \simeq 10^{-18}~J$ \cite{Shaposhnikov:1987pf, Gavela:1993ts, Konstandin:2003dx, Brauner:2011vb}, which is even smaller.


\paragraph{Explanation of the CP violation by the prediction of two new quarks:}
Lee and Yang were the first to postulate back in 1956 that weak interactions violate parity.\footnote{$P$ violation in weak interactions comes from the fact that only left-handed fermions are charged under $SU(2)_L$.} $P$ \cite{Lee:1956qn} They proposed an experiment which validates their prediction one year later \cite{Wu:1957my, Garwin:1957hc}. In 1963, Cabibbo introduced a mixing angle (the largest angle of the CKM matrix) to explain flavor-changing interactions \cite{Cabibbo:1963yz}.  In 1964, Cronin and Val Fitch provided evidence for $CP$ violation in kaon decays \cite{ Christenson:1964fg}. In 1973, Kobayashi and Maskawa predicted the existence of a \textbf{third generation} of quarks in order to explain CP violation \cite{Kobayashi:1973fv}. The first bottomium, $\Upsilon:b\bar{b}$, was detected in 1977 at Fermilab. The top was detected via $t\bar{t}$ pair production in 1995 at Fermilab.  All these discoveries were rewarded with Nobel prizes.

\subsection{Anomaly cancellation}
\label{sec:anomaly_cancellation}

Let us consider a massless Dirac fermion coupled to a classical external $U(1)$ gauge field
\begin{equation}
\mathcal{L} = i\bar{\psi}\slashed{\partial} \psi - e J_V^\mu A_\mu, \qquad {\rm with} \quad J_V^\mu=\bar{\psi}\gamma^\mu \psi.
\end{equation}
The lagrangian looks invariant under the axial symmetry\footnote{\label{footnote:chiral_transf}A chiral transformation rotates the left-handed and right-handed fields in an independent way
\begin{equation}
\psi_{\rm L} \longrightarrow e^{i\alpha}\psi_{\rm L}, \qquad {\rm and} \qquad \psi_{\rm R}\longrightarrow e^{i\beta}\psi_{\rm R}.
\end{equation}
The transformation $\psi \rightarrow e^{i\gamma_5 \theta \psi}$ is a particular chiral transformation called `axial' transformation in which left-handed and right-handed fields rotate in the opposite way.}
\begin{equation}
\psi \xrightarrow{U(1)_{\rm A}}  e^{i\gamma_5 \theta \psi} \psi.
\end{equation}
The expectation value of the corresponding axial current $J^\mu_5 = \bar{\psi} \gamma^\mu \gamma^5 \psi$ in the gauge field background can be expressed in the path integral formalism
\begin{equation}
\left<A| J^{\mu}_5 | A\right>= \frac{\int \mathcal{D} \bar{\psi}\mathcal{D}\psi ~  J^{\mu}_A ~e^{i \int d^4x (i\bar{\psi}\slashed{\partial} \psi - e J_V^\mu A_\mu)} }{\int \mathcal{D} \bar{\psi}\mathcal{D}\psi ~ e^{i \int d^4x (i\bar{\psi}\slashed{\partial} \psi - e J_V^\mu A_\mu )}}.
\end{equation}
Using perturbation theory, we get
\begin{align}
\left<A| J^{\mu}_5 | A\right> = -ie \int d^4 y &\cancel{ \left< 0| T\left[ J_5^\mu(x) J_V^\alpha(y) \right]|0 \right>}  A_\alpha(y) \\
&-\frac{e^2}{2} \int d^4 y_1 d^4 y_2 \left< 0| T\left[ J_5^\mu(x) J_V^\alpha(y_1) J_V^\beta (y_2) \right]|0 \right >  A_\alpha(y_1) A_\alpha(y_2),
\end{align}
where we used that $J_5^\mu(x) J_V^\alpha(y)$ is odd under parity in the first line. Hence, we see that $\left<A| J^{\mu}_5 | A\right>$ is proportional to $\left< 0| T\left[ J_A^\mu(x) J_V^\alpha(y) J_V^\beta (0) \right] |0 \right >$ which via Feynman rules is given by a triangle diagram \cite{AlvarezGaume:2012zz}. Careful computations realized in 1968 by Adler, Bell and \cite{Adler:1969gk,Bell:1969ts} show that
\begin{equation}
\label{eq:ABJ}
\left<A|  \partial_\mu J^{\mu}_5  | A\right> = - \frac{e^2}{16 \pi^2} \epsilon^{\mu \nu \alpha \beta} F_{\mu \nu} F_{\alpha \beta}.
\end{equation}
In absence of mass terms, the chiral transformation is an exact symmetry of the classical lagrangian. However, Eq.~\eqref{eq:ABJ} shows that $U(1)_{\rm A}$ is broken by quantum effects in the presence of gauge fields. If the anomalous symmetry $U(1)_A$ is global it does not lead to any problem. However, if the anomalous symmetry $U(1)_A$ is gauged, then it can lead to inconsistencies (e.g. Lorentz violation). Therefore, we must check that the gauge groups in the SM are anomaly  free.
We can generalize the previous treatment to non-abelian symmetry groups in which case the Noether current reads
\begin{equation}
J_{\mu}^{a} =\sum_\psi \bar{\psi}_i \tau_{ij}^a \gamma^\mu \psi_j,
\end{equation}
where $\tau_{ij}^a$ are the group generators. Now, each vertex of the triangle diagram picks up a group generator and we obtain
\begin{equation}
\partial_\alpha J_\alpha^a(x) =-  \frac{A_{\alpha \beta \gamma}}{64 \pi^2} \epsilon^{\mu \nu \delta \sigma} F^\beta_{\mu \nu} F^\gamma_{\delta \sigma},
\end{equation}
where $A_{\alpha \beta \gamma}$ is a completely symmetry tensor
\begin{equation}
A_{\alpha \beta \gamma} = {\rm tr}\left[ \tau_\alpha^L\,\left\{ \tau_\beta^L\,\tau_\gamma^L\right\}  \right] -  {\rm tr}\left[ \tau_\alpha^R\,\left\{ \tau_\beta^R\,\tau_\gamma^R\right\}  \right] .
\end{equation}
Now, we can check that the $SU(2)_L^3$ anomaly vanishes since $\left\{\tau^a,\,\tau^b\right\}=\tfrac{1}{2}\delta^{ab}$, and so $A^{abc}=\delta^{bc}{\rm tr} [\tau^a] =0$. The $SU(3)_c^3$ anomaly also cancels since right-handed and left-handed have the same color. All the anomalies of the type $SU(N)U(1)^2$ cancel since ${\rm tr}\left[ \tau^a \left\{ 1,1\right\}\right]=2 {\rm tr} \left[ \tau^a \right] = 0$. Therefore, the only potentially dangerous anomalies are the ones of $U(1)_Y$ in the background of $U(1)_Y$, $SU(2)_L$, $SU(3)_c$ and gravity (universal coupling). But the matter content of the SM is exactly such that the anomalies cancel\footnote{Note that the mixed gauge-gravitational anomaly in Eq.~\eqref{eq:mixed-gauge-gravitational-anomaly} is automatically satisfied if the 3 other gauge anomalies, in Eq.~\eqref{eq:U13-anomaly}, \eqref{eq:SU32U1-anomaly} and \eqref{eq:SU22U1-anomaly}, vanish and hypercharge is quantized. See \cite{Lohitsiri:2019fuu} for an elegant proof which uses that Eq.~\eqref{eq:U13-anomaly}, \eqref{eq:SU32U1-anomaly} and \eqref{eq:SU22U1-anomaly} can be recast into the Diophantine equation $x^3+y^3=z^3$ and then applies Fermat's last theorem.}
\begin{align}
&U(1)^3_Y: \qquad \quad & \left( 2Y^3_L - Y^3_e - Y^3_\nu\right) + 3\left(2 Y_Q^3 - Y_u^3 - Y_d^3 \right) = 0,\label{eq:U13-anomaly}\\
&SU(3)^2U(1)_Y: \qquad \quad & 2Y_Q - Y_u - Y_d = 0,\label{eq:SU32U1-anomaly}\\
&SU(2)^2U(1)_Y: \qquad \quad & 2Y_Q - Y_u - Y_d = 0,\label{eq:SU22U1-anomaly}\\
&{\rm Grav}^2U(1)_Y: \qquad \quad & \left( 2Y_L - Y_e - Y_\nu\right) + 3\left(2 Y_Q - Y_u - Y_d \right) = 0. \label{eq:mixed-gauge-gravitational-anomaly}
\end{align}
Hence the SM is anomaly free: the marriage between quarks and leptons is a very peculiar one. Particularly, the electron has exactly the opposite charge as the proton. Actually, we can show that there is one more $U(1)$ symmetry which is anomaly free in the background of the four fundamental forces of Nature: $U(1)_{\rm B-L}$. In contrast, $SU(2)_L^2U(1)_B$ and $SU(2)_L^2 U(1)_L$ are anomalous. As a consequence, $U(1)_B$ and $U(1)_L$ are broken by non-perturbative processes with $\Delta(B+L)\neq 0$, called sphalerons, which can play an important role for generating the observed asymmetry between matter and anti-matter, see Sec.~\ref{sec:Matter-anti-matter-asymmetry} of Chap.~\ref{chap:SM_cosmology}.

The discovery of the chiral anomaly by ABJ  in 1968 was a particularly important result at that time since it explained the decay of the neutral pion, $\pi_0 \rightarrow \gamma \gamma$, from the anomaly $U(1)_{\rm e.m.}^2U(1)_A$. Another triumph of chiral anomalies is the explanation by Weinberg and 't Hooft in 1975 \cite{Weinberg:1975ui,tHooft:1976rip}, of the heaviness of the $\eta'$, with respect to the $\eta$ which has the same flavor content, as resulting from the global anomaly $SU(3)_{c}^2U(1)_A$. In the next section, we discuss an other application of the $SU(3)_{c}^2U(1)_A$ anomaly.

\subsection{Strong CP violation}
The SM Lagrangian contains terms in Eq.~\eqref{eq:line2}, of the form $\tilde{F}_{ \mu \nu}F^{ \mu \nu}$, which are total derivatives, violate $CP$ and are generated by chiral transformations. The one involving the gluons of $SU(3)_c$ can have physical consequences.

\paragraph{Total derivative:} 

The quantity $\tilde{F}_{ \mu \nu}$, called the Hodge-dual of the tensor strength, is defined by
\begin{equation}
\tilde{F}_{ \mu \nu} \equiv  \epsilon_{\mu\nu\alpha\beta} F^{\alpha\beta}.
\end{equation}
We can show that 
\begin{equation}
\label{eq:FFtilde}
\epsilon^{\mu\nu\alpha\beta} F^a_{\mu\nu} F^a_{\alpha\beta} = \partial^\mu K_\mu, \qquad K_\mu = \epsilon_{\mu\nu\alpha\beta} \left(A_{\rm nu}^a F_{\rm \alpha \beta}^a   - \frac{g}{3} f^{abc} A_\nu^a A_\alpha^b A_\beta^c \right),
\end{equation}
where $K_\mu$ is known as the Chern-Simons current. Total derivatives never contribute in perturbation theory since the corresponding Feynman rule contains a factor with the sum of all momenta going into the vertex minus the momenta going out, which is equal to zero.  However, $\theta$ can still have physical consequences through non-perturbative effects. 

\paragraph{$CP$-violation:} 
Furthermore, we can check that the terms in Eq.~\eqref{eq:FFtilde} violate $P$, $T$ and $CP$. This can be easily shown in the electromagnetism case where $F_{\mu\nu}\tilde{F}^{\mu\nu} = 4 \vec{E}.\vec{B}$, the transformation rules of the electric and magnetic fields being $ \vec{E} \xrightarrow{CP,\,T}  \vec{E}$ and $ \vec{B} \xrightarrow{CP,\,T} -\vec{B}$.

\paragraph{Generated by a chiral transformation:} 
\label{par:anomaly_strong}
Using the results from the previous section, e.g. Eq.~\eqref{eq:ABJ}, we now make the observation that the terms in Eq.~\eqref{eq:FFtilde} can be generated after performing a chiral rotation of the fermions
\begin{equation}
\label{eq:anomaly_chiral}
U(1)_A: \qquad \int \mathcal{D} \bar{\psi}\mathcal{D}\psi \xrightarrow{\psi \rightarrow e^{i\gamma_5 \theta } \psi} \int \mathcal{D} \bar{\psi}\mathcal{D}\psi \exp \left( i \theta \int \frac{g^2}{32\pi^2} \epsilon^{ \mu \nu \alpha \beta} F_{\mu \nu}^a F^a_{\alpha \beta} \right),
\end{equation}
where $F_{\mu\nu}^a$ is the field strength for anything under which fermions are charged. Therefore, the terms $F\tilde{F}$ in the SM Lagrangian are in one-to-one connection with the anomaly of the chiral symmetry $U(1)_A$ in the background of the SM gauge fields: $SU(3)_c^2U(1)_A$, $SU(2)_L^2U(1)_A$ and $U(1)_Y^2U(1)_A$.

\paragraph{A physical parameter:} 

For $U(1)_Y$ which is abelian, the term $A_\nu^a A_\alpha^b A_\beta^c$ is absent in Eq.~\eqref{eq:FFtilde} and $\int d^4x~ F\tilde{F} = 0$.\footnote{For $\int d^4x~ F^2$ to converge, we must have $F_{\mu\nu} < \frac{1}{r^2}$. Naively, this would imply $A_\mu < \frac{1}{r}$. However, $A_\mu$ can scale like $A_{\mu } \sim \frac{1}{r}$ if it is pure gauge $U\partial_\mu U$ where $U$ is the gauge matrix. For those fields we have $A_\nu F_{\rho \sigma}< \frac{1}{r^3}$ but $A_\nu A_\rho A_\sigma \sim \frac{1}{r^3}$. So we conclude that only the term $A_\nu A_\rho A_\sigma$ in Eq.~\eqref{eq:FFtilde} can contribute to $\int d^4x ~F\tilde{F}$. I thank Filippo Sala for pointing me this fact.}
For $SU(2)_L$, we can rotate the left-handed fermion fields to move the $\theta_2$ angle in the mass matrix and then rotate the right-handed fields, which are uncharged and therefore not subject to the anomaly, to remove it completely.

However, for $SU(3)_c$, where both left-handed and right-handed quarks are colored, we can not rotate away $\theta_3$. Instead, we can only move it back and forth between the anomaly term and the Yukawa matrix. From performing a chiral rotation to remove the Yukawa phase, we obtain
\begin{equation}
\label{eq:QCD_lagrangian_theta_bar}
\mathcal{L}_\theta = \bar{\theta} \frac{g_s^2}{32\pi^2} F^a_{\mu\nu}\tilde{F}^{a\mu\nu}, \qquad \bar{\theta} \equiv \theta_3 - \theta_{\rm Yuk},
\end{equation}
where $\theta_{\rm Yuk}$ is the phase in the Yukawa matrix (if any)
\begin{equation}
\label{eq:theta_yukawa_matrix}
\theta_{\rm Yuk} \equiv \rm arg \,det \left( Y_d Y_u \right).
\end{equation}
Thus, $\bar{\theta}$ is a basis-independent measure of the strong CP violation and can be physical. Note, that $\theta_{\rm Yuk}$ is not contained among the physical parameters of the CKM matrix in Sec.~\ref{par:CP_violation_CKM}, because the counting $9-5=4$ ignores the quantum anomaly and assumes chiral transformations to be symmetries of the Lagrangian.

\paragraph{Where is the neutron EDM ? :} 

Particularly, $\bar{\theta}$ generates an electron dipole moment (EDM) for the neutron \cite{Crewther:1979pi, Pospelov:2005pr,Dubbers:2011ns,Chupp:2017rkp}
\begin{equation}
d_N \simeq \left(2.5 \times 10^{-16}~\rm e.cm \right) \bar{\theta}.
\end{equation}
The current bound on the neutron EDM  is $d_N \lesssim 3 \times 10^{-26}~\rm e.cm$ \cite{Baker:2006ts, Pendlebury:2015lrz,nEDM:2020crw}, so that 
\begin{equation}
\label{eq:theta_bound}
\bar{\theta} \lesssim 10^{-10}.
\end{equation}
The smallness of $\bar{\theta}$ without any symmetry protection, except $CP$, is known as the strong $CP$ problem, see Sec.~\ref{sec:strong_CP_pb}.

\paragraph{$\theta$-vacua -  dilute instanton gas approximation:} 
Another physical effect of $\bar{\theta}$ is that different values of the angle correspond to different vacua, called \textbf{$\theta$-vacua}. The explicit dependence of the vacuum energy on the $\bar{\theta}$ parameter was first computed in the \textbf{dilute instanton gas approximation}, cf. original works pioneered by Callan, Dashen and Gross in 1976 \cite{Callan:1976je,Callan:1977gz,Callan:1978bm,Bernard:1979qt,Pisarski:1980md,Gross:1980br,Luscher:1981zf,Morris:1984zi} and reviews \cite{Coleman:1978ae,Vainshtein:1981wh,Schafer:1996wv}.
Let us start by discussing properties of the vacuum of $SU(N)$ gauge theories. In order to give a finite action, Yang-Mills fields must go to zero at infinity, up to a pure gauge configuration 
\begin{equation}
A_\mu \rightarrow  i g(x) \partial_\mu g(x)^{-1}, \qquad \textrm{where}\quad g(x) : S^3_\infty \rightarrow  SU(N).
\end{equation}
The set of maps $g(x)$ falls into an infinite number of distinct equivalence classes. Indeed, the third homotopy group of $SU(N)$ is non-trivial
\begin{equation}
\pi_3(SU(N)) = \mathbb{Z},\qquad N \geq 2.
\end{equation}
This means that the connected components of the space of finite-action solutions of Yang-Mills $SU(N)$ theories can be classified with an integer. This integer is known has the \textbf{winding number} or \textbf{Pontryagin number}
\begin{equation}
\label{eq:nu_Pontryagin}
\nu(g) = \frac{1}{24\pi^2} \int_{S^3}d\theta_1 d\theta_2 d\theta_3 \epsilon^{ijk} \textrm{tr} \left( g^{-1} \partial_i g g^{-1} \partial_j g g^{-1} \partial_k g  \right) = \frac{1}{32\pi^2} \int d^4x \,F\tilde{F},
\end{equation}
where $\theta_1$, $\theta_2$ and $\theta_3$ are three angles parameterizing $S^3$.
The field configuration which interpolates between neighboring winding numbers are the configuration which have the lowest Euclidean action
\begin{equation}
\label{eq:euclidean_action_Bogomolny}
S_{\rm E} = \frac{1}{4 g^2} \int d^4x\, F_{\mu\nu}F^{\mu\nu} = \frac{1}{4 g^2} \int d^4x\, \left[F\tilde{F} + \frac{1}{2}(F-\tilde{F})^2\right] = \frac{8\pi^2\nu}{g^2} + \frac{1}{8g^2} \int d^4x(F-\tilde{F})^2.
\end{equation}
They saturate the Bogomol’nyi bound \cite{Bogomolny:1975de} for $\nu = 1$ or $\nu = -1$,
\begin{equation}
\label{eq:BPST_action}
F=\pm \tilde{F} \qquad \implies \qquad S_0 =\frac{8\pi^2 }{g^2}.
\end{equation}
They have been computed by Belavin, Polyakov, Schwarz and Tyupkin in 1975 \cite{Belavin:1975fg} and are known as \textbf{BPST instantons}, 
\begin{equation}
A_\mu(x) = if(r^2) \left( \frac{\tau + i \vec{x} \cdot \vec{\sigma}}{r}\right) \partial_\mu\left( \frac{\tau + i \vec{x} \cdot \vec{\sigma}}{r}\right)^{-1},\qquad f(r^2) =  \frac{r^2}{r^2 + \rho^2}.
\end{equation}
Since the instantons offer the possibility of quantum tunneling between the classical vacua $\left|n\right>$ characterized by the winding number $n$ of gauge fields at infinity, the quantum vacuum is a linear combination $\left|{\rm vac}\right> = \sum_{n=-\infty}^{+\infty} a_n \left|n\right>$. We can easily show that the only linear combination which is invariant under the gauge transformation $\left|n\right> \xrightarrow{g(x)} \left|n+1\right>$ is the so-called \textbf{$\theta$-vacuum}
\begin{equation}
\left|\theta\right> = \sum_{n=-\infty}^{+\infty} e^{-i\theta}\left|n\right>.
\end{equation}
Due to tunneling, the infinite degeneracy between the energy of the classical vacua $\left|n\right>$ is lifted. The energy of the ground state can be computed from the transition amplitude in the stationary limit
\begin{equation}
\left<\theta \right| e^{iHT} \left|\theta\right>= \sum_{i} e^{-E_i T} \left< \theta | E_i\right>\left<E_i |\theta\right> \quad\xrightarrow{T \to\infty}\quad e^{-E_0 T}.
\end{equation}
In quantum field theory, the transition amplitude can be computed in the Feynman path integral formulation
\begin{equation}
\left<\theta \right| e^{iHT} \left|\theta\right> = N \int DA_\mu\, e^{-S_{\rm E}}.
\end{equation}
From Eq.~\eqref{eq:euclidean_action_Bogomolny}, we can see that the path integral is dominated by the BPST instantons, whose action is minimal. They come as two types, the instantons $\nu =1$ and the anti-instantons $\nu =-1$. Suppose we have a gas of $n$ instantons with centers $x_1, x_2, \cdots x_n$. We must integrate over the positions of all the individual instantons, which gives
\begin{equation}
\int_{-T/2}^{T/2}dt_1 \int_{-T/2}^{t_1}dt_2 \cdots \int_{-T/2}^{t_{n-1}}dt_n \int \prod_{i=1}^{n} d^3x_i = \frac{(TV)^{n}}{n!}.
\end{equation}
In addition, we suppose that we have $\bar{n}$ anti-instantons. The path integral over all possible histories gives
\begin{equation}
\label{eq:path_integral_DIGA}
\left<\theta \right| e^{iHT} \left|\theta\right> =  \sum_{n,\bar{n}}e^{i(n-\bar{n})\theta} \frac{(TV K e^{-S_0})^{n}}{n!}\frac{(TV K e^{-S_0})^{\bar{n}}}{\bar{n}!}
\end{equation}
where $S_0$ is defined in Eq.~\eqref{eq:BPST_action} and where $K$ is a prefactor accounting for the quantum fluctuations of the instanton size $\rho$, and which has been shown to contain an IR divergence by t'Hooft in 1976 \cite{tHooft:1976snw}, which he called \textbf{infrared embarrassment}. In Eq.~\eqref{eq:path_integral_DIGA}, we have neglected the overlap between instantons and anti-instantons. This is valid as long as the instanton size $\rho$ is larger than their separation. This is the so-called \textbf{dilute instanton gas approximation} \cite{Callan:1977gz,Coleman:1978ae,Vainshtein:1981wh}.  We obtain the energy of the $\theta$-vacuum
\begin{equation}
E_0(\theta)/V= K e^{-S_0} \cos{\theta}.
\end{equation}
Comparisons with lattice studies show that the dilute instanton gas approximation is valid only for $T\gtrsim 10^6~ \rm GeV$ \cite{Berkowitz:2015aua,Borsanyi:2015cka,diCortona:2015ldu}. Lattice remains the only reliable tool to compute the axion potential for $\Lambda_{\rm QCD} \lesssim T\lesssim 10^6~ \rm GeV$. 

\paragraph{$\theta$-vacua - chiral lagrangian:} 
For $T\lesssim \Lambda_{\rm QCD} $, the axion potential can be computed rather precisely using the \textbf{Chiral lagrangian of QCD} \cite{DiVecchia:1980yfw, diCortona:2015ldu}. We already introduced the chiral Lagrangian of QCD in Eq.~\eqref{eq:chiral_lagrangian} of Sec.~\ref{sec:QCD} in order to compute the mass of the pion. From Eq.~\eqref{eq:theta_yukawa_matrix}, we see that the $\bar{\theta}$ parameter in the QCD Lagrangian in Eq.~\eqref{eq:QCD_lagrangian_theta_bar} can be moved to the quark mass matrix via the chiral transformation
\begin{equation}
\begin{pmatrix}  u \\  d\end{pmatrix} \quad \xrightarrow{U(1)_{\rm A}} \quad e^{i\gamma_5 Q_a\frac{\bar{\theta}}{2}}\begin{pmatrix}  u \\  d\end{pmatrix}, \qquad \textrm{tr}\left(Q_a\right) = 1.
\end{equation}
The chiral Lagrangian of QCD in \eqref{eq:chiral_lagrangian} becomes 
\begin{align}
\mathcal{L}_{\rm eff}\supset \frac{V^3}{2}\textrm{tr}\left[M_a U + M_a^\dagger U^\dagger \right],\qquad V^3 = \frac{F_{\pi}^2m_\pi^2}{m_u + m_d}, \label{eq:eff_lag_mass_pion_2}
\end{align}
with 
\begin{equation}
U = e^{i\Pi/F_{\pi}}, \qquad \Pi =\begin{pmatrix}  \pi^0 & \sqrt{2} \pi^+\\  \sqrt{2}\pi^- & -\pi^0\end{pmatrix} ,
\end{equation}
and
\begin{equation}
M_a = e^{i\gamma_5 Q_a\frac{\bar{\theta}}{2}} M_q e^{i\gamma_5 Q_a\frac{\bar{\theta}}{2}}, \qquad M_q=  \begin{pmatrix}  m_u &0 \\  0 & m_d \end{pmatrix} .
\end{equation}
Keeping only the terms depending on $\bar{\theta}$, we obtain
\begin{align}
V(\bar{\theta}) &= -V^3\left[ m_u \cos\left( \frac{\pi^0}{F_{\pi}} - \frac{\bar{\theta}}{2} \right) + m_d \cos\left( \frac{\pi^0}{F_{\pi}} + \frac{\bar{\theta}}{2} \right)\right]\\
&= -m_\pi^2 F_{\pi}^2 \sqrt{1 - \frac{4m_um_d}{(m_u+m_d)^2} \sin^2 \left( \frac{\bar{\theta}}{2} \right)} \cos{\left(\frac{\pi^0}{F_{\pi}} - \phi_a\right)},\\
\end{align}
where
\begin{equation}
\tan{\phi_a} \equiv \frac{m_u - m_d}{m_u+m_d} \tan{\left( \frac{\bar{\theta}}{2} \right)}.
\end{equation}
In order to minimize the potential, the vacuum expectation value of the meson field is equal to $\left<\pi^0\right> = \phi_a\,F_\pi$. We obtain that the vacuum energy density is a function of the $\bar{\theta}$ parameter \cite{diCortona:2015ldu}
\begin{equation}
\label{eq:pot_theta}
V(\bar{\theta}) = - m_{\rm \pi}^2 f_{\rm \pi}^2 \sqrt{1-\frac{4m_um_d}{(m_u+m_d)}\sin^2 \left( \frac{\bar{\theta}}{2} \right)},
\end{equation}
where $m_{\pi}\simeq 135~$MeV and $F_{\pi} \simeq 92~$MeV. We can see that the vacuum with minimum energy is the $CP$-conserving vacuum $\bar{\theta}=0$, in accordance with the Vafa-Witten theorem \cite{Vafa:1983tf}.

\section{Open problems}

\subsection{Hierarchy problem}
\label{sec:hierarchy_pb}
\begin{figure}[h!]
\centering
\raisebox{0cm}{\makebox{\includegraphics[width=0.8\textwidth, scale=1]{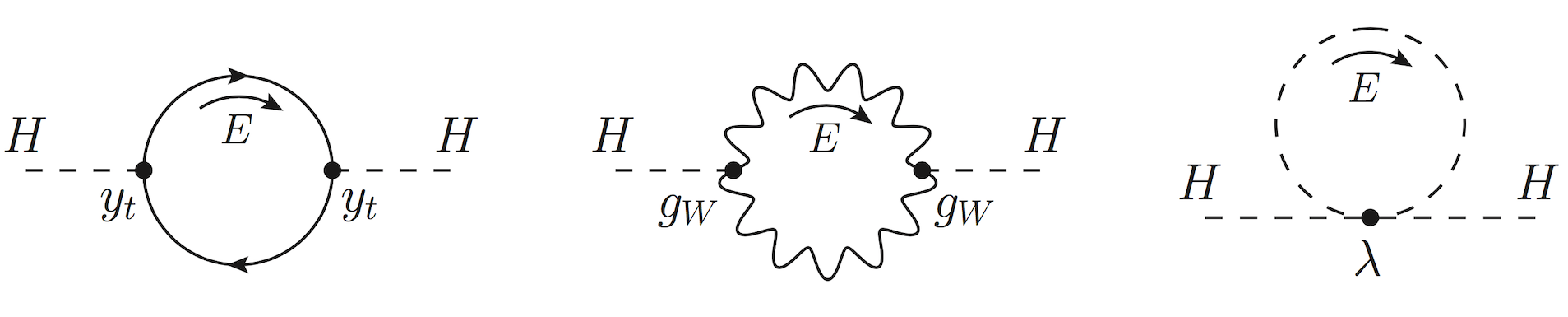}}}
\caption{\it \small 1-loop corrections to the Higgs propagator from the top quark, the weak gauge bosons and the Higgs boson itself. Figure reproduced from \cite{Panico:2015jxa}.}
\label{fig:Higgs_corrections}
\end{figure}


\paragraph{Renormalization:}
The parameters in the SM Lagrangian density in Sec.~\ref{sec:lagrangian}, the gauge couplings $g_i$, the Yukawa couplings $\lambda_i$ or the Higgs potential parameters $\mu^2$ and $\lambda$,  are called \textbf{bare} parameters. They are different from the \textbf{measured} or \textbf{renormalized} parameters due to the presence of quantum corrections. This is the essence of \textbf{renormalization}, initially developed at the end of the 40s by Tomonaga \cite{Tomonaga:1946zz, Tomonaga:1948zz}, Schwinger \cite{Schwinger:1948iu, Schwinger:1948yk, Schwinger:1948yj, Schwinger:1949ra, Schwinger:1951nm} and Feynman \cite{Feynman:1949hz, Feynman:1949zx, Feynman:1950ir} (Nobel prize 1965) but also Bethe \cite{Bethe:1947id}, Pauli \cite{Pauli:1949zm} and Dyson \cite{Dyson:1949ha, Dyson:1949bp}, in order to absorb the troublesome divergent integrals into a rescaling of the masses and coupling constants of the theory.\footnote{They applied it to the get predictions in Quantum Electrodynamics (QED), the theory of light, electrons and positrons. The approach of Tomonaga and Schwinger is based on operators whereas the approach of Feynman is based on diagrams. Nowadays, the value of the magnetic moment of the electron can be computed within the framework of QED up to 4 loops \cite{Laporta:2017okg} and experimental measurements in Laboratoire Kastler-Brossel are able to agree with the theoretical predictions with a precision of one part out of $10^{14}$ \cite{Bouchendira:2010es}} 
The quantum corrections are often UV divergent and then, they are sensitive to any unknown physics above the energy scale at which the SM has been tested experimentally.
For instance, we can show that the quantum corrections to the fermion masses are logarithmically divergent  (e.g. \cite{Schwartz:2013pla})
\begin{equation}
\delta m_{\psi_i}\propto\,\alpha \,  m_{\psi_i} \,\rm log\left(\frac{\Lambda}{m_{\psi_i}}\right) ,
\end{equation}
where $\Lambda$ is a UV cut-off parameterizing the UV divergence.  

\paragraph{Technical naturalness:}
The logarithmic dependence of the fermion mass on the UV scale, instead of a linear one $\delta m_{\psi_i}\propto \Lambda$, originates from symmetry consideration. Indeed, setting the fermion mass to zero introduces a new symmetry in the Lagrangian, the \textbf{chiral symmetry}, defined in footnote \ref{footnote:chiral_transf}.
Quantum corrections $\delta m_{\psi_i}$ to the fermion masses must respect the possibility to enhance the symmetry group when we set $m_{\psi_i}=0$, so they must be proportional to $m_{\psi_i}$. Therefore, from dimensional considerations, the renormalized mass can only depend logarithmically on the scale of new physics $\Lambda$. Whatever the scale of new physics $\Lambda$, the renormalized mass will only differ from the bare mass by tiny logarithmic corrections. 

The quantum corrections to the photon mass are identically zero because of the $U(1)_{\rm e.m.}$ gauge symmetry.
We say that the fermion and vector boson masses are \textbf{protected} from large quantum corrections by chiral and gauge symmetries respectively.\footnote{\label{footnote:technically_natural}This is the idea being \textbf{technical naturalness}, introduced by 't Hooft \cite{tHooft:1979rat}: an observed coupling constant can be much smaller than 1 if a symmetry is restored when the coupling is set to zero.}

\paragraph{The electroweak scale is not natural:}
In contrast, setting to zero the Higgs mass $m_{\rm H}$ does not introduce any additional symmetry and the quantum corrections to the Higgs mass at 1-loop, see Fig.~\ref{fig:Higgs_corrections}, are quadratically sensitive to the scale of new physics $\Lambda$, e.g. \cite{Giudice:2013yca}
\begin{equation}
\delta m_{\rm H}^2= \frac{3}{8\pi^2 }\left( \frac{\Lambda}{v}\right)^2\left( m_{\rm H}^2 + 2m_{\rm W}^2 + m_{\rm Z}^2 - 4m_{\rm t}^2  \right) .
\end{equation}
The incompatibility between the quadratic sensitivity of the Higgs boson mass on the scale of new physics $\Lambda$ and its measured value at $m_{\rm H} \simeq 125~\rm GeV$, is called the \textbf{Naturalness} or \textbf{Hierarchy problem}. 

The existence of physics Beyond the Standard Model (BSM) at an energy scale $\Lambda \gg m_{\rm H}$ is motivated by different open problems, the small neutrino mass problem, the strong CP problem, the flavor hierarchy problem, the dark matter puzzle or the baryonic asymmetry of the universe. Or at least \textbf{quantum gravity} around the \textbf{Planck scale} $\Lambda \simeq M_{\rm pl} \simeq 2.44\times 10^{18}~$GeV, the scale where the strength of gravity becomes comparable to the other forces. In that last case, the hierarchy problem $ M_{\rm pl} \gg m_{\rm H}$ can be rephrased as: why does the classical \textbf{gravity force} between two electrons 
\begin{equation}
F_{\rm Grav} = \frac{G \,m_e^2}{r} \simeq \frac{m_{\rm H}^2}{M_{\rm pl}^2} \frac{Y_e^2}{r},
\end{equation}
is much smaller than the \textbf{Coulomb force}
\begin{equation}
F_{\rm Coul} = \frac{\alpha}{r},
\end{equation}
knowing that \textbf{quantum corrections} to $m_{\rm H}^2$ are of order $M_{\rm pl}^2$ ?

The Naturalness problem suggests the existence of a principle which protects the smallness of the Higgs mass relative to its would-be UV corrections. We cite  below the most studied proposals to make the measured value of the Higgs mass natural.
\paragraph{Supersymmetry (SUSY):} 
In 1967, Coleman and Mandula proved that the largest symmetry group of quantum field theory is the direct product of some internal symmetries and the Poincaré group of spacetime isometries ($4$ transations, $3$ rotations, $3$ boosts) \cite{Coleman:1967ad}. In 1971, Golfand and Likhtman found a loophole and extended the Poincaré group with a two-component weyl spinor operator which transforms \textbf{bosons into fermions}, and vice-versa, and whose anticommutator is a space-time translation \cite{Golfand:1971iw}. 
If the universe is supersymmetric, each SM particle has a \textbf{super-partner} with equal mass but different spin, an idea first proposed by Pierre Fayet in 1977 \cite{Fayet:1977yc}. In 1981, Dimopoulos and Georgi used SUSY to stabilize the EW scale \cite{Dimopoulos:1981zb}. Indeed, due to the relative sign between fermion loops and boson loops, the \textbf{quantum corrections cancel order by order} in perturbation.  However, the non-observation in Nature of those super-particles (sfermions,  gauginos, gravitino) implies that SUSY must be broken at low-energy in order to lift the mass degeneracy.  After SUSY breaking the correction to the Higgs mass from the top $t$ and stop $\tilde{t}$ is \cite{Craig:2013cxa}
\begin{equation}
\label{eq:Higgs_corr_stop}
\delta m_{\rm H}^2 = 
\begin{cases}
-\frac{3\lambda_t^2}{4\pi^2}\left( y_t^2 - y_t^2 \right) =0&\hspace{2em}\textrm{SUSY is not broken},\\
-\frac{3\lambda_t^2}{4\pi^2}\, m_{\tilde{t}}^2 \,{\rm log }\left[ \frac{M_{\rm \cancel{SUSY}}}{m_{\tilde{t}} }\right]  &\hspace{2em}\textrm{SUSY is broken},\\
\end{cases}
\end{equation}
where $M_{\rm \cancel{SUSY}}$ is the SUSY breaking scale and $m_{\tilde{t}}$ is the stop mass.
\\
The discovery of the Higgs at $125~$GeV and the searches on super-partners at LHC have already imposed the presence of a \textbf{fine-tuning} between radiative corrections of the order of 1 out of 100 for the minimal SUSY model \cite{Hall:2011aa, Craig:2013cxa}. 
For instance, latest lower bounds on the stop mass are $m_{\tilde{t}} \gtrsim 1~$TeV \cite{CMS:2017qjo}. Assuming the tree level Higgs mass $m_{\rm H}^2 \simeq m_{\rm Z}^2/2$ and plugging $M_{\rm \cancel{SUSY}} \sim 10~$TeV and $m_{\tilde{t}} \sim 1~$TeV in Eq.~\eqref{eq:Higgs_corr_stop}, we find a needed fine-tuning of $\sim 2\%$ between the different radiative corrections in order to get the correct Higgs mass.

There are many alternatives to reduce the fine-tuning, e.g. next-to-minimal SUSY extensions (NMSSM) that involve an additional SM singlet \cite{Ellwanger:2009dp,Ross:2011xv}, colorless top partners if SUSY is `folded' \cite{Burdman:2006tz}, a focus point on the SUSY RGE \cite{Feng:1999zg,Feng:2012jfa}, or top partners naturally lighter than others sparticle \cite{Craig:2012hc}. The last possibility can arise naturally if SUSY breaking is mediated through a gauged flavor symmetry \cite{Craig:2012di}.  Also, the missing transverse energy signal at LHC can be suppressed due to the sparticles spectrum being compressed \cite{LeCompte:2011cn}, SUSY being `Stealth' \cite{Fan:2011yu}, or $R$-parity being violated \cite{Csaki:2011ge}. 
For lecture notes/books, see e.g. \cite{Wess:1992cp, Martin:1997ns,Weinberg:2000cr,Terning:2006bq, Binetruy:2006ad,Aitchison:2007fn,Peskin:2008nw}. For reviews, see e.g. \cite{Craig:2013cxa, Csaki:2016kln, 1684708,Wulzer:2019max,Lee:2019zbu,RossiaDESYhierarchy,Hebecker:2020aqr}. Particularly, see \cite{Craig:2013cxa} for a review on experimental constraints on SUSY and its theoretical extensions. 

In addition to provide a natural explanation for the stability of the EW scale, SUSY predicts gauge coupling unification \cite{Raby:2006sk} and natural Dark Matter candidates \cite{Jungman:1995df}.

Low-scale SUSY also comes with its own difficulties. In its minimal version, it predicts large flavor and CP violation at odd with observations \cite{Ellis:1981ts,Barbieri:1981gn,Campbell:1983bw,Duncan:1983iq,Gabbiani:1988rb,Dimopoulos:1994gj,Barbieri:1994pv,Barbieri:1995rs}, and dramatic post-BBN entropy injection due to gravitino and moduli decay \cite{Weinberg:1982zq,Coughlan:1984yk,Ellis:1984er, Banks:1993en,Moroi:1993mb,Banks:1995dt,Kawasaki:1994af,Kawasaki:2008qe}. This motivated Arkani-Hamed and Dimopoulos to propose in 2004 that SUSY might be broken at a much higher scale than the TeV one, at the expense of a large fine-tuning \cite{ArkaniHamed:2004yi,Giudice:2004tc,Arkani-Hamed:2004zhs,Wells:2004di,Hall:2009nd,Bagnaschi:2014rsa,Baer:2022naw}. A larger SUSY breaking scale preserves gauge coupling unification \cite{ArkaniHamed:2004yi,Giudice:2004tc,Arkani-Hamed:2004zhs} and is compatible with the Higgs at $125$~GeV \cite{Bagnaschi:2014rsa}. Its realization in which it also explains dark matter and predicts signals at colliders is called \textbf{split SUSY} \cite{ArkaniHamed:2004yi,Giudice:2004tc,Arkani-Hamed:2004zhs}.

\paragraph{Composite Electroweak dynamics:} 
The QCD scale around $200$~MeV arises from dimensional transmutation when the strong force confines, cf. Sec.~\ref{par:color_confinement}. We could expect that the same occurs with the electroweak scale. This is the \textbf{Technicolor} idea introduced at the end of the 70s \cite{Terazawa:1976xx,tHooft:1979rat,Dimopoulos:1979es,Farhi:1980xs}.  We introduce additional massless fermions (technifermions) which are charged under an asymptotically free gauge theory (technicolor) which confines at the electroweak scale. Like in QCD, the global \textbf{chiral} symmetry of the technifermions is \textbf{spontaneously broken} down to the \textbf{vectorial} subgroup when the technifermions condense
\begin{equation}
\label{eq:SSB_technicolor}
SU(2)_L\times SU(2)_R \times U(1)_V \rightarrow  SU(2)_V,
\end{equation}
where $2$ is the number of techniflavors. $U(1)_V$ is the simultaneous vectorial rotation of all the technifermions at the same time. We did not include the axial rotation $U(1)_A$ which is anomalous.
The SM gauge interactions $SU(2)_L\times U(1)_Y$ form a subgroup of the global chiral symmetry group in a way that it gets spontaneously broken to $SU(2)_L\times U(1)_Y \rightarrow U(1)_{\rm e.m.}$ after chiral symmetry breaking. Particularly, three of the four \textbf{Nambu-Golstone bosons} generated in Eq.~\eqref{eq:SSB_technicolor} become the longitudinal components of the SM vector bosons $W^{\pm}$ and $Z$, which then get massive. The SM fermions get their mass from bilinear operators coupling two SM fermions with an operator composed of technifields.

The \textbf{first} advantage of the Technicolor scenario is that there is \textbf{no need} to introduce the Higgs boson responsible for the hierarchy problem. A Higgs-like scalar can still be accommodated as one of the resonance of the strong sector. In that case, the Higgs boson is not sensitive to virtual effects above the compositeness scale, and its mass is natural.

The \textbf{second} advantage of Technicolor is that the electroweak scale is dynamically generated through \textbf{dimensional transmutation}, when the Technicolor gauge coupling runs toward a large value in the infrared, as in QCD, see Sec.~\ref{par:color_confinement}. For reviews on Technicolor, see \cite{Chivukula:2000mb, Lane:2002wv, Hill:2002ap,Shrock:2007km}.

The discovery of the Higgs boson at $125~$GeV, electroweak precision tests as well as the none-observation of exotic resonances at the LHC, have \textbf{excluded} most of the technicolor scenarios, e.g. see \cite{Arbey:2015exa}.
In order to conciliate the composite hypothesis with the LHC constraints, a solution is to introduce a natural little hierarchy between the compositeness scale $f$ and the electroweak scale $v$.
This is the design behind \textbf{Composite Higgs} (CH) scenarios which, for reasons which will become clear soon, are a sort of mix between the Technicolor idea and the SM Higgs mechanism. The composite sector of technicolor is enlarged with a global symmetry group $G$, which gets \textbf{spontaneously broken} to a subgroup $H$ when the strong sector confines
\begin{equation}
\label{eq:SSB_CH}
G \rightarrow H.
\end{equation}
The Higgs field is defined as one of the Nambu-Goldstone bosons of the \textbf{coset} $G/H$. 
In contrast to Technicolor, the breaking pattern in Eq.~\eqref{eq:SSB_CH} preserves the SM EW gauge group which is contained in the unbroken group
\begin{equation}
SU(2)_L \times U(1)_Y \subset H.
\end{equation}
We may ask then, where do the vector boson masses come from? In fact, the gauging of a subgroup of $G$ only, as well as the embedding of SM fermions within large representations of the group $G$, introduce an \textbf{explicit breaking} of $G$. Therefore, loops of SM fermions and gauge fields generate a potential for the Higgs, which in turn can acquire a VEV. 
Consequently, the misalignment between the true vacuum configuration and the SM group $SU(2)_L \times U(1)_Y$, breaks the EW symmetry spontaneously. This is the \textbf{misalignement mechanism} introduced in 1983 by Georgi and Kaplan \cite{Kaplan:1983fs, Kaplan:1983sm, Dugan:1984hq}. The degree of misaligment can be measured by $ \xi \equiv (v/f)^2$. The SM couplings are recovered in the `alignement limit' $\xi \to 0$. In the case of minimal CH models, precision measurements of Higgs couplings to SM vector bosons and fermions impose a little hierarchy between the EW scale $v$ of the compositeness scale $f$ \cite{Aad:2015pla}
\begin{equation}
\label{eq:CH_ATLAS}
\xi \equiv (v/f)^2 \lesssim 0.1, \qquad\rightarrow \qquad f \gtrsim 800~\rm GeV.
\end{equation}
In CH scenario, the SM fermions get their mass from operators which are \textbf{linear} in the SM fermions fields, and not bilinear as in the technicolor way. As a consequence, the physical fermions are linear combinations of elementary states and composite states, hence the name of \textbf{partial compositeness} proposed by D.B Kaplan in 1991 \cite{Kaplan:1991dc} .
For reviews on CH, see \cite{Contino:2010rs,Bellazzini:2014yua, Panico:2015jxa,Wulzer:2019max,RossiaDESYhierarchy}. 

Arrange $\xi$ to be naturally small and compatible with Eq.~\eqref{eq:CH_ATLAS}, is the design of \textbf{Little Higgs}, proposed in 2001 \cite{ArkaniHamed:2001nc,ArkaniHamed:2002qy}. The gauge and fermion sector is extended such that the explicit breaking of $G$, responsible for the generation of the Higgs potential, arises only when \textbf{at least two} couplings are switched on simultaneously in the Lagrangian. As a consequence of this \textbf{collective breaking}, the Higgs potential is only generated at \textbf{two loops}. For reviews on Little Higgs, see \cite{Perelstein:2005ka,Schmaltz:2005ky}.
Another alternative to naturally lower the EW scale $v$ with respect to the confining scale $f$ is the \textbf{Twin Higgs} scenario \cite{Chacko:2005pe,Batell:2022pzc} where a second exact copy of the SM is introduced such that the quadratic dependence on the UV cut-off, in the Higgs potential, accidentally cancels. The Twin Higgs model is a well-known realization of \textbf{neutral naturalness}, an approach for solving the hierarchy problem with top partners which are entirely neutral under the SM gauge groups \cite{Craig:2014aea}.

\paragraph{Large flat extra-dimensions:} 
In a model proposed in 1998 By Antoniadis, Arkani-Hamed, Dimopoulos and Dvali \cite{ArkaniHamed:1998rs, Antoniadis:1998ig}, the SM is confined on a brane \cite{Rubakov:1983bb} whereas the gravity interactions are \textbf{diluted} due to their propagation in large compactified extra-dimensions. If the fundamental gravity scale in the full $4+n$ dimensional space is $M_*$ and that the radii of the $n$ extra dimensions is given by $r$, then the effective gravity scale $M_{\rm pl}$ in 4D is 
\begin{equation}
M_{\rm pl}^2 \sim M_*^{2+n}r^n.
\end{equation}
Assuming that the fundamental scale is the electroweak scale $M_*\sim \rm TeV$ and plugging the value $M_{\rm pl}\simeq 2.44\times 10^{18}~\rm GeV$, we obtain the size of the extra-dimensions
\begin{align}
&r (n=1) \sim 10^{15} ~{\rm mm},\qquad r (n=2) \sim 0.5~{\rm mm}, \qquad \\
&r (n=3) \sim 10^{-6} ~{\rm mm}, \qquad r (n=4) \sim 10^{-8} ~{\rm mm}.
\end{align}
There is no hierarchy problem anymore since the cut-off is now $M_*$.
Newton law has been tested up to a fraction of a millimeter using torsion balance experiment \cite{Tan:2020vpf,Lee:2020zjt} so the size of the extra-dimensions must be $r \lesssim 0.1~\rm mm$. 
Hence $n=1$ is excluded while $n=2$ is under test. 
See \cite{Rattazzi:2003ea,Csaki:2004ay,Cheng:2010pt} for reviews.

\paragraph{Warped extra-dimensions:}
\label{sec:warped_extra_dim_hierarchy}
In 1999, Randall and Sundrum \cite{Randall:1999ee,Randall:1999vf} introduce a warped fifth dimension which \textbf{red-shifts} the Planck scale down to the electroweak scale. 
More precisely, the RS model is a slice of $AdS_5$ which is bounded by two branes, a UV-brane located at $y= 0$ and a IR-brane located at $y=y_{\rm IR}$. The $AdS_5$ metric is given by
\begin{equation}
\label{eq:AdS5}
ds^2 = e^{-2ky} \eta_{\mu\nu} dx^\mu dx^\nu - dy^2,
\end{equation}
where $k$ is the AdS curvature scale and $\mu,\,\nu=1...4$.
Let's assume that  the Higgs field is localized on the $IR$ brane, such that the Higgs action reads
\begin{equation}
S \subset \int d^5x \sqrt{-g}\left[ g^{\mu\nu}\partial_\mu H\partial_\nu H^\dagger - m_{H}^2 \left|H\right|^2 - \lambda \left|H\right|^4 \right]\, \delta\left( y - y_{\rm IR}\right),
\end{equation}
with the metric determinant $\sqrt{-g}=e^{-4ky}$.
Integrating over the extra-dimension
\begin{equation}
S \subset \int d^4x\, e^{-4ky_{\rm IR}} \left[ e^{2ky_{\rm IR}} \eta^{\mu\nu}\partial_\mu H\partial_\nu H^\dagger - m_{H}^2 \left|H\right|^2 - \lambda \left|H\right|^4\right],
\end{equation}
and rescaling $e^{-ky}H \rightarrow H$ to make the kinetic terms canonical, we obtain
\begin{equation}
S \subset \int d^4x  \left[\eta^{\mu\nu}\partial_\mu H\partial_\nu H^\dagger - e^{-2 k y_{\rm IR}}m_{H}^2 \left|H\right|^2 - \lambda \left|H\right|^4 \right].
\end{equation}
The effective Higgs mass is \textbf{redshifted} $m_{H} \rightarrow e^{- k y_{\rm IR}}m_{H}$ with a strength exponentially sensitive to the separation between the two branes. This solves the Hierarchy problem. 
For instance, if the fundamental Higgs mass in $5D$ is near the Planck scale, the effective Higgs mass on the brane can be around the electroweak scale for $k\,y_{\rm IR}\simeq 35$. The stabilization of the position of the IR brane at $k\,y_{\rm IR}\simeq 35$ is realized by promoting the branes inter-distance $y_{\rm IR}$ to a field having a non-trivial potential, this is the \textbf{Goldberger-Wise} mechanism \cite{Goldberger:1999uk,Goldberger:1999un}.
The gravity scales in 5D and 4D are related though $M_{\rm pl}^2=M_{5}^3/2k$, where we can safely choose $M_{5}\sim k \sim M_{\rm pl}$, in contrast to the proposal of large flat extra-dimensions. 
Warped geometry in 5D can be interpreted as \textbf{holographic dual} to Composite Higgs scenarios in 4D, e.g. see \cite{Agashe:2004rs} for one of the first implementation.
See reviews on warped extra-dimensions in \cite{Rattazzi:2003ea, Csaki:2004ay, Sundrum:2005jf,Gherghetta:2010cj}. In Sec.~\ref{sec:holography_confining_phase_PT} of Chap.~\ref{chap:1stOPT}, we introduce first-order cosmological phase transitions in the RS scenario as potential sources of large supercooling.

\paragraph{Cosmological naturalness:} 
The LHC has not discovered any new physics around the $\rm TeV$ scale as predicted by the three traditional solutions to the Hierarchy problem: SUSY, composite Higgs and extra-dimensions.  Taking these results at face value together with the cosmological constant (CC) problem, might reveal the existence of fine tuning in Nature.  This idea is already popular among string theorists, since the only consistent theory of quantum gravity predicts the existence of a vast landscape of values for the Higgs mass $m_{\rm h}$  and for the CC \cite{Susskind:2003kw, Douglas:2003um}.
The question becomes \textbf{how can cosmology select a small Higgs mass and a small CC ?} 
The efforts in recent years which have been developed to address that question can be classified  in three categories \cite{TeresiGGI}.
\begin{itemize}
\item
\textbf{Dynamical} selection. The values of Higgs mass or CC are obtained as the result of a dynamical evolution, e.g. \cite{Abbott:1984qf,Bousso:2000xa,Feng:2000if, Dvali:2003br,Dvali:2004tma,Giudice:2019iwl, Kaloper:2019xfj, Graham:2015cka, Arkani-Hamed:2016rle,Bloch:2019bvc,Csaki:2020zqz,Strumia:2020bdy,TitoDAgnolo:2021nhd,TitoDAgnolo:2021pjo}
\item
\textbf{Statistical} selection. The Higgs mass or CC are linked to a scanning field whose quantum distributions peaks on the SM values, e.g. \cite{Dvali:2005zk,Geller:2018xvz,Cheung:2018xnu,Strumia:2020bdy,Giudice:2021viw}
\item
\textbf{Anthropic} selection. All values of the landscape are populated via eternal inflation, e.g. \cite{Arvanitaki:2016xds,Arkani-Hamed:2020yna}.
\end{itemize}
We review some of the dynamical and anthropic selection ideas below. We study the case of the CC in more details in Sec.~\ref{par:dyn_rel_CC}  of Chap.~\ref{chap:SM_cosmology}.

\paragraph{Cosmological relaxation:} 
The Higgs mass parameter is dynamical and set by a \textbf{rolling axion field} in the \textbf{early universe}. After the field reaches its equilibrium, the value of the electroweak scale has become independent of the UV cut-off. The first proposal of that type is the \textbf{relaxion} scenario (proposed in 2015 \cite{Graham:2015cka}), inspired from Abbot's idea for relaxing the Cosmological Constant (proposed in 1984 \cite{Abbott:1984qf})
\begin{equation}
\label{eq:relaxion_pot}
\mathcal{L} = g \Lambda^3 \phi + \left(\Lambda^2 - g \Lambda \phi \right) h^2 + \Lambda_{b}^4\left(\frac{\left< h \right>}{v}\right)^n \cos{\frac{\phi}{f}}.
\end{equation}
The first term is a \textbf{slope} for the scalar field $\phi$ where $g$ is a constant. The second term is the Higgs mass. The last term are is a back-reaction term from Higgs-dependent \textbf{wiggles}, with potential barrier $\Lambda_{b}$ and decay constant $f$. The scale $\Lambda$ is the UV cut-off of the effective field theory.
The dynamics is engineered such that the Higgs mass is set by the VEV of the relaxion $\phi$ and conversely, the size of the wiggles of $\phi$ are set by the VEV of the Higgs $\left< h \right>$.  Hence, when the Higgs mass turns negative, the Higgs gets a VEV and the relaxion $\phi$ gets a barrier. Shortly after the formation of the wiggles, the relaxion stops due to Hubble friction, and the Higgs gets a final tiny mass $m_{h} \ll \Lambda$. In order to have enough friction and time to relax, in the original proposal \cite{Graham:2015cka} the evolution must occur during inflation. Although see \cite{Fonseca:2019lmc} for revisiting the assumption.

The parameters of the model need to fulfill some conditions 
\begin{align}
&V(\phi) \lesssim V_{\rm inf} \quad \rightarrow \quad \frac{\Lambda^2}{M_{\rm pl}} \lesssim H_I, \qquad  &\text{inflaton potential dominates,} \label{eq:inflaton_pot_dom}\\
&H_I^{-1} \dt{\phi} \gtrsim H_I \quad \rightarrow \quad H_I \lesssim g^{1/3}\Lambda,\qquad &\text{class. roll. > quant. fluct.},  \label{eq:class_vs_quantu} \\
&V' \simeq 0 \quad \text{exists}\quad \rightarrow \quad g \Lambda^3 \lesssim \frac{\Lambda_b^4}{f},\qquad &\text{barrier large enough}, \label{eq:barrier_large}\\
&m_h \simeq 0  \quad \text{is reached} \quad \rightarrow \quad \Delta \phi \sim \frac{\Lambda}{g},\qquad  &\text{field excursion}, \label{eq:field_exc}\\
&H_I \lesssim \Lambda_{\rm b}, \qquad  &\text{barrier is effective}, \label{eq:barrier_form_QCD}
\end{align}
where we used that the velocity of the rolling scalar field is set by the Hubble friction
\begin{equation}
\label{eq:slow_roll_vel_rel}
\ddt{\phi}+3H\dt{\phi}+ V' = 0  \qquad \rightarrow \qquad \dt{\phi} \simeq \frac{g \Lambda^3}{H_I},\qquad \qquad \text{slow-roll velocity}.
\end{equation}
As a consequence, the original relaxion proposal suffers from several issues (see \cite{Fonseca:2019lmc} for a visualization of the allowed parameter space)
\begin{enumerate}
\item
From Eq.~\eqref{eq:inflaton_pot_dom}, Eq.~\eqref{eq:class_vs_quantu} and Eq.~\eqref{eq:barrier_large}, we obtain $\Lambda \lesssim (\Lambda_b^4 \,M_{\rm pl}^3/f)^{1/6}$ where $\Lambda_b$ must be `natural' $\Lambda_b \lesssim \rm TeV$. Hence, assuming $f \gtrsim \rm TeV$, the \textbf{UV cut-off} $\Lambda$ can not be larger than $10^{10}$~GeV. Therefore, the hierarchy problem re-appears if new physics couples to the Higgs above that scale.
\item
From Eq.~\eqref{eq:barrier_large}, the \textbf{linear coupling} $g$ must be very small, e.g. smaller than $g \lesssim 10^{-30}$ for the largest UV cut-off \cite{Fonseca:2019lmc}. Note however, that its smallness is technically natural since it is protected by a shift symmetry $\phi \rightarrow \phi +f$.
\item
From Eq.~\eqref{eq:field_exc}, the relaxion has to perform huge \textbf{field excursion} $\Delta \phi \gg M_{\rm pl}$. This enters in conflict with the Swampland Distance Conjecture  \cite{Brennan:2017rbf, Palti:2019pca,Grana:2021zvf}.
\item
From Eq.~\eqref{eq:field_exc} and Eq.~\eqref{eq:slow_roll_vel_rel}, we can compute the \textbf{number of e-folds} $N_e$
\begin{equation}
\Delta \phi \sim \dt{\phi} \frac{N_e}{H_I} \quad \rightarrow \quad N_e \sim \left(\frac{H_I}{g \Lambda}\right)^2, \qquad \qquad \text{number of e-folds}.
\end{equation} 
The smallness of $g$ implies that the whole relaxation process takes a huge number of e-folds $10^4 \lesssim N_e \lesssim 10^{60}$ \cite{Fonseca:2019lmc}.
\item
If we assume the relaxion to be the \textbf{QCD axion}, then the potential barrier can be generated by QCD instantons, $\Lambda_b \simeq \Lambda_{\rm QCD}$ with $n=1$ in Eq.~\eqref{eq:relaxion_pot}, whenever the de Sitter temperature is not too large cf. Eq.~\eqref{eq:barrier_form_QCD}.
The final $\phi$ position is displaced from its $CP$-conserving minimum by the $\theta$ angle
\begin{equation}
\label{eq:displacement_relaxion}
\theta \sim \frac{g \Lambda^3 f}{\Lambda_b^4}, \qquad \qquad \text{displaced minimun}.
\end{equation}
From Eq.~\eqref{eq:class_vs_quantu}, Eq.~\eqref{eq:displacement_relaxion} and Eq.~\eqref{eq:barrier_form_QCD}, we can check that the neutron EDM constraint $\theta \lesssim 10^{-10}$, together with the astrophysical constraints $f \gtrsim 10^9~$GeV, almost \textbf{exclude} the scenario where the relaxation is the QCD axion since the UV cut-off must be very low\footnote{Note that it can be relaxed to $\Lambda \lesssim 1000~ \text{TeV}$ if the classical rolling constraint in Eq.~\eqref{eq:class_vs_quantu} is dropped. However, this leads to eternal inflation and therefore to a loss of predictivity.} $\Lambda \lesssim 30~ \text{TeV}$ \cite{Graham:2015cka}.
\item
The term $\Lambda_b^4 (\left< h\right>/v)^n \cos{\phi/f}$ is \textbf{unstable} under radiative corrections 
\begin{equation}
\text{e.g.}\quad \epsilon \Lambda_b^4 \cos{\phi/f}, \qquad \epsilon \Lambda_b^3 \, g\phi \,\cos{\phi/f},
\end{equation}
where $\epsilon$ stands for the loop-suppression factors. Those terms should be included and since there are independent of the Higgs VEV, they spoil the relaxion mechanism, unless $\Lambda_b \lesssim v$, which implies new physics at the TeV scale. This is the \textbf{coincident problem}. Ref \cite{Espinosa:2015eda} proposes to solve the issue by introducing, in addition to the initial relaxion field $\phi$ which scans the Higgs mass, a second relaxion field $\sigma$ which scans the potential barrier. This is the \textbf{double scanning mechanism}.
\end{enumerate}
Interestingly, the relaxion can play the role of \textbf{Dark Matter} candidate \cite{Fonseca:2018kqf, Banerjee:2018xmn}. See \cite{Banerjee:2020kww} for a review of the experimental constraints on the relaxion.

Refs.~\cite{Fonseca:2019ypl, Fonseca:2019lmc,Morgante:2021bks} demonstrate that the relaxion can also stop due to the dissipation resulting from \textbf{axion fragmentation}, instead of Hubble friction. Upon decomposing the field into the sum of the homogeneous background plus a perturbation
\begin{equation}
\phi(x, \, t) = \phi_0(t) + \delta \phi(x, \, t), \qquad {\rm with} \quad
\delta \phi(x, \, t) = \int \frac{d^3k}{(2\pi)^3} a_k \, u_k(t) \, e^{ikx} + {\rm h.c.},
\end{equation}
where $\left[a_k,\, a_k^\dagger\right] = (2\pi)^3 \delta^3( k - k')$, then
the equation of motion can be written as
\begin{align}
&\ddt{\phi} + 3H \dt{\phi} + V'(\phi) + \frac{1}{2} V'''(\phi) \int \frac{d^3 k }{(2\pi)^3} \left| u_k\right|^2 = 0, \\
&\ddt{u}_k + 3H\dt{u}_k + \left[ \frac{k^2}{a^2} + V''(\phi)  \right] u_k = 0.
\end{align}
In the relaxion scenario, the second equation reads
\begin{equation}
\ddt{u}_k + \left(k^2 - \frac{\Lambda_b^4}{f^2} \cos{\frac{\dt{\phi}}{f} t} \right) u_k = 0,
\end{equation}
where it is implicitly assumed that $\Lambda_b$ depends on the higgs vev $\left<h\right>$. This is the well-known \textbf{Mathieu equation} \cite{Kofman:1997yn}
\begin{equation}
\ddt{u}_k + \left(A_k - 2q \cos{2z} \right) u_k = 0,
\end{equation}
with $A_k = 4k^2/(\dt{\phi}/f)^2$, $q = 2\Lambda_b^4/\dt{\phi}^2$ and $z = \dt{\phi}/2f$. Its solution has an \textbf{exponential growth} when $k$ is in one of the $n$ \textbf{instability bands}, which in the \textbf{narrow-resonance} limit $q \ll A_k$, lies around $A_k \sim n$. The largest growth, $u_k \propto \exp{qz/2}$, arises in the first instability band $n=1$, given by $ 1 - q \lesssim A_k \lesssim 1+q $, and which can be rewritten as
\begin{equation}
\frac{\dt{\phi}^2}{4f^2} - \frac{\Lambda_b^4}{2f^2} \lesssim k^2 \lesssim \frac{\dt{\phi}^2}{4f^2} + \frac{\Lambda_b^4}{2f^2} \quad \rightarrow \quad \frac{\dt{\phi}}{2f} - \frac{\Lambda_b^4}{2\,f\,\dt{\phi}} \lesssim k \lesssim  \frac{\dt{\phi}}{2f} + \frac{\Lambda_b^4}{2\,f\,\dt{\phi}}.
\end{equation}
 This alternative dissipation mechanism can help the relaxion to stop more efficiently. As a consequence, in some region of the parameter space, only a natural number of e-folds of inflation is required $N_e \sim 10^2$, the process can even occur after inflation. The upper bound on $g$ is relaxed, e.g. $g \sim 10^{-10}$, and the field excursion is never super-planckian. The price to pay is a smaller UV cut-off $\Lambda \lesssim 10^{5}~$GeV.

%
%

\paragraph{Anthropic solution:} 

Different studies suggest that the measured value of the weak scale $v\simeq 246~\rm GeV$ is consubstantial to our existence as observers.
\begin{itemize}
\item
Proton decay arises from dimension 6 operators and yields the lifetime $\tau_p = M^4/m_p^5$. Assuming $M \simeq 10^{16}~\rm GeV$, the weak scale must be smaller than $v \lesssim 400~\rm TeV$ in order for the proton to be stable over the universe age \cite{Agrawal:1997gf,Agrawal:1998xa}.
\item
The pion, mediator of the nuclear force, has a mass proportional to the weak scale $m_{\rm \pi}^2 \propto m_u+m_d \propto v$, see Eq.~\eqref{eq:pion_mass}. An increase of the electroweak scale $v$ by more than 60$\%$ would lead to the instability of complex nuclei, e.g. $^{16}O$ \cite{Damour:2007uv}. The same study finds that if the weak scale is lowered by $60\%$, the mass difference $m_d-m_u$ becomes too small to prevent the instability of hydrogen atoms $p+e \to n+\nu_e$. The bounds are revisited in \cite{Hall:2014dfa}. Those bounds however suppose fixed Yukawa couplings.
\item
The increase of the weak scale $v$ by one or two orders of magnitudes implies that the helium fraction reaches 100$\%$ which make halo cooling less efficient and stars\footnote{Note however the possibility of a universe containing stars, without weak interactions at all \cite{Harnik:2006vj}.} shorter-lived \cite{Hall:2014dfa}. 
\item
If weak interactions whose strength is controlled by $G_F=1/(\sqrt{2}v^2)$ were too small (large) then neutrinos produced in the final stages of stellar collapse might easily escape (get trapped) preventing supernova explosions, which are necessary to produce vital elements as oxygen. The gravitational free-fall time in neutron star is
\begin{equation}
\tau_{\rm fall} \sim \frac{1}{\sqrt{G\rho}} \sim \frac{M_{\rm pl}}{m_n^2}.
\end{equation}
The neutrino free-streaming length is
\begin{equation}
l_\nu \sim \frac{1}{\sigma_{\rm weak} n_n} \sim \frac{v^4}{T^5} \sim \frac{v^4}{m_n^5}.
\end{equation}
The neutrino trapping time is the time it takes for them to diffuse over a distance equal to the star radius $R_s \sim M_{\rm pl}/m_n^2$
\begin{equation}
\tau_{\rm trap} \sim \frac{R_s^2}{l_\nu} \sim \frac{M_{\rm pl}^2 m_n}{v^4}.
\end{equation}
Core-collapse supernova explosion requires the neutrino to be trapped for a time $\tau_{\rm trap}$ comparable to the gravitational time-scale $\tau_{\rm fall}$, which implies that the weak scale must be tuned to 
\begin{equation}
v \sim c_0 \,m_n^{3/4}\, M_{\rm pl}^{1/4},\qquad c_0 \sim 0.01,
\end{equation}
within a factor of few \cite{DAmico:2019hih}. 
\item
The increase of the top mass by less than $5\%$ might lead to an instability of our EW vacuum \cite{Buttazzo:2013uya}.\footnote{The analysis in \cite{Buttazzo:2013uya} shows, first, that the EW vacuum is metastable with a lifetime much longer than the age of the the universe, and second, that the EW vacuum sits at the edge of a region of instability ($+5\%~m_t$).  However, the analysis assumes the validity of the SM up to an energy scale of $10^{10}~$GeV. Hence, any new physics below that scale can potentially alter the statement.} 
\end{itemize}
Hence, the SM seems to sit inside a tiny \textbf{anthropic windows} outside which, a carbon-based life seems impossible.
This motivates the possibility that the solution to the hierarchy problem is not microscopic but \textbf{environmental}.

The multiverse idea, based on eternal inflation and the string landscape, provides a theoretical framework to scan over the values of the observed physical parameters.
The number of different vacua predicted in string theories is of googolplex order $10^{\rm few\,100}$ \cite{Susskind:2003kw, Douglas:2003um}.
In 2000, Bousso, Polchinski and others \cite{Bousso:2000xa,Feng:2000if} propose to use the nucleation of membranes charged under four-form fields in order to scan over the different values of the cosmological constant within that landscape. The membrane nucleation is the exact $(3+1)D$ equivalently of the Schwinger process of $e^+/e^-$ nucleation in $(1+1)D$ \cite{Brown:1987dd,Brown:1988kg}. A generalization of the Bousso-Polchinski approach to a scan over the values of the EW scale is straightforward upon the introduction of an operator - which is non-renormalizable in \cite{Dvali:2003br,Dvali:2004tma} and renormalizable in \cite{Giudice:2019iwl, Kaloper:2019xfj} - coupling the Higgs field with the four-form fields. More details on the relaxation \`a la Bousso-Polchinski can be found in Sec.~\ref{par:dyn_rel_CC} of Chap.~\ref{chap:SM_cosmology} when discussing about the relaxation of the Cosmological Constant.

\subsection{Neutrino oscillations}
\label{sec:neutrino_oscillations}
\paragraph{The discovery of neutrino oscillations}
In $1968$,  Ray Davis et al. \cite{Davis:1968cp} reported the first detection of \textbf{solar neutrinos}, using a tank of Chlorine and the charged-current interaction $\nu_e + n \rightarrow p+e$.
The flux of electron neutrino $\nu_e$ coming from the sun is measured smaller than the prediction by a factor $\sim 1/3$. This was the famous \textbf{solar neutrino problem}.
The same year, Pontecorvo \cite{Pontecorvo:1967fh} suggested that the electron neutrino produced by the sun are transformed in flight into other flavors to which the experiment was not sensitive to. This is the mechanism of \textbf{neutrino oscillation}.
In 2001, the Super-Kamiokande \cite{Fukuda:2001nj} and Sudbury Neutrino Observatory \cite{Ahmad:2001an,Ahmad:2002jz} were able to measure the total flux of neutrinos $\nu_e+\nu_\mu+\nu_\tau$ and found the missing $\sim 2/3$. Super-Kamiokande, which uses water tank, measures the total neutrino flux through the elastic interactions $\nu_{e,\,\mu,\,\tau}+e^- \rightarrow \nu_{e,\,\mu,\,\tau}+e^- $, while Sudbury Neutrino Observatory, which uses heavy water, can also detect neutral current interactions $\nu_{e,\,\mu,\,\tau}+n \rightarrow \nu_{e,\,\mu,\,\tau}+n$.  The initial Davis experiments, which first detected solar neutrinos in 1968, received the Nobel prize in 2002, while the two experiments, Super-Kamiokande and Sudbury Neutrino Observatory, which confirmed the existence of neutrino oscillations and solved the solar neutrino problem, were co-awarded the Nobel Prize $2015$.

In fact, the first evidence for neutrino oscillations was already found a bit earlier in 1998 by Super-Kamiokande in the case of \textbf{atmospheric neutrinos}, generated by the decay $\pi^+ \rightarrow \mu^+ + \nu_\mu$ followed by $\mu^+ \rightarrow e^+ +  \bar{\nu}_\mu + \nu_e$ \cite{Fukuda:1998mi}. Instead of measuring of factor $\nu_\mu/\nu_e$ of order $2$, the collaboration measured $\sim 1.2$, showing the first evidence for neutrino oscillations.

\begin{figure}[h!]
\centering
\raisebox{0cm}{\makebox{\includegraphics[width=0.95\textwidth, scale=1]{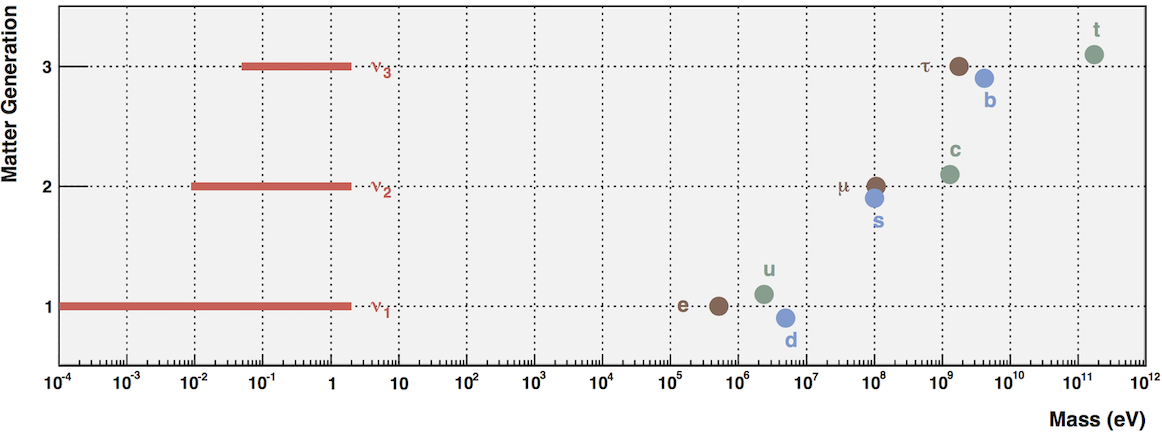}}}
\caption{\it \small Fermion masses in the SM. Figure reproduced from \cite{GomezCadenas:2011it}.}
\label{fig:Fermion_masses}
\end{figure}

At the theoretical level, neutrino oscillations rely on the existence of a mass matrix in the neutrino sector with two properties. First, it is non-diagonal and so contains mixing angles between the different neutrinos flavors. Second, it contains small mass differences $\Delta m$.  
The surprising features of the neutrino matrix are its small entries: the neutrino masses are incredibly small compared to other fermions, see Fig.~\ref{fig:Fermion_masses}.
In what follows, we discuss how to generate a mass matrix for the neutrino sector, with small entries with respect to the electroweak scale.

\paragraph{The see-saw mechanism}
\label{sec:see_saw}
The smallness of the neutrino mass compared to the other fermions calls for an explanation.
Actually, it is straightforward to account for the neutrino masses and oscillations by adding to the SM a dimension 5 operator, called the \textbf{Weinberg operator} \cite{Weinberg:1979sa}
\begin{equation}
\mathcal{L}_{\rm dim-5} = - \lambda_{ij} \left( \bar{L}^i \tilde{H} \right) \left( \tilde{H} L_j\right)^\dagger,
\end{equation}
where $\tilde{H}$ is defined in Eq.~\eqref{eq:Htilde}.
After EWSB, the Weinberg operator provides to the neutrino sector, a non-diagonal mass matrix with flavor mixing and therefore accounts for the oscillations. However, even though the operator respects all the gauge symmetries in the SM, it is not renormalizable. In fact, this term is the low-energy description of a more complete theory which includes right-handed neutrinos $\nu_R^i$, where $i$ indexes the different right-handed neutrino families (generally one, two or three). If we assume that the right-handed neutrinos are neutral under both  the weak and electromagnetic forces, then the most general renormalizable operators are
\begin{equation}
\label{eq:neutrino_mass_matrix_1}
\mathcal{L}_{\rm \nu,\rm \,mass} = - Y_{ij}^e \bar{L}^i H e_R^j - Y_{ij}^\nu \bar{L}^i \tilde{H} \nu_R^j - iM_{\rm ij} \left( \nu_R^i \right)^c \nu_R^j +h.c.
\end{equation}
where $\nu_R^c = \nu_R^T \sigma_2$ denotes the charge conjugate.
We can embed the Weyl spinors  $\nu_R$ and $\nu_L$ into a Dirac notation\footnote{Dirac spinor are usually built with two independent left-handed and right-handed weyl fields $\psi  =\begin{pmatrix}  \psi_L\\ \psi_R \end{pmatrix}$, see Sec.~\ref{sec:lorentz_representations}.}
\begin{equation}
\label{eq:embed_weyl_dirac}
\psi_L =  \begin{pmatrix}  \tilde{\nu}_R \\ \nu_R \end{pmatrix}, \qquad {\rm and} \qquad \psi_R =  \begin{pmatrix}  \nu_L\\ \tilde{\nu}_L \end{pmatrix},
\end{equation}
where $\tilde{\nu}_{L,\,R} = i\sigma_2 \nu_{L,\,R}^*$. Then, the Dirac and Majorana mass terms can be simply written as
\begin{equation}
\label{eq:neutrino_mass_matrix_2}
\mathcal{L}_{\rm \nu,\rm \,mass}  = - m_{ij} \bar{\psi}^i_L \psi_R^j - \frac{M_{ij}}{2} \bar{\psi}_R^i \psi_R^j,
\end{equation}
where 
\begin{equation}
\label{eq:neutrino_dirac_mass}
m_{ij}=\frac{Y_{ij}}{\sqrt{2}} v,
\end{equation}
 after EWSB.
Assuming one flavor generation for simplicity, the mass eigenstates are linear combinations which diagonalize the matrix $ \begin{pmatrix}  0 & m \\ m &M \end{pmatrix}.$ In the limit $M \gg m$, we find 
\begin{equation}
m_{\rm heavy} \simeq M \qquad {\rm and} \qquad m_{\rm light} \simeq \frac{m^2}{M}.
\end{equation}
Hence, the smallness of the neutrino mass results from the heaviness of the Majorana mass $M$, this is the \textbf{see-saw mechanism}, developed at the end of the 70s \cite{Minkowski:1977sc, Ramond:1979py,GellMann:1980vs,Yanagida:1979as,Mohapatra:1979ia}. See \cite{Drewes:2013gca,King:2015aea} for some reviews on the see-saw mechanism. 
The see-saw mechanism predicts the existence of right-handed neutrinos, called \textbf{sterile neutrinos}. The upper bound from neutrinoless double beta decay, presented later in Eq.~\eqref{eq:neutrinolessbetadecaybound}, translates to a lower bound on the Majorana mass of sterile neutrinos
\begin{equation}
\label{eq:lower_bound_majorana_neutrinolessbetadecay}
M \gtrsim 10^{14} ~Y_\nu^2.
\end{equation}  
Interestingly, such particle can explain the matter/anti-matter asymmetry through the \textbf{leptogenesis} mechanism, see Sec.~\ref{par:leptogenesis} of Chap.~\ref{chap:SM_cosmology}. 
Unfortunately, for natural values of the Yukawa coupling $Y_\nu$, the Majorana mass scale is beyond the reach of most of the experiments (colliders, telescopes, labs,..).
On the other hand, Yukawa couplings below $\lesssim 10^{-6}$ and $\lesssim 10^{-10}$, predict sterile neutrinos in the collider energy range $\sim \rm TeV$, and in the $\rm keV$ range, respectively. The latter scenario can explain \textbf{Dark Matter} and can lead to astrophysical signatures \cite{Canetti:2012kh,Canetti:2012zc}. 
For reviews on sterile neutrinos, see \cite{Abazajian:2012ys,Adhikari:2016bei, Boser:2019rta}.

\paragraph{Are neutrinos Majorana or Dirac fermions ?}
Upon taking the charge conjugate of Eq.~\eqref{eq:embed_weyl_dirac}, we find\footnote{We have introduced the charge conjugation operation $\psi_c = -i\gamma_2 \psi^*$. We can check that the Dirac equation $(i\slashed{\partial} - e \slashed{A} - m)\psi = 0$ becomes $(i\slashed{\partial} + e \slashed{A} - m)\psi_c = 0$.}
\begin{align}
-i\gamma_2 \psi^* &= -i  \begin{pmatrix}  0 & \sigma_2 \\ -\sigma_2  & 0 \end{pmatrix}  \begin{pmatrix}  \nu_L\\ i\sigma_2 \nu_{L}^* \end{pmatrix}^*
= \begin{pmatrix}  -\sigma_2\sigma_2^*\nu_L\\ i\sigma_2 \nu_{L}^* \end{pmatrix} = \begin{pmatrix}  \nu_L\\ i\sigma_2 \nu_{L}^* \end{pmatrix}\notag \\
&= \psi,
\end{align}
and same for $\nu_R$. Hence if neutrinos are majorana fermions, i.e. $M \neq 0$, they are their own anti-particle.\footnote{The proposition that fermions could be their own anti-particle was first theorized by Ettore Majorana, for the case of electron and positron, in 1937 \cite{majorana1937teoria}, one year before its disparition under mysterious circumstances \cite{klein2000ettore}.}
In contrast, if neutrinos are Dirac fermions, i.e. $M  = 0$ - if for instance the right-handed neutrino carries lepton number and the latter is conserved - then the right-handed and left-handed neutrinos become components of the same Dirac field $\psi = \left( \nu_L,\, \nu_R \right)$.  Then charge conjugation gives
\begin{align}
-i\gamma_2 \psi^* &= -i  \begin{pmatrix}  0 & \sigma_2 \\ -\sigma_2  & 0 \end{pmatrix}  \begin{pmatrix}  \nu_L\\ \nu_{R} \end{pmatrix}^*
= \begin{pmatrix}  - i\sigma_2 \nu_{R}^*\\ i\sigma_2 \nu_{L}^* \end{pmatrix}= \begin{pmatrix}   \tilde{\nu}_{R} \\ \tilde{\nu}_{L} \end{pmatrix} \notag\\
& \neq \psi,
\end{align}
and neutrinos are different from their anti-particle.
If neutrinos are Dirac fermions, then the tritium bound in Eq.~\eqref{eq:nu_mass_bound_tritium} on the Dirac mass in Eq.~\eqref{eq:neutrino_dirac_mass}, translates to a very tight upper bound on the Yukawa couplings $Y_\nu \lesssim 10^{-12}$. We can understand that this number seems unnatural and motivates alternative like the introduction of a large Majorana mass. But it is actually \textbf{technically natural} in the 't Hooft sense, see footote \ref{footnote:technically_natural}, since it is protected from large quantum corrections by the chiral symmety, see footnote \ref{footnote:chiral_transf}. 

\paragraph{The mixing angles and the PMNS matrix}

The diagonalization of the lepton Yukawa matrix in the presence of right-handed neutrinos modifies the flavor-mixing interaction terms with the $W^{\pm}$, in the same way as the quarks sector studied in Sec.~\ref{par:CKM}
\begin{equation}
\label{eq:CKM_lagrangian}
 \mathcal{L} \supset \frac{g_2}{\sqrt{2}} \left[  W_\mu^- \bar{e}_L^i \gamma^\mu \left( U \right)^{ij} \nu_L^{j} + W_\mu^+ \bar{\nu}_L^i \gamma^\mu \left(U^{\dagger}\right)^{ij} e_L^j \right].
\end{equation}
The matrix $U$ is the analog of the CKM matrix\footnote{The CKM is traditionally defined as $V=\left(U_L^u\right)^\dagger U_L^d$. A pure analog in the lepton sector, with respect to the $SU(2)_L$ charges, would be $U=\left(U_L^\nu \right)^\dagger U_L^e$. However, the PMNS matrix is traditionally defined as its complex conjugate, $U=\left(U_L^e \right)^\dagger U_L^\nu$. } in Eq.~\eqref{eq:neutrino_mass_matrix_1}.  It is called the \textbf{PMNS matrix} for Maki, Nakagawa and Sakata who introduced it in 1962 \cite{Maki:1962mu} in order to explain the neutrino oscillations predicted by Pontecorvo in 1957 \cite{Pontecorvo:1957qd}. As we did for the CKM matrix, we can count how many physical parameters it contains. First of all, as a $U(3)$ matrix, it has $3$ rotation angles and $6$ complex phases. Notice that we can freely rotate the phases of the $3$ charged-leptons, but because of the presence of the Majorana mass term, we can not rotate the phases of the neutrinos. So the number of physical parameters is $3$ rotation angles and $3$ phases. It can be written with an almost identical parameterization to Eq.~\eqref{eq:CKM_matrix}
\begin{equation}
U_{\rm PMNS} = \begin{pmatrix} c_{12} \,c_{13} & s_{12}\, c_{13} & s_{13} \,e^{-i\delta}\\ -s_{12} \,c_{23} - c_{12} \,s_{23} \,s_{13} \,e^{i\delta} & c_{12} \,c_{23} - s_{12} \,s_{23}\, s_{13}\,e^{i\delta} & s_{23} \,c_{13} \\ s_{12}\,s_{23} - c_{12}\,c_{23} \,s_{13} \,e^{i\delta} & -c_{12}\,s_{23}-s_{12}\,c_{23}\,s_{12}\,e^{i\delta} & c_{23}\,c_{12}  \end{pmatrix} ~P
\end{equation}
where $P = \left( 1,\, e^{i\phi_1}, \, e^{i\phi_2} \right)$ accounts for the \textbf{two extra-phases} in the presence of Majorana masses. 

Since two decades, measurements of neutrino oscillations have enabled the relatively precise determination of the neutrino mass differences and mixing angles.
To the best of our knowledge, the physical parameters of the lepton sector are  \cite{Tanabashi:2018oca, Esteban:2018azc}
\begin{align}
& m_e \simeq 0.511~{\rm MeV}, \quad \qquad \qquad m_{\mu} \simeq 106~{\rm MeV}, \quad \qquad \qquad m_{\tau} \simeq 1.78~{\rm GeV}, \quad \\
& \Delta m_{\rm 21}^2 \simeq \left(7.39^{+0.21}_{-0.20}\right)  \times 10^{-5}~{\rm eV^2}, \quad \qquad \Delta m_{\rm 32}^2 \simeq   \left\{
                \begin{array}{ll}
                +\left(2.525^{+0.033}_{-0.031}\right) \times 10^{-3}~{\rm eV^2}, \quad \text{NH} \\
                 -\left(2.512_{-0.031}^{+0.034}\right) \times 10^{-3}~{\rm eV^2}, \quad \text{IH}
                \end{array}
              \right.~, \quad \\
&\theta_{12}= \left(33.82^{+0.78}_{-0.76}  \right)^\circ, \quad \theta_{23}= \left\{
                \begin{array}{ll}
                \left(49.7^{+0.9}_{-1.1}  \right)^\circ,\quad \text{NH} \\
                 \left(49.7^{+0.9}_{-1.0}  \right)^\circ,\quad \text{IH}
                  \end{array}
              \right., \quad \theta_{13}= \left\{
                \begin{array}{ll}
                \left(8.61^{+0.12}_{-0.13}  \right)^\circ,\quad \text{NH} \\
                  \left(8.65^{+0.12}_{-0.13}  \right)^\circ,\quad \text{IH} 
                  \end{array}
              \right.~, \quad\\
&\qquad \qquad \qquad  \quad \qquad  \qquad  \delta= \left\{
                \begin{array}{ll}
                \left(215^{+40}_{-29}  \right)^\circ,\quad \text{NH} \\
                  \left(284^{+27}_{-29}  \right)^\circ,\quad \text{IH} 
                  \end{array}
              \right.,
\end{align}
where NH and IH denote normal ($m_{\nu_e}<m_{\nu_\mu}<m_{\nu_\tau}$) and inverted ($m_{\nu_\tau}<m_{\nu_e}<m_{\nu_\mu}$) hierarchy. The Majorana phases $\phi_1$ and $\phi_2$ have not yet been measured.
Note that the neutrino oscillation dynamics is insensitive to the absolute neutrino mass which is then constrained by other experiments.
For instance, KATRIN experiment \cite{Aker:2019uuj} provides an upper bound on the neutrino mass from measuring the high-energy cut-off of the electron spectrum in tritium decay 
\begin{equation}
\label{eq:nu_mass_bound_tritium}
m_{\rm \beta} \equiv \sqrt{\sum_{k=1}^3 \left| U_{ek} \right|^2 m_k^2} \lesssim 1.1~ \rm eV.
\end{equation}
Other constraints on the absolute neutrino mass come from searches for neutrinoless double beta decay  experiments like GERDA, KamLAND and EXO-200 (which however assumes that neutrino have Majorana masses) \cite{Dolinski:2019nrj}
\begin{equation}
\label{eq:neutrinolessbetadecaybound}
m_{ee} \equiv \left| \sum_{k=1}^3 U_{ek}^2 m_k  \right| \lesssim 0.2-0.4~ \rm eV,
\end{equation}
and cosmological probes (which rely on the $\Lambda$CDM model), CMB + BAO + Lyman-$\alpha$ \cite{Yeche:2017upn}
\begin{equation}
\sum m_{\nu_j} \lesssim 0.12~{\rm eV}.
\end{equation}
See \cite{Arguelles:2022xxa} for an overview of the experimental prospects on neutrino physics and \cite{Lesgourgues:2018ncw,Abazajian:2022ofy} for their importance in cosmology.

\subsection{Flavor hierarchy problem}
\begin{figure}[h!]
\centering
\raisebox{0cm}{\makebox{\includegraphics[width=0.7\textwidth, scale=1]{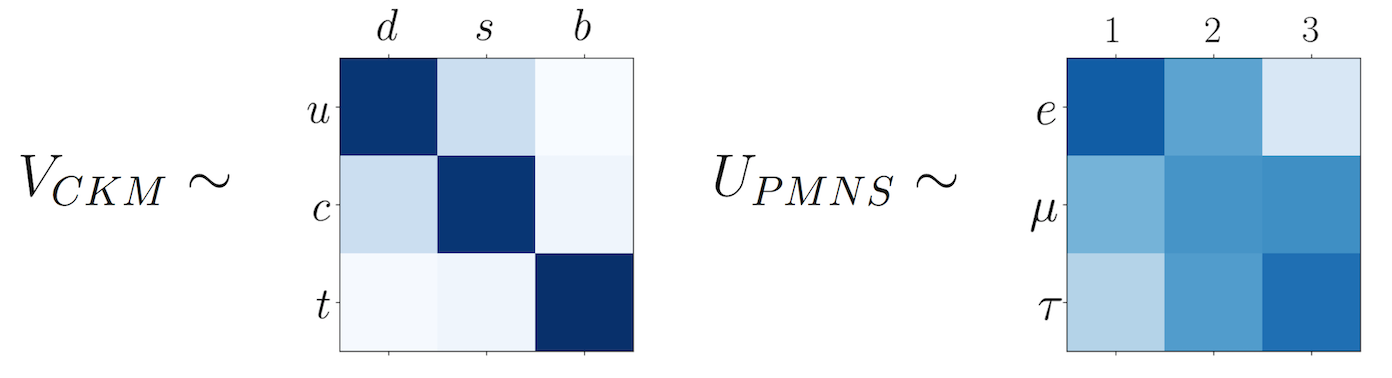}}}
\caption{\it \small Representation of the size of the mixing matrix elements. Lighter colors represent smaller
absolute value with respect to darker ones. Figure reproduced from \cite{QuilezLasanta:2019wgl}.}
\label{CKM_vs_PMNS}
\end{figure}

\paragraph{The Flavor Puzzle:}
In the previous section, Sec.~\ref{sec:neutrino_oscillations}, we have discussed the hierarchy between the neutrino masses and the masses of the other fermions. However, the matter sector suffers from two other (smaller) hierarchies which are sometimes called the \textbf{Flavor Puzzle}.
\begin{enumerate}
\item
There is a hierarchy among the different entries of the CKM matrix of the electroweak interactions in the quark sector. By comparison, the PMNS matrix of the electroweak interactions in the lepton sector has a much weaker hierarchy, see Fig.~\ref{CKM_vs_PMNS}. Particularly, the mixing angles of the CKM matrix, in Eq.~\eqref{eq:CKM_matrix}, satisfy $\theta_{12}\gg \theta_{13},\,\theta_{23}$, such that at first-order we can write it as
\begin{equation}
\left|V_{\rm CKM}\right| = \begin{pmatrix} 1 & \lambda & \lambda^3 \\ \lambda  & 1 & \lambda^2 \\ \lambda^3 & \lambda^2 & 1  \end{pmatrix},
\label{eq:CKM_matrix_wolfenstein_approx}
\end{equation}
with $\lambda = \sin{\theta_{12}} \simeq 0.22$ the Cabibbo angle.
\item
There is a hierarchy among the masses of the different flavors of charged fermions, see Fig.~\ref{fig:Fermion_masses}. The mass ratios in the quark sector seem to follow the rules
\begin{equation}
\frac{m_b}{m_t}\simeq \epsilon^3,\qquad \frac{m_c}{m_t}\simeq \epsilon^4,\qquad \frac{m_s}{m_t}\simeq \epsilon^5, \qquad \frac{m_b}{m_t}\simeq \epsilon^7, \qquad \frac{m_u}{m_t}\simeq \epsilon^8,
\end{equation}
with $\epsilon \simeq \lambda$.
\end{enumerate}
\paragraph{The Froggatt-Nielsen solution:}
A popular explanation is the Froggatt-Nielsen mechanism introduced in 1978 \cite{Froggatt:1978nt} which introduces a new global symmetry $U(1)_{\rm FN}$ and a new field, the \textbf{flavon} $S$. Upon assigning $U(1)_{\rm FN}$ charges to the SM fermions and to the Higgs, we obtain
\begin{equation}
\label{eq:Froggatt_Nielsen}
\mathcal{L}_{\rm FN} = -  y_{ij}^e \left( \frac{S}{M_{\rm FN}} \right)^{n_{ij}^e} \bar{L}_{\rm L}^i H e_{\rm R}^j- y_{ij}^d \left( \frac{S}{M_{\rm FN}} \right)^{n_{ij}^d} \bar{Q}_{\rm L}^i H d_{\rm R}^j-  y_{ij}^u \left( \frac{S}{M_{\rm FN}} \right)^{n_{ij}^u}  \bar{Q}_{\rm L}^i \tilde{H} u_{\rm R}^j  + h.c.,
\end{equation}
where $M_{\rm FN}$ is a heavy scale and the $y$ parameters are numbers of order 1.
The power exponents $n_{ij}^{e,\,d,\,u}$ depend on the $U(1)_{\rm PQ}$ charge assignment.
The Yukawa matrix of the SM is generated when the flavon $S$ acquires a VEV $\left< S\right>$.
The expansion parameter $\lambda$ in the CKM and mass matrix is given by $\lambda = \left<S \right>/M_*$. The different flavor hierarchies $\lambda^n$ are controlled by the charges of the SM fermions under the new $U(1)_{\rm FN}$ symmetry. The non-renormalizable operators in Eq.~\eqref{eq:Froggatt_Nielsen} can be generated after integration of \textbf{messengers fermions} heavier than $M_{\rm FN}$. Generalization to other symmetry groups are possible, e.g. to $SU(3)$ \cite{King:2001uz} or to discrete symmetries \cite{Altarelli:2010gt,King:2013eh}.

\paragraph{The partial-compositeness solution:}
A Yukawa coupling hierarchy can also arise from wave-function localization in Randall-Sundrum-type 5D warped geometry, and their dual description in 4D strongly-coupled conformal Higgs sector \cite{Grossman:1999ra,Gherghetta:2000qt,Huber:2003tu,Agashe:2004cp,Csaki:2008zd,Casagrande:2008hr,Bauer:2009cf}. In the 4D description, the flavors hierarchies are generated by the slow running of the Yukawa couplings, see partial compositeness in \cite{Kaplan:1991dc,Panico:2015jxa} which replaces the old walking technicolor idea.
See \cite{Babu:2009fd, Isidori:2010kg, Grinstein:2017pvg,Grossman:2017thq,Altmannshofer:2022aml,Chauhan:2022gkz} for reviews on flavor physics.

\subsection{Strong CP problem}
\label{sec:strong_CP_pb}

The non-detection of the electric dipole moment of the neutron provides a tight upper bound on the strong CP parameter, $\tilde{\theta} \lesssim 10^{-10}$, cf. Eq.~\eqref{eq:theta_bound}. This motivates the formulation of a principle responsible for its smallness.
In the SM, CP violation by weak interactions generates quantum corrections to  $\bar{\theta}$. However, $\bar{\theta}$ is not proportional solely to the CKM phase, but to the CKM phase times a flavor-neutral combination of Yukawa couplings, which implies a number of loops. This reduces the size of the quantum corrections and make the measured value of $\bar{\theta}$ not surprising from the point of view of the sole SM. However, it becomes a problem in the presence of any new physics which does not respect CP, and which is likely to induce large quantum corrections to $\bar{\theta}$.
Possible solutions to the strong $CP$ problem are listed below.\footnote{I am grateful for having benefited from the great expertise of Pablo Qu\'{i}lez on the strong CP problem and axion physics, to find answers to most of my silly questions.}
\paragraph{The Nelson-Barr solution:}
 $CP$ is asumed to be a fundamental symmetry of nature which is spontaneously broken by a scalar taking a complex vev  \cite{Nelson:1983zb,Barr:1984qx,Barr:1984fh}. For instance in \cite{Bento:1991ez}, new fermionic states are introduced and engineered such that the mass matrix gives a zero $\bar{\theta}=\rm arg \, det \left( Y_u Y_d \right)=0$ while generating a complex phase for the CKM matrix.
 
\paragraph{The Peccei-Quinn solution:}
An additional anomalous $U(1)_{\rm PQ}$ symmetry is added to the SM \cite{Peccei:1977hh, Peccei:1977ur, Wilczek:1977pj, Weinberg:1977ma}. If this $U(1)_{\rm PQ}$ is spontaneously broken, it will generate a pseudo-Nambu-Goldstone boson $a$, called the \textbf{axion}, which after a chiral rotation can be transferred to the $\bar{\theta}$ term. Hence, the $\bar{\theta}$ parameter in the vacuum energy in Eq.~\eqref{eq:pot_theta} is replaced by $\bar{\theta} \rightarrow \bar{\theta}+a/f_a$. Therefore, we have promoted the constant parameter $\bar{\theta}$ to a dynamical degree of freedom $a$ which after minimizing its energy, given by Eq.~\eqref{eq:pot_theta}, relaxes to the CP-preserving vacuum
\begin{equation}
\bar{\theta}+\frac{a}{f_a}=0.
\end{equation}
In fact, after QCD confinement the axion mass eigenstate does not exactly coincide with the $U(1)_{\rm PQ}$ Golstone boson but instead it couples to the SM mesons $\pi^0$ and $\eta^0$. This has great importance for phenomenology since it makes the axion mass eigenstate couple to photons, see e.g. \cite{QuilezLasanta:2019wgl} for the details.
Upon expanding Eq.~\eqref{eq:pot_theta} to quadratic order, we get the well-known formula for the axion mass \cite{Weinberg:1977ma}
\begin{equation}
\label{eq:ma_fa_QCD_axion}
m_a^2 =\frac{\chi}{f_a^2}, \qquad \textrm{with}\quad  \chi = \frac{m_u m_d}{(m_u+m_d)^2} m_\pi^2 F_{\pi}^2 \simeq \left(78~\rm GeV\right)^4.
\end{equation}
$\chi$ is known as the \textbf{susceptibility of the topological charge}. Higher-order corrections lead to $\chi = \left(75.5(5)~\rm GeV\right)^4$ \cite{diCortona:2015ldu}. 
In addition the axion is an excellent \textbf{Dark Matter} candidate \cite{Preskill:1982cy,Abbott:1982af,Dine:1982ah} which continues to receive a lot of interest both theoretically and experimentally, see \cite{Marsh:2015xka,Hook:2018dlk,Irastorza:2018dyq, DiLuzio:2020wdo} for reviews.

\paragraph{The massless quark solution:}
If one quark mass vanishes then we are free to perform a chiral rotation to cancel $\bar{\theta}$. In a sense, this last solution is similar to the axion solution. Indeed, by setting $m_u = 0$, we introduce an additional anomalous $U(1)$ symmetry, which is spontaneously broken when QCD confines. In that case, the axion $a$ would exactly coincide with the $\eta'$ \cite{tHooft:1976rip}. This solution was proposed a long time ago but is no longer acceptable since lattice simulations have found a non-zero mass for the up quark \cite{Aoki:2019cca}.

Instead, we can introduce BSM massless fermions charged under both $SU(3)_c$ and a new confining force, called axicolor \cite{Kim:1984pt,Choi:1985cb}, whose confining scale is much larger than $\Lambda_{\rm QCD}$ \cite{Weinberg:1975gm,Susskind:1978ms,Dimopoulos:1981xc}.  The Goldstone boson which emerges when the chiral symmetry of the exotic fermions is spontaneously broken by axicolor confinement is called the \textbf{dynamical axion} \cite{Kim:1984pt,Choi:1985cb}.  The axion shift symmetry can be preserved until operators of very high dimensions if the axicolored fermions are chiral \cite{Redi:2016esr,Gavela:2018paw}. We say that we have an \textbf{accidental Peccei-Quinn symmetry}.

\paragraph{Axion quality problem:}
One concern about any solutions to the strong $CP$ problem is that any global symmetry is expected to be violated by \textbf{quantum gravity} \cite{Brennan:2017rbf, Palti:2019pca,Grana:2021zvf}. Hence we expect $\bar{\theta}+a/f_a$ to be shifted from its $CP$-preserving minima due to quantum gravity effects \cite{Georgi:1981pu, Dine:1986bg, Kamionkowski:1992mf,Ghigna:1992iv,Holman:1992us,Barr:1992qq,Dobrescu:1996jp}. In order for the Peccei-Quinn solution to remain valid, we must forbid high-dimensional operators up to dimension $12$, assuming their coefficient is $\sim 10^{-2}$ \cite{Kamionkowski:1992mf}. The Nelson-Barr solution also suffers from a quality problem. In order for the mechanism to work, the fermionic mass matrix  must have a special structure, with some zeroes here and there. However, high-dimensional operators are expected to give corrections to the Nelson-Barr mass matrix so that $\bar{\theta}= \rm arg \, det \left( Y_u Y_d \right)\neq 0$. 

A class of solution to the quality problem is to make the QCD axion \textbf{heavy}. A few years after t'Hooft pioneering paper on IR-dominated instantons in QCD, it was shown  that the introduction of heavy colored particles would make the strong coupling constant $\alpha_s$ grow again at a higher scale $\Lambda{'}$ \cite{Holdom:1982ex,Holdom:1985vx,Dine:1986bg,Flynn:1987rs}.\footnote{See $\alpha_s(\mu)$ in Fig.~1 of \cite{Flynn:1987rs} which looks like a `half-pipe' according to Iason Baldes.} The axion mass would then receive a large contribution from UV instantons and could be as large as $(m_a f_a)^2 \sim (\Lambda{'})^4$ \cite{Flynn:1987rs}.
Other proposals are $SU(3)_c$ being the diagonal subgroup of an $SU(3)^N$ gauge group \cite{Agrawal:2017ksf,Agrawal:2017evu,Csaki:2019vte}, mirror QCD \cite{Rubakov:1997vp,Berezhiani:2000gh,Hook:2014cda,Fukuda:2015ana,Dimopoulos:2016lvn,Hook:2019qoh}, enlarged QCD $SU(3+N)$ \cite{Dimopoulos:1979pp,Gherghetta:2016fhp,Gherghetta:2020ofz}, 5D instantons \cite{Gherghetta:2020keg}, and contributions from real \cite{Fischler:1983sc} or virtual \cite{Fan:2021ntg} monopoles. Heavy axion are further motivated by possible signatures in colliders \cite{Bauer:2017ris,Bauer:2018uxu,CidVidal:2018blh} and gravitational wave interferometers \cite{ZambujalFerreira:2021cte,Ferreira:2022zzo}.

\xintifboolexpr { \x = 2}
  {
  }
{
\medskip
\small
\bibliographystyle{JHEP}
\bibliography{thesis.bib}
}

%% file: chap3.tex
\chapterimage{Planck_CMB} 
\chapter{Standard Model of Cosmology}
\label{chap:SM_cosmology}

Humanity's understanding of the universe has evolved significantly over time. During the major part of History, Earth was regarded as the center of all things, with planets and stars orbiting around it.
A profound shift in the understanding of the cosmos was made by Nicolaus Copernicus who suggested that Earth and the other planets in the solar system in fact orbited around the sun. The legend said he released the final version of his book in 1543 on the day of his death \cite{copernicus2010revolutions}. Despite waves of opposition, the heliocentric view of the solar system was finally accepted after the year 1687 when Isaac Newton formulated the law of gravitation to unify terrestrial and celestial mechanics \cite{Newton:1687eqk}. An important step in the history of astronomy was made by Charles Messier with the release in 1781 of a catalog of 100 nebulas and star clusters visible from Paris \cite{messier1781catalogue}. During the next century, in 1864, the catalog got enlarged to 5000 objects including the ones visible from the south hemisphere, by Wiliams Hershell, his sister Carolina and his son John \cite{herschel1864catalogue}.
At that time the study of the universe was restricted to the observation and description of astrophysical objects. Any conjecture on the fundamental origin of those structures was speculative and metaphysical.
Without any doubt, the greatest triumph was made in 1915 when Einstein formulated the theory of general relativity \cite{Einstein:1915ca,Einstein:1916vd}. Before Einstein, space and time were regarded as immutable coordinates introduced to describe the ``real'' physical objects. The universe  itself was not part of the physical world, instead its role was restricted to be the container of the physical world. The masterstroke of Einstein was to understand that this container, space-time, is in fact a physical object subject to its own dynamics like any other. After 1915, it became possible to formulate theoretical models describing the past history of the universe. This task was conducted between 1922 and 1927 by Friedmann and Lemaitre, followed in 1935 by Robertson and Walker \cite{friedmann1922125, Friedmann:1924bb, Lemaitre:1927zz,Lemaitre:1931zza, Lemaitre:1933gd, Robertson:1935zz, Robertson:1936zza,Robertson:1936zz,walker1935riemanntan, walker1937milne}.

In parallel to the theoretical work, three important observations were made by astronomers. First, in 1912 Henrietta Leavitt discovered the relation between the luminosity and the period of Cepheid variables, providing a new way to measure the distance of faraway galaxies \cite{Leavitt:1912zz}. Second, the same year Vesto Slipher measured galactic spectral redshifts which revealed anomalously large recessional velocities  \cite{slipher1913radial,slipher1915spectrographic}. Finally, in 1929 Edwin Hubble used the two previous discoveries to show that the universe is expanding according to Friedmann and Lemaitre prediction, and measured the constant ratio between the recession speed and the redshift, which is today known as the Hubble constant \cite{Hubble:1929ig}.

Upon reversing the arrow of time, the expansion measured by Hubble unavoidably predicts the universe to have started from a dense and hot state, this is the hot big-bang scenario. The term was coined in 1948 by a firm opponent of this theory, Fred Hoyle, who proposed the same year, in parallel of Bondi and Gold, that the universe is instead in a steady state \cite{Hoyle:1948zz,Bondi:1948qk}. Fred Hoyle has never accepted the hot big-bang scenario but instead he has defended the steady state proposal until his death in 2001 \cite{Hoyle:1993qg,Hoyle:1994fw,Hoyle:1994mq,Narlikar:2002bk}.

However, arguments in favour of the big-bang theory became more and more difficult to discredit. In 1948, Alpher, Bethe, Gamow and Herman showed how a hot big-bang would create hydrogen, helium and heavier elements in the correct proportions to explain their abundance in the early universe \cite{Alpher:1948ve,Alpher:1948srz,Alpher:1948yqy,Alpher:1948gsu}. In 1965, Robert Dicke, Jim Peebles and David Wilkinson at Princeton predicted that if the universe would have started with a hot big-bang, then the universe should be filled with microwave radiation \cite{Dicke:1965zz}. The discovery by inadvertence of this cosmic microwave background, just $60$~km away in Bell Labs the same year, by Arno Penzias and Robert Wilson \cite{Penzias:1965wn}, convinced most cosmologists that the hot big-bang model is the best explanation for the origin of the universe.

This new framework brings together the two most fundamental theories of Nature, General Relativity and Quantum Field Theory. In particular, the study of particle physics can be viewed as a proxy to study the early universe, and particle accelerators can be viewed as a place where the physical conditions of the early universe are recreated. The unification of electromagnetism and weak interactions by Glashow in 1961 \cite{Glashow:1961tr}, and Weinberg and Salam in 1967  \cite{Weinberg:1967tq,Salam:1968rm}, implies the existence of a phase transition, taking place $10^{-12}~\rm s$ after the big-bang, during which the Higgs field acquires a vacuum expectation value. The discovery of asymptotic freedom by Gross, Wilczek and Politzer in 1973 \cite{Gross:1973id, Politzer:1973fx} implies that quarks were propagating freely before $10^{-7}~\rm s$.

In 1980, an important breakthrough was made by Guth \cite{Guth:1980zm} who introduced the inflation paradigm to explain the flatness and homogeneity of the universe. Thanks to the works of, among others, Linde, Steinhardt, Starobinsky and Mukhanov \cite{Linde:1981mu, Albrecht:1982wi,Starobinsky:1980te,Starobinsky:1982ee,Mukhanov:1981xt,Mukhanov:1990me}, in the following years it became possible to relate the existence of the large scale structures of the universe to quantum fluctuations produced during an inflationary phase preceding the hot big-bang.

Since then, the interest in cosmology has blossomed due to the multiplication of the observational evidences for dark matter \cite{Bertone:2010zza,Einasto:2013lka,Majumdar:2014wki,Vavilova:2015bed,Profumo:2017hqp,Essig:2019buk}, and to the direct evidence in 1998 for the acceleration of the universe expansion from supernovae data
\cite{SupernovaSearchTeam:1998fmf,SupernovaCosmologyProject:1998vns}, but also thanks to impressive development in the science of telescopes. Today, cosmology has entered an era of precision. The Planck mission from 2009 to 2013, together with other probes such as Type-1a supernovae, Large Scale Structures (LSS), weak gravitational lensing and Lyman-$\alpha$ forests, have measured cosmological parameters like the abundance of dark matter or the slow-roll parameter below the percent level, or the sound horizon size below one part per $10^{4}$. 
The 2nd Gaia data release in 2018, has measured the celestial position and apparent brightness for 1.7 billions of sources, with parallaxes and proper motions available for 1.3 billions of those \cite{Gaia:2018ydn}. The third gravitational-wave transcient catalogue released by LIGO-Virgo-Kagra Collaboration contains 90 compact binary merging events \cite{LIGOScientific:2021psn,LIGOScientific:2021djp,LIGOScientific:2021aug,LIGOScientific:2021sio}. 

The near future of cosmology is really exciting. In the next decade, Planck successor Litebird in space \cite{Suzuki:2018cuy,Hazumi:2019lys,Sugai:2020pjw,LiteBIRD:2020khw} or CMB-S4 on ground \cite{CMB-S4:2016ple,CMB-S4:2017uhf,Abazajian:2019eic,Abazajian:2019tiv,CMB-S4:2020lpa,Abazajian:2022nyh} could detect primordial B-modes and measure the inflation scale, or reveal the existence of light relics contributing to $N_{\rm eff}$ \cite{Dvorkin:2022jyg}.
Galaxy surveys like Euclid \cite{EUCLID:2011zbd,Amendola:2016saw}, LSST \cite{LSSTScience:2009jmu,LSSTDarkEnergyScience:2018jkl,Ferraro:2022cmj} or SKA \cite{Santos:2015gra,Maartens:2015mra} will measure the statistical distribution of LSS up to redshift $z\sim 3$ with such precision that theoretical uncertainties on non-linear scales will become the main obstacles \cite{Sprenger:2018tdb}. 21-cm intensity mapping with SKA might permit to trace LSS back to the reionization era and even the cosmic dawn, up to redshift $z \sim 20$ \cite{Sprenger:2018tdb,Pritchard:2011xb,Cosmology-SWG:2015tjb,Liu:2022iyy}. One will be able to probe the cosmological properties of Dark Energy, Dark Matter \cite{Dvorkin:2022bsc,Chakrabarti:2022cbu}, or neutrinos with an unprecedented level of precision, possibly revealing deviation from $\Lambda$CDM. Planned multi-TeV cosmic-ray telescopes CTA \cite{CTAConsortium:2010umy,CTAConsortium:2013ofs,Silverwood:2014yza,Pierre:2014tra,Ibarra:2015tya,Duangchan:2022jqn}, LHAASO \cite{Addazi:2019tzi} or KM3NeT \cite{KM3Net:2016zxf,Lopez-Coto:2022pff} will strengthen the constraints on WIMP dark matter. The other side of the mass spectrum, the so-called low energy frontier of particle physics \cite{Jaeckel:2010ni,Essig:2013lka}, will continue being probed by new generations of haloscope, helioscopes, light-shining-through the wall experiments \cite{Marsh:2015xka,Hook:2018dlk,Irastorza:2018dyq, DiLuzio:2020wdo}, fifth force experiments  \cite{Berge:2017ovy}, atomic clocks \cite{Arvanitaki:2014faa}, and all sorts of astrophysical observables  \cite{Arvanitaki:2009fg, Caputo:2021eaa}. 
In a bit more than a decade, GW interferometers LISA in space \cite{Klein:2015hvg,Caprini:2015zlo,Tamanini:2016zlh,Bartolo:2016ami,Babak:2017tow,Caprini:2019egz,LISACosmologyWorkingGroup:2022jok} or ET \cite{Punturo:2010zz,Sathyaprakash:2012jk,Regimbau:2012ir,Maggiore:2019uih} and CE \cite{Reitze:2019iox,Evans:2021gyd} on ground will detect thousands of astrophysical sources, and will search for traces of gravitational waves of primordial origins which will open a new avenue of investigation of early universe cosmology.

In this chapter, we introduce the $\Lambda$CDM model (Sec.~\ref{sec:LambdaCDM}), the hot big-bang scenario (Sec.~\ref{sec:hotBB}), the inflation paradigm (Sec.~\ref{sec:inflation_review}), and we discuss possible sources of primordial gravitational waves (Sec.~\ref{sec:GW_cosmology}). Next, we review the three major puzzles in cosmology, which are the cosmological constant problem (Sec.~\ref{sec:CC_pb}), the matter-antimatter asymmetry (Sec.~\ref{sec:Matter-anti-matter-asymmetry}) and the dark matter mystery (Sec.~\ref{sec:DM}). Finally, we close the chapter by presenting three hot open problems which are the fragility of $\Lambda$CDM (Sec.~\ref{sec:fragility_LambdaCDM}), the Hubble tension (Sec.~\ref{sec:H0_tension}) and the $21$~cm anomaly (Sec.~\ref{sec:21cm}). 

For textbooks on general relativity and cosmology, we refer the reader to \cite{Weinberg:1972kfs,Misner:1974qy,Birrell:1982ix,Wald:1984rg,Stewart:1990uf,Schutz:2003nr,Hartle:2003yu, Carroll:2004st,Stephani:2003tm,Stephani:2004ud,Cheng:2005fn,Hobson:2006se,Rindler:2006km,Hoyng:2006zz,Plebanski:2006sd,Groen:2007zz,Woodhouse:2007yq,Mukhanov:2007zz,Chow:2008zz,Parker:2009uva,Griffiths:2009dfa,Ryder:2009zz,taylor2010exploring,Lambourne:2010zz,Baumgarte:2010ndz,Baumgarte:2021skc,Straumann:2013spu,Zee:2013dea,Gourgoulhon:2013gua,Choquet-Bruhat:2014okh,Ashtekar:2015vfx,Williams:2015ofx,Bohmer:2016ome,Deruelle:2018ltn,Bambi:2018drb,Hall:2018jbs,Carlip:2019vrx,Compere:2019qed,Guidry:2019zwb,Soffel:2019aoq,Mayerson:2019auw,Krasnov:2020lku,Ferrari:2020nzo,rovelli_2021},  and
\cite{Hawking:1973uf,Kolb:1990vq, Linde:1990flp,Peebles:1994xt,Roos:1994fz,Hawley:1998du,Liddle:2000cg,Bonometto:2002xz,Narlikar:2002wxx,Dodelson:2003ft, Mukhanov:2005sc,Liebscher:2005it,Weinberg:2008zzc, Belusevic:2008zza, Belusevic:2008zz,Lyth:2009zz,Montani:2009hju,Goodstein:2012zz,lachieze2012theoretical,Peter:2013avv,Lesgourgues:2013sjj,Baumann:2014nda,Lambourne:2015cfz,Bambi:2015mba,Heacox:2015tzw,Jones:2017xsc,Lyth:2017hyk,Saunders:2017akc,Fazio:2018djb,baumann_2022}, respectively.  For books on the historical development of general relativity and cosmology, we call attention to \cite{friedmann1997essais,Hawley:1998du,Renn:2007psn,Kragh:2019teu,Kolata:2020ydd,Peebles2020,kragh2021cosmology}. We also mention various astrophysical complements, e.g. general astrophysics   \cite{ginzburg2013theoretical,padmanabhan2000theoretical1,padmanabhan2000theoretical2,
padmanabhan2002theoretical3,duric2004advanced,makarov2014introduction,carroll2017introduction,longair2010high,weinberg2019lectures}, galaxy formation \cite{chabrier2011structure,lagos2013physics,fox2017gas,mo2010galaxy,
cimatti2019introduction}  and dynamics \cite{longair2007galaxy,binney2011galactic}, 
stars formation \cite{hartmann2000accretion,stahler2008formation,
ward2011introduction,bodenheimer2011principles,ward2011introduction,krumholz2017star}
and dynamics \cite{Raffelt:1996wa,salaris2005evolution,schwarzschild2015structure,
iliadis2015nuclear}, black holes astrophysics \cite{melia2007galactic,frolov2011introduction,meier2012black,romero2013introduction,
haardt2015astrophysical,grenzebach2016shadow,
falanga2016physics,di2019black,latif2019formation,brito2020superradiance}, plasma astrophysics \cite{tucker1975radiation,rybicki1991radiative,melrose1980plasma,draine2011physics,leblanc2011introduction,somov2012plasma,
somov2012plasma,zheleznyakov2012radiation,benz2012plasma,goossens2012introduction,ghisellini2013radiative,chiuderi2014basics,
williams2018plasma,tajima2018plasma,
goedbloed2019magnetohydrodynamics,
kulsrud2020plasma,williams2021introduction}, and astroparticles \cite{kim1993neutrinos,klapdor1997particle,grieder2001cosmic,miroshnichenko2001solar,zuber2003neutrino,
grupen2005astroparticle,sarkar2007particle,ginzburg2013origin,valle2015neutrinos,
gaisser2016cosmic,sigl2016astroparticle,duffy2017astroparticle,de2015introduction,de2018introduction,de2021particle}.

\section{The $\Lambda$CDM cosmological model}
\label{sec:LambdaCDM}
\subsection{A homogeneous and isotropic expanding universe}
At large scales, the distribution of matter and radiation in the universe appears to be \textbf{homogeneous}, \textbf{isotropic} and in \textbf{expansion}. This suggests that the geometry of our universe is given by the Friedmann-Lemaitre-Robertson-Walker (FLRW) metric\footnote{The FLRW metric is a result of various works between 1922 and 1937 \cite{friedmann1922125, Friedmann:1924bb, Lemaitre:1927zz,Lemaitre:1931zza, Lemaitre:1933gd, Robertson:1935zz, Robertson:1936zza,Robertson:1936zz,walker1935riemanntan, walker1937milne}.}
\begin{equation}
\label{eq:metric_FLRW}
ds^2 = dt^2 - a^2(t)\left[ \frac{dr^2}{1-k\,r^2} + r^2 d\theta^2 +r^2 \sin^2 \theta \, d\phi^2\right],
\end{equation}
where $a$ is the scale factor and $k=-1,\,0,\, +1$ for \textbf{open}, \textbf{flat} and \textbf{closed} universes.  The evolution of the geometry of space-time is controlled by the distribution of energy-momentum within it, as stated by the Einstein equation of \textbf{General Relativity} (1915 \cite{Einstein:1915ca,Einstein:1916vd})
\begin{equation}
\label{eq:Einstein_eq_0}
G_{\mu\nu} = \frac{1}{M_{\rm pl}^2}T_{\mu\nu},
\end{equation}
where $G_{\mu\nu}$ and $T_{\mu\nu}$ are the Einstein and energy-momentum tensors and $M_{\rm pl} \simeq 2.44\times 10^{18}~$GeV is the reduced Planck mass.\footnote{ The reduced Planck mass $M_{\rm pl}$ is related to the Newton gravitation constant $G$ through $M_{\rm pl}^2 \equiv 1/8\pi G $.} The Einstein tensor can be built starting from the metric $g_{\mu \nu}$, by means of the Ricci tensor $R_{\mu\nu}$, the Ricci scalar $R$, the Riemann tensor $R^\mu_{\alpha\beta\gamma}$ and the Christoffel symbols $\Gamma^\mu_{\alpha \beta}$, e.g. \cite{Misner:1974qy,Hobson:2006se}
\begin{align}
&G_{\mu\nu}= R_{\mu \nu} - \frac{1}{2}g_{\mu\nu}R, \\
&R_{\mu\nu} = R^\alpha_{\mu\alpha\nu}, \\
&R = g^{\mu\nu}R_{\mu\nu}, \\
&R^\mu_{\alpha\beta\gamma} = \partial_\beta \Gamma^\mu_{\alpha\gamma} - \partial_\gamma \Gamma^\mu_{\alpha \beta} + \Gamma^\mu_{\sigma\beta}\Gamma^{\sigma}_{\alpha \gamma} -  \Gamma^{\mu}_{\sigma\gamma}\Gamma^{\sigma}_{\alpha \beta}, \\
&\Gamma^{\mu}_{\alpha \beta} = \frac{1}{2}g^{\mu\sigma}\left(\partial_\alpha g_{\sigma\beta} + \partial_{\beta} g_{\alpha \sigma} - \partial_{\sigma} g_{\alpha \beta} \right).
\end{align}
From injecting the FLRW metric in Eq.~\eqref{eq:metric_FLRW} in the Einstein equation in Eq.~\eqref{eq:Einstein_eq_0}, we find that the dynamics of the universe is governed by the \textbf{Friedmann equation} (1922),
\begin{equation}
\label{eq:Friedmann_equation}
H^2 + \frac{k}{a^2} = \frac{\rho}{3M_{\rm pl}^2}, \qquad \quad H= \frac{\dt{a}}{a}, \quad 
\end{equation}
where $H$ is the Hubble constant and $\rho = \rho_{M}+\rho_R + \rho_\Lambda$ is the total energy density, sum of a matter, radiation and vacuum component. Assuming that the different fluids do not interact, the conservation of the energy-momentum of each individual component $i$ in the FLRW universe $\partial_\mu T_{\mu \nu}=0$ leads to the \textbf{continuity equation}
\begin{equation}
\dt{\rho}_{i} + 3H(\rho_i + p_i) = 0.
\end{equation}
It encodes how each fluid evolves in the expanding universe
\begin{equation}
\label{eq:energy_density_FLRW}
\rho_i \propto a^{-3(\omega_i+1)},\qquad \omega_i \equiv \frac{p_i}{\rho_i},
\end{equation}
with $\omega_i$ being the equation of state, $\omega_i = 1,\, \frac{1}{3},\, 0,\, -\frac{1}{3}, \,-1$ for kination, radiation, matter, curvature and vacuum energy, respectively. From injecting Eq.~\eqref{eq:energy_density_FLRW} into the Friedmann equation in Eq.~\eqref{eq:Friedmann_equation}, we obtain
\begin{equation}
\label{eq:scale_factor}
a(t) \propto t^{2/3(1+\omega)}.
\end{equation}
We can define the fraction of energy density $\Omega_i = \rho_i/3M_{\rm pl}^2 H^2$, whose sum, $\Omega$, is respectively equal to $<1$, $1$, $>1$ for open, flat and closed universes
\begin{equation}
\label{eq:Omega_i_universe_friedmann}
\Omega = \frac{\rho}{M_{\rm pl}^2 H^2} = \sum_i \Omega_i = 1 + \frac{k}{(a\,H)^2}.
\end{equation}

\subsection{Energy content of the universe}
\label{sec:content_U}

The observation of the Cosmic Microwave Background (CMB) with the telescope \textbf{Planck}, between 2009 and 2013, together with low-redshift probes (mainly the luminosities of Type $Ia$ Supernova (SNe Ia) and the angular scale of Baryonic Acoustic Oscillations, have allowed the precise measurement of the energy content of the universe (per order of dominance) \cite{Aghanim:2018eyx, Tanabashi:2018oca}
\begin{align}
&\Omega_{\Lambda} = 0.685(7), \\
&\Omega_{\rm DM} = 0.1200(12)\,h^{-2} = 0.265(7), \\
&\Omega_{b} = 0.02237(15) \,h^{-2} = 0.0493(6), \\
&\Omega_{\nu} = h^{-2}\sum m_{\nu_j}/93.14\rm \, eV \quad < 0.003~{\rm (CMB+BAO) }\quad > 0.0012 ~\rm (mixing). \\
& \Omega_{\gamma} = 2.473 \times 10^{-5}\,h^{-2} = 5.38(15) \times 10^{-5}, \qquad T_{\gamma} = 2.7255(6)~\rm Kelvin,
\end{align}
where $\Omega_{\Lambda} $ is the \textbf{Dark Energy} (vacuum energy),  $\Omega_{\rm DM}$ is the \textbf{Dark Matter} (non-baryonic cold matter), $\Omega_{b}$ are the \textbf{baryons}, $\Omega_{\nu}$ are the \textbf{neutrinos} and $\Omega_{\gamma}$ are the \textbf{photons}. 

The Hubble parameter, which quantifies the expansion rate of the universe, is measured by Planck satellite to be \cite{Aghanim:2018eyx, Tanabashi:2018oca}
\begin{equation}
H_0 = 100\,h~{\rm km/s / Mpc} \qquad {\rm with} \quad h = 0.674(5).
\end{equation}
The Dark Energy is a fluid with \textbf{constant energy density} and \textbf{negative pressure} $\rho_\Lambda = - p_\Lambda = \rm constant$, responsible for the acceleration of the expansion of the universe, as confirmed in 1998 with SNe Ia measurements \cite{Riess:1998cb,Perlmutter:1998np}. The Dark Matter is a fluid which redshifts as \textbf{cold matter} $\rho_{\rm DM} \propto a^{-3}$.

Finally, the trigonometry in our universe appears to be close to \textbf{euclidean}, implying that the curvature is close to \textbf{flat} \cite{Aghanim:2018eyx, Tanabashi:2018oca} \footnote{Note however that the CMB fit leading to Eq.~\eqref{eq:flatness_planck} has an anomaly in the lensing amplitude $A_{\rm L}$ at $2.8~\sigma$ \cite{Aghanim:2018eyx}
\begin{equation}
A_{\rm L} = 1.180 \pm 0.065.
\end{equation} 
Interestingly, $A_{\rm L}$ and $\Omega_{\rm K}$ are degenerate and the resolution of the lensing anomaly, $A_{\rm L} \to 1$, is possible if our universe is closed at 3.4 standard deviations, $-0.007>\Omega_{\rm K} > -0.095$ \cite{DiValentino:2019qzk} (See also \cite{Park:2017xbl,Handley:2019tkm}). }
\begin{equation}
\label{eq:flatness_planck}
\Omega_{\rm K} = 0.0007 \pm 0.0019.
\end{equation}

\section{The hot big-bang scenario}
\label{sec:hotBB}
\subsection{Thermal equilibrium}

The expansion of the universe implies that if we go backward in times the universe is contracting, and the temperature increases. The Einstein equation even predicts that the size of the universe $a(t)$, cf. Eq.~\eqref{eq:scale_factor}, goes to zero when $t\to 0$, implying that the temperature becomes infinite. This is the \textbf{Hot Big-Bang Scenario}.\footnote{In the standard big-bang scenario, Einstein equations predicts a cosmic singularity at $a \to 0$ and $T \to \infty$. This is the \textbf{Cosmic singularity problem}. Alternatively, in the \textbf{big-bounce scenario}, the initial singularity is replaced by a contracting phase in which the scale factor shrinks to a critical finite size and bounces. However, the inherent difficulty of the big-bounce scenario is the violation of the Null-Energy-Condition in Eq.~\eqref{eq:NEC_violation}. See \cite{Lehners:2008vx,Cai:2014bea,Battefeld:2014uga,Brandenberger:2016vhg,Ijjas:2018qbo,Mironov:2019haz} for reviews on bouncing cosmology.}

At that time the interaction rate of particles $\Gamma$ is larger than the expansion rate and the particles are at \textbf{thermal equilibrium}. Therefore we can describe the thermodynamics quantities of the plasma using \textbf{statistical mechanics}. The number density of particles with degrees of freedom $g$ is given by
\begin{equation}
n = \frac{g}{(2\pi)^3} \int d^3 p ~ f(\vec{p}),
\end{equation}
where $f(\vec{p})$ is the average number of fermions/bosons in the single-particle state of momentum $\vec{p}$
\begin{equation}
f(\vec{p}) = \frac{1}{e^{(E-\mu)/T} \pm 1},
\end{equation}
with $+$ for \textbf{Fermi-Dirac} (FD) statistics (1926) \cite{zannoni1999quantization,dirac1926theory} and $-$ for \textbf{Bose-Einstein} (BE) statistics (1924) \cite{10017606624}. $T$ is the photon temperature and $\mu$ is the chemical potential. The chemical equilibrium for $i+j \leftrightarrow k+l$ is realized when $\mu_i+\mu_j = \mu_k + \mu_l$. A given particle $X$ has the opposite chemical potential as its antiparticle $\bar{X}$, $\mu_X = - \mu_{\bar{X}}$. Therefore, in a symmetric universe where $n_X = n_{\bar{X}}$, we have $\mu_X=0$. Assuming it is the case, the \textbf{number density} at equilibrium is
\begin{equation}
\label{eq:number_density}
n  \simeq \begin{cases}
c\,g \,\frac{\zeta(3)}{\pi^2}\, T^3,  &\hspace{2em} T \gtrsim m,\\
g \left( \frac{m\,T}{2\pi} \right)^{3/2} \, e^{-m/T},  &\hspace{2em} T \lesssim m,
\end{cases}
\end{equation}
where $\zeta(3) \simeq 1.2$, $c = 1$ for BE, and $c= 3/4$ for FD.

The next important thermodynamic quantity is the \textbf{energy density} of particles at thermal equilibrium, which reads
\begin{equation}
\label{eq:energy_density}
\rho = \frac{g}{(2\pi)^3} \int d^3 p ~ E ~ f(\vec{p}) 
\simeq  \begin{cases}
\frac{\pi^2}{30} \,g_{\rm eff} \, T^4,  &\hspace{2em} T \gtrsim m,\\
m \, n,  &\hspace{2em} T \lesssim m,
\end{cases}
\end{equation}
where $g_{\rm eff}$ is the effective number of relativistic degrees of freedom, 
\begin{equation}
g_{\rm eff} = \sum_{i=b} g_i \left(  \frac{T_i}{T}\right)^4 + \frac{7}{8} \sum_{i=f} g_i  \left(  \frac{T_i}{T}\right)^4.
\end{equation}
We have introduced an additional factor $(T_i/T)^4$ in order to account for relativistic species that are not in thermal equilibrium with the photons. 

Similarly we define the \textbf{pressure}
\begin{equation}
P= \frac{g}{(2\pi)^3} \int d^3 p ~ \frac{p^2}{3E} ~ f(\vec{p}).
\end{equation}
which for radiation is simply $P = \rho/3$.

Finally, we define the \textbf{entropy density}\footnote{Upon assuming an adiabatic evolution, the entropy of the universe at the big-bang epoch was carried by the $410$ photons and $340$ neutrinos per $cm^3$ of the observable universe today. This is actually $10^{-122}$ times lower than the maximal entropy of the universe, which is given by the generalization of the Bekeinstein-Hawking formula $S=A^2/4$ for black hole horizons \cite{Bekenstein:1972tm,Hawking:1974sw} to cosmological horizons \cite{Fischler:1998st,Bousso:1999xy,Spradlin:2001pw,Bousso:2002fq}. In comparison, a high-entropy big-bang universe would be born full of black holes.
The fraction of the whole phase space in which our universe sits is one part in $e^{10^{122}}$, which is extremelly fine-tuned if one assumes a uniform distribution. This is the \textbf{entropy problem} raised by Roger Penrose in 1988 \cite{Penrose:1988mg,Barnes:2020kpu}. Inflation, see Sec.~\ref{sec:inflation_review}, can not offer a solution if the distribution of high-entropy gravitational remnants from big-bang is scale-invariant as one could expect \cite{Penrose:1988mg}. This leads to the question of whether `the initial conditions of the universe could have originated from a Poincare recurrence' \cite{Dyson:2002pf}.}
\begin{equation}
\label{eq:entropy_density}
s = \frac{\rho +P}{T} = \frac{2\pi^2}{45} h_{\rm eff}\,T^3,
\end{equation}
with
\begin{equation}
h_{\rm eff} = \sum_{i=b} g_i \left(  \frac{T_i}{T}\right)^3 + \frac{7}{8} \sum_{i=f} g_i  \left(  \frac{T_i}{T}\right)^3.
\end{equation}
If the evolution of the universe is \textbf{adiabatic}, the total entropy is conserved
\begin{equation}
\label{eq:adiab_evolution}
{\rm adiab. ~evolution:} \qquad h_{\rm eff}\,T^3\,a^3 = {\rm constant}.
\end{equation}
Therefore, the scale factor evolves as $a \propto h_{\rm eff}^{-1/3} T^{-1}$. It is sometimes convenient to introduce the \textbf{redshift parameter} $z$ 
\begin{equation}
\label{eq:redshift}
1+z(t) \equiv a_0/a(t),
\end{equation}
which increases with the temperature.
During a radiation-dominated evolution, the time-to-temperature relation is given by
\begin{equation}
t = 2.42~{\rm s} ~ h_{\rm eff}^{-1/2} \left( \frac{1~\rm MeV}{T} \right)^2.
\end{equation}

\subsection{Beyond thermal equilibrium}
\label{eq:Beyond_therm_equil}

In an expanding universe, the thermal equilibrium condition is not always satisfied. A given particle $\chi$ (e.g. Dark Matter, baryons, neutrinos..) interacting with the hot SM plasma with a rate $\Gamma$, is at thermal equilibrium when 
\begin{equation}
\label{eq:equil_expU_cond}
\Gamma \gtrsim H.
\end{equation}

In the SM, the interactions between SM particles are mediated by the gauge bosons of $SU(3)_c\times SU(2)_L\times U(1)_Y$. In the early universe, when the temperature is larger than the mass of the gauge bosons, $T \gtrsim m_A$, the cross-section can be estimated by
\begin{equation}
\label{eq:weak_cross_section_high_T}
\sigma \sim \alpha^2 / T^2
\end{equation}
where $\alpha$ is the typical structure constant of the gauge interaction.
The number of interaction per unit of time is given by
\begin{equation}
\Gamma = n_\chi \sigma \vrel \sim \alpha^2\,T.
\end{equation}
It must be compared with the expansion rate of the universe $H \sim T^2/M_{\rm pl}$. The plasma of particles in the SM are interacting faster than the universe expands when $\Gamma \gtrsim H$ with
\begin{equation}
\frac{\Gamma}{H} \sim \left( \frac{\alpha}{1/30} \right)^2 \frac{10^{15}~ \rm GeV}{T}.
\end{equation}
Hence, the SM particles enter a thermal equilibrium state around $T \sim 10^{15}$ GeV.

If the universe was a hot plasma in the past, it is not the case anymore today. As a consequence of the universe expansion, the particles stop interacting.
Very often\footnote{If particles are very light, like the neutrinos, the freeze-out can occur at a temperature much larger than its mass, cf. paragraph on neutrinos of the present section.}, this occurs when the temperature reaches the mass threshold, $T \lesssim m_\chi$, the thermal excitations of the plasma become too weak to afford for the production of $\chi$, the production rate drops exponentially with the temperature and the particle $\chi$ leaves the thermal equilibrium. Consequently, the total number of particles becomes frozen $n_{\chi}\, a^3 = \rm constant$. It is common to introduce the comoving number density $Y_\chi = n_{\chi}/s$  which is conserved under Hubble expansion. The residual $\chi$ abundance can be computed in the \textbf{instantaneous freeze-out} approximation
\begin{equation}
\label{eq:inst_FO}
\Gamma \simeq H \quad \rightarrow \quad Y_{\rm FO} \simeq \frac{H}{s\,\left< \sigma \vrel \right>} \simeq 0.75 \, \frac{g_{\rm eff}^{1/2}}{h_{\rm eff}}\, \frac{x_{\rm FO}}{M_{\chi} \, M_{\rm pl} \left< \sigma \vrel \right>}
\end{equation}
with the temperature of departure from thermal equilibrium given by
\begin{equation}
\label{eq:inst_FO_x}
x_{\rm FO} \equiv \frac{M_{\chi}}{T_{\rm FO}} \simeq \rm{Log} \left[ 0.192 \tfrac{g_\chi}{\sqrt{g_{\rm SM}^{\rm FO}}} \; M_{\rm pl} \; \MDM \;  \left< \sigma \vrel \right>\; x_{\rm FO}^{1/2} \right] \sim 30.
\end{equation}
More details on thermal freeze-out are given in Sec.~\ref{par:FI_FO_DM} of Chap.~\ref{chap:DM}.

\paragraph{Baryons:}
As an application, we propose to compute the abundance of baryons. The baryon abundance today is set when the proton/anti-proton annihilation process freezes-out, and we can assume the proton-antiproton annihilation cross-section to be \cite{Kolb:1990vq}
\begin{equation}
\sigma_{p\bar{p}\,}\vrel= \frac{c_1}{m_{\pi}^2},
\end{equation}
where $c_1$ is a constant of order $1$ and $m_{\pi}\simeq 135~\rm MeV$ is the pion mass.
From Eq.~\eqref{eq:inst_FO_x}, we get $x_{\rm FO} \simeq 49$ where we have plugged $g_p = 4$, $g_{\rm SM}^{\rm FO}= 10.75$ and $c_1 = 1$.  From Eq.~\eqref{eq:inst_FO} we find
\begin{equation}
\label{eq:baryon_asym_th}
\eta_{\rm B}^{\rm th}  = \left.\frac{n_b}{n_\gamma}\right|_{\rm th} \simeq  1.8 \times 10^{-19} + \left.\frac{n_b -n_{\rm \bar{b}}}{n_\gamma}\right|_{\rm prim}.
\end{equation}
where $ \sim 10^{-19}$ is the expected baryon abundance today if $n_{\rm b}=n_{\rm \bar{b}}$, whereas $\left.\frac{n_b -n_{\rm \bar{b}}}{n_\gamma}\right|_{\rm prim}$ is the primordial baryon asymmetry (which can not annihilate).
The baryon-to-photon\footnote{Note however that the lepton asymmetry $\eta_{\rm \nu}$ is only poorly constrained by BBN and is at most \cite{Mangano:2011ip}
\begin{equation}
\eta_{\rm \nu}^{\rm exp} \lesssim 0.1.
\end{equation} } ratio is precisely inferred from the comparison between the predictions for the abundance of light element produced during the Big-Bang Nucleosynthesis (BBN) and the measurement of these abundances \cite{Fields:2014uja}
\begin{equation}
\label{eq:baryon_asym_BBN}
\eta_{\rm B}^{\rm exp} = \left.\frac{n_b}{n_\gamma}\right|_{\rm exp} \simeq  6.5 \times 10^{-10}.
\end{equation}
From comparing the expected baryon abundance assuming $n_b=n_{\bar{b}}$ in Eq.~\eqref{eq:baryon_asym_th} and the observed abundance in Eq.~\eqref{eq:baryon_asym_BBN}, we understand that it is necessary to introduce a \textbf{primordial baryon asymmetry}, given by Eq.~\eqref{eq:baryon_asym_BBN}, in order to match the observations. 
This is problematic since the observed asymmetry $b/\bar{b}$ can not be explained in the SM of elementary particles. We give more details in Sec.~\ref{sec:Matter-anti-matter-asymmetry}.

\paragraph{Neutrinos:}
\label{par:neutrino_heating}
For very light particles like the neutrinos, thermal freeze-out can occur at a temperature much above the neutrino mass. After that the temperature becomes smaller than the $W^\pm$ gauge boson mass, the cross-section of neutrino annihilation in Eq.~\eqref{eq:weak_cross_section_high_T} becomes $\sigma \sim G_F^2 T^2$ with $G_F^{-1} = \sqrt{2} \,v^2$, such that the neutrino decoupling occurs around $1~\rm MeV$
\begin{equation}
\frac{\Gamma_\nu}{H} \sim \left( \frac{T_{\nu,\, \rm dec}}{1\rm ~MeV} \right)^3.
\end{equation}
The neutrino being relativistic at the time of the decoupling, their abundance today reads
\begin{equation}
\label{eq:neutrino_abundance}
\Omega_\nu h^2 = \frac{\sum m_\nu}{3 M_{\rm pl}^2 (H_0/h)^2} Y_\nu^{\rm dec}\,s_0 \simeq \frac{\sum m_\nu}{94.1(93.1)~\rm eV}
\end{equation}
where $g_\nu = \tfrac{3}{4}\cdot 2$ (only one neutrino flavor), $h_{\rm eff}^{\rm dec}=2+\tfrac{7}{8}(4+6)$ (photons, $e^+/e^-$ and neutrinos) and $s_0$ is the entropy today $s_0 = \frac{2\pi^2}{45}\,(2+\frac{7}{8}\cdot 2 \cdot 3 \cdot \frac{4}{11}) \,T_0^3$. The factor $(T_\nu/T_\gamma)^3 = 4/11$ results from the $e+/e-$ annihilation around $\sim 511/30~\rm keV$, which heats the photon bath but not the neutrino bath.\footnote{The conservation of the entropy during the $e+/e-$ annihilation assuming that the neutrinos have already completely decoupled reads $(2 + \frac{7}{8} \cdot 4 ) T_\nu^3 = 2 T_\gamma^3$.} The number in bracket in Eq.~\eqref{eq:neutrino_abundance} accounts for the correction due to the slight heating of the neutrino bath resulting from $e^+/e^-$ annihilation \cite{Lesgourgues:2018ncw}. Effectively, this is the same as replacing the number of neutrino species $N_\nu=3$ by $N_{\nu,\,\rm eff}=3.045$.

\begin{figure}[h!]
\centering
\raisebox{0cm}{\makebox{\includegraphics[width=0.7\textwidth, scale=1]{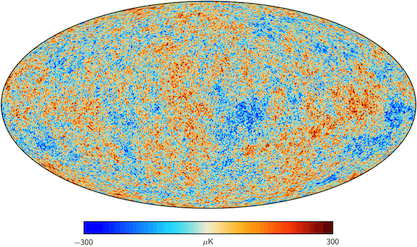}}}
\caption{\it \small  Map of the temperature differences $\delta T\left( \theta, \, \phi \right)/T$ (at the $10^{-5}$ level) on the oldest known optical surface: the background sky when our universe first became transparent to light, $380 \,000$ years after the big-bang. Credits: European Space Agency and Planck Collaboration.}
\label{fig:Planck_2015}
\end{figure}

\paragraph{Photons:}
\label{sec:photon_decoupling}
The photon decouples from the thermal plasma when the rate of \textbf{Compton scattering}
\begin{equation}
e^-+ \gamma \longleftrightarrow e^- + \gamma, \qquad \Gamma_{\gamma} \simeq n_e\,\sigma_T, \qquad \sigma_T \simeq 2 \times 10^{-3}~{\rm MeV^{-2}},
\label{eq:compton_scattering}
\end{equation}
becomes comparable with the expansion rate of the universe
\begin{equation}
\label{eq:inst_FO_decl}
\Gamma_\gamma(T_{\rm dec}) \sim H(T_{\rm dec}).
\end{equation}
Compton scattering becomes inefficient when the electrons \textbf{recombine} with the protons to form hydrogen atoms. The electron density $n_e$ can be computed from the chemical equilibrium condition 
\begin{equation}
\label{eq:electron_proton_equil}
\mu_{e^-} + \mu_{p^+} = \mu_H + \mu_\gamma , \qquad \left(\frac{n_H}{n_e^2}\right)_{\rm eq}= \left( \frac{2\pi}{m_e\,T}  \right)^{3/2}~e^{B_H/T} \qquad B_H \simeq 13.6~\rm eV.
\end{equation}
From injecting Eq.~\eqref{eq:electron_proton_equil} into Eq.~\eqref{eq:compton_scattering} and then into Eq.~\eqref{eq:inst_FO_decl} and using the numerical value of $\eta_{\rm B}$ in Eq.~\eqref{eq:baryon_asym_BBN}, we obtain the temperature at which the photon decouples from the hot plasma
\begin{equation}
\label{eq:decoupling_CMB}
T_{\rm dec} \simeq 0.27~{\rm eV}, \qquad t_{\rm dec} \simeq 380\,000~\rm yrs
\end{equation}
Below that temperature, the universe becomes transparent and photons propagate freely.

The observation of these photons today constitutes the Cosmic Microwave Background (CMB). The CMB was predicted in 1965 by Robert Dicke, Jim Peebles and David Wilkinson at Princeton \cite{Dicke:1965zz} and discovered inadvertently, just $60$~km away in Bell Labs, the same year, by Arno Penzias and Robert Wilson \cite{Penzias:1965wn}. Since then, the CMB has been measured with high precision by the Planck satellite from 2009 to 2013, cf. Fig.~\ref{fig:Planck_2015}. It is isotropic at the $10^{-5}$ level and, for a given direction in the sky, it follows the spectrum of a black-body corresponding to a temperature $T_\gamma = 2.7255~$K. 


\section{Inflation}
\label{sec:inflation_review}

\subsection{The homogeneity problem}

On the one hand, the greatest distance from which an observer at the time of photon decoupling $t_{\rm dec}$ is able to receive signals is given by the \textbf{comoving particle horizon} 
\begin{equation}
\chi_{\rm ph}^{\rm dec}=\int_{0}^{t_{\rm dec}} \frac{dt}{a(t)}.
\end{equation}
This gives the maximal size of the causal patch in the CMB we observed today. Actually, the true causal patch, called the \textbf{sound horizon}, is a bit smaller since the speed of sound is $<1$ and close to $1/\sqrt{3}$.  

On the other hand, the distance a photon has undergone from decoupling until today is $\int_{t_{\rm dec}}^{t_{\rm today}} \frac{dt}{a(t)}$. Therefore, the number of causal patches imprinted on the CMB sphere (or last-scattering-surface) we observe today, is equal to the area of the CMB sphere, divided by the area of an individual causal patch
\begin{equation}
\label{eq:nbr_causal_patches}
\#_{\rm causal~patches} = \frac{4\pi}{\pi}\left(\frac{\int_{t_{\rm dec}}^{t_{\rm today}} \frac{dt}{a(t)}}{\int_{0}^{t_{\rm dec}} \frac{dt}{a(t)}}\right)^2  \simeq 10\,000
\end{equation}
where we computed $a(t)$ from integrating the Friedmann equation in Eq.~\eqref{eq:Friedmann_equation} with the energy content measured by Planck, cf.  Sec.~\ref{sec:content_U} and from assuming a radiation-dominated universe before BBN. Hence, by looking at the celestial sphere, we should observed $10\,000$ causally independent Hubble patches with individual angular size $1^{\circ}$, the age of each patch being $t_{\rm dec}$. However, the temperature of the Cosmic Microwave Background appears to have a striking homogeneity of $\frac{\delta \rho}{\rho} \sim 10^{-5}$. This is the \textbf{homogeneity problem}.

\subsection{The flatness problem}
Another surprise is the observed flatness of the universe at $\Omega_{\rm K} \lesssim 0.1~\%$, cf. Eq.\eqref{eq:flatness_planck}. From Eq.~\eqref{eq:Omega_i_universe_friedmann}, we can see that the curvature  fraction $\Omega_{\rm K}$ should have increased with cosmic time like the square of the \textbf{comoving Hubble radius} $(a\,H)^{-1}$, 
\begin{equation}
\left| \Omega_{\rm K}(a)  \right|=(a\,H)^{-2},
\label{eq:flatness_inflation}
\end{equation}
implying that when the temperature was $T\sim \rm TeV$, the curvature would have been $\Omega_{\rm K}^{\rm TeV} \lesssim 10^{-32}$. This, except if the geometry is rigorously flat $k=0$, requires extreme fine-tuning. This is the \textbf{flatness problem}.

\subsection{The solution: shrinking the comoving Hubble radius}

\paragraph{The inflation idea:}
The homogeneity and flatness problems motivated Alan Guth in 1980 \cite{Guth:1980zm} to postulate the existence of an early \textbf{inflation} period between $t_{\rm start}$ and $t_{\rm end}$, during which the scale factor increases exponentially fast 
\begin{equation}
a(t) \propto e^{Ht} \qquad {\rm \longrightarrow } \qquad \chi_{\rm ph}^{\rm dec}=\int_{t_{\rm start}}^{t_{\rm end}} \frac{dt}{a(t)} \simeq (a\,H)^{-1}_{t_{\rm start}}.
\end{equation}
The comoving Hubble radius $(a\,H)^{-1}$ decreases and, for arbitrary long inflation, $a(t_{\rm start}) \to 0$, the comoving particle horizon $\chi_{\rm ph}^{\rm dec}$ can be made arbitrary large. As a consequence, the whole CMB sphere can be contained in a single inflated causal patch. The homogeneity problem is solved when the comoving Hubble radius $(a\,H)^{-1}$ during inflation has become smaller than the comoving Hubble radius today $(a_0\, H_0)^{-1}$. Also, the flatness problem is solved by stretching the curvature, see Eq.~\eqref{eq:flatness_inflation}.

For pedagogical introductions to inflation, we mention the reviews \cite{Riotto:2002yw,Linde:2005ht,Baumann:2009ds,Baumann:2014nda, Langlois:2010xc,Senatore:2013roa,Burgess:2017ytm,Vazquez:2018qdg} and the reference textbooks \cite{Kolb:1990vq, Liddle:2000cg,Dodelson:2003ft, Mukhanov:2005sc,Weinberg:2008zzc, Lyth:2009zz,Peter:2013avv}. 

\paragraph{Duration of inflation:}
During inflation, a given wave number gets stretched as $k_{\rm phys} = k/a$, where $k$ is the comoving wave number, until it leaves the causal patch when $k > a\,H$ and freezes. The number of e-folds of accelerated expansion between the time when the mode $k$ exits the horizon and the end of inflation is
\begin{equation}
N_k \equiv \log{\frac{a_{\rm end}}{a_k}}, \qquad {\rm with} \quad k=a_k\,H_k.
\end{equation}
We introduce the comoving Hubble horizon today $a_0\,H_0$, such that we can write (e.g. \cite{Liddle:1993fq})
\begin{equation}
N_k = - {\rm ln} \left( \frac{k}{a_0 H_0} \right) + {\rm ln} \left( \frac{a_{\rm end} H_{\rm end}}{a_0 H_0} \right)  -{\rm ln} \left( \frac{H_{\rm end}}{H_k} \right).
\label{eq:efold_nbr_2}
\end{equation}
We expect the last term to be of order 1, however the second term depends strongly on the evolution of the universe after the end of inflation. If we assume that the inflaton behaves like matter (if it oscillates in a quadratic potential), between the end of inflation and the end of the reheating phase, then
\begin{equation}
a_{\rm end} = a_0 \left( \frac{H_0}{H_{\rm eq}} \right)^{2/3} \left( \frac{H_{\rm eq}}{H_{\rm reh}}  \right)^{1/2}\left( \frac{H_{\rm reh}}{H_{\rm end}}  \right)^{2/3},
\label{eq:aend_reh_eq}
\end{equation}
where $H_{\rm reh}$ and $H_{\rm eq}$  are the Hubble parameters at the end of the reheating phase and at matter-radiation equality.\footnote{The Hubble factor at matter-equality can be computed as $H_{\rm eq} \simeq H_0\,z_{\rm eq}^{3/2}$ with $z_{\rm eq} = 3402(26)$ \cite{Tanabashi:2018oca}. } Plugging Eq.~\eqref{eq:aend_reh_eq} into Eq.~\eqref{eq:efold_nbr_2}, we obtain 
\begin{equation}
N_k \simeq 56 - {\rm ln} \left( \frac{k}{a_0 H_0} \right)  + \frac{1}{3} {\rm ln} \left( \frac{T_{\rm reh}}{10^9 ~ {\rm GeV}} \right) + \frac{1}{3} {\rm ln} \left( \frac{H_{\rm end}}{10^{13}~ \rm GeV} \right) - {\rm ln} \left( \frac{H_{\rm end}}{H_k} \right).
\end{equation}
By setting $k \lesssim a_0 H_0$ we obtain the minimal number of e-folds of inflation for two given points within the Hubble patch today to be causally connected.
Taking the lowest inflation scale such that sphaleron transitions are still active and successful baryogenesis is still possible, $T_{\rm reh} \sim 1~\rm TeV$ (assuming instantaneous reheating), we obtain $N_{\rm min} \simeq 32$. If we consider the lowest temperature allowed by BBN, $T_{\rm reh} \sim 10~\rm MeV$, we obtain $N_{\rm min} \simeq 21$. Taking the maximal reheating temperature, $T_{\rm reh}\sim 10^{16}~$GeV, compatible with the non-detection of fundamental $B$-modes in the CMB, see Eq.~\eqref{eq:max_Hinf_Bmodes}, we get $N_{\rm max} \simeq 62$.

\subsection{Slow-roll inflation}
\begin{figure}[h!]
\centering
\raisebox{0cm}{\makebox{\includegraphics[width=0.7\textwidth, scale=1]{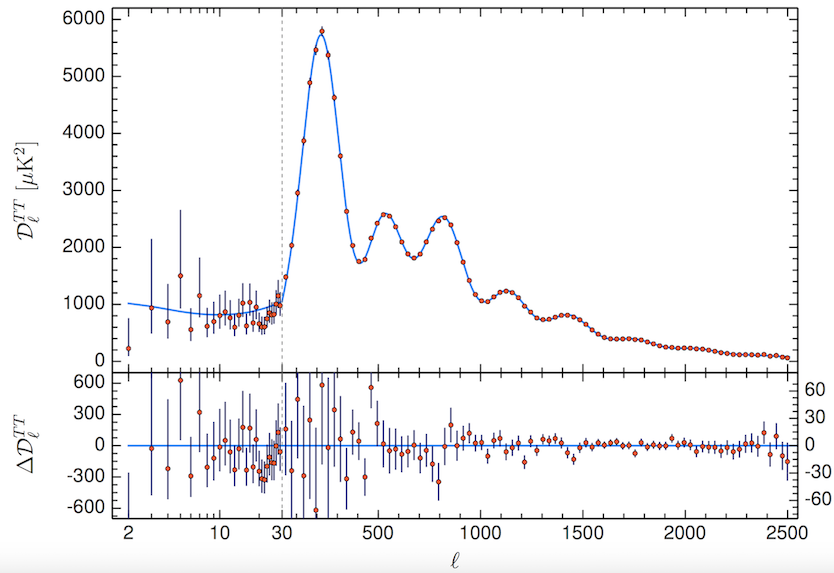}}}
\caption{\it \small Cosmic microwave background temperature power spectrum (CMB). The blue line shows the best fit to the Planck data \cite{Aghanim:2018eyx}. The small (large) multipoles $l$ correspond to the long (small) angular scales: the main acoustic peak at $l=200$ corresponds to $1~$degree on the sky whereas $l=1800$ corresponds to $0.1~$degree. Figure reproduced from \cite{Aghanim:2018eyx}.}
\label{fig:CMB_Planck2018}
\end{figure}

\paragraph{The temperature anisotropies problem: }
\label{par:temp_anisotropy_pb}

The observation of the CMB with the telescopes COBE (1989-1993), WMAP (2001-2010) and Planck (2009-2013) has revealed the presence of anisotropy of temperature at the level of $\delta T/T \sim 10^{-5}$. A theory which solves the horizon and the flatness problem must also be able to explain the presence of these temperature anisotropies. 
The temperature anisotropies in the CMB can be expanded under spherical harmonics as 
\begin{equation}
\frac{\delta T}{T}\left( \theta, \, \phi \right)=\sum_{l=2}^{+\infty} \, \sum_{m=-l}^{+l} a_{lm} \, Y_{lm}\left( \theta , \, \phi \right ),
\end{equation}
where $Y_{lm}\left( \theta , \, \phi \right)$ are spherical harmonics. If the temperature fluctuations are assumed to be Gaussian, as it seems to be the case, the variance $C_l$ of $a_{lm}$, given by 
\begin{equation}
\label{eq:cl}
C_{l} \equiv \frac{1}{2l+1} \sum_{m=-l}^{+l} \left|a_{lm}\right|^2,
\end{equation}
contain all the information about the temperature anisotropies. The $C_l$ have been very well measured by Planck, see Fig.~\ref{fig:CMB_Planck2018}. 

\paragraph{The inflaton:}
Guth's initial 1980 proposal for generating a fast expansion relied on the universe being stuck in a false vacuum of high density \cite{Guth:1980zm}. This idea was also proposed the same year by Kazanas \cite{Kazanas:1980tx}. However, Guth recognized himself that bubbles of true vacuum might never percolate and that the universe might remain stuck in the false vacuum forever \cite{Guth:1981uk,Guth:1982pn}. This led Linde, Albrecht and Steinhardt \cite{Linde:1981mu, Albrecht:1982wi} to propose the \textbf{slow roll} idea the next year. Right after, Starobinsky, Mukhanov and Chibisov showed that quantum fluctuations produced during slow roll generate a spectrum of inhomogeneities possibly accounting for the large scale structures in our universe \cite{Starobinsky:1980te,Starobinsky:1982ee,Mukhanov:1981xt,Mukhanov:1990me}.

The standard lore is to assume the universe to be dominated by a scalar field $\phi$, the \textbf{inflaton}, which slow-roll down its potential $V(\phi)$ under the effect of Hubble friction, and to impose the parameter, e.g. \cite{Baumann:2009ds}
\begin{equation}
\epsilon \equiv -\frac{\dt{H}}{H^2} = - \frac{d \ln{H}}{dN} = \frac{\frac{1}{2}\dt{\phi}^2}{M_{\rm pl}^2H^2},
\end{equation}
to be small.  Deviation from slow-roll is encoded in the second parameter $\eta$ defined through 
\begin{equation}
\frac{d \rm ln \epsilon}{dN}  = 2(\epsilon - \eta), \qquad \text{where} \quad \eta \equiv- \frac{d\,\ln{H_{,\phi}}}{dN}.
\end{equation}
The equation of motion for the slow-rolling scalar field reduces to $\dt{\phi} \simeq V'/3H$.
In the slow-roll approximation $\epsilon$, $|\eta| \ll 1$, the Hubble-slow-roll parameters $\epsilon$ and $\eta$ can be related to the potential-slow-roll parameters
\begin{equation}
\epsilon_v(\phi) \equiv  \frac{M_{\rm pl}^2}{2} \left( \frac{V_{,\phi}}{V} \right)^2, \qquad \eta_v(\phi) \equiv M_{\rm pl}^2 \frac{V_{,\phi\phi}}{V},
\end{equation}
through
\begin{equation}
\epsilon \simeq \epsilon_v, \qquad \eta \simeq \eta_v - \epsilon_v.
\end{equation}
The number of e-folds of accelerated expansion between the time when the mode $k$ exits the horizon at $t_k$ and the end of inflation at $t_e$ is
\begin{equation}
N_{k} = \int_{a_k}^{a_e} d \, \ln{a} =  \int_{t_k}^{t_e} H(t) \,dt = \int_{\phi_k}^{\phi_e} \frac{H}{\dt{\phi}} d\phi = \int_{\phi_k}^{\phi_e}  \frac{1}{\sqrt{2\epsilon}}\frac{d\phi}{M_{\rm pl}}.
\end{equation}

\paragraph{Primordial scalar perturbation:}
Inflation ends when $\epsilon(\phi_e) \approx 1$. However, quantum fluctuation of $\phi (t\,\vec{x})= \bar{\phi}(t) + \delta \phi(t\,\vec{x})$ will make local difference in time when inflation ends, which leads to difference in local density $\delta \rho(t\,\vec{x})$, and therefore, in the CMB temperature $\delta T(t\, \vec{x})$. The inflaton fluctuations are equivalent to the ones of an harmonic oscillator in de Sitter space. In the spatially flat gauge, one finds
\begin{equation}
\Delta_{\delta \phi}^2(k) \equiv \left<0| \delta \phi^2  |0\right> \simeq \left( \frac{H}{2\pi} \right)^2 \Big|_{k=aH}.
\end{equation}
In order to get ride of the gauge dependence, we usually introduce the gauge-invariant curvature perturbation
\begin{equation}
\Delta_{{R}}^2(k) \equiv \frac{H}{\dt{\phi}} \Delta_{\delta \phi}^2  = \frac{1}{2\epsilon} \frac{ \Delta_{\delta \phi}^2}{M_{\rm pl}^2},
\end{equation}
such that the slow-roll prediction reads
\begin{equation}
\Delta_{{R}}^2(k)  \simeq \frac{1}{8\pi^2}\frac{1}{\epsilon} \frac{H^2}{M_{\rm pl}^2}\Big|_{k=aH}.
\end{equation}
Since during inflation, $H$ is almost constant, the spectrum of primordial scalar perturbation is scale-invariant (power-law) with a slight tilt, usually expressed as
\begin{equation}
\Delta_{{R}}^2(k)  \equiv A_s \left( \frac{k}{k_*} \right)^{n_s-1},
\end{equation}
The Planck collaboration \cite{Aghanim:2018eyx} has measured the amplitude of the scalar spectrum $A_s$ and the tilt $n_s$ around the pivot scale $k_* = 0.002~\rm Mpc^{-1}$, with a good precision 
\begin{align}
&A_s = (2.101 \pm 0.034) \times 10^{-9}, \\
&n_s = 0.9649 \pm 0.0042.
\end{align}

\paragraph{Primordial Gravitational Waves}
\label{sec:Primordiale_spectrum_GW}
Independently of the inflaton dynamics, during the accelerated expansion, quantum fluctuations of the tensor modes $h_{ij}$, defined by
\begin{equation}
ds^2 = dt^2 - a^2(\delta_{ij}+ h_{ij}) dx^i dx^j,
\end{equation}
are also generated. The power spectrum of these Gravitational Waves is found to be
\begin{equation}
\Delta_{t}^2(k) \equiv \left<0| h_{\times}^2+h_{+}^2  |0\right> \simeq \frac{2}{\pi^2}\frac{H^2}{M_{\rm pl}^2},
\end{equation}
and again is expected to be scale-invariant and slightly tilted
\begin{equation}
\Delta_{t}^2(k)  \equiv A_t \left( \frac{k}{k_*} \right)^{n_t-1}.
\end{equation}
See \cite{Caprini:2018mtu} for a review.
The fundamental tensor modes are expected to source the so-called fundamental B-modes in the CMB. Their non-detection by neither by Planck 2018 or BICEP2/Keck sets an upper bound on the tensor-to-scalar ratio \cite{Ade:2018gkx, Akrami:2018odb}
\begin{equation}
\label{eq:scalar-to-tensor_bound}
r \equiv \frac{A_t}{A_s} \lesssim 0.055,\qquad (95\%~\rm C.L.)
\end{equation}
which can be recast as an upper bound on the inflation scale
\begin{equation}
H_{\rm inf} \lesssim 6 \times 10^{13}~{\rm GeV}.
\label{eq:max_Hinf_Bmodes}
\end{equation}
The future experiment LiteBird \cite{Matsumura:2013aja,Hazumi:2019lys}, planned to be launched within the next decade, is expected to have a sensitivity $r \lesssim 0.001$.

\paragraph{Slow-roll}
In the slow-roll approximation $\epsilon_v$, $|\eta_v| \ll 1$, the scalar and tensor spectra are given by \cite{Baumann:2009ds}
\begin{equation}
\Delta_{s}^2 \simeq \frac{1}{24\pi^2} \frac{V}{M_{\rm pl}^4} \frac{1}{\epsilon_v}\Big{|}_{k=aH}, \qquad \Delta_{t}^2 \simeq \frac{2}{3\pi^2} \frac{V}{M_{\rm pl}^4}\Big{|}_{k=aH}.
\end{equation}
The scalar and tensor spectral index are
\begin{equation}
n_s -1 \simeq 2\eta_v - 6\epsilon_v, \qquad n_t \simeq -2 \epsilon_v.
\end{equation}
The tensor-to-scalar ratio is 
\begin{equation}
r \simeq 16 \epsilon_v,
\end{equation}
which implies  the relation $r \simeq -8n_t$.

\paragraph{Inflation potentials}
Perhaps the simplest inflation model is $V(\phi)=\frac{1}{2}m^2\phi^2$. It belongs to the class of chaotic inflation models $V(\phi) \propto \phi^n$ introduced by Linde in 1983 \cite{Linde:1983gd}. It has slow-roll parameters \cite{Baumann:2009ds}
\begin{equation}
\epsilon_v =\eta_v = 2 \left(  \frac{M_{\rm pl}}{\phi}\right)^2.
\end{equation}
The slow-roll conditions $\epsilon, |\eta| \lesssim 1 $ require super-Planckian values for the inflaton
\begin{equation}
\phi > \sqrt{2}M_{\rm pl}.
\end{equation}
The possibility to generate inflation from a potential like $m^2\phi^2$ which is omnipresent in particle physics suggests that inflation is not only a peculiar phenomenon desirable for explaining the homogeneity and the flatness of the universe, but that \textit{it is also a natural and inevitable consequence of the chaotic initial conditions in the very early universe} \cite{Linde:1983gd}.
The number of inflation e-folding reads 
\begin{equation}
N(\phi) = \frac{\phi^2}{4M_{\rm pl}^2} - \frac{1}{2},
\end{equation}
and fluctuations in the CMB are created at
\begin{equation}
\phi_{\rm cmb} = 2 \sqrt{N_{\rm cmb}}M_{\rm pl} \sim 15 M_{\rm pl}.
\end{equation}
In order to agree with the observed scalar fluctuation
\begin{equation}
\Delta_{s}^2 = \frac{m^2}{M_{\rm pl}^2}\frac{N_{\rm cmb}^2}{3} \sim 10^{-9},
\end{equation}
we must fix $m \sim 10^{-6}M_{\rm pl}$.
The scalar spectral index is
\begin{equation}
n_s  = 1 + 2 \eta_v -6\epsilon_v = 1 - \frac{2}{N_{\rm cmb}} \simeq 0.96,
\end{equation}
and the tensor-to-scalar ratio is 
\begin{equation}
r = 16 \epsilon_v = \frac{8}{N_{\rm cmb}} \simeq 0.1.
\end{equation}
For other power-law potentials $V(\phi) \propto \phi^n$, it generalizes to \cite{Akrami:2018odb}
\begin{equation}
r  = \frac{16n}{4N_{\rm cmb}+n}.
\end{equation}
The bound on $r$ from Planck data in Eq.~\eqref{eq:scalar-to-tensor_bound}, excludes chaotic inflation models with $n \geq 2$. Other values $n=4/3$, $1$ and $2/3$, motivated by axion monodromy \cite{Silverstein:2008sg,McAllister:2008hb,McAllister:2014mpa} are compatible with Planck.

Instead of chaotic models, constraints from Planck in the $r-n_s$ plane \cite{Akrami:2018odb} favor a class of inflationary models with plateau-like inflaton potentials: natural (axionic) inflation \cite{Freese:1990rb,Adams:1992bn}, symmetry breaking with non-minimal (quadratic) coupling \cite{Salopek:1988qh,Fakir:1990eg,Bezrukov:2007ep,Kaiser:2013sna,Cheong:2021vdb}, hilltop \cite{Boubekeur:2005zm}, $R^2$ \cite{Starobinsky:1980te}, no-scale supergravity \cite{Ellis:2013xoa,Ellis:2013nxa}, $\alpha$-attractors \cite{Kallosh:2013hoa,Kallosh:2013lkr,Ferrara:2013rsa,Kallosh:2013maa,Kallosh:2013yoa,Kallosh:2013daa,Kallosh:2013tua}. For a review of the main inflation models and their constraints from Planck data, see \cite{Martin:2013tda,Martin:2013nzq,Akrami:2018odb}. 

It is worth to mention that in spite of the remarkable success of the inflation paradigm, the fact that Planck data favors plateau-like inflation models over `more natural' chaotic inflation models has raised \textbf{criticisms} \cite{Ijjas:2013vea,Guth:2013sya,Linde:2014nna, Ijjas:2014nta, Ijjas:2015hcc,Chowdhury:2019otk}. In addition, the inflation paradigm could be at odd with \textbf{string theory} considerations \cite{Obied:2018sgi,Agrawal:2018own,Bedroya:2019snp,Bedroya:2019tba}, which instead suggests that our universe could have started from a gas of strings \cite{Agrawal:2020xek}.
%

\section{Gravitational waves of primordial origin}
\label{sec:GW_cosmology}

Gravitational Waves (GW) have been predicted by Einstein in 1916 \cite{Einstein:1916cc,Einstein:1918btx}, one year after publishing his work about the relatistic theory of gravitation \cite{Einstein:1915ca,Einstein:1916vd}). GW have been detected indirectly the first time by Hulse and Taylor in 1974 \cite{Hulse:1974eb} and detected directly the first time by the LIGO collaboration on the 14 September of 2015 \cite{TheLIGOScientific:2016qqj}, one century after their prediction. The GW detected by LIGO were not of primordial origin but they have instead been produced $1.4$ billions years ago when two $\sim 30$ solar masses coalescing black holes merged into each other. Future experiments will have the sensitivity to probe potential GW backgrounds of primordial origin, and GW astronomy is opening a new era of exploration of high-energy physics in the early universe. 

As discussed in Sec.~\ref{sec:photon_decoupling}, the universe only becomes transparent to light when the photons decouple from the electron/proton plasma at the time of recombination $e^-+p^+ \to H+\gamma$. So we can not use the electromagnetic probe to observe the state of the universe when it is younger than $380~000$~yrs. However, GW decouple from the hot plasma much earlier. Indeed, assuming that the interaction rate of gravitons with SM is
\begin{equation}
\Gamma \sim G T^3,
\end{equation}
where $G = 8\pi/M_{\rm pl}^2$ is the Newton constant, we obtain that gravitons decouple around 
\begin{equation}
T \sim M_{\rm pl}.
\end{equation}
Therefore, GW produced in the early universe propagate freely in the universe until today. Such primordial GW carry information about very high-energy physics and probe particle physics at scales that will never be probed by colliders. Current and future GW experiments, LIGO/VIRGO/KAGRA \cite{Aasi:2014mqd}, LISA \cite{Audley:2017drz,LISACosmologyWorkingGroup:2022jok}, Einstein Telescope \cite{Hild:2010id, Punturo:2010zz}, Cosmic Explorer \cite{Evans:2016mbw}, DECIGO, BBO \cite{Yagi:2011wg}, AION/MAGIS \cite{Graham:2016plp,Graham:2017pmn,Badurina:2019hst}, AEDGE \cite{Bertoldi:2019tck}, constitute a new avenue of investigation in particles physics and cosmology. See \cite{Maggiore:2018sht} for a textbook and \cite{Caprini:2018mtu,Kuroyanagi:2018csn,LISACosmologyWorkingGroup:2022jok,Caldwell:2022qsj} for reviews on primordial GW. See \cite{Aggarwal:2020olq} for a review on the GW cosmological sources in the $MHz$ and $GHz$ range and the proposed experiments.

After having introduced GW as linearized solutions of Einstein equations in Sec.~\ref{sec:linerized_GW}, we derive the power of emitted GW by a macroscopic object in Sec.~\ref{sec:energy_GW}, and use it to compute the GW spectrum produced by different cosmological sources in Sec.~\ref{sec:GW_cosmo_signal}.

\subsection{Linearized wave solutions of Einstein equations}
\label{sec:linerized_GW}

Gravitational-waves (GW) are small \textbf{tensor} perturbation around the \textbf{flat-space} metric  \cite{Misner:1974qy,Hobson:2006se, Maggiore:1900zz} 
\begin{equation}
g_{\mu \nu} = \eta_{\mu\nu} + h_{\mu\nu}, \qquad \qquad \left|h_{\mu\nu}\right| \ll 1.
\label{eq:tensor_pert_flat}
\end{equation}
General relativity is invariant under \textbf{any} coordinate transformations 
\begin{equation}
x^\mu \rightarrow x^{'\mu}(x),
\end{equation}
where $x^{'\mu}(x)$ is an arbitrary \textbf{smooth} function of $x$.
In order to preserve the flat-space background in Eq.~\eqref{eq:tensor_pert_flat}, we only perform tiny gauge transformations
\begin{equation}
x^\mu \rightarrow x^{'\mu} = x^\mu + \xi^\mu(x).
\end{equation}
Using the transformation law of the metric
\begin{equation}
g_{\mu\nu}(x) \rightarrow g_{\mu\nu}^{'}(x')  = \frac{\partial x^\rho}{\partial x^{'\mu}}\frac{\partial x^\sigma}{\partial x^{'\nu}}\,g_{\rho \sigma}(x),
\end{equation}
we find the transformation of $h_{\mu\nu}$ at \textbf{linearized} order
\begin{equation}
h_{\mu\nu}(x) \rightarrow h_{\mu\nu}^{'}(x') = h_{\mu\nu}(x)  - (\partial_\mu \xi_\nu + \partial_\nu \xi_\mu).
\end{equation}
For convenient reasons, we introduce the variable
\begin{equation}
\bar{h}_{\mu\nu} = h_{\mu\nu} - \frac{1}{2} \eta_{\mu\nu} h,
\end{equation}
with $h = \eta^{\mu \nu} h_{\mu\nu}$, which transforms as
\begin{equation}
\bar{h}_{\mu\nu}(x) \rightarrow \bar{h}_{\mu\nu}^{'}(x') = \bar{h}_{\mu\nu}(x)  - (\partial_\mu \xi_\nu + \partial_\nu \xi_\mu - \eta_{\mu\nu} \partial_\rho \xi^\rho).
\label{eq:gauge_transf_hbar}
\end{equation}
The linearized equations of motion can be found after injecting Eq.~\eqref{eq:tensor_pert_flat} in the \textbf{Einstein equations} in Eq.~\eqref{eq:Einstein_eq_0}
\begin{equation}
\Box \bar{h}_{\mu\nu} + \eta_{\mu\nu} \partial^\rho \partial^\sigma \bar{h}_{\rho \sigma} - \partial^\rho \partial_\nu \bar{h}_{\mu\rho}  - \partial^\rho \partial_\mu \bar{h}_{\nu \rho} = - \frac{16\pi\,G}{c^4} T_{\mu\nu}.
\end{equation}
From imposing the \textbf{Lorentz gauge} (also called the Hilbert gauge, or the harmonic gauge, or the De Donder gauge)
\begin{equation}
\partial^\nu \bar{h}_{\mu\nu} = 0, \qquad \qquad (\rm Lorentz~gauge)
\label{eq:Lorentz_gauge}
\end{equation}
the equations of motion considerably simplify
\begin{equation}
\Box \bar{h}_{\mu\nu} = - \frac{16\pi\,G}{c^4} T_{\mu\nu}.
\label{eq:GW_wave_equation}
\end{equation}
By looking at Eq.~\eqref{eq:gauge_transf_hbar}, we  observe that even after imposing the Lorentz gauge in Eq.~\eqref{eq:Lorentz_gauge}, we can still perform coordinates transformations $x^\mu \rightarrow x^{'\mu} = x^\mu + \xi^\mu(x)$ with
\begin{equation}
\Box \xi_\mu = 0.
\label{eq:box_xi}
\end{equation}
The Lorentz gauge in Eq.~\eqref{eq:Lorentz_gauge} + the extra constraint in Eq.~\eqref{eq:box_xi} imposes $4+4$ constraints on the $10$ initial independent components of the symmetric $4\times 4$ matrix $g_{\mu\nu}$, which therefore reduce to only $2$ independents parameters. 
A convenient change of coordinates compatible with Eq.~\eqref{eq:box_xi} is the one which imposes the tensor perturbation $h_{\mu\nu}$ to be transverse and traceless
\begin{equation}
h^{0i} = 0, \qquad h^{i}_i = 0, \qquad \qquad (\rm TT~gauge)
\end{equation}
This defines the \textbf{transverse-traceless gauge} or TT gauge. 
Observe that $h^{0i} = 0$ together with the Lorentz gauge implies $\partial_0 h^{00}=0$, which for non-stationary perturbations (GW), reduces to $h^{00}=0$.
Hence we can summarized the Lorentz + TT gauge as follows
\begin{equation}
h^{0\mu} = 0, \qquad h^{i}_i = 0, \qquad \partial^j h_{ij} = 0, \qquad \qquad (\rm Lorentz~+~TT~gauge).
\label{eq:TT_gauge_Lorentz}
\end{equation}
In what follows, we call TT gauge the association of the Lorentz gauge + the TT gauge.
Note that only the spatial components are non-zero and $\bar{h}^{ij}_{\rm TT}=\bar{h}^{ij}$.
Starting from a tensor perturbation with spatial component $h^{ij}$ defined in an arbitrary gauge, we can construct the associated tensor $h^{ij}_{TT}$ in the TT gauge by performing the \textbf{projection}
\begin{equation}
h^{ij}_{\rm TT} = \Lambda_{ij,\,kl}(\hat{\mathbf{k}}) ~h^{kl},
\label{eq:h_TT_proj_def}
\end{equation}
where 
\begin{align}
&\Lambda_{ij,\,kl}(\hat{\mathbf{k}})\equiv P_k^i P_l^j ~ -~ \frac{1}{2} P^{ij} P_{kl} , \\
&P_{ij}(\hat{\mathbf{k}}) \equiv \delta_{ij} ~ - ~ \hat{k}_i \hat{k}_j.
\end{align}
$P_{ij} $ projects spatial tensor components onto the surface orthogonal to the spatial wavevector $k_i$ while $\Lambda_{ij,\,kl} $ subtracts the trace. For any spatial vector $v^i$, we have the relations $k_i \, P^i_j \,v^j = 0$ and $P^i_k \, P_j^k\, v^j = P^i_j\,v^j$, which we can use to check that $h^{ij}_{\rm TT}$  satisfies Eqs.~\eqref{eq:TT_gauge_Lorentz}.
For a plane wave $h^{ij} = A^{ij}\,e^{i\,(\omega t -k\,z)}$ propagating in the $z$ direction, the GW amplitude in the  TT gauge after applying the projection in Eq.~\eqref{eq:h_TT_proj_def}, reads
\begin{equation}
h^{\mu\nu}_{\rm TT} = 
\begin{pmatrix}
0 & 0 & 0 & 0\\
0 & h_+ & h_x & 0\\
0 & h_x & -h_+ & 0 \\
0 & 0 & 0 & 0
\end{pmatrix}\, \exp ~{i\,(\omega t -k\,z)},
\end{equation}
with $h_x = A^{12}$ and $h_+ = \frac{1}{2}(A^{11} - A^{22})$.

\subsection{Energy of gravitational-waves}
\label{sec:energy_GW}

\paragraph{Energy-momentum tensor of gravitational-waves: }

After having defined GW, we now discuss the definition of their energy-momentum tensor. Actually, the question does not make sense since there is no \textbf{local} notion of gravitational energy density in General Relativity. Indeed, we can always transform coordinates to a local \textbf{inertial} frame (a freely falling frame), at a point $p$ where
\begin{align}
&g_{\mu\nu} \big |_p = \eta_{\mu\nu}, \\
&\Gamma_{\mu\nu}^\rho \big|_p = 0.
\end{align}
However, we can show that GW act as a force between \textbf{two or more} nearby test masses. Hence, an energy-momentum tensor of GW can be defined after \textbf{averaging} over a space-time volume of size $L$ larger than the GW wavelength $L \gg \lambda$.
The energy-momentum sourced by GW $t_{\mu\nu}$ (the gravity of gravity) can be found by going to second order in perturbation
\begin{equation}
G^{(1)}_{\mu\nu} = -\frac{8\pi\,G}{c^4} \left( T_{\mu\nu} \,_+\,t_{\mu\nu} \right)
\end{equation}
with 
\begin{equation}
t_{\mu\nu}  \equiv  \frac{c^4}{8\pi\,G} \left< G^{(2)}_{\mu\nu} \right>,
\end{equation}
where $G^{(1)}$ and $G^{(2)}$ are the Einstein tensors at first and second order in $h^{ij}$, and $\left<\cdots\right>$ denotes the space-time average. Expansion at second order in the TT gauge yields  \cite{Misner:1974qy,Hobson:2006se, Maggiore:1900zz} 
\begin{equation}
t_{\mu\nu}  \equiv  \frac{c^4}{32\pi\,G} \left< \partial_\mu h^{\rm TT}_{ij}~\partial_\nu h_{\rm TT}^{ij}\right>.
\end{equation}

\paragraph{Energy flux:}

The outward-propagating GW carries away an energy flux 
\begin{align}
\frac{dE}{dA\,dt} &= +c\,t^{00},\\
&=\frac{c^3}{32\pi\,G} \left<  \dot{h}_{ij}^{\rm TT}~ \dot{h}_{ij}^{\rm TT}  \right>.
\label{eq:Energy_flux}
\end{align}

\paragraph{The quadrupole approximation}

The solution of the gravitational wave equation in Eq.~\eqref{eq:GW_wave_equation}, projected into the TT gauge as prescribed by Eq.~\eqref{eq:h_TT_proj_def}, reads
\begin{align}
\bar{h}^{\rm TT}_{ij}(t,\, \vec{x}) &= \frac{4G}{c^4}~ \Lambda_{ij,\,kl}(\hat{k})~ \int d^3 x^{'} \frac{1}{\left| \vec{x} - \vec{x}^{'}\right|}\, T^{kl} \left( t - \frac{\left| \vec{x} - \vec{x}^{'}\right|}{c},~\vec{x}^{'}  \right) \\
&= \frac{1}{r} \frac{4G}{c^4}~ \Lambda_{ij,\,kl}(\hat{k})~   \int d^3 x^{'} \, \left(T_{kl} (t- \frac{r}{c}, ~\vec{x}^{'}) ~ +~ \frac{  \vec{x}^{'}\cdot \hat{k}}{c} \partial_0 T_{kl}~+~\cdots  \right) 
\label{eq:solution_GW}
\end{align}
where we have assumed the distance to the source $r$ to be larger than the source typical size $r\gg d$, and the motion of the source to be non-relativistic $v\ll c$ (we have kept the first term in the multipole expansion), such that we could write $\left|  \vec{x} - \vec{x}^{'} \right| \simeq r - \vec{x}^{'}\cdot \hat{k}$, and perform an expansion in $\vec{x}^{'}\cdot \hat{k}/c$.
Now comes an interesting relation stemming from successive use of \textbf{integration by part} and \textbf{energy-momentum conservation} $\partial_\mu T^{\mu\nu}=0$
\begin{align}
\int d^3x ~T^{ij}(t,~\vec{x}) 
& = - \int d^3x ~\partial_k\,T^{kj}(t,~\vec{x})~x^i  \notag \\
& =  +  \int d^3x ~\partial_0\,T^{0j}(t,~\vec{x})~x^i \notag \\
& = - \frac{1}{2} \int d^3x ~\partial_0 \partial_k\,T^{0k}(t,~\vec{x})~x^i~x^j \notag \\
& = +\frac{1}{2}  \int d^3x ~\partial_0^2\,T^{00}(t,~\vec{x})~x^i~x^j \label{eq:quadrupole_first_apparition}
\end{align}
This allows to re-express the GW tensor component in Eq.~\eqref{eq:solution_GW} as a function of $T^{00}$ only
\begin{equation}
\bar{h}^{\rm TT}_{ij}(t,\, \vec{x})= \frac{1}{r} \frac{4G}{c^4}~ \Lambda_{ij,\,kl}(\hat{k})~   \int d^3 x^{'} \, \ddt{T}_{00} \left(t- \frac{r}{c}, ~\vec{x}^{'}  \right) .
\end{equation}
Therefore, the GW radiated power in Eq.~\eqref{eq:Energy_flux} at first-order in the multipole expansion (quadrupole) reads
\begin{align}
\left(\frac{dP}{d\Omega}\right)_{\rm quad} &= \frac{r^2\,c^3}{32\pi\,G} \left< \dot{h}_{ij}^{\rm TT}   \dot{h}_{ij}^{\rm TT}   \right> \\
&= \frac{G}{8\pi\,c^5} ~ \Lambda_{ij,\,kl}(\hat{k})~ \left< \dddt{Q}_{ij} ~ \dddt{Q}_{kl}  \right>,
\end{align}
where $Q_{ij}$ is the quadrupole moment of the macroscopic object radiating the GW
\begin{equation}
Q_{ij} \equiv \frac{1}{c^2} \int d^3x ~T^{00}(t,~\vec{x})~\left(x^i~x^j ~-~\frac{1}{3} r^2\,\delta^{ij}\right)
\end{equation}
which is the traceless version of Eq.~\eqref{eq:quadrupole_first_apparition}.
The angular dependence is only contained in  $\Lambda_{ij,\,kl}(\hat{k})$. Using that
\begin{equation}
\int d\Omega~\Lambda_{ij,\,kl}(\hat{k}) = \frac{2\pi}{15} \left( 11\delta_{ik\delta_{jl}} - 4\delta_{ij}\delta_{kl} + \delta_{il} \delta_{jk}    \right),
\end{equation}
we find the \textbf{total radiated power} in the \textbf{quadrupole approximation}
\begin{equation}
P_{\rm quad} = \frac{G}{5c^5}~ \left<  \dddt{Q}_{ij} ~ \dddt{Q}_{kl}    \right>.
\label{eq:quadrupole_formula_final}
\end{equation}

\subsection{Cosmological signals}
\label{sec:GW_cosmo_signal}

\paragraph{Stochastic Gravitational-Waves Background:}

We are motivated by the possibility for future experiments to observe GW of primordial origin. They can be generated during inflation, during preheating, during a first-order phase transition or by a network of cosmic strings.. See \cite{Maggiore:2018sht} for a textbook and \cite{Caprini:2018mtu} for a review on primordial GW. 
Due to causality, a GW signal generated in the early universe must have a correlation length $\Delta t_c$ smaller than the Hubble horizon at the time of production
\begin{equation}
\Delta t_c \lesssim H_*^{-1}.
\end{equation}
The number of causal sources which we observe today can be estimated as the area of the sphere of events at time of GW production $t_*$ divided by the area of an individual causal patch
\begin{equation}
\label{eq:nbr_causal_sources}
\#_{\rm causal~sources} \sim \frac{4\pi}{\pi}\left(\frac{\int_{t_{*}}^{t_{\rm today}} \frac{dt}{a(t)}}{\int_{0}^{t_{*}} \frac{dt}{a(t)}}\right)^2  \sim 10^{29} ~\left( \frac{g_*(T_*)}{106.75}\right)^{\!1/2}
\left( \frac{T}{1~\rm TeV}\right)^2,
\end{equation}
where we computed $a(t)$ from integrating the Friedmann equation in Eq.~\eqref{eq:Friedmann_equation} with the energy content measured by Planck, cf.  Sec.~\ref{sec:content_U} and from assuming a radiation-dominated universe before BBN. Therefore, a GW signal of primordial origin can only be observed today as a superposition of GW generated by an enormous number of causally independent sources. Individual sources can not be resolved but instead we can only observe a \textbf{Stochastic Gravitational-Wave Background (SGWB)}. For most of the cosmological sources, the SGWB is homogeneous, isotropic, gaussian and unpolarized such that we can write \cite{Caprini:2018mtu}
\begin{equation}
\left< h_r(\mathbf{k},\, t )h_{p}^*(\mathbf{q},\, t) \right> = \frac{8\pi^5}{k^3} \delta^{(3)}(\mathbf{k}-\mathbf{q})\, \delta_{rp}\, h_c^2(k,\, t).
\end{equation}
We now use the quadrupole formula in Eq.~\eqref{eq:quadrupole_formula_final} in order to compute the SGWB generated by given cosmological sources in the early universe.

\paragraph{Model-independent:}

The contribution to the total energy density of the universe today from GW generated by sources emitting during $\Delta t$ in the early universe reads
\begin{equation}
\label{eq:GWabundance_with_respect_photons}
\Omega_{\rm GW} = \Omega_{\gamma} ~ \frac{P_{\rm GW} ~ \Delta t ~ H^3}{\rho_{\rm tot}},
\end{equation}
where $\Omega_{\gamma}\simeq 4.2\times 10^{-5}$ is the photon abundance today \cite{Tanabashi:2018oca}, $P_{\rm GW}$ is the radiated GW power per Hubble volume, $H$ and $\rho_{\rm tot}$ are the Hubble scale and total energy density at the time of emission.  We have used that the GW energy density $\rho_{\rm GW}$ redshifts as fast as the photon energy density $\rho_{\gamma}$.  We have assumed GW emission to take place during radiation domination and we have assumed an adiabatic evolution. 
If the source is made of $N$ macroscopic objects per Hubble volume, the power of GW emission can be estimated from the quadrupole formula in Eq.~\eqref{eq:quadrupole_formula_final}
\begin{equation}
P_{\rm GW} = N~G~\dddt{Q}^2, \qquad \text{with}~ \dddt{Q} = E~L^2/\tau^3,
\end{equation}
where $E$, $L$ and $\tau$ are the characteristic energy, length and time scale of the dynamical objects.
This can be recast in the simple form
\begin{equation}
\Omega_{\rm GW} = \Omega_{\gamma} ~ N ~ \left(  \frac{\rho_s}{\rho_{\rm tot}} \right)^2 ~ \left(  L/\tau \right)^6 ~ \left(\Delta t~H\right)~ \left( L~H\right)^4.
\label{eq:SGWB_model_indpt_quadrupole}
\end{equation}
where we have introduced the energy density of the source $\rho_s \equiv E/L^3$ and made use of the Friedmann equation $H^2=\rho_{\rm tot}/3M_{\rm pl}^2$.

\paragraph{First-order phase transition:}
Thanks to the universal formula in Eq.~\eqref{eq:SGWB_model_indpt_quadrupole}, we can recover some of the well-known SGWB of cosmological origin. For instance, the SGWB generated by \textbf{bubbles collision} (scalar field contribution) or \textbf{sound shell overlap} (sound wave contribution) during a first-order phase transition, see \cite{Grojean:2006bp, Caprini:2015zlo,Caprini:2019egz} and Chap.~\ref{chap:1stOPT}. The time scale of the bubble collision / sound shell overlap is given by $\tau \sim \beta^{-1}$ where $\beta$ is the \textbf{time derivative of the tunneling rate}.  The size of the bubbles/sound shells when they collide/overlap is given by $L\sim v_w/\beta^{-1}$ where $v_w$ is the speed of the wall. The number of bubbles/sound shells per Hubble volume is $N \sim (\beta/v_w \, H)^3$. The fraction of kinetic energy density of the source at the time of the phase transition can be expressed in terms of the usual parameters $\alpha$ and $\kappa$
\begin{equation}
\frac{\rho_s\,L^2/\tau^2}{\rho_{\rm tot}}~ = ~ \frac{\kappa ~\alpha}{1+\alpha},
\end{equation}
where $\alpha \equiv \Delta V/\rho_{\rm tot}$ is the vacuum energy (\textbf{latent heat}) fraction and $\kappa = (\rho_s \,L^2/\tau^2)\,/\,\Delta V$ is the \textbf{efficiency} of the energy transfer from vacuum energy to the GW source. The later is either the scalar field kinetic energy (scalar field contribution) or the fluid kinetic energy (sound wave contribution). Applying Eq.~\eqref{eq:SGWB_model_indpt_quadrupole} to the case of 1stOPT gives
\begin{equation}
\label{eq:1stOPT_GW_quadrupole}
\Omega_{\rm GW}  \sim \Omega_{\gamma} ~\left(  \frac{\kappa~\alpha}{1+\alpha} \right)^2  ~v_w^3 ~ \left( \frac{H}{\beta} \right)^2~\left( \Delta t ~\beta  \right),
\end{equation}
where the duration of the emission is usually \textbf{short-lived} $ \Delta t \sim \beta^{-1}$, except for sound-wave contributions of weak phase transitions (low $\alpha$) which can be long-lived $\delta t \sim H^{-1}$ \cite{Caprini:2009yp, Ellis:2020awk}. 
The GW frequency today reads
\begin{equation}
f_0 = \frac{a_*}{a_0} f_* = 1.65 \times 10^{-5}~{\rm Hz}~\left(\frac{T_{*}}{100~\rm GeV}\right) \left( \frac{g_{\rm eff, \,*}}{100} \right)^{1/6} \frac{f_*}{H_*},
\label{eq:GW_frequency_emission_today_1stOPT_intro}
\end{equation}
where the frequency at emission $f_*$ is given by the inverse size of the bubbles/sound shells when they collide/overlap
\begin{equation}
f_* \sim \beta/v_w.
\end{equation}
We have assumed instantaneous reheating after the time of GW emission $t_*$, see Eq.~\eqref{eq:prefactor_Omega_GW} and Eq.~\eqref{eq:GW_frequency_emission_today_1stOPT} in Chap.~\ref{chap:1stOPT} for the effect of the equation of state during reheating on the SGWB from 1stOPT.
For $\beta/H = 100$ and $T_* = 100 $~GeV, Eq.~\eqref{eq:GW_frequency_emission_today_1stOPT_intro} becomes $f_0 \simeq 1.65~\rm mHz$ which corresponds to the sensitivity window of the future LISA\footnote{After an observation time of $T=10~$years, LISA is expected to be able to detect SGWB with abundance $\Omega_{\rm GW} = 10^{-11}$, with signal-to-noise ratio $S=10$, in the range $48~\mu\textrm{Hz} \lesssim f_0 \lesssim 38~\rm mHz$, with a peak sensitiviy around $3.2~\rm mHz$, see App.~\ref{app:sensitivity_curves}.} experiment \cite{Grojean:2006bp, Caprini:2015zlo,Caprini:2019egz,LISACosmologyWorkingGroup:2022jok}. Hence, LISA is an observational window on the electroweak scale for which the existence of new physics can be motivated by the hierarchy problem, cf. Sec.~\ref{sec:hierarchy_pb} in Chap.~\ref{chap:SM_particle}.


\paragraph{Inflation:}
The \textbf{primordial} SGWB generated during inflation, cf. Sec.~\ref{sec:Primordiale_spectrum_GW} and review \cite{Caprini:2018mtu}, is not generated by macroscopic objects and the formula Eq.~\eqref{eq:SGWB_model_indpt_quadrupole} should not by used. However, just as a \textbf{curiosity}, we can check that the correct expression for the inflationary SGWB is recovered with Eq.~\eqref{eq:SGWB_model_indpt_quadrupole}, if we assume that GW are generated by $N=1$ quantum mode of Hubble size $L \sim H^{-1}$ with Hubble dynamical time scale $\tau\sim H^{-1}$, Hubble energy $E\sim H$, and emitting during one Hubble time $\Delta t\sim H^{-1}$. Indeed, we get
\begin{equation}
\Omega_{\rm GW} \sim \Omega_{\gamma}~\left( \frac{H_{\rm inf}}{M_{\rm pl}} \right)^4,
\label{eq:SGWB_inflation_quadrupole}
\end{equation}
which coincides with the well-known result \cite{Caprini:2018mtu}.
The most precise probe of GW generated during inflation is the search for $B$-modes patterns in the CMB.
As already mentioned in Eq.~\eqref{eq:max_Hinf_Bmodes}, the non-detection of primordial $B$-modes by neither Planck 2018 nor BICEP2/Keck implies $H_{\rm inf} \lesssim 6 \times 10^{13}~{\rm GeV}$ \cite{Ade:2018gkx, Akrami:2018odb}.

\paragraph{Cosmic String (global):}
The last major potential source of primordial SGWB are Cosmic Strings (CS). They are topological defects formed after spontaneous breaking of a $U(1)$ symmetry. CS can be either \textbf{global} or \textbf{local} according to whether the $U(1)$ symmetry is accompanied of a gauge boson or not. We devote Chap.~\ref{chap:cosmic_strings} to the precise computation of the SGWB in the presence of a network of cosmic strings and App.~\ref{app:global_strings} for the case of global strings specifically. For global strings, the dominant GW emission results from the oscillations of $N= 1$ \textbf{Hubble-sized loops}, $L\sim \tau \sim H^{-1}$.  Global loops are \textbf{short-lived}, due to the efficient \textbf{Goldstone bosons} emission. More precisely, the global loop lifetime is
\begin{equation}
\delta t \sim H^{-1}~ {\rm log}\left( \eta/H\right),
\end{equation}
where the log term results from the IR divergence of the string tension due to the \textbf{long-range force} mediated by the massless mode and is of order $10^2$. $\eta$ is the typical energy scale in the string (scalar field VEV). The loop energy density goes like 
\begin{equation}
\rho_s \sim \frac{\eta^2 ~ {\rm log}\left( \eta/H\right)}{H^{-2}},
\end{equation}
where $\eta$ is the radial mode VEV. Plugging the previous relations in Eq.~\eqref{eq:SGWB_model_indpt_quadrupole} yields
\begin{equation}
\Omega_{\rm GW} \sim \Omega_{\gamma}~\left( \frac{\eta}{M_{\rm pl}} \right)^4~{\rm log}^3\left( \eta/H\right).
\label{eq:SGWB_quadrupole_global_string}
\end{equation}
See App.~\ref{app:global_strings} for a more precise derivation. Hence, the SGWB from global string receives a $\rm log^3 \simeq 10^6$ enhancement factor with respect to the inflationary spectrum, which bring the spectrum in the LISA band if $\eta > 5\times 10^{14}$~GeV and the SKA band if $\eta > 10^{14}$~GeV. On the other hand, $\eta$ receives an upper bound from the condition that the network should form after inflation and that the non-detection of B-modes translates to $V_{\rm inf} \lesssim 10^{16}~$GeV \cite{Ade:2018gkx, Akrami:2018odb}. Hence, the string scale which will be probed by LISA and SKA corresponds to the interval $10^{14}$~GeV~$\lesssim \eta \lesssim$~ $10^{16}~$GeV. However, in order to not overclose the universe the goldstone mass should be $m_a \lesssim 10^{-18}~\rm eV$ \cite{Gorghetto:2021fsn}.

\paragraph{Cosmic String (local):}

In the case where CS arise from the spontaneous breaking of a local symmetry, there is no massless mode and the strings are \textbf{long-lived}: if formed at time $t_i$, they decay much later, at time $\tilde{t} = t_i/G\mu$ where $\mu \sim \eta^2$ is the string tension. Hence, the dominant population of loops emitting at time $\tilde{t}$ were produced at a much earlier time, $t_i = G\mu~\tilde{t}$, corresponding to one loop-lifetime ago.\footnote{The loop-formation rate scales as $dn/dt \propto 1/t^4$. So the earlier the formation time, the larger the number of loops and the resulting GW signal. This explains why the dominant population of loops emitting at time $\tilde{t}$ are the loops produced the earliest as possible, which corresponds to $t_i = G\mu~\tilde{t}$.} Hence, the SGWB from local strings can be derived from the global case in Eq.~\eqref{eq:SGWB_quadrupole_global_string}, after replacing $N=1$ by $N=(G\mu)^{-3/2}$ and removing the log divergence
\begin{equation}
\Omega_{\rm GW} \sim \Omega_{\gamma}~\frac{\eta}{M_{\rm pl}} 
\label{eq:SGWB_quadrupole_local_string}
\end{equation}
The last expression being independent of the GW frequency at emission $f \sim L^{-1}$, the SWGB from CS is \textbf{flat} in frequency. This is not the case anymore in the presence of a non-standard cosmology. See App.~\ref{app:derivationGWspectrum} for a thorough derivation of the SGWB from CS, in standard and non-standard cosmology.

We conclude the section by emphasizing that searches for primordial GW with current and future experiments offer a fantastic new field of exploration of particle physics and cosmology.


\section{Open problems}

The cosmological model based on a flat universe containing today $69\%$ of Dark Energy, $26\%$ of Dark Matter, $5\%$ baryons, a sub-leading neutrino background, photons at $2.72~\rm K$ and inflationary initial conditions is called the \textbf{$\Lambda$CDM Standard Model of cosmology}.  Thanks to the impressive work of many collaborations of experimentalists, in the last decade, the level of precision at which the parameters of the $\Lambda$CDM model are measured, has been considerably improved. However, big questions remain unanswered. What is the nature of the Dark Energy ? The nature of Dark Matter ? What generates the asymmetry between matter and anti-matter ? 
None of these observations can be consistently justified theoretically within the Standard Models of Cosmology and Elementary particles.
In this section, we give short descriptions about these open questions. We also present the astrophysical small scale problems and the more recent anomalies in Hubble measurement and $21$ cm signal. 

Another open problem in cosmology (or astrophysics) which we only discuss briefly now is the \textbf{Lithium problem}: the theoretical prediction of ${}^{7}$Li is a factor 3 above its determination in the atmosphere of metal-poor halo stars \cite{Lind:2013iza}. Unfortunately, CMB data can not allow to determine the ${}^{7}$Li abundance \cite{Switzer:2005nd}. The publication in 1999 of the NACRE compilation of nuclear reactions \cite{Angulo:1999zz} (previous compilations were \cite{Fowler:1967ty,Fowler:1975kz}) allowed the precise determination of the abundance of light elements \cite{Cyburt:2001pp,Coc:2003ce,Cuoco:2003cu,Descouvemont:2004cw,Cyburt:2004cq, Cyburt:2008kw,Coc:2011az}. In 2013, an update compilation of nuclear rates  \cite{Xu:2013fha} but also the release of Planck data \cite{Planck:2013pxb} allowed to improve the precision of BBN predictions \cite{Coc:2014oia,Cyburt:2015mya,Pitrou:2018cgg,Fields:2019pfx} and to confirm the ${}^{7}$Li problem to an unprecedented level. Proposed solutions are modification of nuclear rates with resonances \cite{Cyburt:2009cf,Boyd:2010kj, Chakraborty:2010zj,Broggini:2012rk}, decay of an exotic particle \cite{Pospelov:2006sc,Jedamzik:2009uy,Cyburt:2010vz,Cyburt:2012kp,Cyburt:2013fda,Poulin:2015woa,Goudelis:2015wpa,Depta:2020zbh}, stellar depletion \cite{Vauclair:1998it, Pinsonneault:1998nf,Pinsonneault:2001ub,Richard:2004pj, Korn:2006tv}, lithium diffusion in the post-recombination universe \cite{Pospelov:2012pa, Kusakabe:2014dta}, or variation of coupling constants \cite{Coc:2006sx,Berengut:2009js,Coc:2012xk}. For reviews on the ${}^{7}$Li problem, see \cite{Iocco:2008va, Pospelov:2010hj, Spite:2012us, Iocco:2012vg,Mathews:2019hbi}.

Another open problem which we briefly discuss now is the \textbf{origin of intergalactic magnetic fields}. Indeed, the non-observation by Fermi telescope of secondary $\gamma-$rays collinear to blazar-produced $\gamma-$rays suggest the presence of intergalactic magnetic fields with strength $B\gtrsim 3\times 10^{-16}$~G \cite{Neronov:1900zz}.  Interestingly, the upcoming Cherenkov Telescope Array (CTA) may push the lower bound to $B\gtrsim 10^{-12}$~G \cite{Korochkin:2020pvg}, closer to the CMB upper bound at $B\lesssim 10^{-9}$~G  \cite{Ade:2015cva} (or even $B\lesssim 10^{-11}$~G \cite{Jedamzik:2018itu}). Primordial magnetic fields can be generated by particular inflation models or by a thermal first-order phase transition \cite{Durrer:2013pga, Subramanian:2015lua, Neronov:2020qrl}. See \cite{Grasso:2000wj,Widrow:2002ud,Vallee:2004osq,Durrer:2013pga,Subramanian:2015lua,Vachaspati:2020blt} for reviews on cosmological magnetic fields.

\subsection{Cosmological constant problem}
\label{sec:CC_pb}

\paragraph{The measurement of the Dark Energy :}
Dark energy is a fluid with equation of state $\rho_\Lambda = - p_{\Lambda} =  \rm constant$. Its precise abundance, $\Omega_\Lambda = 68.5 \pm 0.7 \%$ of the total energy density of the universe, is inferred from two independent observations at low-redshift (when mixed with Planck measurements of the CMB)
\begin{itemize}
\item
\textbf{Type Ia Supernova (SNe Ia)}:  they are produced by the thermonuclear explosion of white dwarfs. Such an explosion arises when a white dwarf accretes gas from a companion star and reaches it Chandrasekhar limit around a mass of $1.44~M_\odot$. Above that mass the electron degeneracy pressure in the core is insufficient to balance the gravitational self-attraction and the star collapses. The supernova explosion results from the outward-going shock wave generated by the bounce of the inner core. Since all SNe Ia involve a white dwarf with a well-defined mass $1.44~M_\odot$, they have the same absolute luminosity and are qualified of \textbf{standard candles} \cite{Phillips:1993ng}. The measure of the luminosity-distance versus redshift allows to measure the expansion rate $H(z)$ and infer $\Omega_\Lambda$.
\item
\textbf{Baryonic acoustic oscillations (BAO)}: As a consequence of the radiation pressure, an acoustic wave expands in the baryon-photon fluid from the moment of the Big-Bang until the time $t_{\rm dec}$ when the photons decouple. After the photons propagate freely, the sound horizon at decoupling stays imprinted in the baryon distribution and, after the non-linear evolution of the matter perturbations, at much later time, it appears as a little bump in the galaxy correlation function. The \textbf{sound horizon at recombination}, also known as the \textbf{BAO} scale, is well measured by Planck from the position of the first peak of the power spectrum of the temperature anisotropies in the CMB, cf. Fig.~\ref{fig:CMB_Planck2018}. The measurements of this \textbf{standard ruler} at different redshifts, using either galaxy redshift surveys or Lyman-$\alpha$ forests, allow to obtain the expansion parameter $H(z)$ and to measure $\Omega_\Lambda$.
\end{itemize}

\paragraph{The Cosmological Constant:}
The measurement of the presence of a fluid with equation of state $\omega = -1$ in the universe corresponds to add a constant $\Lambda$ to the Einstein equations,
\begin{equation}
\label{eq:Einstein_eq_2}
G_{\mu\nu} = \frac{1}{M_{\rm pl}^2}\left( T_{\mu\nu} + \Lambda \, g_{\mu\nu} \right),
\end{equation}
where $g_{\mu\nu}$ is the FLRW metric. This is the \textbf{Cosmological Constant (CC)}, first introduced by Einstein in 1917 as a mistake. From its appearance in Eq.~\eqref{eq:Einstein_eq_2}, the CC coincides with the \textbf{vacuum energy} of our universe.
The measured value of the constant in Planck unit is extremely small
\begin{equation}
\Lambda (= \rho_\Lambda = -p_{\Lambda}) = 3M_{\rm pl}^2 H_0^2\,\Omega_{\Lambda} \simeq 5\times 10^{-121}~M_{\rm pl}^4 \simeq \left( 1~\rm meV \right)^{1/4}.
\end{equation}
The reason why it constitutes $69\%$ of the energy density of the universe in spite of being such a small number is that our universe is mainly made of vacuum.

\paragraph{The SM prediction :}
In the SM of elementary particles, the vacuum energy comes from two contributions.
\begin{itemize}
\item
\textbf{Zero-point energy:}  In Quantum Field Theory, particles are quantum excitations of continuous fields that fill the whole space-time. A characteristic feature of quantum fields is that they constantly produce particles/anti-particles pairs which annihilate after a time $\lesssim \hbar/m$, in accordance with Heisenberg uncertainty principle. From the point of view of Feynman rules, the contribution of quantum fields to the vacuum energy are given  by bubble diagrams.
\item
\textbf{Spontaneous symmetry breaking:} After each phase transition in the universe leading to spontaneous breaking of a symmetry (chiral symmetry when QCD confines, EW symmetry, supersymmetry, Peccei-Quinn symmetry, grand unified symmetry,...), a field operator gets a vev and the vacuum energy gets shifted.
\end{itemize}
Hence, the theoretical prediction of the vacuum energy is\footnote{Note that the naive estimation of the vacuum energy based on introducing a UV cut-off $M$ on the momentum and leading to $\left<\rho\right> = \frac{1}{2}\int_{\left|\vec{k}\right| <M} \frac{d^3\vec{k}}{(2\pi)^3}\, \omega(\vec{k}) \simeq \frac{M^4}{16\pi^2}$, violates Lorentz invariance and does not lead to the correct equation of state $\left<\rho\right>=-\left<p\right>$. Instead, one should use dimensional regularization. An interesting consequence is that the massless fields do not contribute to the vacuum energy. See \cite{Martin:2012bt} for more details.} \cite{Martin:2012bt}
\begin{equation}
\Lambda = \underbrace{\Lambda_{\rm bare}}_{\rm bare~ CC} + \underbrace{\sum_i n_i \frac{m_i^4}{64 \pi^2} \log \frac{m_i^2}{\mu^2}}_{\rm zero-point~ fluctuations} + \rho_{\Lambda}^{\rm QCD}+ \rho_{\Lambda}^{\rm EW} + \underbrace{\cdots}_{\rm other~ possible~ phase~ transitions},
\end{equation}
where $n_i=(-1)^{2s_i}g_i$ where $s_i$ and $g_i$ are respectively the spin and number of degrees of freedom of the particle $i$. $\mu$ is the renormalization scale after dimensional regularization. The sum operates over all the fields in the universe. $\Lambda_{\rm bare}$ is the CC already present in the Lagrangian in the absence of the SM. Note that in the SM the negative contribution from the top/anti-top fluctuations largely dominates over the others. The vacuum energy of QCD is fixed by the gluon correlator \cite{Narison:2011xe}, whereas the vacuum energy shift after EWPT is found from injecting Eq.~\eqref{eq:VEV_Higgs} into Eq.~\eqref{eq:line3}, $\rho_{\Lambda}^{\rm EW} = m_{h}^2\,v^2/8$. Hence, the prediction of the vacuum energy in the SM is
\begin{equation}
\Lambda = \underbrace{\Lambda_{\rm bare}}_{\rm bare~ CC} + \underbrace{-(120~\rm GeV)^4}_{\rm zero-point~ fluctuations}+ \underbrace{(300~\rm MeV)^4}_{\rho_{\Lambda}^{\rm QCD}}+ \underbrace{(105~\rm GeV)^4}_{\rho_{\Lambda}^{\rm EW}}    +\underbrace{\cdots}_{\rm other~ possible~ phase~ transitions}.
\end{equation}
We can see that in order to compensate for the SM contribution  $\rho_{\Lambda}^{\rm SM} =-(95~\rm GeV)^4$, we need to finely tune the bare constant $\rho_{\Lambda}^{\rm bare}$ at the level of $10^{-56}$. Moreover, new physics at larger scales, e.g. GUT or Planck scale, could also contribute to the CC such that the needed fine-tuning could become even worst, e.g. $10^{-120}$. This extreme fine-tuning is the \textbf{Cosmological Constant problem}. In addition, the CC appears to have started dominating the energy density of the universe only recently, around the cosmic age $\sim 9~$Gyrs. This is the \textbf{Coincidence problem} (there is even a triple coincidence $\rho_\Lambda \sim \rho_{\rm mat} \sim \rho_{\rm rad}$ \cite{ArkaniHamed:2000tc}). Finally, another curious fact is the relation $\Lambda \sim {\rm TeV^2}/M_{\rm pl}$, which is reminiscent of the \textbf{WIMP miracle}, which connects the measured Dark Matter abundance to the hierarchy problem, see Sec.~\ref{par:WIMP_miracle}.

\paragraph{Weinberg no-go theorem: }
\hspace{0.5 cm}\say{ \textit{The original symmetry of general covariance, which is always broken by the appearance of any given metric, cannot, without fine-tuning, be broken in such a way as to preserve the subgroup of space-time translations. }}\hspace{0.8 cm} Weinberg (1988 \cite{Weinberg:1988cp}) 

In other words, a given universe with space-time metric $g_{\mu \nu}$, containing an adjusting scalar field $\phi$, is \textbf{invariant under translations} (or under the whole Poincaré symmetry group) if $g_{\mu \nu}$ and $\phi$ are homogeneous and constant, which implies
\begin{enumerate}
\item
$\frac{\partial \mathcal{L}}{\partial \phi} \approx 0 \quad \rightarrow \quad \frac{\partial V}{\partial \phi} \approx 0$: \hspace{0.5 cm}the adjusting field reaches a vev.
\item
$\frac{\partial \mathcal{L}}{\partial g_{\mu\nu}} \approx 0 \quad \xrightarrow{\mathcal{L} \propto \sqrt{-g}\,V(\phi)} \quad V(\phi) \approx 0$: \hspace{0.5 cm} the CC must vanish.
\end{enumerate}
Any dynamical adjustment of the CC to a vanishing value, using the vev of a scalar field $\phi$, needs to satisfy these two last conditions. Since the two conditions are \textbf{independent}, then their simultaneous enforcement requires \textbf{fine-tuning}. This is the \textbf{Weinberg no-go theorem} (1988 \cite{Weinberg:1988cp}). A possible resolution would be the existence of a symmetry such that the two conditions are not independent, but instead, they are related through
\begin{equation}
g_{\mu\nu} \frac{\partial \mathcal{L}}{\partial g_{\mu\nu}} = \kappa \frac{\partial \mathcal{L}}{\partial \phi}, 
\end{equation}
where $\kappa$ is some parameter. We can check, using $\delta \sqrt{-g} = - \frac{1}{2}\sqrt{-g}g^{\alpha \beta}\delta g_{\alpha \beta}$, that this implies
\begin{equation}
\mathcal{L} = \sqrt{-g}\,e^{-2 \kappa \phi} \mathcal{L}_0.
\end{equation} 
Then, the condition of Poincaré invariance can be fulfilled with a \textbf{unique} condition: $\phi \to  \infty$. However, all masses in the Lagrangian, scale as $e^{-\phi}$, and would vanish concertedly with the CC. This corresponds to a scale-invariant universe which is not ours. Any proposed solution to the CC problem much find a may to evade the Weinberg no-go theorem. Either by cutting-off gravity at large scale, adding extra-dimensions, allowing the scalar field to vary in time, or introducing many scalar minima. See \cite{Weinberg:1988cp,Carroll:2000fy,Straumann:2002tv,Copeland:2006wr, Nobbenhuis:2006yf, Bousso:2007gp,Bousso:2012dk, Martin:2012bt,Sola:2013gha,Burgess:2013ara,Padilla:2015aaa,Brax:2017idh} for some reviews on the Cosmological Constant problem and its attempted solutions.\footnote{I thank all the participants of the \href{https://indico.desy.de/indico/category/727/}{\textbf{Workshop Seminar Series on Vacuum Energy}} which has taken place in the DESY theory group between 28/10/2017 and 13/02/2018. An upcoming review written by the participants of the series is in preparation.} We discuss some of them below.

\paragraph{Degravitate the CC in massive gravity: }
The idea is to modify General Relativity at scales larger than $L \gtrsim  1/H_0$ such that the Newton constant $G_{\rm N}$ behaves as a \textbf{high-pass filter} \cite{ArkaniHamed:2002fu, Dvali:2007kt}
\begin{equation}
\label{eq:degravitation}
G_{\rm N}^{-1}G_{\mu \nu} = 8\pi T_{\mu \nu} \quad \rightarrow \quad G_{\rm N}^{-1}\left[ 1 + \left(\frac{m^{2}}{-\Box}\right)^\kappa \right]G_{\mu \nu} = 8\pi T_{\mu \nu},
\end{equation}
where $\frac{1}{2}<\kappa \lesssim 1$. Particularly, $\kappa = 1$ corresponds to massive gravity \cite{deRham:2010tw} whereas $\kappa < 1$ corresponds to having a continuum of massive gravitons.
As a consequence, gravity is \textbf{screened} at large scales and the cosmology becomes insensitive to the vacuum energy. A prediction is a mass for the graviton around $H_0 \sim 10^{-33}~\rm eV$. These effective field theories can find UV completions in higher dimensional models \cite{deRham:2007rw}, see next paragraph. See \cite{Hinterbichler:2011tt, Clifton:2011jh, Joyce:2014kja} for reviews on modified gravity.

\paragraph{Self-tune the CC in 6D Brane-World: }
The claim is that our world could be localized on a (3+1)-brane in a $\gtrsim 6$D space-time. The extra dimensions either have \textbf{infinite-volume} \cite{Dvali:2000xg, Dvali:2002pe, Niedermann:2014bqa} (inspired from DGP model \cite{Dvali:2000hr}, they can be mapped to the case $\kappa \approx 1$ in Eq.~\eqref{eq:degravitation}  \cite{deRham:2007rw}),  or \textbf{finite-volume} and \textbf{supersymmetric} \cite{Aghababaie:2003wz,Burgess:2004ib,Burgess:2013ara,Niedermann:2015via} (sub-milimeter, inspired from ADD model \cite{ArkaniHamed:1998rs}). In those scenarios, the observed CC corresponds to the \textbf{brane tension}. With \textbf{more than two} extra-dimensions, interestingly the brane tension only warps the extra-dimensions but not the brane itself, similarly to a long cosmic string in 4D, which generates a deficit angle instead of curving itself under its tension. Therefore, an observer on the brane sees a flat curvature.
See \cite{Maartens:2010ar,Clifton:2011jh} for reviews on brane-world gravity, and particularly \cite{Niedermann:2016wyv} for recent applications of 6D brane models to the CC problem.

\paragraph{Self-adjust the CC with a scalar field: }
The Weinberg no-go theorem assumes Poincaré invariance, not only at the level of the spacetime metric but also at the level of the self-adjusting scalar field. Hence, it may be interesting to drop the last assumption by allowing the scalar field to \textbf{vary in time}. We can even ask what is the most general covariant theory of gravity giving a viable self-adjusting mechanism?  In order to address this question, in 2011, the authors of Ref. \cite{Charmousis:2011bf} have passed \textbf{Horndeski}'s theory of gravity, the most general covariant theory of gravity with second order equation of motion (first introduced in 1974 in \cite{Horndeski:1974wa} and revisited in \cite{Deffayet:2011gz}), through a \textbf{self-tuning filter}. They have obtained \textbf{Fab Four}, built on four operators, Einstein-Hilbert (George), Gauss-Bonnet (Ringo), double Hodge-dualized Riemann (Paul) and kinetic `braiding' term (John)
\begin{align}
&\mathcal{L}_{\rm George} = V_{\rm George}(\phi) ~R, \\
&\mathcal{L}_{\rm Ringo} = V_{\rm Ringo}(\phi)  \left(R_{\mu \nu \alpha \beta} R^{\mu \nu \alpha \beta} - 4 R_{\mu \nu} R^{\mu \nu} + R^2 \right), \label{eq:Ringo}\\
&\mathcal{L}_{\rm Paul} =V_{\rm Paul}(\phi)~\epsilon^{\mu \nu \lambda \sigma} \epsilon^{\alpha \beta \gamma \delta} R_{\lambda \sigma \gamma \delta}~ \nabla_\mu \phi \nabla_\nu \phi \nabla_\alpha \phi \nabla_\beta \phi, \\
&\mathcal{L}_{\rm John} = V_{\rm John}(\phi) \left( R^{\mu \nu} - \frac{1}{2}g^{\mu\nu}R\right)\nabla_\mu \phi \nabla_\nu \phi, 
\end{align}
A particular example is the k-essence scenario \cite{ArmendarizPicon:2000dh,ArmendarizPicon:2000ah,Erickson:2001bq} where a negative pressure term arises from a non-linear kinetic term during matter-domination.
The Fab Four theories admit a Minkowski vacuum for any value of the CC, even after an instantaneous change of the vacuum energy following a phase transition. 
However, the presence of light scalar field gives rise to \textbf{fifth force} constraints in the solar system such that we must employ a \textbf{screening mechanism}, e.g. Vainshtein mechanism (1972 \cite{Vainshtein:1972sx}), where the fifth force is screened in local environment (at short length scales). Such a possibility is explored in Ref. \cite{Kaloper:2013vta} where the authors show for instance that Ringo in Eq.~\eqref{eq:Ringo} should make himself tiny. They also demonstrate an apparent incompatibility between the requirement of having new effects in the IR, at horizon scale $H^{-1}_0$, and the requirement of maintaining perturbativity in the UV, down to $mm$, the shortest length scale at which the Newtonian law has been tested \cite{Tan:2020vpf,Lee:2020zjt}.

\paragraph{Sequester the CC with global Lagrange multipliers: }
The Einstein equation are left intact locally but are modified \textbf{globally}, in the infinite wavelength limit. In addition to the vacuum energy of the SM, a CC \textbf{counterterm} $\Lambda$ is introduced. The idea of \textbf{sequestering}, first exposed in 2013 in \cite{Kaloper:2013zca, Kaloper:2014dqa}, is to promote the CC counterterm $\Lambda$  as well as the Planck mass $M_{\rm pl}$ to \textbf{global} dynamical degrees of freedom. Their equations of motion are global constraints which enforce the CC counterterm $\Lambda$  to cancel the vacuum energy. Hence, the vacuum energy does not gravitate. A prediction is that the universe should be finite in space-time and therefore, should collapse in the future. Other cosmological consequences are studied in \cite{Coltman:2019mql}. However, we may still wonder what it means for a dynamical variable to be defined globally.
In \cite{Kaloper:2015jra}, a \textbf{local} version of sequestering is developed thanks to the introduction of auxiliary fields.

\paragraph{The Weinberg window: }
It can be shown that the values of the CC compatible with the existence of observers are very restricted and must be contained within the \textbf{Weinberg window}
\begin{equation}
\label{eq:weinberg_window}
-2 \times 10^{-119}~M_{\rm pl}^4 \lesssim \Lambda \lesssim 5\times 10^{-119}~M_{\rm pl}^4.
\end{equation}
The lower bound prevents the Universe to undergo a Big Crunch before galaxy formation, e.g. \cite{Kallosh:2002gg}, while the upper bound forbids the possibility to have a CC domination before the galaxy formation (Weinberg 1988 \cite{Weinberg:1988cp}). This suggests that the CC may have an \textbf{anthropic origin}.

\paragraph{Dynamical relaxation of the CC on a landscape: }
\label{par:dyn_rel_CC}
In 1984, Abbott \cite{Abbott:1984qf} suggested a mechanism of dynamical adjustment of the CC based on a scalar field rolling down a slope with wiggles. This allows to relax the CC \textbf{step by step}, each time the scalar field \textbf{tunnels} from one minimum to the next one. In 1987, Brown and Teitelboim (BT) suggested to use 3-form fields $A^{\mu\nu\alpha}$ whose 4-form strenght tensor $F^{\mu\nu\alpha\beta}$ have the advantage to be \textbf{constant} in $(3+1)-D$, hence contributing to the CC \cite{Brown:1987dd, Brown:1988kg}. In this scenario, the CC is relaxed step by step each time a \textbf{membrane} - the $2D$ object charged under the 3-form gauge field - \textbf{nucleates}. The CC inside the membrane is decreased by one unit of membrane charge $q$. An effect analog to the Schwinger nucleation of $e^-/e^+$ pair in an electric field in $(1+1)-D$. 
However, the step-wise dynamical relaxation of the CC suffers from \textbf{two major flaws}.
\begin{itemize}
\item
In order to be able to predict a CC within the Weinberg window in Eq.~\eqref{eq:weinberg_window}, the vacuum energy difference $\epsilon$ between two neighboring vacua in Abbott's model, or the $2$-brane charge $q$ in BT's model, must be smaller than $\lesssim 10^{-119}$ in Planck unit. This the \textbf{step-size problem}.
\item
The decay rate of the last vacua before entering inside the Weinberg window is ridiculously small. Hence, when the CC is finally adjusted to nearly zero, the universe has undergone an extremely long period of inflation and is nearly empty with no possibility of reheating. This is the \textbf{empty universe problem}.
\end{itemize}
In 2000, Bousso and Polchinski \cite{Bousso:2000xa,Bousso:2012dk} demonstrated that the two problems can be solved after introducing a \textbf{large number} $N$ of four-form fields (such a large number of $4$-form is moreover predicted in string compactification scenario). The parameter space of the $N$ four-forms is a $N-$dimensional grid, whose inter-space is the membrane charge $q$, and in which the Weinberg window is represented by a shell. The trick is that, for a constant shell thickness, the number of grid points within the \textbf{Weinberg shell} grows extremely fast with the number of dimensions $N$. Hence, for $N\gtrsim 100$ we can get \textbf{more than one} vacuum configuration inside the Weinberg shell for membrane charge as large as $q\sim 1$ in Planck unit, which solves the step-size problem. The membrane charge being $q \sim 1$, the vacuum energy difference during the last step is also of order $1$ in Planck unit. The conversion of this large vacuum energy into thermal energy leads to a hot big-bang and the empty universe problem is solved. 

Upon introducing an operator (which is non-renormalizable in \cite{Dvali:2003br,Dvali:2004tma} and renormalizable in \cite{Giudice:2019iwl, Kaloper:2019xfj}) which couples the four-form fields with the Higgs field, the Bousso-Polchinski mechanism can be used to scan over the EW scale. Note that in that case, four-form fluxes act as a connection between the landscape of CCs and the landscape of EW scales. See also \cite{Arkani-Hamed:2020yna} where the authors discuss a possible connection between the CC and EW-scale landscapes in another context.

Now we discuss another possibility to solve the empty universe problem. Introduced in \cite{Steinhardt:2006bf, Graham:2017hfr}, it consists to relax the CC to negative values, $-(\rm meV)^4$, such that the universe crunches and undergoes a bounce, followed by a reheating phase. The inherent difficulty with bouncing cosmology scenarios is that Einstein equations predict an unavoidable singularity at $a\to 0$. 
From a combination between the Friedmann equation and the continuity equation, we can write
\begin{equation}
\label{eq:2ndFriedmann_eq}
\frac{\ddot{a}}{a}-\left(\frac{\dt{a}}{a}\right)^2 = - \frac{1}{2M_{\rm pl}^2}(\rho + p) + \frac{k}{a^2}.
\end{equation}
From this, we can see that a way to allow for a bounce defined by
\begin{equation}
\label{eq:Bounce_condition}
\text{Bounce:} \qquad \dt{a}=0, \quad \ddot{a} > 0,
\end{equation}
and therefore to avoid the singularity at $a\to 0$, is to introduce a fluid which violates the null energy condition (NEC)\footnote{From Eq.~\eqref{eq:2ndFriedmann_eq}, we could think of producing a bounce using the positive curvature $k>0$ of the universe, in light of the recent results \cite{DiValentino:2019qzk}. The inherent difficulty is that the curvature term blueshift as $\propto a^{-2}$ too slowly compared to the other fluids, e.g. radiation $\propto a^{-4}$ or kination $\propto a^{-6}$, such that the curvature would hardly dominate the energy budget when $a \to 0$. Then the issue becomes how to convert, when the universe crunches, fast-blue-shifting fluids having $\omega > -1/3 $ (e.g. matter, radiation, kination) with $\rho \propto a^{-3(1+\omega)}$, into slow-blue-shifting fluids $\omega < -1/3 $ (e.g. dark energy \cite{Barrau:2020nek}), such that the curvature $\omega = 1/3$ can take over when $a \to 0$. This is challenged by the second law of thermodynamics since the faster the blue-shift, the larger the entropy $s \propto (1+\omega) $ (e.g. kination has the largest entropy while dark energy has a vanishing entropy). }
\begin{equation}
\label{eq:NEC_violation}
\textrm{NEC violation:} \qquad  \rho + p <0.
\end{equation} 
Assuming that the bounce can be realized, the negative CC $-(\rm meV)^4$ can be converted to a positive CC $+(\rm meV)^4$ through a phase transition \cite{Graham:2019bfu}.
More popular options for avoiding singularities rely on \textbf{beyond-Einstein} \cite{Ijjas:2015hcc,Ijjas:2016tpn,Ijjas:2018qbo,Ijjas:2019pyf,Rubakov:2014jja,Mironov:2019haz,Sloan:2019jyl} or \textbf{quantum gravity theories} \cite{Bojowald:2007ky}. See  \cite{Lehners:2008vx,Cai:2014bea,Battefeld:2014uga,Brandenberger:2016vhg,Ijjas:2018qbo,Mironov:2019haz} for reviews on bouncing cosmology.

\paragraph{Eternal inflation and multiverse: }

In scenarios where the CC is relaxed by vacuum tunneling, the quantum transition between two vacua is generally \textbf{much longer} that the Hubble time, which implies that there is at most one membrane of new vacuum per Hubble patch. Even if the just nucleated membrane of new vacuum expands at the speed of light, it is not fast enough to catch up with the expansion of the universe. Therefore, the old vacuum can not be entirely converted in the true vacuum. There is always some of the old vacuum left, which \textbf{inflates eternally}. We say that the membranes (or bubbles) do not \textbf{percolate} \cite{Guth:1982pn}.\footnote{However, in case of multiple bubble nucleation per Hubble volume, if the nucleation rate increases fast enough with time, percolation can occur during an inflating period. See Sec.~\ref{par:percolation_temp_inf} in Chap.~\ref{chap:1stOPT} for a quantitative criterium.}
The consequence is the formation of a \textbf{multiverse} containing an infinite number of causal patches corresponding to different \textbf{membrane cascades}. In this picture, our universe would correspond to one of the membrane cascades among those ending up inside the Weinberg window in Eq.~\eqref{eq:weinberg_window}. However, the eternal inflation in a landscape suffers from \textbf{two main issues}, see also \cite{Freivogel:2011eg}.
\begin{itemize}
\item
The different patches in the multiverse are causally disconnected. Hence, the multiverse seems \textbf{not experimentally testable}. See however some attempts \cite{Hook:2019pbh, Hook:2019zxa}.
\item
The multiverse scenario suffers from a loss of predictability. Not only the total number of distinct patches are infinite, but also the number of patches with the correct CC. \textit{Anything that can happen will happen, and it will happen an infinite number of times} \cite{Guth:2007ng}. Hence, the probability that a given patch  has a CC inside the Weinberg window can not be computed, $p = \frac{\infty}{\infty}$. This is the \textbf{measure problem}.
\end{itemize}
Note that the scenario of eternal inflation may be challenged \cite{Guth:2012ww} if the recent claim \cite{DiValentino:2019qzk} that we live in a closed universe is confirmed.

In \cite{Bloch:2019bvc}, eternal inflation and its intrinsic measure problem is evaded by \textbf{crunching} all the patches after some time using a \textbf{supercooled first-order phase transition} with AdS true vacuum. The patches with the largest CC are the ones with the temperature decreasing the fastest and which crunch the fastest. Conversely, regions with a small CC are long-lived.

In this section, we have outlined a set of creative ideas that have been put forward to solve the cosmological constant problem. However, it is fair to say that none of them is satisfactory. This is one of the biggest puzzles in high-energy physics and cosmology. We now move to the Baryon Asymmetry of the Universe (BAU).

\subsection{Matter-anti-matter asymmetry}
\label{sec:Matter-anti-matter-asymmetry}

When Dirac combined Quantum Mechanics and Special Relativity in 1928 in order to generalize the Schrodinger equation to relativistic particles \cite{Dirac:1928hu, Dirac:1928ej}, he predicted the existence of \textbf{antimatter}. Dirac's prediction, followed with the discovery of the positron in 1932 by Anderson after investigation of cosmic rays in a cloud chamber \cite{Anderson:1932zz,Anderson:1933mb}, is a major breakthrough in the history of science.  Dirac received the Nobel prize the next year in 1933.    \\
However, our universe appears to have a large deficit of anti-matter with respect to matter. See e.g. \cite{Canetti:2012zc} for a review. The small fraction of anti-protons and positrons in cosmic rays is well explained by secondary production processes.
The existence of some region of the universe filled with anti-matter is ruled-out  due to the non-observation of gamma-rays coming from matter/anti-matter annihilation, on scales ranging from the solar system, to the Milky-Way, to galaxy clusters, and even to distances comparable to the scale of the present horizon \cite{Cohen:1997ac, Steigman:2008ap}. Anyway, there are no known mechanisms to separate domains.

In Sec.~\ref{eq:Beyond_therm_equil}, we explicitly showed that the baryon-to-light ratio $n_b/n_\gamma$, if one assumes a symmetric universe, would be much smaller than the observed ratio, hence demonstrating the need of a primordial baryon/anti-baryon asymmetry. Inflation is expected to dilute any pre-existing matter asymmetry, so the initial conditions are set to $B=L=0$, where $L=L_e+L_\mu+L_\tau$, and we must provide a mechanism for generating $B\neq 0$.

In 1967, \textbf{Sakharov} gave three conditions that any theory seeking to explain the baryon asymmetry must satisfy \cite{Sakharov:1967dj}
\begin{itemize}
\item 
$B$ violation: otherwise the baryon number is a conserved quantity.
\item
$C$ and $CP$ violation: otherwise for each process $\mathcal{P}$ generating $B$, there would exist a $C$ (or $CP$) conjugate process $\bar{\mathcal{P}}$ which would generate $-B$ with the same probability. 
\item
Out of equilibrium: otherwise the system is time-invariant and $\frac{d}{dt}B(t)=0$.
\end{itemize}

In Sec.~\ref{par:global_sym_SM} of Chap.~\ref{chap:SM_particle}, we discussed that the baryonic and leptonic charges $B$ and $L$ were a priori conserved in the SM at tree level.
But actually, $B$ and $L$ are violated by non-perturbative processes called \textbf{sphalerons}.\footnote{The world `sphalerons' which means `ready to fall' in Greek, was proposed by Klinkhamer and Manton in 1984 \cite{Klinkhamer:1984di}.} This results from the chirality of the weak interactions (only left-handed fermions are charged) which leads to the anomalies $SU(2)_L^2U(1)_B$ and $SU(2)_L^2 U(1)_L$, see Sec.~\ref{sec:anomaly_cancellation} of Chap.~\ref{chap:SM_particle}. As a consequence, we can write
\begin{equation}
\label{eq:sphaleron_B_L}
\partial_\mu j^\mu_B = \partial_\mu j_L^\mu = \frac{n_f}{32\pi^2}\left( g_2^2 W^a_{\rm \mu\nu}\tilde{W}^{a\mu\nu} - g_1^2 B_{\mu\nu} B^{\mu\nu} \right),
\end{equation}
where $ j^\mu_B$ and $j_L^\mu$ are the $B$ and $L$ Noether currents. Note that only $B+L$ is violated while $B-L$ is conserved. 't Hooft was the first to introduce $B+L$ violation by non-perturbative effects in 1976 \cite{tHooft:1976rip}. In his paper, he showed that the tunneling amplitude is extremely suppressed at zero temperature $\mathcal{A} \sim e^{-8\pi^2/g_2^2} \sim e^{-170}$. In 1978, Dimopoulos and Susskind proposed that such procesess are active at high temperature \cite{Dimopoulos:1978kv} and a number of subsequent papers investigated this effect \cite{Christ:1979zm,Manton:1983nd,Klinkhamer:1984di,Kuzmin:1985mm,Arnold:1987mh,Arnold:1987zg}.  Particularly, in 1985, Kuzmin, Rubakov, and Shaposhnikov \cite{Kuzmin:1985mm} computed that sphaleron transitions are at thermal equilibrium in the early universe above $130~$GeV (see \cite{DOnofrio:2014rug} for recent lattice simulations). More particularly, the sphaleron rate per unit of volume goes like
\begin{align}
\label{eq:sphaleron_rate_1}
& \Gamma_{\rm sph}  \sim \left(\frac{E_{\rm sph}}{T} \right)^3 \left( \frac{m_W}{T} \right) T^4 \, e^{-E_{\rm sph}/T} \qquad &T \lesssim 130~{\rm GeV}, \\
 &\Gamma_{\rm sph}  \sim 30\,\alpha_w^5\,T^4,  \qquad &T \gtrsim 130~{\rm GeV},
 \label{eq:sphaleron_rate_2}
\end{align}
where $\alpha_w = g_2^2/4\pi \sim 1/30$ and $E_{\rm sph} = 8\pi v/g_2 \simeq 10~{\rm TeV} $ is the height of the energy barrier between two neighboring values of $B+L$. Hence, above $T\gtrsim 130~\rm GeV$, any $B+L\neq 0 $ will be quickly relaxed to $B+L \simeq 0$, while $B-L\neq 0$ will remain conserved, such that the final baryon asymmetry will be $B_{\rm eq} \simeq - L_{\rm eq}/2$. More precisely, thermal equilibrium of various processes\footnote{Sphalerons transitions only acts on left-handed fermions which explains why $B_{\rm eq}$ is not exactly equal to $-L_{\rm eq}/2$. In addition to be set by $SU(2)$ EW instantons, the baryon number at equilibrium in Eq.~\eqref{eq:equilibrium_B_BmoinsL}, also relies on $SU(3)$ instantons which equilibrate left-handed with right-handed quarks, hypercharge-conserving processes which equilibrate $Y$, and on Yukawa interactions which relate the population of left-handed with right-handed fermions \cite{Buchmuller:2005eh}.} in the SM implies  \cite{Khlebnikov:1988sr}
\begin{equation}
\label{eq:equilibrium_B_BmoinsL}
B_{\rm eq} = \frac{8N_f + 4}{22N_f +13}(B_{\rm eq}-L_{\rm eq}),
\end{equation}
where $N_f = 3$ is the number of flavors.

We list below the main proposals for baryogenesis.\footnote{Recent years have seen the development of a variety of new theoretical ideas for explaining the baryon asymmetry of the universe. They can be very different from the mainstream mechanisms discussed in this book, and we refer to \cite{Elor:2022hpa} for a review.}

\paragraph{Baryogenesis in Grand Unified Theories (GUT):}  
Models which unify strong and electroweak interactions, the simplest being $SU(5)\rightarrow SU(3)_c\times SU(2)_L$ proposed in 1974 by Georgi and Glashow \cite{Georgi:1974sy}\footnote{The GUT based on $SO(10)$ was proposed the same year by Fritzsch and Minkowski \cite{Fritzsch:1974nn}. The embedding of $SU(5)\times U(1)$ in string theory by Antoniadis, Ellis, Hagelin and Nanopoulos in 1987 is one of the first example of Theory of Everything \cite{Ellis:1986mm,Antoniadis:1987dx,Antoniadis:1989zy}.}, put leptons and hadrons in the same multiplet,  unavoidably leading to baryon number violation. Hence, a baryon asymmetry can be created by the \textbf{decay} of a heavy particle $X$ with mass comparable to the GUT scale $M_X \sim 10^{16}$. When the decay occurs \textbf{out of equilibrium}, namely when $\Gamma_X \lesssim H$ for $T \sim M_X$, the baryon asymmetry is given by
\begin{equation}
n_{\rm B}/s \sim  \frac{\kappa}{g_*} \sum_f B_f \frac{\Gamma(X \rightarrow f) - \Gamma(\bar{X} \rightarrow \bar{f})}{ \Gamma_X} ,
\end{equation}
where the sum is performed over all the final state $f$ of the $X$ decay. $f$ has baryon number $B_f$. The total decay rate of $X$ and $\bar{X}$ are equal, $\Gamma_X =\Gamma_{\bar{X}}$, because of CPT invariance and unitary. $\kappa$ accounts for the possible partial wash-out due to inverse-decay if the out-of-equilibrium conditions are not satisfied
\begin{equation}
\label{eq:efficiency_fac_washout_GUT}
\kappa \sim
\begin{cases}
1&\hspace{2em} {\rm if}~\Gamma_{X} \lesssim H\Big|_{T \simeq M_X},\\
\frac{H}{\Gamma_{X} }\Big|_{T \simeq M_{X}} &\hspace{2em}  {\rm if}~\Gamma_{X} \gtrsim H\Big|_{T \simeq M_X}.\\
\end{cases}
\end{equation}
Particularly, if $X$ decays into $f = a, \, b$, then we obtain
\begin{equation}
n_{\rm B}/s \sim \frac{\kappa}{g_*} \left( B_a - B_b  \right) \frac{\Gamma(X \rightarrow a) - \Gamma(\bar{X} \rightarrow \bar{a})}{\Gamma_X},
\end{equation}
where we have used $\Gamma_X = \Gamma(X \rightarrow b)+ \Gamma(X \rightarrow a) $ and idem for $\bar{X}$.
We can explicitly see that we need \textbf{B-violation}, $B_a\neq B_b$, as well as \textbf{CP-violation}, $\Gamma(X \rightarrow a) \neq \Gamma(\bar{X} \rightarrow \bar{a})$, satisfied together to produce a baryon asymmetry.

However in practice, there are some difficulties with GUT baryogenesis. Any \textbf{entropy production} below the GUT scale, due to inflation or the decay of heavy relic, would \textbf{dilute} the baryon asymmetry.
Particularly, the upper bound on the inflationary vacuum energy density, from B-modes non-obervation in the CMB, is very close to the GUT scale, $V_{\rm inf}^{1/4}= (1.88\times 10^{16})(r/0.10)^{1/4}$, with the scalar-to-tensor ratio $r$ constrained to $r\lesssim 0.055$ \cite{Aghanim:2018eyx}. Another flaw is that minimal GUT models conserve $B-L$ such that any $B$-asymmetry produced at the GUT scale is subsequently washed-out by sphaleron transitions. Note also that minimal GUT models, like the simple $SU(5)$, are constrained by the lower bound on the proton decay, $\tau_p / BR(p \rightarrow e^+ \pi^0) > 1.67 \times 10^{34}~$years \cite{Miura:2016krn,Tanabashi:2018oca}. See some of the pioneering articles in 1978 \cite{Yoshimura:1978ex, Dimopoulos:1978kv, Ellis:1978xg, Toussaint:1978br,Weinberg:1979sa, Kolb:1979qa, Fry:1980ph} and some reviews \cite{Riotto:1998bt, Dine:2003ax, Cline:2006ts}.

\paragraph{Affleck-Dine mechanism:} 
A non-zero baryon number can be generated dynamically by the \textbf{coherent oscillation} of a scalar field $\phi$ charged under baryon number, in a potential $V(\phi)$ containing correction terms which explicitly break baryon number. The baryon asymmetry is given by the current
\begin{equation}
n_{\rm B} =  {\rm Im} \left[ \phi^{\dagger} \dot{\phi} \right],
\end{equation}
and its evolution follows from the Klein-Gordon equation for $\phi$
\begin{equation}
\dt{n}_{\rm B} + 3H n_{\rm B} = {\rm Im} \left[ \phi \frac{\partial V}{\partial \phi}   \right].
\end{equation}
Explicit breaking of baryon baryon number can be provided by $V(\phi) \supset \frac{\phi^n}{\Lambda^{n-4}} +\rm h.c.$ with $n$ odd.
The baryon number is transmitted to the SM when the scalar condensate decays . See the seminal article in 1984 \cite{Affleck:1984fy}, a review \cite{Dine:2003ax} and a more recent study involving the QCD axion \cite{Harigaya:2019emn, Co:2019wyp,Gouttenoire:2021jhk} .

\paragraph{Spontaneous Baryogenesis: } 
In 1987, Cohen and Kaplan (and later Nelson) \cite{Cohen:1987vi, Cohen:1991iu} proposed a minimal scenario for generating a baryon asymmetry simply based on
\begin{equation}
\mathcal{L} \supset V(\phi) +  \frac{1}{f} \partial_\mu \phi \, j^\mu_{\rm B} + \mathcal{O}_{\cancel{B}}.
\end{equation}
The first term induces a \textbf{motion of the scalar field} $\dt{\phi} \neq 0$, which thanks to the second term, generates an energy difference between baryons and anti-baryons by an amount $2\dt{\phi}/f$
\begin{equation}
\Delta  \mathcal{H} = \frac{\dt{\phi}}{f} \left( n_b - n_{\bar{b}} \right) = -\mu\,\Delta B, \qquad (\text{assuming }\partial_i\phi =0).
\end{equation}
This is equivalent to give $b$ and $\bar{b}$ a non-vanishing opposite \textbf{chemical potential} $ \mu_{\bar{b}}=-\mu_b =\dt{\phi}/f$. If additionally we switch on some $B$-violation operator $\mathcal{O}_{\cancel{B}}$, then we would start producing a net $B$-number $\Delta B \neq 0$. In the presence of $\dt{\phi} \neq 0$, the $CPT$ symmetry is spontaneously broken which is another Sakharov condition. If the B-violating interactions freeze-out, at $T_{\rm fo}$, before $\phi$ stops rolling, then a residual baryon asymmetry remains
\begin{equation}
n_{\rm B}/s \sim \frac{\dt{\phi}}{f\, g_*\, T_{\rm fo}}.
\end{equation}
Spontaneous baryogenesis can occur during axion inflation either from the operator $\tfrac{\phi}{f}F\tilde{F}$ or $\tfrac{\partial_\mu \phi}{f}\bar{\psi}\gamma^\mu \gamma_5 \psi$ \cite{Adshead:2018oaa, Domcke:2018eki, Domcke:2019mnd}. These two operators are equivalent up to a field rotation, and \cite{Domcke:2020kcp} have explicitly checked that the transport equations for the baryonic charge $n_{\rm B}$ are independent of this choice of representation. See \cite{Abel:2018fqg} for a realization of spontaneous baryogenesis with the relaxion.

Spontaneous breaking of $CPT$ has also been proposed to be generated by the expansion of the universe through the operator \cite{Davoudiasl:2004gf}
\begin{equation}
\sqrt{-g} \frac{(\partial_\mu \mathcal{R})}{M_{\rm pl}} J^\mu_{\rm B},
\end{equation}
where $\mathcal{R}$ is the Ricci scalar curvature whose time derivative
\begin{equation}
\dt{\mathcal{R}}=\sqrt{3}(1-3\omega)(1+\omega)\frac{\rho^{3/2}}{M_{\rm pl}^2},
\end{equation}
is different from zero when the equation of state $\omega$ departs from pure-radiation or pure-inflation. This mechanism is called \textbf{gravitational baryogenesis}. Interestingly, the needed source of $B$-violation $\mathcal{O}_{\cancel{B}}$ can be generated by the Hawking radiation from black holes \cite{Hook:2014mla}.

\paragraph{Leptogenesis:} 
\label{par:leptogenesis} 
This is the successor of GUT baryogenesis, the role of the heavy decaying particle is played by a heavy right-handed neutrino $N_I$ which decays into a left-handed neutrino and a Higgs. Right-handed neutrino are motivated for explaining the neutrino oscillations in the seesaw model, see sec.~\ref{sec:see_saw} of Chap.~\ref{chap:SM_particle}. Since right-handed neutrinos do not carry lepton number, their decay violates the $L$ number. The decay of $N_I$ must occur out of equilibrium, which implies $\Gamma_{N_I} \lesssim H$ when $T \lesssim M_{N_I}$.  Assuming it is the case, then the $L$ asymmetry is given by
\begin{equation}
\frac{n_L}{s} \simeq  \frac{\kappa}{g_*} \sum_\alpha \frac{\Gamma(N_I \rightarrow l_\alpha\, h^\dagger) - \Gamma(N_I \rightarrow \bar{l}_\alpha\, h)}{ \Gamma(N_I \rightarrow l_\alpha\, h^\dagger) + \Gamma(N_I \rightarrow \bar{l}_\alpha\, h)},
\end{equation}
where the sum runs over SM leptons. The needed CP violation, $\Gamma(N_I \rightarrow l_\alpha\, h^\dagger) \neq \Gamma(N_I \rightarrow \bar{l}_\alpha\, h)$, is introduced via complex phases in the neutrino mass matrix. It vanishes at tree level and the leading contribution comes from an interference between the tree-level and one-loop diagrams. $\kappa$ is an efficiency factor which takes into account an eventual wash-out due to inverse decay and other $L$-changing scattering processes. It is given by a similar expression as Eq.~\eqref{eq:efficiency_fac_washout_GUT}.
Then, the $L$ number is subsequently converted into $B$ number via sphaleron transitions, see Eq.~\eqref{eq:equilibrium_B_BmoinsL}.

Leptogenesis was introduced in 1986 by Fukugita and Yanagida the year after sphalerons were shown to reach thermal equilibrium in the early universe  \cite{Fukugita:1986hr}.  See some reviews \cite{Buchmuller:2005eh, Chen:2007fv, Davidson:2008bu, DiBari:2012fz, Fong:2013wr,Garbrecht:2018mrp,Bodeker:2020ghk}. 

At first, it was asserted that successful leptogenesis requires the right-handed neutrinos to be heavier than $m_{\nu_R} \gtrsim 10^9~$GeV \cite{Davidson:2002qv}.  But then, it was shown that it can be relaxed to $m_{\nu_R} \gtrsim 1~$TeV using nearly-degenerate right-handed neutrino masses \cite{Pilaftsis:2003gt}. It has even been reduced to the $\rm keV/GeV$ range by Shaposhnikov et al. \cite{Asaka:2005pn} using $CP$-violating right-handed neutrino oscillations \cite{Akhmedov:1998qx}. In the latter model, called Neutrino Minimal Standard Model, the $\rm KeV$ sterile neutrino can be a Dark Matter candidate \cite{Canetti:2012kh,Canetti:2012zc}. See \cite{Abazajian:2012ys,Adhikari:2016bei, Boser:2019rta} for reviews on sterile neutrinos.

\paragraph{Electroweak baryogenesis:}
 \label{par:EWbaryo} 
 Baryon creation can occur in the vicinity of expanding bubble walls during an electroweak first-order phase transition. When the interactions between particles in the plasma and the Higgs bubble violate $CP$, a $CP$ asymmetry form in the symmetric phase, outside the bubble, which sphalerons convert to $B$-asymmetry. Then, due to the wall motion, the baryons diffuse into the broken phase, inside the bubble, where the sphaleron rate is exponentially suppressed by $\left<\phi\right>/T$. The baryon asymmetry is given by
\begin{equation}
\label{eq:baryon_asym_EW}
n_{\rm B}/s \sim \frac{1}{v_w\, g_* \, T_{\rm nuc}^4} \int_{-\infty}^{\infty} dz ~ \Gamma_{\rm sph}~ \mu_L(z) ~ e^{~c\,\int_{-\infty}^z d\tilde{z}~ \Gamma_{\rm sph}/T_{\rm nuc}^3} ,
\end{equation}
where $z$ is the radius coordinate along the bubble wall profile (broken phase in $z=-\infty$, wall in $z=0$ and symmetric phase in $z=+\infty$) and $v_w$ is the bubble wall velocity. $\mu_L(z)$ is the local chemical potential of the left-handed fermions in the plasma. It can be computed by solving a diffusion equation \cite{Prokopec:2003pj,Prokopec:2004ic,Konstandin:2013caa}.  $\Gamma_{\rm sph}$ is the sphaleron rate given by Eq.~\eqref{eq:sphaleron_rate_1} and Eq.~\eqref{eq:sphaleron_rate_2}. In order to prevent the sphaleron wash-out inside the bubble (due to the exponential term), the nucleation temperature $T_{\rm nuc}$ of the phase transition must satisfy $v/T_{\rm nuc} \gtrsim 1$. For the same reason, the reheating temperature\footnote{For clarity, we omitted the dilution factor $(T_{\rm nuc}/T_{\rm RH})^3$ in Eq.~\eqref{eq:baryon_asym_EW}.} must satisfy $v/T_{\rm RH} \gtrsim 1$. If these last two conditions are satisfied, we obtain 
\begin{equation}
n_{\rm B}/s \sim \frac{\Gamma_{\rm sph}\, \mu_L \, L_w}{v_w\, g_* \, T_{\rm nuc}^4},
\end{equation}
where $L_w$ is the typical wall thickness. 

In the SM, the electroweak phase transition is  a \textbf{cross-over} and needs BSM physics to make it \textbf{first-order}, e.g. supersymmeric extensions \cite{Cline:2000nw}, higher dimensional operator \cite{Grojean:2004xa}, a second scalar \cite{Espinosa:2011ax,Cline:2012hg} or varying Yukawa \cite{Baldes:2016rqn}. 

Also the $CP$ violation from the \textbf{CKM matrix} appears to be to small for successful baryogenesis, since it appears at the $\rm 12^{th}$ order in Yukawa couplings, see Sec.~\ref{par:CP_violation_CKM} in Chap.~\ref{chap:SM_particle} or \cite{Shaposhnikov:1987pf, Gavela:1993ts, Konstandin:2003dx, Brauner:2011vb}. 
It is then necessary to introduce new sources of $CP$ violation, e.g. supersymmeric extensions \cite{Cirigliano:2009yd, Kozaczuk:2012xv}, two-higgs doublet model \cite{Dorsch:2016nrg, Alanne:2016wtx}, strong CP violation \cite{Kuzmin:1992up, Servant:2014bla}, varying Yukawa \cite{Baldes:2016gaf,vonHarling:2016vhf, Bruggisser:2017lhc,Servant:2018xcs} or Composite Higgs scenario \cite{Espinosa:2011eu, Bruggisser:2018mus, Bruggisser:2018mrt}. 

It is commonly assumed \cite{Fromme:2006wx} that successful baryogenesis requires the speed of the wall to be within the range $10^{-3} \lesssim v_{w} < 1/\sqrt{3}$, where $1/\sqrt{3}$ is the speed of sound in a relativistic plasma. However, loopholes to the claim that successfull baryogenesis requires subsonic wall have been proposed \cite{Caprini:2011uz,No:2011fi,Cline:2020jre,Laurent:2020gpg,Dorsch:2021ubz,DeCurtis:2022hlx}. This would offer the possibility to test electroweak baryogenesis models with GW \cite{No:2011fi,Cline:2021iff}. Another proposal based on particle production at the wall boundary allows baryogenesis via relativistic bubble walls $v_w \simeq 1$ \cite{Azatov:2021irb,Baldes:2021vyz}.

See the first implementations of electroweak baryogenesis in 1990 by Cohen, Kaplan and Nelson \cite{Cohen:1990py, Cohen:1990it, Cohen:1993nk}, based on the proposal of Kuzmin, Rubakov, and Shaposhnikov in 1985 \cite{Kuzmin:1985mm}, and some reviews \cite{Riotto:1998bt, Riotto:1999yt, Cline:2006ts,Cline:2018fuq, Morrissey:2012db,Garbrecht:2018mrp,Bodeker:2020ghk}.

\subsection{Dark Matter puzzle}
\label{sec:DM}

So far, DM has only been observed indirectly through its \textbf{gravitational} effects at the scale of individual galaxies, clusters of galaxies, or at larger scales. 
For reviews on DM history, observations, models and searches, we refer to \cite{Jungman:1995df,Bergstrom:2000pn,Bergstrom:2009ib,Bergstrom:2012fi,Olive:2003iq,Bertone:2004pz,Murayama:2007ek,Steffen:2008qp,Hooper:2009zm,Feng:2010gw,Roos:2010wb,Bertone:2010at,Schnee:2011ooa,Peter:2012rz,Zurek:2013wia,Profumo:2013yn,Lisanti:2016jxe,Alexander:2016aln,Bauer:2017qwy,Bertone:2016nfn,Freese:2017idy,Roszkowski:2017nbc,Arcadi:2017kky,Battaglieri:2017aum,Bertone:2018krk,Blanco:2019hah}.  For textbooks, we refer to \cite{Bertone:2010zza,Einasto:2013lka,Majumdar:2014wki,Vavilova:2015bed,Profumo:2017hqp,Essig:2019buk,mambrini2021particles}.

\begin{figure}[h!]
\centering
\raisebox{0cm}{\makebox{\includegraphics[width=0.8\textwidth, scale=1]{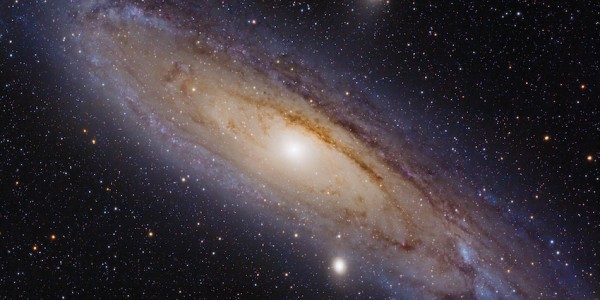}}}
\caption{\it \small Overall view of Andromeda galaxy (M31) taken by Hubble telescope. The visible matter only represents $\sim 20~\%$ of the total mass orbiting around the galaxy, the remaining $\sim 80~\%$ being under the form of Dark Matter.}
\label{fig:hubble-overall-view-of-andromeda-galaxy}
\end{figure}

\begin{figure}[h!]
\centering
\raisebox{0cm}{\makebox{\includegraphics[width=0.6\textwidth, scale=1]{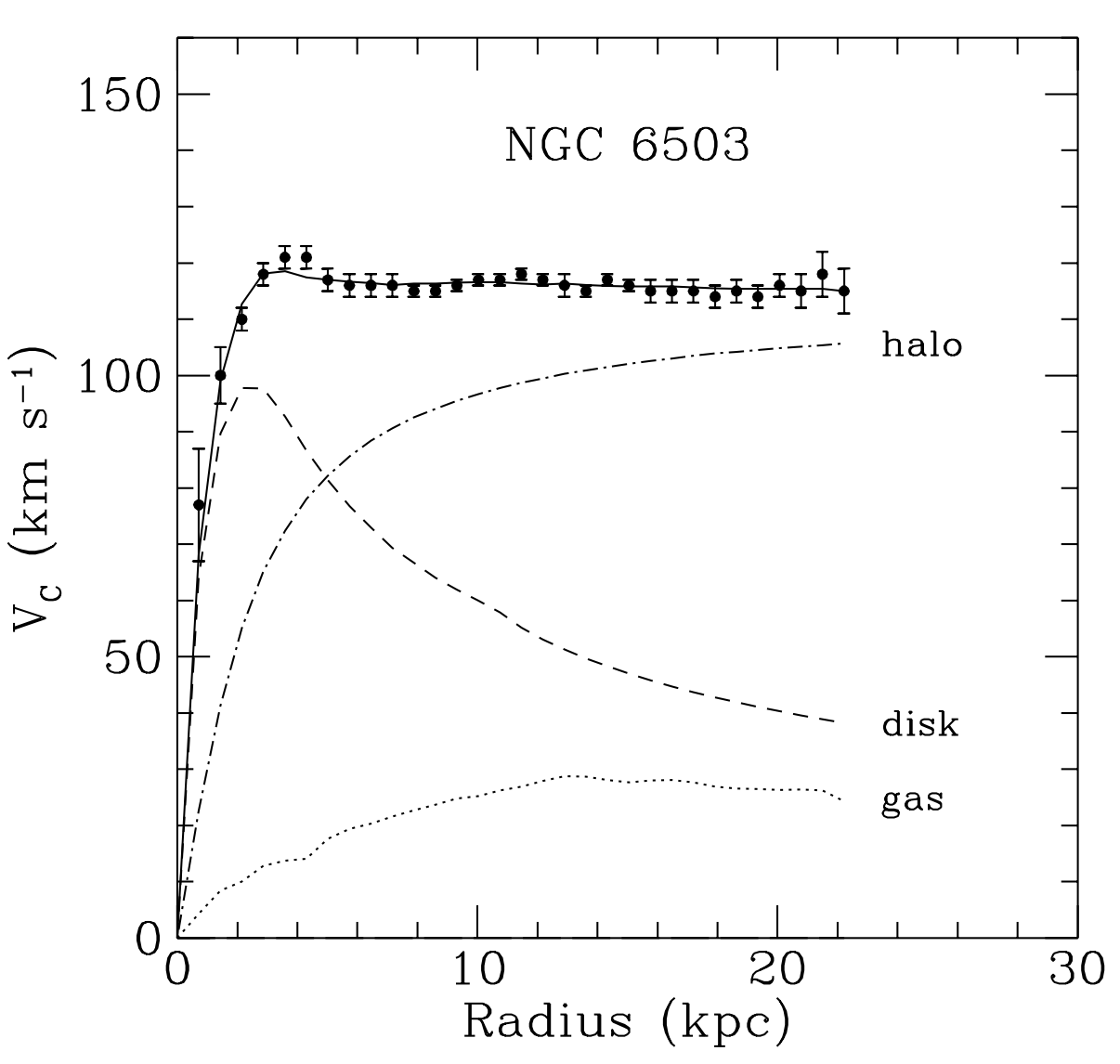}}}
\caption{\it \small The rotation curves of a galaxy (NGC $6503$ here) is well-fitted (solid) upon introducing a spherical dark-halo (dot-dashed) which adds up to the baryonic components measured optically (dashed) and the gas component measured with the $21$~cm line of neutral hydrogen (dotted), cf. Eq.~\eqref{eq:Vcirc}. The fitting parameters are the mass-to-light ratio of the disc, the halo core radius, and the halo asymptotic circular velocity. Figure reproduced from \cite{Begeman:1991iy}.}
\label{fig:NGC6503}
\end{figure}

\paragraph{Galactic scale.}
\label{sec:galactic_scale}
The most famous evidence of dark matter is probably the disagreement between the mass distribution $M_{\rm visible}(R)$ of spiral galaxies based on the observed \textbf{light distribution}, as we can see in Fig.~\ref{fig:hubble-overall-view-of-andromeda-galaxy}, assuming a constant mass-to-light ratio, and the mass distribution infered from the \textbf{circular velocity} $V_{\rm circ}$ of stars and gas. The discordance arises in the outer region and can be accounted for by invoking the presence of a \textbf{spherical halo of dark matter} which extends far beyond the visible galaxy. The discrepancy was first pointed out by Vera Rubin, Kent Ford  and Norbert Thonnard in 1978 \cite{Rubin:1978kmz, Rubin:1980zd} and we refer to \cite{Albada:1986roa} for an old good review written at that epoch.
From Newton's law of gravitation, the circular velocity follows
\begin{equation}
\label{eq:Vcirc}
V_{\rm circ}^2 = \frac{G M_{\rm halo}}{R}+ \frac{G M_{\rm visible}}{R}+ \frac{G M_{\rm gas}}{R},
\end{equation}
where $V_{\rm circ}$ is measured from the Doppler shift of the $H_{\rm \alpha}$ lines for the ionized hydrogen in the inner part, and from the Doppler shift of the $21$-cm lines for the neutral hydrogen in the outer part, respectively. $M_{\rm visible}(R)$ and $M_{\rm gas}(R)$ are infered from the light distribution observed optically and with $21$ cm. Assuming a given mass-to-light ratio of the disc, $M_{\rm halo}$ is the spherical dark matter halo needed to resolve the inconsistency. A rotation curve and its appropriate fit assuming the presence of a spherical dark matter halo is shown in Fig.~\ref{fig:NGC6503}.
Without dark matter halo, we would expect the measured rotation curve to coincide with the one infered from the baryonic mass, namely disk + gas lines, where in that case $V_{\rm circ} \propto 1/R$. Instead, the measured rotation curve does not decrease but quickly flattens, hinting for a conspiracy between baryonic matter and dark matter at intermediate radius and for $M_{\rm halo} \propto R ~ \rightarrow ~ \rho_{\rm halo} \propto 1/R^2$ at large radius.

\begin{figure}[h!]
\centering
\raisebox{0cm}{\makebox{\includegraphics[width=0.6\textwidth, scale=1]{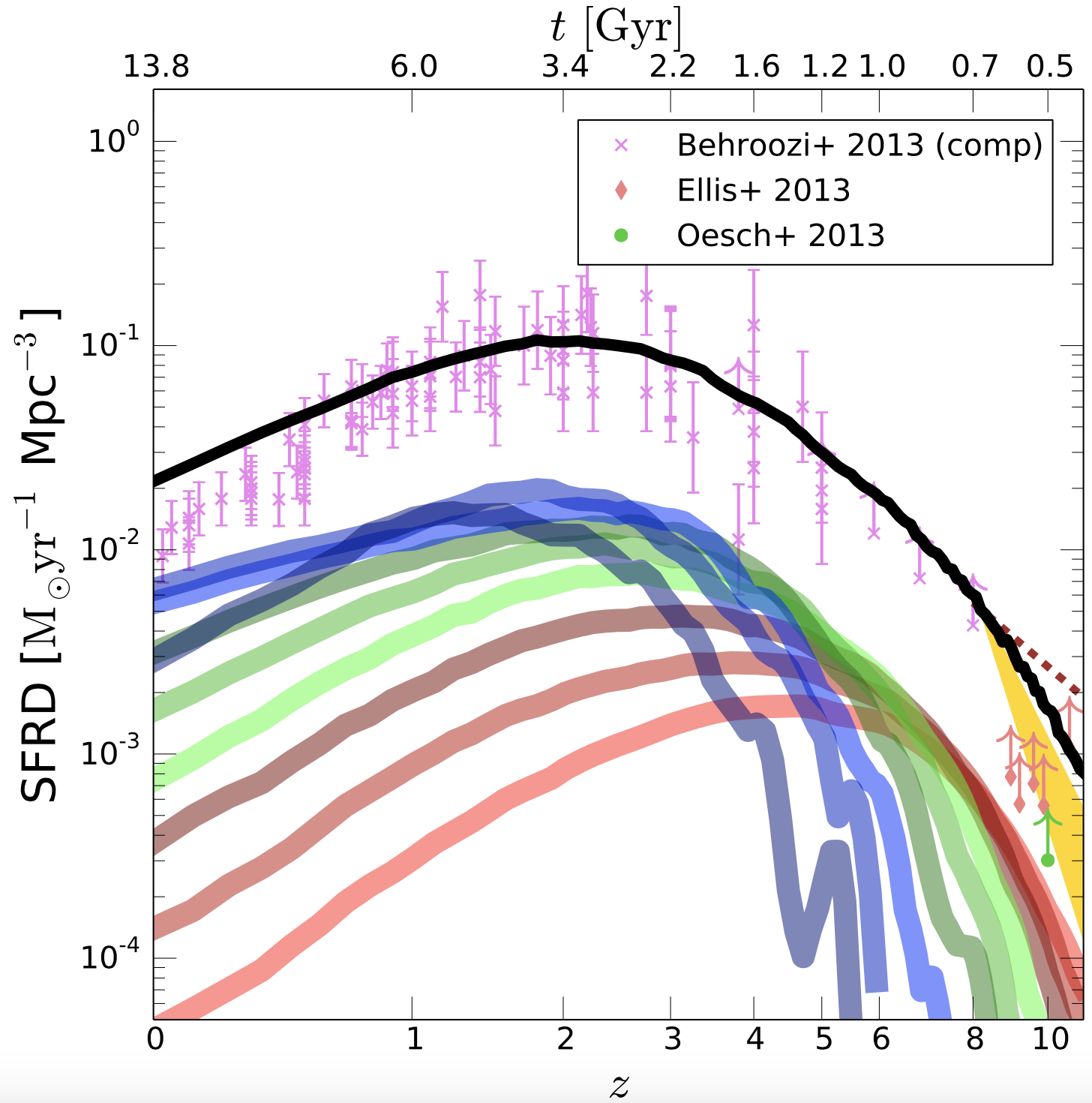}}}
\caption{\it \small  Success of $\Lambda$CDM numerical simulations \cite{Vogelsberger:2014dza} in predicting the Star Formation Rate Density (SFRD), compared to observations (Behroozi et al. 2013a, Ellis et al. 2013, Oesch et al. 2013). Star formation depends on the details of structure growth which is itself sensitive to the dark matter fraction in the universe. The black solid line shows the total SFRD whereas colored thick lines show the contributions from galaxies of different stellar masses ranging from low mass (red lines: $M_*=10^{7},\,10^{7.5},\,10^8~M_{\odot}$), to intermediate mass (green lines: $M_*=10^{8.5},\,10^9,\,10^{9.5}~M_{\odot}$), to the most massive systems (blue lines: $M_*=10^{10},\,10^{10.5},\,10^{11}~M_{\odot}$).}
\label{fig:SFRD_DM_evidence}
\end{figure}

\paragraph{Inter-galactic scale: First dark matter evidence.}
\label{sec:inter_galactic_scale}
Historically, the first proposal for the existence of dark matter goes back to Fritz Zwicky in 1933 and 1937 \cite{Zwicky:1933gu,Zwicky:1937zza} who found a discrepancy between the \textbf{optical mass} of the Coma cluster infered from the measurement of its visible light emission and the \textbf{dynamical mass} measured from the average dispersion velocity of its constituent galaxies. More precisely, the \textbf{Virial theorem} allows to write \cite{Zwicky:1937zza}
\begin{equation}
\left<V_{\rm s}^2 \right> = \frac{5GM}{3R},
\end{equation}
where $\left<V_{\rm s}^2 \right>$ is the average quadratic velocity along the line-of-sight, $R$ and $M$ are the radius and total mass of the Coma cluster. After taking $R \simeq 6.5~\rm Mpc$ and $\left<V_{\rm s}^2 \right> \simeq (700~\rm km/s)^2$, Zwicky computed the total mass $M\simeq 4.5\times 10^{13}~M_{\odot}$, which after division by the measured total luminosity $L = 8.5\times 10^{10}~L_{\rm \odot}$, led him to the \textbf{mass-to-light ratio} of the Coma cluster
\begin{equation}
M/L \sim 500,
\end{equation}
which is two orders of magnitude above the solar neighborhood value. Hence, Zwicky claimed that most of the matter constituting the Coma cluster is \textbf{dark}. His result was not taken seriously during his time, and unfortunately, Zwicky died in 1974, few years before the seminal work from Rubin et al. got published. More precise analysis have later confirmed Zwicky's finding, e.g. \cite{Lokas:2003ks} which finds $M/L \simeq 351$.\footnote{The reason for such a large mass-to-light ratio is not only attributed to dark matter but also to the presence of a lot of gas. More precisely, the study in \cite{Lokas:2003ks} finds in the Coma cluster only $2\%$ of stellar mass against $13\%$ of gas and $85\%$ of dark matter.} \\

\begin{figure}[h!]
\centering
\raisebox{0cm}{\makebox{\includegraphics[width=0.9\textwidth, scale=1]{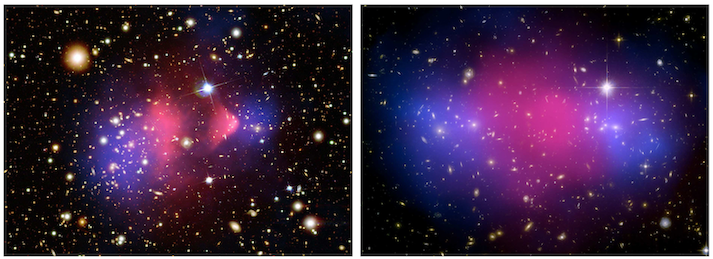}}}
\caption{\it \small The `bullet cluster' and the `baby bullet'. The background pictures show the location of galaxies in yellow. On top, the hot, intra-cluster gas in purple is traced from its X-ray emission whereas the total mass in blue is reconstructed from gravitational lensing measurement. It is found that there is far more mass than that of the galaxies, providing strong evidence for the presence of dark matter. Moreover, the coincident location of dark matter at the same place as the galaxies implies that dark matter has a similarly small interaction cross-section. Figure reproduced from \cite{Massey:2010hh}. For pictures with more quantitative informations, e.g. the gravitation lensing contours, see \cite{Clowe:2006eq}.  Figure credit: Left: X-ray: NASA/CXC/CfA/ M.Markevitch et al.; Lensing Map: NASA/STScI; ESO WFI; Magellan/U.Arizona/ D.Clowe et al. Optical image: NASA/STScI; Magellan/U.Arizona/D.Clowe et al.; Right: NASA/ESA/M.Bradac et al.}
\label{BulletCluster}
\end{figure}

\paragraph{Inter-galactic scale: Structure formation (analytical).} 
Nowadays, the mass of a cluster can be computed by at least three different methods \cite{Bahcall:1998ur}
\begin{itemize}
\item 
Measurement of the dispersion velocities of individual galaxies as initiated by Zwicky.
\item
Measurement of the temperature of the gas, which can be related to the gravitational potential assuming hydrostatic equilibrium \cite{Bertone:2004pz}.
\item
Measurement of the gravitational mass through lensing effects, cf. Fig.~\ref{BulletCluster}.
\end{itemize}
By taking the mean value of the three independent measurements explicited above, we can minimize the uncertainties. Then, we can precisely measure the number of clusters, for each mass range (an histogram), as a function of the redshift. On the theory side, this number of condensed objects for each mass range, as a function of the redshift, can be predicted, assuming a given dark matter abundance, in the Press-Schlechter formalism \cite{Press:1973iz}. The comparison between observations and analytical models allows to measure the dark matter abundance at the inter-galactic scale, e.g. $\Omega_{\rm DM} = 0.22 ^{+0.25}_{-0.1}$ \cite{Bahcall:1998ur} (study performed in 1998).\\

\paragraph{Inter-galactic scale: Structure formation (numerical).}
In the last decade \cite{Banerjee:2022qcb}, the major improvement of the computational power has permitted \textbf{N-body simulations} of large portions of universe, e.g. the 2014 Illustris project \cite{Vogelsberger:2014dza} which covers a volume $(106.5~\rm Mpc)^3$ with mass resolution as low as $10^6~M_{\rm \odot}$ and includes baryonic effects as stellar feeback or supermassive black hole growth, or the 2015 APOSTLE project \cite{Sawala:2015cdf} which simulates the Local group environment. As shown in Fig.~\ref{fig:SFRD_DM_evidence}, the prediction by Illustris of the star formation rate as a function of the redshift is in good agreement with observations. This illustrates the success of the $\Lambda$CDM cosmological model where $26\%$ of the energy density of the universe is contained under the form of cold dark matter and $5\%$ under the form of baryons.\\

\paragraph{Inter-galactic scale: Gravitational lensing.} 
By studying the photon trajectory in Schwarzschild geometry, we find that the \textbf{deflection angle} $\Delta \phi$ of a light ray passing nearby an object of mass $M$ with an impact parameter $b$ is \cite{Hobson:2006se}
\begin{equation}
\Delta \phi = \frac{4GM}{b}.
\end{equation}
When considering full images, bending of light leads to \textbf{gravitational lensing} which can be used to detect the presence of Dark Matter.  By analysing the distortion of the light emitted by the background, we can map the mass distribution of the dark matter in the foreground. The lensing is said \textbf{strong}, when we observe multiple images of the same physical source. If the foreground object responsible for the distorsion is spherically symmetric, we can even observed \textbf{Einstein rings} \cite{Massey:2010hh}. In the \textbf{weak} lensing regime, we can observe magnification and shear.

Well-known examples of weak lensing are the \textbf{bullet cluster} and the \textbf{baby bullet} shown in Fig.~\ref{BulletCluster}. On those images, we can see two clusters after they have collided.  On the one hand, the gas traced by its X-ray emission in purple, is found mostly on the collision site at the center of the image. On the other hand, the two dark matter halos traced by gravitational lensing in blue are found at large distance from the collision site, as so for the galaxies in yellow. We deduce that in contrast to the gas which has strongly interacted during the collision, the dark matter halos have passed through each other. Hence, besides revealing the existence of dark matter, the analysis provides an upper bound on the self-interaction cross-section of dark matter \cite{Kahlhoefer:2013dca,Harvey:2015hha, Robertson:2016xjh}
\begin{equation}
\label{eq:bullet_cluster_bound}
\frac{\sigma}{\MDM}  \lesssim 1 ~\rm cm^2/g.
\end{equation}

\paragraph{Cosmological scale.}
\label{eq:cosmo_scale}

\begin{figure}[h!]
\centering
\raisebox{0cm}{\makebox{\includegraphics[width=0.5\textwidth, scale=1]{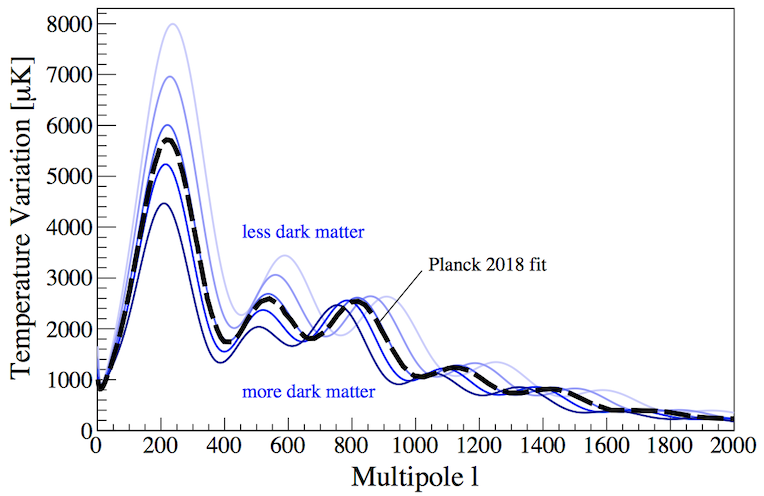}}}
\caption{\it \small Power spectrum of the temperature anisotropies in the Cosmic Microwave Background (CMB), cf. Eq.~\eqref{eq:cl}, for a dark matter density contribution $\Omega_{\rm DM}$ varying between $0.11$ and $0.43$ (blue lines). The dotted black lines show the best fit to the Planck data \cite{Aghanim:2018eyx}. The small (large) multipoles $l$ correspond to the long (small) angular scales: the main acoustic peak at $l=200$ corresponds to $1~$degree on the sky whereas $l=1800$ corresponds to $0.1~$degree. Figure reproduced from \cite{Schumann:2019eaa}.}
\label{fig:Planck_DM_evidence}
\end{figure}


The observation of the temperature anisotropies imprinted in the \textbf{Cosmic Microwave Background (CMB)} , the furthest photons we can observe in the universe, by the satellite Planck from 2009 to 2013, see Fig.~\ref{fig:Planck_2015} and Fig.~\ref{fig:CMB_Planck2018}, provides the most precise measurement of Dark Matter abundance.
As shown in Fig.~\ref{fig:Planck_DM_evidence}, the amplitudes and positions of the peaks of the temperature power spectrum depend crucially on the DM abundance. In the Planck 2018 paper  \cite{Aghanim:2018eyx}, we read the dark matter and baryon density in the universe 
\begin{equation}
\Omega_{\rm DM}h^2 = 0.1200 \pm 0.0011, \quad \Omega_{\rm b}h^2 = 0.02237 \pm 0.00015,
\end{equation}
where $h = 0.674\pm 0.005$ is the Hubble constant in unit of $100~$km/s/Mpc.
In order to further reduce the uncertainties, it is useful to combine the CMB analysis with the measurement of the scale of \textbf{Baryonic Acoustic Oscillations} (which is around $147~$Mpc today) \cite{Aghanim:2018eyx}
\begin{equation}
\Omega_{\rm DM}h^2 = 0.11911 \pm 0.00091, \quad \Omega_{\rm b}h^2 = 0.02242 \pm 0.00014.
\end{equation}

\begin{figure}[h!]
\centering
\raisebox{0cm}{\makebox{\includegraphics[width=0.9\textwidth, scale=1]{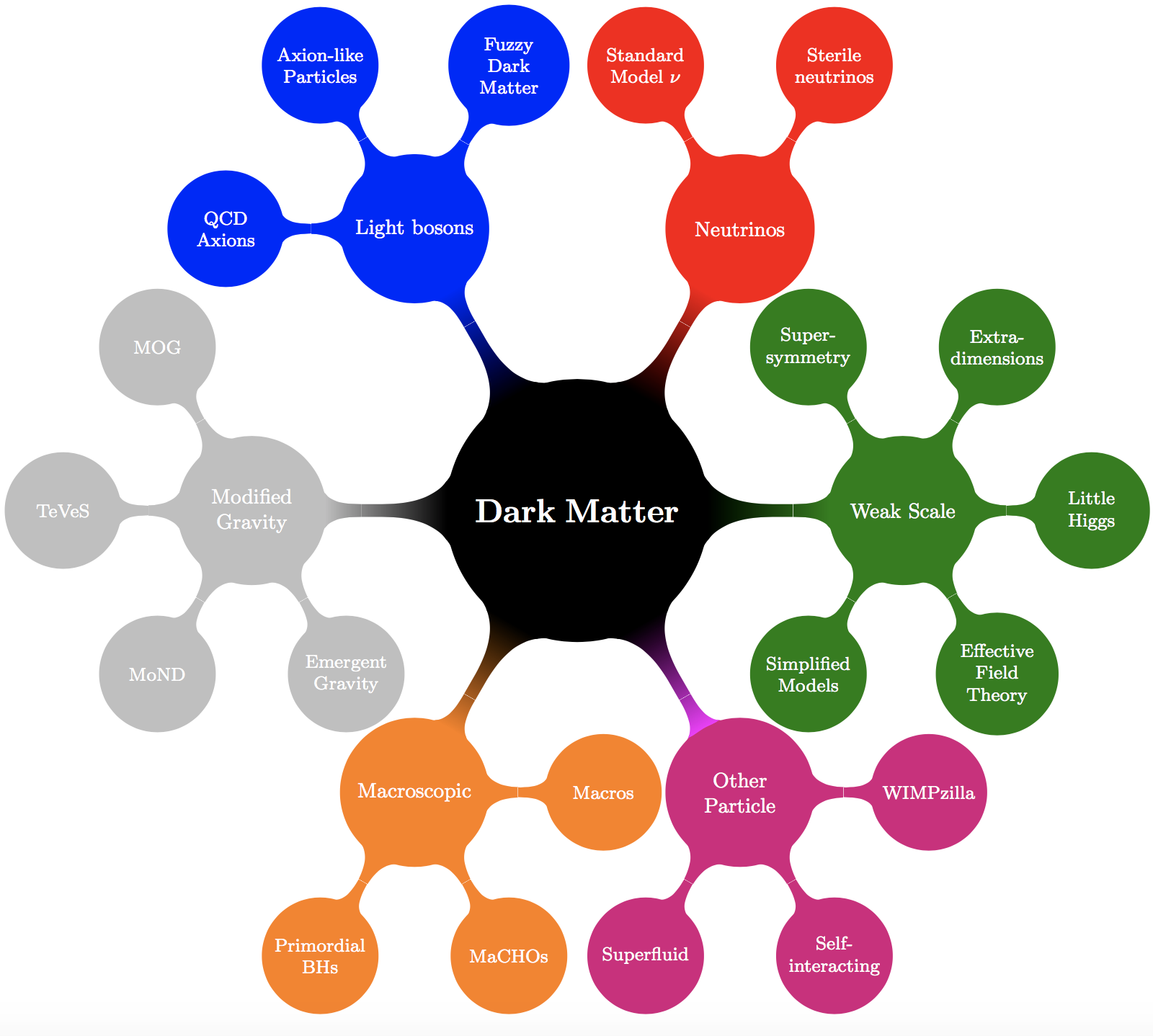}}}
\caption{\it \small Most studied DM candidates. Figure reproduced from \cite{Bertone:2018krk}. }
\label{fig:DM_candidates_landscape}
\end{figure}

\begin{figure}[h!]
\centering

\raisebox{0cm}{\makebox{\includegraphics[width=0.75\textwidth, scale=1]{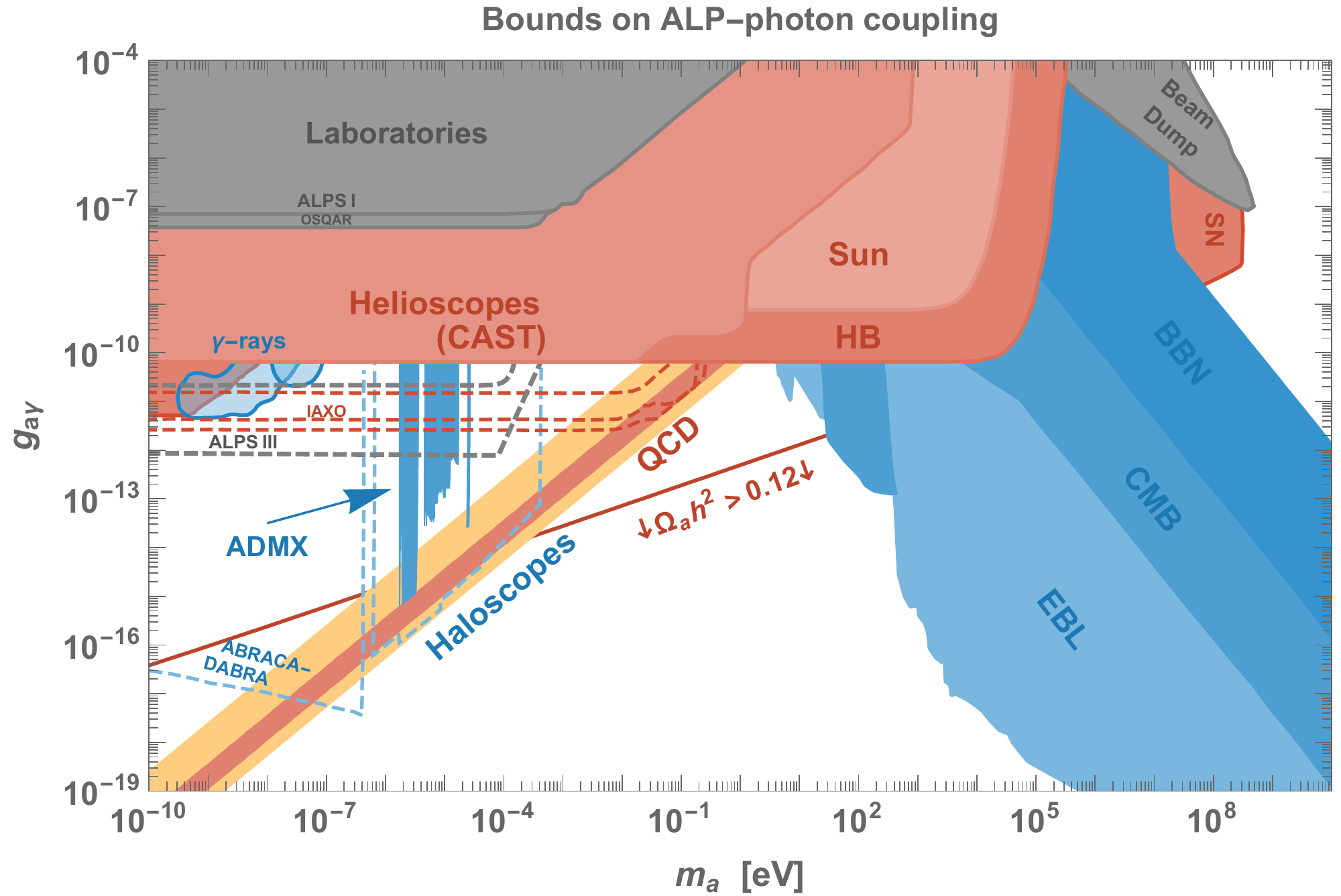}}}
\caption{\it \small Current constraints and prospects on the coupling $g_{a\gamma\gamma}~\rm [GeV]$ between the photon and a CP-odd axion-like particle $\mathcal{L}\supset g_{a\gamma\gamma} a \tilde{F}F$ \cite{DiLuzio:2020wdo}. The red line labeled `QCD' shows where the Peccei-Quinn QCD axion is supposed to lie, based on the $m_a-f_a$ relation in Eq.~\eqref{eq:ma_fa_QCD_axion}, $m_a = (75.5~\rm GeV)^2/f_a$, and $ g_{a\gamma\gamma} = \frac{\alpha_{\rm e.m.}}{2\pi f_a}(1.92 - \rm \frac{E}{N})$ where $1.92$ is induced by the mixing between the Peccei-Quinn axion and the SM $\pi^0$ and $\eta^0$ \cite{QuilezLasanta:2019wgl}, while $\rm E/N$ depends on the UV realization. E.g. in KSVZ model, $\rm E$ and $\rm N$ are the anomaly coefficients of heavy Peccei-Quinn fermions in the background of $U(1)_{\rm e.m.}$ and $SU(3)_c$. The second red line shows the region where the axion abundance overcloses the universe, assuming standard misalignement mechanism and that  the Peccei-Quinn symmetry is broken after inflation.}
\label{fig:axion_landscape_constraints}
\end{figure}

\begin{figure}[h!]
\centering
\raisebox{0cm}{\makebox{\includegraphics[width=0.75\textwidth, scale=1]{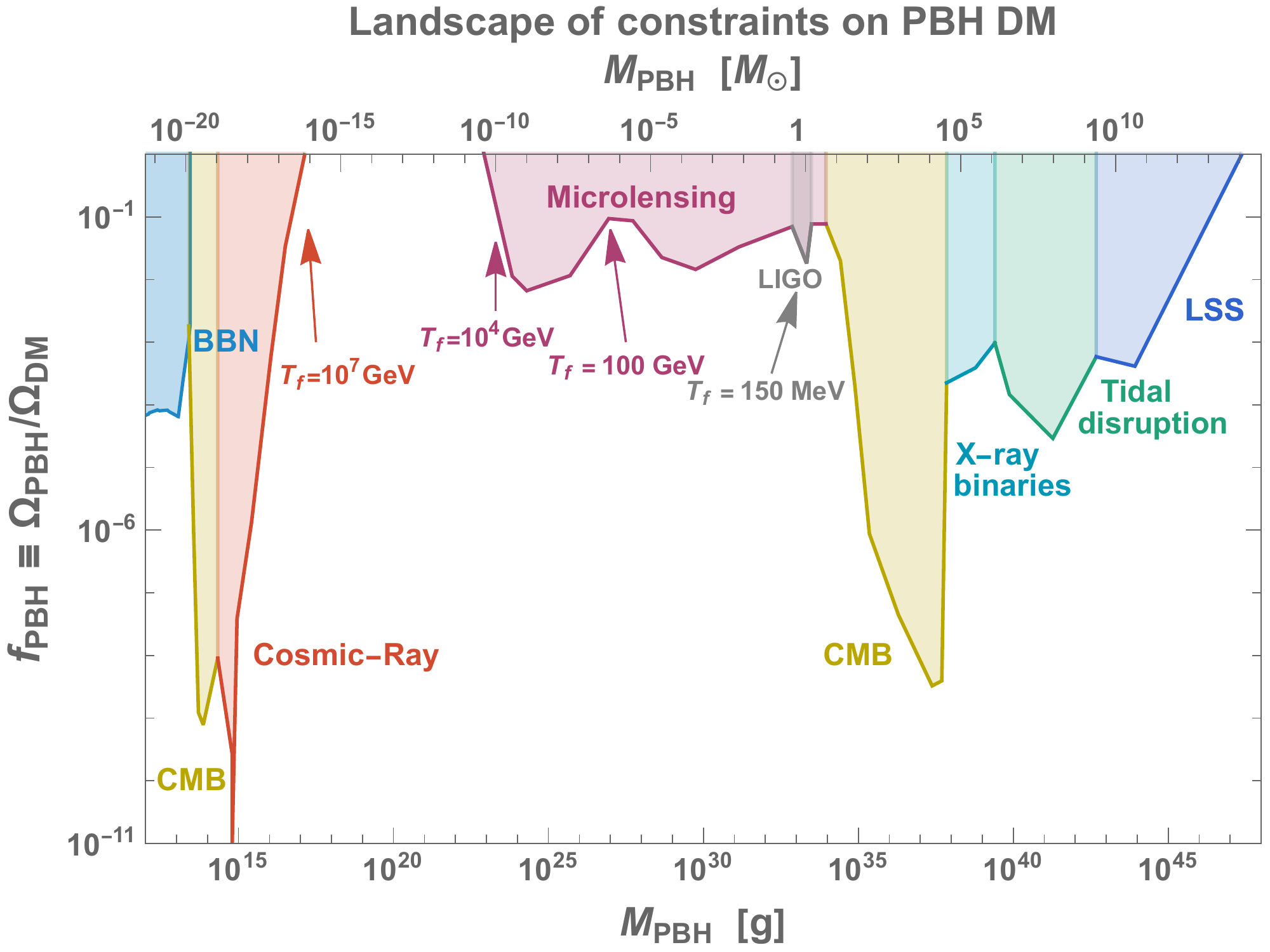}}}
\caption{\it \small Constraints on the PBH fraction of DM  \cite{Carr:2020gox, Green:2020jor,Carr:2020xqk}. The scenario where $100\%$ of DM is made of PBH with a mass within $[10^{17}~{\rm g},~ 10^{23}~\rm g]$ is compatible with experiments. The mechanism producing those BHs would have to take place within the temperature range $[10^{4}~{\rm GeV},~10^7~\rm GeV]$, where we assumed the PBH mass to be the energy contained within the acoustic horizon $c_s H^{-1}$ at the time of their formation, with $c_s=1/\sqrt{3}$. There are claims that BH observed in LIGO/VIRGO data might be of primordial origins \cite{DeLuca:2020qqa, DeLuca:2020sae,Bhagwat:2020bzh,DeLuca:2021wjr,Franciolini:2021tla}. The mechanism responsible for their formation would have to take place around the QCD scale. }
\label{fig:PBH_landscape_constraints}
\end{figure}

\paragraph{Dark Matter candidates.}

There are very numerous DM candidates from masses as small as $10^{-20}~$eV for the ultra-light bosons to masses as large as $10^{-11}~M_{\odot}=10^{46}~\rm eV$ for primordial black holes (PBH). Fig.~\ref{fig:DM_candidates_landscape} shows the most studied DM candidates \cite{Baer:2014eja, Bertone:2018krk}. We discuss some of them below.

\paragraph{Thermal DM:} 
Thermal Dark Matter is motivated by the simplicity of its production mechanism which makes the final abundance only depend on the annihilation cross-section, and by its ubiquity in theories aiming at solving the electroweak hierarchy problem, see Sec.~\ref{sec:hierarchy_pb} in Chap.~\ref{chap:SM_particle}. In view of its importance, we dedicate the whole Chap.~\ref{chap:DM} to thermal DM.

\paragraph{QCD axion:} 
A second well-studied DM candidate is motivated by the strong CP problem, see Sec.~\ref{sec:strong_CP_pb} in Chap.~\ref{chap:SM_particle}. Future experiments aiming at searching for the QCD axion or more generally, for any Axion-Like-Particle, seem promising, e.g Casper, Abracadabra, ALPS II, the IAXO helioscope and a series of haloscopes, see Fig.~\ref{fig:axion_landscape_constraints}. We refer the reader to \cite{Preskill:1982cy,Abbott:1982af,Dine:1982ah} for original articles and \cite{Sikivie:2006ni,Wantz:2009it,Kim:2008hd,Arias:2012az,Kawasaki:2013ae,Ringwald:2014vqa,Graham:2015ouw,Marsh:2015xka,Hook:2018dlk,Irastorza:2018dyq,Sikivie:2020zpn, DiLuzio:2020wdo,Agrawal:2021dbo,Adams:2022pbo} for reviews on the QCD axion and ALPs. See \cite{Agrawal:2022yvu} for a review of the open questions in axion theory.

\paragraph{Primordial black holes:} They are black holes formed during the early universe whose abundance is a substantial fraction of DM. Fig.~\ref{fig:PBH_landscape_constraints} shows the current state of the phenomenological constraints on the fraction of DM contained in PBHs.  The mass window $[10^{17}~{\rm g},~ 10^{23}~\rm g]$ is compatible with 100$\%$ of DM being under the form of PBHs. We call the attention to \cite{Garcia-Bellido:2017fdg, Sasaki:2018dmp,Carr:2020gox, Green:2020jor,Carr:2020xqk} for reviews.

We now discuss astrophysical puzzles in galaxy dynamics which challenge $\Lambda$CDM model. We discuss 5 types of DM candidates whose properties have been shown to possibly resolve some of those issues: self-interacting DM, ultra-light DM, MOND, superfluid DM and baryon-interacting DM.

\subsection{The fragility of $\Lambda$CDM}

\label{sec:fragility_LambdaCDM}

$\Lambda$CDM N-body simulations predict DM halos to follow the Navarro-Frenk-White density profile \cite{Navarro:1995iw} (1995)
\begin{equation}
\label{eq:NFW}
\rho_{\rm NFW}(r) = \frac{\rho_0}{\dfrac{r}{r_{0}}\left(1 + \dfrac{r}{r_{0}} \right)^2}
\end{equation}
Typical parameters are e.g. $ \rho_0=24.42$~GeV/cm$^3$ and $r_{0}=0.184$~kpc in order for the local DM density at the sun position $r_{\odot} = 8.33$~kpc to be $\rho_{\odot} = 0.3 $~GeV/cm$^3$ and for the total DM mass contained within $60~$kpc to be $M_{60} \equiv 4.7\times 10^{11}~M_{\odot}$ \cite{Cirelli:2010xx}. 
However, $\Lambda$CDM N-body simulations disagrees with observations on some points which we review below. See \cite{Famaey:2011kh,Pontzen:2014lma, DelPopolo:2016emo,Bullock:2017xww, Buckley:2017ijx,Tulin:2017ara,Bechtol:2022koa} for reviews on the small-scale problems. See also the rather rich introduction in \cite{Fitts:2018ycl}.

\paragraph{The cusp vs core problem.}
The presence of a cusp in NFW (DM profile peaked as $\rho_{DM} \sim 1/r$ at the center) contradicts the observation of a core in dwarf galaxies ($\rho_{DM} \sim r^{0}$ at the center). More precisely, it has been found that the star rotation velocity in the inner part of the galaxy shows a solid-body behaviour (rises linearly with the radius) hence indicating the presence of a central core in the DM distribution \cite{Moore:1994yx,Flores:1994gz,Battaglia:2008jz,deBlok:2008wp,Walker:2011zu,Amorisco:2011hb,Amorisco:2012rd, Agnello:2012uc,Adams:2014bda,Oh:2015xoa}, although the possibility to distinguish, in the data, a core from a cusp is still debated \cite{CarlosFrenkTalk, stw3004, Oman:2017vkl}. Baryonic physics have been proposed to resolve the conflict: the center density of DM halos can be flattened by repeated gas outflows due to supernova feedback \cite{Navarro:1996bv, Read:2004xc, Pontzen:2011ty, Governato:2012fa} or dynamical friction \cite{1976MNRAS.174...19W, ElZant:2001re,Tonini:2006gwz,RomanoDiaz:2008wz}. 
On the one hand, energy transfer to DM from gas outflows grows with the stellar mass, implying a minimum stellar-to-halo mass for the center halo density to be flattened \cite{Governato:2012fa, Penarrubia:2012bb}. One the other hand, effects of the gas outflow are overcome by the enlarged gravitational potential at larger stellar mass, implying also a maximum stellar-to-halo mass for the center halo density to be flattened \cite{ DiCintio:2013qxa, DiCintio:2014xia}. The dependence of the core creation with the mass in stars has been verified later by hydrodynamical simulations, e.g. NIHAO \cite{Tollet:2015gqa} and FIRE-2 \cite{Fitts:2016usl, Hopkins:2017ycn}, see however \cite{Read:2015sta} who find cores for all stellar masses or \cite{Sawala:2015cdf} who find no cores at all.

\paragraph{The missing satellite problem.}
High-resolution hydrodynamical simulations predict a mass spectrum of DM halos which is self-similar and grows as a power law up to the numerical cut-off, hence predicting thousands of visible MW satellites \cite{Springel:2008cc, Stadel:2008pn, Garrison-Kimmel:2013eoa, Griffen:2016ayh}. However, only $\sim 50$ satellites in the local group (MW+M31) have been observed so far \cite{Tollerud:2008ze, Hargis:2014kaa, Newton:2017xqg}. The missing satellite problem might be explained by inhibition of galaxy formation in low-mass halo in the presence of photoionizing background due to stars and quasars \cite{Bullock:2000wn, Benson:2001au,Somerville_2003,OkamotoMassLoss}, or by ram pressure stripping and tidal effects due to interaction with central hosts \cite{Zolotov:2012xd,Arraki:2012bu}. Hence, DM halos become increasingly inefficient at making galaxies at low masses.

\paragraph{The too-big-to-fail (TBTF) problem.}
The missing satellite problem can be solved by assigning the known MW satellites to the largest DM halos by abundance matching and by attributing the lack of observed satellites to the inefficiency of smaller subhalos to form galaxies. The problem is that the largest DM halos predicted in simulations are too dense compared to the brightest observed satellites. More precisely, $\Lambda$CDM N-body simulations predict at least $10$ subhalos with $V_{\rm max}>25$~km/s in the satellite population of the MW, and we don't observe any of them \cite{BoylanKolchin:2011de, BoylanKolchin:2011dk}. Note the claim of \cite{Purcell:2012kd} that TBTF is not a statement of high statistical significance. Later, the TBTF problem has been found in Andromeda (M31) \cite{Tollerud:2014zha} as well as in satellites `in the field', beyond the virial radius of the MW and Andromeda \cite{Kirby:2014sya}. The existence of TBTF beyond the influence of MW and M31 implies that environment effects, e.g. ram pressure stripping and tidal interactions, which are absent for field satellites, are unlikely to explain the TBTF problem \cite{Garrison-Kimmel:2014vqa}. Another proposal is to lower the center density of DM halo using repeated gas outflows due to supernova feedback \cite{Navarro:1996bv, Read:2004xc, Pontzen:2011ty, Governato:2012fa} or dynamical friction \cite{1976MNRAS.174...19W, ElZant:2001re,Tonini:2006gwz,RomanoDiaz:2008wz}. Evidences of gas outflows include the observation of blue-shifted absorption lines \cite{Heckman:2000sj, Martin:2012mh} or the observation of bulgeless galaxy \cite{Governato:2009bg}. High-resolution hydrodynamics simulations of the Local Group including star formation and stellar/supernova feedback, e.g. APOSTLE \cite{Sawala:2015cdf} or FIRE-2 \cite{Hopkins:2017ycn,LGonFIRE}, find no evidence for missing satellite and TBTF problem, which confirms the results from earlier simulations \cite{Governato:2012fa, Brooks:2012ah, Brooks:2012vi} and seems to clear up the conflict whether SN feedback is likely to inject enough energy into the DM halo \cite{Amorisco:2013uwa} or not \cite{Gnedin:2001ec, GarrisonKimmel:2013aq}. The crucial feature of baryonic feedback is that energy injection in the DM halo is additive over multiple SN burst events occurring during an extended period of time \cite{Pontzen:2014lma}. \\ However, the existence of TBTF problem for satellites in the field \cite{Kirby:2014sya}, where external feedback (environment effects) is absent, and below the critical stellar mass $10^6~M_{\odot}$, where internal feedback (e.g. SN feedback) is inefficient, remains to be explained. A possibility is the ram pressure induced by the flow of gas in filaments and pancakes of the DM cosmic web \cite{BenitezLlambay:2012by}. Another possibility is the presence of biases in the measurement of the rotation curves of ultra-faint dwarfs \cite{papastergis2017testing}. 
Whereas the TBTF problem can be interpreted as a new formulation of the missing satellite problem in light of the abundance matching between brightest dwarfs satellites and densest DM sub-halos, the cusp vs core problem can be uncorrelated, e.g. \cite{Sawala:2015cdf} which solves the TBTF problem with only cuspy profiles. \\

\paragraph{The diversity problem.}
Rotation curves are observed to have a large scatter at small radius for a fixed value of the velocity at large radius $V_{\rm max}$ \cite{Burkert:1995yz, Persic:1995ru, Navarro:2008kc, Oman:2015xda}. DM self-interactions are believed to lead to the formation of a core whose size increases with the strength of the self-interaction cross-section. The size of the core is also tight to the baryon density: an increase of the baryon density enlarges the gravitational well and then decreases the core size. For a given halo mass ($V_{\rm max}$), different baryon contents corresponding to different galaxy formation histories lead to different inner rotation curves. This is how the diversity problem is addressed in \cite{Ren:2018jpt}.  The diversity problem is also well addressed with DM self-interactions replaced by baryonic feedback in \cite{Katz:2016hyb}.

\paragraph{Proposed solution: Self-interacting DM.}
In the text above, we have reviewed solutions to the small-scale problems involving  baryonic physics. 
An alternative approach, proposed by Spergel \& Steinhardt in 1999 \cite{Spergel:1999mh}, is to go beyond the assumption that DM is collisionless and instead to postulate that DM is self-interacting.  The inclusion of DM collisions in N-body simulations, e.g.  \cite{Rocha:2012jg, Peter:2012jh, Vogelsberger:2012ku,Zavala:2012us,Elbert:2014bma,Vogelsberger:2014pda,Fry:2015rta,Dooley:2016ajo,Robertson:2016qef,Elbert:2016dbb,Kummer:2017bhr,Kummer:2019yrb,Robles:2019mfq}, favors a self-interaction cross-section of order  $\sigma/m \sim 1~\rm cm^2/g$. Notice that it is much larger - about $15$ orders of magnitude - than the weak-scale cross section expected for a usual WIMP DM candidate and is close to the upper bound from bullet cluster observation, cf. Eq.~\eqref{eq:bullet_cluster_bound}. 

The presence of collisions generates a cosmological pressure $p$ for the DM fluid, which can be related to the DM energy density $\rho$ and temperature $T$ using the ideal gas law
\begin{equation}
p = \frac{k_{\rm B}T}{m} \rho,
\end{equation}
where $m$ is the DM mass and $k_{\rm B}$ is the Boltzmann constant.
The Navier-Stokes equation at equilibrium reads
\begin{equation}
\frac{dp}{dr} = \frac{k_{\rm  B}T}{m} \frac{d\rho}{dr} = - \rho\frac{d\phi }{dr},
\end{equation}
Everywhere collisions are fast enough, we expect heat flow to homogenize the temperature $T=\rm cst$. This is the so-called \textbf{isothermal} limit for which the previous equation becomes
\begin{equation}
\label{eq:Jeans_eq_iso}
\frac{d\ln \rho}{dr} = - \frac{d(\phi/\sigma_v^2) }{dr},
\end{equation}
where $\sigma_v^2 \equiv \frac{k_{\rm B}T}{m}$ is the mean velocity dispersion squared. The later equation can be trivially integrated to give the density profile of an isothermal self-gravitating sphere \cite{binney2011galactic}. 
\begin{equation}
\label{eq:isothermal_rho}
\rho_{\rm iso}(r) = \rho_0 \exp\left(\frac{\phi(0) - \phi(r)}{\sigma_v^2}\right).
\end{equation}
Gauss's theorem states that at the center of the galaxy the gravitational potential must reach a constant $\phi(r) \to \phi_0$. From Eq.~\eqref{eq:isothermal_rho}, it implies that the density must reach a plateau  $\rho_0$. Hence, while $\Lambda$CDM N-body simulations find the density profile to be peaked at the center $\rho \propto r^{-1}$, cf. Eq.~\eqref{eq:NFW}, the presence of self-interaction leads to a cored profile. The radius $r_0$ at which the density flattens, also called the radius of the core, can be simply estimated by dimensional analysis from Eq.~\eqref{eq:Jeans_eq_iso} together with Poisson's equation $\nabla^2 \phi = 4\pi G\rho$
\begin{equation}
 r_0 \simeq \sqrt{\frac{\sigma^2}{2\pi G \rho_0}}.
\end{equation}
Far from the galactic center where the DM density is weaker, we expect the self-interaction to be negligible and the NFW profile in Eq.~\eqref{eq:NFW} to be a good approximation. However, below the radius $r_1$  where the number of self-interactions per DM particles is larger than 1
\begin{equation}
\frac{\rho(r_1)}{m}\left<\sigma v\right> t_{\rm age}  \equiv 1,
\end{equation}
the collisionless NFW profile should be replaced by the isothermal profile in Eq.~\eqref{eq:isothermal_rho}.
The inner region where self-interactions dominates is also the region where the gravitational potential is dominated by baryons, hence leading to a correlation between the effective stellar radius  and the SIDM core size \cite{Kaplinghat:2013xca}. This analytical model allows to address the cusp vs core problem \cite{Kaplinghat:2013xca}, the diversity problem \cite{Kamada:2016euw,Creasey:2017qxc,Kahlhoefer:2019oyt}, and the regularity problem \cite{Ren:2018jpt}. 

The simplest and first particle physics model of self-interacting DM, proposed in 2000, is a quartic scalar $\lambda \phi^4$  \cite{Bento:2000ah,McDonald:2001vt}. However it is now excluded by invisible constraints of the Higgs boson which leads to DM overabundance \cite{CMS:2016dhk}. Instead the abundance could be depleted further with $3\to 2$ scattering \cite{Carlson:1992fn,deLaix:1995vi,Hochberg:2014kqa,Bernal:2015ova,Pappadopulo:2016pkp,Farina:2016llk}. 
Unfortunately a model with a quartic scalar produces a constant cross-section over all scales, which is disfavored by astrophysical data from galaxies to clusters  \cite{Harvey:2015hha,Kaplinghat:2015aga}. Indeed while $\sigma/m \sim 1~\rm cm^2/g$ is favored in dwarf galaxies, it must be less than $\sigma/m \sim  0.1~ \rm cm^2/g$ on cluster scales \cite{Harvey:2015hha,Kaplinghat:2015aga}. 

A non-trivial velocity-dependence of the self-scattering cross-section can be obtained with Yukawa interactions \cite{Feng:2009hw,Buckley:2009in,Tulin:2012wi,Tulin:2013teo,Schutz:2014nka}. However these models with light mediators face strong bounds from direct and indirect probes of dark matter  \cite{DelNobile:2015uua,Bringmann:2016din,Kahlhoefer:2017umn}, except if DM has a large asymmetric component \cite{Baldes:2017gzu}.  Other proposals for generating velocity-dependant scattering properties are models involving atomic interactions \cite{Cline:2013pca,Boddy:2016bbu}, a Breit-Wigner resonance \cite{Chu:2018fzy,Chu:2019awd}, or strongly-coupled interactions \cite{Boddy:2014yra,Soni:2016gzf,Prilepina:2016rlq,Kribs:2016cew}.

Another solution is to make DM warmer in order to damp structure formation at small scales. Warm DM was first proposed in 2000 \cite{Bode:2000gq,Dalcanton:2000hn,Zentner:2003yd}. However, the resolution of small scale problems which suggests $m_{\rm WDM}\sim 2~$keV \cite{Smith:2011ev,Lovell:2011rd, Anderhalden:2012jc, Lovell:2013ola}, is disfavored by Lyman-$\alpha$ forest observations which imposes $m_{\rm WDM} \gtrsim 3~$keV \cite{Viel:2013apy, Schneider:2013wwa}. Other proposals for damping structure formation at small scales using DM heating is delayed DM kinetic decoupling \cite{Feng:2009hw,vandenAarssen:2012vpm,Ahlgren:2013wba,Laha:2013xua,Bertoni:2014mva,Chu:2014lja,Buckley:2014hja,Tang:2016mot,Agrawal:2017rvu}, dark matter - radiation interactions \cite{Boehm:2014vja,Schewtschenko:2015rno}, or the presence of a phase of DM cannibalism \cite{Carlson:1992fn,deLaix:1995vi,Hochberg:2014kqa,Bernal:2015ova,Pappadopulo:2016pkp,Farina:2016llk}.

We refer to \cite{Tulin:2017ara} for a review on the particle solution to the small scale crisis.

\paragraph{Proposed solution: Ultra-light DM.} 
They are DM candidates with a mass smaller than $m \lesssim 5~\rm eV$. In the galaxy, their particle number  per de Broglie wavelength $\lambda_{\rm dB} = 1/mv$ is much larger than 1 
\begin{equation}
n_{\rm DM}\lambda_{\rm dB}^3~ =~ \frac{\rho_{\rm DM}}{m}\lambda_{\rm dB}^3 ~=~ 10 \left(\frac{\rho_{\rm DM}}{0.3~\rm GeV/cm^3} \right) \left( \frac{5~\rm eV}{m} \right)^4\left( \frac{220~\rm km/s}{v} \right)^3 ~\gg ~1.
\end{equation}
Obviously, Fermi-Dirac statistic forbids ULDM to be a fermion, as already stated by Tremaine and Gunn in 1979 \cite{Tremaine:1979we}.\footnote{Careful analysis of the phase space distribution of DM in Dwarf galaxies implies the lower bound on the mass of fermion DM to be $m_{\rm DM} \gtrsim 1~\rm keV$ \cite{Boyarsky:2008ju}.}
For a large occupation number of a phase-space cell, the particle description breaks down and must be replaced by a wave description. A first possibility is the Schrodinger-Poisson system of equations
\begin{align}
\label{eq:schrodinger_ULDM}
&i\dot{\psi} = - \frac{1}{2m}\nabla^2\psi + m\Phi \psi - \frac{3}{2}iH\psi  +  g_2 |\psi|^2 \psi+  g_3 |\psi|^4\psi + \cdot \\
&\nabla^2 \Phi = 4\pi G(\rho - \overline{\rho}).
\end{align}
where $\Phi$ is the gravitational potential and the $g_i$ are the coupling constants of $i$-body self-interactions. In the literature, the case $g_i = 0$ is called Fuzzy DM (FDM) while models with $g_i \neq 1$ are called self-interacting fuzzy DM (SIFDM). In the long-wavelength limit, the field theory can be replaced by a set of hydrodynamical-like equations using the Madelung decomposition \cite{madelung1927quantum}
\begin{equation}
\psi \equiv \sqrt{\frac{\rho}{m}} e^{i\theta}, \qquad \mathbf{v} = \frac{1}{a m}\mathbf{\nabla}\theta = \frac{1}{2ima}\left(\frac{1}{\psi}\nabla\psi - \frac{1}{\psi^{*}}\nabla \psi^{*}\right).
\end{equation}
The equations of motion for $\psi$ become
\begin{align}
&\dot{\rho} + 3H \rho + \frac{1}{a}\mathbf{\nabla} \cdot (\rho \mathbf{v}) = 0,\\
&\rho\dot{\mathbf{v}} +\rho\, H \mathbf{v} + \frac{\rho}{a}(\mathbf{v} \cdot \mathbf{\nabla}) \mathbf{v} = - \frac{\rho}{a} \nabla \Phi + \nabla\left( P_{\rm int} + P_{\rm QG} \right),
\end{align}
with 
\begin{align}
&P_{\rm int} =  \sum K_j \rho^{(j+1)/j} =  g_2 n^2 +  g_3  n^3+\cdots, \\
& \nabla P_{\rm QG} = \frac{1}{2a^3 m^2} \nabla \left(\frac{\nabla^2 \sqrt{\rho}}{\sqrt{\rho}} \right).
\end{align}
$P_{\rm int} $ is the pressure due to self-interactions with $j$ being the polytropic index. $P\propto n^2$ is generated from 2-body interactions, $P\propto n^3$ from 3-body interactions \cite{Sharma:2018ydn}, and so on. The term $P_{\rm QG}$ which depends on the curvature of the amplitude of the wave function is the quantum pressure. It plays the role of a repulsive interaction which can be understood as arising from Heisenberg uncertainty principle. Interests for Fuzzy DM started in 2000 after that Hu, Barkana and Gruzinov \cite{Hu:2000ke} proposed to use the quantum pressure $P_{\rm QG}$ to address the incompatibility between astrophysics observation and $\Lambda$CDM cosmology at small scales. Indeed, the impact of $P_{\rm QG}$ is to smoothen structures below the de Broglie wavelength
\begin{equation}
\lambda_{\rm dB} = 2~{\rm kpc}\left( \frac{10^{-22}~\rm eV}{m} \right) \left( \frac{10 ~\rm km/s}{v} \right).
\end{equation}
However, $m\sim 10^{-22}~\rm eV$ is in tension with Lyman-$\alpha$ forest observations which imply $m \gtrsim 2 \times 10^{-22}~\rm eV$ \cite{Rogers:2020ltq}, and raise doubts on the possibility for fuzzy DM to address small scale problems. See also the constraints coming from CMB \cite{Hlozek:2014lca} and LSS \cite{Lague:2021frh}. For more details, we refer the reader to existing reviews on astrophysical probes of ULDM \cite{Suarez:2013iw,Hui:2016ltb,Urena-Lopez:2019kud,Niemeyer:2019aqm,Ferreira:2020fam,Marsh:2021lqg}.
Even if the resolution of small scale problems by ULDM becomes completely ruled out in the near future, ULDM remains an interesting scenario due to the ubiquity of ultra light scalars in the string landscape, with possibly axion-like or dilaton-like coupling to SM \cite{Arvanitaki:2009fg,Cicoli:2012sz,Mehta:2020kwu,Mehta:2021pwf,Demirtas:2021gsq}. When composing a significant fraction of DM they can be responsible for variation of coupling constants and give signatures in atomic clocks \cite{Flambaum:2008kr,1407.3493,Arvanitaki:2014faa,VanTilburg:2015oza,Hees:2016gop,Stadnik:2016zkf,Flambaum:2018ssk,Kennedy:2020bac,Berti:2022rwn}, GW interferometers \cite{Stadnik:2014tta,Stadnik:2015xbn,Arvanitaki:2016fyj,Grote:2019uvn}, resonant mass detectors \cite{Arvanitaki:2015iga}, or BBN physics \cite{Stadnik:2015kia,Sibiryakov:2020eir}. Even when they do not constitute DM, the existence of ultra light scalars can be probed by fifth force experiments \cite{Schlamminger:2007ht,Piazza:2010ye,Graham:2015ifn,Berge:2017ovy,Hees:2018fpg} and black hole superradiance \cite{Arvanitaki:2010sy,Arvanitaki:2014wva,Brito:2017zvb,Brito:2017wnc,Stott:2018opm,Berti:2022rwn}.

\paragraph{The regularity problem.}
We now discuss another astrophysical deviation from $\Lambda$CDM whose nature is a bit different from the cusp-core, satellite, too-big-to-fail and diversity problems at small scales $<\rm kpc$. The regularity problem arises at the scale of the whole DM halo, since it deals with the asymptotic region of the rotation curves.
 Analog to Kepler's laws for planets, galaxies have shown to have their own rotation laws. Their stricking feature is their intrinsic small scatter. It seems that the diversity problem, i.e. large scatter of the raising part of the rotation curves, is correlated with the baryon distribution in such a way that tight galactic laws, with a small scatter, emerge. They can be formulated \textbf{globally}, \textbf{centrally} or \textbf{locally} \cite{LelliLPTHE}. \\

\textbf{$\bullet$~ Global law:} a tight relation between the maximal rotation velocity $V_{\rm max}$ and the total baryonic mass $M_{\rm bar}$
\begin{equation}
\label{eq:def_BTFR}
V_{\rm max}^4 \simeq g_{0}\,G\,M_{\rm bar}.
\end{equation}
It has first been known as a tight correlation between the maximum circular velocity and the stellar mass \cite{Tully:1977fu} before being generalized to the total baryonic mass \cite{McGaugh:2000sr}, and so is called the baryonic Tully-Fisher relation (BTFR). Introducing the factors $f_{\rm V} \equiv V_{\rm max}/V_{\rm vir}$ and $f_{\rm b } \equiv M_{\rm bar}/M_{\rm vir}$, $\Lambda$CDM cosmology suggests the relation \cite{McGaugh:2011ac}
\begin{equation}
V_{\rm max}^3 \propto f_{\rm V}^3 f_{\rm b}^{-1} M_{\rm bar},
\end{equation} 
with a large scatter \cite{Lelli:2015wst}. Consistency with the data requires $f_{\rm V}^3 f_{\rm b}^{-1}  \propto V_{\rm max}^{-1}$. Whereas $f_V$ remains close to $1$, the baryon content $f_{\rm b}$ can decrease with the baryonic feedback $\epsilon$, e.g. $f_{\rm b} \propto \epsilon^{-1}$ . We expect baryonic feedback to increase with the stellar mass, hence with $V_{\rm max}$, which is in contradiction with the requirement from the BTFR. In spite of this apparent contradiction, they are many attempts of recovering the BTFR in $\Lambda$CDM either with hydrodynamical simulations \cite{Keller:2016gmw, Ludlow:2016qzh} or analytical modelisations \cite{DiCintio:2015eeq,Desmond:2016azy, Navarro:2016bfs, Ren:2018jpt}, where it is claimed that the scatter from $\Lambda$CDM estimation may be overestimated. 

\textbf{$\bullet$~ Central law:} a tight relation between the central surface stellar density 
\begin{equation}
\Sigma_b(0) \equiv \int_{-\infty}^{+\infty} dz \, \rho_b(\vec{x}),
\end{equation}
where $z$ is the coordinate transverse to the disc, and the dynamical mass \cite{Lelli:2016uea}
\begin{equation}
\Sigma_{\rm dyn}(0) \equiv \frac{1}{2\pi G}\int_0^{\infty} \tfrac{V^2(R)}{R^2}dR =  \int_{-\infty}^{+\infty} dz \, \rho(\vec{x}),
\end{equation}
High-surface density galaxies, dominated by baryon, follow the 1:1 line whereas low-surface galaxies, dominated by DM, deviate from the 1:1 line with a small scatter, showing that DM and baryon are correlated
 \begin{equation}
\Sigma_{\rm dyn}(0) = 
 \begin{cases}
 \Sigma_b(0), \qquad \qquad  ~~\Sigma_b(0) > \frac{a_{0}}{G}, \vspace{0.2cm}\\
 \sqrt{\frac{2}{\pi} \frac{a_0}{G} \Sigma_b(0)},\qquad  \Sigma_b(0) < \frac{a_{0}}{G},
 \end{cases}
 \end{equation}
where the critical acceleration scale $a_{0}$ is measured to be $a_0\simeq 1.2 \times 10^{-8}~ \rm m^2/s$.
The central law is called the central surface density relation (CSDR). 

\textbf{$\bullet$~ Local law:} a tight relation between the radial acceleration traced by rotation curves
\begin{equation}
g_{\rm obs} \equiv \frac{v^2}{R} = \frac{GM(R)}{R},
\end{equation}
and that predicted by the observed distribution of baryons
\begin{equation}
g_{\rm bar} \equiv \frac{GM_b(R)}{R}.
\end{equation}
 Large accelerations are baryon dominated and follow the 1:1 line \cite{McGaugh:2016leg}. However below the critical acceleration scale  $a_{0}\simeq 1.2\times 10^{-8}~ \rm m^2/s$, the observed acceleration $g_{\rm obs}$ scales as the square root of the baryon gravitational potential $g_{\rm bar}$
 \begin{equation}
 \label{eq:def_MDAR}
 g_{\rm obs} = 
 \begin{cases}
 g_{\rm bar}, \qquad \quad  \quad g_{\rm bar} > a_{0}, \\
 \sqrt{g_{\rm bar} a_0},\qquad   g_{\rm bar} < a_{0}.
 \end{cases}
 \end{equation}
Again the small scatter shows a tight correlation between the DM and the baryons. This local relation is called the mass discrepancy-acceleration relation (MDAR) or Milgrom's relation.
It is a powerful statement about how DM is distributed in galaxies: in regions where baryons dominate, the theory behaves like Newtonian theory, and in regions where the DM dominates, the DM mass is uniquely determined by the baryonic distribution. Also surprising is the numerical coincidence between the critical acceleration scale $a_0$ and the current expansion scale of the universe $H_0$
 \begin{equation}
  a_{0}~\simeq~ 1.2 \times 10^{-8}~ \rm m^2/s ~\simeq ~H_0/6.
 \end{equation}

\paragraph{Proposed solution: MOND.}
Probably the simplest explanation for the MDAR in Eq.~\eqref{eq:def_MDAR} (and then BTFR in Eq.~\eqref{eq:def_BTFR} follows trivially) is given by Modified Newtonian Dynamics (MOND) proposed by Milgrom in 1983 \cite{Milgrom:1983ca, Milgrom:1983pn, Sanders:2002pf}
\begin{equation}
\label{eq:Milgrom_law}
\mu\left(\frac{g}{a_0} \right) \vec{g} ~=~ \vec{g}_{\rm N},
\end{equation}
where $\vec{g} \equiv d^2\vec{x}/dt^2$, $g_{\rm N} \equiv GM/r^2$ and
\begin{equation}
 \mu(x) = 
 \begin{cases}
1, \qquad \quad  x \gg 1, \\
 x,\qquad   \quad x \ll 1.
 \end{cases}
\end{equation}
A first Lagrangian formation, called AQUAL, is proposed the following year by Bekenstein and Milgrom \cite{Bekenstein:1984tv}, which we now review.
Newtonian gravity can be derived from the non-relativistic action
\begin{equation}
\label{eq:action_newton}
S_{\rm N} = \int d^3x \,dt \,\left[\frac{\rho \vec{v}^2}{2} - \rho \Phi_{\rm N} - \frac{|\vec{\nabla} \Phi_{\rm N}|^2}{8\pi G} \right].
\end{equation}
Varying $S_{\rm N}$ with respect to space coordinates $\vec{x}$ and with respect to potential $\Phi_{\rm N}$ yields the usual equation of motion and Poisson's equation
\begin{equation}
\label{eq:eom_Poisson_Newton}
\frac{d^2\vec{x}}{dt^2}~=~-\vec{\nabla} \Phi_{\rm N}, \qquad {\rm and} \qquad
\vec{\nabla}^2 \Phi_{\rm N}~=~4\pi G \rho.
\end{equation}
Milgrom's law in Eq.~\eqref{eq:Milgrom_law} can be recovered after replacing the last term of the Newtonian Lagrangian in Eq.~\eqref{eq:action_newton} by some function $F(z)$
\begin{equation}
S_{\rm BM,grav} =  -\frac{a_0}{8\pi G}\int d^3x\, dt \, F\left(\frac{|\nabla\Phi|^2}{a_0^2}\right).
\end{equation}
This is the original Aquadratic Lagrangian (AQUAL) of Bekenstein and Milgrom  \cite{Bekenstein:1984tv}. Varying with respect to $\Phi$ leads to a generalization of Poisson's equation
\begin{equation}
\vec{\nabla} \cdot \left[\mu\left(\frac{|\vec{\nabla} \Phi|}{a_0}\right) \vec{\nabla} \Phi\right] = 4\pi G \rho,
\end{equation}
where $\mu(x) = F^{'}(z)$ and $z=x^2$. Agreement with Milgrom's law in Eq.~\eqref{eq:Milgrom_law} imposes
\begin{equation}
 F(z) = 
 \begin{cases}
z, \qquad \quad~~~  x \gg 1, \\
\frac{2}{3}z^{3/2},\qquad    x \ll 1.
 \end{cases}
\end{equation}
In the low gravity regime, the AQUAL Lagrangian reads
\begin{equation}
\label{eq:BM_action}
|\vec{\nabla} \Phi| \ll a_{0} \qquad \implies \qquad S_{\rm BM} =  -\int d^3x \,dt \,\left[ \rho \Phi + \frac{1}{12\pi G a_0}\left(|\vec{\nabla}\Phi|^2\right)^{3/2}\right],
\end{equation}
Later, other Lagrangian formulations of Milgrom's law have been proposed, either non-relatistic: QMOND \cite{Milgrom:2009ee}, or relativistic: TeVeS \cite{Bekenstein:2004ne} and BMOND \cite{Milgrom:2009gv}. MOND can  be interpreted as a DM fluid which has a dipole interaction with gravity \cite{Blanchet:2006yt,Blanchet:2009zu}. Another MOND proposal is to assume that spacetime and gravity emerge together from the entanglement structure of an underlying microscopic theory \cite{Verlinde:2010hp, Verlinde:2016toy, Lelli:2017sul}. However, MOND without DM has been challenged by measurements of gravitational lensing (e.g. bullet cluster in Fig.~\ref{BulletCluster}) and CMB anisotropies, in particular by the amplitude of the third peak of the CMB power spectrum \cite{Skordis:2005xk,Skordis:2009bf}, even though progress is still on going, e.g. \cite{Skordis:2020eui}. Lower bounds on the speed of gravitational waves following neutron star merger events have also brought additional constraints on MOND theories \cite{Skordis:2019fxt}.   

\paragraph{Proposed solution: Superfluid DM.}
From the particle theory side, the Lagrangian in Eq.~\eqref{eq:BM_action} with a fractional power can be obtained by a complex scalar field dominated by a 3-body interactions
\begin{equation}
\mathcal{L} = -|\partial_\mu \Phi|^2 - m^2|\Phi|^2 - \frac{\lambda}{3}|\Phi|^6.
\end{equation}
This theory is invariant under a global $U(1)$ symmetry with associated conservation of number of particles. We replace $\psi = \sqrt{2m}\,\Phi\, e^{imt}$ such that the Lagrangian in the non-relativistic limit becomes
\begin{equation}
\label{eq:schrodinger_equation_superfluid_1}
\mathcal{L} = \frac{i}{2}\left(\psi \partial_t \psi^{*} - \psi^{*} \partial_t \psi \right) - \frac{\vec{|\nabla }\psi|^2}{2m}- \frac{\lambda}{24m^3}|\psi|^6,
\end{equation}
whose equation of motion  is a non-linear Schrodinger equation
\begin{equation}
-i\partial_t \psi + \frac{\nabla^2 \psi}{2m} - \frac{\lambda}{8m^3}|\psi|^4 \psi = 0.
\end{equation}
The homogeneous solution describes a Bose-Einstein condensate at zero temperature
\begin{equation}
\label{eq:mu_SFDM}
\psi_0 = \sqrt{2m}\, v \, e^{i\mu t},\qquad \mu \equiv  \frac{\lambda v^4}{2m},
\end{equation}
where $\mu$ is the chemical potential. The solution spontaneously breaks the $U(1)$ symmetry because it is associated with a finite number density $n$, given by the time component of the Noether current $J^\mu = \frac{1}{m} \textrm{Im}\left(\psi^* \partial^\mu  \psi \right)$
\begin{equation}
\label{eq:number_density_SFDM}
n =\frac{1}{m}\textrm{Im}\left(\psi^* \dot{  \psi} \right) = 2 \mu v^2 = \left(\frac{8\mu m^2}{\lambda}\right)^{1/2}.
\end{equation}
Perturbations around the homogeneous solution read
\begin{equation}
\psi = \sqrt{2m}(v+\rho) e^{i(\mu t + \phi)},
\end{equation}
here $\rho$ is the perturbation of the order parameter, while $\phi$ is the Goldstone boson.
Substituting into Eq.~\eqref{eq:schrodinger_equation_superfluid_1} gives
\begin{equation}
\label{eq:schrodinger_equation_superfluid_2}
\mathcal{L} = -(\vec{\nabla}\rho)^2 + 2m(v+\rho)^2 \left[ \mu + \dot{\phi} - \frac{(\vec{\nabla}\phi)^2}{2m}\right] - \frac{\lambda}{3}(v+\rho)^6.
\end{equation}
The dispersion relation of the linearized equation of motion is 
\begin{equation}
\label{eq:dispersion_superfluid_phonon}
\omega^2 = c_s^2 k^2 + \mathcal{O}(k^4), \qquad c_s^2 = \frac{\lambda v^4}{m^2} = \frac{2\mu}{m}.
\end{equation}
The linear dispersion of phonons $\phi$ leads to the existence of superfluidity via the Landau criterion \cite{Landau:1941vsj}: an impurity moving at a subsonic velocity $v<c_s$ moves without friction.\footnote{Suppose an impurity of mass $M$ moving with initial velocity $\vec{v}_i$ which produces a phonon of momentum $\vec{k}$ and acquires the final velocity $\vec{v}_f$ after scattering. Energy-momentum conservation reads
\begin{equation}
\frac{Mv_i^2}{2}= \frac{Mv_f^2}{2} + \omega ,\qquad \textrm{and} \qquad M\vec{v}_i = M\vec{v}_f +\vec{k},
\end{equation}
The 2nd equation leads to $\vec{v}_f = \vec{v}_i-\vec{k}/M$, which after injection in the 1st equation leads to the condition that phonon emission can only occur if the initial impurity velocity is supersonic 
\begin{equation}
v_i ~= ~ \frac{c_s+k/2M}{\cos{\theta}} ~\geq~c_s.
\end{equation}
}
Integrating out the heavy mode $\rho$, we obtain an EFT for the phonons \cite{Berezhiani:2015bqa}
\begin{equation}
\label{eq:schrodinger_equation_superfluid_3}
\mathcal{L} = \frac{4}{3}m \left( \mu + \dot{\phi} - \frac{(\vec{\nabla}\phi)^2}{2m}\right) \left(\frac{2m}{\lambda}\left[\mu + \dot{\phi} - \frac{(\vec{\nabla}\phi)^2}{2m}\right]\right)^{1/2},
\end{equation}
Upon adding a baryon-phonon interaction
\begin{equation}
\label{eq:baryon_phonon_int}
\mathcal{L}_{\rm int} = -\frac{\Lambda}{M_{\rm pl}} \phi \rho_b,
\end{equation}
in the limit $\frac{(\vec{\nabla}\phi)^2}{2m} \gg \dot{\phi},\, \mu$ we obtain the Bekenstein-Milgrom Lagrangian Eq.~\eqref{eq:BM_action} with the fractional $3/2$ power and the correct sign in front of the kinetic term if 
\begin{equation}
\lambda~<~0.
\end{equation}
However because of Eq.~\eqref{eq:dispersion_superfluid_phonon}, $\lambda~<~0$ leads to an imaginary frequency and to a destruction of the condensate. More precisely, in the small perturbation regime the Lagrangian in Eq~\eqref{eq:schrodinger_equation_superfluid_3} becomes
\begin{equation}
\label{eq:linearized_SFDM_lagrangian}
\mathcal{L}_{\rm int} = \frac{n}{2}\left( \frac{4}{3}\mu +2\dot{\phi} + \frac{1}{2\mu}\left(\dot{\phi}\right)^2 - \frac{1}{m}\left(\vec{\nabla}\phi\right)^2\right).
\end{equation}
From Eq.~\eqref{eq:mu_SFDM}, $\lambda < 0$ leads to $\mu<0$ which implies a negative pressure (constant term in Eq.~\eqref{eq:schrodinger_equation_superfluid_3}) and to a ghost-like kinetic term. Negative $\lambda$ corresponds to an attractive interaction which makes the BEC unstable against collapse. It is therefore needed to add repulsive interactions to stabilise the BEC. We refer to \cite{Berezhiani:2015bqa,Khoury:2016ehj} for more details on this point. Superfluid DM is shown to provide successful model of galaxy clusters and galaxy rotation curves \cite{Hodson:2016rck,Berezhiani:2017tth}. The dissipationless nature of subsonic motion can be used to explain existing astrophysical puzzles \cite{Berezhiani:2019pzd}. Dynamics explaining dark energy can be added \cite{Khoury:2018vdv,Ferreira:2018wup}. Superfluid DM has also received criticisms \cite{Lisanti:2018qam,Lisanti:2019nmn,Mistele:2022vhh}. For more details, we refer the reader to the reviews \cite{Ferreira:2020fam,Khoury:2021tvy}.

\paragraph{Proposed solution: environment-dependent baryon-DM interactions.}
Another proposal to explain BTFR and MDAR is that baryons and DM interact together with a cross-section scaling as the inverse of the DM density and squared velocity \cite{Famaey:2017xou, KhouryTalkIAS,Famaey:2019baq} 
\begin{equation}
\sigma_{\rm int} \propto \frac{1}{\rho v^2}.
\end{equation}
We now propose to quickly review the reasoning. The phase space density $f(t,\vec{r},\vec{v})$ of the DM component obeys the Boltzmann transport equation
\begin{equation}
\frac{\partial f}{\partial t} + \vec{v}\cdot \frac{\partial f}{\partial \vec{r}} + \vec{g} \cdot\frac{\partial f}{\partial \vec{v}} = I\left[f,\,f_b\right],
\end{equation}
where $ I\left[f,\,f_b\right]$ accounts for interactions between DM particles and baryons.
The 0th, 1st and 2nd velocity moments of the Boltzmann equation, known as the continuity, Jeans and heat equations, read
\begin{align}
&\frac{\partial \rho}{\partial t} + \vec{\nabla} \cdot (\rho \vec{u}) = 0, \\
&\rho \left( \frac{\partial}{\partial t} + \vec{u} \cdot \vec{\nabla}\right) u^i + \partial_j P^{ij} = \rho \vec{g}, \\
&\frac{3}{2}\left(\frac{\partial}{\partial t} + \vec{u} \cdot \vec{\nabla}\right) T + \frac{m}{\rho}P^{ij} \partial_i u_j + \frac{m}{\rho} \vec{\nabla} \cdot \vec{q} = \dot{\varepsilon},
\end{align}
where $\vec{u} \equiv \left< \vec{v}\right>$ is the bulk velocity\footnote{We define $\left<A\right> = \frac{1}{n}\int d^3v \, f(t,\vec{r}, \vec{v})A$}, $P^{ij} \equiv \rho \left<\left(v^i-u^i\right)\left(v^j - u^j\right)\right> $ is the pressure tensor, $T\equiv \frac{m}{3}\left< \left|\vec{v} - \vec{u}\right|^2 \right>$ is the local DM temperature, and $\vec{q} \equiv \frac{1}{2}\rho\left<(\vec{v}-\vec{u})|\vec{v}-\vec{u}|^2\right>$ is the heat flux. The local heat rate $\dot{\varepsilon}$ is due to interaction with baryons. For the sake of simplicity, we assume that the DM bulk velocity is zero and that the  DM distribution is described by a Maxwell-Boltzmann distribution,
\begin{equation}
\vec{u}\simeq 0, \qquad f(\vec{r}, \vec{v}) \simeq f_{\rm MB}(\vec{r}, \vec{v}).
\end{equation} 
This implies $P^{ij} = \rho v^2 \delta^{ij}$ and $v=\sqrt{T/M}$. The heat flux can be obtained by perturbing the M-B distribution
\begin{equation}
\vec{q} \simeq - \kappa \vec{\nabla}T, \qquad \textrm{with} \quad \kappa = \frac{3}{2}n \frac{l^2}{t_{\rm relax}}, 
\end{equation}
where $\kappa$ is the thermal conductivity with $l$ the system size and $t_{\rm relax}$ the time scale for DM particle to lose memory of its initial velocity. For dilute system like galaxies, the relaxation time is set by the dynamical time or Jeans time, $t_{\rm relax} \sim t_{\rm dyn} \sim 1/\sqrt{G\rho}$, so we have $l \sim v \,t_{\rm dyn}  \sim r$ and 
\begin{equation}
\kappa ~\sim ~n \,r \,v .
\end{equation}
The heating rate can be written
\begin{equation}
\dot{\varepsilon} = -\Gamma_{\rm int} \epsilon, \qquad \Gamma_{\rm int}  = n_b \sigma_{\rm int} v,\qquad \epsilon \simeq \frac{m m_b}{m+m_b}v^2,
\end{equation}
where $\sigma_{\rm int}$ is the DM-baryon scattering cross-section and $\epsilon$ is the exchanged energy per scattering. We now show that the following ansatz reproduces the TBFR \cite{Famaey:2017xou, KhouryTalkIAS,Famaey:2019baq}
\begin{equation}
\label{eq:master_relation_Khoury}
\sigma_{\rm int} = \frac{C a_0 m_b}{n \epsilon} \simeq \frac{m+m_b}{m}\frac{C a_0}{n v^2}, 
\end{equation}
with $a_0 \simeq 1.2 \times 10^{-8}~{\rm m/s^2}$ and $C=\mathcal{O}(1)$.
Poisson and approximated Jeans and heat equations read
\begin{align}
&\vec{\nabla}\cdot \vec{g} = -4\pi G(\rho + \rho_{b})\quad \implies \quad g = - \frac{G\left(M+M_b\right)}{r^2}, \\
&\frac{1}{\rho}\partial \left(\rho v^2\right) = g,  \\
&  \vec{\nabla} \cdot \left(\frac{\rho r v}{a_0} \vec{\nabla} \left( v^2\right) \right) = C v \rho_b.
\end{align}
The heat equation in spherical coordinate gives 
\begin{equation}
\frac{1}{r^2}\frac{d}{dr}\left(\rho v r^3 \frac{dv^2}{dr} \right) = C a_0 v \rho_b, \qquad \implies \qquad \rho r \frac{d v^2}{dr} \sim C a_0 \frac{M_b}{4\pi r^2}.
\end{equation}
The Jeans equation in DM-dominated region $M(r) \gg M_b(r)$ reads
\begin{equation}
\rho(r) \sim \frac{v^2(r)}{2\pi G r^2},
\end{equation}
which after substituting in the previous equation gives the BTFR, cf. Eq.~\eqref{eq:def_BTFR}
\begin{equation}
v^{4}(r) \sim C a_0 G M_b(r),
\end{equation}
up to a log factor which we neglected. We refer the reader to the original articles \cite{Famaey:2017xou, KhouryTalkIAS,Famaey:2019baq} for more details. These are exploratory works and further studies are needed to find a particle physics model giving rise to the master DM-baryon scattering cross-section in Eq.~\eqref{eq:master_relation_Khoury}. A proposal is superfluid DM. Indeed, upon canonicalizing the gradient term of the linearized lagrangian in Eq.~\eqref{eq:linearized_SFDM_lagrangian}, the baryon-phonon interaction in Eq.~\eqref{eq:baryon_phonon_int} becomes
\begin{equation}
\label{eq:baryon_phonon_int_cano}
\mathcal{L}_{\rm int} = -\frac{\Lambda}{M_{\rm pl}} \phi \rho_b \qquad \implies \qquad  \mathcal{L}_{\rm int}^{\rm canon.} = - \frac{2m}{n}\frac{\Lambda}{M_{\rm pl}} \tilde{\phi} \rho_b ,
\end{equation}
which has the $1/n$ behavior of the master relation in Eq.~\eqref{eq:master_relation_Khoury}.

In the remaining part of this chapter, we discuss two other cosmological (or astrophysical) problems which should not be put on the same footing as the previous ones and may end up being attributed to experimental issues.

\subsection{The Hubble tension}

\label{sec:H0_tension}

The $H_0$ tension results from the $\sim 5 ~\sigma$ discordance between the measurements of the Hubble parameter today $H_0$ using late time observables and the measurements using early time observables.
\begin{enumerate}
\item
\textbf{Late time:} The expansion parameter is inferred from the luminosity-distance versus redshift relation of Cepheid-calibrated\footnote{Cepheids are stars whose brightness oscillates with a period precisely correlated with its amplitude through the period-luminosity relation (PLR). They are more accurate standard candles than SNe Ia, but since they are less bright, they can not be observed at high redshift in contrast to SNe Ia. They are useful for calibrating the SNe Ia.} \textbf{Supernova Ia}, also known as the local distance ladder. Particularly, the
SHOES collaboration has determined \cite{Riess:2019cxk}
\begin{equation}
H_0 = 74.03 \pm 1.42 ~ \rm km/s/Mpc.
\end{equation}
Another solid low-redshift observable is the \textbf{time-delays of strongly-lensed quasars} such as measured by H0LiCOW \cite{Wong:2019kwg}
\begin{equation}
H_0 = 73.3^{+1.7}_{-1.8} ~\rm km/s/Mpc.
\end{equation}
\item
\textbf{Early time:} the high-precision measurement of the \textbf{CMB}, by Planck leads to (Planck 2018 \cite{Aghanim:2018eyx})
\begin{equation}
H_0 = 67.3 \pm 0.6 ~\rm km/s/Mpc.
\end{equation}
This is supplemented at high redshift by the observation of the \textbf{Baryonic Acoustic Oscillations} in galaxy redshift surveys and Lyman-$\alpha$ forests (also known as inverse distance ladder) calibrated against \textbf{Big-Bang Nucleosynthesis}  (deuterium + helium abundance) \cite{Schoneberg:2019wmt} (see also  \cite{Abbott:2017smn})
\begin{equation}
H_0 = 68.3^{+1.1}_{-1.2}~{\rm km/s/Mpc}.
\end{equation}
\end{enumerate}
The robustness of the Hubble tension lies on the fact that it is verified by different independent observables, see \cite{Verde:2019ivm} for a review on the different $H_0$ measurements.
Therefore, proposed solutions which focus on only one probe, like dimming the supernovas \cite{Csaki:2001yk} or introducing a fifth force which increases the gravity on the Cepheids in order to change the luminosity-profile relation and bias the SNe Ia measurement, are difficult to reconcile with other late-time probes.

The Hubble tension can be recast as a mismatch between $H_0$ at low redshift, and the sound horizon at high redshift \cite{Bernal:2016gxb}, see however \cite{Jedamzik:2020zmd}.
Early time observables, BAO and CMB, measure the Hubble parameter\footnote{The location of the first peak of the CMB power spectrum is given by $l \simeq \pi/\theta(z)$.} $H_0$  through the \textbf{sound horizon angle} \cite{Schoneberg:2019wmt}
\begin{equation}
\theta_s(z)\Big|_{\rm obs} \equiv \frac{r_s}{D_{\rm A}} = \frac{\int_{z_{\rm D}}^{\infty} d\tilde{z} ~c_s(\tilde{z})~ H(\tilde{z})^{-1}}{\int_0^z d\tilde{z} ~ H(\tilde{z})^{-1}} \simeq \frac{\int_{z_{\rm D}}^{\infty} d\tilde{z} ~c_s(\tilde{z}) \left[  \Omega_r/\Omega_m(1+\tilde{z})^4) + (1+\tilde{z})^3\right]^{-1/2}}{\int_0^z d\tilde{z} ~ \left[  \Omega_\Lambda/\Omega_m + (1+\tilde{z})^3  \right]^{-1/2}} 
\label{eq:sound_horizon_angle}
\end{equation}
where $c_s(z)  = 1/\sqrt{3(1+3\rho_b/4\rho_r)}$ is the speed of sound. So the sound horizon is sensitive to the ratios $\Omega_r/\Omega_m$ and $\Omega_\Lambda/\Omega_m$. Hence, by increasing $N_{\rm eff}$, we must increase $\Omega_m$ while keeping $\Omega_\Lambda/\Omega_m$ constant which increases $H_0$ and solves the discrepancy between Planck and SNe Ia.
But, this enters in tension with BAO+BBN  \cite{Schoneberg:2019wmt}. 
A possible solution is to increase $N_{\rm eff}$ between BBN and CMB \cite{Bernal:2016gxb, Binder:2017lkj, Bringmann:2018jpr,Hooper:2019gtx}. 
Moreover, just by looking at the CMB fit alone, it can be shown that a period of non-standard cosmology hoping for solving the Hubble tension must occur one or two decades of scale factor evolution prior to recombination \cite{Hojjati:2013oya, Aylor:2018drw}.
Nonetheless, to increase $N_{\rm eff}$ has also effects on the CMB at the perturbation level (like Silk damping and neutrino drag at high $l$) which must be counterbalanced with other ingredients.
Particularly, the CMB is sensitive to the so-called \textbf{damping angle}
\begin{equation}
\theta_{\rm D} \simeq \frac{r_{\rm D}}{\int_0^z d\tilde{z} ~ H(\tilde{z})^{-1}} \qquad \textrm{where} \qquad r_{\rm D}^2 \propto \int_{z_{\rm D}}^{\infty} \frac{dz}{H(z)} \frac{1}{a\,\sigma_{T}\,n_e}.
\end{equation}
While the sound horizon scale $r_s$ in Eq.~\eqref{eq:sound_horizon_angle} was linear in $H_{\rm D}$, the damping scale $r_{\rm D}$, which is the variance of a \textbf{random walk}, is a \textbf{quadratic} object.
As a consequence, the decrease of $7\%$ of $r_{\rm s}$ needed to solve the Hubble tension (as well as the decrease of $D_{\rm A}$ by the same amount in order to keep $\theta_s$ constant in Eq.~\eqref{eq:sound_horizon_angle}), implies a decrease of $3.5\%$ of $r_{\rm D}$ and therefore an increase of $3.5\%$ of $\theta_{\rm D}$. This is the $\mathbf{\theta_{\rm D}/\theta_{\rm s}}$ \textbf{problem} \cite{Knox:2019rjx}.

A proposed solution is to increase $N_{\rm eff}$ with new species which also have self-interactions in order to reduce their free-streaming length, e.g. self-interacting neutrinos \cite{Kreisch:2019yzn}, self-interacting dark radiation \cite{Aloni:2021eaq},

An other popular solution \cite{Poulin:2018cxd, Agrawal:2019lmo,Lin:2019qug,Smith:2019ihp,Berghaus:2019cls} is a \textbf{frozen scalar field} behaving as dark energy, reaching $10\%$ of the total energy density at redshift $z \sim 3000$, and which subsequently \textbf{dilutes faster than matter} when oscillating in its potential, and which could solve the Hubble tension at the background and perturbation level.
This solution may be challenged by LSS contraints on the matter power spectrum which break the degeneracy between the sound horizon $r_s$ and the Hubble constant $H_0$ \cite{Hill:2020osr,DAmico:2020ods,Ivanov:2020ril}, or not \cite{Niedermann:2020qbw, Smith:2020rxx,Chudaykin:2020igl}.


They are many other proposals, e.g. primordial magnetic fields \cite{Jedamzik:2020krr}, early AdS phase \cite{Ye:2020btb}, time-variation of fine-strucutre constant and electron mass \cite{Hart:2017ndk,Hart:2021kad}, Newton constant $G_{\rm N}$ \cite{Lin:2018nxe,Ballesteros:2020sik,Braglia:2020iik,Braglia:2020auw,Ballardini:2020iws,Abadi:2020hbr}, so instead of citing them all we refer to the reviews \cite{Knox:2019rjx,DiValentino:2021izs,Schoneberg:2021qvd}.

In the forthcoming years, we expect others probes of $H_0$ to enter the same precision level as CMB or SNe Ia.
This is the case of \textbf{Tip of Red Giant Branch (TRGB)} measurement \cite{Freedman:2019jwv}
\begin{equation}
H_0 = 69.8\pm 1.7 ~\rm km/s/Mpc,
\end{equation}
which currently, appears to sit halfway between CMB and SNe Ia. 

Another interesting promising probe is the detection of gravitational waves emited by \textbf{binary neutron stars (BNS) mergers}, also known as \textbf{standard sirens}. It is claimed that a sample of 50 BNS, detectable by LIGO and VIRGO within a decade would be able to arbitrate the $H_0$ tension between CMB and local distance ladder \cite{Chen:2017rfc, Feeney:2018mkj}. 

Finally, another exciting possibility for measuring the current expansion rate of the universe is the direct observation of the time evolution of the cosmological redshift of distant sources due to the accelelation of the universe expansion, also known as \textbf{redshift drift} \cite{sandage1962change,mcvittie1962appendix,Loeb:1998bu}. Even though the level of spectral drifts is of the order of $10^{-10}/10~\rm yr$, ultra stable and ultra precise spectroscopy coupled with ultra large telescopes like ELT or SKA would be able to detect them
\cite{Balbi:2007fx,Liske:2008ph,Bolejko:2019vni}.

\subsection{The 21-cm anomaly}
\label{sec:21cm}

The $21$~cm line ($1420~{\rm MHz}$) arises from the hyperfine splitting of the $1S$ ground state of hydrogen due to the interaction between the magnetic moments of the proton and the electron. The spin temperature $T_{\rm S}$ determines the relative abundance of the triplet state relative to the singlet
\begin{equation}
\frac{n_{1}}{n_{0}} = \frac{3}{1} \exp \left( -\frac{E_{21}}{T_{\rm S}}\right),
\end{equation}
with $E_{21} = 2\pi/\lambda_{21} = 0.068~\rm K$. 

By passing through the hydrogen clouds, the CMB photons may subtract or add energy to it by absorption or emission. The resulting intensity at a given frequency $\nu$ and a given redshift $z$ is found after integrating the radiative transfer equation \cite{Rybicki:2004hfl}
\begin{equation}
\label{eq:radiative_transfer_theory}
I_{\rm CMB}(\tau) = I_{\rm H}\left(1-  e^{-\tau}\right) + I_{\rm CMB}(0)  e^{-\tau},
\end{equation}
where $I_{\rm CMB}(0)$ is the black body CMB intensity in a transparent universe, $I_{\rm H}$ is the intensity emitted by the medium considered as a black body with temperature $T_{\rm S}$, and $\tau$ is the optical depth, found after integrating the inverse mean free path over the line of sight \cite{Madau:1996cs}
\begin{equation}
\tau = \int_{\rm l.o.s} ds \frac{h \nu}{4\pi}\left[n_0 B_{01} - n_1B_{10} \right] \simeq  \frac{3\lambda_{21}^2 n_{H} A_{10}}{16T_{\rm S}\,H(z)}, \qquad A_{10} \simeq 1.85 \times 10^{-15}~{\rm s^{-1}}.
\end{equation}
We have used the relation between the Einstein coefficients, $g_1B_{10}=g_{0}B_{01}$ and $\frac{2h\nu_{21}^3}{c^3}B_{10}=A_{10}$, and approximated the neutral hydrogen column length by the Hubble horizon $\int ds \simeq H(z)^{-1}$.
We recall that Eq.~\eqref{eq:radiative_transfer_theory} is evaluated for a particular frequency $\nu$ and a particular redshift $z$. We assume the Rayleigh-Jeans limit, $\nu \ll T_{\rm S},\, T_{\rm CMB} $, such that we can approximate $ I(\nu,\,T,\,z) \simeq \frac{2\nu^2}{c^2}k_B T(z)$.  The contrast temperature observed today, between the spin temperature $T_{\rm S}$ and the CMB temperature $T_{\rm CMB}$\footnote{In the context of $21$~cm forest astronomy, the brightness temperature of the $21$~cm line is defined relative to bright radio sources, e.g. active galactic nuclei or gamma-ray bursts, instead of relative to the CMB. $21$~cm forests are a promising probe of the Epoch of Reionization and might be detected with future experiments SKA \cite{Santos:2015gra,Maartens:2015mra, Sprenger:2018tdb,Pritchard:2011xb,Cosmology-SWG:2015tjb,Liu:2022iyy}.}, also known as the differential \textbf{brightness temperature}, reads
\begin{align}
\delta T_b &= \frac{T_{\rm S} - T_{\rm CMB}}{1+z}   \left(1-  e^{-\tau_\nu}\right) \\
&\simeq {27~\rm mK} ~ x_{H}(1+\delta_b) \left(\frac{\Omega_b h^2}{0.022}\right) \left(\frac{0.14}{\Omega_m h^2} \frac{1+z}{10} \right)^{1/2} \frac{T_{\rm S}-T_{\rm CMB}}{T_{\rm S}} \label{eq:brightness_temp},
\end{align}
where $x_{H}$ is the neutral-to-ionized hydrogen fraction and $\delta_b$ is a density fluctuation parameter (important for $21$ cm forests).

The spin temperature results from three competing processes
\begin{equation}
T_{\rm S}^{-1} = \frac{T_{\rm CMB}^{-1} + x_c T_{\rm K}^{-1} + x_{\alpha} T_{\rm {\rm C}}^{-1}}{1+ x_{c} + x_{\alpha}},
\end{equation}
\begin{enumerate}
\item
The absorption/emission of 21 cm photons from/to the CMB at temperature $T_{\rm CMB}$.
\item
The collision with other hydrogen atoms, electrons and protons which have a kinetic temperature $T_{\rm K}$. 
\item
The resonant scattering of Lyman-$\alpha$ photons with color temperature $T_{\rm C}$ which pump the 21 cm transition through an intermediate state (the $1^{\rm st}$ excited hydrogen level).
\end{enumerate}
\begin{figure}[h!]
\centering
\raisebox{0cm}{\makebox{\includegraphics[width=1\textwidth, scale=1]{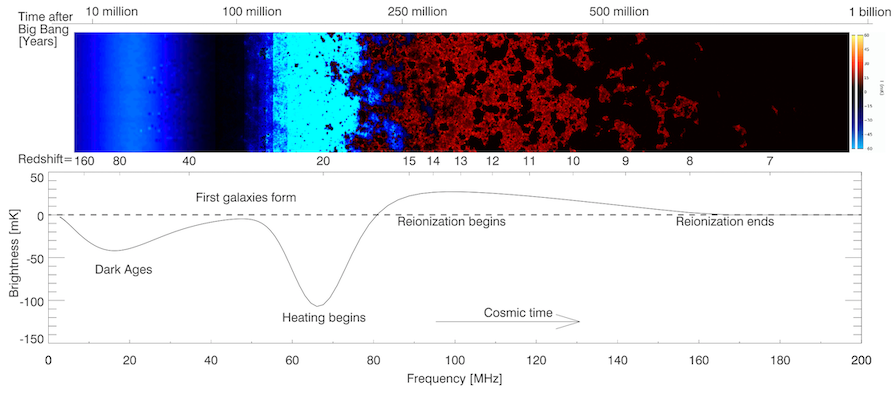}}}
\caption{\it \small The 21 cm cosmic hydrogen signal $\delta T_b \propto T_{\rm S} - T_{\rm CMB}$. \textbf{Top:} The fluctuation of the brightness temperature $\delta T_b$. \textbf{Bottom:} The sky-averaged signal $\delta \bar{T}_b$. Figure reproduced from \cite{Pritchard:2011xb}.}
\label{fig:21cm}
\end{figure}
The evolution of the 21 cm cosmic hydrogen signal $\delta T_b$ with redshift is summarized on Fig.~\ref{fig:21cm}.
\begin{itemize}
\item
\textbf{200 $\lesssim$ z $\lesssim$ 1100}: The Compton scattering between CMB photons and residual electrons allows $T_{\rm K}=T_{\rm CMB}$. Moreover, the high gas density leads to $T_{\rm S}=T_{\rm K}$ and there is no $21$ cm signal $\delta T_{b} = 0$.
\item
\textbf{40 $\lesssim$ z $\lesssim$ 200}: The gas cools adiabatically $T_{\rm K} \propto (1+z)^2$ leading to $T_{\rm K} \lesssim T_{\rm CMB}$ while collisional coupling sets $T_{\rm S} < T_{\rm CMB}$, leading to a first absorption signal $\delta T_{b} < 0$.
\item
\textbf{25 $\lesssim$ z $\lesssim$ 40}: The spin temperature decouples from the gas, whose density decreases due to Hubble expansion and $T_{\rm S} \sim T_{\rm CMB}$. The 21 cm signal switches off $\delta T_{b} \approx 0$.
\item
\textbf{15 $\lesssim$ z $\lesssim$ 25}: As the first stars form, they emit Lyman-$\alpha$ photons such that $T_{\rm S} \simeq T_{\rm C}$ with $T_{\rm C} \simeq T_{\rm K}$. This results in an absorption depth $\delta T_b \lesssim 0$. 
\item
\textbf{8 $\lesssim$ z $\lesssim$ 15}: Continuated heating drives $T_{\rm K} > T_{\rm CMB}$ giving rise to an emission signal $\delta T_b > 0$.
\item
\textbf{ z $\lesssim$ 8}: A significant fraction of the hydrogen atoms become ionized ($x_{H} \to 0$ in Eq.~\eqref{eq:brightness_temp}) such that $\delta \bar{T}_b \to  0$.
\end{itemize}
For reviews on 21 cm cosmology, see \cite{Madau:1996cs, Furlanetto:2006jb, Pritchard:2011xb,Furlanetto:2019dhm}.
In 2018, upon analysing EDGES data, the authors of \cite{Bowman:2018yin} have claimed to have found a $21$ cm absorption depth in the CMB\footnote{The EDGES absorption depth has been observed at $78~\rm MHz$ which corresponds to the low energy tail of the CMB, whose peak frequency is much larger: $160~\rm GHz$ (Wien's law).}, due to the Epoch of Reionization, peaked at the frequency $(21~{\rm cm})^{-1}/(1+z)$ with $z \simeq 17$ 
\begin{equation}
\label{eq:EDGES_result}
\delta T_b\big|_{\rm EDGES} = -0.500^{+200}_{-500} ~ {\rm mK} 
\qquad \rm (99\%~C.L.).
\end{equation}
However, the $\Lambda$CDM cosmology predicts a baryon temperature $T_{\rm K}\simeq 7~$K and a ionized fraction $x_{\rm H} \simeq 1$ \cite{Chluba:2010ca,AliHaimoud:2010dx}, which from imposing $T_{\rm S} \gtrsim T_{\rm K}$ in Eq.~\eqref{eq:brightness_temp}, leads to
\begin{equation}
\delta T_b|_{\rm \Lambda CDM} = -0.22 ~ {\rm mK}.
\end{equation}
This missing factor two between the observed value and the theoretical prediction is the \textbf{21 cm anomaly}. We now briefly discuss the proposed solutions.
\begin{enumerate}
\item
A first explanation is a lower baryon temperature than expected from $\Lambda$CDM. From Eq.~\eqref{eq:brightness_temp} and Eq.~\eqref{eq:EDGES_result}, we deduce the baryon temperature inferred by EDGES result 
\begin{equation}
T_{\rm baryon}\big|_{\rm EDGES}  \lesssim T_{\rm S}\big|_{\rm EDGES} \simeq 3.3^{+2}_{-1.6} ~ \rm K,
\end{equation}
such that
\begin{equation}
\Delta T_{\rm baryon} = T_{\rm baryon}\big|_{\rm EDGES} -T_{\rm baryon}\big|_{\rm \rm \Lambda CDM} \lesssim -3.7^{+2}_{-1.6} ~ \rm K,
\end{equation}
with $T_{\rm baryon}\big|_{\rm \rm \Lambda CDM} \simeq 7~$K \cite{Chluba:2010ca,AliHaimoud:2010dx}.
Hence, the $21$ cm anomaly predicts the baryons at redshift $z=17$ to be at least $1.7~\rm K$ cooler (at $99\%~C.L.$) than in the $\Lambda$CDM scenario. Such a cooling can arise in the presence of long range interactions between baryons and DM when the latter is millicharged \cite{Tashiro:2014tsa,Munoz:2018pzp,Barkana:2018lgd} (note however \cite{Munoz:2015bca}). However, in \cite{Barkana:2018qrx,Berlin:2018sjs,Kovetz:2018zan,Falkowski:2018qdj,Creque-Sarbinowski:2019mcm}, it was shown that this possibility is severly constrained by colliders, CMB, BBN, cooling of SN1987A, stars and DM overclosure. This possibility can be revived if a small fraction of DM is milicharged under both SM and dominant DM \cite{Liu:2019knx} or if the milicharged DM forms composites blobs with higher millicharge \cite{1848079}.  Also, upon asking minimal baryon heating due to DM annihilation, generic upper bound on DM annihilation cross-section can be obtained \cite{DAmico:2018sxd, Liu:2018uzy}, which allows to exclude models of thermal WIMP DM lighter than $\MDM \lesssim 10~$GeV. A lower baryon temperature can also be obtained with the presence of primordial magnetic fields \cite{Natwariya:2020mhe}.
\item
The second possibility is to inject photons in the Rayleigh-Jeans tail of the CMB via invoking additional radio sources \cite{Feng:2018rje}, e.g. black holes population \cite{Ewall-Wice:2018bzf} or hidden dark photons from DM decay, oscillating into photons \cite{Pospelov:2018kdh}.
\end{enumerate}
In the meantime, the proper interpretation of the data has raised doubts. Particularly, the authors of \cite{Hills:2018vyr} find that the conclusion depends on the modelization of the foreground, while the authors of \cite{Bradley:2018eev} suggest the absorption depth to be due to the presence of an unsuspected resonant cavity in the EDGES antenna. The interpretation of EDGES anomaly in terms of systematic effects receives further support in December 2021 after that another radiometer, SARAS 3 \cite{Singh:2021mxo}, did not detect any sign of the profile found by EDGES \cite{Bowman:2018yin} in their data.

We conclude the chapter by saying that the measurement of the cosmological parameters by Planck satellite has considerably transformed the field of Cosmology into a science of precision. However, the big questions remains unresolved. The future of the field is promising thanks to planned experiments like GW interferometers, multi-TeV telescopes, direct-detection experiments, observation of $21$-cm forests, B-modes searches... We know move to the chapter on thermal Dark Matter.


%

\xintifboolexpr { \x = 2}
  {
  }
{
\medskip
\small
\bibliographystyle{JHEP}
\bibliography{thesis.bib}
}

%% file: chap4.tex
\chapterimage{Dark_matter_distribution} 
\chapter{Thermal Dark Matter}
\label{chap:DM}

Since there are zillions of Dark Matter models and thus different types of production mechanisms characterized by different masses and couplings, see Fig.~\ref{fig:DM_candidates_landscape}, instead of listing the different possible candidates, we prefer to focus on one of the best motivated and most studied scenario: the one in which Dark Matter is thermally produced. For reviews on DM history, observations, models and searches, we refer to \cite{Jungman:1995df,Bergstrom:2000pn,Bergstrom:2009ib,Bergstrom:2012fi,Olive:2003iq,Bertone:2004pz,Murayama:2007ek,Steffen:2008qp,Hooper:2009zm,Feng:2010gw,Roos:2010wb,Bertone:2010at,Schnee:2011ooa,Peter:2012rz,Zurek:2013wia,Profumo:2013yn,Lisanti:2016jxe,Alexander:2016aln,Bauer:2017qwy,Bertone:2016nfn,Freese:2017idy,Roszkowski:2017nbc,Arcadi:2017kky,Battaglieri:2017aum,Bertone:2018krk,Blanco:2019hah}.  For textbooks, we call attention to \cite{Bertone:2010zza,Einasto:2013lka,Majumdar:2014wki,Vavilova:2015bed,Profumo:2017hqp,Essig:2019buk,mambrini2021particles}.
The results of the present chapter, in particular Sec.~\ref{sec:heavy_WIMP} which deals with Sommerfeld effects and the unitarity limit, are crucial for Chap.~\ref{chap:secluded_DM} on Homeopathic Dark Matter. All the figures in the chapter are mine (Fig.~\ref{fig:feyn_XX_ZZ}, \ref{FI_VS_FO_Boltzmann}, \ref{FI_VS_FO}, \ref{fig:DMabundance_WIMP}, \ref{fig:sommerfeld-ladder-feynman} and \ref{fig:sommerfeld_factor}).

\section{Production mechanism}
\subsection{The Boltzmann equation}
\paragraph{The production rate of DM:}
The numerous, compelling evidences for the presence of dark matter in our universe raise the question of its nature and of its production mechanism. Probably the most straightforward assumption is that DM is a particle which was at thermal equilibrium with the SM in  the early universe. An example of quantum process involving both DM and SM is a pair of SM particles which annihilate into a pair of DM and anti-DM particles\footnote{At least at first approximation, the SM should be neutral with respect to the DM quantum charge (if any), hence the SM can only produce $DM\bar{DM}$ but not $DMDM$ or $\bar{DM}\bar{DM}$.} with a cross-section $\sigma_{\mathsmaller{ \rm SM SM \rightarrow DM \bar{DM}}}$. Due to this interaction, DM can be thermally produced out of the primordial SM plasma with a rate per unit of volume
\begin{equation}
\gamma_{\mathsmaller{ \rm SM SM \rightarrow DM \bar{DM}}} = n_{\rm SM}^2 \left< \sigma_{\mathsmaller{  \rm SM SM \rightarrow DM \bar{DM}}} \vrel \right>,
\end{equation}
where $\vrel$ is the relative velocity between the incoming particles and $\left<\,.\,\right>$ denotes the average over the thermal distributions of the incoming particles
\begin{equation}
\label{eq:thermal_average_sigmav}
\left< \sigma \vrel \right> = \frac{\int \sigma \, \vrel \, e^{-E_1/T}e^{-E_2/T} d^3 p_1 \, d^3 p_2}{\int  e^{-E_1/T}e^{-E_2/T} d^3 p_1 \, d^3 p_2} .
\end{equation}
\paragraph{The destruction rate of DM:}
DM production is not the only process happening, but instead it competes with the inverse process where DM annihilates into SM and whose rate per unit of volume is
\begin{equation}
\gamma_{\mathsmaller{ \rm  DM \bar{DM} \rightarrow SM SM}} = n_{\rm DM} n_{\rm \bar{DM}}\left< \sigma_{\mathsmaller{ \rm  DM \bar{DM} \rightarrow SM SM}}\, \vrel\right>.
\end{equation}
The change per unit of time in the number of DM particles in a comoving volume is simply the production rate minus the destruction rate
\begin{equation}
\label{eq:Boltzmann_eq_1}
\frac{1}{a^3}\frac{d(n_{\rm DM}^3 a^3)}{dt} = \gamma_{\mathsmaller{ \rm SM SM \rightarrow DM \bar{DM}}} - \gamma_{\mathsmaller{ \rm DM \bar{DM} \rightarrow SM SM}},
\end{equation}
where $a$ is the scale factor of the expanding universe.
The SM number density can be safely fixed to their thermal equilibrium value
\begin{equation}
n_{\rm SM} = n_{\rm SM}^{\rm eq}.
\end{equation}
At this stage, it is useful to note that DM is at thermal equilibium when its destruction rate is equal to its production rate, hence leading to a relation between the thermally averaged cross-sections of the direct and inverse process (often called the `detailed balance condition')
\begin{equation}
\label{eq:detailed_balance}
\left( n_{\rm SM}^{\rm eq}\right)^2 \left< \sigma_{\mathsmaller{ \rm SM SM \rightarrow DM \bar{DM}}} \,\vrel \right> = n_{\rm DM}^{\rm eq} n_{\rm \bar{DM}}^{\rm eq}  \left< \sigma_{\mathsmaller{ \rm  DM \bar{DM} \rightarrow SM SM}}\, \vrel\right>.
\end{equation}
Upon injecting the equilibrium condition \eqref{eq:detailed_balance} in the the out-of-equilibrium equation \eqref{eq:Boltzmann_eq_1}, we obtain
\begin{equation}
\frac{1}{a^3}\frac{d(n_{\rm DM} a^3)}{dt} =   -\left< \sigma\, \vrel\right>\left( n_{\rm DM}n_{\rm \bar{DM}}  - n_{\rm DM}^{\rm eq}n_{\rm \bar{DM}}^{\rm eq}  \right),
\end{equation}
where $\left< \sigma\, \vrel\right>$ is short for $\left< \sigma_{\mathsmaller{ \rm  DM \bar{DM} \rightarrow SM SM}}\, \vrel\right>$.
\paragraph{Two convenient changes of variable:}
If we assume that the universe follows an adiabatic expansion, cf. Eq.~\eqref{eq:adiab_evolution} in Chap.~\ref{chap:SM_cosmology}, then the SM entropy per unit of comoving volume is conserved
\begin{equation}
\frac{d(s a^3)}{dt} = 0,
\end{equation}
where $s$ is the entropy density.
Therefore, it is convenient to trade the variable $n_{\rm DM}a^3$ for $Y_{\rm DM} \equiv n_{\rm DM}/s$ with 
\begin{equation}
\frac{1}{a^3}\frac{d(n_{\rm DM}^3 a^3)}{dt}  = s \frac{dY_{\rm DM}}{dt}.
\end{equation}
A second advantageous choice is to replace the cosmic time $t$ by the inverse temperature $x \equiv \MDM/T$ using $Hdt = da/a$ and 
\begin{equation}
d(h_{\rm eff}\, T^3\,a^3) = 0 \quad \rightarrow \quad \frac{da}{a} = \left(1+\frac{T}{3 h_{\rm eff}}\frac{d h_{\rm eff}}{dT} \right)\frac{dx}{x}.
\end{equation}
Finally, we obtain the so-called Boltzmann equation for annihilating DM
\begin{equation}
\label{eq:Boltzman_eq_bef_final}
\frac{dY_{\rm DM}}{dx}= -\frac{s \left< \sigma\, \vrel\right>}{x \, H} \left( Y_{\rm DM}^2 - \left(Y_{\rm DM}^{\rm eq}\right)^2 \right).
\end{equation}
For simplicity, we have omitted the factor $ \left(1+\frac{T}{3 h_{\rm eff}}\frac{d h_{\rm eff}}{dT} \right)$ on the right hand side, and we have supposed DM to be symmetric $n_{\rm DM} = n_{\rm \bar{DM}}$. 

It is often convenient to extract the temperature dependence from the cross-section by introducing $ \left< \sigma\, \vrel\right> = \sigma_0~ x^{-n}$, such that Eq.~\eqref{eq:Boltzman_eq_bef_final} becomes
\begin{equation}
\label{eq:Boltzman_eq_final}
\frac{dY_{\rm DM}}{dx}= -\frac{\lambda}{x^{2+n}}\left( Y_{\rm DM}^2 - \left(Y_{\rm DM}^{\rm eq}\right)^2 \right),\qquad \lambda = M_{\rm pl}\,\MDM\, \sigma_0\,\sqrt{8\pi^2 g_{*}/45}
\end{equation}
with $g_*^{1/2} \equiv \frac{h_{\rm eff}}{g_{\rm eff}^{1/2}} \left(1+\frac{1}{3}\frac{T}{h_{\rm eff}}\frac{d h_{\rm eff}}{dT} \right)$.
Then, the DM abundance today is given by 
\begin{equation}
\label{eq:final_DM_abundance_exact}
\Omega_{\rm DM}h^2 = 2_{\mathsmaller{\rm DM\neq \bar{DM}}} \frac{h^2\, s_{0} \, \MDM}{3 M_{\rm pl}^2 H_0^2}Y_{\rm DM}^{\infty},
\end{equation}
where $s_0$ is the entropy today and $H_0/h=100$~km/s/Mpc. We can either choose the PDG value $s_0=2 891.2~\rm cm^{-3}$ \cite{Tanabashi:2018oca} or the SM prediction $s_0=2913~\rm{cm^{-3}}$ which uses $N_{\rm eff} \simeq 3.045$ \cite{Mangano:2005cc,deSalas:2016ztq,Escudero:2020dfa}. The factor $2$ stands for DM as being counted as $\Omega_{\rm DM} + \Omega_{\rm \bar{DM}}$. It must be removed if DM is its own anti-particle. Note that Eq.~\eqref{eq:final_DM_abundance_exact} can also be written as
\begin{equation}
2_{\mathsmaller{\rm DM\neq \bar{DM}}}\MDM Y_{\rm DM}^{\infty} = \frac{\Omega_{\rm DM}}{\Omega_{\rm DM}+\Omega_{\rm b}}\frac{3g_{\rm eff,0}}{4h_{\rm eff,0}} T_{\rm eq} \simeq 0.43 ~\rm eV,
\end{equation}
where $T_{\rm eq} \simeq 0.8~\rm eV$ is the temperature at matter-radiation equality, $g_{\rm eff,0}\simeq 3.36$, $h_{\rm eff,0}\simeq 3.94$, $\Omega_{\rm DM} \simeq 0.26$ and $\Omega_{\rm b} \simeq 0.048$.

\subsection{Freeze-in versus Freeze-out}
\label{sec:FI_vs_VO}
\begin{figure}[h!]
\centering{\begin{tikzpicture}
\begin{feynman}
    \vertex (a1) {\(\bar{X}\)};
    \vertex[right=2.5cm of a1] (a2);
    \vertex[right=2cm of a2] (a3){\(V\)};

    \vertex[below=2cm of a1] (b1) {\(X\)};
    \vertex[right=2.5cm of b1] (b2);
    \vertex[right=2cm of b2] (b3){\(V\)};

    \diagram*{
       {
      },
      (a2) -- [fermion] (a1),
      (a3) --  [boson] (a2),
      (b2) -- [fermion] (a2),
      (b1) -- [fermion] (b2),
      (b2) -- [boson] (b3),
    };

\end{feynman}
\end{tikzpicture}
}
\caption{Annihilation of a pair a DM particles $X$ into a pair of hidden vector bosons $V$. The Feynman diagrams are designed with Tikz-Feynman \cite{Ellis:2016jkw}.}
\label{fig:feyn_XX_ZZ}
\end{figure}
\paragraph{A simple DM model:}
We now propose to compute the DM abundance today in a simple model, where DM is a dirac fermion $X$ which couples to a vector boson $V_\mu$ with a coupling constant $\alpha_{\rm D}=\gD^2/4\pi$
\begin{equation}
\label{eq:simple_model_interaction}
\mathcal{L} \supset \gD \bar{X} \gamma^\mu X \, V_\mu.
\end{equation}
We suppose that the vector boson $V_\mu$ interacts with the SM through a kinetic coupling with the hypercharge SM boson $B_\mu$
\begin{equation}
\mathcal{L} \supset \epsilon \,F_{\rm D}\,F_{\rm Y},
\end{equation}
where $\epsilon$ is the coupling constant, $F_{\rm D\,\mu\nu} = \partial_\mu V_\nu - \partial_\nu V_\mu$ and $F_{\rm Y\,\mu\nu} =\partial_\mu B_\nu - \partial_\nu B_\mu$.
We suppose $\MDM > \mV$ such that $X$ can annihilate with its antiparticle $\bar{X}$ into a pair of $V_\mu$, cf. Fig.~\ref{fig:feyn_XX_ZZ}. The corresponding cross-section can easily be computed with FeynCalc \cite{Shtabovenko:2020gxv}
\begin{equation}
\label{eq:cross_section_toy_model_Z0}
\sigma \vrel \simeq \left\{
                \begin{array}{ll}
                 \dfrac{\pi \alpha_D^2}{\MDM^2}, \qquad  \qquad S \simeq 4\MDM^2, \\ \\
                  \dfrac{\pi \alpha_D^2}{S}, \qquad \qquad  S \gg 4\MDM^2,
                \end{array}
              \right.
\end{equation}
where $S=-(p_1+p_2)^2$ is the Mandelstram variable. The thermal average over the thermal distribution of the incoming particles in Eq.~\eqref{eq:thermal_average_sigmav} can be computed with the Gelmini \& Gondolo formula \cite{Gondolo:1990dk}
\begin{equation}
\label{eq:Gelmini_Gondolo}
\left< \sigma \vrel \right> =  \frac{1}{8 \MDM^4 T K_2^2\left(\frac{\MDM}{T}\right)}\int_{4\MDM^2}^{\infty}  (S-4\MDM^2) \sqrt{S} K_{1}\left(\frac{\sqrt{S}}{T}\right) \, \sigma \, dS,
\end{equation}
where $K_\alpha$ are the hyperbolic Bessel functions of the second kind.\footnote{ We can show that it is a relatively good approximation to simply use Eq.~\eqref{eq:cross_section_toy_model_Z0} with $S \simeq 4\MDM^2 + 9 T^2$, instead of integrating out Eq.~\eqref{eq:Gelmini_Gondolo} for each temperature.}
\begin{figure}[h!]
\centering
\raisebox{0cm}{\makebox{\includegraphics[width=0.9\textwidth, scale=1]{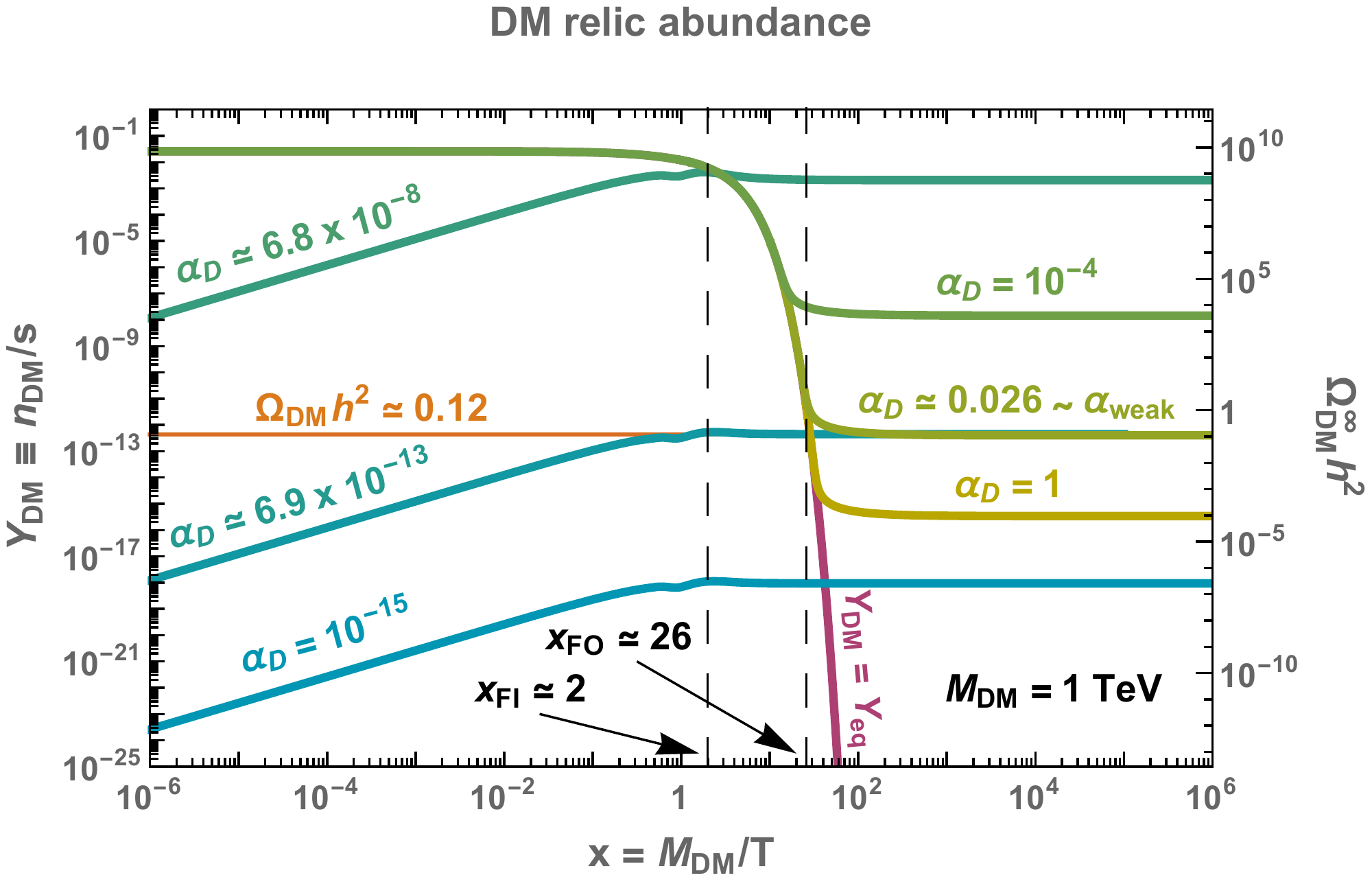}}}
\caption{\it \small Solutions of the Bolzmann equations for different coupling constant $\alpha_D$. On the one hand, for small coupling constant, namely for $\alpha_{\rm D} \lesssim 6.8\times 10^{-8}$ here in the $U(1)_D$ model (blue shaded lines), the dark matter particle never reaches thermal equilibrium but is instead constantly produced until its production rate becomes smaller than the expansion rate.  This is the `freeze-in' mechanism. On the other hand, for larger coupling constant (gold shaded lines), the dark matter thermalized with the SM. Its abundance today depends on the amount of self-annihilation at the time when it becomes non-relativistic. This is the `freeze-out' mechanism. }
\label{FI_VS_FO_Boltzmann}
\end{figure}
\paragraph{Solutions of the Boltzmann equation (numerical):}
The solutions of the Boltzmann equation in Eq.~\eqref{eq:Boltzman_eq_final} are shown in Fig.~\ref{FI_VS_FO_Boltzmann}, and are compared with the thermal abundance in purple. We can see two different regimes
\begin{enumerate}
\item
\textbf{Freeze-in regime} (blue): The DM-SM interaction strength $\alpha_D$ is so small that DM never reaches thermal equilibrium. Instead, it is constantly produced until the temperature drops below the DM mass, around $x_{\rm FI} \simeq 2$, where then the DM production becomes exponentially suppressed by a Boltzmann factor $\exp\left(-\MDM/T\right)$. Assuming $\MDM =1~\rm TeV$,  the abundance of DM fits the Planck data for $\alpha_D \simeq 6.9 \times 10^{-13}$. The corresponding DM candidates are called \textbf{Feebly-Interacting Dark Matter (FIMP)} \cite{Hall:2009bx, Bernal:2017kxu}.
\item
\textbf{Freeze-out regime} (yellow): When the DM-SM coupling is larger than $\alpha_D \gtrsim 6.8 \times 10^{-8}$, DM reaches thermal equilibrium. As explained in the previous paragraph, when the temperature becomes smaller than the DM mass, DM production becomes exponentially suppressed. Therefore, DM annihilates massively (drop of the purple line) until its annihilation rate $\Gamma_{\rm ann} \simeq n_{\rm DM} \left< \sigma \vrel \right>$ becomes smaller than the expansion rate of the universe $H$ around $x_{\rm FO} \simeq 30$ (final plateau). For a mass close to the TeV scale, the needed coupling to have the correct relic abundance is close to the weak coupling $\alpha_W \simeq 1/30 \simeq 0.03$. Such DM candidates, supposedly charged under $SU(2)_L$ are called \textbf{Weakly-Interacting Dark Matter (WIMP)}. The possible connection between the measured DM abundance and the hierarchy problem which motivates new physics around the TeV scale has been called the \textbf{WIMP miracle}. In the literature however, the term WIMP is sometimes enlarged to any \textbf{frozen-out thermal DM}: a particle which was once at thermal equilibrium with the SM and whose abundance is set via the freeze-out mechanism. We give more details in Sec.~\ref{sec:thermal_DM}.
\end{enumerate}
\begin{figure}[h!]
\centering
\raisebox{0cm}{\makebox{\includegraphics[width=0.9\textwidth, scale=1]{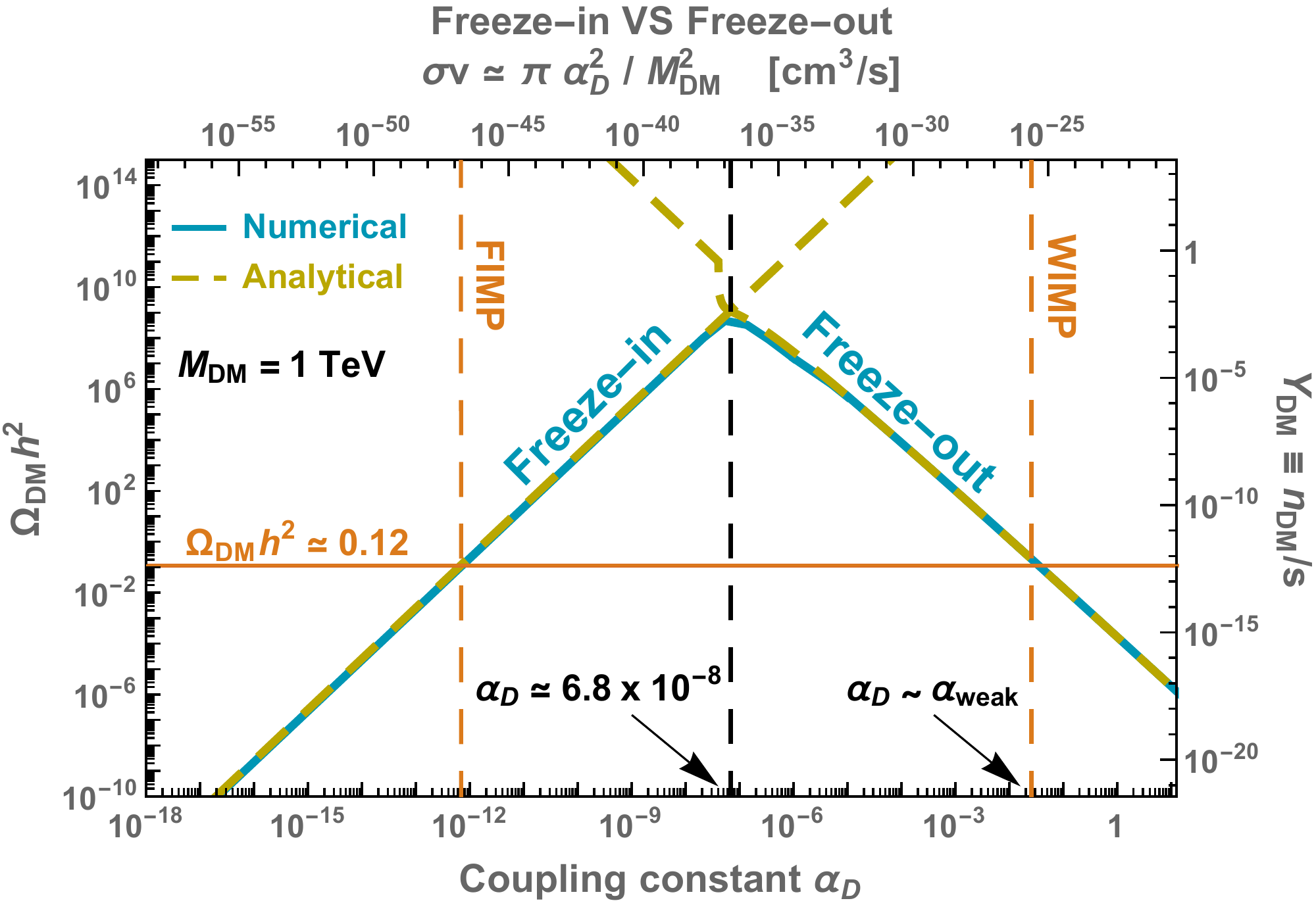}} }
\caption{\it \small   Dark matter abundance as a function of the coupling constant $\alpha_D$. At small coupling constant, the abundance is set by freeze-in, and the corresponding DM particle is called a \textbf{FIMP}. At larger coupling constant, the abundance is set by freeze-out and  the corresponding DM particle is called a \textbf{WIMP}.}
\label{FI_VS_FO}
\end{figure}
The transition between the FI and the FO regime can be appreciated in Fig..~\ref{FI_VS_FO}. The DM abundance grows with the coupling constant in the FI regime, reaches a maximum and switches to the FO regime where, this time, the abundance decreases with the coupling constant.

\paragraph{Solutions of the Boltzmann equation (analytical):}
\label{par:FI_FO_DM}
The DM abundance can also be derived analytically after integrating Eq.~\eqref{eq:Boltzman_eq_final} upon assuming $Y_{\rm DM} \ll Y_{\rm DM}^{\rm eq}$ in the FI regime and upon assuming $Y_{\rm DM} \gg Y_{\rm DM}^{\rm eq}$  in the FO regime
\begin{equation}
\label{eq:analytical_solution_Boltzmann}
Y_{\rm DM} \simeq \left\{
                \begin{array}{ll}
                  \dfrac{\lambda }{x^{2+n}}\left(Y_{\rm DM}^{\rm eq}(x_{\rm FI})\right)^2/2,    \qquad  \qquad \text{[freeze-in]}, \\ \\
                \dfrac{(n+1)\,x_{\rm FO}^{n+1}}{\lambda}, \qquad  \qquad  \qquad \quad \text{[freeze-out]},
                \end{array}
              \right.
\end{equation}
where $\lambda$ is defined in Eq.~\eqref{eq:Boltzman_eq_final}.
We numerically find the `freeze-in' temperature $x_{\rm FI}\simeq 2$. The freeze-out temperature $x_{\rm FO}$ which at first-order obeys the equation $n_{\rm DM} \left< \sigma \vrel \right> \simeq H$, is precisely computed semi-analytically in \cite{Kolb:1990vq}. It is found to be solution of the root equation
\begin{equation}
\label{eq:xFO_KolbTurner}
x_{\rm FO} = \ln\left(0.19\, (n + 1)\,\frac{g_{\rm DM}}{\sqrt{g_{\rm eff}}}\,M_{\rm pl} \, \MDM \, \sigma_0 \right)+\left(n+\frac{1}{2}\right)\ln {\, x_{\rm FO} },
\end{equation}
where $g_{\rm DM}$ is the effective number of degrees of freedom present in $g_{\rm DM}^{\rm eq}$ (e.g. $n_{\rm DM} = \frac{3}{4}\cdot 4$ for a Dirac fermion). See Lee and Weinberg 1977 \cite{Lee:1977ua} for a pioneering work on freeze-out calculation.

\subsection{Exceptions}

We discuss thermal production scenarios which deviate from the vanilla freeze-out and freeze-in studied in the previous section. 
\paragraph{Resonant DM.}
In 1990, Griest and Seckel \cite{Griest:1990kh} studied three exceptions to the freeze-out calculation: resonance, threshold and co-annihilation. Annihilation takes place near a pole of the cross-section $S = m_V^2$, where $m_V$ is the mediator mass. The depletion of the DM abundance is enhanced and the cross-section at zero-temperature $\left<\sigma v\right>_0$ is potentially much smaller than $4 \times 10^{-26}~ \rm cm^3/s$, cf. Eq.~\eqref{eq:DM_abundance_FO}. The maximal effect is obtained when the freeze-out temperature is $x_{\rm FO} = (3u-2)/(2u(1-u))$ where $u\equiv (2\MDM/m_V)^2$ \cite{Griest:1990kh}. For $x_f =25$, we get $u \simeq 0.98$.

\paragraph{Forbidden DM.}
The second exception of Griest and Seckel \cite{Griest:1990kh} is when annihilation products are heavier than DM $m_V > \MDM$. Hence, DM annihilation at zero-temperature is forbidden \cite{DAgnolo:2015ujb,DAgnolo:2020mpt}. During freeze-out, the kinematic threshold is reached by the tail of the thermal distribution $S > 4 m_V^2$ or by $3$-to-$2$ scatterings \cite{Cline:2017tka}, which allows thermal DM to be in the $\rm MeV$ - $10~\rm GeV$ mass range, and can become a target of low-recoil direct detection experiments \cite{Dolan:2014ska,Krnjaic:2015mbs,Knapen:2017xzo}.

\paragraph{Co-annihilating DM.} 
The last exception of Griest and Seckel \cite{Griest:1990kh} is when DM has partner(s) whose mass $m_{\rm partner}$ satisfies $m_{\rm partner} -\MDM \lesssim T_{\rm FO}$. Hence the freeze-out of DM and of its partner(s) are inter-connected and a system of Boltzmann equations must be solved. The co-annihilating specie(s) can be either parasitic or symbiotic according to whether they impede or improve the efficiency of DM annhilation \cite{Profumo:2017hqp}. A well-known application is the computation of the relic abundance of the lightest supersymmetric particle (LSP) when the next-to-LSPs  are close in mass \cite{Edsjo:1997bg}.

\paragraph{Asymmetric DM.}
Some mechanism violates the DM quantum number and generates a non vanishing DM asymmetry $ n_{\rm DM} \gg n_{\rm \overline{DM}}$ in the very early universe such that the DM comoving abundance is later set by $n_{\rm DM}/s$. It is interesting to imagine that the same mechanism generating the baryon number, see Sec.~\ref{sec:Matter-anti-matter-asymmetry} in Chap.~\ref{chap:SM_cosmology}, could also generate the DM number in such a way that $n_{\rm B}=n_{\rm DM}$. Hence the observed relation $\Omega_{\rm DM} = 5\Omega_{\rm b}$ would imply that $\MDM \simeq 5~\rm GeV$. In this work, we do not discuss further the possibility that DM is asymmetric and we instead refer the interested reader to existing reviews \cite{Kaplan:2009ag,Petraki:2013wwa,Zurek:2013wia}. 

\paragraph{Self-annihilating DM.} 
If DM self-annihilates with itself according to \cite{Carlson:1992fn}
\begin{equation}
\underbrace{\chi \cdots \chi}_{n} \to \chi \chi,\qquad n\geq 3,
\end{equation}
then DM freezes-out through a cannibalistic phase during which the comoving number of particle $N$ decreases but the comoving entropy $S \propto a^3 T^{1/2} e^{-\MDM/T}$ remains constant. The exothermic cannibal scatterings make the temperature decrease much slower than in the standard case
\begin{equation}
\frac{T}{\MDM}~ \simeq ~\frac{1}{3\log{a}}.
\end{equation}
$n\geq 3$ scatterings are generally in competition with $n=2$ scatterings \cite{Pappadopulo:2016pkp,Farina:2016llk}. 3-to-2 scattering arises naturally for Strongly-Interacting Massive Particle (SIMP) in pion EFT from the Wess-Zumino-Witten term \cite{Hochberg:2014kqa}. The fact that the confining scale required to reproduce the correct DM abundance is close to the QCD scale has been called the SIMP miracle \cite{Hochberg:2014dra}
\begin{equation}
\begin{cases}
\Gamma_{3\to 2} = n_{\rm DM}^2 \left<\sigma v^2\right>_{3\to 2}\Big|_{T=T_{\rm FO}} = H(T_{\rm FO}), \\
\left<\sigma v^2\right>_{3\to 2} \sim \alpha_{\rm s}^3/\MDM^5
\end{cases}
\qquad 
\implies 
\qquad 
\MDM \sim M_{\rm pl}^{1/3}T_{\rm eq}^{2/3}\alpha_{\rm s} \sim 100 ~\rm MeV.
\end{equation}
In the original SIMP model \cite{Hochberg:2014kqa}, it is assumed that DM and SM are still at thermal equilibrium when DM self-annihilation freezes-out. Instead if DM and SM thermally decouple earlier, then the DM mass prediction becomes dependent on the DM-SM coupling $\epsilon$ instead of the self-coupling $\alpha_{\rm s}$. Such scenario has been called Elastically Decoupling Relic (ELDER) \cite{Kuflik:2015isi,Kuflik:2017iqs} and its realization in $U(1)_{\rm D}$ model kinetically coupled to $U(1)_Y$, cf. App.~\ref{app:U1D_mixing} in Chap.~\ref{chap:secluded_DM}, has been called Kinetically Decoupling Relic (KINDER) \cite{Fitzpatrick:2020vba,Fitzpatrick:2021cij}.

\paragraph{Semi-annihilating DM.}
DM particles $\psi_i$ annihilate through $\psi_i \psi_j \to \psi_k \phi$ where $\phi$ is unstable \cite{DEramo:2010keq}. It has a specific indirect detection signature \cite{DEramo:2012fou}, see Sec.~\ref{sec:uni_bound}, and can evade the unitarity bound on the mass of DM \cite{Kim:2019udq,Kramer:2021hal,Bian:2021vmi}.

\paragraph{Homeopathic DM.} The standard lore is that the temperature in the universe evolves adiabatically such that the comoving entropy $S=s a^3$ is conserved. In presence of a mechanism increasing the entropy of the universe by $S \to D S$, the DM relic abundance gets diluted by $1/D$, cf. Chap.~\ref{chap:secluded_DM}
\begin{equation}
S \to D S \qquad \implies \qquad \Omega_{\rm DM} \to \frac{\Omega_{\rm DM}}{D}. 
\end{equation}
McDonald was the first in 1989 to propose the idea to dilute DM with entropy injection using the decay of a massive particle \cite{McDonald:1989jd}. This has motivated numerous studies \cite{Giudice:2000ex,Patwardhan:2015kga,Berlin:2016vnh,Berlin:2016gtr,Bramante:2017obj,Hamdan:2017psw,Allahverdi:2018aux,Cirelli:2018iax,Contino:2018crt,Chanda:2019xyl,Contino:2020god,Asadi:2022vkc}. Entropy can also be injected by the reheating phase following a supercooled first-order phase transition (1stOPT) \cite{Konstandin:2011dr,Hambye:2018qjv,Baldes:2018emh}.

\paragraph{Bubble-wall-interacting DM.} Interactions between the thermal plasma and bubble walls during a 1stOPT can drastically change the DM abundance calculation. E.g. gluon string formation followed by fragmentation \cite{Baldes:2020kam,Baldes:2021aph}, see Chap.~\ref{chap:SC_conf_PT},  DM filtering \cite{Baker:2019ndr,Chway:2019kft}, DM squeezing \cite{Asadi:2021pwo,Asadi:2021yml}, DM production by wall-wall collisions \cite{Falkowski:2012fb} or plasma-wall scattering \cite{Azatov:2021ifm}.

\section{The WIMP paradigm}
\label{sec:thermal_DM}

Assuming that the universe evolves adiabatically, the freeze-out abundance of a symmetric particle is given by Eq.~\eqref{eq:final_DM_abundance_exact} and Eq.~\eqref{eq:analytical_solution_Boltzmann}
\begin{equation}
\label{eq:DM_abundance_FO}
\frac{\Omega_{\rm DM}^{\rm FO}h^2}{0.1186} \simeq \frac{3.8 \times 10^{-9}~\rm GeV^{-2}}{\left< \sigma \vrel \right>} \simeq \frac{4.4 \times 10^{-26}~ \rm cm^3/s}{\left< \sigma \vrel \right>},
\end{equation}
where we have set $\MDM=1~\rm TeV$ and $n=0$. We have included the factor $2$ in Eq.~\eqref{eq:final_DM_abundance_exact} due to DM being non-self-conjugate.

\subsection{Motivations}

We here discuss the two main reasons why the `frozen-out thermal dark matter', has been the subject of such extensive studies for the last decades.
\paragraph{1) A generic prediction:}
A first remarkable feature of the freeze-out mechanism is that the final DM abundance in Eq.~\eqref{eq:DM_abundance_FO} does not depend on the DM mass.\footnote{A slight dependence of the DM abundance on the DM mass arises from the change in effective number of relativistic degrees of freedom, $\Omega_{\rm DM} \propto x_{\rm FO}^{n+1} g^{-1/2}_{*}(\MDM/x_{\rm FO})$, cf. Eq.~\eqref{eq:analytical_solution_Boltzmann}, and from the logarithmical dependence of $x_{\rm FO}$ on $\MDM$, cf. Eq.~\eqref{eq:xFO_KolbTurner}. See e.g. \cite{Steigman:2012nb} for a plot showing this dependence.} Hence, the prediction for the DM abundance is very generic, and only depends on the value of the annihilation cross-section which must be of order $10^{-26}~\rm cm^3/s$. More precise values depending on the number of fermionic degrees of freedom and on the velocity-dependence of the cross-section are given in Tab.~\ref{table:cross-section-mass-uni}.

\paragraph{2) The WIMP miracle:}
\label{par:WIMP_miracle}
If we assume the generic expression $\left<\sigma \vrel\right> \simeq \frac{\pi \alpha^2}{\MDM^2}$ for the DM anihilation cross-section, then we recover the DM value $10^{-26}~\rm cm^3/s$ with $\MDM \sim \rm TeV$ and $\alpha \sim \alpha_W \simeq 1/30$, see Fig.~\ref{FI_VS_FO_Boltzmann} and Fig.~\ref{FI_VS_FO}, which are exactly the values hinted by the `Naturalness' or `Hierarchy' problem, see Sec.~\ref{sec:hierarchy_pb} in Chap.~\ref{chap:SM_particle}. The common point of most of the solutions to the Naturalness problem is the prediction of new physics at the TeV scale, e.g. susy partners in Supersymmetry, resonances and top partners for Composite Higgs, Kaluza-Klein modes for models with extra-dimensions. This often comes along with the prediction of a Dark Matter candidate if one of the new particles is protected against decay with a symmetry, e.g. $R$-parity in SUSY or $Z_3$-symmetry in warped GUT scenario \cite{Agashe:2004ci,Agashe:2004bm}, which are both needed for proton stability, or the KK parity in models with universal extra-dimensions \cite{Servant:2002aq,Hooper:2007qk}, or the T-parity in little Higgs models \cite{Birkedal:2006fz}. The connection between the measured DM abundance and solutions to the Hierarchy problem is often called the \textbf{WIMP miracle}.

\subsection{The WIMP abundance}
\label{sec:wimp_abundance}
We now study how the abundance of thermal DM varies with respect to its mass $\MDM$, for a fixed coupling.
We assume that the Dark Matter freezes-out after self-annihilation through weak interactions with the generic cross-section \cite{Baltz:2004tj}
\begin{equation}
\label{eq:WIMP_cross_section}
\sigma \vrel \simeq \kappa\, \pi\, \alpha_{\rm w}^2 \frac{ \MDM^2}{(S-m_{\rm Z'}^2)^2+m_{\rm Z'}^2\Gamma_{\rm Z'}^2},
\end{equation}
with $\alpha_{\rm w} \equiv g_{\rm w}^2/4\pi \simeq 1/30$ and where $\kappa$ is a model-dependent parameter\footnote{In the case where DM would be neutrinos (excluded by various bounds), we have $\sigma \vrel \simeq \dfrac{c_2}{2\pi} G_{\rm F}^2 \MDM^2$, with $c_2\simeq 5$ \cite{Kolb:1990vq} and $G_{\rm F} \equiv \sqrt{2}g_{\rm w}^2/8m_{\rm W}^2\simeq 1.166\times 10^{-5}~\rm GeV^{-2}$, which corresponds to Eq.~\eqref{eq:WIMP_cross_section} with $\kappa \simeq 4$.} which we allow to vary within the range $10^{-4}\lesssim \kappa \lesssim 10^4$. The vector boson $Z'$ can be either the SM boson\footnote{The case where DM annihilates directly into the SM gauge boson $Z_0$ is well constrained by direct detection and collider bounds \cite{Escudero:2016gzx,Arcadi:2017kky}. Note however the possibility to revive this scenario using DM dilution via entropy injection \cite{Chanda:2019xyl}.} $Z_0$ or an hidden vector boson. For simplicity, we assume its decay width to be $\Gamma_{\rm Z'} \simeq m_{\rm Z'}$. The model in Sec.~\ref{sec:FI_vs_VO} whose cross-section is given in Eq.~\eqref{eq:cross_section_toy_model_Z0} is encompassed by Eq.~\eqref{eq:WIMP_cross_section} in the limit $\MDM \gtrsim m_{\rm Z'}$.

\begin{figure}[h!]
\centering
\raisebox{0cm}{\makebox{\includegraphics[width=0.9\textwidth, scale=1]{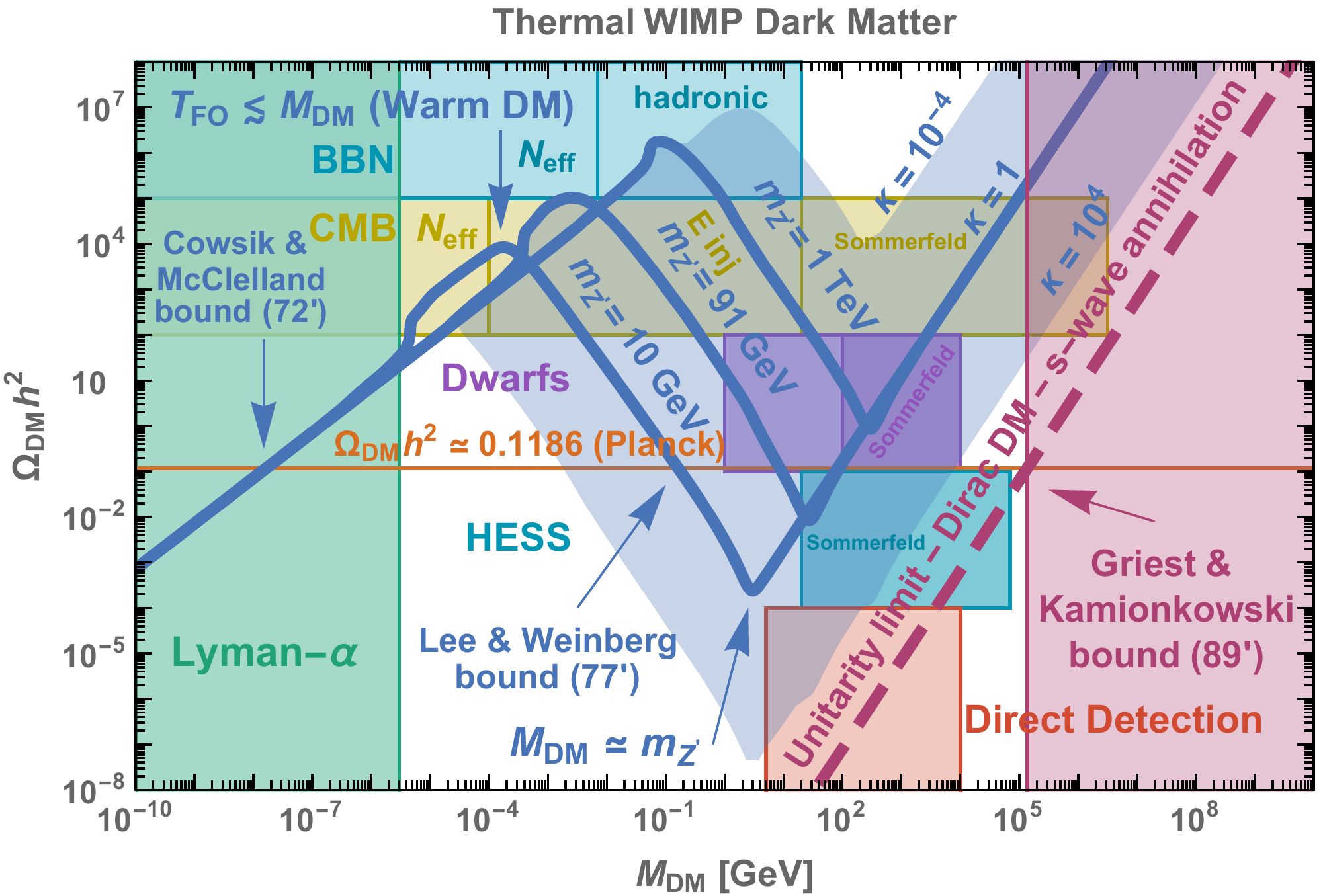}}}
\caption{\it \small Abundance of WIMP thermal Dark Matter with generic annihilation cross-section defined in Eq.~\eqref{eq:WIMP_cross_section}. Thermal Warm DM is contrained by Lyman-$\alpha$ forests  \cite{Viel:2005qj,Baur:2015jsy,Irsic:2017ixq}. Thermal Cold DM is constrained by BBN \cite{Jedamzik:2004ip,Jedamzik:2009uy, Hisano:2011dc,Kawasaki:2015yya,Hufnagel:2017dgo,Hufnagel:2018bjp,Depta:2019lbe,Depta:2020zbh}, CMB \cite{Slatyer:2015jla, Slatyer:2015kla,Elor:2015bho}, indirect detection experiments, among gamma-ray from Dwarfs by Fermi-Lat \cite{Ackermann:2015zua}, gamma-ray from Galactic Center by HESS \cite{Abdallah:2016ygi}, or neutrino by ANTARES \cite{Albert:2016emp,ANTARES:2016obx,ANTARES:2022aoa}, and direct detection constraints from XENON1T \cite{XENON:2018voc,XENON:2019gfn}.  The BBN bounds from photo-dissociation due to hadronic injection stops around the deuterium threshold $2.22~\rm MeV$. The CMB bounds due to energy injection stops at the WDM limit, when the DM abundance becomes independent of the annihilation cross-section below $T_{\rm FO}\lesssim \MDM$ (here we considered the case $m_{Z'}\simeq 10~\rm GeV$). Lower masses are constrained by the $N_{\rm eff}$ bound. The regions labeled "Sommerfeld" include the Sommerfeld enhancement factor of the annihilation cross-section, cf. Sec.~\ref{sec:heavy_WIMP}. For all the constraints, we have assumed $\Omega_{\rm DM}h^2 \simeq 0.12$.
Masses larger than $\MDM \gtrsim 100~\rm TeV$ are constrained by unitarity, see Sec.~\ref{sec:uni_bound}.  The requirement of not overclosing the universe implies all sorts of bounds. The Cowsik-McClelland upper bound on hot DM, $\MDM \lesssim 15~\rm eV$ \cite{Cowsik:1972gh}. The Lee-Weinberg lower bound on cold DM with weak-scale mediator, $\MDM \gtrsim 1~\rm GeV$ \cite{Lee:1977ua,Hut:1977zn,Sato:1977ye,Vysotsky:1977pe}. The Griest-Kamionkowski upper bound due to unitarity, $\MDM \lesssim 100 ~\rm GeV$, see Sec.~\ref{sec:uni_bound}.}
\label{fig:DMabundance_WIMP}
\end{figure}

The DM abundance computed from Eq.~\eqref{eq:final_DM_abundance_exact} in the freeze-out regime is displayed in Fig.~\ref{fig:DMabundance_WIMP} for a wide range of masses.  We can see three different behaviors.
\begin{enumerate}
\item
$ \MDM \gtrsim T_{\rm FO} $ and $\MDM \gtrsim m_{Z'}$: this is the case already studied earlier in Sec.~\ref{sec:FI_vs_VO}. The annihilation cross-section is $\sigma \vrel = \pi \alpha_D^2 /\MDM^2$ such that the DM abundance, see Eq.~\eqref{eq:DM_abundance_FO}, is $\Omega_{\rm DM} \propto \MDM^{2}$. This is the \textbf{vanilla WIMP scenario} where $\MDM \sim \rm TeV$ for electroweak-like coupling. 
The mass can not be larger than $\sim 100~$TeV due to unitarity, see \ref{sec:uni_bound}.
\item
$ \MDM \gtrsim T_{\rm FO} $ and $\MDM \lesssim m_{Z'}$: the annihilation cross-section goes like $\sigma \vrel \propto \MDM^2$ such that the abundance, set by the FO mechanism in Eq.~\eqref{eq:DM_abundance_FO}, decreases like $\Omega_{\rm DM} \propto \MDM^{-2}$.

\item
$ \MDM \lesssim T_{\rm FO} $:  this is the \textbf{Warm Dark Matter scenario}, which we discuss in the next section.
\end{enumerate}

\subsection{Minimal WIMP under pressure}

\paragraph{Indirect detection.}
Indirect detection experiments search for the products of the self-annihilation or decay of dark matter particles in region of high dark matter density, e.g. the milky-way galaxy or its satellites, the dwarf galaxies. We refer to App.~\ref{app:Gamma-ray from DM annihilation} of Chap.~\ref{chap:secluded_DM} for a computation of the cosmic-ray flux at Earth. DM could be detected indirectly through an excess of cosmic  rays. A major difficulty inherent in such searches is that various astrophysical sources can mimic the signal expected from DM. In Fig.~\ref{fig:DMabundance_WIMP}, we show DM constraints resulting from measurements of gamma rays fluxes from Dwarfs by Fermi-lat \cite{Ackermann:2015zua}, gamma rays from Galactic Center by HESS \cite{Abdallah:2016ygi}, and neutrinos by ANTARES \cite{Albert:2016emp,ANTARES:2016obx,ANTARES:2022aoa}. There is a large incertitude on the cosmic ray flux produced in the galactic center according to whether the DM profile forms a core or a cusp \cite{Cirelli:2010xx}, or even a spike \cite{Gondolo:1999ef}. The latter possibility proposed by Silk and Gondolo in 1999 is the potential accumulation of DM in presence of the adiabatic growth of the super-massive black hole in the center.\footnote{Think of what would happen to Earth trajectory if the mass of our Sun would be slowly growing with time.} For reviews on indirect detection, we call attention to \cite{Cirelli:2010xx, Bringmann:2012ez,Cirelli:2012tf,Conrad:2014tla,Gaskins:2016cha,Hooper:2018kfv,Slatyer:2017sev,Slatyer:2021qgc,Strigari:2018utn,Ando:2022kzd,Boddy:2022knd}.

\paragraph{Direct detection.}
Direct detection experiments aim to observe low-energy recoils (typically a few keVs) of nuclei induced by interactions with particles of dark matter.  The non-detection of would-be scintillation light  or would-be phonons induced by such recoils implies an upper bound on the DM-proton cross-section.  In order to limit interferences from cosmic rays, such experiments are conducted deep underground where the background is minimized.  In Fig.~\ref{fig:DMabundance_WIMP}, we show constraints on DM from XENON1T \cite{XENON:2018voc,XENON:2019gfn}.
For reviews on direct detection, we refer to \cite{Baudis:2012ig,Saab:2012th,Cushman:2013zza,Liu:2017drf,Lin:2019uvt,Schumann:2019eaa,Billard:2021uyg}. In recent years, a lot of experiments with lower energy recoils have been proposed which allow to search for sub-GeV DM. We refer the interested reader to the dedicated reviews \cite{Dolan:2014ska,Krnjaic:2015mbs,Knapen:2017xzo}. 

\paragraph{Collider.}
Another approach is the production of DM in a laboratory. Particle colliders like the Large Hadron Collider (LHC) could produced DM in collisions of the proton beams. Since DM have negligible interactions with normal visible matter, it may be detected indirectly as missing energy and momentum that escape the detectors.
For reviews on DM searches at colliders, we  draw attention to \cite{Kane:2008gb,Kahlhoefer:2017dnp,Buchmueller:2017qhf,Morgante:2018tiq,Boveia:2018yeb,Ilten:2018crw,Criado:2021trs},

\paragraph{BBN and CMB.}
An alternative approach to the detection of dark matter particles in nature is to look for deviation from the $\Lambda$CDM expectation in BBN and CMB observables.
At large DM mass, the DM annihilation cross-section is constrained from the amount of energy injection at the BBN \cite{Jedamzik:2004ip,Jedamzik:2009uy, Hisano:2011dc,Kawasaki:2015yya,Hufnagel:2017dgo,Hufnagel:2018bjp,Depta:2019lbe,Depta:2020zbh} or CMB epoch \cite{Zavala:2009mi,Hannestad:2010zt,Slatyer:2015jla, Slatyer:2015kla,Elor:2015bho}. At smaller DM masses or in presence of light mediators, the dark sector could contribute to the effective number of degrees of freedom in the early Universe, potentially detectable in BBN or CMB observations \cite{Ackerman:2008kmp,Baumann:2015rya,Chacko:2015noa,Chacko:2016kgg,CMB-S4:2016ple,Ko:2016fcd,Ko:2016uft,Banerjee:2016suz,Brust:2017nmv}. We show the BBN and CMB constraints on thermal DM in Fig.~\ref{fig:DMabundance_WIMP}.
In the next Sec.~\ref{sec:WDM}, we discuss other cosmological probes of DM, at later times, in large scale structure observables.

\paragraph{This thesis.}

Under the various contraints, it seems that minimal WIMP models are under pressure. The parameter space can however be re-opened using secluded models with an heavy/feebly interacting mediator \cite{Pospelov:2007mp,Pospelov:2008jd,Ackerman:2008kmp,Hambye:2008bq,Arkani-Hamed:2008hhe,Chu:2011be,Berlin:2016gtr,Cirelli:2016rnw,Cirelli:2018iax,Hambye:2019dwd}.
In this work, we study the possibility to relax the DM overclosure bound above $\sim 100$ TeV, by diluting the DM abundance using entropy injection. We explore two directions.
\begin{enumerate}
\item
Entropy injection due to the decay of the DM mediator, see Chap.~\ref{chap:secluded_DM}.  The universe is reheated after a matter era.
\item
Entropy injection due to a supercooled 1st order cosmological phase transition, see Chap. \ref{chap:SC_conf_PT}. The universe is reheated after a vacuum-dominated era.
\end{enumerate}
Such scenarios involve non standard cosmologies and therefore can be probed using the associated imprint on the GW spectrum from Cosmic Strings if such GW are observed with future GW detectors, see Chap.~\ref{chap:cosmic_strings} and Chap.~\ref{chap:DM_GW_CS}.


\subsection{Warm Dark Matter}
\label{sec:WDM}

In the previous section we have discussed impact of DM in BBN and CMB observables. DM could also lead to deviation from $\Lambda$CDM much later, during structure formation. We will discuss the possibility that DM is warm  (WDM), meaning that it was relativistic during a substantial amount of time in the early universe and structure formation is damped at small scales.
The lowest scale at which structures can form is limited by the comoving distance over which DM has moved during cosmologicall history, the so-called free-streaming length $\lambda_{\rm FS}$.

\paragraph{Free-streaming length.}
On general grounds, after being produced the velocity of DM evolves according to 
\begin{equation}
\label{eq:velocity_DM_evol}
v(a)= \begin{cases}
1, \qquad \qquad \qquad a <a_{\rm NR} \\
\left(\frac{a_{\rm NR}}{a}\right)^{\frac{1}{2}},  \qquad a_{\rm NR}< a < a_{\rm KD}, \\
\frac{a_{\rm KD}}{a}, \qquad \qquad \quad  a_{\rm KD}< a,
\end{cases}
\end{equation}
where $a_{\rm NR}$ and $a_{\rm KD}$ are the scale factors when DM becomes non-relativistic and decouples kinetically from the thermal bath, respectively.
As a result of their thermal velocity, the WDM particles slow down the growth of structures at a scale smaller than their \textbf{free-streaming length}, $\lambda_{\rm FS}$, defined by \cite{Kolb:1990vq}
\begin{equation}
\label{eq:free_stream_1}
\lambda_{\rm FS}(t) = \int_0^{t} \frac{v(t')}{a(t')}dt'.
\end{equation}
We are interested in its value just after matter-radiation equality  at $t_{\rm eq}$ when structures start to form
\begin{align}
\lambda_{\rm FS}(t_{\rm eq}) &= \int_0^{t_{\rm NR}} \frac{dt}{a(t)}  + \int_{t_{\rm KD}}^{t_{\rm NR}} \frac{a^{\frac{1}{2}}_{\rm NR}}{a^{\frac{3}{2}}(t)}  dt  +  \int_{t_{\rm KD}}^{t_{\rm eq}} \frac{a_{\rm KD}}{a^2(t)}  dt \\
&\simeq 2  \frac{t_{\rm NR}}{a_{\rm NR}}\left[ 1   + 2  \left(\frac{t_{\rm KD}}{t_{\rm NR}}\right)^{\!\frac{1}{4}}   \left(  1 - \left( \frac{t_{\rm NR}}{t_{\rm KD}}\right)^{\frac{1}{4}} + \frac{1}{4} \ln\frac{t_{\rm eq}}{t_{\rm KD}}\right)\right],\\
&\simeq 2\frac{t_{\rm eq}}{a_{\rm eq}} \frac{T_{\rm eq}}{T_{\rm NR}}\left[  1 + 2\left( \frac{T_{\rm NR}}{T_{\rm KD}} \right)^{\frac{1}{2}}\left(  1 - \left( \frac{T_{\rm KD}}{T_{\rm NR}}\right)^{\frac{1}{2}} + \frac{1}{2} \ln\frac{T_{\rm KD}}{T_{\rm eq}}\right)\right],
\label{eq:free_stream_2}
\end{align}
where we used $a(t) = (t/t_{\rm NR})^{\frac{1}{2}} a_{\rm NR}$.
We now consider two possible scenarios of warm dark matter: relativistic freeze-out and delayed kinetic decoupling. 

\paragraph{Relativistic freeze-out.}
When $ \MDM \lesssim T_{\rm FO} $, the DM freezes-out while being relativistic, cf. left part of Fig.~\ref{fig:DMabundance_WIMP}, similar to SM neutrinos which decouple from the $e^{\pm}, \, \gamma$ bath around $T_{\rm dec, \,\nu} \sim 1~\rm MeV$. In that case the abundance grows linearly with the DM mass, cf. Eq.~\eqref{eq:neutrino_abundance} in Chap.~\ref{chap:SM_cosmology}
\begin{equation}
\label{eq:WDM_abundance}
\Omega_{\mathsmaller{\rm DM}}h^2 \simeq \frac{M_{\mathsmaller{\rm DM}}}{94~\rm eV} \frac{11}{4} \left(\frac{T_{\mathsmaller{\rm DM}}}{T_{\gamma}}\right)^3.
\end{equation}
The ratio of temperatures accounts for any possible asymmetric entropy injection in one sector and not the other.
We suppose that kinetic decoupling takes place  during the relativistic phase 
\begin{equation}
T_{\rm KD} ~>~T_{\rm NR}, 
\end{equation}
such that there is no extra heating due to interaction with the thermal bath in the non-relativistic phase, cf. second line in Eq.~\eqref{eq:velocity_DM_evol}. This implies that the DM  temperature can evolve differently from the SM temperature due to non-adiabatic processes in one of the two sectors, see  e.g. neutrinos heating in Sec.~\ref{par:neutrino_heating} of Chap.~\ref{chap:SM_cosmology}. Hence we denote the DM temperature with a superscript $`{{\rm DM}}'$.
Fixing $T_{\rm KD}^{\mathsmaller{\rm DM}}=T_{\rm NR}^{\mathsmaller{\rm DM}}$ in Eq.~\eqref{eq:free_stream_2} leads  to
\begin{align}
\lambda_{\rm FS}(t_{\rm eq}) \simeq 2\frac{t_{\rm eq}}{a_{\rm eq}} \frac{T_{\rm eq}^{\mathsmaller{\rm DM}}}{T_{\rm NR}^{\mathsmaller{\rm DM}}}\left[  1 + \ln\frac{T_{\rm NR}^{\mathsmaller{\rm DM}}}{T_{\rm eq}^{\mathsmaller{\rm DM}}}\right].
\label{eq:free_stream_3}
\end{align}
If the DM is thermally produced then the average DM momentum follows
\begin{equation}
\left< p\right>= 3.15 ~T_{\mathsmaller{\rm DM}}
\end{equation}
and DM becomes non-relativistic at the temperature\footnote{WDM bounds from structure-formation can also be applied to non-thermal DM candidate. For those cases, the thermal relation $\left< p\right>\Big|_{\rm thermal} = 3.15 ~T_{\mathsmaller{\rm DM}}$ and the temperature Eq.~\eqref{eq:WDM_NR} when DM becomes non-relativistic must be replaced accordingly.}
\begin{equation}
\label{eq:WDM_NR}
T_{\rm NR}^{\mathsmaller{\rm DM}} \simeq M_{\mathsmaller{\rm DM}}/3.
\end{equation}
Upon plugging $a_{\rm eq}^{-1} = 3402(26)$, $t_{\rm eq}=51.1(8)~{\rm kyrs}$ and $T_{\rm eq} \simeq 0.8~{\rm eV}$ \cite{Tanabashi:2018oca} into  Eq.~\eqref{eq:free_stream_3}, we obtain
\begin{equation}
\lambda_{\rm FS}(t_{\rm eq}) \simeq 0.69~{\rm Mpc} ~ \left( \frac{3~\rm keV}{M_{\mathsmaller{\rm DM}}} \right) \left( \frac{T^{\mathsmaller{\rm DM}}_{\rm eq}}{T_{\rm eq}}  \right)\left[ 1+ 0.12 \ln\left(\frac{M_{\mathsmaller{\rm DM}}}{3~\rm keV}\right)\left(\frac{T_{\rm eq}}{T^{\mathsmaller{\rm DM}}_{\rm eq} }\right)\right].
\end{equation}
We can use Eq.~\eqref{eq:WDM_abundance} to relate the DM temperature $T_{\rm eq}^{\mathsmaller{\rm DM}}$ to the photon temperature $T_{\rm eq}$
\begin{equation}
\label{eq:TDM_vs_TSM_eq}
T^{\mathsmaller{\rm DM}}_{\rm eq} = \left(\frac{4}{11} \frac{94 ~ \rm eV}{M_{\mathsmaller{\rm DM}}} \Omega_{\mathsmaller{\rm DM}} h^2 \right)^{1/3}T_{\rm eq},
\end{equation}
so that we finally obtain\footnote{The full expression which keeps track of the log is
$\lambda_{\rm FS}(t_{\rm eq}) \simeq 0.098~{\rm Mpc} ~ \left( \frac{\Omega_{\mathsmaller{\rm DM}}h^2}{0.12} \right)^{1/3}\left( \frac{3~\rm keV}{M_{\mathsmaller{\rm DM}}} \right)^{4/3} \left[ 1 + 0.097\ln{\left( \frac{0.12}{\Omega_{\mathsmaller{\rm DM}}}\right)^{1/3}\left( \frac{M_{\mathsmaller{\rm DM}}}{3~\rm keV} \right)^{4/3}} \right]$}
\begin{equation}
\lambda_{\rm FS}(t_{\rm eq}) \simeq 0.098~{\rm Mpc} ~ \left( \frac{\Omega_{\mathsmaller{\rm DM}}h^2}{0.12} \right)^{1/3}\left( \frac{3~\rm keV}{M_{\mathsmaller{\rm DM}}} \right)^{4/3}.
\end{equation}
As a result, the measurement of the \textbf{matter power spectrum} using Lyman-$\alpha$ forests provides a lower bound on the mass of thermal WDM \cite{Viel:2005qj,Baur:2015jsy,Irsic:2017ixq}
\begin{equation}
\lambda_{\rm FS}(t_{\rm eq}) \lesssim 0.1~{\rm Mpc}\qquad \rightarrow \qquad M_{\mathsmaller{\rm WDM}}\Big|_{\rm ly-\alpha} \gtrsim 3~ \rm keV.
\end{equation}
Potential candidates for WDM are sterile neutrinos, see \cite{Abazajian:2012ys,Adhikari:2016bei, Boser:2019rta} for reviews. Note that the previous bounds only applies for thermal DM, namely when the relativistic momentum obeys $\left< p \right> = 3.15~T$. Otherwise, they must be derived case by case, e.g resonant production of sterile neutrino from active-sterile neutrino mixing \cite{Shi:1998km, Laine:2008pg}, hidden sector hotter than the SM \cite{Hambye:2020lvy}, WDM production from primordial black hole evaporation \cite{Baldes:2020nuv, Auffinger:2020afu}, or FIMP DM \cite{DEramo:2020gpr}.

\paragraph{Delayed kinetic decoupling.}
In presence of large interactions with the thermal bath, after being produced the temperature of DM continues to follow the temperature of the plasma until the momentum transfer rate $\Gamma$ between DM and the thermal bath
\begin{equation}
\label{eq:kinetic_coupling_rate}
\Gamma = n_{\rm rad} \left< \sigma_{\rm rad-DM} \vrel \right> \frac{T}{\MDM},
\end{equation}
falls below the Hubble rate. This defines the \textbf{kinetic decoupling} temperature $T_{\rm KD}$.
We suppose that kinetic decoupling is so delayed that it takes place deep inside the non-relativistic phase
\begin{equation}
T_{\rm KD} ~ \ll ~T_{\rm NR},
\end{equation}
so that Eq.~\eqref{eq:free_stream_2} becomes
\begin{equation}
\lambda_{\rm FS}(t_{\rm eq}) \simeq 4\frac{t_{\rm eq}}{a_{\rm eq}}\frac{T_{\rm eq}}{T_{\rm NR}} \left( \frac{T_{\rm NR}}{T_{\rm KD}} \right)^{\frac{1}{2}} \ln\frac{T_{\rm KD}}{T_{\rm eq}}, \qquad T_{\rm NR} \simeq \MDM/3.
\end{equation}
Plugging $a_{\rm eq}^{-1} = 3402(26)$, $t_{\rm eq}=51.1(8)~{\rm kyrs}$ and $T_{\rm eq} \simeq 0.8~{\rm eV}$ \cite{Tanabashi:2018oca}, we obtain
\begin{equation}
\lambda_{\rm FS}(t_{\rm eq}) \simeq 2.5~{\rm kpc} \left(\frac{\rm keV}{T_{\rm KD}} \right)^{1/2}  \left(\frac{\rm MeV}{\MDM} \right)^{1/2} \ln\frac{T_{\rm KD}}{T_{\rm eq}}.
\end{equation}
Suppression of structure formation at the $\rm kpc$ scale is too small for being probed by Lyman-$\alpha$ observations. However it translates into a suppression of the number of halos with mass smaller than \cite{Chen:2001jz,Hofmann:2001bi,Loeb:2005pm,Green:2005fa,Profumo:2006bv,Bertschinger:2006nq,Bringmann:2009vf,Hooper:2007tu}
\begin{align}
M_{\rm cut} &\simeq \frac{4\pi}{3}\rho_{\rm M,0} \lambda_{\rm FS}^3(t_{\rm eq}) \\
&\simeq  3 \times 10^3~{M_{\odot}} \left(\frac{T_{\rm KD}}{\rm keV} \right)\left( \frac{\MDM}{\rm MeV} \right),
\label{eq:M_cut_WDM}
\end{align}
where $\rho_{\rm M}$ is the cosmological matter density today.  Using Press-Schechter formalism, Ref.~\cite{Hooper:2007tu} found a value for $M_{\rm cut}$ which is larger than Eq.~\eqref{eq:M_cut_WDM} by 4 orders of magnitude. 

We now evaluate the kinetic decoupling temperature in two different scenarios.
A first possibility is that DM $X$ remains coupled to dark radiation $V_\mu$ through Thomson scattering $XV\to XV$ with cross-section $\sigma_{\rm T} = 8\pi \alpha_{\rm D}^2/(3\MDM^2)$, e.g. \cite{Cirelli:2016rnw}. Injecting in Eq.~\eqref{eq:kinetic_coupling_rate} gives
\begin{equation}
T_{\rm kd} \simeq 1~{\rm MeV} \left(\frac{0.03}{\alpha_{\rm D}} \right)\left(\frac{\MDM}{1~\rm TeV} \right)^{3/2}\left(\frac{g_*}{100} \right)^{1/4}.
\end{equation}
A second possibility is an interaction between DM and SM radiation. We consider the existence of a kinetic mixing $\epsilon$ between $U(1)_{\rm D}$ and $U(1)_{\rm Y}$. Then electrons $e$ with electric charge $q$ acquire a $U(1)_{\rm D}$ charge $\epsilon q$, cf. App.~\ref{app:U1D_mixing} in Chap.~\ref{chap:secluded_DM}. Dark matter particles  $X$ can be be kept in kinetic equilibrium through $e\,X \to e\, X$ scattering with t-channel dark photon $V_\mu$ exchange. At temperature $m_{\rm V} \gtrsim T \gtrsim m_e$, the cross-section reads \cite{Feng:2009mn}
\begin{equation}
\left<\sigma_{\rm el} \vrel \right> \sim \frac{\epsilon^2 \alpha_{\rm EM}\alpha_{\rm D}T^2}{m_{\rm V}^4}, 
\end{equation}
and the resulting kinetic decoupling temperature is
\begin{equation}
T_{\rm kd} \simeq \textrm{Max}\left[m_e,~1~{\rm MeV} \left(\frac{10^{-3}}{\epsilon} \right)^{1/2}\left(\frac{0.03}{\alpha_{\rm D}} \right)^{1/4}\left(\frac{m_{\rm V}}{30 ~\rm MeV} \right)\left(\frac{\MDM}{1~\rm TeV} \right)^{1/4}\left(\frac{g_*}{10.75} \right)^{1/8}\right].
\end{equation}
Below $T< m_e$, the electron number density is Boltzmann suppressed and kinetic coupling can not be maintained. Kinetic decoupling temperature lower than a $\rm keV$ has been considered using interaction of DM with various dark fermion\footnote{DM interactions with SM neutrinos, proposed in \cite{vandenAarssen:2012vpm} are severely constrained \cite{Ahlgren:2013wba,Laha:2013xua}.} \cite{Hooper:2007tu,vandenAarssen:2012vpm,Dasgupta:2013zpn,Bringmann:2013vra,Bertoni:2014mva,Ko:2014bka,Cherry:2014xra,Chu:2014lja,Buckley:2014hja,Binder:2016pnr,Bringmann:2016ilk,Tang:2016mot,Agrawal:2017rvu}. It has been shown that the suppression of the number of halos lighter than Eq.~\eqref{eq:M_cut_WDM} can solve the \textbf{missing satellite}, \textbf{cusp-core} and \textbf{too-big-too-fail} problems \cite{Hooper:2007tu,vandenAarssen:2012vpm,Dasgupta:2013zpn,Bringmann:2013vra,Bertoni:2014mva,Ko:2014bka,Cherry:2014xra,Chu:2014lja,Buckley:2014hja,Binder:2016pnr,Bringmann:2016ilk,Tang:2016mot,Agrawal:2017rvu}. We refer to Sec.~\ref{sec:fragility_LambdaCDM} in Chap.~\ref{chap:SM_cosmology} for a review of small-scale problems.

Finally, let's mention that a tight coupling between DM and a dark radiation component leads to acoustic oscillations and diffusion damping for DM, analogous to the baryon-photon fluid, modifying the linear matter power spectrum. We refer to \cite{Boehm:2000gq,Boehm:2001hm,Boehm:2004th,Cyr-Racine:2012tfp,Cyr-Racine:2013fsa,Buckley:2014hja} for dedicated studies.  We also call attention to \cite{Cyr-Racine:2015ihg,Vogelsberger:2015gpr} in which the fundamental parameters of DM at the Lagrangian level are mapped into physical effective parameters that shape the linear matter power spectrum.

\section{Heavy WIMP}
\label{sec:heavy_WIMP}

Assuming a typical DM annihilation cross-section of the form $\sigma_{\rm ann} \vrel = \pi \alpha^2/\MDM^2$, we can see that the condition of satisfying the correct DM relic abundance in Eq.~\eqref{eq:DM_abundance_FO} leads to
\begin{equation}
\alpha \simeq 0.3 \left(\frac{\MDM}{10~\rm TeV}\right).
\end{equation}
Hence, the larger the DM mass, the larger the coupling constant, which leads to two difficulties.
\begin{enumerate}
\item
The validity of perturbation theory breaks down. Non-perturbative effects have to be taken into account: the Sommerfeld enhancement of the annihilation cross-section and the formation of bound-states.
\item
The unitarity of the theory is violated when the probability, $P_{\rm ann}\propto \alpha^2$, for a given pair of DM particles to annihilate is larger than 1.
\end{enumerate}

\subsection{Breakdown of perturbation theory}
\paragraph{An infinite number of exchanged mediator:}
In order to simplify what follows in the present paragraph, we replace the charged fermion DM $X$ by a charged scalar $X$ and the vector boson $V_\mu$ by a scalar boson $\phi$, and the interaction in Eq.~\eqref{eq:simple_model_interaction} becomes
\begin{equation}
\mathcal{L} \supset g \, \phi\, X^\dagger X.
\end{equation}
We now want to study what happens when the coupling constant $\alpha = g^2/(4\pi)^2$ becomes large, which corresponds to the mass of the corresponding thermal Dark Matter candidate being close to the unitarity bound. 

\begin{figure}[h!]
\centering
\begin{equation}
\begin{tikzpicture}
\begin{feynman}
    \vertex (a1) {\(\bar{X}\)};
    \vertex[right=1cm of a1] (a2);
    \vertex[right=1cm of a2] (a3);
    \vertex[right=1cm of a3] (a4);
    \vertex[right=1cm of a4] (a5) {\(V\)};

    \vertex[below=1cm of a1] (b1) {\(X\)};
    \vertex[right=1cm of b1] (b2);
    \vertex[right=1cm of b2] (b3);
    \vertex[right=1cm of b3] (b4);
    \vertex[right=1cm of b4] (b5) {\(V\)};

    \diagram*{
       {[edges=boson]
        (b2) -- (a2),
        (b3) -- (a3),
      },
      (b3) -- [boson, edge label'=\;\;...] (a3),
      (a2) --  (a1),
      (a3) -- [fermion] (a2),
      (a4) --  (a3),
      (a4) -- [boson] (a5),
      (b1) --  (b2),
      (b2) -- [fermion] (b3),
      (b3) --  (b4),
      (b4) -- [boson] (b5),
      (b4) -- [fermion] (a4),
    };

\end{feynman}
\end{tikzpicture}
\end{equation}
\caption{\it \small Infinite ladder of mediator exchanges, which become relevant in the long range / small velocity / large coupling limit. }
\label{fig:sommerfeld-ladder-feynman}
\end{figure}
Using Feynman rules, we can estimate the amplitude of the double boson exchange relative to the amplitude of the single boson exchange, see Fig.~\ref{fig:sommerfeld-ladder-feynman}
\begin{equation}
\frac{\alpha\,m_X^2}{(p_1-p_3)^2 - \mV^2} + \int \frac{dk^4}{(2\pi)^4}\frac{\alpha\,m_X^2}{k^2 - \mV^2}\frac{\alpha\,m_X^2}{(p_3-p_1+k)^2 - \mV^2}\frac{1}{(p_1-k)^2 - m_X^2}\frac{1}{(p_2+k)^2 - m_X^2}.
\end{equation}
In the non-relativistic limit, the typical exchanged momentum and energy between the two incoming particles are of the order of the Bohr momentum $|\vec{k}| \sim \mu \alpha$ and of the would-be bound-state energy $k_0 \sim \mu \alpha^2/2$ \cite{Hoyer:2014gna, Petraki:2015hla}. We have introduced the reduced mass $\mu = m_X/2$. Also, the external momenta are of order $p_\mu = \left(m_X+\mu \vrel^2/2, ~\pm \mu \vec{v}_{\rm rel}\right)$, with $\mu \vrel^2/2 \ll \mu \vec{v}_{\rm rel}$, where we have introduced the relative velocity $\vrel$.  Hence, we can write
\begin{equation}
\frac{\alpha\,m_X^2}{(\mu \alpha)^2 + \mV^2} +2\pi^2\frac{ (\mu \alpha^2/2)}{2\pi}  \frac{(\mu \alpha)^3}{(2\pi)^3} \frac{\alpha\,m_X^2}{(\mu \alpha)^2 + \mV^2}\frac{\alpha\,m_X^2}{(\mu \alpha)^2 + \mV^2}\frac{1}{(\mu \vrel)^2}\frac{1}{(\mu \vrel)^2}.
\end{equation}
which in the limit $\mu \alpha \gg \mV$, leads to
\begin{equation}
\frac{1}{\alpha} + \frac{1}{\alpha} \frac{1}{8\pi^2}\left( \frac{\alpha}{\vrel}  \right)^4.
\end{equation}
We can see that in the \textbf{long range} / \textbf{small velocity} / \textbf{large coupling} limit, $\mu \alpha \gg \mV$ and $\alpha \gg \vrel$, the perturbation theory breaks down and we must account for all the ladder diagrams.

The relevance of the infinite ladder of mediator exchanges can be understood as bound states (BS) being present in the theory.
Non-perturbative effects arise when those bound states start to play a role in the dynamics. Hence, we must account for the presence of non-perturbative effects when the following condition are satisfied.

\paragraph{BS existence:}
The BS are present in the theory when the range of the Dark force $1/\mV$, where $\mV$ is the mediator mass, is larger than the size of the would-be bound state system $(\mu \alpha)^{-1}$, $\mu$ being the reduced mass $\MDM/2$ of the 2-body system
\begin{equation}
\text{BSE:} \qquad \frac{\alpha \MDM}{2\mV} \geq 0.84.
\label{eq:BSE_condition}
\end{equation}
The numerical factor $0.84$ is found after numerically solving the Schr\"{o}dinger equation in the presence of a Yukawa potential \cite{Petraki:2016cnz}.

\paragraph{BS formation 1:}
Those bound states can form when the kinetic energy of the system in its center of mass $\mu \vrel^2/2$ is less than the ground state binding energy $\mu \alpha^2/2$
\begin{equation}
\label{eq:BSF}
\text{BSF 1:} \qquad \alpha \geq \vrel.
\end{equation}
Otherwise the system remains unbound and the two particles just pass through each other. An equivalent formulation of Eq.~\eqref{eq:BSF} is to say that the incoming wave packets can form BS only if their spatial extension, the De Broglie wavelength $(\mu \vrel)^{-1}$, is larger than the typical bound state size $(\mu \alpha)^{-1}$.

\paragraph{BS formation 2:}
In order to relax to the ground state of negative energy $-\mu \alpha^2/2$, the system with initial kinetic energy $+\mu \vrel^2/2$, must emit of vector boson of energy $\mu \vrel^2/2 + \mu \alpha^2/2 \lesssim \mu \alpha^2$. This is impossible if the mediator is too massive, hence the second BS formation condition
\begin{equation}
\label{eq:BSF2}
\text{BSF 2:} \qquad  \mV \leq \MDM(\vrel^2 +\alpha^2)/4 \qquad \rightarrow \qquad \frac{M_{\rm DM} \alpha^2}{4 \mV} \gtrsim 1.
\end{equation}

\paragraph{Coulomb approximation:}
We can neglect the mass of the mediator when the range of the Dark force $1/\mV$ is larger than the De Broglie wavelength $(\mu \vrel)^{-1}$. This is the Coulomb approximation
\begin{equation}
\label{eq:coulomb_limit}
\text{Coulomb:} \qquad \frac{\MDM \vrel}{2 \mV} \geq 1.
\end{equation}

\subsection{Sommerfeld enhancement}
\label{sec:som_enh}

The enhancement of the annihilation cross section at low velocities due to the presence of long-range interactions was first proposed by Sommerfeld in 1931 \cite{Sommerfeld:1931qaf}.
Sommerfeld enhancement became the center of the attention in the DM community \cite{Hisano:2004ds,Cirelli:2007xd,Cirelli:2008jk,Cirelli:2008pk,Cholis:2008wq,Cholis:2008vb,Kamionkowski:2008gj,ArkaniHamed:2008qn,Donato:2008jk,Lattanzi:2008qa,March-Russell:2008klu,Bedaque:2009ri,Finkbeiner:2010sm,Slatyer:2011kg,Liu:2013vha} in 2008 after the data release from PAMELA satellite \cite{PAMELA:2008gwm,Chang:2008aa} showing an overabundant spectrum of cosmic ray positrons with energies between 10 GeV and a few TeV. The positron excess was later confirmed by Fermi-LAT \cite{Fermi-LAT:2011baq} and AMS-02 \cite{AMS:2013fma}.
Today, the DM explanation of the positron excess seems ruled out by CMB distortion \cite{Zavala:2009mi,Slatyer:2009yq,Hannestad:2010zt}, measurements of gamma-ray \cite{Profumo:2009uf,Belikov:2009cx,Kawasaki:2009nr,Cirelli:2009dv,Abazajian:2011ak} and
neutrino fluxes \cite{Spolyar:2009kx,Buckley:2009kw,Hisano:2009fb}, and the asphericity of DM halo \cite{Feng:2009hw,Buckley:2009in}. Instead, an astrophysical origin seems more likely, e.g. pulsars \cite{Hooper:2008kg,Yuksel:2008rf,Profumo:2008ms} or supernova remnants \cite{Blasi:2009hv,Blasi:2009bd}.

On the good side, the enthusiasm for the PAMELA excess has motivated important theoretical works on the theory of Sommerfeld enhancement \cite{ArkaniHamed:2008qn,Iengo:2009ni,Iengo:2009xf,Cassel:2009wt,Slatyer:2009vg,Blum:2016nrz}.
This has allowed to push computations of the thermal relic abundance \cite{Dent:2009bv,Feng:2010zp,Hryczuk:2010zi} and of the cosmic-ray flux  \cite{Pieri:2009zi,Cirelli:2010xx,Das:2016ced,Boddy:2017vpe,Cirelli:2016rnw,Cirelli:2018iax} to the multi-TeV DM mass region, and to more precisely determine the parameter space of neutralino DM \cite{Hryczuk:2011vi,Hryczuk:2011tq,Beneke:2014gja,Beneke:2014hja,Beneke:2019qaa,Beneke:2020vff,Rinchiuso:2018ajn,Baumgart:2018yed}.

%
%
%

\paragraph{Warm-up with classical gravity:}
Before giving more details about the nature of quantum interactions in the long range / small velocity / large coupling limit, we first introduce its analog in classical gravity, as proposed in \cite{ArkaniHamed:2008qn}.
Consider a point particle in free fall nearby a star of radius $R$. Neglecting gravity, only particles with impact parameter $b$ smaller than $R$ can hit the star. Hence, the cross-section for the particle to crash on the star is just the geometrical cross-section $\sigma_0 = \pi R^2$. 
However, after including the \textbf{long range} Newton's force, particles can crash on the star for larger impact parameter. Then, the cross-section is $\pi b_{\rm max}^2$, where $b_{\rm max}$ is the largest impact parameter leading to a crash. 
The \textbf{conservation of angular momentum} allows to write
\begin{equation}
\label{eq:ang_mom_Arkani}
m\,v\,b_{\rm max} = m\, v(R) \, R \quad \rightarrow \quad v(R)= \frac{b_{\rm max}}{R} v,
\end{equation}
where $v$ is the velocity at infinity. Together with \textbf{conservation of energy}, Eq.~\eqref{eq:ang_mom_Arkani} gives
\begin{align}
\frac{1}{2}m\,v^2 &= \frac{1}{2}m \,v(R)^2 - \frac{G\,M\,m}{R}, \\
v^2 &= v^2 \frac{b_{\rm max}^2}{R^2}  - 2\frac{G\,M}{R}, \\
1 &= \frac{b_{\rm max}^2}{R^2} - \left( \frac{v_{\rm esc}}{v} \right)^2,
\end{align}
where $v_{\rm esc}^2 = 2G_{\rm N} M / R$ is the escape velocity from the surface of the star.
Hence, we get
\begin{equation}
\label{eq:final_answer_classGrav_Arkani}
\sigma = \pi^2 \, b_{\rm max} = \sigma_0  \left( 1 + \frac{v_{\rm esc}^2}{v^2} \right).
\end{equation}
We conclude that a long-range Newton's interaction \textbf{enhances} the cross-section $\sigma v$ by a factor growing as $1/v$. This is a classical counterpart of the \textbf{Sommerfeld enhancement} which we discuss in the next section.

\paragraph{Quantum Mechanics:}
We now discuss the effects of the presence of a long range interaction, in the small velocity / large coupling limit, in Quantum Mechanics\footnote{`Small velocity' implies that we can safely work in the non-relativistic limit.}. Suppose that a non-relativistic particle is moving along the $z$ direction with the wavefunction
\begin{equation}
\psi_{k}^{(0)}(\vec{r}) = e^{ikz},
\end{equation}
and can be annihilated and converted into an other state due to a short-range interaction at the origin: $H_{\rm ann}=U_{\rm ann}\delta(\vec{r})$ \cite{ArkaniHamed:2008qn}. It is easy to convince oneself that the rate of the process will be proportional to the density of presence at the origin $|\psi_{k}^{(0)}(\vec{r})|^{2}=1$.
If we now add a long-range interaction $V(\vec{r})$, the wavefunction of the particle will be distorted and this will change the annihilation rate. More precisely, the wavefunction $\psi_{k}(\vec{r})$ obeys the Schrodinger equation
\begin{equation}
\label{schrodinger_long_short_range}
\left[ -\frac{1}{2M}\nabla^2 + V(\vec{r}) + U_{\rm ann}\delta(\vec{r}) \right] \psi_{k} = \epsilon_{k} \;\psi_{k}.
\end{equation}
Since the annihilation by the short range potential $H_{\rm ann}$ takes place locally, in $r=0$, the effective effect of the long-range force is to modify the value of the wave-function at the origin. This changes the cross-section 
\begin{equation}
\sigma = \sigma_{0}~ S_{\rm ann},
\end{equation}
by a factor called the \textbf{Sommerfeld factor}
\begin{equation}
\label{eq:sommerfeld_fac}
S_{\rm ann} = \frac{|\psi_{k}(0)|^{2}}{|\psi_{k}^{(0)}(0)|^{2}}.
\end{equation}
$\sigma_{0}$ is the perturbative cross-section, computed when taking only into account the short-range potential.

\paragraph{Hulthen potential:}

The Sommerfeld enhancement with a Yukawa potential has no analytical solution and must be computed numerically, see App.~\ref{app:Sommerfeld_enhancement_num}.
However, an analytical solution exists when replacing the Yukawa potential by the Hulth\'en potential 
	\begin{equation}
	\label{eq:Hulthen_potential}
	 V_{H} = - \frac{\alpha\, m_{*}\,e^{-m_*r}}{(1-e^{-m_*r})}, \qquad  \mathrm{where} \qquad  m_*= \frac{\pi^2}{6} \,\mV.
	\end{equation}
We report here the resulting Sommerfeld enhancement factor for arbitrary $l$-wave process~\cite{Cassel:2009wt} (see also \cite{Iengo:2009ni})
\begin{equation}
\label{eq:sommerfeld_enhancement_Hulthen}
S_{H}^{(l)} = \left| \frac{\Gamma(a^-)\Gamma(a^+)}{\Gamma(1+l+2i\omega))} \frac{1}{l!}   \right|^2,
\end{equation}
with $a^{\pm} \equiv 1+ l + i \omega(1 \pm \sqrt{1-x/\omega})$, $x \equiv 2\alpha/v_{\mathsmaller{\rm rel}}$ and $\omega \equiv \MDM v_{\mathsmaller{\rm rel}} / (2 m_{*})$.
Explicitly, for s-wave annihilation one finds~\cite{Slatyer:2009vg}
	\begin{equation}
	S_{H}^{(l=0)} = \frac{2\pi\alpha}{v_{\mathsmaller{\rm rel}}} \frac{\sinh (\pi \MDM v_{\mathsmaller{\rm rel}}/m_*)}{\cosh{(\pi\MDM v_{\mathsmaller{\rm rel}} /m_*)} - 		\cosh{ \pi \sqrt{\MDM^2 v_{\mathsmaller{\rm rel}}^2/m_*^2  -  4\MDM \alpha/m_*}}}.
	\end{equation}
	
\paragraph{Coulomb potential:}


In the Coulomb limit, $\MDM \vrel/2 \mV>1$,  the potential in Eq.~\eqref{eq:Hulthen_potential} reduces to the Coulomb potential, and the associated Sommerfeld enhancement factor in Eq.~\eqref{eq:sommerfeld_enhancement_Hulthen} becomes
\begin{equation}
S_{\ann}^{\rm C} = \frac{2\pi\alpha}{	\vrel}\frac{1}{1-e^{2\pi \alpha/	\vrel}}~ P_l \qquad{\rm with } \quad 
 P_l  =
 \begin{cases}
1 &\hspace{2em}\textrm{if } l=0,\\
 \prod_{k=1}^{l}\left[ 1 + \frac{1}{k^2}\frac{\alpha^2 }{\vrel^2}\right] &\hspace{2em}\textrm{if } l>0.
\end{cases}
\end{equation}

\begin{figure}[h!]
\centering
\raisebox{0cm}{\makebox{\includegraphics[width=0.7\textwidth, scale=1]{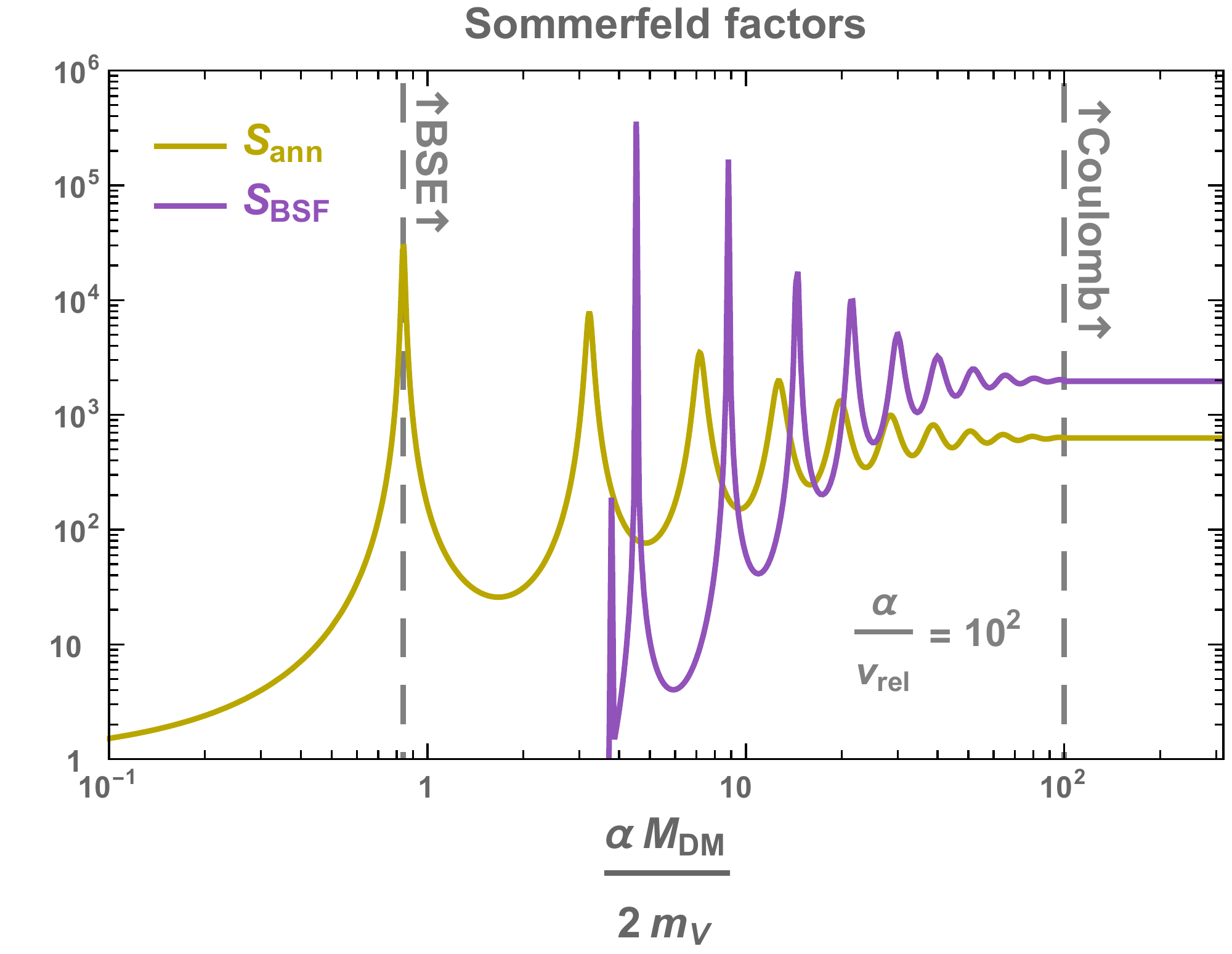}}}
\caption{\it \small Sommerfeld enhancement factor (gold), cf. Sec.~\ref{sec:som_enh} and Bound-State-Formation factor (purple), cf. Sec.~\ref{sec:BSF}. We assume a Dirac fermion interacting with a vector mediator and only include the capture in the ground state. With dashed vertical lines, we highlight the BSE and Coulomb regime, defined in Eq.~\eqref{eq:BSE_condition} and Eq.~\eqref{eq:coulomb_limit}.}
\label{fig:sommerfeld_factor}
\end{figure}

\subsection{Bound-state-formation}
\label{sec:BSF}

The presence of a long-range force also allows for the formation of bound-states. Significant theoretical works on bound state formation (BSF) were accomplished by Petraki et al. \cite{Petraki:2015hla,Petraki:2016cnz}, which were then applied to study various DM models with a particular focus on Higgs portal \cite{Harz:2017dlj,Harz:2018csl,Harz:2019rro,Oncala:2018bvl,Oncala:2019yvj,Oncala:2021tkz,Oncala:2021swy}. BSF opens additional annihilation channels which deplete further the relic-abundance during freeze-out \cite{vonHarling:2014kha,Binder:2018znk,Binder:2019erp,Binder:2020efn,Binder:2021vfo,Binder:2021otw,Garny:2021qsr,Bollig:2021psb} and add features to the cosmic-ray flux at Earth \cite{Pearce:2015zca,Baldes:2020hwx,Geller:2020zhq,Eiger:2020anw}.
The maximal mass of thermal DM compatible with the unitarity limit was recomputed more precisely accounting for those non-perturbative effects in \cite{vonHarling:2014kha,Baldes:2017gzw,Cirelli:2018iax,Smirnov:2019ngs}. The theory of BSF has been applied to the study of various topics such as colored dark sectors \cite{ElHedri:2016onc,ElHedri:2017nny,Mitridate:2017izz,Mitridate:2017oky,DeLuca:2018mzn,Gross:2018zha,Contino:2018crt,Redi:2018muu,Harz:2018csl,Becker:2022iso}, asymmetric DM \cite{Baldes:2017gzw,Baldes:2017gzu,Mahbubani:2019pij}, or dark nucleosynthesis \cite{Mahbubani:2020knq}. 

\paragraph{The capture into the ground-state :} 
When the conditions Eq.~\eqref{eq:BSF} and Eq.~\eqref{eq:BSF2} are realized the incoming particles are slow enough to get trapped in a bound state and the mediator is light enough to be emitted.

We will consider only capture into the ground state, with quantum numbers $\{n \ell m\} = \{100\}$. Capture to other bound states is possible but is always subdominant either with respect to capture to the ground state or to the DM annihilation~\cite{Petraki:2015hla, Petraki:2016cnz}. For BSF via emission of a vector boson, the bound-state angular momentum $\ell$ differs by one unit with respect to that of the scattering state. Therefore, the capture into the $\{100\}$ state is a $p$-wave process. This is important for the velocity dependence of $\sigma_\BSF$ at small DM velocities, as we discuss below. The BSF cross-sections can be expressed as
\beq
\sigma_{\BSF} \vrel = \frac{\pi\aD^2}{\MDM^2}  S_{\BSF} \Pi_{\BSF},
\label{eq:sigmaBSF}
\eeq
where  $\Pi_{\BSF}$ includes both the phase-space suppression due to $\mV >0$, and the enhancement due to the third polarization of the vector boson, $\Pi_{\BSF} = s_{\rm ps}^{1/2} (3-s_{\rm ps})/2$, with $s_{\rm ps} = 1-16 \mV^2/[\MDM (\vrel^2 + \aD^2)]^2$. 
$S_\BSF$ can be computed analytically only in the Coulomb regime  \cite{Petraki:2015hla, Petraki:2016cnz}, which is specified by the condition $\MDM \vrel/2 \lesssim \mV$, cf. \eqref{eq:coulomb_limit}
\beq
S_{\BSF}^{\rm C} =  
\frac{2\pi / \epsilon_v}{1-e^{-2\pi/ \epsilon_v}} \,
\frac{2^9}{3}\,\frac{e^{-4 {\rm arctan}(\epsilon_v)/\epsilon_v}}{(1+\epsilon_v^2)^2} \,,
\label{eq:SBSF_Coulomb}
\eeq
where $\epsilon_v = \vrel/\alpha$. The BSF factor in the Yukawa regime is plotted in Fig.~\ref{fig:sommerfeld_factor}.\footnote{I thank Kallia Petraki for providing the tables for computing the BSF factor. See \cite{Petraki:2015hla, Petraki:2016cnz} for the derivation.}

\paragraph{The role of the DM velocity:} 

In the light mediator / large coupling limit, $2 \mV/\MDM \lesssim \vrel \lesssim \aD$, both $S_{\ann}^{\rm C}$ and $S_{\BSF}^{\rm C}$ scale as $\sim 1/\vrel$, and their ratio is
$S_{\BSF}^{\rm C}/S_{\rm ann}^{\rm C} \simeq 3.13$, so that BSF dominates over the direct DM annihilation, cf. Fig.~\ref{fig:sommerfeld_factor}.

However, away from the Coulomb limit $\vrel \lesssim 2 \mV/\MDM$, $S_{\rm ann}$ and $S_{\BSF}$ depend also on $\aD \MDM/\mV$, and display a resonant structure along this parameter, as mentioned above~\cite{Petraki:2016cnz}. In this regime, and away from resonances, the $1/\vrel$ behavior in the Coulomb regime switches to $\vrel^{2\ell}$, so that $S_{\rm ann}$ saturates to a constant value while $S_{\BSF}$ becomes insignificant, cf. Fig.~\ref{fig:sommerfeld_factor}.

Close to resonances, the cross sections instead grow with decreasing velocity as $1/\vrel^2$, i.e.~faster than in the Coulomb limit. 
This scaling can eventually cause an unphysical violation of the partial-wave unitarity bound on the total inelastic cross section, $\sigma_{\rm inel}^{(\ell)} \vrel < 4 \pi (2\ell+1)/(\vrel \MDM^2)$, cf. Sec.~\ref{sec:uni_bound}. The unitarity limit is also seemingly violated for very large $\aD$. 
To ensure that our computations do not violate unitarity, we impose the unitarity bound for $\ell=0$ and $\ell=1$, as a hard cutoff on the annihilation and BSF cross-sections respectively.

\paragraph{The role of BSF in depleting the DM abundance:}
The formation of bound states (BSF),
\begin{equation}
\begin{tikzcd}[column sep=4em,row sep=-0.5em]
& \qquad (\chi\overline{\chi})_{\rm BS}^{\updownarrows} + \DP,  \\
\chi+\overline{\chi}  \arrow[ur, shorten >=-5ex, pos=1.4, "\sigma_{\rm BSF}/4"] \arrow[dr, shorten >=-5ex, swap, pos= 1.5, "3\,\sigma_{\rm BSF}/4" ]   \\
& \qquad (\chi\overline{\chi})_{\rm BS}^{\upuparrows} + \DP, 
\end{tikzcd}
\end{equation}
is either followed by their decay into vector bosons $\DP$ (we can neglect the capture into excited states which is suppressed by $1/n^2$ \cite{vonHarling:2014kha}),
\begin{equation}
\hspace{-1cm}
\begin{tikzcd}[column sep=3em,row sep=-0.25em]
& (\chi\overline{\chi})_{\rm BS}^{\updownarrows} \arrow[r, "\Gamma_{\updownarrows}"] & \DP \DP,  \\
&  (\chi\overline{\chi})_{\rm BS}^{\upuparrows} \arrow[r, "\Gamma_{\upuparrows}"] & \DP \DP \DP,
\end{tikzcd}
\end{equation}
or by their photo-ionization due to ambient $\DP$,
\begin{equation}
(\chi\overline{\chi})_{\rm BS} + \DP \xrightarrow{\sigma_{\rm ion}} \chi+\overline{\chi}.
\end{equation}
BSF counts as a second process participating in the DM depletion only if bound states decay before they get ionized, $\Gamma_{\rm dec} \gtrsim \Gamma_{\rm ion}$.
On one side, since the capture into the ground state dominates, the corresponding decay rates are \cite{Stroscio:1975fa}
\begin{equation}
\Gamma_{\updownarrows} = \alpha^{5}\mu \quad \text{and} \quad \Gamma_{\upuparrows}=c_{\alpha}\alpha^{5}\mu,
\end{equation}
with $\mu \equiv m_{\chi}/2$ and $c_{\alpha}=0.12\alpha$.
On the other side, the photo-ionization rate can be evaluated using the Milne relation
\begin{equation}
\frac{\sigma_{\rm ion}}{\sigma_{\rm BSF}} = \frac{\mu^{2}v_{\rm rel}^{2}}{2\omega^{2}},
\end{equation}
with $\omega = \Delta + \mu v_{\rm rel}^{2}/2$ and $\Delta=\alpha^{2}\mu/2$. Indeed, the thermally averaged photo-ionization rate can be expressed as
\begin{equation}
\Gamma_{\rm ion} = \frac{2\cdot 4\pi}{(2\pi)^{3}}\int_{0}^{\infty} \frac{d\omega \; \omega^{2}}{e^{\omega/T}-1}\sigma_{\rm ion} =  \frac{\mu}{\pi^{2}}\int_{0}^{\infty} \frac{ d\omega \,\omega}{e^{\Delta/T}e^{\omega/T}-1}\, \sigma_{\rm BSF}.
\end{equation}
One can show \cite{vonHarling:2014kha} that the BS decay rate $\Gamma_{\rm dec} $ is faster than the photo-ionization rate $\Gamma_{\rm ion} $ when the temperature is lower than the BS binding energy $\Delta=\alpha^{2}\mu/2 $,
\begin{equation}
\label{eq:decVSion}
 \Gamma_{\rm dec} \gtrsim \Gamma_{\rm ion} \quad \to \quad T \lesssim \alpha^{2}\mu/2.
\end{equation}
Upon plugging the freeze-out temperature, $T\simeq \MDM/30$, we conclude that BSF participates in the DM depletion if 
\begin{equation}
\text{BSF depletes DM if:} \qquad \alpha \gtrsim 0.3.
\end{equation}
Hence, for large coupling $\alpha \gtrsim 0.3$, in the coulomb limit where $S_{\BSF}^{\rm C}/S_{\rm ann}^{\rm C} \simeq 3.13$, we expect BSF to additionally deplete the DM abundance by a factor $3$ related to the annihilation channel alone. Note however situations where DM depletion through BSF can be more efficient, e.g. thermal field theory corrections \cite{Binder:2018znk,Binder:2019erp,Binder:2020efn,Binder:2021vfo,Binder:2021otw}, co-annihilation \cite{Garny:2021qsr}, non-thermal DM \cite{Bollig:2021psb}.

\subsection{The unitary bound}
\label{sec:uni_bound}

\begin{table}[h!]
\centering
\hspace{-1cm}
\begin{tabular}{c|cc}
& \begin{tabular}[c]{@{}c@{}} Annihilation cross-section \\ which reproduces the \\correct DM abundance \\$\left< \sigma v_{\mathsmaller{\rm rel}} \right>_{\mathsmaller{\rm FO}}$\end{tabular} & \begin{tabular}[c]{@{}c@{}}Maximal DM \\ mass compatible \\with unitarity\\ $\MDM^{\rm uni}$\end{tabular}  \\ \\[-1em]\hline\\[-0.75em]
\begin{tabular}[c]{@{}c@{}} Majorana and \\ $n=0$ (s-wave)\\ $\quad$ \end{tabular} & $2.4 \times 10^{-26}~\rm cm^3/s$                                           & $138~\rm TeV$                                     \\
\begin{tabular}[c]{@{}c@{}} Majorana and \\ $n=-1/2$ (s-wave + Sommerfeld) \\ $\quad$\end{tabular}  & $1.1 \times 10^{-26}~\rm cm^3/s$ 1                                           & $197~\rm TeV$  \\
\begin{tabular}[c]{@{}c@{}} Dirac and \\ $n=0$ (s-wave) \\ $\quad$\end{tabular} & $4.9 \times 10^{-26}~\rm cm^3/s$                                           &$96~\rm TeV$                                      \\
\begin{tabular}[c]{@{}c@{}} Dirac and \\ $n=-1/2$ (s-wave + Sommerfeld) \\ $\quad$\end{tabular} & $2.4 \times 10^{-26}~\rm cm^3/s$                                            & $137~\rm TeV$                                                                                             \\[0.25em] \hline 
\end{tabular}
\caption{\it \small Precise values of WIMP annihilation cross-section which reproduce the correct DM abundance and the corresponding upper bound on the mass derived from unitarity. We did not include the effect of delayed annihilation and bound states formation which are sub-dominant \cite{vonHarling:2014kha}. Note that the similarity of $\MDM^{\rm uni}$ for Majorana and $n=0$ with $\MDM^{\rm uni}$ for Dirac and $n=-1/2$ comes from the almost exact cancellation between the Dirac-to-Majorana factor $2$ and the velocity dependence of the non-perturbative (Sommerfeld-enhanced) cross-section $\left<\sigma v_{\rm rel}\right>\propto 1/v_{\rm rel}$.}
\label{table:cross-section-mass-uni}
\end{table}

The unitarity of the S-matrix $S^{\dagger}S$ is just the matrix form of the conservation of the occupation probability of a quantum state $\sum_{f} P_{i\to f}$. We can split the channels into elastic and inelastic scatterings $P_{\rm elast} + P_{\rm inelast} = 1$. Thanks to the orthonormality properties of the spherical harmonics, the conservation of the probability is even satisfied for each partial wave $J$,  $P_{\rm elast}^{(J)} + P_{\rm inelast}^{(J)} = 1$.\footnote{Partial waves are  defined as eigenstates of the square of the angular momentum $\bold{J}^2$.} Next, the inelastic cross-section of a 2-to-2 process can be decomposed into partial waves with each coefficient being a function of $P_{\rm inelast}^{(J)}$
\begin{equation}
\sigma_{\rm ine} = \sum \sigma_{\rm ine}^{(J)} \quad \text{with} \quad \sigma_{\rm ine}^{(J)} = \frac{\pi (2J+1)}{p_{i}^2} P_{\rm inelast}^{(J)}.
\end{equation}
This is how in 1990, Griest and Kamionkowski  converted the unitarity requirement to an upper-bound on the annihilation cross-section  \cite{Griest:1989wd} (see App.~\ref{app:unitarity_bound} for a re-derivation) 
\begin{equation}
\label{eq:sigmav_uni}
P_{\rm inelast}^{(J)} \lesssim 1 \qquad \rightarrow \qquad \sigma_{\rm ine}^{(J)} \lesssim \frac{\pi (2J+1)}{p_{i}^2}. 
\end{equation}
In the early universe, the momentum of the particle $i$ can be written as
\begin{equation}
p_{i}^{2} = E_i^{2} - \MDM^{2} = (\gamma_i^2 -1)\MDM^{2} = \frac{\MDM^{2}v_i^{2}}{(1-v_i^{2})} \approx \MDM^{2}v_{i}^{2},
\end{equation}
Then the relative velocity $\vrel$ is easily related to the individual velocity $v_i$ in the center of mass 
\begin{equation}
\vrel^2 = (\vec{v}_1-\vec{v}_2)^2 =  4 v_i^2,
\end{equation}
hence leading to
\begin{equation}
\sigma_{\rm ine}^{(L)}\vrel\leq \frac{4\pi(2J+1)}{\MDM^{2} \vrel}.
\end{equation}
This leads to an upper bound on the mass of thermal DM. Indeed, assuming that DM freezes-out by annihilating through the partial wave $J$, we can write
\begin{equation}
\left< \sigma v_{\rm rel}  \right>_{\rm FO} \leq \left< \sigma \vrel \right>_{\rm uni}^{\rm max} \quad \rightarrow \quad \MDM^2 \leq \frac{4\pi(2J+1)}{\left< \sigma\vrel  \right>_{\rm FO} } \left<\frac{1}{\vrel}\right>,
\end{equation}
where the required annihilation cross-section in order to obtain the correct DM relic abundance, $\left< \sigma v_{\rm rel}  \right>_{\rm FO}$, can be computed from Eq.~\eqref{eq:final_DM_abundance_exact} and  Eq.~\eqref{eq:analytical_solution_Boltzmann}.
From using $\left<1/\vrel\right>=\sqrt{x_{\rm FO}/\pi}$ \cite{Griest:1989wd}, we obtain
\begin{equation}
\MDM \lesssim \left\{
                \begin{array}{ll}
                 138 \sqrt{2J+1}~{\rm{TeV} }\, f(n), \quad \text{Majorana}, \\
                 96 \sqrt{2J+1}~{\rm{TeV}}\, f(n), \qquad \text{Dirac},
                \end{array}
              \right.
\end{equation}
where $f(n)$ is a weak function of $n$, where $n$ encodes the velocity dependence of the cross-section $\sigma = \sigma_0 \,x^{-n}$, normalized at $f(n=0) = 1$ (e.g. we find $f(n=-1/2)\simeq \sqrt{2}$). We give some values of the maximal thermal DM mass allowed by unitarity, for different velocity dependence $n$, in table~\ref{table:cross-section-mass-uni}.

\begin{subappendices}

\chapterimage{overfilledglass_2} 

\section{Unitary bound on cross-sections}
\label{app:unitarity_bound}
Clarification of the computation of the unitary bound on $2\to2$ cross-section in \cite{Griest:1989wd} using results from \cite{Pilkuhn:2005cwa}.

\subsection{Partial-wave expansion of the cross-section}

We consider the scattering $1 + 2 \rightarrow 3 + 4$. The transition probability amplitude between the two asymptotic states $\left<i \right|S \left|f \right> = \left< p_{1}, \lambda_{1}; p_{2}, \lambda_{2} \right| S \left|p_{3}, \lambda_{3}; p_{4}, \lambda_{4} \right>$
can be decomposed as
\begin{equation}
S_{if} = \delta_{if} + i(2\pi)^{4} \delta^{(4)}(P_{i} - P_{f}) T_{if}
\label{s_matrix}
\end{equation}
where the matrix element $T_{if}$ contains the non-trivial part. 
The cross-section reads
\begin{equation}
d\sigma(ab \rightarrow cd) = \frac{1}{2E_{1}2E_{2} v_{\rm rel}} d \; \rm{ Lips}(s; P_{3}, P_{4})
\end{equation}
where 
\begin{equation}
2E_{1}2E_{2} \; v_{\rm rel}=\sqrt{(p_{1}\cdot p_{2})^{2} - m_{1}m_{2}} \equiv 2 \lambda^{1/2}(s, m_{1}^{2}, m_{2}^{2})
\end{equation}
In the rest frame of particule $a$, we have $v_{\rm rel} = p_{2}/E_{2} = v_{2}$ while in the center of mass frame we have 
\begin{equation}
v_{\rm rel} = \frac{p \sqrt{s}}{E_{1}E_{2}}.
\end{equation}
We notice that the center of mass initial $p_{i}$ and final $p_{f}$ momenta can be expressed as
\begin{equation}
p_{i} = (4s)^{-1/2} \lambda^{1/2} (s, m_{1}^{2}, m_{2}^{2}), \quad p_{f} = (4s)^{-1/2} \lambda^{1/2} (s, m_{3}^{2}, m_{4}^{2})
\label{initial_final_momenta}
\end{equation}
The Lorentz-invariant phase space is
\begin{align}
d \; \rm{Lips}(s, P_{3}, P_{4}) &= \frac{d^{3}p_{3}}{(2\pi)^{3}}\frac{1}{2E_{3}}\frac{d^{3}p_{4}}{(2\pi)^{3}}\frac{1}{2E_{4}}  (2\pi)^{4} \; \delta(P_{1}+P_{2} - P_{3}-P_{4}) \\
 &= \frac{1}{16\pi^{2}} \frac{1}{E_{3}E_{4}} p^{2} dp \; d\Omega \; \delta(s^{1/2} - E_{1} - E_{2} ) .
\end{align}
Now using  
\begin{align}
E &= (m_{3}^{2} + p^{2})^{1/2} + (m_{4}^{2} + p^{2})^{1/2} \\
dE &= (E_{3}^{-1} + E_{4}^{-1})\,p\,dp = E_{3}^{-1}E_{4}^{-1} E \, p \, dp,
\end{align}
the LIPS becomes
\begin{equation}
d \; \rm{Lips}(s, P_{1}, P_{2}) = \frac{1}{16\pi^{2}}\frac{p}{E} \delta(s^{1/2} - E) dE d\Omega 
\end{equation}
and 
\begin{equation}
\boxed{
d \; \rm{Lips}(s, P_{1}, P_{2}) = \frac{1}{16\pi^{2}} \frac{p}{\sqrt{s}} d\Omega
}
\label{Lips}
\end{equation}
Then the differential cross section reads
\begin{align}
d\sigma(12 \rightarrow 34) &= \frac{p_{f}}{4p_{i}s} \sum_{f} \left|\frac{T_{if}}{4\pi} \right|^{2} d\Omega \\
&= \frac{\pi p_{f}}{2p_{i}s} \sum_{f} \left|\frac{T_{if}}{4\pi} \right|^{2} d \cos \theta
\end{align}
Now, we expand $T_{if}(\theta)$ in terms of Legendre polynomials of $\cos \theta$:
\begin{equation}
T_{if}(\theta) = 8 \pi s^{1/2} \sum^{\infty}_{L=0} (2L+1) P_{L}(\cos \theta) T_{if,L}(s) .
\label{Legendre_polynomials}
\end{equation}
Using the orthogonality relation 
\begin{equation}
\frac{1}{2} \int dx P_{L'}(x)P_{L}(x) (2L+1) = \delta_{LL'}
\label{orthogonality_relation}
\end{equation}
we obtain the cross section in terms of partial waves $\sigma = \sum \sigma_{L}$
\begin{equation}
\sigma_{L} = 4\pi(2L+1)\sum_{f}\frac{p_{f}}{p_{i}} \left| T_{if, L} \right|^{2},
\end{equation}
which averaged over polarizations becomes
\begin{equation}
\boxed{
\sigma_{L} = \frac{4\pi(2L+1)}{(2s_{1}+1)(2s_{2}+1)}\sum_{\lambda}\sum_{f}\frac{p_{f}}{p_{i}} \left| T_{if, L} \right|^{2}.
}
\label{sigma_L}
\end{equation}

\subsection{Unitarity of the partial-wave expansion}
\begin{figure}[htp]
\centering
\raisebox{17pt}{\resizebox{0.6\textwidth}{!}{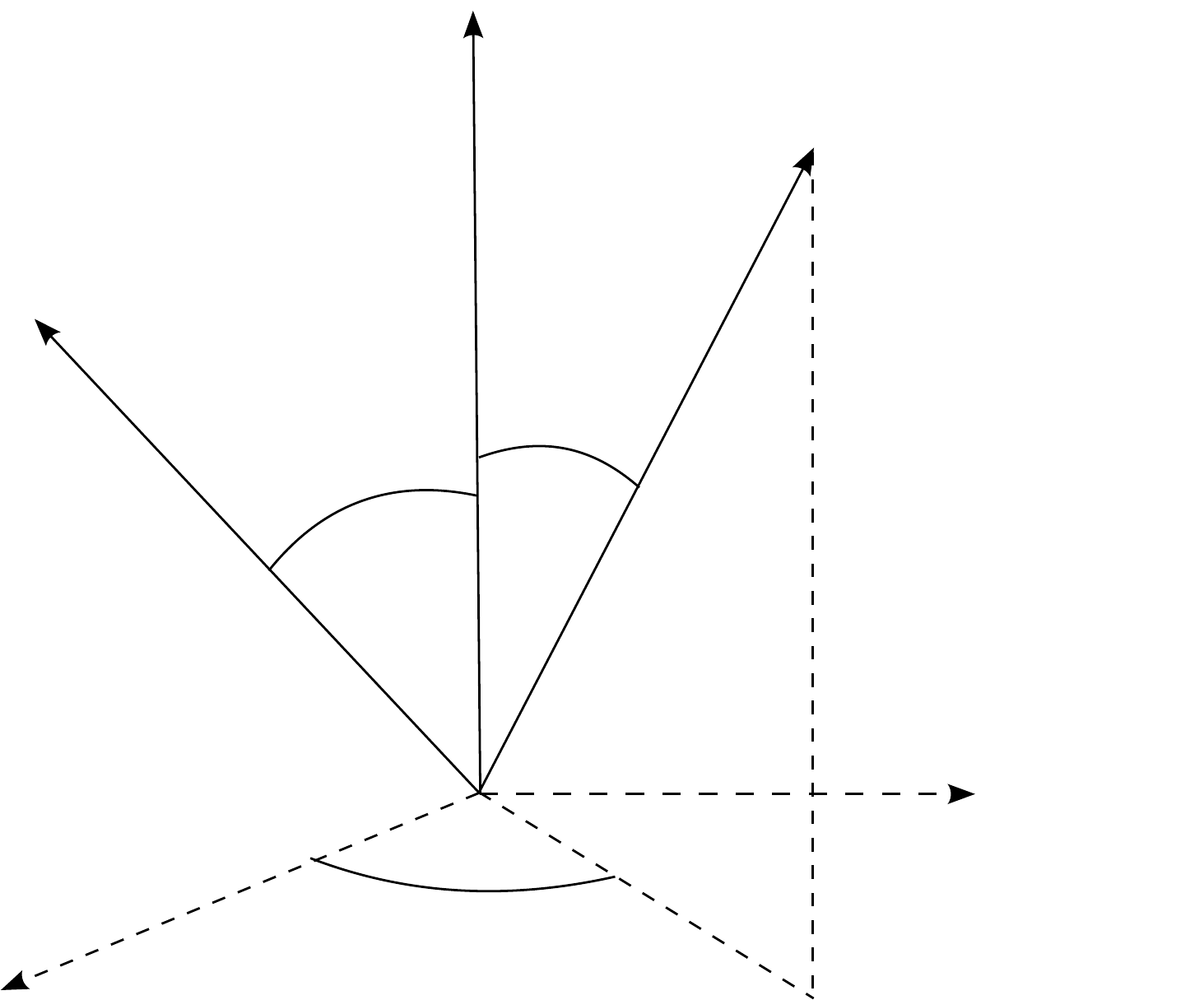}}
\caption{\it \small Solid angles $\Omega$, $\Omega'$ and $\Omega''$ which appear in the unitarity equation Eq.~\eqref{unitary_equation_L}.}
\label{unitary_bound_momenta}
\end{figure}

The unitarity means that the Hamiltionian should be Hermitian $H^{\dagger} = H$. Since the $S$-matrix is $S= e^{-iHt}$, it implies
\begin{equation}
S^{\dagger}S = 1
\end{equation}
and yields to 
\begin{equation}
-i(2\pi)^{4} \delta^{(4)}(P_{i}-P_{f})(T_{if} - T_{fi}^{*}) = \sum_{n} \left( \prod^{n}_{k=1} \int \frac{d^{3}p_{k}}{(2\pi)^{3}}\frac{1}{2E_{k}} \right) T_{ik} T_{fk}^{*} (2\pi)^{4} \delta^{(4)}(P_{i}-P_{k}) (2\pi)^{4} \delta^{(4)}(P_{f}-P_{k}),
\end{equation}
where we used Eq.~\eqref{s_matrix}. 
We get the generalized optical theorem
\begin{equation}
\boxed
{
-i(T_{if} - T_{fi}^{*}) = \sum_{n} \left( \prod^{n}_{k=1} \int \frac{d^{3}p_{k}}{(2\pi)^{3}}\frac{1}{2E_{k}} \right) T_{ik} T_{fk}^{*} (2\pi)^{4} \delta^{(4)}(P_{i}-P_{k})
}
\end{equation}
which holds order by order.
Remark: taking $i=f=A$, we obtain 
\begin{equation}
\rm{Im} \; T(A \rightarrow A) = m_{1} \sum_{X} \Gamma(A \rightarrow X) = m_{1} \Gamma_{\rm tot}
\end{equation}
if $A$ is a $1$-particle state and
\begin{equation}
\rm{Im} \; T(A \rightarrow A) = 2E_{\rm cm}\,p_{\rm cm} \sum_{X}\sigma(A \rightarrow X)
\end{equation}
if $A$ is a $2$-particles state. 
In the case of the scattering $12 \rightarrow 34$, by using Eq.~\eqref{Lips} we get
\begin{align}
-i(T_{if}(\Omega) - T_{fi}^{*}(\Omega)) &= \sum_{k} \int d \; \rm{Lips} (s; k) T_{ik} T_{fk}^{*} \\
&= \sum_{k} \frac{1}{16\pi^{2}} \frac{p_{k}}{\sqrt{s}} \int d \Omega' T_{ik}(\Omega') T_{fk}^{*}(\Omega'') 
\label{unitary_equation_L}
\end{align}
where $\Omega$ is the solid angle between $p_{1}$ and $p_{3}$, $\Omega'$ is the solid angle between $p_{1}$ and $p_{k}$ and $\Omega''$ is the solid angle between $p_{3}$ and $p_{k}$. See Fig.~\ref{unitary_bound_momenta} for a visual representation.

Now using the expansion in Legendre polynomials in Eq.~\eqref{Legendre_polynomials}, the following property of Legendre polynomials
\begin{equation}
\int d\phi^{'} P_{L}(\cos \theta'') = 2\pi P_{L}(\cos \theta) P_{L}(\cos \theta') 
\end{equation}
as well as the orthogonality relation in Eq.~\eqref{orthogonality_relation}, we can write
\begin{align}
-8\pi s^{1/2} &\sum_{L} (2L+1) P_{L}(x) i (T_{if,L} - T_{fi,L}^{*}) \\
=& \sum_{k} p_{k} \; 4 s^{1/2} \int d x' \sum_{L'} (2L'+1) P_{L'}(x') T_{ik, L'} \sum_{L}(2L+1) 2\pi P_{L}(x) P_{L}(x') T^{*}_{fk,L} \\
=& 16 \pi s^{1/2} \sum_{k} p_{k} \sum_{L'} (2L'+1)P_{L'}(x) T_{ik,L'}T^{*}_{fk,L'},
\end{align}
with $x=\cos \theta$ and $x'=\cos \theta'$. 
Finally, we get
\begin{equation}
 T_{if,L} - T_{fi,L}^{*} = 2i \sum_{k} p_{k} \;T_{ik,L} T_{fk,L}^{*}.
\end{equation}
Using matrix notation, this can be rewritten as
\begin{equation}
\boxed
{
T_{L} - T_{L}^{\dagger} = 2i T_{L} \tilde{p} T_{L}^{\dagger}
}
\end{equation}
with $\tilde{p} =$ diag$(...p_{k}...)$ where $p_{k}$ is the $3$-momentum of the intermediate state $k$.
Defining $S_{L} = 1 + 2i \tilde{p}^{1/2} T_{L} \tilde{p}^{1/2}$, we see that the partial-wave unitarity can also be written $S_{L}S_{L}^{\dagger}=1$ or 
\begin{equation}
\left| S_{el, L} \right| + \sum_{f} \left|S_{i\neq f,L}\right|^{2} = 1,
\end{equation}
where $S_{el, L}$ stands for the elastic channel, $i=f$. If we define $S_{el, L} = \eta_{L} e^{2i\delta_{L}}$, where $\delta_{L}$ is a real phase shift and $\eta_{L}$ is an inelaticity factor, $0 \leq \eta_{L} \leq 1$. Then $\left| S_{el,L} \right|^{2} = \eta_{L}^{2}$, and $\sum_{f} \left|S_{i\neq f, L} \right|^{2} = 1 - \eta_{L}^{2}$. Finally, using $T_{el, L} = (S_{el, L} - 1)/2ip$ and $T_{f\neq i, L} = S_{f\neq i, L}/2i(p_{i}p_{f})^{1/2}$ and the standard formula for the unpolarized cross section in terms of partial waves $\sigma = \sum \sigma_{L}$, where $\sigma_{L}$ is given in Eq.~\eqref{sigma_L}, we find 
\begin{align}
\sigma_{r,L} &= 4\pi \frac{2L+1}{(2s_{1}+1)(2s_{2}+1)} \sum_{\lambda} \sum_{f\neq i} \frac{p_{f}}{p_{i}} \left| T_{if,L} \right|^{2} \\
&= \frac{\pi (2L+1) (1-\eta_{L})}{p_{i}^{2}}.
\end{align}
Here $\sigma_{r,L}$ is the \textit{reaction} cross section, that is, the total cross section minus the elastic piece. It has a maximum when $\eta_{L} = 0$, so we conclude that 
\begin{equation}
\sigma_{L}(a+b \rightarrow c+d) \leq \pi (2L+1)/p_{i}^{2}.
\end{equation}
In the early Universe,
\begin{equation}
p_{i}^{2} = E^{2} - m_{X}^{2} = \frac{m_{X}^{2}v_{\rm rel}^{2}}{4(1-v_{\rm rel}^{2}/4)} \approx \frac{m_{X}^{2}v_{\rm rel}^{2}}{4},
\end{equation}
where we used $v_{i}=p_{i}/E$ and $v_{\rm rel} = 2v_{i}$. \\
So $\sigma_{L}v_{\rm rel} \leq (\sigma_{L})_{\rm max}  \;v_{\rm rel}$ where
\begin{equation}
\boxed
{
(\sigma_{L})_{\rm max} v_{\rm rel} \approx \frac{4\pi(2L+1)}{m_{X}^{2} v_{\rm rel}}
}
\end{equation}

\chapterimage{ladder_mediator_exchange} 

\section{Computation of the Sommerfeld factor}
\label{app:Sommerfeld_enhancement_num}

\subsection{The Schr\"{o}dinger equation}

The non-relativistic potential between the DM wave-functions can be computed as the Fourier transform
	\begin{equation}
	V(r) = -\frac{1}{4i\MDM^2} \int \frac{d\vec{k}^3}{(2\pi)^3} \mathcal{W}(\vec{k}) \, e^{-i\vec{k}\vec{r}},
	\label{eq:def_NR_pot_fourier_transform}
	\end{equation}
of the four-point function amplitude $\mathcal{W}(\vec{k})$ with one-boson mediator exchange~\cite{Petraki:2015hla, Oncala:2018bvl}. We are interested in the situations where the computation in Eq.~\eqref{eq:def_NR_pot_fourier_transform} leads to the static Yukawa potential (long range force)
\begin{equation}
\label{yukawa_potential}
V(\bold{r}) = -\alpha\frac{e^{-\mV r}}{r}.
\end{equation}
The DM wave-functions $\psi_k(r)$ are solutions of the Schr\"{o}dinger equation, which upon neglecting the short range potential reads
\begin{equation}
\left[ -\frac{1}{2\mu}\nabla^2 -\alpha\frac{e^{-\mV r}}{r} \right] \psi_{k} = \frac{k^{2}}{2\mu} \psi_{k},
\end{equation}
where $\mu$ is the reduced mass of the $X\overline{X}$ system $\mu = m_{\chi}/2$. \\
We expand $\psi_{k}(r)$ on Legendre polynomials
\begin{align}
&\psi_{k}(r) = \frac{1}{k}\sum_{i} i^{l}(2l+1)e^{i\delta_{l}}P_{l}(\cos{\theta}) R_{kl}(r), \\
&R_{kl}(r) = \frac{\chi_{kl}(r)}{r},
\end{align}
and we introduce dimensionless variables
\begin{align}
&x = \kappa r,\\
&\kappa = \mu \alpha = (\rm Bohr \; radius)^{-1},\\
&\xi = \kappa/\mV = \frac{\alpha m_{X}}{2 \mV} =(\rm Force~ range \;/\; BS~size), \\
&\zeta = \kappa/k = \alpha/\vrel = (\rm De~Brogglie~\lambda \;/\; BS~size),
\end{align}
such that the Schr\"{o}dinger equation becomes
\begin{equation}
\label{schrodinger_eq_yukawa}
\chi^{''}(x)+\left[-\frac{l(l+1)}{x^{2}}+\frac{1}{\zeta^{2}}+\frac{2e^{-x/\xi}}{x} \right] \chi(x) = 0.
\end{equation}

\subsection{Coulomb potential}

The solution for a massless mediator $\xi = \infty$ is found to be (see $\S$.36 of \cite{landau2013quantum}, see also \cite{Petraki:2015hla,Petraki:2016cnz})
\begin{align}
&R_{kl}(r) = \frac{C_{k}}{(2l+1)!}(2kr)^{l}e^{ikr}F(i/k+l+1,2l+2,-2ikr),
\end{align}
with the confluent hypergeometric function being such that
\begin{equation}
F(\alpha,\gamma,0) = 1.
\end{equation}
If we normalize at infinity with the condition
\begin{equation}
\psi_{k}(r) \rightarrow  \frac{\sin(kr+\delta)}{r},
\end{equation}
the normalization constant $C_{k}$ is found to be
\begin{equation}
C_{k} = \sqrt{\frac{\pi}{2}} \frac{2\sqrt{k}}{\sqrt{e^{2\pi/k}-1}}\prod^{l}_{s=1}\sqrt{s^{2}+\frac{1}{k^{2}}}.
\end{equation}
Since $R_{kl}(r)$ vanishes for $l\neq0$, the Sommerfeld enhancement factor, defined in Eq.~\ref{eq:sommerfeld_fac}, is just
\begin{equation}
S_{\rm ann} = \left| \frac{R_{k,l=0}(0)}{k} \right|^{2},
\end{equation}
which can be expressed as
\begin{equation}
S_{\rm ann}(\zeta) = \frac{2\pi\zeta}{1-e^{-2\pi\zeta}}.
\end{equation}

\subsection{Yukawa potential}

The case with finite vector boson mass does not have any analytical solution. We need to numerically integrate the Schr\"{o}dinger equation Eq.~\eqref{schrodinger_eq_yukawa} with the boundary conditions
\begin{align}
\label{bc_def1_schrodinger}
&\chi(x) \underset{\substack{x\to0}}{\longrightarrow}  0, \\
&\chi(x) \underset{\substack{x\to\infty}}{\longrightarrow} \sin(x/\zeta+\delta) .
\label{eq:BC2_yukawa_sommerfeld}
\end{align}
The second boundary condition in Eq.~\eqref{eq:BC2_yukawa_sommerfeld} is difficult to impose numerically. Instead, it is preferable to rely on an different formulation of the Sommerfeld enhancement factor, but equivalent to Eq.~\eqref{eq:sommerfeld_fac} \cite{ArkaniHamed:2008qn}
\begin{equation}
\tilde{S}_{\rm ann} = \frac{|\tilde{\chi}(\infty)|^{2}}{|\tilde{\chi}(0)|^{2}},
\end{equation}
where $\tilde{\chi}(x)$ is solution of Eq.~\eqref{schrodinger_eq_yukawa} with boundary conditions
\begin{equation}
\label{bc_def2_schrodinger}
\begin{split}
&\tilde{\chi}_(x) \underset{\substack{x\to0}}{\longrightarrow}  1, \\
&\tilde{\chi}^{'}(x) \underset{\substack{x\to\infty}}{\longrightarrow} \frac{i}{\zeta} \tilde{\chi}(x).
\end{split}
\end{equation}
The latter boundary conditions in Eq.~\eqref{bc_def2_schrodinger} are easier to implement numerically than the former in Eq.~\eqref{bc_def1_schrodinger} and \eqref{eq:BC2_yukawa_sommerfeld}. 
\paragraph{Proof:}
The equivalence between 
\begin{equation}
\label{sommerfeld_enhancement_factor_def1_and_2}
S_{\rm ann} = |\psi(0)|^{2} = \left|\frac{1}{k}\frac{\chi(0)}{dr} \right|^{2} , \qquad {\rm and} \qquad
\tilde{S}_{\rm ann} = \frac{|\tilde{\chi}(\infty)|^{2}}{|\tilde{\chi}(0)|^{2}},
\end{equation}
 can be demonstrated by using the conservation of the Wronskian
\begin{equation}
W(r) = \chi_{1}^{'}(r)\chi_{2}(r) - \chi_{1}(r)\chi_{2}^{'}(r),
\end{equation}
where $\chi_{1}$ and $\chi_{2}$ are solutions of the Schr\"{o}dinger equation in Eq.~\eqref{schrodinger_eq_yukawa} with boundary conditions
\begin{align}
\begin{split}
&\chi_{1}(x) \underset{\substack{x\to0}}{\longrightarrow}  0, \\
&\chi_{1}(x) \underset{\substack{x\to\infty}}{\longrightarrow} \sin(x/\zeta+\delta), \\
\end{split}
\begin{split}
&\chi_{2}(x) \underset{\substack{x\to0}}{\longrightarrow}  A, \\
&\chi_{2}(x) \underset{\substack{x\to\infty}}{\longrightarrow} \cos(x/\zeta+\delta).
\end{split}
\end{align}
First, notice that  $\tilde{\chi}$ introduced  Eq.~\eqref{bc_def2_schrodinger} can be written as $\tilde{\chi}(x) = \frac{1}{A}(\chi_{2}(x)+i\chi_{1}(x))$, such that
\begin{equation}
\tilde{S}_{\rm ann}=\frac{1}{|A|^{2}}.
\end{equation}
Second, notice that $\chi$, introduced in Eq.~\eqref{bc_def1_schrodinger} and \eqref{eq:BC2_yukawa_sommerfeld}, is simply $\chi_1$, such that we can compute $S_{\rm ann}$ from the conservation of the Wronskian
\begin{equation}
W(0) = \chi_{1}^{'}(0) A = W(\infty) = - k \quad \Rightarrow \quad S_{\rm ann}=\frac{1}{|A|^{2}}.
\end{equation}
Hence, the two definitions in Eq.~\eqref{sommerfeld_enhancement_factor_def1_and_2} are equivalent.\\

\end{subappendices}


%

\xintifboolexpr { \x = 2}
  {
  }
{
\medskip
\small
\bibliographystyle{JHEP}
\bibliography{thesis.bib}
}

%% file: Figures/unitary_bound_momenta.pdf_tex.tex
\begingroup%
  \makeatletter%
  \providecommand\color[2][]{%
    \errmessage{(Inkscape) Color is used for the text in Inkscape, but the package 'color.sty' is not loaded}%
    \renewcommand\color[2][]{}%
  }%
  \providecommand\transparent[1]{%
    \errmessage{(Inkscape) Transparency is used (non-zero) for the text in Inkscape, but the package 'transparent.sty' is not loaded}%
    \renewcommand\transparent[1]{}%
  }%
  \providecommand\rotatebox[2]{#2}%
  \ifx\svgwidth\undefined%
    \setlength{\unitlength}{421.30475883bp}%
    \ifx\svgscale\undefined%
      \relax%
    \else%
      \setlength{\unitlength}{\unitlength * \real{\svgscale}}%
    \fi%
  \else%
    \setlength{\unitlength}{\svgwidth}%
  \fi%
  \global\let\svgwidth\undefined%
  \global\let\svgscale\undefined%
  \makeatother%
  \begin{picture}(1,0.83002558)%
  \Large{
    \put(0,0){\includegraphics[width=\unitlength]{Figures/unitary_bound_momenta.pdf}}%
    \put(0.41506441,0.77231795){\color[rgb]{0,0,0}\makebox(0,0)[lb]{\smash{$p_{1}$}}}%
    \put(0.056372,0.57570844){\color[rgb]{0,0,0}\makebox(0,0)[lb]{\smash{$p_{3}$}}}%
    \put(0.69264511,0.66682014){\color[rgb]{0,0,0}\makebox(0,0)[lb]{\smash{$p_{k}$}}}%
    \put(0.28942616,0.3311454){\color[rgb]{0,0,0}\makebox(0,0)[lb]{\smash{$\theta$}}}%
    \put(0.41794162,0.3829352){\color[rgb]{0,0,0}\makebox(0,0)[lb]{\smash{$\theta'$}}}%
    \put(0.36017485,0.10455035){\color[rgb]{0,0,0}\makebox(0,0)[lb]{\smash{$\phi'$}}}%
    \put(-0.00093185,0.04673296){\color[rgb]{0,0,0}\makebox(0,0)[lb]{\smash{$x$}}}%
    \put(0.70978541,0.10369885){\color[rgb]{0,0,0}\makebox(0,0)[lb]{\smash{$y$}}}%
    }
  \end{picture}%
\endgroup%

%% file: chap5.tex
\chapterimage{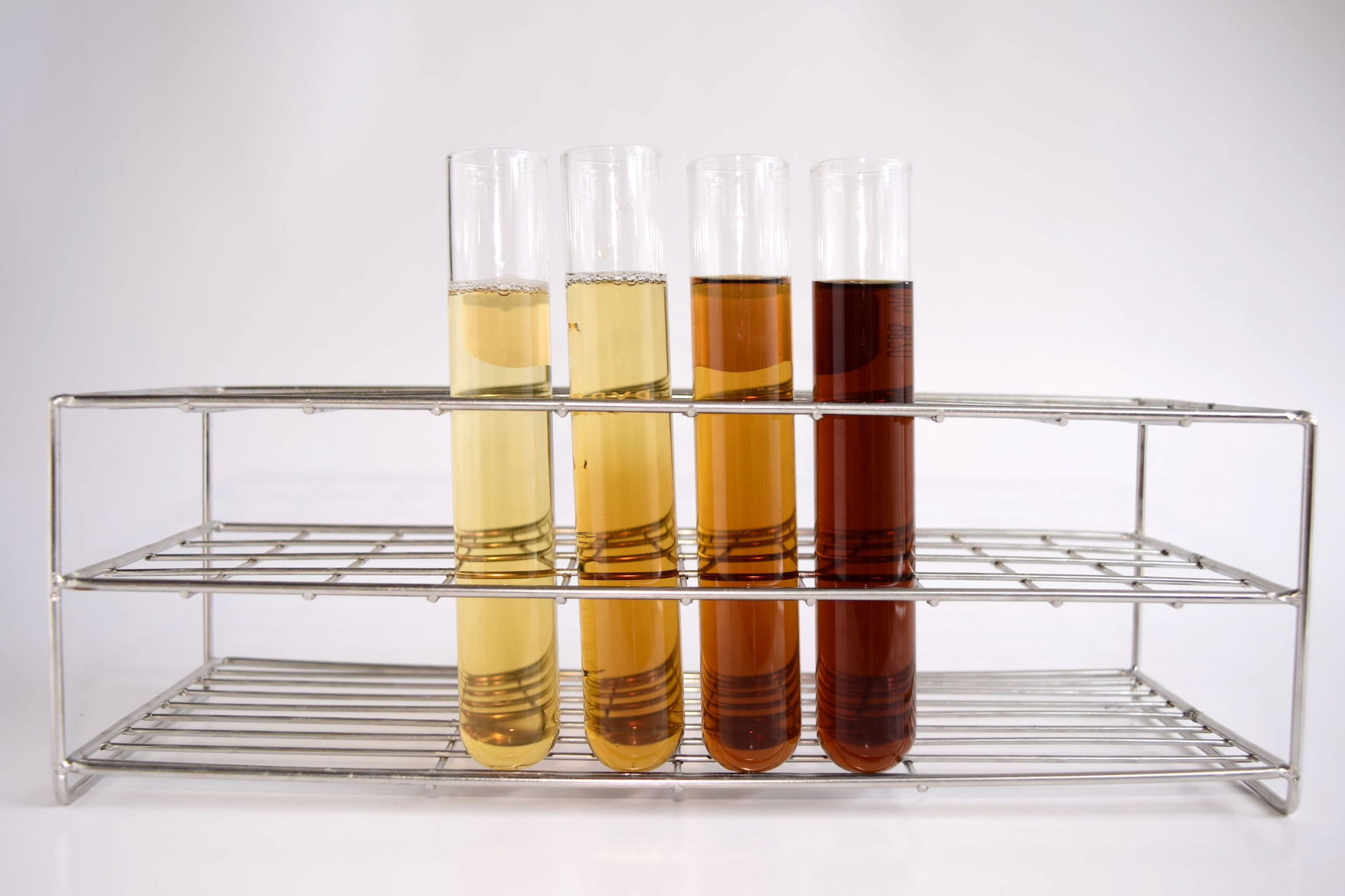} 
\chapter{Homeopathic Dark Matter}
\label{chap:secluded_DM}

\begin{tikzpicture}[remember picture,overlay]
\node[text width=18cm,text=black,minimum width=\paperwidth,
minimum height=7em,anchor=north]%
 at (7,4) {This chapter is based on \cite{Cirelli:2018iax}.};
 \draw (-2.0,2) -- (5,2);
\end{tikzpicture}

\section{Introduction}

The much anticipated physics beyond the Standard Model (BSM) has so far not been discovered at experiments, that have then pushed the scale of many related theories beyond the TeV range.
How can we gain an experimental access to such energy scales?
High energy cosmic rays currently constitute a privileged access to that realm, as opposed to the more subtle one offered by various precision measurements.
Many telescopes are indeed now observing such cosmic rays~({\sc Hess~II, Hawc, Veritas, Magic, Taiga, Antares, IceCube, Baikal, Ams, Calet, Dampe}), and many are planned for the near future~({\sc Cta, Lhaaso, Km3net, Herd, Iss-Cream}).
Using this data optimally to advance our knowledge of fundamental physics is therefore of pressing importance.

Dark matter (DM) is arguably the strongest evidence for BSM physics that can be potentially probed by these observations.
New physics (NP) sectors at scales beyond 10-100 TeV, addressing other open questions in our understanding of Nature, have long been known to 
provide DM candidates in that same mass range, see for example~\cite{Dimopoulos:1996gy,Antipin:2014qva}. Moreover, 
the empty-handed searches for weakly interacting massive particles (WIMPs) 
motivate exploring new regimes of WIMP models, including DM masses larger than about 10 TeV, see e.g.~\cite{Antipin:2015xia,DelNobile:2015bqo}.
But masses much larger than this are challenged, in the wide class of DM models where thermal freeze-out sets the relic abundance, by the so-called unitarity bound~\cite{Griest:1989wd}: the DM annihilation cross section is bounded from above by the unitarity of the $S$ matrix, and this translates into an upper bound on the DM mass of $O(100)$~TeV, see e.g.~\cite{vonHarling:2014kha,Baldes:2017gzw}.

However, the injection of entropy in the SM bath after DM freeze-out can dilute its abundance, and thus open the possibility of obtaining thermal relic DM beyond the unitarity bound. Such a dilution of relics takes place in well motivated NP models, see e.g.~\cite{deCarlos:1993wie,Banks:1993en,Moroi:1999zb}. Its consequences for DM have been studied since a long time~\cite{McDonald:1989jd}, also in relation with heavy DM~\cite{Giudice:2000ex} (in the rest of this chapter, `heavy DM' will be used as a synonym of  DM with a mass at and beyond 10-100 TeV). 
Recent years have seen a growing interest toward the possibility to evade the Griest and Kamionkowski unitarity bound \cite{Griest:1989wd} on the mass of thermal DM.
For example, DM dilution after a matter era \cite{Giudice:2000ex,Patwardhan:2015kga,Berlin:2016vnh,Berlin:2016gtr,Bramante:2017obj,Hamdan:2017psw,Allahverdi:2018aux,Cirelli:2018iax,Contino:2018crt,Chanda:2019xyl,Contino:2020god,Asadi:2022vkc}, or having a dark sector being much cooler than SM \cite{Heurtier:2019eou,Hambye:2020lvy}, DM becoming heavy only after freezing-out \cite{Davoudiasl:2019xeb}, DM annihilating with one spectator field \cite{Kramer:2021hal,Bian:2021vmi} or with many of them \cite{Kim:2019udq}, DM forming an extended object which undergoes a second annihilation stage \cite{Harigaya:2016nlg, Mitridate:2017oky,Geller:2018biy}. Mechanisms involving phase transitions include the possibility of a short inflationary stage associated with gluon-string-mediated quark-wall interactions \cite{Baldes:2020kam,Baldes:2021aph}, perturbative DM mass generation~\cite{HKonstandin:2011dr,Hambye:2018qjv,Baldes:2018emh}, DM filtered \cite{Baker:2019ndr,Chway:2019kft} or squeezed-out \cite{Asadi:2021pwo,Asadi:2021yml} by non-relativistic bubble wall motion, DM produced by elastic bubble-bubble collisions \cite{Falkowski:2012fb} or perturbative plasma interactions with relativistic walls \cite{Azatov:2021ifm}. We refer the reader to \cite{Carney:2022gse} for a review on heavy DM.

These DM scenarios offer therefore an important physics goal for high energy cosmic ray telescopes, which could test BSM physics at an energy scale where no other experiment has direct access. 
However, in the standard WIMP-inspired DM scenarios, this is complicated by the treatment of electroweak (EW) radiation at large DM masses. The SM spectra from DM annihilations are indeed governed by an expansion in $\sim \alpha_w/\pi \, \log^2(\MDM/m_W)$, that should be resummed when this number is order one (analogoulsy to what is done for QCD radiation), which happens at $\MDM \approx 100$ TeV.
At present this resummation constitutes a technical challenge, and limits the reliability of the interpretation of high energy cosmic ray data in terms of DM.\footnote{
For example, two of the tools often used for this purpose, {\sc Pppc4dmid}~\cite{Cirelli:2010xx} and Pythia~\cite{Sjostrand:2014zea}, respectively include EW radiation only at first-order~\cite{Ciafaloni:2010ti} and lack radiation processes among gauge bosons like $W^* \to W \gamma$~\cite{Christiansen:2014kba}, and thus cannot be completely trusted for heavy DM.}

\medskip

As pointed out recently in~\cite{Berlin:2016vnh,Berlin:2016gtr,Cirelli:2016rnw}, a class of DM models that allows to evade the unitarity limit consists of DM annihilating into mediators belonging to a dark sector that themselves decay into SM particles. If their lifetime is long enough and they are sufficiently heavy, the mediators may temporarily dominate the energy density of the Universe, so that, when they eventually decay, they inject significant entropy in the SM bath and thus dilute any relic, including DM. A frozen-out overabundant DM can then be diluted to the observed density. This allows for a smaller cross section at freeze-out time, and therefore a larger DM mass, relaxing the unitarity limit.

We observe that the same class of models circumvents also the challenge of reliably computing the SM spectra from heavy DM annihilations. Indeed, the energy scale relevant for computing the SM spectra is now the mass of the mediator $\mV$, rather than $\MDM$. For mediators lighter enough than $100$~TeV, $\alpha_w/\pi \log^2(\mV/m_W) \ll 1$, so one can use the spectra with EW corrections computed as in~\cite{Cirelli:2010xx,Ciafaloni:2010ti} (see \cite{Bauer:2020jay} for a more recent study) and then boost them from the mediator rest frame to the DM one. 

From the two points of view we just discussed, these \textit{secluded DM} models appear to constitute an ideal target for telescopes observing high energy cosmic rays, and would allow them to reliably test for the first time annihilating DM beyond 100 TeV. However, to our knowledge,
only two experimental analyses of secluded DM models have been performed, which are specific to neutrinos from DM in the Sun and reach DM masses up to 10~TeV~\cite{Adrian-Martinez:2016ujo,Ardid:2017lry}.
Are current and planned telescopes sensitive to DM models of this kind?
In this study we show this to be the case, thus enriching the physics case of these experiments, and opening a new window on BSM theories at the high energy frontier.

The discussion is organized as follows. In section~\ref{sec:inj_entropy} we present a quantitative model-independent study of dilution in secluded DM models, and of the limits from Big-Bang Nucleosynthesis (BBN).
For concreteness, we then consider a specific model of DM charged under a dark $U(1)$, which we discuss in section~\ref{sec:U1model}.
We study its signals in section~\ref{sec:pheno}, where we analyse the constraints from observations of CMB, 21cm, neutrinos, gamma rays, antiprotons, electrons and positrons. In section~\ref{sec:outlook} we summarise our results and indicate possible future directions.

\section{Relaxing the unitarity bound by injecting entropy}
\label{sec:inj_entropy}

We consider the dark photon, as the mediator of the DM which is charged under $U(1)_{D}$. The dark photon is produced from the annihilation of dark matter at freeze-out. When long-lived and heavy, the dark photon dominates the energy density of the universe before it decays into SM radiation. During the decay, non-relativistic degrees of freedom held in the dark photon are converted into relativistic degrees of freedom held in the radiation, hence creating entropy.

\subsection{The start of the matter era}
The temperature $T_{\rm dom}$ at which the heavy relic dominates the energy density of the universe must satisfy $\rho_{\rm rad} \simeq \rho_{V}$ (already introduced in Eq.~\eqref{eq:Tdom_def})
\begin{equation}
\frac{\pi^2}{30}g_{\rm SM} T_{\rm dom}^4 \simeq m_V f_{\mathsmaller{\rm V}}  \frac{2\pi^2}{45}g_{\rm SM}T_{\rm dom}^3 \quad \rightarrow \quad T_{\rm dom} \simeq \frac{4}{3} \; f_{\mathsmaller{\rm V}}  \;m_V.
\end{equation}
$\fV$ is the ratio of the mediator number density over the SM entropy density before decay
\begin{equation}
\fV \equiv \frac{n_V^{\rm  before}}{\sSM^{\rm before}}
= \frac{45 \, \zeta(3)}{2\pi^{4}}\frac{\gD^{\rm before}}{\gSM^{\rm before}} r_{\rm before}^{3}
= \frac{45 \, \zeta(3)}{2\pi^{4}} \frac{\gtildeD}{\gtildeSM} \, \rtilde^3 \simeq 0.0169 \left(\frac{\gtildeD}{6.5}\right)  
\rtilde^3  \,,
\label{eq:fV}
\end{equation}
Our discussion here will encompass both possibilities, and only assume that the dark sector and the SM were {\it not} in kinetic equilibrium when the SM plasma had a temperature $\TSM < \TtildeSM$.\footnote{
$\TtildeSM \gtrsim$~few~MeV could be justified by the requirement to not ruin BBN, depending on the specific model. It is not our purpose here to elaborate on this observation.}
The subsequent evolution of $\TD$ and $\TSM$ is controlled by the conservation of the entropy density of the two sectors separately; their ratio reads
\begin{equation}
\label{eq:r_ratio_decay}
r \equiv \frac{\TD}{\TSM}
=\left( \frac{\gSM}{\gD}
\frac{\gtildeD}{\gtildeSM}\right)^{\!\!\frac{1}{3}}\,
\rtilde\,,
\end{equation}
where $\gSM$ and $\gD$ are the numbers of relativistic degrees of freedom in the SM and in the dark sector, and the $\sim$ on top of a symbol denotes that the corresponding quantity is evaluated at $\TSM = \TtildeSM$.
$\rtilde = 1$ corresponds to the case where the SM and the dark plasma were in thermal equilibrium for $\TSM > \TtildeSM$.

\subsection{The end of the matter era}

\begin{figure}[!t]
\begin{center}
\includegraphics[width=0.48\textwidth]{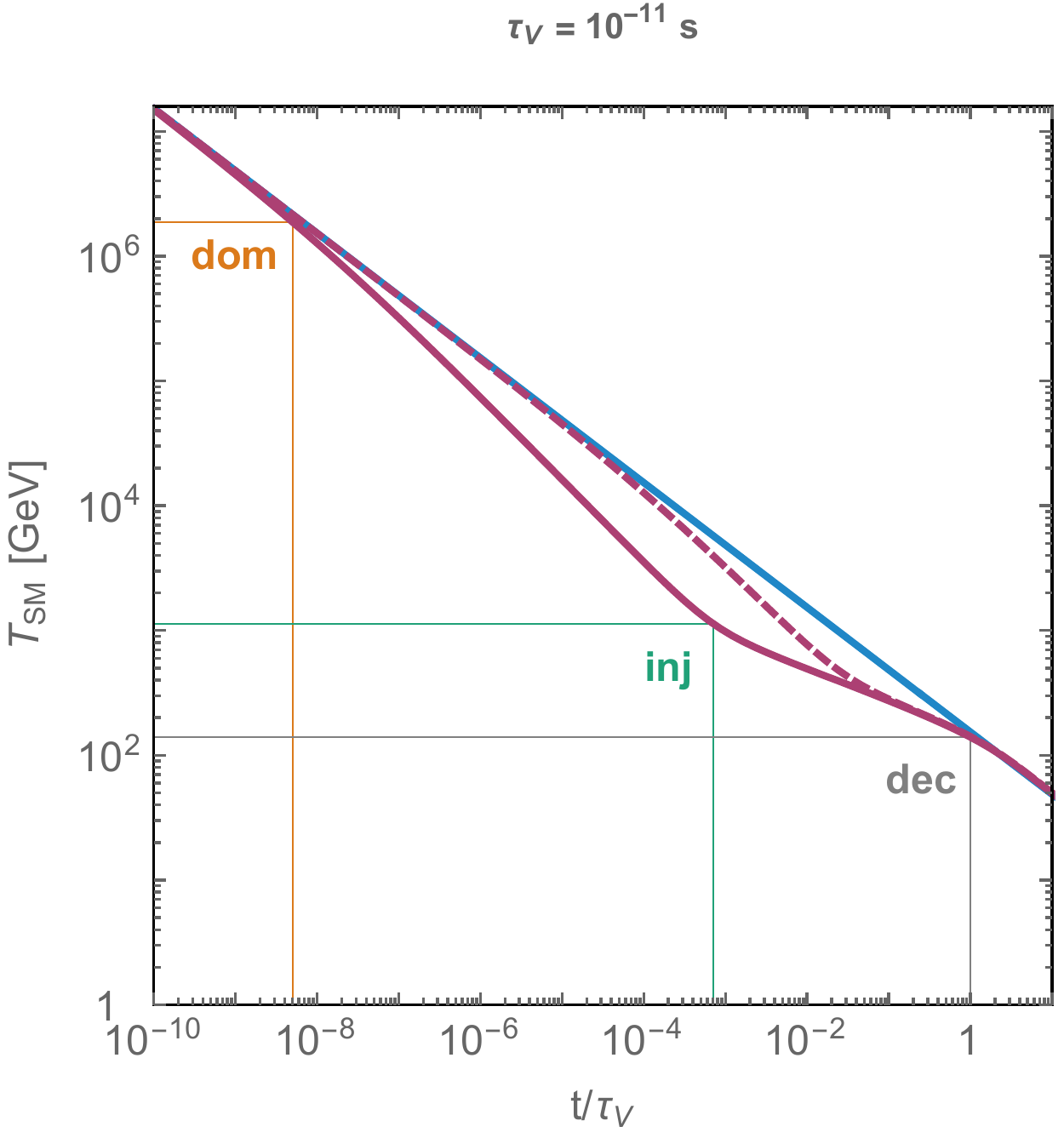}
\includegraphics[width=0.48 \textwidth]{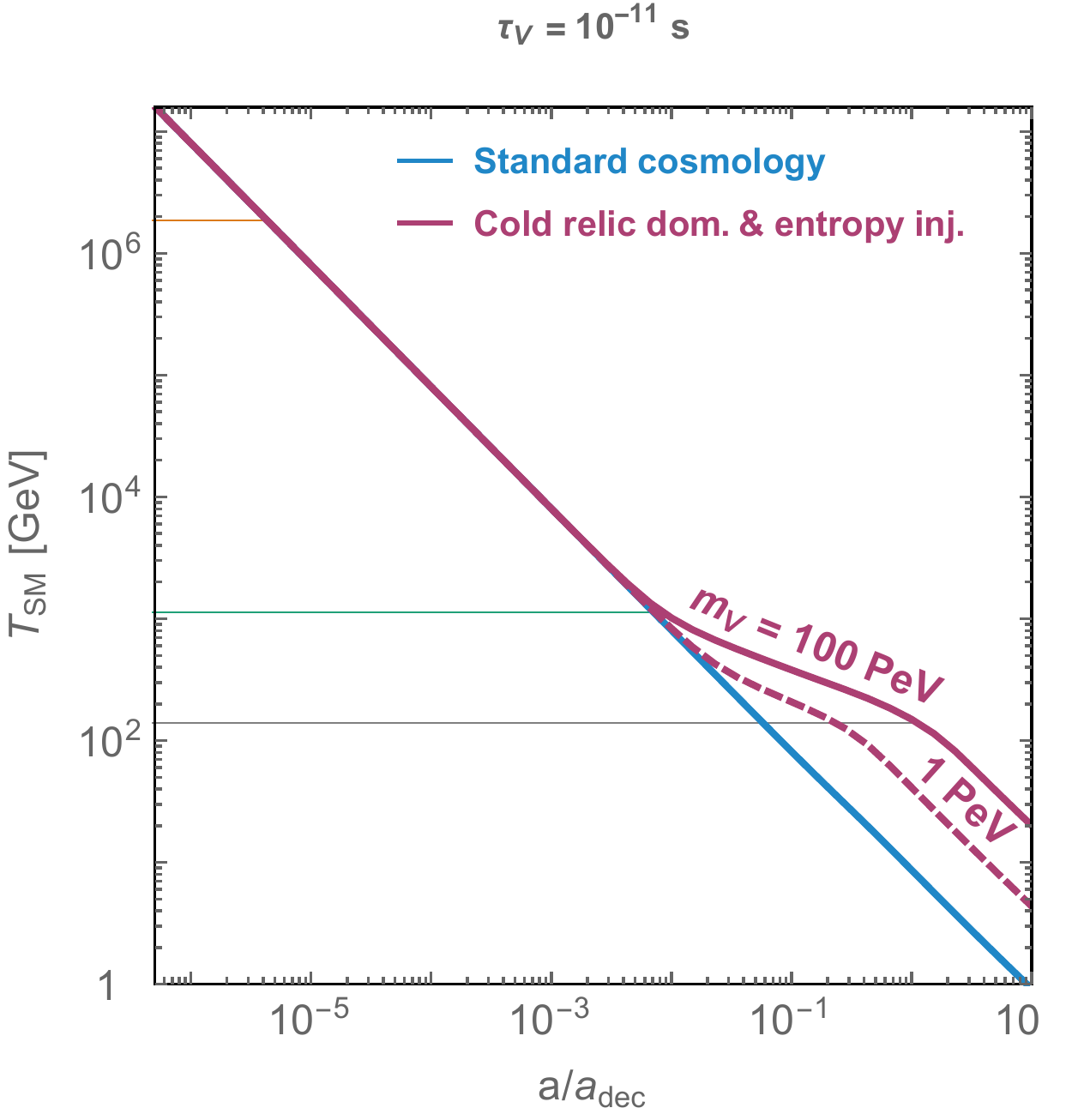}
\end{center}
\caption{\it \small
Evolution of the SM temperature with time (left) and with scale factor (right), for $\mV = 100~\text{PeV}$ (continuous purple) and 1~PeV (dashed purple), computed after integrating the full Friedmann equation. The evolution of the SM temperature in the absence of dilution is shown as a continuous blue line for comparison. 
Just after the cold relic dominates the energy density of the universe (`dom'), the temperature decreases with time as $T \propto t^{-2/3}$, i.e. faster than in a radiation-dominated Universe for which $T \propto t^{-1/2}$.
However, when entropy injection starts to become significant (`inj'),
the temperature decreases more slowly $T \propto t^{-1/4}$.
The faster redshift during the period of matter domination is counterbalanced by the slower redshift during the period of entropy injection such that the temperature of the SM at the time of decay (`dec') is independent of the mass of the relic and is the same as in standard cosmology.
Concerning the scale factor, the temperature starts to depart from the standard evolution,  $T \propto a^{-3/8}$ instead of $T \propto a^{-1}$, only after the entropy injection has started.
The three moments described above, `dom' `dec' and `inj', are indicated for the case $\mV=100$~PeV with thin lines, obtained analytically as described in the text.
}
\label{fig:temperature_evolution}
\end{figure}

In order to preserve the light element abundance, the mediator must decay before BBN starts. This places an upper limit on the possible dilution. The study in \cite{Jedamzik:2006xz} shows that the lifetime of an abundant particle, decaying $30\%$ or more hadronically, must be smaller than $0.03$~s.
In order to incorporate the modification of the cosmology due to the mediator dominating the energy density of the universe and injecting entropy, we rewrite the upper bound on the lifetime, $\tauV < 0.03$ s, as a lower bound on the SM temperature at the time of decay, $\TSMdecay > 5$ MeV. Therefore, we shall now describe the evolution of the SM temperature in our non-standard cosmology.

The evolution of the SM temperature $\TSM(z)$ in the presence of the cold relic is driven by a combination of redshift and entropy injection
\begin{equation}
\label{eq:T_vs_z}
\TSM(z) = \frac{\TSM^{\rm before}}{z}  \left( \frac{\gSM(\TSM^{\rm before})}{\gSM\left(\TSM(z)\right)} \right)^{\!1/3}  \left(  \frac{\SSM(x)}{\SSM^{\rm before}}  \right)^{\!1/3},
\end{equation}
where $z(x) \equiv a(x)/a^{\rm before}$, $x=t/\tauV$ and~\cite{Scherrer:1984fd}
\begin{equation}
\label{eq:entropy_ratio}
\frac{\SSM(x)}{\SSM^{\rm before}} = \left(1 + 
\frac{4}{3}
\frac{\fV \mV}{(\gSM^{\rm before})^{1/3} \, \TSM^{\rm before}}
\int_{0}^{x} d\tilde{x} 
\, \gSM^{1/3}(\tilde{x} ) \, z(\tilde{x} ) \, e^{-\tilde{x} }  
\right)^{3/4}.
\end{equation}
Integrating 
the Friedmann equation assuming that the cold relic dominates the universe energy density, gives
\begin{equation}
\label{eq:z_vs_x_matter_domination}
z(x)\simeq\left( \frac{3}{2}\frac{x}{x_\text{before}}\right)^{\!2/3},
\end{equation}
where $x_\text{before} = \GammaV \sqrt{3\MPl^{2}/\rho_V^{\rm before}}$ and $\rho_V^{\rm before} = \mV \fV (2\pi^{2}/45) \gSM^{\rm before}T_{\rm SM, before}^{3}$ is the energy density in the mediators before decay.
For $x \leq 1$, for simplicity we use $e^{-\tilde{x}} \simeq 1$ in eq.~\eqref{eq:entropy_ratio}.
Using eq.~\eqref{eq:z_vs_x_matter_domination}, eq.~\eqref{eq:T_vs_z} then becomes
\begin{equation}
\label{eq:T_vs_x}
\TSM(x) \simeq 1.25\,\sqrt{\MPl\GammaV}\;\frac{(\gSMdec)^{\!1/12}}{\gSM^{\!1/3}}
\frac{x_\text{inj}^{5/12}}{x^{2/3}}\left( 1 + \Big( \frac{x}{x_{\rm inj}} \Big)^{\! 5/3}  \right)^{\!1/4} ,
\end{equation}
where 
\begin{equation}
x_\text{inj} \simeq 1.43 \left(\frac{\MPl^2}{\gSMdec \, \fV^{4} \, \mV^4 \, \tauV^2} \right)^{1/5}\simeq  \left(  \frac{T_{\rm dec}^{\rm after}}{T_{\rm dom}} \right)^{4/5}, \qquad T_{\rm inj}\simeq T_{\rm dec}^{\rm after} \left(  \frac{T_{\rm dom}}{T_{\rm dec}^{\rm after}} \right)^{1/5}.
\end{equation}
$x_\text{inj}$ corresponds to the time where entropy injection becomes significant, which we define as $\SSM(x_\text{inj})/\SSM^{\rm before} = 2^{3/4}$, see eq.~\eqref{eq:entropy_ratio}. Note that eq.~\eqref{eq:T_vs_x} implies $\TSM \propto x^{-2/3}$ for $x \ll x_\text{inj}$, as expected for a matter-dominated Universe, and $\TSM \propto x^{-1/4}$ for $x \gg x_\text{inj}$, as expected for a matter-dominated Universe during a period of entropy dilution (in which $T \propto a^{-3/8}$).
We emphasise that, in deriving eq.~\eqref{eq:T_vs_x}, we have assumed a matter dominated universe. That is an excellent approximation for $x \ll 1$. However, around $x\simeq 1$, the radiation energy content of the universe is non-negligible, and therefore our result for the temperature at time of decay, $\TSMdecay$, is a priori less accurate.
We also determine the temperature evolution by solving the Friedmann equations (with decay) numerically, and show our results in fig.~\ref{fig:temperature_evolution}.
Our analytical determination of $x_{\rm inj}$ and $\TSM^{\rm inj}$ reproduces very well the point where the numerical solution changes slope, as also shown in fig.~\ref{fig:temperature_evolution}.
Setting $x=1$ in eq.~\eqref{eq:T_vs_x} and assuming a significant entropy injection $x_{\rm inj}\ll1$, we get the well known temperature below which standard cosmology is recovered~\cite{Scherrer:1984fd}
\begin{equation}
\label{eq:TSMdecay}
\TSMdecay \approx 1.25 \frac{(\gSMdec)^{\!1/12}}{\gSM^{\!1/3}}\sqrt{\MPl\GammaV}.
\end{equation}
We first note that $\TSMdecay$ reproduces fairly well its numerical determination, see fig.~\ref{fig:temperature_evolution}.
$\TSMdecay$ is independent of the mass of the relic (except via the dependence of $\GammaV$ on $\mV$), its number of degrees of freedom or its initial temperature. %
Furthermore, under the very good approximation where $\gSMdec=\gSM$, $\TSMdecay$  is almost equal to the temperature of a radiation-dominated universe with age $\tauV$ (set by $H(\TSM)=1/2\tauV$),
$
T_{\mathsmaller{\rm rad}}(\tauV) \simeq  1.23 \sqrt{\MPl\GammaV}/\gSM^{1/4}
$.\footnote{This corroborates the validity of the approximation, common in the literature, where it is assumed that the cold relic decays instantaneously when $H = \Gamma_{V}$.}
The net result is that if a cold relic has dominated the energy density of the universe and has subsequently injected entropy during its decay, then the temperature of the universe at decay time, $\TSMdecay$, corresponds roughly to that in the standard cosmology (namely without early matter domination nor entropy injection) at time $t=\tauV$. The only difference with respect to the standard case is that the universe has expanded more: this is the dilution effect.

\subsection{Dilution by entropy injection}
\label{sec:dilution_DarkU1}

\paragraph{The dilution factor:}
We first review the dilution via entropy injection. When a cold relic (in our case, the mediator) decays into SM particles, its energy density is transferred to the SM bath. The non-relativistic degrees of freedom of the mediator are converted into relativistic degrees of freedom of the SM. This heat production increases the total entropy of the Universe and dilutes the frozen-out comoving DM number density $n_{X}/s_{\rm tot}$.
We shall use the superscripts `before' and `after' to denote the times before $V$ decays but after DM freezes-out ($t \ll \tauV$),  and after $V$ decays ($t \gtrsim \tauV$), respectively. 
Then the comoving DM density today $n^{0}_{X}/s_{\rm tot}^{0}$ is related to the comoving DM density at freeze-out $n^\FO_{X}/s^\FO_{\rm tot}$ through\footnote{
As is conventional, the uppercase letters $S$ and $N$ refer to the comoving entropy and number densities whereas the lowercase $s$ and $n$ refers to the entropy and number densities, such that $S=s a^{3}$ and $N=n a^{3}$ where $a$ is the scale factor.
}
\beq
\frac{n^{0}_{X}}{s_{\rm tot}^{0}}
= \frac{n^{0}_{X}}{\sSM^{0}}
= \frac{N^{\rm after}_{X}}{\SSM^{\rm after}} 
= \frac{1}{\DSM}\frac{N^{\rm before}_{X}}{\SSM^{\rm before}} 
=   \frac{1}{\DSM}  \frac{n^\FO_{X}}{\sSM^\FO} 
=   \frac{1}{D}  \frac{n^\FO_{X}}{s_{\rm tot}^\FO} \,,
\label{eq:dilution_factor}
\eeq
where we used $N_X^{\rm after} = N_X^{\rm before}$ and defined the dilution factors
\begin{equation}
\label{eq:dilution_factor_definition}
\DSM \equiv \SSM^{\rm after}/\SSM^{\rm before}
\qquad \text{and} \qquad
D \equiv \SSM^{\rm after}/S_{\rm tot}^{\rm before} \,,
\end{equation}
which are of course related via
\begin{equation}
D = \frac{\DSM}{1+ (\gtildeD/\gtildeSM) \rtilde^3} \,. 
\label{eq:D_and_DSM}
\end{equation}
$D = 1$ corresponds to no dilution of the DM density.

\paragraph{The instantaneous approximation:}
If we assume that the decay occurs instantaneously when $H\simeq \Gamma_V$, then we can neglect the universe expansion and the energy density is conserved through the decay. Therefore, we can deduce the temperature of the universe just before the decay $T_{\rm dec}^{\rm before}$ 
\begin{align}
\rho_{\rm dec}^{\rm before} = \rho_{\rm dec}^{\rm after} \quad &\rightarrow \quad m_V \; f_{\mathsmaller{\rm V}}  \; \frac{2\pi^2}{45}g_{\rm SM} \left(T_{\rm dec}^{\rm before} \right)^{3} = \frac{\pi^2}{30}g_{\rm SM} \left(T_{\rm dec}^{\rm after} \right)^{4}, \\
 \quad &\rightarrow \quad  \frac{4}{3}\;m_V\;f_{\mathsmaller{\rm V}}  = \dfrac{\left(T_{\rm dec}^{\rm after}\right)^4}{\left(T_{\rm dec}^{\rm before}\right)^3}, \\
  \quad &\rightarrow \quad  \left(T_{\rm dec}^{\rm before}\right)^3 = \dfrac{\left(T_{\rm dec}^{\rm after}\right)^4}{T_{\rm dom}}.
\end{align}
We deduce the dilution factor
\begin{equation}
\label{eq:dilution_factor}
D_{\rm SM} \equiv \frac{S^{\rm after}}{S^{\rm before}} \simeq \left( \frac{T_{\rm dec}^{\rm after}}{T_{\rm dec}^{\rm before}} \right)^3 \simeq \frac{T_{\rm dom}}{T_{\rm dec}^{\rm after}} \simeq \frac{4}{3} \,f_{\mathsmaller{\rm V}}  \,\left( \frac{\pi^2 g_{\rm SM}}{90} \right)^{\!1/4}\frac{m_V}{\sqrt{\Gamma_V m_{\rm pl}}}.
\end{equation}
where $S$ is the total entropy $S = s \; a^3$ and $s$ is the comoving entropy.

\paragraph{The Friedmann equation with decay term:}

A more precise computation of the dilution factor involves the second law of thermodynamics together with the resolution of the Friedmann equation in the presence of decaying matter. We refer the reader to \cite{Scherrer:1984fd} for the derivation and we just report here the expression for the dilution factor 
($\MPl \simeq 2.4\times10^{18}$ GeV, $\GammaV=1/\tauV$)
\begin{equation}
\DSM = 
\left[ 1 + 0.77 \; (\gSMdec)^{1/3} \; \fV^{4/3} 
\left(\frac{ \mV^2 }{\GammaV\,\MPl}\right)^{\!\! \frac{2}{3}} \right]^{\!\! \frac{3}{4}},
\label{eq:dilution_factor_expression_VD_dom_approx}
\end{equation}
which assumes that the mediator dominates the energy density of the Universe when it starts to decay. By comparing Eq.~\eqref{eq:dilution_factor} and Eq.~\eqref{eq:dilution_factor_expression_VD_dom_approx}, we conclude that the error done by using the instantaneous approximation is only $7$~\%.

\paragraph{The effect of the dark sector temperature on the dilution.}
From eq.~(\ref{eq:dilution_factor_expression_VD_dom_approx}) we see that for a given mass and lifetime of the mediator, the dependence of the dilution factor on the thermodynamics of the dark sector -- its degrees of freedom and its temperature -- is dominantly encoded in $\fV$. We neglect the dependence on $\rtilde$ introduced by $\gSMdec$, because it is much milder than that arising from $\fV$, and because here we do not aim at an extremely precise study but rather at a simple analytical understanding.
We may then rewrite eq.~(\ref{eq:dilution_factor_expression_VD_dom_approx}) as
\beq
\DSM \simeq \left[1 + \left(\DSMbar^{4/3}-1 \right) \, \rtilde^4 
\right]^{3/4},
\label{eq:DSM_rtilde}
\eeq
where $\DSMbar$ encapsulates all the dependence on the model parameters $\mV$, $\GammaV$, $\gtildeD$ and $\gtildeSM$, %
\begin{equation}
\DSMbar \equiv 
\left[1+ 
\left(0.23 
\:(\gSMdec)^{1/4} 
\: \frac{\gtildeD}{\gtildeSM} 
\frac{\mV}{\sqrt{\GammaV \MPl}}
\right)^{4/3}
\right]^{3/4}
\,,
\label{eq:DSM_rtilde1}
\end{equation}
where $\DSMbar = \DSM^{\rtilde=1}$. For large values of $\DSMbar$ and/or the temperature ratio $\tilde{r}$, eq.~\eqref{eq:DSM_rtilde} implies the simple scaling 
$\DSM \propto \rtilde^3$,  with $D$ consequently saturating to a constant value, $D \simeq (\gtildeSM/\gtildeD) \DSMbar \simeq 0.23 (\gSMdec)^{1/4}  \mV  /  \sqrt{\GammaV \MPl}$ [cf.~eq.~\eqref{eq:D_and_DSM}].

\paragraph{Cross-section.} Ultimately, we are interested in the couplings needed to  reproduce the correct relic abundance, since they determine the DM signals. Let $\langle \sigma \vrel \rangle^\FO$ be the thermally averaged effective annihilation cross-section times relative velocity around the time of freeze-out, which is of course determined by these couplings.\footnote{
We use the specification ``$\FO$" because if $\sigma \vrel$ depends on $\vrel$, as is the case with Sommerfeld enhanced processes that are relevant for heavy DM, then its value during freeze-out is of course different than at later times, when the DM indirect detection signals are generated. Moreover, we use the term ``effective annihilation cross-section" because the DM density may be depleted via processes other than direct annihilation -- in particular, the formation and decay of unstable bound states -- as we shall see in section~\ref{sec:U1model}.}  
Large dilution implies that a smaller $\langle \sigma \vrel \rangle^\FO$ is needed. 
From eq.~(\ref{eq:DSM_rtilde}), it is evident that the dilution is larger the warmer the dark sector is, since more energy is stored into the mediators and transferred to the SM once they decay. However, $r_\FO \neq 1$ also affects $\langle \sigma \vrel \rangle^\FO$ directly, i.e.~even in the absence of any significant dilution.  We shall now incorporate this effect.

In the presence of a dark plasma, the Hubble parameter is
$ H/H_\mathsmaller{\rm SM} = \sqrt{1+ (\gD / \gSM) \,r^4}$, 
where $H_\mathsmaller{\rm SM}$ is the Hubble parameter in the absence of the dark sector. The comoving DM energy density today is
\beq
\frac{\MDM n_X^0}{\sSM^0} 
= \frac{\MDM}{\DSM} \frac{n_X^\FO}{\sSM^\FO }
\simeq  
\frac{\MDM }{\DSM \, \sSM^\FO} \ 
\frac{H^\FO}{\langle \sigma \vrel \rangle^\FO} 
=
\frac{\MDM  H_{\mathsmaller{\rm SM}}^\FO}{\sSM^\FO}
\frac{\sqrt{1+ (\gD^\FO / \gSM^\FO) \,r_\FO^4}}{\DSM} 
\frac{1}{\langle \sigma \vrel \rangle^\FO}
\,,
\eeq
where in the first step we used eq.~\eqref{eq:dilution_factor}, and in the second step we used the approximate freeze-out condition 
$n_X^\FO \langle \sigma \vrel \rangle^\FO \simeq H^\FO$. The combination 
$\MDM H_{\mathsmaller{\rm SM}}^\FO / \sSM^\FO$ is proportional to 
$\MDM /\TSM = r_\FO x^\FO$, where $x^\FO \equiv \MDM/\TD^\FO$ (not to be confused with $x=t/\tauV$ defined earlier) itself depends on $r_\FO$. From the above, we deduce that 
\beq
\frac
{\langle \sigma \vrel \rangle^\FO} 
{\langle \sigma \vrel \rangle^\FO_{\mathsmaller{\rm SM}}}
\simeq 
\frac{\xfo}{\xfoSM}\,
\frac{r_\FO \sqrt{1+ (\gD^\FO/\gSM^\FO)r_\FO^4}}{\DSM} \,,
\label{eq:sigmav_fo}
\eeq
where $\xfoSM$ and $\langle \sigma \vrel \rangle^\FO_{\mathsmaller{\rm SM}}$ denote respectively the time parameter $\xfo$ and the cross-section needed to establish the observed DM density if DM were annihilating directly into SM particles (in which case we identify $\TD$ with $\TSM$ in the definition of $\xfo$). $\DSM$ can be computed from  eqs.~\eqref{eq:DSM_rtilde} and \eqref{eq:DSM_rtilde1}, and the dark-to-SM temperature ratio at freeze-out $r_\FO$ is related to $\rtilde$ via eq.~\eqref{eq:r_ratio_decay}.

\begin{figure}
\centering
\includegraphics[height=0.45\textwidth]{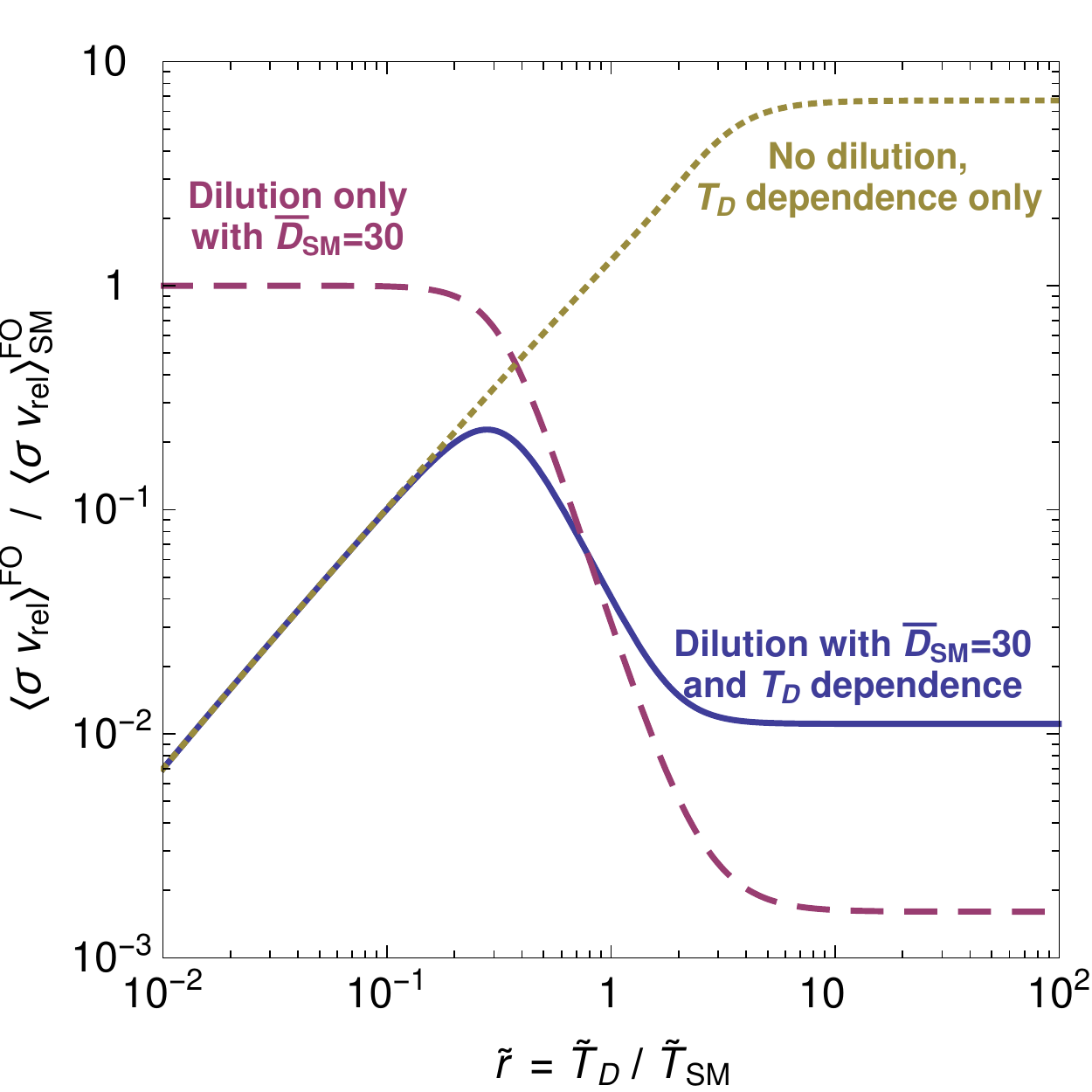}
~~~
\includegraphics[height=0.45\textwidth]{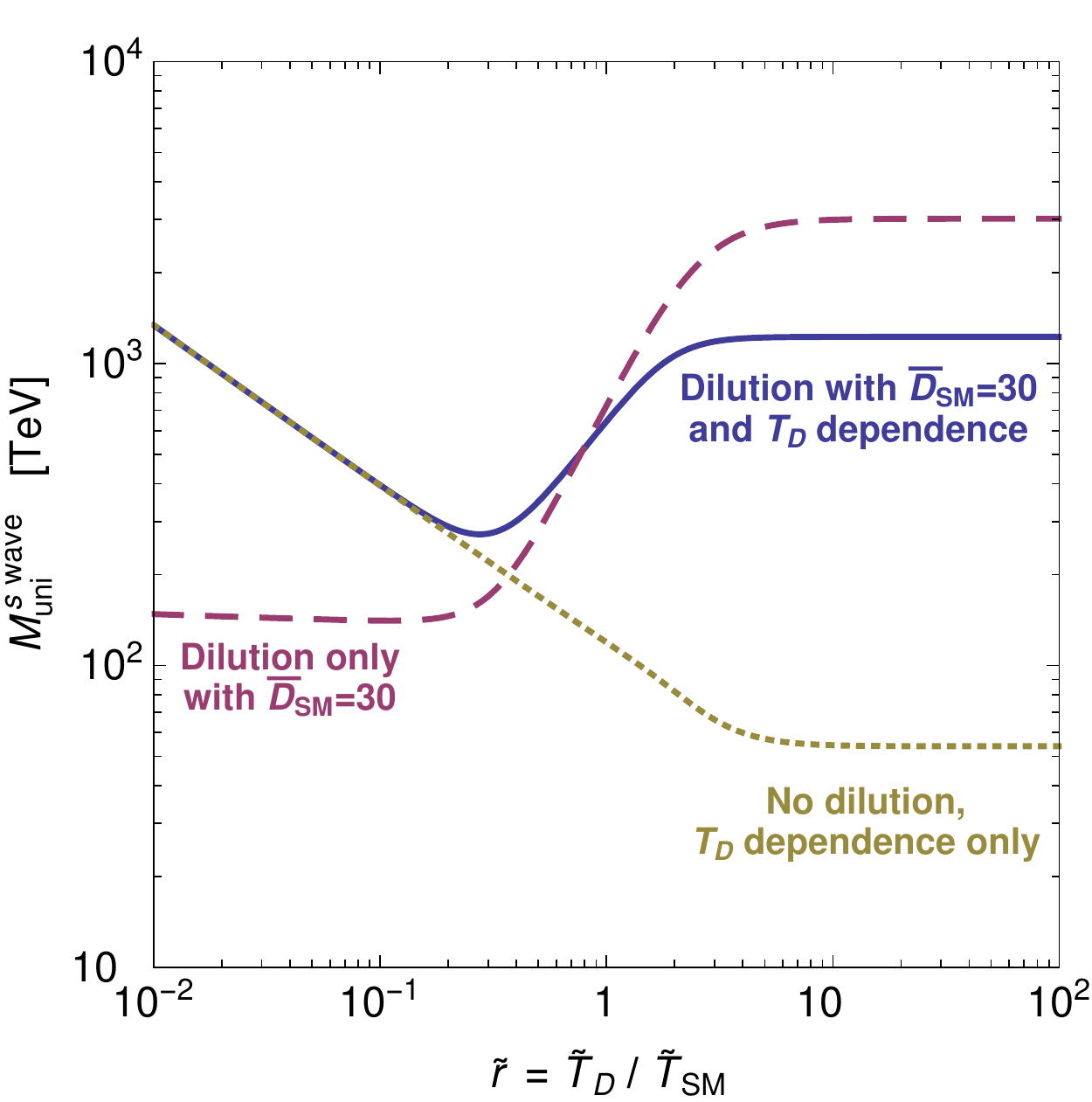}
\caption{\it \small \label{fig:sigmaFOandMuni} \it \small
Dependence of the effective annihilation cross-section at freeze-out required to attain the observed DM density (left) and of the $s$-wave unitarity limit on the mass of thermal relic DM (right), on the dark-to-SM temperature ratio and on the dilution due to the decay of the dark mediators. 
$\langle \sigma \vrel\rangle^\FO$ is normalized to the corresponding value for DM of the same mass annihilating directly into SM particles with no dilution occurring, $\langle \sigma \vrel\rangle_{\mathsmaller{\rm SM}}^\FO$.  The continuous blue lines display the combined effect of the dilution and the temperature ratio.
To ease its understanding we also show the effect without dilution, namely $D=1$ (dotted yellow), and of the dilution only, namely we retain the dependence on $\rtilde$ of $\langle \sigma \vrel\rangle_{\mathsmaller{\rm SM}}^\FO$ (dashed purple).
Note that higher partial waves may contribute significantly to the depletion of DM in the early Universe, thereby raising the unitarity bound on the DM mass. 
For definiteness, we have used $\gtildeSM = 106.75$, $\gtildeD = 6.5$, $\gD^\FO =3$ (cf.~sec.~\ref{sec:U1model}) and for the left panel we assumed $\xfoSM =29$.
}
\end{figure}

Let us now discuss the features of eq.~\eqref{eq:sigmav_fo}. For small $\rtilde$, the dilution factor is close to unity and eq.~\eqref{eq:sigmav_fo} implies $\langle \sigma \vrel \rangle_\FO \propto r_\FO$. 
This is expected since for smaller $\tilde{r}$ the DM freeze-out is happening earlier, thus $\TSM^\FO$ is larger. This implies that
\begin{itemize}
\item[$\circ$] the dilution due to the subsequent Hubble expansion is larger, so that the cross-section must be reduced by $r_\FO^3$ in order to compensate for the deficit in the DM abundance,
\item[$\circ$] the Hubble expansion rate during freeze-out is faster, so that the cross-section must be enhanced by $r_\FO^{-2}$ in order to keep $\TD^\FO$ and then the DM abundance fixed.
\end{itemize} 
As a result $\langle \sigma \vrel \rangle_\FO \propto r_\FO$.
For larger $\rtilde$, the dilution factor becomes more significant, thus suppressing $\langle \sigma \vrel \rangle_\FO$. However, for sufficiently large $\rtilde$, the dark sector energy density dominates the Universe and the Hubble expansion rate becomes independent of $r_\FO$, resulting in the suppression of $\langle \sigma \vrel \rangle_\FO$ to saturate to the value
\beq
\left[\frac
{\langle \sigma \vrel \rangle^\FO} 
{\langle \sigma \vrel \rangle^\FO_{\mathsmaller{\rm SM}}}
\right]_{\min} 
\simeq
\frac{\xfo}{\xfoSM}
\frac{1}{\DSMbar} \, \frac{\gtildeD/\gtildeSM}{\sqrt{\gD^\FO/\gSM^\FO}}
\simeq
4.35~\frac{\xfo}{\xfoSM}
\frac{\sqrt{\GammaV \MPl}}{\mV}
\frac{(\gSM^\FO/\gD^\FO)^{1/2}}{(\gSMdec)^{1/4}} \,.
\label{eq:sigmaFO_Min}
\eeq
The interplay between $\rtilde$ and $\DSMbar$ in determining $\langle \sigma \vrel \rangle^\FO$ is depicted in fig.~\ref{fig:sigmaFOandMuni} (left panel).

\subsection{Impact on unitary bound  \label{sec:UniBound}}

\begin{figure}[t!]
\begin{center}
\includegraphics[width=0.48\textwidth]{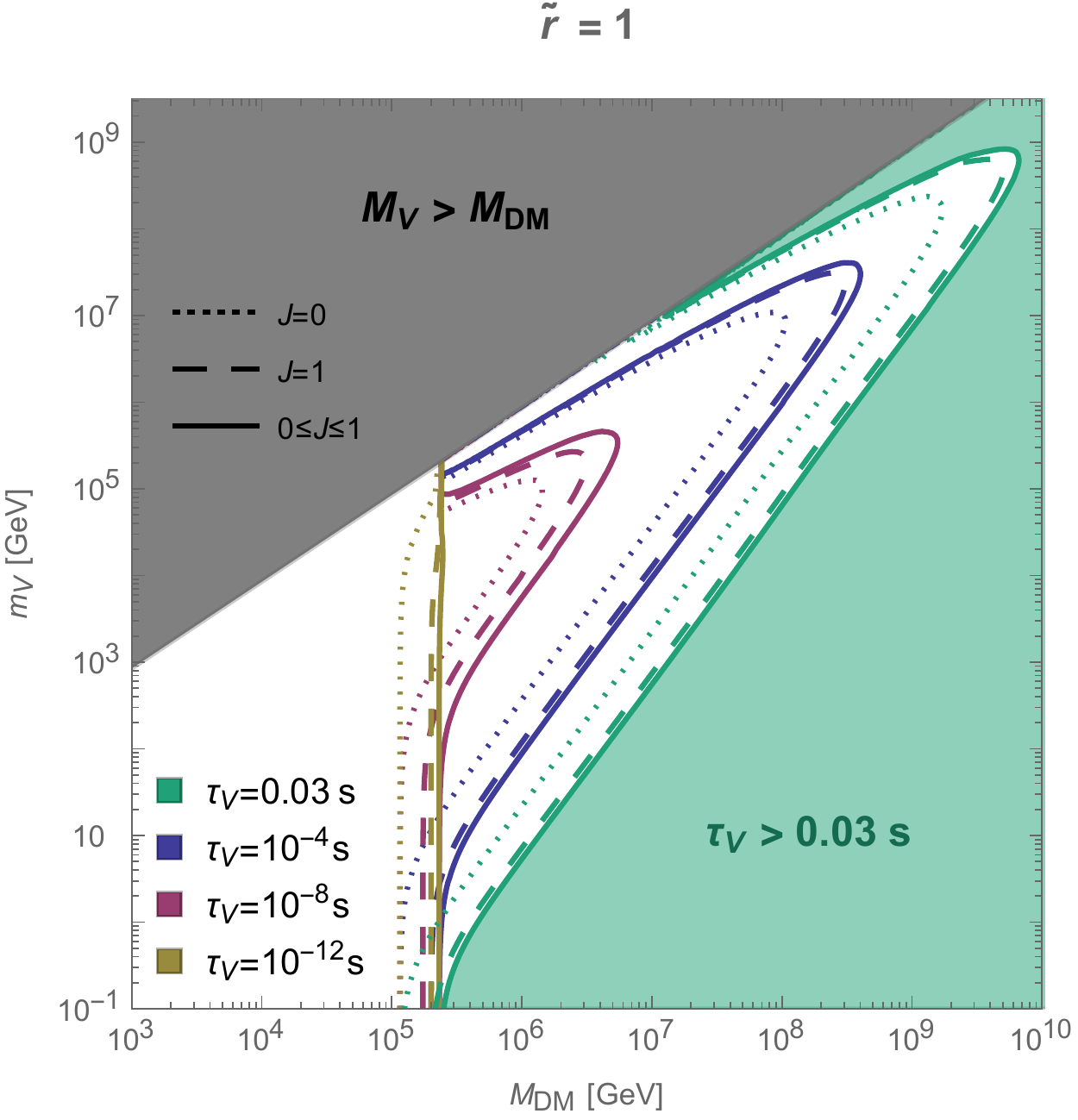}
~~~
\includegraphics[width=0.48\textwidth]{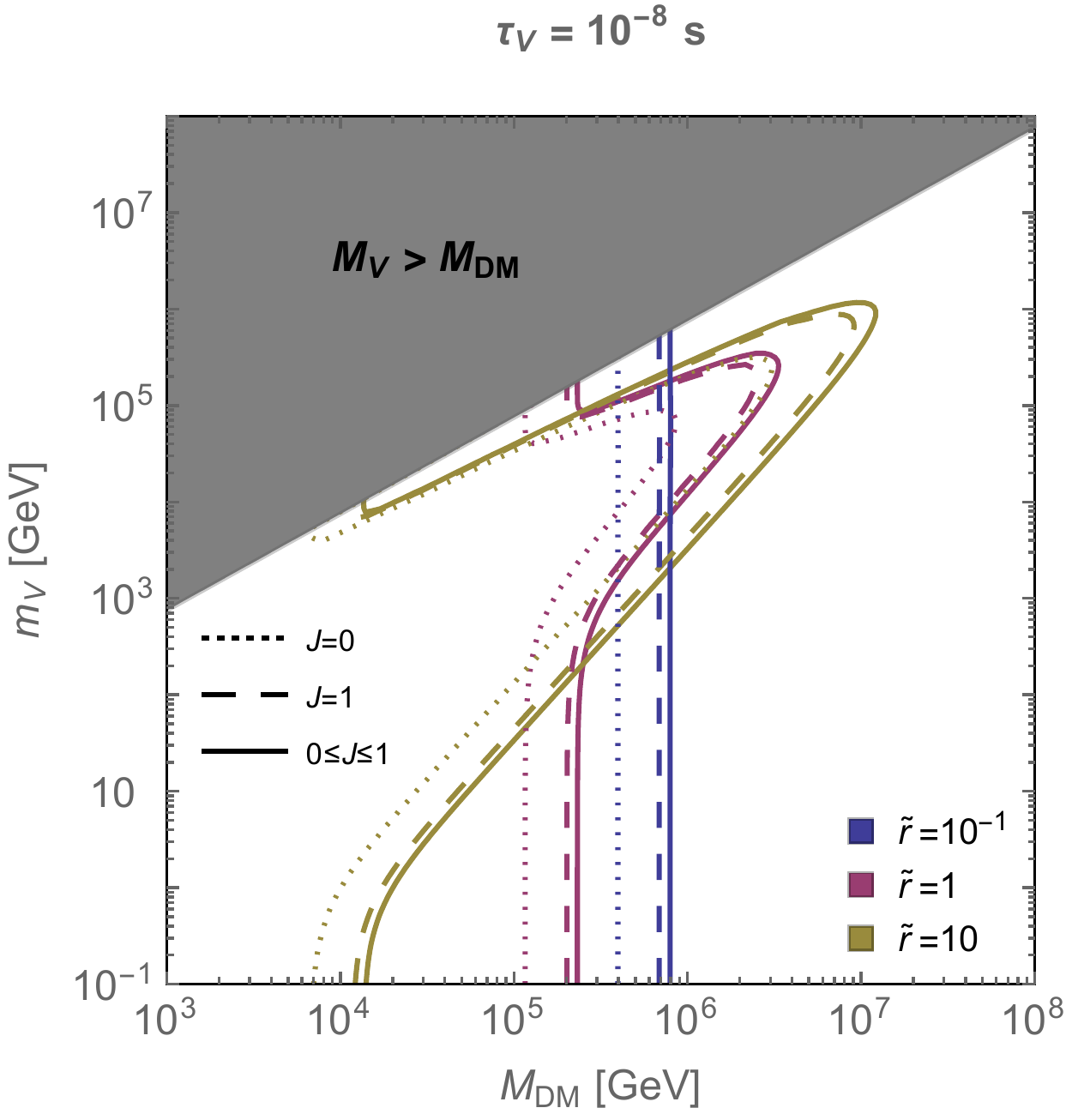}
\end{center}
\caption[]{\it \small
The dependence of the upper bound on the mass of thermal relic DM implied by unitarity, $\MDM \leqslant \Muni$ (coloured lines), on the mediator mass $\mV$ and lifetime $\tauV$, and the dark-to-SM temperature ratio $\tilde{r}$. For definiteness, we have used $\gtildeSM = 106.75$, $\gtildeD = 6.5$ and $\gD^\FO =3$ (cf.~sec.~\ref{sec:U1model}).
\emph{Left:} The dark plasma and the SM are assumed to have the same temperature at early times, $\rtilde =1$. Large $\mV$ and/or $\tauV$ imply that the cosmological energy density carried by the mediators before their decay is large, leading to significant dilution of the frozen-out DM abundance upon their decay and raising $\Muni$. However, a lifetime longer than $0.03$~s
is in conflict with BBN (green-shaded area). Also, the dilution factor gets exponentially suppressed with $\mV$ when $\mV$ is close to $\MDM$ (footnote 4), thus bringing $\Muni$ back to its value in the standard scenario (for the appropriate $\rtilde$).
\emph{Right:} For fixed $\tauV$ and small $\mV$ the dilution is negligible; then, larger $\rtilde$ implies less efficient depletion of DM (due to less Universe expansion from DM freeze-out, which happens later, till today) and more stringent $\Muni$. 
However, for larger $\mV$, the dilution due to the mediator decay is significant and increases with $\tilde{r}$, thereby raising $\Muni$. Note that for sufficiently large $\rtilde$, $\Muni$ becomes independent o{\tiny }f $\rtilde$ (see text for discussion).
}
\label{fig:UnitaryLimit}
\end{figure}

\paragraph{Sketch (neglecting $r$):}
In this paragraph, we first drop the dark-to-SM temperarure ratio $r$.
From comparing  Eq.~\eqref{eq:sigmav_fo} with the thermally averaged unitarity cross-section, 
$
\langle\sigma \vrel \rangle^\FO =
\langle\sigma_{\rm uni} \vrel \rangle
=4 (2J+1) \sqrt{\pi x^\FO} / \Muni^2
$, cf. Sec.~\ref{sec:uni_bound}, we deduce that the upper bound on the DM mass from unitarity is increased by a factor $\sqrt{D}$
\begin{equation}
\label{eq:max_mass_uni}
\MDM \lesssim 137~\rm{TeV}\;\sqrt{D}
\end{equation}
with $D$ in Eq.~\eqref{eq:dilution_factor} and where $137$~TeV is the standard unitarity bound on Sommerfeld-enhanced Dirac DM, annihilating via s-wave, cf. table~\ref{table:cross-section-mass-uni}. 

Plugging $m_V = 100~\rm{PeV}$, $\Gamma_V=(0.1~\rm{s})^{-1}$, $g_{\rm SM} = 106.75$ and $f_{\mathsmaller{\rm V}} = 10^{-2}$ (cf. Eq.~\eqref{eq:DP_comoving_number}), we get the maximal dilution factor compatible with BBN
\begin{equation}
D \simeq 6.2\times 10^8 ~ \left( \frac{f_{\mathsmaller{\rm V}} }{10^{-2}} \right)\left( \frac{m_V}{100~\rm{PeV}} \right) \left( \frac{\tau_V}{0.1~\rm{s}} \right)^{\! 1/2}.
\end{equation}

Therefore, for the purpose of setting BBN constraints, we impose $\tauV <0.03$ s for all values of the mass $\mV$, i.e. even for those leading to matter domination and entropy injection.
Therefore, the maximal DM mass compatible with both unitarity and BBN is 
\begin{equation}
\label{eq:maximal_DM_mass}
\MDM^{\rm max} \simeq 3.5~\rm{EeV} \left( \frac{f_{\mathsmaller{\rm V}} }{10^{-2}} \right)^{1/2}\left( \frac{m_V}{100~\rm{PeV}} \right)^{\! 1/2} \left(  \frac{\tau_V}{0.1~\rm{s}} \right)^{\! 1/4}.
\end{equation}
The maximal DM mass in Eq.~\eqref{eq:maximal_DM_mass} can be visualized in Fig.~\ref{fig:UnitaryLimit}.

\paragraph{Dependence on the dark-to-SM temperature ratio $r$:}

In the previous paragraph, we have been sloppy about the dependence on the dark-to-SM temperature ratio $r$. If we now re-introduce this dependence, we obtain
\begin{equation}
\Muni \simeq \MuniSMs 
\left(\frac{\xfoSM}{\xfo} \right)_{\rm \! \! uni}^{1/4} 
\left[
\frac{\DSM}
{r_\FO \sqrt{1+ (\gD^\FO/\gSM^\FO)r_\FO^4}}
\right]^{1/2}  
\times \left\{
\begin{alignedat}{10}
\sqrt{2J+1},	& \quad \text{solely}~J
\\
J_{\max}+1,		& \quad 0\leqslant J \leqslant J_{\max}
\end{alignedat}
\right.
\,,
\label{eq:Muni}
\end{equation}
where $\MuniSMs \simeq 135$~TeV is the $s$-wave unitarity limit in the absence of any dilution and if DM is annihilating directly into the SM plasma.
The parameters $\xfo$ and $\xfoSM$ at the unitarity limit, appearing in  eq.~\eqref{eq:Muni}, can be determined by the standard formula\footnote{\label{foot:xfo}
For $\langle \sigma \vrel \rangle	= \sigma_0 x^{1/2}$, 
the freeze-out temperature can be estimated by 
(see e.g.~\cite{Kolb:1990vq}, and ref.~\cite{Baldes:2017gzw} for generalization to dark sector freeze-out)
\beq 
\xfo \simeq \ln \left[
0.095
\frac{g_{\mathsmaller{\rm DM}}}{\sqrt{\gSM^\FO}} 
\frac{ r_\FO^2 }{ \sqrt{1+ (\gD^\FO/\gSM^\FO) r_\FO^4}}
M_{\rm Pl} \MDM \sigma_0 
\right]  \,,
\nonumber
\eeq
where $g_{\mathsmaller{\rm DM}}$ are the DM degrees of freedom and $M_{\rm Pl}$ is the reduced Planck mass.
\label{footnote:xfoKolbTurner}
}, using the unitarity cross-section; we find 
$\xfoSM \simeq 31 + (1/2) \ln [(2J+1) / \gSM^\FO]$ and 
$\xfo - \xfoSM =-\ln(\Muni/\MuniSM) 
+ \ln (r_\FO^2 / \sqrt{1+ (\gD^\FO/\gSM^\FO)r_\FO^4})$.

We recall that $r_\FO$ and $\DSM$ can be computed using eqs.~\eqref{eq:r_ratio_decay}, \eqref{eq:DSM_rtilde} and \eqref{eq:DSM_rtilde1}. This result is consistent with previous results in the limit of no dilution and $\rtilde = 1$~\cite{vonHarling:2014kha,Baldes:2017gzw}.
Evidently, $\Muni$ depends on the thermodynamics of the dark sector, its temperature and its degrees of freedom; in the presence of dilution, it also depends on the properties of the mediator, its lifetime and mass. In figs.~\ref{fig:sigmaFOandMuni} (right panel) and \ref{fig:UnitaryLimit}, we illustrate these relations.

To conclude this section, we find it interesting to report the maximal mass that DM could in principle reach in such models, $\Muni^\text{max}$, being agnostic on the way this limit would be realised. We obtain $\Muni^\text{max}$ from eq.~\eqref{eq:Muni} upon maximising $\DSM$ (we recall that EeV $= 10^3$~PeV)
\beq
\Muni^\text{max} \simeq 1.1~\text{EeV} \, \Big(\gD^\FO \frac{\mV}{100~\text{PeV}} \left( \frac{\tauV}{0.03~\text{s}}\right)^{\!\frac{1}{2}}\Big)^{\!\frac{1}{2}}
\frac{r_\FO}{\left( 1+0.009\,\gD^\FO\,r_\FO^4\right)^{\!\frac{1}{4}}}
\times \left\{
\begin{alignedat}{10}
\sqrt{2J+1},	& \quad \text{solely}~J
\\
J_{\max}+1,		& \quad 0\leqslant J \leqslant J_{\max}
\end{alignedat}
\right.\,,
\label{eq:Muni_Max}
\eeq
where for simplicity of the exposition we have omitted the weak log dependence.\footnote{For completeness, it amounts
to multiplying eq.~(\ref{eq:Muni_Max}) by 
\beq
\left[1 - 0.025 \log \left( r_\FO^{-2} \; \gD^\FO \, \frac{\mV}{\text{100~\text{PeV}}} \, \left(\frac{\tauV}{0.03~\text{s}}\right)^{\!1/2} \left( 1+0.009\,\gD^\FO\,r_\FO^4  \right)^{1/2} \right)\right]^{-1/4}
\eeq
}
In the dark $U(1)$ model, with $\gtildeD = 6.5$, $\rtilde=1$, $\gD^\FO=3$, $s+p$ wave, this translates to
\beq
\Muni^\text{max} \simeq 2.4~\text{EeV}\,\left(\frac{\mV}{100~\text{PeV}}\right)^{\!1/2}\left( \frac{\tauV}{0.03~\text{s}}\right)^{\!1/4},
\label{eq:mDMmax_darkU1}
\eeq
where we have normalized $\mV$ to a value which guarantees that freeze-out happens when $V$ is relativistic (otherwise eq.~(\ref{eq:mDMmax_darkU1}) does not hold).

\section{The dark $\mathbf{U(1)}$ model as a case of study} \label{sec:U1model}

\subsection{The Lagrangian}
We consider Dirac fermionic DM $X$, charged under a dark gauge group $U(1)_D$, under which all the SM particles are neutral. We assume that the dark sector communicates with the SM via kinetic mixing between $U(1)_D$ and the hypercharge group $U(1)_Y$. The Lagrangian then reads
\beq
\mathcal{L}= - \frac{1}{4} {\FD}_{\mu\nu} \FD^{\mu\nu} - \frac{\epsilon}{2\,c_{w}} {\FD}_{\mu\nu} \FY^{\mu\nu}
 + \bar{X}(i \slashed{D} - \MDM)X,
 \label{eq:L}
 \eeq
where $D_\mu = \partial_\mu + i \gD V_\mu'$ is the covariant derivative, with $V_\mu'$ being the dark gauge field. (We reserve the symbol $V^\mu$ for the state arising after the diagonalisation of the kinetic terms, as described below.)  $F_{\mathsmaller{D},\mathsmaller{Y}}$ are the dark and hypercharge field strength tensors, and $c_w$ is the cosine of the weak angle. We define the dark fine structure constant $\aD~\equiv~\gD^2/(4 \pi)$.

We further assume that the dark photon $V^\mu$ obtains a mass $\mV$, either via the St\"{u}ckelberg 
or the Higgs mechanisms. We shall remain agnostic about which of the two is realized, and only note that in case of the Higgs mechanism, the extra scalar can be decoupled (and thus made irrelevant for the phenomenology discussed in this study) by choosing its charge to be much smaller than one.
Dark photons described by eq.~(\ref{eq:L}) and with a St\"{u}ckelberg mass arise, for example, in string theory constructions, where the predicted ranges for $\epsilon$ and $\mV$ include those of interest in our study, see e.g.~\cite{Jaeckel:2010ni,Essig:2013lka}.

\subsection{Dark photon}
\label{sec:darkphoton}

The dark photon interacts with SM particles via its kinetic mixing $\epsilon$ with the hypercharge gauge boson. Upon diagonalisation of the kinetic and mass terms\footnote{
This has an effect also on the mass and couplings of the $Z$ boson, that is phenomenologically irrelevant for the small values of $\epsilon$ that we will be interested in, see e.g.~\cite{Curtin:2014cca} for the related constraints.}, 
its couplings may be written as
\beq
\mathcal{L} \supset
\epsilon\,e \left( \frac{1}{1-\big(\frac{\mV}{\mZ}\big)^{\!2}}\,J^\mu_{\rm em}
-\frac{1}{c_w^2} \frac{\big(\frac{\mV}{\mZ}\big)^{\!2}}{1-\big(\frac{\mV}{\mZ}\big)^{\!2}}\,J^\mu_{Y}
+O(\epsilon^2) \!\!\right) \! V_\mu
\label{eq:L_DP}
\eeq
where $J^\mu_{\rm em}$ and $J^\mu_Y$ are the usual electromagnetic and hypercharge currents, and where we have denoted by $V$ the mass eigenstate dominated by the gauge eigenstate $V'$.
Equation~\eqref{eq:L_DP} makes transparent both the $\epsilon$-suppression of the $V$ couplings to SM particles, and the physical limits where $V$ couples as the photon ($\mV \ll \mZ$) or as the hypercharge gauge boson ($\mV \gg \mZ$). The expansion of eq.~\eqref{eq:L_DP} in $\epsilon$ is not valid for $\mV \simeq \mZ$, where $V$ couples as the $Z$ boson. In our study we use the full expressions valid also in that limit.

\begin{figure}[t]
\begin{center}
\includegraphics[width=0.6 \textwidth]{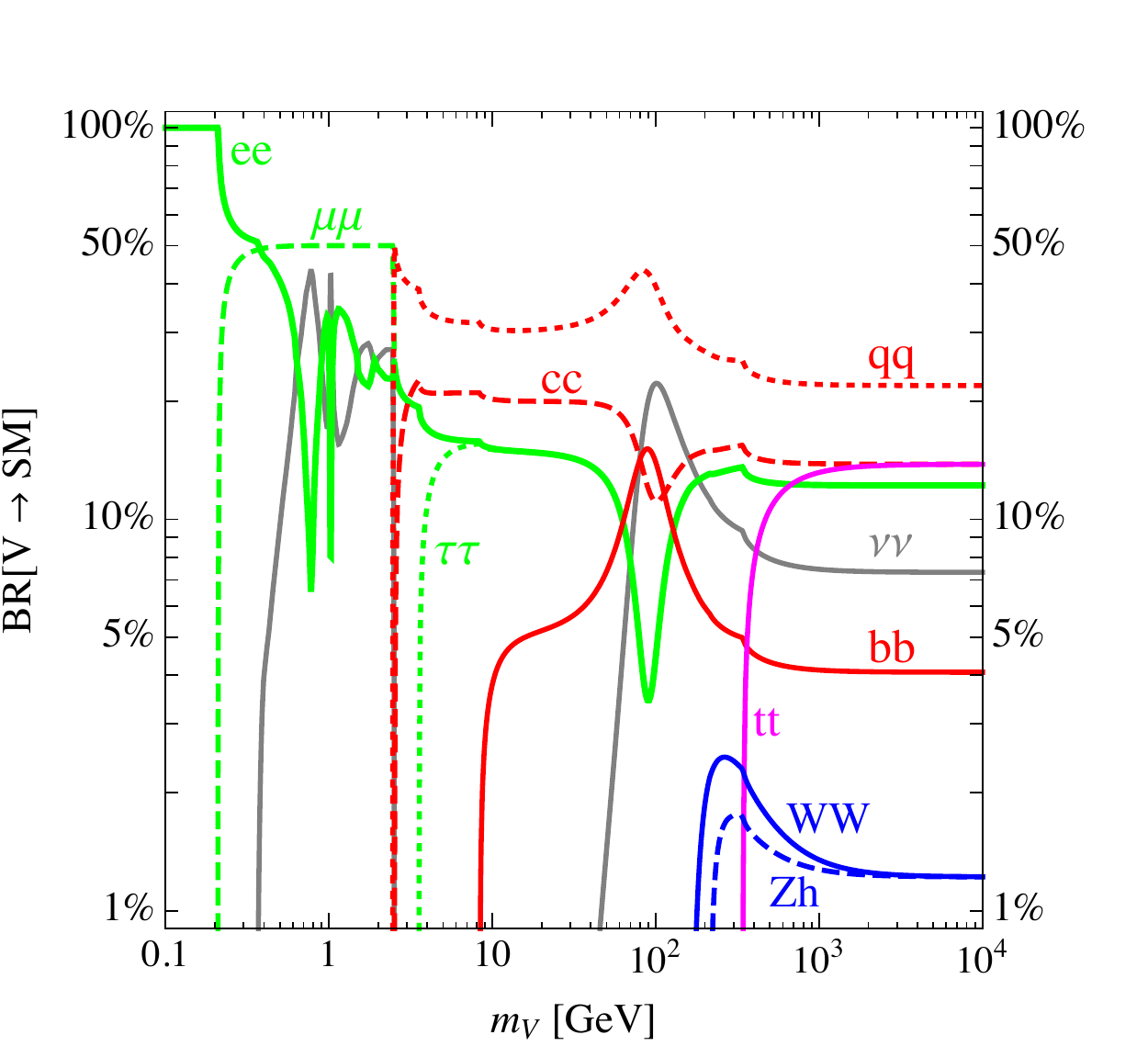}
\end{center}
\caption{\it \small Branching ratios of the dark photon.  This figure updates figure 4 of~\cite{Cirelli:2016rnw}.}
\label{fig:BRs}
\end{figure}

In this study we will be interested in dark photon masses larger than 100~MeV, so that tree-level decays to at least a pair of SM particles are always open. We compute the decay widths of the dark photon to all possible two-body final states, and present them in appendix~\ref{app:DP}. For $\mV > \Lambda_{\rm QCD}$, the hadronic decay channels are open, but for $\mV < {\rm few~GeV}$ they cannot be described perturbatively.
In that interval, which for definiteness we take to be $350~{\rm MeV} < \mV < 2.5~{\rm GeV}$, we determine the total decay width of $V$ from measurements of $e^+ e^- \to$~hadrons at colliders (see e.g.~\cite{Curtin:2014cca,Buschmann:2015awa}). Following~\cite{Cirelli:2016rnw}, we assume this width to consist half of $\mu$ and half of $\nu$ pairs, since these constitute the dominant final states from hadronic decays. 
We show the resulting branching ratios in fig.~\ref{fig:BRs} and refer to appendix~\ref{app:DP} for more details. 

Throughout this work, we will be interested in the region $\MDM > \mV$. The phenomenology of DM is then driven by its annihilations into dark photons, whose tree-level cross section scales as $\aD^2$. The tree-level cross section for DM annihilation into SM pairs scales as $\epsilon^2\,\aD \aSM$, where $\aSM$ generically stands for an SM EW coupling, and is therefore negligible for the values of $\epsilon$ we will be interested in.

\subsection{DM relic abundance and dilution}
\label{sec:DMabundance}

\begin{figure}[t]
\centering
\includegraphics[width=0.49\textwidth]{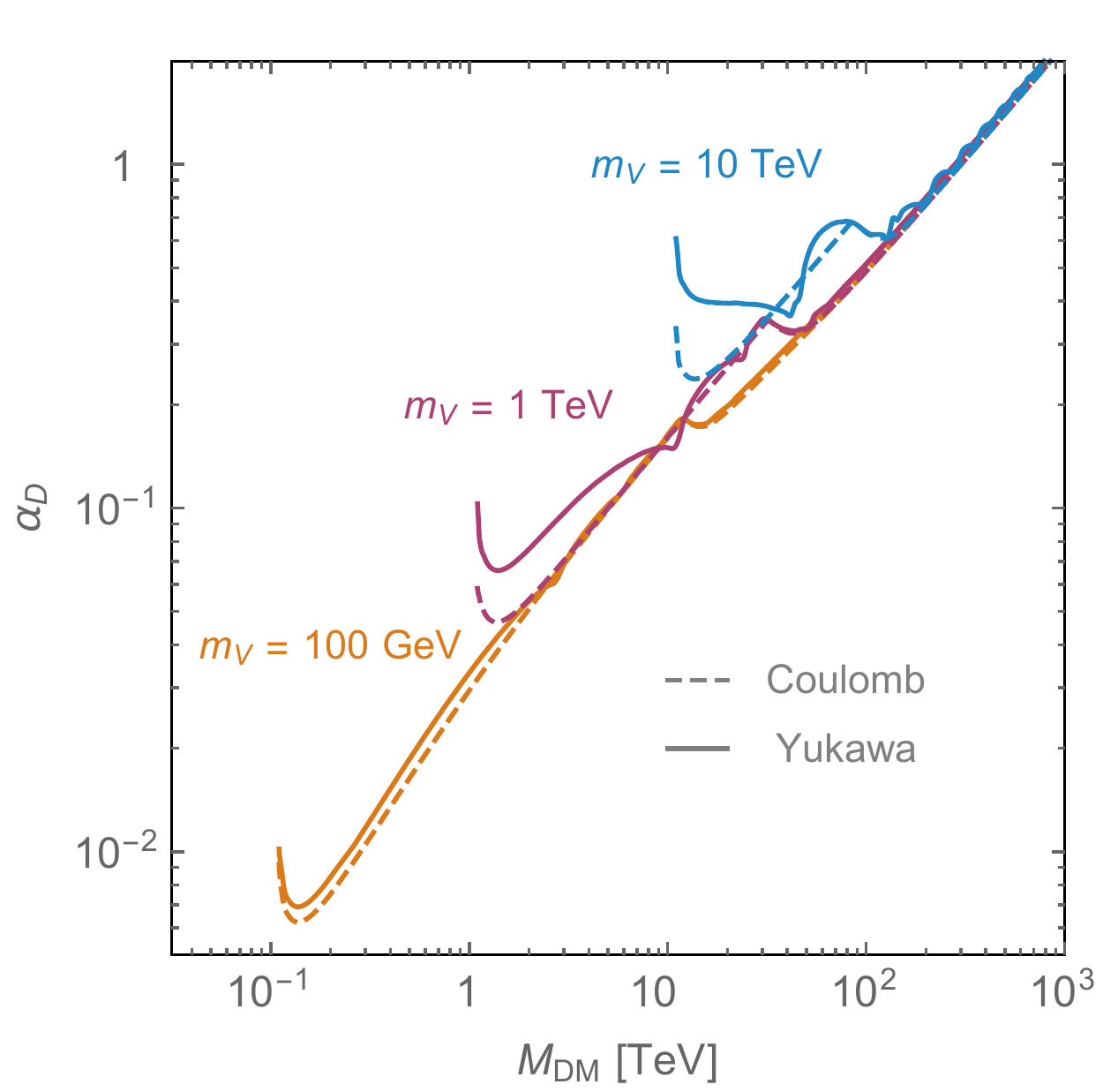}\;
\includegraphics[width=0.49\textwidth]{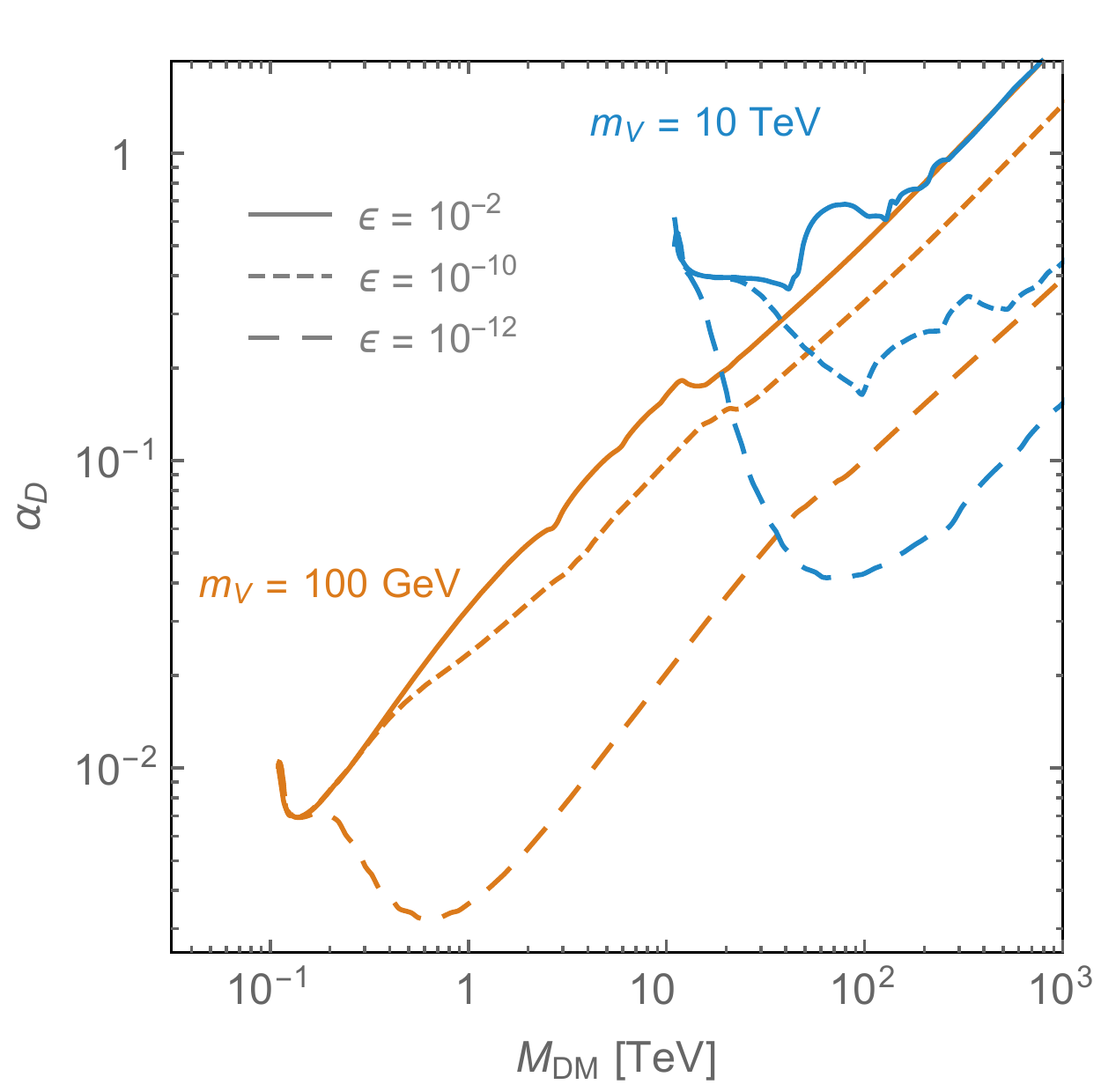}
\caption{\it \small \label{fig:alphaD}
Values of $\aD$ that yield the observed DM abundance as a function of the DM mass, for some reference values of the dark photon mass. Left: our results including the full Yukawa potential (continuous) are compared with the ones in the Coulomb approximation (dashed).
Right: comparison of cases with different kinetic mixing~$\epsilon$ and therefore with different dilution.
The dilution is exponentially suppressed when $\mV$ is close to $\MDM$ (see App. A of \cite{Cirelli:2018iax}).}
\end{figure}

For each values of the masses $\MDM$ and $\mV$, we compute\footnote{
We use the latest value of the DM relic abundance today measured by Planck~\cite{Patrignani:2016xqp},  $\Omega_{\rm DM}h^{2}=0.1186 \, \pm \, 0.0020$ and the value of the effective number of neutrinos computed in~\cite{Mangano:2005cc}, $N_{\rm eff}=3.046$.}
the value of the dark fine structure constant $\aD$ which leads to the correct value of the DM relic abundance today. 
We use the procedure in~\cite{vonHarling:2014kha} which takes into account both the Sommerfeld enhancement of the direct annihilation and BSF. The novelties of the present work with respect to the most recent analogous computations~\cite{vonHarling:2014kha,Cirelli:2016rnw} are the following:
\begin{itemize}
\item[$\circ$] 
We take into account the full Yukawa potential for computing the Sommerfeld enhancement and BSF, as opposed to the Coulomb limit only.
\footnote{The Sommerfeld factors with the full Yukawa potential have been computed by Kallia Petraki.}
This is important at large values of the dark photon mass, as seen in the left panel of fig.~\ref{fig:alphaD} (we remind that the cross section relevant for the signals scales as $\aD^3$ away from resonances).
\item[$\circ$] We study the impact of the entropy dilution due to the dark photon decay, whose effect is shown in the right panel of fig.~\ref{fig:alphaD} for some reference values of the kinetic mixing~$\epsilon$ (note $\GammaV \propto \epsilon^2$ and see eq.~(\ref{eq:dilution_factor_expression_VD_dom_approx}) for the dilution).
\end{itemize}
Equipped with the dark coupling constant that gives the observed DM relic density, we next compute DM annihilation and BSF rates relevant for the cosmic ray signals.

\subsection{DM signals}
\label{sec:DMsignals}

\begin{figure}[!t]
\begin{center}
\includegraphics[width=0.48 \textwidth]{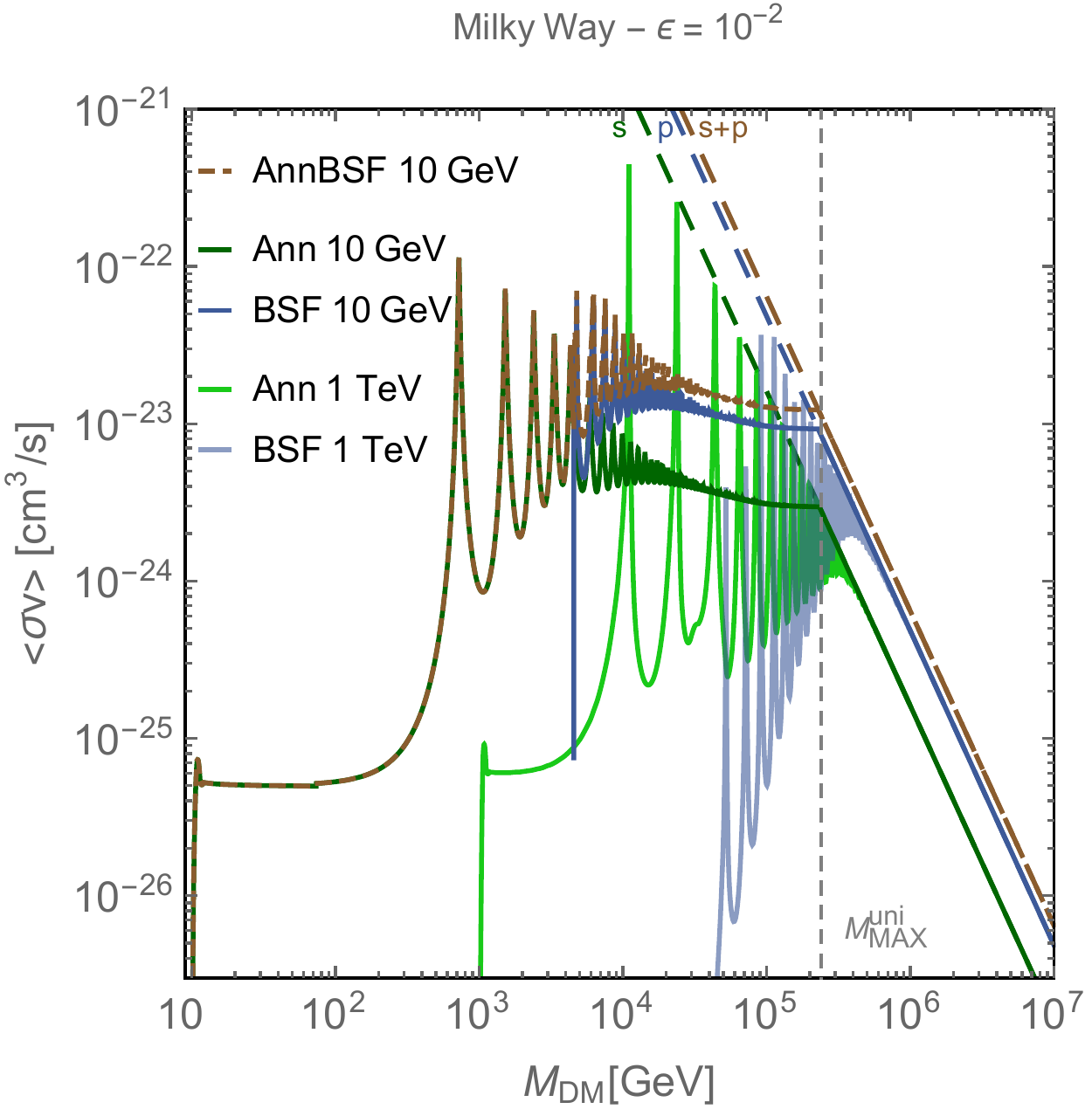} \;
\includegraphics[width=0.48 \textwidth]{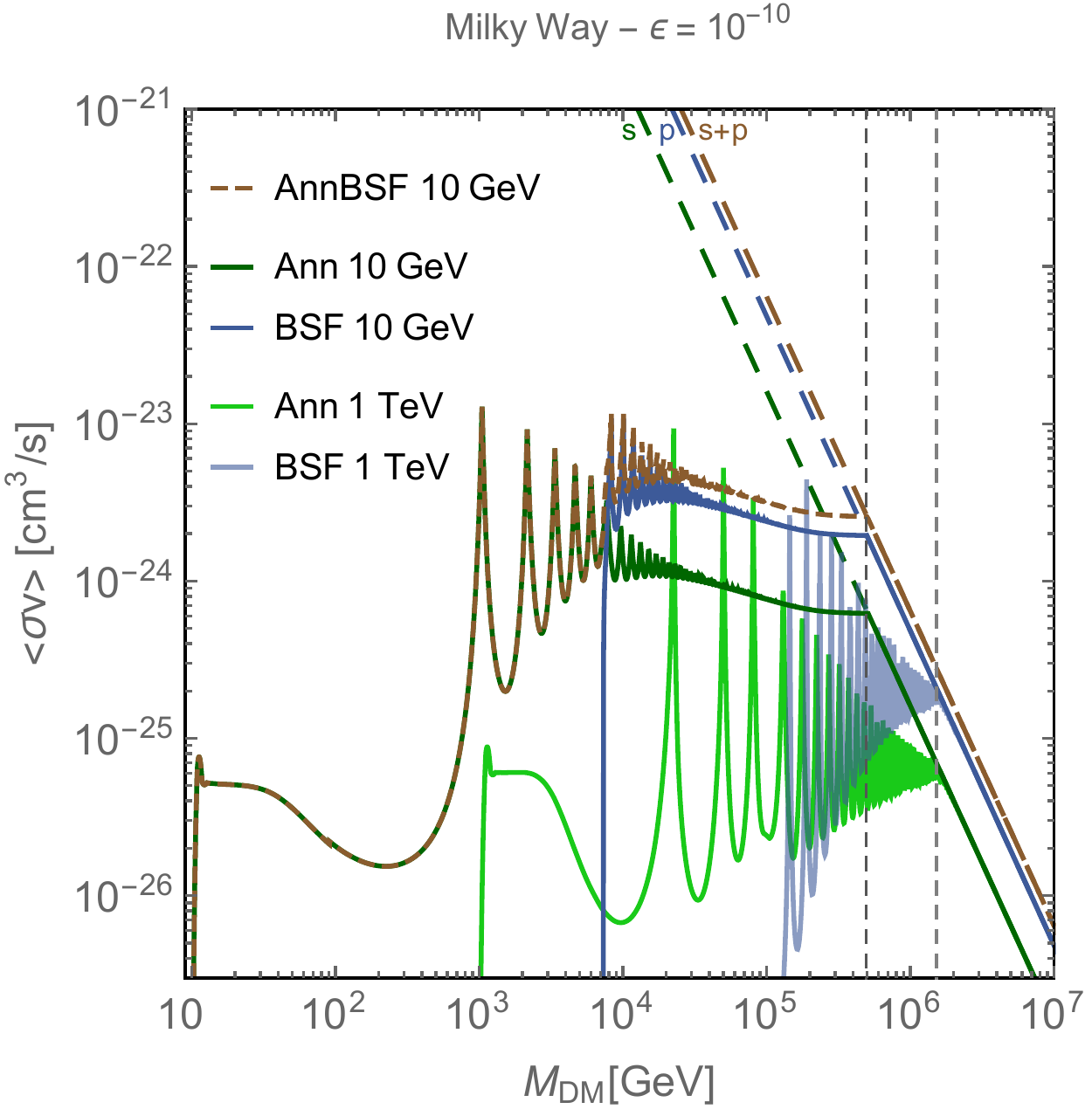} 
\end{center}
\begin{center}
\includegraphics[width=0.48 \textwidth]{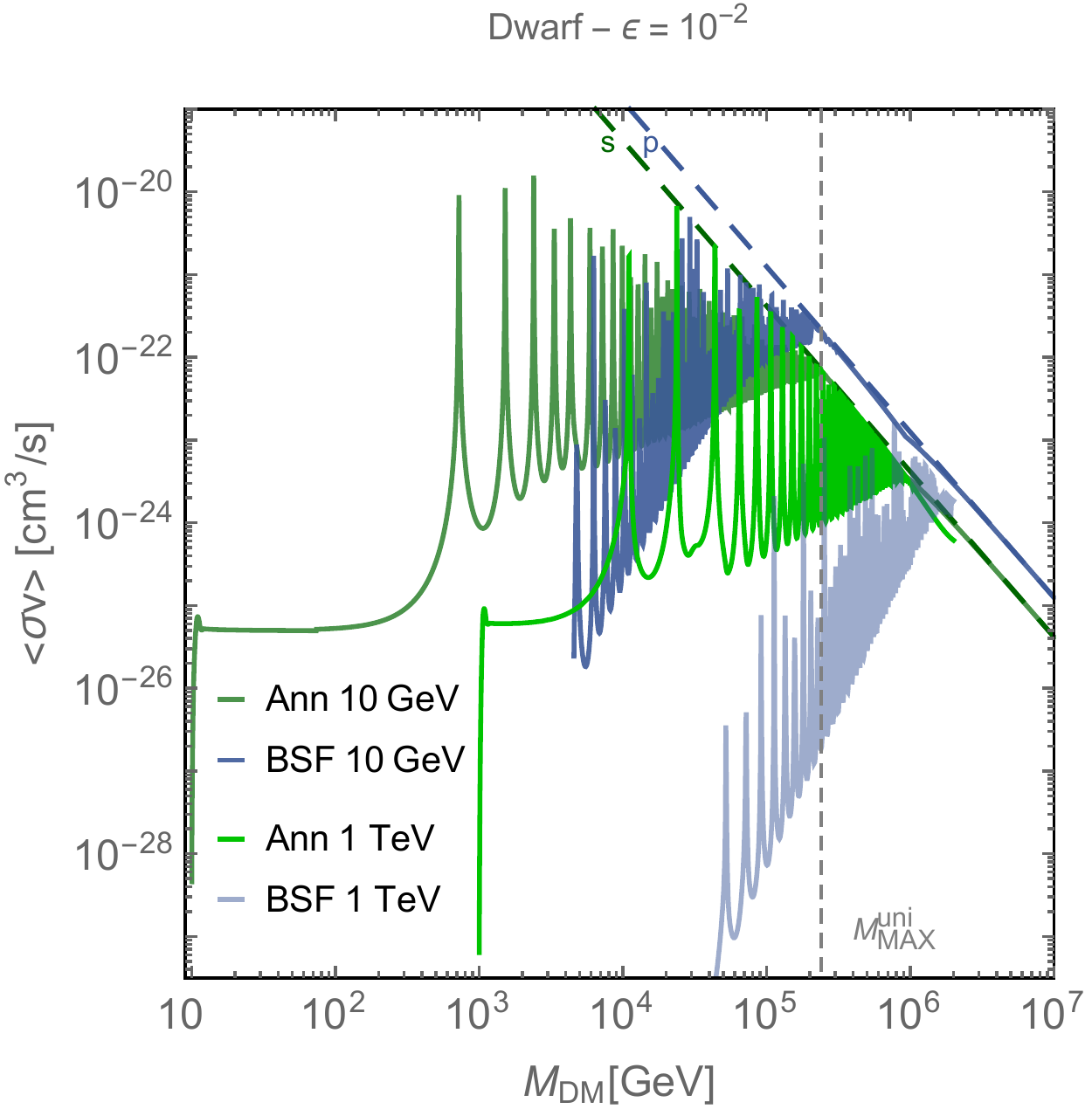} \;
\includegraphics[width=0.48 \textwidth]{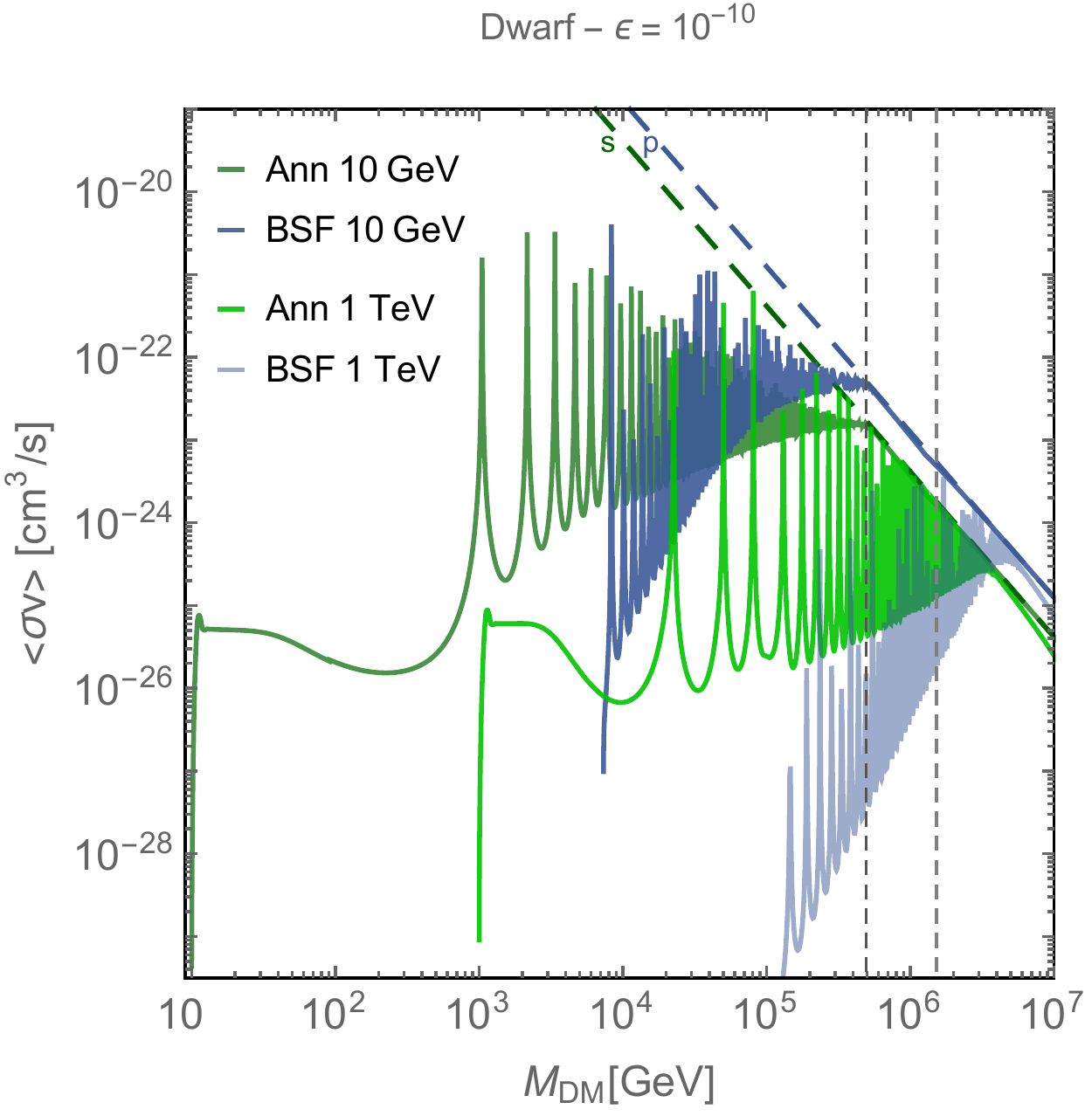} 
\end{center}
\caption{\it \small \label{fig:sigmav_MW_unitarity_mVD_10GeV}
Cross-section averaged over the velocity distribution of the DM in the Milky Way and in Dwarfs versus the dark photon mass. The dashed straight lines are the maximum cross-section values allowed by unitarity for $s$, $p$ and $s+p$ partial waves. The first column is without entropy dilution whereas in the second column, entropy dilution is present with kinetic mixing $\epsilon=10^{-10}$. For $\mV$ close to $\MDM$, the dilution is exponentially suppressed (see App. A of \cite{Cirelli:2018iax}).}
\end{figure}

We take both  the direct DM annihilation and BSF into account in the computation of the DM relic abundance (and of the unitarity bound), and in the estimation of the DM signals during the cosmic microwave background (CMB), 21 cm, at reionization, in the Milky Way (MW) and in dwarf spheroidal galaxies (dSph).

\paragraph*{\bf MW and dSph.}
For the MW and dSph we compute $S_{\rm ann}$ and $S_{\BSF}$ numerically as in~\cite{Petraki:2016cnz}. We then assume a Maxwellian DM velocity distribution $f(\mathbf{v}) = N \, \theta (v - v_{\rm esc}) \, e^{-v^2/v_0^2}$ for each of the interacting particles, where $N$ is fixed by imposing $\int d^3 v \,f(\mathbf{v}) = 1$.  The velocity $\mathbf{v}$ and the distribution $f(\mathbf{v})$ should not be confused with $\vrel$ and the distribution over $\vrel$; the latter is derived from
$\int f(v_1) f(v_2) \, d^3v_1 d^3v_2$ after carrying out the integration over the mean velocity $(\mathbf{v_1}+\mathbf{v_2})/2$. We choose $v_{\rm esc} = 533$~km/s and $v_0 = 220$~km/s for the MW~\cite{Piffl:2013mla}, and $v_{\rm esc} = 15$~km/s and $v_0 = 10$~km/s for dSph's~\cite{McConnachie:2012vd,Burkert:2015vla}.

The resulting cross sections $\langle \sigma_{\ann} \vrel \rangle$ and $\langle \sigma_{\BSF} \vrel \rangle$ are shown in fig.~\ref{fig:sigmav_MW_unitarity_mVD_10GeV} as a function of $\MDM$, for some reference values of $\mV$. We have fixed $\aD$ to the value that reproduces the correct DM relic abundance, and which depends on $\MDM$ and very mildly on $\mV$ (see section~\ref{sec:DMabundance}).
Figure~\ref{fig:sigmav_MW_unitarity_mVD_10GeV} displays the features discussed in the previous subsection.
Going from smaller to larger $\MDM$, the total cross section first coincides with the tree-level one, then the Sommerfeld enhancement of $\sigma_{\ann}\vrel$ becomes relevant, and for even larger $\MDM$ also BSF becomes important. Both $\sigma_{\ann}\vrel$ and $\sigma_{\BSF}\vrel$ display resonances, and reach the Coulomb limit at a larger value of $\MDM$, which is different between the MW and dSph because the two systems are characterised by different DM velocities. The change of slope around $\MDM \simeq 200$~TeV reflects our prescription for not violating the unitarity limit.

\paragraph*{\bf CMB and 21 cm.}
The evolution of the DM temperature implies that, for the values of the parameters of interest for this study, at redshifts relevant for CMB constraints ($z \approx 600$, see e.g.~\cite{Finkbeiner:2011dx}) one has $\vrel \ll 10^{-10}$~\cite{Cirelli:2016rnw}. 
Therefore
\begin{itemize}
\item[i)] 
the capture into the ground state is negligible because of the $\vrel^2$ suppression, 
and
\item[ii)] 
the Sommerfeld enhancement of the annihilation processes is well within a saturated regime, so we do not need to compute the velocity average.
\end{itemize}
At even smaller redshifts, inhomogeneities grow and DM structures start to form, so that the DM velocity acquires a component that depends on the gravitational potential of such structures, that could be larger than the value of $\vrel$ relevant for CMB. However, at redshifts relevant for the 21~cm signal ($z \approx 17$, see e.g.~\cite{Furlanetto:2015apc}), the largest DM halos formed are expected to have masses several orders of magnitude smaller than MW-size galaxies, thus inducing low DM velocities. In addition, the dominant part of the signal from DM annihilations comes from the smallest halos, see e.g.~\cite{Lopez-Honorez:2016sur,Liu:2016cnk}.
Therefore, the DM velocities relevant for the 21~cm limits are not expected to be so large that they invalidate i) and ii).

The small values of the velocity also imply that the numerical calculation of $S_{\ann}$ is impractical.
Therefore, to derive the CMB and 21cm constraints in section~\ref{sec:ID_early}, we work with the Hulth\'en potential $V_H = - \aD m_* e^{-m_* r}/(1-e^{-m_* r})$, that allows for an analytic solution $S_{\ann}^H$  (see e.g.~\cite{Cassel:2009wt,Slatyer:2009vg}).
$S_{\ann}^H$ approximates well $S_{\ann}$ off-resonance, it correctly reproduces the fact that the resonances become denser at larger values of $\aD\MDM/\mV$, and we choose $m_* = 1.68 \,\mV$ such that the position of the first resonance coincides with the position of the first resonance in the standard Yukawa case.
For both the CMB and 21cm constraints we work with $v = 10^{-11}$.\footnote{We have explicitly verified that the regions excluded by CMB do not change for smaller $v$ values, and that those excluded by 21cm observations and allowed by unitarity do not change up to $v \simeq 10^{-8}$~(see section~\ref{sec:ID_early}).}

\section{Phenomenology \label{sec:pheno}}

In this Section\footnote{Sec.~\ref{sec:pheno} has been mostly written by Filippo Sala, apart from Sec.~\ref{sec:ID_local_pbar} and Sec.~\ref{sec:ID_local_epem} which have been mostly written by Marco Cirelli.} we present the constraints on the parameter space of the model imposed by the different signals that we consider.
For the ease of the reader, we first present a summary of the results in fig.~\ref{fig:ID_summary}, where each panel corresponds to decreasing values of $\epsilon$, i.e.~increasing dilution ($\GammaV \propto \epsilon^2$ and see eq.~(\ref{eq:dilution_factor_expression_VD_dom_approx}) for the dilution). A few comments are in order.
\begin{itemize}
\item[$\diamond$]
The various indirect detection probes are complementary and collectively cover very large portions of the parameter space. It is remarkable that some of them (neutrinos, $\gamma$-ray and CMB searches) are sensitive to DM with a mass beyond $O(10)$~TeV.
Multimessenger astronomy is therefore a concrete possibility also for the quest of heavy DM, adding motivation to indirect detection in this mass range.
\item[$\diamond$] The fact that the {\sc Antares} and {\sc Hess} regions end abruptly at $\MDM \simeq 100$~TeV is only a consequence of the largest masses considered in the papers of the experimental collaborations. Our study provides a strong motivation for these experiments to extend their current searches to heavier DM masses, as this would likely test unexplored regions of parameter space.
In addition, we point out that  our analysis is partly a reinterpretation of the studies performed by the collaborations themselves and as such suffers from the fact that the SM spectra from cascade decays predicted in secluded models differ from the ones used by these collaborations. The above difficulties would be circumvented if public access to the data was possible.

\item[$\diamond$] As expected, the dilution weakens the sensitivities of the various indirect detection probes, since the cross sections needed to obtain the correct DM abundance are smaller than in the standard case.
It is interesting to note, however, that for increasing dilution the parameter space accessible to indirect detection shrinks because of BBN constraints. This gives a well-defined window of parameter space to aim at, with current and future telescopes.\footnote{Another challenge to test large $\MDM$ with gamma ray telescopes is posed by the non-transparency of the Milky Way to photons energies of hundreds of TeV, see e.g.~\cite{Blanco:2017sbc} for a recent related study.}
From the model-building point of view, this shows explicitly that the upper limit on dilution sets a lower limit on the communication with the SM of secluded DM models.
\end{itemize}
Collider and direct detection searches have instead no power in testing these models, because of the large DM mass and of the small communication with the SM, parameterized by $\epsilon$. We now move to  illustrate in more details the phenomenology of this model.

\subsection{Constraints on the kinetic mixing}
\label{sec:kinetic_mixing}

For $\epsilon \lesssim 10^{-9}$ and $\mV \gtrsim$ GeV, the $\epsilon - \mV$ parameter space is constrained by the requirement that decays of the dark photon do not spoil BBN.
We use here the results of the study~\cite{Jedamzik:2006xz}, that derived BBN constraints on neutral particles decaying at early times. 
It found that, for particles decaying more than 30\% hadronically (like the dark photon in the mass range of our interest), all constraints evaporate for lifetimes of the particle shorter than 0.03 seconds.
Lifetimes as large as $10^2$ seconds become allowed as soon as the abundance of the dark photon, at the time of its decay, drops below $O(0.1)$ times the critical density.
As presented in section~\ref{sec:dilution_DarkU1}, the region we are interested in is that where dark photons dominate the energy density of the Universe just before decaying, to realise sufficient entropy injection. Therefore, we conservatively label as `disfavoured by BBN'  the regions where
$\tauV > 0.03$~sec.
Note that for mediator masses below the GeV range, where hadronic decay modes close, the BBN constraints weaken significantly~\cite{Hufnagel:2018bjp}.
A more detailed analysis, e.g. with a proper mass dependence of the constraints, goes beyond the purpose of this work.

For smaller $\mV$ and/or larger $\epsilon$, the existence of dark photons is severely constrained also by DM Direct Detection, observations of the supernova SN1987A, beam dump experiments, and neutrino limits from the Sun. As that region is not the focus of interest of this work, we do not discuss these constraints here, and refer the interested reader to the discussion in~\cite{Cirelli:2016rnw} for the first three, and to~\cite{Adrian-Martinez:2016ujo,Ardid:2017lry} for constraints from DM annihilation in the Sun.

\subsection{DM constraints from the Early Universe}
\label{sec:ID_early}

\subsubsection{CMB}
DM annihilations at the time of CMB inject energy in the SM bath and could therefore alter the observed CMB spectrum, resulting in stringent limits~\cite{Ade:2015xua}.
Such limits are driven by the ionizing power of the SM final state $i$, which can be encoded in efficiency factors $f^i_{\rm eff}(\MDM)$ that we take from~\cite{Slatyer:2015jla,Slatyer:2015kla}. Since these depend mostly on the total amount of energy injected, such limits do not depend on the number of steps between the DM annihilation and the final SM products~\cite{Elor:2015bho}.
Following also our discussion in section~\ref{sec:DMsignals}, we then place limits as follows
\beq
\sum_{i=e\bar{e}, u\bar{u},\dots} \!\! \langle \sigma_{\rm ann}^H v\rangle \, {\rm BR}(V\to i) \, f^i_{\rm eff} < 8.2 \times 10^{-26} \,\frac{{\rm cm}^3}{\rm sec}\,\frac{\MDM}{100~{\rm GeV}}\,,
\label{eq:CMB}
\eeq
where on the left-hand side the dependence on $\MDM$ and $\mV$ is implicit.
The resulting exclusion is displayed in light yellow in figure~\ref{fig:ID_summary} for various possibilities of DM dilution.

\subsubsection{21 cm}
$\Lambda$CDM predicts an absorption signal in the radio band at 21 cm, associated with the relative occupation number of the singlet and triplet hyperfine states of the hydrogen atom at $z\simeq 17$, see e.g.~\cite{Furlanetto:2015apc}.
The {\sc Edges} collaboration has recently reported~\cite{Bowman:2018yin} the observation of such a signal, although with an amplitude larger than expected in $\Lambda$CDM.
Because extra injection of energy in the SM bath would make that amplitude smaller than expected, this observation allows to place limits on DM annihilations and/or decays, that turn out to be competitive with CMB ones~\cite{DAmico:2018sxd,Liu:2018uzy,Cheung:2018vww,Clark:2018ghm,Mitridate:2018iag}.
Here we do not aim at explaining why the amplitude is observed to be larger than predicted, but rather we simply impose a limit on the DM annihilation following ref.~\cite{DAmico:2018sxd}, as
\beq
\sum_{i=e\bar{e}, u\bar{u},\dots} \!\! \langle \sigma_{\rm ann}^H v\rangle \, {\rm BR}(V\to i) \, f^i_{\rm eff} < 6.3 \times 10^{-26} \,\frac{{\rm cm}^3}{\rm sec}\,\frac{\MDM}{100~{\rm GeV}}\,,
\label{eq:21cm}
\eeq
where we have taken into account that DM in our model is not self-conjugate.
Among the limits presented in~\cite{DAmico:2018sxd}, we have chosen the more conservative one, derived by imposing that DM does not reduce the standard amplitude by more than a factor of 4 (`$T_{21} \gtrsim 50$~mK'), and using the more conservative boost due to substructures (`Boost~1').

\begin{figure}[p]
\begin{center}
\includegraphics[width=0.43\textwidth]{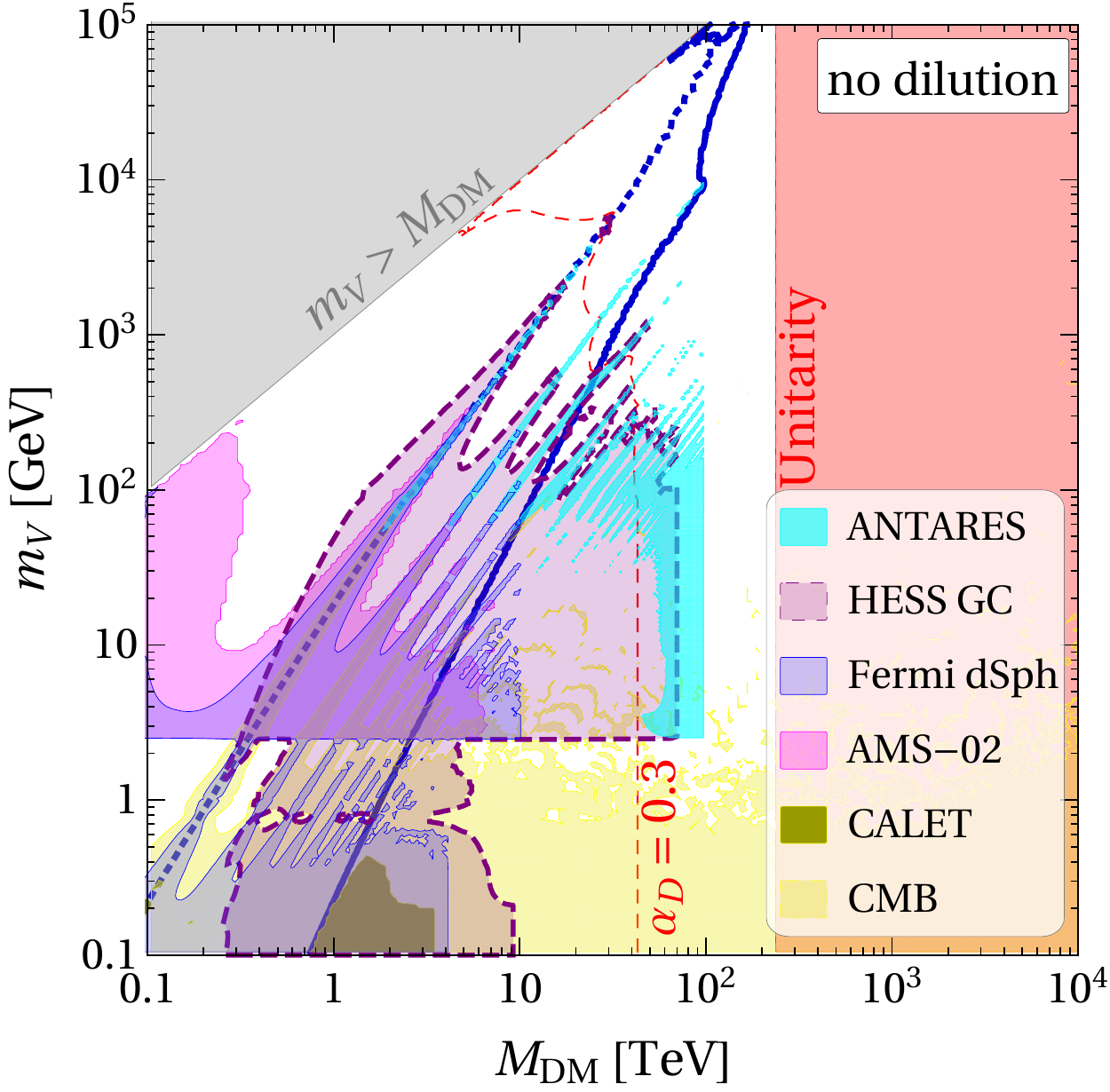} \quad
\includegraphics[width=0.43\textwidth]{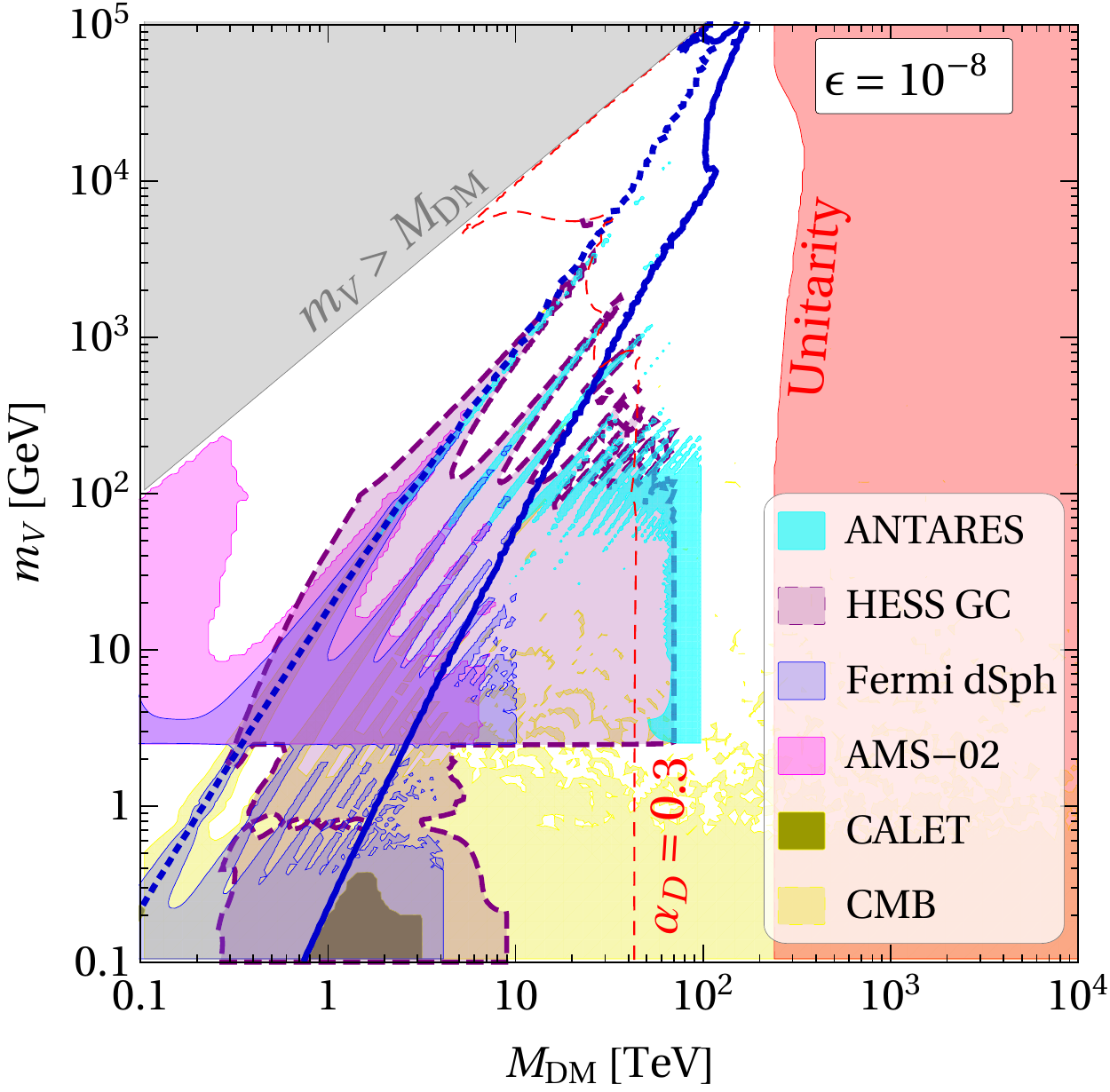}\\
\includegraphics[width=0.43\textwidth]{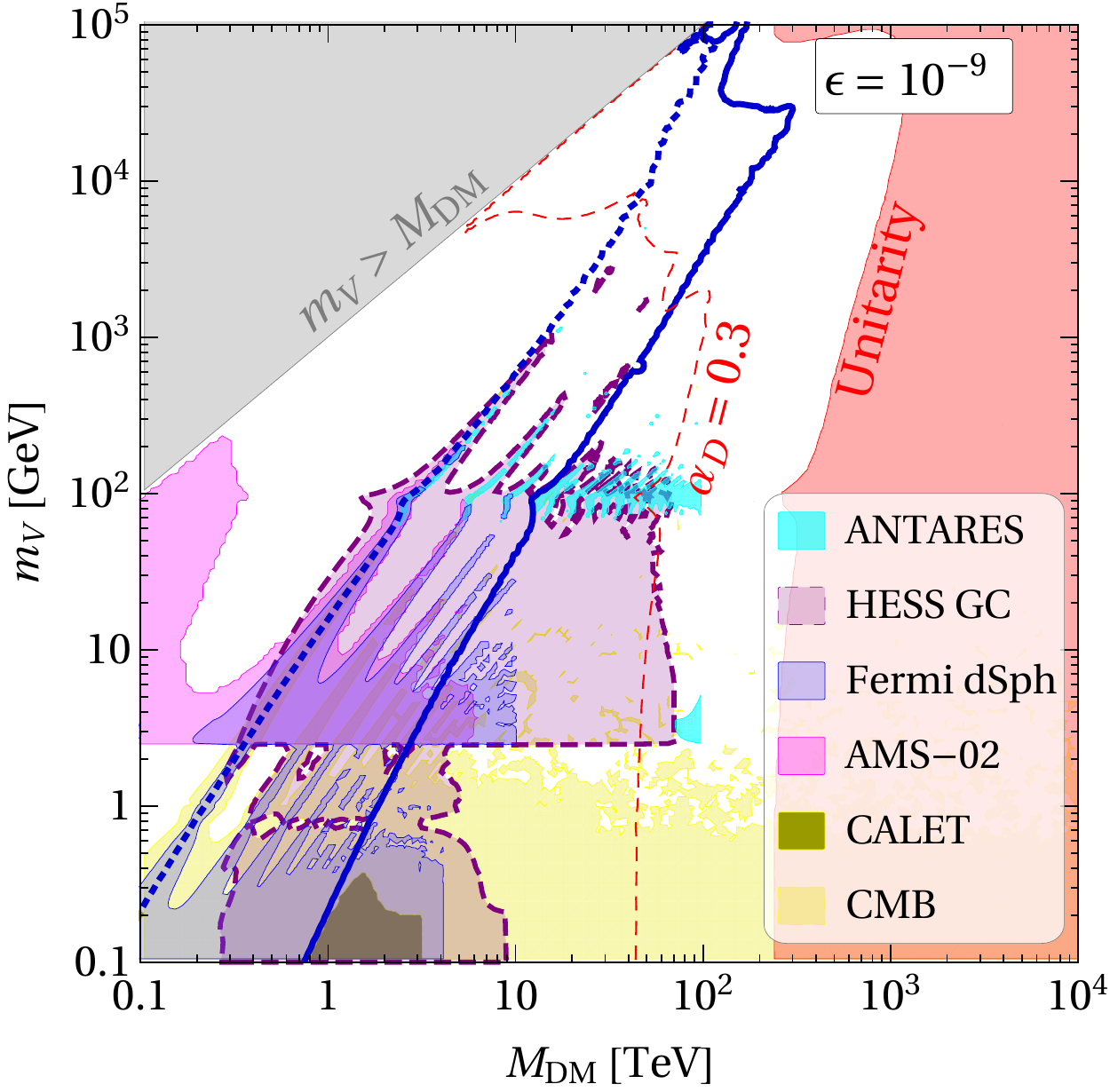}\quad
\includegraphics[width=0.43\textwidth]{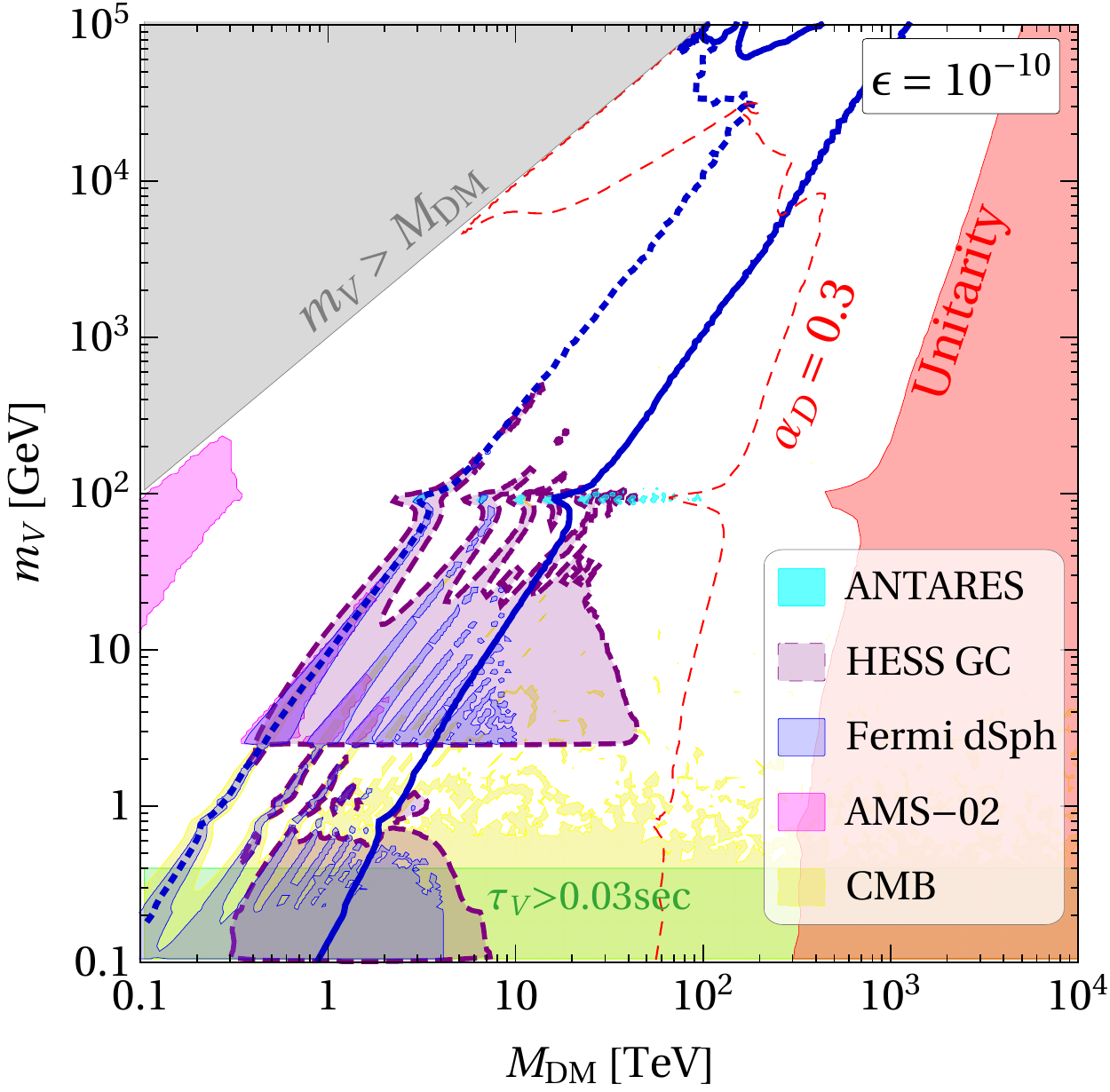}\\
\includegraphics[width=0.43\textwidth]{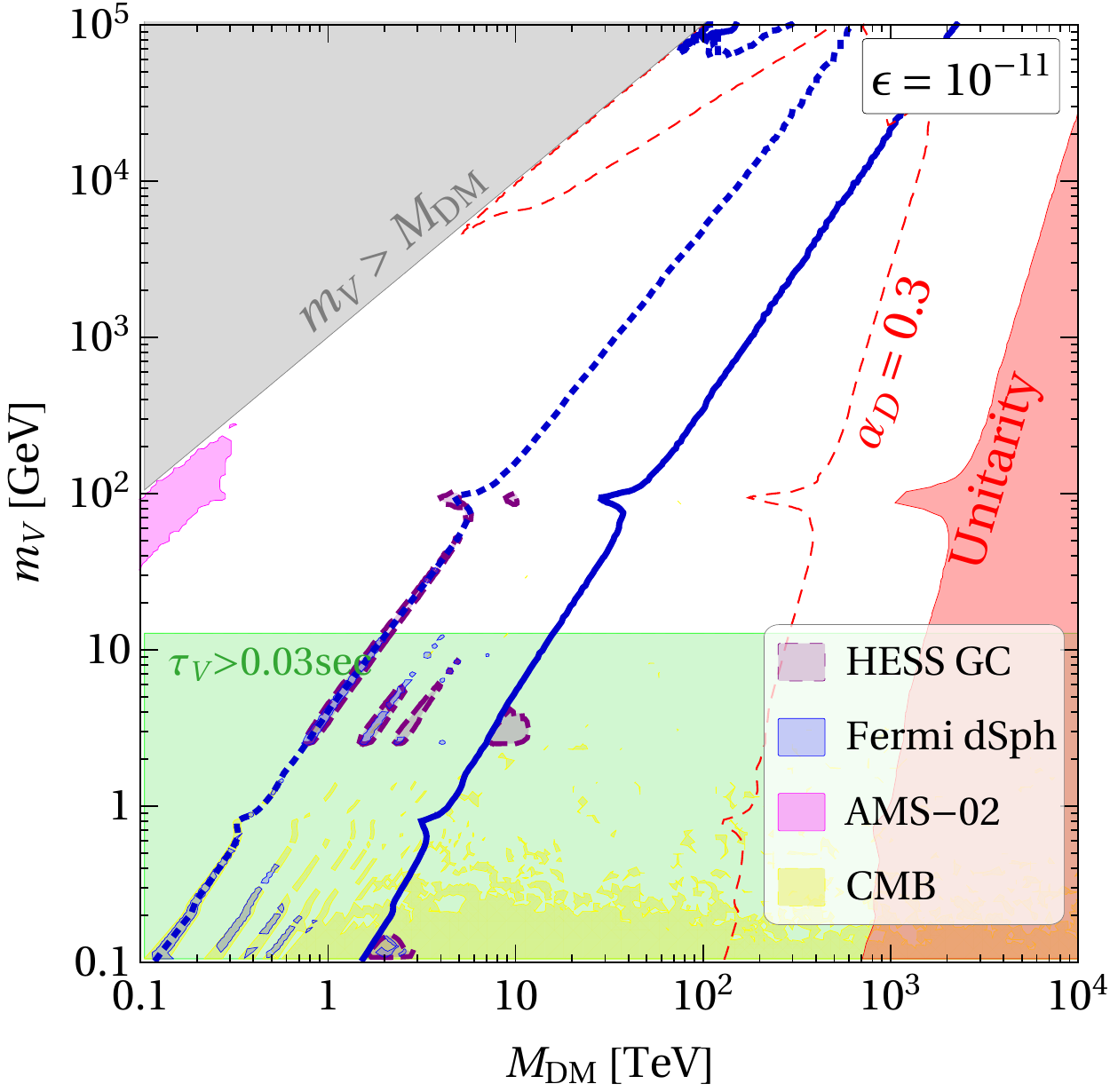}\quad
\includegraphics[width=0.43\textwidth]{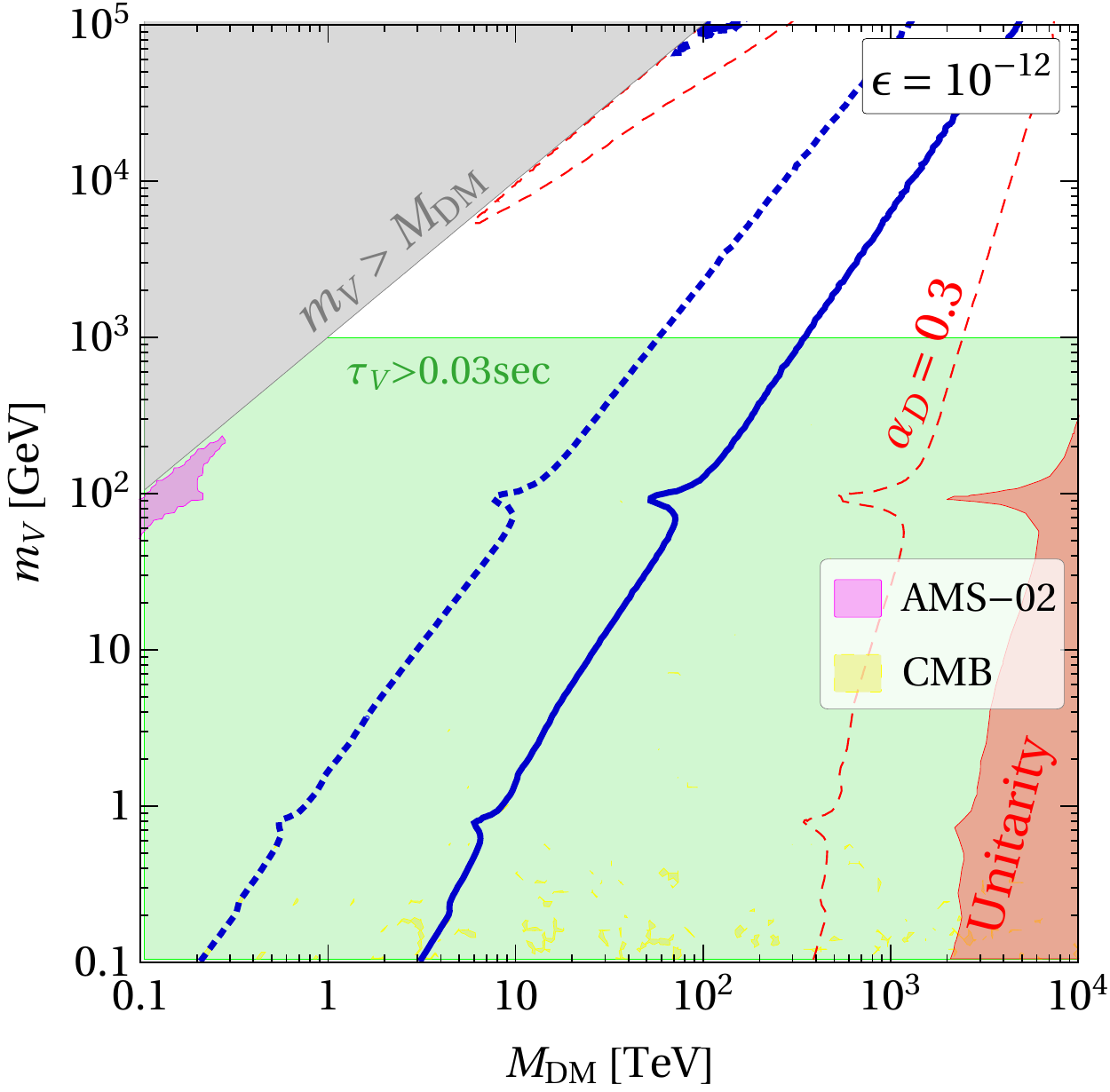}
\end{center}
\caption{\it \small  \label{fig:ID_summary} Indirect detection limits.
Shaded: excluded by {\sc Antares} (cyan), {\sc Ams} (magenta), {\sc Calet} (ocra), {\sc Planck} (light yellow), {\sc Fermi} (blue), within {\sc Hess} reach (magenta), violates s+p wave unitarity bound (red), disfavoured by BBN (green), DM annihilation to dark photons kinematically forbidden (gray). On the right of the lines: bound states exist (blue dotted), bound states can form (blue), $\aD > 0.3$ (red dashed, as a rough indication of where higher order in $\aD$ are expected to become important). From top-left to bottom-right: no DM dilution, and then decreasing values of $\epsilon$ corresponding to increasing DM dilution. 
Note the wider extension in $\MDM$ and $\mV$ in the bottom right panel.
}
\end{figure}

It is important to stress that the 21~cm limit in eq.~(\ref{eq:21cm}) is derived assuming an immediate absorption of the energy injected from the SM products of DM annihilations (so that, again, the number of steps in the cascade does not matter). This is however not the case especially when they are very energetic, thus for large DM masses.
The effects of delayed energy deposition have been taken into account e.g. in~\cite{DAmico:2018sxd,Cheung:2018vww}, however the resulting limits have not been reported for DM masses beyond~10 TeV.
Up to those masses and choosing as an example the $b\bar{b}$ final state, the limits derived as in eq.~(\ref{eq:21cm}) are stronger than those that take into account delayed energy deposition by a factor of a few ($\MDM \simeq 10$~GeV) to a few tens ($\MDM \simeq 10$~TeV)~\cite{DAmico:2018sxd}.
We expect a difference in the same ballpark also in our model, because of the extra softening of the spectra from the extra-steps in the cascade, and because the dominant SM final states (see figure~\ref{fig:BRs}) result in an electron and photon spectra of energies much lower than $\MDM$, in qualitative agreement with $b\bar{b}$.
Such differences in the limits (between eq.~(\ref{eq:21cm}) and those that include delayed energy depositions) are of the same order of the uncertainties induced by assumptions like the boost model and the precise value of the bound to apply, see e.g.~\cite{DAmico:2018sxd,Cheung:2018vww}.
In Fig.~8 of \cite{Cirelli:2018iax}, we show that the $21$~cm-constraints only slightly improves the CMB ones.
Waiting for further experimental confirmations of the signal, and for a precise assessment of the 21~cm limit to be settled, we refrain from putting the 21~cm limits on the same footing of the CMB ones, and we show only the latter ones in the summary in Fig.~\ref{fig:ID_summary}.

\subsection{DM constraints from the Local Universe}
\label{sec:ID_local}

Dark Matter interactions (annihilation, formation and decay of bound states) in the local Universe may induce signals at space- and ground-based telescopes, whose observations therefore put limits on the parameter space of our model.
Here we study constraints from telescopes observing neutrinos (section~\ref{sec:ID_local_nu}), gamma rays (\ref{sec:ID_local_gamma}), antiprotons (\ref{sec:ID_local_pbar}) and electrons and positrons (section~\ref{sec:ID_local_epem}).

In our model, DM interactions result into 2 (from annihilations and from decays of $B_{\uparrow\downarrow}$) or 3 (from decays of $B_{\uparrow\uparrow}$) energetic dark photons, themselves decaying to SM pairs according to the branching ratios in Fig.~\ref{fig:BRs}.
The resulting `one-step' SM energy spectra are therefore moved to lower energies compared to the standard case of direct DM annihilation into SM pairs.
This constitutes an important input in deriving the constraints discussed in this section, and will be discussed case by case.
The additional and much less energetic $V$, that is emitted to form the bound state, will be discussed separately in section~\ref{sec:ID_DPemittedBS}.

\subsubsection{Neutrinos}
\label{sec:ID_local_nu}

The {\sc Antares} collaboration derived constraints from the non-observation of excesses in muon neutrino fluxes coming from the Milky Way halo~\cite{Albert:2016emp}.
Those limits were cast assuming direct DM annihilations into various SM pairs, and therefore cannot be directly applied to our case.
We circumvent this limitation following~\cite{Baldes:2017gzu}, that observed that the {\sc Antares} limits are driven by the most energetic part of the neutrino spectra from DM. Therefore they can roughly be applied to one-step signals that result in quarks (up to bottom), because their zero- and one-step neutrino spectra are very similar for neutrino energies close to $\MDM$.
They can also be applied to one-step signals that result in neutrinos, as is the case in our model for $\mV$ larger than a few tens of GeV, with a further caveat that we now discuss.
The resulting `one-step' neutrino spectrum is spread to energies lower than $\MDM$, while direct DM annihilation into neutrino pairs would induce a spike at $E_\nu \simeq \MDM$. However, the {\sc Antares} finite energy resolution of O(50\%) causes a sizeable fraction of `one-step' neutrinos to fall in the same energy range of those originating from a spike.
To summarize, we exclude a point in our parameter space if it does not respect the limit
\beq
\langle \sigma_{\rm tot} \vrel \rangle {\rm BR}(V \to \bar{f} f) < 2\,C\,\langle \sigma_{{\rm DM\,DM} \to f \bar{f}}\, \vrel \rangle_{\rm limit}^{\textsc{Antares}}
\label{eq:ID_ANTARES}
\eeq
for at least one of the two final states i) $\bar{f} f = \bar{b} b +\bar{c} c +\dots +\bar{u} u$, that we compare with the $\bar{b} b$ limit from {\sc Antares} because of the very similar spectrum, and ii) $\bar{f} f = \bar{\nu}_\mu \nu_\mu$, where we include the effect of oscillations ({\sc Antares} presents limits for DM annihilation into $\bar{\nu}_\mu \nu_\mu$, so that the effective flux of muon neutrinos at the telescope is reduced to roughly 40\% of the initial one).
The factor of 2 in eq.~(\ref{eq:ID_ANTARES}) accounts for the fact that, in our model, DM is not self-conjugate.
We also choose $C=1$ for quark final states and $C=2$ for neutrino ones, to very roughly account for the neutrinos that will be properly recognized to have $E_\nu < \MDM$.\footnote{This last prescription for neutrinos, together with the precise values of the dark photon branching ratios, are the only aspects in which this analysis departs from the one in~\cite{Baldes:2017gzu}.}

The resulting limits are shown in cyan in figure~\ref{fig:ID_summary} for various values of the kinetic mixing $\epsilon$ (and therefore of DM dilution).
We find also an excluded region for $\mV < 2.5$~GeV, but we do not show it as it is an artifact of the way we modeled dark photon decays in that mass region, see~\cite{Baldes:2017gzu} for more details. 
Our limits can be considered conservative in the sense that one should have summed over the neutrino fluxes from all final states (this would be possible if {\sc Antares} data were public), and aggressive in the sense that {\sc Antares} assumes a DM density profile which is rather peaked towards the Galactic Center.
We do not show the analogous {\sc Icecube} limits~\cite{Aartsen:2017ulx} because they are not provided for $\MDM > 1$~TeV, and are weaker than those of {\sc Antares}.
The latter aspect could be caused by the fact that {\sc Icecube} does not perform as good as {\sc Antares} for neutrinos coming from the Galactic Center, which drive the exclusions (see also~\cite{ElAisati:2017ppn} for a derivation of DM limits from IceCube data).

\subsubsection{Gamma rays}
\label{sec:ID_local_gamma}

We consider first constraints from the {\sc Fermi} satellite observations of several dwarf spheroidal galaxies (dSphs)~\cite{Ackermann:2015zua}, which are the strongest ones on our model among those derived from {\sc Fermi} data (see~\cite{Baldes:2017gzu}).
We use the results of~\cite{Profumo:2017obk}, which derived limits on secluded DM models starting from public {\sc Fermi} data from dSphs observations, and presented them for different assumptions on the decay products of the mediator and of its mass. In particular, we exclude a point in our parameter space if it does not satisfy
\beq
\langle \sigma_{\rm tot} \vrel \rangle {\rm BR}(V \to \bar{f} f) < 2\,\langle \sigma_{{\rm DM\,DM} \to f \bar{f}}\, \vrel \rangle_{\rm limit}^\gamma
\label{eq:ID_Gammas}
\eeq
for at least one of the final states $\bar{f} f = \bar{b} b, \bar{q}q, \tau^+\tau^-,\dots$, and where again the factor of 2 accounts for the fact that, in our model, DM is not self-conjugate.
The resulting exclusions are shown in dark blue in figure~\ref{fig:ID_summary}, for various values of $\epsilon$. We find that these exclusions do not significantly depend on the assumption on $\mV$, among the two presented in~\cite{Profumo:2017obk}.
As a cross-check we note that the excluded regions are in very good agreement with those derived in~\cite{Cirelli:2016rnw}, for the same model in the case of no dilution, and which made direct use of {\sc Fermi} data.

Moving to higher energy gamma rays, we now consider observations of gamma rays from the GC with the {\sc Hess} telescope~\cite{Abdallah:2016ygi}.
We use the results in~\cite{Profumo:2017obk} that, analogously to what done for {\sc Fermi} dSphs, derived {\sc Hess} limits on secluded DM models.
An important caveat is that they are derived from the public data used in the {\sc Hess} analysis~\cite{Abramowski:2011hc} (corresponding to 112 hours of observation time), but presented assuming a rescaling of the 112h limits with the observation time of 254h on which~\cite{Abdallah:2016ygi} is based (because these extra data are not public).
Therefore we only treat these limit as an indication of the sensitivity, to this model, that could be achieved by the {\sc Hess} telescope. In this spirit, we extend them up to $\MDM = 70$~TeV, because this is the maximal DM mass reached in the {\sc Hess} analysis~\cite{Abdallah:2016ygi}.
We derive the sensitivities as in eq.~(\ref{eq:ID_Gammas}), where now of course $\langle \sigma_{{\rm DM\,DM} \to f \bar{f}}\, \vrel \rangle_{\rm limit}^\gamma$ are the exclusions given in~\cite{Profumo:2017obk},  and show them in purple in figure~\ref{fig:ID_summary}.
We stress that the {\sc Hess} sensitivity, from observations of the GC, would completely evaporate if the DM density profile has a core of size larger than~500~pc, and that it would be substantially weakened for cores down to a few tens of pc~\cite{HESS:2015cda} (both possibilities are currently allowed both by simulations~\cite{DiCintio:2013qxa,Marinacci:2013mha,Tollet:2015gqa} and by data~\cite{Nesti:2013uwa,Cole:2016gzv,Wegg:2016jxe}).

Other ground-based telescopes have observed high energy gamma rays and cast bounds on annihilations of heavy DM, e.g. {\sc Veritas}~\cite{Archambault:2017wyh}, {\sc Magic}~\cite{Ahnen:2017pqx} and {\sc Hawc}~\cite{Albert:2017vtb,Abeysekara:2017jxs}.
We do not show the resulting limits here because they are derived assuming no steps in the DM annihilations, and because they are anyway weaker than the {\sc Hess} sensitivity that we show.

\subsubsection{Antiprotons}
\label{sec:ID_local_pbar}

We now turn to the constraints imposed by galactic antiproton observations. The method follows closely the previous analyses\footnote{See also \cite{Boudaud:2019efq,Genolini:2021doh,Calore:2022stf}.} in~\cite{Boudaud:2014qra,Giesen:2015ufa,Cirelli:2016rnw,Baldes:2017gzu}. For completeness, however, we briefly review it here.
 
Consistently with the previous studies, we use the {\sc Ams} measurement~\cite{Aguilar:2016kjl} of the antiproton to proton ratio over the range from 1 to 450 GeV. The data are rather well explained by astrophysical antiproton sources, at least within the uncertainties related to $\bar p$ production in the interstellar medium, to their propagation process in the galactic environment and to the modulation of their flux due to solar activity~\footnote{Possible claimed excesses at low energies~\cite{Hooper:2014ysa, Cuoco:2016eej, Cui:2016ppb, Huang:2016tfo, Feng:2017tnz, Arcadi:2017vis} are now shown to have a limited statistical significance~\cite{Reinert:2017aga}.}. 
As for the DM contribution, we adopt an Einasto profile for the distribution in the MW galactic halo and we set the local dark matter density to $\rho_{sun}=0.42$~GeV/cm$^3$~\cite{Pato:2015dua}. We choose a `MAX' propagation scheme, as it is the one favored by the fit of the astrophysical background.
Upon imposing that the DM contribution does not worsens the astrophysics-only fit by more than $\Delta \chi^2 =9$, we obtain conservative DM bounds.

The excluded regions are highlighted in magenta in fig.~\ref{fig:ID_summary}. They consist of an area below $M_{\rm DM} \lesssim 1$ TeV, where the DM $\bar p$ contribution falls into the relatively low energy and high precision portion of the {\sc Ams} data, and a series of funnels at $M_{\rm DM} \gtrsim 1$ TeV, corresponding to the resonances of the annihilation and BSF cross-sections. The excluded regions terminate for $m_{\rm V} \lesssim 2.5$ GeV, consistently with the fact that the BRs of a light $V$ into protons and antiprotons are inhibited (see the discussion in sec.~\ref{sec:darkphoton} and appendix~\ref{app:DP}).

\subsubsection{Electrons and positrons}
\label{sec:ID_local_epem}

We derive constraints also from the measurements of the `all-electron flux' (i.e.: $e^++e^-$) by the {\sc Calet} experiment~\cite{Adriani:2017efm} onboard the International Space Station, which cover the range 10 GeV $-$ 3 TeV. Measurements of the same observable have also been produced by the {\sc Dampe} satellite~\cite{Ambrosi:2017wek}, by {\sc Ams}~\cite{Aguilar:2014fea} and by {\sc Hess}~\cite{HESSICRC2017}. The {\sc Calet} data agree well with {\sc Ams} but the former extend to higher energies: since our focus is on very heavy DM, we use {\sc Calet}. The data by {\sc Hess}, which are an update of former measurements~\cite{Aharonian:2008aa,Aharonian:2009ah} with extended energy range, have so far been presented only at conferences. They display large systematic uncertainties and we checked that they are currently non competitive (in terms of constraining power), so that we do not use them here.
Finally, the data by {\sc Dampe} are somewhat higher in normalization then those by {\sc Calet} and {\sc Ams}, in the range 50$-$1000 GeV, and have a harder spectrum. As this might be due to energy miscalibration in {\sc Dampe}, we use {\sc Calet} only.

The astrophysical fluxes of $e^++e^-$ are poorly modeled at the high energies under consideration. We therefore decide to proceed, as customary in these case, with two methods: a conservative one and a more aggressive one. In the conservative approach, we assume no astrophysical background, we compute the DM $e^++e^-$ and we impose that it does not exceed any {\sc Calet} data point by more than 2$\sigma$. This yields the excluded regions in ocra in fig.~\ref{fig:ID_summary}, which sit in the ballpark of $M_{\rm DM} \sim$ few TeV and $m_{\rm V} \lesssim$ 500 MeV. They shrink and disappear as soon as the dilution parameter $\epsilon \lesssim 10^{-10}$. 
In the aggressive approach, we include an astrophysical $e^++e^-$ background flux modeled with a power-law fitted on the {\sc Calet} data over the whole energy range. We find $E^3 \, \Phi^{e^++e^-}_{\rm astro} = 3.4 \times 10^{-2} E^{-0.18} \, {\rm GeV}^2/({\rm cm}^{2} \, {\rm s}\, {\rm sr})$, with $\chi^2 = 16.4$ for 38 d.o.f., in reasonably good agreement with the analogous power law fit quoted in \cite{Adriani:2017efm}.
Then, in analogy with the procedure followed for antiprotons, we impose that the DM contribution does not worsens the modeled-astrophysics-only fit by more than $\Delta \chi^2 =9$ and we obtain the regions drawn in yellow in fig.~10 of \cite{Cirelli:2018iax}. Not surprisingly, these regions are much more extended than those derived in the conservative approach. We stress, however, that these regions should be interpreted more as a {\it reach} of the $e^++e^-$ measurements {\it under the assumption} that the astrophysical flux is well modeled by the {\it empirical} fit described above. In order to be conservative, we do not show them in the summary of Figure~\ref{fig:ID_summary}.

\subsubsection{On galactic subhalos and on DM limits from clusters of galaxies.}

In all the limits presented in this section~\ref{sec:ID_local}, we have not included a possible `boost' of the DM signals from the existence of DM clumps. We summarize their possible impact here.

In principle, the presence of DM subhalos is indeed expected to enhance the annihilation flux~\cite{Silk:1992bh,Bergstrom:1998jj,Ullio:2002pj,Berezinsky:2003vn}.
However, in the central regions of the Milky Way subhalos are typically tidally disrupted by both the GC and the Galactic disk, so that such a boost turns out to be negligible for signals, e.g.~in gamma rays, originating from the inner kiloparsecs (see~\cite{Stref:2016uzb} for a recent assessment).
Boost factors up to a few tens of percent are possible for signals coming from dwarf spheroidal galaxies, even though, for conservative assumptions on the mass index, they become again negligible~\cite{Moline:2016pbm}.
Boosts of signals coming from our neighborhood in the MW, relevant for DM searches in charged cosmic rays, are found not to exceed a few tens of percent for conservative assumptions~\cite{Lavalle:2006vb,Lavalle:1900wn,Pieri:2009je,Stref:2016uzb}.

On the basis of the above, we find it justified to neglect the impact of DM subhalos as far as limits are concerned. On the other hand, we underline that the interplay of a possible DM signal in charged cosmic rays with gamma observations could provide precious indications on DM subhalo existence and properties.

DM subhalos might have a more dramatic impact on DM signals from galaxy clusters~\cite{Moline:2016pbm,Stref:2016uzb}, leading to an interesting interplay with other signals from heavy annihilating DM~\cite{Murase:2012xs}.
However, DM limits from galaxy clusters are either negligible with respect to all those considered here~\cite{Abramowski:2012au}, or at most comparable for aggressive assumptions on the DM subhalo properties~\cite{Ackermann:2015fdi}. Therefore, while a possible ID signal from a cluster could in principle be ascribed to annihilating DM, we do not discuss these limits further here.

We finally comment that additional DM subhalos could arise in cosmological histories with a matter-dominated phase after DM (chemical and thermal) decoupling, see~\cite{Gelmini:2008sh,Erickcek:2011us,Erickcek:2015jza} for quantitative studies of this possibility. The investigation of the formation and survival of these subhalos in our model, while interesting, goes beyond the purpose of this study. In the spirit of being conservative on this matter we neglect it in our ID constraints.

\subsection{On the dark photon emitted during the formation of bound states}
\label{sec:ID_DPemittedBS}

So far we have not discussed the signals from the low-energy dark photon emitted during BSF, which could potentially strengthen the indirect detection constraints. This $V$ carries away the binding energy $E_B = \aD^2 \MDM/4$, where we neglect the kinetic energy $\vrel^2 \MDM/4$ because $\aD \gtrsim \vrel$ when BSF is important.
It thus produces a spectrum of SM final states analogous to the one produced by a DM of mass $E_B/2$ annihilating directly into SM pairs (the same SM pairs that $V$ decays into). However, DM with such low mass would have number density larger by $\MDM/(E_B/2) = 8/\aD^2$. 

We may then recast the existing upper bounds, $\overline{\sigma v}_\mathsmaller{\text{rel}}$, on (non-self-conjugate) DM annihilating into SM pairs as follows
\beq
\langle \sigma_{\BSF} \vrel \rangle (\MDM, \mV) < 
\Big(\frac{8}{\aD^2}\Big)^2   \times 
\overline{\sigma v}_\mathsmaller{\text{rel}} (M = \aD^2 \MDM /8)\,.
\label{eq:BSFphoton_RecastID}
\eeq
Even for large values of the dark fine structure constant, the prefactor above is rather large ($64/\aD^4 \gtrsim 10^3$ for $\aD \lesssim 0.5$), and the existing indirect detection bounds (see e.g.~\cite{Ade:2015xua,Ackermann:2015zua} for some strongest ones) do not yield any further constraints on our parameter space -- except perhaps on top of the resonances. We shall therefore not consider this low-energy dark photon further, but instead we refer to \cite{Baldes:2020hwx} for a dedicated study.

Note that the CMB and 21~cm constraints scale proportionally to the DM mass [cf.~eqs.~\eqref{eq:CMB} and \eqref{eq:21cm}], which partly mitigates the loosening of the constraints seen in \eqref{eq:BSFphoton_RecastID}. However, as noted earlier, BSF via vector emission is a $p$-wave process, and becomes entirely negligible for the very low $\vrel$ during CMB.

\section{Summary and Outlook}
\label{sec:outlook} 

High energy telescopes constitute our unique direct access to energies much above a TeV, therefore they are in a privileged position to test BSM physics at those scales. Annihilating heavy Dark Matter is a particularly motivated example of such BSM physics.
The existing and upcoming telescope data at unprecedented energies, together with the growing theoretical interest in DM with masses beyond 10-100 TeV, make it particularly timely to explore the associated phenomenology.
In this study we performed a step in this direction.

We pointed out that models where DM pairs annihilate mostly in `dark' mediators, themselves decaying into SM particles, allow to naturally evade the unitarity limit on the DM mass and, at the same time, to reliably compute cosmic ray signals. They therefore overcome the two main obstacles to the use of data of high energy telescopes to test annihilating heavy Dark Matter.
Encouraged by this observation, we then studied in detail several aspects of the cosmological history and of the indirect detection signals of these models, which we summarise as follows:
\begin{enumerate}
\item We computed the dilution of relics induced by i) the entropy injection from decays of the mediators after DM freeze-out, and ii) the possibility that the SM and dark sector were not in thermal equilibrium at early times.
We then determined how i) and ii) affect a) the thermal freeze-out cross section and, in turn, the maximal DM mass compatible with unitarity, see Figures~\ref{fig:sigmaFOandMuni} and \ref{fig:UnitaryLimit}; b) the maximal dilution compatible with BBN constraints, see Section~\ref{sec:UniBound}. 
Our study is the first, to our knowledge, to systematically explore the combination of these effects, therefore we reported ready-to-use analytical results (on dilution, freeze-out cross section, BBN limits, maximal DM mass allowed by unitarity) throughout Section~\ref{sec:dilution_DarkU1}.

\item As a case study we then considered a model of fermion DM charged under a spontaneously broken dark $U(1)$, where the role of the mediator is played by the associated massive vector. 
We computed the values of the dark gauge coupling that results in the correct DM abundance, over a wide mass range and for different values of the kinetic mixing and therefore of dilution (see Figure~\ref{fig:alphaD} right). Our freeze-out computation  improves over previous ones in that it includes long-range interactions beyond the Coulomb approximation, which is particularly relevant for heavy DM and mediator masses (see Figure~\ref{fig:alphaD} left).
The resulting cross sections relevant for DM indirect detection are shown in Figure~\ref{fig:sigmav_MW_unitarity_mVD_10GeV}.

\item We finally studied the indirect detection phenomenology of the Dark $U(1)$ model, both with and without dilution, see Figure~\ref{fig:ID_summary} for a summary.
We found the encouraging result that multimessenger searches for heavy DM are possible, offering different lines of attack to test these models.
Our findings also give reasons to think that some telescopes (especially neutrino and gamma ray) have an unexplored potential to test annihilating DM with a mass of  $O(100)$~TeV. This motivates an experimental effort in this direction and, importantly, the grant of public access to experimental data in some form.
The latter would also overcome the problem, in interpreting experimental results, that some DM models (e.g. secluded ones) result in SM spectra different from the benchmarks used by the collaborations.

\end{enumerate}

Our results open the doors to many possible future investigation.
Concerning high energy cosmic rays, it would be interesting to determine the sensitivity of future experiments like {\sc Cta \cite{Siqueira:2019wdg}, Lhaaso, Taiga, Km3net, Herd, Iss-Cream} on secluded DM models, and to explore whether annihilations of heavy secluded DM could explain the highest energy neutrinos observed by {\sc Icecube}.
It would also be intriguing to study the interplay of indirect detection signals with other effects of entropy injections, e.g. on the baryon asymmetry.
On a more theory side, two exciting directions of exploration regard the connection, of the small portal of these models with the SM and/or of the heavy DM mass, with the solution of other problems of the SM.
We plan to return to some of these aspects in future work.

\begin{subappendices}

\section{$U(1)_{D}$ coupled to hypercharge $U(1)_{Y}$.}
\label{app:U1D_mixing}
\subsection{The Lagrangian}

We consider the Dark matter charged under $U(1)_{D}$ and interacting through a Dark mediator $V_{\mu}$, the Dark photon, with mass $m_{V}$. The Dark photon is kinematically coupled to the $U(1)_{Y}$ gauge boson $B_{\mu}$. If the temperature is below $100$ $\rm GeV$, then the electroweak symmetry is broken

\begin{align}
L = & \quad \bar{\psi}_{D}i\slashed{D}\psi_{D} - m_{DM}\bar{\psi}_{D}\psi_{D} \\ 
 &- \frac{1}{4} F_{D\mu\nu}F_{D}^{\mu\nu} -\frac{1}{2}m_{V}^{2}V_{\mu}V^{\mu} \\  
 &-\frac{1}{4} W^{3}_{\mu\nu}W^{3\mu\nu} -\frac{1}{4} B_{\mu\nu}B^{\mu\nu} -\frac{1}{2}m_{Z}^{2}Z_{\mu}Z^{\mu}  \\  
 & + \frac{\epsilon}{c_{w}} B_{\mu\nu}F_{D}^{\mu\nu} 
\label{lagrangian_U(1)_D_model}
\end{align}

where 
\begin{align}
&F_{D}^{\mu\nu} =  \partial ^{\mu} V^{\nu} - \partial ^{\nu} V^{\mu}, \qquad W^{3\mu\nu} =  \partial^{\mu} W^{3\nu} - \partial^{\nu} W^{3\mu}, \\
&B^{\mu\nu} =  \partial ^{\mu} B^{\nu} - \partial ^{\nu} B^{\mu}, \qquad Z_{\mu} = c_{w} W_{3\mu} - s_{w} B_{\mu}, 
\end{align}
with $D_{\mu} = \partial_{\mu} + i g_{D} V_{\mu}$,  $c_{w} = \cos \theta_{w}$, $s_{w} = \sin \theta_{w}$, and $\theta_{w}$ the weak mixing angle. $W^{3}_{\mu}$ is the $SU(2)_{L}$ gauge boson associated with the generator $T_{3}$. The coupling constant $\epsilon$ is supposed to be small: $\epsilon \ll 1$. 
For later use, we defined the Dark fine structure constant $\alpha_{D} = g_{D}^{2}/4\pi$.

\subsection{Gauge eigenstates versus mass eigenstates}
\paragraph{Definitions.}

The gauge eigenstates are the vector fields which transform under the adjoint representation of their gauge group. In our case they are $V_{\mu}$, $W^{3}_{\mu}$ and $B_{\mu}$ or $V_{\mu}$, $Z_{\mu}$ and  $A_{\mu}$ modulo a rotation. If we define
\begin{equation}
G_{\mu} = \left( \begin{array}{c}
V_{\mu} \\
B_{\mu} \\
W^{3}_{\mu}
\end{array} \right) , \qquad
H_{\mu} = \left( \begin{array}{c}
V_{\mu} \\
Z_{\mu} \\
A_{\mu}
\end{array} \right)
\end{equation}
then the orthogonal matrix $R$ which rotates $H_{\mu}$ into $G_{\mu} = R \cdot H_{\mu}$, is equal to
\begin{equation}
R= \left( \begin{array}{ccc}
1 & 0 & 0 \\
0 & -s_{w} & c_{w} \\
0 & c_{w} & s_{w} 
\end{array} \right).
\label{R-matrix}
\end{equation}
Now, let's rewrite the kinetic terms as
\begin{equation}
-\frac{1}{4}B^{\mu\nu}B_{\mu\nu} = -\frac{1}{2} \left( \partial^{\mu}B^{\nu}\partial_{\mu}B_{\nu} -  \partial^{\mu}B^{\nu}\partial_{\nu}B_{\mu} \right) =  B_{\mu} k^{\mu\nu} B_{\nu}
\end{equation}
with $k^{\mu\nu} = \frac{\eta^{\mu\nu}\partial^{2} - \partial^{\mu}\partial^{\nu}}{2}$.\\
We use it to write the Lagrangian as
\begin{equation}
L = \quad \bar{\psi}_{D}i\slashed{D}\psi_{D}  + G_{\mu}^{t} K k^{\mu\nu} G_{\nu} - \frac{1}{2} H_{\mu}^{t} M \eta^{\mu\nu} H_{\nu} 
\end{equation}
with 
\begin{equation}
K =  \left( \begin{array}{ccc}
1 & \frac{\epsilon}{c_{w}} & 0 \\
\frac{\epsilon}{c_{w}} & 1 & 0 \\
0 & 0 & 1 
\end{array} \right)
\end{equation}
and
\begin{equation}
M = \left( \begin{array}{ccc}
m_{V}^{2} & 0 & 0 \\
0 & m_{Z}^{2} & 0 \\
0 & 0 & 0 
\end{array} \right).
\end{equation}
We define the mass eigenstates as the fields with a \textbf{kinetic} matrix $K$ equal to the \textbf{identity} matrix and with a \textbf{mass} matrix $M$ which is \textbf{diagonal}. 

Let's explain how we compute them.
\begin{enumerate}
\item
First, we find new fields $\tilde{G}_{\mu}$, linear combinations of $G_{\mu}$, with \textbf{canonical} kinetic terms, i.e.  the kinetic matrix is the \textbf{identity} matrix $I$. We are looking for the transformation matrix $P_{G}$ $\in$ $M(3,\mathbb{R})$ defined as $G_{\mu} = P_{G} \tilde{G}_{\mu}$ and which fulfills
\begin{equation}
P_{G}^{t} K P_{G} = I. 
\label{P_{G}}
\end{equation}
The Lagrangian becomes
\begin{equation}
L =\quad \bar{\psi}_{D}i\slashed{D}\psi_{D}  + \tilde{G}_{\mu}^{t} k^{\mu\nu} \tilde{G}_{\nu} - \frac{1}{2} H_{\mu}^{t} M \; \eta^{\mu\nu} H_{\nu}.
\end{equation}
Introducing $\tilde{H}_{\mu}$ defined by $\tilde{G}_{\mu} = R \tilde{H}_{\mu}$ with $R$ defined in Eq.~\eqref{R-matrix} allows us to write
\begin{equation}
L =\quad \bar{\psi}_{D}i\slashed{D}\psi_{D}  + \tilde{H_{\mu}}^{t} k^{\mu\nu} \tilde{H_{\nu}} - \frac{1}{2} \tilde{H_{\mu}}^{t} P^{t}_{H} M  P_{H} \; \eta^{\mu\nu} \tilde{H_{\nu}} 
\end{equation}
with 
\begin{equation}
P_{H} = R^{-1} P_{G} R.
\label{P_{H}}
\end{equation}
\item
Second, we find new fields $\tilde{\tilde{H}}_{\mu}$, linear combinations of $\tilde{H}_{\mu}$, with still a \textbf{canonical} kinetic matrix but with a \textbf{diagonal} mass matrix. These are the mass eigenstates. One needs to find the transformation matrix $Q$ $\in$ $M(3, \mathbb{R})$ defined by $\tilde{H}_{\mu} = Q \tilde{\tilde{H}}_{\mu}$ and which verifies
\begin{equation}
\begin{array}{ll}
Q^{-1}Q = 1 \quad \rightarrow \quad Q \in O(3), \\
Q^{t}P^{t}_{H} M P_{H} Q = D. 
\end{array}
\label{Q}
\end{equation} 
with $D$ a diagonal matrix. 

Indeed, the lagrangian becomes
\begin{align}
L = &\quad \bar{\psi}_{D}i\slashed{D}\psi_{D}  + \tilde{\tilde{H}}_{\mu} Q^{-1}Q k^{\mu\nu} \tilde{\tilde{H}}_{\mu} - \frac{1}{2} \tilde{\tilde{H}}_{\mu} Q^{t}P^{t}_{H} M P_{H} Q \; \eta^{\mu\nu} \tilde{\tilde{H}}_{\mu}, \\
= &\quad \bar{\psi}_{D}i\slashed{D}\psi_{D}  + \tilde{\tilde{H}}_{\mu} \left( k^{\mu\nu} - \frac{1}{2} D \; \eta^{\mu\nu} \right) \tilde{\tilde{H}}_{\mu}.
\end{align}
\end{enumerate}

\paragraph{Computation of the mass eigenstates.}

Let's now apply the recipe. First, we need to solve Eq.~\eqref{P_{G}} in order to find $P_{G}$ $\in$ $M(3, \mathbb{R})$ that we choose to parametrize as
\begin{equation}
P_{G} = \left( \begin{array}{ccc}
R_{1}\cos \theta_{1} & R_{1}\sin \theta_{1}  & 0 \\
R_{2}\cos \theta_{2} & R_{2}\sin \theta_{2} & 0 \\
0 & 0 & 1
\end{array} \right).
\end{equation}
since only the restriction to $(V, B)_{\mu}$ is not diagonal. 

Eq.~\eqref{P_{G}} yields to
\begin{align}
&R_{1}^{2}\cos^{2} \theta_{1} + R_{2}^{2} \cos^{2} \theta_{2} + 2\frac{\epsilon}{c_{w}} R_{1}R_{2} \cos \theta_{1} \cos{\theta_{2}} = 1, \\
&R_{1}^{2}\sin^{2} \theta_{1} + R_{2}^{2} \sin^{2} \theta_{2} + 2\frac{\epsilon}{c_{w}} R_{1}R_{2} \sin \theta_{1} \sin{\theta_{2}} = 1, \\
&R_{1}^{2}\sin 2\theta_{1} + R_{2}^{2} \sin 2\theta_{2} + 2\frac{\epsilon}{c_{w}} R_{1} R_{2} \sin(\theta_{1}+\theta_{2}) = 0.
\end{align}
We can check that $\theta_{1}=0$ and $R_{1}=R_{2}=R$ are allowed by the last equations which simplify to
\begin{align}
&R^{2} \left( \cos^{2} \theta_{2} + 1 + 2\frac{\epsilon}{c_{w}} \cos \theta_{2} \right) = 1, \\
&R^{2} \sin^{2} \theta_{2} = 1, \\
& \cos \theta_{2} + \frac{\epsilon}{c_{w}} = 0.
\end{align}
We finally get
\begin{equation}
P_{G} = \left( \begin{array}{ccc}
\frac{1}{\sqrt{1-\left( \frac{\epsilon}{c_{w}} \right) ^2}} & 0 & 0 \\
-\frac{1}{\sqrt{1-\left( \frac{\epsilon}{c_{w}} \right) ^2}}\frac{\epsilon}{c_{w}} & 1 & 0 \\
0 & 0 & 1
\end{array} \right),
\end{equation}
from which and from Eq.~\eqref{P_{H}} we compute $P_{H}$
\begin{equation}
P_{H} = \left( \begin{array}{ccc}
\frac{1}{\sqrt{1-\left( \frac{\epsilon}{c_{w}} \right) ^2}} & 0 & 0 \\
\frac{t_{w} \epsilon}{\sqrt{1-\left( \frac{\epsilon}{c_{w}} \right) ^2}}  & 1 & 0 \\
-\frac{\epsilon}{\sqrt{1-\left( \frac{\epsilon}{c_{w}} \right) ^2}}  & 0 & 1
\end{array} \right).
\end{equation}
One deduces the new mass matrix
\begin{equation}
P_{H}^{t} M P_{H} = \left( \begin{array}{ccc}
\frac{m_{V}^{2} + t_{w}^{2}\epsilon^{2}m_{Z}^{2}}{1-\left( \frac{\epsilon}{c_{w}} \right) ^2} & \frac{t_{w}\epsilon}{\sqrt{1-\left( \frac{\epsilon}{c_{w}} \right) ^2}}m_{Z}^{2} & 0 \\
\frac{t_{w}\epsilon}{\sqrt{1-\left( \frac{\epsilon}{c_{w}} \right) ^2}}m_{Z}^{2} & m_{Z}^{2} & 0 \\
0 & 0 & 0
\end{array} \right).
\end{equation}
The last step is to diagonalize it. 

We find the diagonal mass matrix
\begin{equation}
D = \left( \begin{array}{ccc}
m_{Z}^{2}\delta^{2} \left( 1 + \epsilon^{2} \frac{1 - \left( \frac{\delta}{c_{w}} \right) ^{2}}{1-\delta^{2}} \right) & 0 & 0 \\
0 & m_{Z}^{2} \left( 1 + \epsilon^{2} \frac{t_{w}^{2}}{1-\delta^{2}} \right) & 0 \\
0 & 0 & 0
\end{array} \right) +O(\epsilon^{4}),
\end{equation}
and the matrix with the eigenvectors as its columns at leading orders in $\epsilon$
\begin{equation}
Q = \left( \begin{array}{ccc}
-1 & \frac{t_{w}\epsilon}{1-\delta^{2}} & 0 \\
\frac{t_{w}\epsilon}{1-\delta^{2}} & 1 & 0 \\
0 & 0 & 1
\end{array} \right) +O(\epsilon^{2})
\end{equation}
with $\delta \equiv m_{V}/m_{Z}$. \\
When $\epsilon \to 0$, the matrix $D$ becomes
\begin{equation}
\lim_{\epsilon \to 0} D  = \left( \begin{array}{ccc}
m_{V}^{2} & 0 & 0 \\
0 & m_{Z}^{2} & 0 \\
0 & 0 & 0
\end{array} \right)
= M.
\end{equation}
From this, we can identifies the first, second and third coordinate of $\tilde{\tilde{H}}_{\mu}$ to be the new Dark photon $\tilde{\tilde{V}}_{\mu}$, the new $U(1)_{Y}$ boson $\tilde{\tilde{Z}}_{\mu}$ and the new photon $\tilde{\tilde{A}}_{\mu}$ and we write
\begin{equation}
\tilde{\tilde{H}}_{\mu} = \left( \begin{array}{c}
\tilde{\tilde{V}}_{\mu} \\
\tilde{\tilde{Z}}_{\mu} \\
\tilde{\tilde{A}}_{\mu}
\end{array} \right).
\end{equation}
We finally deduce the transformation between the gauge eigenstates $H_{\mu}$ and the mass eigenstates $\tilde{\tilde{H}}_{\mu}$ at leading orders in $\epsilon$
\begin{equation}
H_{\mu}  = P_{H}Q \tilde{\tilde{H}}_{\mu}
\quad
\rightarrow
\quad
\left( \begin{array}{c}
V_{\mu} \\
Z_{\mu} \\
A_{\mu}
\end{array} \right)
=
\left( \begin{array}{ccc}
-1 & \frac{1}{1-\delta^{2}}t_{w}\epsilon & 0 \\
\frac{\delta^{2}}{1-\delta^{2}}t_{w}\epsilon & 1 & 0 \\
\epsilon & 0 & 1
\end{array} \right)
\left( \begin{array}{c}
\tilde{\tilde{V}}_{\mu} \\
\tilde{\tilde{Z}}_{\mu} \\
\tilde{\tilde{A}}_{\mu}
\end{array} \right).
\end{equation}

\subsection{Dark photon interaction with SM}

Now let's look at how the interaction terms between gauge fields and SM matter fields are modified.
\begin{align}
L_{int} &= e J_{em}^{\mu} A_{\mu} + \frac{g}{c_{w}} J_{z}^{\mu} Z_{\mu} \\
&= e J_{em}^{\mu} \left( \epsilon \tilde{\tilde{V}}_{\mu} + \tilde{\tilde{A}}_{\mu} \right) + \frac{g}{c_{w}} J_{z}^{\mu} \left( \frac{\delta^{2}}{1-\delta^{2}} t_{w} \epsilon \tilde{\tilde{V}}_{\mu} + \tilde{\tilde{Z}}_{\mu} \right) + O(\epsilon^{2}) \\
&= e J_{em}^{\mu} \tilde{\tilde{A}}_{\mu} + \frac{g}{c_{w}} J_{z}^{\mu} \tilde{\tilde{Z}}_{\mu} 
+ \epsilon \left( e J_{em}^{\mu} + g \frac{s_{w}}{c_{w}^{2}} \frac{\delta^{2}}{1-\delta^{2}} J_{z}^{\mu} \right) \tilde{\tilde{V}}_{\mu} + O(\epsilon^{2})
\end{align}
We can express $J_{Z}^{\mu}$ as $J_{Z}^{\mu} = c_{w}^{2} J_{em} - J_{Y}$ and rewrite the interaction terms between the Dark photon $\tilde{\tilde{V}}_{\mu}$ and the SM matter fields as
\begin{equation}
\epsilon \left[ \left( e + g s_{w} \frac{\delta^{2}}{1-\delta^{2}} \right) J_{em}^{\mu} - g \frac{s_{w}}{c_{w}^{2}} \frac{\delta^{2}}{1-\delta^{2}} J_{Y} \right ] \tilde{\tilde{V}}_{\mu} + O(\epsilon^{2}).
\end{equation}
Using $e=\frac{gg'}{\sqrt{g^{2}+g^{'2}}}$, $c_{w} = \frac{g}{\sqrt{g^{2}+g^{'2}}}$ and $s_{w} = \frac{g'}{\sqrt{g^{2}+g^{'2}}}$, we finally get
\begin{equation}
\epsilon \left[ e \frac{1}{1-\delta^{2}} J_{em}^{\mu} - \frac{e}{c_{w}^{2}} \frac{\delta^{2}}{1-\delta^{2}} J_{Y} \right ] \tilde{\tilde{V}}_{\mu} + O(\epsilon^{2}).
\end{equation}
This can be rewritten as 
\begin{equation}
\sum_{f} g_{f} \tilde{\tilde{V}}_{\mu} \bar{f} \gamma^{\mu} f , \qquad g_{f} = \epsilon e \left[ Q_{f} \frac{1}{1-\delta^{2}}  - \frac{Y_{f}}{c_{w}^{2}} \frac{\delta^{2}}{1-\delta^{2}}  \right ] + O(\epsilon^{2}).
\label{V_D_ff_lag}
\end{equation}
The mixing term between $U(1)_{Y}$ and $U(1)_{D}$ in the Lagrangian in Eq.~\eqref{lagrangian_U(1)_D_model} couples the Dark photon to SM fermions.

\chapterimage{decay_DP} 

\section{Dark Photon Decay Widths}
\label{app:DP}

We have computed the decay width  of the dark photon $V$ into pairs of SM fermions, at tree-level, in the small $\epsilon$ limit. The final expression reads
\begin{equation}
\Gamma(V \rightarrow  \bar{f}f) = \frac{N_{f}}{24\pi} \mV \sqrt{1-4\delta_{f}^{-2}} \left[ g_{f_{L}}^{2}+g_{f_{R}}^{2} - \delta_{f}^{-2} (g_{f_{L}}^{2} + g_{f_{R}}^{2} - 6g_{f_{L}}g_{f_{R}}) \right]
\label{eq:VD_decay_rate_ff}
\end{equation}
with 
\begin{align}
&g_{f} = \epsilon e \left[ Q_{f} \frac{1}{1-\delta_{Z}^{2}}  - \frac{Y_{f}}{c_{w}^{2}} \frac{\delta_{Z}^{2}}{1-\delta_{Z}^{2}}  \right ] + O(\epsilon^{2}), \\
&\delta_{f} = \mV/m_{f},  \quad \delta_{Z}=\mV/m_{Z}, \\
&e=\frac{g_{2}g_y}{\sqrt{g_{2}^{2}+g_y^{2}}}, \quad c_{w} = \frac{g_{2}}{\sqrt{g_{2}^{2}+g_y^{2}}}.
\end{align}
$N_{f}$ stands for the color number and is equal to $3$ for  quarks and $1$ for leptons.
$Q_{f}$, $Y_{f}$, $g_{2}$ and $g_y$ are respectively the electric charge of the fermion $f$ in unit of $e$, the hypercharge of $f$, the gauge coupling constant of $SU(2)_{L}$ and that of $U(1)_{Y}$.
We have also computed the decay widths of the dark photon, at tree-level, into $Zh$
\begin{equation}
\Gamma_{V\rightarrow Zh} = \epsilon^{2} \left(\frac{g_y}{g_{2}} \right)^{2} \frac{g_y^{2}+g_{2}^{2}}{192\pi} \mV \left( \frac{\delta_{Z}^{2}}{1-\delta_{Z}^{2}} \right)^{2} \lambda_{Zh,V}^{3/2} \left( 1 + 12 \lambda_{Zh,V}^{-1} \delta_{Z}^{-2} \right)
\label{eq:VD_decay_rate_Zh}
\end{equation}
where the 2-body phase space function $\lambda_{Zh,V}$ is defined as
\begin{equation}
\lambda_{Zh,V} = \left(1 - \delta_{Z}^{-2} - \delta_{h}^{-2} \right)^{2} - 4 \left(\delta_{Z}\delta_{h}\right)^{-2}
\end{equation}
with  $\delta_{h}=\mV/m_{h}$, and into $WW$
\begin{equation}
\Gamma_{V\rightarrow WW} = \epsilon^{2} \left(\frac{g_y}{g_{2}} \right)^{2} \frac{g_y^{2}+g_{2}^{2}}{192\pi} \mV \left( \frac{\delta_{Z}^{2}}{1-\delta_{Z}^{2}} \right)^{2}  \left( 1 - 4\delta_{W}^{-2} \right)^{3/2} \left( 1 + 20\delta_{W}^{-2} + 12 \delta_{W}^{-4} \right).
\label{eq:VD_decay_rate_WW}
\end{equation}
with $\delta_{W} = \mV/m_{W}$.
We have verified that these expressions agree both with the Goldstone equivalence theorem \cite{Chanowitz:1985hj}
\begin{align}
&\lim_{m_{V_{D}} \gg m_{Z},m_{h}} \Gamma_{V_{D}\rightarrow  Zh} =  \Gamma_{V_{D}\rightarrow  \pi^{3} h} \\
&\lim_{m_{V_{D}} \gg m_{W}} \Gamma_{V_{D}\rightarrow  W^{-}W^{+}} =  \Gamma_{V_{D}\rightarrow  \pi^{-}\pi^{+}}.
\end{align}
and with the existing literature, e.g. \cite{Bai:2016vca} and \cite{Ekstedt:2016wyi}.

In our calculations, we use the expressions above in the region $\mV < 350$~MeV and $\mV > 2.5$~GeV. For  350~MeV$ < \mV < 2.5$~GeV we use the perturbative tree-level expression in eq.~(\ref{eq:VD_decay_rate_ff}) for the width into leptons, and for the width into hadrons we instead use measurements of $R(s) = \sigma(e^+e^- \to {\rm hadrons})/\sigma(e^+e^- \to \mu^+\mu^-)$, that we extract from~\cite{Curtin:2014cca}, and where $\sqrt{s}$ is the energy of the collision in the center of mass frame.
These scattering processes are dominated by a virtual photon exchange in $s$-channel, therefore the quantum numbers of the final state coincide with those of the dark photon, making their use justified for our purpose. Therefore we write
\beq
\Gamma(V \to {\rm hadrons}) = R(s = \mV^2)\times \Gamma(V \to \mu^+\mu^-),
\label{eq:Width_hadrons}
\eeq
where we take $ \Gamma(V \to \mu^+\mu^-)$ from eq.~(\ref{eq:VD_decay_rate_ff}).
The hadronic decay chains in $e^+ e^- \to$~hadrons end dominantly in charged pions~\cite{Whalley:2003qr,Buschmann:2015awa}, that have BR$(\pi^\pm \to \mu\nu_\mu)>99.9\%$. Therefore, following~\cite{Cirelli:2016rnw}, we assume for simplicity that the $V$ final states, for 350 GeV~$< \mV < 2.5$~GeV, consist $50~\%$ of $\mu\bar{\mu}$ and $50~\%$ of $\nu\bar{\nu}$.

\chapterimage{Fermi_LAT_sky_map-s} 

\section{Gamma ray from DM annihilation}
\label{app:Gamma-ray from DM annihilation}

Let's compute the contribution to the detected gamma spectrum due to the Dark Matter annihilation in the Milky way. The number of DM pairs in the volume $d^3r$ at position $\vec{r}$ in the MW is 
\begin{equation}
d^{3}r \frac{1}{2} \left( \frac{\rho_{DM}(\vec{r})}{\MDM} \right)^{2}
\end{equation}
where $\rho_{DM}(\vec{r})$ is the DM energy density at $\vec{r}$. We have assumed self-conjugate DM. In the opposit case where particles can only interact with anti-particles, the $1/2$ must be replaced by $1/4$.

\begin{figure}[h!]
\centering
\begin{tikzpicture}[line width=1.5 pt, scale=2]
%

\begin{feynman}
		\draw[fermion](145:.3cm) -- (145:1);
			\node at (145:1.15) {$\overline{DM}$ };
		\draw[fermion](215:1) -- (215:.3cm);
			\node at (215:1.2) {$DM$ };
		\draw[fermion](35:.3cm) -- (35:1);
			\node at (35:1.15) {$f$};
		\draw[fermion](-35:1) -- (-35:.3cm);
			\node at (-35:1.2) {$\bar{f}$};
		\draw[fill=black] (0,0) circle (.3cm);
		\draw[fill=white] (0,0) circle (.29cm);
		\begin{scope}
	    	\clip (0,0) circle (.3cm);
	    	\foreach \x in {-.9,-.8,...,.3}
				\draw[line width=1 pt] (\x,-.3) -- (\x+.6,.3);
	  	\end{scope}
\end{feynman}
\end{tikzpicture}	
\caption{\label{DM_annihilation_into_SM_f} DM annihilation into SM}
\end{figure}

If we suppose that a pair of DM annihilates into a pair of SM particles $f$ as shown in Fig.~\ref{DM_annihilation_into_SM_f} with a cross-section $\left< \sigma v \right>_{f}$, then multiplying the last expression by $\left< \sigma v \right>_{f}$ gives the number of DM annihilations into $f$ per second 
\begin{equation}
d^{3}r \frac{1}{2} \left( \frac{\rho_{DM}(\vec{r})}{\MDM} \right)^{2} \sum_{f} \left< \sigma v \right>_{f}.
\end{equation}
Now multiplying by the number of photons produced per second, in energy bin $\Delta E$ centered on $E$, per DM annihilation into SM particles $f$,  
\begin{equation}
\Delta E \frac{dN_{f\rightarrow\gamma}}{dE},
\end{equation}
we get the number of photons produced in volume $d^3r$, at position $\vec{r}$, in energy bin $\Delta E$ centered on $E$
\begin{equation}
d^{3}r \frac{1}{2} \left( \frac{\rho_{DM}(\vec{r})}{\MDM} \right)^{2} \sum_{f} \left< \sigma v \right>_{f} \Delta E \frac{dN_{f\rightarrow\gamma}}{dE}.
\end{equation}

\begin{figure}[htp]
\centering
\resizebox{0.8\textwidth}{!}{\includegraphics[width=1\textwidth]{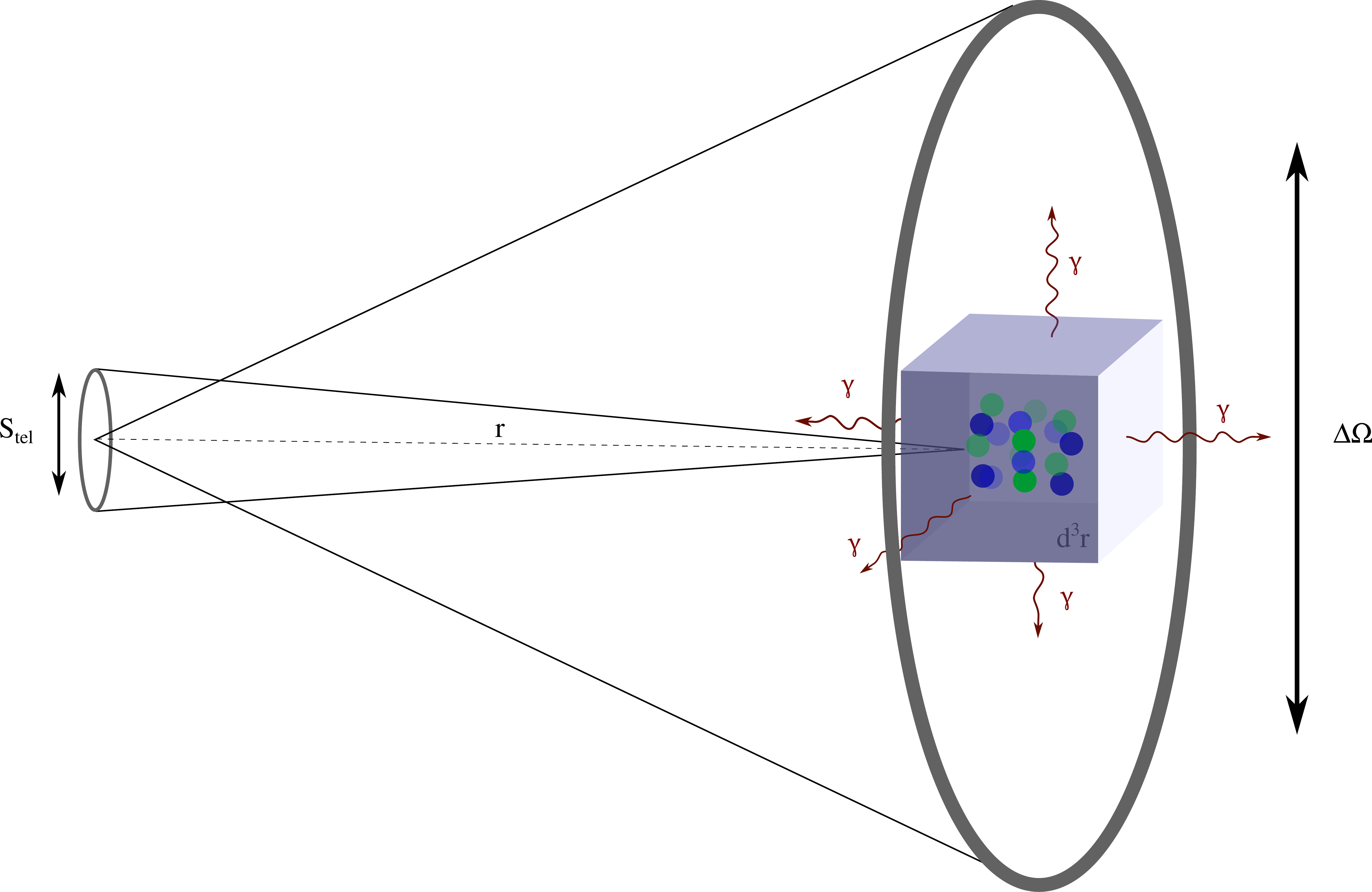}}
\caption{\it \small The detected $\gamma$-ray flux is the fraction $\dfrac{\Delta\Omega}{4\pi}\dfrac{S_{\rm tel}}{4\pi r^2}$ of the total emission.}
\label{gamma_fluxes_angle_solid} 
\end{figure}

However, as described in Fig.~\ref{gamma_fluxes_angle_solid} a telescope with a surface $S_{\rm tel}$, receiving signals within a solid angle $\Delta\Omega$ and within an energy bin $\Delta E$ detects only a fraction 
\begin{equation}
\frac{\Delta\Omega}{4\pi}\frac{S_{\rm tel}}{4\pi r^2}
\end{equation}
of the total number of photons produced. So this telescope detects a number of photons per second which is
\begin{equation}
S_{tel} \Delta \Omega \Delta E \frac{d\phi_{\gamma}}{d\Omega dE} =  \int d^{3}r \frac{1}{2} \left( \frac{\rho_{DM}(\vec{r})}{\MDM} \right)^{2} \sum_{f} \left< \sigma v \right>_{f} \Delta E \frac{dN_{f\rightarrow\gamma}}{dE}  \frac{\Delta\Omega}{4\pi}\frac{S_{\rm tel}}{4\pi r^2} .
\end{equation}
where $\phi_{\gamma}$ is the integrated $\gamma$ flux in cm$^{2}$/s. \\
Finally the detected gamma flux per solid angle unit and per energy unit, assuming self-conjugate DM, is 
\begin{equation}
\label{detected_gamma_spectrum_1}
\frac{d\phi_{\gamma}}{d\Omega dE} =  \frac{1}{2}  \frac{r_{\odot}}{4\pi} \left( \frac{\rho_{\odot}}{\MDM} \right)^{2}  J_{\rm ann} \sum_{f} \left< \sigma v \right>_{f} \frac{dN_{f\rightarrow\gamma}}{dE}
\end{equation}
where
\begin{equation}
J_{\rm ann} = \int \frac{ds}{r_{\odot}} \left( \frac{\rho_{DM}(\vec{r})}{\rho_{\odot}} \right)^{2}
\end{equation}
is the so-called $J_{\rm ann}$ factor for annihilation, $r_{\odot} = 8.33$~kpc is the Earth distance to the GC and $\rho_{\odot} = 0.3$~GeV/cm$^3$ is the supposed DM energy density at the Earth position. \\
A similar treatment for DM decay leads to
\begin{equation}
\label{detected_gamma_spectrum_1}
\frac{d\phi_{\gamma}}{d\Omega dE} =   \frac{r_{\odot}}{4\pi} \left( \frac{\rho_{\odot}}{\MDM} \right)  J_{\rm dec} \sum_{f} \Gamma_{f} \frac{dN_{f\rightarrow\gamma}}{dE}
\end{equation}
where
\begin{equation}
J_{\rm dec} = \int \frac{ds}{r_{\odot}} \left( \frac{\rho_{DM}(\vec{r})}{\rho_{\odot}} \right).
\end{equation}

\end{subappendices}

%

\xintifboolexpr { \x = 2}
  {
  }
{
\medskip
\small
\bibliographystyle{JHEP}
\bibliography{thesis.bib}
}

%% file: chap6.tex
\chapterimage{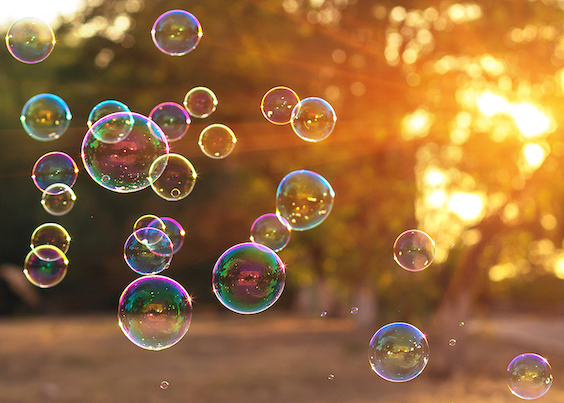} 

\chapter{First-order Cosmological Phase Transition}
\label{chap:1stOPT}

As first observed by Linde and his supervisor in 1972 \cite{Kirzhnits:1972ut}, at high temperature in the Early universe, the Higgs field had a thermal mass and the electroweak (EW) symmetry was effective (we say ``restored'').  The electromagnetism and the weak interaction constituted a single electroweak interaction, and electroweak and Higgs bosons were continuously created and absorbed in the plasma. This raises the question on how the Higgs field acquired the non-zero vacuum expectation value $\left<H^\dagger H \right> \neq 0$ responsible for the masses of the fermions and electroweak bosons which we observe today ? (Sec~\ref{sec:EWSB} in Chap.~\ref{chap:SM_particle}) The nature of the electroweak phase transition (EWPT) was first investigated using lattice simulations at the end of the 90s \cite{Kajantie:1996mn, Rummukainen:1998as, Csikor:1998eu}. The authors showed that in the SM the transition is a cross-over, meaning that the order parameter $\left<H^\dagger H \right>$ varies smoothly with the temperature\footnote{Technically speaking, there is no gauge-invariant order parameter which can distinguish the symmetric high temperature phase and the low temperature Higgs phase \cite{Kajantie:1996mn, Banks:1979fi,Fradkin:1978dv}.  Except along the 1stOPT line in the phase diagram, there is no breaking or restoration of the gauge symmetry across the phase transition. For instance, the gauge boson masses themselves receive finite-temperature corrections and varies smoothly with T.}, if $m_{\rm Higgs} \gtrsim 72$~GeV. In contrast, the scenario where $\left<H^\dagger H \right>$ would manifest a jump is called a first-order phase transition (1stOPT). Also, the chiral phase transition of QCD is known to be a cross-over in the SM, except if, at the time of the phase transtion, the baryon chemical potential is large \cite{Stephanov:2007fk} or if there are more than $N_f \gtrsim 3$ massless quarks, as shown by Pisarski and Wilczek in 1983 \cite{Pisarski:1983ms}.

1stOPT have been the subject of a lot of interest because they can  explain the baryon asymmetry (Sec.~\ref{sec:Matter-anti-matter-asymmetry}), they can generate GW which are potentially detectable by future interferometers like LISA, see \cite{Witten:1984rs, Hogan:1986qda,Kamionkowski:1993fg} for pionnering works and \cite{Caprini:2015zlo, Caprini:2018mtu, Caprini:2019egz} for reviews, they can generate magnetic fields \cite{Vachaspati:1991nm,Zhang:2019vsb,Ellis:2019tjf}, topological defects \cite{Kibble:1976sj, Kibble:1995aa, Borrill:1995gu}, primordial black holes~\cite{Kodama:1982sf,Hawking:1982ga,Jung:2021mku, Garriga:2015fdk,Deng:2016vzb,Deng:2017uwc,Maeso:2021xvl, Gross:2021qgx,Baker:2021nyl,Baker:2021sno,Kawana:2021tde, Liu:2021svg,Hashino:2021qoq},  they can set the abundance of dark matter~\cite{Konstandin:2011dr,Falkowski:2012fb,Hambye:2013dgv,Baker:2016xzo,Baker:2017zwx,Baker:2018vos,Hambye:2018qjv,Bai:2018dxf,Heurtier:2019beu,Baker:2019ndr,Azatov:2021ifm,Baldes:2020kam,Hong:2020est,Asadi:2021yml,Asadi:2021pwo,Baldes:2021aph}, and offer a new access to the scale of supersymmetry breaking~\cite{Craig:2020jfv}. 
We refer to \cite{Mazumdar:2018dfl,Hindmarsh:2020hop} for other reviews on 1stOPT.

It is rather straightforward to make the EWPT first-order by adding new physics in the scalar sector. Specific realizations include: $|H|^6$ operator \cite{Grojean:2004xa,Bodeker:2004ws,Delaunay:2007wb, Huang:2015izx,Huang:2016odd, deVries:2017ncy, Chala:2018ari, Ellis:2018mja, Ellis:2019oqb}, extra singlet models \cite{Anderson:1991zb, Choi:1993cv, Espinosa:1993bs, Profumo:2007wc, Espinosa:2011ax, Ellis:2018mja, Beniwal:2018hyi}, two-Higgs double models \cite{Branco:2011iw, Basler:2016obg}, models with spontaneous breaking of conformality due to 1-loop quantum correction \cite{Meissner:2006zh, Iso:2009ss, Iso:2009nw, Iso:2017uuu, Hambye:2018qjv, Baldes:2018emh, Marzo:2018nov, Ellis:2019oqb, Brdar:2019qut}, models with warped extra-dimensions \cite{Creminelli:2001th, Randall:2006py, Nardini:2007me, Hassanain:2007js, Konstandin:2010cd, Konstandin:2011dr, Bunk:2017fic, Dillon:2017ctw, Megias:2018sxv, vonHarling:2017yew}, or strong dynamics \cite{Bruggisser:2018mus, Bruggisser:2018mrt, Baratella:2018pxi, Agashe:2019lhy,DelleRose:2019pgi,vonHarling:2019gme,Baldes:2021aph}, the last two being related by holography. Generically, we can say that models with polynomial potentials typically lead to weakly first-order phase transitions while models with nearly-conformal potentials can naturally induce strongly first-order phase transitions with large amount of supercooling. We mention \cite{Caprini:2019egz} and \cite{Schmitz:2020rag} for reviews on SM extensions leading to strong 1stOPT.

The present chapter is a review on 1stOPT and not my research. I believe it might be useful for students and researchers desiring learning about this topic.
In Sec.~\ref{sec:bubble_nucleation}, we show how to compute the effective potential in a given model, and how to compute the tunneling temperature. In Sec.~\ref{sec:bubble_propagation}, we present different methods to compute the velocity of the bubble wall. In Sec.~\ref{sec:GW_generation}, we present the GW spectrum resulting from bubble collisions, sound-waves and turbulence. Finally, in Sec.~\ref{sec:supercool_potential} we focus on the class of models giving the largest GW signal:  models with nearly-conformal potential, which are responsible for supercooled 1stOPT.

All the figures in the chapter are mine (Fig.~\ref{fig:Veff_dim6}, \ref{fig:O3_different_methods_comparison}, \ref{fig:thick_vs_thin_wall}, \ref{fig:Dim6_Tn_Lambda}, \ref{fig:DeltaP_VS_gamma_T_vs_R}, \ref{eq:DeltaV_alpha_contours_v_LO}, \ref{fig:OmegaGW_scalarField_Litterature}, \ref{fig:SC_PT_constraints}, \ref{fig:detonation_deflagration_profile}, \ref{fig:efficiency_sw_RMS_fluid_velocity_vs_vw}, \ref{fig:SW_PT_constraints}, \ref{fig:CW_gX_Tnuc_alpha_beta} and \ref{fig:light_dilaton_msigma_Tnuc_alpha_beta}).

\section{Bubble nucleation}
\label{sec:bubble_nucleation}
\subsection{Effective potential at finite temperature}
\label{sec:effective_potential}

\paragraph{Tree level: }
We start with the following classical effective potential for the SM Higgs
\begin{equation}
V(H) = m^2  \left| H \right|^2 + \lambda \left|  H  \right|^4 + \Lambda^{-2} \left| H \right|^6,
\end{equation}
with $H^T = (\chi_1 + i \chi_2, \, \varphi + i \chi_3)/\sqrt{2} $. It is the SM Lagrangian, defined in Sec~\ref{sec:EWSB} of Chap.~\ref{chap:SM_particle}, augmented with the non-renormalizable operator $\left| H \right|^6$ \cite{Grojean:2004xa,Bodeker:2004ws,Delaunay:2007wb, Huang:2015izx,Huang:2016odd, deVries:2017ncy, Chala:2018ari, Ellis:2018mja, Ellis:2019oqb}.  The energy scale $\Lambda$ sets the range of validity of the model. After expanding the Higgs around its VEV $\varphi = \phi + h$,  we obtain
\begin{equation}
V_{\rm tree}(\phi) = \frac{m^2}{2}\phi^2 + \frac{\lambda}{4}\phi^4 + \frac{1}{8}\frac{\phi^6}{\Lambda^{2}}.
\end{equation}

\paragraph{One-loop finite-temperature effective potential: }
The quantum corrections at one-loop were first computed in 1973 by Coleman and E.J. Weinberg at zero temperature  \cite{Coleman:1973jx} and a few months later by Dolan and Jackiw at finite-temperature \cite{Dolan:1973qd}. We call attention to the reviews \cite{Quiros:1999jp, Delaunay:2007wb} and textbooks \cite{Kapusta:2006pm,Bellac:2011kqa} for more details. The 1-loop corrections read
\begin{equation}
V_{\rm eff}(\phi, \, T) = V_{\rm tree}(\phi) + \Delta V_1(\phi, \, T),
\end{equation}
with
\begin{equation}
\Delta V_1(\phi, \, T) = \sum_{i = \{h,\chi,W,Z,t\}} \frac{n_i\,T}{2} \sum_{n=-\infty}^{+\infty} \int \frac{d^3 \vec{k}}{(2\pi)^3} {\rm log} \left[ \vec{k}^2 + \omega_n^2  + m_i^2(\phi)   \right],
\label{eq:one-loop-eff-pot}
\end{equation}
where $k_E = (\omega_n,\, \vec{k})$ is the euclidean loop 4-momentum, $\omega_n$ are the Matsubara frequencies in the imaginary time formalism, $\omega_n = 2 n \pi T$ for bosons (periodic on the euclidean time circle) and $\omega_n = (2 n+1) \pi T$ for fermions (anti-periodic on the euclidean time circle). We assumed the Landau gauge ($\xi = \infty$ in $R_\xi$ gauge\footnote{The gauge-independence of the energy value at the extrema of the effective potential is guaranteed by the so-called Nielsen identity (1975 \cite{Nielsen:1975fs,Fukuda:1975di})
\begin{equation}
\frac{\partial V_{\rm eff}(\phi, \, \xi)}{\partial \xi} = - C(\phi, \, \xi) \frac{\partial V_{\rm eff}(\phi, \, \xi)}{\partial \phi}. 
\end{equation} 
It says that the gauge dependence of the effective action is equivalent to a nonlocal field redefinition.
As a consequence, the position of the minima are gauge-dependent, e.g. \cite{DiLuzio:2014bua, Andreassen:2014gha}, but the nucleation rate as well as the energy profile of the nucleated bubbles are gauge-independent, e.g. \cite{Metaxas:1995ab, Plascencia:2015pga, Espinosa:2016nld}. I thank Thomas Konstandin for related discussions.}) where the ghosts decouple and the Goldstones are included.\footnote{Note that away from the zero-temperature minimum, longitudinal modes of gauge bosons and Goldstones are independent of each other.} The number of degrees of freedom is then $n_{\{h, \chi, W, Z, t\}} = \{1,3,6,3,-12\}$. The masses $m_i^2(\phi) = \partial^2 V(\phi)/ \partial \chi_i^2$ read
\begin{align}
&m_h^2(\phi) = m^2 + 3 \lambda \phi^2 + \frac{15}{4} \frac{\phi^4}{\Lambda^2}, \\
&m_\chi^2(\phi) =  m^2 +  \lambda \phi^2 + \frac{3}{4} \frac{\phi^4}{\Lambda^2},  \\
&m_W^2(\phi) = \frac{g_2^2}{4}\phi^2,\quad m_Z^2(\phi) = \frac{g_1^2 + g_2^2}{4} \phi^2, \quad m_t^2(\phi) = \frac{y_t^2}{2}\phi^2.
\end{align}
We can decompose the one-loop correction as a sum of a zero-temperature term and a T-dependent one which vanishes when $T\to 0$ 
\begin{align}
&\Delta V_1^0(\phi) = \sum_{i = \{h,\chi,W,Z,t\}} \frac{n_i}{2} \int \frac{d^4 k_E}{(2\pi)^4}\, {\rm log} \left[ k_E^2 + m_i^2(\phi) \right], \label{eq:CW_pot_before_int}\\
&\Delta V_1^T(\phi, \, T) =  \sum_{i = \{h,\chi,W,Z,t\}} \frac{n_i \, T^4}{2 \pi^2} \int_0^\infty dk \,k^2\,  {\rm log} \left[ 1 \mp {\rm exp} ~{-\sqrt{k^2 + m_i^2(\phi)/T^2}} \right] .\label{eq:finite_temp_before_fct_J}
\end{align}

\paragraph{Zero-temperature corrections: }

Upon imposing a hard cut-off $k_E < \Lambda_{\rm UV}$ and integrating over $k_E$, Eq.~\eqref{eq:CW_pot_before_int} gives the so-called Coleman-Weinberg potential
\begin{equation}
\Delta V_1^0(\phi) =  \sum_{i = h,\chi,W,Z,t} n_i \frac{m_i^4(\phi)}{64\pi^2} \left[ {\rm log} ~\frac{m_i^2(\phi)}{\Lambda_{\rm UV}^2}  + \Lambda_{\rm UV}^2 m_i^2(\phi) \right],
\end{equation}
The UV cut-off $\Lambda_{\rm UV}$ disappears after renormalization. For doing this, we introduce counterterms
\begin{equation}
V_{\rm ct} = \delta V_0 + \frac{\delta m^2}{2}\phi^2 + \frac{\delta \lambda}{4} \phi^4,
\end{equation}
and we impose the minimum of the potential to correspond to the Higgs VEV $v = 246~$GeV, and the second derivative to the Higgs mass $m_{\rm Higgs} = 125$~GeV
\begin{align}
&\left.\frac{d (\Delta V_1^0 + V_{ct}) }{d \phi} \right|_{\phi=v} = 0, \\
&\left.\frac{d^2 (\Delta V_1^0 + V_{ct}) }{d \phi^2} \right|_{\phi=v} = m_{\rm Higgs}^2.
\end{align}
We obtain
\begin{equation}
\Delta V_1^0(\phi) =  \sum_{i = \{h,\chi,W,Z,t\}} n_i \,\frac{1}{64\pi^2} \left[ m_i^4(\phi) \left( {\rm log} ~\frac{m_i^2(\phi)}{m_i^2(v)} - \frac{3}{2}\right)  + 2 m_i^2(v) m_i^2(\phi) \right].
\end{equation}

\paragraph{Finite temperature corrections: }

The finite-temperature contributions in Eq.~\eqref{eq:finite_temp_before_fct_J} can be re-expressed in term of the $J_{b,f}$ functions
\begin{align}
\Delta V_1^T(\phi, \, T) &=  \sum_{i = \{h,\chi,W,Z,t\}} \frac{n_i \, T^4}{2 \pi^2} \int_0^\infty dk \, k^2 \, {\rm log} \left[ 1 \mp {\rm exp} ~{-\sqrt{k^2 + m_i^2(\phi)/T^2}} \right] \\
&\equiv  \sum_{\rm i = {bosons}} \frac{n_i T^4}{2\pi^2} J_b\left( \frac{m_i^2(\phi)}{T^2} \right)  +   \sum_{\rm i = {fermions}} \frac{n_i T^4}{2\pi^2} J_f\left( \frac{m_i^2(\phi)}{T^2} \right).
\label{eq:Jb_Jf}
\end{align}
It is sometimes convenient to Taylor-expand $J_{b,f}$ in the high-temperature limit
\begin{align}
&J_b(x) \underset{x \to 0}{=} -\frac{\pi^4}{45} + \frac{\pi^2}{12}x - \frac{\pi}{6} x^{3/2} - \frac{x^2}{32} {\rm log} \frac{x}{a_b} + \mathcal{O}\left( x^3 \log{\frac{x^{3/2}}{\rm cst.}} \right), \label{eq:HT_exp_jb}\\
&J_f(x) \underset{x \to 0}{=}  \frac{7\pi^4}{360} - \frac{\pi^2}{24}x - \frac{x^2}{32} \log{\frac{x}{a_f}}  + \mathcal{O}\left( x^3 \log{\frac{x^{3/2}}{\rm cst.}}\right),
\end{align}
with $\log{a_b} \simeq 5.4076$ and $\log{a_f} \simeq 2.6350$.
In order to speed-up the numerical computation, it may be useful to use the fitting formula \cite{Li:2014wia}\footnote{Note that the Eq.~(3.9) of \cite{Li:2014wia} has a typo.}
\begin{equation}
J_{B/F}(x) = t_{B(F)} f_{B(F)}^{\rm HT}(x) + (1- t_{B(F)}) f^{\rm LT}(x) 
\end{equation}
with
\begin{align}
&J_{B}^{\rm HT}(x) = -\frac{\pi^{4}}{45}+\frac{\pi^2}{12}x-\frac{\pi}{6}x^{3/2}-\frac{1}{32}x^2\left[\log{x}-\log{a_b}\right], \\
&J_{F}^{\rm HT}(x) = -\frac{7\pi^{4}}{360}+\frac{\pi^2}{24}x+\frac{1}{32}x^2\left[\log{x}-\log{a_f}\right],
\end{align}
\begin{equation}
J^{\rm LT}(x) = -\sqrt{\frac{\pi}{2}} x^{3/4} e^{-\sqrt{x}} \left( 1 + \frac{15}{8}x^{-1/2} + \frac{105}{128}x^{-1} \right),
\end{equation}
\begin{align}
&t_{B}(x)=e^{-(x/6.3)^4},\\
&t_{F}(x)=e^{-(x/3.25)^4}.
\end{align}

\paragraph{Daisy corrections: }

Perturbative expansion at finite-temperature suffers from infrared divergences coming from the so-called daisy diagrams. At order $\lambda^N$, they can be described as $N$ independent daughter loops (or petals) attached to $1$ mother loop. Their contributions to the effective potential read \cite{Carrington:1991hz}
\begin{equation}
\Delta V_{\rm daisy}^{N} = -\frac{1}{2} T\, \sum_n \, \frac{d^3\vec{k}}{(2\pi)^3}\,\left[-\frac{\Pi_1}{\omega_n^2+\vec{k}^2 + m^2}\right]^N,
\label{eq:daisy_petals}
\end{equation}
where $\Pi_1$ the 1-loop self-energy of petals and $\vec{k}$ is the $3$-momentum running in the mother loop. We can see that for massless particles $m=0$ and zero Matsubara modes $n=0$, each individual daisy diagram with $N\gtrsim 2$ is IR-divergent. In order to cure those divergences, we must resum all of them, which gives
\begin{equation}
\Delta V_{\rm daisy} =  \sum_{N=1}^{\infty} V_{\rm daisy}^{N} = \frac{1}{2} T\, \sum_n \, \frac{d^3\vec{k}}{(2\pi)^3}\, {\rm ln} \left[ 1 + \frac{\Pi_1}{\omega_n^2 + \vec{k}^2 + m^2}  \right].
\label{eq:daisy_before_int}
\end{equation} 
By comparing the last equation \eqref{eq:daisy_before_int} with Eq.~\eqref{eq:one-loop-eff-pot}, we see that the effect of resumming all the daisy diagrams is a temperature-dependent shift of the square mass by
\begin{equation}
m_b^2(\phi) \to m_b^2(\phi) + \Pi_b(T),
\end{equation}
where $\Pi_b(T)$ is the self-energy of the bosonic field $b$ in the IR limit, $\omega = \vec{p}=0$. Since fermions have no zero Matsubara modes, they are free from IR-divergences.  In the SM + $\left| H \right|^6$ theory, these shifts, also known as Debye masses, read \cite{Carrington:1991hz, Delaunay:2007wb} 
\begin{align}
&\Pi_{h,\chi_i}(T) = \frac{T^2}{4 v^2}\left( m_h^2 + 2m_W^2 + m_Z^2 + 2m_t^2  \right) - \frac{3}{4}T^2 \frac{v^2}{\Lambda^2}, \\
&\Pi_W(T) = \frac{22}{3} \frac{m_W^2}{v^2} T^2,
\end{align}
while the shifted masses $\Pi_{Z/\gamma}(T)$ of $Z$ and $\gamma$ are eigenvalues of the following mass matrix
\begin{equation}
\begin{pmatrix}
\frac{1}{4}g_2^2 \phi^2 + \frac{11}{6}g_2^2 T^2 & -\frac{1}{4} g_1^2 g_2^2 \phi^2 \\
-\frac{1}{4}g_1^2 g_2^2 \phi^2 & \frac{1}{4} g_1^2 \phi^2 + \frac{11}{6}g_1^2 T^2
\end{pmatrix}.
\end{equation}
Note that the daisy resummation being based on a high-T expansion, it should not be relevant at large $m_i/T$. So, in order to prevent the non-physical contributions from gauge boson at large $m_i/T$, it may be useful to introduce the cut-off \cite{Baldes:2018emh}
\begin{equation}.
\Pi_i(T) \rightarrow \Pi_i(T)~ C(m_i/T), \qquad \text{with} \quad C(x)=x^2/2~K_{2}(x),
\end{equation}
$K_{2}(x)$ being the modified Bessel function of the second kind.

Upon performing the integration in Eq.~\eqref{eq:daisy_before_int} while keeping only the dominant contribution from the zero Matsubara mode, and upon summing over the longitudinal modes of all the vector bosons in the theory, we obtain
\begin{equation}
\Delta V_{\rm daisy} = \sum_{i=\{h,\chi,W,Z,\gamma\}} \frac{\bar{n}_i T}{12\pi} \left[ m_i^3(\phi) - (m_i^2(\phi) + \Pi_i(T))^{3/2}  \right],
\end{equation} 
with $\bar{n}_{\{h,\chi,W,Z,\gamma\}} = \{1,3,2,1,1\}$. The non-integer power $3/2$ gives evidence of the breakdown of perturbation theory.

\begin{figure}[h]
\centering
\begin{adjustbox}{max width=1.2\linewidth,center}
\raisebox{0cm}{\makebox{\includegraphics[width=0.55\textwidth]{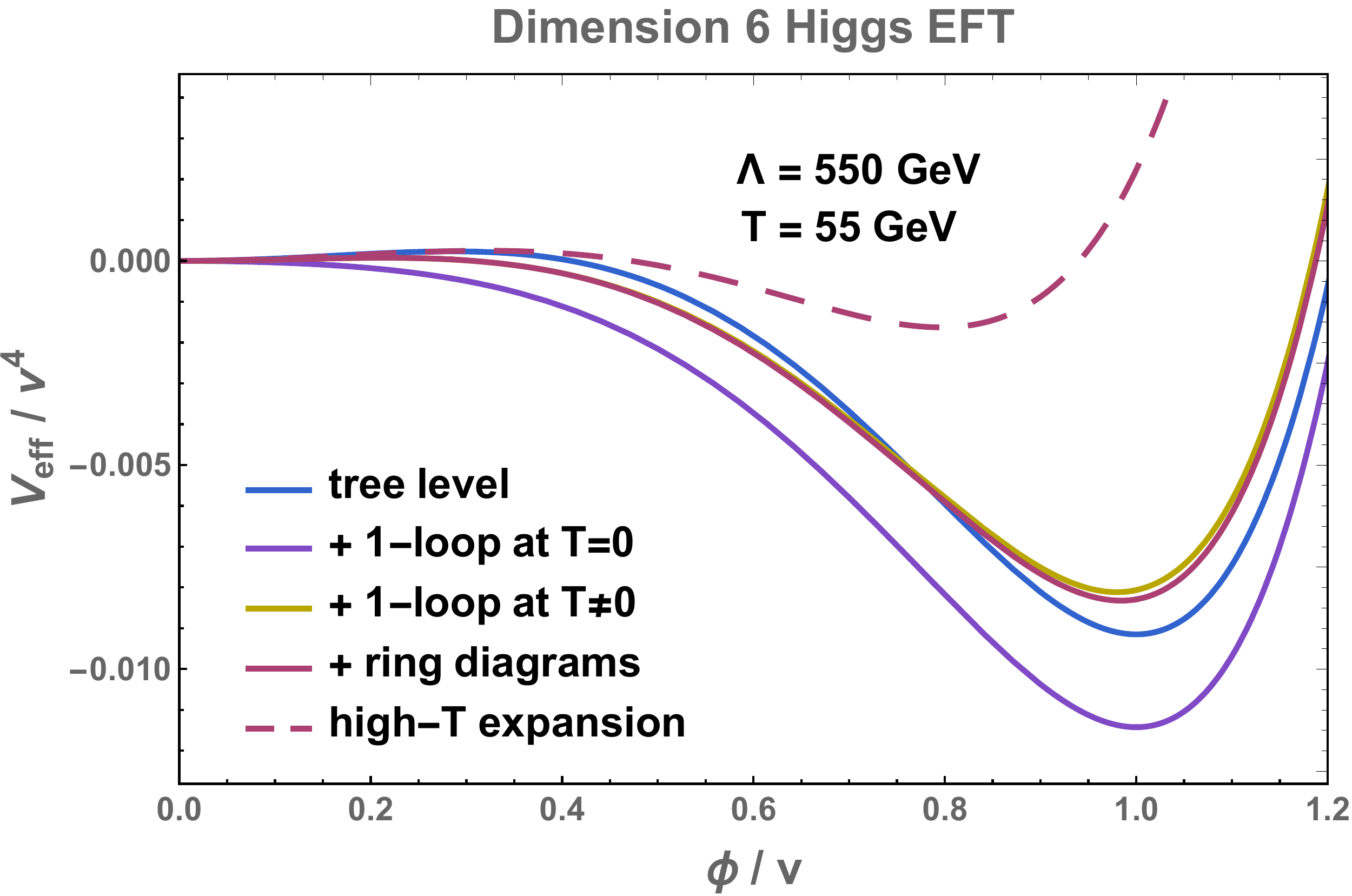}}}
\raisebox{0cm}{\makebox{\includegraphics[width=0.55\textwidth]{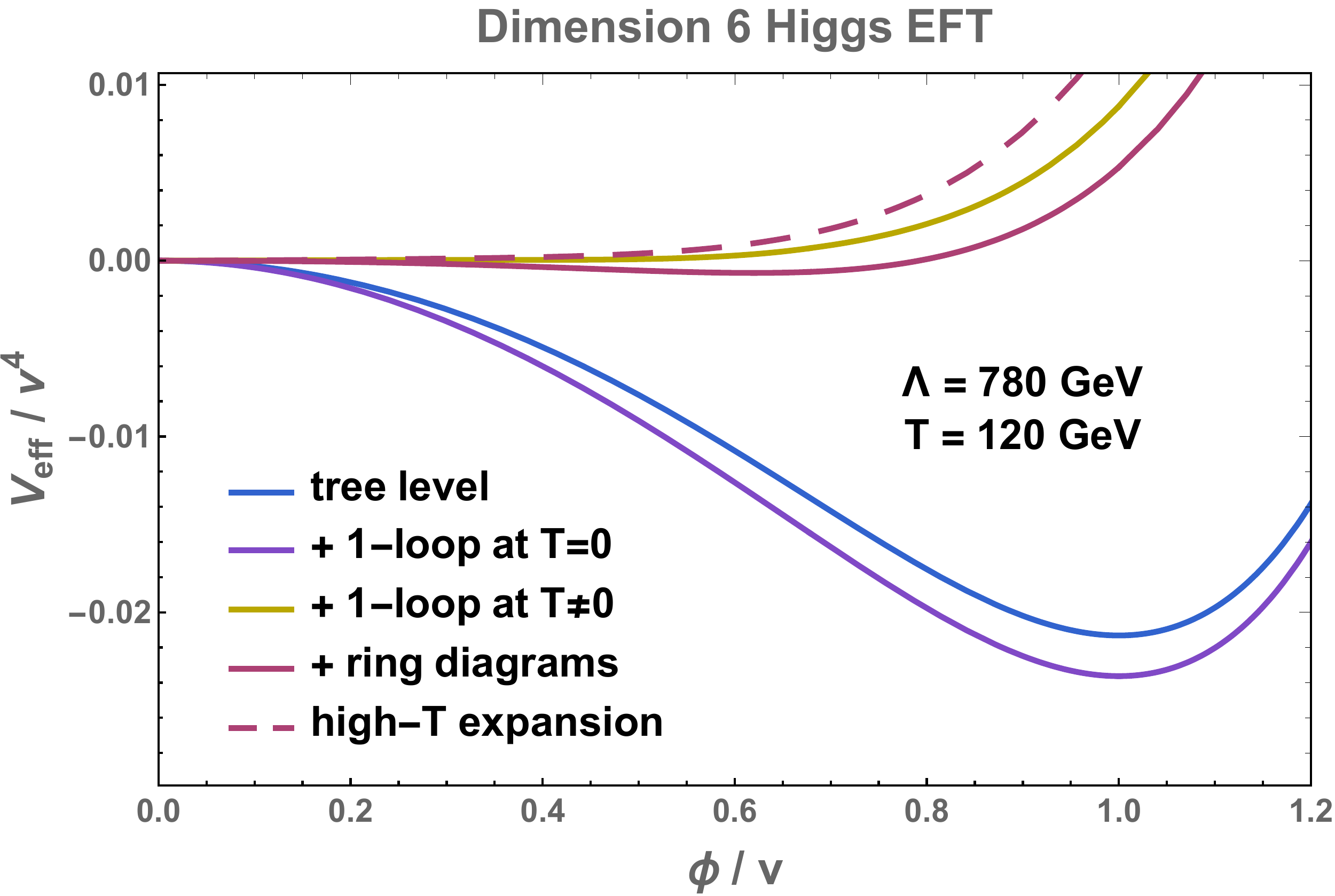}}}
\end{adjustbox}
\caption{\it \small  Effective potential in the Landau gauge at one-loop and finite-temperature of the SM + dimension 6 EFT, for two different choices of temperature and cut-off. }
\label{fig:Veff_dim6}
\end{figure}

\paragraph{Total: }

We conclude that the total effective potential, in the Landau gauge, at one-loop and finite-temperature reads
\begin{align}
V_{\rm eff}(\phi,\, T) =  &\frac{m^2}{2}\phi^2 + \frac{\lambda}{4}\phi^4 + \frac{1}{8}\frac{\phi^6}{\Lambda^{2}} \label{eq:effective_potential_total}\\
&+\sum_{i = \{h,\chi,W,Z,t\}} n_i \,\frac{1}{64\pi^2} \left[ m_i^4(\phi) \left( {\rm log} ~\frac{m_i^2(\phi)}{m_i^2(v)} - \frac{3}{2}\right)  + 2 m_i^2(v) m_i^2(\phi) \right]  \notag \\
&+  \sum_{\rm i = {bosons}} \frac{n_i T^4}{2\pi^2} J_b\left( \frac{m_i^2(\phi)}{T^2} \right)  +   \sum_{\rm i = {fermions}} \frac{n_i T^4}{2\pi^2} J_f\left( \frac{m_i^2(\phi)}{T^2} \right)  \notag \\
&+ \sum_{i=\{h,\chi,W,Z,\gamma\}} \frac{\bar{n}_i T}{12\pi} \left[ m_i^3(\phi) - (m_i^2(\phi) + \Pi_i(T))^{3/2}  \right] \notag ,
\end{align}
with $n_{\{h, \chi, W, Z, t\}} = \{1,3,6,3,-12\}$, $\bar{n}_{\{h,\chi,W,Z,\gamma\}} = \{1,3,2,1,1\}$ and
\begin{align}
m^2 = \frac{m_H^2}{2} - \frac{3}{4} \frac{v^4}{\Lambda^2}, \\
\lambda =  \frac{1}{2} \frac{m_H^2}{v^2} - \frac{3}{2} \frac{v^2}{\Lambda^2}.
\end{align}
We show the different contributions to the effective potential in Fig.~\ref{fig:Veff_dim6}.

Note that the high-temperature expansion of the effective potential reads \cite{Anderson:1991zb}
\begin{equation}
V_{\rm eff}(\phi,\, T) ~ \overset{ T \gg  v }{\longrightarrow} ~ D(T^2 - T_0^2)\phi^2 -ET\phi^3 + \frac{\lambda(T)}{4}\phi^4+ \frac{1}{8}\frac{\phi^6}{\Lambda^{2}},\label{eq:V_tot_SM_dim6_HT}
\end{equation}
with 
\begin{align}
D &= \frac{2m_W^2+m_Z^2 + 2m_t^2}{8v^2}, \\
E &= \frac{2m_W^3+m_Z^3}{4\pi v^3}, \\
T_0^2 &= \frac{m_H^2- 8 B v^3}{4D}, \\
B &= \frac{3}{64\pi^2 v^4} (2m_W^4 + m_Z^4 -4 m_t^4), \\
\lambda(T) &= \lambda - \frac{3}{16\pi^2 v^4} \left( 2m_W^4 ~\textrm{log} ~\frac{m_W^2}{A_B T^2} + m_Z^4 ~\textrm{log} ~\frac{m_Z^2}{A_B T^2} - 4 m_t^4 ~\textrm{log}~ \frac{m_t^2}{A_F T^2} \right)
\end{align}
where $v = 246$~GeV, $\textrm{log}~A_B = \textrm{log}~a_b - 3/2$ and $\textrm{log}~A_F = \textrm{log}~a_f - 3/2$.
As shown in Fig.~\ref{fig:Veff_dim6}, the high-temperature expansion of the effective potential is only a good approximation at small $\phi/T$.


\subsection{Tunneling rate}

As shown in Fig.~\ref{fig:Veff_dim6}, in the presence of finite-temperature corrections, the effective potential can manifest a barrier separating the false vacuum (in $\phi_{\rm FV} \approx 0$) and the true vacuum (in $\phi_{\rm TV}  \approx v $). Below the critical temperature $T_c$, the potential energy in the true minimum becomes lower than the potential energy in the false vacum, which therefore becomes \textbf{metastable}. The decay rate of the false vacuum was first computed by Coleman and Callan in 1977 at zero temperature \cite{Coleman:1977py, Callan:1977pt} and extended to finite-temperature by Linde in 1980 \cite{Linde:1980tt, Linde:1981zj}.\footnote{Gravitational corrections were computed for the first time in 1980 by Coleman and De Luccia \cite{Coleman:1980aw}.} We refer the reader to Coleman's lecture for more details \cite{Coleman:1985rnk}. Based on those results, the decay rate can be expressed as
\begin{equation}
\Gamma(T) \simeq {\rm max} \left[   T^4\left( \frac{S_3/T}{2\pi} \right)^{3/2} {\rm exp} \left(-S_3/T\right) , \quad R_0^{-4} \left( \frac{S_4}{2\pi} \right)^2 {\rm exp} \left(-S_4 \right)  \right],
\label{eq:tunneling_rate}
\end{equation}
where $S_3$ and $S_4$ are the $3$- and $4$-dimensional Euclidean action of the $O(3)$- and $O(4)$-symmetric tunneling solutions. The former is \textbf{thermally-induced} while the later is purely \textbf{quantum}. $R_0$ is the bubble radius. The tunneling rate via $O(3)$ bounce dominates for bubble radius larger than the imaginary time periodicity $R \gtrsim 1/T$, while the tunneling rate for $O(4)$ bounce dominates in the opposite case \cite{Linde:1980tt, Linde:1981zj}.

The $S_3$ and $S_4$ bounce actions read
\begin{align}
&\frac{S_{3}}{T}=4\pi\int dr~r^2~\left[\frac{1}{2}\phi^{'}(r)^2+V\left(\phi(r)\right)\right],\label{eq:O3_bounce_action} \\ &S_{4}=2\pi^2\int dr~r^3~\left[\frac{1}{2}\phi^{'}(r)^2+V\left(\phi(r)\right)\right],\label{eq:O4_bounce_action} 
\end{align}
where $V(\phi)$ is the effective potential discussed in Sec.~\eqref{sec:effective_potential}, and $\phi(r)$ is the field configuration which interpolates between the two asymptotic vacuum. Extremization of the action leads to the \textbf{Euclidean} equation of motion
\begin{equation}
\phi''(r) + \frac{d-1}{r}\phi'(r) = \frac{dV}{d\phi},
\label{eq:bounce_eomotion}
\end{equation}
with \textbf{boundary conditions}
\begin{equation}
\phi'(0)=0, \qquad \textrm{and} \qquad  \lim_{r \to \infty} \phi(r) = 0.
\label{eq:BC_bounce}
\end{equation}
$O(3)$ bounces have $d=3$ dimensions and are described by the coordinate $r = \vec{x}$, while $O(4)$ bounces have $d=4$ dimensions and are described by the coordinate $r= \sqrt{\vec{x}^2 + t^2}$.

It is worth mentionning the existence of a \textbf{stochastic approach} of quantum tunneling, introduced by Linde et al. in 1991  \cite{Ellis:1990bv, Linde:1991sk}. According to this picture, one treats the field $\phi$ in the false vacuum state as a \textbf{random gaussian variable}, which can move over the barrier to the true vacuum after developing large quantum fluctuations \cite{Ellis:1990bv, Linde:1991sk,Braden:2018tky,Hertzberg:2019wgx, Blanco-Pillado:2019xny}. This approach does not rely on classically forbidden tunneling path as in Eq.~\eqref{eq:bounce_eomotion}, but instead it relies on a \textbf{real-time description} which has the advantage to be extendable to time-dependent scenarios.

\paragraph{Numerical solutions:}
The bounce solution $\phi(r)$ is equivalent to the trajectory of a classical point particle in an inverted potential $-V(\phi)$.
Finding the bounce solution $\phi(r)$ can be realized numerically by guessing an initial field value $\phi(0)$ between the zero of the potential $\phi_{\rm zero}$, defined by $V(\phi_{\rm zero}) = V (\phi_{\rm FV})$, and the true vacuum $\phi_{\rm TV}$. If the field trajectory $\phi(r)$ \textbf{overshoots} the false vacuum $\phi_{\rm FV}$, we must decrease $\phi(0)$, while if it \textbf{undershoots} the false vacuum, we must increase $\phi(0)$.

Once the numerical bounce solution is obtained, we can check a posteriori that it corresponds to a stationary point of the action.  To do so, let us first define a one-parameter family of functions
\begin{equation}
\phi_\lambda(x) = \bar{\phi}(x/\lambda)
\end{equation}
where $\bar{\phi}(x)$ is the bounce solution. The corresponding $O(d)$ bounce action reads
\begin{align}
S_d(\lambda) &= \lambda^{d-2} \int d x^d \, \frac{1}{2} (\partial_\mu \bar{\phi})^2 + \lambda^d  \int d x^d \, V(\bar{\phi}) \\
&= \lambda^{d-2} K + \lambda^d U
\end{align}
Since $\bar{\phi}$ is solution of the equation of motion, $S_d(\lambda)$ must be stationary for $\lambda =1$, which implies \cite{Coleman:1985rnk}
\begin{equation}
K  = -\frac{d}{d-2} U.
 \label{eq:check_T_V_bounce}
\end{equation}
The last equality allows to quantify the degree of error of the numerical solution. For instance, we can define the \textbf{error parameter} $\epsilon$
\begin{equation}
\epsilon = \frac{(d-2) K  + d \, U}{(d-2) K  - d \, U}.
\end{equation}

Over the last decade, various numerical packages have been released to the community: CosmoTransitions \cite{Wainwright:2011kj}, AnyBubble \cite{Masoumi:2016wot}, BubbleProfiler \cite{Athron:2019nbd}, SimpleBounce \cite{Sato:2019wpo}, FindBounce \cite{Guada:2018jek, Guada:2020xnz}, as well as semi-analytical approaches \cite{Espinosa:2018hue, Espinosa:2018voj,Espinosa:2018szu}.

\begin{figure}[h!]
\centering
\raisebox{0cm}{\makebox{\includegraphics[width=0.9\textwidth, scale=1]{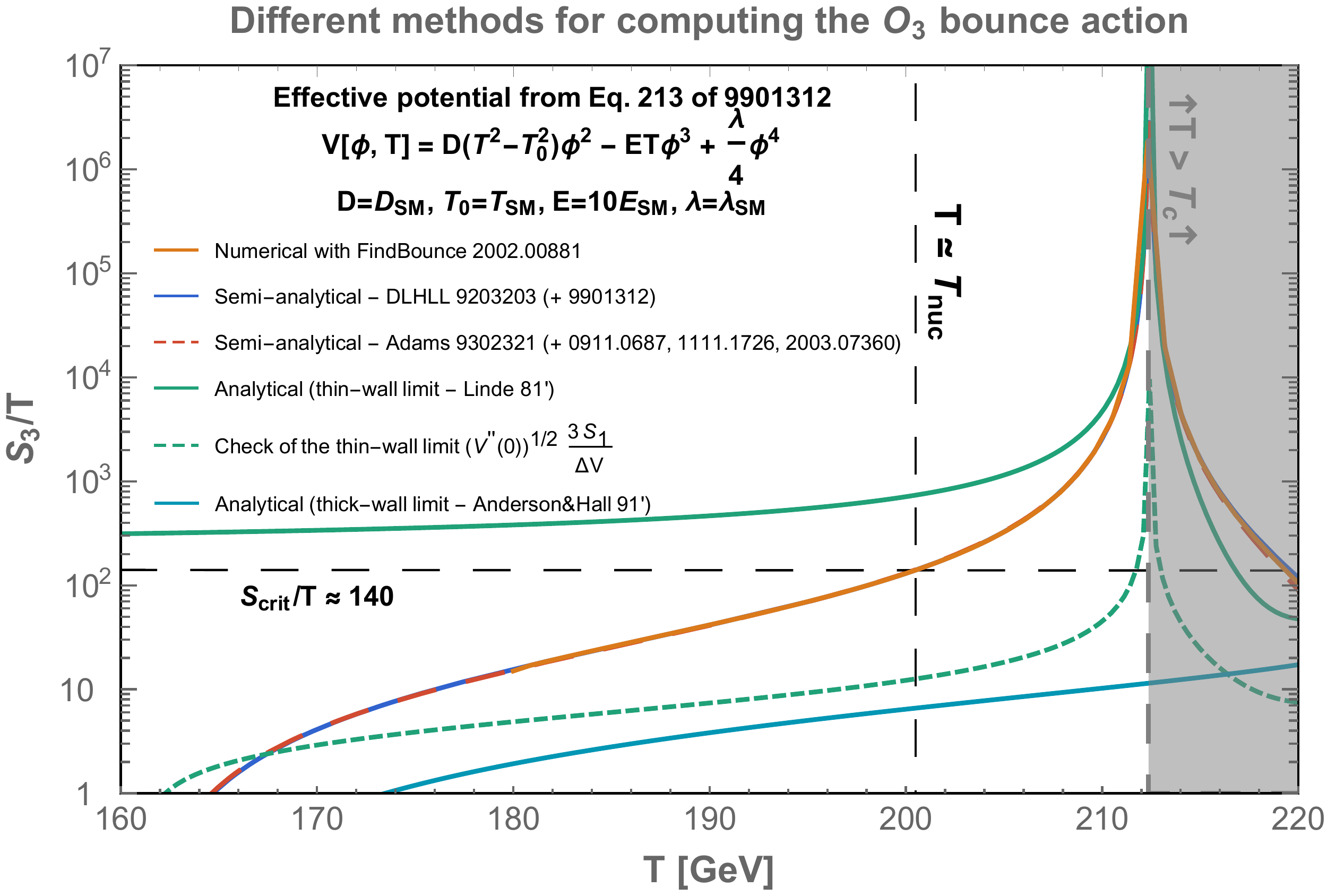}} }
\caption{\it \small We consider the high-temperature expansion of the effective potential of the SM EFT in Eq.~\eqref{eq:V_tot_SM_dim6_HT}, with $\Lambda \to 0$ and $E \to 10 E$. The semi-analytical solutions for the $O_3$-bounce action, in Eq.~\eqref{eq:S3_semi_analytical_Adams} and Eq.~\eqref{eq:S3_semi_analytical_DLHLL}, are very good approximations of the solution from the numerical package FindBounce \cite{Guada:2018jek, Guada:2020xnz}. On the other side, the thin-wall, cf. Eq.~\eqref{eq:S3_thin_before_R_inj}, and thick-wall limit, cf. Eq.~\eqref{eq:S_3_thick_wall}, are very bad approximation for the potential which we consider. More precisely, the thin-wall limit ceases to be valid when the quantity in Eq. 3.7 of \cite{Linde:1981zj} (dashed green line) is smaller than $\sim 10^2$.}
\label{fig:O3_different_methods_comparison}
\end{figure}

\paragraph{Semi-analytical solutions:}
For a polynomial potential of the type
\begin{equation}
V(\phi)=a\phi^2 - b \phi^3 + \frac{\lambda}{4} \phi^4,
\end{equation}
semi-analytical expressions for the $O_3$- and $O_4$-bounce actions exist (Adams 1993 \cite{Adams:1993zs}, see also \cite{Kehayias:2009tn,Caprini:2011uz,Ellis:2020awk} for more recent use)
\begin{align}
&S_3 = \frac{8\pi b}{\lambda^{3/2}} \frac{8\sqrt{\delta}}{81(2-\delta)^2}( \beta_1 \delta + \beta_2 \delta^2 + \beta_3 \delta^3 ), \label{eq:S3_semi_analytical_Adams} \\
&S_4 = \frac{4\pi^2}{3\lambda}\frac{1}{(2-\delta)^3} (\alpha_1 \delta + \alpha_2 \delta^2 + \alpha_3 \delta^3),
\end{align}
with
\begin{align}
& \delta \equiv \frac{2 \lambda a}{b^2},\\
& \beta_1 = 8.2938, \quad \beta_2 = -5.5330, \quad \beta_3 = 0.8180,\\
& \alpha_1 = 13.832, \quad \alpha_2 = -10.819, \quad \alpha_3 = 2.0765.
\end{align}
Another semi-analytical expression for $S_3$ which we can be found in the litterature is the one that DLHLL derived in 1992 \cite{Dine:1992wr} (see also \cite{Quiros:1999jp})
\begin{equation}
S_3 = \frac{13.72\,a^{3/2}}{b^2} f(\delta/2),\label{eq:S3_semi_analytical_DLHLL}
\end{equation}
with
\begin{equation}
f(x)=1 + \frac{x}{4} \left[ 1 + \frac{2.4}{1-x} + \frac{0.26}{(1-x)^2}\right].
\end{equation}
In Fig.~\ref{fig:O3_different_methods_comparison} we show that the two semi-analytical expressions in Eq.~\eqref{eq:S3_semi_analytical_Adams} and Eq.~\eqref{eq:S3_semi_analytical_DLHLL} fit very well the solution of the package FindBounce \cite{Guada:2018jek, Guada:2020xnz}.
As a toy model of the effective potential we consider the high temperature expansion of  the SM EFT in Eq.~\eqref{eq:V_tot_SM_dim6_HT}, with $\Lambda \to \infty$, where we modified $E \to 10 E$ in order to increase the barrier strength and to make the transition strongly first order.

 In the next section, we provide analytical approximations of the bounce action in the so-called thin-wall and thick-wall limit.
 
\subsection{Thin-wall and thick-wall limits}
\label{sec:thick_wall_thin_wall_formula}

\begin{figure}[h!]
\centering
\raisebox{0cm}{\makebox{\includegraphics[width=\textwidth, scale=1]{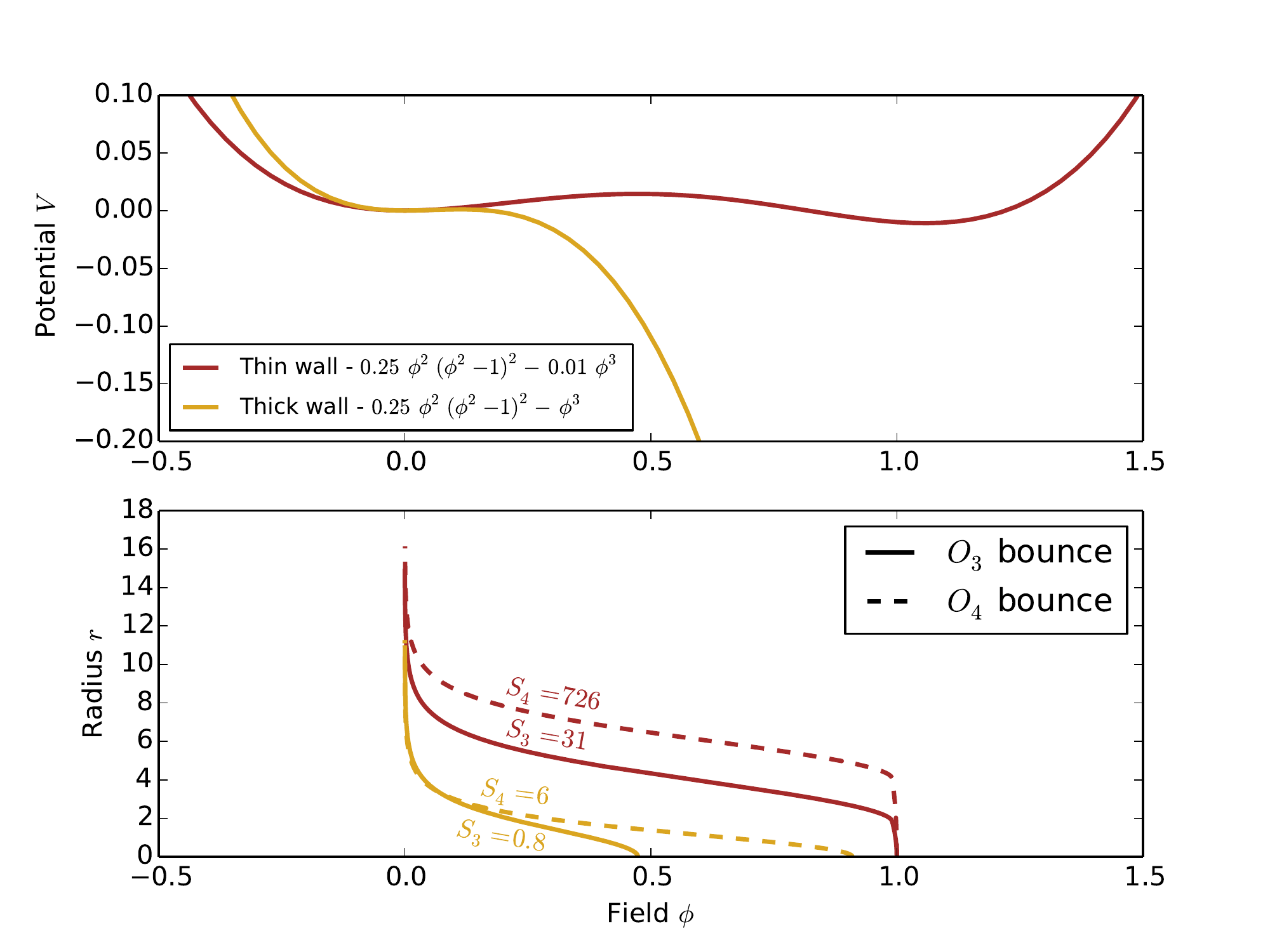}} }
\caption{\it \small   \textbf{Top:} Thin-walled potential (Large barrier compared to the potential energy difference between the minima) versus thick-walled potential (small barrier compared to the potential energy difference between the two minima). The true minimum of the thick-walled potential does not appear on the plot. \textbf{Bottom:} Corresponding $O_3$- and $O_4$- symmetrical bounce profile $\phi(r)$, solution of the Euclidean equation of motion in Eq.~\eqref{eq:bounce_eomotion}. Computed with CosmoTransition \cite{Wainwright:2011kj}.}
\label{fig:thick_vs_thin_wall}
\end{figure}

\paragraph{Thin-wall limit:}

The first class of potential where the bounce action can be estimated analytically are potentials where the extrema are \textbf{nearly-degenerate} and separated by a \textbf{large barrier}. The wall thickness is as thin as possible in order to minimize the region of space sitting on top of the potential barrier. This is the \textbf{thin-wall} limit (red line in Fig.~\ref{fig:thick_vs_thin_wall}). In order for the trajectory to be able to reach the false vacuum, the initial field value $\phi(0)$ has to start close to the true vacuum $\phi_{\rm TV}$ where the potential is very flat. Hence, the particle sits there for a long time before evolving significantly. Hence, we can neglect the damping term in the equation of motion in Eq.~\eqref{eq:bounce_eomotion}
\begin{equation}
\frac{d^2 \phi}{dr^2} = V'(\phi)
\end{equation}
which, after multiplying both sides by $d\phi/dr$ and using the first boundary condition in Eq.\eqref{eq:BC_bounce}, leads to
\begin{equation}
\frac{d\phi}{dr}=\sqrt{2(V(\phi)-V(\phi_{\rm TV}))} 
\end{equation}
Therefore, the $O(4)$-symmetric bounce action can be estimated as
\begin{align}
S_4 &= 2\pi^2 \int dr \, r^3 \, \left[ \frac{1}{2} \left(  \frac{d\phi}{dr} \right)^2 + V(\phi) \right] \\
& = \underbrace{- 2 \pi^2 \frac{R^4}{4} \Delta V}_{r \ll R}  + \underbrace{2 \pi^2 R^3 S_1 }_{r \sim R}
\label{eq:S4_thin_before_R_inj}
\end{align}
where $\Delta V>0$ is the potential energy difference and where $S_1$ is the action of a one-dimensional bounce
\begin{equation}
S_1 = \int_0^\infty dr \left[ \frac{1}{2}\left( \frac{d\phi}{dr} \right)^2 + V(\phi) \right] = \int_{\phi_{\rm TV}}^{\phi_{\rm FV}} d\phi\sqrt{2(V(\phi)-V(\phi_{\rm TV}))} .
\end{equation}
Physically, $S_1$ corresponds to the energy of the bubble wall per unit of area (the \textbf{wall tension}).
The \textbf{critical radius} $R_c$ of the bubble at nucleation is the one which minimizes the action
\begin{equation}
\frac{\delta S_4}{\delta R} = 0 \quad \rightarrow \quad -2 \pi^2\, R_c^3 \, \Delta V + 6 \pi^2 \,R_c^2 \, S_1 =0.
\end{equation}
By injecting the solution $R_c$ into Eq.~\eqref{eq:S4_thin_before_R_inj}, we obtain \cite{Coleman:1977py}
\begin{equation}
R_c = \frac{3 S_1}{\Delta V} \qquad \textrm{and} \qquad S_4 = \frac{27 \pi^2 \, S_1^4}{2 \,\Delta V^3} .
\end{equation}
It is very straightforward to generalize this result to the $O(3)$-symmetric bounce action in Eq.~\eqref{eq:O3_bounce_action} and we find \cite{Linde:1981zj}
\begin{equation}
R_c = \frac{2 S_1}{\Delta V} \qquad \textrm{and} \qquad S_3 = \frac{16 \pi \,S_1^3}{3 \,\Delta V^2} .
\label{eq:S3_thin_before_R_inj}
\end{equation}
Whether the tunneling completes via $O(3)$ or $O(4)$ bounce depends on the ratio
\begin{equation}
\frac{S_4}{S_3/T} = \frac{81 \pi}{32}\frac{S_1}{\Delta V} T \sim 3 \,R_c\, T.
\end{equation}

\paragraph{Thick-wall limit:}

\begin{figure}[h!]
\centering
\raisebox{0cm}{\makebox{\includegraphics[width=0.7\textwidth, scale=1]{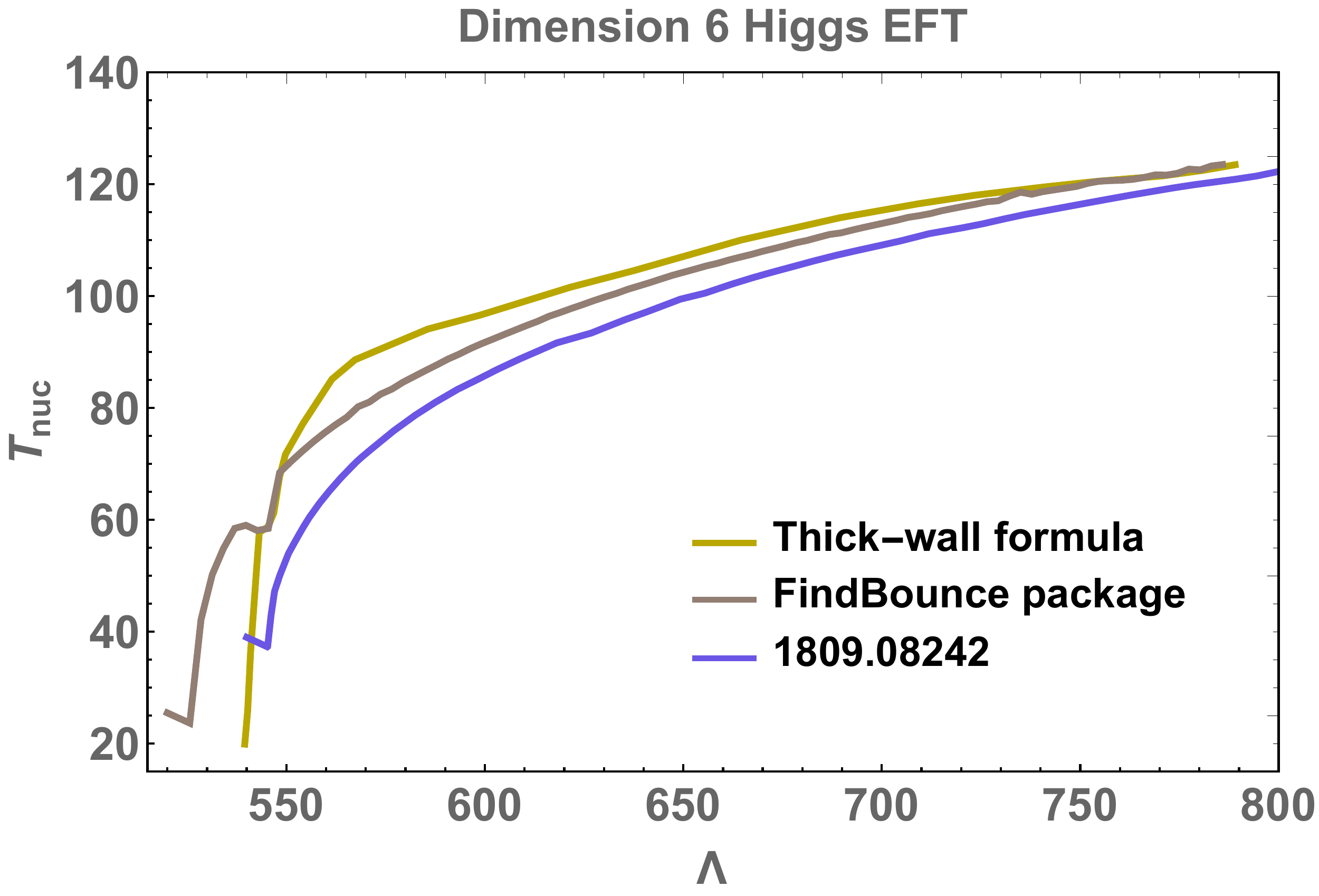}} }
\caption{\it \small   Nucleation temperature, assuming $O_3$ tunneling, in the dimension 6 Higgs EFT model, which effective potential $V_{\rm eff}$ is given by Eq.~\eqref{eq:effective_potential_total}. We compare the results obtained with the (easy-to-use) FindBounce package \cite{Guada:2018jek, Guada:2020xnz}, the results obtained from the thick-wall formula in Eq.~\eqref{eq:S_3_thick_wall} and the results from \cite{Ellis:2018mja}.}
\label{fig:Dim6_Tn_Lambda}
\end{figure}

When the \textbf{potential barrier is small} compared to the potential energy difference, the wall thickness is as large as possible in order to minimize the gradient energy. This is the \textbf{thick-wall} limit (yellow line in Fig.~\ref{fig:thick_vs_thin_wall}). The $O(4)$-symmetric bounce action reads
\begin{equation}
S_4 = \pi^2\, R^3 \,\delta R\, \left( \frac{\delta \phi}{\delta R} \right)^2 + \frac{\pi^2}{2}\, R^4 \,\bar{V}
\end{equation}
where $\delta \phi$ is the field excursion while $\delta R$ is the wall thickness. In the thick-wall limit, we can set $\delta R = R$. The quantity $\bar{V}<0$ is the potential energy averaged in the bubble volume. We approximate it as $\bar{V} = V(\phi_*) - V(\phi_{\rm FV})$ where $\phi_*$ is the initial field value, at the center of the bubble.\footnote{In the thick-wall limit, the initial field value $\phi_*$ is generally close to the zero of the potential $V(\phi_{\rm zero}) = V (\phi_{\rm FV})$.} Additionally, we assume (without lack of generality) that the false vacuum energy and the corresponding field value (at asymptotic distance) are zero, $V(\phi_{\rm FV})=0$ and $\phi_{\rm FV}=0$, such that we can write $\bar{V} = V(\phi_*) $ and $\delta \phi = \phi_*$. Hence, we obtain
\begin{equation}
S_4 = \pi^2 R^2 \,\delta R\, \phi_*^2 + \frac{\pi^2}{2} \,R^4 \,V(\phi_*),
\label{S_4_thick_wall_before_R_inj}
\end{equation}
where $\phi_*$ is understood as the value which minimizes the bounce action.
Upon solving for the \textbf{critical bubble radius} $R_c$, solution of $\delta S_4 / \delta R = 0$, and reinjecting into Eq.~\eqref{S_4_thick_wall_before_R_inj}, we obtain \cite{Anderson:1991zb, Randall:2006py, Nardini:2007me}
\begin{equation}
R_c^2 = \frac{\phi_*^2}{-V(\phi_*)} \qquad \textrm{and}\qquad S_4 = \frac{\pi^2}{2} \frac{\phi_*^4}{-V(\phi_*)}.
\label{S_4_thick_wall}
\end{equation}
Repeating straightforwardly the same steps for the $O(3)$-symmetrical bounce leads to
\begin{equation}
R_c^2 = \frac{\phi_*^2}{-2\,V(\phi_*)} \qquad \textrm{and}\qquad S_3 = \frac{4\pi}{3} \frac{\phi_*^3}{\sqrt{-2\,V(\phi_*)}}.
\label{eq:S_3_thick_wall}
\end{equation}
Whether the tunneling completes via $O(3)$ or $O(4)$ bounce depends on the ratio
\begin{equation}
\frac{S_4}{S_3/T} = \frac{3 \pi}{4}\frac{\phi_*}{\sqrt{-2 V(\phi_*)}} T \sim 2 \,R_c\, T.
\end{equation}
In Fig.~\eqref{fig:Dim6_Tn_Lambda}, we compare the nucleation temperature $T_{\rm nuc}$, discussed in the next section in Sec.~\ref{par:nucleation_temp}, in the dim 6 Higgs EFT model, computed using the thick-wall formula for $S_3$ to the one computed using a public numerical package\footnote{I thank Victor Guada and Jorinde Van de Vis for useful discussions regarding the computation of the bounce action with existing numerical packages.}, as well as to the one which can be found in the literature. The agreement is rather acceptable.

\subsection{Temperature at which the phase transition completes}
\label{sec:nucleation_temp_VS_percolation}

\paragraph{Nucleation temperature:}
\label{par:nucleation_temp}

At first approximation, we can consider the phase transition to complete when the number of bubbles nucleated per Hubble volume and per Hubble time is of order $1$
\begin{equation}
\label{eq:nucTempInstApprox}
\Gamma(T_{\rm nuc}) \simeq H(T_{\rm nuc})^4.
\end{equation}

\paragraph{Percolation temperature:}
\label{par:percolation_temp}

In this paragraph we give a more thorough analysis of the temperature for which regions of false vacuum are entirely converted into true vacuum following the early work of \cite{Guth:1979bh,Guth:1981uk,Enqvist:1991xw,Enqvist:1991xw,Turner:1992tz} and the more recent of work of \cite{Ellis:2018mja}. Starting from $\Gamma(t') \, a(t')^3 $ which is the number of bubbles formed per unit of time and per unit of comoving volume and from $\frac{4\pi}{3} \, \left( \int_{t'}^t \frac{d \tilde{t}}{a(\tilde{t})}  \right)^3$ which is the comoving volume of a bubble which has been nucleated at $t'$ and has expanded at the speed of light until $t$, we deduce the volume fraction converted into the true vacuum at time $t$, including \textbf{multiple counting} when bubbles overlap \footnote{Note that we have neglected the initial size of the bubble just after it nucleates. Including it would give $ I(t) =  \int_{t_c}^t dt' \, \Gamma(t') \, \frac{4\pi}{3} \, \left( \int_{t'}^t \frac{a(t')}{a(\tilde{t})} \, d \tilde{t} + R(t') \right)^3 = \frac{4\pi}{3} \int_{T_{\rm nuc}}^{T_c} \frac{dT'}{T'} \, \frac{\Gamma(T') }{T'^3 \, H(T')} \, \left( \int_{T_{\rm nuc}}^{T'} \frac{d \tilde{T}}{H(\tilde{T})} + R(T') \, T'\right)^3$ where $R'$ is the bubble radius at nucleation. The correction is of order $R(T') \, H' \sim \frac{\text{TeV}}{\, \MPl}$ and is completely negligible. }
\begin{equation}
 I(t) =  \int_{t_c}^t dt' \, \Gamma(t') \, a(t')^3 \, \frac{4\pi}{3} \, \left( \int_{t'}^t \frac{d \tilde{t}}{a(\tilde{t})}  \right)^3.
\end{equation}
We can also interpret $I(t)$ as the average number of bubbles inside which, a given point is contained. It can be larger than $1$ because it includes overlapping. However, multiple counting can been subtracted by \textbf{exponentiation}.\footnote{The probability $P(t)$ that a given point remains in the false vacuum is the $N \to \infty$ limit of the product $\prod_{n=1}^N \left(1-dt\frac{dI(t_n)}{dt}\right)$ where $dt = \frac{t-t_c}{N}$ and $1-dt\frac{dI(t_n))}{dt}$ is the probability of remaining in the false vacuum between $t_n = t_c+(t-t_c)\frac{n}{N}$ and $t_{n+1}$.} Hence, we define 
\begin{equation}
P(t) \equiv e^{-I(t)}
\end{equation}
which is the probability that a given point  remains in the false vacuum at time $t$. 
From this, we can define the percolation to occur when 
\begin{equation}
P(t) \lesssim 1/e \, \rightarrow \, I(t) \gtrsim 1 .
\label{eq:percolation_criterium}
\end{equation}

\paragraph{Do bubbles percolate during an inflating period?}
\label{par:percolation_temp_inf}
During inflation it can happen that the probability $1-P(t)$ that a given point has been converted to the true vacuum tends to $1$, whereas the physical volume remaining in the false vacuum (here given per unit of comoving volume)
\begin{equation}
\mathcal{V}_{\mathsmaller{\rm false}}(t) \equiv a(t)^3 \, P(t),
\end{equation}
goes to  infinity \cite{Guth:1982pn}.
In order to avoid this to occur\footnote{In the study \cite{Barir:2022kzo}, the authors explore the alternative possibility that a 1stOPT occuring during inflation completes after the end of inflation. They show that a continuous nucleation of bubbles during inflation would generate a large spectrum of primordiale scalar fluctuation which could source observable GW at second order in perturbations.}, we add to Eq.~\eqref{eq:percolation_criterium} a second condition which is that the decrease of the probability to remain in the false vacuum $P(t)$ drops \textbf{faster} than the increase of the false vacuum volume due to inflation \cite{Turner:1992tz} 
\begin{equation}
\frac{d\mathcal{V}_{\mathsmaller{\rm false}}(t)}{dt} \lesssim 0.
\label{eq:percolation_criterium_false_vac_volum}
\end{equation}

Upon using the adiabatic relations $H \,dt = -dT/T$ and $d(T \, a)=0$, we can re-express the criteria in Eq.~\eqref{eq:percolation_criterium} and in Eq.~\eqref{eq:percolation_criterium_false_vac_volum} solely in terms of the temperature. Hence, the \textbf{percolation temperature} $T_{\rm per}$ is obtained from
\begin{equation}
\label{eq:percoTemp}
\frac{4\pi}{3} \int_{T_{\rm per}}^{T_c} \frac{dT'}{T'} \, \frac{\Gamma(T') }{T'^3 \, H(T')} \, \left( \int_{T_{\rm per}}^{T'} \frac{d \tilde{T}}{H(\tilde{T})} \right)^3 \simeq 1,
\end{equation}
after checking that 
\begin{equation}
\label{eq:percoTemp_PhysVolDec}
3+T_{\rm per}\frac{dI(T)}{dT}\Big|_{T_{\rm per}} < 0.
\end{equation}
Note however that the distinction between nucleation temperature and percolation temperature is relevant only for tuned polynomial potential in the supercool regime \cite{Ellis:2018mja}. Particularly, the distinction is not relevant for supercooled phase transitions which are naturally generated by non-polynomial potential, see Refs.~\cite{Baldes:2018emh,Baldes:2021aph} and Sec.~\ref{sec:supercool_potential}.

\section{Bubble propagation}
\label{sec:bubble_propagation}

The vacuum energy difference between the broken phase and the symmetric phase acts as a negative pressure which accelerates the bubble wall. If coupled to the scalar field driving the PT, the particles entering the broken phase acquire a mass term proportional to the scalar VEV. Hence, the scalar field gradient along the wall profile induces a force acting on the particles in the plasma.  The terminal velocity of the bubble wall results from the equilibrium between the accelerating vacuum pressure and the retarding friction pressure. In this section we present different formalism which can be found in the literature for computing the bubble wall velocity. In Sec.~\ref{sec:eom_scalar_field_wall}, we derive the equilibrium condition between the accelerating pressure and the friction pressure from the conservation of the energy-momentum tensor of the system plasma+scalar field. In Sec.~\ref{sec:lte_pressure}, we derive the retarding pressure at local-thermal-equilibrium in the field-theoretic approach. In Sec.~\ref{sec:close_lte_pressure}, we present a field-theoretic computation of the friction pressure in the presence of out-of-equilibrium conditions. In Sec.~\ref{sec:ballistic_pressure}, we compute the friction pressure in the limit where collisions are neglected. We discuss the important NLO correction in gauge coupling constant in Sec.~\ref{sec:NLO_pressure}. Finally, we deduce the speed of the bubble wall in Sec.~\ref{sec:speed_wall}.
Part of the results of this section are relevant for the work on string fragmentation in supercooled confinement in Chap.~\ref{chap:SC_conf_PT}, where the new effects which we point out, string fragmentation and deep-inelastic-scattering depends on the value of the bubble wall Lorentz factor $\gamma_{\rm wp}$.

\subsection{Equation of motion for the scalar field}
\label{sec:eom_scalar_field_wall}

The dynamics of the scalar field during the bubble propagation can be derived from the conservation of the energy-momentum tensor
\begin{equation}
\partial^\mu T_{\mu \nu}^{\rm total} = \partial^\mu \left( \sum_i T_{i,\,\mu \nu}^{\rm plasma} + T_{\mu \nu}^
\phi \right) = 0,
\label{eq:Tmunu_conservation_wall}
\end{equation}
where $i$ runs over the species in the plasma. The covariant formulation of the energy-momentum tensor of the plasma in terms of the one-particle state occupation number $f_i(\vec{p}, \, x)$ reads
\begin{equation}
T_{i,\,\mu \nu}^{\rm plasma} = \int \frac{d^3p}{(2\pi)^3}  \frac{p_\mu p_\nu}{E_i} f_i(\vec{p},\, x).
\label{eq:Tmunu_plasma_f}
\end{equation}
$f_i(\vec{p}, \, x)$ is solution of the Boltzmann equation (Liouville equation + collision)
\begin{equation}
\frac{df}{dt} = {\rm coll}\quad \rightarrow \quad \partial_t f + \dt{\vec{x}} \cdot \partial_{\vec{x}} +  \dt{\vec{p}} \cdot \partial_{\vec{p}} = \rm coll,
\label{eq:Liouville_equation}
\end{equation}
which follows from the time conservation of the number of particles $N = \int\, d^3r \,d^3p\, \,  f(\vec{r},\, \vec{p},\,t) $ up to the collision term $``{\rm coll}''$. Now, supposing that the wall velocity is constant $v_w = \rm cst$ (the wall has reached equilibrium), we have
\begin{align}
&\partial_t f =v_w \,\partial_z f \quad (=0 \textrm{ in the wall frame}),\\
&\dt{z} =  \frac{p_z}{E}, \\
&\dt{p}_z = - \frac{\partial_z m(z)^2}{2E} \quad (\textrm{the problem is time-invariant: } \dt{E}=0) .
\end{align}
we can rewrite Eq.~\eqref{eq:Liouville_equation} as (e.g. \cite{Moore:1995si})
\begin{equation}
\left( \frac{p_z}{E} \partial_z ~ -~ \frac{\partial_z m^2}{2E} \partial_{p_z} \right) f(p_z,\, z) = \rm coll.
\label{eq:Liouville_equation_massaged}
\end{equation}
We deduce the divergence of Eq.~\eqref{eq:Tmunu_plasma_f}
\begin{align}
\partial^\mu T_{i,\, \mu \nu}^{\rm plasma} + {\rm coll_i} &= \int \frac{d^3 p }{(2\pi)^3}  p_\nu \left( \frac{1}{E} p^\mu \partial_\mu f_i(\vec{p},\, x)  - \frac{ f_i(\vec{p},\, x)}{2E^3} \partial_\mu m^2 \right) \\
&= \frac{1}{2}\partial_\mu m^2 \int \frac{d^3 p }{(2\pi)^3}  p_\nu \left( -\frac{1}{E}  \,\partial_{p_\mu} f_i(\vec{p},\, x)  -  \partial_{p_\mu}\left(\frac{1}{E}\right)  \, f_i(\vec{p},\, x)   \right)\\
&=  \frac{1}{2}\partial_\nu m^2 \int \frac{d^3 p }{(2\pi)^3}\, \frac{1}{E} \, f_i(\vec{p},\, x).
\label{eq:divergence_Tmunu_plasma_f}
\end{align}
where in the second line we have used the Boltzmann equation in Eq.~\eqref{eq:Liouville_equation_massaged} whereas in the last line we have performed an integration by part.
In addition, the energy-momentum tensor of the classical field background is 
\begin{equation}
T_{\mu \nu}^\phi = \partial_\mu \phi \partial_\nu \phi - g_{\mu\nu} \left( \frac{1}{2}\partial_\rho\phi \partial^\rho \phi - V_0(\phi)  \right),
\end{equation}
where $V_0$ is the effective potential at zero temperature. Its divergence reads
\begin{equation}
\partial^\mu T_{\mu \nu}^\phi = \partial_\nu \phi \left( \Box \phi + \frac{d V_0}{d\phi}    \right).
\label{eq:divergence_Tmunu_classical_field_background}
\end{equation}
Upon plugging Eq.~\eqref{eq:divergence_Tmunu_classical_field_background} and Eq.~\eqref{eq:divergence_Tmunu_plasma_f} into Eq.~\eqref{eq:Tmunu_conservation_wall}, we obtain the equation of motion for the scalar field at constant bubble wall velocity
\begin{equation}
\Box \phi  +  \frac{dV_0}{d\phi}  + \sum_i \frac{dm_i^2}{d\phi} \int \frac{d^3p}{(2\pi)^3} \frac{1}{2E} \, f_i(\vec{p},\, x) = 0.
\label{eq:Higgs_profile_thermal_0}
\end{equation}
Note that the collision terms are absent since they cancel out when they are summed over all the species \cite{Konstandin:2014zta}.
The last term can be expressed as
\begin{align}
 \frac{dm^2}{d\phi} \int \frac{d^3p}{(2\pi)^3} \frac{1}{2E} \, f(\vec{p},\, x) &=  -\frac{1}{3} \frac{dm^2}{d\phi} \int \frac{d^3p}{(2\pi)^3} \frac{p}{2}\cdot \frac{d}{dp} \frac{f_i(\vec{p},\, x) }{E} \\
 &= -\frac{1}{3} \frac{dm^2}{d\phi} \int \frac{d^3p}{(2\pi)^3}  \frac{p}{2} \cdot\frac{dE}{dp} \cdot\frac{dm^2}{dE}\cdot \frac{d\phi}{dm^2}\cdot\frac{d}{d\phi} \frac{f_i(\vec{p},\, x) }{2E} \\
 &= - \frac{d P_{\rm plasma}}{d\phi},
\end{align}
where $P_{\rm plasma}$ is the thermal pressure 
\begin{equation}
P_{\rm plasma}= \int \frac{d^3 p}{(2\pi)^3}\frac{p^2}{3E} \sum_i f_i(\vec{p},\, x)  ,
\label{eq:pressure_def_stat_physics}
\end{equation}
In the presence of collisions, $f_i(\vec{p},\, x)$ can be highly out-of-equilibrium. 
We now express Eq.~\eqref{eq:scalar-time-like} in spherical coordinates (we use the $SO(3)$ symmetry to reduce the 3 Cartesian coordinates to the radial $r$ coordinate\footnote{Note that we could also use the $SO(3,1)$ symmetry of Eq.~\eqref{eq:scalar-time-like} which allows to trade $r$ and $t$ for the time-like light-cone coordinate $s = \sqrt{t^2-r^2}$ \cite{Jinno:2019bxw}
\begin{equation}
\frac{\partial^2 \phi}{\partial s^2} +\frac{3}{s}\frac{\partial \phi}{\partial s} +\frac{\partial V}{\partial \phi} = \frac{d P_{\rm plasma}}{d\phi}.
\label{eq:scalar-time-like}
\end{equation}
In this coordinate, the trajectory of the bubble wall, $r = v_w\,t$, satisfies $s = t/\gamma_w$.})
\begin{equation}
\frac{\partial^2 \phi}{\partial t^2} - \frac{1}{r^2}\frac{\partial}{\partial \phi}\left[ r^2\frac{\partial \phi}{\partial r}\right] +\frac{\partial V}{\partial \phi} =  \frac{d P_{\rm plasma}}{d\phi}.
\label{eq:scalar-spherical}
\end{equation}
Going in the wall frame, we can drop the time derivative in the previous Eq.~\eqref{eq:scalar-spherical}. Then, upon multiplying by $d\phi/dz$ and integrating over $z$ across the wall, we find the equilibrium condition
\begin{equation}
\Delta V_{\rm vac}  ~=~ \Delta P_{\rm plasma}  , \qquad \qquad \rm (equilibrium~ condition)
\label{eq:pressure_equi_bubble_wall}
\end{equation}
where $\Delta V_{\rm vac} $ and $\Delta P_{\rm plasma} $ are respectively the difference of effective potential at zero-temperature (vacuum energy difference) and the difference between the pressure from plasma effects between the two phases. In order to write Eq.~\eqref{eq:pressure_equi_bubble_wall}, we used that the wall profile flattens at infinity $\int \phi'' \phi' \, dz = \left[  \frac{\phi^{'2}}{2} \right]_{-\infty}^{+\infty}$ and we have neglected the Laplace pressure $\Delta P_{\rm Lapace}$ due to the surface tension of the bubble \cite{de2013capillarity} which goes to zero at large radius/time
\begin{equation}
\Delta P_{\rm Laplace}(t) = \int_{\rm across~the~wall} dz ~\frac{2}{r} \left( \frac{d \phi}{dr} \right)^2 \quad \underset{r \to \infty}{\xrightarrow{\hspace*{0.5cm}}} \quad 0.
\end{equation}
We can generally expect the retarding pressure from plasma effects $P_{\rm plasma}$ to be an increasing function of the wall velocity $v_w$ such that the wall accelerates under the vacuum pressure $\Delta V$ until it compensates with $P_{\rm plasma}$.

\subsection{Friction pressure at local thermal equilibrium}
\label{sec:lte_pressure}

\paragraph{Energy-momentum conservation in the stationary regime.}
In this section, we derive the friction pressure at local thermal equilibrium.
 Under the assumption that the bubble wall reaches a constant velocity, the energy-momentum tensor in the bubble frame is time-invariant and its conservation law in the bubble frame reads \cite{Mancha:2020fzw} 
\begin{align}
&\Delta \left<T_{\rm z0}^p \right> + \Delta \left<  T_{\rm z0}^b \right>=0,\label{eq:stressenergyconserv1}\\
&\Delta \left<T_{\rm zz}^p \right> + \Delta \left<  T_{\rm zz}^b \right>=0.\label{eq:stressenergyconserv2}
\end{align}
where the subscripts $p$ and $b$ denote the plasma and bubble components, respectively.
We can write the stress-energy tensor of the plasma in the plasma frame as
\begin{equation}
\left<(T_{00}^{p})_{\rm plasma~frame}\right> = \rho_p, \qquad \left<(T_{zz}^{p})_{\rm plasma~frame}\right> = P_p, 
\end{equation}
such that after a Lorentz boost to the bubble frame, it becomes
\begin{equation}
\left<T_{zz}^{p}\right> = (\gamma^2-1)(\rho_p+P_p) +P_p,\qquad \left<T_{z0}^{p}\right> = \beta \gamma^2 (\rho_p + P_p),
\end{equation}
where $\gamma=1/\sqrt{1-\beta^2}$.
For the bubble components, we obtain
\begin{equation}
\left<T_{zz}^{b}\right> \equiv P_b, \qquad \left<T_{z0}^{b}\right> = 0,
\end{equation}
where we have set $\rho_{\rm b} \, + \, P_{\rm b} = 0$ since a scalar condensate has no entropy.
Eq.~\eqref{eq:stressenergyconserv1} and \eqref{eq:stressenergyconserv2} integrated along the z direction become
\begin{align}
&\Delta (\beta\, \gamma^2\, T\,S)=0, \label{eq:TmunuCons1}\\
&\Delta((\gamma^2-1)\,T\,S ) +\Delta P_p + \Delta P_b = 0,\label{eq:TmunuCons2}
\end{align}
where $S = (\rho_p + P_p)/T$ is the entropy of the plasma. 

\paragraph{First energy-momentum conservation equation:}
The first equation implies that whenever $\Delta(TS) \neq 0$, the velocity of the plasma can not be uniform along the fluid profile. The temperature and velocity of fluid just outside the bubble $T_+,\,\beta_{+}$ and just inside the bubble $T_-,\,\beta_{-}$ must be determined by solving the two equations in Eq.~\eqref{eq:TmunuCons1} and Eq.~\eqref{eq:TmunuCons2}. We refer to \cite{Balaji:2020yrx,Ai:2021kak} for such a treatment. Instead, here we follow \cite{Mancha:2020fzw}  and make the approximation
\begin{equation}
\Delta T \simeq 0.
\end{equation}
This is the \textbf{local-thermal-equilibrium} (l.t.e.) condition.

\paragraph{Second energy-momentum conservation equation:}
Eq.\eqref{eq:TmunuCons2} implies that the pressure difference in the plasma frame at l.t.e. reads \cite{Mancha:2020fzw}\footnote{I thank Bogumi\l{}a \'{S}wie\.{z}ewska for a valuable correspondence.} 
\begin{equation}
\label{eq:pressure_local_thermal_1}
\Delta P \equiv \Delta (P_p + p_b) = -(\gamma^2-1)\,T\,\Delta S 
\end{equation}
Hence, at local thermal equilibrium, the pressure difference is given by the entropy difference between the two phases $\Delta S$ (in the plasma frame). 

\paragraph{Entropy in quantum field theory.}
The entropy is related to the component of the energy-momentum tensor which for a free scalar field, reads
\begin{equation}
T_{\rm \mu\nu} = \partial_\mu \phi \partial_\nu \phi + g_{\mu\nu} \mathcal{L}_\phi
\end{equation}
with
\begin{equation}
\mathcal{L}_\phi = - \frac{1}{2} g_{\mu\nu} \partial_\mu \phi \partial_\nu \phi -\frac{1}{2} m^2 \phi^2.
\end{equation}
The thermal average, after renormalization, reads 
\begin{align}
\left< T_{\mu\nu} \right>_{\rm ren} = -\eta_{\mu\nu} \frac{m^4}{64\pi^2} \left[\textrm{ln} \left(\frac{m^2}{4\pi\mu^2}\right) + \gamma_{\rm E} - \frac{3}{2} \right] + \frac{\eta_\mu\nu}{6\pi^2\beta^5m}\left[\partial_z J_{\rm B}(b,\,z) \right]_{z=\beta m} \\
+ \delta_\mu^0\delta_\nu^0 \left( \frac{2}{3\pi^2 \beta^5 m} \left[\partial_z J_{\rm b}(6,\,z) \right]_{z=\beta m}  + \frac{m}{2\pi^2 \beta^3} \left[\partial_z J_{\rm b}(4,\,z) \right]_{z=\beta m} \right),
\end{align}
with 
\begin{equation}
J_{\rm b/f}(n,\,z) \equiv \int_0^\infty dx~x^{n-2}~{\rm ln} \left( 1 \mp e^{-\sqrt{x^2+z^2}} \right).
\end{equation}
We deduce the change in entropy between the massless phase to the massive phase
\begin{equation}
T \Delta S = \frac{2\pi^2}{45 \beta^4} - \frac{2}{3\pi^2\beta^5 m}\left[\partial_z J_{\rm b}(6,\,z) \right]_{z=\beta m}  - \frac{m}{2\pi^2 \beta^3} \left[\partial_z J_{\rm b}(4,\,z) \right]_{z=\beta m} .
\label{eq:pressure_local_thermal_2}
\end{equation}

\paragraph{Results.}
Since $\Delta S$ is independent of $\gamma$, this implies
\begin{equation}
\Delta P \propto \gamma^2.
\end{equation}
Hence a bubble can not run-away to an arbitrary large $\gamma$ value at local thermal equilibrium. However, we generally expect the local thermal equilibrium condition to be quickly violated as soon as $\Gamma \lesssim \rm few$, due to the mass of particles varying faster than the collision time scale as the wall passes by.  In the following sections, we discuss different possibilities to account for the deviation from local thermal equilibrium.
In \cite{Mancha:2020fzw}, the temperature and the fluid velocity is assumed to be constant across the wall. Even though this contradicts energy-momentum conservation \cite{Espinosa:2010hh}, it can be a good approximation in the limit where particles of the plasma interact strongly with each others \cite{Lewicki:2022nba}.
In \cite{Balaji:2020yrx,Ai:2021kak}, the results of \cite{Mancha:2020fzw} and \cite{Espinosa:2010hh} are merged to compute the wall velocity at local equilibrium with varying temperature and fluid velocity across the wall.

\paragraph{Bubble wall velocity.}
The pressure difference defined from Eq.~\eqref{eq:pressure_local_thermal_1} accounting for both scalar field and plasma contribution coincides with the effective potential at finite temperature $V_{\rm eff}(\phi,\,T)$ defined in Eq.~\eqref{eq:effective_potential_total}. Hence the bubble velocity can be computed from 
\begin{equation}
\label{eq:bubble_wall_velocity_lte}
\Delta V_{\rm eff}(T) = -(\gamma^2 \,-\, 1)\,T\,\Delta S.
\end{equation}
The friction pressure and bubble velocity will be shown in Fig.~\ref{fig:DeltaP_VS_gamma_T_vs_R} and  Fig.~\ref{eq:DeltaV_alpha_contours_v_LO}.

\subsection{Friction pressure close to local thermal equilibrium}
\label{sec:close_lte_pressure}
\paragraph{A thermal field theoretic approach:}

At thermal equilibrium, the thermal pressure $P_{\rm plasma}$ is directly related to the finite-temperature corrections $\Delta V_{\rm plasma} $ to the effective potential \cite{Bellac:2011kqa}
\begin{equation}
\Delta V_{\rm plasma} = -P_{\rm plasma}
\end{equation}
Therefore, splitting the particle distribution functions into an equilibrium part plus some deviation $\delta f_i$  yields
\begin{equation}
\Box \phi  +  \frac{dV(\phi,\,T)}{d\phi}  + K(\phi) = 0,
\label{eq:Higgs_profile_thermal}
\end{equation}
where $V(\phi,\,T)$ is the \textbf{effective potential at finite temperature}, introduced in Sec.~\ref{sec:effective_potential}, and gives the \textbf{force} driving the wall. The last term encodes the deviation from local thermal equilibrium 
\begin{equation}
K(\phi) \equiv \sum_i \frac{dm_i^2}{d\phi} \int \frac{d^3p}{(2\pi)^3} \frac{1}{2E} \, \delta f_i(\vec{p},\, x),
\label{eq:friction_term_wall}
\end{equation}
and is responsible for \textbf{friction}.
Upon considering the Ansatz 
\begin{equation}
\phi(z) = \frac{\phi_0}{2} \left[ 1 + {\rm tanh} \left(\frac{z}{L_{\rm w}}\right)   \right]
\end{equation}
for the wall profile in the wall frame ($\phi_0$ being the scalar VEV ) and taking the moments
\begin{align}
&\int_{-\infty}^{\infty} dz \left[ \rm l.h.s ~ of ~ Eq.~\eqref{eq:Higgs_profile_thermal}\right]\times \phi' = 0, \\
&\int_{-\infty}^{\infty} dz \left[ \rm l.h.s ~ of ~ Eq.~\eqref{eq:Higgs_profile_thermal}\right]\times (2\phi - \phi_0) \phi' = 0,
\end{align}
we obtain
\begin{align}
\frac{\Delta  V_T}{T^4} &= F,\label{eq:system_wall_Eq_1}\\
-\frac{2}{15(T\, L_{\rm w})^2} \left( \frac{\phi_0}{T} \right)^3 + \frac{W}{T^5} &= G,
\label{eq:system_wall_Eq_2}
\end{align}
where $\Delta  V_T$ denotes the difference in finite-temperature effective potential between the two phases and $W$ is the integral
\begin{equation}
W \equiv  \int_0^{\phi_0}  \frac{d V(\phi,\,T)}{d\phi} \,(2\phi \,- \, \phi_0)\,d\phi.
\end{equation}
$\Delta  V_T$ measures the \textbf{pressure difference} while $W$ measures the \textbf{pressure gradient}. The functions $F$ and $G$ measure the deviation from local thermal equilibrium and can be computed in close-to-equilibrium \textbf{thermal field theory} \cite{Moore:1995si,Moore:2000wx,John:2000zq,Huber:2013kj, Konstandin:2014zta, Dorsch:2018pat,Friedlander:2020tnq,Dorsch:2021nje}.
The resolution of the Eq.~\eqref{eq:system_wall_Eq_1} and Eq.~\eqref{eq:system_wall_Eq_2} allows to determine the \textbf{wall velocity} $v_w$ and the \textbf{wall thickness} $L_{\rm w}$.\footnote{Since plasma effects enters the scalar field equation of motion in Eq.~\eqref{eq:Higgs_profile_thermal}, it was expected that the wall thickness would depend on the plasma properties and wall velocity.}

\paragraph{An hydrodynamic approach:}

We can also choose to describe the friction term in Eq.~\eqref{eq:friction_term_wall} in a fluid approach, for instance, starting from the ansatz \cite{Ignatius:1993qn,Megevand:2009ut, Megevand:2009gh,Espinosa:2010hh,Leitao:2015ola,Leitao:2015fmj,Megevand:2017vtb}
\begin{equation}
K(\phi) = \eta\, u^\mu\,\partial_\mu\phi,
\end{equation}
where $\eta$ is a model-dependent parameter encoding the friction efficiency ($[\eta]\rm = energy$). For instance in the SM, we expect $\eta \approx 3\,T$ \cite{Espinosa:2010hh}.
The fluid velocity $v$, the temperature $T$ and the scalar profile $\phi$ obey to the following combined set of equations 
\begin{align}
&\partial_z^2 \phi - \frac{\partial V(\phi,\,T)}{\partial \phi} + \eta\, u^\mu\,\partial_\mu\phi = 0, \\
& \partial_z\left[ \omega \gamma^2 v\right] = 0, \\
& \partial_z\left[ \frac{1}{2}(\partial_z \phi)^2 + \omega \gamma^2 v^2 + p \right] = 0,
\end{align}
where $\omega = \rho+p$ is the fluid enthapy. The last two equations results from the conservation of the components $T_{z0}$ and $T_{00}$ of the energy-momentum tensor. See Ref.~\cite{Espinosa:2010hh} for a determination of the wall velocity in this approach.

\subsection{Friction pressure in the ballistic approximation}
\label{sec:ballistic_pressure}

\paragraph{Collisionless regime:}

For  fast moving wall, we can expect  large departure from local thermal equilibrium to take place, such that close-to-equilibrium thermal field techniques or fluid approaches, as mentioned in the previous section, break down. This is particularly true when the time scale, $L_w/\gamma_w\,v_w$, over which the background change, is much shorter than the  collision time scale, $1/\Gamma_{\rm coll} \sim 1/ \alpha^2\,T$, where $T$ is the plasma temperature and $\alpha$ is the typical coupling constant responsible for the interactions
\begin{equation}
\frac{L_w}{\gamma_w\,v_w} \quad  \ll  \quad \frac{1}{\alpha^2\,T}, \qquad \qquad \textrm{(ballistic regime)}.
\label{eq:out_of_eq_cond_wall}
\end{equation}
$v_w$ and $\gamma_w$ denotes the wall velocity and Lorentz factor with respect to the bubble center.
In that regime, we can neglect the collision term $``\rm coll''$ in the Boltzmann equation in Eq.~\eqref{eq:Liouville_equation_massaged} which becomes
\begin{equation}
\left( \frac{p_z}{E} \partial_z ~ -~ \frac{\partial_z m^2}{2E} \partial_{p_z} \right) f(p_z,\, z)=0,
\label{eq:Liouville_equation_massaged_2}
\end{equation}
and we can consider the particles as evolving independently from each others, this is the \textbf{ballistic regime}. 

\paragraph{Solution of the Liouville equation:}
The solution of the Liouville equation in Eq.~\eqref{eq:Liouville_equation_massaged_2} is any function of the form
\begin{equation}
\label{eq:solution_liouville}
f(\vec{r},\, \vec{p},\,t)  = f(p_z^2 + m(z)^2, \,\vec{p}_\perp).
\end{equation}
Its exact form is determined using the \textbf{boundary conditions} at $z \to \pm \infty$, where local thermal equilibrium is well-satisfied. Indeed, far from the wall, the collision term $``\rm coll''$ is large such that local thermal equilibrium is satisfied and the occupation number is
\begin{equation}
\label{eq:liouville_BC}
f({z \to \pm \infty}) = \frac{1}{e^{\beta p_\mu u^\mu} - 1 } = \frac{1}{e^{\beta \gamma (E - v p_z)} \mp 1},
\end{equation}
with a minus/plus sign for bosons/fermions.
The second term is Lorentz-invariant whereas the third one is expressed in the wall frame, with $v\equiv v_w$ and $\gamma \equiv \gamma_w$.

From Eq.~\eqref{eq:solution_liouville} and Eq.~\eqref{eq:liouville_BC}, we obtain
\begin{equation}
f = 
\begin{cases}
\dfrac{1}{e^{\beta \gamma \left(E - v \sqrt{p_z^2 + m(z)^2} \right)}\mp 1}  &\hspace{2em}\textrm{if coming from the symmetric phase},\\
\dfrac{1}{e^{\beta \gamma \left(E + v \sqrt{p_z^2 + m(z)^2 - m_0^2} \right)}\mp 1}  &\hspace{2em}\textrm{if coming from the broken phase},\\
\end{cases}
\end{equation}
where $m_0$ is the mass of the particle deep inside the broken phase (inside the bubble).

The \textbf{physical intuition} behind the final solutions is that $E = \sqrt{p_z^2 + p^2_\perp + m(z)^2}$ and $p_\perp$ are conserved across the bubble wall, since time- and $\perp$- translation are symmetries of the problem. This implies that $\sqrt{p_z^2 + m(z)^2}$ is conserved. Note that $p_z$ is not conserved since invariance under $z$-translation is broken (spontaneously) by the wall. Conservation of the number of particles passing through the wall implies that $f$ must depend on $\sqrt{p_z^2 + m(z)^2}$ and not on $p_z$.

\begin{figure}[h!]
\centering
\raisebox{0cm}{\makebox{\includegraphics[width=0.9\textwidth, scale=0.5]{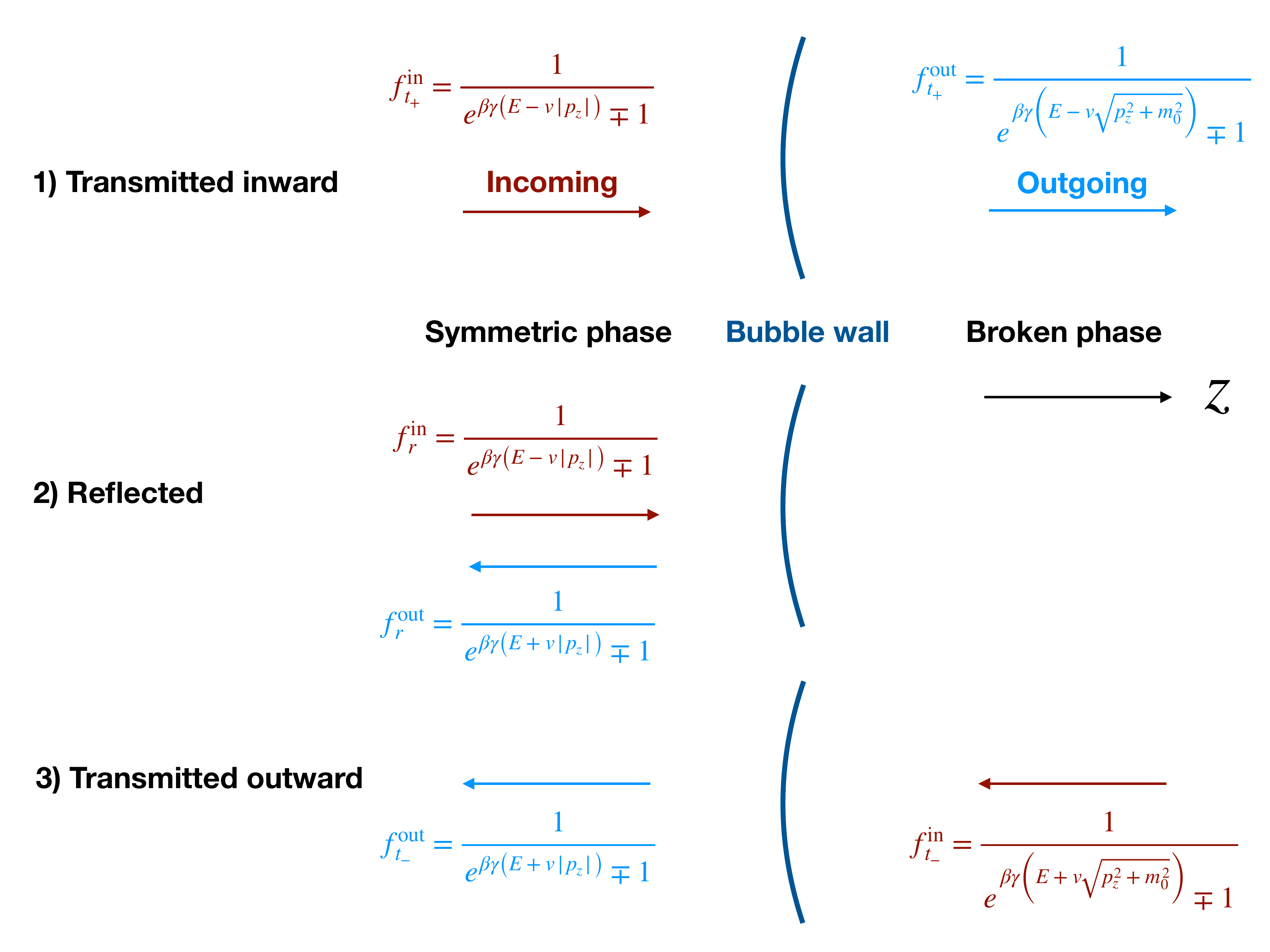}} }
\caption{\it \small  Three contributions to the ballistic pressure. The occupancy functions $f$ are solutions of the Liouville equation is Eq.~\eqref{eq:solution_liouville} with the boundary condition in Eq.~\eqref{eq:liouville_BC}. We distinguish incoming and outgoing particles, with respect to the bubble wall, with red and blue colors respectively.}
\label{fig:ballistic_pressure_contributions}
\end{figure}

\paragraph{Three contributions to the pressure:}
In the ballistic approach, the \textbf{local} pressure due to plasma effects, which enters the bubble wall equilibrium condition in Eq.~\eqref{eq:pressure_equi_bubble_wall}, is
\begin{equation}
P_{\rm plasma} = \int \frac{d^3p}{(2\pi)^3} \frac{p_z^2}{E} \, f(\vec{r},\, \vec{p},\,t),
\end{equation}
and it receives \textbf{three contributions}, as displayed in Fig.~\ref{fig:ballistic_pressure_contributions} and detailed below \cite{Liu:1992tn, Dine:1992wr,Moore:1995si,Mancha:2020fzw}.
\begin{enumerate}
\item
First, there is the contribution due to the particles going from the \textbf{symmetric} to the \textbf{broken} phase, which reads
\begin{multline}
\Delta P_{t+} =  \int_{p_z > m} \frac{d^3p}{(2\pi)^3} \dfrac{1}{e^{\beta \gamma (\sqrt{p_r^2 + p_z^2} - v p_z)} \mp 1}\left( \frac{p_z^2}{\sqrt{p_r^2 + p_z^2}} \right) \\-   \int_{p_z > 0} \frac{d^3p}{(2\pi)^3} \dfrac{1}{e^{\beta \gamma (\sqrt{p_r^2 + p_z^2+m^2} - v \sqrt{p_z^2+m^2})} \mp 1}\left( \frac{p_z^2}{\sqrt{p_r^2 + p_z^2+m^2}} \right).
\end{multline}
This is the difference between the local pressure in the deep symmetric phase minus the pressure in the deep broken phase, due to particles going in the $+z$ direction. We have assumed the $z$-axis pointing in the broken phase direction. We have denoted $m$ as the mass deep inside the broken phase ($m\equiv m_0$).
\item
Second, we have the contribution coming from the particles getting \textbf{reflected}, which reads
\begin{equation}
\Delta P_{r}  = \int_{-m < p_z < m} \dfrac{d^3p}{(2\pi)^3} \frac{1}{e^{\beta \gamma (\sqrt{p_r^2 + p_z^2} - v p_z)} \mp 1}\left( \frac{p_z^2}{\sqrt{p_r^2 + p_z^2}} \right) .
\end{equation}
We have defined $\Delta P_{r} = P^{|p_z|<m}_{\rm sym} - P_{\rm turn}^{p_z=0}$ where $P^{|p_z|<m}_{\rm sym}$ is the pressure in the deep symmetric phase resulting from the particles whose momentum is insufficient to climb up to the potential energy of the wall, $|p_z|<m$, considered either before ($p_z>0$) or after reflection ($p_z<0$). $P_{\rm turn}^{p_z=0}$ is the pressure at the turning-point position and is identically zero since the phase space of $p_z = 0$ is zero. 
\item
Finally, there is the contribution due to the particles passing from the \textbf{broken} to the \textbf{symmetric} phase
\begin{multline}
\Delta P_{t-} =   \int_{p_z < 0} \frac{d^3p}{(2\pi)^3} \dfrac{1}{e^{\beta \gamma (\sqrt{p_r^2 + p_z^2+m^2} + v \sqrt{p_z^2})} \mp 1}\left( \frac{p_z^2}{\sqrt{p_r^2 + p_z^2+m^2}} \right) \\ -\int_{p_z < -m} \frac{d^3p}{(2\pi)^3} \dfrac{1}{e^{\beta \gamma (\sqrt{p_r^2 + p_z^2} + v \sqrt{p_z^2 - m^2})} \mp 1}\left( \frac{p_z^2}{\sqrt{p_r^2 + p_z^2}} \right) .
\label{eq:contribution_3_ballistic_pressure}
\end{multline}
This is the difference between the local pressure in the deep broken phase minus the pressure in the deep symmetric phase, due to particles going in the $-z$ direction.  We can show that this last contribution is \textbf{suppressed} like $\Delta P_{t-} \propto e^{-\beta \gamma m}$ or stronger \cite{Mancha:2020fzw}, so most of the time it is negligible.
\end{enumerate}
Upon changing variable $p_r\,dp_r = E\,dE$, integrating over $E$ between $p_z$ and $\infty$, and performing a shift $p_z^2 \to p_z^2 - m^2$, we obtain
\begin{align}
&\Delta P_{t+} = \mp \frac{1}{4\pi^2} \frac{m^3}{\beta\gamma\,M_-^3}\, \Big[ \int_{M_-}^\infty dx \, x^2\, \textrm{log}\left[ 1\mp e^{-x} \right] - \int_{0}^\infty dx \, x^2\, \textrm{log}\left[ 1\mp e^{-\sqrt{x^2+M_-^2}} \right] \Big], \label{eq:Pt+_log} \\
&\Delta P_{r} = \mp \frac{1}{4\pi^2} \frac{m^3}{\beta\gamma}\, \Big[ \frac{1}{M_-^3}\int_{0}^{M_-} dx \, x^2 \, \textrm{log}\left[ 1\mp e^{-x} \right]   + \frac{1}{M_+^3}\int_{0}^{M_+} dx \, x^2 \, \textrm{log}\left[ 1\mp e^{-x} \right] \Big] ,\label{eq:Pr_log} \\
&\Delta P_{t-} = \mp \frac{1}{4\pi^2} \frac{1}{\beta^4\gamma^4}\,\int_{0}^\infty dx \, x\left[\sqrt{x^2+(\beta\,\gamma \,m)^2} ~-~ x\right]\, \textrm{log}\left[ 1\mp e^{-\sqrt{x^2+(\beta\,\gamma \,m)^2} - v\,x} \right],\label{eq:Pt-_log}
\end{align}
where $M_- = m\,\beta\,\gamma\,(1-v)$ and $M_+ = m\,\beta\,\gamma\,(1+v)$. 
\begin{figure}[h!]
\centering
\raisebox{0cm}{\makebox{\includegraphics[width=0.7\textwidth, scale=0.5]{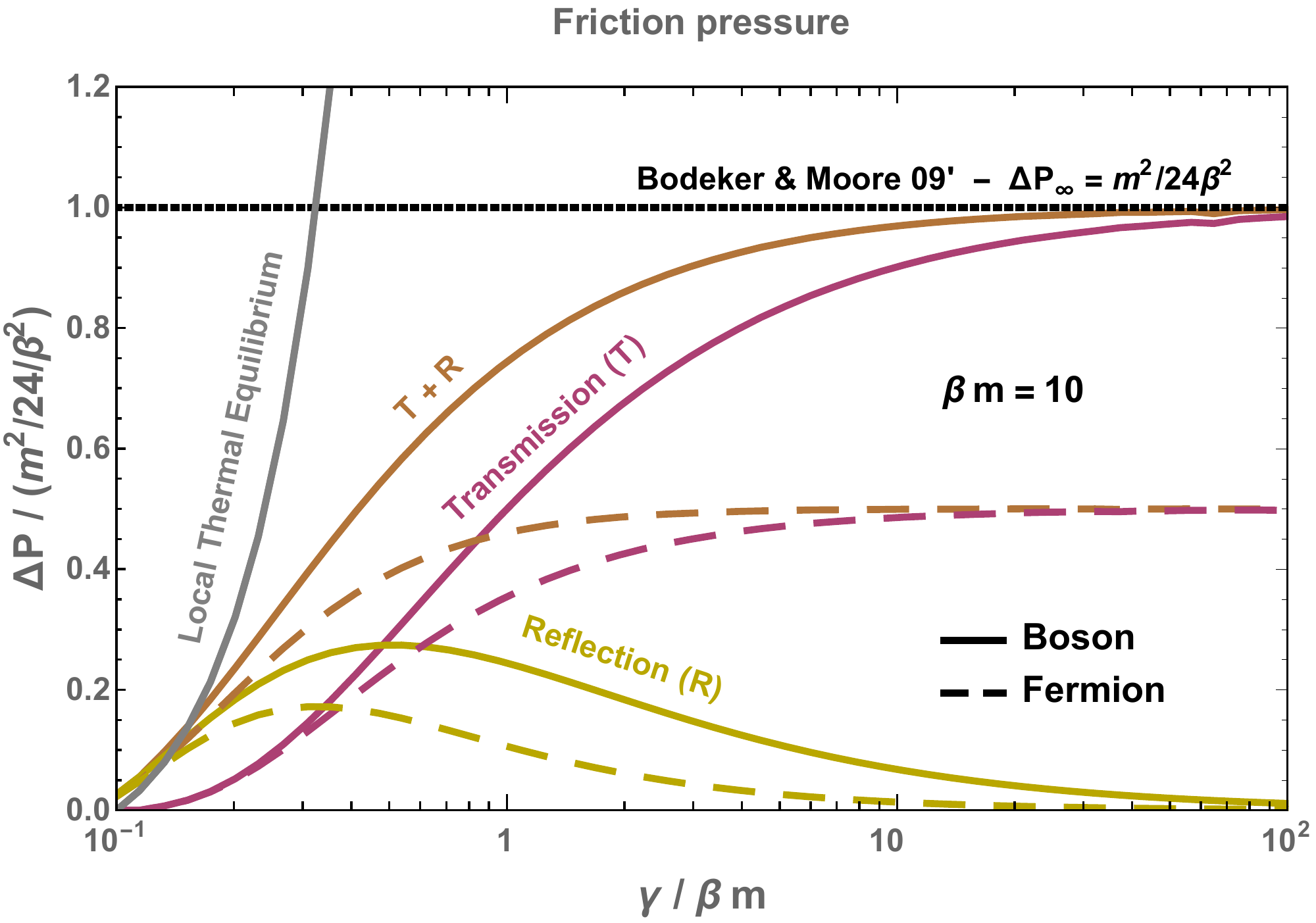}} }
\caption{\it \small   Retarding friction pressure on the bubble wall versus the Lorentz factor of the wall. The pressure computed using local thermal equilibrium (gray lines), cf. Eq.~\eqref{eq:pressure_local_thermal_1}, increases as $\Delta P_{\rm lte} \propto \gamma^2$, implying that the bubble wall would never run away. However, for $\gamma \gtrsim 1$, local thermal equilibrium is violated. Instead, for relativistic wall velocities the ballistic approximation (brown lines) is more trustable. We decompose the later into the contribution from particles entering the bubble (purple lines, Eq.~\eqref{eq:Pt+_log} and case 1 in Fig.~\ref{fig:ballistic_pressure_contributions}) and the contribution from the particles reflected off the wall (yellow lines, Eq.~\eqref{eq:Pr_log} and case 2 in Fig.~\ref{fig:ballistic_pressure_contributions}), the particles being either bosons (solid) or fermions (dashed), both having a unique degree of freedom. The third contribution to the ballistic pressure, due to particles going from the broken phase to the symmetric phase (Eq.~\eqref{eq:Pt-_log} and case 3 in Fig.~\ref{fig:ballistic_pressure_contributions}), is suppressed like $e^{-\beta \gamma m}$ and is not represented. We can see that the ballistic pressure saturates to the Bodeker and Moore value from 2009 \cite{Bodeker:2009qy} at large $\gamma$ (the asymptotic pressure receives an extra factor $1/2$ for fermions).}
\label{fig:DeltaP_VS_gamma_T_vs_R}
\end{figure}
We compare the different contributions to the ballistic pressure in Fig.~\eqref{fig:DeltaP_VS_gamma_T_vs_R}. As expected the dominant contribution comes from particles entering the bubbles ($t+$) and the contribution coming from particles going out of the bubble ($t-$) is negligible (not shown on the plot). The contribution from the \textbf{reflected} particles is \textbf{relevant} for $\gamma \lesssim \beta \, m$.

\paragraph{Total ballistic pressure:}
Upon combining the three contributions\footnote{In order to get Eq.~\eqref{eq_pressure_ballistic_bosons} and Eq.~\eqref{eq_pressure_ballistic_fermions}, we have used the relation 
\begin{equation}
\int_0^M dx \, x^2 \,{\textrm{log}}(1-e^{-x}) = -\frac{\pi^4}{45} + M^2 \,\textrm{Li}_2(e^{-M}) + 2M\, \textrm{Li}_3(e^{-M})  + 2\, \textrm{Li}_4(e^{-M}),
\end{equation}
and
\begin{equation}
\int_0^M dx \, x^2 \,{\textrm{log}}(1+e^{-x}) = \frac{7\pi^4}{360} + M^2 \,\textrm{Li}_2(-e^{-M})+ 2M\, \textrm{Li}_3(-e^{-M}) + 2\, \textrm{Li}_4(-e^{-M}).
\end{equation}
}, we obtain the \textbf{total pressure} in the ballistic approximation \cite{Mancha:2020fzw}
\begin{multline}
\Delta P_{\rm ballistic}^b= \frac{\pi^2}{90\beta^4}\left[ 4(\gamma^2-1)+1 \right] - \frac{1}{4\pi^2\beta^4\gamma^4} \Big[ - \frac{J_{b}(\beta \gamma(1-v)m) }{(1-v)^3} \\+ \frac{\big[M_+^2 \,\textrm{Li}_2(e^{-M_+}) + 2M_+\, \textrm{Li}_3(e^{-M_+})  + 2\, \textrm{Li}_4(e^{-M_+})\big]}{(1+v)^3}\\
+ \int_{0}^\infty dx \, x\left[\sqrt{x^2+(\beta\,\gamma \,m)^2} ~-~ x\right]\, \textrm{ln}\left[ 1- e^{-\sqrt{x^2+(\beta\,\gamma \,m)^2} - v\,x} \right] \Big],
\label{eq_pressure_ballistic_bosons}
\end{multline}
for \textbf{bosons} and 
\begin{multline}
\Delta P_{\rm ballistic}^f= \frac{7}{8}\frac{\pi^2}{90\beta^4}\left[ 4(\gamma^2-1)+1 \right] + \frac{1}{4\pi^2\beta^4\gamma^4} \Big[ - \frac{J_{f}(\beta \gamma(1-v)m) }{(1-v)^3} \\
+\frac{\big[ M_+^2 \,\textrm{Li}_2(-e^{-M_+})+ 2M_+\, \textrm{Li}_3(-e^{-M_+}) +  2 \textrm{Li}_4(-e^{-M_+}) \big]}{(1+v)^3} \\
+ \int_{0}^\infty dx \, x\left[\sqrt{x^2+(\beta\,\gamma \,m)^2} ~-~ x\right]\, \textrm{ln}\left[ 1+ e^{-\sqrt{x^2+(\beta\,\gamma \,m)^2} - v\,x} \right] \Big],
\label{eq_pressure_ballistic_fermions}
\end{multline}
for \textbf{fermions}.\footnote{Ref. \cite{Mancha:2020fzw} only studied the boson case.}
The functions $J_{b}$ and $J_f$ where already defined in Eq.~\eqref{eq:Jb_Jf} and ${\rm Li_s}(z) = \sum_{n=1}^{\infty} (z^n/n^s)$ are the polylogarithm functions. 

\paragraph{Ultra-relativistic limit:}

In contrast to the l.t.e pressure presented in Sec.~\ref{sec:lte_pressure}, the ballistic pressure saturates at large $\gamma$. In this regime, the only relevant contribution to the ballistic pressure arises from particles entering the bubble, cf.$  $ Eq.~\eqref{eq:Pt+_log}, which in the large $\gamma$ limit, reads
\begin{align}
\Delta P_{t+} &= \mp \frac{1}{4\pi^2}~ \frac{m^3}{\beta \gamma \,M_-^3}~\int_{M_-}^{\infty} dx \, x\left(x\,-\,\sqrt{x^2 \, - \, M_{-}^2} \right) ~{\rm log} \left[1 \, \mp\, e^{-x} \right]  \\
&\underset{\gamma \to \infty}{\xrightarrow{\hspace*{0.5cm}}}~ \frac{1}{24} \frac{m^2}{\beta^2}  ~\begin{cases}
1&\hspace{2em}\textrm{bosons},\\
1/2 &\hspace{2em}\textrm{fermions}.\\
\end{cases}
\end{align}
This is the result found in 2009 by Bodeker\&Moore \cite{Bodeker:2009qy}.

\paragraph{Non-relativistic limit:}


At small $v$ the ballistic pressure does not vanish, see Fig.~\ref{fig:DeltaP_VS_gamma_T_vs_R}, in contrast to the pressure at local thermal equilibrium in Eq.~\eqref{eq:pressure_local_thermal_1}, which scales as $P_{\rm lte} \propto v^2$ at small $v$. It was expected that the ballistic approximation breaks down at small $v$ where we can no longer neglect particle interactions.
A possible prescription for an easy-to-use retarding pressure, which vanishes at small $v$ and saturates at large $v$ is
\begin{equation}
P_{\rm lte+ballistic}(v) \, =\, \textrm{Min} \left[\Delta P_{\rm lte}(v),~ P_{\rm ballistic}(v)\right]
\label{eq:pressure_lte_ballistic}
\end{equation}
where $P_{\rm lte}$ is the l.t.e pressure defined in Eq.~\eqref{eq:pressure_local_thermal_1} and Eq.~\eqref{eq:pressure_local_thermal_2}, and $P_{\rm ballistic}$ is the ballistic pressure defined in Eq.~\eqref{eq_pressure_ballistic_bosons} and Eq.~\eqref{eq_pressure_ballistic_fermions}.

\subsection{Friction pressure at NLO}
\label{sec:NLO_pressure}

\paragraph{Transition splitting:}

In the previous section, we have presented the retarding pressure in the ballistic approximation where particle interactions were neglected. In this section we introduce a possible correction which arises in the presence of a finite gauge coupling constant. The correction term, which is called NLO pressure, arises from the possibility for the incoming particle to radiate a soft boson which gets a mass in the broken phase \cite{Bodeker:2017cim}
\begin{equation}
\mathcal{P}_{1\to 2} =\sum_{a} \nu_a \int \frac{d^3 p}{(2\pi)^3}f_a(p)\,\frac{p_z}{p_0} \times \sum_{bc} \int dP_{a\to bc} \times (p_{z,s} - k_{z,h} - q_{z,h}),
\end{equation}
where $h,s$ stands for the `Higgsed' and the symmetric phases. $p$ and $q$ are the momenta of the incoming particle before and after the splitting while $k$ is the momentum of the radiated boson. We summed over all the species $a$ likely to participate in the process, $\nu_a$ being their number of degrees of freedom. The differential splitting probability is given by
\begin{equation}
\int dP_{a\to bc} \equiv \int  \frac{d^3 k}{(2\pi)^3 2k^0}  \frac{d^3 q}{(2\pi)^3 2q^0} \left< \phi | \mathcal{T} |k,q\right> \left< k,q | \mathcal{T} |\phi \right>,
\end{equation}
with the transition element
\begin{align}
 \left< k,q | \mathcal{T} |p \right> &= \int d^4x  \left< k,q | \mathcal{H}_{\rm int} |p \right>,\\
 &= (2\pi)^3 \,\delta (\vec{p}_\perp - \vec{k}_\perp -\vec{q}_\perp) \, \delta (p_0 - k_0 -q_0) \,\mathcal{M},
\end{align}
where
\begin{equation}
\label{eq:M_matrix}
\mathcal{M} \equiv \int dz ~ \chi_k^*(z) \chi_q^*(z) \chi_k(z) V(z)\chi_p(z).
\end{equation}
We obtain \cite{Bodeker:2017cim}
\begin{multline}
\mathcal{P}_{1\to 2} =\sum_{a,bc} \nu_a \int \frac{d^3 p}{(2\pi)^3 2p_0}f_a(p) ~\frac{d^3 k}{(2\pi)^3 2k^0} ~ \frac{d^3 q}{(2\pi)^3 2q^0}~ [1\pm f_k][1\pm f_q] ~(p_{z,s} - k_{z,h} - q_{z,h})\\
\times  (2\pi)^3 \,\delta (\vec{p}_\perp - \vec{k}_\perp -\vec{q}_\perp) \, \delta (p_0 - k_0 -q_0) ~|\mathcal{M}|^2.
\end{multline}
Now we assume $p_z \simeq p_0$, $q_z \simeq q_0 \simeq p_0$ and $k_z \simeq k_0 -\dfrac{m_V^2(z)+k_{\perp}^2}{2k_0}$, from which we get
\begin{equation}
\label{eq:exchanged_momentum_NLO_BM}
p_{z,s} - k_{z,h} - q_{z,h} \simeq \frac{m_V^2(z)+k_{\perp}^2}{2k_0},
\end{equation}
and
\begin{equation}
\mathcal{P}_{1\to 2} =\sum_{a,\,bc} \nu_a \int \frac{d^3 p}{(2\pi)^3 (2p_0)^2}f_a(p) \frac{d^2 k_\perp}{(2\pi)^2 }  \frac{d k_0}{(2\pi) 2k^0} [1\pm f_k][1\pm f_{p-k}] \frac{m_V^2(z)+k_{\perp}^2}{2k_0}~|\mathcal{M}|^2.
\label{eq:P1to2_interm}
\end{equation}

\paragraph{WKB approximation:}
Next, we make use of the WKB approximation, 
\begin{equation}
\chi_k(z) \simeq {\rm exp } \left(i \int_0^z k_z(z') dz' \right) \simeq e^{ik_0 z} \exp \left(  - \frac{i}{2k_0} \int_0^z(m^2(z')+k^2_\perp)~ dz' \right),
\end{equation}
which allows to write the product of wave functions in terms of a phase-dependent quantity $A$.
\begin{equation}
\chi_a(z)\chi_b^*(z)\chi_c^*(z) = {\rm  exp} \left( \frac{i}{2p^0} \int_0^z ~A(z')~dz' \right)
\label{eq:M_matrix_1p5}
\end{equation}
with 
\begin{equation}
-A = m_a^2 - \frac{m_b^2 + k_\perp^2}{x} -\frac{m_c^2 + k_\perp^2}{1-x} \simeq \dfrac{k^2_{\perp}+m_b^2}{x}.
\end{equation}
We have introduced the variable $x \equiv k^0 /p^0$ and assumed $x \ll 1$ in the last equality.
We can now split the integral over $z$ across the wall in Eq.~\eqref{eq:M_matrix} into a contribution from the broken phase and a contribution from the symmetric phase. We assume that the vertices $V$ and WKB phases $A$ on each side of the wall are $z$-independent and we denote them by ($V_h$, $A_h$) and ($V_s$, $A_s$), such that we obtain
\begin{equation}
\mathcal{M} \simeq V_s \int_{-\infty}^0 dz\,{\rm exp} \left[ iz\frac{A_s}{2p^0} \right] + V_h \int_{0}^{\infty} dz\,{\rm exp} \left[ iz\frac{A_h}{2p^0} \right] = 2ip^0 \left(  \frac{V_h}{A_h} - \frac{V_s}{A_s} \right).
\label{eq:M_matrix_2}
\end{equation}

\paragraph{Radiation of a soft transverse boson:}
It can be shown \cite{Bodeker:2017cim} that the most important process contributing to the pressure at large $p_0$ is $X(p) \to V_T(k)~ X(q)$ where $V_T$ is a transverse vector boson. The corresponding vertex function is phase-independent, $V_h=V_s$, and equal to
\begin{equation}
|V^2| = 4\,g^2\,C_2[R]\,\frac{1}{x^2}\,k_\perp^2,
\end{equation}
where $g$ is the gauge coupling constant and $C_2[R]$ is the second casimir of the representation $R$ of $SU(N)$ 
\begin{equation}
C_2[R] = \left\{
			\begin{array}{ll}
                 \frac{N^2-1}{2N} \quad \text{if R= fundamentale,} \\
                 N \quad \text{if R= adjoint.}
                \end{array}
              \right.
\end{equation}
Therefore, Eq.\eqref{eq:M_matrix_2} becomes 
\begin{equation}
|\mathcal{M}|^2 \simeq 16\,g^2 \,C_2[R] \,p_0^2 \,\frac{m_V^4}{k_\perp^2\,(k_\perp^2+m_V^2)^2}.
\end{equation}
where we have replaced $m_V \equiv m_b$.
The $k_\perp$ integral becomes
\begin{equation}
\int \frac{d^2 k_\perp}{(2\pi)^2} ~\frac{1}{k_\perp^2(k_\perp^2+m_V^2)} = \frac{\ln\left( 1+\frac{m_V^2}{k_*^2} \right)}{4\,\pi \,m_V^2},
\end{equation}
where $k_*$ is the IR cut-off on $k_{\perp}$. 

\paragraph{Final NLO pressure:}
Finally, injecting the last two equations into Eq.~\eqref{eq:P1to2_interm} yields
\begin{equation}
\mathcal{P}_{1\to 2} =\sum_{a,\,bc} \nu_a \int \frac{d^3 p}{(2\pi)^3 }f_a(p)\frac{d k_0}{(2\pi) k_0^2} [1\pm f_k][1\pm f_{p-k}] ~ g^2 \,C_2[R] \, m_V^2\,\frac{\ln\left( 1+\frac{m_V^2}{k_*^2} \right)}{4\pi}.
\end{equation}
The Pauli blocking or Bose enhancing factor $1 \pm f_{p-k}$ is of order $1$, while $1\pm f_k$ sums to $1$ when considering both absorption and emission processes.
Hence, the result simplifies to
\begin{equation}
\mathcal{P}_{1\to 2} =\sum_{a} \nu_a\,b_a\, \frac{\zeta(3)}{\pi^2} T^3 \,\gamma_{\rm wp}\, 8\pi\,\alpha \,C_2[R]\,  m_V~\frac{\ln\left( 1+\frac{m_V^2}{k_*^2} \right)}{k_*/m_V},
\label{eq:BM_result}
\end{equation}
where $b_a = 1~(3/4)$ for bosons (fermions) and $\alpha \equiv g^2/4\pi$. The Lorentz factor $\gamma_{\rm wp}$ between the the wall and the plasma comes from $d^3p$. Taking $k_* \sim m_V$, we obtain

\begin{equation}
\gamma \Delta P_{\rm NLO} \equiv \mathcal{P}_{1\to 2} \simeq \sum_{a} \nu_a\,b_a\, \frac{\zeta(3)}{\pi^2} \,T^3 \,\gamma_{\rm wp}\,4\pi\, \alpha \,C_2[R]\,  m_V.
\label{eq:p12}
\end{equation}

\paragraph{Beyong NLO:}

While this chapter was completed, this paper \cite{Hoeche:2020rsg} came out, where the authors aim at improving the B\&M17 result \cite{Bodeker:2017cim} in Eq.~\eqref{eq:BM_result} by resumming soft boson emission processes to all orders in perturbation theory and by obtaining a gauge-invariant matrix element. The authors find that the exchanged momentum $\Delta p \sim m_V$ in Eq.~\eqref{eq:exchanged_momentum_NLO_BM} becomes $\Delta p \sim \gamma\, T$, which implies that the NLO pressure $ \mathcal{P}_{1\to 2} \sim \alpha\,\gamma_{\rm wp}\,m_V\,T^3$ becomes  $ \mathcal{P}_{1\to 2} \sim \alpha\,\gamma_{\rm wp}^2\,T^4$. As pointed out in \cite{Vanvlasselaer:2020niz}, the result from \cite{Hoeche:2020rsg} does not vanish as it should in the symmetry restoration limit $\left<\phi\right> \to 0$. 

Later in \cite{Gouttenoire:2021kjv}, together with Ryusuke Jinno and Filippo Sala, we showed that by breaking energy-momentum conservation, (or by preventing the onshellness of the momenta at the splitting vertex), the presence of the wall boundary prevents the trivial applicability of the Ward identity at the level of the single splitting vertex function. Without any additional ingredient standing for the presence of the wall in the Feynmann rules, gauge invariance is not expected, which is at odd with the finding of \cite{Hoeche:2020rsg}. Instead, the B\&M17 result is not gauge invariant but assumes physical polarizations. In \cite{Gouttenoire:2021kjv}, we show that the average exchanged momentum in which both real and virtual leading-log emissions are resummed is identical to the naive expectation using the perturbative splitting probability as in the original B\&M17 paper  \cite{Bodeker:2017cim}. Additionally, we compute the IR cut-off $k_*$ which regulates the logarithmic divergence, we point out that a dominant fraction of radiated gauge bosons are too soft to enter the broken phase and discuss their fate, we compute effects from wall finite thickness and we discuss various approximations entering the computation.

\begin{figure}[htp!]
\centering
\raisebox{0cm}{\makebox{\includegraphics[width=0.7\textwidth]{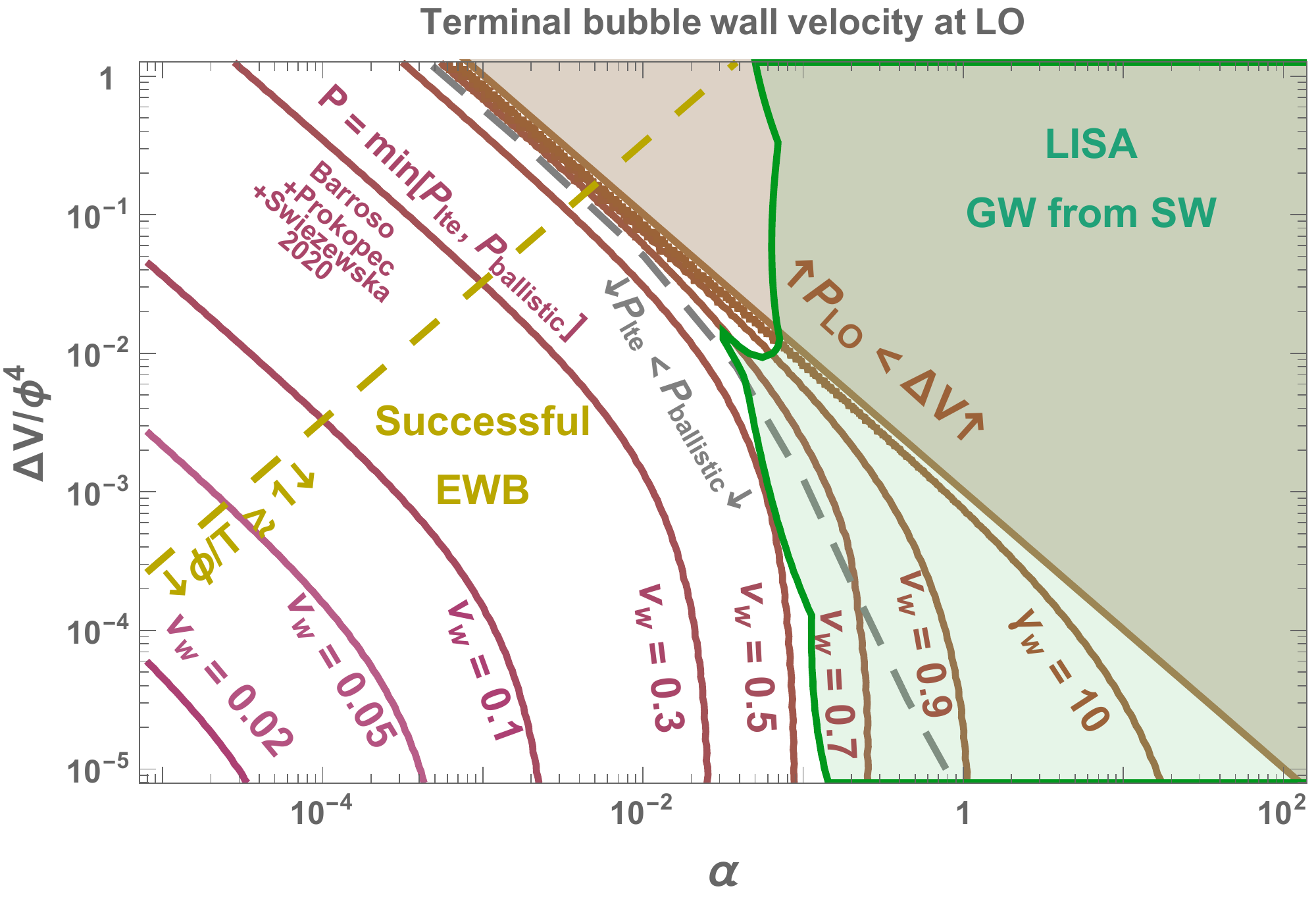}} }
\raisebox{0cm}{\makebox{\includegraphics[width=0.7\textwidth]{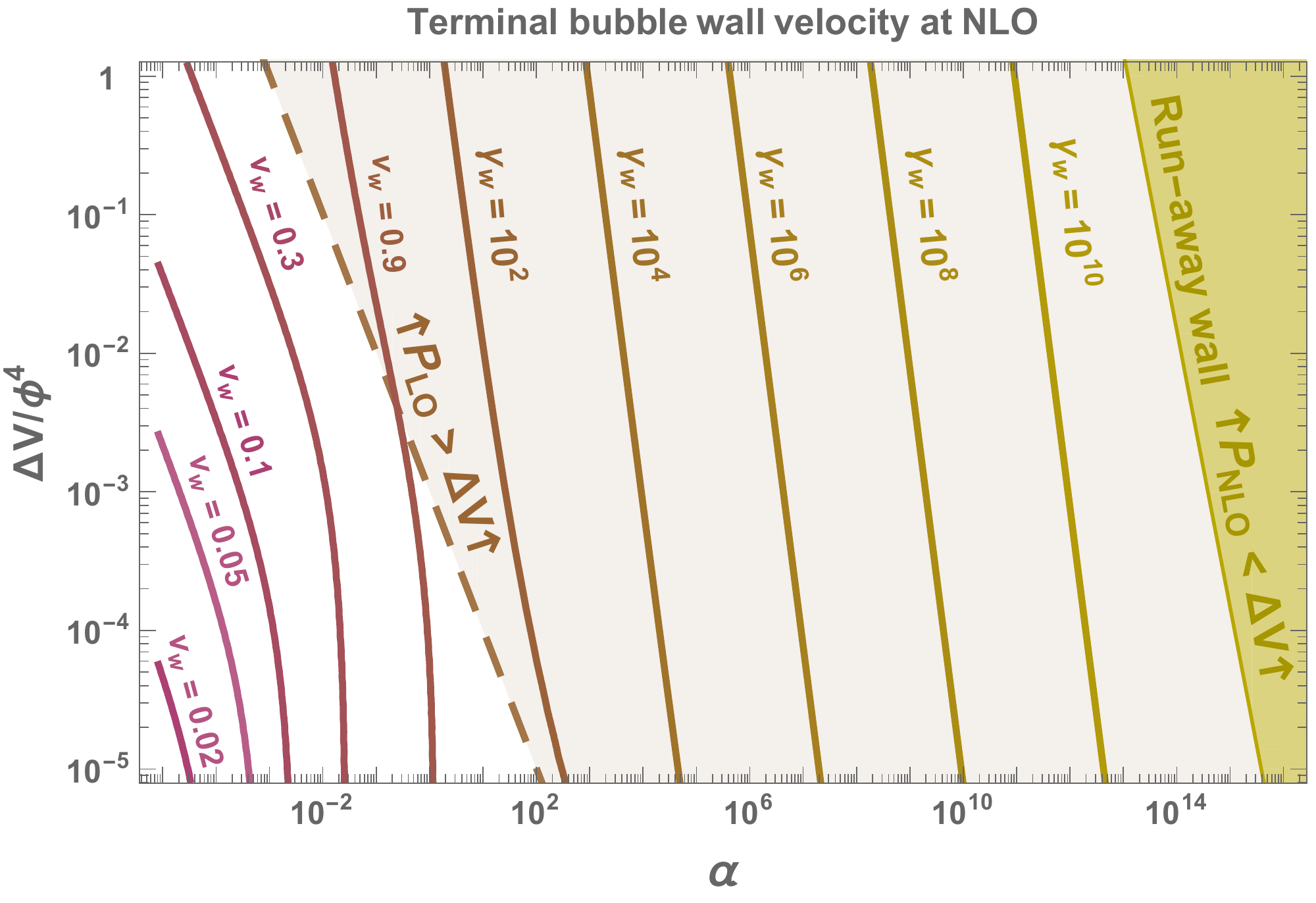}} }
\caption{\it \small    \textbf{Top:} Bubble wall velocity at equilibrium $\Delta P_{\rm LO}(v)=\Delta V_{\rm vac}$ with $\Delta P_{\rm LO}$ defined in Eq.~\eqref{eq:pressure_lte_ballistic_sum_species}, in a minimal extension of the SM including the contributions from $t,\,W^\pm,\,Z$ and $h$. The reach of LISA on the GW spectrum from sound-waves (SW), see Sec.~\ref{sec:GW_SW_turb} requires large $\alpha$. We might worry that the corresponding large wall velocity $v_w \gtrsim 0.5$ could jeopardize successful EW baryogenesis (EWBG). However, successful EWBG only requires the fluid velocity at the wall $v_+$ to be sub-sonic $v_+\lesssim 0.5$, which can be much smaller than $v_w$ for strong phase transition (large $\alpha$) \cite{No:2011fi}. Note also, the claim that EWBG can also work for super-sonic $v_+$ in \cite{Cline:2020jre,Laurent:2020gpg,Dorsch:2021ubz,DeCurtis:2022hlx}.
\textbf{Bottom:} We include the NLO pressure $P_{\rm NLO}(v)=\Delta V_{\rm vac}$ using the trick explained along Eq.~\eqref{eq:PLO_ballistic_sum_species_plus_NLO} for $v_{w} \gtrsim 0.9$. In the yellow region, the vacuum energy is larger than the NLO pressure at the time of bubble collisions, such that bubbles never stop accelerating until they collide, cf. Eq.~\eqref{eq:run_away_wall_cond}.  }
\label{eq:DeltaV_alpha_contours_v_LO}
\end{figure}

\subsection{Speed of the wall}
\label{sec:speed_wall}

\paragraph{At LO:}

As already discussed at the end of Sec.~\ref{sec:ballistic_pressure}, a possible prescription for the pressure at LO in gauge coupling constant, which presents the correct behavior at small $v$ (goes to zero) and at large $v$ (saturates) is
\begin{equation}
\Delta P_{\rm LO}(v)  =  \sum_a\, \nu_a ~\textrm{Min} \left[\Delta P_{\rm lte}^{b/f}(v),~ \Delta P_{\rm ballistic}^{b/f}(v)\right]
\label{eq:pressure_lte_ballistic_sum_species}
\end{equation}
where we summed over all the species in the plasma with $\nu_a$ degrees of freedom. $\Delta P_{\rm lte}^{b/f}$ is the l.t.e pressure defined in Eq.~\eqref{eq:pressure_local_thermal_1} and Eq.~\eqref{eq:pressure_local_thermal_2}. $\Delta P_{\rm ballistic}^{b/f}$ is the ballistic pressure given by Eq.~\eqref{eq_pressure_ballistic_bosons} and Eq.~\eqref{eq_pressure_ballistic_fermions}. We might worry that for intermediate bubble wall velocities (with a range of values might be large), Eq.~\eqref{eq:pressure_lte_ballistic_sum_species} is too simplistic and instead one should use a close-to-equlibrium approach, along the line of Sec.~\ref{sec:close_lte_pressure}. However, such computations go beyond the scope of this chapter whose goal is not to perform thorough computations but instead to introduce different approaches available in the litterature.

The bubble wall reaches a finite velocity $v$ when the retarding pressure $\Delta P_{\rm LO}(v)$ compensates for the accelerating vacuum pressure $\Delta V_{\rm vac}$, cf. Eq.~\eqref{eq:pressure_equi_bubble_wall}. In the top panel of Fig.~\ref{eq:DeltaV_alpha_contours_v_LO}, we give the bubble wall velocity with respect to the almost model-independent variables 
\begin{equation}
\alpha \equiv \frac{\Delta V_{\rm vac}}{\rho_{\rm rad}}, \qquad \text{and} \qquad \Delta V_{\rm vac}/\phi^4,
\end{equation}
where $\phi = 246~$GeV. In the brown region, we have 
\begin{equation}
\Delta V_{\rm vac} ~>~ \Delta P_{\rm LO}(v=1),
\end{equation}
and the bubble is predicted to run-away (to keep accelerating until collisions with neighboring bubbles) at LO in the gauge coupling constant $g$. 

\paragraph{We add the NLO term:}

In the bottom panel of Fig.~\ref{eq:DeltaV_alpha_contours_v_LO}, we add the pressure resulting from NLO effets (in gauge coupling constant $g$), $\gamma \Delta P_{\rm NLO} \sim \gamma \, g^2 \, f\, \Tnuc^3$, computed in the previous Sec.~\ref{sec:NLO_pressure}. As a result, the bubble wall reaches a terminal velocity given by
	\begin{equation}
	\label{eq:gamma_max_NLO}
	\gamma_{\rm eq}^{\rm NLO} = \frac{\Delta V - P_{\rm LO}}{P_{\rm NLO}} \sim \frac{\Delta V}{g^2 \, f \, \Tnuc^3} \sim 10^{18} \left( \frac{f}{\text{TeV}} \frac{\text{MeV}}{\Tnuc} \right)^3,
	\end{equation}
where $f$ is the scalar field vev.

Since the NLO pressure is only known in the large $\gamma$ limit, cf. Eq.~\eqref{eq:p12}, we propose to define the LO+NLO pressure with the following trick
\begin{equation}
\Delta P_{\rm LO+NLO}(v)  = \Delta P_{\rm LO}(v)\left[ 1 ~+~ \frac{\Delta P_{\rm NLO}(\gamma)}{\Delta P_{\rm LO}(v=1)}\right],
\label{eq:PLO_ballistic_sum_species_plus_NLO}
\end{equation}
where $P_{\rm LO}(v)$ is the LO ballistic pressure defined in Eq.~\eqref{eq:pressure_lte_ballistic_sum_species} and $P_{\rm NLO}(\gamma)$ is NLO pressure in the large $\gamma$ limit, defined in Eq.~\eqref{eq:p12}.

\paragraph{When do bubbles run-away ? }
\label{eq:run_away_bubbles_criterium}

The energy gained upon formation of a bubble of radius $R$ is
\beq
\label{eq:Ebubble}
E_\text{bubble} \simeq \frac{4}{3}\, \pi \,R^3 ~\Delta V_{\rm vac },
\eeq
where $ \Delta V_{\rm vac }$ is the vacuum energy of the transition.
As the bubble grows, the later energy is stored in the bubble wall given by
\beq
\label{eq:Ewall}
E_\text{wall} \simeq 4 \, \pi\, R^2 \,\gamma~ \sigma,
\eeq
where $\sigma$ is the surface energy of the wall (wall tension) in the wall frame. After equating Eq.~\eqref{eq:Ebubble} and Eq.~\eqref{eq:Ewall}, we obtain the wall Lorentz factor at collision in the run-away regime
\begin{equation}
\label{eq:gamma_run_away_0}
\gamma_{\rm coll} = \frac{\Delta V_{\rm vac }\, R_{\rm coll}}{3\,\sigma},
\end{equation}
where $R_{\rm coll}$ is the bubble radius at collision. Hence, the wall Lorentz factor grows linearly with the bubble radius (see e.g. the review in \cite{Ellis:2019oqb}).
The wall tension decomposes as the product
\begin{equation}
\sigma = L_{\rm w} \, E_{\rm w},
\end{equation}
of $L_{\rm w}$, the wall thickness, and $E_{\rm w}$, the averaged energy density in the wall. $E_{\rm w}$ is the kinetic energy density of the field oscillating around the true minima. From conservation of energy, we expect
\begin{equation}
 E_{\rm w} \simeq \Delta V_{\rm vac },
\end{equation}
such that 
\begin{equation}
\label{eq:gamma_run_away}
\gamma_{\rm coll} = \frac{R_{\rm coll}}{3\,L_{\rm w}}.
\end{equation}

In order for the wall to run-away (i.e. $\dt{\gamma} \neq 0$) at the collision time, we need to lower the friction pressure as much as possible and therefore to lower the nucleation temperature. Hence, situations where run-away walls are expected correspond to phase transitions with a lot of supercooling, meaning that
\begin{equation}
\alpha \gg 1, \qquad \qquad (\rm supercooled ~phase ~transition).
\end{equation}
Typically, supercooled phase transitions are generated by shallow $(T=0)$-potentials, namely potentials where the curvature $f$ close to the true vacuum is much larger than the curvature close to the false vacuum, $f\,\exp{-c/\epsilon}$ where $\epsilon \ll 1$ and $c=O(1)$.  In that case, the tunneling exit point is very close to the false minimum and the bounce action is only sensitive to the scale $f\,\exp{-c/\epsilon}$ \cite{Konstandin:2011dr}. Hence, we expect the wall thickness (or equivalently the bubble size at nucleation) to be
\begin{equation}
L_{\rm w} \simeq \Tnuc^{-1}.
\end{equation} 
Later in Sec.~\ref{sec:supercool_potential}, we study in detail two classes of potential potentially leading to large amount of supercooling: the Coleman-Weinberg potential and the light-dilaton potential.

Finally, Eq.~\eqref{eq:gamma_run_away} becomes\footnote{Note that the difference between Eq.~\eqref{eq:gamma_run_away_2} and \cite{Caprini:2019egz} is due to the choice of the authors of \cite{Caprini:2019egz} of setting $L_w \sim f$ instead of $L_w \sim \Tnuc$. For typical potentials leading to large supercooling, the later choice is the correct one, see App.~\ref{app:wall_profile}.}
\begin{equation}
\label{eq:gamma_run_away_2}
\gamma_{\rm coll}  \simeq ~\Tnuc\, \beta^{-1},
\end{equation}
where $\beta$ is the rate of change of the tunneling rate (inverse of the bubble propagation time).

Upon comparing Eq.~\eqref{eq:gamma_run_away_2} with Eq.~\eqref{eq:gamma_max_NLO}, we find that collisions occur before the terminal velocity is reached (run-away bubbles) when 
	\begin{equation}
	\gamma_{\rm coll}  \lesssim \gamma_{\rm eq} \quad \Longrightarrow \quad T_{\rm nuc} \lesssim T_{\rm eq}  \equiv \left( \frac{f^4}{ M_{\rm pl} }  \frac{
\beta}{H} \right)^{1/3} \simeq 10~\text{MeV} \, \left(\frac{f}{\text{TeV}} \right)^{4/3}  \left( \frac{ \beta/H}{10} \right)^{1/3}.
\label{eq:run_away_wall_cond}
	\end{equation}
If $\Tnuc \lesssim T_{\rm eq}$, then most of the vacuum energy is used for accelerating the bubble walls and the GW spectrum is dominated by the scalar field contribution. If $\Tnuc \gtrsim T_{\rm eq}$, then the terminal velocity is reached and most of the vacuum energy is converted into thermal and kinetic energy of the plasma via the friction, leading to a GW spectrum dominated by sound waves and turbulence contributions (see e.g. the review in \cite{Ellis:2019oqb}).

\paragraph{Bubble wall velocity from heating and cooling rates.}
We have derived the bubble wall velocity from the equilibrium condition between the friction pressure and the expanding pressure, cf. Eq.~\eqref{eq:pressure_equi_bubble_wall}. We would like to mention a completely different approach - and as far as we can tell which gives a different result - for deriving the bubble wall velocity, which does not even need the computation of the friction pressure \cite{Witten:1984rs,Asadi:2021pwo}. Instead the wall velocity is found as resulting from a detailed balance between the rates at which the bubble wall heats and cools the surrounding plasma.
On the one hand, the plasma around the wall boundary is heated due to the latent heat conversion at a rate
\begin{equation}
\dot{T}_{\rm heat} \sim l\,\Lambda\, \left(-\frac{d R}{dt}\right) \times \frac{dT}{d\rho} \sim \Lambda\, \frac{d R}{dt} ,
\end{equation}
where $-dR/dt$ is the outward-oriented wall velocity and $\Lambda^{-1}$ is the wall thickness.
On the other hand, due to the temperature gradient the same plasma cools down at a rate
\begin{equation}
\dot{T}_{\rm cooling} \sim -K \nabla^2 T \sim \Lambda^{-1} \frac{T_c-T}{\Lambda^2},
\end{equation}
where we assume that the transport coefficient $K$ and the gradient length scale are both given by $\Lambda^{-1}$. The wall velocity results from the balance between the two thermal rates \cite{Witten:1984rs,Asadi:2021pwo}
\begin{equation}
\dot{T}_{\rm heat} \sim \dot{T}_{\rm cooling} \quad \implies \quad \dot{R} \simeq \frac{T - T_c}{T_c}, \label{eq:radius_time}
\end{equation}
where we have assumed $T_c \simeq \Lambda$.

\section{GW generation}
\label{sec:GW_generation}

In Sec.~\ref{sec:GW_cosmology} of Chap.~\ref{chap:SM_cosmology}, we have introduced Gravitational Waves (GW) as linearized solutions of the Einstein equation. We have discussed the possibility that the universe today is filled with a Stochastic Gravitational-Wave Background (SGWB) from primordial origin, which we have computed in the quadrupole approximation. In the present section, we focus on the case where the SGWB is sourced by bubble dynamics during a cosmological 1stOPT.

\subsection{The GW spectrum for a generic source}

\paragraph{The GW from the energy-momentum tensor correlator:}

The energy fraction contained in SGWB today is given by its energy fraction at the time of its emission $*$ redshifted until today as a radiation fluid
\begin{equation}
h^2 \Omega_{\rm GW}(k) = \frac{h^2}{\rho_c} \, \left(  \frac{a_p}{a_0} \right)^4 \, \left(\frac{d\rho_{\rm GW}}{d\log k} \right)_*.
\end{equation}
$\rho_c$ is the critical density today and $\rho_{\rm GW}$ is the energy density of GW \cite{Misner:1974qy,Maggiore:1900zz}
\begin{equation}
\rho_{\rm GW}  = \frac{ \left<  \dot{h}_{\rm ij}(\vec{x},\, \eta)  \dot{h}_{\rm ij}(\vec{x}, \, \eta)  \right>}{32 \pi G a^2}.
\end{equation}
$\eta $ is the conformal time and $h_{\rm ij}(k, \, t)$ is the tensor spatial perturbation of the FLRW metric
\begin{equation}
ds^2 = dt^2 - a(t)^2(\delta_{\rm ij} + h_{\rm ij})d\vec{x}^2 = a(t)^2(d\eta^2 - (\delta_{\rm ij} + h_{\rm ij})d\vec{x}^2 )
\end{equation}
in the transverse-traceless (TT) gauge, so that $\partial_{\rm i} h_{\rm ij} =0$ and $h_{\rm ii}=0$.
We define the Fourier modes as $h_{\rm ij}(\vec{k},\, \eta) = \int \frac{ d\vec{k}^3}{(2\pi)^3} h_{\rm ij}(\vec{k},\, \eta) e^{- i \vec{k}\cdot \vec{x}} $ and the characteristic amplitude $h(k, \, \eta) $ from the correlation function of the metric perturbation
\begin{equation}
\left<  \dot{h}_{\rm ij}(\vec{k},\, \eta)  \dot{h}_{\rm ij}(\vec{q}, \, \eta)  \right> = (2\pi)^3 \delta^{(3)}(\vec{k} - \vec{q}) \, | \dot{h}(k, \, \eta) | ^2.
\end{equation}
from which we deduce
\begin{equation}
 \left<  \dot{h}_{\rm ij}(\vec{x},\, \eta)  \dot{h}_{\rm ij}(\vec{x}, \, \eta)  \right> = \int d\ln{k} \, \frac{k^3}{2\pi^2} \, | \dot{h}(k, \, \eta) | ^2
\end{equation}
and
\begin{equation}
\left(\frac{d\rho_{\rm GW}}{d\log k} \right)_* = \frac{ k^3 | \dot{h}(k, \, \eta) | ^2}{64 \pi^3 G a^2} = \frac{ k^5 | h'(k, \, \eta) | ^2}{64 \pi^3 G a^2} .
\end{equation}
The dot stands for the derivative with respect to the conformal time $\eta$ and the prime for the derivative with respect to the dimensionless conformal time $\tilde{\eta} \equiv k \eta$. We define the conformal Hubble factor $\mathcal{H} \equiv \dot{a}/a = H/a$.
The linearised Einstein equation in the FLRW background leads to \cite{Caprini:2018mtu}
\begin{equation}
\ddot{h}_{\rm ij}(k, \, \eta) + 2 \mathcal{H} \dot{h}_{\rm ij}(k, \, \eta) +k^2 h_{\rm ij}(k, \, \eta) = 16 \pi G a^2 \, \Pi_{\rm ij}^{\mathsmaller{\rm TT}}(\vec{k}, \, \eta))
\end{equation}
where $\Pi_{\rm ij}^{\mathsmaller{\rm TT}}(\vec{k}, \, \eta)$ is the transverse and traceless part of the anisotropic stress. It is easier to compare the different terms of the equation if we use the dimensionless conformal time $\tilde{\eta} = k \eta$
\begin{equation}
{h}''_{\rm ij}(k, \, \eta) + 2 \frac{\mathcal{H}}{k} h'_{\rm ij}(k, \, \eta) +h_{\rm ij}(k, \, \eta) = \frac{16 \pi G a^2}{k^2} \, \Pi_{\rm ij}^{\mathsmaller{\rm TT}}(\vec{k}, \, \tilde{\eta})
\end{equation}
For sub-Hubble modes $k \gg \mathcal{H}$, we can neglect the first derivative and the solution reads
\begin{equation}
h_{\rm ij}(k, \, \eta)  =  \frac{16 \pi G }{k^2} \int_{\tilde{\eta}_{\rm in}}^{\tilde{\eta}_{\rm fi}} d\tilde{\eta}' \, G(\tilde{\eta} - \tilde{\eta}') \, a^2 \, \Pi_{\rm ij}^{\mathsmaller{\rm TT}}(\vec{k}, \, \tilde{\eta}'))
\end{equation}
where $G(\tilde{\eta} - \tilde{\eta}')= \cos(\tilde{\eta}' - \tilde{\eta})$ is the Green function of the operator $\left[\frac{d}{d\tilde{\eta}^2} + 1\right]$ with the boundary conditions $G(0)=0$ and $G'(0)=0$.
We define $\tilde{\Pi}_{\rm ij}(\vec{k}, \, \eta))$ as $\Pi_{\rm ij}^{\mathsmaller{\rm TT}}(\vec{k}, \, \eta)) = \rho_s \, \tilde{\Pi}_{\rm ij}(\vec{k}, \, \eta))$ where $\rho_s$ is the energy density of the source, e.g. due to the gradient of the scalar field or the compressional and turbulent motion of the plasma. Then
\begin{multline}
\left<  {h}'_{\rm ij}(\vec{k},\, \tilde{\eta})  {h}'_{\rm ij}(\vec{q}, \, \tilde{\eta})  \right> = \left( \frac{16\pi G }{k^2} \right)^2  \rho_s^2  \int_{\tilde{\eta}_{\rm in}}^{\tilde{\eta}_{\rm fi}}  d\tilde{\eta}_1 \int_{\tilde{\eta}_{\rm in}}^{\tilde{\eta}_{\rm fi}}  d\tilde{\eta}_2 \cos(\tilde{\eta}_1 - \tilde{\eta}) \cos(\tilde{\eta}_2 - \tilde{\eta})  \\  
\, a(\tilde{\eta}_1)^2 a(\tilde{\eta}_2)^2 \, \tilde{\Pi}_{\rm ij}(\vec{k}, \, \tilde{\eta}_1) \,  \tilde{\Pi}_{\rm ij}(\vec{q}, \, \tilde{\eta}_2) 
\end{multline}
which after averaging over time becomes
\begin{multline}
\left<  {h}'_{\rm ij}(\vec{k},\, \tilde{\eta})  {h}'_{\rm ij}(\vec{q}, \, \tilde{\eta})  \right> \simeq \frac{1}{2} \left( \frac{16 \pi G a^2 }{ k^2} \right)^2 \rho_s^2 \int_{\tilde{\eta}_{\rm in}}^{\tilde{\eta}_{\rm fi}}  d\tilde{\eta}_1 \int_{\tilde{\eta}_{\rm in}}^{\tilde{\eta}_{\rm fi}}  d\tilde{\eta}_2 \, \cos(\tilde{\eta}_1 - \tilde{\eta}_2) \\ \,a(\tilde{\eta}_1)^2 a(\tilde{\eta}_2)^2 \, \tilde{\Pi}_{\rm ij}(\vec{k}, \, \tilde{\eta}_1) \,  \tilde{\Pi}_{\rm ij}(\vec{q}, \, \tilde{\eta}_2)
\end{multline}
We define the unequal-time correlator of the transverse-traceless energy stress tensor as
\begin{equation}
\left<   \tilde{\Pi}_{\rm ij}(\vec{k},\, \tilde{\eta}_1)  \tilde{\Pi}_{\rm ij}(\vec{q}, \, \tilde{\eta}_2)  \right> = (2\pi)^3 \delta^{(3)}(\vec{k} - \vec{q}) \,  \tilde{\Pi}(\vec{k},\, \tilde{\eta}_1,\, \tilde{\eta}_2)  
\end{equation}
such that we can write
\begin{equation}
\left(\frac{d\rho_{\rm GW}}{d\log k} \right)_* = \frac{ 2 G }{\pi a^2} \,  \rho_s^2 \, k \, \int_{\tilde{\eta}_{\rm in}}^{\tilde{\eta}_{\rm fi}}  d\tilde{\eta}_1 \int_{\tilde{\eta}_{\rm in}}^{\tilde{\eta}_{\rm fi}}  d\tilde{\eta}_2 \, \cos(\tilde{\eta}_1 - \tilde{\eta}_2)\, a(\tilde{\eta}_1)^2 a(\tilde{\eta}_2)^2\, \tilde{\Pi}(\vec{k},\, \tilde{\eta}_1,\, \tilde{\eta}_2)  .
\end{equation}
For short-lasting source we can approximate $\tilde{\eta}_{\rm in} \simeq \tilde{\eta}_{\rm fi}$ and we obtain (e.g. \cite{Jinno:2016vai})
\begin{equation}
\left(\frac{d\rho_{\rm GW}}{d\log k} \right)_* = \frac{ 2 G a^2}{ \pi } \,  \rho_s^2 \,   k \,\int_{\tilde{\eta}_{\rm in}}^{\tilde{\eta}_{\rm fi}}  d\tilde{\eta}_1 \int_{\tilde{\eta}_{\rm in}}^{\tilde{\eta}_{\rm fi}}  d\tilde{\eta}_2 \, \cos(\tilde{\eta}_1 - \tilde{\eta}_2)\, \tilde{\Pi}(\vec{k},\, \tilde{\eta}_1,\, \tilde{\eta}_2)  .
\end{equation}
After its emission at time $*$, the GW energy density redshifts as radiation, such that its abundance today reads
\begin{equation}
h^2 \Omega_{\rm GW}(k) = \frac{h^2}{\rho_c} \, \left(  \frac{a_{*}}{a_0} \right)^4 \,  a_*^2 \, ( \rho_{\rm tot}^{*})^2 \, K^2\,\frac{ 2 G }{ \pi } \, k \, \int_{\tilde{\eta}_{\rm in}}^{\tilde{\eta}_{\rm fi}}  d\tilde{\eta}_1 \int_{\tilde{\eta}_{\rm in}}^{\tilde{\eta}_{\rm fi}}  d\tilde{\eta}_2 \, \cos(\tilde{\eta}_1 - \tilde{\eta}_2)\, \tilde{\Pi}(\vec{k},\, \tilde{\eta}_1,\, \tilde{\eta}_2)  .
\end{equation}
We introduced the fraction of  total energy density that gets converted into the  energy of the source
\begin{equation}
K \equiv \frac{\rho_s^{*}}{\rho_{\rm tot}^{*}},
\end{equation}
where $\rho_{\rm tot}^{*}$ is the total energy density just before the phase transition (in the symmetric phase).
For convenience, we can define the dimensionless quantity
\begin{equation}
\Delta(k)= \frac{ 3 }{ 4\pi^2 } \,  a_*^2 \, \int_{\tilde{\eta}_{\rm in}}^{\tilde{\eta}_{\rm fi}}  d\tilde{\eta}_1 \int_{\tilde{\eta}_{\rm in}}^{\tilde{\eta}_{\rm fi}}  d\tilde{\eta}_2 \, \cos(\tilde{\eta}_1 - \tilde{\eta}_2)\,  \frac{k}{\beta} \, \left( \beta^3 \, \tilde{\Pi}(\vec{k},\, \tilde{\eta}_1,\, \tilde{\eta}_2)  \right).
\end{equation}
where $\beta$ is the time variation of the nucleation rate
\begin{equation}
\beta \equiv \frac{dS}{dt}\Big|_{*}= -H_* T_* \frac{dS}{dT} \Big|_{*} ,
\end{equation}
such that the SGWB abundance today becomes
\begin{equation}
\Omega_{\rm GW}(k)\,h^2  = h^2 \frac{\rho_{\rm tot}^{*}}{\rho_c} \, \left(  \frac{a_*}{a_0} \right)^4  \,  \left(\frac{H_*}{\beta} \right)^2\, K^2  \, \Delta(k) .
\end{equation}
with $H_*^2 = 8\pi G \rho_{\rm tot}^{*}/3$. The GW emission time denoted by $*$ coincides with the bubble percolation time 
\begin{equation}
a_{\rm *} = a_{\rm perc},
\end{equation}
which can be different from the nucleation temperature, see discussion Sec.~\ref{sec:nucleation_temp_VS_percolation}, and the end of reheating $a_{\rm reh}$.

\paragraph{The GW propagation from the percolation epoch to today:}
Upon assuming an adiabatic evolution from today $a_0$ up to reheating $a_{\rm reh}$, i.e. $h_{\rm eff}\,T^3\,a^3 = \rm cst$, and possibly assuming the domination of the universe by a scalar field redshifting as $\rho \propto a^{-n}$ between the end of reheating $a_{\rm reh}$ and the time of bubble nucleation $a_*$, we obtain
\begin{equation}
 \Omega_{\rm GW}(k)\,h^2  = F_{\rm gw,\,0}\,h^2 \, \left(\frac{H_*}{\beta} \right)^2\, K^2  \, \Delta(k) ,
\end{equation}
with 
\begin{align}
F_{\rm gw,\,0} \,h^2 &\equiv \Omega_{\rm \gamma}\,h^2 \,\left(\frac{h_{\rm eff,\, 0}}{h_{\rm eff,\, reh}}\right)^{4/3}\frac{g_{\rm eff,\,reh}}{g_{\gamma,\,0}} \left[\frac{\rho_{\rm reh}}{\rho_*\,(1+\alpha)}\right]^{\frac{4-n}{n}} , \notag \\
&\simeq 1.657 \times 10^{-5}~\left( \frac{100}{g_{\rm eff, \,reh}} \right)^{1/3} \left[\frac{\rho_{\rm reh}}{\rho_*\,(1+\alpha)}\right]^{\frac{4-n}{n}},
\label{eq:prefactor_Omega_GW}
\end{align}
where $\rho_{\rm reh}/\rho_* = g_{\rm eff,\, reh}/ g_{\rm eff,\,*} ~T_{\rm reh}^4/T_{*}^4$.
 The last factor $\left[\cdots\right]^{\frac{4-n}{n}}$ accounts for the extra-redshift\footnote{Eq.~\eqref{eq:prefactor_Omega_GW} follows from $F_{\rm gw,\,0} = (a_{\rm reh}/a_0)^4 (a_*/a_{\rm reh})^4 (\rho_{\rm reh}/\rho_{c}) (\rho_{\rm tot}^{*}/\rho_{\rm reh})$ with $\rho_{\rm tot}^{*}/\rho_{\rm reh} = (a_{\rm reh}/a_{*})^n$.} of the universe as $\rho \propto a^{-n}$ between bubble nucleation and reheating, and only differs from unity if the lifetime of the scalar field driving the PT is longer than a Hubble time. For a scalar field oscillating in a potential $V\propto \phi^{2p}$, the equation of state averaged over the oscillations is $\bar{\omega}=(p-1)/(p+1)$ \cite{Turner:1983he}, so the resdhift parameter is $n=3(1+\bar{\omega})=6p/(1+p)$. Note that for instantaneous reheating, $a_{\rm reh} = a_{*}$, the reheating temperature reads
 \begin{equation}
 T_{\rm reh} = \left(\frac{g_{\rm eff *}}{g_{\rm reh}}\right)^{1/4}\left(1 + \alpha \right)^{1/4} T_{*}.
 \end{equation}
We made used of $g_{\rm eff, \,p}=h_{\rm eff, \,p}$, $g_{\gamma,\,0}=2$, $h_{\rm eff,\, 0}=3.94$ (which assumes $N_{\rm eff} \simeq 3.045$  \cite{Mangano:2005cc,deSalas:2016ztq,Escudero:2020dfa}) and $\Omega_{\rm \gamma}\,h^2 \simeq 2.473 \times 10^{-5}$ \cite{Tanabashi:2018oca}.

From doing similar maths, the GW frequency today reads
\begin{equation}
f_0 = \frac{a_*}{a_0} f_* = 1.65 \times 10^{-5}~{\rm Hz}~\left(\frac{T_{\rm reh}}{100~\rm GeV}\right) \left( \frac{g_{\rm eff, \,reh}}{100} \right)^{1/6} \frac{f_*}{H_*} \left[\frac{\rho_*\,(1+\alpha)}{\rho_{\rm reh}}\right]^{\frac{n-2}{2n}},
\label{eq:GW_frequency_emission_today_1stOPT}
\end{equation}

\subsection{Contribution from the scalar field }

\paragraph{The envelope approximation:}

Bubble collisions resulting from first-order phase transition are known for producing GW for 4 decades (Witten 1984 \cite{Witten:1984rs,Hogan:1986qda}). The first computations of the GW spectrum have been carried out in the early 90s \cite{ Kosowsky:1991ua, Kosowsky:1992vn, Kamionkowski:1993fg}, with the simulation of the collision of two bubbles on a lattice and the analytical computation of the GW spectrum. Along these lines, the authors introduced the \textbf{envelope approximation} where GW are sourced by the anisotropic stress-tensor of the scalar field gradient localised on infinitely thin shells in which the collided portions have been removed.
Since then, other analytical studies \cite{Caprini:2007xq, Huber:2008hg,Weir:2016tov,Jinno:2016vai,Konstandin:2017sat} have refined the results found earlier.
In the envelope approximation, the stochastic GW background (SGWB) sourced by the gradient energy of the scalar field reads (LISA paper 2015 \cite{Caprini:2015zlo}, derived by Huber+ in 2008 \cite{Huber:2008hg})
\begin{equation}
\Omega_{\rm \phi}  h^2 = F_{\rm gw}\,h^2\,  \left(\frac{H_*}{\beta} \right)^2\, \left( \frac{\kappa_{\phi} \alpha}{1+\alpha} \right)^2  \,  \left( \frac{0.11 v_{\rm w}^3}{0.42 + v_{\rm w}^2}\right) \; S_{\phi}(f_*),
\label{eq:Omega_GW_bubble_collision_envelope}
\end{equation} 
where $S_{\phi}(f_*)$ describes the spectral shape
\begin{align}
\label{eq:spectral_shape_scalar}
&S_{\phi}(f_*) = \frac{3.8 (f_*/f_{\phi})^{2.8}}{1+2.8(f_*/f_{\phi})^{3.8}}, \\
&\frac{f_{\phi}}{H_*} = \left( \frac{0.62}{1.8-0.1v_w + v_w^2} \right) \left( \frac{\beta}{H_*} \right).
\end{align}
The prefactor $F_{\rm gw}\,h^2$ is given in  Eq.~\eqref{eq:prefactor_Omega_GW}. $f_*$ denotes the frequency at emission and is related to the frequency today $f_0$ through Eq.~\eqref{eq:GW_frequency_emission_today_1stOPT}.
The parameters $\alpha$, $\beta$ and $\kappa_{\phi}$ are defined by
\begin{align}
&\alpha \equiv \frac{\Delta V}{\rho_{\rm tot}^*} , \label{eq:alpha_definition}\\
&\kappa_{\phi} \equiv \frac{\dt{\phi}^2/2+(\nabla{\phi})^2/2}{\Delta \theta}, \\
&\beta \equiv \frac{dS}{dt}\Big|_{*}= -H_* T_* \frac{dS}{dT} \Big|_{*} ,\label{eq:beta_definition}\\
\end{align}
where $\rho_{\rm tot}^*$ is the total energy density in the symmetric phase (sometimes noted $e_+$) and where $\Delta V$ is the zero-temperature energy difference between the false vacuum and the true vacuum.  
 
\begin{figure}[h]
\centering
\raisebox{0cm}{\makebox{\includegraphics[width=0.9\textwidth]{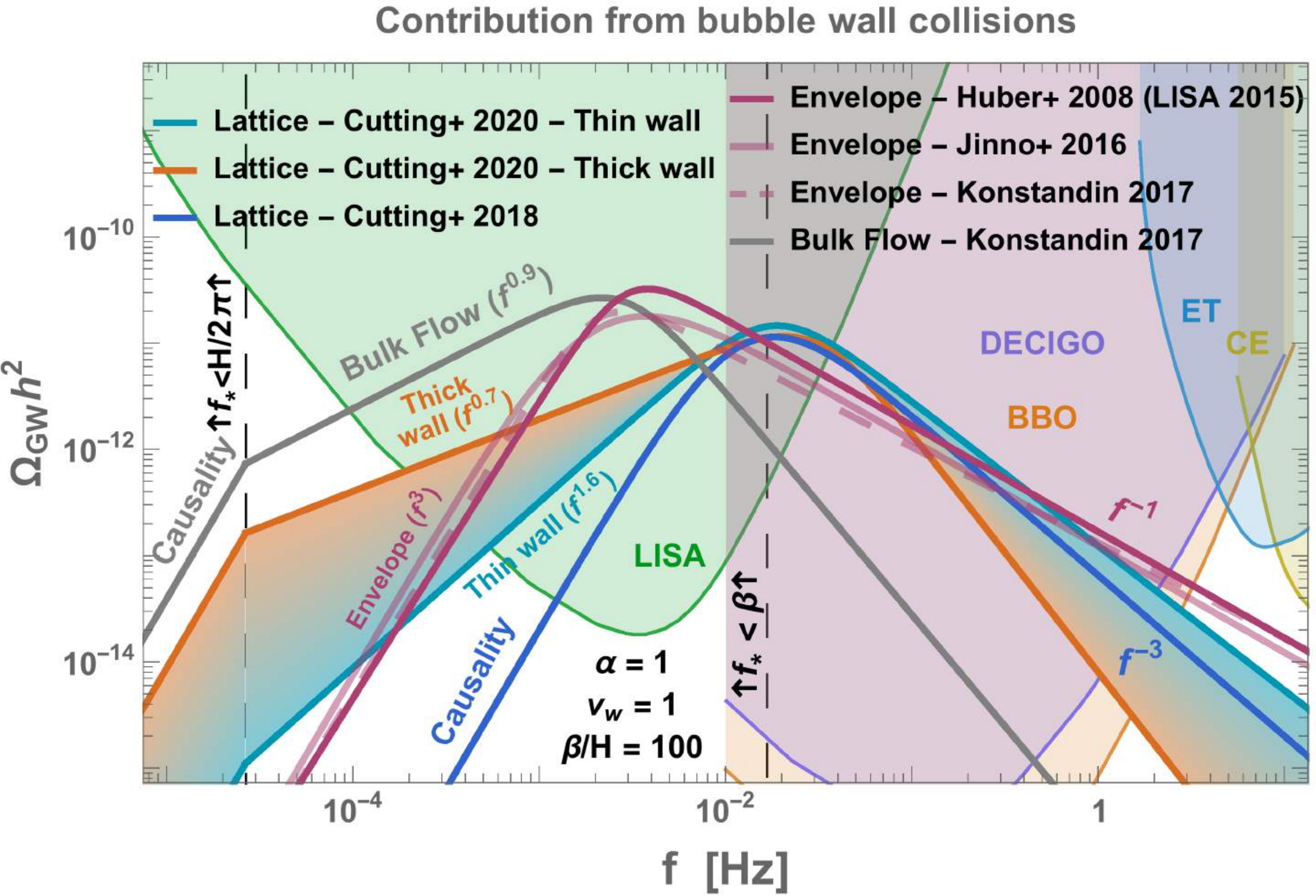}} }
\caption{\it \small    Comparison of the different expression for the SGWB sourced by bubble collisions (scalar field contribution), available in the litterature. The envelope approximation predicts a $f^3$ and $f^{-1}$ scaling in the IR and the UV respectively, c.f Huber-Konstandin 2008 \cite{Huber:2008hg} (version in the LISA paper 2015 \cite{Caprini:2015zlo}), Jinno-Takimoto 2016 \cite{Jinno:2016vai} and Konstandin 2017 \cite{Konstandin:2017sat}. First deviations from the envelope approximation are found by Lattice simulations run in vacuum by Cutting+ in 2018 \cite{Cutting:2018tjt}: a $f^{-1.5}$ at higher frequency and a larger peak frequency. Later in 2020, new deviations from the envelope approximation are found by Cutting+ \cite{Cutting:2020nla}: a dependence on the wall profile (thick versus thin wall) and an enhancement in the IR (see footnote~\ref{footnote:CuttingPrivateCom}) due to the shells of transverse-traceless stress-energy tensor still propagating after the collision (especially for thick walls). This confirms the predictions from the analytical bulk flow model (Jinno-Takimoto 2017 \cite{Jinno:2017fby} and Konstandin 2017 \cite{Konstandin:2017sat}). Note that whether the IR slope at $ f \lesssim \beta$ is enhanced or not, at much lower frequency, $f \lesssim H/2\pi$, the slope must converge to $k^3$ due to causality \cite{Durrer:2003ja,Caprini:2009fx, Cai:2019cdl,Hook:2020phx}. }
\label{fig:OmegaGW_scalarField_Litterature}
\end{figure}

The IR slope is $f^3$ (fixed by causality \cite{Durrer:2003ja,Caprini:2009fx, Cai:2019cdl,Hook:2020phx}) while the UV slope is $f^{-1}$. The peak frequency is given by $f_* \simeq \beta/5$.
In purple lines of Fig.~\ref{fig:OmegaGW_scalarField_Litterature}, we compare the formula of Huber+ 2008 for $v_w=1$ to more recent studies, Jinno+ 2016 \cite{Jinno:2016vai} and Konstandin 2017 \cite{Konstandin:2017sat}, and show that there are only little differences. 

\paragraph{Beyond the envelope:}
The envelope approximation was first challenged by the lattice simulation of Cutting+ in 2018 \cite{Cutting:2018tjt}, which showed that the peak frequency is $5$ times higher and the UV slopes is steeper $(f^{1.5})$ (blue line in Fig.~\ref{fig:OmegaGW_scalarField_Litterature}).
Later, Cutting+ 2020 \cite{Cutting:2020nla} showed that the IR and UV slopes depend on the wall profile (orange-to-blue gradient in Fig.~\ref{fig:OmegaGW_scalarField_Litterature}) and that the IR slope is softer\footnote{The dynamical range in Cutting+ 2018 \cite{Cutting:2018tjt} was not large enough to estimate the spectral index of the IR slope. Therefore, the authors simply imposed $f^3$ in order to respect causality. In Cutting+ 2020 \cite{Cutting:2020nla}, both the kinetic $\dt{\phi}^2/2$ and gradient energy $(\nabla{\phi})^2/2$ are considered (while only gradient energy was considered in 2018), which allows to reduce the artificial oscillations in the IR slope and to see the hint of the shallower power law \cite{CuttingPrivateCom}. \label{footnote:CuttingPrivateCom}} due to the long-lasting free propagation of the shells of anisotropic energy-momentum tensor after the collision. This confirms the analytical work of Jinno+ 2017 \cite{Jinno:2017fby} and Konstandin 2017 \cite{Konstandin:2017sat} (bulk flow model, see gray line of Fig.~\ref{fig:OmegaGW_scalarField_Litterature}). Note that Cutting+ 2020 \cite{Cutting:2020nla} finds that the IR enhancement is stronger for thick-walled bubbles, since for thin-walled bubbles, after collision the scalar field can be trapped back in the false vacuum. Hence, instead of propagating freely, the shells of energy-momentum tensor dissipates via multiple re-bounces of the walls, see \cite{Konstandin:2011ds, Jinno:2019bxw,Lewicki:2019gmv,Gould:2021dpm}. Note that whether the IR slope at $ f \lesssim \beta$ is enhanced or not, at much lower frequency, $f \lesssim H/2\pi$, the slope must converge to $k^3$ due to causality \cite{Durrer:2003ja,Caprini:2009fx, Cai:2019cdl,Hook:2020phx}. Finally, it has been shown that scalar field oscillations after bubble collisions can also contribute to the GW spectrum \cite{Child:2012qg}, however the time scale is set by the scalar mass and the signal is Planck-suppressed $\propto \beta/m_\phi$ \cite{Cutting:2018tjt}.

\paragraph{Only relevant for runaway bubbles:}
Following the study of Bodeker\&Moore in 2017, which shows the presence of a NLO contribution to the retarding pressure scaling linearly with $\gamma$, cf. Sec.~\ref{sec:NLO_pressure}, in most of the models studied in the litterature, bubbles do not run-away and the scalar field contribution to the GW signal is \textbf{irrelevant}. Instead, most of the vacuum energy is converted into \textbf{fluid kinetic energy}, see next section.
As already discussed along Eq.~\eqref{eq:run_away_wall_cond}, only phase transitions with a lot of \textbf{supercooling}, $\alpha \gg 1$, (e.g. $T_{\rm nuc} \lesssim 10$~MeV for EWPT) have run-away bubbles\footnote{Or strong phase transitions without gauge bosons.}, such that the released vacuum energy is contained in the kinetic+gradient energy of the scalar field, or equivalently in the wall kinetic energy, and not in the fluid motion.

The energy transfer between the vacuum energy and the kinetic energy of the bubble wall, is given by \cite{Ellis:2019oqb}
\begin{equation}
\kappa_\phi = 
\begin{cases}
\frac{\gamma_{\rm eq}}{\gamma_{\rm coll}} \left[ 1 - \left(\frac{\alpha_\infty}{\alpha} \right)^2  \right]  ~\simeq~ 0, &\hspace{2em}\gamma_{\rm coll} > \gamma_{\rm eq}, \qquad \rm (constant~wall~velocity~at~coll.)\\
1 - \frac{\alpha_\infty}{\alpha}~ \simeq ~1, &\hspace{2em}\gamma_{\rm coll} \leq \gamma_{\rm eq},\qquad \rm (run-away~wall~at~coll.)
\end{cases}
\end{equation}
where $\alpha_\infty \equiv \Delta P_{\rm LO}/\rho_{\rm rad}(\Tnuc)$. We recall that $\gamma_{\rm eq}$ and $\gamma_{\rm coll}$ are the Lorentz factors when bubbles reach finite velocity and when bubbles collide, respectively. They are given by Eq.~\eqref{eq:gamma_max_NLO}  and Eq.~\eqref{eq:gamma_run_away_2}, which we report here
\begin{align}
&\gamma_{\rm eq} \simeq \frac{\Delta V}{g^2\,f\,T_{\rm nuc}^3},\\
&\gamma_{\rm coll} \simeq  T_{\rm nuc} \, \beta^{-1}.
\end{align}

\begin{figure}[h]
\centering
\begin{adjustbox}{max width=1.2\linewidth,center}
\raisebox{0cm}{\makebox{\includegraphics[width=0.6\textwidth]{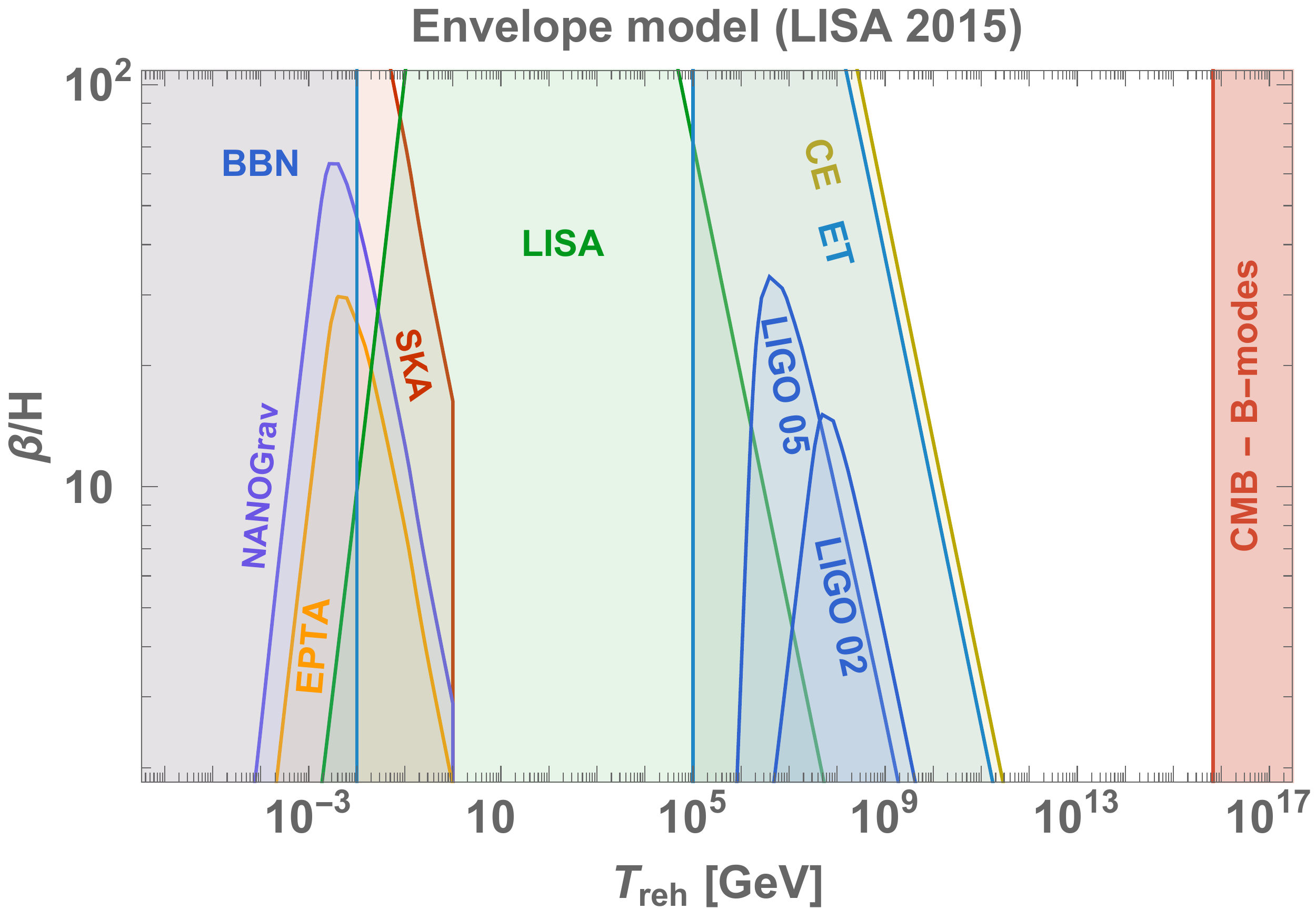}}}
\raisebox{0cm}{\makebox{\includegraphics[width=0.6\textwidth]{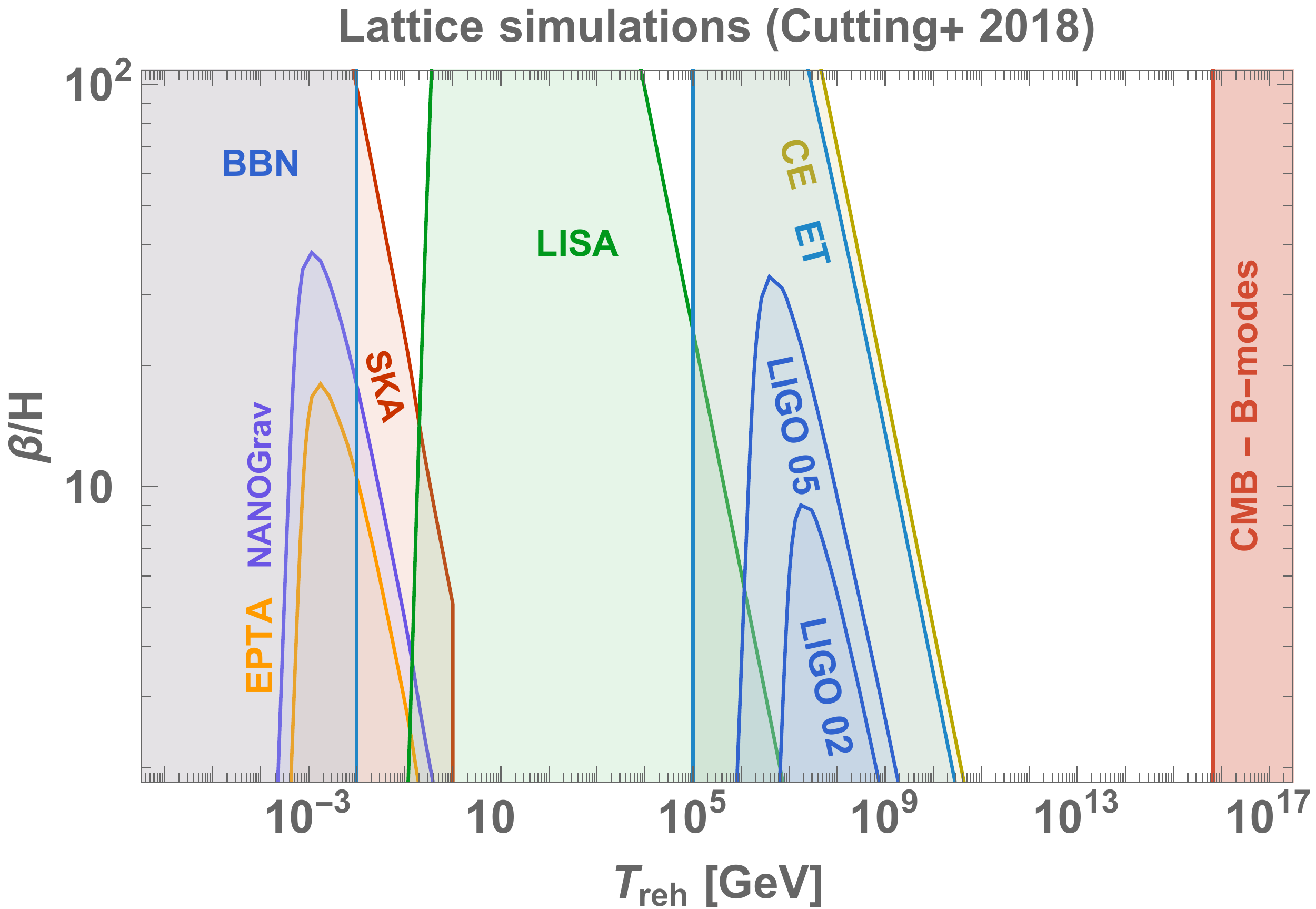}}}
\end{adjustbox}
\begin{adjustbox}{max width=1.2\linewidth,center}
\raisebox{0cm}{\makebox{\includegraphics[width=0.6\textwidth]{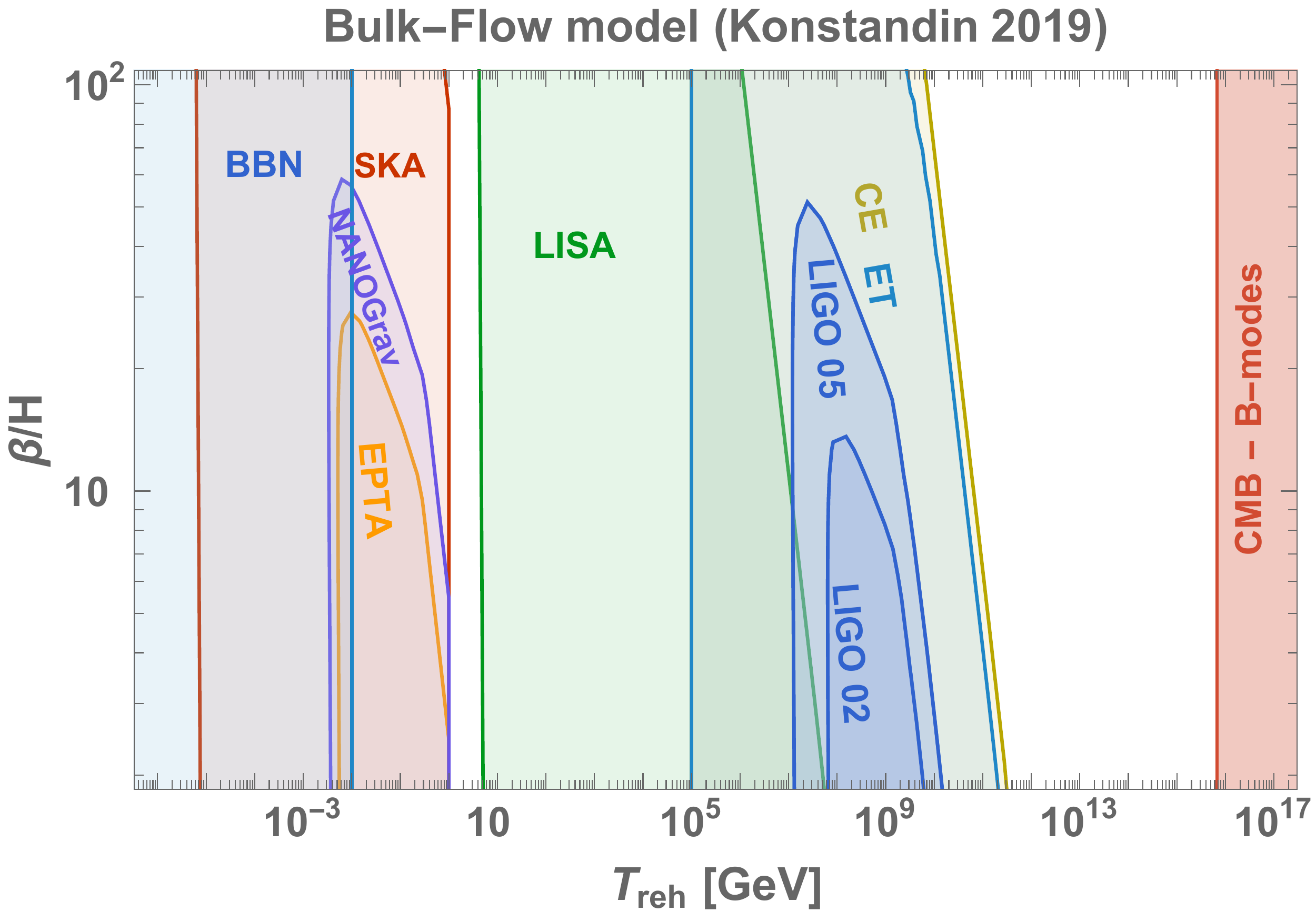}}}
\raisebox{0cm}{\makebox{\includegraphics[width=0.6\textwidth]{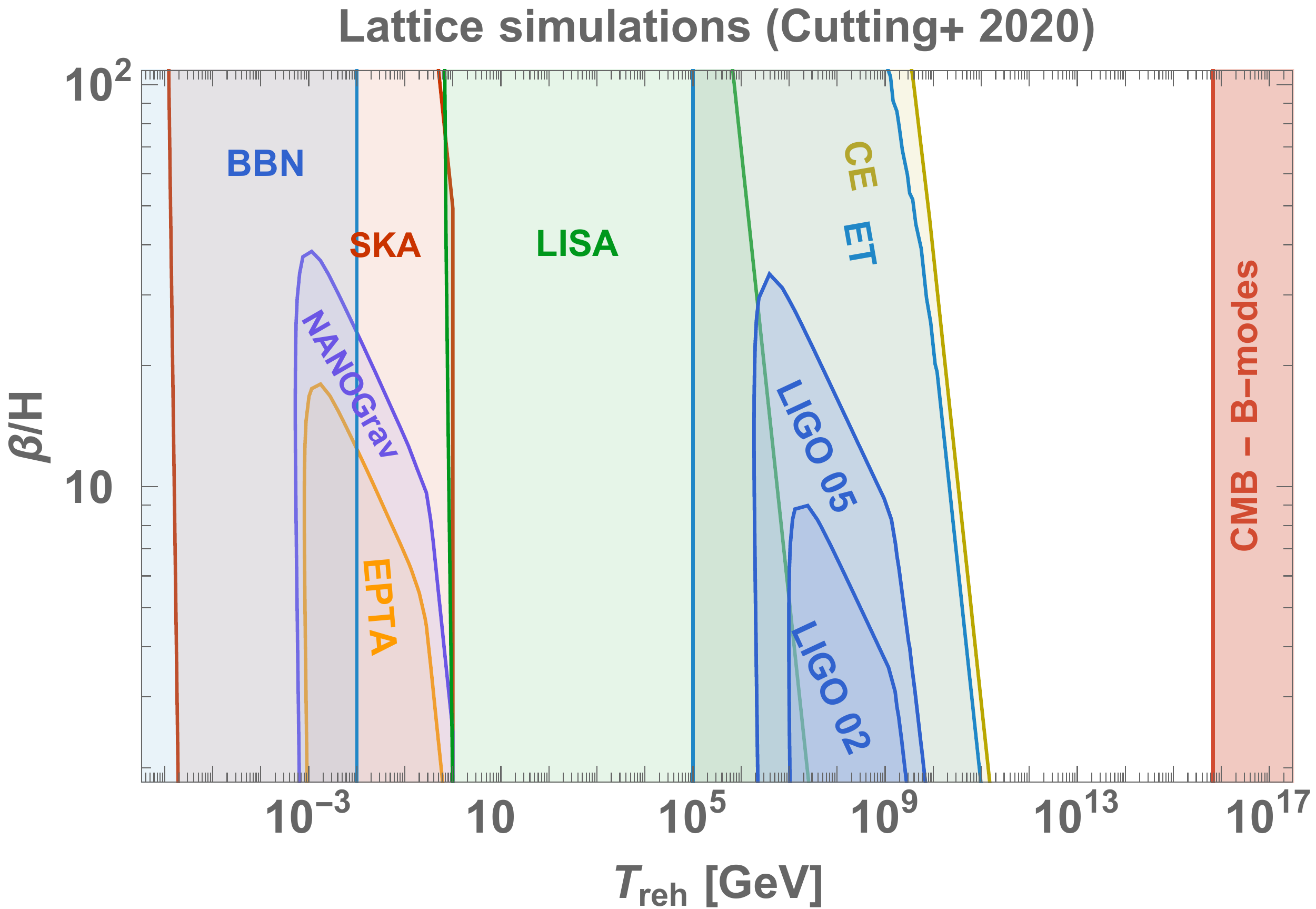}}}
\end{adjustbox}
\caption{\it \small  Sensitivity regions of various experiments on vacuum-dominated first-order phase transitions with $\alpha \gg 1$ (Supercooled phase transitions).  For such PT, we expect $\beta/H \sim 10$. GW are sourced by the scalar field gradient during bubble collisions, details on the different models/simulations are given in Fig.~\ref{fig:OmegaGW_scalarField_Litterature} and in text. The power-law integrated sensitivity curves of the different experiments are computed in App.~\ref{app:sensitivity_curves}.}
\label{fig:SC_PT_constraints}
\end{figure}

\paragraph{Constraints}

We show the currents and future constraints on the GW signal generated by bubble collision during a supercooled first-order phase transition in Fig.~\ref{fig:SC_PT_constraints}. We consider pulsar-timing arrays EPTA \cite{Lentati:2015qwp}, (current), NANOGrav \cite{Arzoumanian:2018saf} (current) and SKA \cite{Breitbach:2018ddu} (planned), Space-based interferometer LISA \cite{Audley:2017drz} (planned), Earth-based interferometers LIGO O2 \cite{Aasi:2014mqd} (current), LIGO O5 (planned), Einstein Telescope \cite{Hild:2010id, Punturo:2010zz} (waiting for approval) and Cosmic Explorer \cite{Evans:2016mbw} (waiting for approval). We fixed the signal-to-noise ratio SNR = 10 and the observation time to $T=10$~years, except for LIGO O2 ($T=268~$days) and LIGO O5 ($T=1~$year). See App.~\ref{app:sensitivity_curves} for details about the computation of the power-law-integrated-sensitivity curves.


\subsection{Contributions from sound waves and turbulence}
\label{sec:GW_SW_turb}

In the presence of an interaction between the plasma and the scalar field driving the phase transition, the expanding bubble wall, if subsonic, generates a shock wave in front of the wall (deflagration fluid profile) and, if supersonic, generates a rarefaction wave behind the wall (detonation fluid profile)  \cite{Espinosa:2010hh}.
The first studies of the GW spectrum sourced by sound shells goes back to the 90s \cite{Kamionkowski:1993fg}. Since then, the hydrodymical simulations have been performed  \cite{Hindmarsh:2013xza,Giblin:2014qia, Hindmarsh:2015qta, Hindmarsh:2017gnf,Cutting:2019zws}, complemented by analytical investigations of the GW production from sound shell overlap \cite{Hindmarsh:2016lnk, Hindmarsh:2019phv}

As sound waves propagate, the flow becomes turbulent and non-linearities forms, triggering energy cascades to smaller scales and leading to another source of GW production \cite{Kosowsky:2001xp, Caprini:2006jb, Caprini:2009yp, Gogoberidze:2007an,Kahniashvili:2008pf, Kahniashvili:2008pe, Kahniashvili:2009mf, Kalaydzhyan:2014wca, Pen:2015qta,Niksa:2018ofa,Pol:2019yex}. 
 
The stochastic GW background sourced by sound waves, based on the lattice work of \cite{Hindmarsh:2013xza, Hindmarsh:2015qta, Hindmarsh:2017gnf} and sourced by turbulence, cf. LISA paper 2019 \cite{Caprini:2019egz}, reads
\begin{align}
& h^2\Omega_{\rm sw}= F_{\rm gw}\,h^2 ~0.049 \,\left(H_* \tau_{\rm sw}\right)\,  \left(\frac{H_* R_*}{c_s} \right)\, K_{\rm sw}^2  \, S_{\rm sw}(f) , \label{eq:GW_from_sw_formula}\\
&h^2\Omega_{\rm turb}=  F_{\rm gw}\,h^2 ~ 11.9 \, \left(1-H_* \tau_{\rm sw}\right)\,   \left(\frac{H_* R_*}{c_s} \right)\, K_{\rm turb}^{3/2}   \, S_{\rm turb}(k) .\label{eq:GW_from_turb_formula}
\end{align} 
with the spectral shapes
\begin{align}
&S_{\rm sw}(f_*) = \left( \frac{f_*}{f_{\rm sw}}  \right)^3  \left( \frac{1}{1 + \frac{3}{4} (f_*/f_{\rm sw})^2}  \right)^{7/2}, \\
&S_{\rm turb}(f_*) = \frac{(f_*/f_{\rm turb})^3}{(1+(f_*/f_{\rm turb}))^{11/3}(1+8\pi f/H_*)}, \\
&\frac{f_{\rm sw}}{H_*} = \frac{1.57}{H_*R_*}, \\
&\frac{f_{\rm turb}}{H_*} =\frac{2.38}{H_*R_*} .
\end{align}
The prefactor $F_{\rm gw}\,h^2$ is given in  Eq.~\eqref{eq:prefactor_Omega_GW}. $f_*$ denotes the frequency at emission and is related to the frequency today $f_0$ through Eq.~\eqref{eq:GW_frequency_emission_today_1stOPT}. The radius at collision is related to the duration of the phase transition $\beta^{-1}$ though
\begin{equation}
R_*  =  \frac{(8\pi)^{1/3}}{\beta}~{\rm Max}(v_w,\,c_s).
\end{equation}
$K_{\rm sw} $ and $K_{\rm turb} $ are the efficiency of energy transfer from the total energy density (in the symmetric phase) $\rho_{\rm tot}^*$ to the energy density of sound waves and turbulence.

We discuss $K_{\rm turb} $ in the next paragraph and dedicate the entire next section, Sec.~\ref{sec:energy_budget_sw}, to $K_{\rm sw} $, mainly based on the work of \cite{Espinosa:2010hh, Giese:2020rtr,Giese:2020znk}. The result for $K_{\rm sw} $ is given by Eq.~\eqref{eq:K_sw_cs_model} for arbitrary speed of sound $c_s$ and more specifically by Eq.~\eqref{eq:K_sw_Bag_model} in the case of the well-studied Bag model in which $c_s = 1/\sqrt{3}$.

$\tau_{\rm sw}$ is the life-time of sound-waves, and will also be defined later in Eq.~\eqref{eq:life_time_sound_wave} after having introduced the averaged fluid velocity $U_f$.

\paragraph{Efficiency transfer to turbulence:}
For turbulence, lattice simulations at small fluid velocity $U_f \lesssim 0.05$ have found  \cite{Hindmarsh:2015qta,Hindmarsh:2017gnf}
\begin{equation}
K_{\rm turb} \equiv  \epsilon_{\rm turb} \, K_{\rm sw} , \qquad \textrm{with}~ \epsilon_{\rm turb} \simeq 5\%.
\end{equation}
The precise value of $ \epsilon_{\rm turb}$ as well as its dependence on $\alpha$ remains to be determined. As hinted by recent simulations Cutting+ 19 \cite{Cutting:2019zws}, we expect $ \epsilon_{\rm turb}$ to increase with $\alpha$. See also the recent works \cite{Niksa:2018ofa, Pol:2019yex}.
The life-time of sound-waves $\tau_{\rm sw}$ coincides with the time after which non-linearities start to form and turbulence motion takes place. This explains the prescription of \cite{Ellis:2020awk} of adding the factor $(1-H_* \tau_{\rm sw})$ in Eq.~\eqref{eq:GW_from_turb_formula}.

In the next section, we discuss how to compute the efficiency of the energy transfer to sound waves kinetic energy $K_{\rm sw}$, needed in order to compute the GW signal in Fig.~\ref{fig:SW_PT_constraints}.

%

\subsection{Energy transfer to sound-waves }
\label{sec:energy_budget_sw}

\paragraph{Sound-waves as a source of GW:}
The energy momentum tensor of the fluid at first-order in the fluid gradient $\partial_\mu u_\nu$ expansion (perfect fluid approximation) is
\begin{equation}
T_{\mu\nu} = w ~u_\mu u_\nu - g_{\mu\nu}~p 
\end{equation}
where $u_\mu = \left(\gamma,~\gamma\,\vec{v}\right)$ is the fluid 4-velocity and $w=e+p$ is the fluid enthalpy density. $e$ and $p$ are the total energy and pressure densities ($e \equiv \rho_{\rm tot}$). GW are sourced by the transverse-traceless-spatial part of $T_{\mu\nu}$, cf. Sec.~\ref{sec:linerized_GW}, which for one bubble, has only one component along $rr$, and we get
\begin{align}
K_{\rm sw} = \frac{\rho_{\rm fl}}{\rho_{\rm tot}^*}, \qquad \rho_{\rm fl} = \left< w\, \gamma^2 \, v^2 \right>,
\label{eq:K_sw}
\end{align}
with 
\begin{equation}
\left< w\, \gamma^2 \, v^2 \right> = \frac{3}{\xi_w^3} \int d\xi \, \xi^2 \, v^2 \, \gamma^2 \, w.
\end{equation}
We have assumed that the profiles $v(\xi)$ and $w(\xi)$ are self-similar and only depend on $\xi = r/t$ where $r$ is the bubble radius and $t$ is the nucleation time. $\rho_{\rm tot}^*$ is the total energy density in the symmetric phase far from the wall (sometimes noted $e_n \equiv \rho_{\rm tot}^*$).

\begin{figure}[h!]
\centering
\begin{adjustbox}{max width=1.\linewidth,center}
\raisebox{0cm}{\makebox{\includegraphics[width=0.5\textwidth]{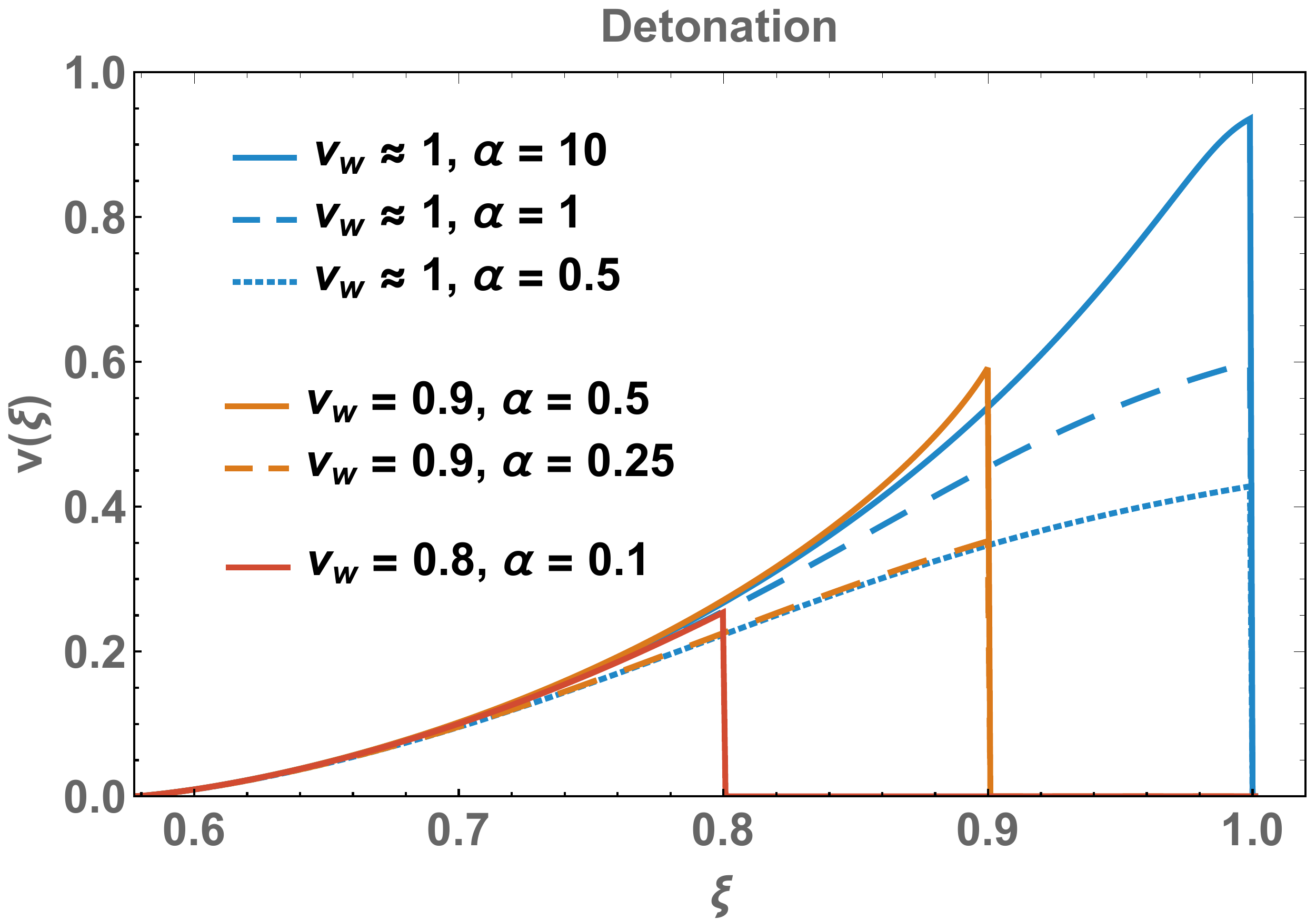}}}
\raisebox{0cm}{\makebox{\includegraphics[width=0.5\textwidth]{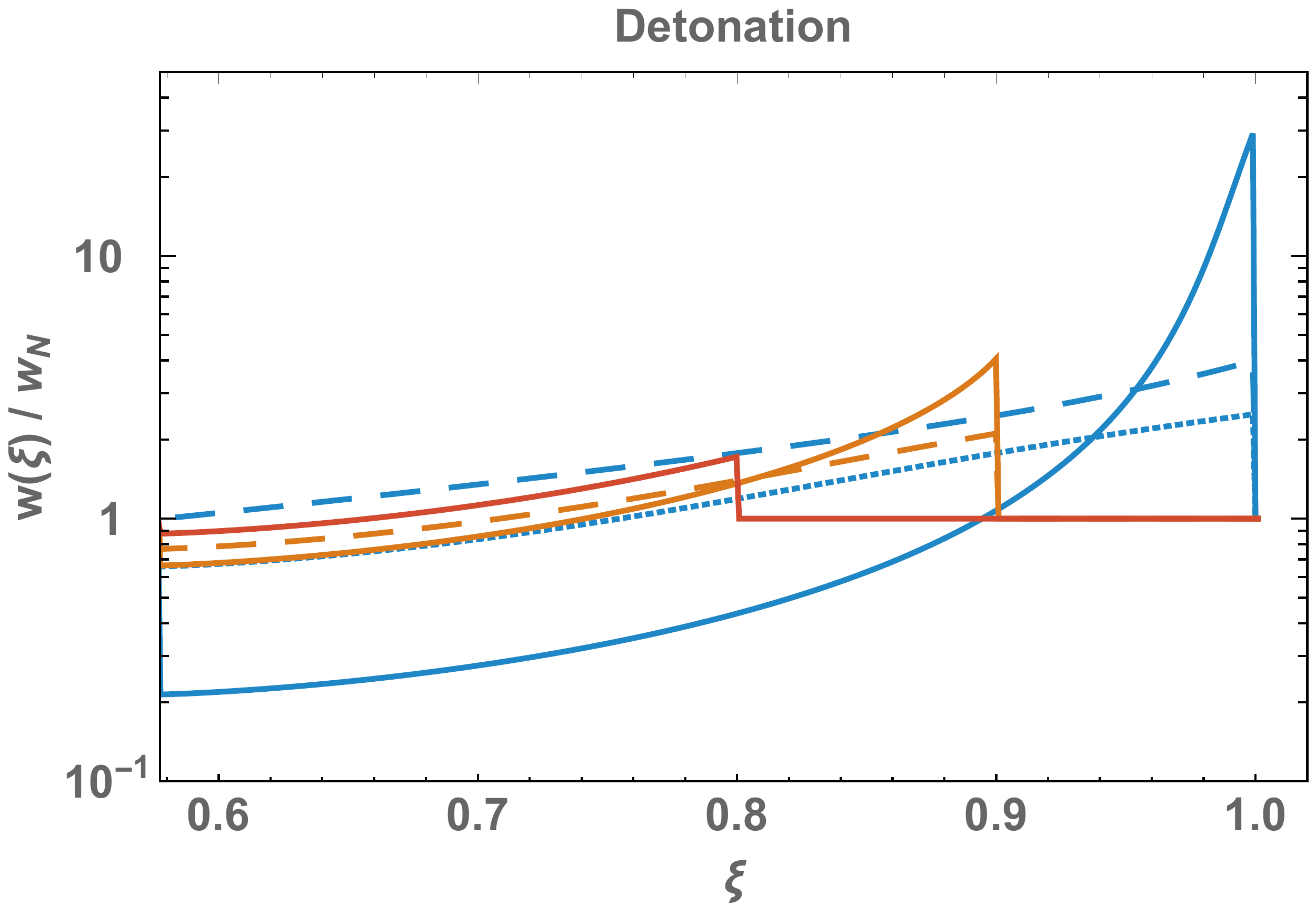}}}
\end{adjustbox}
\begin{adjustbox}{max width=1.\linewidth,center}
\raisebox{0cm}{\makebox{\includegraphics[width=0.5\textwidth]{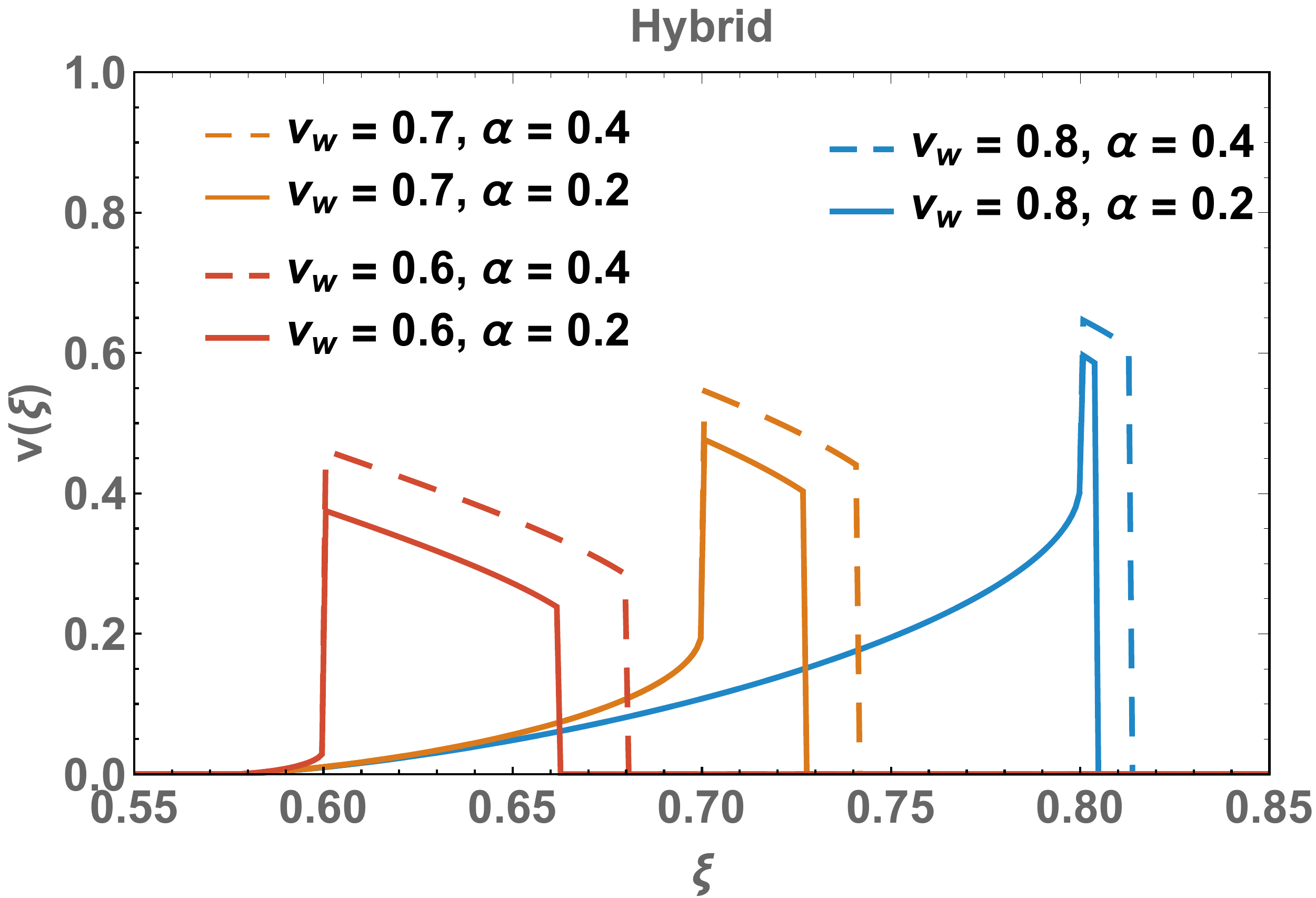}}}
\raisebox{0cm}{\makebox{\includegraphics[width=0.5\textwidth]{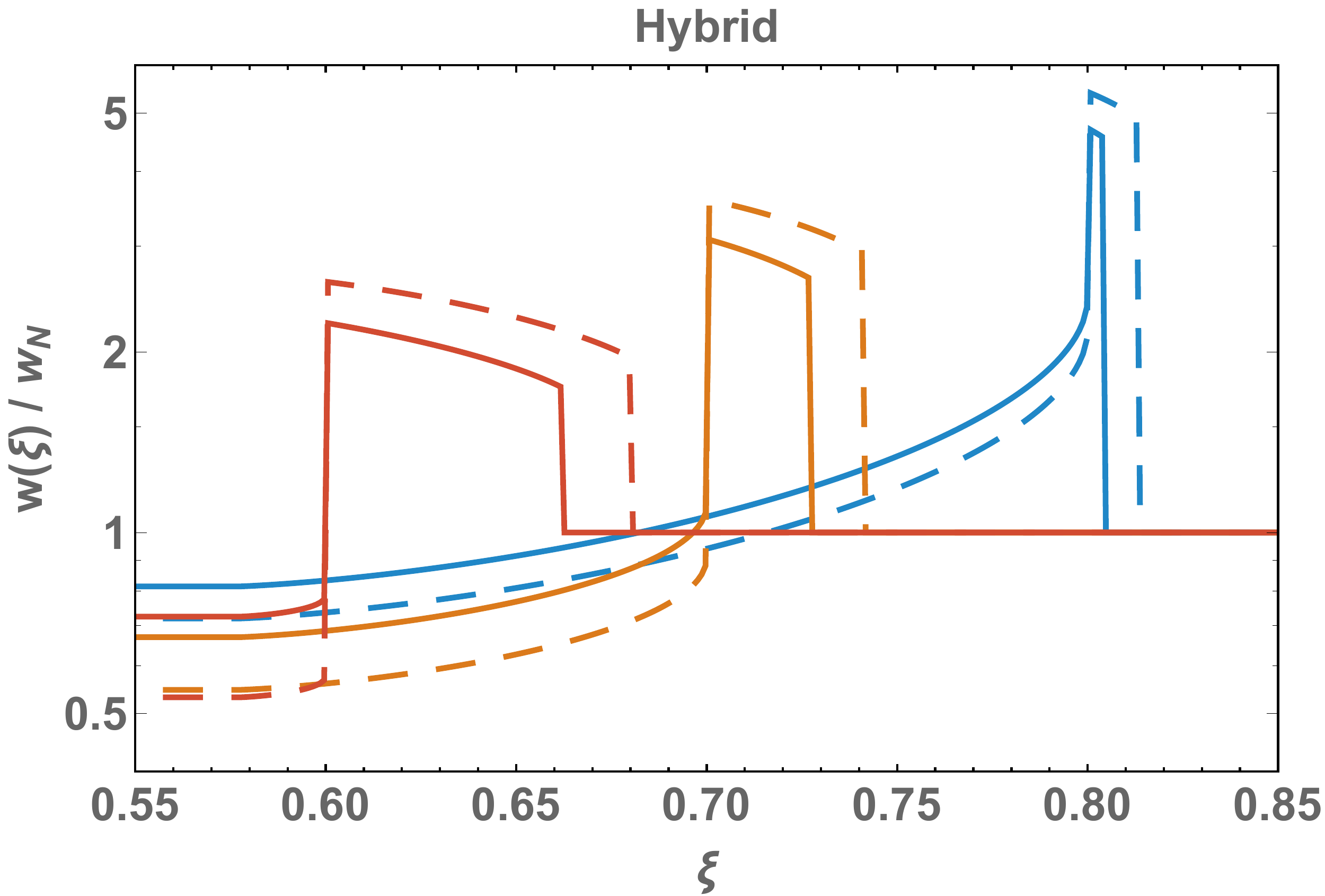}}}
\end{adjustbox}
\begin{adjustbox}{max width=1.\linewidth,center}
\raisebox{0cm}{\makebox{\includegraphics[width=0.5\textwidth]{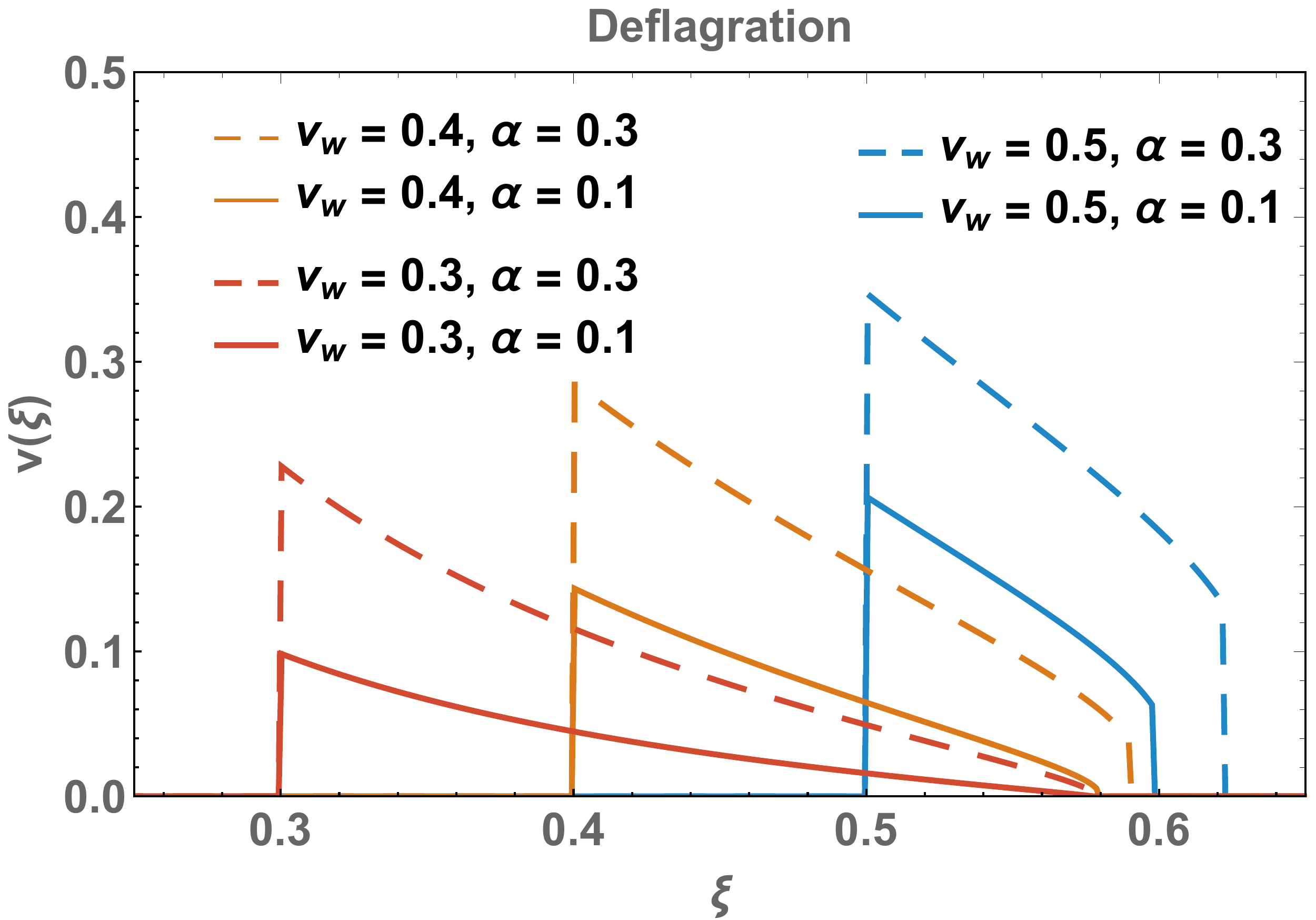}}}
\raisebox{0cm}{\makebox{\includegraphics[width=0.5\textwidth]{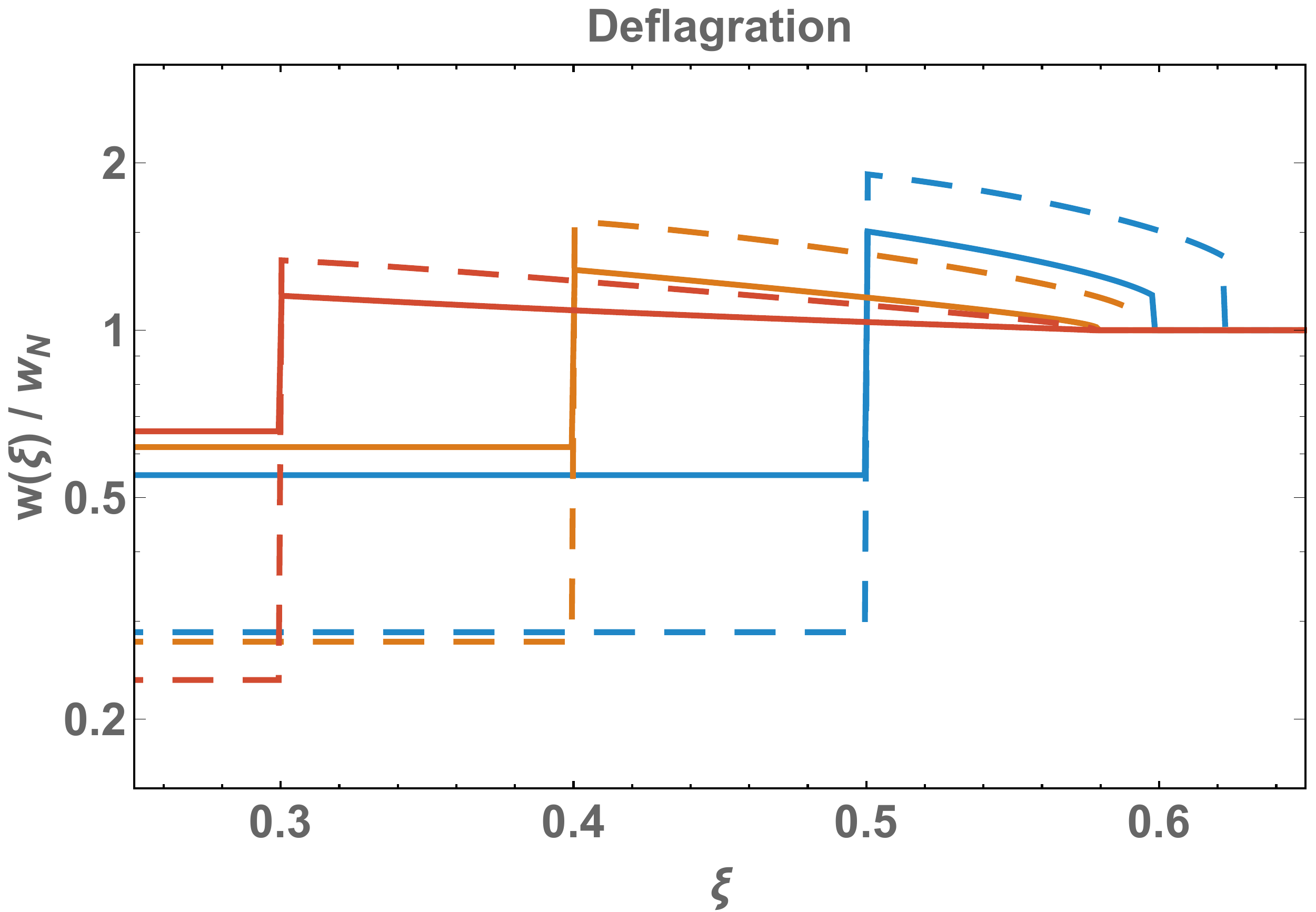}}}
\end{adjustbox}
\caption{\it \small  Velocity (in the plasma frame) and enthalpy profile found after integration of the hydrodynamic equations in Eq.~\eqref{eq:hydro_eq_v} and Eq.~\eqref{eq:hydro_eq_w} with the matching conditions in Eq.~\eqref{eq:vp_vm_bag}. Detonations, hybrids and deflagrations are defined by the boundary conditions $v_+ = v_w$ (only exists if $v_w> \xi_J$), $v_-= c_s$ (only exists if $\xi_J> v_w> c_s$) and $v_- = v_w$ (only exists if $c_s>v_w$), respectively. See \cite{Espinosa:2010hh} and \cite{Ellis:2020awk} for details about the computation of the hydrodynamic profile.}
\label{fig:detonation_deflagration_profile}
\end{figure}

\paragraph{The velocity and enthalpy density profile of the sound-shell:}
The velocity and enthalpy profile of the fluid can be computed using local conservation of the energy-momentum tensor \cite{Espinosa:2010hh}
\begin{equation}
\partial_\mu T^{\mu\nu}= u^\nu~ \partial_\mu(u^\mu~ w) + u^\mu ~w ~\partial_\mu u^\nu - \partial^\nu p = 0.
\label{eq:energy_momentum_conservation_fluid}
\end{equation}
which for a self-similar profile described by $\xi =r/t$ and a constant speed of sound $c_s^2 = dp/de$, becomes
\begin{align}
2\frac{v}{\xi} = \gamma^2(1-v\,\xi) \left[ \frac{\mu^2}{c_s^2} - 1 \right] \partial_\xi v,  \label{eq:hydro_eq_v}\\
\frac{\partial_v w}{w} = 2w(1+c_s^2) \frac{v}{\xi} \frac{\xi-v}{(\xi-v)^2 - c_s^2(1-\xi v)^2},\label{eq:hydro_eq_w}
\end{align}
where $\mu$ is the boosted fluid velocity
\begin{equation}
\mu(\xi,\,v) = \frac{\xi-v}{1-\xi\,v}.
\end{equation}
Additionally due the change of pressure between the two phases, the velocity in the \textbf{symmetric} phase $v_+$ and in the \textbf{broken} phase $v_-$ at the  wall position $\xi_w \equiv v_w$, in the wall frame, do not coincide, but can be related using the two matching equations \cite{Espinosa:2010hh}
\begin{align}
&\partial_z T^{zz} = 0, \quad \rightarrow \quad \omega_+v_+^2\gamma_+^2 +p_+ = \omega_-v_-^2\gamma_-^2 + p_-, \label{eq:matchin_eq_wall_1}\\
&\partial_z T^{z0} = 0,\quad \rightarrow \quad  \omega_+ v_+ \gamma_+^2 = \omega_- v_- \gamma_-^2 \label{eq:matchin_eq_wall_2}.
\end{align}
In the \textbf{Bag} model where the energy and pressure densities read \cite{Espinosa:2010hh}
\begin{align}
&e_+ = a_+\,T_+^4 \, +\, \epsilon, \qquad e_- = a_-\,T_-^4, \label{eq:Bag_model_e} \\ 
&p_+ = a_+\,T_+^4/3 \, -\, \epsilon, \qquad p_- = a_-\,T_-^4/3, \label{eq:Bag_model_p}
\end{align}
we obtain 
\begin{equation}
v_+ = \frac{1}{1+\alpha_+} \left[\left(\frac{v_-}{2} + \frac{1}{6v_-}\right) \pm \sqrt{\left(\frac{v_-}{2}+\frac{1}{6v_-}\right)^2 + \alpha_+^2 + \frac{2}{3}\alpha_+ - \frac{1}{3}}    \right],
\label{eq:vp_vm_bag}
\end{equation}
with
\begin{equation}
\alpha_+ \equiv \frac{\epsilon}{a_+\,T_+^4}.
\end{equation}
Note that the value of $\alpha_+$ at the wall is different from the input value $\alpha_N$ far ahead from the wall, which we simply denote by $\alpha$. They are related through
\begin{equation}
\alpha \equiv \alpha_N = \alpha_+\,\frac{w(\xi = v_w^+)}{w_N}.
\end{equation}
Depending on the value of $v_w$, the fluid profile behaves in three different ways: deflagration (shock-front, $v_w < c_s$ and $v_- = v_w$), hybrid (shock-front and rarefaction wave, $c_s<v_w<\xi_J$, $v_-= c_s$) and detonation (rarefaction wave, $v_w> \xi_J$, $v_+ = v_w$).  The transition between deflagration and hybrid occurs at the speed of sound, which in the Bag model reads
\begin{equation}
c_s^2 = 1/3,
\end{equation}
while the transition between detonation and hybrid occurs at the Jouguet velocity $\xi_J$ 
\begin{equation}
\xi_J =  \frac{1+\sqrt{\alpha_+(2+3\alpha_+}}{\sqrt{3}(1+\alpha_+)},
\end{equation}
obtained after setting $v_-=c_s$ in Eq.~\eqref{eq:vp_vm_bag}. We compute some profile solutions\footnote{I thank Felix Giese and Jorinde Van de Vis for useful discussions regarding the computation of the fluid profile.} for given values of $v_w$ and $\alpha$ and plot them in Fig.~\ref{fig:detonation_deflagration_profile}.

\paragraph{The potential energy sourcing the bubble expansion:}

The first thermodynamics law at constant volume (take a volume much larger than the bubble) in the presence of a heat transfer $\delta (e_QV)$ reads
\begin{equation}
\delta (e_QV) = d(eV)  + pdV  \qquad \overset{ V=\rm cst}{\xrightarrow{\hspace*{0.5cm}}} \qquad \delta e_Q = de = \frac{dw}{1+c_s^2} ,
\end{equation}
where in the last term we have introduced the speed of sound
\begin{equation}
c_s^2 = \frac{dp}{de}.
\end{equation}
Hence, we can identify 
\begin{equation}
e_Q = w/(1+c_s^2),
\end{equation}
as the thermal energy density of the fluid.
First, observe that the component $T_{00}$ of the energy-momentum tensor can be split into
\begin{align}
T_{00} = w\gamma^2-p &= w\gamma^2 v^2 +e \\
&= w\gamma^2 v^2+ \frac{w}{1+c_s^2}+\frac{c_s^2}{1+c_s^2}\left(e - \frac{p}{c_s^2}\right).
\end{align}
Second, note that the integral of $T^{00}$ around a single bubble
\begin{equation}
E = 4\pi \int_0^R dr \, r^2\, T_{00},
\end{equation}
must be constant for $R$ much larger than the bubble radius.  As a result, we get the following energy budget\footnote{These lines are inspired from Appendix B.2 of \cite{Hindmarsh:2019phv} where the energy budget is realized in the restricted case $c_s^2=1/3$.}
\begin{equation}
 e_{\rm K} \, +\, \Delta e_{\rm Q} =-\Delta e_{\rm pot},
 \label{eq:energy_budget_interp_trace_anomaly}
\end{equation}
where
\begin{align}
&\Delta e_{\rm pot} = \frac{3}{\xi_{\rm max}^3} \int_0^{\xi_{\rm max}} d\xi\, \xi^2\,(e_{\rm pot, \,\rm n} \,-\,e_{\rm pot}), \\
&e_{\rm K} = \frac{3}{\xi_{\rm max}^3} \int_0^{\xi_{\rm max}} d\xi\, \xi^2\,w\,\gamma^2\,v^2, \\
&\Delta e_{\rm Q} = \frac{3}{\xi_{\rm max}^3} \int_0^{\xi_{\rm max}} d\xi\, \xi^2\,(\frac{w_{\rm n}}{1+c_s^2}\,-\,\frac{w}{1+c_s^2}), 
\end{align}
and where $e_{\rm pot}$ is defined thought the pseudotrace $\bar{\theta}$ of the energy-momentum tensor
\begin{equation}
\label{eq:potential_energy_source_bubble_expansion}
e_{\rm pot} = \frac{c_s^2}{1+c_s^2}\bar{\theta}, \qquad {\rm with } \qquad \bar{\theta} \equiv \left(e - \frac{p}{c_s^2}\right).
\end{equation}
The upper integral boundary is $\xi_{\rm max} = {\rm max}(v_w,\, \xi_{\rm sh})$, with $\xi_{\rm sh}$ being the outermost position of the velocity profile. Quantities with the subscript $n$ are evaluated in the symmetric phase, far from the wall. 
Eq.~\eqref{eq:energy_budget_interp_trace_anomaly} allows to interpret the \textbf{pseudotrace anomaly} $ \dfrac{c_s^2}{1+c_s^2}\bar{\theta}$ as the \textbf{potential energy} available for the transformation. As the bubble expands, it gets converted into the fluid  kinetic energy density $w\,\gamma^2\,v^2$ and the fluid  thermal energy density $w/(1+c_s^2)$.

In the Bag model where, by definition, the energy and pressure densities read Eq.~\eqref{eq:Bag_model_e} and Eq.~\eqref{eq:Bag_model_p}, the potential energy which drives the phase transition coincides with the \textbf{vacuum energy} 
\begin{equation}
\Delta e_{\rm pot} = \epsilon, \qquad {\rm (Bag~ model)}
\end{equation}

However for generic models, the pressure and the energy density are functions of the type $a+b\,T+c\,T^2+c\,T^3+d\,T^4+\cdots$. Therefore, the net distinction between a purely relativistic component and a purely vacuum component, characteristic of the Bag model, is not possible anymore. Generally, the pressure is given by the effective potential at finite-temperature $p = -V_{\rm eff}(\phi,\,T)$ and the energy follows from the identity $e \equiv T \frac{\partial p}{\partial T} - p$, such that the potential energy in Eq.~\eqref{eq:potential_energy_source_bubble_expansion} driving the bubble expansion reads 
\begin{equation}
\Delta e_{\rm pot} \simeq e_{\rm pot, \, s} - e_{\rm pot, \, b},
\end{equation}
with 
\begin{align}
e_{\rm pot, \, i} = \left[V_{\rm eff} - \frac{T}{1+\frac{1}{c_s^2}} \frac{dV_{\rm eff}}{dT}\right]_{\rm phase~i}.
\end{align}
Note that the speed of sound
\begin{equation}
c_{s}^2 = \frac{dp/dT}{de/dT},
\end{equation}
depends on the phase (symmetric s or broken b). Idem for the temperature $T_s$ or $T_b$.

\paragraph{Efficiency parameter for arbitrary $c_s$}

We have seen in Eq.~\eqref{eq:energy_budget_interp_trace_anomaly} that the pseudotrace $\bar{\theta}$ pops up naturally when computing the energy budget of the bubble expansion for  arbitrary sound of speed c$_s$.

Additionally, the authors of \cite{Giese:2020rtr,Giese:2020znk} have shown that, for arbitrary $c_s$, the pseudotrace\footnote{Before the pseudo-trace $\bar{\theta}$, the simple trace (or trace anomaly) $\theta \equiv e - 3p$ was first advocated by \cite{Hindmarsh:2013xza, Hindmarsh:2015qta, Hindmarsh:2017gnf}.} is the good quantity which appears in the matching equations at the wall position $T_{\mu\nu}^+=T_{\mu\nu}^-$, in Eq.~\eqref{eq:matchin_eq_wall_1} and Eq.~\eqref{eq:matchin_eq_wall_2}, after performing a Taylor expansion in $T_- = T_+ +\delta T$.

For a \textbf{detonation} profile\footnote{For the deflagration case, we refer to \cite{Giese:2020znk}.}, this motivates the re-writing of the energy transfer parameter $K_{\rm sw}$ in Eq.~\eqref{eq:K_sw}, as
\begin{equation}
K_{\rm sw} = \frac{\rho_{\rm fl}}{e_+} = \Gamma ~ \alpha_{\bar{\theta}} ~ \kappa_{\bar{\theta}}, \qquad {(\rm arbitrary~ c_s~model)}
\label{eq:K_sw_cs_model}
\end{equation}
with
\begin{align}
&\textrm{pseudotrace:}\qquad &\bar{\theta}\equiv e - p/c_s^2, \\
&\textrm{pseudotrace difference:} &D\bar{\theta}=\bar{\theta}_+(T_+) -\bar{\theta}_-(T_+),\\
&\textrm{speed of sound:}\qquad &c_s^2 \equiv \left. \frac{dp_-/dT}{de_-/dT}\right|_{T_+}, \\
&\textrm{adiabaticity parameter:}\qquad &\Gamma = w_+/e_+, \\
&\textrm{strength parameter:}\qquad &\alpha_{\bar{\theta}} = \frac{D\bar{\theta}}{4w_+}, \\
&\textrm{efficiency parameter:}\qquad &\kappa_{\bar{\theta}}(c_s,\,\alpha_{\bar{\theta}}, \, v_w) = \frac{4\rho_{\rm fl}}{D\bar{\theta}}, \label{eq:kappa_bar_theta}\\
&\textrm{fluid kinetic energy density:}  \qquad &\rho_{\rm fl} = \left< w\, \gamma^2 \, v^2 \right>.
\end{align}
$e_+$, $w_+$ and $T_+$ are the energy, enthalpy density and temperature in the symmetric phase at the wall position, which for the detonation case, coincide with the quantities in the symmetric phase, far from the wall, $e_n$, $w_n$, $T_n$.
This allows to cast all the model-dependence within the three quantities $c_s$, $\alpha_{\bar{\theta}}$, $\xi_w$ such that the function $\kappa_{\bar{\theta}}(c_s,\,\alpha_{\bar{\theta}}, \, v_w)$ is model-independent and provided in Appendix A of \cite{Giese:2020rtr} in the detonation case.

In the Bag model, we have
\begin{align}
&c_s^2 = 1/3, \\
&\Gamma = \frac{4}{3}\frac{1}{1+\alpha_\epsilon}, \\
&D\bar{\theta} = 4\epsilon,  \\
&\alpha_{\bar{\theta}}= \frac{3}{4}\alpha_\epsilon, \\
&\kappa_{\bar{\theta}} = \kappa_{\epsilon}, \\
\end{align}
or equivalently
\begin{equation}
K_{\rm sw} = \frac{\rho_{\rm fl}}{\rm e_n} = \frac{ \alpha_{\epsilon}  \kappa_{\epsilon}}{1+\alpha_{\epsilon}}, \qquad {(\rm Bag~model)},
\label{eq:K_sw_Bag_model}
\end{equation}
where\footnote{In the bag model, which is defined by the energy $e$ and pressure $p$ density being given by a radiation component $\propto T^4$ plus a vacuum component $\propto T^0$, we have $\alpha_\epsilon \equiv \frac{\Delta \theta}{e_{+}}$ where $\theta =\frac{1}{4}(e - 3p)$ is the trace anomaly, $e_{+}$ energy density in the symmetry phase, and $\Delta$ being the phase difference \cite{Espinosa:2010hh}. From using the thermodynamic identity $e = T \frac{\partial p}{\partial T} - p$, we obtain $\theta = -p  + \frac{T}{4} \frac{\partial p}{\partial T}$. } 
\begin{align}
&\alpha_\epsilon \equiv \frac{\epsilon}{a_+\,T_+^4}, \\
&\kappa_{\epsilon}(\alpha, \, v_w)\equiv \frac{\rho_{\rm fl}}{\epsilon},\label{eq:kappa_epsilon}
\end{align}
are the usual $\alpha$ and $\kappa$ parameters  \cite{Espinosa:2010hh}, which however are only relevant for the (restricted) Bag model. The function $\kappa_{\rm \epsilon}(\alpha,\, v_w)$ is provided in Appendix A of \cite{Espinosa:2010hh} for detonation, deflagration and hybrid profiles.

For general models, the pressure is given by the effective potential at finite-temperature $p = -V_{\rm eff}(\phi,\,T)$ and the energy follows from the identities $e \equiv T \frac{\partial p}{\partial T} - p$, such that the parameters of the \textbf{arbitary c$_s$ model}, $c_s$, $\alpha_{\bar{\theta}}$ and $\Gamma$, reduce to 
\begin{align}
&c_s^2 =  \left[\frac{dV_{\rm eff}/dT}{Td^2V_{\rm eff}/dT^2}\right]_{\phi=\phi_{\rm true},\,T_+},\\
&\alpha_{\bar{\theta}} = \frac{1}{4} \left[1-(1+\frac{1}{c_s^2})\frac{DV_{\rm eff}}{T\partial DV_{\rm eff}/ \partial T} \right]_{T_+}, \\
&\Gamma = \left[\frac{T dV_{\rm eff}//dT}{TdV_{\rm eff}/dT - V_{\rm eff}}\right]_{\rm \phi=0,\,T_+}, \\
&DV_{\rm eff} = V_{\rm eff}(\phi=0,\,T_+) - V_{\rm eff}(\phi=\phi_{\rm true},\,T_+). 
\end{align}

\begin{figure}[h]
\centering
\begin{adjustbox}{max width=1.2\linewidth,center}
\raisebox{0cm}{\makebox{\includegraphics[width=0.55\textwidth]{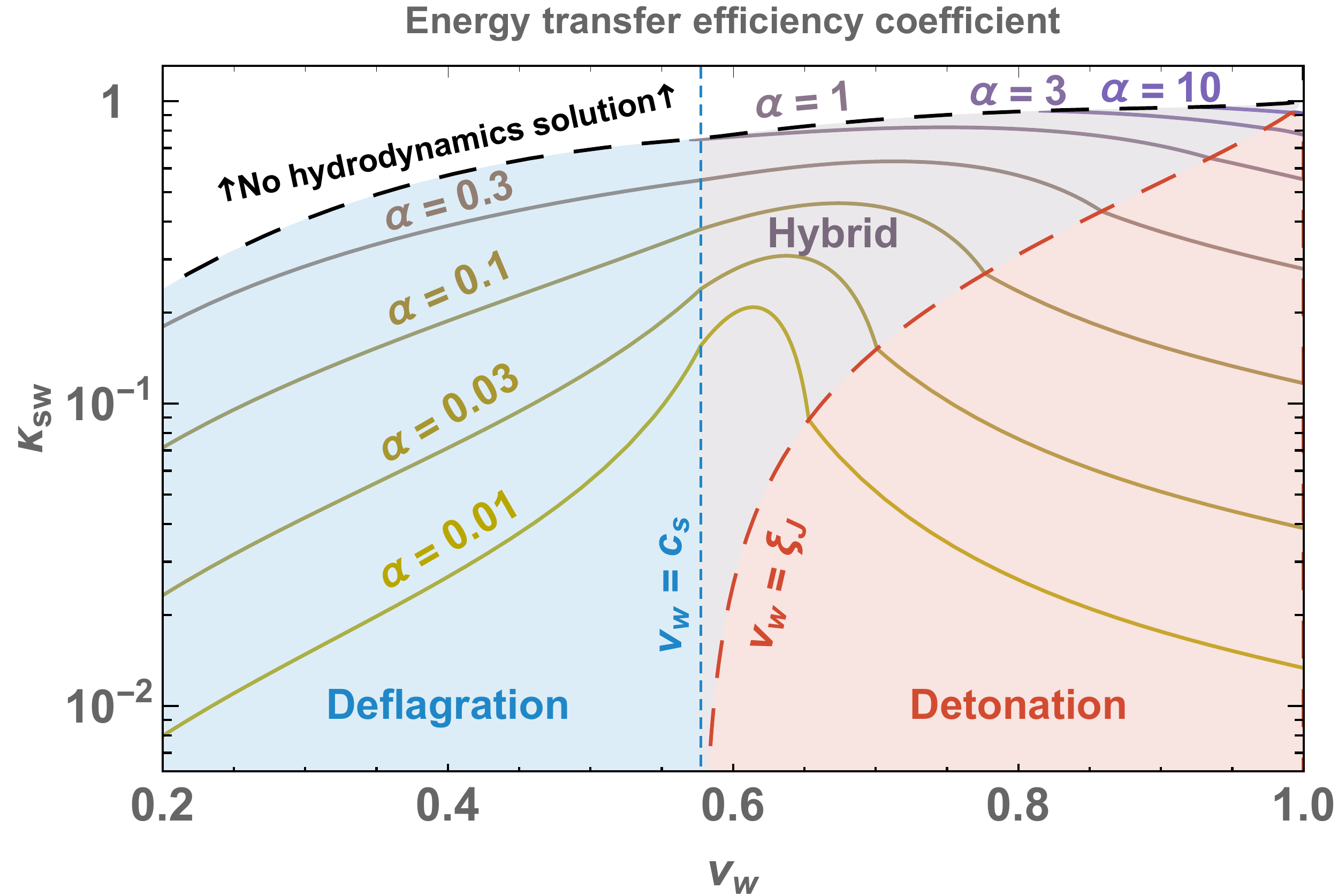}} }
\raisebox{0cm}{\makebox{\includegraphics[width=0.55\textwidth]{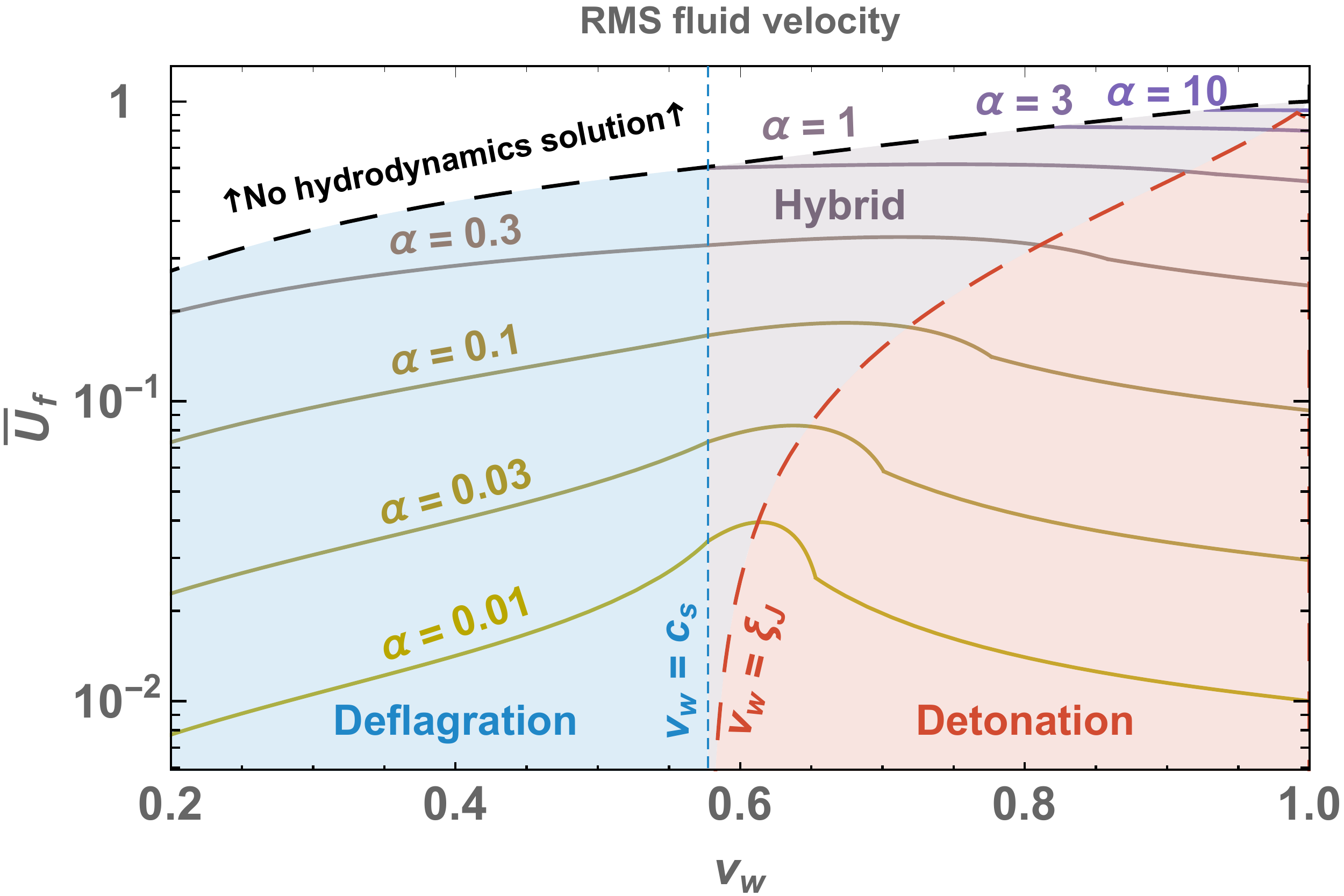}}}
\end{adjustbox}
\caption{\it \small   \textbf{Left}: Efficiency coefficient $\kappa_{\rm \epsilon}$ in the Bag model, defined in Eq.~\eqref{eq:kappa_epsilon}, of the energy transfer from vacuum energy to fluid bulk motion. The function is given in App.~A of \cite{Espinosa:2010hh}. \textbf{Right}: Enthalpy-averaged root-mean-square fluid velocity in the bubble-center frame, defined in Eq.~\eqref{eq:fluid_velocity_avg}, in the case of the Bag model. }
\label{fig:efficiency_sw_RMS_fluid_velocity_vs_vw}
\end{figure}

\begin{figure}[h]
\centering
\begin{adjustbox}{max width=1.2\linewidth,center}
\raisebox{0cm}{\makebox{\includegraphics[width=0.6\textwidth]{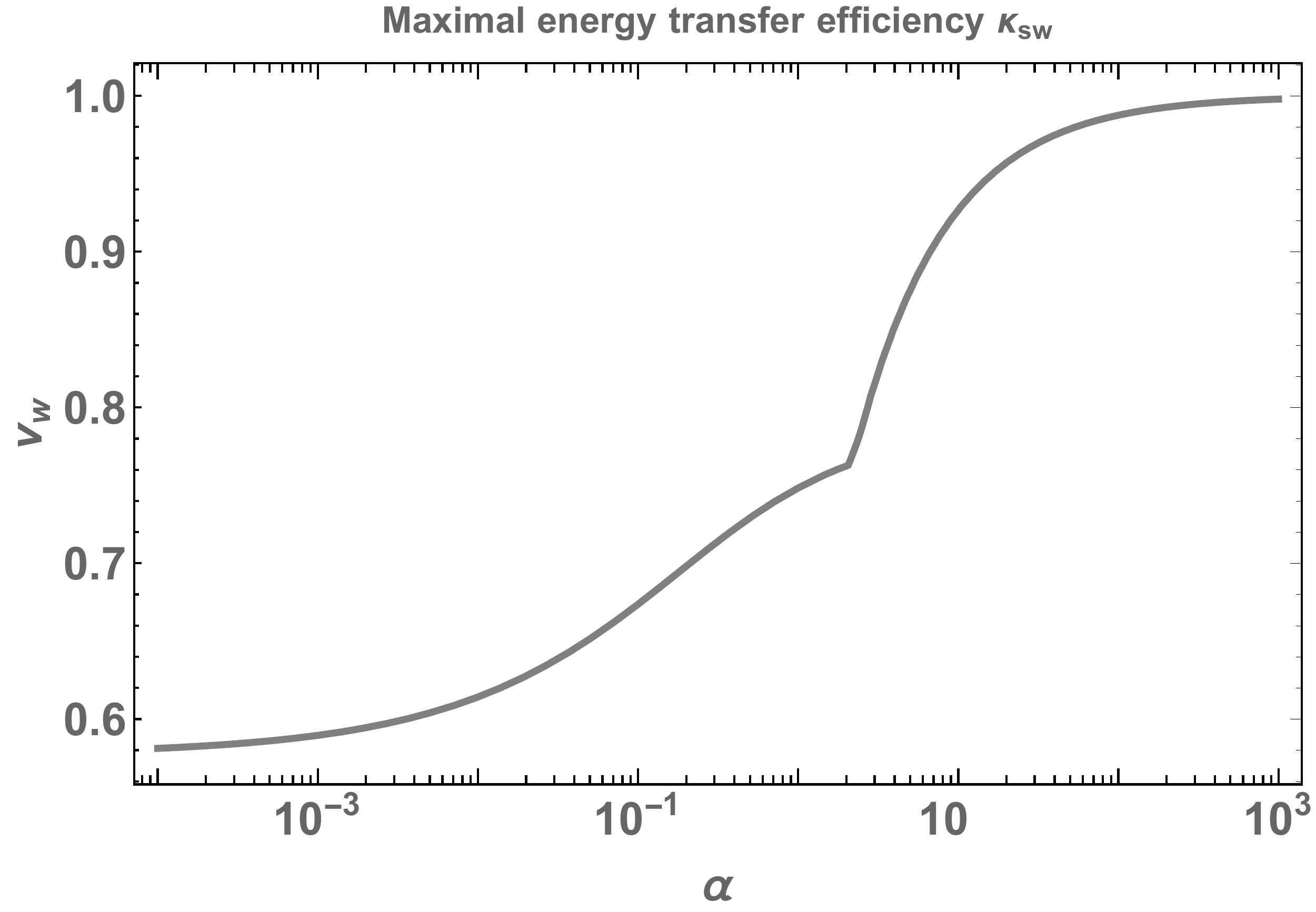}}}
\raisebox{0cm}{\makebox{\includegraphics[width=0.6\textwidth]{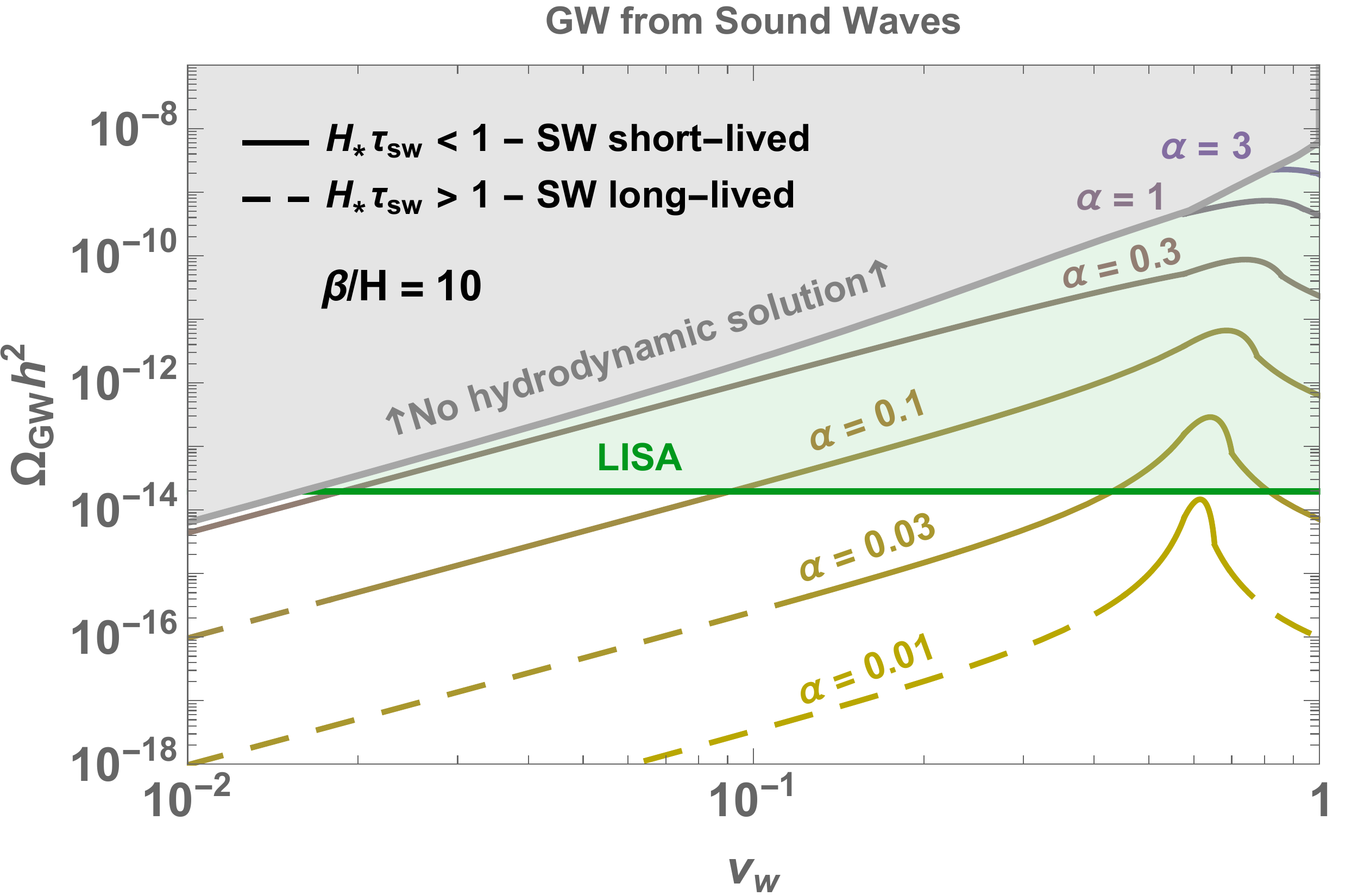}}}
\end{adjustbox}
\begin{adjustbox}{max width=1.2\linewidth,center}
\raisebox{0cm}{\makebox{\includegraphics[width=0.6\textwidth]{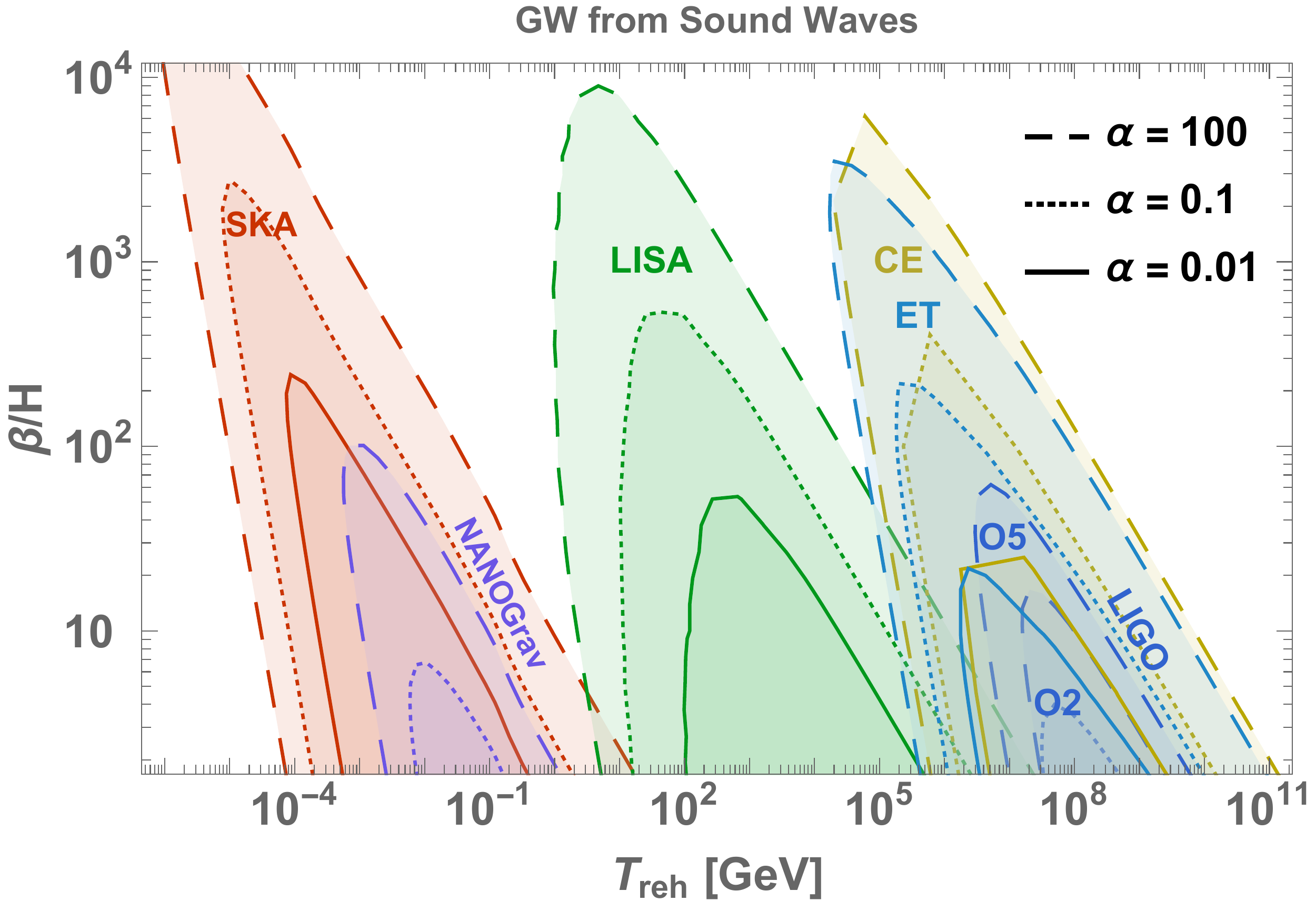}}}
\raisebox{0cm}{\makebox{\includegraphics[width=0.6\textwidth]{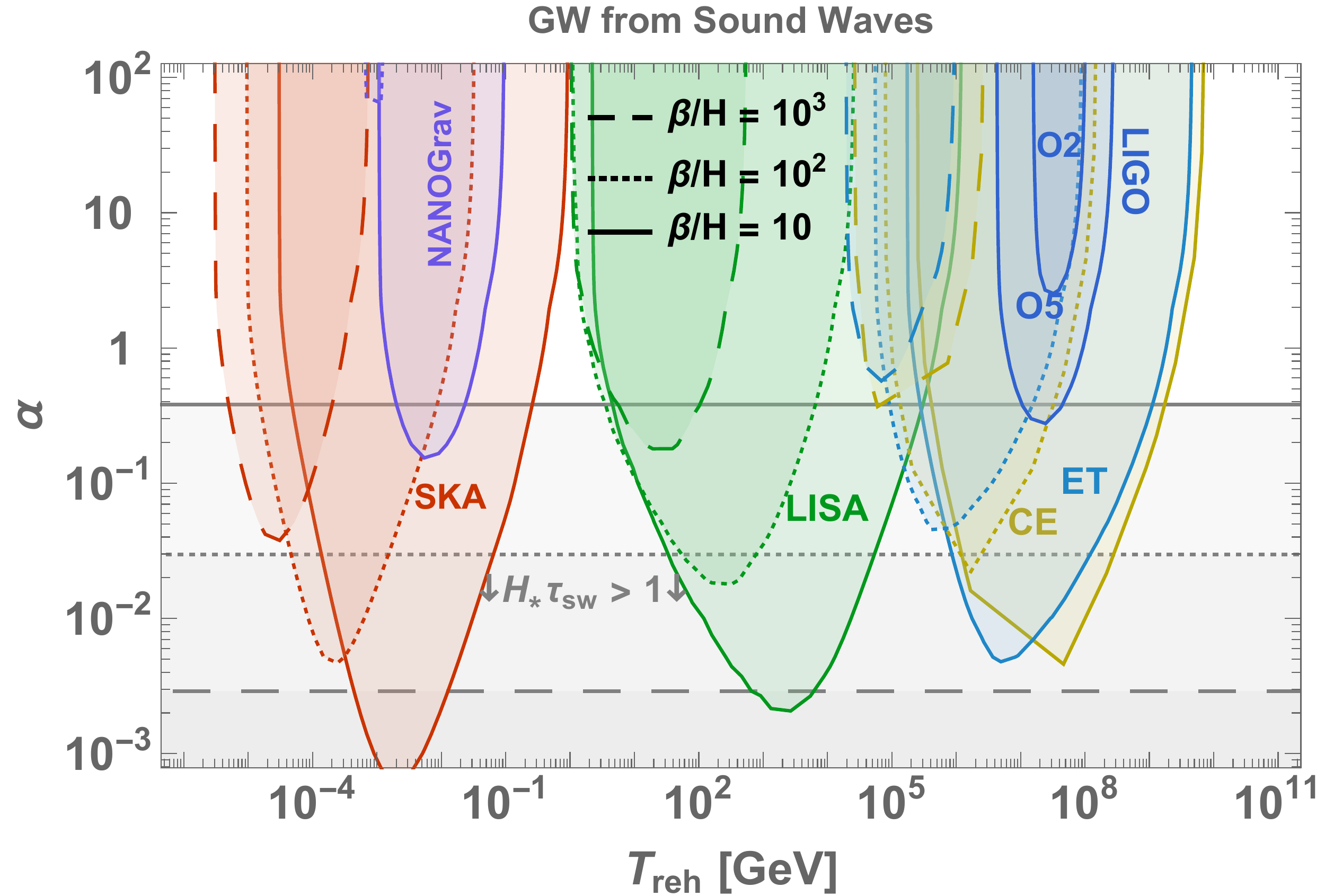}}}
\end{adjustbox}
\caption{\it \small  Reach of current and future experiments on thermal 1stOPT. GW are sourced by the acoustic waves, cf. formula of LISA paper 2019 in Eq.~\eqref{eq:GW_from_sw_formula}. \textbf{Top left:} Wall velocity which maximizes the energy transfer efficiency, see left panel of Fig.~\ref{fig:efficiency_sw_RMS_fluid_velocity_vs_vw}. \textbf{Top right:} Dependence of the peak amplitude of the GW signal on the wall velocity. In the gray region, corresponding to $\alpha> \frac{1}{3}(1-v_w)^{-13/10}$, the fluid profile has no hydrodynamic solution \cite{Espinosa:2010hh}. \textbf{Bottom:} The wall velocity $v_w$ is chosen in order to maximize the GW signal. We consider the Bag model, such that $c_s = 1/\sqrt{3}$, $\alpha=\alpha_{\epsilon}$ and $\kappa_{\rm sw} = \kappa_{\epsilon}$, cf. definitions around Eq.~\eqref{eq:K_sw_cs_model}. The contribution from turbulence is not included.  }
\label{fig:SW_PT_constraints}
\end{figure}

\paragraph{Average fluid velocity}
Another averaged quantity which we can compute once we know the profile $v(\xi)$ and $w(\xi)$ is the enthalpy-averaged root-mean-square velocity of the fluid
\begin{equation}
\bar{\gamma}_f^2\,\bar{U}_f^2 \equiv \frac{1}{\bar{w}V} \int_V T_{ii}\,d^3x = \frac{\left< w\, \gamma^2 \, v^2 \right>}{\bar{w}}
\label{eq:fluid_velocity_avg}
\end{equation}
The average fluid velocity in Eq.~\eqref{eq:fluid_velocity_avg} can be related to the efficiency parameter $\kappa_{\bar{\theta}}$ in Eq.~\eqref{eq:kappa_bar_theta} and $\kappa_{\epsilon}$ in Eq.~\eqref{eq:kappa_epsilon}
\begin{align}
\bar{\gamma}_f^2\,\bar{U}_f^2 =\begin{cases}
\kappa_{\bar{\theta}} \alpha_{\bar{\theta}} &\hspace{2em}\textrm{(Arbitrary}~ c_s~\rm model),\\
 \frac{3}{4} \kappa_{\epsilon} \alpha_{\epsilon} &\hspace{2em}\textrm{(Bag~model:}~c_s^2 = 1/3),\\
\end{cases}
\end{align}
or equivalently\footnote{Note the difference with the litterature \cite{Hindmarsh:2013xza, Hindmarsh:2015qta, Hindmarsh:2017gnf,Ellis:2020awk,Caprini:2019egz} where the authors do the approximations $\bar{\gamma}_f^2\,\bar{U}_f^2 \rightarrow \bar{U}_f^2$ and $\frac{\bar{w}}{\bar{e}} = \frac{4}{3(1+\alpha_\epsilon)}  \rightarrow  \frac{4}{3}$, which implies $\bar{U}_f^2 = \dfrac{ \kappa_{\bar{\theta}} \alpha_{\bar{\theta}}}{1+ \kappa_{\bar{\theta}} \alpha_{\bar{\theta}}} \rightarrow \frac{3}{4}\dfrac{\kappa_\epsilon \alpha_\epsilon}{1+\alpha_\epsilon}$ .}
\begin{align}
\bar{U}_f^2 =\begin{cases}
\dfrac{ \kappa_{\bar{\theta}} \alpha_{\bar{\theta}}}{1+ \kappa_{\bar{\theta}} \alpha_{\bar{\theta}}}&\hspace{2em}\textrm{(Arbitrary}~ c_s~ \rm model),\\
 \dfrac{3 \kappa_{\epsilon} \alpha_{\epsilon}/4}{1+ 3\kappa_{\epsilon} \alpha_{\epsilon}/4} &\hspace{2em}\textrm{(Bag~model:}~c_s^2 = 1/3),\\
\end{cases}
\end{align}
We plot the functions $\kappa_\epsilon$ and $\bar{U}_f$ in the case of the Bag model in Fig.~\ref{fig:efficiency_sw_RMS_fluid_velocity_vs_vw}.

\paragraph{Life-time of sound-waves and onset of turbulence:}

The averaged fluid velocity is an important quantity since it allows to determine the lifetime $\tau_{\rm sw}$ of the sound-waves, after which shocks form and non-linearities develop \cite{Pen:2015qta,Hindmarsh:2017gnf}
\begin{equation}
\tau_{\rm sw} \sim L_f / \bar{U}_f,
\label{eq:life_time_sound_wave}
\end{equation}
where $L_f$ is the characteristic length of the fluid flow. It is approximately given by the  mean bubble separation at percolation $R_*$, such that  \cite{Ellis:2020awk}
\begin{equation}
H \tau_{\rm sw} \sim \frac{H\,R_*}{\bar{U}_f} \simeq \frac{(8\pi)^{1/3} \textrm{Max}(v_w,\,c_s)}{\bar{U}_f} \left( \frac{H}{\beta} \right)
\end{equation}

\paragraph{Constraints on GW from sound-waves:}

We show the current and future constraints on the GW signal generated by acoustic waves during a thermal first-order phase transition in Fig.~\ref{fig:SW_PT_constraints}. We consider pulsar-timing arrays NANOGrav \cite{Arzoumanian:2018saf} (current) and SKA \cite{Breitbach:2018ddu} (planned), Space-based interferometer LISA \cite{Audley:2017drz} (planned), Earth-based interferometers LIGO O2 \cite{Aasi:2014mqd} (current), LIGO O5 (planned), Einstein Telescope \cite{Hild:2010id, Punturo:2010zz} (waiting for approval) and Cosmic Explorer \cite{Evans:2016mbw} (waiting for approval). We fixed the signal-to-noise ratio SNR = 10 and the observation time to $T=10$~years, except for LIGO O2 ($T=268~$days) and LIGO O5 ($T=1~$year). See App.~\ref{app:sensitivity_curves} for details about the computation of the power-law-integrated-sensitivity curves.

Note with bottom-right panel of Fig.~\ref{fig:SW_PT_constraints} that except for the poorly motivated case where $\beta/H \sim 10$ and $\alpha \ll 1$, the life-time of sound-waves corresponding to the reach of experiments is never longer than a Hubble time $H_* \tau_{\rm sw} <1$. This is in contrast to what was initially claimed \cite{Caprini:2009yp,Caprini:2015zlo}. See the extensive discussion in \cite{Ellis:2020awk}.

\section{Supercooling from a nearly-conformal sector}
\label{sec:supercool_potential}

In the previous sections of the chapter, we have discussed the rate of bubble nucleation, the speed of the bubble wall and the GW generation due to bubble collisions. We know study in details two classes of potential leading to a supercooled first-order phase transition: the Coleman-Weinberg potential in the weakly-coupled scenario and the light-dilaton potential in the strongly-coupled scenario. For both cases, we compute the bounce action, the nucleation temperature and the GW parameters $\alpha$ and $\beta$. The results for the light-dilaton potential are relevant for Chap.~\ref{chap:SC_conf_PT} about string fragmentation in supercooled confinement and follow-ups \cite{Baldes:2021aph}.

\subsection{Weakly-coupled scenario: the Coleman-Weinberg potential}

\paragraph{The lagrangian:}

We extend the SM with a new gauge group $SU(2)_X$ whose generators define three dark gauge bosons $X$. We also add a scalar $\phi$ transforming as a doublet of $SU(2)_X$. We suppose that the tree level Lagrangian is \textbf{scale-invariant}, so that the bare mass terms are forbidden and the only interaction terms between $\phi$ and the SM Higgs $H$ which we can write are  \cite{Hambye:2018qjv, Baldes:2018emh}
\begin{equation}
\mathcal{L} \supset -V =   -\lambda_{\rm h}H^{4}  -\lambda_{\rm \sigma} \phi^{4} -\lambda_{\rm h\sigma} \phi^{2} H^{2}.
\end{equation}
with
\begin{equation}
H=\exp\left(i \sigma^{i} G^{i}/v \right) \begin{pmatrix}
           0 \\
           h/\sqrt{2}
         \end{pmatrix},
         \qquad
\phi=\exp\left(i \sigma^{i} G_{X}^{i}/v \right) \begin{pmatrix}
           0 \\
         \sigma/\sqrt{2}
         \end{pmatrix},
         \qquad
i \in {1,2,3}.
\end{equation}
$G^{i}$ and $G_{X}^{i}$ are the Golsdstone bosons whereas $h$ and $\sigma$ are the massive states after taking the vevs $\left<h\right>=v$ and $\left<\sigma\right>=f$.

\paragraph{Radiatively-induced symmetry breaking:}
In the absence of bare mass term, spontenous symmetry breaking of $SU(2)_X$ can occur due to radiative effects \cite{Gildener:1976ih}. Indeed, the integration of the $\beta$ function 
\begin{equation}
\beta_{\lambda_{\rm \sigma}} \equiv \frac{d\lambda_{\rm \sigma}}{d\ln{\mu}} \simeq \frac{1}{(4\pi)^{2}}\frac{9 g_{x}^4}{8},
\end{equation}
gives the one-loop Coleman-Weinberg potential  \cite{Hambye:2018qjv, Baldes:2018emh}
\begin{equation}
\label{eq:1loopRadiativePotentialDarkScalar}
V_{\rm 1-loop}(\sigma) \simeq\beta_{\lambda_{\rm \sigma}}  \frac{\sigma^4}{4} \left( \ln{\frac{\sigma}{f}} - \frac{1}{4}\right),
\end{equation}
which has a minimum at $\left<\sigma\right>=f$. \newline
As a consequence
\begin{enumerate}
\item
The $SU(2)_X$ gauge bosons receive the mass\footnote{The Dark gauge bosons transform as a triplet under a custodial $SO(3)$ symmetry so their masses are degenerate and DM has 9 degrees of freedom.}
\begin{equation}
\label{eq:DM_mass_definition}
M_{X} = g_{X} ~ f/2
\end{equation}
where $g_{X}$ is the $SU(2)_X$ coupling constant. 
\item
The higgs gets the vev
\begin{equation}
v = f \sqrt{\frac{-\lambda_{\rm h\sigma}}{2\lambda_{\rm h}}}.
\end{equation}
\end{enumerate}
The mass matrix of the $\sigma-h$ system is
\begin{equation}
\frac{\partial V}{\partial h \partial \sigma}|_{(v,~\omega)} = \begin{pmatrix}
           2\lambda_{\rm h} v^2 & \lambda_{\rm h\sigma} v f \\
          \lambda_{\rm h\sigma} v f &   \beta_{\lambda_{\rm \sigma}} f^2 - \lambda_{\rm h\sigma}^{2}\frac{f^2}{4 \lambda_{\rm h}}
         \end{pmatrix}.         
\end{equation}
In the limit of small $h-\sigma$ mixing $\lambda_{\rm h\sigma}\ll 1$, the mass eigenvalues reduce to
\begin{align}
m_{h}^2 \simeq 2\lambda_{\rm h}v^2, \\
m_{\sigma}^2 \simeq  \beta_{\lambda_{\rm \sigma}} \, f^2.
\end{align}

\paragraph{Finite-temperature contributions:}

At finite temperature the potential for $\sigma$ receives 1-loop thermal corrections from the $3$ gauge bosons + the Daisy contribution from the $3$ longitudinal modes, cf. Sec.~\ref{sec:effective_potential}
\begin{equation}
\label{eq:ThermalPotentialDarkScalar}
V_{T}(\sigma,~T) = V_{\rm 1-loop}^{T}+V_{\rm Daisy} =  \frac{9 T^{4}}{2\pi^2} J_{B}\left(\frac{M_{X}^2}{T^2}\right) + \frac{T}{4\pi} \left[ M_{X}^3 - \left( M_{X}^{2} + \Pi_{X}\right)^{3/2} \right].
\end{equation}
where the dark gauge boson thermal mass
\begin{equation}
\Pi_{X} = \frac{5}{6}g_{X}^{2}T^{2} ~C(M_{X}/T),
\end{equation}
is cutted-off with $C(x)=x^2/2~K_{2}(x)$ to prevent non-physical contributions from gauge boson at large $M_{X}/T$ \cite{Baldes:2018emh}.

At high temperature, the scalars get positive thermal masses such that the minimum of the potential is in $\left<\sigma\right>=\left<h\right>=0$. When the radiation energy density $\pi^2 g_{*} T^4/30$ gets smaller than the vacuum energy
\begin{equation}
V_{\Lambda}\simeq \frac{9M_{X}^4}{8(4\pi)^2},
\end{equation}
a period of inflation starts, this is the \textbf{supercooling} period. It happens when the temperature reaches the critical temperature
\begin{equation}
T_{\rm inf}  \simeq \frac{M_{X}}{8.4} \left( \frac{116}{g_*}\right)^{1/4}.
\end{equation}

\paragraph{Bounce action:}

\begin{figure}[h]
\centering
\begin{adjustbox}{max width=1.2\linewidth,center}
\raisebox{0cm}{\makebox{\includegraphics[width=0.55\textwidth]{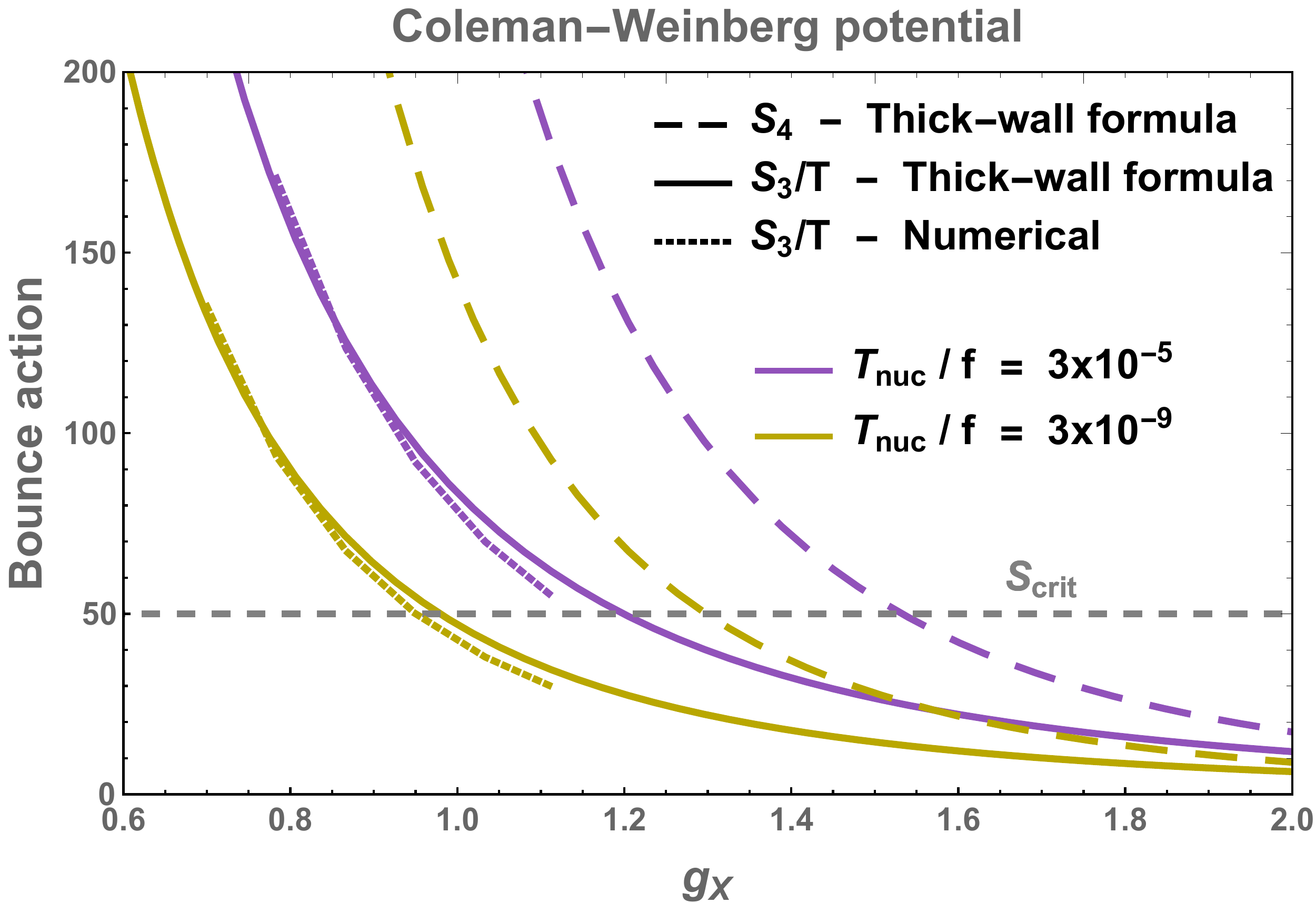}}}
\raisebox{0cm}{\makebox{\includegraphics[width=0.55\textwidth]{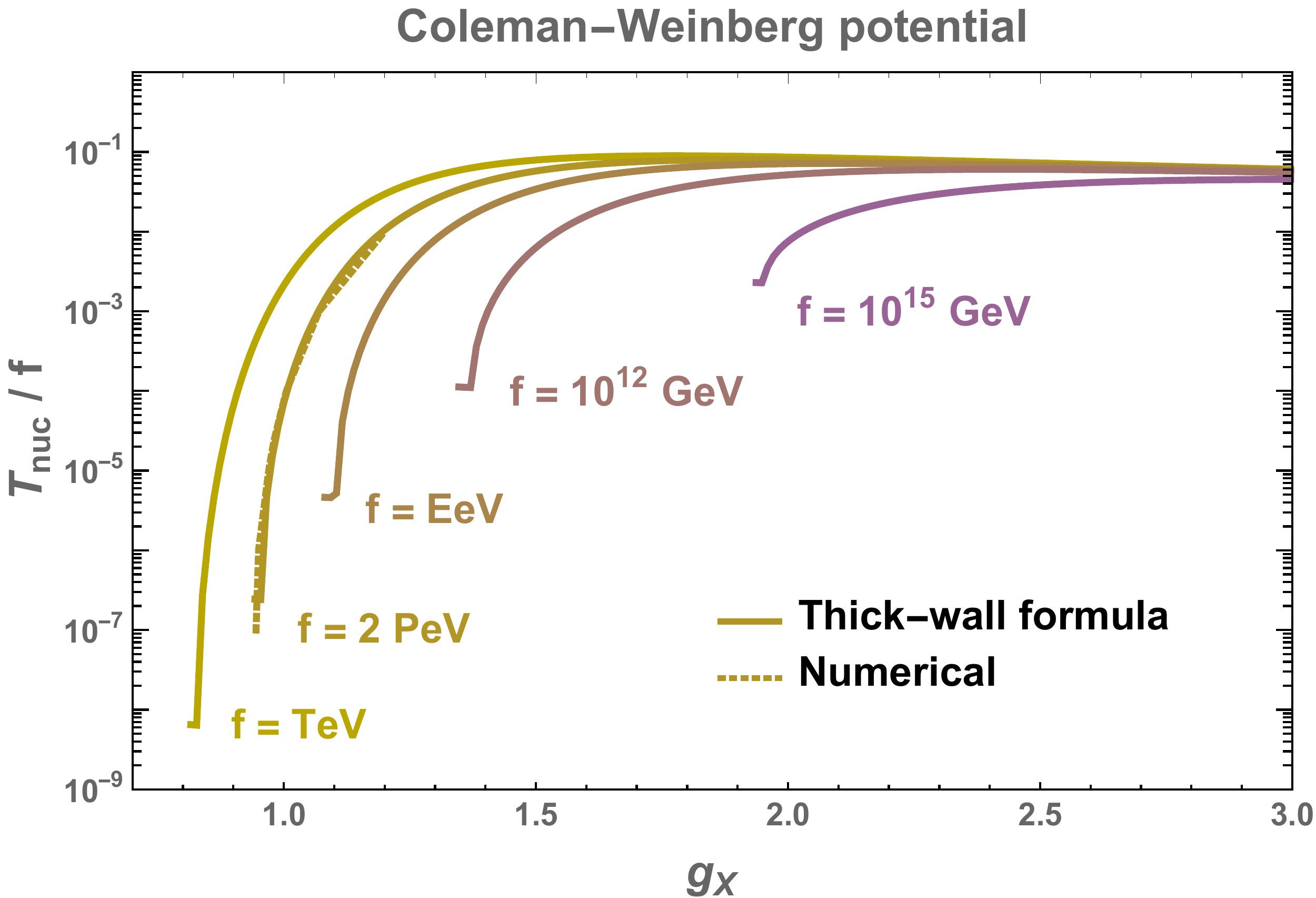}}}
\end{adjustbox}
\begin{adjustbox}{max width=1.2\linewidth,center}
\raisebox{0cm}{\makebox{\includegraphics[width=0.55\textwidth]{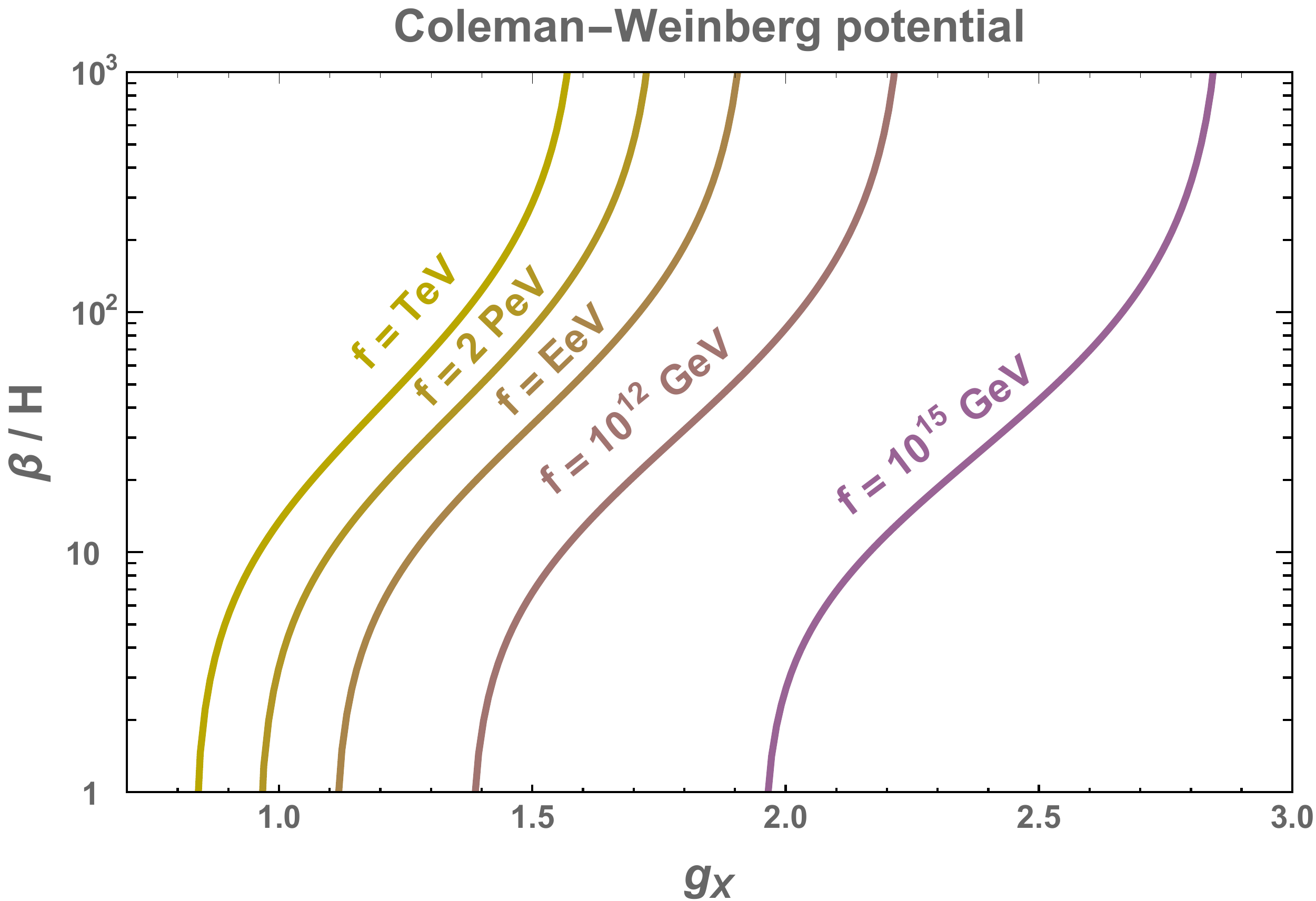}}}
\raisebox{0cm}{\makebox{\includegraphics[width=0.55\textwidth]{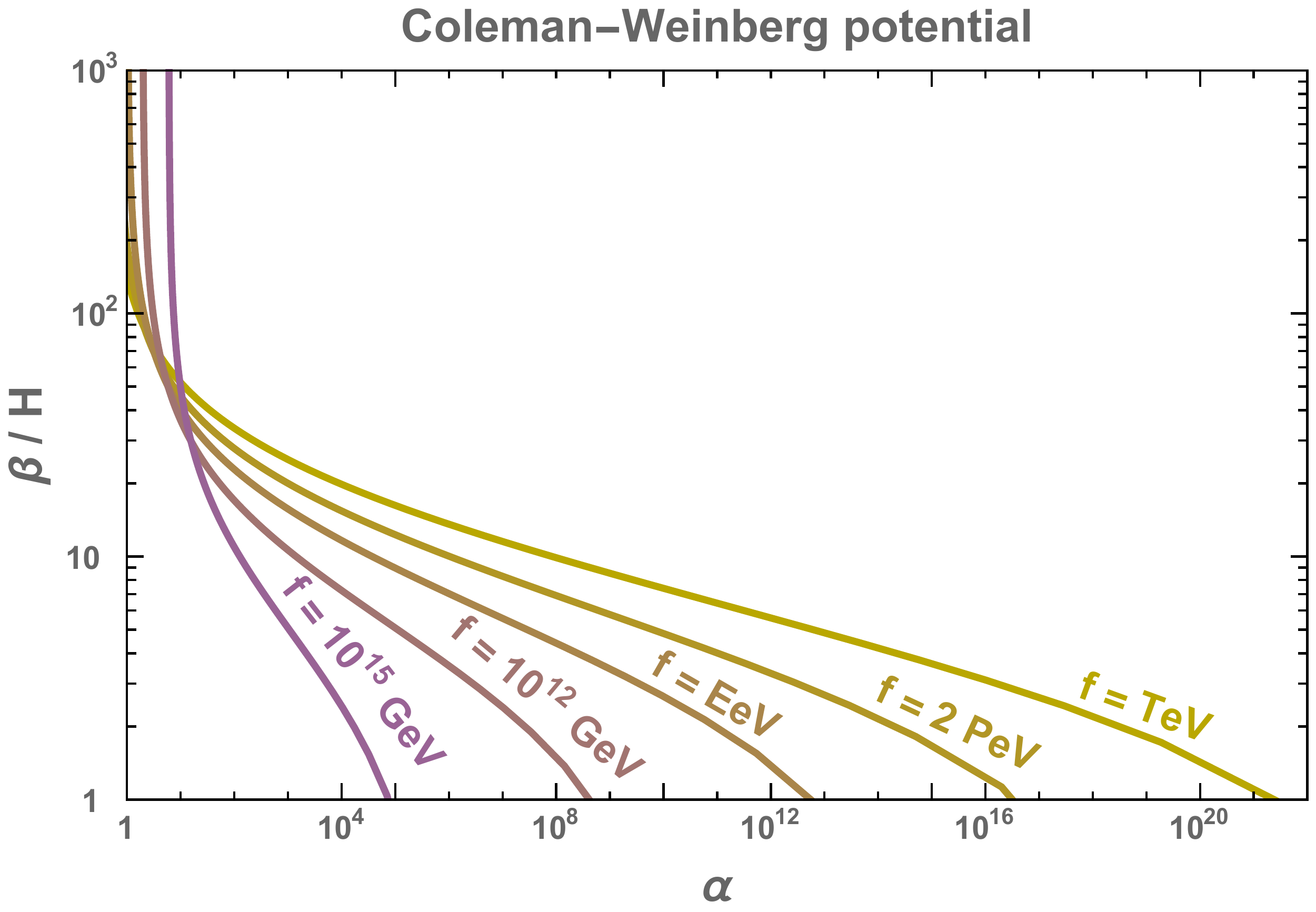}}}
\end{adjustbox}
\caption{\it \small  \textbf{Top left:} We compare the $O_3$- and $O_4$- bounce action computed with the thick-wall formula, cf. Eq.~\eqref{eq:S3overT_cw} and Eq.~\eqref{eq:S4_cw}, to the $O_3$-bounce action computed with our own shooting method. \textbf{Top right:} Nucleation temperature computed with the thick-wall formula for $S_3/T$.  The end points correspond to the last possible nucleation before eternal inflation, cf. Eq.~\eqref{eq:minimal_temp_S3_CW}. \textbf{Bottom:} GW parameters $\alpha$ and $\beta$.}
\label{fig:CW_gX_Tnuc_alpha_beta}
\end{figure}

Close to the false vacuum the Coleman-Weinberg potential is very flat such that the thermal barrier in Eq.~\eqref{eq:ThermalPotentialDarkScalar} is well approximated by its high-temperature expansion, cf. Eq.\eqref{eq:HT_exp_jb}
\begin{equation}
V_{T}(\sigma,~T) \quad  \overset{ T \gg  \sigma }{\longrightarrow} \quad  \frac{m_{\rm eff}(T)^2}{2} - \frac{\lambda_{\rm eff} }{4}\sigma^4,
\label{eq:effective_pot_cw_HT}
\end{equation}
with 
\begin{align}
&m_{\rm eff}(T)^2 = \frac{3}{16}\,g_X^2\,T^2, \\
&\lambda_{\rm eff}  =  -\beta_{\lambda_{\rm \sigma}}\left( \ln{\frac{\sigma}{f}} - \frac{1}{4} \right), \qquad \beta_{\lambda_{\rm \sigma}} \simeq \frac{1}{(4\pi)^{2}}\frac{9 g_{x}^4}{8}.
\end{align}
We now use the thick-wall formula introduced in Sec.~\ref{sec:thick_wall_thin_wall_formula} in order to evaluate the $O_3$- and $O_4$- bounce actions.
Upon injecting Eq.~\eqref{eq:effective_pot_cw_HT} into Eq.~\eqref{eq:S_3_thick_wall}, we obtain
\begin{equation}
\frac{S_3}{T} = \frac{16 \pi}{3\,\lambda_{\rm eff} } \frac{m_{\rm eff}}{T}, \qquad \text{and}  \qquad \phi_* = 2\frac{m_{\rm eff}}{\sqrt{\lambda_{\rm eff}}}.
\end{equation}
We recall that $\phi_*$ is the field value after tunneling. In terms of the physical parameters, the $O_3$-bounce action reads
\begin{equation}
\frac{S_3}{T} \simeq  \frac{A}{\textrm{log}\left(\frac{M}{T}\right)} \qquad \text{with} \quad A= \frac{1018}{g_X^3} \quad \text{and} \quad M = 5.65\,f.
\label{eq:S3overT_cw}
\end{equation}
We now repeat the same steps in the $O_4$ case, so that we inject Eq.~\eqref{eq:effective_pot_cw_HT} into Eq.~\eqref{S_4_thick_wall}. However, by doing that we get
\begin{equation}
S_4 = 0,
\end{equation}
which means that in the presence of the thermal barrier, there is no $O_4$ bounce. However, if we neglect the thermal barrier and set $m_{\rm eff} =0$ in Eq.~\eqref{eq:effective_pot_cw_HT}, then the thick-wall formula in Eq.~\eqref{S_4_thick_wall} gives
\begin{equation}
S_4 \simeq \frac{2 \pi^2}{\lambda_{\rm eff}}.
\label{eq:S4_cw}
\end{equation}
Note that due to the scale-invariance of  $V = -\lambda_{\rm eff} \sigma^4/4$, no exit point $\phi^*$ of the tunneling is singled out in Eq.~\eqref{S_4_thick_wall} which leads to a degenerate family of instantons with arbitrary $\phi^*$ \cite{Espinosa:2019hbm}. They are the so-called Fubini bounces and their tunneling actions can be computed exactly \cite{Fubini:1976jm,Lipatov:1976ny}
\begin{equation}
S_4 = \frac{8\pi^2}{3\lambda_{\rm eff}},
\end{equation}
which confirms the good convenience of the thick-wall formula.
We conclude that tunneling occurs via $O_3$ when
\begin{equation}
\frac{S_3}{T} \lesssim S_4 \qquad \rightarrow \qquad m_{\rm eff} \lesssim T \qquad \rightarrow \qquad g_X \lesssim 2.7. \label{eq:comp_S3_S4_cw}
\end{equation}
In the top left panel of Fig.~\ref{fig:CW_gX_Tnuc_alpha_beta}, we compare the $O_3$- and $O_4$- bounce actions computed with the thick-wall formula and we confirm the previous estimate in Eq.~\eqref{eq:comp_S3_S4_cw}.
We also compare the analytical $O_3$-bounce action to the one computed numerically. For this, we implemented our own undershoot-overshoot algorithm\footnote{I thank Iason Baldes, Sebastian Bruggisser, Benedict Von Harling and Oleksii Matsedonskyi for useful discussions related to the numerical computations of the bounce action.}. Note that in order to improve the convergence of the code, it can be useful to modify the potential as follows
\begin{align}
&\tilde{V}(\sigma)=V(\sigma)-V(0), \\
&\tilde{V}(\sigma)=0 \qquad \sigma<0, \\
&\tilde{V}(\sigma)=V(\sigma)-V(0) - \sigma^4 + f^4 \qquad \sigma>f.
\end{align}
We numerically find that the size of the bounce is of order $R \sim 10/T$.

\paragraph{Nucleation temperature:}
The nucleation happens when the tunneling rate in Eq.~\eqref{eq:tunneling_rate} becomes comparable to the Hubble expansion rate per Hubble volume
\begin{equation}
\label{eq:nucleation_eq}
\frac{S_{3}(T_{\rm nuc})}{T_{\rm nuc}} \simeq 4 \ln{\frac{T_{\rm nuc}}{H(T_{\rm nuc})}} + \frac{3}{2} \ln \frac{S_3/T}{2\pi}
\end{equation}
with
\begin{equation}
H^2(T) = H^2_{\Lambda}+H^2_{\rm rad} = \frac{1}{(4\pi)^2}\frac{9M_{X}^4}{24 M_{\rm pl}^2}+\frac{\pi^2 g_* T^4}{90M_{\rm pl}^2}, \quad g_* = 116.75.
\end{equation}
From injecting Eq.~\eqref{eq:S3overT_cw} into Eq.~\eqref{eq:nucleation_eq}, we obtain
\begin{equation}
\Tnuc = \sqrt{ H_{\Lambda}\,M} ~\left( \frac{2\pi}{S_3/\Tnuc} \right)^{3/16}~\exp \left( \frac{1}{2} \sqrt{-A + \left( \ln \frac{M}{H_{\Lambda}} + \frac{3}{8} \ln \frac{S_3/\Tnuc}{2\pi} \right)^2} \right).
\end{equation}
Neglecting the $S_3/2\pi$ terms, we conclude that there is no nucleation solution when \cite{DelleRose:2019pgi}
\begin{equation}
A \gtrsim  \ln \frac{M}{H_{\Lambda}},
\end{equation}
and that the minimal nucleation temperature is
\begin{equation}
T_{\rm nuc}^{\rm min} \simeq  \sqrt{ H_{\Lambda}\,M}  \simeq 0.1~\left(\frac{f}{M_{\rm pl}}\right)^{1/2}~f.
\label{eq:minimal_temp_S3_CW}
\end{equation}
In top right panel of Fig.~\ref{fig:CW_gX_Tnuc_alpha_beta}, we show the nucleation temperature computed with the thick-wall formula.

\paragraph{Catalysis by QCD effects:}
The period of super-cooling ends when the phase transition occurs and the vacuum energy becomes zero. If the tunneling rate is large enough, the phase transition happens by bubble nucleation, otherwise it can be trigged by the QCD phase transition, see the pioneering paper \cite{Witten:1980ez} and more recent works \cite{Iso:2017uuu, Hambye:2018qjv}.\footnote{For studies of the QCD effects on a 1stOPT arising from a strongly-coupled sector, see the pioneering paper \cite{vonHarling:2014kha} in which the gluon condensate induces a new term in the dilaton potential, and more recent works \cite{Baratella:2018pxi,Bloch:2019bvc}. We discuss catalysis by QCD effects in confining PT at the end of Sec.~\ref{par:QCD_effect_light_dilaton}.} Indeed when quarks condensates, the Yukawa coupling induces a linear term for the Higgs potential and the latter gets a temperature dependent vev \cite{Witten:1980ez}. Then $\left<h\right>$ induces a negative mass term for the dark scalar $m_{\sigma}^{2}=\lambda_{\rm h\sigma}\left<h\right>^2/2$ and $\sigma$ starts rolling down as soon as the Higgs-mediated quarks condensate contribution becomes comparable to the thermal mass $m_{\sigma}^{2}(T)=\frac{3}{16}g_{X}^2T^2$. If $-\lambda_{\rm h\sigma}$ is large enough, it happens immediately at the QCD phase transition when $T_{\rm QCD}=85\MeV$\footnote{This is the temperature at which the QCD phase transition occurs when the quarks are massless.} otherwise it happens at \cite{Hambye:2018qjv}
\begin{equation}
T_{\rm end}^{\rm QCD}\approx \frac{0.1 \left<h\right>}{M_{X}/\rm TeV}
\end{equation}
Then the temperature at the end of super-cooling is
\begin{equation}
T_{\rm end}=\text{Max}\left(T_{\rm nuc},\; \text{Min}\left(T_{\rm end}^{\rm QCD},\; T_{\rm QCD}\right)\right).
\end{equation}

\paragraph{Catalysis by gravity effects:}

When the bubble size at nucleation $R_0 \sim 10/T$ is of the same order as the Hubble horizon $H^{-1}$\cite{Coleman:1980aw,Joti:2017fwe}, the bounce action receives important correction in $R_0H$. Intuitively, when $T \lesssim H$, the quantum fluctuations induced by the de Sitter temperature $T_{\rm dS} = H/2\pi$ becomes larger than the thermal barrier induced by the plasma temperature $T$ such that we expect the tunneling to be dominated by gravity effects when
\begin{equation}
T \lesssim H.
\end{equation}
See \cite{Lewicki:2021xku} for a related study.

\paragraph{Catalysis by primordial black holes:}

The idea that the bounce action of 1stOPT could be enhanced by the presence of a black hole (BH) is credited to Hiscock in 1987 \cite{Hiscock:1987hn}. More recently, the idea has been the object of numerous studies
\cite{Gregory:2013hja,Burda:2015isa,Burda:2015yfa,Burda:2016mou,Rajantie:2016hkj,Canko:2017ebb,Gorbunov:2017fhq,Mukaida:2017bgd,Kohri:2017ybt,Oshita:2018ptr,Dai:2019eei, El-Menoufi:2020ron}.
The optimum BH mass $M_{\rm PBH}$ is a compromise between the enhancement of the bounce action (which decreases with $M_{\rm PBH}$) and the BH lifetime though evaporation (which increases with $M_{\rm PBH}$ and which should not be much smaller than the false vacuum lifetime). For instance, the maximal effect on the metastability of the electroweak vacuum is reached for BH masses around $M_{\rm PBH} \sim 10^4-10^8$~$M_{\rm pl} \sim 1~\textrm{g}-1~\textrm{kg} $, depending on the Higgs potential parameters \cite{Burda:2016mou}, while the nucleation rate of electroweak-like 1stOPT could be boosted in the presence of BH lighter than $M_{\rm PBH} \lesssim 10^{15}$~$M_{\rm pl} \sim 10^9~\rm g$ \cite{El-Menoufi:2020ron}. The impact of seed BH on the nucleation rate of nearly-conformal potential remains to be investigated.

\paragraph{Catalysis by topological defects:}

In 1981, Steinhardt showed that cosmological 1stOPT could be seeded by monopoles \cite{Steinhardt:1981ec,Steinhardt:1981mm}. If the monopole core is made of the false vacuum and that the later is sufficiently metastable, then the monopole is unstable under fluctuations of the core radius, causing it to expand. This effect called \textbf{monopole dissociation} has been explored further in \cite{Hosotani:1982ii,Kumar:2009pr,Kumar:2010mv,Agrawal:2022hnf}. Other works have studied the possibility to catalyze cosmological phase transitions with cosmic strings \cite{Yajnik:1986tg,Yajnik:1986wq,Dasgupta:1997kn,Kumar:2008jb,Lee:2013zca} and domain walls \cite{Blasi:2022woz}.

\paragraph{Reheating:}
The reheating temperature can be shown to be 
\begin{equation}
T_{\rm RH} = T_{\rm inf} ~ \text{Min}\left(1,\; \frac{\Gamma}{H}\right)^{1/2}
\end{equation}
After the period of super-cooling, the dark scalar field oscillates around its minima and reheates the Universe by transfering its energy into SM. However the dark scalar being coupled to the Higgs, a part of the energy will go into oscillation of the Higgs around its minima. The equation of motion of the scalar fields are
\begin{align}
\ddot{\sigma}+(3H+\Gamma_{\sigma}) \dot{\sigma} &= -\frac{\partial V}{\partial \sigma} \\
\ddot{h}+(3H+\Gamma_{h}) \dot{h} &= -\frac{\partial V}{\partial h} 
\end{align}
Which at first-order around the minima becomes
\begin{align}
\ddot{h}+(3H+\Gamma_{h}) \dot{h} &= m_{h}^2(h-v) + \lambda_{h \sigma} (\sigma - \sigma_{0})\\
\ddot{\sigma}+(3H+\Gamma_{\sigma}) \dot{\sigma} &= m_{\sigma}^2(\sigma-\sigma_{0}) + \lambda_{h \sigma} (h-v)\\
\end{align}
with $\lambda_{h\sigma} = \partial^2 V/\partial h \partial \sigma = - m_{h}^2 (v/f)^2$. In the limit $\Gamma_{\sigma},~ H \ll \Gamma_{h} \ll M_{\sigma}$, one gets the frequency of the normal modes $\omega \simeq m_{h} \pm i(3H+\Gamma_h)/2$ and  $f \simeq m_{\sigma} \pm i(3H+\Gamma)/2$ with \cite{Hambye:2018qjv}.
\begin{equation}
\Gamma = \Gamma_{h} \sin^2{\theta} + \Gamma_{\sigma} \cos^2{\theta}
\end{equation}
where $\theta=-v/f$.

\paragraph{GW signal: }

In order to estimate the GW signal, we first need to compute the parameters $\alpha$ and $\beta$, cf. Eq.~\eqref{eq:alpha_definition} and Eq.~\eqref{eq:beta_definition}
\begin{align}
\alpha &= \frac{\Delta V}{\rho_{\rm rad}(\Tnuc)} \simeq 2\times 10^{-4} \frac{100}{g_*} \left( \frac{M_X}{\Tnuc} \right)^4, \\
\beta/ H & \simeq - 4 + T \frac{d(S_3/T)}{dT} \Big|_{\Tnuc} = - 4 + \dfrac{S_3/T}{\text{log} \frac{M}{T}} \Big|_{\Tnuc}  .
\end{align}
Note the non-polynomial relation $\beta/H \propto 1/\text{log}~\alpha $ characteristic of nearly-conformal potential. In contrast, for polynomial potentials in the thin-wall limit, we expect $\beta / H \propto \alpha^{-2}$.
We show $\alpha$ and $\beta$ in Fig.~\ref{fig:CW_gX_Tnuc_alpha_beta}.

\subsection{Strongly-coupled scenario: the light-dilaton potential}

\paragraph{The dilaton potential:}
In this section we suppose that spontaneous symmetry breaking arises from the condensation of a nearly-conformal sector, the broken symmetry being scale invariance. If the source of explicit breaking of scale invariance is small, the spontaneous breaking of scale invariance generates a pseudo Nambu-Goldstone boson, the dilaton which we parametrize as \cite{Goldberger:2008zz}
\begin{equation}
\dilaton(x) = \dilatonvev e^{\frac{\sigma(x)}{\dilatonvev}},
\end{equation}
where $\sigma(x)$ transforms non-linearly under the scale transformation $\sigma(x)  \rightarrow \sigma{\lambda x} + \log{\lambda}$. 
We are interested in the case where the dilaton $\dilaton$ is the lightest state of the nearly-conformal sector such that we can integrate out heavier degrees of freedom and describe the confined phase with the dilaton only  \cite{Bruggisser:2018mrt,Baldes:2021aph}. We suppose that the Higgs VEV is small compared to the one of the dilaton so that we neglect the field excursion in the Higgs direction. The only operator invariant under dilatations $x_{\mu} \rightarrow x_{\mu} / \lambda $, $\dilaton \rightarrow  \lambda  \dilaton$ is $\dilaton^4$ and the dilaton has vanishing vev. Spontaneous breaking of scale invariance is only possible after adding an explicit source of scale invariance breaking \cite{ContinoPomarolRattazzi, Chacko:2012sy, Bellazzini:2013fga, Coradeschi:2013gda,Megias:2014iwa, Pomarol:2019aae}. So we add a slightly relevant operator
\begin{equation}
\label{eq: explicit_source_scale_break}
\epsilon \, \mathcal{O}_{\epsilon},
\end{equation}
with scaling dimension $d = 4+\gamma_{\epsilon} \lesssim 4$ and the potential for the dilaton becomes
 \begin{equation}
V_{\dilaton}^{\mathsmaller{\rm T=0}}(\dilaton)  = c_{\dilaton} g_{\dilaton}^2 \dilaton^4 - \epsilon(\dilaton)\dilaton^4.
\end{equation}
where $g_{\dilaton}$ is the typical coupling constant of the strong sector and $c_{\dilaton} $ is an order $1$ constant.
The source of explicit breaking of scale invariance in Eq.~\eqref{eq: explicit_source_scale_break} leads to the running of the coupling constant $\epsilon$
\begin{equation}
\frac{\partial \epsilon}{\partial \log \mu}\simeq \gamma_{\epsilon} \epsilon
\end{equation}
where $\mu$ is the renormalization scale. This yields a non-zero vev $\dilatonvev$ for the dilaton $\chi$
 \begin{equation}
\frac{dV}{d\chi}\Big|_{\chi=\dilatonvev}=0 \quad \rightarrow \quad  \epsilon(\dilaton)=\frac{ c_{\dilaton} g_{\dilaton}^2}{1+\gamma_{\epsilon}/4}\left(\frac{\dilaton}{\dilatonvev}\right)^{\gamma_{\epsilon}},
\end{equation} 
and also implies a none-vanishing mass for the dilaton  $m_{\dilaton}$ whose smallness is tied to the smallness of the anomalous dimension $\gamma_{\epsilon}$
\begin{equation}
\frac{d^2V}{d\sigma^2}\Big|_{\sigma = 0}=m_{\dilaton}^2 \quad \rightarrow \quad \gamma_{\epsilon} \simeq -\frac{1}{4 c_{\dilaton}}\frac{m_{\dilaton}^2}{g_{\dilaton}^2\dilatonvev^2}.
\end{equation}
The dilaton potential reduces to \cite{Bruggisser:2018mrt,Baldes:2021aph}
 \begin{equation}
 \label{eq:zero_temp_pot_dil}
V_{\dilaton}^{\mathsmaller{\rm T=0}}(\dilaton)  =  c_{\dilaton} \,g_{\dilaton}^2 \, \dilaton^4 \, \left(1- \frac{1}{1+\gamma_{\epsilon}/4}\left(\frac{\dilaton}{\dilatonvev}\right)^{\gamma_{\epsilon}} \right),
\end{equation}
with
\begin{equation}
\label{eq:gammaepsilonVSmdilaton_march}
\gamma_{\epsilon} \simeq -\frac{1}{4 }\frac{y_{\dilaton}^2}{ c_{\dilaton} \,g_{\dilaton}^2},
\end{equation}
and $y_{\dilaton} \equiv m_{\dilaton}/\dilatonvev$.
From Eq.~\eqref{eq:gammaepsilonVSmdilaton_march}, the conformal limit $|\gamma_{\epsilon}| \ll 1$ coincides with the light dilaton limit $m_\dilaton/f \ll 1$. Later in Eq.~\eqref{eq:g_chi_holography} we will deduce from holography that $g_\chi = 4\pi/N$. Hence, at fixed $|\gamma_{\epsilon}|$ the light dilaton limit coincides with the large $N$ limit.

Note that in the conformal limit $|\gamma_{\epsilon}| \ll 1$, the dilaton potential at zero-temperature reduces to the Coleman-Weinberg potential in Eq.~\eqref{eq:1loopRadiativePotentialDarkScalar}
\begin{equation}
V_{\dilaton}^{\mathsmaller{\rm T=0}}(\dilaton)   \overset{\gamma_{\epsilon}\ll 1}{=}  -\gamma_{\epsilon} \, c_{\dilaton} \, g_{\dilaton}^2\, \dilaton^4 \, \textrm{log} \left( \frac{\dilaton}{\dilatonvev} \right) \simeq \frac{m_\dilaton^2}{f^2} \frac{\dilaton^4}{4} \, \textrm{log} \left( \frac{\dilaton}{\dilatonvev} \right) .
\label{eq:dilaton_potential_taylor_cw_march}
\end{equation}
However, the different nature of the thermal corrections between the weakly-coupled and the strongly-coupled scenario - presence of the thermal mass in $T^2$ at the tunneling point in the former case and absence of the finite-temperature contributions at the tunneling point in the latter case - make the bounce action behaving differently, see below.

\paragraph{Non-canonical kinetic terms:}
Note that dual theories in 5D motivate the presence of non-canonical kinetic terms, cf. Eq.~\eqref{eq:kin_5D_AdS}
\begin{equation}
\mathcal{L}_{\rm K} = \frac{\kappa^2}{2} \partial_{\mu} \chi\partial^{\mu} \chi .
\end{equation}
After rescaling
\begin{equation}
\label{eq:non_can_kin}
 \frac{\kappa^2}{2} \partial_{\mu} \chi \partial^{\mu} \chi  \overset{\chi \rightarrow \chi/ \kappa}{\xrightarrow{\hspace*{1.5cm}}}  \frac{1}{2} \partial_{\mu} \chi \partial^{\mu} \chi,
\end{equation}
the dilaton potential becomes
 \begin{equation}
 \label{eq:zero_temp_pot_dil_kappa}
V_{\dilaton}^{\mathsmaller{\rm T=0}}(\dilaton)  = \frac{ c_{\dilaton} \,g_{\dilaton}^2}{\kappa^4} \, \dilaton^4 \, \left(1- \frac{1}{1+\gamma_{\epsilon}/4}\left(\frac{\dilaton}{\kappa\,\dilatonvev}\right)^{\gamma_{\epsilon}} \right),
\end{equation}
with
\begin{equation}
\label{eq:gammaepsilonVSmdilaton_march_kappa}
\gamma_{\epsilon} \simeq -\frac{\kappa^2}{4 }\frac{y_{\dilaton}^2}{ c_{\dilaton} \,g_{\dilaton}^2},
\end{equation}
and $y_{\dilaton} \equiv m_{\dilaton}/\dilatonvev$. In the rest of the chapter, we keep $\kappa = 1$, and we refer to \cite{Baldes:2021aph} for the study of $\kappa \neq 1$.

\paragraph{Holography - the confined phase:}
\label{sec:holography_confining_phase_PT}
The strongly-coupled 4D theory explicited above can be mapped to a weakly-coupled 5D theory with $AdS_5$ metric  (see the AdS-CFT correspondence pioneering articles \cite{Maldacena:1997re,Witten:1998qj} and some reviews about holography \cite{Rattazzi:2000hs, Rattazzi:2003ea, Gherghetta:2010cj})
\begin{align}
ds^2 &=e^{-2ky}  \eta_{\mu\nu} dx^\mu dx^\nu + dy^2 , \\
&= \frac{L^2}{z^2} \left(\eta_{\mu\nu} dx^\mu dx^\nu + dz^2 \right), 
\end{align}
where $k=1/L$ is the AdS curvature and $z =L\, e^{ky}$ is the coordinate along the fifth dimension. The AdS/CFT correspondence relate the AdS curvature in Planck unit to the rank of the 4D gauge group
\begin{equation}
\left(M_5L \right)^3 = \frac{N^2}{16\pi^2}.
\end{equation}
We consider the Randall-Sundrum scenario \cite{Randall:1999ee,Randall:1999vf}, which is a slice of $AdS_5$ bounded by two branes, a UV-brane located at $y=0$ and a IR-brane located at $y_{\rm IR}$. The use of the RS scenario for solving the hierarchy problem, is discussed in Sec.~\ref{sec:warped_extra_dim_hierarchy}. The position of the IR brane along the 5th dimension $y_{\rm IR}$ defines the radion field 
\begin{equation}
\mu =\frac{1}{z_{\rm IR}} = \frac{ e^{-k\,y_{\rm IR}}}{L},
\end{equation}
whose kinetic term is 
\begin{equation}
\mathcal{L}_{\rm K} = 12 (M_5L)^3(\partial \mu)^2 = 12 \frac{N^2}{16\pi^2}(\partial \mu)^2.
\label{eq:kin_5D_AdS}
\end{equation}
Upon adding a field $\phi_{\rm GW}$ of mass $m$ in the 5D bulk, the radion field receives a stabilizing potential (Goldberger-Wise mechanism \cite{Goldberger:1999uk})
\begin{equation}
V_{\rm GW}(\mu) =  v_{\rm IR}^2 \,\mu^4 \left[ (4+2\gamma_{\epsilon})\left(1- \frac{v_{\rm UV}}{v_{\rm IR}} \left( \frac{\mu}{\mu_0} \right)^{\gamma_{\epsilon}}\right)^2  + \delta \right] ,
\label{eq:GW_potential}
\end{equation}
with 
\begin{equation}
\gamma_{\epsilon}  =  \sqrt{4+m^2/k^2} - 2 \simeq m^2/k^2,
\label{eq:gamma_epsilon_5D_GW}
\end{equation}
and where $\mu_0$ is the UV scale, $\delta$ is a dimensionless parameter, $v_{\rm UV}$ and $v_{\rm IR}$ are the vevs in Planck unit of the Goldberger-Wise field $\phi_{\rm GW}$ on the Planck and IR branes. 

The quantity $\gamma_{\epsilon}$ in Eq.~\eqref{eq:gamma_epsilon_5D_GW}, related to the mass $m$ of the light 5D field, is the holographic dual of the anomalous dimension $\gamma_{\epsilon}=d-4$ of the slightly relevant 4D scalar operator in Eq.~\eqref{eq: explicit_source_scale_break}. The 5D radion potential in Eq.~\eqref{eq:GW_potential} is the holographic dual of the 4D light-dilaton potential in Eq.~\eqref{eq:zero_temp_pot_dil}.  Note however that the 5D and 4D potentials do not exactly coincide since in addition to the term $\left(\dilaton/\dilatonvev\right)^{\gamma_{\epsilon}}$, the 5D potential contains a term $\left({\dilaton/\dilatonvev}\right)^{2\gamma_{\epsilon}}$, which can generate a barrier at zero-temperature. 

Upon matching the non-canonical kinetic terms in 5D and 4D
\begin{equation}
\mathcal{L}_{\rm K,\,5D}= 12 \frac{N^2}{16\pi^2}(\partial \mu)^2 \qquad \textrm{and} \qquad \mathcal{L}_{\rm K,\,4D}=  \frac{\kappa^2}{2} \frac{1}{g_{\dilaton}^2} \partial_{\mu} \dilaton \partial^{\mu} \dilaton ,
\end{equation}
we obtain
\begin{equation}
\label{eq:g_chi_holography}
\kappa = \sqrt{24} \qquad \text{and} \qquad g_{\dilaton} = \frac{4\pi}{N}.
\end{equation}
The interaction coupling constant scales like $g_{\dilaton} \propto 1/N$, which corresponds to the glueball scaling at large N \cite{Witten:1979kh}. Finally, the AdS/CFT dictionary also identifies the 5D Kaluza-Klein resonances to the bound-states of the 4D theory (e.g. \cite{Rattazzi:2003ea}) whose masses are
\begin{equation}
m_{\rm kk} \simeq (n+\frac{1}{4})\, \pi \mu, \qquad n=1,\,2,\cdots
\end{equation}
and which after rescaling of the kinetic terms $\frac{\kappa^2}{g_{\dilaton}^2}\frac{1}{2}(\partial \mu)^2 \rightarrow \frac{1}{2}(\partial \dilaton)^2$ become
\begin{equation}
m_{\rm kk} \simeq (n+\frac{1}{4}) \,\frac{\pi\, g_{\dilaton}}{\kappa} \dilaton,
\end{equation}
in conformity with the rescaling of the dilaton mass in Eq.~\eqref{eq:gammaepsilonVSmdilaton_march_kappa}.

\paragraph{Holography - the deconfined phase:}
In the 5D perspective, at high enough temperature the IR brane is replaced by a black-hole horizon at position $z_h$. The corresponding metric called $AdS_5$-Schwarzschild is solution of the Einstein equation with compact Euclidean time \cite{Creminelli:2001th}
\begin{equation}
ds^2 = -\frac{L^2}{z^2} \left( - f(z)dt^2 + dx_i^2 + \frac{1}{f(z)} dz^2 \right), \qquad f(z) = 1 - \left(\frac{z}{z_h}\right)^4.
\end{equation}
and its free energy density is \cite{Creminelli:2001th}
\begin{equation}
F_{\rm BH}(T_{\rm BH}) = 2\pi^4 (M_5L)^3 \left(3T_{\rm BH}^4 - 4T\,T_{\rm BH}^3\right) = \frac{\pi^2}{8}N^2 \left(3T_{\rm BH}^4 - 4T\,T_{\rm BH}^3\right),
\label{eq:free_energy_bh}
\end{equation}
where $T_{\rm BH}  = 1/\pi z_h $ is the black-hole temperature. The free energy density of the deconfined phase is minimized when $T=T_{\rm BH}$ due to the existence of a conical singularity when $T \neq T_{\rm BH}$. Note that the deconfined phase and the confined phase coincide when both the black hole horizon and the IR-brane are sent to infinity, which corresponds to $T_{\rm BH}=0=\mu$.
Hence, the total free energy density can be constructing by analytically extending the zero-temperature potential in Eq.~\eqref{eq:GW_potential} to negative value of the field $\mu$ using Eq.~\eqref{eq:free_energy_bh} with $\mu = - T_{\rm BH}$ \cite{Creminelli:2001th}. Using holography, the finite-temperature corrections in the 4D theory can be incorporated by extending the zero-temperature potential in Eq.~\eqref{eq:zero_temp_pot_dil} to negative value of the field $\chi$ using Eq.~\eqref{eq:free_energy_bh} with $\dilaton = - T_{\rm BH}$.

We refer to \cite{Creminelli:2001th} for the first study of the phase transition in Randall-Sundrum scenarios and \cite{Randall:2006py, Nardini:2007me, Hassanain:2007js, Konstandin:2010cd, Konstandin:2011dr, Bunk:2017fic, Dillon:2017ctw, Megias:2018sxv, vonHarling:2017yew, Baratella:2018pxi,Fujikura:2019oyi} for later works on holographic phase transitions. For studies of the dual 4D confinement, we call attention to \cite{Bruggisser:2018mus, Bruggisser:2018mrt, Baratella:2018pxi, Agashe:2019lhy,DelleRose:2019pgi,vonHarling:2019gme}.

\paragraph{The deconfined phase in 4D:}
We now describe another approach for modelling the free energy between the deconfined phase and the confine phase, borrowed from \cite{Randall:2006py,Bruggisser:2018mrt}, which relies on the usual thermal corrections in 4D QFT.
In the 4D perspective, the effective potential receives finite-temperature corrections generated by the particles in the plasma.
\begin{enumerate}
\item
When $T<g_{\dilaton} \, \dilaton$, the finite-temperature effects are the standard SM thermal corrections, which we neglect.
\item
When $T>g_{\dilaton} \, \dilaton$, we expect the existence of heavier bound states as well as eventually their deconfined constituents and the EFT which only describes the SM, the Higgs and the dilaton, breaks done. 
\end{enumerate}
We expect the free energy density of the CFT phase to scale like
\begin{equation}
\label{eq:free_energy_CFT}
F_{\mathsmaller{\rm CFT}}(\dilaton=0) \simeq -b\, N^2\, T^4.
\end{equation}
Instead of looking for a finer description of the strong sector which would be very model dependent, we assume that we can approximate the deconfined phase by a $\mathcal{N}=4$ $SU(N)$ large $N$ super-YM dual to an AdS-Schwarzschild space-time where in that particular case, we have\footnote{Note that the free-energy of an interacting gluons gas is lower than the free-energy of a none-interacting gluons gas, $F= -\frac{\pi^2}{90} 2N^2 T^4$, by a factor $5$, in conformity with the intuition that switching on the strong interaction would stabilize the phase.} \footnote{We neglect the number of degrees of freedom $\gSM$ of the SM and of the techni-quarks $g_{\mathsmaller{\rm q}}$ compared to the effective number of the techni-gluons $ \frac{45}{4} N^2 $, namely, $-\frac{\pi^2}{90} (\gSM  + g_{\mathsmaller{\rm q}} + \frac{45}{4} N^2 ) T^4 \simeq -\frac{\pi^2}{8} N^2 T^4$.}, cf. Eq.~\eqref{eq:free_energy_bh}
\begin{equation}
\label{eq:free_energy_CFT_b}
b = \frac{\pi^2}{8}.
\end{equation}
Without a precise UV description, we are unable to determine the potential for $0<\dilaton<T/g_{\dilaton}$, so we assume that the dilaton still exists, that its zero-temperature potential still holds, and that it receives thermal corrections, coming from the $N$ CFT bosons  \cite{Randall:2006py,Bruggisser:2018mrt}
 \begin{equation}
 \Delta V_{T}^{\rm 1-loop}( \dilaton) = \sum_{\rm\mathsmaller{CFT}~ bosons} \frac{n T^4}{2 \pi^2} J_B\left( \frac{m_{\rm \mathsmaller{CFT}}^2}{T^2} \right)
 \label{eq:finite_temp_dilaton}
\end{equation}
with mass
\begin{equation}
\label{eq:m_CFT_march}
m_{\mathsmaller{\rm CFT}} = g_{\dilaton} \dilaton, \qquad  \textrm{with} \quad g_{\dilaton} = 4\pi/N.
\end{equation}
In order to recover the free energy eq.~\eqref{eq:free_energy_CFT} in the deconfined phase, we set
\begin{equation}
\sum_{\mathsmaller{\rm CFT}~\rm bosons} n = \frac{45 N^2}{4}.
\label{eq:number_dof_CFT}
\end{equation}
Finally the total potential for the dilaton is
\begin{equation}
\label{eq:dilaton_pot_tota_march_1}
V_{\rm tot}(\dilaton) =  V_{\dilaton}(\dilaton) + \Delta V_{T}^{\rm 1-loop}( \dilaton) 
\end{equation}
where $V_{\dilaton}(\dilaton)$ and $ \Delta V_{T}^{\rm 1-loop}( \dilaton) $ are given by Eq.~\eqref{eq:zero_temp_pot_dil_kappa} and Eq.~\eqref{eq:finite_temp_dilaton}.
The zero-temperature energy difference between the false vacuum and the true vacuum reads
\begin{equation}
\label{eq:vacuum_energy_difference}
\Delta V \simeq \frac{m_{\dilaton}^2 \dilatonvev^2}{16}.
\end{equation}
According to Eq.~\eqref{eq:HT_exp_jb}, the high-temperature expansion of Eq.~\eqref{eq:dilaton_pot_tota_march_1} reads 
\begin{equation}
\label{eq:dilaton_pot_tota_march_2}
V_{\rm tot}(\dilaton)  \simeq  -\frac{\pi^2}{8}N^2 T^4 +15 \pi^2T^2 \dilaton^2   + \frac{m_\dilaton^2}{f^2} \frac{\dilaton^4}{4} \, \textrm{log} \left( \frac{\dilaton}{\dilatonvev} \right) + \cdots,
\end{equation}
where we have assumed the conformal limit $|\gamma_\epsilon| \ll 1$ in Eq.~\eqref{eq:dilaton_potential_taylor_cw_march} and used Eq.~\eqref{eq:m_CFT_march}. From Eq.~\eqref{eq:dilaton_pot_tota_march_2}, we can see that in the large $N$ limit, the large number of degrees of freedom in the deconfined phase leads to a large negative free energy at $\dilaton = 0$. However, the $1/N$ scaling of the strong coupling $g_\chi$ implies a smaller contribution to the thermal mass $m_{\mathsmaller{\rm CFT}}$ in Eq.~\eqref{eq:m_CFT_march}, which then becomes negligible. This is an important difference with respect to the weakly-coupled Coleman-Weinberg scenario in Eq.~\eqref{eq:effective_pot_cw_HT} for which the thermal mass plays a significant role in the bounce action. While the bounce action of the weakly-coupled scenario was dominated by $O_3$, we will now see that in the large N strongly-coupled scenario\footnote{We recall that due to Eq.~\eqref{eq:gammaepsilonVSmdilaton_march}, the large $N$ limit coincides with the light-dilaton limit $m_\dilaton/f \ll 1$ and with the conformal limit $|\gamma_\epsilon| \ll 1$.}, the bounce action can be dominated by $O_4$.

\paragraph{Bounce action:}

\begin{figure}[h]
\centering
\begin{adjustbox}{max width=1.2\linewidth,center}
\raisebox{0cm}{\makebox{\includegraphics[width=0.55\textwidth]{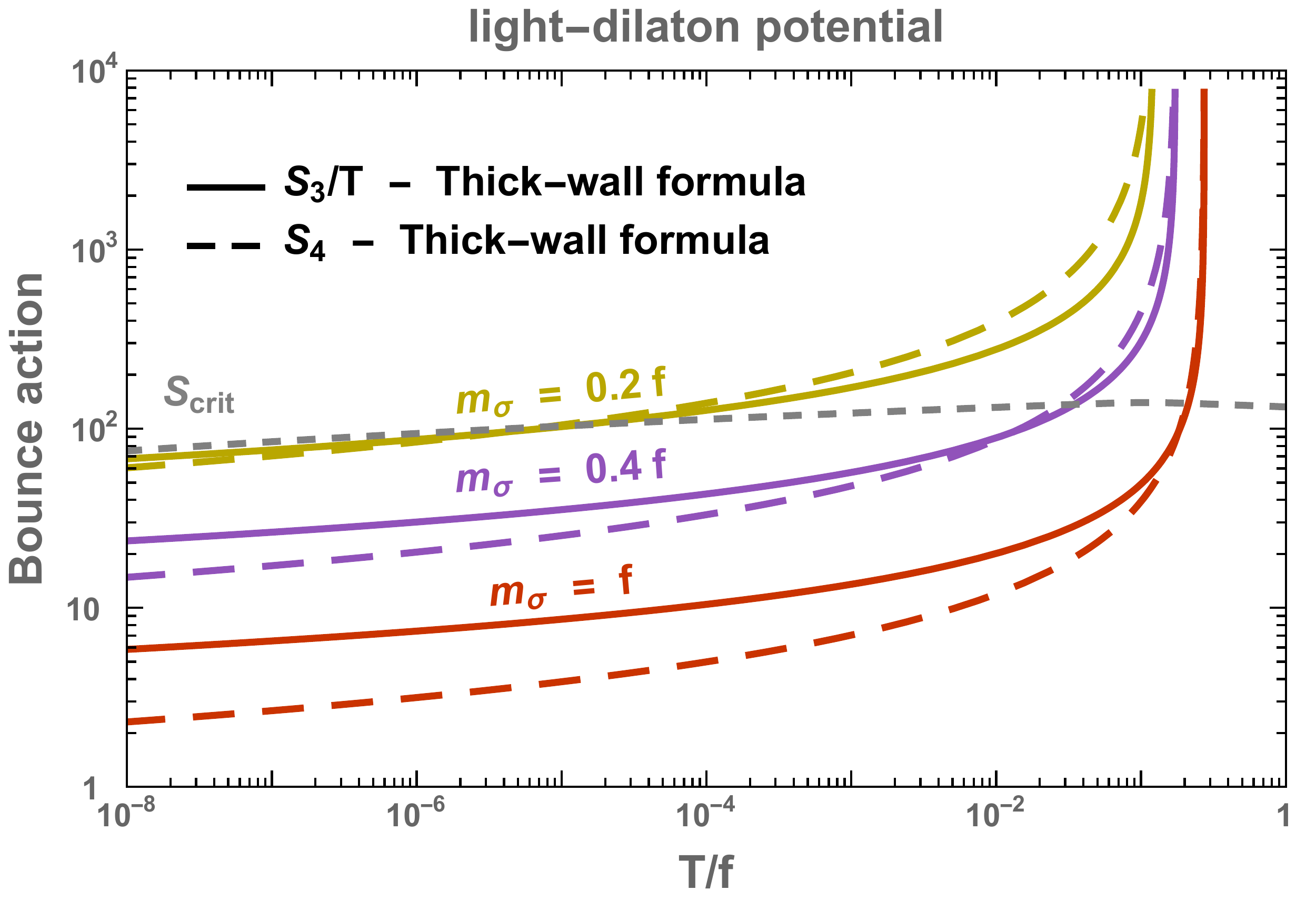}}}
\raisebox{0cm}{\makebox{\includegraphics[width=0.55\textwidth]{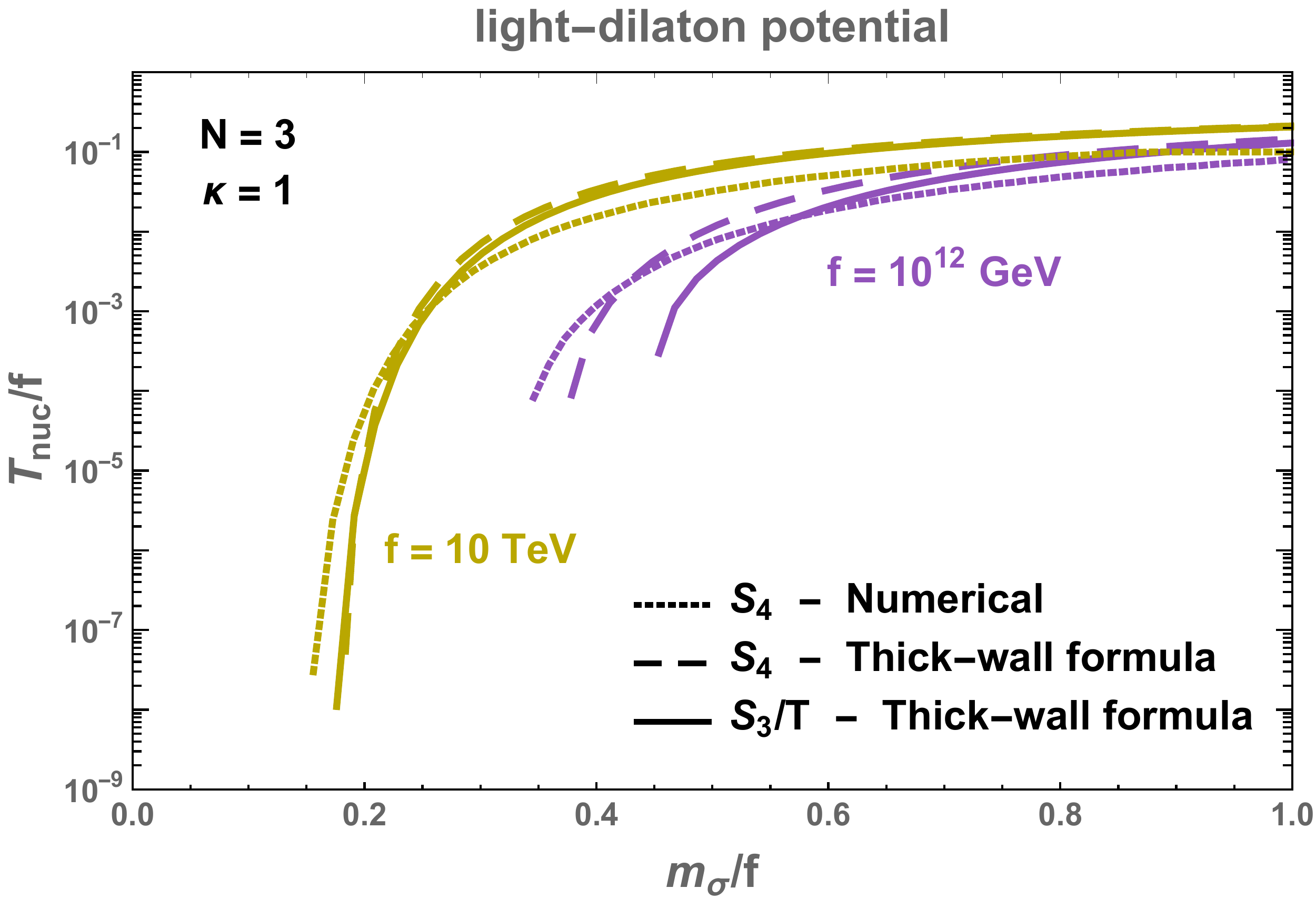}}}
\end{adjustbox}
\begin{adjustbox}{max width=1.2\linewidth,center}
\raisebox{0cm}{\makebox{\includegraphics[width=0.55\textwidth]{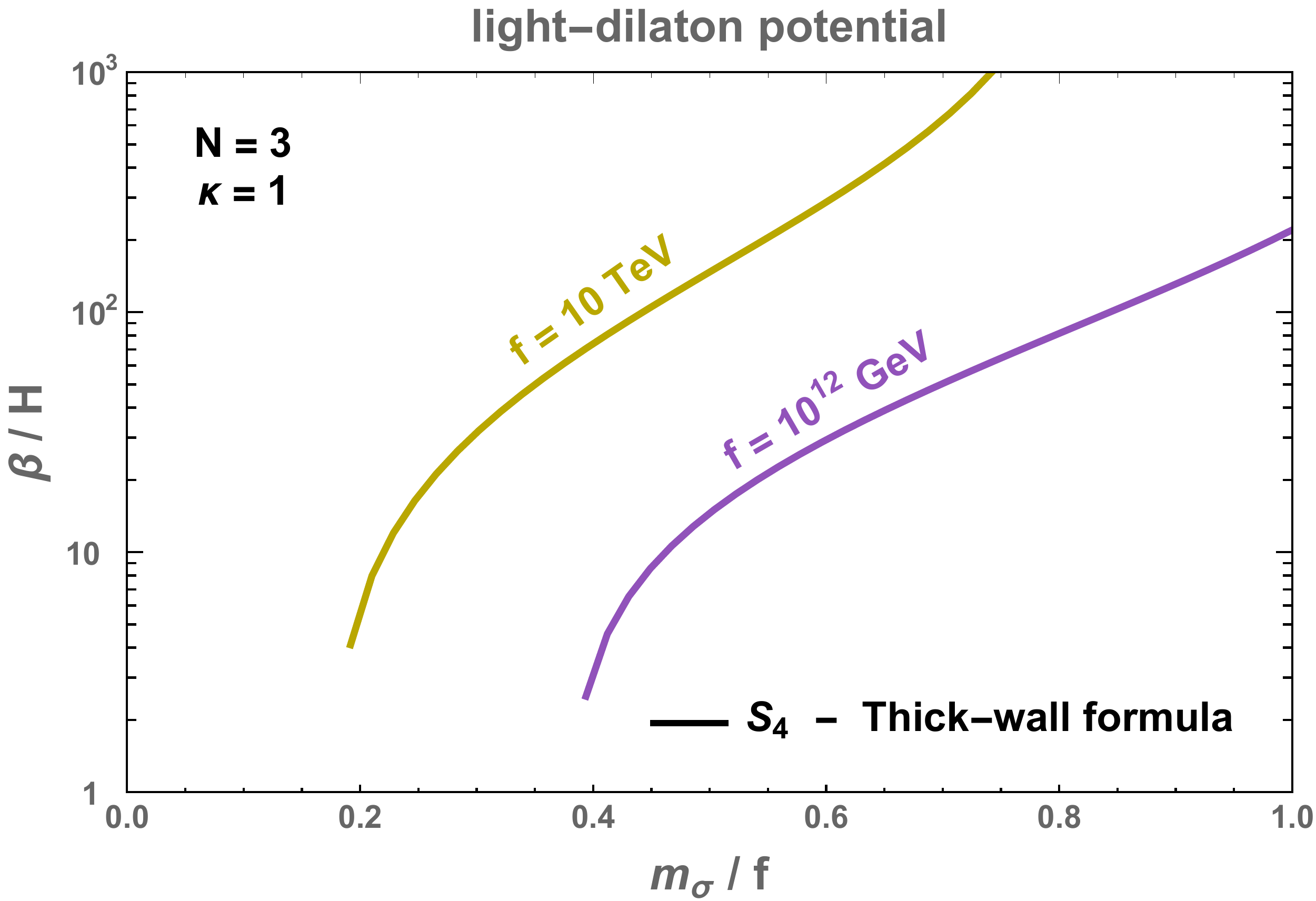}}}
\raisebox{0cm}{\makebox{\includegraphics[width=0.55\textwidth]{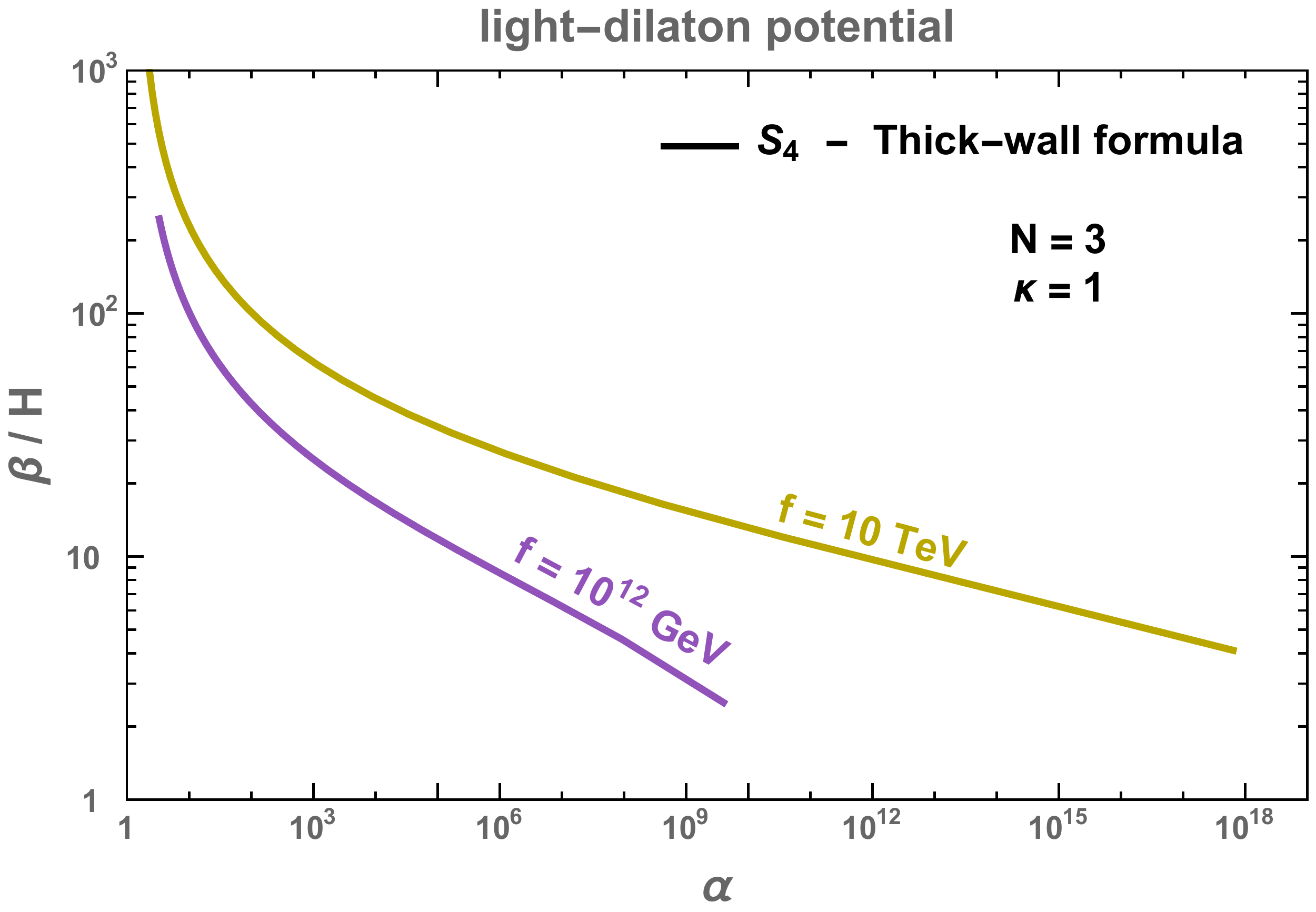}}}
\end{adjustbox}
\caption{\it \small  \textbf{Top left:} We compare the $O_3$- and $O_4$- bounce action computed with the thick-wall formula, cf. Eq.~\eqref{eq:S3_thickwall_light_dilaton} and Eq.~\eqref{eq:S4_thickwall_light_dilaton}, to the $O_4$-bounce action computed with our own shooting method. \textbf{Top right:} Nucleation temperature computed with the thick-wall formula for $S_4$. The end points correspond to the last possible nucleation before eternal inflation, cf. Eq.~\eqref{eq:minimal_msigma_S4_light_dilaton}. \textbf{Bottom:} GW parameters $\alpha$ and $\beta$.}
\label{fig:light_dilaton_msigma_Tnuc_alpha_beta}
\end{figure}

We numerically checked that the bounce action in the small $T$ and small $|\gamma_{\epsilon}|$ limit, is insensitive to the shape of the free energy between the thermal CFT phase and the confined phase. This is due to the thermal corrections being absent at the tunneling point which makes the bounce action only depend on the depth of the thermal CFT free energy and not on its exact shape.\footnote{See also \cite{Agashe:2019lhy,DelleRose:2019pgi} where the authors show that the contribution to the bounce action coming from the free energy in the thermal CFT phase is negligible.} In other world, the bounce action can simply be determined from the boundary conditions \cite{Konstandin:2010cd, Agashe:2019lhy,DelleRose:2019pgi}
\begin{equation}
\dilaton'(0) = 0, \qquad \dilaton'(\dilaton\to 0 ) = \sqrt{b}\,T^2.
\end{equation}
However, we will not use these boundary conditions but instead we will rely on the thick-wall formula.\footnote{For small $|\gamma_{\epsilon}|\ll 1$, the barrier is very small compared to the potential energy difference such that the bounce can be estimated with the thick-wall formula. }
For $O_3$-bounce, Eq.~\eqref{eq:S_3_thick_wall} with $V(\phi_{\rm FV}) = -bN^2T^4$ and Eq.~\eqref{eq:dilaton_potential_taylor_cw_march} reads\footnote{We assume that the kinetic terms are canonically normalized $(\kappa = 1)$.}
\begin{equation}
S_3 =c_* ~\underset{\dilaton_*}{\textrm{Min}} \frac{4\pi}{3} \frac{\dilaton^3_*}{\sqrt{-2\left(b \,N^2\,T^4 - \lambda_*\, \dilaton_*^4\right)}},
\end{equation}
with 
\begin{equation}
\lambda_* =  \left|\gamma_{\epsilon}\right|\, c_\dilaton \,g_\dilaton^2\, \textrm{log} \frac{f}{\dilaton_*},
\end{equation}
which gives
\begin{align}
&S_3/T  =c_* ~ \frac{2\pi}{3^{1/4}} \frac{(b\,N^2)^{1/4}}{\lambda_*^{3/4}}, \label{eq:S3_thickwall_light_dilaton}\\
&\chi_{*} =  \left(\frac{3\,b\,N^2}{\lambda_*}\right)^{1/4}~T.
\end{align}
$c_*$ is a coefficient which we add and which we fit on the numerical solution. We set
\begin{equation}
c_* \simeq 2,
\end{equation}
for both the $O_3$- and $O_4$- bounce actions.
The thick-wall formula for the $O_4$-bounce in Eq.~\eqref{eq:S_3_thick_wall}, reads
\begin{equation}
S_4 = c_* ~\underset{\dilaton_*}{\textrm{Min}} \frac{\pi^2}{2} \frac{\dilaton^4_*}{-\left(b \,N^2\,T^4 - \lambda_*\, \dilaton_*^4\right)},
\end{equation}
which gives
\begin{align}
&S_4  = c_* ~\frac{\pi^2}{2} \frac{1}{\lambda_*}, \label{eq:S4_thickwall_light_dilaton}\\
&\chi_{*} =   \frac{T}{T_c}~f.
\end{align}
where $T_c$ is the critical temperature, defined when the two minima of the free energy are equal
\begin{equation}
b\,N^2\,T_c^4 = \frac{m_{\dilaton}^2\,f^2}{16}, \qquad \rightarrow \quad T_c = \left(\frac{m_{\dilaton}^2 \,f^2}{16\,b\,N^2}\right)^{1/4} = \left( \frac{\left|\gamma_{\epsilon}\right|\,c_{\dilaton}\,g_{\dilaton}^2}{4\,b\,N^2} \right)^{1/4}~f
\end{equation}
We conclude that 
\begin{equation}
\frac{S_4}{S_3/T} = \frac{T}{b\,N^2\, \lambda_* } = \frac{f^2}{m_{\dilaton}^2}\frac{4}{b\,N^2} \frac{1}{\textrm{log} \frac{T_c}{T}}.
\end{equation}
Hence, at low temperature we expect the $O_4$ bounce to overtake over the $O_3$ bounce, see top left panel of Fig.~\ref{fig:light_dilaton_msigma_Tnuc_alpha_beta}. We refer to App.~B of \cite{Baldes:2021aph} for more details on the computation of the bounce action in the light-dilaton model.

\paragraph{Nucleation temperature:}
The nucleation happens when the tunneling rate in Eq.~\eqref{eq:tunneling_rate} becomes comparable to the Hubble expansion rate per Hubble volume.  For $O_4$-dominance, the nucleation temperature is solution of 
\begin{equation}
\label{eq:nucleation_eq_S_4_light_dilaton}
S_{4}(T_{\rm nuc}) \simeq 4 \ln{\frac{R_0^{-1}}{H(T_{\rm nuc})}} + \frac{1}{2} \ln \frac{S_4}{2\pi}
\end{equation}
where $R_0$ is the bubble radius which according to the thick-wall formula in Eq.~\eqref{S_4_thick_wall}, reads
\begin{equation}
R_0 = \left(\frac{f/m_{\rm \dilaton}}{\sqrt{b}\,N\,\textrm{\rm log} \frac{T_c}{T}}\right)^{1/2}~T^{-1} \sim T^{-1}.
\end{equation}
The Hubble parameter is, cf. Eq.~\eqref{eq:number_dof_CFT} and Eq.~\eqref{eq:vacuum_energy_difference}
\begin{equation}
H^2(T) = H^2_{\Lambda}+H^2_{\rm rad} = \frac{m_{\dilaton}^2\,f^2/16}{3 M_{\rm pl}^2}+\frac{\pi^2 g_* T^4}{90M_{\rm pl}^2}, \quad g_* = 106.75 +\frac{45 N^2}{4} .
\end{equation}
From rewriting the $S_4$ bounce action in Eq.~\eqref{eq:S4_thickwall_light_dilaton} as 
\begin{equation}
S_4  = \frac{A}{\textrm{log}~T_c/T} \qquad \textrm{with} \quad  A= c_*~\frac{\pi^2}{2}\frac{1}{ \left|\gamma_{\epsilon}\right|\, c_\dilaton \,g_\dilaton^2} = c_*~2\pi^2 \frac{f^2}{m_{\dilaton}^2}.
\label{eq:S_4_light_dilaton_A_T_c}
\end{equation}
From injecting it into Eq.~\eqref{eq:nucleation_eq_S_4_light_dilaton}, we obtain
\begin{equation}
\Tnuc = \sqrt{ H_{\Lambda}\,T_c} ~\left( \frac{2\pi}{S_4} \right)^{1/4}~\exp \left( \frac{1}{2} \sqrt{-A + \left( \ln \frac{T_c}{H_{\Lambda}} + \frac{1}{2} \ln \frac{S_4}{2\pi} \right)^2} \right).
\end{equation}
Neglecting the $S_4/2\pi$ terms, we conclude that there is no nucleation solution when
\begin{equation}
A \gtrsim  \ln \frac{T_c}{H_{\Lambda}},
\label{eq:minimal_msigma_S4_light_dilaton}
\end{equation}
and that the minimal nucleation temperature is
\begin{equation}
T_{\rm nuc}^{\rm min} \simeq  \sqrt{ H_{\Lambda}\,T_c} \simeq 0.1~\left( \frac{m_{\dilaton}}{f}  \right)^{3/4}\left(\frac{f}{M_{\rm pl}}\right)^{1/2}~f.
\label{eq:minimal_nucleation_temp_S4_light_dilaton}
\end{equation}
In top right panel of Fig.~\ref{fig:light_dilaton_msigma_Tnuc_alpha_beta}, we show the nucleation temperature computed with the thick-wall formula. 
Note that from inverting Eq.~\eqref{eq:S_4_light_dilaton_A_T_c}, we can write
	\begin{equation}
	\label{eq:analyticalTnuc}
	\left(\frac{T_{\rm nuc}}{T_c} \right)^4 \simeq \text{Exp}\left( -c_{*}\, \frac{2\, \pi^2}{ S_{\mathsmaller{\rm crit}} } \frac{1}{ \kappa^2 y_{\dilaton}^2} \right),
	\end{equation}
where $y_\dilaton \equiv m_\dilaton/\dilatonvev$.  The advantage of Eq.~(\ref{eq:analyticalTnuc}) is to make explicit that the nucleation temperature is exponentially suppressed for small dilaton mass $m_{\dilaton}$, or equivalently for small anomalous coupling $|\gamma_{\epsilon}|$. Note also that from using Eq.~\eqref{eq:gammaepsilonVSmdilaton_march_kappa} and $g_{\dilaton} = 4\pi/N$, we can rewrite Eq.~\eqref{eq:analyticalTnuc} as 
	\begin{equation}
	\label{eq:analyticalTnuc_N}
	\left(\frac{T_{\rm nuc}}{T_c} \right)^4 \simeq \text{Exp}\left( -c_{*}\, \frac{N^2}{ 32\,c_{\dilaton}\,\left|\gamma_{\epsilon}\right|\,S_{\mathsmaller{\rm crit}} } \right),
	\end{equation}
and we recover that at fixed $\gamma_{\epsilon}$, the nucleation temperature scales like $\Tnuc/f \propto e^{-N^2}$  \cite{Baratella:2018pxi}.

\paragraph{Catalysis by QCD effects:}
\label{par:QCD_effect_light_dilaton}

With Eq.~\eqref{eq:minimal_msigma_S4_light_dilaton}, we have concluded that there is no nucleation solution for too small dilaton mass $m_{\dilaton}$, such that the universe gets stuck in the false vacuum and inflates eternally.
We know discuss the possibility, first pointed out by \cite{vonHarling:2017yew}, that the QCD PT triggers the light-dilaton or holographic PT. In 4D, it comes from the possibility for the gluon condensate to generate a potential for the dilaton.
If the CFT couples to the $SU(3)_c$ of the SM, we expect it to modify the QCD beta function \cite{vonHarling:2017yew,Baratella:2018pxi}
\begin{equation}
\frac{dg_3}{d\ln{\mu}} = \beta_{3} = \frac{g_3^2}{16\pi^2} \left( - \frac{11}{3}N_c + \frac{2}{3}N_f + \alpha N  \right).
\end{equation}
Hence, in the presence of the deconfined CFT phase, we expect the QCD scale $\Lambda_{\rm QCD}$ to be modified as follows \cite{vonHarling:2017yew,Baratella:2018pxi}
\begin{equation}
\Lambda_{\rm QCD}^{\rm dec} = \Lambda_{\rm QCD} \left( \frac{\Lambda_{\rm QCD}}{f} \right)^{\frac{n}{1-n}}, \qquad \textrm{with} \quad n= \frac{\beta_{3}(\mu<f) - \beta_3(\mu>f)}{\beta_3(\mu<f)},
\end{equation}
where $\beta_{3}(\mu<f) $ and $\beta_{3}(\mu>f) $ are the QCD beta functions without and with the $\alpha N$ correction, respectively. Note that for $0<n<1$, we have $\Lambda_{\rm QCD}^{\rm dec} < \Lambda_{\rm QCD}$.
Additionally, when the QCD phase transition takes place, we expect it to generate a potential for the dilaton with a negative sign 
\begin{equation}
V_{\rm QCD}(\dilaton) \simeq  - \Lambda_{\rm QCD}(\dilaton)^4 \qquad \textrm{with} \quad  \Lambda_{\rm QCD} \sim \dilaton^n,
\end{equation}
Note that this provides an elegant solution to the graceful exit problem of eternal inflation. See \cite{Bloch:2019bvc,Fujikura:2019oyi} for other applications of the same mechanism.

\begin{subappendices}

\chapterimage{LISA-GW} 

\section{Sensitivity curves of GW detectors}
\label{app:sensitivity_curves}

\subsection{The signal-to-noise ratio}
The total output of a detector is given by the GW signal plus the noise, $ h(t)+n(t)$ where the level of noise $n(t)$ is measured by its noise spectral density $S_n(f)$ \cite{Caprini:2018mtu}.
\begin{equation}
\left< \tilde{n}^*(f) \tilde{n}(f')  \right> \equiv \delta(f-f')\, S_n(f).
\end{equation}
We define the detector sensitivity $\Omega_{\rm sens}(f)$ as the magnitude of the SGWB energy density which would mimick the noise spectral density $S_n(f)$
\begin{equation}
\Omega_{\rm sens}(f) = \frac{2\pi^2}{3H_0^2}\,f^3\,S_n(f).
\end{equation}
The capability of an interferometer to detect a SGWB of energy density $\Omega_{\rm GW}(f)$ after an observation time $T$ is measured by the signal-to-noise ratio (SNR) \cite{Maggiore:1999vm}
\begin{equation}
\label{eq:SNR_def_Sam}
{\rm SNR} = \sqrt{T \int_{f_{\rm min}}^{f_{\rm max}} df \, \left[ \frac{\Omega_{\rm GW}(f)}{\Omega_{\rm sens}(f)}\right]^2}.
\end{equation}
\subsection{The power-law integrated sensitivity curve}
Assuming a power law spectrum
\begin{equation}
\label{eq:SGWB_power_law_Sam}
\Omega_{\rm GW}(f) = \Omega_\beta\left( \frac{f}{f_{\rm ref}}\right)^\beta,
\end{equation}
with spectral index $\beta$, amplitude $\Omega_\beta$ and reference frequency $f_{\rm ref}$, we deduce from Eq.~\eqref{eq:SNR_def_Sam} the amplitude $\Omega_\beta$ needed to reach a given SNR after a given observation time $T$
\begin{equation}
\Omega_\beta = \frac{{\rm SNR}}{\sqrt{T}}\left( \int_{f_{\rm min}}^{f_{\rm max}} df \left[ \frac{h^2}{h^2\Omega_{\rm sens}(f)} \left( \frac{f}{f_{\rm ref}} \right)^\beta \right]^2  \right)^{-1/2},
\end{equation}
which upon re-injecting into Eq.~\eqref{eq:SGWB_power_law_Sam} gives
\begin{equation}
h^2 \Omega_{\rm GW}(f) = f^\beta \frac{{\rm SNR}}{\sqrt{T}}\left(\int_{f_{\rm min}}^{f_{\rm max}} df \left[ \frac{f^\beta}{h^2\Omega_{\rm sens}(f)} \right]^2 \right)^{-1/2}.
\end{equation}
For a given pair $({\rm SNR}, \, T)$, one obtains a series in $\beta$ of power-law integrated curves. One defines the power-law integrated sensitivity curve $\Omega_{PI}(f)$ as the envelope of those functions \cite{Thrane:2013oya}
\begin{equation}
\Omega_{PI}(f) \equiv \underset{\beta}{\rm max}  \left[ f^\beta \frac{{\rm SNR}}{\sqrt{T}} \left( \int_{f_{\rm min}}^{f_{\rm max}} df \left[ \frac{f^\beta}{h^2\Omega_{\rm sens}(f)}  \right]^2 \right)^{-1/2}   \right].
\end{equation}
Any SGWB signal $\Omega_{\rm GW}(f)$ which lies above $\Omega_{PI}(f)$ would gives a signal to noise ratio $>{\rm SNR}$ after an observation time $T$.
\subsection{Results}
For the purpose of our study, we computed the power-law integrated sensitivity curve $\Omega_{PI}(f)$, starting from the noise spectral density in \cite{Hild:2010id} for ET, \cite{Evans:2016mbw} for CE and \cite{Yagi:2011wg} for BBO/DECIGO. 
For pulsar timing arrays EPTA, NANOGrav and SKA, we directly took the sensitivity curves from \cite{Breitbach:2018ddu}. The signal-to-noise ratio can be improved by using cross-correlation between multiple detectors, e.g. LIGO-Hanford, LIGO-Livingston, VIRGO but also KAGRA which may join the network at the end of run O3, which began on the 1$^{\rm st}$ of April 2019, or LIGO-India which may be operational for run O5 \cite{Aasi:2013wya}. We computed the SNR for LIGO from the expression \cite{Thrane:2013oya}
\begin{equation}
\textrm{SNR} = \left[ 2T \int_{f_{\rm min}}^{f_{\rm max}} df \frac{\Gamma^2(f) S_{\rm h}^2(f)}{S_{n}^1(f)S_n^2(f)}  \right]^{1/2},
\end{equation}
where $S_n^1$ and $S_n^2$ are the noise spectral densities of the detectors in Hanford and in Livingston for the runs \href{https://dcc.ligo.org/LIGO-T1500293/public}{O2}, \href{https://dcc.ligo.org/LIGO-T1800044/public}{O4} or \href{https://dcc.ligo.org/LIGO-T1800042-v4/public}{O5} and $\Gamma(f)$ is the \href{https://dcc.ligo.org/public/0022/P1000128/026/figure1.dat}{overlap function} between the two LIGO detectors which we took from \cite{Abadie:2011fx}. The GW power spectral density $S_h(f)$ is related to the GW energy density through
\begin{equation}
S_h(f) = \frac{3H_0^2}{2\pi^2}\frac{\Omega_{\rm GW}(f)}{f^3}.
\end{equation}

We fixed the signal-to-noise ratio $\rm SNR = 10$ and the observational time $T =$ 268 days for LIGO O2, 1 year for LIGO O4 and O5, and 10 years for other sensitivity curves.

As this study was completed, Ref.~\cite{Mingarelli:2019mvk} appeared, where the sensitivity curves may differ from us by a factor of order 1.

\end{subappendices}


%

\xintifboolexpr { \x = 2}
  {
  }
{
\medskip
\small
\bibliographystyle{JHEP}
\bibliography{thesis.bib}
}

%% file: chap7.tex
\chapterimage{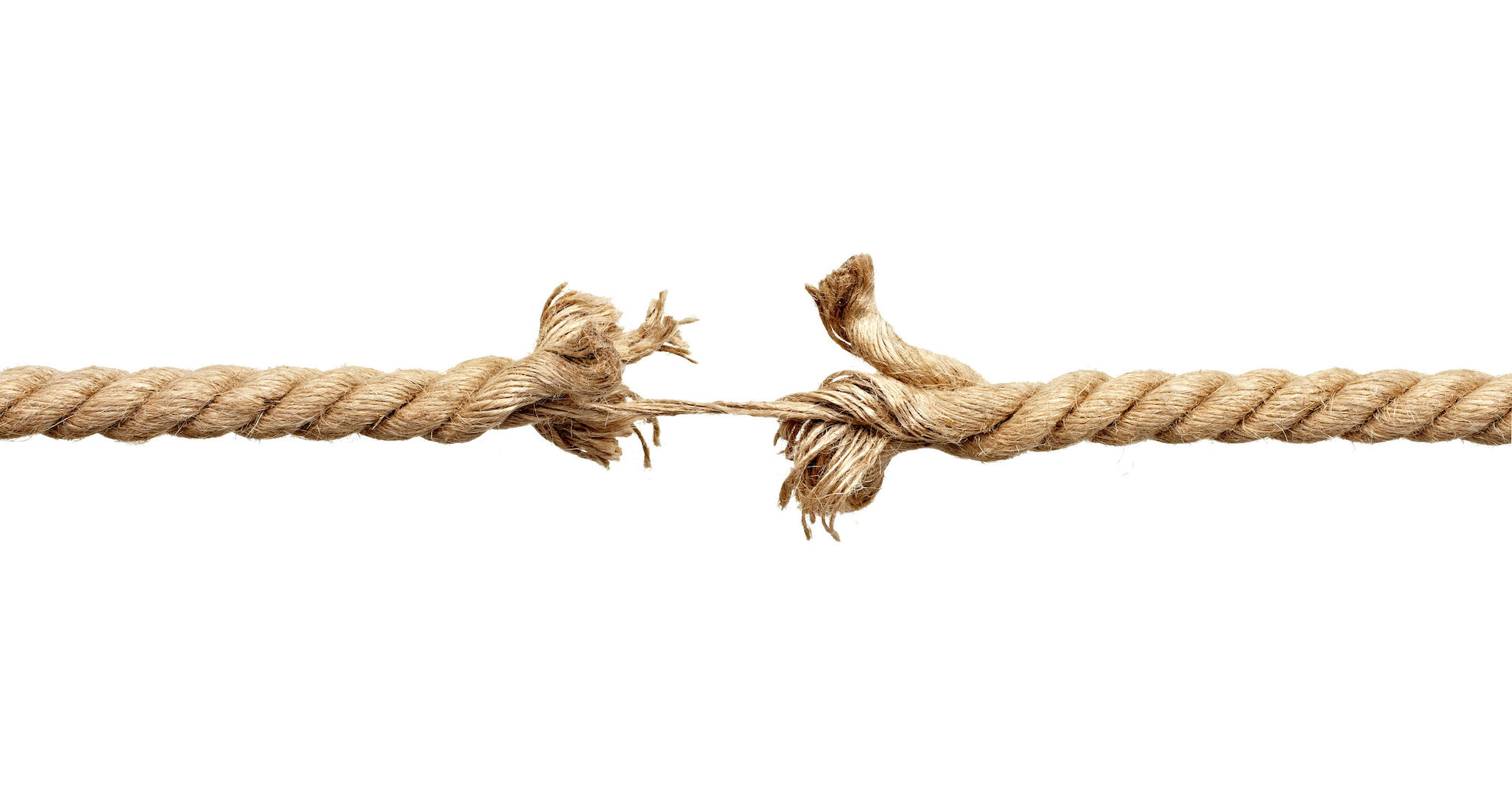} 


\chapter{String Fragmentation in Supercooled Confinement and implications for Dark Matter}
\label{chap:SC_conf_PT}

\begin{tikzpicture}[remember picture,overlay]
\node[text width=18cm,text=black,minimum width=\paperwidth,
minimum height=7em,anchor=north]%
 at (7,3.5) {This chapter is based on \cite{Baldes:2020kam}.};
 \draw (-2.0,1.8) -- (5,1.8);
\end{tikzpicture}

\section{Introduction}

The possible existence of new confining sectors is motivated by most major failures of our understanding of Nature at a fundamental level.
First, the stability of particle Dark Matter can be elegantly achieved as an accident if it is a composite state of a new strongly-coupled sector, similarly to proton stability in QCD, see e.g.~\cite{Antipin:2015xia}.
The hierarchy problem of the Fermi scale is solved via dimensional transmutation by new confining gauge theories, whose currently most appealing incarnation is that of composite Higgs models~\cite{Contino:2010rs,Panico:2015jxa}. Analogous composite pictures can UV-complete~\cite{Geller:2014kta,Barbieri:2015lqa,Low:2015nqa} twin-Higgs scenarios~\cite{Chacko:2005pe}, and so ameliorate also the little hierarchy problem. A rationale to understand the SM hierarchies of masses and CKM mixing angles is provided by partial compositeness of the SM fermions~\cite{Kaplan:1991dc}. Finally, new confining sectors play crucial roles in addressing the strong CP problem~\cite{Rubakov:1997vp,Redi:2016esr}, the baryon asymmetry~\cite{Konstandin:2011ds,Servant:2014bla}, etc.

Given their ubiquity, it makes sense to look for predictions of confining sectors that do not depend on the specific way they address a given SM issue.
Cosmology naturally offers such a playground, in association with the confinement phase transition (PT) in the early universe.
The low-density QCD phase transition would for example be strongly first-order if the strange or more quarks had smaller masses~\cite{Pisarski:1983ms}, with associated signals in gravitational waves~\cite{Witten:1984rs,Helmboldt:2019pan}. New confining sectors could also well feature a similar PT.
In addition, the confinement transition could be supercooled, a property that for example arises naturally in 5-dimensional (5D) duals of 4D confining theories~\cite{Creminelli:2001th,Randall:2006py,Nardini:2007me}.

Generically, supercooling denotes a PT in which bubble percolation occurs significantly below the critical temperature.
Here we are interested in the case where a cosmological PT becomes sufficiently delayed so that the radiation energy density becomes subdominant to the vacuum energy.
The universe then experiences a stage of inflation until the PT completes~\cite{Kolb:1979bt}. This implies a dilution of any pre-existing relic, such as dark matter (DM), the baryon or other asymmetries, topological defects, and gravitational waves, see e.g.~\cite{Hambye:2018qjv,Barreiro:1996dx,Easther:2008sx}.

In this paper we point out an effect that, to our knowledge, had been so far missed: when the fundamental quanta of the strong sector enter the expanding bubbles of the confined phase, their relevant distance can be much larger than the inverse of the confinement scale, thus realising a situation whose closest known analogues are perhaps QCD jets in particle colliders or cosmic ray showers.
We anticipate that our attempt to model this phenomenon implies an additional production mechanism of any composite resonance --- string fragmentation followed by deep inelastic scattering --- which introduces a mismatch between the dilution of composite and other relics. This opens new model building and phenomenological avenues, which we begin exploring here in a model independent manner for the case of composite DM. The application of our findings to a specific model, namely composite dark matter with dilaton mediated interactions, will appear elsewhere~\cite{Baldes:2021aph}.

\section{Synopsis}
\label{sec:synopsis}
Due to the numerous effects which will be discussed in the following sections, it is perhaps useful for the reader that we
summarise the overall picture in a few paragraphs. We begin in the deconfined phase in which the techniquanta $\TC$ of the new strong sector (which we will call quarks and gluons) are in thermal equilibrium. Their number density normalised to entropy takes a familiar form	
	\begin{equation}
	\label{eq:quark_density}
	Y_\TC^{\rm eq} = \frac{45\, \zeta(3) \, g_\TC}{2\pi^4 g_s},
	\end{equation}
where $g_\TC$ ($g_{s}$) are the degrees of freedom of the quarks and gluons (entropic bath) respectively.
Next a period of supercooling occurs, in which the universe finds itself in a late period of thermal inflation, which is terminated by bubble nucleation. As is known from previous studies, such a phase will dilute the number density of primordial particles. The dilution factor is given by
	\begin{equation}
	D^{\rm SC} = \left( \frac{T_{\rm nuc}}{T_{\rm start}} \right)^3 \frac{T_{\rm RH}}{T_{\rm start}},
	\end{equation}
where $T_{\rm nuc}$ is the nucleation temperature, $T_{\rm start} \propto f$ is the temperature at which the thermal inflation started, $T_{\rm RH}$ is the temperature after reheating, and $f$ is the energy scale of confinement. We assume reheating to occur within one Hubble time, so that $T_{\rm RH} \propto f$.
The supercooled number density of quarks and gluons then becomes
	\begin{equation}
	Y_\TC^{\SC} =  D^{\rm SC}~Y_\TC^{\rm eq} \propto \left( \frac{T_{\rm nuc}}{f} \right)^3.
	\label{eq:SCintro}
	\end{equation}
	For completeness, the details entering Eq.~(\ref{eq:SCintro}) will be rederived in Sec.~\ref{sec:supercool_relics}. 
	
	When the fundamental techniquanta are swept into the expanding bubbles, they experience a confining force. Because $f \gg T_{\rm nuc}$ in the supercooled transition, the distance between them is large compared to the size of the composite states  $\psi$ (which we will equivalently call `hadrons').
The field lines attached to a quark or gluon then find it energetically more convenient to form a flux tube oriented towards the bubble wall, rather than directly to the closest neighbouring techniquantum, which is in general much further than the wall (see Fig.~\ref{fig:wall_diagram}).
The string or flux tube connecting the quark or the gluon and the wall then fragments, producing a number of hadrons inside the wall.
Additionally, because of charge conservation, techniquanta must be ejected outside the wall to compensate (see Fig.~\ref{fig:string_breaking}).
The process is conceptually analogous to the production of a pair of QCD partons at colliders, and we model it as such. The details are explained in~Sec.~\ref{sec:inside_bubble}.
The result is an increase of the yield of composite particles, compared to the naive estimate following directly from Eq.~\eqref{eq:SCintro}, by a string fragmentation factor $K^{\rm string}$,
	\begin{equation}
	Y_\psi^{\rm \SC+ string } =  K^{\rm string}  D^{\rm SC} ~Y_\TC^{\rm eq}
	\propto \left( \frac{ \Tnuc}{ f } \right)^3 \times  \text{logs}{\left(\frac{ \gwp \Tnuc }{ f }\right)},
	\end{equation}
where $\gwp > f/\Tnuc \gg 1$ is the Lorentz factor of the bubble wall at the time the quarks enter.

The Lorentz factor is estimated in Sec.~\ref{sec:wall_speed}.
In Sec.~\ref{sec:our_picture_relevant} we show that our picture can be relevant already for $\Tnuc/\Tstart\lesssim 1$.
The quarks ejected from the bubbles are treated in detail in Sec.~\ref{sec:outside_bubble}. 
We find they enter neighbouring bubbles and confine there into hadrons.
Acting as a cosmological catapult, string fragmentation at the wall boundary gives a large boost factor to the newly formed hadrons, such that their momenta in the plasma frame can be $\gg f$.

The composite states and their decay products can next undergo scatterings with other particles they encounter, e.g~with particles of the preheated `soup' after the bubbles collide.
Since the associated center-of-mass energy can be much larger than $f$, the resulting deep inelastic scatterings (DIS) increase the number of hadrons.
We explore this in detail in Sec.~\ref{sec:DIS}. The resulting effect on the yield can be encapsulated in a factor $K^{\rm DIS}$, and reads
	\begin{equation}
	Y_\psi^{\rm \SC+string+\DIS }
	=  K^{\rm DIS} D^{\rm SC}~Y_\TC^{\rm eq}
	\propto \left( \frac{ \Tnuc}{ f } \right)^3 \gwp
	\xmapsto{\rm if~runaway} \left( \frac{ \Tnuc}{ f } \right)^4 \frac{M_\text{Pl}}{m_*}\,,
	\label{eq:Yintro_SC+str+DIS}
	\end{equation}
	where $M_\text{Pl}$ is the Planck mass and $m_* = g_*f$ is the mass scale of hadrons. The last proportionality holds in the regime of runaway bubble walls, relevant for composite DM.

Finally the late-time abundance of the long-lived and stable hadrons, if any, evolves depending on their inelastic cross section in the thermal bath $\langle \sigma v_{\mathsmaller{\rm rel}}  \rangle$, and on $Y_\psi^{\rm \SC+string+\DIS }$ as an initial condition at $\TRH$. We compute it in Sec.~\ref{sec:DMabundance} by solving the associated Boltzmann equations.

By combining all the above effects we arrive at an estimate of the final relic abundance of the composite states. 
Our findings impact their abundance by several orders of magnitude, as can be seen in Fig.~\ref{fig:compositeDM_generic} for the concrete case where the relic is identified with DM.
The formalism leading to this estimate can readily be adapted for other purposes. For example, if $\psi$ instead decays out-of-equilibrium, it could source the baryon asymmetry. The estimate of $Y_\psi^{\rm \SC+string+\DIS}$ would then act as the first necessary step for the determination of the baryonic yield.

\section{Supercooling before Confinement}
\label{sec:supercool_relics}
\subsection{Strongly coupled CFT}

Although striving to remain as model independent as possible in our discussion, we shall be making a minimal assumption that the confined phase of the strongly coupled 
theory can be described as an EFT with a light scalar $\chi$, e.g.~a dilaton.
The scalar VEV, $\langle \chi \rangle$, then parametrizes the local value of the strong scale. It can be thought of as a scalar condensate of the strong sector, such as a glueball- or pion-like state. The scalar VEV at the minimum of its zero-temperature potential is identified with $\langle \chi \rangle = f$, where $f$ is the confinement energy scale, while $\langle \chi \rangle = 0$ at large enough temperatures.
In order to have strong supercooling, we require the approximate (e.g.~conformal) symmetry to be close to unbroken, thus justifying the lightness of the associated pseudo-Nambu-Goldstone boson (e.g.~the dilaton~\cite{Bardeen:1985sm}). That supercooling occurs with a light dilaton is known from a number of previous studies~\cite{Creminelli:2001th, Randall:2006py, Nardini:2007me}, see~\cite{Paterson:1980fc, Bruggisser:2018mus, Bruggisser:2018mrt, Baratella:2018pxi, Agashe:2019lhy, DelleRose:2019pgi, vonHarling:2019gme,Bloch:2019bvc,Azatov:2020nbe} for studies in a confining sector and 
\cite{Hassanain:2007js, Konstandin:2010cd, Konstandin:2011dr, vonHarling:2017yew,Fujikura:2019oyi,Bunk:2017fic, Dillon:2017ctw, Megias:2018sxv,Megias:2020vek, Agashe:2020lfz} for studies of holographic dual 5D warped extra dimension models.

\subsection{Thermal history}
\label{sec:thermal_history}
The vacuum energy before the phase transition is given by
	\begin{equation}
	\label{eq:lambda_vac}
	\Lambda_{\rm vac}^{4} \equiv c_{\rm vac} \, f^{4},
	\end{equation}
with some model dependent $c_{\rm vac} \sim \mathcal{O}(0.01)$ constant. The radiation density is given by
	\begin{equation}
	\rho_{\rm rad} = \frac{g_{R}\pi^{2}}{30}T^{4},
	\end{equation}
where $g_{R}$ counts the effective degrees of freedom of the radiation bath. We define $g_{R} \equiv g_{Ri} \; (g_{Rf})$ in the deconfined (confined) phase. Now consider the case of strong supercooling. The universe will enter a vacuum-dominated phase at a temperature
	\begin{equation}
	T_{\rm start} = \left( \frac{ 30\, c_{\rm vac} }{  g_{Ri} \, \pi^{2}} \right)^{\! 1/4}f,
	\label{eq:Tstart}
	\end{equation}
provided the phase transition has not yet taken place beforehand. The vacuum domination signals a period of late-time inflation. The phase transition takes place at the nucleation temperature, $T_{\rm nuc}$, when the bubble nucleation rate becomes comparable to the Hubble factor.
Following the phase transition, the dilaton undergoes oscillations and decay, reheating the universe to a temperature
	\begin{equation}
	T_\RH = \left( \frac{g_{Ri}}{g_{Rf}} \right)^{1/4}T_{\rm start},
	\end{equation}     
At this point the universe is again radiation dominated. We have assumed the decay to occur much faster than the expansion rate of the universe such that we can neglect a matter-dominated phase~\cite{Hambye:2018qjv}.

\subsection{Dilution of the degrees of freedom}
\label{sec:dilution}
Now consider some fundamental techniquanta of the strong sector, e.g. techniquarks or technigluons (for simplicity we always refer to them as quarks and gluons). Prior to the phase transition the number density of techniquanta follows a thermal distribution for massless particles
	\begin{equation}
	n_\TC^{\rm eq} = g_\TC \frac{\zeta(3)}{\pi^{2}}T^{3},
	\end{equation}
where $g_\TC$ denotes the degrees of freedom of the quanta under consideration.
The entropy density is given by
	\begin{equation}
	s = \frac{2\pi^{2}g_s}{45}T^{3},
	\end{equation}
where $g_s$ are the total entropic degrees of freedom.\footnote{In a picture with $N_f$ flavours of quarks in fundamental representations of an $SU(N)$ confining gauge group, one has $g_q = 2 N_f N$, $g_g = 2 (N^2-1)$, $g_\TC = g_g + 3g_q/4$, $g_s = g_g + 7g_q/8$.
}
The number density normalized to entropy before the phase transition,
	\begin{equation}
	Y_\TC^{\rm eq} = \frac{45 \zeta(3) \, g_\TC}{2\pi^4 g_s},
	\label{eq:Yeqi}
	\end{equation}
remains constant up to the point when the phase transition takes place. 
The entropy density then increases during reheating giving 
	\begin{equation}
	\label{eq:qyield}
	Y_\TC^{\SC} = D^{\rm SC} ~Y_\TC^{\rm eq},
	\end{equation}  
when we find ourselves back in the radiation-dominated phase.
The dilution factor from the additional expansion during the vacuum-dominated phase can be derived by finding the increase in entropy between $\Tnuc$ and $T_{\rm RH}$. It reads
	\begin{equation}
	\label{eq:DSC}
	D^{\rm SC}
	\equiv \left( \frac{T_{\rm nuc}}{T_{\rm start}} \right)^3 \left( \frac{T_{\rm RH}}{T_{\rm start}} \right)
	\simeq \frac{g_{Ri}}{c_\text{vac}^{3/4} g_{Rf}^{1/4} } \, \Bigg(\frac{\Tnuc}{f}\Bigg)^3\,.
	\end{equation}
If the quarks and gluons were non-interacting following the phase transition, the yield today would be given by the above formula. (In the presence of interactions the above would be taken as an initial condition at $T_{\rm RH}$ for the Boltzmann equations describing the effects of number changing interactions between reheating and today.) The picture would then be analogous to that studied, in a theory without confinement, in~\cite{Hambye:2018qjv}.
The picture is completely changed, however, for supercooled confining phase transitions, which we elucidate next.

\section{Confinement and String Fragmentation}
\label{sec:inside_bubble}

\subsection{Where does confinement happen?}
\label{sec:when_confinement}

\begin{figure}[t]
\centering
\hspace{-1cm}
\raisebox{0cm}{\makebox{\includegraphics[height=0.35\textwidth, scale=1]{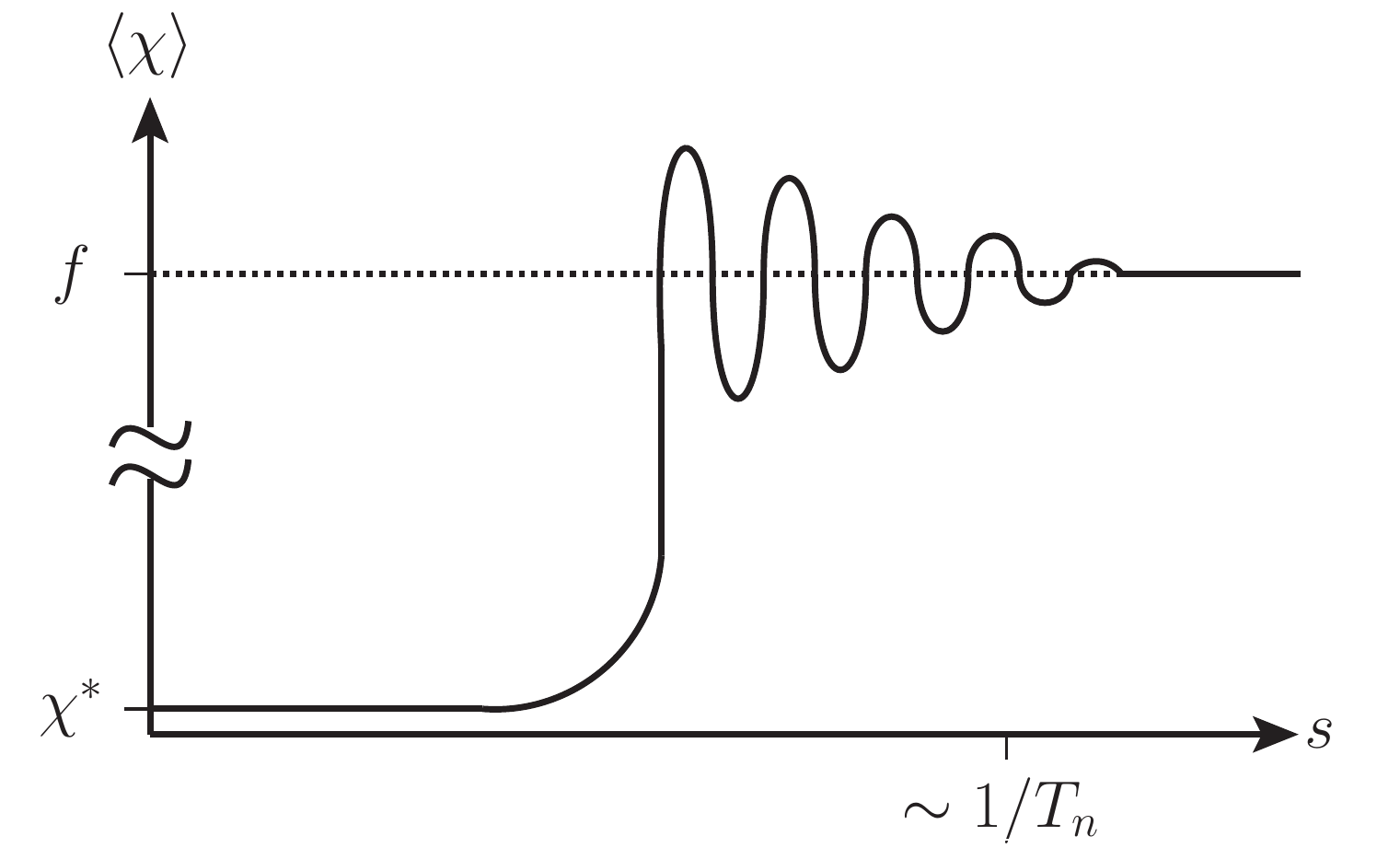}}}
\caption{\it \small A typical wall profile found in close-to-conformal potentials. After nucleating by tunneling to the exit point, $\chi^{\ast} \ll f$, the field rolls down and undergoes damped oscillations around the minimum of its potential. The typical wall thickness is $L_w \sim 1/\Tnuc$.}
\label{fig:wall_profile} 
\end{figure}

\paragraph{Bubble wall profile.}
The expanding bubble is approximately described by the Klein-Gordon equation~\cite{Jinno:2019bxw}
	\begin{equation}
	\label{eq:KG}
	\frac{ d^{2} \chi }{ d s^{2} } +  \frac{3}{s}\frac{d \chi}{ ds}  + \frac{d V }{ d\chi}   = 0,
	\end{equation}
where $s^{2} =t^2- r^{2}$ is the light-cone coordinate and $V$ is the scalar potential.
A sketch of a typical bubble profile for close-to-conformal potentials is shown in Fig.~\ref{fig:wall_profile}.
The key point here is that the wall thickness is 
	\beq
	L_{\rm w} \lesssim \frac{1}{\Tnuc},
	\label{eq:Lw}
	\eeq
as shown by numerical computations and analytical estimates, see App.~\ref{app:wall_profile} for a calculation in an explicit example.

\paragraph{Confinement time scale.}
The techniquanta (quarks and gluons) constitute a plasma with temperature of order $\Tnuc$ before entering the bubble. Once they enter the bubbles, they could in principle either confine in a region close to the bubble wall where $\langle \chi \rangle \ll f$,
or approach as free particles the region where $\chi$ has reached its zero-temperature expectation value $\langle \chi \rangle = f$.
To determine this, let us define a `confinement rate' and a `confinement length' as
	\beq
	\Gamma_\text{conf} = L_\text{conf}^{-1} = n_\TC\, v_\TC\, \sigma_\text{conf},
	\label{eq:confinement_rate}
	\eeq
where $n_\TC$ and $v_\TC$ are, respectively, the number density and the relative M{\o}ller velocity of the techniquanta $v_\TC \equiv \left[|\bold{v}_1-\bold{v}_2|^2 - |\bold{v}_1\times\bold{v}_2|^2  \right]^{1/2}$~\cite{moller1945general}, and $\sigma_\text{conf}$ is a `confining cross section'. 
We want to compare $L_\text{conf}$  with the length of the bubble wall, defined as the distance over which $\chi$ varies from its value at the exit point, $\langle \chi \rangle = \chi^* \ll f$, to $\langle \chi \rangle =f$.
Of course we need to perform this comparison in the same Lorentz frame, so we emphasise our definition of $L_w$ as the bubble-wall length in the bubble-wall frame, and $L_{\rm p} = L_{\rm w}/\gwp$ as the bubble-wall length in the frame of the center of the bubble, which coincides with the plasma frame, and where $\gwp$ is the boost factor between the two frames.
Let us now move to the confinement timescale of Eq.~(\ref{eq:confinement_rate}).
Since we expect confinement to happen `as soon as possible', we assume the related cross section to be close to the unitarity limit \cite{Griest:1989wd},
	\beq
	\sigma_\text{conf} \sim \frac{4\pi}{\Tnuc^2}\,.
	\label{eq:confinement_cross_section}
	\eeq
Since $n_\TC^2 v_\TC$ is Lorentz invariant~\cite{moller1945general, Gondolo:1990dk},
one then has that $n_\TC \,v_\TC$ transforms under boosts as $n_\TC^{-1}$.
The boost to apply in this case is $\gwp$, because by definition the string forms after confinement, so we can treat the plasma frame as the center-of-mass frame of the techniquanta.
Combining this with the Lorentz invariance of the cross section, we obtain
	\beq
	\label{eq:conf_rate}
	\Gamma_{\text{conf},w}
	= n_{\TC,\, \rm w}\,v_{\TC,\, \rm w}\,\sigma_\text{conf}
	=  \frac{ n_{\TC,\, \rm p}\,v_{\TC, \rm \,p} }{ \gwp }\,\sigma_\text{conf}
	\sim  \frac{ 4 \pi \,\Tnuc }{ \gwp },
	\eeq
where in the last equality we have used that the average relative speed and density of the techniquanta in the plasma frame satisfy, respectively, $v_{\TC,\,p} \simeq 1$ and $n_{\TC,\,p} \sim \Tnuc^3$, because they are relativistic.
This in turn implies
	\beq
	\label{eq:Lconf}
	L_{\text{conf, w}}
	\sim \frac{\gwp}{4\pi}L_{\rm w}.
	\eeq
	
\paragraph{Confinement takes place deep inside the bubble.}
For the regimes of supercooling we are interested in, the phase transition is of detonation type and the Lorentz factor $\gwp$ is orders of magnitude larger than unity.
Therefore, $L_{\text{conf},\,w} \gg L_{\rm w}$ such that confinement does not happen in the outermost bubble region where $\langle \chi \rangle \ll f$.
This conclusion is solid in the sense that it would be strengthened by using a confinement cross section smaller than what assumed in Eq.~(\ref{eq:confinement_cross_section}), which is at the upper end of what is allowed by unitarity. The end effect of the above discussion, is that for practical purposes, we can consider the wall profile to be a step-like function between the deconfined phase, $\langle \chi \rangle=0$, and confined phase, $\langle \chi \rangle=f$. Furthermore, as we shall discuss below, the quarks will not confine directly in pairs but rather form fluxtubes pointing toward the bubble wall as they penetrate the $\langle \chi \rangle=f$ region of the bubble.

\paragraph{The ballistic approximation is valid.}
Equation~\eqref{eq:Lconf}, together with the large wall-Lorentz-factors encountered in this study, implies that we can safely neglect the interactions between neighbouring techniquanta during the time when they cross the bubble wall. This is the so-called ballistic regime, see e.g.~\cite{Mancha:2020fzw}, which will be useful for deriving the friction pressure in Sec.~\ref{sec:wall_speed}.

\subsection{Fluxtubes attach to the wall following supercooling}
\label{sec:string_breaking}

\paragraph{A hierarchy of scale.}
Upon entering the region $\langle \chi \rangle = f$ of expanding bubbles, the techniquanta experience a confinement potential much stronger than in the region close to the wall.
This can be easily understood by taking the long-distance potential of the Cornell form~\cite{Eichten:1974af, Thorn:1979gv, Greensite:1988tj, Poulis:1996iv, Ko:1999yx, Greensite:2001nx, Bardakci:2002xi, Greensite:2003xf, Greensite:2003bk,Trawinski:2014msa}
	\begin{equation}
	\label{eq:stringenergy}
	E_\TC = c_\TC \,f^{2} \, d_{c},
	\end{equation}
where $d_c$ is the techniquanta seperation in their `center of interaction frame' (or equivalently `string center of mass frame')\footnote{Lattice simulations find that the QCD potential at $d_c \gtrsim$~fm saturates to a constant, a behavior which is interpreted in terms of pair creation of quarks from the vacuum, see e.g. the recent~\cite{Bulava:2019iut}. Therefore this realises an outcome that, for our purposes, coincides with having $E_\TC \propto d_c$ to larger distances. Lattice simulations with quarks only as external sources~\cite{Bali:2000gf}, so without sea quarks (`quenched'), find that the linear regime of the QCD Cornell potential extends up to the maximal distances probed, namely $d_c\simeq 3$~fm in the results reported in~\cite{Bali:2000gf}.
}, and $c_\TC$ is an adimensional constant\footnote{$c_\TC$ does not hide any `coupling dimension', indeed in units where $\hbar \neq 1$, $[f] = (\text{energy}/\text{distance})^{\!\frac{1}{2}}$.
}, $c_{q\bar{q}} \simeq 10$ in QCD~\cite{Trawinski:2014msa}.
A crucial point regarding the string energy in this context, besides the fact it grows proportionally to $\chi^2$, is that the inter-quanta distance is large compared to the natural confinement scale, i.e. $d_{c} \gg f^{-1}$, due to the supercooling.
Indeed the distance between quanta outside the wall, in the plasma and wall frames respectively, scales as $d_p \sim \Tnuc^{-1}$ and $d_w \sim \gwp^{-1/3} \Tnuc^{-1}$. Since $\gwp \ll (f/\Tnuc)^3$ (see Sec.~\ref{sec:wall_speed}) and $d_c \geq d_w$ (because $d_c = d_p$ outside the wall, and because the quarks and gluons cannot be accelerated upon entering so $d_w$ is Lorentz contracted with respect to $d_p$), one ends up with $d_c \gg f^{-1}$.
What happens then to the techniquanta and to the fields connecting them?

\begin{figure}[t]
\begin{minipage}[t]{0.6\linewidth}
    \centering
    \includegraphics[height=7 cm]{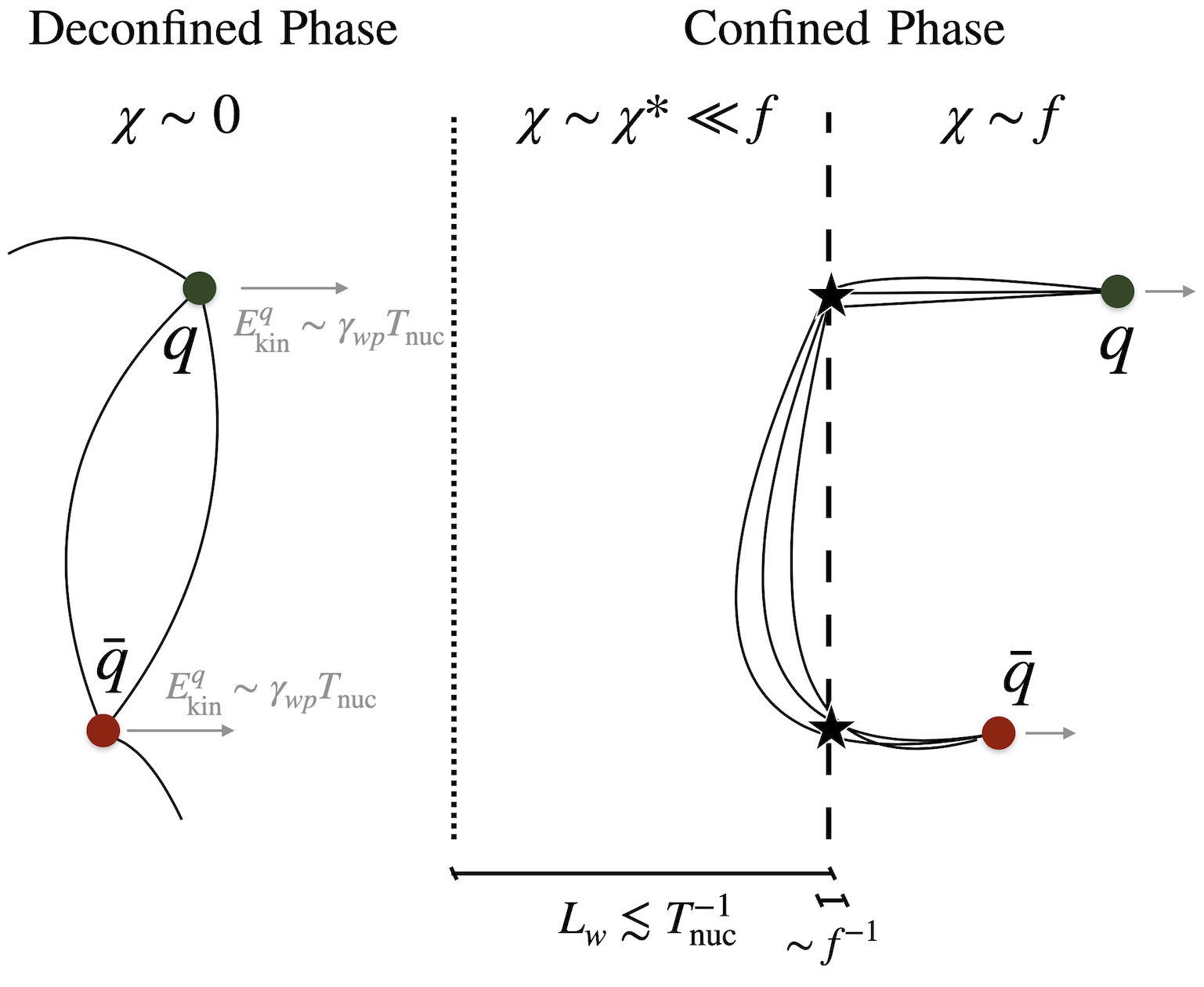}
    \caption{\it \small  Quarks entering the bubble as seen in the frame of the bubble wall, together with the associated field lines and quantities defined in the text. The rest energy of the string is minimized if the fluxtubes in the region $\chi = f$ point to the bubble wall, rather than if they point to the closest color charge.}
    \label{fig:wall_diagram}
\end{minipage}
\hspace{0.02\linewidth}
\begin{minipage}[t]{0.38\linewidth} 
    \centering
    \includegraphics[height=7 cm]{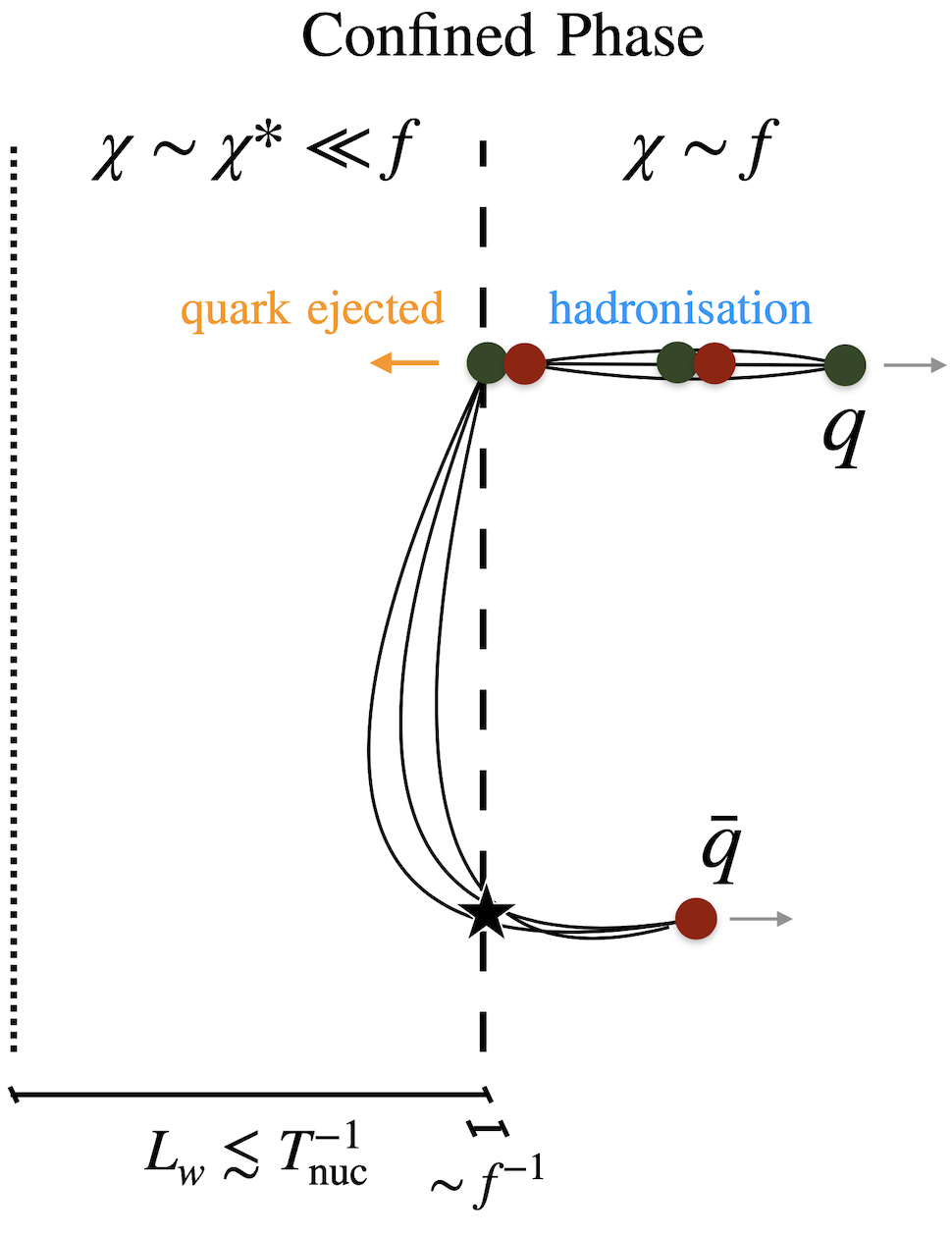}
    \caption{\it \small The string inside the wall breaks, producing hadrons (Sec.~\ref{sec:fragmentation_multiplicity_energy}), and a quark is ejected from the wall (Sec.~\ref{sec:ejected}).}
    \label{fig:string_breaking}
\end{minipage}        
\end{figure}  

\paragraph{Flux tubes minimize their energy.}

In a picture without hierarchy of scales, the fields would compress in fluxtubes connecting different charges, `isolated' in pairs or groups to form color-singlets. Here, we argue that the fluxtubes have another option, which is energetically preferable: that of orienting themselves towards the direction of minimal energy, i.e. as perpendicular as possible to the bubble-wall\footnote{We wish to express our gratitude to Benedict von Harling, Oleksii Matsedonskyi, and Philip Soerensen, for discussions which lead us to develop the picture we employ in this paper.}, and to keep a `looser' connection in the outer region where $\chi \ll f$.
Indeed, a straight-line connection between techniquanta would result in a much longer portion of fluxtubes in the region $\langle \chi \rangle = f$, with respect to our picture of fluxtubes perpendicular to the wall.
Via Eq.~\eqref{eq:stringenergy}, this would in turn imply a much higher cost in energy, disfavoring that option.
We stress that, in our picture, the fluxtubes are still connecting techniquanta in such a way to form an overall color singlet, just these fluxtubes minimise their length in the region $\chi = f$, and partly live in a region $\chi \simeq \chi^* \ll f$.
This picture is visualised in Fig.~\ref{fig:wall_diagram}. Note the nearest neighbour quark from the plasma may also be located outside the bubble.

\paragraph{Condensed matter analogy.}
An interesting analogue to the picture above is the vortex string of magnetic flux in the Landau-Ginzburg model of superconductivity. To match onto confinement dynamics a dual superconductor is pictured, in which the external colour-electric field --- rather than the magnetic field --- is expelled by the Meissner effect~\cite{Ripka:2003vv}. Here the bubble of confining phase corresponds to the superconductor from which the colour-electric field is expelled. Quarks entering the bubble then map onto magnetic monopoles being fired into a regular superconductor.

\subsection{String energy and boost factors}
\label{sec:boosts}

To possibly be quantitative on the implications of the picture we just outlined, we first need to determine the string energy and the Lorentz boosts among the frames of the plasma, wall and center-of-mass of the string.

\paragraph{String end-points.}
Let us define as $\TC_i$ the quark or gluon that constitutes an endpoint, inside the bubble, of a fluxtube pointing towards the wall, and $\bigstar$ the end-point of the fluxtube on the wall. The energy of the incoming techniquantum in the wall frame is $E_{i,\text{w}} = 3 \gwp \Tnuc$, where for simplicity we have averaged over their angle with respect to the wall. We assume $\bigstar$ to be at rest or almost, and to carry some $\mathcal{O}(1)$ fraction of the inertia of the string. Hence the respective four-momenta are
\beq
p_{i,\text{w}} = 
\begin{pmatrix}
 3\,\gwp \Tnuc \\
\sqrt{9\, \gwp^2 \Tnuc^2 - m_i^2}
\end{pmatrix},
\quad
p_{\bigstar,w} = 
\begin{pmatrix}
m_f \\
\epsilon f
\end{pmatrix},
\qquad \epsilon \ll 1,
\qquad m_i \simeq m_f = q f, \quad q\leq \frac{1}{2}\,.
\label{eq:momenta_star}
\eeq

\paragraph{String center-of-mass.}
Then we define the center-of-mass of the string as the one of $\TC_i$ and $\bigstar$, and find
\beq
\ECM = |p_{\bigstar,w} + p_{i,\text{w}}|
\simeq \sqrt{3\, \gwp \, \Tnuc\, f}\,,
\label{eq:ECMstring}
\eeq
where the second expression is valid up to relative orders $(\gwp f/\Tnuc)^{-1} \ll 1$.
By employing a Lorentz boost between the wall and center-of-mass frames, and imposing $\vec{p}_{i,c} = - \vec{p}_{\bigstar,c}$, we find
\beq
\gwc \simeq \sqrt{3\,\gwp\frac{\Tnuc}{f}}\,.
\label{eq:gwc}
\eeq
On the right-hand side of the equations above we have omitted a factor of $\sqrt{2 (q-\epsilon)}$, in~(\ref{eq:ECMstring}), and of $1/\sqrt{2 (q-\epsilon)}$, in~(\ref{eq:gwc}), because for simplicity we take these to be $\approx 1$ from now on (as per the benchmark $q=1/2$, $\epsilon=0$).
Finally we determine the boost between the center-of-mass frame of the string and the plasma frame as
\beq
\gcp \simeq\frac{\gwp}{2\gwc} = \frac{1}{2} \sqrt{\frac{\gwp}{3} \frac{f}{\Tnuc}},
\label{eq:gcp}
\eeq
which is valid up to a relative order $(\gwp f/\Tnuc)^{-1} \ll 1$.

\subsection{Hadrons from string fragmentation: multiplicity and energy}
\label{sec:fragmentation_multiplicity_energy}

The fluxtubes connecting a quark or gluon to the wall will fragment and form hadrons, singlet under the new confining gauge group. We would now like to determine:
\begin{itemize}
\item The number of hadrons formed per fluxtube.
\item The momenta of said hadrons.
\end{itemize}
\paragraph{Collider analogy.}
We start by noticing that the process of formation of a fluxtube, in our picture, is analogous to two color charges in an overall-singlet state, $\TC_i$ and $\bigstar$, moving apart with a certain energy $\ECM$, where $\ECM =  \sqrt{3\,\gwp\, \Tnuc\, f}$ in the modelling of Sec.~\ref{sec:boosts}.
This physical process appears entirely analogous to what would happen in a collider that produces a pair of techniquanta of the new confining force, starting from an initial singlet state. In light of this observation, we then decide to model the process
by analogy with a very well-studied process observed in Nature, that of QCD-quark pair production at electron-positron colliders, where the analogy lies also in the fact that the initial state electron-positron pairs is in a color singlet state.
Needless to say, a BSM confining sector needs not behave as QCD in terms of number and momenta of hadrons produced per scattering, see e.g.~\cite{Strassler:2008bv}.
However, QCD constitutes a well studied and tested theory, so that we find it reasonable to use it as our benchmark. Moreover, we anticipate from Sec.~\ref{sec:DIS} that our final result for the cosmological abundance of hadrons, in the assumption of efficient-enough interactions between them and the SM, will only depend on the initial available energy $\ECM$. This suggests that, within that assumption, our final findings hold for confining sectors that distribute this energy over a number of hadrons different from~QCD.

\begin{figure}[t]
\centering
\begin{adjustbox}{max width=1.2\linewidth,center}
\includegraphics[width= 0.5\textwidth]{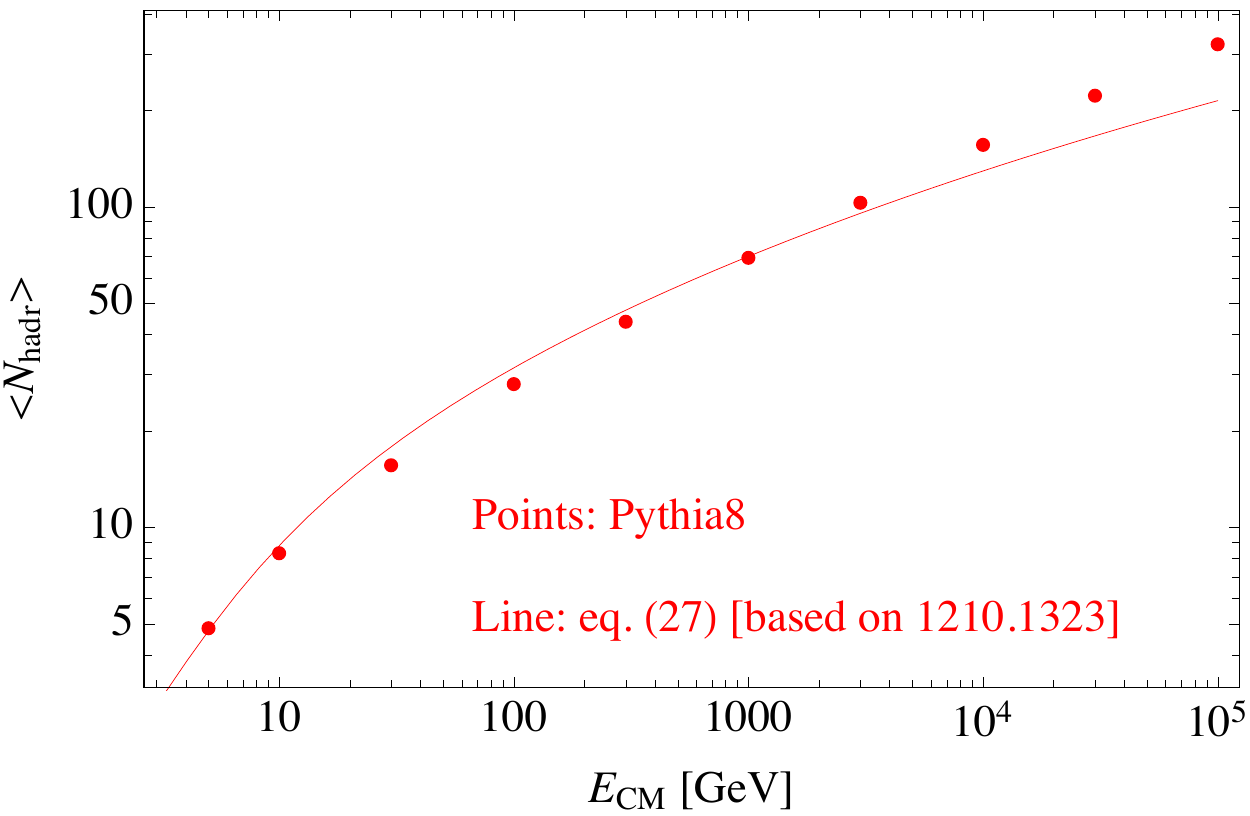}
\;
\includegraphics[width= 0.5\textwidth]{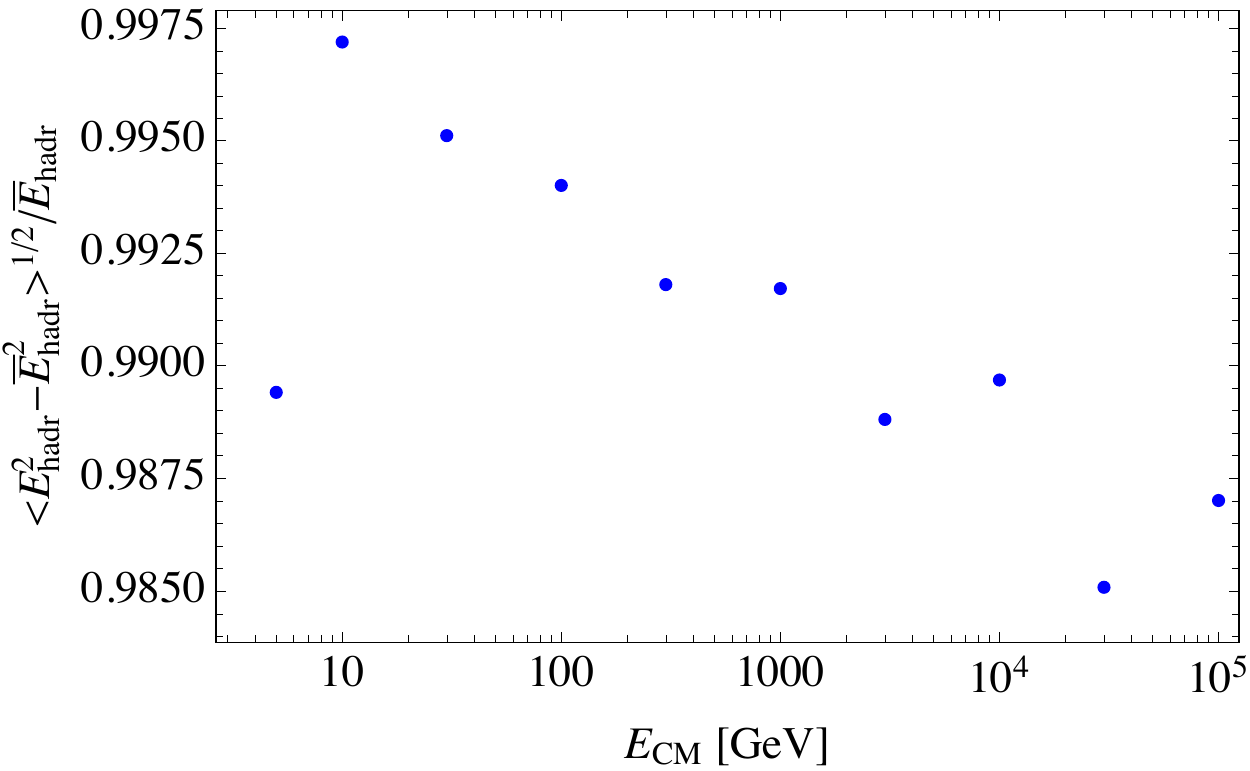}
\end{adjustbox}
\caption{\it \small Left: average hadron multiplicity per single QCD scattering $e^-e^+ \to q\bar{q}$.
Right: Root square mean of the hadron energy per single QCD scattering $e^-e^+ \to q\bar{q}$, where $\bar{E}_\text{hadr} = \ECM/\langle N_\text{hadr}\rangle$ is the average hadron energy per scattering.
Dots in both plots are extracted via MadAnalysis v1.8.34~\cite{Conte:2012fm} by simulations with MadGraph v2.7.0~\cite{Alwall:2014hca} plus Pythia v8.2~\cite{Sjostrand:2014zea}, the line in the left-hand plot displays Eq.~(\ref{eq:Npsi_string}).
Results are expressed as a function of the center-of-mass energy of the scattering in GeV, to export them to our cosmological picture we simply substitute $\mathrm{GeV}\simeq 4\pi f_\pi$ by $m_* = g_* f$, and use $\ECM = \sqrt{3\, \gwp \, \Tnuc\, f}$.}
\label{fig:pythia} 
\end{figure}

\paragraph{Numerical simulations.}
We use Pythia v8.2~\cite{Sjostrand:2014zea} interfaced to MadGraph v2.7.0~\cite{Alwall:2014hca} to simulate the process $e^-e^+ \to q\bar{q}$ for different center-of-mass energies, and MadAnalysis v1.8.34~\cite{Conte:2012fm} to extract from these simulations both the total number of hadrons produced per scattering and their energy distribution. We thus recover known QCD results and display them in Fig.~\ref{fig:pythia}.
We translate them to our picture by replacing the units of a GeV$\simeq 4 \pi f_\pi$ used by Pythia, with the generic mass of a composite state $m_* = g_* f$, where $1\leq g_* \leq 4\pi $ is some strong effective coupling.
These results can be summarised as follows:
\begin{itemize}
\item The number of hadrons produced per fluxtube grows logarithmically in $\ECM$.
\item The distribution of hadron energies is such that its root square mean coincides, to a percent level accuracy, with the average energy per hadron
\beq
\bar{E}_\text{hadr} = \frac{\ECM}{\langle N_\text{hadr}\rangle}\,.
\label{eq:energy_per_hadron}
\eeq
This will support, in Sec.~\ref{sec:DIS_hadron_abundance}, our simplifying assumption that all hadrons produced by the string fragmentation carry an energy of order $\bar{E}_\text{hadr}$.
\end{itemize}

\medskip
\paragraph{Results from the literature.}
The multiplicity of QCD hadrons from various scattering processes has been the object of experimental and theoretical investigation, since the late 1960s~\cite{Feynman:1969ej}.
We now leverage such studies both to check the results of our simulation and to obtain analytical control over them.
Collider studies have typically focused on the multiplicity of charged QCD resonances per scattering, $\langle n_\text{ch}\rangle $. In particular, works such as~\cite{GrosseOetringhaus:2009kz,Kumar:2012hx} have carried out the exercise of collecting the most significant measurements of $\langle n_\text{ch}\rangle$ and `filling' the missing phase space --- not covered by detectors --- with the output of MC programs, thus obtaining a full-phase-space quantity. We take as our starting point the result provided in~\cite{Kumar:2012hx} from $pp$ collisions, which reads
\beq
\langle n_\text{ch}\rangle (\ECM) = a + b \log \frac{\ECM}{m_*} + c \log^2 \frac{\ECM}{m_*} + d \log^3 \frac{\ECM}{m_*},
\label{eq:ncharged_s}
\eeq
with $(a,b,c,d) = (0.95,0.37,0.43,0.04)$. Here, as already explained, we substituted the normalisation of a GeV with $m_* = g_* f$. 
\paragraph{Our modelling.}
To obtain the total number of hadrons from $e^+e^-$ collisions we proceed as follows.
First, most hadrons coming out from hard scatterings consist in the lightest ones, i.e. the pions.
Second, the total number of pions produced is very well approximated by $3 \langle n_\text{ch}\rangle /2$, because of isospin conservation. By the first argument, this coincides with very good approximation to the total number of hadrons produced. 
Third, the multiplicity of composite states from $e^+e^-$ collisions has been found to roughly match the one from $pp$ collisions, upon increasing the $e^+e^-$ energy by a factor of 2, see e.g.~Sec.~2.2 in~\cite{Rosin:2006av}\footnote{\label{foot:Npsi_ee_vs_pp}This is qualitatively understood by the fact that, in purely leptonic initial states, there is more energy available to produced hadrons, while in the case with protons in the initial state much energy is carried over by the initial hadron remnant.
}.
We then model the total number of composite states produced, per string fragmentation, as
\beq
N_\psi^\text{string}(\ECM)
\simeq
\frac{3}{2} \langle n_\text{ch}\rangle (2 \ECM) \exp(-3 m_*/\ECM) + 1\,,
\label{eq:Npsi_string}
\eeq
where we have multiplied by an exponential and added one to smoothen $N_\psi^\text{string}(\ECM)$ to 1 as  $\ECM \to m_*$, because this physical regime was not taken into account in\cite{Kumar:2012hx}.
In the left-hand panel of Fig.~\ref{fig:pythia} one sees that Eq.~(\ref{eq:Npsi_string}) reproduces the results of our Pythia simulation for $\ECM$ smaller than a few TeV rather well. This was to be expected since Eq.~(\ref{eq:ncharged_s}) was determined in~\cite{Kumar:2012hx} from fits to data up to that energy. It is not the purpose of this paper to improve on this fit, as stated above, we simply use the above results as a check of our Pythia simulation.

\subsection{Enhancement of number density from string fragmentation}
\label{sec:string_summary}

\paragraph{Production of composite states.}
Prior to (p)reheating, we then have a yield of composite states given by the yield of strings, which can be estimated from Eq.~(\ref{eq:qyield}), multiplied by the number of composite states per string
\begin{equation}
K^{\rm string} = 
\left\{  \begin{array}{ll}
             \frac{\frac{3}{4}g_q N_\psi^\text{string}(\ECM) + g_g (N_\psi^\text{string}(\ECM)-1)}{g_\TC} & \quad \text{heavy~composite~state}, \\
               N_\psi^\text{string}(\ECM) & \quad \text{light~composite~state}\,,
                \end{array}
              \right.
\label{eq:Kstring}
\end{equation}
where $N_\psi^\text{string}(\ECM)$ is given by Eq.~\eqref{eq:Npsi_string} and $\ECM = \sqrt{3\, \gwp \, \Tnuc\, f}$ in Eq.~\eqref{eq:ECMstring}.
We have distinguished the cases where the composite state of interest is heavier or lighter than the glueballs (e.g. the analogous of a proton or a pion in QCD). In the former case, the $-1$ we added to the factor multiplying $g_g$ accounts for the fact that, if the final composite states produced by string fragmentation do not undergo other additional interactions, then glueballs decay to the light composite states and do not contribute to the final yield of any heavy composite state of quarks. 
The yield of composite states $\psi$ then reads
\begin{align}
	Y_{\psi}^{\rm \SC + string} & =  \,Y_\TC^{\rm eq} ~ D^{\rm SC}  ~ K^{\rm string}
	\propto \left( \frac{ \Tnuc}{ f } \right)^3 \times  \text{logs}{\left( \frac{ \gwp\Tnuc }{ f } \right)}.
	\label{eq:Ystring} 
	\end{align}
	The appearance of $Y_\TC^{\rm eq}$ in Eq.~(\ref{eq:Ystring}) accounts for string formation from both quarks and gluons. Hence, not only is the number of $\psi$'s enhanced by the string fragmentation, relative to the case with no confinement, but also by the possibility of gluons to form strings. $K^{\rm string}$ and $Y_{\psi}^{\rm \SC + string} $ are plotted in Fig.~\ref{fig:contributions}.

\paragraph{Hadrons are highly boosted in the plasma frame.}
The hadrons formed after string fragmentation schematically consist of two equally abundant groups.  
Hadrons in the first group, which for later convenience we call `Population A', move towards the bubble wall with an average energy
\beq
E_\text{A,p} \simeq 2 \gcp \frac{ \ECM}{N_\psi^\text{string}(\ECM)} \simeq \frac{\gwp f}{N_\psi^\text{string}(\ECM)},
\label{eq:EAp}
\eeq
 where we have boosted the energy per hadron of Eq.~(\ref{eq:energy_per_hadron}) to the plasma frame with the $\gcp$ of Eq.~(\ref{eq:gcp}), and also used Eqs.~(\ref{eq:ECMstring}) and (\ref{eq:Kstring}).
  We conclude from Eq.~\eqref{eq:EAp} that the newly formed hadrons have large momenta in the plasma frame. The formation of a gluon string between the incoming techniquanta and the wall acts as a cosmological catapult which propels the string fragments in the direction the wall is moving.
Hadrons in the second group move, in the wall frame, towards the bubble wall center, and their energy in the plasma frame is negligible compared to~(\ref{eq:EAp}).
Note that if only one hadron is produced on average per every string, then it would roughly be at rest in the center-of-mass frame of the string, with an energy (mass) of order $\ECM$. In the plasma frame, its energy would then read $E_\text{p} \simeq \gcp \ECM \simeq \gwp f/2$. As we will see in~Sec.~\ref{sec:DIS}, the impact of this hadron on the final yield would then be captured by our expressions.

Following this first stage of string fragmentation, the composite states, and/or their decay products, can undergo further interactions with remnant particles of the bath, preheated or reheated plasma, and among themselves.
Such interactions may change the ultimate yield of the relic composite states.
Before taking these additional effects into account in Sec.~\ref{sec:DIS}, in the next sections we complete the modelling we proposed above, by describing the behaviour of the ejected quarks and deriving the Lorentz factor of the wall, $\gwp$.

\subsection{Ejected quarks and gluons and their energy budget}
\label{sec:ejected}
So far we dealt with what happens inside the bubble wall.
The process we described apparently does not conserve color charge: we started with a physical quark or gluon with a net color charge entering the bubble, and we ended up with a system of hadrons which is color neutral. Where has the color charge gone?

\paragraph{The necessity of ejecting a quark or gluon.}
To understand this, it is convenient to recall the physical modelling behind the process of string fragmentation that converts the initial fluxtube into hadrons, see e.g.~the original Lund paper~\cite{Andersson:1983ia}.
When the fluxtube length, in its center-of-mass frame, becomes of order $f^{-1}$, the string breaks at several points via the nucleation of quark-antiquark pairs from the vacuum. Now consider, in our cosmological picture, the quark-antiquark pair nucleated closest to the bubble wall. One of the two --- say the antiquark --- forms a hadron inside the wall. The only thing that can happen to the quark is for it to be ejected from the wall, because of the lack of charge partners inside the wall.
This process, somehow reminiscent of black hole evaporation, thus allows for charge to be conserved.
The momentum of the ejected quark, in the wall frame, has to be some order-one fraction of the confinement scale $f$, because that is the only energy scale in the process. For definiteness, in the following we will take this fraction to be a half.
This picture is visualized in Fig.~\ref{fig:string_breaking}, and it is analogous if $\TC_i$ is a gluon instead of a quark.
 
\medskip
\paragraph{Energy of the ejected quark or gluon.}
One then has one ejected quark (at least) or gluon per fluxtube, thus per quark or gluon that initially entered. Therefore, the number of techniquanta outside the bubble wall does not diminish upon expansion of the bubble.
This population of ejected techniquanta is energetically as important as that of hadrons inside the bubble.
Indeed the energy of an ejected quark or gluon (or quark pair), in the plasma frame, reads
\beq
E_\text{ej,p}\simeq \gwp f.
\label{eq:energy_ejected}
\eeq
This is of the same order as the total energy in the hadrons from the fragmentation of a single string,
\beq
E^\text{tot}_\text{A,p} = \frac{N_\psi^\text{string}(\ECM)}{2} E_\text{A,p} \simeq \gwp \frac{f}{2},
\label{eq:ECMp}
\eeq
obtained by multiplying $E_\text{A,p}$ of Eq.~(\ref{eq:EAp}) times half of the total number of hadrons produced per string (i.e.~we included only the energetic ones).
The population of ejected techniquanta cannot therefore be neglected in the description of the following evolution of this cosmological system.

\section{Bubble wall velocities}
\label{sec:wall_speed}

The wall boost in the plasma frame, $\gwp$, affects many key properties of our scenario, from the ejection of techniquanta to the number and energy of the hadrons produced by string fragmentation.
It is the purpose of this section to study the possible values it can take over the~PT.

\begin{figure}[t]
\begin{center}
\includegraphics[width=.65\textwidth]{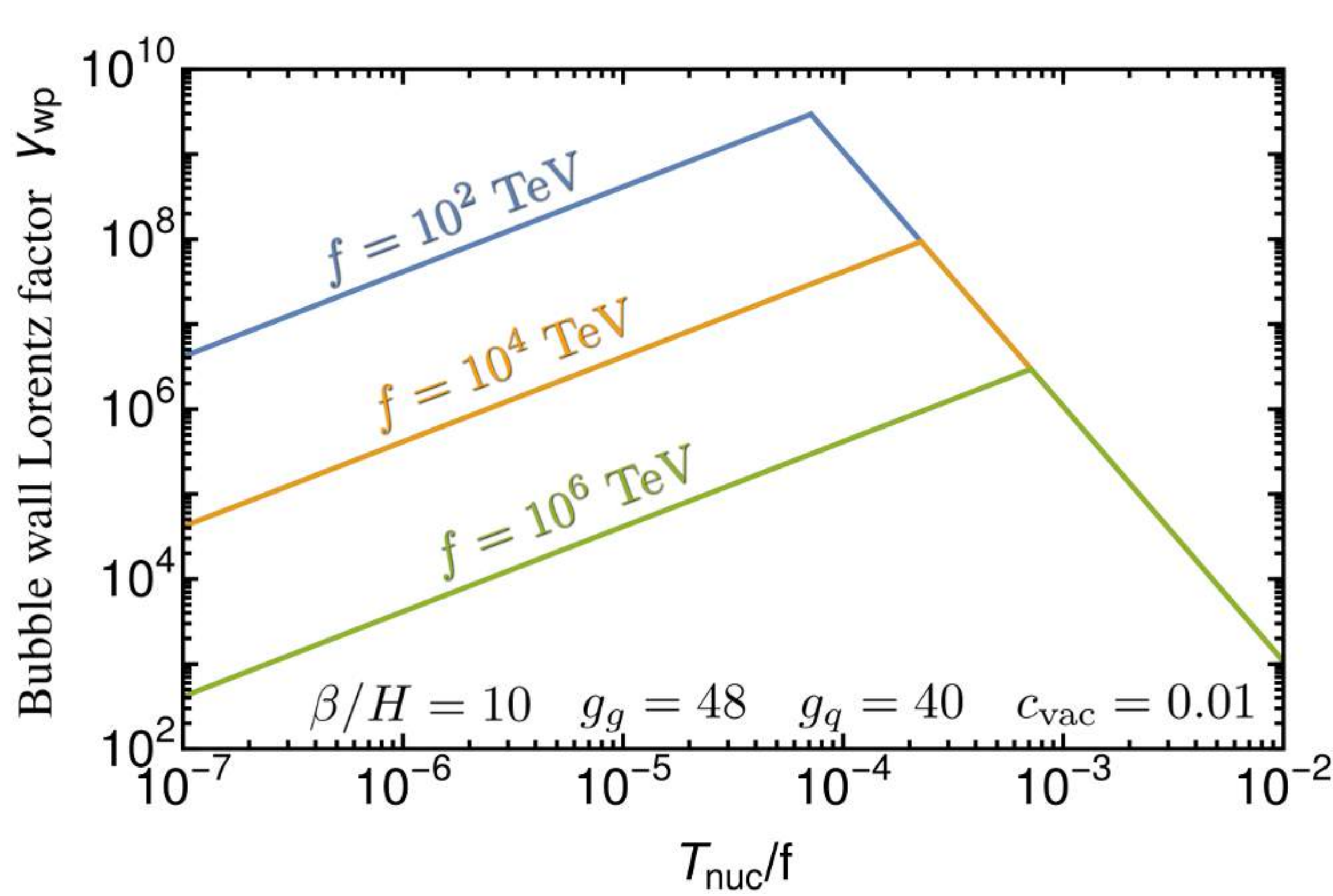}
\caption{
\label{fig:gwp} 
\it \small
The Lorentz factor of the wall at bubble percolation for various values of $f$ and amounts of supercooling, assuming $\beta/H=10$. For extreme supercooling (on the left side of the plot) $\gwp$ is in the runaway regime. In this regime, larger $f$ or smaller $\Tnuc$ leads to a smaller distance over which the bubble can accelerate. The former because of the smaller Hubble horizon and the latter due to the larger bubble size at nucleation. Therefore $\gwp$ decreases for more supercooling.
}
\end{center}
\end{figure}

\paragraph{Final results.}
As bubbles are nucleated and start to expand, $\gwp$ starts growing as well.
If nothing slows down the bubble-wall acceleration, then $\gwp$ keeps growing until its value at the time of bubble-wall collision, $\gwp^\text{runaway}$.  Sources of friction that could prevent this runaway regime are given by the equivalent, in this scenario, of the so-called leading order (LO) and next-to-leading order (NLO) contributions of~\cite{Bodeker:2009qy} and~\cite{Bodeker:2017cim} respectively.
We find it convenient to report right away our final result for the maximal possible value of $\gwp$,
\begin{equation}
\gwp^\text{max} \simeq \text{Min}\Big[
1.7 \,\frac{10}{\beta/H} \Big(\frac{0.01}{c_\text{vac}}\Big)^{\!\frac{1}{2}}\, \frac{\Tnuc}{f} \frac{\MPl}{f},~
1.0\times 10^{-3}\frac{c_\text{vac}}{0.01}\frac{80}{g_\TC} \Big(\frac{f}{\Tnuc}\Big)^{\!3}
\Big],
\label{eq:gwp_max}
\end{equation}
where the first entry is associated to $\gwp^\text{runaway}$, and the second to the boost as limited by the LO pressure, $\gwp^\text{LO}$. $\gwp^\text{LO}$ is always smaller than $\gwp^\text{NLO}$ in the parameter space of our interest, so that $\gwp^\text{NLO}$ does not enter Eq.~\eqref{eq:gwp_max}.
We learn that in the regime of very strong supercooling and/or of very large confinement scale $f$, which will be the most relevant one for the DM abundance, bubble walls run away. The behaviour of $\gwp$ is illustrated in Fig.~\ref{fig:gwp}.

\paragraph{The impact on GW.}
The behaviour of $\gwp$ also has important consequences for the gravitational wave signal from the phase transition~\cite{Caprini:2015zlo,Caprini:2019egz}. If $\gwp^\text{max}=\gwp^\text{runaway}$ then the vacuum energy is converted into kinetic energy of the bubble walls \cite{Ellis:2019oqb}. The gravitational wave (GW) spectrum sourced by scalar field gradient is traditionally computed in the envelope approximation \cite{Kosowsky:1992vn,Huber:2008hg, Jinno:2016vai}. However, the latest lattice results \cite{Cutting:2020nla, Lewicki:2020jiv} suggest an enhancement of the GW spectrum at low frequency due to the free propagation of remnants of bubble walls after the collision, the IR slope $\propto k^3$ becoming close to $\propto k^1$. This confirms the predictions from the analytical bulk flow model \cite{Jinno:2017fby, Konstandin:2017sat}. Note that  the IR-enhancement is stronger for thick-walled bubbles \cite{Cutting:2020nla}, which is the case relevant for nearly-conformal potential leading to strong supercooling, and thus for the PT considered here. (Instead, for thin-walled bubbles, after collision the scalar field can be trapped back in the false vacuum \cite{Konstandin:2011ds, Jinno:2019bxw,Lewicki:2019gmv}. Instead of propagating freely, the shells of energy-momentum tensor remain close to the collision point and dissipate via multiple bounces of the walls.)
Irrespectively of whether the IR slope at $ f \lesssim \beta$ is $\propto k^3$ or $\propto k^1$, at much lower frequency, $f \lesssim H$, the slope must converge to $k^3$ due to causality \cite{Durrer:2003ja,Caprini:2009fx, Cai:2019cdl}.
Oscillations of the condensate following the PT can provide an additional source of GW~\cite{Child:2012qg}. However, instead of $\beta^{-1}$ the time scale is set by the inverse scalar mass $\sim f^{-1}$ and the signal is Planck-suppressed $\propto \beta/f$ \cite{Cutting:2018tjt}.

If instead, $\gwp^\text{max} = \gwp^\text{NLO}$, the vacuum energy is converted into thermal and kinetic energy of the particles in the plasma already prior to the bubble wall collision. The contribution from sound waves or turbulence~\cite{Caprini:2015zlo,Caprini:2019egz}, however, in supercooled transitions is not yet clearly understood. Indeed, current hydrodynamical simulations, which aim to capture the contribution of the bulk motion of the plasma to the gravitational wave signal, do not yet extend into the regime in which the energy density in radiation is subdominant to the vacuum~\cite{Cutting:2019zws}. And analytical studies of shock-waves in the relativistic limit have just started \cite{Jinno:2019jhi}. In any case, we expect supercooled transitions to provide promising avenue for detection in future GW observatories.

\medskip
\noindent
We now proceed to a detailed derivation of Eq.~(\ref{eq:gwp_max}).

\paragraph{Linear growth.}
The energy gained upon formation of a bubble of radius $R$ is
$E_\text{bubble} = \frac{4}{3} \pi R^3 \Delta V_\text{vac}$,
where $\Delta V_\text{vac}$ is the difference between the vacuum energy density outside and inside the bubble.
The energy lost upon formation of a bubble of radius $R$ is $E_\text{wall} \simeq 4 \pi R^2 \gwp \sigma_w$,
where $\sigma_w$ is the surface energy density of the wall (surface tension) in the wall frame.
If a bubble nucleates and expands, its energy $E_\text{bubble}$ is transferred to the wall energy $E_\text{wall}$.
As soon as a nucleated bubble contains the region $\chi \simeq f$, neither $\Delta V_\text{vac}$ nor $\sigma_w$ change upon bubble expansion. Indeed both are a function of the bubble wall profile, which does not change in that regime (also see Fig.~\ref{fig:wall_profile}).
We thus recover the well-known property that $\gwp$ grows linearly in $R$,
\beq
\gwp =\frac{R}{R_0} \sim \Tnuc \,R\,,
\label{eq:gwp_growth}
\eeq
where $R_0$ is a normalisation of the order of the minimal radius needed for a bubble to nucleate, and where in the second relation we have used $R_0 \gtrsim L_w \sim \Tnuc^{-1}$ because we assumed the nucleated bubble to contain the region $\chi \simeq f$. A more precise treatment can be found, e.g.~in the recent~\cite{Ellis:2019oqb}, which confirms the parametric dependence of Eq.~(\ref{eq:gwp_growth}). 

\paragraph{At collision time.}
In a runaway regime, i.e.~for small enough retarding pressure on the bubble walls, $\gwp$ at collisions then reads
\beq
\gwp^\text{runaway}
\sim \Tnuc \, \beta^{-1}
\simeq 1.7 \, \frac{10}{\beta/H}\,\Big(\frac{0.01}{c_\text{vac}}\Big)^{\!\frac{1}{2}}\frac{\Tnuc}{f}\frac{\MPl}{f},
\label{eq:gwp_runaway}
\eeq
where $\beta^{-1}$ is the average radius of bubbles at collision, $H \simeq \Lambda_\text{vac}^2/(\sqrt{3} \MPl)$, and the value $\beta/H \simeq 10$ is a benchmark typical of supercooled phase transitions~\cite{Randall:2006py,Konstandin:2010cd,Konstandin:2011dr,Baldes:2018emh,Bruggisser:2018mrt,Megias:2018sxv,Baldes:2021aph}, which we employ from now on.

The bubbles swallow most of the volume of the universe, and thus most techniquanta, when their radius is of the order of their average radius at collision $\beta^{-1}$. Therefore, in the regime of runaway bubble walls, the relevant $\gwp$ for all the physical processes of our interest (hadron formation from string fragmentation, quark ejection, etc.) will be some order one fraction of $\gwp^\text{runaway}$.
For simplicity, in the runaway regime we will then employ the simplifying relation $\gwp =\gwp^\text{runaway}$.
This will not only be a good-enough approximation for our purposes, but it will also allow to clearly grasp the parametric dependence of our novel findings.
Moreover, a more precise treatment, to be consistent, would need to be accompanied by a more precise solution for $\gwp$ than that of Eq.~(\ref{eq:gwp_growth}), i.e.~we would need to specify the potential driving the supercooled PT and solve for $\gwp$.
As the purpose of this paper is to point out effects which are independent of details of the specific potential, we leave a more precise treatment to future work.

\subsection{LO pressure}
\label{sec:PLO}

\paragraph{Origin.}
By LO pressure we mean the pressure from the partial conversion --- of the quark's momenta before entering the bubbles --- into hadron masses~\cite{Bodeker:2009qy}, plus that from the ejection of quarks.
We use the subscript LO in reference to~\cite{Bodeker:2009qy,Bodeker:2017cim}, because this pressure is of the form $\mathcal{P_{\rm LO}} \sim \Delta m^2 T^2$, where $\Delta m $ is the rest energy of the flux tube between the incoming techni-quanta and the wall.
However, in contrast to \cite{Bodeker:2009qy,Bodeker:2017cim}, here the pressure arises from non-perturbative effects.
 
\paragraph{Momentum transfer.}
The momentum exchanged with the wall, upon hadronization of a single entering quark plus the associated quark ejection, reads in the wall frame
\beq
\Delta p_\LO
= E_\text{in} - \sqrt{E_\text{in}^2 - \Delta m^2_\text{in}} + E_\text{ej}
\simeq f\,,
\label{eq:Deltap_LO}
\eeq
where $E_\text{in} \simeq 3\,\gwp \Tnuc$ is the energy of the incoming quark, $\Delta m^2_\text{in}$ is the fraction of that energy that is converted into `inertia' of the string, and $E_\text{ej} \simeq f/2$ is the energy of the ejected quark or gluon.
In the second equality, we have used $\Delta m^2_\text{in} \simeq \ECM^2 \simeq 3\,\gwp\, \Tnuc \, f$ from Eq.~(\ref{eq:ECMstring}) and $\gwp \gg f/\Tnuc$. Note that $\Delta p_\LO$ is independent of $p_{\rm in}$.

\paragraph{Pressure.}
In light of Sec.~\ref{sec:when_confinement}, we can safely consider a collision-less approach and neglect the interactions between neighboring quarks.
The associated pressure is given by
\beq
\mathcal{P}_\LO
= \sum_a g_a \int\frac{d^3 p_\text{in}}{(2 \pi)^3} \frac{1}{e^{|p_\text{in}|/\Tnuc}\pm 1}\,\Delta p_\LO,
\eeq
where $g_a$ is the number of internal degrees of freedom of a given species $a$ of the techniquanta.
Upon using Eq.~(\ref{eq:Deltap_LO}), we get
\beq
\mathcal{P}_\LO
\simeq  \frac{\zeta(3)}{\pi^2}\,g_\TC \,\gamma_{\rm wp}\,\Tnuc^3 f\,,
\label{eq:LOpressure}
\eeq
where we remind that $g_\TC =g_g + \frac{3g_q}{4}$.
This result can be understood intuitively from $\mathcal{P}_\LO \sim n_{\TC,w} \Delta p_\LO$, where $\gamma_{\rm wp}$ enters through $n_{\TC,w}$~\cite{Bodeker:2009qy}. Note that, in the absence of ejected particles, the pressure would have been a half of our result in Eq.~(\ref{eq:LOpressure}).

\paragraph{Terminal velocity.}
The resulting upper limit on $\gwp$ is obtained by imposing that the LO pressure equals that of the internal pressure from the difference in vacuum energies,
\beq
\mathcal{P}_\text{expand} = c_\text{vac} f^4\,,
\label{eq:pressure_expansion}
\eeq
and reads
\beq
\gwp^\text{LO} =  c_\text{vac} \frac{\pi^2}{\zeta(3)}\frac{1}{g_\TC} \Big(\frac{f}{\Tnuc}\Big)^{\!3}\,.
\label{eq:P_LO}
\eeq
We finally remark that $\mathcal{P}_\LO$ grows linearly in $\gwp$, unlike in `standard' PTs where it is independent of the boost. The reason lies in the fact that the effective mass $\Delta m^2_\text{in}$ grows with $\gwp$, whereas in `standard' PTs it is constant in $\gwp$. Our results then imply that, in confining phase transitions, the LO pressure is in principle enough to ensure the bubble walls do not runaway asymptotically. This is to be contrasted with non-confining PTs, where the asymptotic runaway is only prevented by the NLO pressure.\footnote{In our scenario, bubble walls can still run away until collision for some values of the parameters, and we anticipate they will. Unlike in non-confining PTs, the scaling of our LO pressure with $\gwp$ implies they could not runaway indefinitely if there were no collisions.}

\subsection{NLO pressure}
\label{sec:P_NLO}
\paragraph{Origin.}
The NLO pressure comes from the techniquanta radiating a soft gluon \cite{Bodeker:2017cim} which itself forms a string attached to the wall in the broken phase.  

\paragraph{Result.}
We derive it in detail in Sec.~\ref{sec:NLO_pressure}. We find
\begin{equation}
\mathcal{P}_\NLO
\simeq  \big(g_g C_2[g]+ \frac{3}{4}g_q C_2[q]\big) \frac{8 \zeta(3)}{\pi} \frac{g_\text{conf}^2}{4\pi} \epsilon_\text{ps}\,\frac{\log\big(1+\frac{m_g^2}{k_*^2}\big)}{k_*/m_g}\,\gwp \Tnuc^3 m_g\,,
\label{eq:NLOpressure}
\end{equation}
where $C_2[g,q]$ are the second Casimirs of the representations of gluons and quarks under the confining group (if $SU(N)$, $C_2[g] = N$, $C_2[q] = (N^2-1)/2N$), $g_\text{conf}$ is the gauge coupling of the confining group, $\epsilon_\text{ps} \leq 1$ encodes the suppression from phase-space saturation of the emitted soft quanta $g$, important for large coupling $g_\text{conf}$, $m_g$ is an effective mass of the soft radiated gluons responsible for this pressure, and $k_*$ the IR cut-off on the momentum radiated in the direction parallel to the wall.
\paragraph{Vector boson mass.}
As we model the masses of our techniquanta as the inertia that their fluxtube would gain inside the bubble, these masses increase with increasing momentum of the techniquanta, in the wall frame. The NLO pressure is caused by emission of gluons `soft' with respect to the incoming quanta. Their would-be mass $m_g$ upon entering the wall cannot, therefore, be as large as that of the incoming quanta that emit them, $\Delta m_\text{in} \simeq \sqrt{3\,\gwp\, \Tnuc\, f}$.
At the same time, the effective gluon mass should at least allow for the formation of one hadron inside the wall, therefore we assume it to be of the order of the confinement scale, $m_g \sim f$. The fact that $m_g$ does not grow with $\gwp$ while $\Delta m_\text{in}$ does, is the reason why unlike in non-confining phase transitions, we find here that $\mathcal{P}_\NLO$ and $\mathcal{P}_\LO$ have the same scaling in $\gwp$ and in the amount of supercooling.

\paragraph{NLO pressure is sub-leading.}
By making the standard~\cite{Bodeker:2017cim} choice $k_* \simeq m_g$, and assuming $\epsilon_\text{ps} g_\text{conf}^2 <1 $, we then find that $\mathcal{P}_\NLO \ll \mathcal{P}_\LO$ in the entire parameter space of our interest. Thus, for simplicity, we do not report the NLO limit on $\gwp^{\rm max}$ in Eq.~(\ref{eq:gwp_max}).

Recently, Ref.~\cite{Hoche:2020ysm} performed a resummation of the log-enhanced radiation that leads to the scaling $\mathcal{P}_\NLO \propto g_\text{conf}^2 \gwp^2 \Tnuc^4$. By using the analogue of that result for confining theories, we find that $\mathcal{P}_\NLO$ dominates over $\mathcal{P}_\LO$ in some region of parameter space, and therefore that the values of the parameters for which bubble walls run away slightly change.
Still, even by using that resummed result, we find that the region relevant for DM phenomenology corresponds to the region where bubble walls run away, so that the difference between the results of~\cite{Bodeker:2017cim} and~\cite{Hoche:2020ysm} does not impact the DM abundance. As observed in~\cite{Vanvlasselaer:2020niz}, the pressure as determined in~\cite{Hoche:2020ysm} does not tend to zero when the order parameter of the transition goes to zero, casting a shadow on that result. As pointed out in \cite{Gouttenoire:2021kjv}, this might result from the violation of energy-momentum conservation due to the presence of the wall not being correctly accounted.  Therefore, both for this issue as well as for the limited impact on the DM abundance that we will discuss later, we content ourselves with a treatment analogous to~\cite{Bodeker:2017cim} in our paper.

\paragraph{Summary and runaway condition.}
At small supercooling (i.e. not too small $\Tnuc/f$) the bubble wall velocity reaches an equilibrium value set by the LO pressure.
At larger supercooling bubble walls collide before reaching their terminal LO velocity, and $\gwp$ is set by the runaway value Eq.~\eqref{eq:gwp_runaway}. By comparing Eq.~\eqref{eq:P_LO} with Eq.~\eqref{eq:gwp_runaway}, we find that bubble walls run away for
\beq
\frac{\Tnuc}{f}
\lesssim 1.2 \times 10^{-4}
\left(\frac{80}{g_\TC}\frac{\beta/H}{10}\,\frac{f}{\text{PeV}}\right)^{\!\frac{1}{4}}
\left(\frac{c_\text{vac}}{0.01}\right)^{\!\frac{3}{8}}
\label{eq:run-away_cond}
\eeq
The bubble wall Lorentz factor is plotted in Fig.~\ref{fig:gwp} against the amount of supercooling.

\subsection{Ping-pong regime}
\label{sec:pingpong}
\paragraph{Condition to enter.}
For even a single hadron to form inside the bubble, one needs $\ECM \geq m_\pi$, where $\pi$ is the lightest hadron of the new confining sector (e.g. a pseudo-goldstone boson). Via Eq.~(\ref{eq:ECMstring}), this implies
\beq
\gwp \gtrsim \gwp^\text{enter} = \frac{m_\pi^2}{3\,\Tnuc\,f}\,.
\label{eq:gamma_enter_def}
\eeq
\paragraph{Contribution to the pressure.}
For $\gwp \lesssim \gwp^\text{enter}$, which holds at least in the initial stages of the bubble expansion, the quarks and gluons are reflected and induce a pressure
\beq
\mathcal{P}_\text{refl} \sim n_{\TC,w} \times \Delta p_{\TC,w} \sim \Tnuc^3 \gwp \times \gwp \Tnuc \sim \gwp^2 \Tnuc^4\,.
\label{eq:pressure_reflectedquarks_init}
\eeq
This is to be compared with Eq.~(\ref{eq:pressure_expansion}), $\mathcal{P}_\text{expand} = c_\text{vac} f^4$, which implies the bubble wall could in principle be limited by this pressure to $\gwp \sim (f/\Tnuc)^2$. Nevertheless, as $(f/\Tnuc)^2 \gg \gwp^\text{enter}$, this pressure ceases to exist at an earlier stage of the expansion, namely once $\gwp = \gwp^\text{enter}$. Hence the maximum Lorentz factor remains encapsulated by Eq.~(\ref{eq:gwp_max}).
\paragraph{Ping-pong regime.}
In some extreme regions of parameter space, however, one could have $\gwp^\text{max} < \gwp^\text{enter}$, so that all techniquanta in the plasma are reflected at least once before entering a bubble. We leave a treatment of this `ping-pong' regime to future work.

\section{Amount of supercooling needed for our picture to be relevant}
\label{sec:our_picture_relevant}

\paragraph{Intuition about the limit of no supercooling.}
In the limit of no supercooling, one does not expect the fluxtubes to attach to the bubble wall, but rather to connect the closest charges that form a singlet and induce their confinement.
In other words, in the limit of no supercooling one expects the picture of confinement to be the one of `standard phase transitions'.
By continuity, there should exist a value of $\Tnuc$, smaller than $f$, such that the our picture ceases to be valid, and one instead recovers the more familiar confinement among closest color charges.
We now wish to determine it.
In order to do so, we note that the absence of ejected techniquanta is a necessary condition for the above to hold, therefore we now phrase the problem in terms of absence of ejected techniquanta.

\paragraph{Rate of detachment of $\bigstar$.}
We propose and analyse some effects that could lead to fluxtubes detaching from the bubble walls without ejecting particles.
To take place, these effects need to happen before the end-point of the fluxtube on the wall, $\bigstar$, ceases to exist, i.e.~when the string breaking inside the bubble has already taken place and a quark is ejected.
So we start by computing the rate $\Gamma_{\text{det}\bigstar}$ of detachment of $\bigstar$, the point where the fluxtubes is attached to the wall, from the wall itself. To estimate it, we again borrow the modelling of the classic paper on string fragmentation~\cite{Andersson:1983ia}.

The distances between the several points of breaking of a given string (that connects in our case $\TC_i$ and $\bigstar$) are space-like. In the frame of each point of breaking, that breaking is itself the first to happen, a time of order $N/f$ after the string formation (we adopt the scaling for strong sector gauge groups $SU(N)$~\cite{Casher:1978wy,Armoni:2003nz}).
This time therefore also applies to the outermost breaking point in our picture, i.e. that closest to the wall, whose frame approximately coincides with the wall frame. We remind the reader that the outermost breaking is the one that nucleates the quark or gluon that is eventually ejected. The rate we need can therefore be estimated as the inverse of the nucleation time of the outermost pair,
\beq
\Gamma_{\text{det}\bigstar\text{,w}} = \tau_{\text{det}\bigstar\text{,w}}^{-1} \simeq f/N\,.
\label{eq:rate_detachment}
\eeq

We now enumerate and model effects that could lead to fluxtubes detaching from the bubble walls without ejecting techniquanta, and compare their time scales with Eq.~(\ref{eq:rate_detachment}).
\begin{enumerate}
\item
\textbf{Flux lines overlap.} 
The faster a bubble-wall, the denser and thus the closer together in the wall frame are the quarks and gluons entering it. Eventually, they could get closer than the typical transverse size of a fluxtube $d_\text{tr} \simeq f^{-1}$~\cite{Bicudo:2017uyy}. When that happens, the fluxtubes between different color charges have a non-negligible overlap. We expect that in this situation it will not be clearly preferable energetically for these strings to attach directly to the wall. Thus there would be no ejected techniquanta.
This situation is of course realised also in the case of small supercooling $f/\Tnuc$, in addition to and independently of the case of fast bubble-walls.

We then obtain a rate of `string breaking by fluxtube overlap', $\Gamma_\text{overlap}$, as follows.
We define an effective associated cross section as the area of a circle on the wall, centered on any $\bigstar$ and with radius $d_\text{tr}$,
\beq
A_\text{overlap} = \pi d_\text{tr}^2 \simeq \pi f^{-2}\,.
\eeq
The associated rate then reads
\beq
\Gamma_\text{overlap} = A_\text{overlap} v \,n_{\TC,\text{w}} 
\simeq \frac{\gwp \zeta(3) g_\TC}{\pi} \frac{\Tnuc^3}{f^{2}}  \,,
\eeq
where $n_{\TC,\text{w}} = \gwp n_{\TC,\text{p}}$ is the density of techniquanta in the wall frame, $g_\TC = g_g + 3g_{q}/4$, and we have used that they are relativistic $v = 1$. The condition of no ejected techniquanta then reads
\beq
\Gamma_\text{overlap} > \Gamma_{\text{det}\bigstar\text{,w}}
\Rightarrow
\gwp \gtrsim \frac{2.6}{g_\TC N}\Big(\frac{f}{\Tnuc}\Big)^3\,.
\label{eq:noejquarks_detach}
\eeq

\item
The entire fluxtube connecting real color charges, so including its portion in the region $\chi \simeq \chi^* \ll f$ (see Fig.~\ref{fig:wall_diagram}), could enter the region $\chi =f$ before its portions in the region $\chi = f$ break and form hadrons, and eject particles. We see two ways this could happen.
\begin{enumerate}
\item[2.1] \textbf{Attractive interaction between neighboring flux lines.}  
The points $\bigstar$ are not static, because they move by the force exerted by the part of the string which is outside the wall, in the layer where $\langle \chi \rangle \simeq \chi^*$. Defining $y_\bigstar$ as the transverse distance, on the wall, between two $\bigstar$ points connected by a fluxtube, one has
\beq
\frac{d^2 y_\bigstar}{dt^2}
= \frac{F}{m_\bigstar}
\sim -\frac{d E_{q\bar{q}}/dy}{f}
\simeq -c_{q\bar{q}}\frac{{\chi^*}^2}{f}
\sim -c_{q\bar{q}} \frac{\Tnuc^2}{f}\,,
\eeq
where, consistently with our previous treatments, we have assigned to $\bigstar$ an inertia $m_{\bigstar} \sim f$.
If $y_\bigstar$ goes to zero in a time shorter than the breaking time $\tau_{\text{det}\bigstar\text{,w}} \sim N f^{-1}$, then the two fluxtubes connect and become fully contained in the region $\chi = f$ before they break and form hadrons, and thus there are no ejected techniquanta.
To determine this condition, we assume initially static points $\bigstar$, and thus we only need the initial distance between them $y_\bigstar (t=0) \simeq (\gwp n_{\TC,\text{p}})^{-1/3}$. We then obtain
\beq
y_\bigstar(t=\tau_{\text{det}\bigstar,\text{w}})
\simeq (\gwp n_{\TC,\text{p}})^{-1/3} - c_{q\bar{q}}\frac{\Tnuc^2}{f} \frac{\tau_{\text{det}\bigstar\text{,w}}^2}{2} \,.
\eeq
The resulting condition for no ejected quarks reads 
\beq
y_\bigstar(t=\tau_{\text{det}\bigstar\text{,w}}) < 0
\Rightarrow
\gwp \gtrsim \frac{ 6.6 \times 10^{-2}}{g_\TC  N^{6} }  \left( \frac{ 10 }{ c_{q\bar{q}} } \right)^{3}  \Big(\frac{f}{\Tnuc}\Big)^9\,.
\label{eq:noejquarks_classical}
\eeq

\item[2.2]
\textbf{Limit of no distortion of the flux lines.}  
When the string portion in the region $\chi \simeq \chi^* \ll f$ has a small enough length $d_\bigstar$, the possibility that it is pulled inside the region
$\chi = f$ could be energetically more convenient than the one of our picture, where it stays outside and instead energy goes in increasing the length of the strings that are perpendicular to the wall.
The energy price, for the string portion in the region $\chi \simeq \chi^* \ll f$ to enter the region $\chi = f$, reads in the wall frame
\beq
\Delta E_\text{pull-in,w}
\simeq c_{q\bar{q}} (f^2 - {\chi^*}^2)d_\bigstar
\simeq c_{q\bar{q}} f^2 d_\bigstar\,,
\label{eq:price_pulledin}
\eeq
where we stress that the length of the string portion $d_\bigstar$ is transverse to the bubble-wall velocity and therefore is not Lorentz contracted in the process of being pulled into the bubble. In the wall frame, it reads $d_\bigstar \simeq (\gwp n_{\TC,\text{p}})^{-1/3}$
The transition between $\chi \simeq \chi^* \ll f$ and $\chi = f$ is exponentially fast in the proper coordinate $s$ (see App.~\ref{app:wall_profile}), and happens over an interval (a distance, in the wall frame) $L_\text{f} \sim f^{-1}$.
The energy price of Eq.~(\ref{eq:price_pulledin}) should therefore be compared with the one to stretch two strings, inside the wall, by an amount $L_\text{f}$:
\beq
\Delta E_\text{stretch,w}
\simeq 2 c_{q\bar{q}} f^2 L_\text{f}/\gwc
\sim 2 c_{q\bar{q}} f \Big(\frac{f}{3\,\gwp \Tnuc}\Big)^{1/2},
\label{eq:Lf_introduce}
\eeq
where we have used that the string length in the expression for $E_{q\bar{q}}$, Eq.~(\ref{eq:stringenergy}), has to be evaluated in the string center-of-mass frame, and that $\gwc \simeq \sqrt{3\,\gwp \Tnuc/f}$ from Eq.~(\ref{eq:gwc}).
Therefore, it is energetically more convenient to pull the fluxtube inside the region $\chi \simeq f$, and so to have no ejected quarks, if
\beq
\Delta E_\text{pull-in} < \Delta E_\text{stretch}
\Rightarrow
\gwp \lesssim 0.035 \, g_\TC^2 \, \Big(f\,L_f\Big)^{\!6}\, \Big(\frac{\Tnuc}{f}\Big)^{\!3}\,.
\label{eq:noejquarks_pulled}
\eeq
Contrary to the previous two possibilities to have no ejected quarks, Eqs.~(\ref{eq:noejquarks_detach}) and (\ref{eq:noejquarks_classical}), the possibility in Eq.~(\ref{eq:noejquarks_pulled}) imposes an upper limit on $\gwp$. We anticipate that, in the regimes of supercooling interesting for our work $\Tnuc/f \ll 1$, Eq.~(\ref{eq:noejquarks_pulled}) cannot be satisfied consistently with $\gwp > 1$, so that it is not relevant for our work. 
\end{enumerate}

\end{enumerate}
%

\begin{figure}[t]
\begin{center}
\begin{adjustbox}{max width=1\linewidth,center}
\includegraphics[width=.5\textwidth]{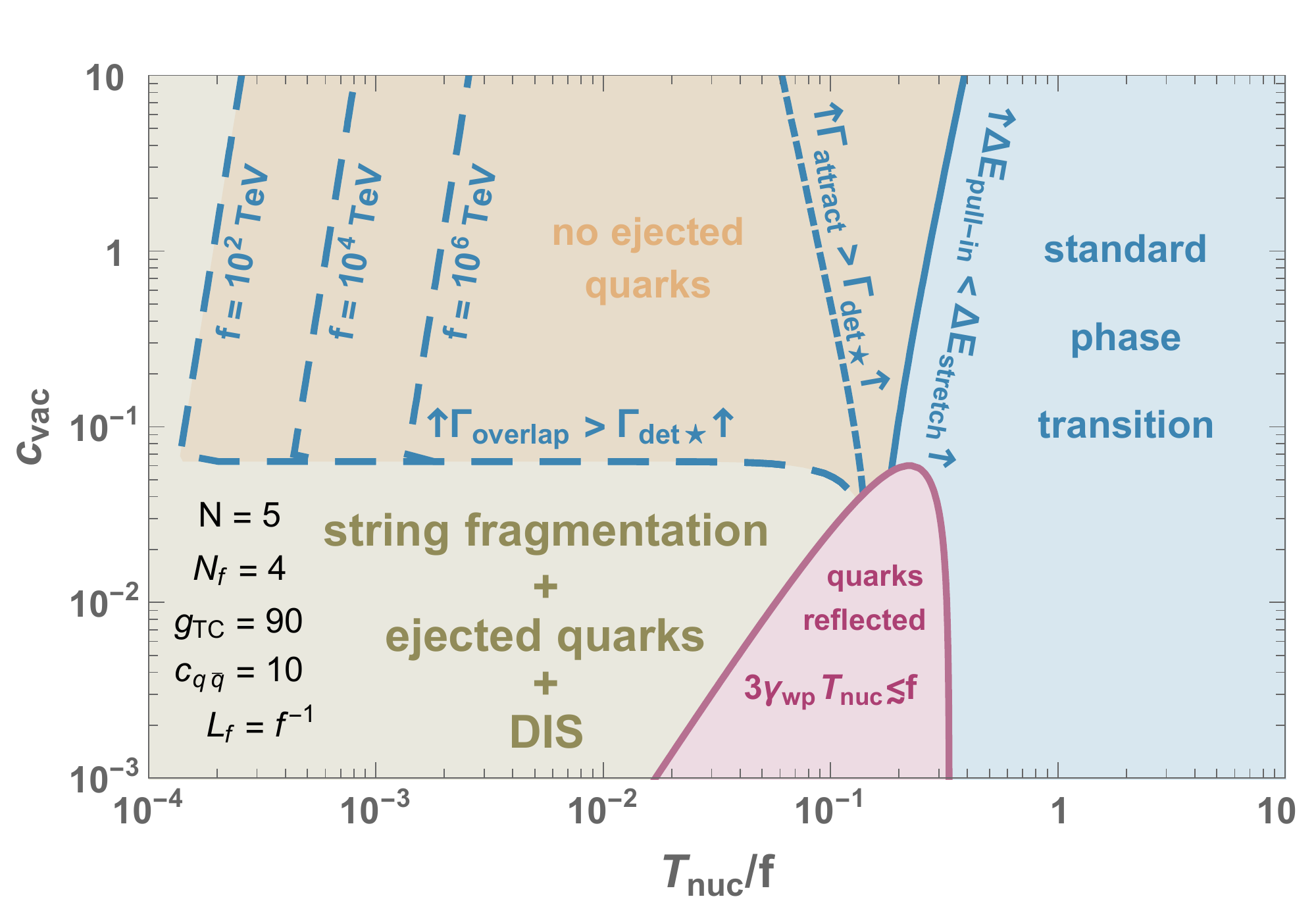}
\includegraphics[width=.5\textwidth]{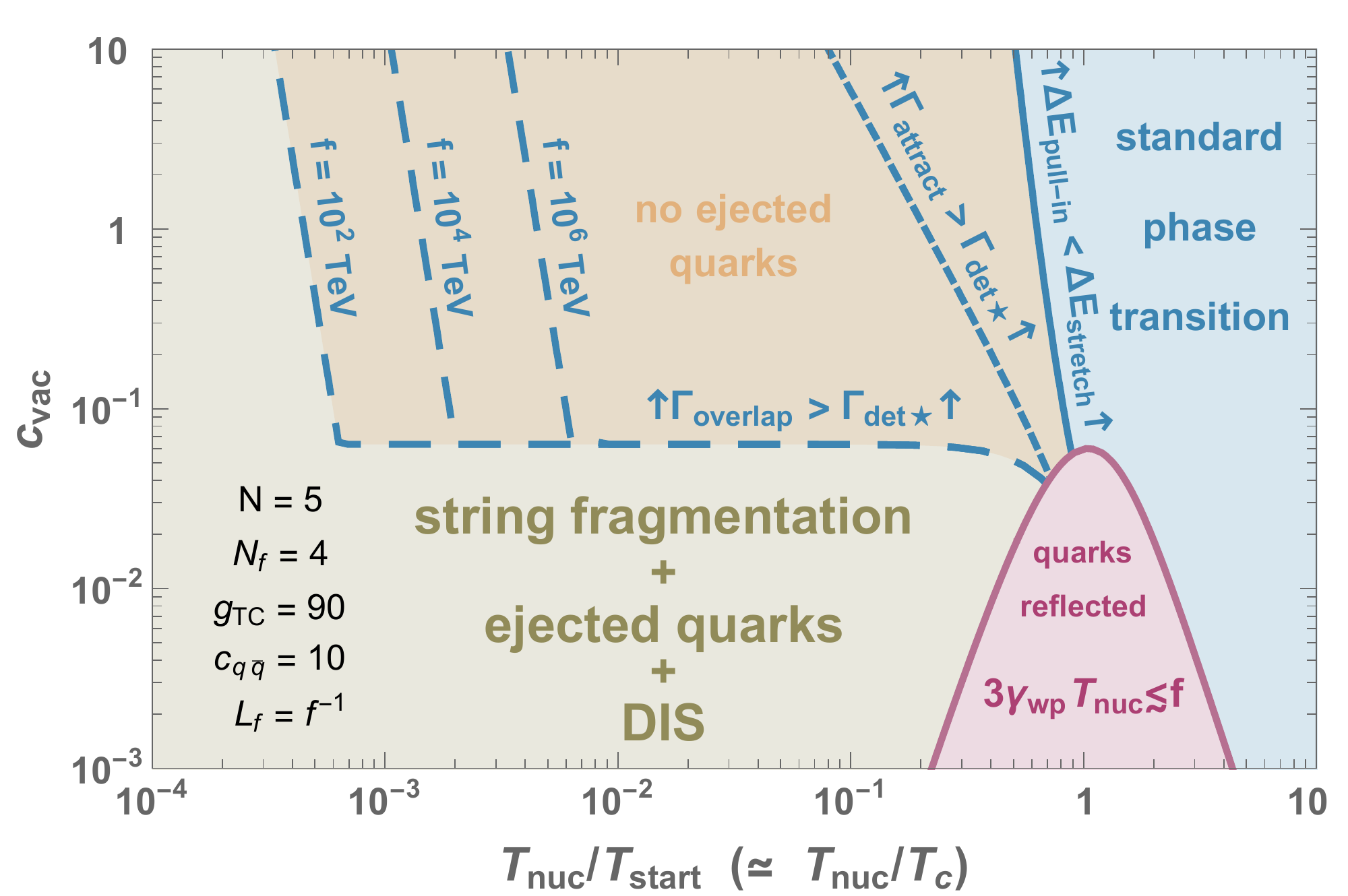}
\end{adjustbox}
\caption{
\label{fig:StdPictureRecovered} 
\it \small  \newline
\textbf{Blue Region}: The incoming techniquanta confine with their neighbours as in the standard picture of phase transitions that are not supercooled. \newline
\textbf{Olive Region}: All three inequalities, \eqref{eq:noejquarks_detach}, \eqref{eq:noejquarks_classical}, \eqref{eq:noejquarks_pulled}, are violated and the new effects pointed out in this study, i.e. string fragmentation, ejection of techniquanta and deep inelastic scattering, should be taken into account. 
\newline 
\textbf{Orange Region}: At least one but not all of the inequalities above hold, therefore there are no ejected techniquanta.
The dynamics taking place in this region remains to be investigated.
\newline
\textbf{Purple Region}: Quarks are too weakly energetic to enter the bubbles, see.~\ref{sec:pingpong}. \newline 
\textbf{Left of solid line}: Eq.~\eqref{eq:noejquarks_pulled}
is violated and it is energetically favourable for the flux lines to be distorted.  \newline 
\textbf{Left of dotted line}: Eq.~\eqref{eq:noejquarks_classical} is violated we can neglect the attractive interactions between neighboring flux lines.  \newline
\textbf{Left of dashed lines}: Eq.~\eqref{eq:noejquarks_detach} is violated and we can neglect the overlap of neighbouring flux lines.  \newline
The two plots only differ through their horizontal axis, see Sec.~\ref{sec:thermal_history} for the definitions of $c_{\rm vac}$ and $T_{\rm start}$, and App.~\ref{app:wall_profile} for that of $T_c$. 
To avoid the unphysical values $\gwp < 1$, we have added 1 to Eq.~\eqref{eq:gwp_max}.
}
\end{center}
\end{figure}

\paragraph{\bf Summary of required supercooling.}

In the regime where $\Tnuc \gtrsim f$, we expect that neither ejection of techniquanta nor string fragmentation should take place, and that the standard picture of quarks and gluons confining with their neighbors should be recovered (which we dub the `standard phase transition'). More precisely, if any of Eqs.~(\ref{eq:noejquarks_detach}), (\ref{eq:noejquarks_classical}) and (\ref{eq:noejquarks_pulled}) hold, we depart from our picture in at least one regard.
By demanding none of these inequalities hold, we expect the new effects of our study, namely flux line attached to the wall, string fragmentation, quark ejection and deep inelastic scattering, to take place. In the non-runaway regime, we require
		\begin{align}
	c_{\rm vac} & \lesssim  \frac{0.32}{N} \quad \text{and} \\
	 \frac{\Tnuc}{f} & \lesssim  \text{Min}\Big[  0.19 \left(\frac{5}{N}\right) \left( \frac{0.01}{c_{\rm vac}} \right)^{1/6} \left( \frac{10}{c_{q\bar{q}}} \right)^{1/2}, \,  \frac{ 0.12 }{ fL_{f} }  \left( \frac{c_{\rm vac}}{0.01} \right)^{1/6} \left( \frac{90}{g_{\rm TC}} \right)^{1/2}  \Big], \nonumber
	\end{align}
for our picture to hold. In the runaway regime, we instead require
\begin{align}  
 	\frac{\Tnuc}{f} &  \lesssim \nonumber \\
	\text{Min}\Big[ & 6.1 \times 10^{-5} \left( \frac{\beta/H}{10}\right)^{1/4}  \left( \frac{c_{\rm vac}}{0.01} \right)^{1/8} \left( \frac{90}{g_\TC} \right)^{1/4} \left( \frac{5}{N} \right)^{1/4} \left( \frac{ f }{ 10 \; \mathrm{TeV} } \right)^{1/4},   \nonumber \\
	& 6.4\times 10^{-3} \left( \frac{\beta/H}{10}\right)^{1/10} \left( \frac{c_{\rm vac}}{0.01} \right)^{1/20} \left( \frac{10}{c_{q\bar{q}}} \right)^{3/10} \left( \frac{90}{g_\TC} \right)^{1/10} \left( \frac{5}{N} \right)^{3/5} \left( \frac{ f }{ 10 \; \mathrm{TeV} } \right)^{1/10}   , \nonumber \\
	& \frac{ 1.2 \times 10^6 }{ (fL_{f})^{3} } \left( \frac{10}{\beta/H} \right)^{1/2}\left( \frac{0.01}{c_{\rm vac}} \right)^{1/4} \left( \frac{90}{g_\TC} \right) \left( \frac{ 10 \; \mathrm{TeV} }{ f } \right)^{1/2}     \Big].
\end{align}
for our picture to hold. Here we have used $\gwp$ in Eq.~\eqref{eq:gwp_max}. The conditions are visually summarised in Fig.~\ref{fig:StdPictureRecovered}.

 In light of this figure, we conclude that some new effects pointed out in our study are also relevant in confining phase transitions where $\Tnuc \sim T_{\rm start} \sim T_{c}$ (see App.~\ref{app:wall_profile} for the definition of the critical temperature $T_c$), e.g.
\cite{Schwaller:2015tja, Aoki:2017aws,Aoki:2019mlt, Bigazzi:2020avc,Ares:2020lbt,Huang:2020mso, Halverson:2020xpg}, provided $c_{\rm vac}$ is small enough.
A possible impact on the QCD phase transition, e.g. \cite{Buballa:2003qv,Fukushima:2010bq,Fukushima:2013rx,Herzog:2006ra,Schwarz:2009ii,Schettler:2010dp,Alho:2013hsa,Ahmadvand:2017xrw,Ahmadvand:2017tue,Chen:2017cyc}, remains to be investigated.

\paragraph{Averaged quantities only. }
We conclude this section by also stressing that all the conditions above refer to averaged quantities, and therefore do not take into account the leaks from tails of distributions. These leaks could for example imply that there are a few strings that hadronise without ejecting particles, even if all conditions Eqs.~(\ref{eq:noejquarks_detach}), (\ref{eq:noejquarks_classical}) and (\ref{eq:noejquarks_pulled}) are violated.
As these strings constitute a small minority of the total ones, these effects have a negligible impact on the phenomenology we discuss.
They could however be important in studying other situations of supercooled confinement.
Though certainly interesting, the exploration of these effects goes beyond the scope of this paper.

\section{Ejected quarks and gluons}
\label{sec:outside_bubble}

\subsection{Density of ejected techniquanta}
\label{sec:nej}

In the wall frame, since we have one ejected quark or gluon per each incoming one, we find
\beq
n_\text{ej,w} = n_{\TC,\text{w}}(r_\text{ej}) = \gwp(r_\text{ej}) n_{\TC,\text{p}}\,,
\eeq
where $n_{\TC,\text{p}} = g_\TC \zeta(3) \Tnuc^3/\pi^2$ is the density of the diluted bath in the plasma frame. The density of ejected techniquanta then depends on the time passed since bubble wall nucleation, or equivalently on the bubble radius at the time of ejection $r_\text{ej}$, via $\gwp(r)$ (see Sec.~\ref{sec:wall_speed}). 
In the plasma frame, and at a given distance $D$ from the center of the bubble, we then have\footnote{The factor $2$ arises when we boost the quark current $(\gwp\, n_{\TC,\text{p}} , \,  \gwp\, \vec{\beta} \,n_{\TC,\text{p}})$, with $\vec{\beta} = \vec{e}_r$, from the wall to the plasma frame.}
\beq
n_\text{ej,p}(D)
= 2\gwp^2(r_\text{ej}) \Big(\frac{r_\text{ej}}{D}\Big)^{\!2} n_{\TC,\text{p}}\,,
\label{eq:nejp}
\eeq
where we have included the surface dilution from the expansion between the radius at which a given quark has been ejected, $r_\text{ej}$, and the radius $D$ where we are evaluating $n_\text{ej,p}$. 
\paragraph{Radial dependence. }
It is convenient to express $n_\text{ej,p}$ as a function of the radial distance $x$ from the bubble wall in the plasma frame, where for definiteness $x= 0$ denotes the position of the wall and $x = L_\text{ej,p}$ the position of the techniquanta ejected first (which constitute the outermost layer). In order to do so, we determine the relation between the position $x$ of a quark and the radius $r_\text{ej}(x)$ when it has been ejected.
We assume that the bare mass of the quarks is small enough such that they move at the speed of light, like the gluons.
The wall at $x = 0$, instead, moves at a speed $v_\text{wall} \simeq 1-1/(2\gwp^2)$ (we have used the relativistic limit $\gwp \gg 1$), dependent on its radius. 
The coordinate $x$ of a given layer of ejected particles can then be found by integrating the difference between the world line of an ejected particle and that of the wall,
\beq
x
= \int_{t_\text{ej}}^{t_\mathsmaller{D}} \!\!\! dt(1-v_\text{wall}) 
\simeq \int_{t_\text{ej}}^{t_\mathsmaller{D}} \frac{1}{2 \gwp^2(t)}
\simeq \frac{1}{2 \Tnuc^2} \Big(\frac{1}{t_\text{ej}} - \frac{1}{t_\mathsmaller{D}}\Big)\,,
\label{eq:rej}
\eeq 
where we defined $t_\mathsmaller{D}$ and $t_\text{ej}$ as the times when the bubble radius is respectively $D$ and $r_\text{ej}$, and we used $\gwp(t) \simeq \Tnuc t$, cf. Eq~(\ref{eq:gwp_growth}), valid up to relative orders $1/\gwp^2 \ll 1$. 
It is convenient to rewrite Eq.~(\ref{eq:rej}) as
\beq
r_\text{ej} (x) \simeq \frac{D }{1 + 2\,\Tnuc^2\,D\, x}\,.
\label{eq:gwp_x_runaway}
\eeq
We finally obtain
\beq
n_\text{ej,p}(x)
= \frac{2\,\gwp^2 (x)}{\big(1+2\,\Tnuc^2 \,D \,x\big)^{\!2}} n_{\TC,\text{p}}
\simeq  \frac{2\,\Tnuc^2\,D^2}{ \big(1+2\,\Tnuc^2 \,D \,x\big)^{\!2} } n_{\TC,\text{p}}\,,
\label{eq:nejpx}
\eeq
where the last equality is valid as long as the bubbles run away, i.e. as long as Eq.~(\ref{eq:gwp_growth}) $\gwp \simeq \Tnuc \,r$ holds.

\begin{figure}[t]
\begin{center}
\includegraphics[width=.65\textwidth]{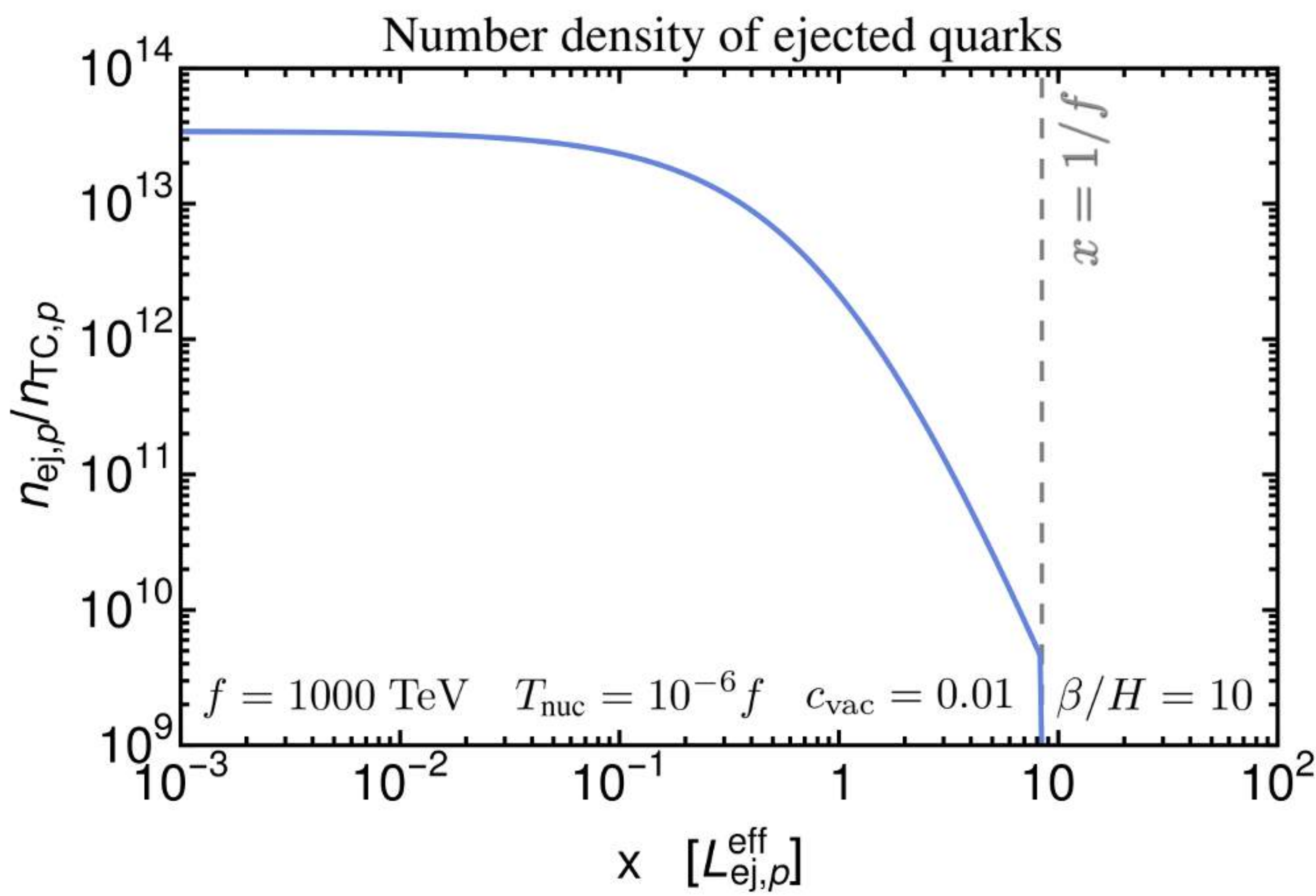}
\caption{
\label{fig:plasmaprofile} 
\it \small
The density of the ejected quarks in front of the bubble wall as a function of the distance $x$ in front of the bubble wall, Eq.~(\ref{eq:nejpx}), for an example parameter point. Here we have used the relation $\gwp \simeq D\Tnuc$. The distance to the outermost techniquanta $L_\text{ej,p} \approx 1/f$ is also shown. 
}
\end{center}
\end{figure}

\paragraph{Thickness of the layer of ejected techniquanta. }
Our result Eq.~\eqref{eq:nejpx} implies that the highest density, of ejected techniquanta, is located in the shell within a distance of the bubble wall 
\beq
L_\text{ej,p}^{\rm eff} \simeq \frac{1}{2\, \Tnuc^2\,D}\,.
\label{eq:Lej_eff}
\eeq
The density of ejected quarks $n_\text{ej,p}(x)$ extends to $x = L_\text{ej,p}$, i.e. to the outermost ejected layer, that we now show to be much larger than $L_\text{ej,p}^{\rm eff}$.
Indeed, $L_\text{ej,p}$ can be related to the time $t_\text{first}$ of ejection of the first techniquanta (corresponding to $\gwp \simeq m_\pi/\Tnuc$, Eq.~\eqref{eq:gamma_enter_def}).  Using $t_i \gg t_\text{first}$ and $t_\text{first} \sim m_\pi/\Tnuc^2$, we find
\beq
L_\text{ej,p} \simeq \frac{1}{t_{\rm first}\,2\,\Tnuc^2} \sim \frac{1}{f}\,,
\label{eq:Lej_velocity}
\eeq
where for simplicity we have assumed $m_\pi \simeq f$ as in QCD.
As long as $L_\text{ej,p} \gg L_\text{ej,p}^{\rm eff}$, as it holds for our estimate Eq.~\eqref{eq:Lej_velocity}, the value of $L_\text{ej,p}$ does not affect any of the results of this paper.
\footnote{One could easily envisage situations in which $m_\pi$ differs sizeably from $f$, e.g. because pions are much lighter or because of a possible dependence of the mass of the lightest resonances on the number of colours $N$. The exploration of if and how this possibility would affect our results (for example the conclusion that $L_\text{ej,p} \gg L_\text{ej,p}^{\rm eff}$), while certainly interesting, goes beyond the purposes of this paper.
}
The density profile of Eq.~(\ref{eq:nejpx}) is shown in Fig.~\ref{fig:plasmaprofile}.

\paragraph{Sanity check. }
As a check of our result Eq.~\eqref{eq:nejpx}, we verify that one has one ejected quark or gluon per each one that entered the bubble. Indeed, we compute
\beq
4 \pi D^2
\int_{0}^{L_\text{ej,p}}\!\!\!dx \,n_\text{ej,p}(x)
= \frac{4}{3} \pi D^3 n_{\TC,\text{p}}\,,
\label{eq:cons_number}
\eeq
where we have assumed $D \gg f/\Tnuc^2$, i.e. we have placed ourselves deep in the regime where hadrons can form inside bubbles (see Eq.~(\ref{eq:gamma_enter_def})).
Equation~(\ref{eq:cons_number}) guarantees that the number of ejected techniquanta in the layer of thickness $L_\text{ej,p}$ is equal to the total number of techniquanta that entered the bubble up to radius $D$.

\medskip
\paragraph{Interactions between ejected quarks. }
Let us finally comment why, we think, interactions among the ejected techniquanta cannot much alter their density.
The density of the particles in the incoming bath does not change out of their own interactions.
In the wall frame, both the density and the relative momentum of the ejected techniquanta are of the same order of those of the particles in the incoming bath. Therefore, we analogously expect that the density of the ejected techniquanta would also not change after ejection.
Since what will matter for the following treatment is the energy in the ejected techniquanta, rather than how this energy is spread among the various degrees of freedom, we content ourselves with this qualitative understanding and leave a more precise treatment to future work.

\subsection{Scatterings of ejected quarks and gluons before reaching other bubbles}
\label{sec:energy_transfer}

Before possibly reaching other expanding bubble-walls and their ejected techniquanta, ejected quarks and gluons could undergo scatterings with particles from the supercooled bath at temperature $\Tnuc$, and with techniquanta ejected from other bubbles. In this section we study the effects of these scatterings.

\paragraph{Ejected techniquanta are energetic. }
As soon as a bubble occupies an order one fraction of its volume at collision, the total energy in ejected particles is much larger than that in the supercooled bath outside the bubble.
Indeed, we have seen that for each quark or gluon in the supercooled bath that enters a bubble, there is at least an ejected one, and that the energy ejected per each incoming particle is much larger than the energy per each particle in the bath, $E_\text{ej,p}\simeq \gwp f \gg \Tnuc$, Eq.~(\ref{eq:energy_ejected}).
Assuming the degrees of freedom in quarks and gluons are not an extremely small fraction of those in the diluted medium, then the diluted medium outside the bubbles does not have enough energy to act as a bath for the ejected particles.
This implies that most ejected particles keep most of their energy upon passing through the supercooled bath.

\paragraph{Energy transfer between ejected techniquanta and diluted bath. }
By reversing the logic above, the ejected particles can deposit in the supercooled bath an energy much larger than its initial one.
Pushing this to the extreme, the ejected techniquanta could make the bath move away from the bubble wall, thus making our treatment so far valid only in the first stages of bubble expansion.
In order to assess this, we estimate
the rate of transferred energy between ejected techniquanta and particles from the bath outside the bubbles,
\beq
\Gamma_\text{ej-bath}
=  n_\text{ej} \int_{\Delta E_\text{min}}^{\Delta E_\text{max}} \!\! \!\!\!d \Delta E\,\frac{d\sigma v}{d \Delta E}\, \Delta E
\simeq n_\text{ej} \int_{-s}^{t_\IR} \!\! dt\frac{d\sigma}{d t}\, \sqrt{-t}\,,
\eeq
where $n_\text{ej}$ is the density of ejected techniquanta, $\Delta E$ is the energy transferred per single scattering, and where in the second equality we have taken the limit of relativistic particles $v \simeq 1$ and small energy transfer per single scattering $\Delta E$, so that the Mandaelstam variable $t$ can be expressed as $t \simeq -\Delta E^2$.
The quantity $d\sigma/dt$ depends on the specific model under consideration, in particular it depends both on whether the ejected particle is a quark or a gluon, and on the identity of the scatterer in the bath outside the bubbles. For definiteness, we model it as the cross section for fermion-fermion scattering mediated by a light vector with some effective coupling $\sqrt{4 \pi \alpha_\text{eff}}$,
\beq
\frac{d\sigma}{dt} = \frac{4\pi \alpha_\text{eff}^2}{t^2}\,.
\label{eq:dsigmadt}
\eeq
We then obtain
\beq
\Gamma_\text{ej-bath}
=  n_\text{ej} \frac{8 \pi\alpha_\text{eff}^2}{\sqrt{-t_\IR}}.
\label{eq:transfer_energy_rate}
\eeq
$\Gamma_\text{ej-bath}$ is of course not Lorentz invariant, it depends on the frame via the density of ejected techniquanta $n_\text{ej}$ determined in Sec.~\ref{sec:nej}.

\paragraph{Impact on diluted bath. }
The average energy transferred to a particle in the diluted bath at position $D$, when this particle goes across the layer of ejected techniquanta (so before it reaches the wall and initiates the processes described in Sec.~\ref{sec:inside_bubble}), then reads
\beq
Q_\text{ej-bath} \equiv
\int_{0}^{L_\text{ej,p}}\!\!\!dx\, \Gamma_\text{ej-bath,p} (x)\,,
\label{eq:Qeji}
\eeq
where we remind that the spatial coordinate $x$ is the distance between a given layer of ejected techniquanta and the wall at $x=0$. 
Upon use of Eqs.~(\ref{eq:transfer_energy_rate}) and \eqref{eq:nejpx}, we can then evaluate the average energy transferred to an incoming particle from the diluted bath, Eq.~(\ref{eq:Qeji}), as
\beq
Q_\text{ej-bath}
\simeq \frac{8 \pi\alpha_\text{eff}^2}{3}\frac{D \,n_{\TC,\text{p}}}{\sqrt{-t_\IR}}\,.
\label{eq:Qejbath}
\eeq
Note that the product $D \, n_{\TC,\text{p}}$ is Lorentz-invariant, so that $Q_\text{ej-bath}$ is indeed a Lorentz-invariant quantity.
To learn whether particles from the diluted bath are prevented from entering the wall, because of the interaction with the ejected techniquanta, we compare the energy they exchange with them upon passing their layer with their initial energy in the wall frame\footnote{Had we chosen another frame, we would have had to include the wall velocity in the condition.},
$E_{i,w}\simeq 3\,\gwp \Tnuc$,
\beq
\frac{Q_\text{ej-bath}}{E_{i,w}}
\simeq  \frac{8 \zeta(3)}{9\pi} \alpha_\text{eff}^2 \,g_\TC \frac{\Tnuc D}{\gwp} \frac{\Tnuc}{\sqrt{-t_\IR}}\,. 
\label{eq:QoverEi}
\eeq
The novel physical picture we described in Secs.~\ref{sec:inside_bubble} and~\ref{sec:outside_bubble} is valid as long as $Q_\text{ej-bath}/E_{i,w} \ll 1$.
As seen in Sec.~\ref{sec:wall_speed}, $\gwp$ initially grows linearly with the bubble radius, $\gwp \simeq \Tnuc\,D$, until the retarding pressure possibly becomes effective. It will turn out in Sec.~\ref{sec:DMabundance} that the runaway regime of linear growth is the one relevant for the phenomenology we will discuss. In that regime, the condition $Q_\text{ej-bath}/E_{i,w} \ll 1$ translates into
$\Tnuc/\sqrt{-t_\IR}~\ll~1$.
\paragraph{IR cut-off. }
The quantity $-t_\IR$ is the IR cutoff of the scattering, $-t_\IR \equiv m_V^2$, with $m_V$ some effective mass of the mediator responsible for the interactions that exchange momentum. 
In the absence of mass scales, which is the case for example for the SM photon and for the gluons, the effective mass $m_V$ is equal to the plasma mass of these particles in the thermal bath.
If the only bath was the diluted one, one would have $m_{V,\text{therm}}^2 \sim \alpha_\text{eff}\, n_{\TC,\text{p}}/\langle E_{\TC,\text{p}}\rangle \sim \Tnuc^2$ (see e.g.~\cite{Kapusta:2006pm}).
However, the process of our interest here happens in the much denser bath of ejected techniquanta, $n_\text{ej,p} \gg n_{\TC,\text{p}}$, so that we indeed expect $m_{V,\text{therm}}^2 \gg \Tnuc^2$, so that $Q_\text{ej-bath}/E_{i,w} \ll 1$ and our picture so far is valid.
More precisely, the screening mass for non-equilibrium systems scales as~\cite{Arnold:2002zm} ($f(p)$ is the non-equilibrium phase space distribution of the particles in the system)
\beq
m_{V,\text{therm}}^2
\simeq g_\TC \alpha_\text{eff} \int \frac{f(p)}{|p|}
\sim  \frac{n_\text{ej,p}}{\langle E_\text{ej,p}\rangle}
\sim \frac{\gwp}{f/\Tnuc} \Tnuc^2 \gg \Tnuc^2\,,
\label{eq:mV}
\eeq
where we have used $\langle E_\text{ej,p} \rangle\sim \gwp f$ and $n_\text{ej,p}  \sim  (D/L_\text{ej,p}) n_{\TC,\text{p}} \sim D^2 \Tnuc^5 \sim \gwp(D)^2 \Tnuc^3$.
Equations~\eqref{eq:QoverEi} and \eqref{eq:mV} teach us that, in the regions of parameter space where $\gwp \gg f/\Tnuc$, the energy received by each particle in the diluted bath, from scatterings with the ejected techniquanta, is much smaller than their energy in the wall frame $E_{i,w}\simeq 3\,\gwp \Tnuc$.\footnote{One could be worried that in the outer shell of size $L_{\rm ej,p} \sim  1/f$, the thermal mass $m_{V,\text{therm}}$ is much smaller than its value in the densest region in $x\simeq 0$, explicited in Eq.~\eqref{eq:mV}, such that $ \frac{Q_\text{ej-bath}}{E_{i,w}}$ becomes larger than $1$.
We can check that it is not the case by including the $x$-dependence of $m_{V,\text{therm}}$, Eq.~\eqref{eq:mV}, in the integral in Eq.~\eqref{eq:Qeji} 
\begin{equation}
\label{eq:fullint}
Q_\text{ej-bath, p} =
\int_{0}^{L_\text{ej,p}}\!\!\!dx\,\, n_\text{ej, p} \frac{8 \pi\alpha_\text{eff}^2}{\sqrt{n_\text{ej,p}/\langle E_\text{ej,p}\rangle}}\,,
\end{equation} 
where $n_\text{ej,p}$ is defined in Eq.~\eqref{eq:nejpx}, $L_\text{ej,p} \sim 1/f$ in Eq.~\eqref{eq:Lej_velocity}, $D \sim \gwp / \Tnuc$ in Eq.~\eqref{eq:gwp_growth}, and $E_\text{ej,p} \sim \langle E_\text{ej,p} \rangle\sim \gwp f$. 
 We compute the integral in Eq.~(\ref{eq:fullint}) and obtain
\begin{equation}
\frac{Q_\text{ej-bath}}{E_{i,w}} \simeq 
 \left\{
                \begin{array}{ll}
                 \frac{8\sqrt{2 \zeta(3)}}{9} \alpha_\text{eff}^2\, \sqrt{g_\TC \frac{f/\Tnuc}{\gwp}} \ll 1, \quad \text{if}~ \gwp \gtrsim f/\Tnuc \\
                  \frac{8\sqrt{2 \zeta(3)}}{3} \alpha_\text{eff}^2\, \sqrt{g_\TC \frac{\gwp}{f/\Tnuc}} \ll 1, \quad \text{if}~ \gwp \lesssim f/\Tnuc,
                \end{array}
              \right.
\end{equation}

This confirms that the energy of incoming particles, $E_{i,w} \simeq 3 \gwp \Tnuc$, is not affected by the shell of ejected~quarks.
  }
Since $E_{i,w}$ was the crucial input quantity for our treatment in Sec.~\ref{sec:inside_bubble}, the picture that emerged there is not affected by these scatterings.

\medskip

\paragraph{Energy transferred to techniquanta ejected from other bubbles. }
Finally, before ejected techniquanta can possibly enter another expanding bubble, they also have to pass through the layer of the techniquanta ejected from that other bubble. To investigate this, one can use the result derived above, Eq.~(\ref{eq:Qejbath}), with the specification that now $D$ is the maximal radius reached on average by expanding bubbles, because the shells of ejected quarks and gluons meet just before the bubble walls do.
We then find that the average energy transferred is much smaller than the energy of an ejected techniquanta in the plasma frame $\simeq \gwp f$,
\beq
\frac{Q_\text{ej-ej}}{\gwp f}
\simeq \frac{8 \zeta(3)}{3\pi} \alpha_\text{eff}^2 \,g_\TC \frac{\Tnuc}{f} \frac{\Tnuc D}{\gwp} \frac{\Tnuc}{\sqrt{-t_\IR}}
\ll 1\,.
\label{eq:QoverEej}
\eeq
Hence, for the purpose of determining the average energy of ejected quarks when they enter another bubble, one can safely ignore the interactions between the two shells.

\subsection{Ejected techniquanta enter other bubbles (and their pressure on them)}
\paragraph{Ejected techniquanta are squeezed. }
In the plasma frame, all ejected techniquarks are contained within a shell of length given by Eq.~(\ref{eq:Lej_velocity}) $L_\text{ej,p} \sim 1/f$, and most of them lie within a length given by Eq.~(\ref{eq:Lej_eff}) $L_\text{ej,p}^\text{eff} \sim 1/(\Tnuc^2 D) \ll 1/f$.
In the frame of the wall of the bubble they are about to enter, these lengths are further shrunk, so that ejected techniquarks are closer to each other than $1/f$ by several orders of magnitude.
Therefore we expect no phenomenon of string fragmentation when they enter other bubbles. So each ejected particle, upon entering another bubble, forms a hadron with one or more of its neighbours. This also implies there is no further ejection of other techniquanta.
Each of these hadrons carries an energy equal to that of the techniquanta that formed it, of order $\gwp f$ in the plasma frame.

\paragraph{Contribution to the retarding pressure. }
This conversion of ejected techniquanta into hadrons results in another source of pressure on the bubble walls, that acts for the relatively short time during which the bubble wall swallows the layer of ejected techniquanta.
In the frame of the bubble wall that they are entering, the energy of each ejected quark or gluon reads $E_\text{ej,w2} \simeq 2 \gwp^2 f$.
We then proceed analogously to what done in Sec.~\ref{sec:PLO}, and compute
\beq
\Delta p_\text{LO}^\text{ej}
= E_\text{ej,w2}-\sqrt{E_\text{ej,w2}^2 - \Delta m^2_\text{in}}
\simeq \frac{f}{4\,\gwp^2}\,,
\eeq
\beq
P_\text{LO}^\text{ej}
\simeq n_\text{ej,w2}\,\Delta p_\text{LO}^\text{ej}
\simeq  \frac{\zeta(3)}{\pi^2}\,g_\TC \,\gamma_{\rm wp}\Tnuc^3 f\,,
\eeq
where we have used $g_\TC = g_g + 3g_q/4$, $\Delta m^2_\text{in} \simeq f^2$, $n_\text{ej,w2} \simeq 2 \gwp n_\text{ej,p}$ and, for simplicity, the peak value $n_\text{ej,p} \simeq 2\,\gwp^2 n_{\TC,\text{p}}$ of Eq.~(\ref{eq:nejpx}).
The population of techniquanta ejected from other bubbles thus exert, on a given bubble wall, a pressure comparable to that exerted by the techniquanta incoming from the bath at LO, cf. Eq.~(\ref{eq:LOpressure}). 
Therefore, the pressure from ejected techniquanta does not alter the picture described so far --- a fortiori --- because it is exerted only just before bubble walls collide and not throughout their entire expansion.

\subsection{Ejected techniquanta heat the diluted SM bath}
\label{sec:SMbathheated_byejecta}
In Sec.~\ref{sec:energy_transfer} we found that the scatterings between ejected techniquanta and the diluted bath do not quantitatively change the picture of string fragmentation described in Sec.~\ref{sec:inside_bubble}.
These scatterings may however affect the properties of the particles, in the diluted bath, that do not confine.
These particles include all the SM ones that are not charged under the new confining group, so that for simplicity we denote them as `SM'.
By a derivation analogous to the one that lead us to Eq.~(\ref{eq:Qejbath}), we find that the average energy they exchange with the ejected quarks reads
\beq
Q_{\text{ej-}\SM}
\simeq \frac{8 \pi\alpha_\SM^2}{3}\frac{D \,n_{\TC,\text{p}}}{\sqrt{-t_\IR}}
\sim \alpha^2_\SM \,g_\TC \Big(\frac{\gwp}{f/\Tnuc}\Big)^{\!\frac{1}{2}} f\,,
\label{eq:QejSM}
\eeq
where we have used $\Tnuc D \simeq \gwp$ and $-t_\IR \sim \Tnuc^3 \gwp/f$, cf. Eq.~(\ref{eq:mV}).
We have denoted by $\alpha_\SM$ an effective coupling between SM particles and the techniquanta, which is model-dependent.

Now assume the techniquarks carry SM charges, e.g. as expected in composite Higgs models.
Then, in the wall frame, the fractional change of energy is of course similar to that derived in Eq.~\eqref{eq:QoverEi} for the incoming techniquanta. However the incoming techniquanta next undergo string fragmentation, and Eq.~\eqref{eq:QoverEi} does not affect that energy balance for $\gwp \gg f/\Tnuc$. In other words, string fragmentation renders this energy transfer irrelevant for the techniquanta, while the SM particles neutral under the confining group just proceed undisturbed so they keep track of it.
In particular, $Q_{\text{ej-}\SM}$, is much larger than the latter energy in the plasma frame $\sim \Tnuc$, and may even be slightly larger than the confinement scale $f$.\footnote{As already anticipated, in the regime of interest for DM phenomenology we will find that bubble walls run away, so that $\gwp^\text{max}$ is (much) smaller than $\sim 10^{-3} (f/\Tnuc)^3$, see Eq.~(\ref{eq:gwp_max}). 
Note also that, for Eq.~\eqref{eq:QejSM} only, $g_\TC = 3g_q/4$, i.e. the gluon contribution to heating the SM is negligible because they cannot carry SM charge.}

This need not be the case, however, as the new techniquanta may be very weakly interacting with the SM.
As they cannot interact too weakly, otherwise our assumption of instantaneous reheating would not hold, for simplicity we ignore this case in what follows and we assume that some techniquarks carry SM charges.

\section{Deep Inelastic Scattering in the Early Universe}
\label{sec:DIS}

The physical picture described so far results in a universe that, before (p)reheating from bubble wall collisions, contains three populations of particles.
\begin{itemize}
\item {\bf Population A.} Arises from hadronisation following string fragmentation. It consists of $N_\psi^\text{string}/2$ hadrons per quark or gluon in the initial bath, each on average with energy
\beq
E_A \simeq \frac{\gwp f}{N_\psi^\text{string}(\ECM)},
\label{eq:EA}
\eeq
in the plasma frame, and of roughly the same number of hadrons with much smaller energy. (The latter can be thought as coming from the half of the string closer to the center of the bubble wall.)
The physics resulting in this population is described in Sec.~\ref{sec:inside_bubble}, see Eq.~(\ref{eq:EAp}) for $E_A$ and Eq.~(\ref{eq:Npsi_string}) for  $N_\psi^\text{string}(\ECM = \sqrt{3\, \gwp \, \Tnuc\, f})$.
\item {\bf Population B.} Comes from the hadronisation of the ejected techniquanta. This population consists in $\sim$ one hadron per quark or gluon in the initial bath, each with energy
\beq
E_B \simeq \gwp f\,.
\label{eq:EB}
\eeq
So this population carries an energy of the same order of that of population A. Its physics is described in Sec.~\ref{sec:outside_bubble}, the energy $E_B$ is that of the initial quark or gluon, Eq.~(\ref{eq:energy_ejected}).
\item {\bf Population C.} Consists of the particles that do not feel the confinement force, that we denote `SM' for simplicity, each with a model-dependent energy given by Eq.~(\ref{eq:QejSM}), and whose total energy is much smaller than that in populations A and B.
\end{itemize}
The direction of motion of all these populations points, on average, out of the centers of bubble nucleation.

Hadrons from both populations A and B have large enough energies, in the plasma frame, that showers of the new confining sector are induced when they (or their decay products) scatter with the other particles in the universe and/or among themselves.
These deep inelastic scatterings (DIS):
\begin{itemize}
\item Increase the number density of composite states.
\item Decrease the momentum of each of these states with respect to the initial one $|\vec{p}_\psi|$.
\end{itemize}
Hence, such effects need to be taken into account to find the yield of any long-lived hadron.

The evolution of our physical system would require solving Boltzmann equations for the creation and dynamics of populations A, B and C in a universe in which preheating is occurring, and of the interactions of populations A, B and C among themselves and with the preheated particles produced from bubble wall collisions.
While certainly interesting, such a refined treatment goes beyond the purpose of this paper. 
In this Section, we aim rather at a simplified yet physical treatment, in order to obtain an order-of-magnitude prediction for the yield of long-lived hadrons.

\subsection{Scatterings before (p)reheating}
\label{sec:scatters_before_preheating}
We begin by considering the interactions among populations A, B and C.
\paragraph{Number densities of scatterers. }
Let us define $L_X$, with $X=A,B,C$, the effective thickness of the shells containing populations A, B, and C respectively.
For example, $L_\text{B,p} = L_\text{ej,p}^\text{eff}$ of Eq.~(\ref{eq:Lej_eff}).
We know that population A(B) consists on average of $N_\psi^\text{string}/2$ hadrons (one hadron) per each quark or gluon in the initial diluted bath, and that population C is the initial diluted SM population.
By conservation of the number of particles, we then obtain the number densities
\beq
n_A \simeq \frac{K^\text{string}}{2} \times \frac{D}{3 L_A} n_\TC,
\qquad
n_B \simeq \frac{D}{3 L_B} n_\TC,
\qquad
n_C \simeq \frac{D}{3 L_C} n_\SM,
\eeq
where $D$ is the average radius of a bubble at collision and we have used $L_X \ll D$.

\medskip
\paragraph{Energy transferred between scatterers. }
We now determine the average momentum, transferred to a particle from population $X$, upon going across a shell of population $Y$. 
In order to do so, we use our result Eq.~(\ref{eq:transfer_energy_rate}) for the rate of transferred energy and compute
\beq
Q_{\text{Y} \to \text{X}}
\simeq \Gamma_{\text{Y} \to \text{X}} L_Y
\simeq n_\text{Y} L_\text{Y} \frac{8\pi \alpha_\text{X-Y}^2}{\sqrt{-t_\IR}}
\simeq \frac{8\zeta(3)}{3\pi} \alpha_\text{X-Y}^2 g_\text{Y} \frac{\Tnuc}{\sqrt{-t_\IR}} \,\gwp \Tnuc\,,
\label{Q_YX}
\eeq
where $\alpha_\text{X-Y}$ is the effective interaction strength of the scatterings of interest,
$g_\text{Y}$ the number of degrees of freedom in density of population Y (where we include a factor of $N_\psi^\text{string}/2$ for Y = A), and we have used the relation $\Tnuc D = \gwp$ valid in the runaway regime.
We conclude that:
\begin{itemize}
\item {\bf Populations A and B.} The energies of the hadrons of population A and B in the plasma frame, respectively $\gwp f/N_\psi^\text{string}$ and $\gwp f$, are both much larger than the energy they can exchange with any of the other baths among A,B,C, by a factor that scales parametrically as $f/\Tnuc$ or larger (because for all populations we have $-t_\IR \sim n/\langle E\rangle > \Tnuc^2$, see the discussion in Sec.~\ref{sec:energy_transfer}).
Therefore these elastic scatterings are not effective in reducing the energy of the hadrons of either population A or population B.
\item {\bf Population C.} On the contrary, $Q_{\text{A,B} \to \text{C}}$ can be of the same order of the energy of each particle in population C, Eq.~(\ref{eq:QejSM}), which therefore are significantly slowed down by these interactions.
Importantly for our treatment, this does not alter the fact that population C was energetically subdominant with respect to populations A and B.
\end{itemize}

\paragraph{No significant DIS between populations A, B and C.}
Finally, we determine whether any of the scatterings among particles in populations A,B,C could result in significant hadron production, via deep inelastic scattering. A single scattering event potentially results in a shower of the new confining sector if the exchanged momentum is larger than the confinement scale, $t^2 > f^2$.
This condition is allowed by kinematics, because the center-of-mass energy of the scatterings between any of the populations above is much larger than $f$.
A significant amount of DIS happens if the DIS scattering rate $\Gamma_{\text{Y} \to \text{X}}^\text{DIS}$ of a particle from population $X$, upon going across a shell of population $Y$, is much larger than the inverse of the length of the shell $Y$.
We then compute
\beq
\Gamma_{\text{Y} \to \text{X}}^\text{DIS} L_\text{Y}
\simeq n_\text{Y} \sigma_\text{X-Y} v\, L_\text{Y}
\simeq \frac{4 \zeta(3)}{3 \pi} \alpha_\text{X-Y}^2 g_\text{Y} \frac{\gwp}{(f/\Tnuc)^2}\,,
\label{eq:DIS_XY}
\eeq
where again we have used the runaway relation $\Tnuc D = \gwp$ and, for definiteness, we have assumed the scattering cross section has the form of Eq.~(\ref{eq:dsigmadt}).
Therefore, no significant DIS happens in the regions where $\gwp \ll (f/\Tnuc)^2$.
This condition will turn out to be always satisfied in the parameter space of our interest, so we can ignore the DIS among populations A, B and C in what follows.

\subsection{Scatterings with the (p)reheated bath}
\label{sec:DIS_preheated}
By \textit{preheating}, we intend the stage between the time when bubble walls collide and start to produce particles (e.g. from the resulting profile of the condensate), and the \textit{reheating} time when these particles have thermalised into a bath.
We now discuss the scatterings of populations A and B with the particles produced at preheating, that we have assumed to be efficient. 
The contribution of population C to the final yield of hadrons is subdominant with respect to the one of populations A and B because, as seen in Secs.~\ref{sec:SMbathheated_byejecta} and \ref{sec:scatters_before_preheating}, the total energy in population C is much smaller than that in populations A and B.

\paragraph{Energy of the (p)reheated bath. }
The preheated particles are produced with energies, in the plasma frame, of the order of the mass of the scalar condensate,\footnote{In the picture we have in mind, non-perturbative effects such as Bose enhancement or parametric resonance (see e.g.~\cite{Kofman:1997yn}) are not relevant: the first because the SM particles are interacting, thus they exchange momentum and do not occupy the same phase space cells; the second because the variation of their masses from the dilaton's oscillations is smaller than their mass at the minimum. Note that, unlike what occurs in many inflationary scenarios, we expect only a small hierarchy $ T_{\rm RH} \lesssim \langle E_\text{prh} \rangle $.}
\beq
\langle E_\text{prh} \rangle \simeq m_\chi < f\,.
\eeq
Their total energy scales as
\beq
E_\text{prh}^\text{tot} \sim f^4 V\,,
\eeq
with V the volume of a large enough region of the universe.
For comparison, the total energy in populations A and B scales as
\beq
E_\text{A,B}^\text{tot} \sim \gwp f \Tnuc^3 V\,,
\eeq
which is much smaller than $f^4 V$ because $\gwp \ll (f/\Tnuc)^3$, Eq.~(\ref{eq:gwp_max}).
So the preheated particles can act as a thermal bath for all the other populations A, B and C, because the energy of A, B, and C is subdominant in the energy budget of the universe.

\paragraph{Inelastic versus elastic scattering. }
Scatterings of hadrons (or their decay products) with the preheated bath will, therefore, eventually slow down and thermalise populations A and B.
However, these scatterings can also exchange energies much larger than $f$,  thus inducing deep inelastic scatterings. Indeed their center-of-mass energy squared reads
\beq
s_\text{A,B} \simeq 2\, m_\chi E_\text{A,B}\,,
\label{eq:s_DIS}
\eeq
where $E_\text{A} \simeq \gwp f /N_\psi^\text{string}(\ECM)$ and $ E_\text{B} \simeq \gwp f$.
Eq.~(\ref{eq:s_DIS}) is the result of our simplifying assumption to neglect masses and to average to zero scattering angles with particles in a bath: define $p_E= E(1,\,\hat{E})$, $p_\text{prh} = m_\chi(1,\,\hat{m})$, then $s = (p_E+p_\text{prh})^2 \simeq 2 E \,m_\chi (1- \hat{E} \cdot \hat{m}) \simeq 2 E\, m_\chi$.
We now determine if those center of mass energies are entirely available for particle production via DIS, or if instead they are reduced by several low-momentum-exchange interactions.
In order to do so, we evaluate the rate of energy loss of a particle from population A or B, $\Gamma^\text{loss}_\text{A,B}$, as the ratio between the rate of energy it exchanges with the preheated bath, that we evaluate analogously to Eq.~(\ref{eq:transfer_energy_rate}), and its initial energy $E_\text{A,B}$.
We then compare this quantity with the rate for a deep inelastic scattering to happen with the full energy available $s_\text{A,B}^{1/2}$,
\beq
\frac{\Gamma^\text{loss}_\text{A,B}}{\Gamma^\text{DIS}_\text{A,B}}
\simeq \frac{n_\text{prh} 8 \pi\alpha_\text{eff}^2/(E_{A,B} \sqrt{-t_\IR})}{n_\text{prh} 4 \pi\alpha_\text{eff}^2/s_{A,B}}
\simeq \frac{m_\chi}{\sqrt{-t_\IR}}
\sim \frac{1}{\sqrt{c_\text{vac}}}\frac{m_\chi^2}{f^2}\,.
\label{eq:loss_over_DIS}
\eeq
In the last equality, we have again used the screening mass for non-equilibrium systems~\cite{Arnold:2002zm}
\beq
-t_\IR
\sim \frac{n_\text{prh}}{\langle E_\text{prh} \rangle}
\sim c_\text{vac} \frac{f^4}{m_\chi^2}\,,
\eeq
where we have used that by conservation of energy $n_\text{prh} \sim \rho_\RH/\langle E_\text{prh} \rangle$, and where we have expressed the energy density of the reheated bath $\rho_\RH$ using the results of Sec.~\ref{sec:thermal_history}.

We conclude that, if
\beq
\Big(\frac{m_\chi}{f}\Big)^2 \ll c_\text{vac}^{1/2}\,,
\label{eq:condition_full_DIS}
\eeq
the full center-of-mass energies $s_\text{A,B}$ are available for deep inelastic scattering, i.e. populations A and B do not lose a significant amount of their energy via interactions with the preheated bath.
For simplicity, in what follows we assume this model-dependent property to hold.

\subsection{Enhancement of hadron abundance via DIS}
\label{sec:DIS_hadron_abundance}

\paragraph{The picture: a cascade of DIS. }
The number of composite states arising from a hard scattering depends on how the strings fragment, so on the same physics that set the abundance of the composite states when the techniquanta cross the bubble walls, discussed in Sec.~\ref{sec:fragmentation_multiplicity_energy}.
Each scattering, depending on its center-of-mass energy, produces a number $N_\psi^\text{string}$ of hadrons $\psi$, that we model in the same was as in Eq.~(\ref{eq:Npsi_string}).
Given the large initial energies $s_\text{A,B}$, the daughter hadrons typically still have enough energy to themselves induce further deep inelastic scatterings with the particles in the preheated bath, and hence additional hadron production.
Analogously, SM particles produced in such DIS typically have large enough energies to also initiate showers of the new confining force with their subsequent scatterings.
This process iterates until the average energy of scatterings drops below the confinement scale.

\paragraph{Number of hadrons produced per scattering. }
For reasons given in Sec.~\ref{sec:fragmentation_multiplicity_energy}, together with simplicity, we assume that the available energy $\sqrt{s}$ at each scattering splits equally among all the outcoming particles.  We then write the average of this number as
\beq
N^\DIS(s) =  N_\psi^\text{string}(\sqrt{s}/2)\,,
\label{eq:Npsi_NSM}
\eeq
where the factor of 2 in the argument of $N_\psi^\text{string}$ arises because Eq.~(\ref{eq:Npsi_string}), which defines $N_\psi^\text{string}$, assumes that $\sqrt{s}$ is the center of mass energy of the scattering of two particles neutral under the new confining force.
If a hadron is included among the two scatterers, then QCD studies find that the final number of hadrons can be obtained by just halving the energy in the center of mass frame~\cite{Rosin:2006av}, also see footnote~\ref{foot:Npsi_ee_vs_pp}.\footnote{
Note that if a hadron instead decays to two SM particles before it scatters, which is model-dependent, then $\sqrt{s}/2$ is again the good argument for the function $N_\psi^\text{string}$, because then one has two particles each with half the initial energy, but both neutral under the new confining force. In this case, however, Eq.~\eqref{eq:Npsi_NSM} becomes $N^\DIS(s) = 2 N_\psi^\text{string}(\sqrt{s}/2)$.
When iterating the treatment to many scatterings, we find that this extra factor of~2 does not impact the final abundance of hadrons, which can be understood by thinking that the same initial energy is spread faster to zero.}

\paragraph{Energies of produced hadrons. }
Explicitly, we assume $E'_\text{com} = \sqrt{s}/N^\DIS$, where $E'_\text{com}$ is the energy of any outgoing particle (SM and/or composite) in the center-of-mass frame of the scattering.
To iterate to many scatterings, we write $E'_\text{com}$ in the plasma frame as
$E' = \gamma'\,E_{\rm com}'(1 -   \hat{v}'\cdot \hat{v})$, where $\gamma'$ and $\hat{v}'$ are the associated Lorentz boost and its direction, and $\hat{v}$ is the direction of motion of the outgoing particle in the center-of-mass frame of the scattering.
By averaging $\hat{v}'\cdot \hat{v}$ to zero for simplicity, we obtain
\beq
E' = \gamma'\,E_{\rm com}'\,.
\eeq
We then determine $\gamma'$ by observing that the energy of each particle, in the center-of-mass frame of the scattering, is both $E_{\rm com} = \sqrt{s}/2$ and $E_{\rm com} = \gamma' E_\text{prh} (1 + \hat{v}' \cdot \hat{E}_{\rm com})$, where $E_\text{prh}$ is the energy in the plasma frame of the particles in the preheated bath. By averaging $\hat{v}' \cdot \hat{E}_{\rm com}$ to zero for simplicity, we obtain the Lorentz boost
\beq
\gamma' \simeq \frac{\sqrt{s}}{2 \langle E_\text{prh}\rangle}\,.
\eeq
Using Eq.~(\ref{eq:s_DIS}) for $s$ we finally obtain
\beq
E'_\text{A,B} \simeq \frac{1}{N^\DIS} E_\text{A,B}\,.
\eeq
(If we did not average over angles, we would have obtained $E'_\text{A,B} = (E_\text{A,B}/N^\DIS)(1-\hat{v}'\cdot \hat{v})(1-\hat{E} \cdot \hat{m})/(1+\hat{v}' \cdot \hat{E}_{\rm com})$).
So, after a hard scattering the energy of each outgoing particle in the plasma frame is roughly the initial energy divided by a factor $N^\DIS$. The subsequent $s$ is then reduced by the same factor, ensuring a convergence of $N^\DIS(s)$ to unity, via Eq.~(\ref{eq:Npsi_string}), after only a few iterations.
This also teaches us that the average energy of the particles, produced this way, quickly decreases to values lower than about $m_*$.

\paragraph{Number of hadrons produced by a chain of DIS. }
Let us now estimate the yield of final hadrons by following the above arguments. Assuming interactions are fast enough, also those following the first one happen with preheated particles of the same average energy $\langle E_\text{prh}\rangle$.
Now define the number of states (both composite and not) $N_{k}$ produced at the $k^\text{th}$ interaction. This can be expressed as
\beq
N_k(s)
\simeq N^\DIS
\Big(\frac{s}{
N_{k-1}
\times N_{k-2}
\times \cdots
\times N_{1}}\Big),
\eeq
where we remind the reader that the function $N^\DIS$ is obtained from Eqs.~(\ref{eq:ncharged_s}) and (\ref{eq:Npsi_NSM}).
Starting from a single resonance produced from the fragmentation of strings between quanta inside the bubble, after this chain of scattering processes one obtains a total number of resonances given by the product $\prod_k N_{k}(s)$.
We find numerically that this product can be expressed as
\beq
K^\DIS_\text{A,B}
 \simeq \frac{s_\text{A,B}}{m_*^2}\,.
 \label{eq:KDIS_AB}
\eeq
In other words, the iterative process we described converts the initial available energy into the rest mass of hadrons $m_*$.
Since our aim here is not to achieve a more precise treatment, we refrain from refining the assumption that the momenta are distributed evenly among the particles coming out of a scattering process. 
In the same spirit of building a physically-clear picture without drowning in model-dependent details, we do not cover here the possibility that every scattering produces, in addition to the composite states, a comparable or larger amount of SM particles. (That would result in $N^\DIS > N_\psi^\text{string}$ and in a faster degrowth of the available scattering energy to $m_*$ at each step.) In addition to simplicity, this can be justified by observing that, in the limit of large number of degrees of freedom in the dark sector, our assumption that they carry SM charges will make their production dominant with respect to the one of SM particles.

\paragraph{Additional comments.}
We conclude our derivations with two comments concerning its validity.
\begin{itemize}
\item
If the full center-of-mass energies are not available for DIS, i.e. if Eq.~(\ref{eq:condition_full_DIS}) does not hold, then one could use the same result  $K^\DIS_\text{A,B}$ of Eq.~(\ref{eq:KDIS_AB}), upon substituting $s_\text{A,B} = 2E_\text{A,B} m_\chi$ with the largest energy for which $\Gamma^\text{loss}_\text{A,B} \ll \Gamma^\text{DIS}_\text{A,B}$, that can be derived via Eq.~(\ref{eq:loss_over_DIS}).

\item We have ignored the production of heavy particles from the collisions of bubble walls~\cite{Hawking:1982ga, Watkins:1991zt, Konstandin:2011dr, Braden:2014cra}.
This is justified as it has been shown that it only occurs when the minima of the potential are nearly degenerate and seperated by a sizable barrier~\cite{Falkowski:2012fb, Katz:2016adq}, which is not the case for the close-to-conformal potentials we have in mind. 
Hence we expect only particles lighter than the scalar condensate to be produced during reheating following the wall collision.
\end{itemize}

\subsection{DIS summary}
\label{sec:DIS_summary}

The yield of hadrons, resulting from the processes of deep inelastic scattering described above, receives contributions from:
\begin{itemize}
\item {\bf Population A.} That is, the hadrons produced from string fragmentation as described in Sec.~\ref{sec:inside_bubble}.
Their contribution reads
\beq
Y^{\SC+\rm string+\DIS }_\text{A}
\simeq \frac{1}{2}  K^\DIS_\text{A}  N_\psi^\text{string}(\ECM) D^\SC Y_\TC^{\rm eq}
\simeq \frac{\gwp f m_\chi}{m_*^2}\,D^\SC Y_\TC^{\rm eq}\,,
\label{eq:KDIS_popA}
\eeq
where we have used $K^\DIS_\text{A} = s_\text{A}/m_*^2$, cf. Eq.~\eqref{eq:KDIS_AB}, with $s_\text{A} \simeq 2 m_\chi E_\text{A} \simeq 2 m_\chi \gwp f/N_\psi^\text{string}(\ECM)$, cf. Eqs.~\eqref{eq:s_DIS} and \eqref{eq:EA}.
Note that the above expression captures also the regime where each string fragmentation produces on average one hadron, because the energy of that single hadron is  roughly $\gwp f/2$, see the related discussion in Sec.~\ref{sec:string_summary}.

\item {\bf Population B.} That is, the hadrons produced out of the techniquanta ejected from the bubbles, described in Sec.~\ref{sec:outside_bubble}.
Their contribution reads
\beq
Y^{\SC+\rm string+\DIS }_\text{B}
\simeq K^\DIS_\text{B} D^\SC Y_\TC^{\rm eq}
\simeq 2\,\frac{\gwp f m_\chi}{m_*^2}\,D^\SC Y_\TC^{\rm eq}\,,
\eeq
where we have used $K^\DIS_\text{B} = s_\text{B}/m_*^2$, cf. Eq.~\eqref{eq:KDIS_AB}, with $s_\text{B} \simeq 2\,m_\chi E_\text{B} \simeq 2\,m_\chi \gwp f$, cf. Eqs.~\eqref{eq:s_DIS} and \eqref{eq:EB}.
\end{itemize}

Thus, the combined contribution to the total hadron yield is given by
\beq
	\label{eq:DISrelic}
	Y^{\SC+\rm string+\DIS }
	\simeq K^\DIS D^\SC \,Y_\TC^{\rm eq}
	\simeq  3\,\frac{\gwp\, f\, m_\chi}{m_*^2}\,D^\SC\, Y_\TC^{\rm eq}\,,
\eeq
where we have defined
\beq
K^\DIS = \frac{1}{2}K^\DIS_\text{A} N_\psi^\text{string}(\ECM)  + K^\DIS_\text{B}\,.
\label{eq:KDIS_definition}
\eeq
Note finally that, in the regime of runaway bubble-walls, one obtains the parametric scaling $Y^{\SC+\rm string+\DIS} \propto (\Tnuc/f)^3 \gwp$. Which is much larger than the simple supercooling dilution, $\sim (\Tnuc/f)^3$, in the regions of parameter space where our analysis holds, namely for $\gwp > f/\Tnuc$.

\section{Supercooled Composite Dark Matter}
\label{sec:DMabundance}

\subsection{Initial condition for thermal evolution}

Finally, all unstable resonances decay either to SM or to the long-lived or stable hadrons, which we take to form DM.
To obtain the yield of any such hadron $i$ at the onset of reheating, one should use the expression
	\beq
	\label{eq:DISrelic_i}
	Y_i^{\SC+\rm string+\DIS } =  \mathrm{BR}_i\,K^\DIS  D^\SC Y_\TC^{\rm eq}\,,
	\eeq
where $K^\DIS$, $D^\SC$ and  Y$_\TC^{\rm eq}$ are defined, respectively, in Eqs.~\eqref{eq:KDIS_definition}, \eqref{eq:DSC} and~\eqref{eq:Yeqi}.
$\mathrm{BR}_i$ is a pseudo-branching ratio, of the energy available to the confining techniquanta, into $\psi_i$ particles.
	Estimates of $\mathrm{BR}_i$ for the cases where $\psi_i$ is a meson and a baryon are given in App.~\ref{app:brstring}, which show a broad range of underlying-model dependent values are possible, albeit with a large uncertainty.
	For example, in a QCD-like theory where $\psi_i$ is a baryon with mass $\sim 4\pi f$ and the pions have mass $\sim f$, one obtains values $\mathrm{BR}_i \sim 10^{-6}$, while larger values $\mathrm{BR}_i$ are obtained for baryon-pion mass rations closer to one, or if $\psi_i$ is a meson. Hence, we will take $\mathrm{BR}_i$ to be a free parameter.

		\begin{figure}[t]
\centering
\begin{adjustbox}{max width=1.\linewidth,center}
\includegraphics[width= 0.5\textwidth]{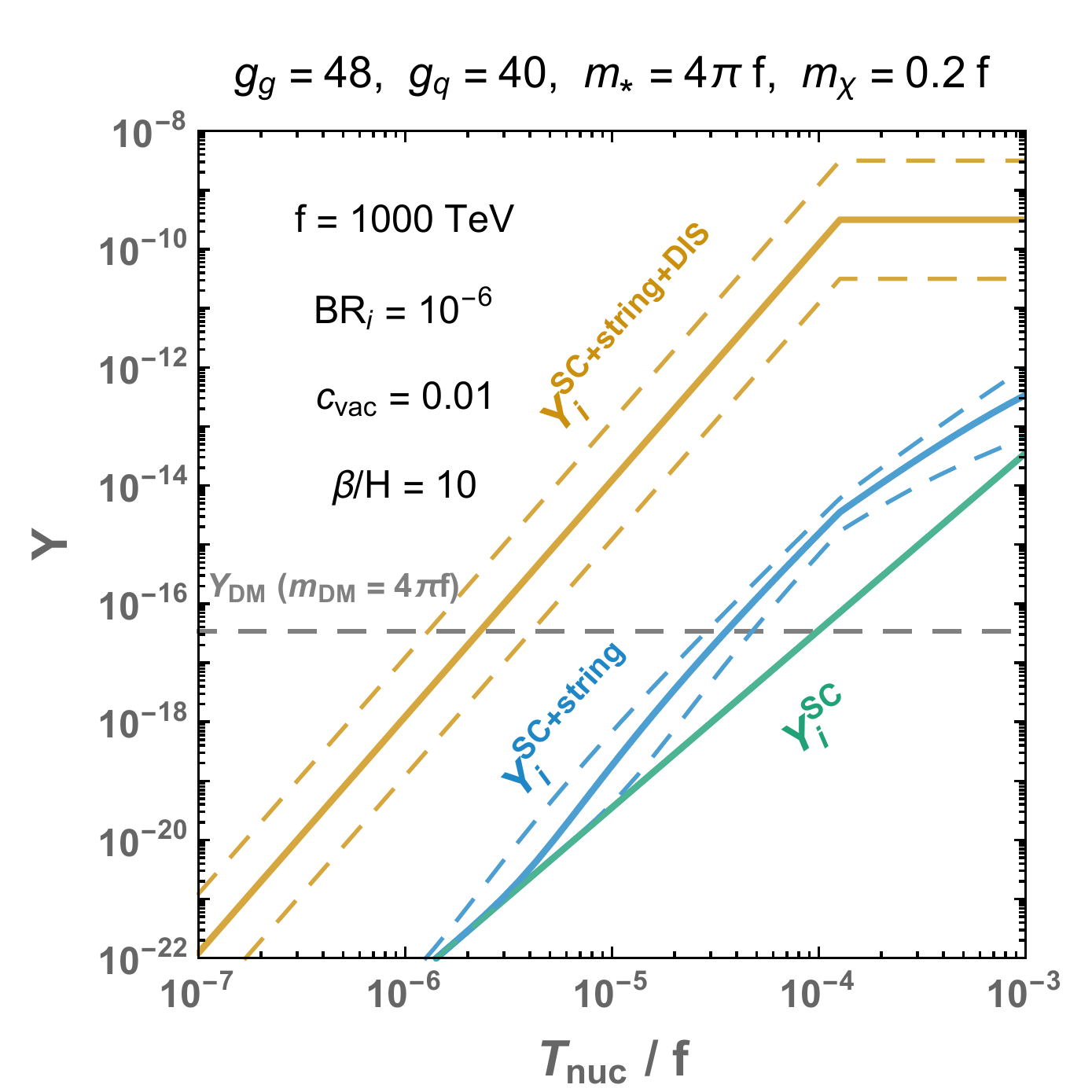}
\includegraphics[width= 0.5\textwidth]{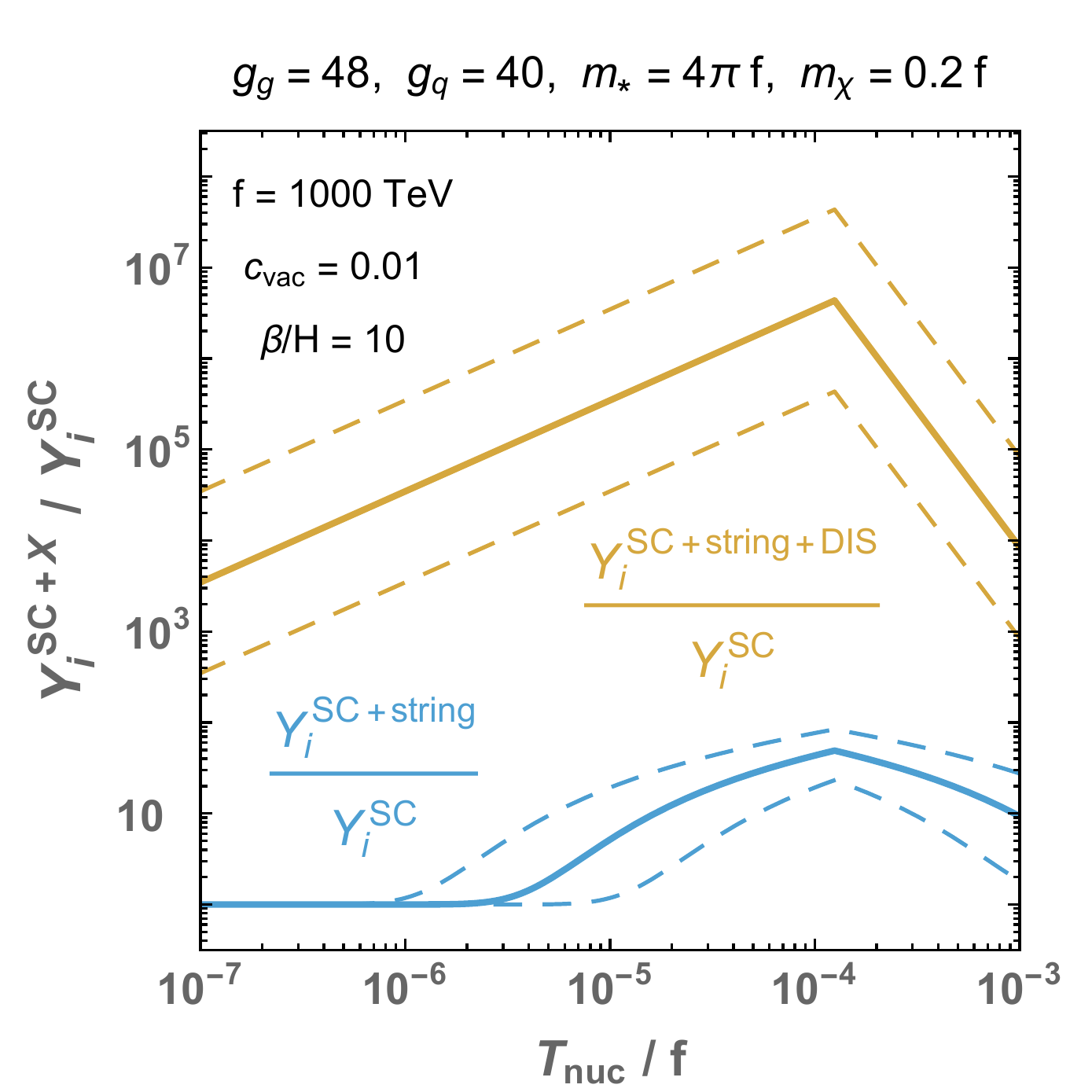}
\end{adjustbox}
\caption{\it \small Left: the yields following supercooling Eq.~(\ref{eq:SC_relic_i}), string fragmentation Eq.~\eqref{eq:SC_and_string_relic_i} and deep inelastic scattering Eq.~\eqref{eq:DISrelic_i}. The yield matching the observed relic abundance of DM for $\mDM = 4\pi\,f$ is also shown. The dashed lines show the effect of varying $\gwp$ by an order of magnitude in either direction around Eq.~(\ref{eq:gwp_max}). This illustrates the sensitivity of the yield to our determination of $\gwp$. Right: ratios of the same yields. The peak corresponds to the maximum $\gwp$. }
\label{fig:contributions} 
\end{figure}

	 For completeness, the supercooling plus string  and supercooling yields read
	\begin{eqnarray}
	Y_i^{\SC+\rm string} &=&  \mathrm{BR}_i\,K^{\text{string}}  D^\SC Y_\TC^{\rm eq}, \label{eq:SC_and_string_relic_i}\\
	Y_i^\SC &=&  \mathrm{BR}_i\, D^\SC \frac{\frac{3}{4}g_q}{g_\TC} Y_\TC^{\rm eq}\,, \label{eq:SC_relic_i}
	\end{eqnarray}
	where $K^{\text{string}}$ is defined in Eq.~\eqref{eq:Kstring}. We have included a factor $\frac{3}{4}g_q/g_\TC$ in $Y_i^\SC$ to account for the fact that, in the case of no string fragmentation nor DIS, gluons do not contribute to the final abundance of heavy composite states of quarks. It would be absent if one was interested in light composite states of quarks. The yield of the various contributions is shown in Fig.~\ref{fig:contributions}.

It will turn out that the measured DM abundance is achieved in the regime of runaway bubble walls. In that regime, the resulting expression for the DM yield has a simple parametric form that eventually results in the DM abundance being independent of the DM mass, if it is to match onto observation $Y_{\rm DM}  \simeq 0.43 \, \mathrm{eV}/\MDM$~\cite{Aghanim:2018eyx}, which we find convenient to report here.
By using~Eqs.~(\ref{eq:Yeqi}), (\ref{eq:DSC}), (\ref{eq:gwp_runaway}), and (\ref{eq:DISrelic}),  with $g_{Rf} = g_\SM = 106.75$, we find
\beq
Y_{i,\text{runaway}}^{\SC+\rm string+\DIS } 
\simeq 0.43\,\frac{\text{eV}}{m_*}
\times
\frac{\mathrm{BR}_i}{10^{-6}}
\, \frac{g_\TC}{120}
\,\left(\frac{0.01}{c_\text{vac}}\right)^{\!\frac{5}{4}}
\,\frac{m_\chi/f}{0.2}
\,\frac{4\pi}{g_*}
\,\left(\frac{\Tnuc/f}{10^{-5.7}}\right)^{\!4}\,.
\label{eq:DISrelic_runaway}
\eeq

\subsection{Thermal contribution}
\label{sec:sub-thermal-abundance}

To complete our discussion, we must still determine the effects on the yield of any DM interactions with the thermal bath after supercooling, DIS, and reheating.
The importance of thermal effects following reheating was already pointed out in~\cite{Hambye:2018qjv} (therein dubbed the subthermal contribution). Following the phase transition and particle production through DIS, the SM bath and the DM have returned to kinetic equilibrium. The scattering energy is now insufficient to break the resonances, but these may still annihilate into SM particles or be produced in the inverse process. Thus, just after the reheating, the DM abundance evolves according to the well known Boltzmann equation~\cite{Kolb:1990vq}
	\begin{align}
	\label{eq:first_case}
	\frac{dY_{\mathsmaller{\rm DM}}}{dx} = -   \sqrt{ \frac{ 8\pi^2 \gSM }{ 45 }} \frac{ M_{\rm pl} \MDM \, \left<\sigma v_{\mathsmaller{\rm rel}} \right>}{x^2} \, \left( Y_{\mathsmaller{\rm DM}}^2 -  Y_{\mathsmaller{\rm DM}}^{\mathrm{eq} \, 2} \right),
	\end{align}
where we use $x\equiv\MDM/T$ as the time variable, and $M_{\rm pl}$ is the reduced Planck mass. For simplicity we only consider velocity independent cross sections here. As an intitial condition we take the relic abundance at the reheat temperature, $Y_{\mathsmaller{\rm DM}}(T_{\rm RH}) = Y_{\mathsmaller{\rm DM}}^{\SC+\rm string+\DIS }$, estimated following string fragmentation and DIS enhancement in Eq.~\eqref{eq:DISrelic}. 
For our plots we solve the Boltzmann equation numerically. If the cross section and reheating temperatures are sufficiently large the system will be driven back into equilibrium. The relic density is then largely set by freezeout dymanics, albeit with somewhat different initial conditions. On the other hand, if the cross section and reheat temperatures are small enough, the relic density is set by dilution, string fragmentation and DIS, with only negligible thermal corrections following reheating.
Using the dilution mechanism of the PT, of course, we can avoid the usual unitarity constraint on the maximum thermal relic DM mass~\cite{Griest:1989wd} (see e.g.~\cite{vonHarling:2014kha,Baldes:2017gzw} for recent appraisals).

\begin{figure}[t!]
\begin{center}
\begin{adjustbox}{max width=1.0\linewidth,center}
\includegraphics[width=.5\textwidth]{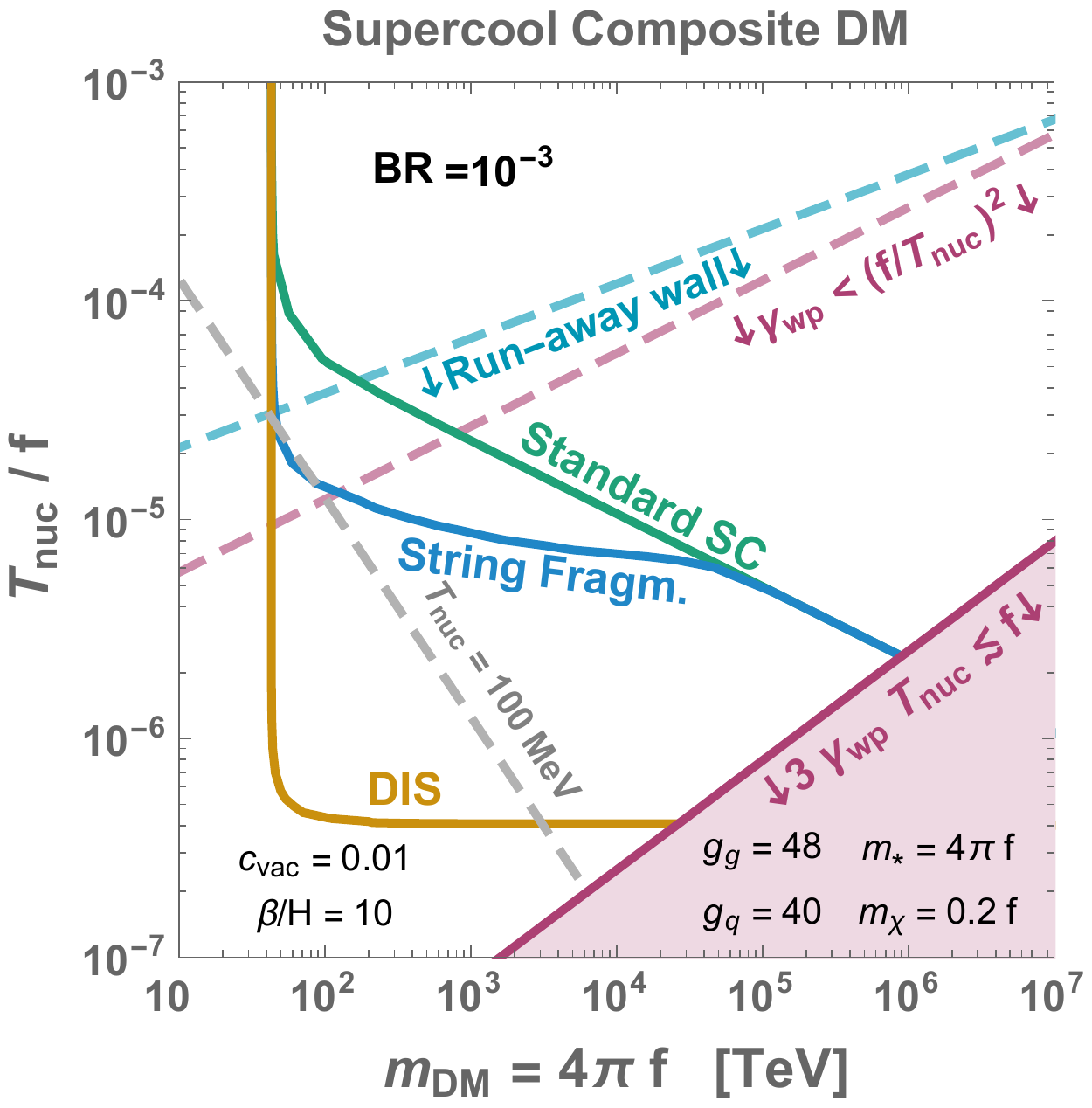} \quad
\includegraphics[width=.5\textwidth]{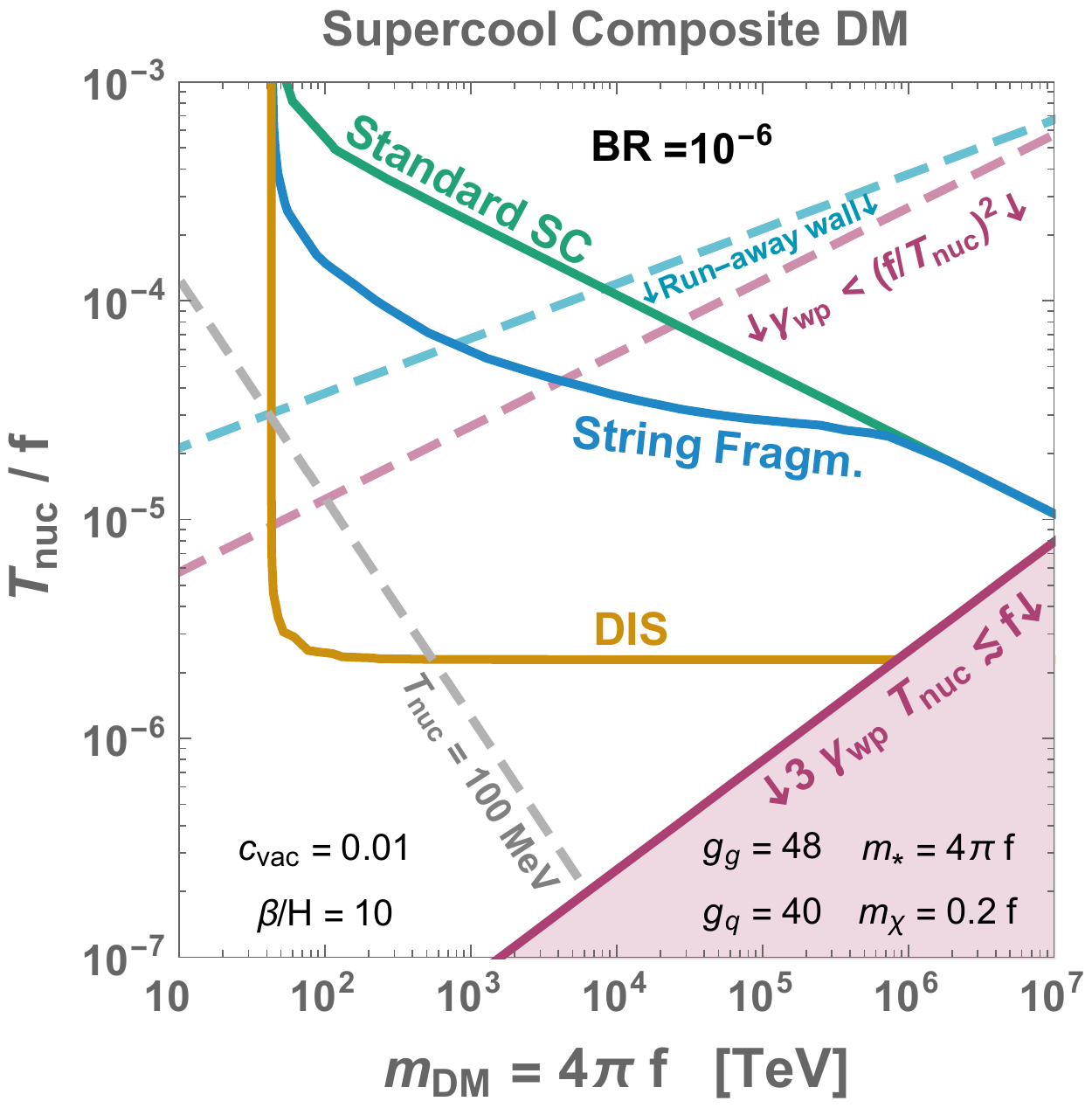}
\end{adjustbox}
\caption{
\label{fig:compositeDM_generic} 
\it \small
Solid lines: supercooling $\Tnuc/f$ and DM mass $\MDM$ required to obtain the observed DM abundance. The parameters chosen imply a reheating temperature $T_\RH \simeq 0.13~f$, see Sec.~\ref{sec:thermal_history}.
All lines include the thermal contribution discussed in Section~\ref{sec:sub-thermal-abundance}.
The line with initial condition $Y^{\rm \SC+string+\DIS}_i$ of Eq.~(\ref{eq:DISrelic_i}) corresponds to the yellow contour.
For comparison, we show in green (blue) the contour that one would obtain by skipping directly from the supercooling (supercooling plus string fragmentation) step to the thermal corrections, respectively Eqs.~\eqref{eq:SC_relic_i} and~\eqref{eq:SC_and_string_relic_i}.
All contours converge at some $\MDM$ where thermal effects following reheating become dominant, of the order of $\MDM \approx 100 \, \mathrm{TeV}$ because we fixed $\langle \sigma v_{\mathsmaller{\rm rel}}\rangle =4\pi/\MDM^2$. Below this mass, the relic density is necessarily suppressed compared to the observed DM density, due to efficient DM annihilation after reheating. 
In the purple region $3\,\gwp\,\Tnuc \lesssim f$ the quarks are reflected by the first wall they encounter, but may enter the bubbles in following stages of their evolution, and the DM abundance lines ignore possible modifications arising from this `ping-pong' effect.
They also ignore that, for values of $\gwp$ only slightly larger than $f/\Tnuc$ and depending on other model-dependent parameters, the energetics of our treatment may be more complicated, see Eqs.~(\ref{eq:QoverEi}) and~(\ref{eq:mV}).
The dashed gray line delimits the area $\Tnuc < O(100) \, \mathrm{MeV}$ where the supercooled phase transition could happen because of QCD dynamics.
The dashed light blue line indicates the regimes where bubble walls run away, cf. Eq.~\eqref{eq:run-away_cond}. The dashed purple line indicates the regime where $\gwp < (f/\Tnuc)^2$, and the fact it lies above the horizontal part of the DIS line confirms that our treatment has been consistent when ignoring the DIS of Eq.~(\ref{eq:DIS_XY}).
}
\end{center}
\end{figure}

\subsection{Dark matter relic abundance}

We now combine all our results together and determine the amount of supercooling required to match the observed relic abundance $Y_{\rm DM}  \simeq 0.43 \, \mathrm{eV}/\MDM$. Examples are shown in Fig.~\ref{fig:compositeDM_generic} for some representative choices of the parameters. From these figures we can draw a number of conclusions. 

\begin{enumerate}
\item[i)] If we assume $\left<\sigma v_{\mathsmaller{\rm rel}} \right> \propto 1/\MDM^{2}$, thermal effects will necessarily dominate if the DM is light enough. This occurs because $T_{\rm RH}$ cannot realistically be arbitrarily suppressed below $f$, for sensible choices of $g_{Ri}$ and $g_{Rf}$. This regime corresponds to the point in which the contours turn vertical in Fig.~\ref{fig:compositeDM_generic}. At which value of $\MDM$ this occurs depends on the precise choice for $\left<\sigma v_{\mathsmaller{\rm rel}} \right> $.
For definiteness, in Fig.~\ref{fig:compositeDM_generic} we choose $\langle \sigma v_{\mathsmaller{\rm rel}} \rangle = 4\pi/\MDM^2$ as typical of baryon scatterings in a strongly coupled sector.
Thermal effects can of course be further suppressed if we depart from the efficient reheating assumption made here~\cite{Hambye:2018qjv}.

\item[ii)] String fragmentation and DIS lead to large corrections to the composite DM relic density, compared to the naive supercooling dilution. This implies a mismatch between the relic abundances of primordial elementary and composite relics alluded to before. Whether the composite or elementary relic would have the greater abundance depends on the details of confinement (for elementary relics BR$_i=1$). If the composite relic is say, a light meson which is produced abundantly, the multiplicative DIS process can be highly efficient in populating these states following the PT. This implies we require much more supercooling to match onto the observed DM relic abundance. On the other hand, if the composite relic is some heavy state, perhaps a baryon, it could be produced in a highly suppressed rate both in string fragmentation and DIS. In this latter case, the required amount of supercooling to match onto the DM relic density is also reduced. The two cases are illustrated with two different assumptions for the branching ratios in Fig.~\ref{fig:compositeDM_generic}.\footnote{
We checked that the DIS line is unaffected if we use the treatment of the NLO pressure of~\cite{Hoche:2020ysm}, instead of the one of~\cite{Bodeker:2017cim} that we have employed in this paper, cf. Sec.~\ref{sec:P_NLO} and  Sec.~\ref{sec:NLO_pressure}. On the other hand, the run-away wall dashed line, and hence the string fragmentation line, could be affected by this choice. For simplicity as well as in light of the criticism of~\cite{Hoche:2020ysm} appeared in~\cite{Vanvlasselaer:2020niz,Azatov:2021ifm,Gouttenoire:2021kjv}, we employed the results of~\cite{Bodeker:2017cim} in this paper.}

\item[iii)] In some cases, we find $T_{\rm nuc} \lesssim 100$ MeV, as delineated in Fig.~\ref{fig:compositeDM_generic}. Thus QCD effects could assist in completing the PT~\cite{Witten:1980ez,Iso:2017uuu,vonHarling:2017yew,Hambye:2018qjv}. On the other hand, if QCD effects help the transition to occur, they can also suppress the eventual gravitational wave signature~\cite{Baldes:2018emh} (simply because the QCD effects increase the tunneling probability and thus will act to shorten the timescale of the PT). The details will depend on the physics entering the effective potential of the scalar $\chi$ and need to be studied in a model dependent way.

\end{enumerate}

Together with the gravitational wave signal from the PT, there may also be model dependent collider, direct, and indirect detection signatures associated with the DM from the strongly coupled sector. We will investigate these further, together with their interplay with the novel string fragmentation and DIS effect, in a concrete realisation of such a confining sector in a companion paper~\cite{Baldes:2021aph}.

\section{Discussion and Outlook}
\label{sec:outlook} 
The possible existence of a new confining sector of Nature is motivated by several independent problems of the Standard Model of particle physics and by cosmology.
This encourages the identification of predictions of confining sectors, that are independent of the specific problem they solve.
One such prediction is the possibility that the finite temperature phase transition in the early universe, between the deconfined and confined phase, is supercooled.
This possibility has received a lot of attention in recent years, see e.g.~\cite{Iso:2017uuu,vonHarling:2017yew,Hambye:2018qjv,Baldes:2018emh,Bruggisser:2018mrt, Baratella:2018pxi, Agashe:2019lhy, DelleRose:2019pgi, Ellis:2019oqb,vonHarling:2019gme,Baldes:2020kam,Baldes:2021aph}.

In this paper, we have pointed out and modelled a novel dynamical picture taking place in every supercooled confining phase transition, that (to our knowledge) had been missed in the literature.
This novel picture stems from the observation that, when fundamental techniquanta of the confining sector are swept into expanding bubbles of the new confining phase, the distance between them is large with respect to the confinement scale.
Therefore the energy of the fluxtubes connecting techniquanta is so large that string breaking produces many hadrons per fluxtube, with large momenta in the plasma (CMB) frame, in a sense analogously to QCD hadrons produced in electron-positron collisions at colliders.
These hadrons and their decay products subsequently undergo scatterings with other particles in the universe, with center-of-mass energies much larger than both the confinement scale and the temperature that the universe reaches after reheating.
The dynamics just described is partly pictured in Figs.~\ref{fig:wall_diagram}~and~\ref{fig:string_breaking}.

The processes of string fragmentation and `deep inelastic scatterings in the sky', synthetised above, have a plethora of implications.
A key quantity to study them is the pressure on the bubble walls induced by this novel dynamics, which we have determined in Sec.~\ref{sec:wall_speed}, see Eq.~(\ref{eq:gwp_max}) and Fig.~\ref{fig:gwp} for the resulting bubble-wall velocities.
An interesting aspect of our findings is that the so-called `leading-order' pressure is proportional to the boost factor of the bubble wall, unlike in the case of non-confining supercooled PTs~\cite{Bodeker:2009qy,Bodeker:2017cim}.

We then quantified the values of supercooling below which one recovers the `standard phase transition', where confinement happens between nearest charges.
By relying on the modelling we proposed in Sec.~\ref{sec:our_picture_relevant} we found, interestingly,
that the PT does not proceed in the `standard' way already for minor supercooling, i.e. if bubbles are nucleated and expand just after vacuum energy starts to dominate.
Our proposed dynamics should not only be employed in the large supercooling region, but also in the minor supercooling one depending on the value of another model-dependent parameter, see Fig.~\ref{fig:StdPictureRecovered}.
The regimes in between these regions (one being the `ping-pong' regime of Sec.~\ref{sec:pingpong}) will be studied in future work, to not charge this paper with too much content.

Next, we have focussed on the implications of our dynamical picture for the abundance of long-lived or stable particles that are composite states of the new confining sector. They are summarised in the Synopsis, Sec.~\ref{sec:synopsis}, and a quantitatively accurate expression of the final yield of a given composite particle is given in Eq.~(\ref{eq:DISrelic_runaway}), for concreteness in the regime where bubble walls run away.
Compared to the simple dilution of relics induced by supercooling of non-confinement transitions, these processes enhance their abundance by parametrically large factors. Therefore they have to be taken into account whenever a property of the universe, e.g.~the DM and/or the baryon abundance, depends on the final yield of hadrons. As an example, their dramatic impact on the abundance of supercooled composite DM can be seen in Fig.~\ref{fig:compositeDM_generic}.

Concerning DM in particular, this study constitutes a novel production mechanism of DM with mass beyond the unitarity bound~\cite{Griest:1989wd}. It would be interesting and timely to study its experimental signals, given the new wave of telescopes that is starting to take data of high-energy neutrinos and gamma rays (e.g.~KM3NeT, LHAASO, CTA) and given their potential in testing heavy DM, e.g.~see~\cite{Cirelli:2018iax}.
One such study will appear in a forthcoming publication~\cite{Baldes:2021aph}.

During the course of carrying out this study we have made a number of simplifications, for the purpose of obtaining a general and clear enough picture of the physics involved.
For example, the various populations of particles created by this novel dynamics, such as the ejected techniquanta and the hadrons that follow the bubble walls, could be better described by Boltzmann equations, by the use of simulations etc., rather than with our simple treatment that focused on their average properties.

%

Finally, this dynamics opens broader and exciting avenues of investigation, that we think deserve exploration.
For example, it would be interesting to study  its interplay with recent interesting ideas regarding phase transitions~\cite{Creminelli:2001th,Randall:2006py,Konstandin:2011ds,Konstandin:2011dr,
Falkowski:2012fb,Ipek:2018lhm,Bai:2018dxf,Baker:2019ndr,Chway:2019kft,Bloch:2019bvc,DelleRose:2019pgi,Kitajima:2020kig}, or its impact on the production of gravitational waves in supercooled confining phase transitions.
As for the latter, our study of the bubble wall Lorentz factor in~Sec.~\ref{sec:wall_speed} constitutes a necessary first step.

\begin{subappendices}

\chapterimage{wall_profile3}

\section{Wall profile of the expanding bubbles}
\label{app:wall_profile}
\FloatBarrier

\begin{figure}[t]
\centering
\begin{adjustbox}{max width=1.2\linewidth,center}
\raisebox{0cm}{\makebox{\includegraphics[ width=0.5\textwidth, scale=1]{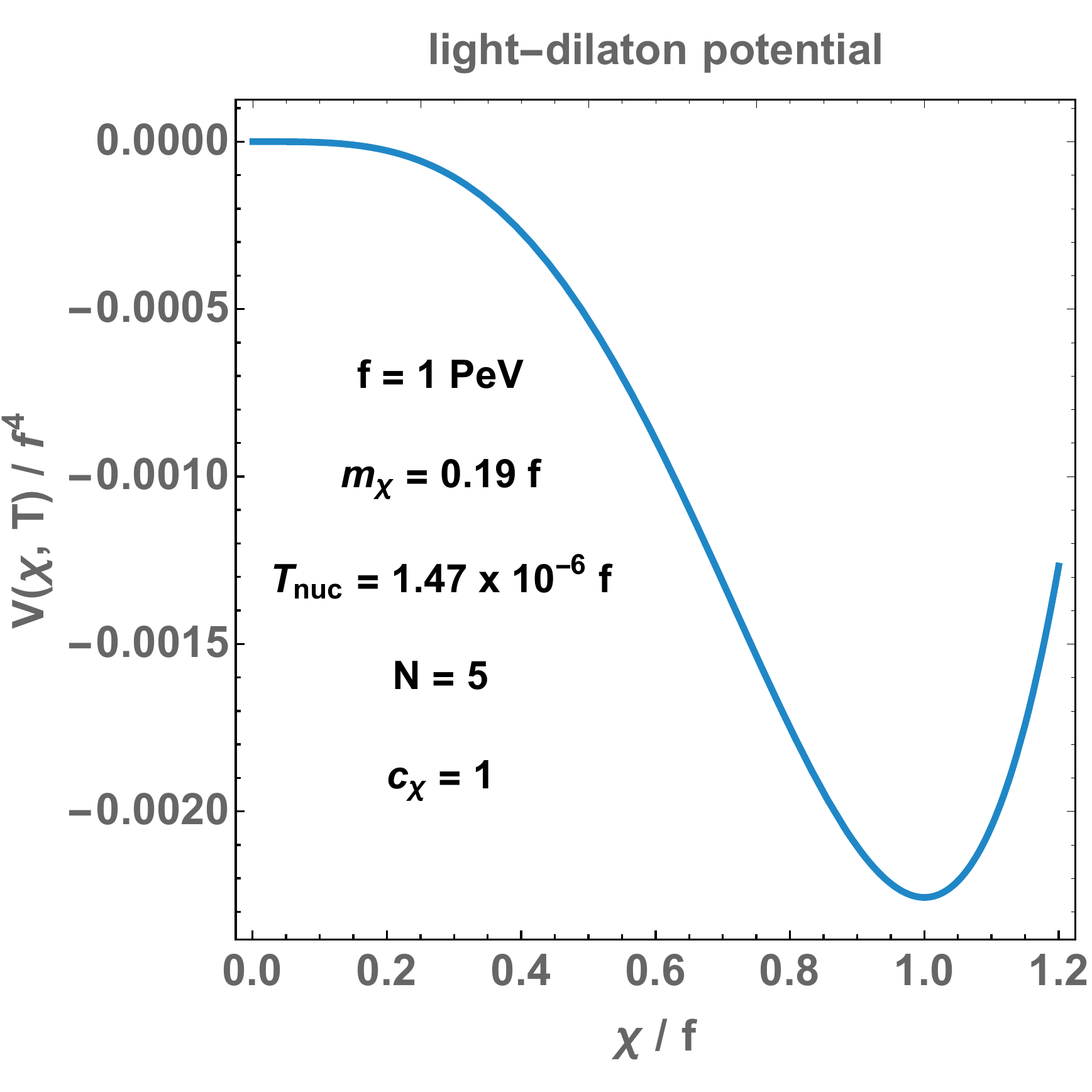}}}
\raisebox{0cm}{\makebox{\includegraphics[ width=0.5\textwidth, scale=1]{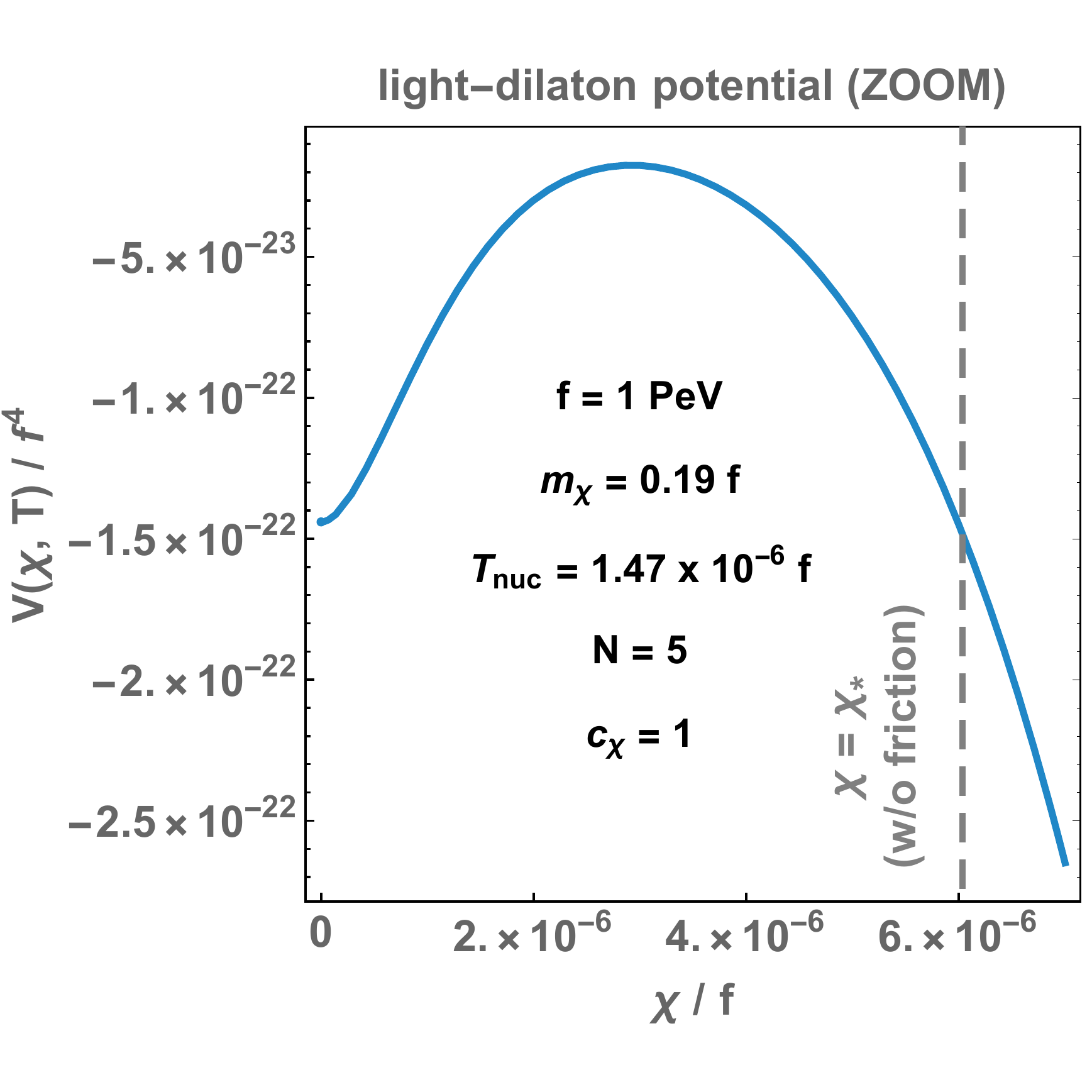}}}
\end{adjustbox}
\caption{\it \small \textbf{Left:} \label{fig:lightdilatonpot} light-dilaton potential with temperature corrections in Eq.~\eqref{eq:total_dilaton_potential}. \textbf{Right:} Zoom on the thermal barrier. The tunneling point $\chi_*$, in the case where the friction term in the Euclidean equation of motion is neglected,  is also shown.}
\end{figure}

\subsection{The light-dilaton potential}
\paragraph{Zero-temperature.}

In this section we suppose that confinement occurs from the condensation of a nearly-conformal strongly-interacting sector, when an approximate scale-invariance gets spontaneously broken. If the source of explicit breaking is small, the spontaneous breaking of scale invariance generates a pseudo Nambu-Goldstone boson, the dilaton which we parameterize as \cite{Goldberger:2008zz}
\begin{equation}
\dilaton(x) = \dilatonvev e^{\frac{\sigma(x)}{\dilatonvev}},
\end{equation}
where $f$ is the confining scale and where $\sigma(x)$ transforms non-linearly $\sigma(x)  \rightarrow \sigma{(\lambda x)} + \log{\lambda}$ under the scale transformation $x  \rightarrow \lambda x$. 
Its potential is given by \cite{Bruggisser:2018mrt}
 \begin{equation}
 \label{eq:zero_temp_pot_dil}
V_{\dilaton}^{\mathsmaller{\rm T=0}}(\dilaton)  =c_{\dilaton}\, g_{\dilaton}^2 \, \dilaton^4 \, \left[1- \frac{1}{1+\gamma_{\epsilon}/4}\left(\frac{\dilaton}{\dilatonvev}\right)^{\gamma_{\epsilon}} \right],
\end{equation}
with 
\begin{equation}
\label{eq:gammaepsilonVSmdilaton}
\gamma_{\epsilon} \simeq -\frac{1}{4 }\frac{m_{\dilaton}^2}{c_{\dilaton}\,g_{\dilaton}^2~f^2} <1,
\end{equation}
where $m_{\dilaton}$ is the dilaton mass, and $c_{\dilaton}$ is a constant of order $1$, which we fix to $c_{\chi} =1$.
The dilaton coupling constant $g_{\dilaton}$ is chosen to reproduce the glueball normalization
\begin{equation}
g_{\dilaton} \simeq \frac{4\pi}{N},
\end{equation}
with $N$ being the rank of the confining gauge group.
 The validity of the EFT relies on the smallness of the parameter $|\gamma_{\epsilon}| \ll 1$ (here taken negative) which controls the size of the explicit breaking of scale invariance, and thus of the dilaton mass.
 
Note that in the limit where $|\gamma_{\epsilon}| \ll 1$, the dilaton potential at zero-temperature reduces to the Coleman-Weinberg potential \cite{Coleman:1973jx}, i.e. 
\begin{equation}
V_{\dilaton}^{\mathsmaller{\rm T=0}}(\dilaton)   \overset{|\gamma_{\epsilon}|\ll 1}{=}  -\gamma_{\epsilon} \, c_{\dilaton}\,g_{\dilaton}^2\, \dilaton^4 \, \textrm{log} \left( \frac{\dilaton}{\dilatonvev} \right).
\label{eq:dilaton_potential_taylor_cw}
\end{equation}

\paragraph{Thermal corrections.}
To model thermal effects, we follow \cite{Randall:2006py,Bruggisser:2018mrt}, and consider the finite-temperature corrections generated by the particles charged under the confining force (the CFT bosons)
 \begin{equation}
  V_T( \dilaton,\, T) = \sum_{i \in \rm\mathsmaller{CFT}~ bosons} \frac{n T^4}{2 \pi^2} J_B\left( \frac{m_{i}^2}{T^2} \right), \qquad \textrm{with} \quad m_{i} \simeq g_{\dilaton }\,\chi.
 \label{eq:finite_temp_dilaton}
\end{equation}
The total number of CFT bosons $n$ is fixed to\footnote{The effective number of gluons in the deconfined phase $45N^2/4$ being different from $2(N^2-1)$ is a property valid at thermal equilibrium. It results from the peculiar strongly-coupled dynamics of the CFT. However, due to the large wall Lorentz factor, the CFT gas entering the wall can be considered as collisionless, cf. Sec.~\ref{sec:when_confinement}. This is why in the main text we consider the number of gluons entering the wall as $g_g = 2(N^2-1)$.}
\begin{equation}
\sum_{\mathsmaller{\rm CFT}~\rm bosons} n =\frac{45 N^2}{4} \equiv  \tilde{g}_{g},
\label{eq:number_dof_CFT}
\end{equation}
in order to recover the free energy of $\mathcal{N}=4$ $SU(N)$ large $N$ super-YM dual to an AdS-Schwarzschild space-time \cite{Creminelli:2001th}
\begin{equation}
\label{eq:free_energy_CFT}
V_T( 0,\, T) \simeq -b\, N^2\, T^4, \qquad \textrm{with} \quad b= \frac{\pi^2}{8}.
\end{equation}
By writing Eq.~\eqref{eq:free_energy_CFT}, we have neglected the contribution from the fermions present in the plasma.
For simplicity, we suppose that the dilaton degree of freedom $\chi$ still exists in the deconfined phase, such that the total potential for the dilaton is
\begin{equation}
\label{eq:total_dilaton_potential}
V_{\rm tot}(\dilaton,\, T) =  V_{\dilaton}(\dilaton) + V_T( \dilaton,\, T) ,
\end{equation}
where $V_{\dilaton}(\dilaton)$ and $ V_T(\dilaton, \, T) $ are given by Eq.~\eqref{eq:zero_temp_pot_dil} and Eq.~\eqref{eq:finite_temp_dilaton}. We plot the potential in Fig.~\ref{fig:lightdilatonpot}.
The supercooling stage starts when the energy density becomes vacuum-dominated
\begin{equation}
\frac{\pi^2}{30}g_{\rm Ri} T_{\rm start}^4 \simeq c_{\rm vac} f^4 \quad \implies \quad T_{\rm start}\simeq \left(\frac{30 c_{\rm vac}}{g_{\rm Ri} \pi^2}  \right)^{\! 1/4} f, \label{eq:Tstart_app}
\end{equation}
with $c_{\rm vac}=\dfrac{m_{\sigma}^2}{16f^2}$ and $g_{\rm Ri}= g_{\rm SM}+ g_{\rm TC}$ where (see Eq.~\eqref{eq:number_dof_CFT})
\begin{equation}
g_{\rm TC} = g_{\rm q} + \tilde{g}_{g} \simeq \frac{45N^2}{4}. \label{eq:gTC_app}
\end{equation}

\subsection{The wall profile}
\paragraph{Space-like region: the bounce profile.}
We solve the tunneling temperature by solving the equation 
\begin{equation}
\label{eq:nucTempInstApprox}
\Gamma(\Tnuc) \simeq H(T_{\rm nuc})^4.
\end{equation}
with \cite{Coleman:1977py, Callan:1977pt}
\begin{equation}
\Gamma(\Tnuc) = R_{0} ^{-4} \left( \frac{S_4}{2\pi} \right)^2 {\rm exp} \left(-S_4 \right) ,
\end{equation}
where $R_{0} \sim 1/\Tnuc$ is the bubble radius at nucleation and $S_4$ is the $O_4$-bounce action 
\begin{equation}
S_{4}=2\pi^2\int dr~r^3~\left[\frac{1}{2}\phi^{'}(r)^2+V\left(\phi(r)\right)\right],
\end{equation}
which we compute from solving the Euclidean equation of motion ($d=4$)
\begin{equation}
\phi''(s) + \frac{d-1}{s}\phi'(s) = \frac{dV}{d\phi},
\label{eq:bounce_eom}
\end{equation}
with boundary conditions
\begin{equation}
\phi'(0)=0, \qquad \textrm{and} \qquad  \lim_{r \to \infty} \phi(r) = 0.
\label{eq:BC_bounce}
\end{equation}
$s = \sqrt{\vec{r^2}+t_E^2} = \sqrt{\vec{r^2}-t^2}$ is the space-like light-cone coordinate and $t_E = i\,t$ is the Euclidean time.

\begin{figure}[t]
\centering
\begin{adjustbox}{max width=1.2\linewidth,center}
\raisebox{0cm}{\makebox{\includegraphics[ width=0.5\textwidth, scale=1]{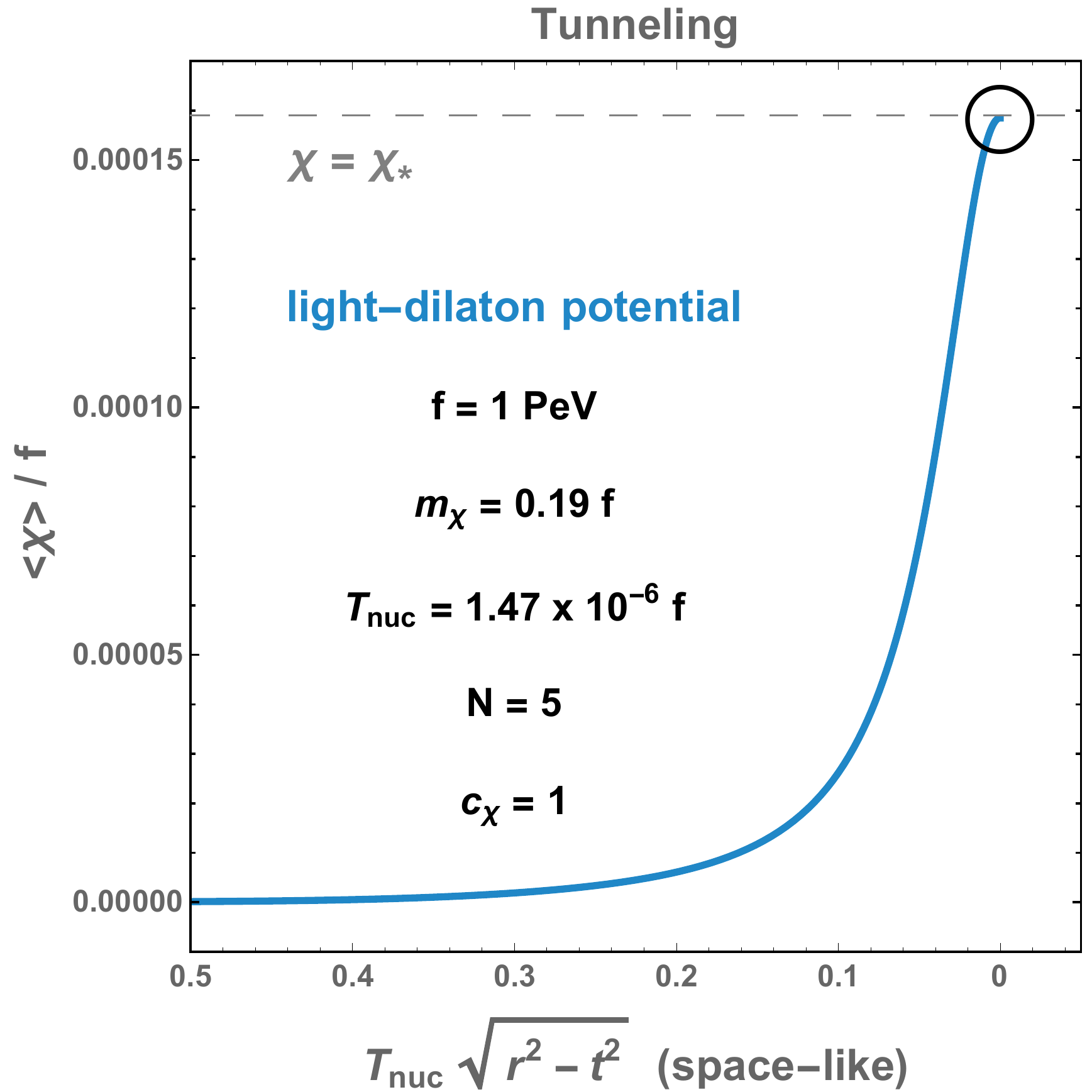}}}
\raisebox{0cm}{\makebox{\includegraphics[ width=0.5\textwidth, scale=1]{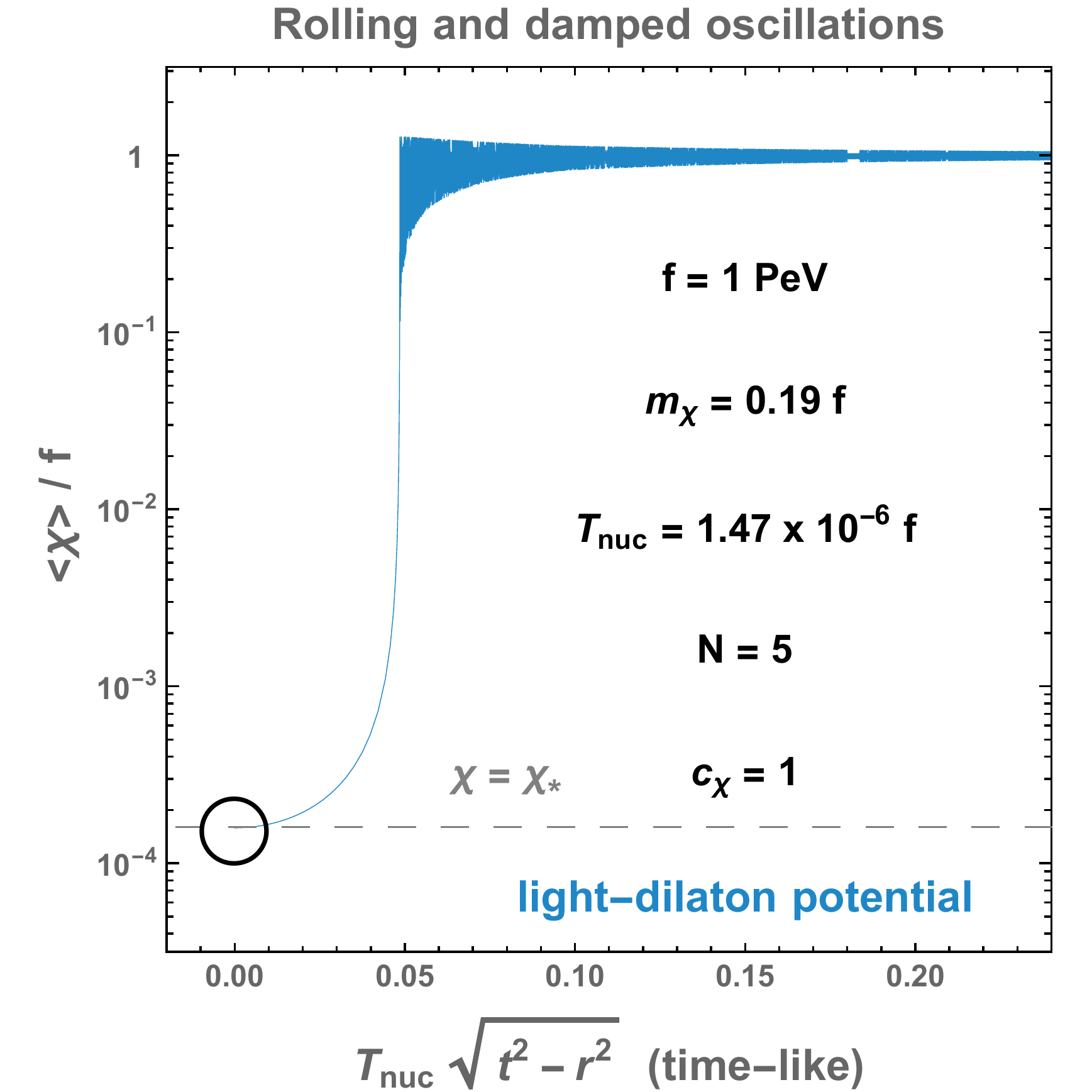}}}
\end{adjustbox}
\caption{\it \small \label{fig:scalarTimeLike} \textbf{Left:} Bounce profile at nucleation. It interpolates between the false vacuum $\left<\dilaton\right> = 0$ outside the bubble and the release point $\left<\dilaton\right>_{*}$ at the center of the bubble. \textbf{Right:} Evolution of the scalar field after tunneling. First, the scalar field rolls along the shallow part of the nearly-conformal potential and then realized oscillations with period $\sim \dilatonvev^{-1}$ with a damping time $\lesssim T_{\mathsmaller{\rm nuc}}^{-1}\gg\dilatonvev^{-1}$. Taking into account the decay of the scalar field would reduce the damping time after the first oscillation. The full bubble wall profile can be obtained after connecting the two figures through the two black circles.}
\end{figure}

We plot the bounce profile in the left-hand panel of Fig.~\ref{fig:scalarTimeLike} for given parameters relevant for the study. 
The value at the center of the bubble --- the tunneling point $\dilaton_*$ --- can be estimated analytically by energy conservation between $\chi=\chi_*$ and the false vacuum in $\chi =0$ if we neglect the friction term in the equation of motion in Eq.~\eqref{eq:bounce_eom},
\begin{equation}
V_{\rm tot}(\dilaton_*) \simeq V_{\rm tot}(0), \qquad \rightarrow \qquad \frac{\chi_{*}}{f} \simeq \frac{1}{\sqrt{2}\,{\rm log}^{1/4}(f/\chi_{*} )}\frac{T}{T_c}.
\end{equation}
Here (coincidence numeric) $T_c$ is the critical temperature, defined when the two minima of the free energy are equal
\begin{equation}
\label{eq:Tc_def}
c_{\rm vac} f^4 + V_T(f,\,T_c) - V_T(0,\,T_c) \equiv 0,
\end{equation} 
Note that for confining phase transition with $m_i(f) \gtrsim f$, the quantity $V_T(f,\,T_c)$ in Eq.~\eqref{eq:Tc_def} vanishes\footnote{
We recall that the thermal functions in Eq.~\eqref{eq:finite_temp_dilaton} verify the property $\lim_{x\to\infty} J_{\rm B/F}(x) = 0$. From using Eq.~\eqref{eq:Tstart_app}, we observe that
\begin{equation}
4.0~ \frac{m_i(f)}{f} \left(\frac{0.1}{c_{\rm vac}} \right)^{1/4}\left( \frac{g_{\rm Ri}}{80} \right)^{1/4} \gg 1 \qquad \implies \qquad m_i(f)/T_{\rm start} \gg 1. \label{eq:TC_Tstart_condition}
\end{equation}
The connection between $T_c$ and $T_{\rm start}$ in Eq.~\eqref{eq:TC_Tstart} applies for all phase transitions satisfying Eq.~\eqref{eq:TC_Tstart_condition}.} and $T_c$ is related to the temperature at which supercooling starts $T_{\rm start}$ in Eq.~\eqref{eq:Tstart_app} through
\begin{equation}
T_{\rm c} \simeq 3^{1/4} \left( \frac{g_{\rm Ri}}{g_\TC} \right)^{1/4} T_{\rm start}. \label{eq:TC_Tstart}
\end{equation}
$g_{\rm Ri}$ is the total number of relativistic d.o.f in the symmetric phase while $g_\TC$ only counts those which are involved in the phase transition ($g_\TC <g_{\rm Ri}$).
In the scenario studied in this appendix, upon assuming $g_{\rm Ri}\simeq g_{\rm TC}$ with $g_{\rm TC}$ given in Eq.~\eqref{eq:gTC_app}, we get
\begin{equation}
 T_c = \left(\frac{m_{\dilaton}^2 \,f^2}{16\,b\,N^2}\right)^{1/4} = \left( \frac{\left|\gamma_{\epsilon}\right|\,c_{\dilaton}\,g_{\dilaton}^2}{4\,b\,N^2} \right)^{1/4}~f. \label{eq:Tc_app}
\end{equation}

The tunneling point $\chi_*$ in absence of friction is shown in Fig.~\ref{fig:lightdilatonpot}, while the tunneling point from numerically solving the bounce equation is visible in Fig.~\ref{fig:scalarTimeLike}.
Plugging the numbers chosen for making the plots, we find $\chi_{*}/f \simeq 6.0 \times 10^{-6}$ for the analytical value and $\chi_{*}/f \simeq 1.6 \times 10^{-4}$ for the numerical value. This difference was expected since the analytical estimate neglects the friction term in Eq.~\eqref{eq:bounce_eom}.

\paragraph{Time-like region: rolling and damped oscillations.}

As soon as the bubble expands, the scalar field starts to roll toward the true vacuum $\left<\dilaton\right> = f$ and realize damped oscillations. The field dynamic is captured by the Klein-Gordon equation for an inhomogeneous field
	\begin{equation}
	\square \phi -\frac{\partial V}{\partial \phi} = 0.
	\end{equation}
We first use the $SO(3)$ symmetry to reduce the 3 Cartesian coordinates to the radial $r$ coordinate
	\begin{equation}
	\frac{\partial^2 \phi}{\partial r^2} +\frac{2}{r}\frac{\partial \phi}{\partial r} - \frac{\partial^2 \phi}{\partial t^2} -\frac{\partial V}{\partial \phi} = 0.
	\end{equation}
We used the Minkowski metric since we can neglect the universe expansion during the time of bubble propagation.
We then use the $SO(3,1)$ symmetry which reduces $r$ and $t$ to the time-like light-cone coordinate $s = \sqrt{t^2-r^2}$ only \cite{Jinno:2019bxw}
	\begin{equation}
	\label{eq:scalar-time-like}
	\frac{\partial^2 \phi}{\partial s^2} +\frac{3}{s}\frac{\partial \phi}{\partial s} +\frac{\partial V}{\partial \phi} = 0.
	\end{equation}
Note the opposite sign in front of the potential $V$ between the space-like (or Euclidean) equation of motion in Eq.~\eqref{eq:bounce_eom} and the time-like (or Minkowskian) equation of motion in Eq.~\eqref{eq:scalar-time-like}.
Here the damping is purely geometrical, reminiscent of the $SO(3,1)$ symmetry and we do not consider the damping due to the dilaton decay or due to the interaction with the plasma (see e.g. \cite{Dorsch:2018pat}). In right panel of Fig.~\ref{fig:scalarTimeLike}, we display the scalar field profile obtained after integration of the time-like equation in Eq.~\eqref{eq:scalar-time-like}, using the initial condition $\chi(s=0) = \chi_*$ given by the bounce solution in Eq.~\eqref{eq:bounce_eom}. 

\paragraph{The full bubble wall profile.}
The full bubble wall profile is obtained after matching the profile in the space-like region, left panel of Fig.~\ref{fig:scalarTimeLike}, with the profile in the time-like region, right panel of Fig.~\ref{fig:scalarTimeLike}.
One can see that the first confining scale, that the incoming techniquanta are subject to upon entering the wall, is the exit scale
\beq
\chi_* \gtrsim \Tnuc\,.
\eeq
Our explicit computation also shows that the length of the section of the bubble wall where $\langle \chi \rangle = \chi_*$, in the wall frame, satisfies
\beq
L_\text{w} \lesssim \Tnuc^{-1}\,,
\eeq
as we assumed in Eq.~(\ref{eq:Lw}) in the main text.
Then, $\langle \chi \rangle$ transits to its zero-temperature value $f$ over a length, in the wall frame, of order $f^{-1}$.

\chapterimage{dices} 
\section{Example estimates of the string to DM branching ratio}
\label{app:brstring}

In Sec.~\ref{sec:string_breaking}, we have discussed that, after supercooling, the quarks enter inside the confined phase, with a typical seperation $\sim \Tnuc^{-1}$, much larger than the confining scale $f$, such that a highly energetic fluxtube forms.
We have shown that this string, which is unstable under quark-anti-quark pair nucleation, breaks into $K_{\rm string}$ pieces.
The dynamics of strings is then also relevant in the processes of deep inelastic scatterings of section~\ref{sec:DIS}.
In this section, we estimate the branching ratio of a string to a given hadron $i$, introduced in~Eq.~\eqref{eq:DISrelic_i}, in two different cases.
First, when $i$ is a light meson, in which case we expect the yield of $i$ to be independent of its mass and given by a combinatoric factor implying the number of flavors. Second, when $i$ is a heavy baryon in which case one expects the yield to be Boltzmann suppressed.

\subsection{Light meson -- Combinatorics}

In the limit of large string energy, $\ECM \gg f$, one expects the fragmentation of the string to be democratic with respect to the different bound-states if they are light enough. In that case, the string-to-i branching ratio is given by a combinatoric factor depending on the number of flavors $N_{f}$ and the number of quark constituents (either $2$ for meson and $N_\TC$ for baryons). In the particular case of a light meson $q_1 \bar{q}_2$, one obtains
\begin{equation}
\mathrm{Br(\text{string} \rightarrow} i)  = \begin{dcases} 1/N_f^2, \qquad ~\text{if } q_1 = q_2, \\
2/N_f^2 ,  \qquad  ~\text{if } q_1 \neq q_2.
\end{dcases}
\label{eq:combinatorics}
\end{equation}

\subsection{Heavy baryon -- Boltzmann suppression}

For this example a useful model for us will be the thermal one~\cite{Chliapnikov:1994qc,Becattini:1995if,Pei:1996kq,Chliapnikov:1999qi}, which was able to fit LEP data of particle yields up to a $10\%$ error~\cite{Andronic:2008ev}, even with an initial state far from thermal equilibrium. In this model, the yield of heavy mesonic or baryonic resonances is suppressed by a Boltzmann factor~\cite{Chliapnikov:1994qc,Becattini:1995if,Pei:1996kq,Chliapnikov:1999qi}, in which the strong scale plays the usual role of temperature. The yield of heavy resonances can be modelled by
\begin{equation}
\langle N_{i} \rangle \sim A_i  \frac{ (2J_i + 1) }{ \mathrm{Exp}\left[ M_{i}/B_i \right]} ,
\end{equation}
where $M_i$ and $J_i$ are the mass and spin of the state $i$ respectively. Here $A_i$ is an overall normalisation, which will depend on whether the particle is a pseudoscalar meson, vector meson, or baryon etc. In QCD it was found to differ by $\lesssim 10$ between vector mesons, tensor mesons, and baryons~\cite{Chliapnikov:1999qi}.
For these particles $B_{i}$ was found to be a common factor between the groups, $B_{i} \equiv B \sim 150$ MeV~\cite{Chliapnikov:1999qi}.
Note the pseudoscalar mesons in QCD, however, which are lighter, follow a softer spectrum.

Following the above discussion, we shall construct a toy model for the baryonic particle yield from our string fragmentation. In order to retain some simplicity in our model we will consider all particles to share a common $B_i = m_* = g_* f$. In our toy model we consider $SU(N_{c})$ theories, with techniquarks in the fundamental representation, in which baryons will contain $N_{c}$ quarks. Mesons on the other hand will contain a techni-quark-anti-quark pair independent of $N_c$. In order to take into account the reduced probability of creating a baryon as opposed to a meson it is therefore suitable to include an additional suppression in the prefactor $A_{i} $ for baryons~\cite{Mitridate:2017oky}
\begin{equation}
p_{\mathcal{B}i} = \begin{dcases} \frac{1}{1+2^{N_{c}-1}/N_c}, \qquad ~\text{if $i$ is a baryon}, \\
1 , \qquad \qquad \qquad \qquad \text{if $i$ is a meson}.
\end{dcases}
\end{equation}
Other than this we take a common $A_{i} = p_{\mathcal{B}i}A$.
Applying energy conservation, we thus find the average number of  the composite state $i$ produced per string breaking to be
\begin{equation}
\langle N_{i} \rangle
\simeq  \, \frac{p_{\mathcal{B}i}(2J_i + 1) }{ \mathrm{Exp}\left[ M_{i}/m_\pi \right] } \, \left( \sum_k  \frac{p_{\mathcal{B}j} (2J_k + 1) }{ \mathrm{Exp}\left[ M_{k}/m_\pi \right] } \frac{M_k}{m_\pi} \right)^{-1} \langle N_\psi \rangle
\equiv   \mathrm{BR}_i \,\langle N_\psi \rangle,
\end{equation}
where the sum runs over all the states in the spectrum, and we remind that $\pi$ denotes the lightest composite state(s). In this case it is clearly possible to have a highly suppressed $\mathrm{BR}_i$, e.g. $\mathrm{BR}_i \simeq 10^{-6}$ for $m_i = m_* \simeq 4 \pi f$, $m_\pi \simeq f$, $N_c=10$, $N_\pi = 3$.

\end{subappendices}

%

\xintifboolexpr { \x = 2}
  {
  }
{
\medskip
\small
\bibliographystyle{JHEP}
\bibliography{thesis.bib}
}

%% file: chap8.tex
\chapterimage{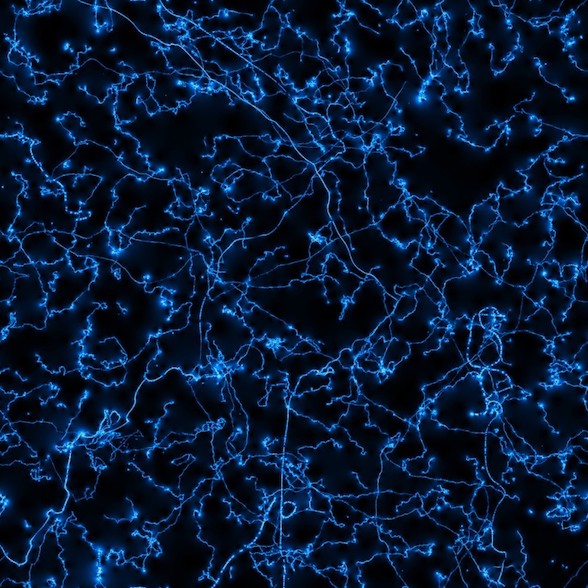} 

\chapter{Gravitational Waves from Cosmic Strings}
\label{chap:cosmic_strings}

\begin{tikzpicture}[remember picture,overlay]
\node[text width=18cm,text=black,minimum width=\paperwidth,
minimum height=7em,anchor=north]%
 at (7,4) {This chapter is based on \cite{Gouttenoire:2019kij}.};
 \draw (-2.0,2) -- (5,2);
\end{tikzpicture}

\section{Introduction}

The Standard Model of particle physics needs to be completed 
to address observational facts such as the matter antimatter asymmetry and the dark matter of the universe, as well as the origin of inflation. These, together with a number of other fundamental theoretical puzzles associated with e.g. the flavour structure of the matter sector and the ultra-violet properties of  the Higgs scalar field, motivate extensions of the Standard Model featuring new degrees of freedom and new energy scales. In turn, such new physics can substantially impact the expansion history in the early universe and leads to deviations  with respect to the standard cosmological model. Any deviations in the Friedmann equation occurring at  temperatures above the MeV remain to date  essentially unconstrained.

In the standard cosmological model,   primordial inflation is  followed by a long period of radiation domination until the more recent transitions to matter and then dark energy domination. Evidence for this picture comes primarily from observations of the Cosmic Microwave Background (CMB) and the successful predictions of Big-Bang Nucleosynthesis (BBN), which on the other hand, do not allow to test  cosmic temperatures above ${\cal O}$(MeV). 

An exciting prospect for deciphering the pre-BBN universe history and therefore high energy physics unaccessible by particle physics experiments, comes from the possible detection of a stochastic background of gravitational waves (SGWB), originating either from cosmological phase transitions, from cosmic strings or from inflation \cite{Caprini:2018mtu}.

Particularly interesting are cosmic strings (CS), which act as a long-lasting source of gravitational waves (GW) from the time of their production, presumably very early on, until today. The resulting frequency spectrum therefore encodes information from the almost entire cosmic history of our universe, and could possibly reveal precious details about the high energy particle physics responsible for a modified universe expansion.

There has been a large literature on probes of  a non-standard cosmology through the nearly-scale invariant primordial GW spectrum generated during inflation \cite{ Giovannini:1998bp, Giovannini:2009kg, Riazuelo:2000fc, Sahni:2001qp, Seto:2003kc, Tashiro:2003qp, Nakayama:2008ip, Nakayama:2008wy, Durrer:2011bi, Kuroyanagi:2011fy, Kuroyanagi:2018csn, Jinno:2012xb, Lasky:2015lej, Li:2016mmc, Saikawa:2018rcs, Caldwell:2018giq, Bernal:2019lpc, Figueroa:2019paj, DEramo:2019tit}. In contrast, little efforts have been invested to use the scale-invariant GW spectrum generated by CS  \cite{Cui:2017ufi, Cui:2018rwi, Auclair:2019wcv, Guedes:2018afo, Ramberg:2019dgi, Chang:2019mza}  while there has been intense activity on working out predictions for the SGWB produced by CS in standard cosmology \cite{Vilenkin:1981bx, Hogan:1984is, Vachaspati:1984gt, Accetta:1988bg, Bennett:1990ry, Caldwell:1991jj, Allen:1991bk, Battye:1997ji, DePies:2007bm, Siemens:2006yp, Olmez:2010bi, Regimbau:2011bm, Sanidas:2012ee, Sanidas:2012tf, Binetruy:2012ze, Kuroyanagi:2012wm, Kuroyanagi:2012jf,Sousa:2016ggw,Sousa:2020sxs,Figueroa:2020lvo,Hindmarsh:2021vih}.

In this study, we propose to use the detection of a SGWB from local cosmic strings to test the 
existence of alternative stages of cosmological expansion between the end of inflation and the end of the radiation era.
Particularly well-motivated is a stage of early-matter domination era induced by  a heavy cold particle dominating the universe and decaying before BBN. Another possibility is a stage of kination triggered by the fast rolling evolution of a scalar field down its potential, e.g. \cite{Spokoiny:1993kt, Joyce:1996cp} for the pioneering articles. See \cite{Gouttenoire:2021wzu,Gouttenoire:2021jhk} for a detailed discussion of kination occuring during the standard radiation era. Finally, supercooled confining phase transitions \cite{Creminelli:2001th, Randall:2006py,Nardini:2007me,Konstandin:2011dr,vonHarling:2017yew,Iso:2017uuu,Bruggisser:2018mrt,Baratella:2018pxi,Agashe:2019lhy,DelleRose:2019pgi,vonHarling:2019gme,Baldes:2020kam,Baldes:2021aph} can induce some late short stages of inflation inside a radiation era.  The latter were motivated at the TeV scale but the properties of the class of scalar potentials naturally inducing a short  inflationary era can be applied to any other scale. We will consider these various possibilities and their imprints on the GW spectrum from cosmic strings.

The dominant source of GW emission from a cosmic string network comes from loops which are continuously formed during the network fragmentation.
We thus primarily need to compute the loop-production efficiency during the non-standard  transition eras.
This is crucial for a precise prediction of the turning-point frequency as a signature of the non-standard era.
The temperature of the universe at the end of the non-standard era can be deduced from the measurement of these turning point frequencies.

The observational prospects for measuring the SGWB from cosmic string networks at LISA was recently reviewed in \cite{Auclair:2019wcv}.
Besides, the  effect  of particle production on the loop distribution and thus on the SGWB was recently discussed \cite{Matsunami:2019fss,Auclair:2019jip} where it was however concluded that the expected cutoff is outside the range of current and planned detectors (see also \cite{Kawasaki:2011dp}). 
Our study integrates these  recent developments and goes beyond in  several directions:

\begin{itemize}

\item We go beyond the so-called  scaling regime by computing  the time evolution of the string network parameters (long string mean velocity and correlation length) and thus the loop-production efficiency during modifications of the equation of state of the universe. Including these transient effects results in a turning-point frequency smaller by ${\cal O}$(20) compared to the prediction from the scaling regime.\footnote{ The turning-point frequency can even be smaller by ${\cal O}$(400) if in a far-future, a precision of the order of $1\%$ can be reached in the measurement of the SGWB, cf. Eq.~\eqref{turning_point_general_scaling_app}.} As a result, the energy scale of the universe associated with the departure from the standard radiation era that can be probed is correspondingly larger than the one predicted from scaling networks.

\item We investigate a large variety of non-standard cosmologies, in particular models where a non-standard era can be rather short inside the radiation era, due for instance to some cold particle temporarily dominating the energy density (short matter era, see Fig.~\ref{intermediate_matter_spectrum}) \cite{Gouttenoire:2019rtn} or some very short stage of inflation (for a couple of efolds) due to a high-scale supercooled confinement phase transition,  see Fig.~\ref{figure_spect_inflation}.
Such inflationary stages occurring at scales up to $10^{14}$ GeV could be probed, see Fig.~\ref{fig:contour_power_inflation2}. Even 1 or 2 e-folds could lead to observable features, see Fig.~\ref{figure_spect_inflation}. The study of GW signatures of cosmological histories containing a kination era is given in \cite{Gouttenoire:2021wzu,Gouttenoire:2021jhk}.

\item  We also consider longer low-scale inflation models. For instance,  an intermediate inflationary era lasting for ${\cal O}(10)$ efolds, the SGWB from cosmic strings completely looses its scale invariant shape and has a peak structure instead,  see Fig.~\ref{fig:beautifulpeaks}.  A TeV scale inflation era can lead to a broad peak either in the LISA or BBO band or even close to the SKA band, depending on the precise value of the string tensions $G\mu$, and the number of efolds $N_e$.

\item We include high-frequency cutoff effects from particle production which can limit observations for small value of the string tension $G\mu \lesssim 10^{-15}$ and high-frequency cutoff from thermal friction,  see Fig.~\ref{sketch_scaling} and top left panel of Fig.~\ref{ST_vos_scaling}.

{  \item We discuss how to read information about the small-scale structure of CS from the high-frequency tail of the GW spectrum, see App.~\ref{sec:study_impact_mode_nbr}.}

\item We discuss the comparison between local and global string networks, see App.~\ref{app:global_strings}.

\end{itemize}

The plan of the chapter is the following.
 In Sec.~\ref{cs_chapter}, we recap the key features of CS networks, their cosmological evolution, decay channels and the pulsar timing array constraints on the string tension. 
 Sec.~\ref{sec:GWCS} reviews the computation of the SGWB from Nambu-Goto CS. We first discuss the underlying assumptions on the small-scale structure and  on the loop distribution and then derive the master formula of the GW frequency spectrum.
 An important discussion concerns the non-trivial frequency-temperature relation and how it depends on the cosmological scenario. 
 Sec.~\ref{sec:VOS} is devoted to the derivation  of the loop production efficiency beyond the scaling regime, taking into account transient effects from the change in the equation of state of the universe. We apply this to predict the SGWB in the standard cosmological model in Sec.~\ref{sec:standard}. We then move to discuss non-standard cosmological histories, a short intermediate matter era inside the radiation era in  Sec.~\ref{sec:interm_matter}, and an intermediate inflationary era in Sec.~\ref{sec:inflation}.
 We discuss the specific spectral features in each of these cases and their observability by future instruments. 
 We conclude in Sec.~\ref{sec:conclusion}.
 Additional details are moved to appendices, such as non-GW constraints on the string tension $G\mu$ in App.~\ref{app:phenoCS}, a step-by-step derivation of the GW spectrum as well as the values of its slopes in App.~\ref{app:derivationGWspectrum}, the formulae of the various turning-point frequencies in 
App.~\ref{derive_turning_points},  the derivation of the equations which govern the evolution of the long-string network in the Velocity-dependent One-Scale (VOS) model in App.~\ref{sec:VOS_proof},  a discussion of the extensions to the original VOS model in App.~\ref{app:VOScalibration}, 
  the prediction of the GW spectrum from global strings in App.~\ref{app:global_strings}, and the calculation of the integrated power-law sensitivity curves for each experiment in App.~\ref{app:sensitivity_curves}.

\section{Recap on Cosmic Strings}\label{cs_chapter}
Cosmic strings have been the subject of numerous studies since the pioneering paper \cite{Kibble:1976sj}, see  \cite{Hindmarsh:1994re, Vilenkin:2000jqa, Vachaspati:2015cma} for reviews. 

\subsection{Microscopic origin of Cosmic Strings}
\label{intro_cs}

\paragraph{Fundamental objects:} A cosmic string is an extended object of cosmological size. It can be a fundamental object in superstring theories \cite{Witten:1985fp, Dvali:2003zj,Copeland:2003bj, Polchinski:2004ia, Sakellariadou:2008ie, Davis:2008dj, Sakellariadou:2009ev, Copeland:2009ga}. In superstring theory, strings are the fundamental mandatory new degrees of freedom, which are usually unstable and decay before they could stretch to cosmological scales. However, the so-called $F$-, $D$-, and $(p,q)$-strings can grow large and have cosmological consequences. 
\paragraph{Topological defects:} Cosmic strings can also be topological solitons in classical field theories. It is a classical field configuration which arises whenever there is a symmetry breaking $G\rightarrow H$ with a non-trivial homotopy group\footnote{The existence of other topological defects, domain walls, monopoles, or textures also relies on the non triviality of an homotopy group, $\pi_n(G/H)\neq \textrm{Id}$ for $n=0,2$ or $3$ respectively.} $\pi_1(G/H)\neq \textrm{Id}$, where `$\textrm{Id}$' denotes the identity map, i.e. one can find at least two non-equivalent maps from the circle to the vacuum manifold $G/H$. For example, any theory with spontaneous breaking of a $U(1)$ symmetry \cite{Buchmuller:2013lra, Dror:2019syi} has a string solution, since $\pi_1(U(1)) = \mathbb{Z}$. 
More complex vacuum manifolds with string solutions can appear in various grand unified theories \cite{Jeannerot:2003qv, Sakellariadou:2007bv}, e.g. $SO(10) \rightarrow SO(5)\times \mathbb{Z}_2$.
However, we emphasize that there is no string solution in the Standard Model (SM) because 
the symmetry breaking pattern 
\begin{equation}
SU(2)_L\times U(1)_Y\rightarrow U(1)_\textrm{EM},
\end{equation}
has a trivial homotopy group\footnote{The homotopy group of a coset can be easily computed via the so-called fundamental theorem: $\pi_n(G/H)\cong\pi_{n-1}(H)$ \cite{Vilenkin:2000jqa}.}
\begin{equation}
\pi_1(SU(2)\times U(1)/U(1))\cong \pi_0(U(1))\cong \textrm{Id}.
\end{equation}
Thus, the existence of cosmic strings requires physics Beyond the Standard Model (BSM). In this study, we will assume a generic BSM scenario giving rise to cosmic strings without specifying a particular model.
\paragraph{Abelian-Higgs string:} We now present the standard example of field theories with a string-liked solution: the abelian-Higgs (AH) model. It is a field theory with a complex scalar field $\phi$ charged under a $U(1)$ gauge interaction mediated by a gauge field $A_\mu$. Note that the symmetry can also be global in the absence of the gauge boson. The resulting strings solutions corresponding to local and global symmetries are called local and global strings, respectively. The AH lagrangian in $d+1$ space-time dimensions is described by 
\begin{equation}
\mathcal{L}=-\frac{1}{4}F_{\mu\nu}^2+|D_\mu\phi|^2-\frac{1}{4}\lambda(|\phi|^2-\eta^2)^2\hspace{2em}\textrm{for }\mu,\nu=0,1,2,\dotsc,d\textrm{ },
\end{equation}
where $D_\mu = \partial_\mu -i\,e\,A_\mu$ and where $\lambda$ and $\eta$ are parameters of the scalar potential generating the spontaneous symmetry breaking (SSB). $\eta$ is the vacuum expectation value (VEV) of the scalar field. The energy of the system can be derived as the sum of the energy in the scalar and gauge field
\begin{equation}
E=\int d^{d}x \left(-\frac{1}{4}F_{\mu\nu}^2+|D_\mu\phi|^2+V(\phi)\right).
\end{equation}
It has been shown that a stable field configuration with finite energy density exists, known as the Nielsen-Olesen vortex \cite{Nielsen:1973cs} which are lines where the scalar field sits on the top of its `mexican hat potential' $V(\phi)$ and approaches its VEV at large distance. When following a closed path around the string, the phase of the complex scalar field returns to its original value after winding around the mexican an integer $n$ number of times. Hence, the angular component of the gradient of the scalar field diverges at large distance
\begin{equation}
\int_{\delta}^{R} d^2 r   \left| \frac{1}{r}\frac{\partial \phi }{ d \theta} \right|^2 =  \int 2\pi r d r \, n \frac{\eta^2}{r^2} = 2 \pi \eta^2\, n \log{\frac{R}{\delta}} \underset{R\to \infty}{\longrightarrow} \infty
\end{equation}
where $\delta \sim m_{\phi}^{-1}=(\sqrt{\lambda} \eta)^{-1}$ or $m_{A}^{-1}=(\sqrt{2} e\eta)^{-1}$ is the string core width. $m_{\phi}^{-1}$ and $m_{A}^{-1}$ are the length scales below which the scalar and gauge field gradients are respectively confined. For local strings, the divergence can be exactly canceled by a judicious gauge choice. However, for global strings, which are a particular realization of the AH model in the absence of gauge field or in the presence of a vanishing gauge coupling $e=0$, the logarithmic dependence of the energy with the infrared cut-off can not be avoided. Physically, the logarithm divergence is due to the existence of a long-range interaction mediated by the massless Goldstone mode (which is the complex phase of $\phi$). Therefore, the energy per unit of length, also known as the `string tension' reads
\begin{equation}
\mu\approx2\pi \eta^2\,n\,\times
\begin{cases}
1&\hspace{2em}\textrm{for local strings},\\
\ln\left(\frac{R}{\delta}\right)&\hspace{2em}\textrm{for global strings},\\
\end{cases}
\label{tension_string_exp}
\end{equation}
where the infrared cut-off $R$ can be either the string inter-distance or the causal horizon $H^{-1}$ and where the string core width $\delta$ is given by the inverse mass of the radial mode $m_{\phi}^{-1}$. The interaction between strings receives an attractive contribution from the vortex core-overlap and a repulsive contribution (attractive for vortex-anti-vortex interactions) from the circulation of the magnetic lines outside the vertex core. Strings where the attractive interaction dominates are called `type I` whereas the other ones are called `type II', in analogy with vortices in superconductors. Type II with winding number $n>1$ are unstable against decay to $n'<n$. Conversely, type I strings with high winding numbers are stable, but it has been shown in numerical simulations that string interactions could quickly decrease the winding number via `peeling' \cite{Laguna:1989hn}. 
\paragraph{Nambu-Goto string (local):} Masses of the underlying fields determine the size of the core width. However, for local strings this can be rather small compared to the cosmological scale $H^{-1}$ such that an observer is unable to resolve any physics within the core. Hence, it seems reasonable to consider only the effective description of theories in the zero-width limit. For the local abelian-Higgs model, the appropriate effective action is the Nambu-Goto (NG) action \cite{Vilenkin:2000jqa,Maeda:1987pd,Forster:1974ga}, which strictly speaking is defined by the area of the world-sheet swept out by the motion of a featureless one-dimensional object,
\begin{equation}
\mathcal{S}_\textrm{NG}=-\mu\int dt dl \left(1 - v_{\perp}^2  \right)^{1/2},
\end{equation}
where $d\tau =  \left(1 - v_{\perp }\right)^2 dt $ is the proper time of a string segment of length $dl$. It can also be written in a covariant form
\begin{equation}
\mathcal{S}_\textrm{NG}= -\mu\int d^2 \zeta \sqrt{-\gamma},
\label{NG_action}
\end{equation}
where $\gamma \equiv \det{\gamma_{ab}}$ is the determinant of the metric $\gamma_{ab}$ induced on the 2D world-sheet, starting from the 4D bulk metric $g_{\mu \nu}$,
\begin{equation}
\gamma_{ab} = g_{\mu \nu} \frac{\partial x^{\mu}}{\partial \zeta^a} \frac{\partial x^{\nu}}{\partial \zeta^b,}.
\end{equation}
The coordinates $x^\mu(\zeta_a)$, with $\mu =1,\,2,\,3,\,4$ and $a =1,\,2$, describe the embedding of the 2D world-sheet in the 4D space-time. 
\paragraph{Nambu-Goto string (global):} However, global strings whose core extends due to the long-range interaction mediated by the massless Goldstone boson $\theta$, complex phase of the scalar field $\phi = \eta\, e^{i\theta}$, can not be completely described by the NG action. One must account for the interaction between the string and the Golstone field, the so-called Kalb-Ramond action \cite{Witten:1985fp,Vilenkin:1986ku}
\begin{equation}
\mathcal{S}_\textrm{KR}= -\mu\int d^2 \zeta \sqrt{-\gamma}+\frac{1}{6}\int H^2 d^4x + 2\pi \eta \int B_{\mu\nu} d\sigma ^{\mu\nu},
\end{equation}
where $B_{\mu\nu}$ is the antisymmetric field, defined as the hodge-dual of the Goldstone boson
\begin{equation}
 \frac{1}{2} \epsilon_{\mu \nu \lambda \rho} \partial^{\nu} B^{\lambda \rho} \equiv \eta \, \partial_\mu \mathcal{\theta},
\end{equation} 
$H^{\mu\nu\lambda} \equiv \partial^{\mu}B^{\nu\lambda}+\partial^{\lambda}B^{\mu\nu}+\partial^{\nu}B^{\lambda\mu}$ is the field strength tensor of $B_{\mu\nu}$ and $d\sigma^{\mu\nu} \equiv \epsilon^{a b}x^{\mu}_{,a}x^{\nu}_{,b}d^2\zeta$ is the world-sheet area element.

\subsection{Cosmic-string network formation and evolution}

\paragraph{Kibble mechanism:} The formation of cosmic strings occurs during a cosmological phase transition associated with spontaneous symmetry breaking, occurring at a temperature, approximately given by the VEV acquired by the scalar field
\begin{equation}
\label{eq:network_formation}
T_{\rm p}\sim 10^{11} ~ \text{GeV}  \left( \frac{G\mu}{10^{-15}} \right)^{1/2}.
\end{equation} 
CS are randomly distributed and form a network characterized by its correlation length $L$, which can be defined as
\begin{equation}
L \equiv \sqrt{\mu / \rho_\infty},
\end{equation}
where $\mu$ is the string tension, the energy per unit length, and  $\rho_\infty$ is the energy density of long strings. More precisely, long strings form infinite random walks \cite{Scherrer:1986sw} which can be visualized as collections of segments of length $L$.

\paragraph{Loop chopping:}  Each time two segments of a long string cross each other, they inter-commute, with a probability $P$ and form a loop. Loop formation is the main energy-loss mechanism of the long string network.
In numerical simulations \cite{Shellard:1987bv} and analytical modelling \cite{Eto:2006db}, the probability of inter-commutation has been found to be $P=1$ but in some models it can be lower. This is the case of models with extra-dimensions \cite{Dvali:2003zj, Jackson:2004zg}, strings with junctions \cite{Copeland:2006if} or peeling \cite{Laguna:1989hn}, or the case of highly relativistic strings \cite{Achucarro:2006es}. 
\begin{figure}[]
				\centering
				\includegraphics[width=\textwidth]{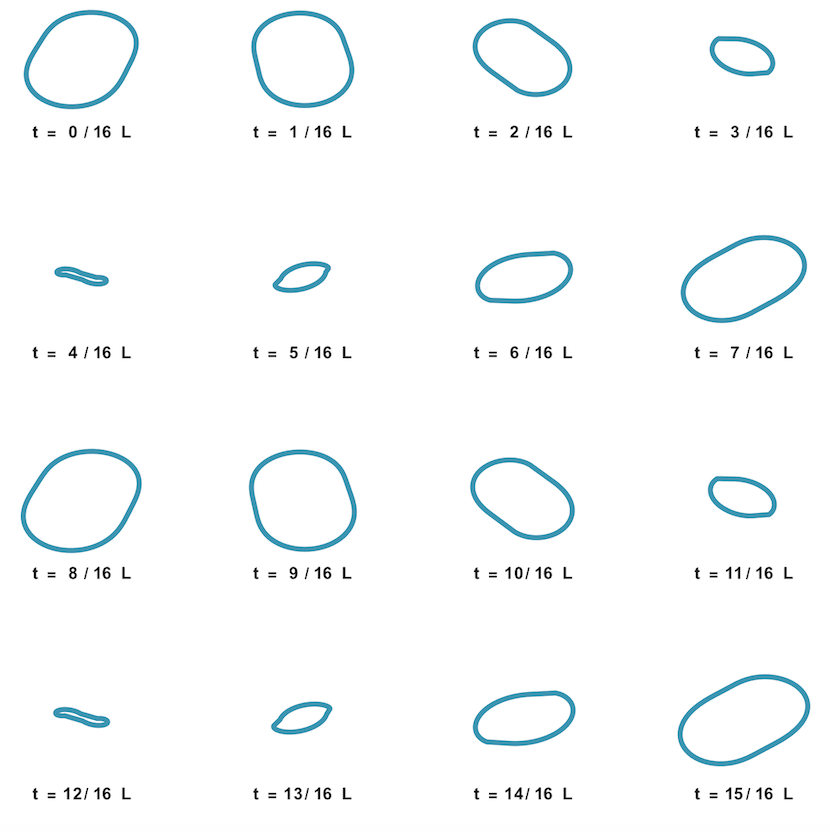}
				\caption{\it \small \label{fig:loopTraj} Loop trajectory found after numerically integrating (c++) the Nambu-Goto equation in flat space. The loop initial conditions correspond to the parameters $\phi =0$ and $\kappa =0.5$ in Sec.~6.2.4 of \cite{Vilenkin:2000jqa}. Under the effect of its tension, the loop oscillates: it contracts while converting its potential energy into kinetic energy and vice-versa. We can see that the loop oscillation period is $L/2$. This figure is not included in \cite{Gouttenoire:2019kij} but instead it has been realized during the redaction of the manuscript.}
\end{figure}

\paragraph{Scaling regime:} Just after the network is formed, the strings may interact strongly with the thermal plasma such that their motion is damped. When the damping stops, cosmic strings oscillate and enter the phase of \textit{scaling} evolution.
During this phase, the network experiences two competing dynamics:
\begin{enumerate}
\item Hubble stretching: the correlation length scale stretches due to the cosmic expansion, $L\sim a$.
\item Fragmentation of long strings into loops: a loop is formed after each segment crossing. Right after their formation, loops evolve independently of the network, they oscillate under the effect of the tension, see Fig.~\ref{fig:loopTraj} and start to decay through gravitational radiation and/or particle production. 
\end{enumerate}
It is known since a long time ago \cite{Kibble:1984hp, Bennett:1987vf, Bennett:1989ak, Albrecht:1989mk, Allen:1990tv}, that out of the two competing dynamics, Hubble expansion and loop fragmentation, there is an attractor solution, called the \textit{scaling regime}, where the correlation length scales as the cosmic time, 
\begin{equation}
L\sim t .
\label{eq:scaling_def}
\end{equation}
Note however that in the case of global-string network, it has been claimed that the scaling property in Eq.~\eqref{eq:scaling_def}, is logarithmically violated due to the dependence of the string tension on the Hubble horizon \cite{Klaer:2017ond, Gorghetto:2018myk,Gorghetto:2020qws, Gorghetto:2021fsn, Kawasaki:2018bzv, Vaquero:2018tib, Buschmann:2019icd, Martins:2018dqg}. Notet that an opposite conclusion has been drawn in \cite{Hindmarsh:2019csc,Hindmarsh:2021vih}.

\paragraph{Proof:} We know show that the scaling regime in Eq.~\eqref{eq:scaling_def} is an attractor solution, following the approach of \cite{Vilenkin:2000jqa}. We start by evaluating the loop chopping rate. In a correlation volume $L^3$, a segment of length $L$ must travel a distance $L$ before encountering another segment. So the collision rate, per unit of volume, is $\tfrac{v}{L} \cdot \tfrac{1}{L^3} \sim \tfrac{1}{L^4}$. At each collision forming a loop, the network looses a loop energy $\mu \, L$. Hence, the master equation for the energy density $\rho_\infty \sim \mu/L^2$ reads
\begin{equation}
\dt{\rho}_\infty = -2\frac{\dot{a}}{a} \rho_\infty - \frac{1}{L^4}\cdot \mu \, L.
\end{equation}
Plugging the ansatz $L = \xi t$ in the above equation leads to
\begin{equation}
\dt{\xi} =  \frac{1}{2t} \left( 1- \xi  \right).
\end{equation}
Clearly $\xi =1$ is a fix point. If the density of long string is too high, $\xi <1$, then one has $\dt{\xi}>0$ and the loop production becomes more efficient. If the density of long string is too low $\xi >1$, then one has $\dt{\xi}<0$ and the loop production becomes less efficient. 
The exact coefficient $\xi = L/t$ can be computed numerically or analytically, e.g. within the Velocity-dependent One-Scale model presented in sec.~\ref{sec:VOS}. 

\paragraph{Number of strings:} During the scaling regime, the number of strings per Hubble patch is conserved
\begin{equation}
\frac{\rho_{\infty} H^{-3}}{\mu L} = \rm constant.
\end{equation}
Moreover, the energy density of the long-string network, which scales as $\rho_\infty\sim  \mu/t^2$, has the same equation-of-state as the main background cosmological fluid $\rho_\textrm{bkg}\sim a^{-n}$, 
\begin{equation}
\frac{\rho_\infty}{\rho_\textrm{bkg}}\sim \frac{a^{n}}{t^2} \sim \textrm{constant},
\end{equation}
where we used $a=t^{2/n}$. Hence, the long-string energy density redshifts as matter during matter domination and as radiation during radiation domination. The scaling regime allows cosmic strings not to dominate the energy density of the universe, unlike other topological defects.
The scaling property of a string network has been checked some fifteen years ago in numerical Nambu-Goto simulations \cite{Ringeval:2005kr, Vanchurin:2005pa, Martins:2005es, Olum:2006ix} and more recently with larger simulations \cite{BlancoPillado:2011dq}. 
During the scaling regime, the loop production function is scale-free, with a power-law shape, meaning that loops are produced at any size between the Hubble horizon $t$ and the scale $\sim \Gamma \, G \mu \, t,$ below which the strings have been smoothened by the gravitational backreaction and there is no further segment crossing.
\paragraph{A scale-invariant  SGWB:} An essential outcome is the scale-invariance of the Stochastic GW Background generated by loops during the scaling regime \cite{Vilenkin:1981bx, Hogan:1984is, Vachaspati:1984gt, Accetta:1988bg, Bennett:1990ry, Caldwell:1991jj,Allen:1991bk, Battye:1997ji, DePies:2007bm, Siemens:2006yp, Olmez:2010bi, Regimbau:2011bm, Sanidas:2012ee, Sanidas:2012tf, Binetruy:2012ze, Kuroyanagi:2012wm, Kuroyanagi:2012jf}.  We construct the GW spectrum in Sec.~\ref{sec:SGWB} and give more details in App.~\ref{app:derivationGWspectrum}. Remarkably, the spectrum generated by loops produced during radiation domination is flat, $\propto f^0$, whereas an early matter domination or an early kination-domination era turns the spectral index from $f^0$ to respectively $f^{-1/3}$ or $f^1$. As recently pointed out by \cite{Blasi:2020wpy}, in the presence of an early matter, the slope $f^{-1}$ predicted by \cite{Cui:2017ufi, Cui:2018rwi}, is changed to $f^{-1/3}$ due to the high-k modes. We give more details on the impact of high-k modes on the GW spectrum in the presence of a decreasing slope due to an early matter era, a second period of inflation, particle production, thermal friction or network formation in App.~\ref{sec:study_impact_mode_nbr}.  Hence, the detection of the SGWB from CS by LIGO \cite{Aasi:2014mqd}, DECIGO, BBO \cite{Yagi:2011wg}, LISA \cite{Audley:2017drz,LISACosmologyWorkingGroup:2022jok}, Einstein Telescope \cite{Hild:2010id, Punturo:2010zz} or Cosmic Explorer \cite{Evans:2016mbw} would offer an unique observation window on the  equation of state of the Universe at the time when the CS loops responsible for the detected GW are formed.

\subsection{Decay channels of Cosmic Strings}	

Cosmic strings can decay in several ways, as we discuss below.

\paragraph{GW radiation from long strings:}
Because of their topological nature,  straight infinitely-long strings are stable against decay. However, 
small-scale structures of wiggly long strings can generate gravitational radiation. Intuitively, a highly wiggly string can act as a gas of small loops. The GW emission from long strings can be neglected compared to the GW emission from loops, as loops live much longer than a Hubble time \cite{Allen:1991bk, Vilenkin:2000jqa}. Indeed, the GW signal emitted by loops is enhanced by the large number of loops (continuously produced).  Nambu-Goto numerical simulations have shown that the loop energy density is at least $100$ times larger than the long-string energy density \cite{Blanco-Pillado:2013qja}. Only for global strings where loops are short-lived due to efficient Goldstone production, the GW emission from long strings can give a major contribution to the SGWB \cite{Krauss:1991qu, JonesSmith:2007ne, Fenu:2009qf, Figueroa:2012kw}. In what follows, we only consider the emission from loops.
	
\begin{figure}[]
				\centering
				\includegraphics[width=9.5cm]{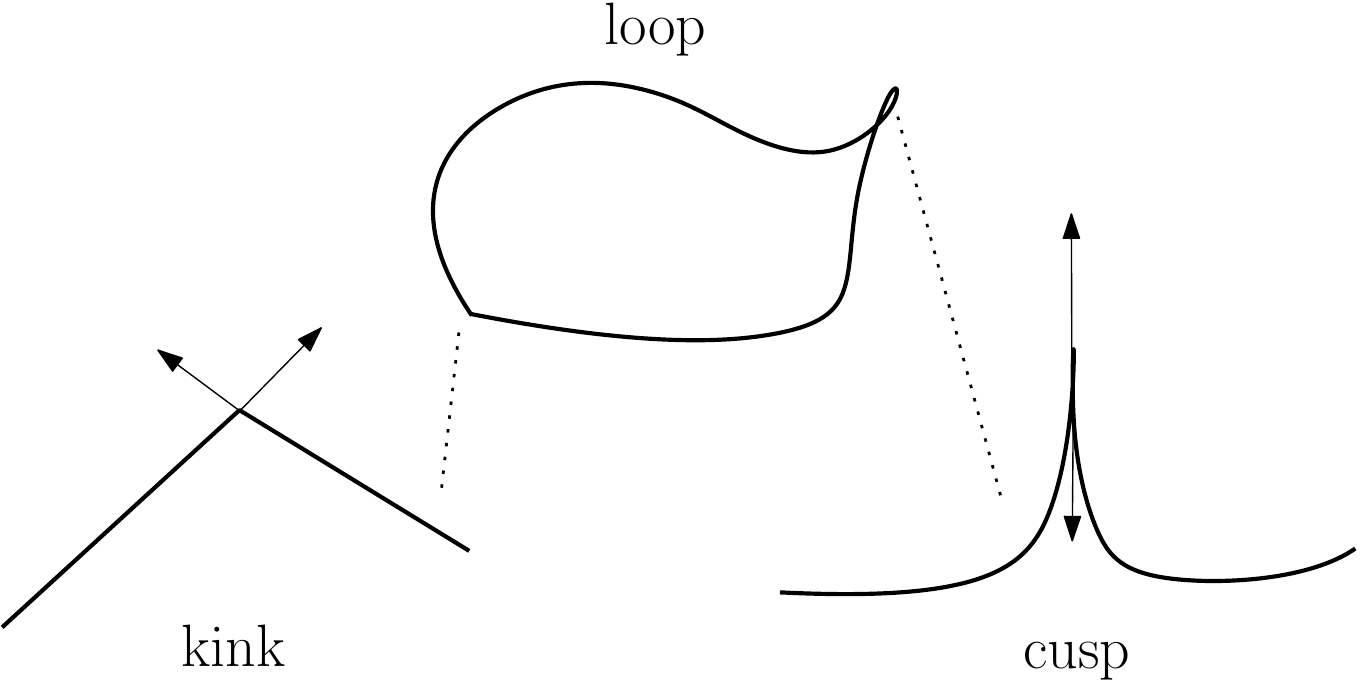}
				\caption{\it \small Cartoon showing the geometry of a kink and a cusp which are singular structures formed on loops. The arrows denote the tangent vectors of the string segments.}
				\label{kink_cusp_cartoon}
\end{figure}

\paragraph{GW radiation from loops (local strings):} 
\label{sec:masslessradiation}
In contrast to long strings, loops do not contain any topological charge and are free to decay into GW. The GW radiation power is found to be \cite{Vilenkin:2000jqa}
\begin{equation}
P_\textrm{GW}=\Gamma G \mu^2,
\label{eq:power_GW_0}
\end{equation}
where the total GW emission efficiency $\Gamma$ is determined from Nambu-Goto simulations, $\Gamma\simeq 50$ \cite{Blanco-Pillado:2017rnf}. Note that the gravitational power radiated by a loop is independent of its length. This can be understood from the quadrupole formula $P = G/5 (Q''')^2$ \cite{maggiore2008gravitational,Vachaspati:1984gt} where the triple time derivative of the quadrupole, $Q''' \propto {\rm mass \, (length)^2 / (time) ^3 \propto \mu}$, is indeed independent of the length. 
The resulting GW are emitted at frequencies \cite{Kibble:1982cb,Hindmarsh:1994re}
\begin{equation}
\label{eq:GWspecfreq_0}
\tilde{f}=\frac{2k}{l}, \qquad  k\in\mathbb{Z}^{+},
\end{equation}  
corresponding to the proper modes $k$ of the loop.  The tilde is used to distinguished the frequency emitted at $\tilde{t}$ from the frequency today 
\begin{equation}
f = a(\tilde{t})/a(t_0)~ \tilde{f}.
\end{equation}
 The frequency dependence of the power spectrum $P_\textrm{GW}(k)$ relies on the nature of the loop small-scale structures \cite{ Damour:2001bk, Ringeval:2017eww}, e.g. kinks or cusps, cf. Fig.~\ref{kink_cusp_cartoon}.
More precisely, the spectrum of the gravitational power emitted from one loop reads
\begin{equation}
\label{eq:one-loop-spectrum}
P_\textrm{GW}^{(k)}= \Gamma^{(k)} G \mu^2, \qquad \text{with} \quad \Gamma^{(k)} \equiv \frac{\Gamma \, k^{-n} }{ \sum_{p=1}^{\infty}p^{-n}} \ , \  \ n= 
\begin{cases}
4/3 & \mbox{cusps } \\
5/3 & \mbox{kinks} \\
2 &  \mbox{kink-kink collisions }\\
\end{cases}
\end{equation}
where the spectral index $n={4/3}$ when the small-scale structure is dominated by cusps \cite{Vachaspati:1984gt, Burden:1985md,Olmez:2010bi}, $n={5/3}$ for kink domination \cite{Olmez:2010bi}, or $n={2}$ for kink-kink collision domination \cite{Damour:2001bk, Ringeval:2017eww}. { A discussion on how to read information about the small-scale structure of CS from the GW spectrum, is given in App.~\ref{sec:study_impact_mode_nbr}. In particular, we show that the high-frequency slope of the GW spectrum in the presence of an early matter era, a second period inflation, particle production or network formation, which is expected to be $f^{-1}$ from the fundamental, $k=1$, GW spectrum alone, is actually given by $f^{1-n}$.}
Immediately after a loop gets created, at time $t_i$ with a length $\alpha\,t_i$, its length $l(\tilde{t})$ shrinks through emission of GW with a rate $\Gamma G \mu$
\begin{equation}
\label{eq:CSlength0}
l(\tilde{t}) = \alpha t_i -\Gamma G \mu(\tilde{t}-t_i).
\end{equation}
Consequently, the string lifetime due to decay into GW is given by 
\begin{equation}
\label{eq:GWlifetime}
\tau_{\rm GW} = \frac{\alpha \, t_i }{\Gamma G\mu}.
\end{equation}
The superposition of the GW emitted from all the loops formed since the creation of the long-string network generates a Stochastic GW Background. 
Also, cusp formations can emit high-frequency, short-lasting GW bursts \cite{Damour:2000wa, Damour:2001bk, Siemens:2006yp, Olmez:2010bi, Ringeval:2017eww}. If the rate of such events is lower than their frequency, they might be subtracted from the SGWB.

\paragraph{Goldstone boson radiation (global strings):} For global strings, the massless Goldstone  particle production is the main decay channel. The radiation power has been estimated \cite{Vilenkin:2000jqa}
\begin{equation}
P_{\rm Gold}=\Gamma_{\rm Gold}\, \eta^2,
\label{eq:power_goldstone}
\end{equation}
where $\eta$ is the scalar field VEV and $\Gamma_{\rm Gold} \approx 65$ \cite{Vilenkin:1986ku, Chang:2019mza}. We see that the GW emission power in Eq.~\eqref{eq:power_GW_0} is suppressed by a factor $G\mu$ with respect to the Goldstone emission power in Eq.~\eqref{eq:power_goldstone}. Therefore, for global strings, the loops decay into Goldtone bosons after a few oscillations before having the time to emit much GW \cite{Vilenkin:2000jqa, Saurabh:2020pqe}. However, as shown in App.~\ref{app:global_strings}, the SGWB from global string is detectable for large values of the string scale, $\eta \gtrsim 10^{14}$~GeV. Other recent studies of GW spectrum from global strings in standard and non-standard cosmology include \cite{Bettoni:2018pbl, Ramberg:2019dgi, Chang:2019mza,Gorghetto:2021fsn}. 

A well-motivated example of global string is the axion string coming from the breaking of a $U(1)$ Peccei-Quinn symmetry \cite{Vilenkin:1982ks, Vilenkin:1984ib, Vilenkin:1986ku, Sikivie:1982qv}. Its phenomenological interest relies on the possibility to generate the correct Dark Matter abundance from Goldstone production. However, the value of the energy of the emitted Goldstone, needed to compute the axion relic abundance, is still matter of debate \cite{Gorghetto:2018myk,Gorghetto:2020qws}. Particularly, the precise dependence of the emitted mean energy with the infrared cut-off $H^{-1}$ is unknown and is not tractable in numerical simulations due to the too large hierarchy between the string core size and the Hubble horizon. Hence, in the case of post-inflationnary Peccei-Quinn breaking, it is currently not understood whether the axion abundance today coming from the decay of the axionic string network supersedes the abundance coming from the vacuum misalignement or not. Note, however 
\cite{Hindmarsh:2019csc} which claims that the dependence of the emitted energy with the cut-off is an artefact. Note also the possibility that $1-10\%$ of the axion abundance would come from heavy-radial-mode production followed by its decay, and then would be hot \cite{Saurabh:2020pqe}. Ref.~\cite{Ramberg:2019dgi}  shows the detectability of the GW from the axionic network of QCD axion Dark Matter (DM), after introducing an early-matter era which dilutes the axion DM abundance and increases the corresponding Peccei-Quinn scale $\eta$.

\paragraph{Massive particle radiation:} 
When the string curvature size is larger than the string thickness, one expects the quantum field nature of the CS, like the possibility to radiate massive particles, to give negligible effects and one may instead consider the CS as an infinitely thin 1-dimensional classical object with tension $\mu$: the Nambu-Goto (NG) string. 
However, due to the presence of small-scale structures on the strings, regions with curvature comparable to the string core size can develop and the Nambu-Goto  approximation breaks down. In that case, massive radiation can be emitted during processes known as cusp annihilation \cite{BlancoPillado:1998bv} or kink-kink collisions \cite{Matsunami:2019fss}. We discuss massive particle emission in more details in Sec.~\ref{sec:massive_radiation}.

\subsection{Constraints on  the string tension $G \mu$ from GW emission}
\label{subsec:Gmuconstraints}

The observational signatures of Nambu-Goto cosmic strings are mainly gravitational. The GW emission can be probed by current and future pulsar timing arrays and GW interferometers, while the static gravitational field around the string can be probed by CMB, 21 cm, and lensing observables, see app.~\ref{app:phenoCS} for more details on non-GW probes.
The strongest constraints come from pulsar timing array EPTA, $G \mu \lesssim 8 \times 10^{-10}$ \cite{Lentati:2015qwp}, and NANOGrav,  $G \mu \lesssim 5.3 \times 10^{-11}$ \cite{Arzoumanian:2018saf}. Comparison  with the theoretical predictions from the SGWB from cosmic strings leads to  $G\mu \lesssim 2 \times 10^{-11}$ 
\cite{Blanco-Pillado:2017rnf, Cui:2018rwi} or $G\mu \lesssim 10^{-10}$ \cite{Ringeval:2017eww}, even though it can be relaxed to $G\mu \lesssim 5 \times 10^{-7}$ \cite{Sanidas:2012ee}, after taking into account uncertainties on the loop size at formation and on the number of emitting modes. Note that it can also be strengthened by decreasing the inter-commutation probability \cite{Sakellariadou:2004wq, Damour:2004kw, Binetruy:2012ze}. 

By using the EPTA sensitivity curve derived in \cite{Breitbach:2018ddu},
we obtain the upper bound on $G\mu$, one order of magnitude higher, $2 \times 10^{-10}$, instead of $2 \times 10^{-11}$, cf. Fig.~\ref{ST_vos_scaling}.  This bound 
becomes {$\sim5 \times 10^{-11}$} by using the NANOGrav sensitivity curve derived in \cite{Breitbach:2018ddu}.
Another large source of uncertainty is the nature of the GW spectrum generated by a loop, which depends on the assumption on the loop small-scale structure (e.g. the number of cusps, kinks and kink-kink collisions per oscillations) \cite{Binetruy:2012ze,Ringeval:2017eww}. For instance, the EPTA bound can be strengthened to $G\mu \lesssim 6.7 \times 10^{-14}$ if the loops are very kinky \cite{Ringeval:2017eww}. 
CS can also emit highly-energetic and short-lasting GW bursts due to cusp formation \cite{Damour:2000wa, Damour:2001bk, Siemens:2006yp, Olmez:2010bi, Ringeval:2017eww}. From the non-observation of such events with LIGO/VIRGO \cite{Abbott:2017mem, Abbott:2019prv}, one can constrain $G\mu \lesssim 4.2 \times 10^{-10}$ with the loop distribution function from \cite{Lorenz:2010sm}. However, the constraints are completely relaxed with the loop distribution function from \cite{Blanco-Pillado:2013qja}.

\section{Gravitational waves from cosmic strings}
\label{sec:GWCS}

In the main text of this study, we do not consider the case of global strings where the presence of a massless Goldstone in the spectrum implies that particle production is the main energy loss so that GW emission is suppressed~\cite{Vilenkin:2000jqa}. However, we give an overview of the GW spectrum from global strings in App.~\ref{app:global_strings}, which can be detectable for string scales $\eta \gtrsim 10^{14}~$GeV. Other studies of the sensitivity  of  next generation GW interferometers to GW from global strings are \cite{Bettoni:2018pbl, Ramberg:2019dgi, Chang:2019mza}.

There has been a long debate in the community whether local cosmic strings mainly loose their energy via GW emission or by particle production. We summarise the arguments and clarify the underlying assumptions below.

\subsection{Beyond the Nambu-Goto approximation}
\label{sec:NG}

\paragraph{Quantum field string simulations:}

Quantum field string (Abelian-Higgs) lattice simulations run by Hindmarsh et al. \cite{Vincent:1997cx, Hindmarsh:2008dw, Hindmarsh:2017qff} have shown that decay into massive radiation is the main energy loss and is sufficient to lead to scaling. Then, loops decay within one Hubble time into scalar and gauge boson radiation before having the time to emit GW. It is suggested that the presence of small-scale structures, kinks and cusps, at the string core size are responsible for the energy loss into particle production. In these regions of large string curvature, the Nambu-Goto  approximation, which considers CS as infinitely thin 1-dimensional classical objects, is no longer valid.

However, Abelian-Higgs simulations run by \cite{Moore:1998gp, Olum:1999sg, Moore:2001px} have claimed the opposite result, that energy loss into massive radiation is exponentially suppressed when the loop size is large compared to the thickness of the string.

\paragraph{Small-scale structure:}
	
At formation time, loops are not smooth but made of straight segments linked by kinks \cite{Blanco-Pillado:2015ana}. Kinks are also created in pairs after each string intercommutation, see \cite{Matsunami:2019fssVIDEO} or Fig.~$2.1$ in \cite{Rocha:2008de}. The presence of straight segments linked by kinks prevents the formation of cusps. However, backreaction from GW emission smoothens the shapes, hence allowing for the formation of cusps \cite{Blanco-Pillado:2015ana} (see Fig.~\ref{kink_cusp_cartoon}). Because of the large hierarchy between the gravitational backreaction scale and the cosmological scale $H$, the effects of the gravitational backreaction on the loop shape are not easily tractable numerically. The  effects of backreaction from particle emission are shown in \cite{Matsunami:2019fssVIDEO}. Nevertheless, it has  been proposed since long \cite{Quashnock:1990wv} that the small-scale structures are smoothened below the gravitational backreaction scale $\sim \Gamma \, G \mu \, t,$. Particularly, based on analytical modelling on simple loop models, it has been shown in \cite{Blanco-Pillado:2018ael,Blanco-Pillado:2019nto} that due to gravitational backreaction, kinks get rounded off, become closer to cusps and then cusps get weakened. In earlier works, the same authors \cite{Wachter:2016rwc, Wachter:2016hgi} claimed that whether the smoothening has the time to occur within the loop life time strongly depends on the initial loop shape. In particular, for a four-straight-segment loop, the farther from the square shape, the faster the smoothening, whereas for more general loop shapes, the smoothening may not always occur. 

To summarise the last two paragraphs, the efficiency of the energy loss into massive radiation depends on the nature of the small-scale structure, which can be understood as a correction to the Nambu-Goto approximation. The precise nature of the small-scale structure, its connection with the gravitational backreaction scale and the conflict between Nambu-Goto and Abelian Higgs simulations remain to be explained. Moreover, the value of the gravitational backreaction scale itself, see Sec.~\ref{sec:Ringeval} is matter of debate.  
For our study, we follow the proposal of \cite{Auclair:2019jip} for investigating how the GW spectrum is impacted for two benchmark scenarios: when the small-scale structures are dominated by cusps or when they are dominated by kinks. We give more details in the next paragraph. {
In App.~\ref{sec:study_impact_mode_nbr}, we show that if the high-frequency slope of the fundamental, $k=1$, GW spectrum is $f^{-1}$, as expected in the presence of an early matter era or in the presence of an Heavide cut-off in the loop formation time, then the existence of the high-$k$ modes, turns it to $f^{-1} \rightarrow f^{1-n}$, where $n$, defined in Eq.~\eqref{eq:one-loop-spectrum}, depends on the small-scale structure. We can therefore read information about the small-scale structure of CS from the high-frequency GW spectrum.}

\paragraph{Massive radiation emission:}
\label{sec:massive_radiation}

In the vicinity of a cusp, the topological charge vanishes where the string cores overlap. Hence, the corresponding portions of the string can decay into massive radiation. The length of the overlapping segment has been estimated to be $\sqrt{r\,l}$ \cite{BlancoPillado:1998bv, Olum:1998ag} where $r \simeq \mu^{-1/2}$ is the string core size and $l$ is the loop length. Hence, the energy radiated per cusp formation is $\mu \sqrt{r l}$, from which we deduce the power emitted from a loop
\begin{equation}
P_{\rm cusp}^{\rm part} \simeq N_{\rm c} \frac{\mu^{3/4}}{l^{1/2}},
\label{eq:power_cusp}
\end{equation}
where $ N_{\rm c}$ is the average number of cusps per oscillation, estimated to be $ N_{\rm c} \sim 2$  \cite{Blanco-Pillado:2015ana}. Note that the consideration of pseudo-cusps, pieces of string moving at highly relativistic velocities, might also play a role \cite{Elghozi:2014kya, Stott:2016loe}.

Even without the presence of cusps, Abelian-Higgs simulations \cite{Matsunami:2019fss} have shown that kink-kink collisions produce particles with a power per loop
\begin{equation}
P_{\rm kink}^{\rm part} \simeq  N_{\rm kk} \frac{\epsilon}{l},
\label{eq:power_kink}
\end{equation}
where $ N_{\rm kk}$ is the average number of kink-kink collisions per oscillation. 
Values possibly as large as $N_{\rm kk} \sim O(10^3)$ have been considered in \cite{Ringeval:2017eww} or even as large as $10^6$ for the special case of strings with junctions 
\cite{Binetruy:2010cc}, due to kink proliferations \cite{Binetruy:2010bq}.
In contrast to the cusp case, the energy radiated per kink-kink collision, $\epsilon$, is independent of the loop size $l$ and we expect $\epsilon \sim \mu^{1/2}$.

Upon comparing the power of GW emission in Eq.~\eqref{eq:power_GW_0} with either Eq.~\eqref{eq:power_cusp} or Eq.~\eqref{eq:power_kink}, one expects gravitational production to be more efficient than particle production when loops are larger than \cite{Auclair:2019jip}
\begin{equation}
\label{eq:length_cusps}
l \gtrsim l_c  \equiv \beta_c \, \frac{\mu^{-1/2}}{(\Gamma G \mu)^2},
\end{equation}
for small-scale structures dominated by cusps, and
\begin{equation}
\label{eq:length_kinks}
l \gtrsim l_k  \equiv \beta_k \, \frac{\mu^{-1/2}}{\Gamma G \mu},
\end{equation}
for kink-kink collision domination. $\beta_{\rm c}$ and $\beta_{\rm k}$ are numbers which depend on the precise refinement. We assume $\beta_{\rm c}, \, \beta_{\rm k} \sim O(1)$. Therefore, loops with length smaller than the critical value in Eq.~\eqref{eq:length_cusps} or Eq.~\eqref{eq:length_kinks} are expected to decay into massive radiation before they have time to emit GW, which means that they should be subtracted when computing the SGWB. Equations (\ref{eq:length_cusps}) and (\ref{eq:length_kinks}) are crucial to determine the cutoff frequency, as we discuss in Sec.~\ref{UVcutoff}.

The cosmological and astrophysical consequences of the production of massive radiation and the corresponding constraints on CS from different experiments are presented in Sec.~\ref{sec:particle_prod_pheno}

\subsection{Assumptions on the loop distribution}
\label{sec:Ringeval}

The SGWB resulting from the emission by CS loops strongly relies on the distribution of loops. In the present section, we introduce the loop-formation efficiency and discuss the assumptions on the loop-production rate, inspired from Nambu-Goto simulations. The loop-formation efficiency is computed later, in Sec.~\ref{sec:VOS}.

\paragraph{Loop-formation efficiency:}
The SGWB resulting from the emission by CS loops strongly relies on the assumption for the distribution of loops which we now discuss. 
The equation of motion of a Nambu-Goto string in a expanding universe implies the following evolution equation for the long string energy density, cf. Sec.~\ref{sec:VOS_proof}
\begin{equation}
\label{eq:longstringdensity_eq}
\frac{d\rho_{\infty}}{dt}= -2H(1+\bar{v}^2) \rho_{\infty} - \left.\frac{d\rho_{\infty}}{dt}\right|_\textrm{loop},
\end{equation}
where $\bar{v}$ is the long string mean velocity. The energy loss into loop formation can be expressed as \cite{Vilenkin:2000jqa}
\begin{equation}
\label{eq:energyloss_loops}
\left.\frac{d\rho_{\infty}}{dt}\right|_\textrm{loop} \equiv \mu \int_{0}^{\infty} l f(l,t)dl \equiv \frac{\mu}{t^3} \tilde{C}_{\rm eff},
\end{equation}
with $f(l,t)$ the number of loops created per unit of volume, per unit of time $t$ and per unit of length $l$ and where we introduced the loop-formation efficiency $\tilde{C}_{\rm eff}$.
 The loop-formation efficiency $\tilde{C}_{\rm eff}$ is related to the notation introduced in \cite{Cui:2017ufi, Cui:2018rwi} by
\begin{equation}
\tilde{C}_{\rm eff} \equiv \sqrt{2} \,C_{\rm eff}.
\end{equation}
In Sec.~\ref{sec:VOS}, we compute the loop-formation efficiency $C_{\rm eff}$ as a function of the long string network parameters $\bar{v}$ and $L$, which themselves are solutions of the Velocity-dependent One-Scale (VOS) equations.

\paragraph{Only loops produced at the horizon size contribute to the SGWB:}
\label{sec:mainAssumptions}
As pointed out already a long time ago by \cite{Bennett:1989ak, Quashnock:1990wv} and more recently in large Nambu-Goto simulations \cite{Blanco-Pillado:2013qja}, the most numerous loops are the ones of the size of the gravitational backreaction scale 
\begin{equation}
\Gamma G \mu \times t,
\end{equation} 
which acts as a cut-off below which, small-scale structures are smoothened and such that smaller loops can not be produced below that scale. However, it has been claimed that only large loops are relevant for GW \cite{Hogan:2006we, Siemens:2006yp, Blanco-Pillado:2013qja}. In particular, Nambu-Goto  numerical simulations realized by Blanco-Pillado et al. \cite{Blanco-Pillado:2013qja} have shown that a fraction $\mathcal{F}_{\alpha}\simeq 10\%$ of the loops are produced with a length equal to a fraction $\alpha \simeq 10\%$ of the horizon size, and with a Lorentz boost factor $\gamma \simeq \sqrt{2}$. The remaining $90\%$ of the energy lost by long strings goes into highly boosted smaller loops whose contributions to the GW spectrum are sub-dominant. Under those assumptions, the number of loops, contributing to the SGWB, produced per unit of time can be computed from the total energy flow into loops in Eq.~\eqref{eq:energyloss_loops}
\begin{equation}
\label{eq:loop-formation_rate}
\frac{dn}{dt_i} = \frac{\mathcal{F}_{\alpha}}{\gamma\,\mu\,\alpha\,t_i} \left.\frac{d\rho_{\infty}}{dt}\right|_\textrm{loop},
\end{equation}
with $\mathcal{F}_{\alpha} = 0.1$, $\gamma = \sqrt{2}$ and $\alpha = 0.1$.  In an appendix of \cite{Gouttenoire:2019kij}, we discuss the possibility to define the loop-size as a fixed fraction of the correlation length $L$ instead of a fixed fraction of the horizon size $t$. Especially, we show that the impact on the GW spectrum is negligible. The latter can be recast as a function of the loop-formation efficiency $\tilde{C}_{\rm eff}$ defined in Eq.~\eqref{eq:energyloss_loops}
\begin{equation}
\label{eq:LoopProductionFctBody2}
\frac{dn}{dt_i}=\mathcal{F}_{\alpha} \frac{\tilde{C}_{\rm eff}(t_i)}{\gamma \,\alpha \, t_i^4}.
\end{equation}
This is equivalent to choosing the following monochromatic horizon-sized loop-formation function
\begin{equation}
f(l,\,t_i) = \frac{\tilde{C}_{\rm eff}}{\alpha\,t_i^4} \delta(l-\alpha t_i).
\end{equation}
The assumptions leading to Eq.~\eqref{eq:LoopProductionFctBody2} are the ones we followed for our study and which are also followed by \cite{Cui:2017ufi, Cui:2018rwi}. Our results strongly depend on these assumptions and would be dramatically impacted if instead we consider the model discussed in the next paragraph.

\begin{figure}[]
\centering
\raisebox{0cm}{\makebox{\includegraphics[width=0.7\textwidth, scale=1]{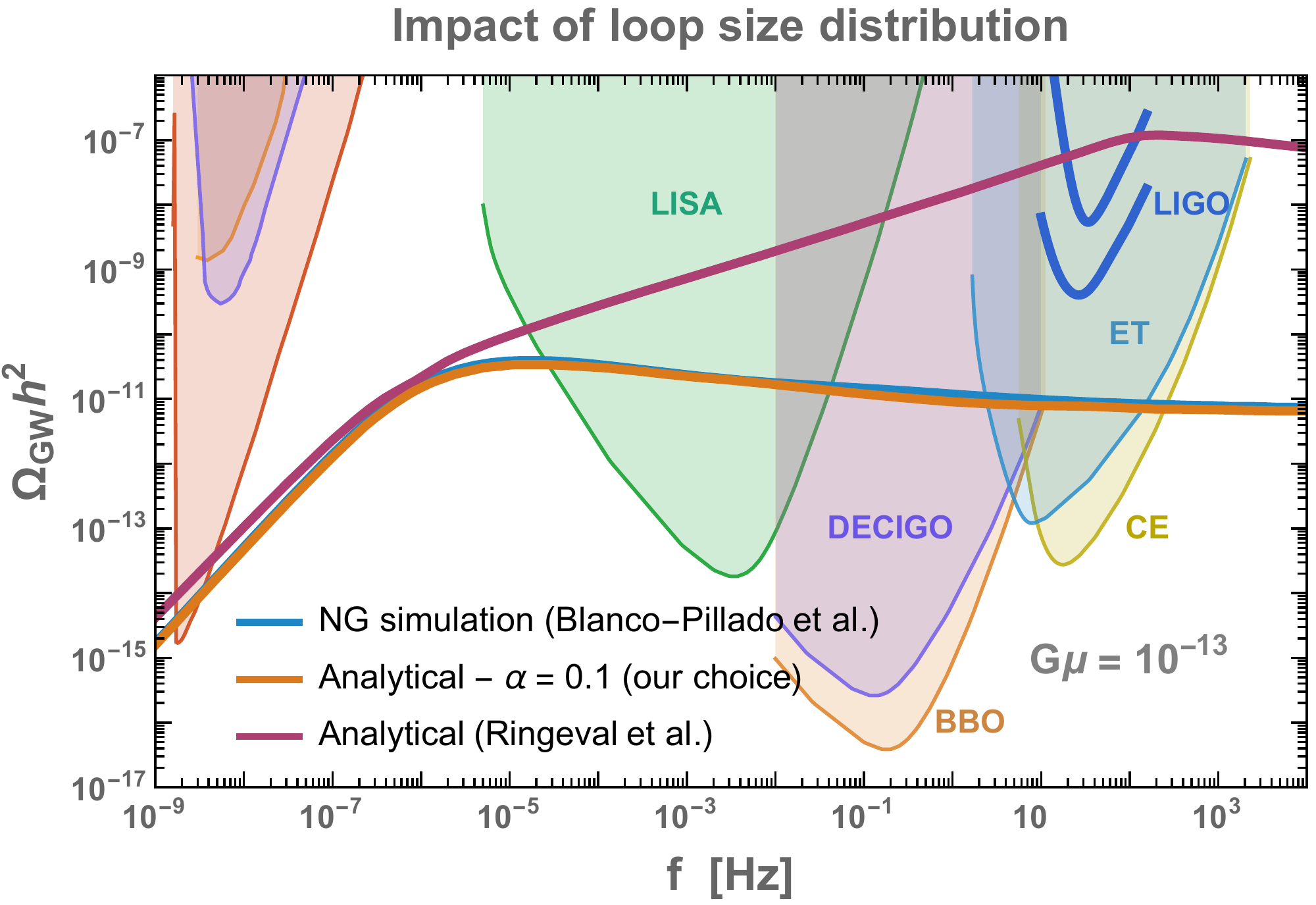}}}
\caption{\it \small  We compare the GW spectrum from CS computed in the NG simulation of Blanco-Pillado et al. \cite{Blanco-Pillado:2013qja, Blanco-Pillado:2017oxo} to the one computed analytically by Ringeval et al. \cite{Lorenz:2010sm, Ringeval:2017eww, Auclair:2019zoz}. In the later case, the cut-off on the loop size is smaller. This acts as a second population of small loops which enhances the GW spectrum at high frequency. The analytical modeling we consider in our work is fitted on the work of Blanco-Pillado et al and consider a monochromatic loop size distribution peaked on $\alpha =0.1$. This figure is not included in \cite{Gouttenoire:2019kij} but instead it has been done during the redaction of the manuscript.}
\label{fig:Impact_loop_size_distrib_Ringeval_vs_BlancoPillado}
\end{figure}

\paragraph{A second population of smaller loops:}
The previous assumption - that the only loops relevant for the GW signal are the loops produced at horizon size - which is inspired from the Nambu-Goto numerical simulations of Blanco-Pillado et al. \cite{Blanco-Pillado:2013qja, Blanco-Pillado:2017oxo}, is in conflict with the results from Ringeval et al. \cite{Lorenz:2010sm, Ringeval:2017eww, Auclair:2019zoz}. In the latter works, the loop production function  is derived analytically starting from the correlator of tangent vectors on long strings, within the Polchinski-Rocha model \cite{Polchinski:2006ee, Polchinski:2007rg, Dubath:2007mf, Rocha:2007ni}. In the Polchinski-Rocha model, which has been tested in Abelian-Higgs simulations \cite{Hindmarsh:2008dw}, the gravitational back-reaction scale, i.e. the lower cut-off of the loop production function, is computed to be 
\begin{equation}
\Upsilon (G \mu)^{1+2\chi} \times t,
\end{equation} 
with $\Upsilon \simeq 10$ and $\chi \sim 0.25$.  Consequently, the gravitational back-reaction scale in the Polchinski-Rocha model is significantly smaller than the usual gravitational back-reaction scale, commonly assumed to match the gravitational radiation scale, $ \Gamma G \mu \, t$. Therefore, the model of Ringeval et al. predicts the existence of a second population of smaller loops which enhances the GW spectrum at high frequency by many orders of magnitude, see Fig.~\eqref{fig:Impact_loop_size_distrib_Ringeval_vs_BlancoPillado}. However, as raised by \cite{Blanco-Pillado:2019vcs}, the model of Ringeval et al. predicts the amount of long-string energy converted into loops, to be $\sim 200$ times larger than the one computed in the numerical simulations of Blanco-Pillado et al. \cite{Blanco-Pillado:2013qja}. These discrepancies between Polchinski-Rocha analytical modeling and Nambu-Goto numerical simulations remain to be understood. \\

\newpage
\hspace{-1cm}
\begin{minipage}[c]{15cm}
\vspace{0.cm}
\subsection{The gravitational-wave spectrum}
\label{sec:SGWB}
For our study, we compute the GW spectrum observed today generated from CS as follows (see app.~\ref{app:derivationGWspectrum} for a derivation)
	\begin{equation}
	\Omega_{\rm{GW}}(f)\equiv\frac{f}{\rho_c}\left|\frac{d\rho_{\rm{GW}}}{df}\right|=\sum_k{\Omega^{(k)}_{\rm{GW}}(f)},
	\label{eq:SGWB_CS_Formula}
		\vspace{-0.25cm}
	\end{equation}
	where
	\begin{equation}
\Omega^{(k)}_{\rm{GW}}(f)=\frac{1}{\rho_c}\cdot\frac{2k}{f}\cdot\frac{\mathcal{F}_{\alpha}\,\Gamma^{(k)}G\mu^2}{\alpha(\alpha+\Gamma G \mu)}\int^{t_0}_{t_{\rm osc}}d\tilde{t}~ \frac{C_{\rm{eff}}(t_i)}{\,t_i^4}\left[\frac{a(\tilde{t})}{a(t_0)}\right]^5\left[\frac{a(t_i)}{a(\tilde{t})}\right]^3\Theta(t_i-t_{\rm osc})\Theta(t_i-\frac{l_*}{\alpha}),
		\vspace{-0.25cm}
	\label{kmode_omega}
	\end{equation}
	\vspace{-0.25cm}
	with
	\begin{eqnarray*}
	    \, \Theta &\equiv& \textrm{Heaviside function,}\\
		\mu, \, G, \, \rho_c &\equiv& \textrm{string tension, Newton constant, critical density,}\\
		 \, a &\equiv& \textrm{scale factor of the universe}\\
		 &&\textrm{(we solve the full Friedmann equation for a given energy density content)},\\
		 \, k &\equiv& \textrm{proper mode number of the loop (effect of high-k modes are discussed in App.~\ref{sec:study_impact_mode_nbr}.} \\
		 && \textrm{For technical reasons, in most of our plots, we restrict to $2\times 10^4$ modes),}\\
		\Gamma &\equiv& \textrm{gravitational loop-emission efficiency, }~(\Gamma \simeq 50 ~ \text{\cite{Blanco-Pillado:2017oxo}})\\
		\Gamma^{(k)} &\equiv& \textrm{Fourier modes of $\Gamma$, dependent on the loop small-scale structures,}\\
		&&(\Gamma^{(k)}\propto k^{-4/3}\textrm{ for cusps, e.g. \cite{Olmez:2010bi})} ,\\
		\mathcal{F}_{\alpha} &\equiv& \textrm{fraction of loops formed with size $\alpha$ ($\mathcal{F}_{\alpha}\simeq 0.1$), cf. Sec.~\ref{sec:mainAssumptions}},\\ 
		C_\textrm{eff} &\equiv& \textrm{loop-production efficiency, defined in Eq.~\eqref{eq:loopEfficiency}, }\\
		&& \textrm{($C_\textrm{eff} $ is a function of the long-string mean velocity $\bar{v}$ and correlation length $\xi$,}\\
		&&  \textrm{both computed upon integrating the VOS equations, cf. Sec.~\ref{sec:VOS})}\\
		\alpha &\equiv& \textrm{loop length at formation in unit of the cosmic time,}~(\alpha \simeq 0.1)\\
		&& \textrm{(we consider a monochromatic, horizon-sized loop-formation function, cf. Sec.~\ref{sec:mainAssumptions}),}\\
		\tilde{t} &\equiv& \textrm{the time of GW emission},\\
		 f &\equiv& \textrm{observed frequency today}\\
		 &&\textrm{(related to frequency at emission $\tilde{f}$ through $f\,a(t_0)=\tilde{f}\,a(\tilde{t})$},\\
		 &&\textrm{related to loop length $l$ through $\tilde{f}=2k/l$},\\
		 &&\textrm{related to the time of loop production $t_i$ through $l = \alpha t_i -\Gamma G \mu(\tilde{t}-t_i)$)},\\
		 t_i &\equiv& \textrm{the time of loop production},\\
		 &&\textrm{(related to observed frequency and emission time $\tilde{t}$ through}\\
		 &&\textrm{$t_i (f, \, \tilde{t})= \frac{1}{\alpha+\Gamma G \mu} \left[ \frac{2k}{f}\frac{a(\tilde{t})}{a(t_0)} + \Gamma G \mu \, \tilde{t} \right]$\big),}\\
		t_0 &\equiv& \textrm{the time today},\\
		t_{\rm osc} &\equiv& \textrm{the time at which the long strings start oscillating, $t_{\rm osc} =\textrm{Max}[t_{\rm fric}, \, t_F]$,}\\
		&&\textrm{$t_F$ is the time of CS network formation, defined as $\sqrt{\rho_{\rm tot}(t_F)}\equiv\mu$ where $\rho_{\rm tot}$ is}\\
		&&\textrm{the universe total energy density. In the presence of friction, at high temperature},\\
		&& \textrm{the string motion is damped until the time $t_{\rm fric}$, computed in app.~\ref{sec:thermal_friction},}\\
		l_* &\equiv&  l_{\rm c}\textrm{ for cusps and } l_{\rm k} \textrm{ for kinks in Eq.~\eqref{eq:length_cusps} and Eq.~\eqref{eq:length_kinks}} \\
		&&\textrm{(critical length below which the emission of massive radiation}  \\
		&&\textrm{is more efficient than the gravitational emission, cf. Sec.~\ref{sec:massive_radiation})}.
	\end{eqnarray*}
\end{minipage}

\begin{figure}[]
\centering
\raisebox{0cm}{\makebox{\includegraphics[width=0.9\textwidth, scale=1]{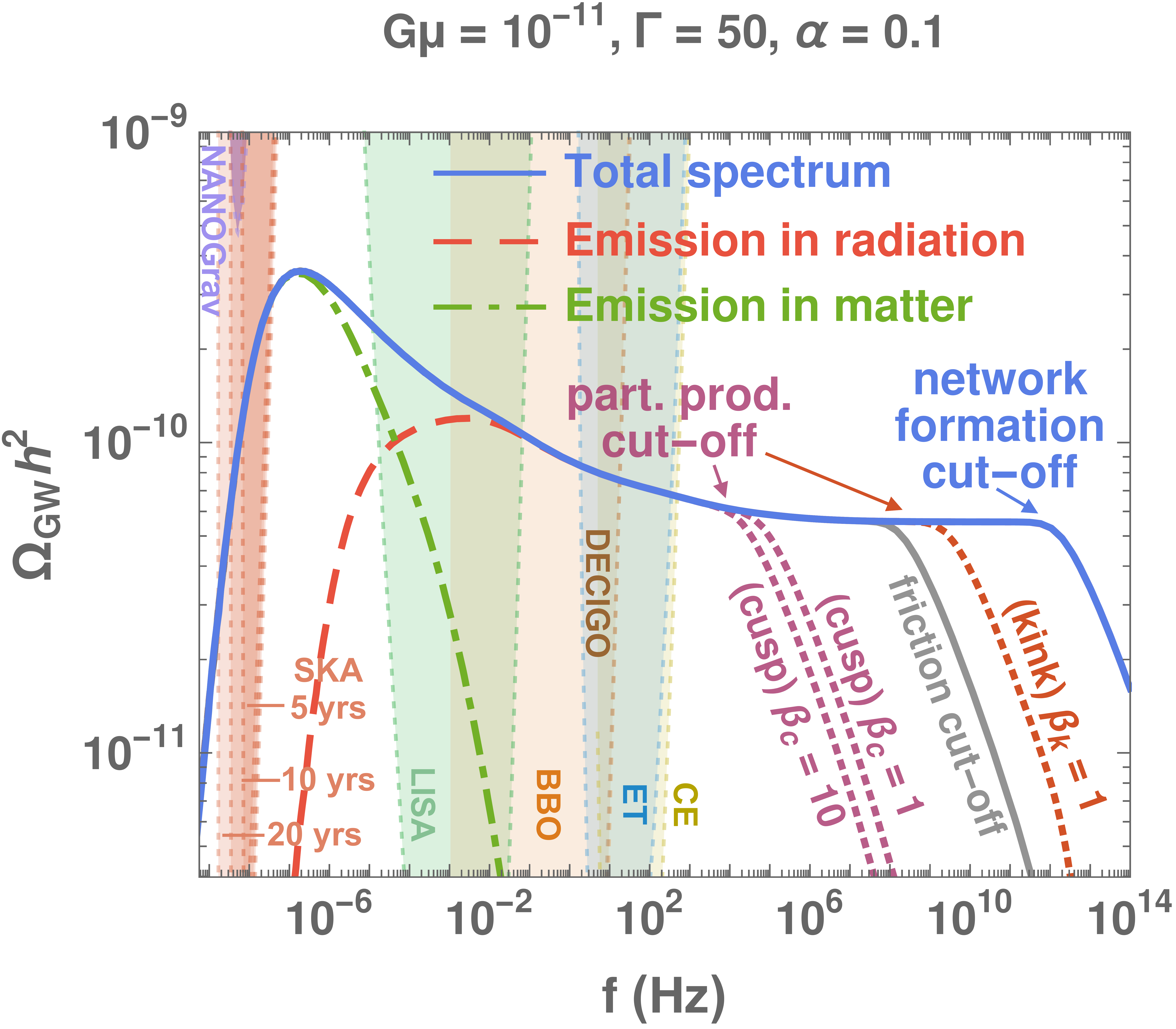}}}
\caption{\it \small GW spectrum from the scaling cosmic-string network evolving in a standard cosmology.
Contributions from GW emitted during radiation and matter eras are shown with red and green dashed lines respectively.
The high-frequency cut-offs correspond to either the time of formation of the network, cf. Eq.~\eqref{eq:network_formation}, the time when friction-dominated dynamics become irrelevant, cf.  App.~\ref{sec:thermal_friction}, or the time when gravitational emission dominates over massive particle production, for either kink or cusp-dominated small-scale structures, cf. Sec.~\ref{sec:massive_radiation}. The cut-offs are described by Heaviside functions in the master formula in Eq.~\eqref{eq:SGWB_CS_Formula}.  In App.~\ref{sec:study_impact_mode_nbr}, we show that the slopes beyond the high-frequency cut-offs are given by $f^{-1/3}$. Colored regions indicate the integrated power-law sensitivity of future experiments, as described in app.~\ref{app:sensitivity_curves}.}
\label{sketch_scaling}
\end{figure}

\paragraph{A first look at the GW spectrum:}
Fig.~\ref{sketch_scaling} shows the GW spectrum computed with Eq.(\ref{eq:SGWB_CS_Formula}). The multiple frequency cut-offs visible on the figure, follow from the Heaviside functions in Eq.~\eqref{eq:SGWB_CS_Formula}, which subtract loops formed before network formation, cf. Eq.~\eqref{eq:network_formation}, or when thermal friction freezes the network, cf. App.~\ref{sec:thermal_friction}, or which subtract loops decaying via massive particle emission from cusps and kinks instead of GW, cf. Sec.~\ref{sec:massive_radiation}. We indicate separately the contributions from the emission occurring before and after the matter-radiation equality. {One can see that loops emitting during the radiation era contribute to a flat spectrum whereas loops emitting during the matter era lead to a slope decreasing as $f^{-1/3}$. Similarly, the high-frequency cut-offs due to particle production, thermal friction, network formation, but also due to a second period of inflation (discussed in Sec.~\ref{sec:inflation}), give a slope $f^{-1/3}$.  In App.~\ref{sec:study_impact_mode_nbr}, we show that the presence of high-frequency modes are responsible for changing the slope $f^{-1}$, expected from the $(k=1)$-spectrum, to $f^{-1/3}$. }

\paragraph{Impact of the cosmology on the GW spectrum:} 
In a nutshell, the frequency dependence of the GW spectrum receives two contributions, a red-tilt coming from the redshift of the GW energy density and a blue-tilt coming from the loop-production rate $\propto t_i^{-4}$. On the one hand, the higher the frequency the earlier the GW emission, so the larger the redshift of the GW energy density and the more suppressed the spectrum. On the other hand, high frequencies correspond to loops formed earlier, those being more numerous, this increases the GW amplitude. Interestingly, during radiation-domination the two contributions exactly cancel such that the spectrum is flat.
As explained in more details in App.~\ref{sec:quadrupole_formula}, the flatness of the GW spectrum during radiation is intimately related to the independence of the GW emission power on the loop length. In the same appendix, we show that a change in the equation of state of the universe impacts the GW spectrum if it modifies at least one of the two following redshift factors: the redshift of the number of emitting loops and the redshift of the emitted GW.

For instance, when GW emission occurs during radiation but loop formation occurs during matter, the loop density redshifts faster. Then, the larger the frequency, the earlier the loop formation, and the more suppressed the GW spectrum (as $f^{-1}$ for $k=1$ and as $f^{-1/3}$ when taking into account high-k modes). Conversely, if loop formation occurs during kination, the loop density redshift slower and the GW gets enhanced at large frequency (as $f^1$).

\subsection{The frequency - temperature relation}
\paragraph{Relation between frequency of observation and temperature of loop formation:}
\label{sec:turning_point_scaling}
In app.~\ref{derive_turning_points}, we derive the relation between a detected frequency $f$ and the temperature of the universe when the loops, mostly responsible for $f$, are formed
\begin{align}
f=(6.7\times10^{-2}\textrm{ Hz})\left(\frac{T}{\textrm{GeV}}\right)\left(\frac{0.1\times 50 \times10^{-11}}{\alpha\,\Gamma G\mu}\right)^{1/2}\left(\frac{g_*(T)}{g_*(T_0)}\right)^{1/4}.
\label{turning_point_general_scaling}
\end{align}
We emphasize that Eq.~\eqref{turning_point_general_scaling} is very different from the relation obtained in the case of GW generated by a first-order cosmological phase transition. In the latter case, the emitted frequency corresponds to the Hubble scale at $T_p$ \cite{Caprini:2018mtu}
\begin{equation}
f = (19 \times 10^{-3}~\text{mHz}) \left( \frac{T_p}{100~\text{GeV}} \right)\left( \frac{g_*(T_p)}{100} \right)^{1/4}.
\end{equation}
In the case of cosmic strings, instead of being set by the Hubble scale at the loop-formation time $t_i$, the emitted frequency is further suppressed by a factor $(\Gamma G \mu)^{-1/2}$, which we now explain.
From the scaling law  $\propto t_i^{-4}$ of the loop-production function in Eq.\eqref{eq:LoopProductionFctBody2}, one can understand that the most numerous population of emitting loops at a given time $\tilde{t}$ is the population of loops created at the earliest epoch. They are the oldest loops\footnote{Note that they are also the smallest loops, with a length given by the gravitational radiation scale $\Gamma G \mu\,t$.}. Hence, a loop created at time $t_i$ contributes to the SGWB much later, at a time given by the loop half-lifetime $\tilde{t}_{\rm M}=\alpha\,t_i/2\Gamma G \mu$, cf. Eq.~\eqref{eq:GWlifetime}. Therefore, the emitted frequency is dispensed from the redshift factor $a(\tilde{t}_{\rm M})/a(t_i)=(\tilde{t}_{\rm M}/t_i)^{1/2} \sim  (\Gamma G \mu)^{-1/2}$, and so, is higher. See app.~\ref{derive_turning_points} and its Fig.~\ref{cartoon_loop_maximal_decay} for more details.

\paragraph{The detection of a non-standard cosmology:}
During a change of cosmology, e.g a change from a matter to a radiation-dominated era, the long-string network evolves from one scaling regime to the other. The response of the network to the change of cosmology is quantified by the VOS equations, which are presented in Sec.~\ref{sec:VOS}. As a result of the transient evolution towards the new scaling regime, the turning-point frequency Eq.~\eqref{turning_point_general_scaling_app} associated to the change of cosmology  is lower in VOS than in the scaling network. The detection of a turning-point in a GW spectrum from CS by a future interferometer would be a smoking-gun signal for non-standard cosmology. Particularly, in Fig.~\ref{fig:contour_int}, we show that LISA can probe a non-standard era ending around the QCD scale, ET/CE can probe a non-standard era ending around the TeV scale whereas DECIGO/BBO can probe the intermediate range. We focus on the particular case of a short matter era in Sec.~\ref{sec:interm_matter} and a short inflation era in Sec.~\ref{sec:inflation}, respectively. In the latter case, the turning-point frequency is even further decreased due to the string stretching which we explain in the next paragraph.

\paragraph{The detection of a non-standard cosmology (intermediate-inflation case):}
If the universe undergoes a period of inflation lasting $N_e$ e-folds, the correlation length of the network is stretched outside the horizon. After inflation, the network achieves a long transient regime lasting $\sim N_e$ other e-folds until the correlation length re-enters the horizon. Hence, the turning-point frequency in the GW spectrum, cf. Eq.~\eqref{turning_point_inf}, receives a $\exp{N_e}$ suppression compared to Eq.~\eqref{turning_point_general_scaling} due to the duration of the transient. We give more details in Sec.~\ref{sec:inflation}.

		\paragraph{Cut-off frequency from particle production:}
\label{UVcutoff}
As discussed in the Sec.~\ref{sec:massive_radiation}, particle production is the main decay channel of loops shorter than
\begin{align}
l_*=\beta_m\frac{\mu^{-1/2}}{(\Gamma G\mu)^m},
\end{align}
where $m=1$ or $2$ for loops kink-dominated or cusp-dominated, respectively, and $\beta_m\sim\mathcal{O}(1)$.
The corresponding characteristic temperature above which loops, decaying preferentially into particles, are produced, is
\begin{equation}
T_* \simeq \beta_m^{-1/2}\, \Gamma^{m/2} \,\sqrt{\alpha}\,(G\mu)^{(2m+1)/4} \, M_{\rm pl} \simeq
\begin{cases}
\text{(0.2 EeV)} ~ \sqrt{\dfrac{\alpha}{0.1}} \,\sqrt{\dfrac{1}{\beta_c}} \left( \dfrac{G\mu}{10^{-15}}  \right)^{3/4}&\textrm{\hspace{0.5em}for kinks},\\[1em]
\text{(1 GeV)} ~ \sqrt{\dfrac{\alpha}{0.1}} \,\sqrt{\dfrac{1}{\beta_k}}\left( \dfrac{G\mu}{10^{-15}}  \right)^{5/4}&\textrm{\hspace{0.5em}for cusps}.\\
\end{cases}
\label{UVcutoff_T_app}
\end{equation}
We have used $l_*=\alpha \, t_i$, $H=1/(2t_i)$ and $\rho_{\rm rad}=3 M_{\rm pl}^2 H^2$. Upon using the frequency-temperature correspondence in Eq.~\eqref{turning_point_general_scaling}, we get the cut-off frequencies due to particle production 
\begin{equation}
f_* \simeq
\begin{cases}
\text{(1 GHz)} ~  \sqrt{\dfrac{1}{\beta_c}} \,\left( \dfrac{G\mu}{10^{-15}} \right)^{1/4}&\textrm{\hspace{0.5em}for kinks},\\[1em]
\text{(31 Hz)} ~ \sqrt{\dfrac{1}{\beta_k}} \left( \dfrac{G\mu}{10^{-15}}  \right)^{3/4}&\textrm{\hspace{0.5em}for cusps}.\\
\end{cases}
\label{UVcutoff_f_app}
\end{equation}
and which we show in most of our plots with dotted red and purple lines. Particularly, in Fig.~\ref{ST_vos_scaling}, we see that particle production in the cusp-dominated case would start suppressing the GW signal in the ET/CE windows for string tension lower than $G\mu \lesssim 10^{-15}$. However, in the kink-dominated case, the spectrum is only impacted at frequencies much higher than the interferometer windows.  In App.~\ref{sec:study_impact_mode_nbr}, we show that the slope of the GW spectrum beyond the high-frequency cut-off $f_*$ is given by $f^{-1/3}$.

\subsection{The astrophysical foreground}

Crucial for our analysis is the assumption that the stochastic GW foreground of astrophysical origin can be substracted.  

LIGO/VIRGO has already observed three binary black hole (BH-BH) merging events \cite{TheLIGOScientific:2016qqj, Abbott:2016nmj, TheLIGOScientific:2016pea} during the first $4$-month observing run O1 in 2015, and seven additional BH-BH \cite{Abbott:2017vtc, Abbott:2017gyy, Abbott:2017oio, LIGOScientific:2018mvr} as well as one binary neutron star (NS-NS) \cite{TheLIGOScientific:2017qsa} merging events during the second $9$-month observing run O2 in 2017. And more events might still be discovered in the O2 data \cite{Venumadhav:2019lyq}.
According to the estimation of the NS merging rate following the detection of the first (and unique up to now) NS-NS merger event GW170817, NS-NS stochatisc background may be detectable after a $20$-month observing run with the expected LIGO/VIRGO design sensistivity in $2022+$ and in the most optimistic scenario, it might be detectable after $18$-month of the third observing run O3 who began in April, $1^{\rm st}$, 2019 \cite{ Abbott:2017xzg}.
Hence, one might worry about the possibility to distinguish the GW SGWB sourced by CS from the one generated by the astrophysical foreground.
However, in the BBO and ET/CE windows, the NS and BH foreground might be substracted with respective reached sensibilities $\Omega_{\rm GW} \simeq 10^{-15}$ \cite{Cutler:2005qq} and $\Omega_{\rm GW} \simeq 10^{-13}$ \cite{Regimbau:2016ike}.
In the LISA window, the binary white dwarf (WD-WD) foreground dominates over the NS-NS and BH-BH foregrounds \cite{Farmer:2003pa, Rosado:2011kv, Moore:2014lga}. The WD-WD galactic foreground, one order of magnitude higher than the WD-WD extragalactic \cite{Kosenko:1998mv}, might be substracted with reached sensibility $\Omega_{\rm GW} \simeq 10^{-13}$ at LISA \cite{Adams:2010vc, Adams:2013qma}. Hence, in the optimistic case where the foreground can be removed and the latter sensibility are reached one might be able to distinguished the signal sourced by CS from the one generated by the astrophysical foreground. Furthermore, the GW spectrum generated by the astrophysical foreground increased with frequency as $f^{2/3}$ \cite{Zhu:2012xw}, which is different from the GW spectrum generated by CS during radiation (flat), matter $f^{-1/3}$, inflation $f^{-1/3}$ or kination $f^1$. We refer to \cite{Baldes:2021aph,Lewicki:2021kmu} for the study of the sensitivity of future GW observatories when assuming that astrophysical foregrounds can not be subtracted.

\section{The Velocity-dependent One-Scale model}
\label{sec:VOS}

The master formula (\ref{eq:SGWB_CS_Formula}) crucially depends on the loop-production efficiency encoded in $C_{\rm eff}$. In this section, we discuss its derivation within the framework of the Velocity-dependent One-Scale (VOS) model.

\subsection{The loop-production efficiency}
\label{sec:loopProductionFct}

In a correlation volume $L^3$, a segment of length $L$ must travel a distance $L$ before encountering 
another segment. $L$ is the correlation length of the long-string network.  The collision rate, per unit of volume, is $\tfrac{\bar{v}}{L} \cdot \tfrac{1}{L^3} \sim \tfrac{\bar{v}}{L^4}$ where $\bar{v}$ is the long-string mean velocity. At each collision forming a loop, the network looses a loop energy $\mu \, L = \rho_{\infty} \, L^3$. Hence, the loop-production energy rate can be written as \cite{Kibble:1984hp}
\begin{equation}
\label{eq:energylossloop}
\left.\frac{d\rho_{\infty}}{dt}\right|_\textrm{loop}=\tilde{c}\, \bar{v}\frac{\rho_{\infty}}{L},
\end{equation}
where one can compute $\tilde{c}=0.23 \pm 0.04$ from Nambu-Goto simulations in expanding universe \cite{Martins:2000cs}. $\tilde{c}$ is the only free parameter of the VOS model. 
Hence, the loop-formation efficiency, defined in Eq.~\eqref{eq:energyloss_loops}, can be expressed as a function of the long-string parameters, $\bar{v}$ and $\xi \equiv L/t$,
\begin{equation}
\label{eq:loopEfficiency}
\tilde{C}_\textrm{eff} \equiv \sqrt{2}\,C_\textrm{eff}(t)=\frac{\tilde{c}\,\bar{v}(t)}{\xi^3(t)}.
\end{equation}
In app.~\ref{app:VOScalibration}, we discuss how our results are changed when considering a recent extension of the VOS model with more free parameters, fitted on Abelian-Higgs field theory numerical simulations \cite{Correia:2019bdl}, and taking into account the emission of massive radiation. Basically, the loop-formation efficiency $C_{\rm eff}$ is only decreased by a factor $\sim 2$.
In the following, we derive $\bar{v}$ and $\xi$ as solutions of the VOS equations.

\subsection{The VOS equations}

The VOS equations describe the evolution of a network of long strings in term of the mean velocity $\bar{v}$ and the correlation length $\xi = L/t$ \cite{Martins:1995tg, Martins:1996jp, Martins:2000cs, martins2016defect}. The latter is defined through the long string energy density $\rho_{\infty} \equiv \mu/L^2$.
Starting from the equations of motion of the Nambu-Goto string in a FRW universe, we can derive the so-called VOS equations (see app.~\ref{sec:VOS_proof} for a derivation)
\begin{align}
\label{eq:VOS_eq_body}
&\frac{dL}{dt}=HL \,( 1+ \bar{v}^2)+\frac{1}{2}\tilde{c}\,\bar{v}, \\
&\frac{d\bar{v}}{dt}=(1-\bar{v}^2)\left[\frac{k(\bar{v})}{L}-2H\bar{v}\right],
\end{align}
where
\begin{equation}
 k(\bar{v})=\frac{2\sqrt{2}}{\pi}(1-\bar{v}^2)(1+2\sqrt{2}\bar{v}^3)\frac{1-8\bar{v}^6}{1+8\bar{v}^6},
\end{equation}
is the so-called momentum parameter and is a measure of the deviation from the straight string, for which $ k(\bar{v})=1$ \cite{martins2016defect}.
The first VOS equation describes the evolution of the long string correlation length under the effect of Hubble expansion and loop chopping.
The second VOS equation is nothing more than a relativistic generalization of Newton's law where the string is accelerated by its curvature $1/L$ but is damped by the Hubble expansion after a typical length $H^{-1}$.

Numerical simulations \cite{Ringeval:2005kr, Vanchurin:2005pa, Martins:2005es, Olum:2006ix, BlancoPillado:2011dq} have shown that a network of long strings is first subject to a transient regime before reaching a scaling regime, in which the long string mean velocity $\bar{v}$ is constant and the correlation length grows linearly with the Hubble horizon $L = \xi \, t$. The values of the quantities $\bar{v}$ and $\xi $ depend on the cosmological background, namely the equation of state of the universe. Hence, when passing from a cosmological era 1 to era 2, the network accomplishes a transient evolution from the scaling regime 1 to the scaling regime 2. We  use the VOS equations to compute the time evolution of $\bar{v}$ and $\xi $ during the change of cosmology and then compute their impact on the CS SGWB.

\subsection{Scaling regime solution and beyond}
\label{sec:scalingVSvos}

\begin{figure}[h!]
\centering
\raisebox{0cm}{\makebox{\includegraphics[width=0.32\textwidth, scale=1]{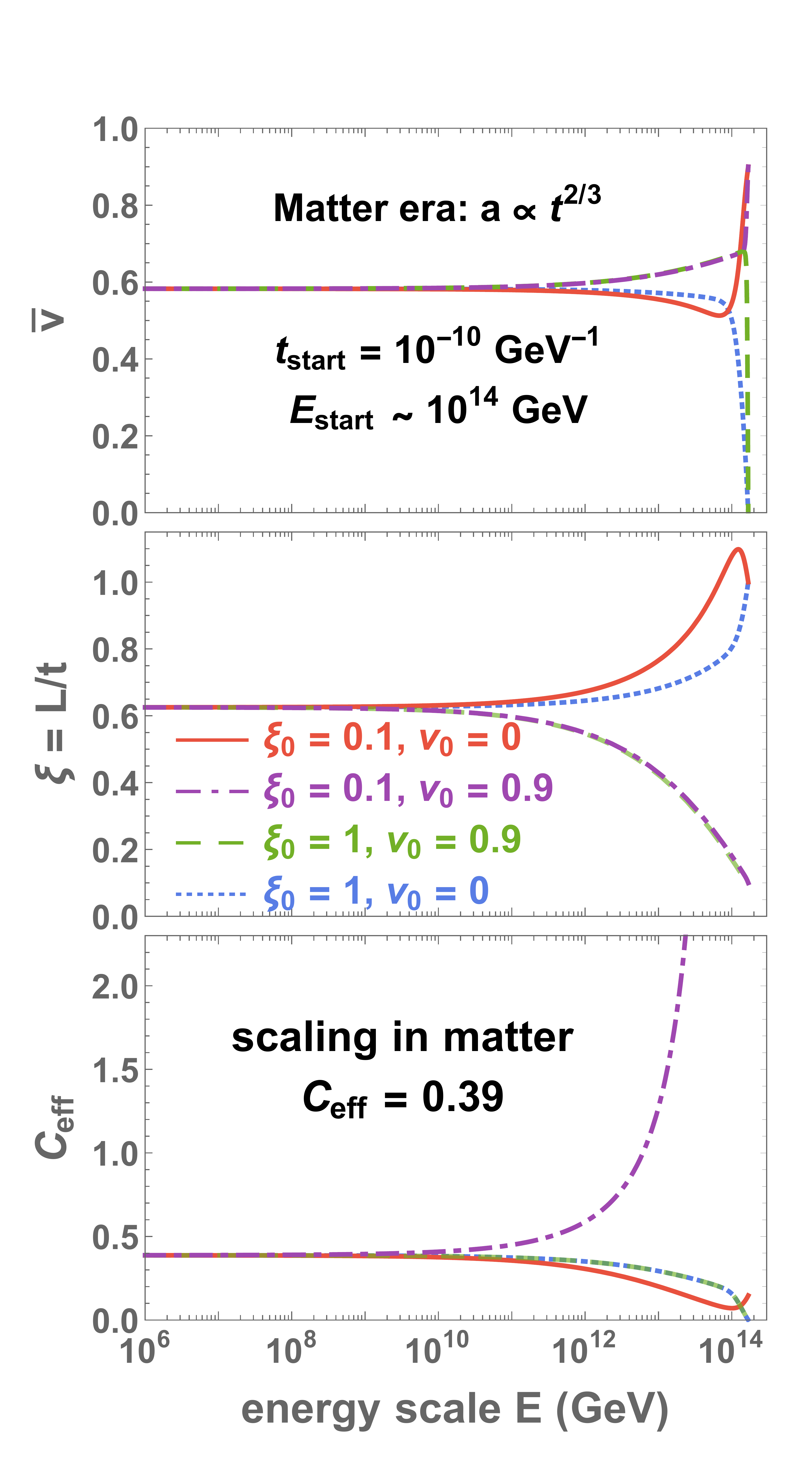}}}
\raisebox{0cm}{\makebox{\includegraphics[width=0.32\textwidth, scale=1]{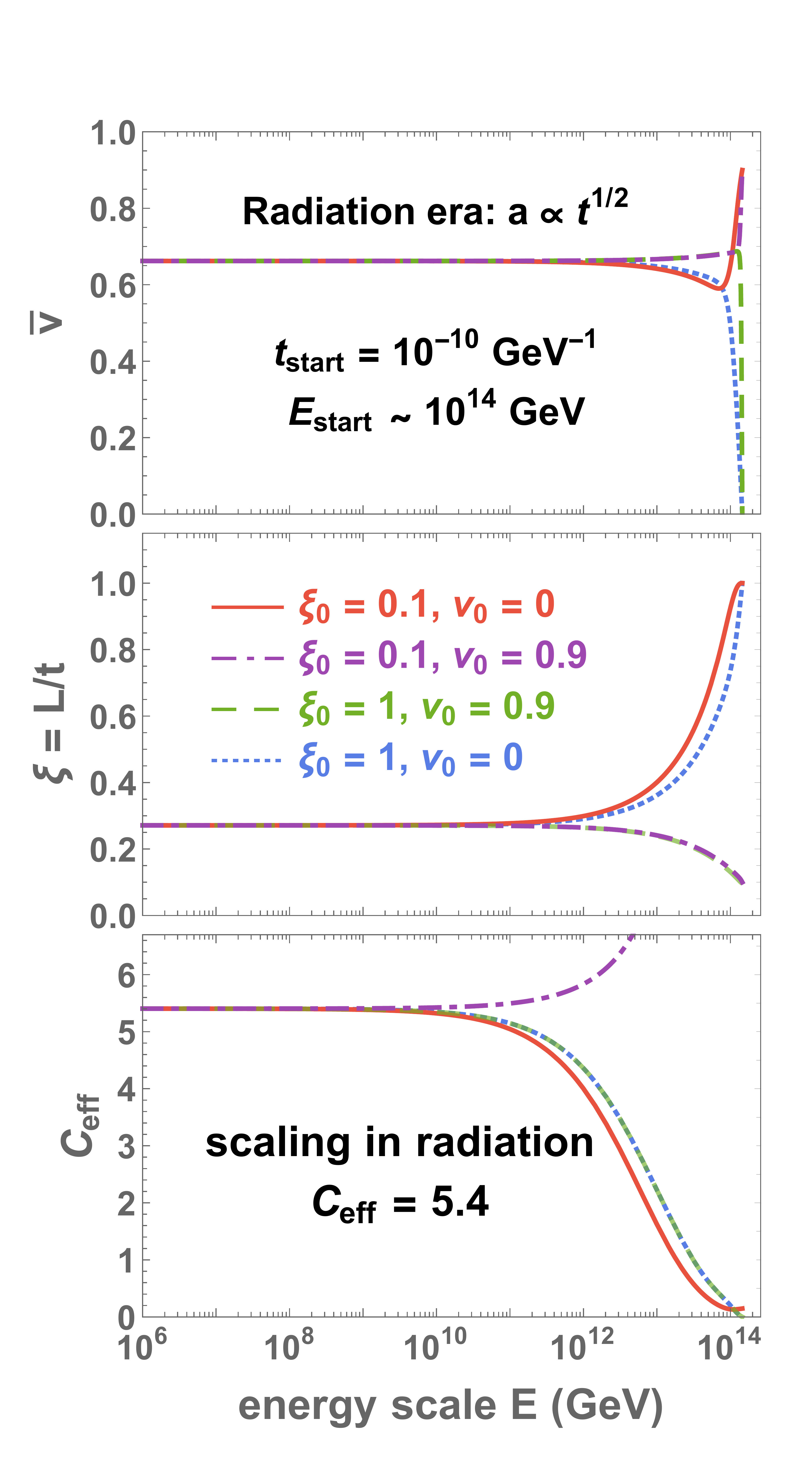}}}
\raisebox{0cm}{\makebox{\includegraphics[width=0.32\textwidth, scale=1]{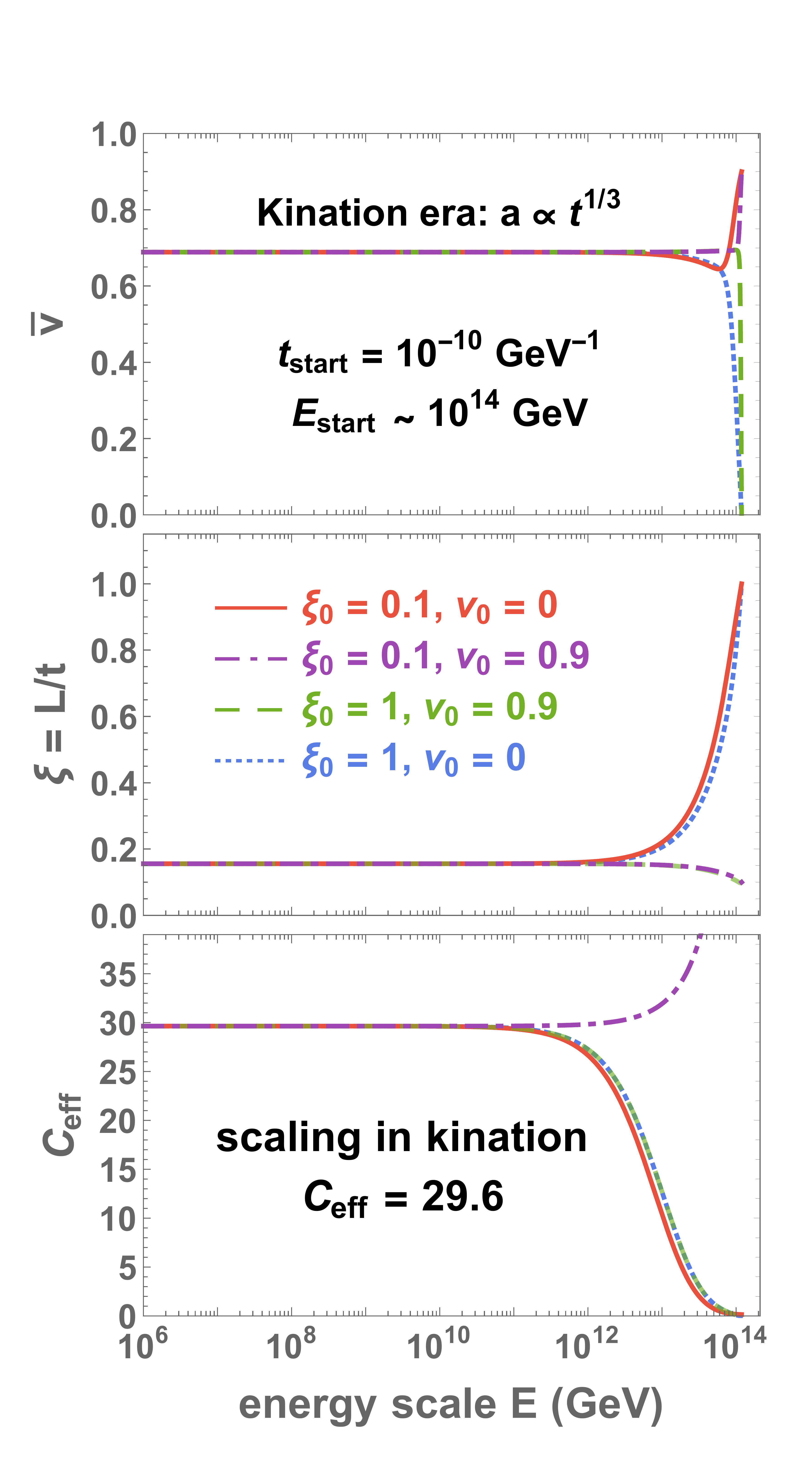}}}
\caption{\it \small Cosmic-string network evolving in the one-component universe with energy density $\rho\sim a^{-n}$ where $n=$ 3, 4 and 6 correspond to matter, radiation and kination, respectively. The long-string-network mean velocity $\bar{v}$, the correlation length $\xi$ and the corresponding loop-production efficiency $C_\textrm{eff}$ reach the scale-invariant solutions after the Hubble expansion rate has dropped by 2 orders of magnitude, independently of the initial conditions.}
\label{scaling_evolution}
\end{figure}

\paragraph{Scaling solution vs VOS solution:}
Fig.~\ref{scaling_evolution} shows the evolutions of $\xi,~\bar{v}$, and $C_\textrm{eff}$, from solving the VOS equations in Eq.~\eqref{eq:VOS_eq_body} with three equations of state, matter, radiation and kination. Regardless of the initial-condition choice, the network approaches a scaling solution where all parameters become constant. The energy scale of the universe has to decrease by some 4 orders of magnitude before reaching the scaling regime after the network formation. For a cosmological background evolving as $a\propto t^{2/n}$ with $n \geq 2$, the scaling regime solution is
\begin{equation}
\label{eq:VOS_scaling_solution}
\xi=\textrm{ constant} \textrm{\hspace{1.5em}and\hspace{1.5em}}\bar{v}=\textrm{ constant},
\end{equation}
with
\begin{equation}
\label{eq:VOS_scaling_solution2}
\textrm{with\hspace{1.5em}}\xi=\frac{n}{2}\sqrt{\frac{k(\bar{v})[k(\bar{v})+\tilde{c}]}{2(n-2)}}\quad\textrm{and\hspace{1em}}\bar{v}=\sqrt{\frac{n}{2}\frac{k(\bar{v})}{[k(\bar{v})+\tilde{c}]}\left(1-\frac{2}{n}\right)}.
\end{equation}
In order to fix the notation used in our plots, we define
\begin{itemize}
\item
\textbf{(Instantaneous) scaling network}: The loop-formation efficiency $C_{\rm eff}$, defined in Eq.~\eqref{eq:loopEfficiency}, is taken at its steady state value, given by Eq.~\eqref{eq:VOS_scaling_solution2}. In particular for matter, radiation and kination domination, one has 
\begin{equation}
\label{eq:Ceff_scaling}
C_\textrm{eff} \simeq 0.39,~5.4,~29.6 \qquad \text{for} \quad n=3,~4,~6.
\end{equation}
During a change of era $1\to 2$, $C_{\rm eff}$ is assumed to change instantaneously from the scaling regime of era $1$ to the scaling regime of era $2$. This is the assumption adopted in \cite{Cui:2017ufi, Cui:2018rwi}.
\item
\textbf{VOS network}: The loop-formation efficiency $C_{\rm eff}$, defined in Eq.~\eqref{eq:loopEfficiency}, is computed upon integrating the VOS equations in Eq.~\eqref{eq:VOS_eq_body}. During a change of cosmology, the long-string network experiences a transient regime.
\end{itemize}

\paragraph{Beyond the scaling regime in standard cosmology:}

In Fig.~\ref{ST_vos_scaling} and Fig.~\ref{ST_vos_scaling_ceff}, we compare the GW spectra and the $C_\textrm{eff}$ evolution, obtained with a scaling and VOS network. They are quite similar. The main difference arises from the change in relativistic degrees of freedom near the QCD confining temperature and from the matter-radiation transition. In contrast, predictions differ significantly when considering non-standard cosmology.

\paragraph{Beyond the scaling regime in non-standard cosmology:}

In Fig.~\ref{intermediate_matter_spectrum}, in dashed vs solid, we compare the loop-production efficiency factor $C_\textrm{eff}$ and the corresponding GW spectra for  a scaling network and for  a VOS network. The VOS frequency of the turning point due to the change of cosmology is shifted to a lower frequency by a factor $\sim 22.5$ with respect to the corresponding scaling frequency.\footnote{ The turning-point frequency can even be smaller by ${\cal O}$(400) if in a far-future, a precision of the order of $1\%$ can be reached in the measurement of the SGWB, cf. Eq.~\eqref{turning_point_general_scaling_app}.}. The shift results from the extra-time needed by the network to achieve its transient evolution to the new scaling regime. In the rest of this work, we go beyond the instantaneous scaling approximation used in \cite{Cui:2017ufi,Cui:2018rwi}.

\section{Standard cosmology}
\label{sec:standard}
\subsection{The cosmic expansion}
The SGWB from CS, cf. master formula in Eq.~\eqref{eq:SGWB_CS_Formula}, depends on the cosmology through the scale factor $a$. We compute the later upon integrating the Friedmann equation
\begin{equation}
H^2=\frac{\rho}{3 M_{\rm pl}^2},\label{friedmann_eq}
\end{equation}
for a given energy density $\rho$. In the standard $\Lambda$CDM scenario, the universe is first dominated by radiation, then a matter era, and finally the cosmological constant so that we can write the energy density as 
\begin{equation}
\rho_\textrm{ST,0}(a)=\rho_{r,0}\,\Delta_R(T(a),T_0)\left(\frac{a}{a_0}\right)^4+\rho_{m,0}\left(\frac{a}{a_0}\right)^3+\rho_{k,0}\left(\frac{a}{a_0}\right)^2+\rho_{\Lambda,0},
\end{equation}
where $r,m,k$ and $\Lambda$ denote radiation, matter, curvature, and the cosmological constant, respectively. We take $\rho_i=\Omega_i h^2 \, 3 M_{\rm pl}^2 H_0^2$, where $H_0=100$~km/s/Mpc, $\Omega_{r}h^2 \simeq 4.2\times 10^{-5}$, $\Omega_{\rm m}h^2 \simeq 0.14$, $\Omega_k\simeq 0$, $\Omega_{\Lambda}h^2 \simeq 0.31$ \cite{Tanabashi:2018oca}. The presence of the function
\begin{equation}
\Delta_R=\left(\frac{g_*(T)}{g_*(T_0)}\right)\left(\frac{g_{*s}(T_0)}{g_{*s}(T)}\right)^{4/3},
\label{eq:DeltaR}
\end{equation}
comes from imposing the conservation of the comoving entropy $g_{*s}\,T^3\,a^3$, where the evolutions of $g_*$ and $g_{*,s}$ are taken from appendix C of \cite{Saikawa:2018rcs}.
We discuss the possibility of adding an extra source of energy density in the next sections, intermediate matter in Sec.~\ref{sec:interm_matter} and intermediate inflation in Sec.~\ref{sec:inflation}.

\subsection{Gravitational wave spectrum}

Fig.~\ref{ST_vos_scaling} shows the dependence of the spectrum on the string tension.  The amplitude decreases with $G\mu$ due to the lower energy stored in the strings. Moreover, at lower $G\mu$, the loops decaying slower, the GW are emitted later, implying a lower redshift factor and a global shift of the spectrum to higher frequencies.
The figure also shows how the change in SM relativistic degrees of freedom introduces a small red-tilt which suppresses the spectrum by a factor $ \Delta_R^{-1} \sim 2.5$ at high frequencies.
We find that the amplitude of the GW spectrum at large frequency, assuming a standard cosmology, is given by 
\begin{equation}
\label{eq:lewicki_formula_GW_spectrum_radiation}
\Omega_{\rm GW}h^2 \simeq 15\pi\, \Delta_{R} \,\Omega_{r}h^2 \,C_{\rm eff}(n=4)\, \mathcal{F}_\alpha \,\left( \alpha\, G \mu /\Gamma\right)^{\! 1/2},
\end{equation}
where $\Omega_{r}h^2 \simeq 4.2\times 10^{-5}$ is the present radiation energy density of the universe \cite{Tanabashi:2018oca}.  
We provide an intuitive derivation based on the quadrupole formula in App.~\ref{sec:quadrupole_formula}.

\subsection{Deviation from the scaling regime}
Fig.~\ref{ST_vos_scaling_ceff} shows how the loop-formation efficiency $C_{\rm eff}$ varies during the change of SM relativistic degrees of freedom and the matter-radiation equality, upon solving the VOS equations, cf. Sec.~\ref{sec:VOS}. We see the associated corrections to the spectrum in Fig.~\ref{ST_vos_scaling}, and which were already pointed out in \cite{Auclair:2019wcv}. The spectrum is enhanced at low frequencies because more loops are produced than when assuming that the matter era is reached instantaneously, cf. Fig.~\ref{ST_vos_scaling_ceff}.

\subsection{Beyond the Nambu-Goto approximation}
\begin{figure}[h!]
\centering
\raisebox{0cm}{\makebox{\includegraphics[width=0.495\textwidth, scale=1]{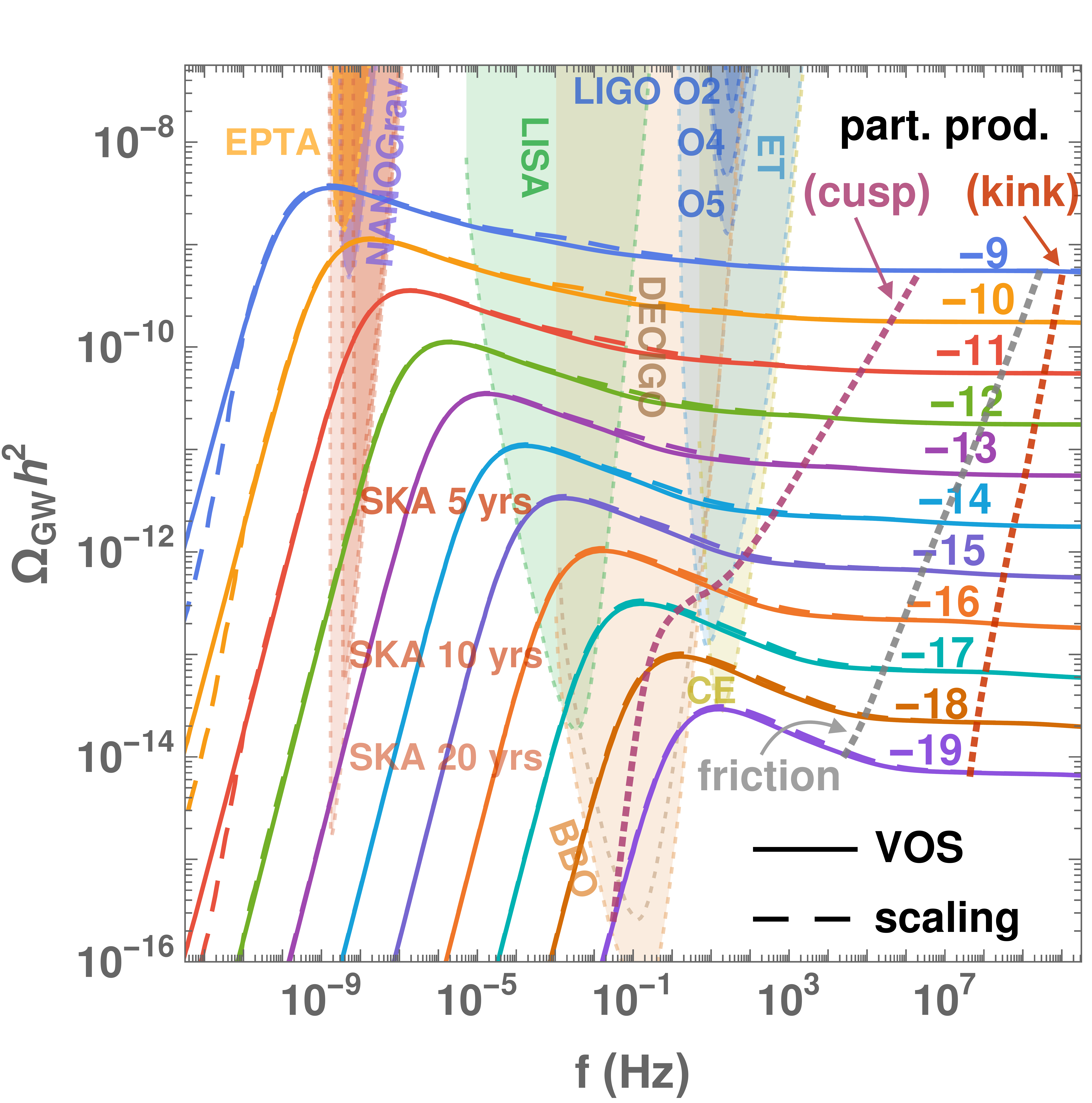}}}
\raisebox{0cm}{\makebox{\includegraphics[width=0.49\textwidth, scale=1]{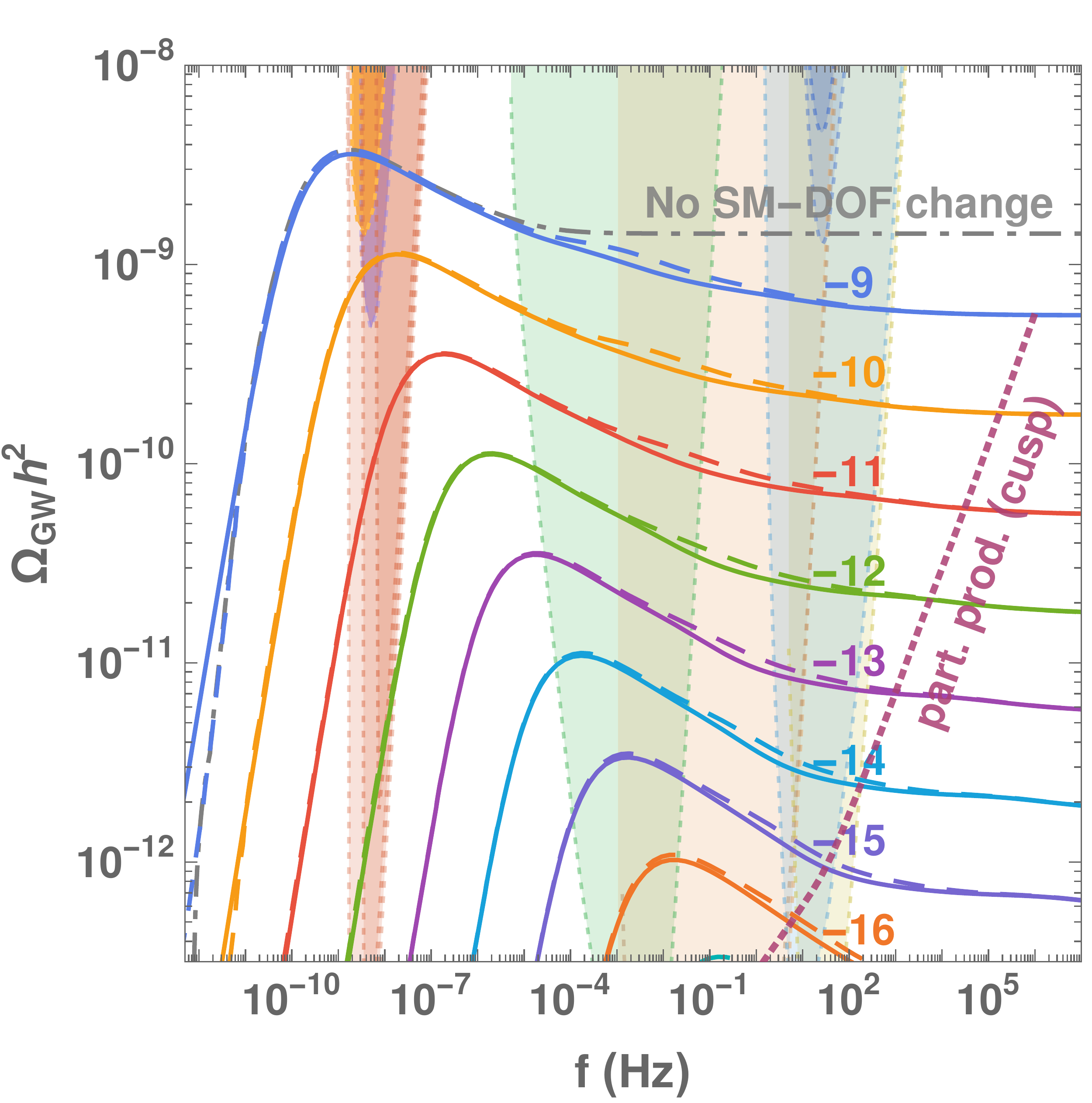}}}
\caption{\it \small \textbf{Left:} GW spectra from cosmic strings assuming either the scaling or VOS network, cf. Sec.~\ref{sec:scalingVSvos}, evolving in the standard cosmological background.  Each line corresponds to string tension $G\mu = 10^{x}$, where $x$ is specified by a number on each line. Dotted lines show the spectral cut-offs expected due to particle production, cf. Sec.~\ref{UVcutoff} and thermal friction, cf. Sec.~\ref{sec:thermal_friction}, which depend on the nature of the loop small-scale structures: cusp or kink-dominated. \textbf{Right:} The zoom-in plot of the left panel shows the effects from the change of SM degrees of freedom on the scaling and VOS networks.}
\label{ST_vos_scaling}
\end{figure}

\begin{figure}[h!]
\centering
\raisebox{0cm}{\makebox{\includegraphics[width=0.49\textwidth, scale=1]{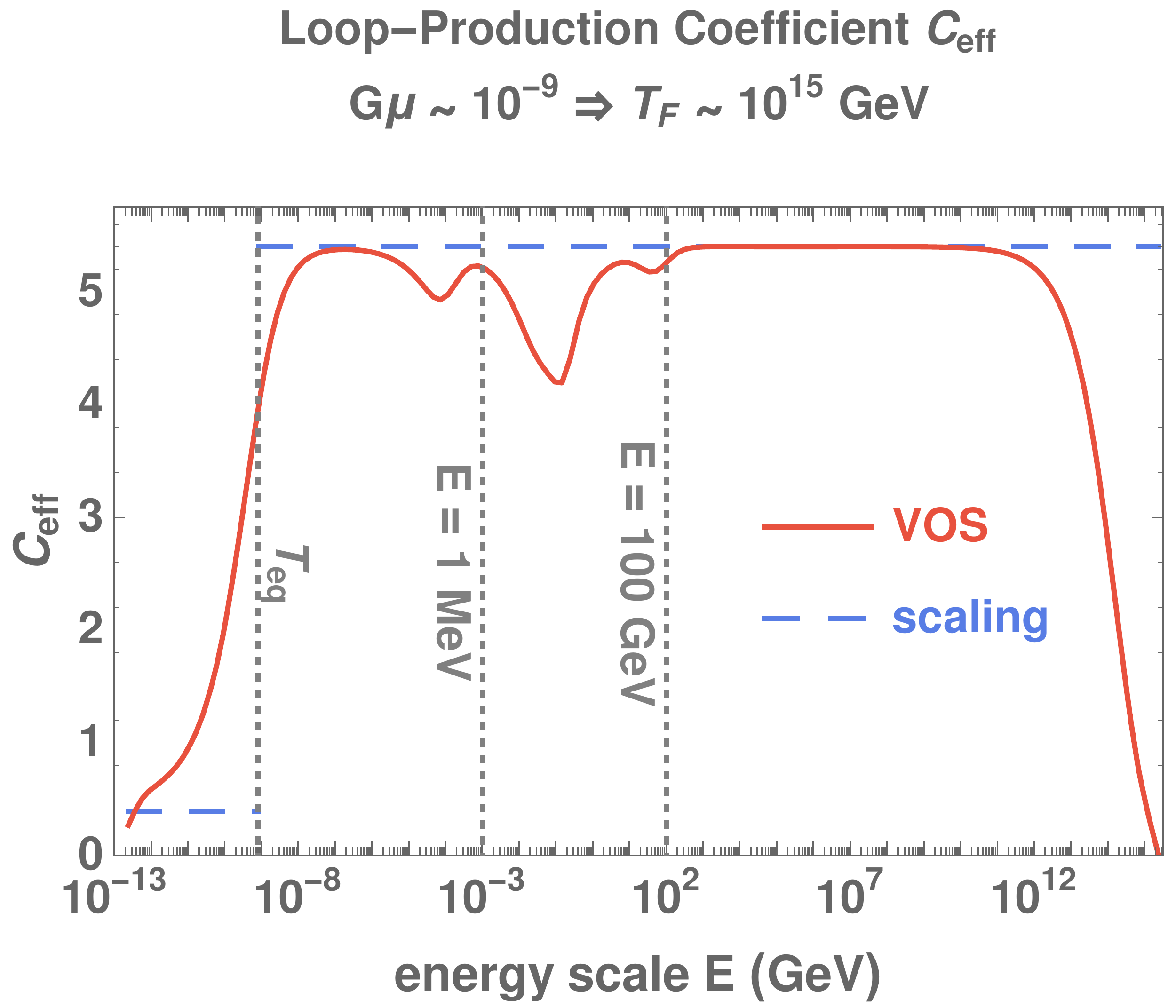}}}
\caption{\it \small  Comparison of the loop-production efficiency under the scaling assumption, where the attractor solution of the VOS equations is assumed to be reached instantaneously, and under the VOS assumptions, where one integrates the VOS equations. A standard cosmology is assumed.}
\label{ST_vos_scaling_ceff}
\end{figure}

Fig.~\ref{ST_vos_scaling} shows the possibility of a cut-off at high frequencies due to particle production, for two different assumptions regarding the loop small-scale structures: cusps or kinks domination, cf. Sec.~\ref{sec:massive_radiation}. Above these frequencies, loops decays into massive radiation before they have time to emit GW. For kinky loops, the cut-off is outside any future-planned observational bands, while for cuspy loops, the cut-off might be in the observed windows for $G\mu \lesssim 10^{-15}$.


\FloatBarrier
\section{Intermediate matter era}
\label{sec:interm_matter}

\subsection{The non-standard scenario}
\begin{figure}[h!]
\centering
\raisebox{0cm}{\makebox{\includegraphics[width=0.49\textwidth, scale=1]{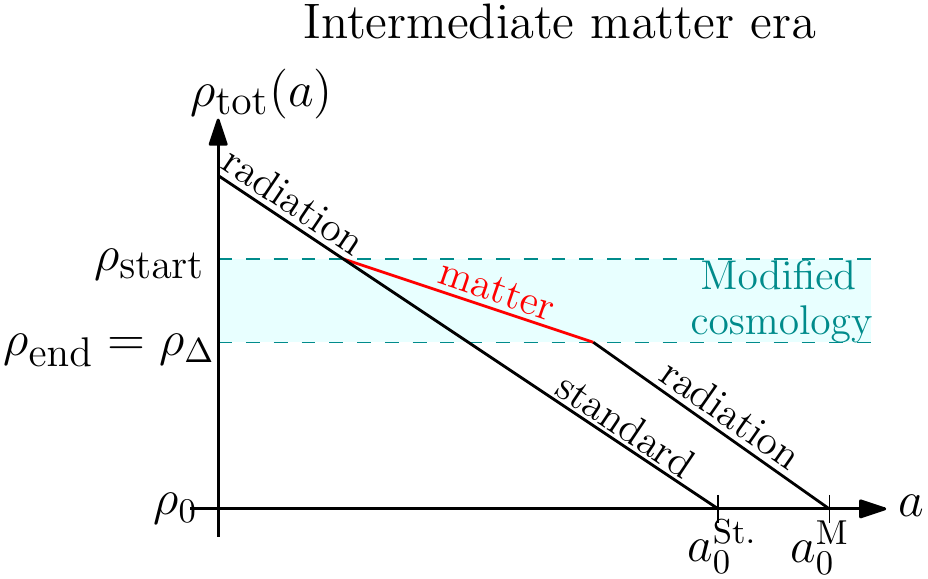}}}
\raisebox{0cm}{\makebox{\includegraphics[width=0.49\textwidth, scale=1]{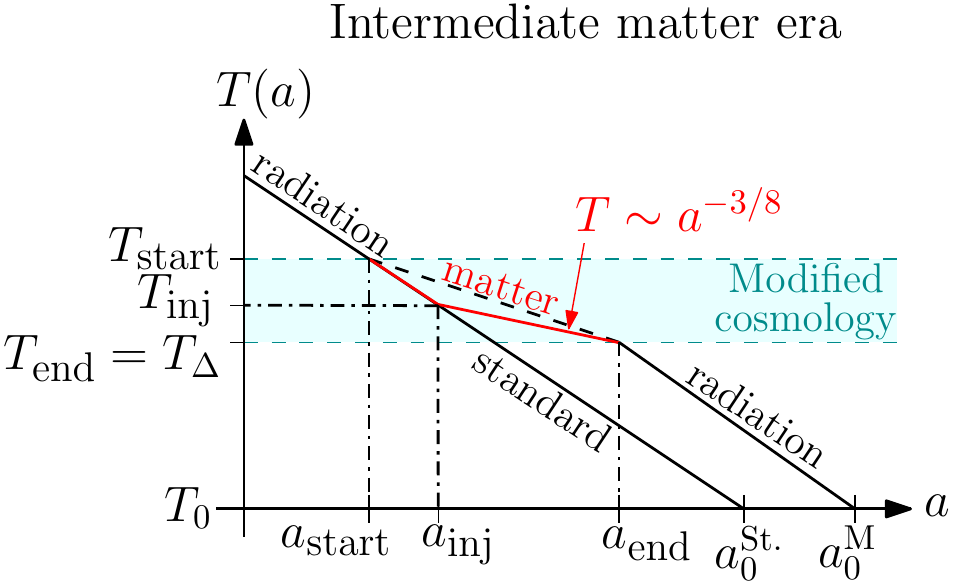}}}
\caption{\it \small Evolution of  the total energy density \textbf{(left)} and  the temperature \textbf{(right)} assuming the presence of an intermediate matter era. $T_{\rm inj}$ and $a_{\rm inj}$ are the temperature and scale factor at which the entropy injected by the decay of the matter content into radiation, starts to be effective, cf. Fig.~2 in \cite{Cirelli:2018iax}. St: standard; M: matter.}
\label{fig:md_intermediate_diag}
\end{figure}

In this section, we consider the existence of an early-intermediate-matter-dominated era, following an earlier radiation era and preceeding the standard radiation era. The intermediate matter-dominated era starts when the matter energy density $\rho_{\rm matter} \propto a^{-3}$ takes over the radiation energy density $\rho_{\rm radiation} \propto a^{-4}$ and ends when the matter content decays into radiation, cf. Fig.~\ref{fig:md_intermediate_diag}. The energy density profile is illustrated in Fig.~\ref{fig:md_intermediate_diag}
and can be written as
\begin{equation}
\rho_\textrm{tot}(a)=\begin{cases}
\rho^\textrm{st}_\textrm{rad}(a)+\rho_\textrm{late}(a)&\textrm{for }\rho>\rho_\textrm{start},\\[0.5em]
\rho_\textrm{start}\  \over{a_\textrm{start}}{a}^n+\rho_\textrm{late}(a)&\textrm{for }\rho_\textrm{start}>\rho>\rho_\textrm{end},\\[0.5em]
\rho_\textrm{end}   \Delta_R(T_\textrm{end},T) \ \over{a_\textrm{end}}{a}^4+\rho_\textrm{late}(a)\hspace{2em}&\textrm{for }\rho<\rho_\textrm{end}.
\end{cases}
\label{ns_cosmo_define_function}
\end{equation}
where
\begin{eqnarray*}
		\rho_\textrm{start}, \rho_\textrm{end} &\equiv& \textrm{ the starting and ending energy density of the non-standard cosmology},\\
		\rho_\textrm{late} &\equiv& \textrm{ the standard-cosmology energy density dominating at late times,}\\
		&&  \textrm{ e.g. the standard matter density, and cosmological constant}.\\
		\Delta_R &\mbox{is}& \mbox{given in Eq.}(\ref{eq:DeltaR}).
	\end{eqnarray*} 

\subsection{Impact on the spectrum}
\begin{figure}[h!]
\centering
\raisebox{0cm}{\makebox{\includegraphics[height=0.54\textwidth, scale=1]{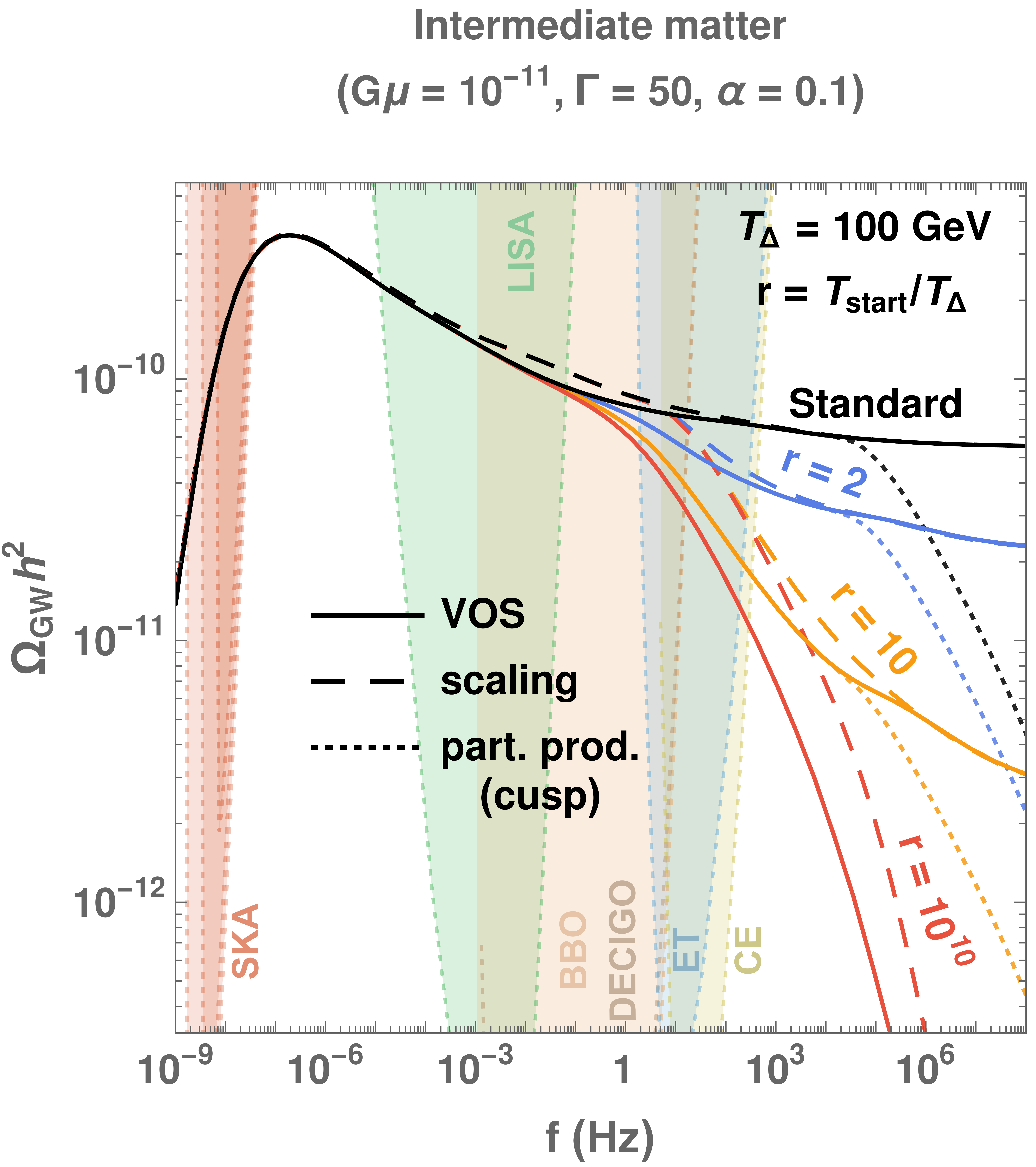}}}
\raisebox{0cm}{\makebox{\includegraphics[height=0.4\textwidth, scale=1]{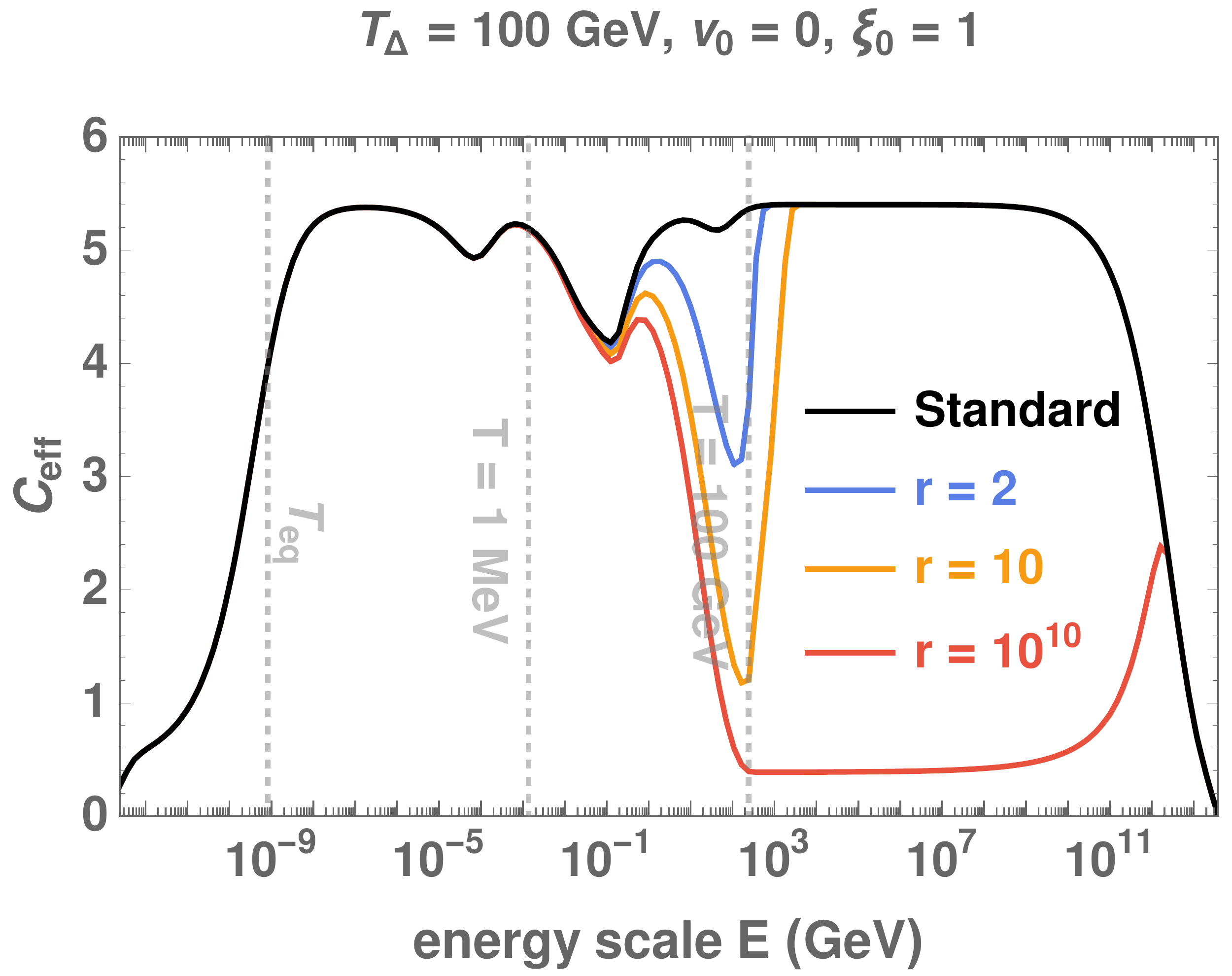}}}
\caption{\it \small \label{fig:VOSvsScaling_Ceff}  GW spectrum from an intermediate matter era starting at the temperature $T_{\rm start}$ and ending at $T_{\Delta}$. \textbf{Left:} The dashed-lines assume that the scaling regime in matter era switches instantaneously to the scaling regime in radiation era, meaning that $C_{\rm eff}$ varies discontinuously, whereas the plain lines incorporate the transient behavior solution of the VOS equations and shown on the right panel.  The cut-offs due to particle production, cf. Sec.~\ref{UVcutoff}, are shown with dotted lines.   \textbf{Right:} Time evolution of the loop-production efficiency $C_{\rm eff}$ after solving the VOS equations, cf. Sec.~\ref{sec:scalingVSvos}. }
\label{intermediate_matter_spectrum}
\end{figure}

\paragraph{A low-pass filter:}
In the left panel of Fig.~\ref{intermediate_matter_spectrum}, we show that an intermediate matter era blue-tilts the spectral index of the spectrum. Furthermore, at higher frequencies, corresponding to loops produced during the radiation era preceding the matter era, the spectrum recovers a flat scaling but is suppressed by the duration $r$ of the matter era
\begin{equation}
r=\frac{T_\textrm{start}}{T_{\rm \Delta}},
\end{equation}
where $T_{\rm \Delta} =T_{\rm end}$. By suppressing the high-frequency part of the spectrum, an early matter era acts on the CS spectrum as a low-pass filter.
The negative spectral index and the suppression can be understood from Fig.~\ref{fig:md_intermediate_diag}. Indeed, the universe, in the presence of an intermediate matter era, has expanded more than the standard universe. Hence at a fixed emitted frequency, loops are produced later and so are less numerous, implying less GW emission.
In the right panel of Fig.~\ref{intermediate_matter_spectrum}, we show that for short intermediate matter era, $r=2$ or $r=10$, the scaling regime in the matter era, which is characterized by $C_{\rm eff} = 0.39$, cf. Eq.~\eqref{eq:Ceff_scaling}, is not reached. 

\paragraph{A turning-point:}
\label{sec:turning_point_general}
A key observable is the frequency above which the GW spectrum differs from the one obtained in standard cosmology. This is the so-called \textit{turning-point} frequency $f_{\Delta}$.
It  corresponds to the redshifted-frequency emitted by the loops created during the change of cosmology at the temperature $T_{\Delta}$. In the instantaneous scaling approximation, cf. dashed line in Fig.~\ref{fig:VOSvsScaling_Ceff}, the turning-point frequency $f_{\Delta}$ is given by the $(T,\,f)$-correspondence relation 
\begin{align}
f_\Delta^{\textrm{scaling}}=(4.5\times10^{-2}\textrm{ Hz})\left(\frac{T_\Delta}{\textrm{GeV}}\right)\left(\frac{0.1\times 50 \times10^{-11}}{\alpha\,\Gamma G\mu}\right)^{1/2}\left(\frac{g_*(T_\Delta)}{g_*(T_0)}\right)^{1/4}.
\label{eq:turning_point_scaling}
\end{align}
However, the deviation from the scaling regime during the change of cosmology, cf. Sec.~\ref{sec:scalingVSvos}, implies a shift to lower frequencies of the $(T,\,f)$-correspondence, by a factor $\sim 22.5$, cf. solid vs dashed lines in Fig.~\ref{fig:VOSvsScaling_Ceff}. The correct $(T,\,f)$-correspondence when applied to a change of cosmology is
\begin{align}
f_\Delta^{\textrm{VOS}}=(2\times10^{-3}\textrm{ Hz})\left(\frac{T_\Delta}{\textrm{GeV}}\right)\left(\frac{0.1\times 50 \times10^{-11}}{\alpha\,\Gamma G\mu}\right)^{1/2}\left(\frac{g_*(T_\Delta)}{g_*(T_0)}\right)^{1/4}.
\label{turning_point_general}
\end{align}
We fit the numerical factor in Eq.~\eqref{turning_point_general} (but also in Eq.~\eqref{turning_point_inf}) by imposing\footnote{The coefficient in Eq.~\eqref{turning_point_general} has been fitted upon considering the matter case $\Omega_\textrm{NS}= \Omega_\textrm{matter}$. Note that the turning-point in the kination case is slightly higher frequency by a factor of order 1, cf. Fig.~\ref{fig:VOSvsScaling_Ceff}.} the non-standard-cosmology spectrum $\Omega_\textrm{NS}$ to deviate from the standard-cosmology one $\Omega_\textrm{ST}$ by $10\%$ at the turning-point frequency,
\begin{align}
\left|\frac{\Omega_\textrm{NS}(f_\Delta)-\Omega_\textrm{ST}(f_\Delta)}{\Omega_\textrm{ST}(f_\Delta)}\right|\simeq 10\%.
\label{10per_criterion}
\end{align}
We are conservative here. Choosing $1\%$ instead of $10\%$ would lead to a frequency shift of the order of  ${\cal O}(400)$, cf. Eq. \eqref{turning_point_general_scaling_app}.
Note that our Eq.~\eqref{turning_point_general} is numerically very similar to the one in \cite{Cui:2017ufi,Cui:2018rwi,Auclair:2019wcv} although an instantaneous change of the loop-production efficiency $C_{\rm eff}$ at $T_{\rm \Delta}$ is assumed in \cite{Cui:2017ufi,Cui:2018rwi,Auclair:2019wcv}. This can be explained if  in Ref.~\cite{Cui:2017ufi,Cui:2018rwi,Auclair:2019wcv},  the criterion in Eq.~\eqref{10per_criterion} is smaller than the percent level.

\subsection{Constraints}

\begin{figure}[h!]
			\centering
			\raisebox{0cm}{\makebox{\includegraphics[height=0.49\textwidth, scale=1]{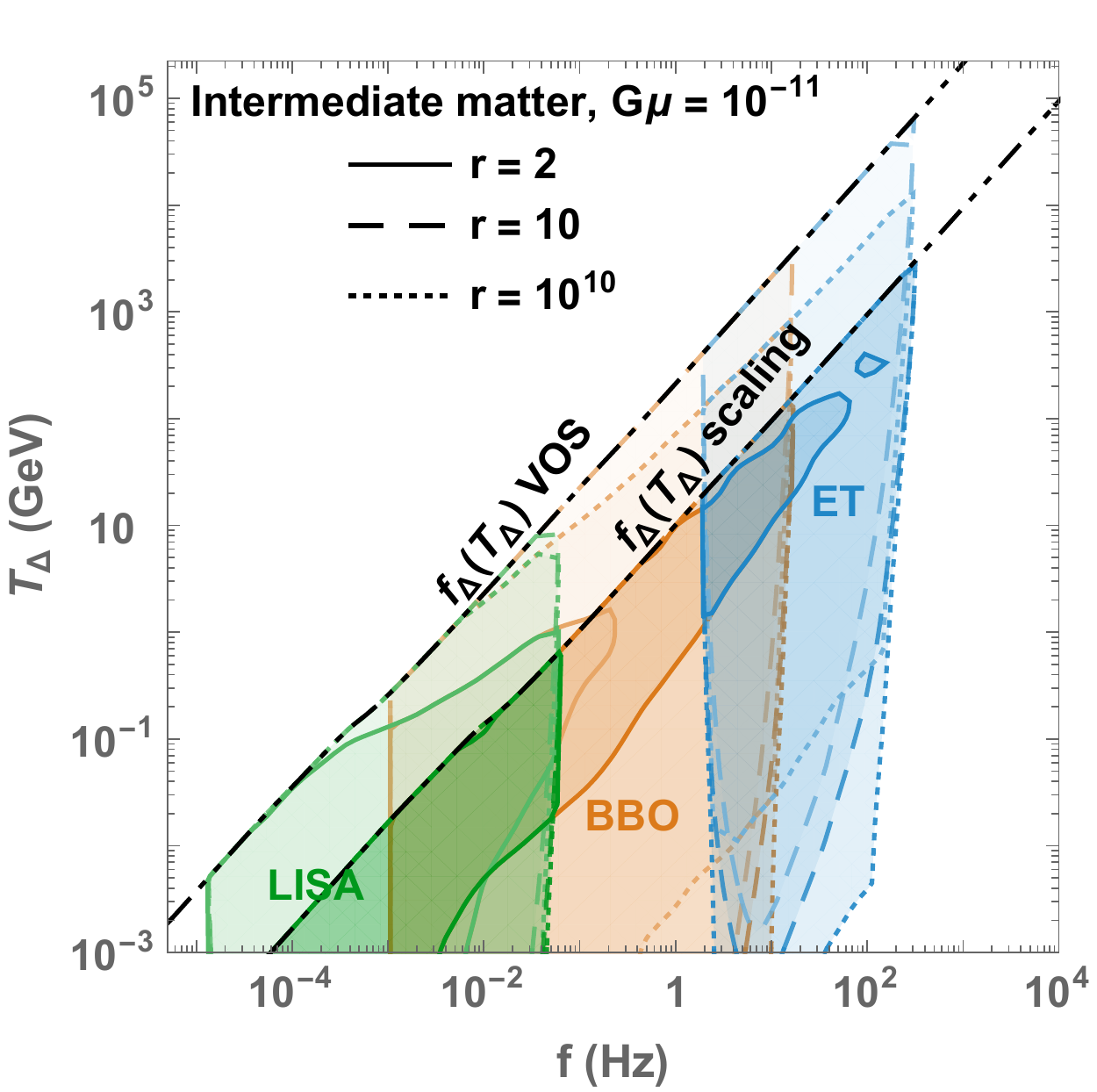}}}
			\raisebox{0cm}{\makebox{\includegraphics[height=0.49\textwidth, scale=1]{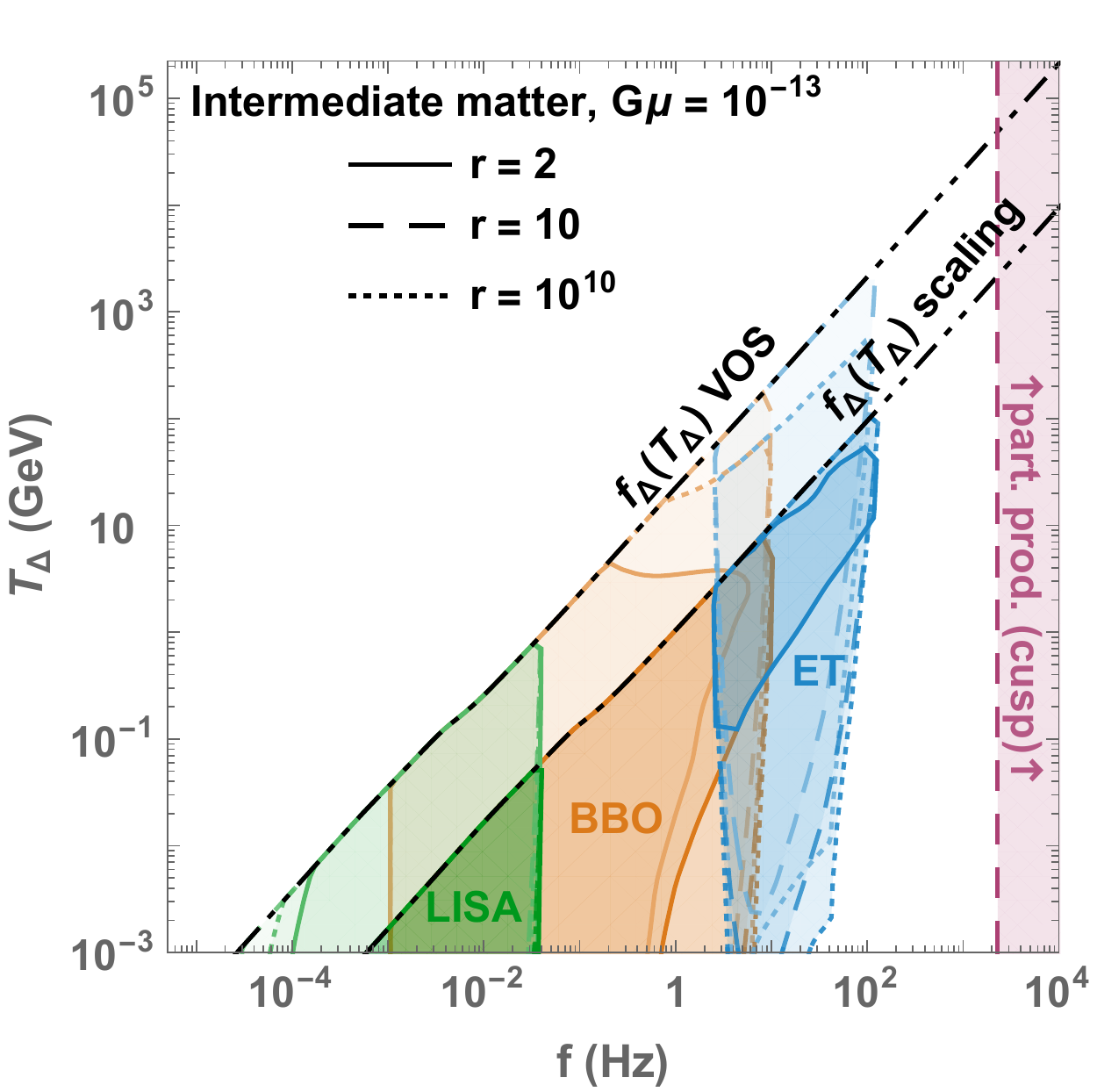}}}\\
			\raisebox{0cm}{\makebox{\includegraphics[height=0.49\textwidth, scale=1]{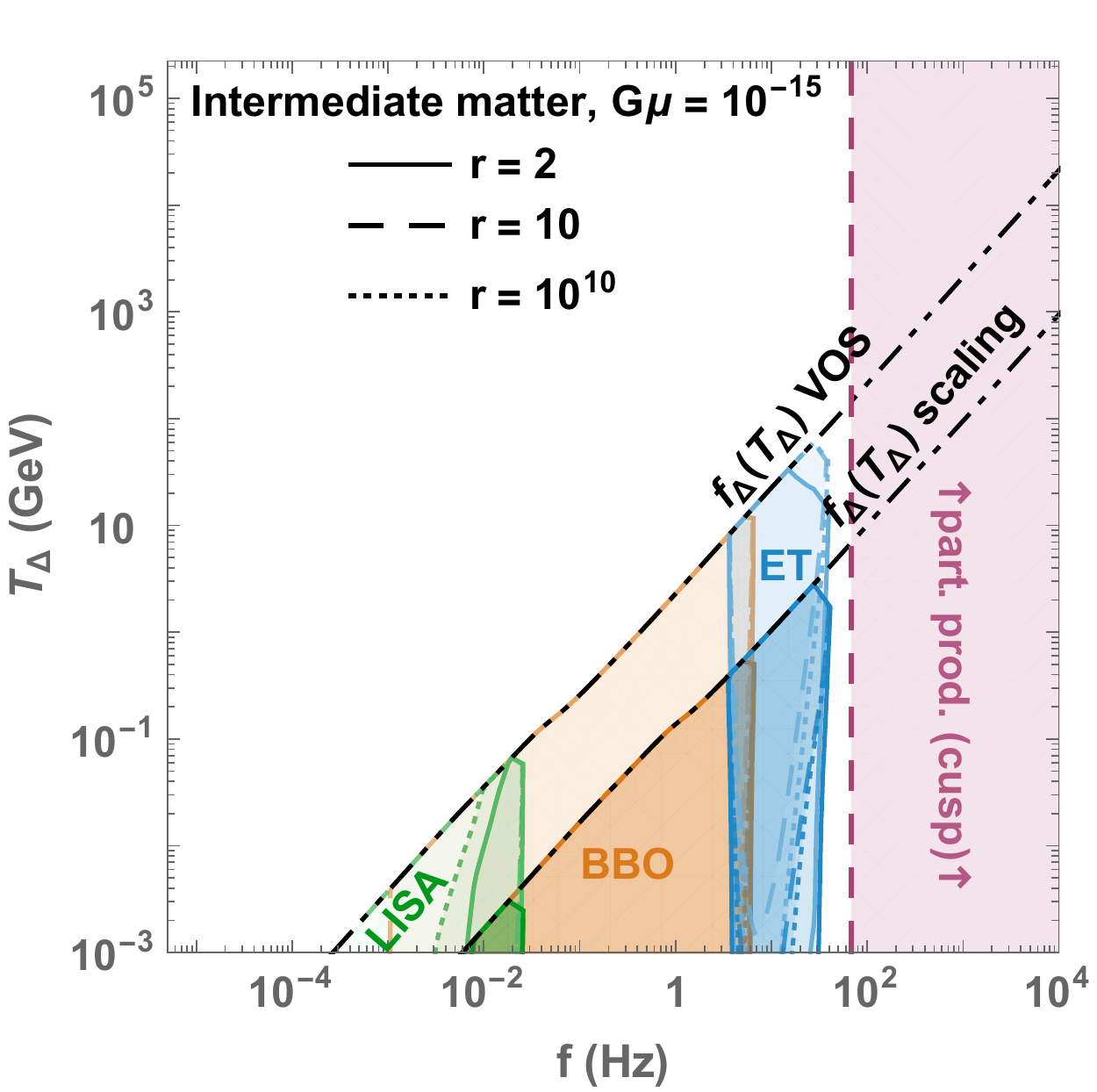}}}
			\hfill
			\caption{\it \small The colored regions show the detectability of the spectral suppression, cf. \textit{spectral-index prescription (Rx 2)} in Sec.~\ref{sec:triggerMatterGW}., due to a NS intermediate matter era with duration $r=T_\textrm{start}/T_\Delta$, assuming scaling and VOS networks, cf. Sec.~\ref{sec:scalingVSvos}. Limitation from particle production, cf. Sec.~\ref{UVcutoff}, is shown in purple.}
			\label{fig:contour_int}
		\end{figure}

In Fig.~\ref{fig:contour_int}, we show the constraints on the presence of an early-intermediate-non-standard-matter-dominated era starting at the temperature $r\,T_{\Delta}$ and ending at the temperature $T_{\Delta}$.
Matter eras as short as $r=2$ and ending at temperature as large as $100$~TeV could be probed by GW interferometers. We assume that an early-matter era is detectable if the spectral index is smaller than $-0.15$, cf. \textit{spectral-index prescription (Rx 2)} in Sec.~\ref{sec:triggerMatterGW}.
In the next chapter, Chap.~\ref{chap:DM_GW_CS}, we provide model-independent constraints on the abundance and lifetime of an unstable particles giving rise to such a non-standard intermediate matter era.

\section{Intermediate inflation}
\label{sec:inflation}

\subsection{The non-standard scenario}
Next, we consider the existence of a short inflationary period with a number of e-folds
\begin{align}
N_e \equiv \log\left(\frac{a_\textrm{start}}{a_\textrm{end}}\right),
\end{align}
smaller than $N_e \lesssim 20 \ll 60$, in order not to alter the predictions from the first inflation era regarding the CMB power spectrum. On the particle physics side, such a short inflationary period can be generated by a highly supercooled first-order phase transition.
It was stressed that nearly-conformal scalar potentials naturally lead to such short, with $N_e \sim 1-15$,  periods of inflation \cite{Konstandin:2011dr,vonHarling:2017yew,Bruggisser:2018mrt}. Those are well-motivated in new strongly interacting composite sectors arising at the TeV scale, as invoked to address the Higgs hierarchy problem and were first studied in a holographic approach \cite{Creminelli:2001th, Randall:2006py} (see also the review \cite{Caprini:2019egz}). As the results on the scaling of the bounce action for tunnelling and on the dynamics  of the phase transitions do essentially not depend on the absolute energy scale, but only on the shallow shape of the scalar potential describing the phase transition, those studies can thus be extended to a large class of confining phase transitions arising at any scale. In this section, we will take this inflationary scale as a free parameter.

\begin{figure}[h!]
\centering
\raisebox{0cm}{\makebox{\includegraphics[width=0.7\textwidth, scale=1]{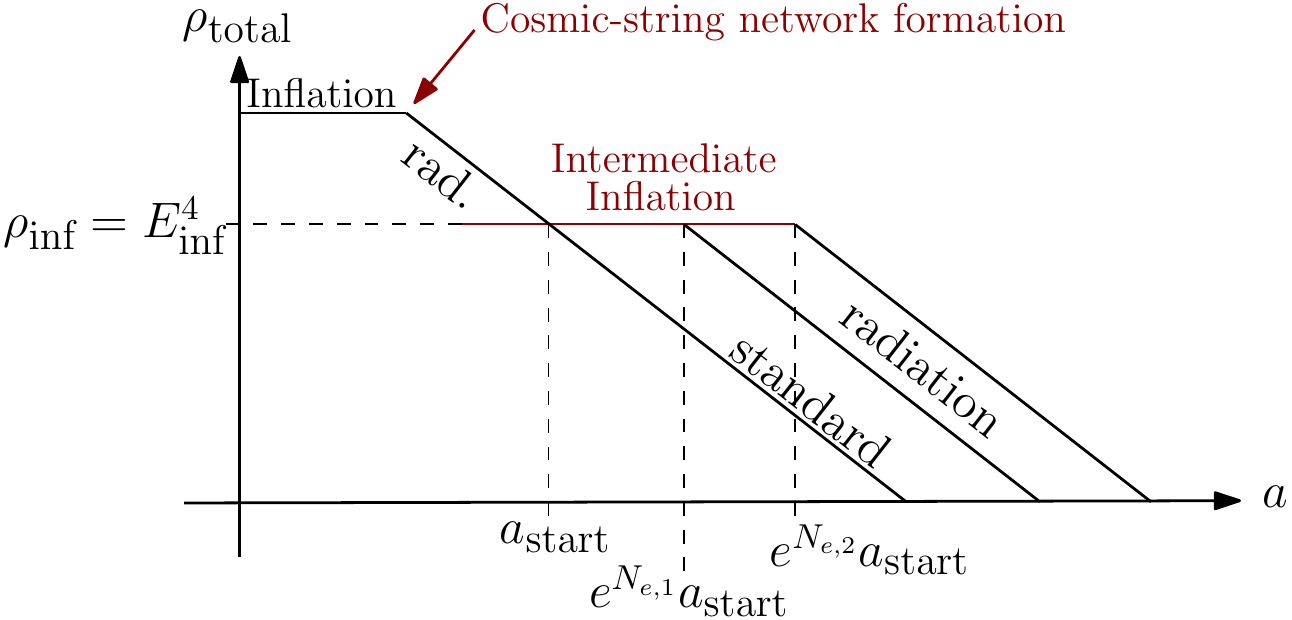}}}
\caption{\it\small Evolution of the total energy density assuming  the presence of an intermediate inflationary era characterised by the energy density $\rho_\textrm{inf}$, for two different durations (number of efolds), $N_{e,1}$ and $N_{e,2}$.}
\label{figure_inter_inflation}
\end{figure}

 We define the energy density profile as, cf. Fig.~\ref{figure_inter_inflation}
\begin{equation}
\rho_\textrm{tot}(a)=\begin{cases}
\rho^\textrm{st}_\textrm{rad}(a)+\rho_\textrm{late}(a)&\textrm{for }\rho>\rho_\textrm{inf},\\[0.5em]
\rho_\textrm{inf}=E^4_\textrm{inf}\hspace{2em}&\textrm{for }\rho=\rho_\textrm{inf},\\[0.5em]
\rho_\textrm{inf}\Delta_R(T_\textrm{end},T)\over{a_\textrm{end}}{a}^4+\rho_\textrm{late}(a)\hspace{2em}&\textrm{for }\rho<\rho_\textrm{inf},
\end{cases}
\label{inf_define_function}
\end{equation}
where $\rho_\textrm{inf}$ is the total energy density of the universe during inflation and $E_\textrm{inf}\equiv \rho_\textrm{inf}^{1/4}$ is the corresponding energy scale.  The function $\Delta_R$ is defined in (\ref{eq:DeltaR}).
%

%
\begin{figure}[h!]
\centering
\raisebox{0cm}{\makebox{\includegraphics[width=0.9\textwidth, scale=1]{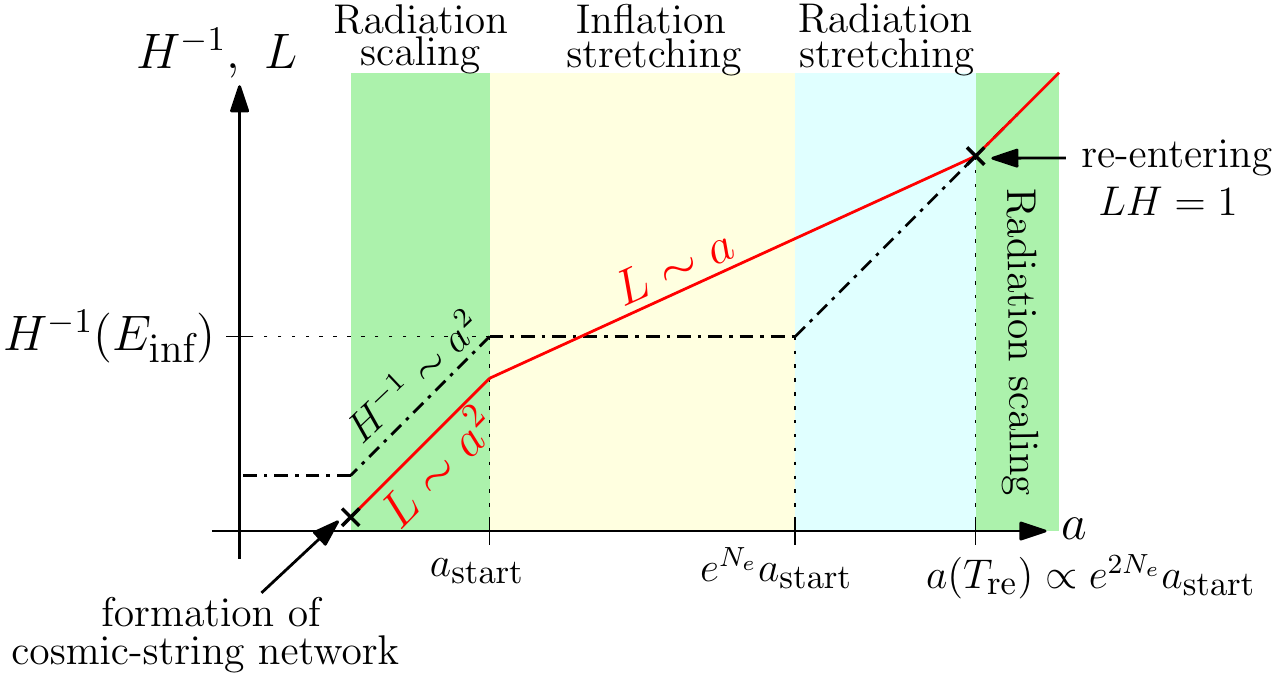}}}
\caption{\it\small After its formation, before inflation, the network enters the scaling regime with $L \sim a^2$ due to loop formation. During the $N_e$ e-folds of inflation, the network correlation length gets stretched out of the horizon by the rapid expansion and loop formation stops, thus $L\sim a$. After inflation, during radiation, the correlation length starts to re-enter the horizon and scales again as $L \sim a^2$.
}
\label{fig:stringsandinflation}
\end{figure}

\begin{figure}[h!]
\centering
\raisebox{0cm}{\makebox{\includegraphics[width=0.7\textwidth, scale=1]{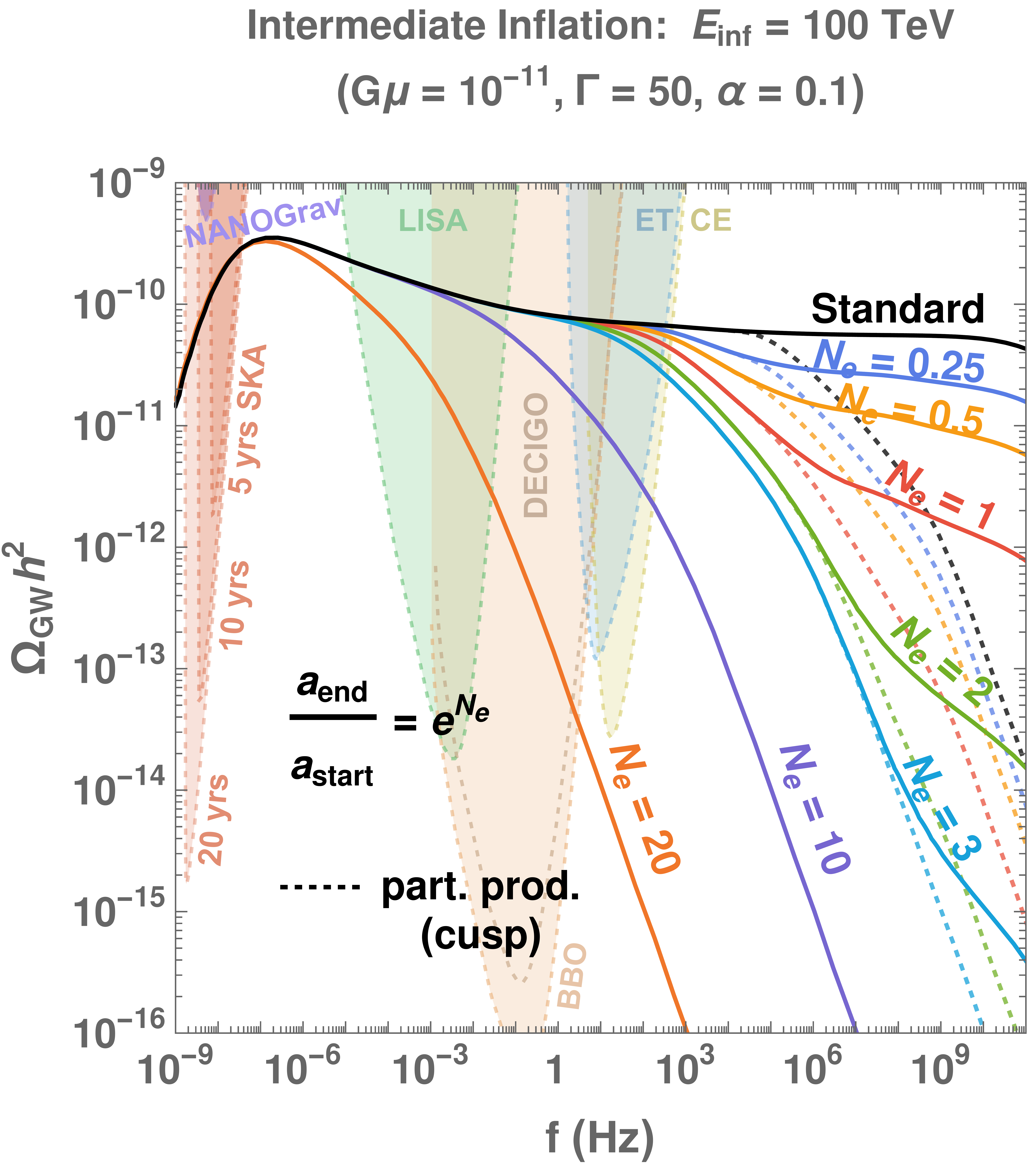}}}
\raisebox{0.65cm}{\makebox{\includegraphics[width=0.49\textwidth, scale=1]{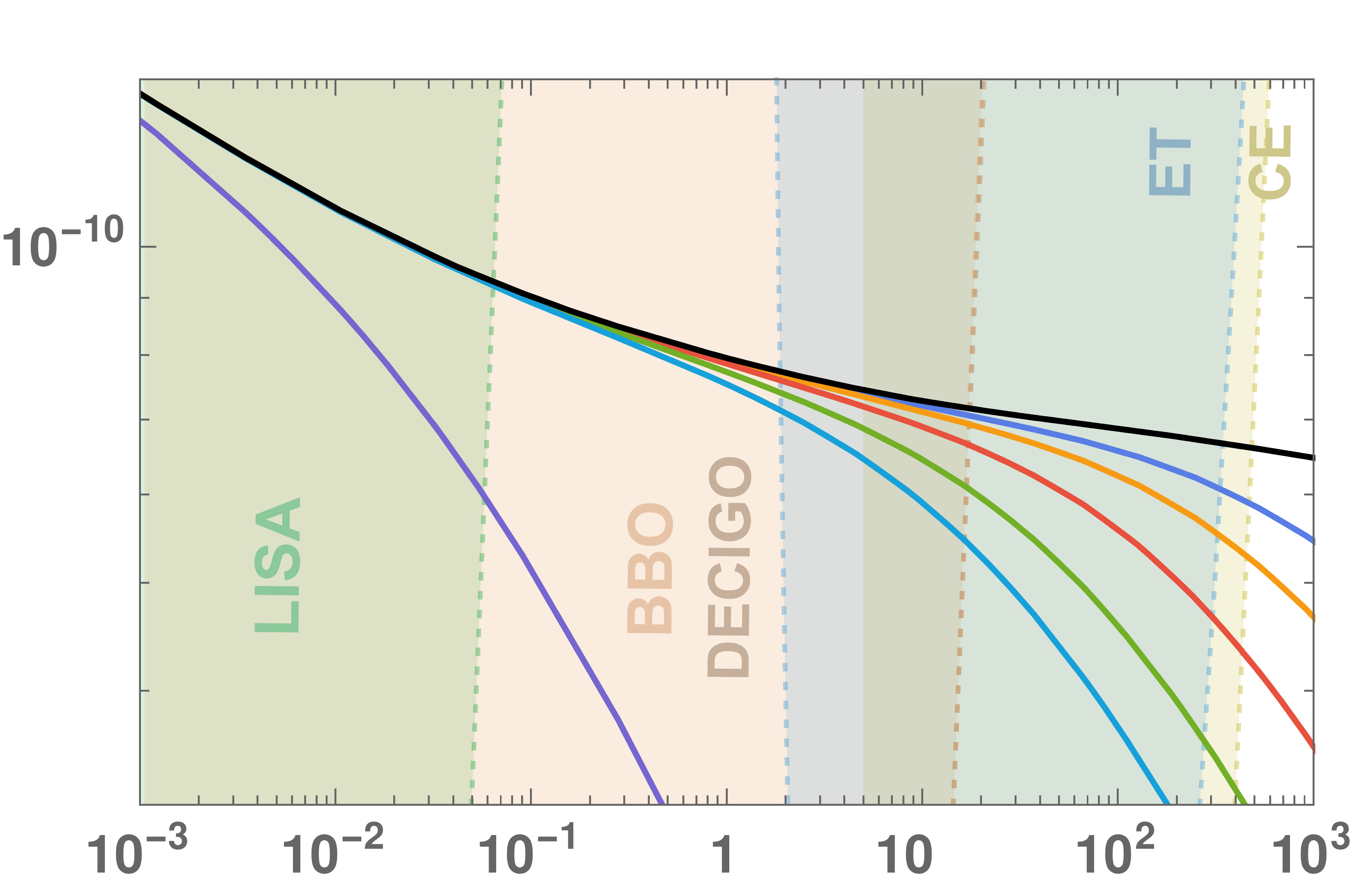}}}
\raisebox{0cm}{\makebox{\includegraphics[width=0.49\textwidth, scale=1]{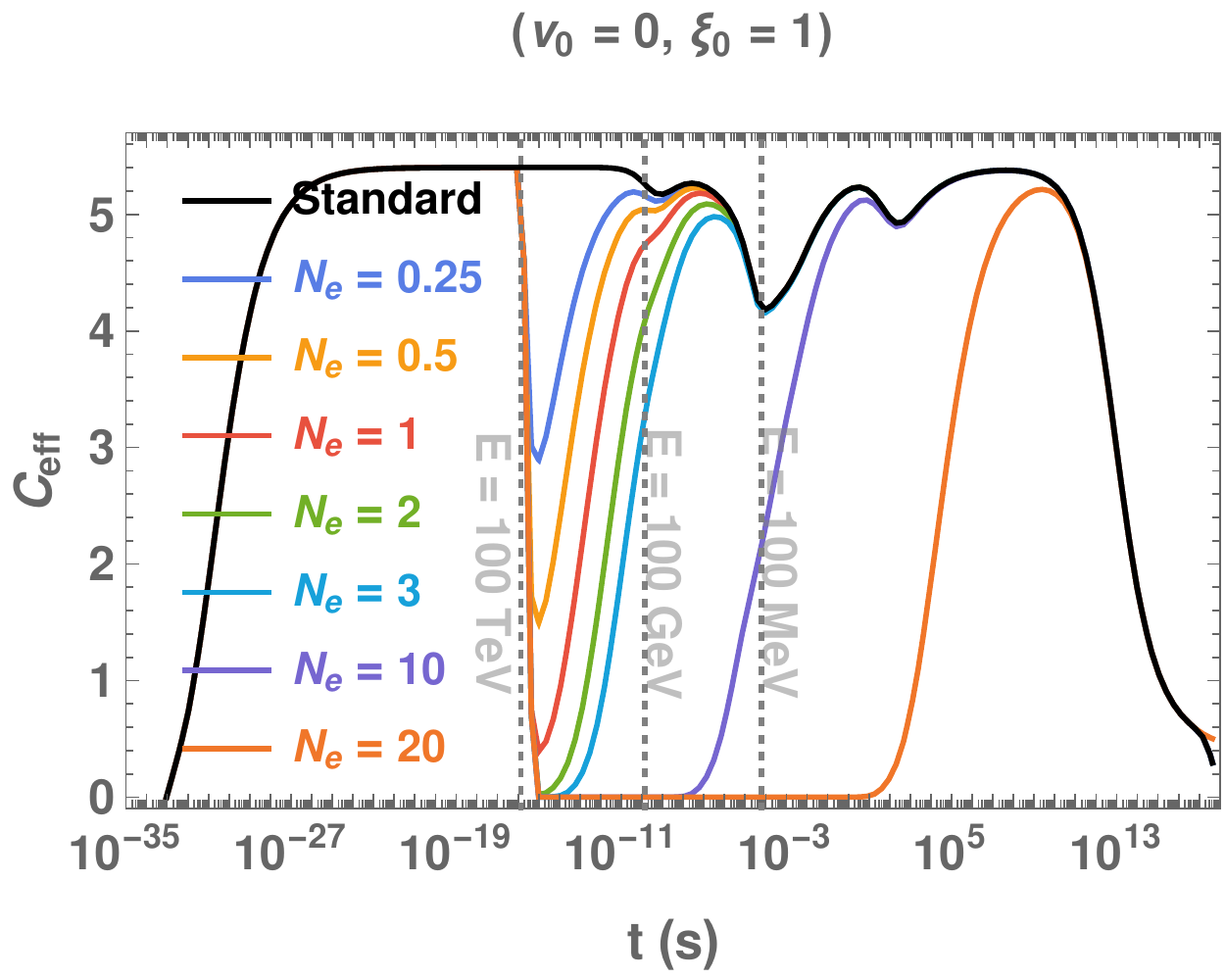}}}
\caption{\it \small \textbf{Top:} GW spectra from cosmic strings assuming either the scaling or the VOS network, evolved in the presence of a non-standard intermediate inflation era. Inflation directly affects the VOS parameters by stretching the strings beyond the horizon.  The transition between the $f^{-1/3}$ scaling after the turning point, to the $f^{-1}$ scaling at even larger frequencies, is an artefact due to total number of modes $k$ being fixed to $2\times 10^4$, see Fig.~\ref{fig:beautifulpeaks} for an extrapolation of the $f^{-1/3}$ behavior to arbitrary large frequencies and App.~\ref{sec:study_impact_mode_nbr} for more details. \textbf{Bottom:} The loop-production is suppressed and only becomes significant again when the correlation length re-enters the horizon. Limitations due to particle production, cf. Sec.~\ref{UVcutoff}, are shown with dotted lines.}
\label{figure_spect_inflation}
\end{figure}

\subsection{The stretching regime and its impact on the spectrum}
\label{sec:turning_point_inf}
Fig.~\ref{figure_spect_inflation} shows how the fast expansion during inflation suppresses the GW spectrum for frequencies above a turning-point frequency $f_\Delta$ which depends on the number of e-folds. The larger the number of e-folds, the lower $f_\Delta$.  Indeed, during inflation, the loop-production efficiency $C_{\rm eff} \propto \xi^{-3}$ is severely suppressed, cf. Fig.~\ref{figure_spect_inflation}, by the stretching of the correlation length $\xi$ beyond the Hubble horizon, and loop production freezes \cite{Guedes:2018afo}. 
After the end of inflation, one must wait for the correlation length to re-enter the horizon in order to reach the scaling regime again. The duration of the transient regime receives an enhancement factor $\exp{N_e}$. As a result, the turning-point frequency $f_\Delta$ receives a suppression factor $\exp{N_e}$ as derived below:
\begin{equation}
f_\Delta=(1.5\times10^{-4}\textrm{ Hz})\left(\frac{T_\textrm{re}}{\textrm{GeV}}\right)\left(\frac{0.1\times 50 \times 10^{-11}}{ \alpha \,\Gamma G\mu}\right)^{1/2}\left(\frac{g_*(T_\textrm{re})}{g_*(T_0)}\right)^{1/4},
\label{turning_point_inf}
\end{equation}
with $T_\textrm{re}$ the temperature at which the long-string network re-enters the Hubble horizon
\begin{equation}
T_\textrm{re}\simeq \frac{E_\textrm{inf}}{(0.1)\,g_*^{1/4}(T_\textrm{re})\,\exp(N_e)},
\label{eq:Tre-enter}
\end{equation}
where $(0.1)$ is the typical correlation length before the stretching starts. Note that the numerical factor in Eq.~\eqref{turning_point_inf} comes from the demanded precision of 10\% deviation, cf. Eq.~\eqref{10per_criterion}. It can be lower by a factor $\sim 300$  if the 1\% precision is applied, as shown in Eq.~\eqref{turning_point_general_scaling_app_inf}.

\begin{figure}[h!]
\centering
\raisebox{0cm}{\makebox{\includegraphics[width=0.7\textwidth, scale=1]{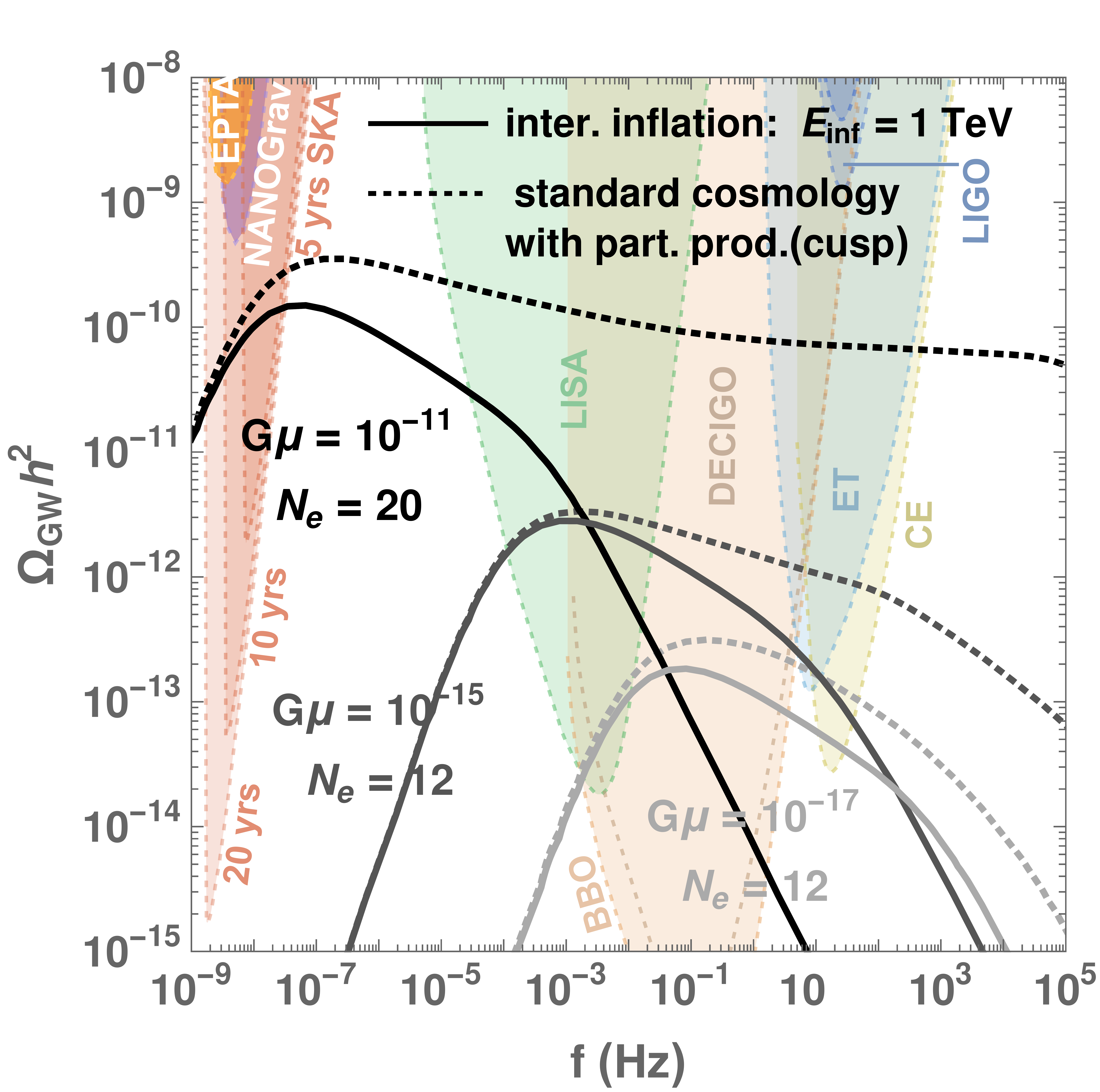}}}
\caption{\it\small In the case an intermediate inflationary era lasting for ${\cal O}(10)$ efolds, the SGWB from cosmic strings completely looses its scale invariant shape and has instead a peak structure. A TeV scale inflation era can lead to broad peaks either in the LISA or BBO band or even close to the SKA band, depending on the value of the string tensions $G\mu$, and the number of efolds $N_e$. At low $G\mu \lesssim 10^{-17},$ the spectrum manifests a peak structure even in standard cosmology because of the emission of massive particles at large frequencies, cf. Sec.~\ref{sec:massive_radiation}.  Here we extrapolate the $f^{-1/3}$ behavior to arbitrary large frequencies, which is equivalent to sum over an infinite number of proper modes $k$, see App.~\ref{sec:study_impact_mode_nbr}.  }
\label{fig:beautifulpeaks}
\end{figure}

Fig.~\ref{fig:beautifulpeaks} shows how a sufficiently long period of intermediate inflation can lead to SGWB with peak shapes in the future GW interferometer bands.
We emphasize that the change of the GW spectrum from CS in the presence of a non-standard matter-dominated era, a short inflation, and particle production look similar. Therefore, the question of how disentangling each effect from one another deserves further studies.

Interestingly, in contrast with the SGWB which is dramatically impacted by an intermediate period of inflation, the short-lasting GW burst signals \cite{Damour:2000wa, Damour:2001bk, Siemens:2006yp, Olmez:2010bi, Ringeval:2017eww} remain preserved if the correlation length re-enters the horizon at a redshift higher than $\sim 5 \times 10^4$ \cite{Cui:2019kkd}. Indeed, the bursts being generated by the small scale structures, they have higher frequencies and then are emitted later than the SGWB, cf. Fig.~2 in \cite{Ringeval:2017eww}.


\paragraph{Derivation of the turning-point formula (inflation case):}
		Let us review the chronology of the network in the presence of an intermediate-inflation period (see figure~\ref{fig:stringsandinflation}) in order to derive Eq.~\eqref{turning_point_inf}.
		In the early radiation era, the network has already been produced and reached the scaling regime before inflation starts. 
		The correlation length scale is of order $(0.1)t$ or equivalently
		\begin{align}
		L_\textrm{start} H_\textrm{start} \sim \mathcal{O}(0.1),
		\end{align}
		where $L$ is the correlation length of strings, and $H$ is the Hubble rate.
		When inflation begins, it stretches cosmic strings beyond the horizon with
		\begin{align}
		L\propto a \textrm{\hspace{1em} leading to \hspace{1em}} LH\gg1,
		\end{align} 
		within a few e-folds. 
		Later, the late-time energy density takes over inflation, but the network is still in the stretching regime $L\propto a$, i.e.
		\begin{align}
		\label{eq:network_radiation_length}
		LH\propto t^{(2-n)/n}\textrm{\hspace{1em} during the era with }\rho\propto a^{n}.
		\end{align} 		
		For $n>2$, the Hubble horizon will eventually catch up with the string length, allowing them to re-enter, and initiate the loop production.
		We consider the case where the universe is radiation-dominated after the inflation period and define the temperature $T_\textrm{re} $ of the universe when the long-string correlation length $L$ re-enters the horizon
		\begin{align}
		L_\textrm{re} H_\textrm{re} = 1,
		\end{align}
		where $L_\textrm{re}$ and $H_\textrm{re}$ are the correlation length and Hubble rate at the re-entering time. We can use Eq.~\eqref{eq:network_radiation_length} to evolve the correlation length, starting from the start of inflation up to the re-entering time
		\begin{align}
		1=L_\textrm{re} H_\textrm{re}&=\left(\frac{t_\textrm{re}}{t_\textrm{end}}\right)^{-1/2}L_\textrm{end}H_\textrm{end},\\
		&=\left(\frac{t_\textrm{re}}{t_\textrm{end}}\right)^{-1/2}\left(\frac{a_\textrm{end}}{a_\textrm{start}}\right)L_\textrm{start}H_\textrm{start},\\
		&\simeq \left(\frac{T_\textrm{re}}{T_\textrm{end}}\right)e^{N_e}(0.1)
		\end{align}
		We have used $t\propto T^{-2}$ during the radiation era and introduced the number $N_e$ of inflation e-folds.
		Finally, we obtain the re-entering temperature in terms of the number of e-folds $N_e$ and the inflationary energy scale $E_\textrm{inf}$ as
		\begin{align}
		\label{eq:Tre}
		T_\textrm{re}\simeq \frac{E_\textrm{inf}}{(0.1)\,g_*^{1/4}(T_\textrm{re})\,\exp(N_e)}.
		\end{align}
After plugging Eq.~\eqref{eq:Tre} into the VOS turning-point formula Eq.~\eqref{turning_point_general}, with $T_{\Delta}=T_{\rm re}$, and adjusting the numerical factor with the GW spectrum computed numerically, we obtain the relation in Eq.~\eqref{turning_point_inf} between the turning-point frequency and the inflation parameters $N_e$ and $E_{\rm inf}$.


\begin{figure}[h!]
			\centering
			\raisebox{0cm}{\makebox{\includegraphics[height=0.47\textwidth, scale=1]{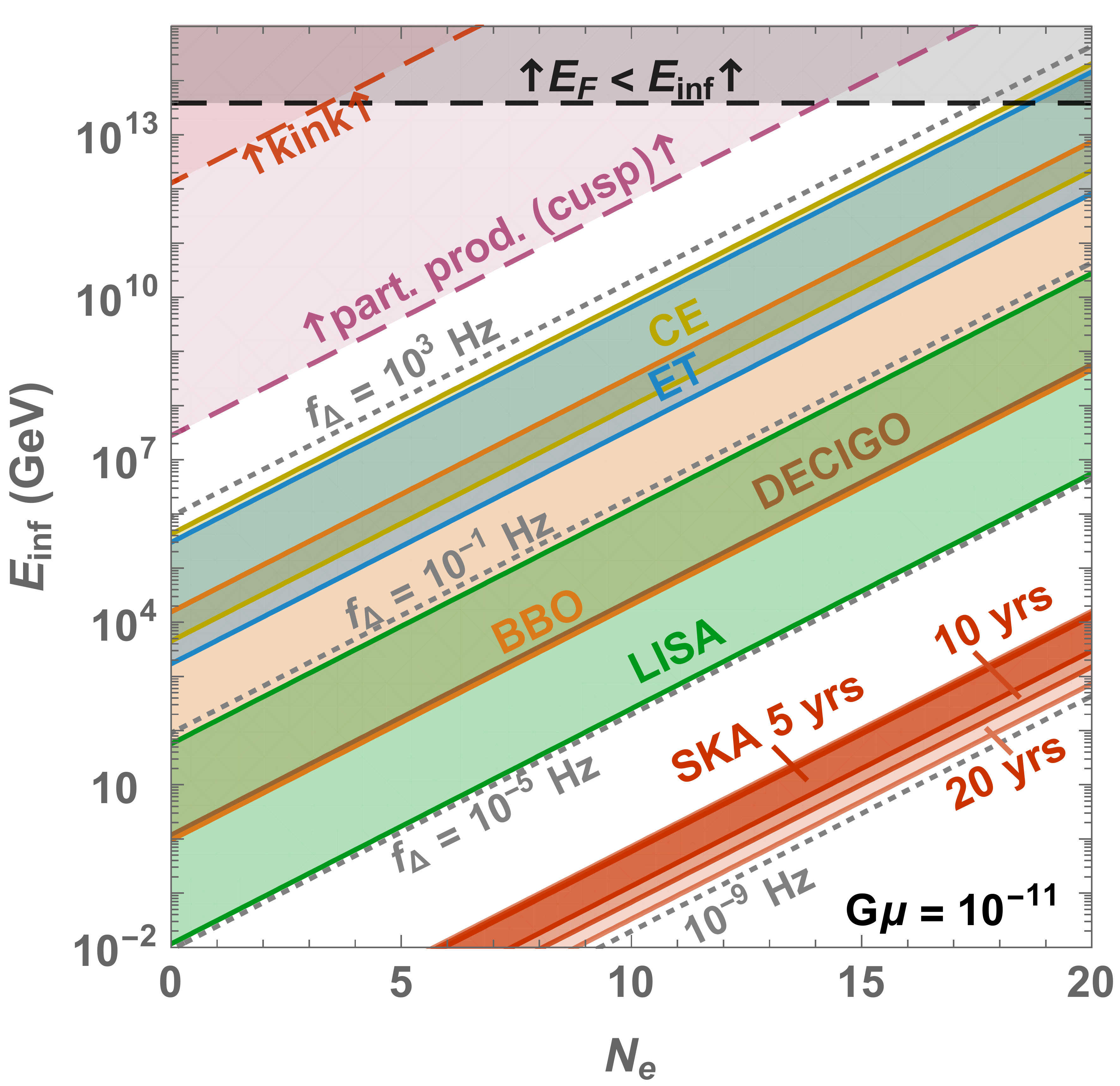}}}
			\raisebox{0cm}{\makebox{\includegraphics[height=0.47\textwidth, scale=1]{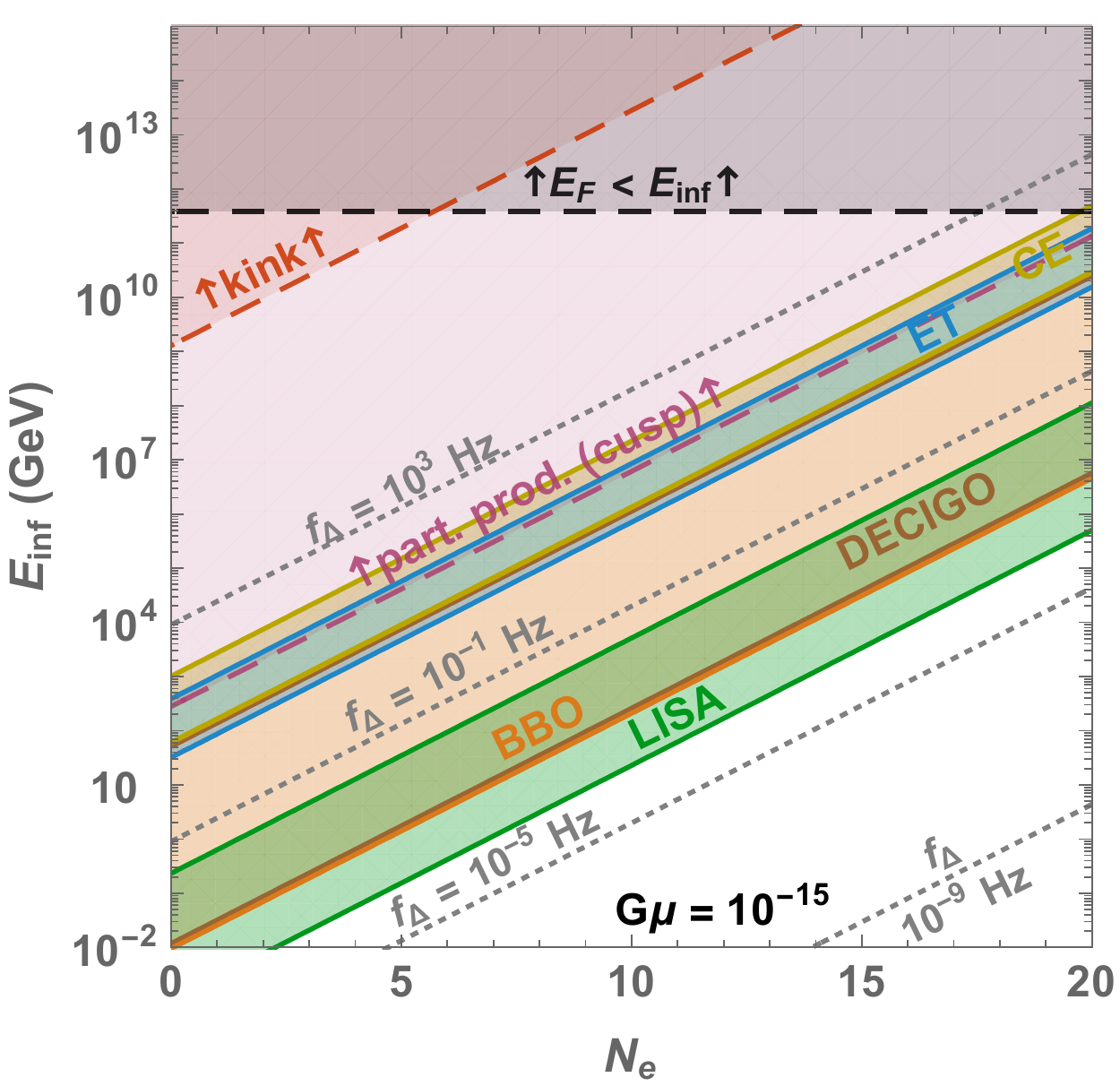}}}
			\hfill
			\caption{\it \small  Reach of future GW interferometers for probing an intermediate-inflation period with an energy scale $E_{\rm inf}$, lasting $N_e$ efolds. Colored regions correspond to the turning-points with amplitude higher than each power-law-sensitivity curve, cf. \textit{turning-point prescription (Rx 1)} in Sec.~\ref{sec:triggerMatterGW}. Gray dotted lines are turning-points, cf. Eq.~\eqref{turning_point_inf}, for given frequencies. Red and purple dashed lines are limitations from particle production, cf. Sec.~\ref{UVcutoff}.}
			\label{fig:contour_power_inflation1}
		\end{figure}
\begin{figure}[h!]
			\centering
			\raisebox{0cm}{\makebox{\includegraphics[height=0.495\textwidth, scale=1]{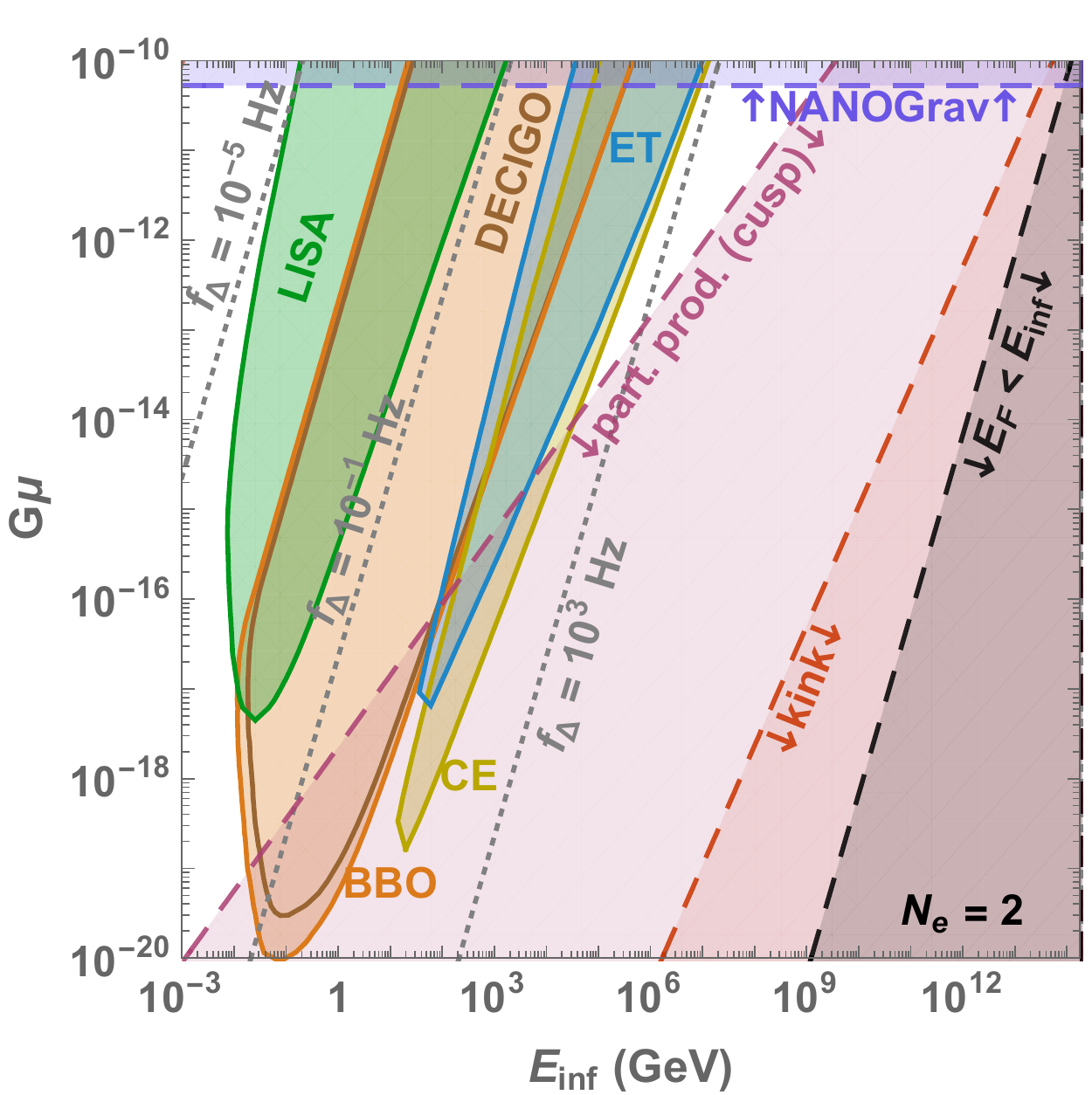}}}
			\raisebox{0cm}{\makebox{\includegraphics[height=0.495\textwidth, scale=1]{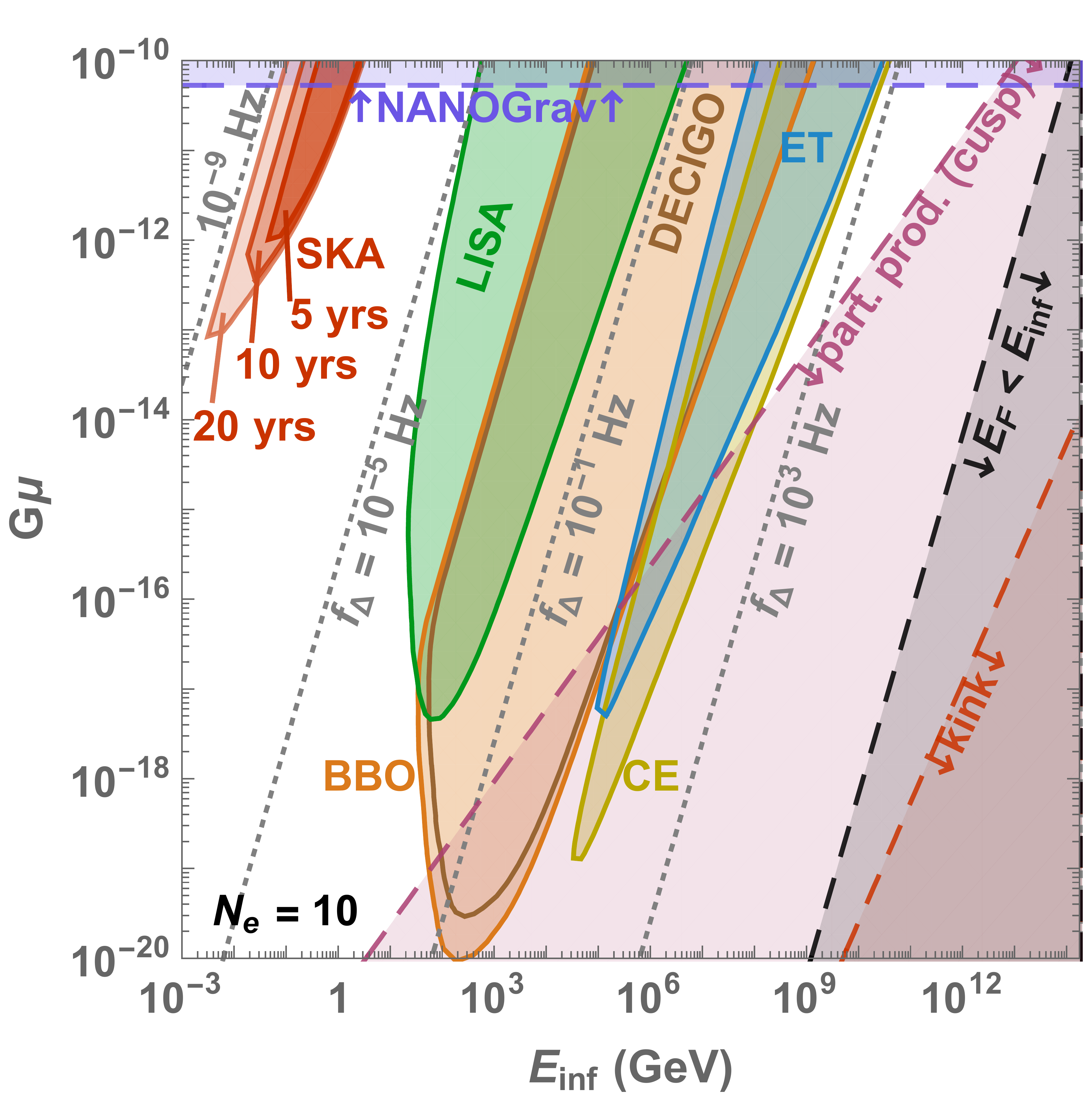}}}\\
			\raisebox{0cm}{\makebox{\includegraphics[height=0.495\textwidth, scale=1]{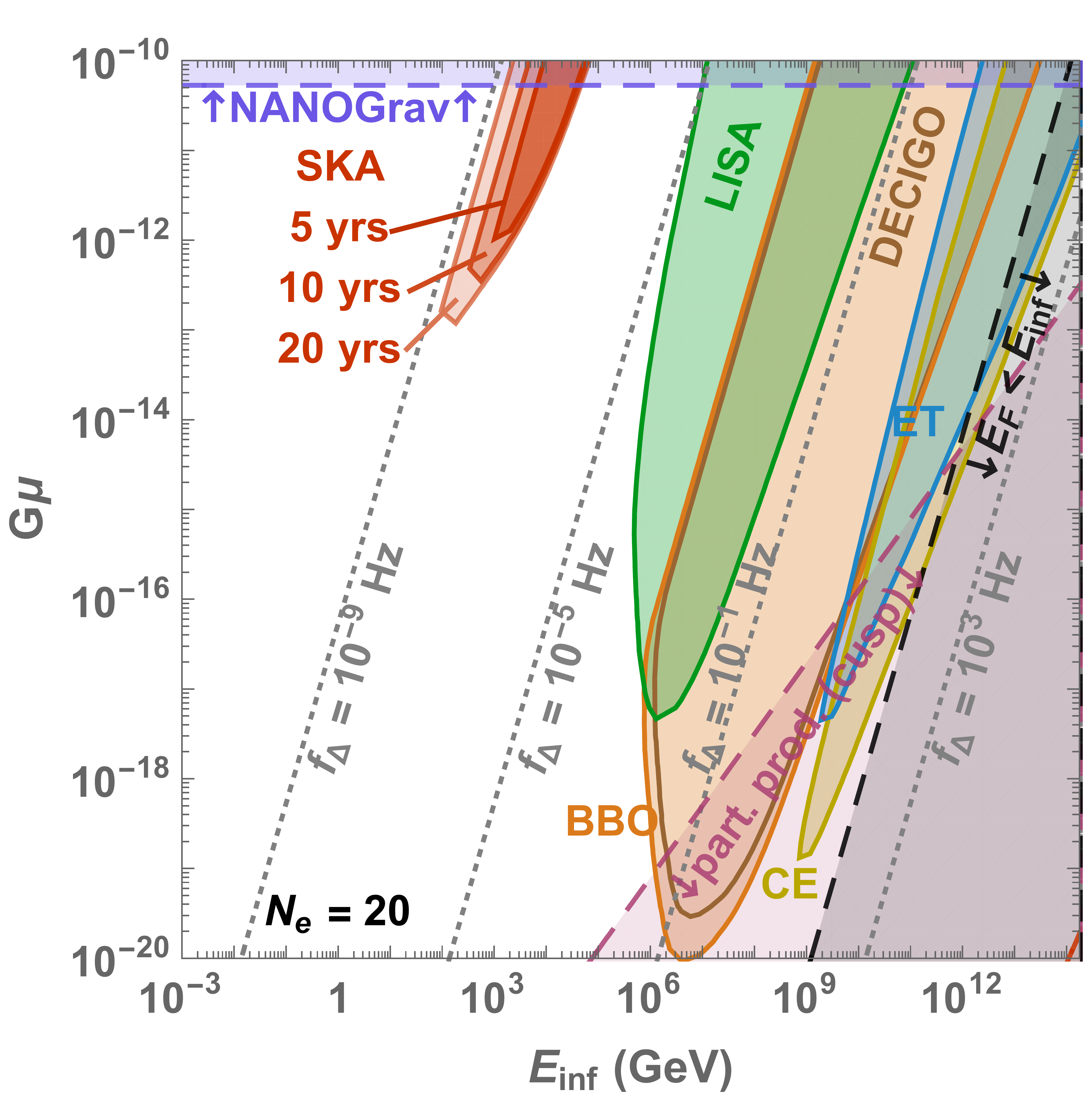}}}
			\hfill
			\caption{\it \small Constraints on intermediate inflation from CS detection by future GW observatories. The longer the intermediate inflation, the later the correlation length re-enters the horizon, the more shifted to lower frequencies the turning-point and the larger the inflation scale which we can probe.   Colored regions correspond to the turning-points with amplitude higher than each power-law-sensitivity curve, cf. \textit{turning-point prescription (Rx 1)} in Sec.~\ref{sec:triggerMatterGW}. The bound $E_F<E_\textrm{inf}$, where $E_F\sim m_{pl}\sqrt{G\mu}$ is the network-formation energy scale, guarantees that the CS network forms before the intermediate-inflation starts. Red and purple dashed lines are limitations from particle production, cf. Sec.~\ref{UVcutoff}.}
			\label{fig:contour_power_inflation2}
		\end{figure}
%
\begin{figure}[h!]
			\centering
			\raisebox{0cm}{\makebox{\includegraphics[height=0.425\textwidth, scale=1]{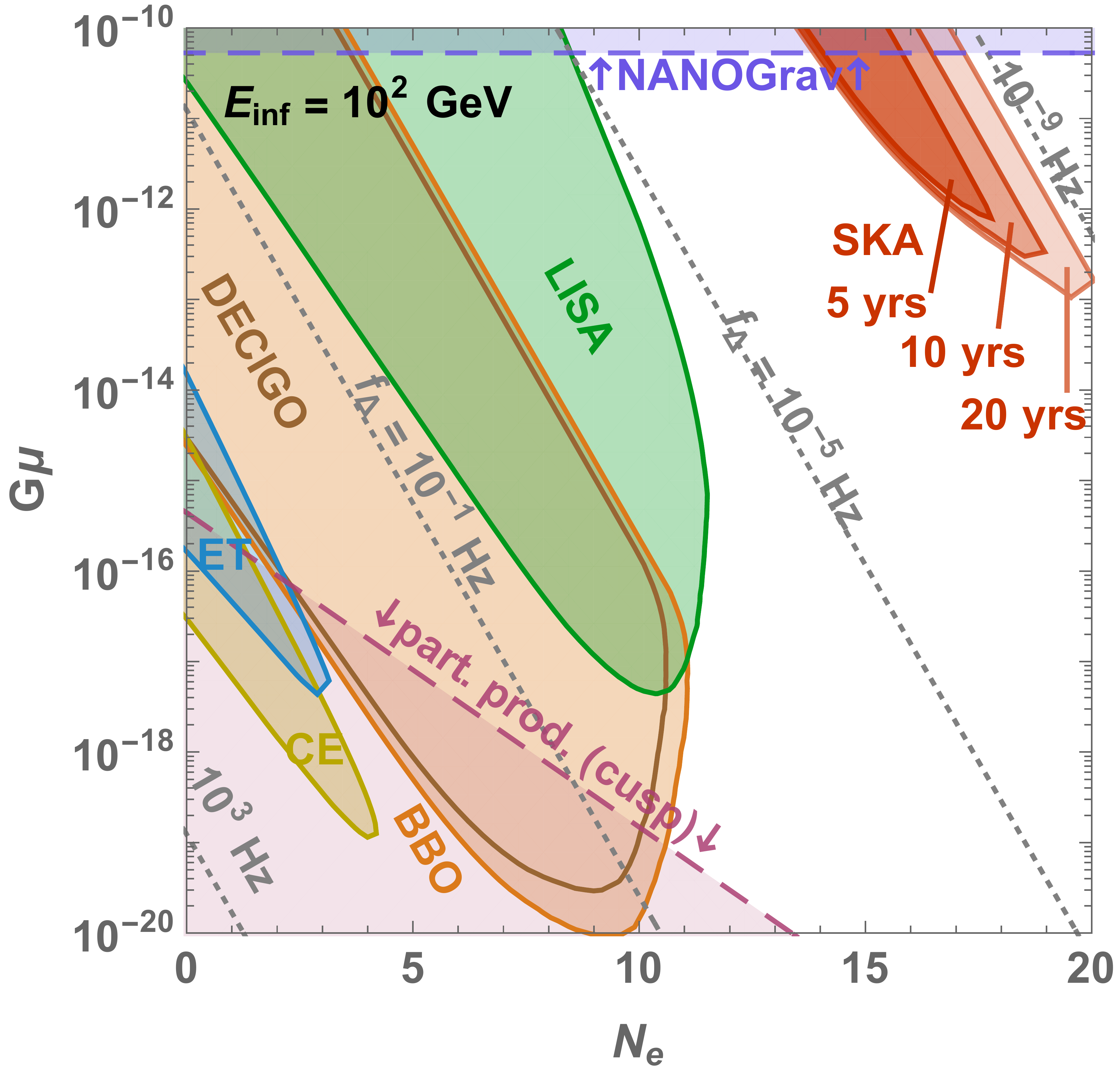}}}
			\raisebox{0cm}{\makebox{\includegraphics[height=0.425\textwidth, scale=1]{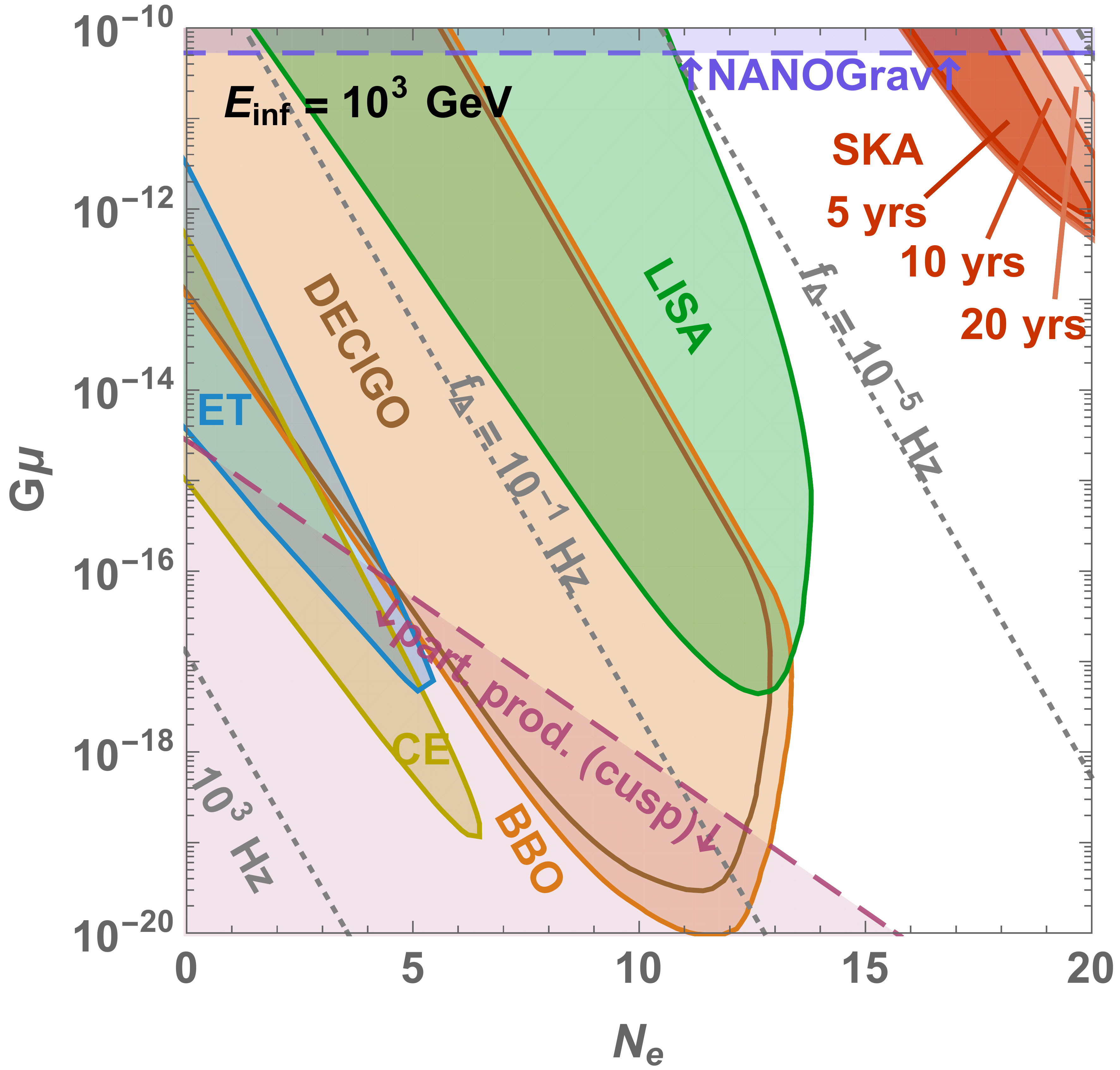}}}\\
			\raisebox{0cm}{\makebox{\includegraphics[height=0.425\textwidth, scale=1]{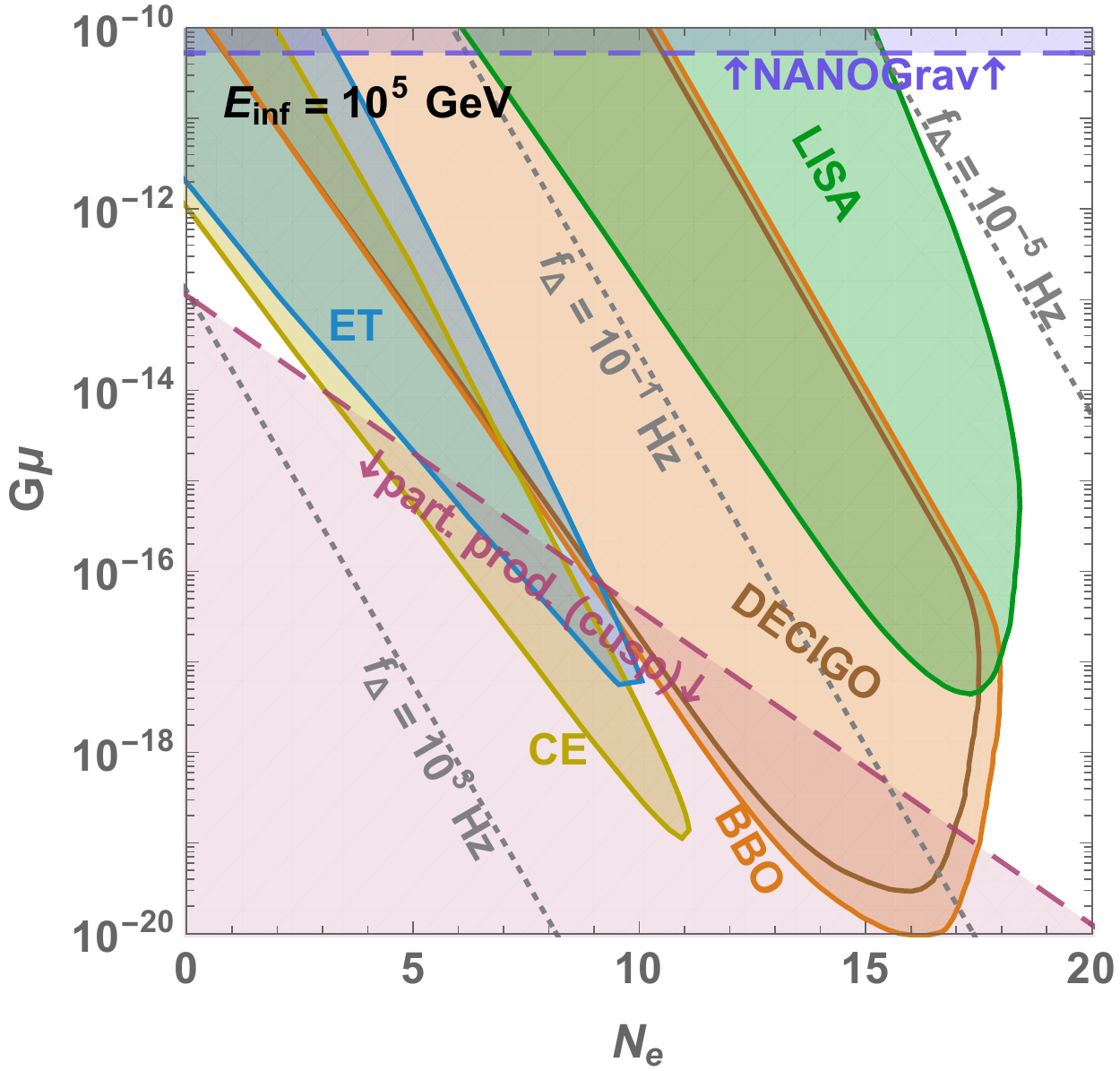}}}
			\raisebox{0cm}{\makebox{\includegraphics[height=0.425\textwidth, scale=1]{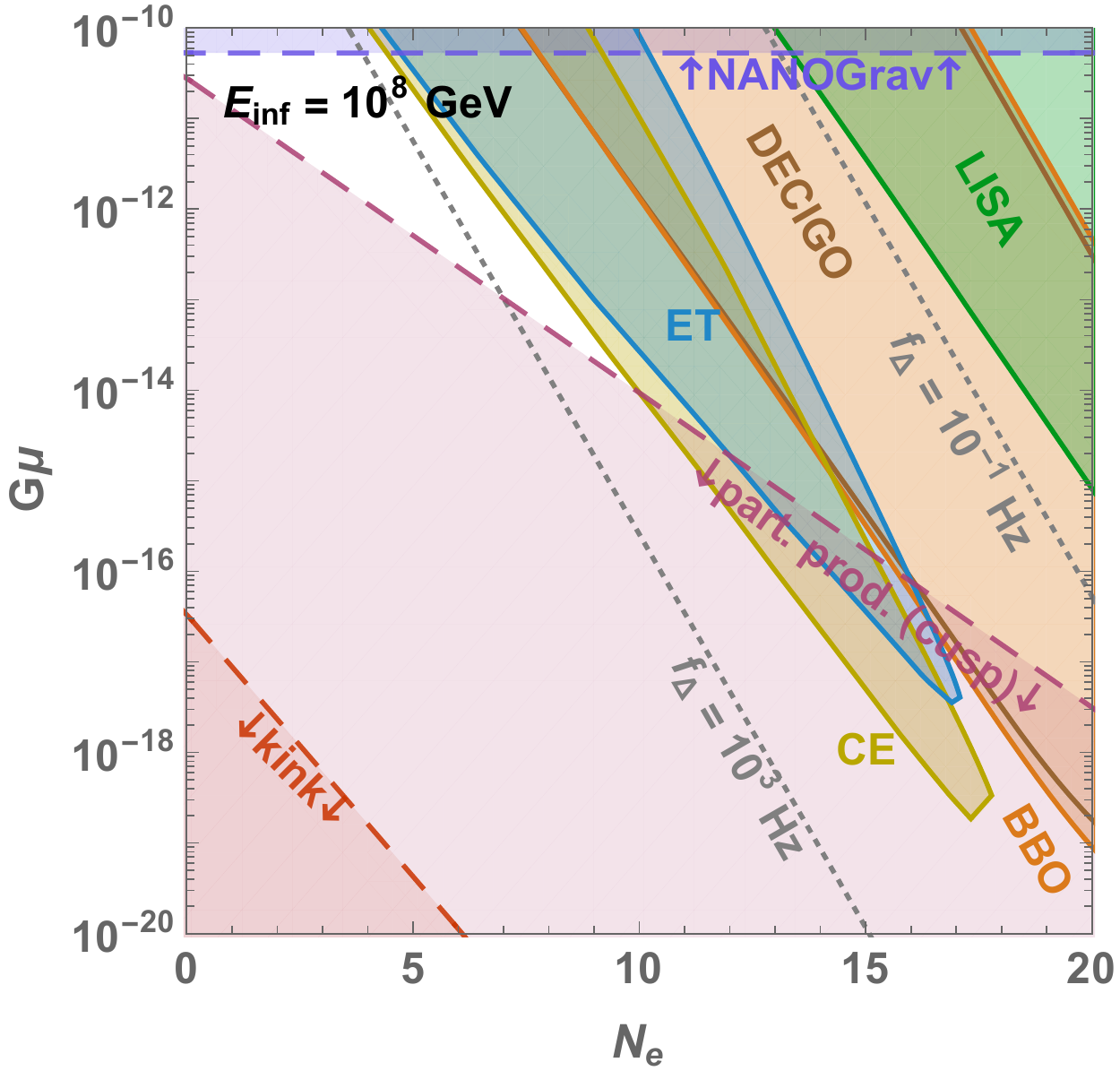}}}\\
			\raisebox{0cm}{\makebox{\includegraphics[height=0.425\textwidth, scale=1]{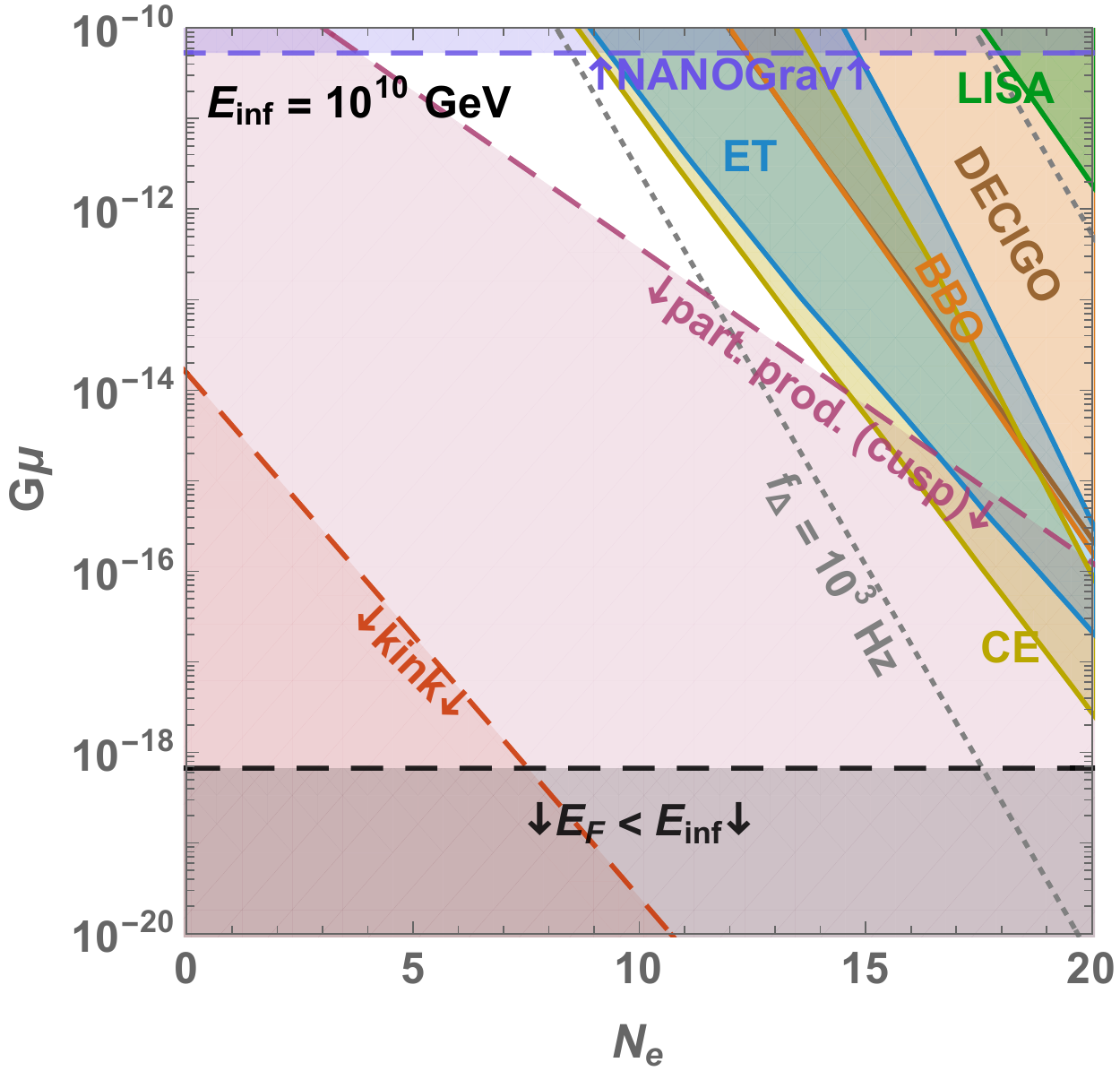}}}\\
			\hfill
		\caption{\it \small Prospect constraints on intermediate inflation if a GW interferometer detects a SGWB from CS with tension $G\mu$. The freezing of the long-string network due to the stretching of the correlation length outside the horizon allows to probe large inflationary scale $E_{\rm inf}$ for large number of efolds $N_e$.   Colored regions correspond to the turning-points with amplitude higher than each power-law-sensitivity curve, cf. \textit{turning-point prescription (Rx 1)} in Sec.~\ref{sec:triggerMatterGW}.  Red and purple dashed lines are limitations from particle production, cf. Sec.~\ref{UVcutoff}.}
			\label{fig:contour_power_inflation3}
		\end{figure}
\FloatBarrier

\subsection{Model-independent constraints}
\label{sec:model_indpt_inflation_GWCS}

In Sec.~\ref{sec:inflation}, we derived the imprint of a short period of inflation on the GW spectrum from CS.
In Figs.~\ref{fig:contour_power_inflation1}, \ref{fig:contour_power_inflation2} and \ref{fig:contour_power_inflation3}, we show the corresponding constraints - assuming one of the mentioned experiments detect a GW spectrum from CS in the future -  on an intermediate short inflation period in the planes $E_\textrm{inf}-N_e$, $G\mu-E_\textrm{inf}$, and $G\mu-N_e$, respectively. We follow the \textit{turning-point prescription (Rx 1)} defined in Sec.~\ref{sec:triggerMatterGW}, which constrains a non-standard cosmology by using the detectability of the turning-point frequency defined by Eq.~\eqref{turning_point_inf}. 
The longer the intermediate inflation, the later the correlation length re-enters the horizon, the latter the long-string network goes back to the scaling regime, the lower the frequency of the turning-point and the larger the inflationary scale which can be probed.
The detection of a GW spectrum generated by CS by future GW observatories would allow to probe an inflationary energy scale $E_\textrm{inf}$ between $10^{-2}$~GeV and $10^{13}$~GeV assuming a number of e-folds $N_e \lesssim 20$.

\begin{figure}[t!]
\centering
\raisebox{0cm}{\makebox{\includegraphics[width=0.565\textwidth, scale=1]{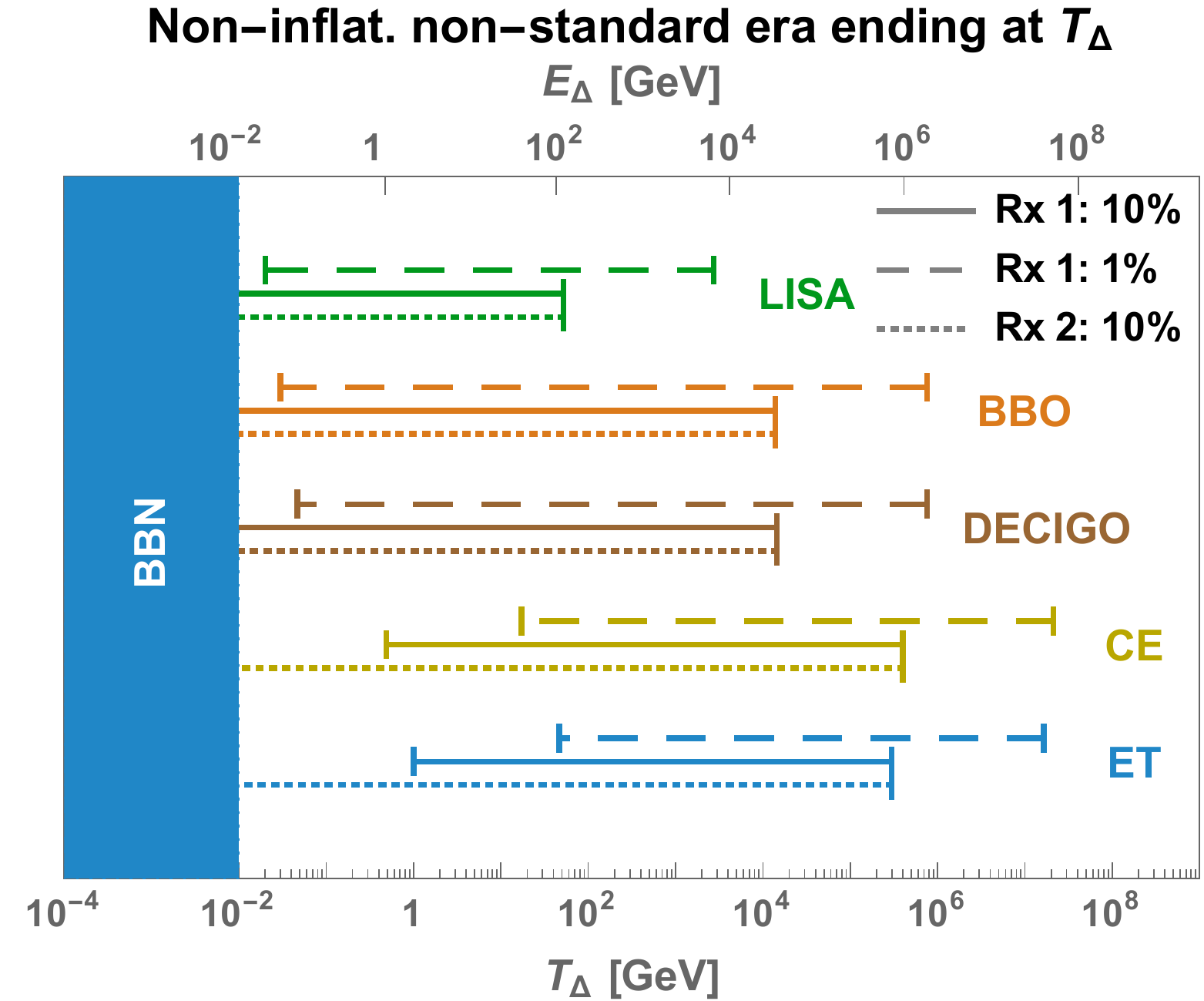}}}\\[0.5em]
\raisebox{0cm}{\makebox{\includegraphics[width=0.55\textwidth, scale=1]{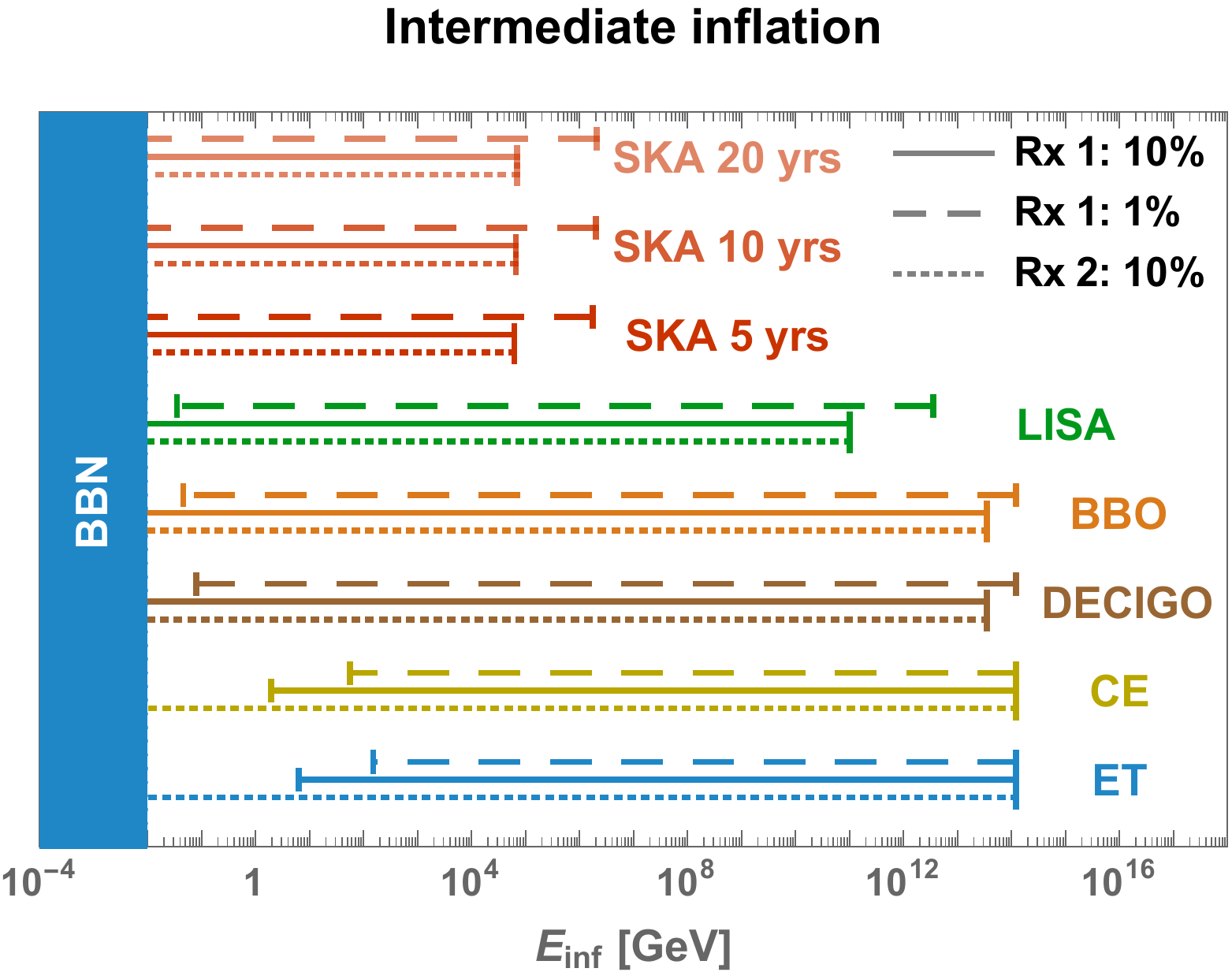}}}
\caption{\it \small \textbf{Top:}  Sensitivity to the energy scale $E_\Delta$ of the universe at the end of any non-inflationary non-standard era  for each future GW interferometer. The connection to $E_\Delta$ is given by the observation of the turning-point frequency defined in Eq.~(\ref{turning_point_general_scaling_app}). The width of the bands includes varying the string tension for $G\mu<10^{-10}$. The dotted, dashed and solid lines correspond to different observational prescriptions defined in Sec.~\ref{sec:triggerMatterGW}. \textbf{Bottom:} Sensitivity to the energy scale $E_{\rm inf}$ of an intermediate inflationary era  for each future GW interferometer as well as for future radio telescope SKA. The connection to $E_{\rm inf}$ is given by the observation of the turning-point frequency defined in Eq.~(\ref{turning_point_general_scaling_app_inf}). The width of the band also includes varying the number of efolds of inflation $N_e$ up to 20.}
\label{fig:summary1}
\end{figure}

\section{Summary and conclusion}
\label{sec:conclusion}

In standard cosmology, the GW spectrum generated by a network of Nambu-Goto cosmic strings (and mainly due to emission by loops)  is nearly scale-invariant. 
Its potential observation by third-generation interferometers would be a unique probe of new effects beyond the standard models of particle physics and cosmology. 
Such opportunity  was pointed out in \cite{Cui:2017ufi, Cui:2018rwi,Auclair:2019wcv,Guedes:2018afo,Chang:2019mza}. 

Deriving firm conclusions is still premature as
theoretical predictions of the GW spectrum from CS are subject to a number of large uncertainties. 
Still, we feel that the extraordinary potential offered by future GW observatories to probe 
high energy physics has not yet been explored, and in a series of papers, we are starting to scrutinise 
 how much can be learnt, even if only in the far-future, after those planned GW observatories will have reached their expected long-term sensitivity and the astrophysical foreground will have been fully understood.

Deviations in the cosmological history  with respect to standard cosmology not only change the redshifting factor of  GW but also modify the time of loop formation and the loop-production efficiency.
We presented  predictions for the resulting GW spectra under a number of assumptions which we have comprehensively  reviewed. 

We extend previous works in several directions, as listed in the introduction.

A particular feature of gravitational waves from cosmic strings is the relation between the observed frequency and the GW production mechanism. In contrast with short-lasting cosmological sources of  gravitational waves, such as phase transitions, where the frequency is simply related to the Hubble radius at the time of GW emission, for cosmic strings the time of the dominant GW emission is much later than the time of  loop production, by a factor $\sim 1/(G\mu)$, such that the observed frequency is higher due to a smaller redshift.
We stressed that a given interferometer may be sensitive to very different energy scales, depending on the nature and duration of the non-standard era, and the value of the string tension. 
This goes  against usual paradigms. For instance, 
it is customary to talk about LISA as a window on the EW scale \cite{Grojean:2006bp,Caprini:2019egz}. This does not apply for GW from cosmic strings, as LISA could either be a window on a non-standard matter era at the  QCD scale or on a 10 TeV inflationary  era, meaning that the GW observed in the LISA band have been emitted by loops that were created at the QCD epoch, or at a 10 TeV epoch depending on the nature of the new physics responsible for the non-standard era.
Interestingly, the Einstein Telescope and Cosmic Explorer offer a window of observation on the highest scales, up to $10^{14}$ GeV inflationary eras. They can also be windows on the EW and TeV scales, as will be discussed in more details in \cite{Gouttenoire:2019rtn}.
BBO/DECIGO could probe new physics in an intermediate range.  Finally, radio telescope SKA may be sensitive to a TeV scale inflationary era. 
The goal of this study was to stress 
the very rich variety of spectral shapes that can be obtained by combining different physical effects. In particular, we showed how peaked shapes can arise naturally in a large variety of models.

We apply these findings to probe well-motivated particle physics scenarios in \cite{Gouttenoire:2019rtn}. Particularly generic are intermediate matter eras triggered by cold  heavy particles arising in UV completions of the Standard Model. In  \cite{Gouttenoire:2019rtn}, we show on specific models how a new  uncharted particle theory space can be probed from analysis of SGWB from CS.

Finally, one important  question  will be to work out how to distinguish a stage of matter or inflationary expansion, which both lead to a suppression of the GW spectrum, from the cutoff induced by particle production from small loops. Both predict a cutoff at high frequencies and lead to similar spectra.
Interestingly, particle production by cosmic string networks can be probed through cosmic rays and bring complementary non-gravitational information on the SGWB.
Besides, the complementarity between different GW instruments will be crucial here as the detection of the low-frequency peak of the spectrum (due to the transition from the standard radiation to the standard matter era) can enable to probe the string tension and to break the degeneracy between different spectral predictions.
The possibility to reconstruct the spectral shape of a SGWB was  analysed in \cite{Caprini:2019pxz} using LISA data only.
In the case of a SGWB generated by CS, which can span more than twenty decades in frequency, it will be crucial to use data from different interferometers  (and even from radio telescopes) to probe the full spectrum.

\begin{subappendices}

\chapterimage{CS_lensing_2} 

\section{Constraints on cosmic strings from BBN, gravitational lensing, CMB and cosmic rays}
\label{app:phenoCS}

By confronting our theoretical predictions for the GW spectrum from CS with the sensitivity curves of EPTA \cite{Lentati:2015qwp} and NANOGrav  \cite{Arzoumanian:2018saf} (which we take from \cite{Breitbach:2018ddu}), we derived the respective bounds
$G \mu \lesssim 2 \times 10^{-10}$ (EPTA) and $G \mu \lesssim 5 \times10^{-11}$ (NANOGrav),
as discussed in Sec.~\ref{subsec:Gmuconstraints}.\blfootnote{A galaxy lensed by a cosmic string. Image credits: \cite{articleCSlensing}}
For this reason, we only considered  in our analysis $G \mu$ values smaller than $5\times 10^{-11}$.
In Sec.~\ref{sec:GW_BBN}, we give the constraints on the string tension from not changing the expansion rate of the universe at BBN. They are much weaker than the ones from Pulsar Timing Arrays but they can become relevant in the presence of kination.

In Sec.~\ref{sec:gravitational_lensing} and Sec.~\ref{sec:temperature_anisotropy}, we give bounds from gravitational lensing and CMB observables. They are also much weaker than the ones from Pulsar Timing Arrays but they have the strong advantage to be independent of our assumptions for the theoretical prediction of the GW background. 
Finally, in Sec.~\ref{sec:particle_prod_pheno}, we discuss the possibility of probing CS from the massive particle production in the presence of kinks and cusps.

\subsection{GW constraints from BBN}

\label{sec:GW_BBN}
As a sub-component of the total energy density of the universe, the amount of GW can impact the expansion rate of the universe which is strongly constrained by BBN and CMB. More precisely, any non-standard energy density can act as an effective number of neutrino relics
\begin{equation}
N_{\rm eff} = \frac{8}{7}\left( \frac{\rho_{\rm tot}-\rho_{\gamma}}{\rho_\gamma} \right)\left( \frac{11}{4} \right)^{4/3},
\end{equation}
which is constrained by CMB measurements \cite{Aghanim:2018eyx} to $N_{\mathsmaller{\rm eff}} = 2.99_{-0.33}^{+0.34}$ and by BBN predictions \cite{Mangano:2011ar, Peimbert:2016bdg} to $N_{\mathsmaller{\rm eff}} = 2.90_{-0.22}^{+0.22}$ whereas the SM prediction \cite{Mangano:2005cc, deSalas:2016ztq, Escudero:2020dfa} is $N_{\mathsmaller{\rm eff}}  \simeq 3.045$.
Using $\Omega_{\gamma} h^2 \simeq 2.47\times 10^{-5}$ \cite{Tanabashi:2018oca}, we obtain the following bound on the GW spectrum 
\begin{equation}
\int^{f_{\rm high}}_{f_\textrm{BBN}}\frac{df}{f}h^2 \Omega_\textrm{GW}(f) \leq 5.6 \times 10^{-6} ~ \Delta N_\nu,
\end{equation}
where $f_{\rm high}$ is the frequency today of the first GW produced,  $f_\textrm{BBN}$ is the frequency today of the GW produced at BBN, and we set $\Delta N_\nu \leq 0.2$. The value of $f_\textrm{BBN}$ depends on the source of GW.
For CS, the temperature at BBN, $T_\textrm{CMB}\simeq 1$ MeV, translates via Eq.~\eqref{fdeltaApp} to the frequency 
\begin{equation}
f_\textrm{BBN} \simeq 8.9 \times 10^{-5}~\textrm{Hz}~\left(\frac{0.1\times 50 \times 10^{-11}}{\alpha \Gamma G \mu}\right)^{1/2}.
\end{equation}
\begin{figure}[h!]
\centering
\raisebox{0cm}{\makebox{\includegraphics[width=0.5\textwidth, scale=1]{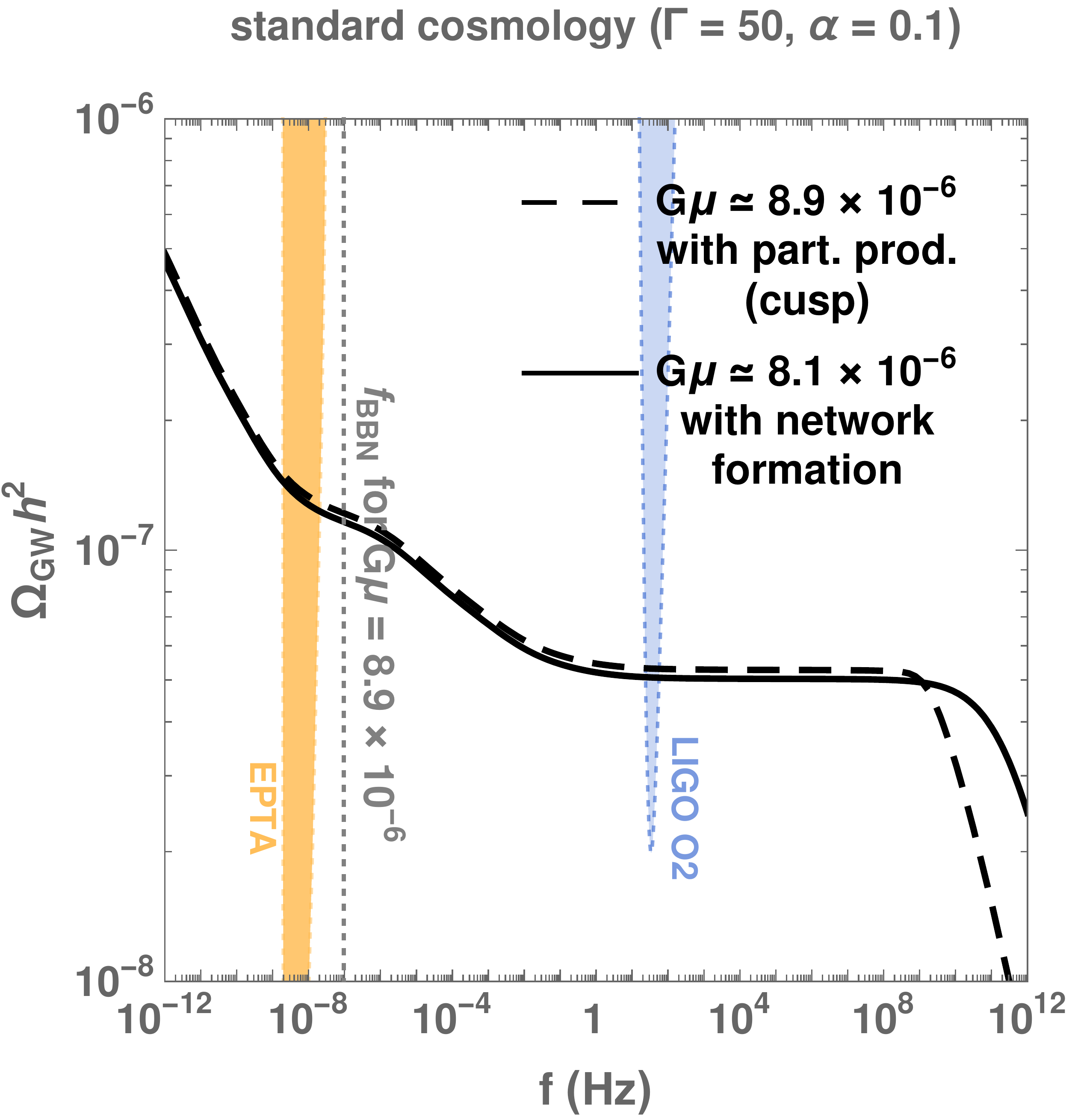}}}
\caption{\it \small 
Two GW spectra which saturate the BBN bounds, assuming a VOS string network, cf. Sec.~\ref{sec:scalingVSvos}, evolving in standard cosmology. The solid line assumes a cut-off due to network formation whereas the dashed line assumes a cut-off due to particle production from cusps. The dotted vertical line is the frequency emitted when BBN starts. We compare the BBN bounds to the bounds from EPTA and LIGO O2. }
\label{excessive_GW}
\end{figure}
In Fig.~\ref{excessive_GW}, we show the GW spectra which saturate the BBN bound for two different high-frequency cut-offs.
We can see that the lower the cut-off, the higher the upper bound on $G\mu$ due to less GW present at the time of BBN. Assuming the presence of the cut-off due to particle production from cusps, we obtain 
\begin{equation}
\text{BBN:}  \qquad h^2\Omega_\textrm{GW}(f)\lesssim 8.9\times 10^{-6}.
\end{equation}
We expect the BBN bounds to become softer in the presence of non-standard matter or inflation era but tighter in the presence of an early kination era. For instance, scenarios of inflation followed by a stiff equation of state (e.g. quintessential inflation \cite{Peebles:1998qn}) are dramatically jeopardized by the BBN bounds \cite{Figueroa:2019paj}.
Similarly, in the case of CS, we find that the maximally allowed string tension is $G\mu \simeq 3.9 \times 10^{-15},~3.8 \times 10^{-17},~\textrm{and }2.9 \times 10^{-20}$ for long-lasting kination era ending at temperature $T_\Delta = 100$ TeV, $1$ TeV, and $1$ GeV, respectively.

\subsection{Gravitational lensing}
\label{sec:gravitational_lensing}
The presence of energy confined within the core of CS affects the spacetime around them. The metric near a CS is locally flat but globally conical \cite{Vilenkin:1981zs}. Photons from a distant celestial object travelling in the vicinity of a CS are subject to gravitational lensing effects. The corresponding constraint $G\mu \lesssim 3 \times 10^{-7}$ has been derived from the search of gravitational lensing signatures of CS in the high-resolution wide-field astronomical surveys GOODS \cite{Christiansen:2008vi} and COSMOS \cite{Christiansen:2010zi}. It has been claimed that constraints from gravitational lensing surveys at radio frequencies like LOFAR and SKA could reach $G\mu \lesssim 10^{-9}$ \cite{Mack:2007ae}. 

\subsection{Temperature anisotropies in the CMB}
\label{sec:temperature_anisotropy}
There are two possible effects from CS on  temperature fluctuations in the CMB:
\begin{enumerate}
\item 
CS moving through the line-of-sight can induce Doppler shifts on the photons coming from the last scattering surface, known as the Kaiser-Stebbins-Gott effect \cite{Gott:1984ef, Kaiser:1984iv,Bouchet:1988hh}, potentially leaving line-like discontinuities in the CMB.
\item 
A CS moving in the primordial plasma leaves surdensity perturbations, the so-called wakes \cite{Silk:1984xk}, possibly imprinted in the CMB temperature anisotropy.
Due to the stochastic behavior of the Kibble mechanism, these perturbations are decoherent and give rise to a CMB spectrum without acoustic peaks \cite{Pogosian:1999np}.
\end{enumerate}
Lattice numerical computation of the temperature anisotropy in Abelian-Higgs \cite{Lizarraga:2014xza, Lizarraga:2016onn}, Nambu-Goto \cite{Pogosian:1999np, Battye:2010xz, Charnock:2016nzm} or global strings \cite{Lopez-Eiguren:2017dmc} have constrained the string tension to $G\mu \lesssim \, \text{few} \times 10^{-7}$ \cite{Ade:2013xla}. Constraints of the same magnitude can be found from non-gaussianities \cite{Ringeval:2012tk, Ade:2013xla,Ciuca:2019oka}. Also, the same signatures as in the CMB can be imprinted in the $21$~cm power spectrum, and an experiment with a collecting area of $10^4-10^6$~km$^2$ might constrain $G\mu \lesssim 10^{-10}-10^{-12}$ \cite{Khatri:2008zw}.


\subsection{Non-gravitational radiation}

\label{sec:particle_prod_pheno}

As discussed in Sec.~\ref{sec:massive_radiation}, the presence of small-scale structures on local strings, cusps and kinks, invalidates the Nambu-Goto approximation and implies the radiation of massive particles. Therefore, CS have been proposed as a possible mechanism for generating non-thermal Dark Matter  \cite{Jeannerot:1999yn, Matsuda:2005fb, Cui:2008bd, Long:2019lwl}.

At a cusp, the string can reach ultrarelativistic velocities. Therefore, CS have been pointed \cite{MacGibbon:1989kk, Bhattacharjee:1998qc, Berezinsky:1998ft, Berezinsky:2011cp} as a possible candidate for the detection of ultra-high energy cosmic rays \cite{bird1994cosmic} above the Greisen-Zatsepin-Kuzmin (GZK) cut-off, around $10^{20}$~eV \cite{PhysRevLett.16.748, zatsepin1966upper,HiRes:2007lra}, even though the expected flux at earth is found to be too small for being detected \cite{Bhattacharjee:1989vu,Srednicki:1986xg,Gill:1994ic}.

More precisely, upon introducing a coupling between the SM and the dark $U(1)'$ from which the CS result, e.g. a Higgs portal or a kinetic mixing, an effective interaction between SM particles and the CS arises \cite{Hyde:2013fia}. In that case, the formation of cusps and kinks on the string radiate SM particles \cite{Long:2014mxa}. The expected gamma-ray flux at the earth is too low to be observed by Fermi-Lat \cite{Long:2014lxa}, even when assuming that all the massive particles radiated by CS subsequently decay into gamma-ray \cite{Auclair:2019jip}. Note however \cite{Auclair:2021jud} which finds particle production to be within the reach of Fermi-Lat for CS modelled according to Ringeval et al. \cite{Lorenz:2010sm, Ringeval:2017eww, Auclair:2019zoz}. For cusp domination and for large coupling between the SM and $U(1)'$, the flux of high-energy neutrino might be measured by the future experiments SKA and LOFAR for $G\mu \sim [10^{-14},\, 10^{-16}]$ \cite{Long:2014lxa}. Also, the distortions in the CMB may be detected by the future telescope PIXIE for $G\mu \sim [10^{-12},\, 10^{-14}]$ \cite{Long:2014lxa}. Finally, depending on the magnitude of the SM-$U(1)'$ coupling, the  BBN constraints can already exclude values of string tensions between $10^{-8} \gtrsim G\mu \gtrsim 10^{-14}$ \cite{Long:2014lxa}. 

Constraints from particle emission apply on an interval of values for $G\mu$, and not as upper bound like for gravitational emission ~\cite{Bhattacharjee:1989vu}.
For longer lifetimes $\propto (\Gamma G \mu)^{-1}$, there are more loops and we expect a larger flux of emitted particles while  gravitational emission grows with $G\mu$. At small $G\mu$, loops decay preferentially into particles, cf. sec.~\ref{sec:massive_radiation}. In that case, the expected flux of emitted particles increases with the string tension which controls the power of the particle emission. Therefore, there exists a value of $G\mu$ for which the expected flux of emitted particles is maximal. This is the value of $G\mu$ when particle production is as efficient as gravitational production. For example for loops created at the recombination time, the value of $G\mu$ maximizing the cosmic ray production is $10^{-18}$ \cite{Laliberte:2019kpb}.

\paragraph{Superconducting Cosmic Strings:} another possibility for generating large particle production is to couple the CS with electromagnetic charge carriers and to spontaneously break electromagnetic gauge invariance inside the vortex \cite{Witten:1984eb}. Upon moving through cosmic magnetic fields, Superconducting Cosmic Strings (SCS) are able to develop a large electric current $\mathcal{I}$. The formation of cusps on SCS is expected to emit bursts of electromagnetic radiation \cite{Vilenkin:1986zz, Spergel:1986uu, Copeland:1987yv, BlancoPillado:2000xy}, up to very high energies, set by the string tension $ \sqrt{\mu}\sim 10^{13}~\textrm{GeV}\,\sqrt{G\mu/10^{-15}}$, hence leading to high-energy gamma-rays \cite{Babul:1987lza, Berezinsky:2001cp, Cheng:2010ae}. 
Hence, SCS could be an explanation for the observed gamma-ray bursts at high redshifts, which depart from the predictions from star-formation-history \cite{Cheng:2010ae}.
However, the expected photon flux at earth is larger in the radio band than in the gamma-ray band 
\cite{Vachaspati:2008su, Zadorozhna:2009zza, Cai:2011bi} (but also mostly generated by kinks instead of cusps \cite{Cai:2012zd}). Thus, it has been proposed that SCS could be 
an explanation for the Fast-Radio-Burst events \cite{Yu:2014gea, Ye:2017lqn, Brandenberger:2017uwo} for string tensions in the range $G\mu \sim [10^{-12},\, 10^{-14}]$ and string currents $\mathcal{I}\sim [10^{-1}, \, 10^2]~\text{GeV}$ \cite{Ye:2017lqn}.
Electromagnetic emission from SCS lead to CMB distortions \cite{Sanchez:1988ek, Sanchez:1990kj, Tashiro:2012nb}. A next-generation telescope like PIXIE \cite{Kogut:2011xw} would exclude string tensions $G\mu \sim 10^{-18}$, for string currents as low as $\mathcal{I} \sim 10^{-8}~$GeV \cite{Tashiro:2012nb}.  Also, electromagnetic radiation, by increasing the ionization fraction of neutral hydrogen, can affect the CMB temperature and polarization correlation functions at large angular scales, leading to the contraint $\mathcal{I}\lesssim 10^{7}$~GeV \cite{Tashiro:2012nv}. Note that ionization of neutral hydrogen has been studied in \cite{Laliberte:2019kpb} in the case of non-superconducting strings. Additionally, the radio emission from SCS can increase the depth of the 21 cm absorption signal, and EDGES data excludes the SCS tension $G\mu \sim 10^{-13}$ for string currents as low at $\mathcal{I}\sim 10$~GeV. Finally, emission of boosted charge carriers from SCS cusps moving in a cosmic magnetic field $B$, has been studied in \cite{Berezinsky:2009xf}, and provide a possible explanation for high-energy neutrino above $10^{20}$~eV, for $G\mu \sim [10^{-14},\,10^{-20}]$.

\section{Derivation of the GW spectrum from CS}
\label{app:derivationGWspectrum}

In this appendix we provide the steps leading to Eq.~(\ref{kmode_omega}).

\subsection{From GW emission to detection}
The GW energy density spectrum today is defined as 
\begin{equation}
\Omega_{\rm GW}(f) = \frac{f}{\rho_{c}}\, \left| \frac{d \rho_{\rm GW}(f, \, t_0)}{df} \right|.
\end{equation}
After emission, the GW energy density redshifts as radiation,  $\rho_{\rm GW}\propto a^{-4}$, so the GW energy density per unit of frequency redshifts as
\begin{equation}
\frac{d \rho_{\rm GW}(f, \, t_0)}{df} =\frac{d \rho_{\rm GW}(\tilde{f}, \, \tilde{t})}{d\tilde{f}}  \, \left(\frac{a(\tilde{t})}{a(t_0)}\right)^3
\end{equation}
where the frequency at emission $\tilde{f}$ is related to the frequency today $f$ through 
$$\tilde{f}=\frac{a(t_0)}{a(\tilde{t})} f.$$  
\subsection{From loop production to GW emission}
After its formation at $t_{i}$, a loop shrinks through emission of GW with a rate $\Gamma G \mu$ so that its length evolves as, cf.~Sec.~\ref{sec:masslessradiation}
\begin{equation}
\label{eq:CSlength_app}
l(t) = \alpha t_i -\Gamma G \mu(t-t_i),
\end{equation}
where $\alpha$ is the length at formation in units of the horizon size.
The resulting GW are emitted at a frequency $\tilde{f}$ corresponding to one of the proper modes of the loop, i.e. 
\begin{equation}
\label{eq:GWspecfreq_app}
\tilde{f}=\frac{2k}{l}, \qquad  k\in\mathbb{Z}^{+}.
\end{equation}  
The GW energy rate emitted by one loop through the mode $k$ is, cf. Sec.~\ref{sec:masslessradiation}
\begin{equation}
\frac{dE_{\rm GW}^{(k)}}{dt}=\Gamma^{(k)}\, G \mu^2, \qquad \text{with} \quad\sum_k\Gamma^{(k)} = \Gamma,
\end{equation}
where 
\begin{equation}
\label{eq:Gamma_k}
\Gamma^{(k)}= \frac{\Gamma \, k^{-4/3} }{ \sum_{p=1}^{\infty}p^{-4/3}} \simeq \frac{\Gamma \, k^{-4/3} }{ 3.60},
\end{equation}
which assumes that the GW emission is dominated by cusps.
The GW energy density spectrum resulting from the emission of all the decaying loops until today is
\begin{equation}
\frac{d\rho_{\rm GW}(\tilde{f}, \, \tilde{t})}{d\tilde{f}} = \int_{t_F}^{t_0} d\tilde{t} \, \frac{dE_{\rm GW}}{d\tilde{t}} \, \frac{d n(\tilde{f}, \, \tilde{t})}{d\tilde{f}},
\end{equation}
where $d n(\tilde{f}, \, \tilde{t})/d\tilde{f}$ is the number density of loops emitting GW at frequency $\tilde{f}$ at time $\tilde{t}$ and $t_0$ is the age of the universe today. Loops start being created at time of CS network formation $t_F$,  after the damped evolution has stopped, cf. Sec.~\ref{sec:VOS_proof}.  \\
\subsection{The loop production}
In Sec.~\ref{sec:mainAssumptions}, we assume the loop-formation rate to be
\begin{equation}
\label{eq:LoopProductionFctBody_app}
\frac{dn}{dt_i}=(0.1)\frac{C_{\rm eff}(t_i)}{\alpha \, t_i^4},
\end{equation}
where $C_{\rm eff}(t_i)$ is the loop-formation efficiency.
We deduce the loop number density per unit of frequency
\begin{align}
\label{eq:loop_density_Jac}
\frac{d n(\tilde{f}, \, \tilde{t})}{d\tilde{f}} &=\left[\frac{a(t_i)}{a(\tilde{t})}\right]^3  \, \frac{dn}{dt_i} \cdot \frac{dt_i}{dl} \cdot \frac{dl}{d\tilde{f}} \\
&=\left[\frac{a(t_i)}{a(\tilde{t})}\right]^3  \,  \sum_{k} (0.1)\frac{C_\textrm{eff}(t_i)}{t_i^4} \cdot \frac{1}{\alpha \, (\alpha+\Gamma G\mu)} \cdot \frac{2k}{f^2} \left[\frac{a(\tilde{t})}{a(t_o)}\right]^2.
\end{align}
\subsection{The master equation}
Finally, we get the GW energy density spectrum
\begin{align}
\label{eq:master_eq_app}
\Omega_{\rm GW}(f) &=\sum_{k} \Omega_{\rm GW}^{(k)}(f) \notag \\ 
&= \sum_{k}\frac{1}{\rho_{c}}\, \frac{2k}{f} \, \frac{\mathcal{F}_{\alpha}\, \Gamma^{(k)} \, G\mu^2}{\alpha \, (\alpha+\Gamma\,G\mu)} \int_{t_{\rm osc}}^{t_{0}} d\tilde{t} \; \frac{C_\textrm{eff}(t_{i})}{t_{i}^4}\, \left[\frac{a(\tilde{t})}{a(t_0)}\right]^5 \, \left[\frac{a(t_i)}{a(\tilde{t})}\right]^3 \, \theta(t_i-t_{\rm osc})\,\theta(t_i-\frac{l_*}{\alpha}).
\end{align}
The first Heaviside function stands for the time $t_{\rm osc}$ at which long-strings start oscillating, either just after formation of the long-string network or after that friction becomes negligible, cf. Sec.~\ref{sec:thermal_friction}. The second Heaviside function stands for the energy loss into particle production which is more efficient than GW emission for loops of length smaller than a characteristic length $l_*$, which depends on the string small-scale structure, cf. sec~\ref{sec:massive_radiation}.
The time $t_i$ of formation of the loops, which emit at time $\tilde{t}$ and which give the detected frequency $f$, can be determined from Eq.~\eqref{eq:CSlength_app} and Eq.~\eqref{eq:GWspecfreq_app}
\begin{equation}
\label{eq:def_t_i}
t_i (f, \, \tilde{t})= \frac{1}{\alpha+\Gamma G \mu} \left[ \frac{2k}{f}\frac{a(\tilde{t})}{a(t_0)} + \Gamma G \mu \, \tilde{t} \right].
\end{equation}
Note that the contribution coming from the higher modes are related to the contribution of the first mode by
\begin{equation}
\label{eq:mode_k_vs_mode_1}
 \Omega_{\rm GW}^{(k)}(f) = k^{-4/3} \,  \Omega_{\rm GW}^{(1)}(f/k).
\end{equation}

\subsection{The GW spectrum from the quadrupole formula}
\label{sec:quadrupole_formula}

\paragraph{In standard cosmology:}
The scaling behavior $\Omega_{\rm GW} \propto  \sqrt{G \mu } \times f^0$, e.g. discussed along Eq.~\eqref{eq:lewicki_formula_GW_spectrum_radiation}, can be understood qualitatively from the quadrupole formula for the power emission of GW \cite{maggiore2008gravitational,Vachaspati:1984gt} 
\begin{equation}
\label{eq:GW_power_quadrupole}
P_{\rm GW}\sim N_{\rm loop} \, \frac{G}{5} \left(Q^{'''}_{\rm loop}\right)^2,
\end{equation}
where the triple derivative of the quadrupole of a loop is simply the string tension
\begin{equation}
\label{eq:GW_quadrupole}
Q^{'''}_{\rm loop} \sim \textrm{mass}\times \textrm{length}^2/\textrm{time}^{3} \sim \mu.
\end{equation}
 During the scaling regime, the number of loops formed at time $t_i$ scales as $t_i^{-3}$. Hence, the number of loops formed at time $t_i$, evaluated at a later time $\tilde{t}$ is 
\begin{equation}
\label{eq:loop_number_naive}
N_{\rm loop} \sim  \left(\frac{\tilde{t}}{t_i} \right)^3 \left(\frac{t_i }{ \tilde{t} }\right)^{\! 3/2},
\end{equation}
where the second factor accounts for the redshift as $a^{-3}$ of the loops between $t_i$ and $\tilde{t}$ during radiation.
Since GW redshift as radiation, their energy density today is
\begin{equation}
\label{eq:GW_amplitude_naive}
\Omega_{\rm GW} \sim \Omega_{\rm rad}~ \frac{\rho_{\rm GW}(\tilde{t})}{\rho_{\rm rad}(\tilde{t})} \sim  \Omega_{\rm rad}  \left( G \mu \right)^2 \left(\frac{ \tilde{t} }{t_i }\right)^{\! 3/2},
\end{equation}
where we assumed radiation-domination at $\tilde{t}$
\begin{equation}
\label{eq:rho_rad_radiation}
\rho_{\rm rad}(\tilde{t}) \sim G^{-1} \,\tilde{H}^2 ~\frac{\rho_{\rm rad}(\tilde{t})} {\rho_{\rm tot}(\tilde{t}) } ~\sim
G^{-1} ~\tilde{t}^{-2},
\end{equation}
and where we used that the energy density of GW at $\tilde{t}$ is
\begin{equation}
\rho_{\rm GW}(\tilde{t}) \sim \left( P_{\rm GW} ~\tilde{t} \right) / \,\tilde{t}^{\,3}.
\end{equation}
with the GW power $ P_{\rm GW}$ defined in Eq.~\eqref{eq:GW_power_quadrupole}.
From Eq.~\eqref{eq:loop_number_naive}, one can see that, at a fixed formation time $t_i$, the later the GW emission, the more numerous the loops. Hence, the dominant contribution to the SGWB from a given population of loops formed at $t_i$ occurs after one loop-lifetime, cf. Eq.~\eqref{eq:GWlifetime}, at 
\begin{equation}
\label{eq:main_emission_naive}
\tilde{t}_{\rm M} \sim  \frac{\alpha \, t_i }{\Gamma G\mu}.
\end{equation}
Upon plugging Eq.~\eqref{eq:main_emission_naive} into Eq.~\eqref{eq:GW_amplitude_naive}, one gets
\begin{equation}
\label{eq:GW_amplitude_final_naive}
\Omega_{\rm GW} \propto  \sqrt{G\mu}\times f^0.
\end{equation}
From Eq.~\eqref{eq:GW_amplitude_naive}, we can see that the GW spectrum during radiation is set by a combination of the strength of the GW emission from loops, $(G\mu)^2$, and the loop-lifetime $\tilde{t}_{\rm M}/t_i$, cf. Eq.~\eqref{eq:main_emission_naive}. Both are set by the triple derivative of the loop-quadrupole $Q^{'''}_{\rm loop}\sim \mu$. Hence we understand that the flatness in frequency during radiation is closely linked to the independence of the triple derivative of the loop-quadrupole \footnote{ In contrast, the GW spectrum generated by domain walls during radiation is not flat since in that case the triple derivative of the wall-quadrupole depends on the emission time: $Q^{'''}_{\rm DW}\sim \sigma\,\tilde{t}$, where $\sigma$ is the wall energy per unit of area. Hence, the energy density fraction in GW before wall annihilation \cite{Saikawa:2017hiv} increases with time $\Omega_{\rm GW}^{\rm DW} \sim \left( G\, \sigma \,\tilde{t} \right)^2$.}, cf. Eq.~\eqref{eq:GW_quadrupole}, on the loop length, and therefore on the frequency. \\

\paragraph{In non-standard cosmology:}
\label{sec:quadrupole_formula_NS}
In presence of non-standard cosmology, possibly being different during loop formation $a(t_i)\propto t_i^{2/m}$ and GW emission $a(\tilde{t})\propto \tilde{t}^{2/n}$, one must modify how the number of emitting loops, c.f. Eq.~\eqref{eq:loop_number_naive}, and the GW energy density, c.f. Eq.~\eqref{eq:rho_rad_radiation}, get redshifted. Namely, Eq.~\eqref{eq:loop_number_naive} and Eq.~\eqref{eq:rho_rad_radiation}, here denoted by $\rm Std.$, become
\begin{equation}
\label{eq:loop_number_naive_NS}
N_{\rm loop} \sim \left(\frac{\tilde{t}}{t_i} \right)^3\left(\frac{a(t_i)}{a(\tilde{t})}\right)^3\propto N_{\rm loop}^{\rm Std.}\times \frac{t_i^{(\frac{6}{m}-\frac{3}{2})}}{\tilde{t}^{(\frac{6}{n}-\frac{3}{2})}},
\end{equation}
and
\begin{equation}
\label{eq:rho_rad_radiation_NS}
\frac{\rho_{\rm rad}(\tilde{t})} {\rho_{\rm tot}(\tilde{t}) }\sim \left( \frac{a(\tilde{t})}{a(t_{\rm end})} \right)^{n-4} \propto ~\tilde{t}^{\,\frac{2(n-4)}{n}},
\end{equation}
where $t_{\rm end}$ is the ending time of the non-standard cosmology. 
The GW spectrum, c.f. Eq.~\eqref{eq:GW_amplitude_naive}, depends on the combination of the loop- and GW-redshift factors in Eq.~\eqref{eq:loop_number_naive_NS} and Eq.~\eqref{eq:rho_rad_radiation_NS}. Upon plugging the scaling $t_i \propto f^{\frac{n}{2-n}}$ and $\tilde{t}_M \propto f^{\frac{n}{2-n}}$ (which are themselves deduced from $t_i \propto a(\tilde{t})/f$, c.f. Eq.~\eqref{eq:GWspecfreq_app}, and Eq.~\eqref{eq:main_emission_naive}), we obtain
\begin{equation}
\label{eq:spectral_index_omega_GW}
\Omega_{\rm GW} \propto f^{-2\left(\frac{3n+m(n-7)}{m(n-2)}\right)}.
\end{equation}
When loop formation at $t_i$ occurs during matter/kination but GW emission at $\tilde{t}_{\rm M}$ occurs during radiation, $(m,\, n) = (3, \, 4)/(6, \, 4)$, we find that the GW spectrum scales like $f^{-1}$/$f^1$. { In Sec.~\ref{sec:study_impact_mode_nbr}, we show that the presence of high-frequency modes $k\gg 1$ turns the $f^{-1}$ behavior to $f^{-1/3}$. }

\subsection{Impact of the high-frequency proper modes of the loop}
\label{sec:study_impact_mode_nbr}
	
			  \begin{figure}[h!]
				\centering
				\centering
			\includegraphics[width=0.45\textwidth]{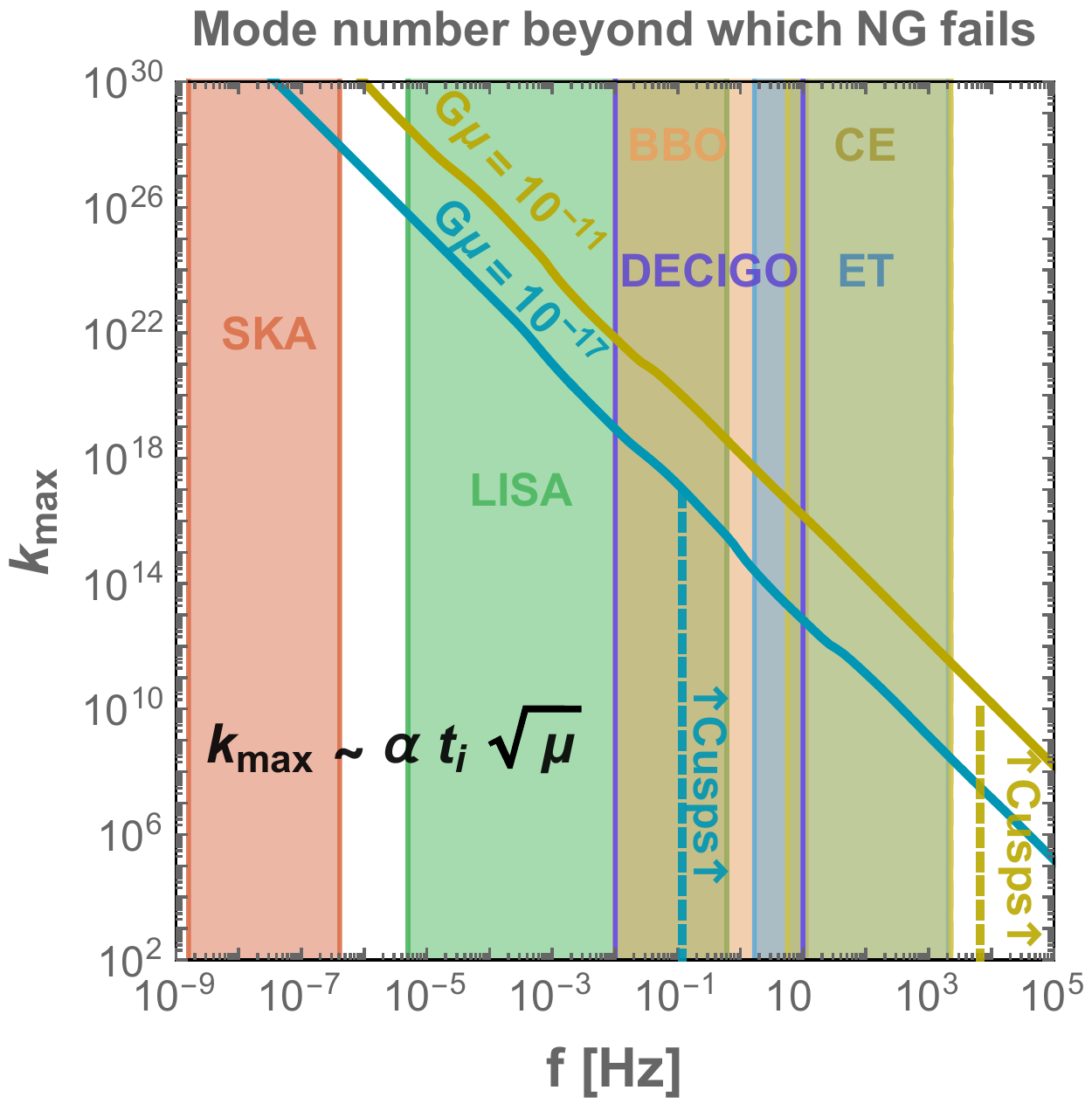}
			\includegraphics[width=0.45\textwidth]{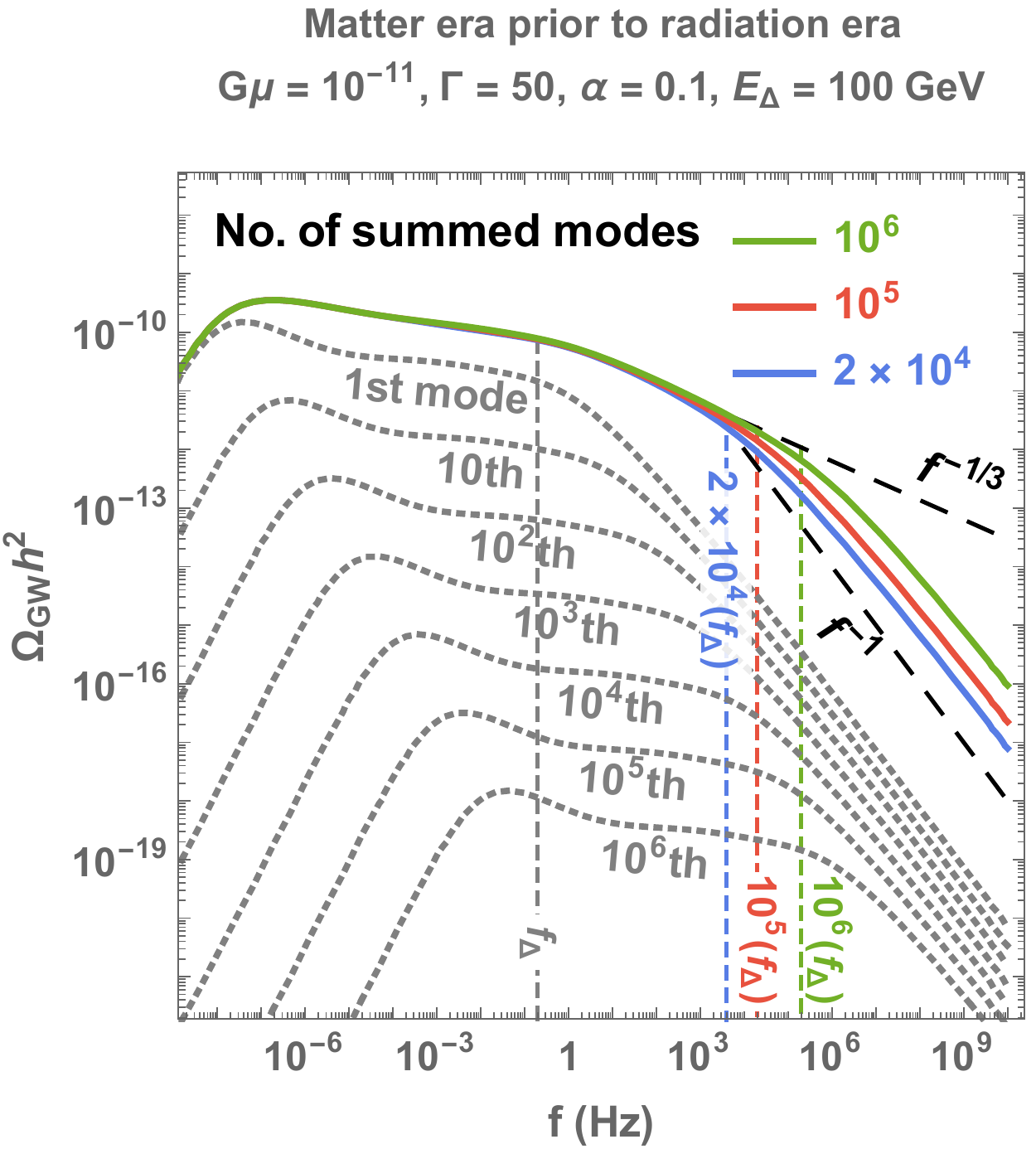}
				\caption{\it \small \textbf{Left:} Maximal mode number $k_{\rm max}$ beyond which we can not trust the Nambu-Goto approximation anymore. It occurs when the wavelength of the oscillation, given by $\alpha \, t_i/k$ where $t_i$ is the Hubble horizon when the loop forms, becomes of the order of the loop thickness $\mu^{-1/2}$. We can see that in the different interferometer windows, $k_{\rm max}$ is extremely large, often much larger than the maximal mode number tractable numerically $\sim 10^6$. \textbf{Right:} Decomposition of a GW spectrum under the contributions coming from the different proper modes of the loop. We can see that high-k modes are responsible for the change of slope $f^{-1/3} \, \rightarrow \,f^{-1}$ between the physical turning point frequency $f_{\rm \Delta}$ and a second, artificial, turning-point $f_{\rm max}$, given by $f_{\rm max} = k_{\rm max}\,f_{\rm \Delta}$, cf. Eq.~\eqref{eq:second_turning_point}, where $k_{\rm max}$ is the total number of modes chosen for doing the computation, here $2\times 10^4$, $10^5$ and $10^6$. Except when explicitly specified, for technical reasons we fix $k_{\rm max}=2 \times 10^4$ modes in all the plots of our study. }
				\label{fig:max_mode_number}
			\end{figure}

\paragraph{The motivation:}
When computing the GW spectrum from cosmic strings, given by Eq.~\eqref{eq:master_eq_app}, we are confronted with an infinite sum over the proper modes $k$ of the loop. The number of modes before we violate the Nambu-Goto approximation is very large, as shown in the left panel of Fig.~\ref{fig:max_mode_number}. 
In what follows, we study the impact of the high-frequency modes on the GW spectrum.
From Eq.~\eqref{eq:GWspecfreq_app}, Eq.~\eqref{eq:Gamma_k} and Eq.~\eqref{eq:master_eq_app}, we can see that the spectrum for the $k^{\rm th}$ mode is related to the fundamental spectrum $k=1$ through Eq.~\eqref{eq:mode_k_vs_mode_1}, which we rewrite here
\begin{equation}
\label{eq:mode_k_vs_mode_1_(2)}
\Omega_{\rm GW}^{(k)}(f) = k^{-\delta} \,  \Omega_{\rm GW}^{(1)}(f/k),
\end{equation}
In this study, we fix $\delta = 4/3$ since we assume that the small-scale structure is dominated by cusps. However, the results of the present section apply to any small-scale structure described by $\delta$.

\paragraph{Case of a fundamental spectrum with a flat slope:}
At first, if we assume that the one-mode spectrum is flat, $\Omega_{\rm GW}^{(1)}(f) \propto f^0$, then the total spectrum is a simple rescaling of the fundamental spectrum by the Riemann zeta function
\begin{equation}
\Omega_{\rm GW}(f) =\zeta\left(\delta\right) \,\Omega_{\rm GW}^{(1)}(f),
\end{equation}
where in particular, $\zeta(4/3)=\sum_k k^{-4/3} \simeq 3.60$.

\paragraph{Case of a fundamental spectrum with a slope $f^{-1}$:}
Now, we consider the case where the fundamental spectrum has a slope $f^{-1}$, as expected in thepresence of an early matter era, cf. Eq.~\eqref{eq:spectral_index_omega_GW}, but also, in the presence of high-frequency cut-offs. The high-frequency cut-offs in the spectrum are described by Heaviside functions in the master formula in Eq.~\eqref{eq:master_eq_app}, of the type $\Theta\left( t_i \, - \, t_{\rm \Delta}  \right)$, where $t_{\rm \Delta}$ is the cosmic time when loop formation starts, assuming it is suppressed before on. The time $t_{\rm \Delta}$ can correspond to either the time of formation of the network, cf. Eq.~\eqref{eq:network_formation}, the time when friction-dominated dynamics become irrelevant, cf.  App.~\ref{sec:thermal_friction}, the time when gravitational emission dominates over massive particle production, cf. Sec.~\ref{sec:massive_radiation}, or the time when the string correlation length re-enters the Hubble horizon after a short period of second inflation, cf. Sec.~\ref{sec:inflation}.  The slope of the $k=1$ spectrum beyond the cut-off frequency can be read from Eq.~\eqref{eq:master_eq_app} after injecting Eq.~\eqref{eq:def_t_i} and $t_i=t_{\rm \Delta}$, where we find
\begin{equation}
\label{eq:first_mode_spectrum}
\Omega_{\rm GW}^{(1)}(f) =  \Omega_\Delta \Theta\left( -f + f_\Delta  \right) \,+\, \Omega_\Delta \,  \frac{f_\Delta }{f} \,\Theta\left( f - f_\Delta  \right).
\end{equation}
The fundamental spectrum is flat until $f_\Delta $ and then shows a slope $f^{-1}$ beyond.
The total spectrum, summed over all the proper modes, can be obtained from Eq.~\eqref{eq:mode_k_vs_mode_1_(2)} and Eq.~\eqref{eq:first_mode_spectrum}
\begin{equation}
\label{eq:spectrum_slope_m1}
\Omega_{\rm GW}(f)  = \sum_{k=1}^{\rm k_{\Delta }} \,\frac{\Omega_\Delta }{k^{\delta}}\, k \,\frac{f_\Delta }{f}  \,+\, \sum_{k=k_\Delta }^{\rm k_{\rm max}} \frac{\Omega_\Delta  }{k^{\delta}},
\end{equation}
where $k_{\rm max}$ is the maximal mode, chosen arbitrarily, and $k_\Delta $ is the critical mode defined such that modes with $k<k_\Delta $ have a slope $f^{-1}$ whereas modes with $k>k_\Delta $ have a flat slope. For a given frequency $f$, the critical mode number $k_\Delta$ obeys
\begin{equation}
\label{eq:critical_mode}
k_\Delta   \simeq  \frac{f}{f_\Delta }.
\end{equation}
We now evaluate Eq.~\eqref{eq:spectrum_slope_m1} in the large $k_\Delta $ limit, while still keeping $k_\Delta < k_{\rm max}$
\begin{equation}
\label{eq:spectrum_slope_m1_(2)}
\Omega_{\rm GW}^{1\ll k_\Delta <k_{\rm max}}(f)  \simeq  \Omega_\Delta \frac{f_\Delta }{f} k_\Delta ^{2-\delta}  \,+\, \frac{1}{\delta-1} \frac{\Omega_\Delta }{k_\Delta ^{\delta-1}} ,
\end{equation}
where we have used the asymptotic expansion of the Euler-Maclaurin formula for the first term and the asymptotic expansion of the Hurwitz zeta function for the second term. Finally, after injecting Eq.~\eqref{eq:critical_mode}, we get
\begin{equation}
\label{eq:spectrum_slope_m1_(3)}
\Omega^{1\ll k_\Delta <k_{\rm max}}_{\rm GW}(f)  \simeq \frac{\delta}{\delta-1}\Omega_\Delta  \left( \frac{f_\Delta }{f} \right)^{\delta - 1} \propto \begin{cases}
f^{-1/3} & \mbox{for cusps ($\delta=4/3$) } \\
f^{-2/3} & \mbox{kinks ($\delta = 5/3$)} \\
f^{-1}  &  \mbox{kink-kink collisions ($\delta = 2$) }\\
\end{cases}
\end{equation}
We conclude that the spectral index beyond a high-frequency turning point $f_\Delta $ due to an early matter era, a second inflation era, particle production, thermal friction domination, or the formation of the network, is modified by the presence of the high-k modes in a way that depends on the small-scale structure. Particularly, if the small-scale structure is dominated by cusps, we find a slope $-1/3$. We comment on the possibility to get information about the nature of the small-structure from detecting a GW spectrum from CS with a decreasing slope. The study \cite{Blasi:2020wpy} was the first one to point out the impact of the high-frequency modes on the value of a decreasing slope.

\paragraph{Impact of fixing the total number of proper modes:}
For technical reasons we are unavoidably forced to choose a maximal number of modes $k_{\rm max}$. We now study the dependence of the GW spectrum on the choice of $k_{\rm max}$. The evaluation of Eq.~\eqref{eq:spectrum_slope_m1} for $k_\Delta  > k_{\rm max}$ leads to
\begin{equation}
\label{eq:spectrum_slope_m1_(3)}
\Omega^{1\ll k_{\rm max} <k_\Delta }_{\rm GW}(f) = \zeta\left(\delta-1\right) \,\Omega_\Delta \,  \frac{f_\Delta }{f}.
\end{equation}
Hence, in addition to the initial physical turning point $f_\Delta $, where the slope changes from flat to $f^{-1/3}$, there is a second artificial turning point $f_{\rm max}$ given by
\begin{equation}
\label{eq:second_turning_point}
f_{\rm max} = k_{\rm max}\,f_{\Delta },
\end{equation}
where the slope changes from $f^{-1/3}$ to $f^{-1}$.  We show the different behaviors in the right panel of Fig.~\ref{fig:max_mode_number}.

\paragraph{Case of a fundamental spectrum with a slope $f^{+1}$:}
As last, we comment on the case where the fundamental spectrum has a slope $f^1$, as in the case of an early kination era, cf. Eq.~\eqref{eq:spectral_index_omega_GW}. Repeating the same steps as in Eq.~\eqref{eq:spectrum_slope_m1}, we obtain
\begin{equation}
\label{eq:spectrum_slope_p1}
\Omega_{\rm GW}(f)  = \zeta\left(\delta+1\right)\,\Omega_\Delta \,\frac{f}{f_\Delta } ,
\end{equation}
hence the slope of the full spectrum is the same as the slope of the fundamental spectrum.

\section{Derivation of the frequency - temperature relation}
\label{derive_turning_points}
In this appendix, we compute the correspondence between an observed frequency $f$ and the temperature $T$ of the universe when the loops responsible for that frequency have been formed. 

		\begin{figure}[h!]
				\centering
				\centering
			\includegraphics[width=0.6\textwidth]{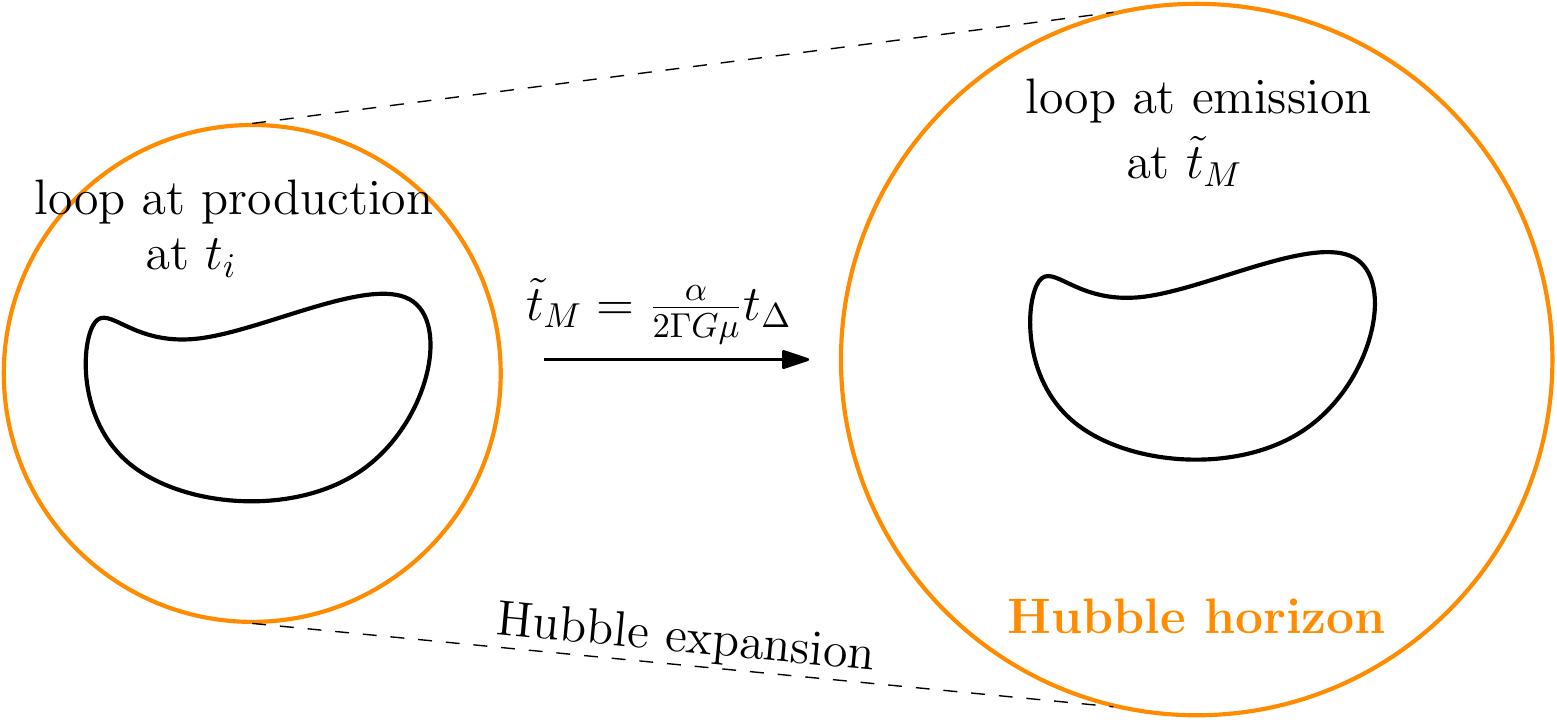}
				\caption{\it \small Loops produced at time $t_i$ contribute to the GW spectrum much later, when they have accomplished half of their lifetime, at $\tilde{t}_{\rm M}\simeq\alpha\, t_i/  (2\Gamma G \mu)$. Hence GW emitted from cosmic-string loops are exempt from a redshift factor $a(\tilde{t}_{\rm M})/a(t_i)$ so have much higher frequency than GW produced from other sources at the same energy scale.}
				\label{cartoon_loop_maximal_decay}
			\end{figure}
\subsection{In standard cosmology}
According to the scaling of the loop-formation rate ${dn}/{dt_i}\propto t_i^{-4}$, the main contribution to the GW emission at time $\tilde{t}$ comes from the loops created at the earliest epoch. Correspondingly, loops created at $t_i$ contribute to the spectrum as late as possible, at the \textit{main emission time} $\tilde{t}_{\rm M}$. The latest emission time is set by the loop lifetime $\alpha \, t_i/ \Gamma G \mu$, where $\alpha$ is the loop-length at formation in horizon unit, cf. Eq.~\eqref{eq:GWlifetime}. Hence, a loop produced at time $t_i$ mainly contributes to the spectrum, much later cf. figure~\ref{cartoon_loop_maximal_decay}, at a time
\begin{equation}
\label{eq:t_M}
\tilde{t}_{\rm M} \simeq \frac{\alpha \, t_i}{ 2 \Gamma G \mu},
\end{equation}
where the factor $1/2$ is found upon maximizing the loop-formation rate ${dn}/{dt_i}\propto t_i^{-4}$ and upon assuming $\alpha\gg \Gamma G\mu$.
The loop length after half the loop lifetime, in Eq.~\eqref{eq:t_M}, is equal to half the length at formation $\alpha\, t_i/2$, cf. Eq.~\eqref{eq:CSlength_app}. Hence the emitted frequency is set by
\begin{eqnarray}
\label{eq:tdelta_fdelta_eq_line1}
\alpha \, t_i &\simeq &\frac{4}{f}\frac{a(\tilde{t}_M)}{a(t_0)},\\
&\simeq & \frac{4}{f}\frac{a(\tilde{t}_M)}{a(t_\textrm{eq})}\frac{a(t_\textrm{eq})}{a(t_0)} \\
&\simeq& \frac{4}{f}\left(\frac{\tilde{t}_M}{t_\textrm{eq}}\right)^{1/2}\left(\frac{\,t_\textrm{eq}}{t_0}\right)^{2/3}
\label{tdelta_fdelta_eq}
\end{eqnarray}
where we used $f \,a(t_0)/a(\tilde{t}) = 2k/l$ and only considered the first Fourier mode $k=1$, cf. Eq.~\eqref{eq:GWspecfreq_0}. By merging Eq.~\eqref{eq:t_M} and Eq.~\eqref{tdelta_fdelta_eq}, we obtain the relation between an observed frequency $f$ and the time $t_i$ of loop formation
\begin{equation}
f\simeq \sqrt{\frac{8z_\textrm{eq}}{\alpha\Gamma G\mu}}\left(\frac{t_\textrm{eq}}{t_i}\right)^{1/2}t_0^{-1},
\end{equation}
where the redshift at matter-radiation equality is  $z_{\rm eq} = \Omega_{\rm C}/\Omega_{\gamma} \simeq 3360$, and  $t_{\rm eq} \simeq 51.8$~kyrs (from integrating Eq.~\eqref{friedmann_eq}) and $t_0\simeq 13.8~$Gyrs \cite{Tanabashi:2018oca}.
Finally, using entropy conservation, we obtain the relation between the frequency $f$ at observation and the temperature $T$ of the universe when the corresponding loops are formed
		 \begin{align}
		 \nonumber
		f &\simeq \sqrt{\frac{8}{z_\textrm{eq}\alpha\Gamma G\mu}}\left(\frac{g_*(T)}{g_*(T_0)}\right)^{1/4}\left(\frac{T}{T_0}\right) \, t_0^{-1}\\
		&\simeq (6.7\times10^{-2}\textrm{ Hz})\left(\frac{T}{\textrm{GeV}}\right)\left(\frac{0.1\times50\times10^{-11}}{\alpha\Gamma G\mu}\right)^{1/2}\left(\frac{g_*(T)}{g_*(T_0)}\right)^{1/4}.
		\label{fdeltaApp}
		\end{align}
\subsection{During a change of cosmology}

The derivation of (\ref{fdeltaApp})  does not take into account the time-variation of  $C_{\rm eff}$.
It assumes that loops are produced and decayed during the scaling regime in the radiation era.
An observable to test the non-standard cosmology is 
 the frequency $ f_{\Delta}$ of the \textit{turning-point} defined as the frequency at which the GW spectrum starts to deviate from the standard-cosmology behavior and the spectral index changes.
We obtain different fitted values for this turning point frequency depending on the prescription. We quote below different expressions, depending whether we assume that the spectrum can be measured with a $10\%$ precision, and $1\%$ respectively. We compare the predictions obtained using a scaling and VOS network:
\begin{align}
f_{\Delta}\simeq \textrm{ Hz} \left(\frac{T_{\Delta}}{\textrm{GeV}}\right)\left(\frac{0.1\times50\times10^{-11}}{\alpha\Gamma G\mu}\right)^{1/2}\left(\frac{g_*(T_{\Delta})}{g_*(T_0)}\right)^{1/4} \times 
\begin{cases}
2 \times 10^{-3} &\textrm{for VOS}, 10 \% \\
45 \times 10^{-3} &\textrm{for scaling}, 10 \% \\
0.04 \times 10^{-3} &\textrm{for VOS}, 1 \% \\
15 \times 10^{-3} &\textrm{for scaling}, 1 \% \\
\end{cases}
\label{turning_point_general_scaling_app}
\end{align}
 Therefore, the turning point frequency is lower in VOS than in scaling  by a factor $\sim 22.5$ if we define the turning-point frequency by  an amplitude deviation of $10\%$ with respect to standard cosmology, and by a factor $\sim 375$ for a deviation of $1\%$. The loops contributing to this part of the spectrum have been formed at the time of the change of cosmology. When the cosmology changes, the network achieves a transient evolution in order to reach the new scaling regime. The long-string network needs extra time to transit from one scaling regime to the other, hence the shift in the relation between observed frequency and temperature of loop formation at the turning-point, cf. Sec.~\ref{sec:turning_point_general}.

\subsection{In the presence of an intermediate inflation period}
		The above derivation of the relation between the observed frequency and the time of loop production assumes that cosmic-string loops are constantly being produced throughout the cosmic history.
		It does not apply if the network experiences an intermediate era of inflation.
		This case is discussed in Sec.~\ref{sec:turning_point_inf} and the turning-point formulae are, for a given precision
		\begin{align}
f_{\Delta}\simeq \textrm{ Hz} \left(\frac{T_{\Delta}}{\textrm{GeV}}\right)\left(\frac{0.1\times50\times10^{-11}}{\alpha\Gamma G\mu}\right)^{1/2}\left(\frac{g_*(T_{\Delta})}{g_*(T_0)}\right)^{1/4} \times 
\begin{cases}
1.5 \times 10^{-4} &\textrm{for } 10 \% \\
5 \times 10^{-6} &\textrm{for } 1 \% \\
\end{cases}
\label{turning_point_general_scaling_app_inf}
\end{align}

\subsection{Cut-off from particle production}
The cutoff frequency due to particle production is given in Sec.~\ref{UVcutoff}.

\section{Derivation of the VOS equations}
\label{sec:VOS_proof}

\subsection{The Nambu-Goto  string in an expanding universe}
The Velocity-dependent One-Scale equations (VOS) in Eq.~\eqref{eq:VOS_eq_body}, describe the evolution of a network of long strings in term of the mean velocity $\bar{v}$ and the correlation length $\xi = L/t$, see the original papers \cite{Martins:1995tg, Martins:1996jp, Martins:2000cs} or the recent review \cite{martins2016defect}. The set of points visited by the Nambu-Goto  string during its time evolution form a 2D manifold, called the world-sheet, described by time-like and space-like coordinates $\mathtt{t}$ and $\sigma$. The embedding of the 2D world-sheet in the 4D space-time is described by $x^\mu(\mathtt{t},\, \sigma)$ where $\mu =1,\,2,\,3,\,4$. The choice of the word-sheet coordinates being arbitrary, we have two gauge degrees of freedom which we can fix by imposing $\mathbf{\dot{x}}\cdot \mathbf{x'} = 0$ and $\mathtt{t}=\tau$ where $\tau$ is the conformal time of the expanding universe. The dot and prime denote the derivatives with respect to the time-like and space-like world-sheet coordinates, $ \mathbf{\dot{x}}\equiv d\mathbf{x}/d\mathtt{t} $ and  $ \mathbf{x'}\equiv d\mathbf{x}/d\sigma $. Then, the equations of motion of the Nambu-Goto string in a FRW universe are \cite{Turok:1984db}
\begin{eqnarray}
\mathbf{\ddot{x}}+2\mathcal{H}(1-\mathbf{\dot{x}}^2)\mathbf{\dot{x}}&=&\frac{1}{\epsilon}\left(\frac{\mathbf{x'}}{\epsilon}\right)',\\
\dot{\epsilon}+2\mathcal{H}\,\mathbf{\dot{x}}^2\,\epsilon&=&0,
\end{eqnarray}
where $\mathcal{H}\equiv \dot{a}/a=Ha$ and $\epsilon\equiv \sqrt{\mathbf{x'}^2/(1-\mathbf{\dot{x}}^2})$ is the coordinate energy per unit of length.
\subsection{The long-string network}
The macroscopic evolution of the long string network can be described by the energy density 
\begin{equation}
\label{vos_long_E}
\rho_{\infty}=\frac{E}{a^3}=\frac{\mu}{a^2(\tau)}\int \epsilon \,d\sigma\equiv\frac{\mu}{L^2},
\end{equation}
 and the root-mean-square averaged velocity
\begin{equation}
\bar{v}^2\equiv\langle\mathbf{\dot{x}}^2\rangle=\frac{\int\mathbf{\dot{x}}^2\epsilon \,d\sigma}{\int\epsilon \,d\sigma},
\label{vos_avg_v}
\end{equation}
where we recall that $\mu$ is the CS linear mass density.
\subsection{VOS 1: the correlation length}
Differentiating Eq.~\eqref{vos_long_E} gives the evolution of the energy density in an expanding universe
\begin{eqnarray}
\frac{d\rho_{\infty}}{dt}&=&\frac{d\rho_{\infty}}{d\tau}\cdot\frac{d\tau}{dt}=\frac{1}{a}\cdot\frac{d\rho_{\infty}}{d\tau},\\
&=&\frac{\mu}{a}\left[\frac{d}{d\tau}\left(\frac{1}{a^2}\right)\int\epsilon \,d\sigma + \frac{1}{a^2}\int\frac{d\epsilon}{d\tau}\,d\sigma\right],\\
&=&-2\frac{\mu}{a^3}\mathcal{H}\left[\int\epsilon \,d\sigma + \int\mathbf{\dot{x}}^2\epsilon \,d\sigma\right],\\
&=&-2H\rho_{\infty} \,( 1+ \bar{v}^2).
\label{E_network}
\end{eqnarray}
Moreover, after each string crossing, the network transfers energy into loops with a rate given by Eq.~\eqref{eq:energylossloop} and we get
\begin{equation}
\frac{d\rho_{\infty}}{dt}=-2H\rho_{\infty} \, ( 1+ \bar{v}^2)-\tilde{c} \, \bar{v}\frac{\rho_{\infty}}{L},
\end{equation}
which after using Eq.~\eqref{vos_long_E}, leads to the first VOS equation
\begin{equation}
\label{eq:VOS_L}
\text{VOS 1:} \qquad \frac{dL}{dt}=HL \,( 1+ \bar{v}^2)+\frac{1}{2}\tilde{c}\,\bar{v}.
\end{equation}
We neglect the back-reaction on long strings from gravitational emission which is suppressed with respect to the loop-chopping loss term by $O(G\mu)$. The case of global strings, for which however, the back-reaction due to particle production may play a role, is considered in App.~\ref{sec:VOS_global}.

\subsection{Thermal friction}
\label{sec:thermal_friction}

In addition to the Hubble friction, there can be friction due to the interactions of the strings with particles of the plasma, leading to the retarding force \cite{Vilenkin:1991zk}
\begin{equation}
F = \rho \, \sigma \, \bar{v} = \beta \, T^3\, \bar{v},
\end{equation}
where $\rho \sim T^4$ is the plasma energy density and $\sigma \sim T^{-1}$ is the cross-section per unit of length. The effect of friction is to damp the string motion and to suppress the GW spectrum \cite{Garriga:1993gj}. For gauge strings, a well-known realisation of friction is the interaction of charged particles with the pure gauge fields existing outside the string, the so-called Aharonov-Bohm effect \cite{Aharonov:1959fk}. In such a case, the friction coefficient $\beta$ is given by \cite{Vilenkin:1991zk}
\begin{equation}
\beta=2\pi^{-2}\zeta (3)\sum_i g_i \sin^2(\pi \nu_i),
\end{equation}
with
\begin{eqnarray*}
i &\equiv&\textrm{\hspace{1em} relativistic particle species ($m_i \ll T$),}\\
 g_i &\equiv& \textrm{\hspace{1em} number of relativistic degrees of freedom of $i$} \times \begin{cases}3/4 \textrm{ (fermion),}\\1 \textrm{ (boson),} \end{cases} \\
2\pi\,\nu_i \equiv e_i\,\Phi &\equiv& \textrm{\hspace{1em}  phase-shift of the wave-function of particle $i$ when transported on a}\\
&& \textrm{\hspace{1em}  close path around the string. $e_i$ being its charge under the associated }\\
&& \textrm{\hspace{1em}  gauge group and $\Phi$ the magnetic field flux along the string.}
\end{eqnarray*}
The friction term in the first VOS equation, Eq.~\eqref{eq:VOS_L}, becomes
\begin{equation}
2H\bar{v}^2\longrightarrow\frac{\bar{v}^2}{l_d} \equiv 2H\bar{v}^2+\frac{\bar{v}^2}{l_f},
\label{VOS_frictionlength_intro}
\end{equation}
where we introduced a friction length due to particle scattering $l_f \equiv \mu/(\sigma\, \rho) = \mu/(\beta T^3)$ \cite{Martins:1995tg, Martins:1996jp}, and the associated effective friction length $l_d$. At large temperature, the large damping due to the frictional force prevents the CS network to reach the scaling regime until it becomes sub-dominant when $2H \lesssim 1/l_f$, so after the time
\begin{align}
\label{eq:eq_friction}
t_\textrm{fric} &\simeq (2.5\times 10^{-5})~\left(\frac{106.75}{g_*(t_\textrm{fric})} \right)^{3/2}~\beta^2 ~(G\mu)^{-2}~ t_{\rm pl},\\
&\simeq (1.4\times10^{-4})~\left(\frac{g_*(t_\textrm{F})}{106.65} \right)^{1/2}\left(\frac{106.75}{g_*(t_\textrm{fric})} \right)^{3/2}~\beta^2 ~(G\mu)^{-1} ~ t_{F},
\end{align}
where $t_{\rm pl} \equiv \sqrt{G}$ and where the network formation time $t_F$ is the cosmic time when the energy scale of the universe is equal to the string tension $\rho_{\rm tot}^{1/2}(t_F) \equiv \mu$. 
For friction coefficient $\beta = 1$, the friction becomes negligible at the temperatures $T_* \simeq 4$~TeV for $G\mu=10^{-17}$, $T_* \simeq 400$~TeV for $G\mu=10^{-15}$, $T_* \simeq 40$~PeV for $G\mu=10^{-13}$, hence respectively impacting the SGWB only above the frequencies $20$~kHz, $200$~kHz, $2$~MHz, cf. Eq.~\eqref{turning_point_general}, which are outside the GW interferometer windows, cf. Fig.~\ref{ST_vos_scaling}. 

\begin{figure}[h!]
\centering
\raisebox{0cm}{\makebox{\includegraphics[width=0.49\textwidth, scale=1]{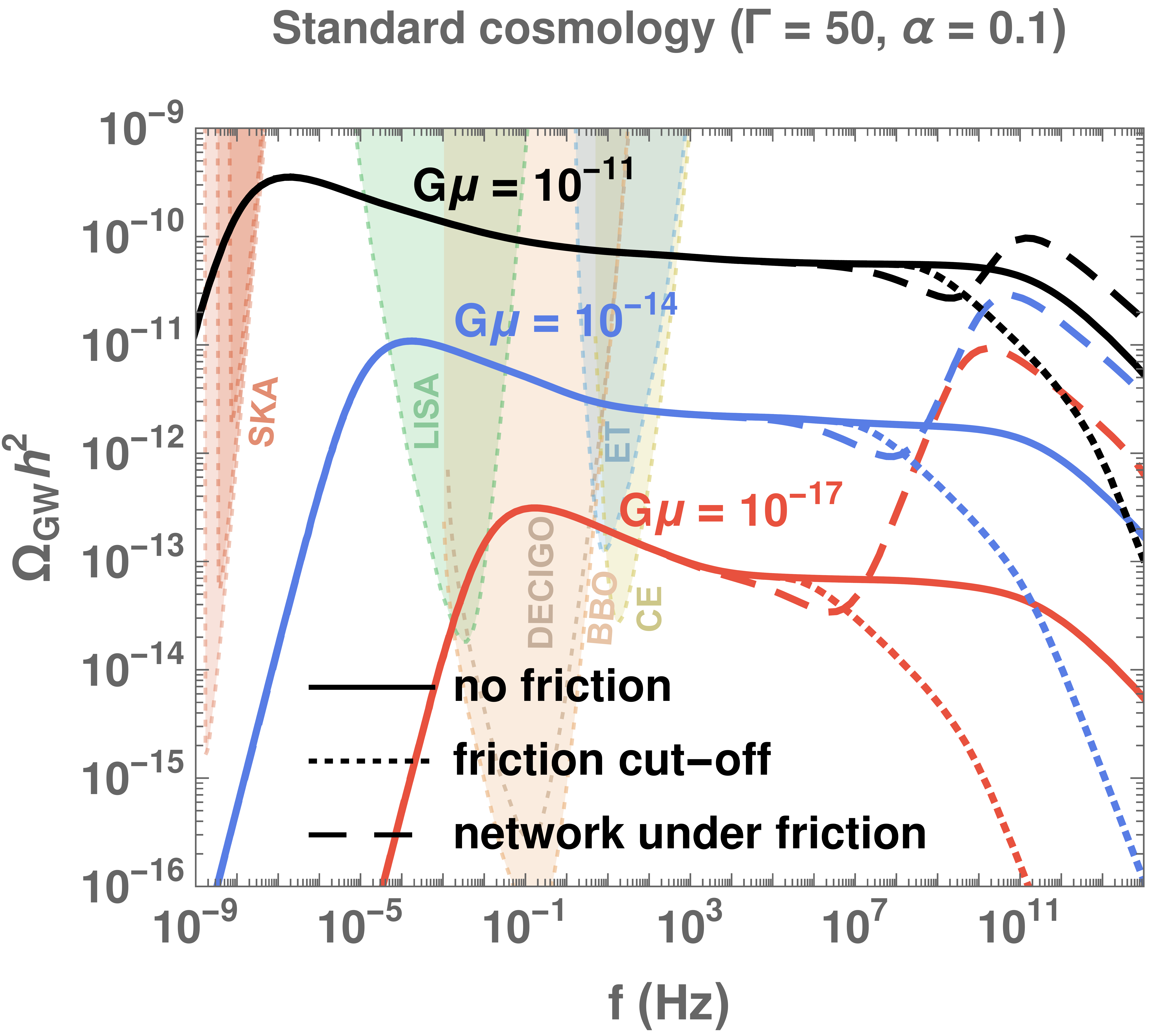}}}
\caption{\it \small GW spectrum from CS assuming no thermal friction (solid lines), thermal friction only at the level of the long-string network,  i.e. upon including eq. \eqref{VOS_frictionlength_intro} in the VOS equations (dashed lines) or thermal friction taken at the loop level, i.e. by removing GW emissions anterior to $t_{\rm fric}$ defined in Eq.~\eqref{eq:eq_friction} (dotted lines). See text for more details. A standard cosmology is assumed. } 
\label{fig:friction_spectrum}
\end{figure}

In Fig.~\ref{fig:friction_spectrum}, we show the impact of thermal friction on the GW spectrum from CS in two different ways.
\begin{itemize}
\item [$\diamond$]
\textbf{Network under friction}\textit{ (dashed lines in Fig.~\ref{fig:friction_spectrum}) :} the thermal friction is only taken into account at the level of the long-string network. Concretely, by simply including the friction term in Eq.~\eqref{VOS_frictionlength_intro} in the VOS equations. The GW peak at high frequency is due to the loop over-production by the frozen network, followed by a fast relaxation (with a little oscillatory behavior) to the scaling regime once friction becomes negligible with respect to Hubble expansion. This approach is insufficient since it assumes that the GW power emitted by loops is still given by $\Gamma G\mu^2$ with $\Gamma \simeq 50$ and therefore it does not take into account the damping of the oscillations of the loops under which we expect $\Gamma \rightarrow 0$.
\item [$\diamond$]
\textbf{GW emission cut-off}\textit{ (dotted lines in Fig.~\ref{fig:friction_spectrum}) :}  The damping of the loop oscillations is now taken into account by discarding all GW emissions happening earlier than $t_{\rm fric}$ in Eq.~\eqref{eq:eq_friction}, when thermal friction is larger than Hubble friction. Technically, the time $t_{\rm osc}$ of first loop oscillations in Eq.~\eqref{kmode_omega} is set equal to $t_\textrm{fric}$ in Eq.~\eqref{eq:eq_friction}. 
\end{itemize}
In many of our plots, e.g. Fig.~\ref{sketch_scaling} or Fig.~\ref{ST_vos_scaling}, we show the GW spectrum in the presence of thermal friction with a gray line, computed according to the second prescription above, entitled `GW emission cut-off'.  Note that in most cases, the effect of friction manifests itself at very high frequencies, outside the observability band of planned interferometers. It could however become relevant if those high frequencies could be probed in future experiments.

\subsection{VOS 2: the mean velocity}
Differentiating Eq.~\eqref{vos_avg_v} gives the evolution of the averaged velocity, which constitutes the second VOS equation
\begin{equation}
\label{eq:VOS_v}
\text{VOS 2:}  \qquad \qquad \frac{d\bar{v}}{dt}=(1-\bar{v}^2)\left[\frac{k(\bar{v})}{L}-\frac{\bar{v}}{l_d}\right],
\end{equation}
with
\begin{equation}
k(\bar{v}) \equiv \frac{\left<(1-\mathbf{\dot{x}}^2)\,(\mathbf{\dot{x}} \cdot \mathbf{u}) \right>}{\bar{v}\,(1-\bar{v}^2)},
\end{equation}
where $\mathbf{u}$ is the unit vector aligned with the radius of curvature $\propto {d^2\mathbf{x}}/{d\sigma^2}$. $k(\bar{v})$ indicates the degree of wiggliness of the string. More precisely, $k(\bar{v})=1$ for a straight string and $k(\bar{v})\lesssim 1$ once we add small-scale structures.
We use the results from numerical simulations \cite{Martins:2000cs}
\begin{equation}
\label{eq:momentum_operator}
 k(\bar{v})=\frac{2\sqrt{2}}{\pi}(1-\bar{v}^2)(1+2\sqrt{2}\bar{v}^3)\frac{1-8\bar{v}^6}{1+8\bar{v}^6}.
\end{equation}
Eq.~\eqref{eq:VOS_v} is a relativistic generalization of Newton's law where the string is accelerated by its curvature $1/L$ but is damped by the Hubble expansion and plasma friction after a typical length $1/l_d$.
{The Eq.~\eqref{eq:VOS_v} neglects the change in long string velocity $\bar{v}$ due to loop formation as proposed in \cite{Avelino:2020ubr}.}

\section{Extension of the original VOS model}
\label{app:VOScalibration}

\subsection{VOS model from Nambu-Goto  simulations}

In our study, we describe the evolution of the long-string network through the VOS model, defined by the equations in Eq. \eqref{eq:VOS_eq_body}. The only free parameter of the model is the loop-chopping efficiency $\tilde{c}$, which is computed to be 
\begin{equation}
\textbf{NG:}\qquad \tilde{c}=0.23\pm 0.04
\end{equation}
from Nambu-Goto  network simulations in an expanding universe \cite{Martins:2000cs}.
\subsection{VOS model from Abelian-Higgs simulations}
Abelian-Higgs (AH) field theory simulations in both expanding and flat spacetime suggest a larger value \cite{Moore:2001px, Martins:2003vd}
\begin{equation}
\label{eq:AH_original}
\textbf{AH:}\qquad \tilde{c}=0.57\pm 0.04.
\end{equation}
Indeed, in Abelian-Higgs simulations, no loops are produced below the string core size so the energy loss into loop formation is lower. Consequently, the loop-chopping efficiency must be increased to maintain scaling.
\FloatBarrier	
\begin{figure}[h!]
			\centering
			\raisebox{0cm}{\makebox{\includegraphics[height=0.53\textwidth, scale=1]{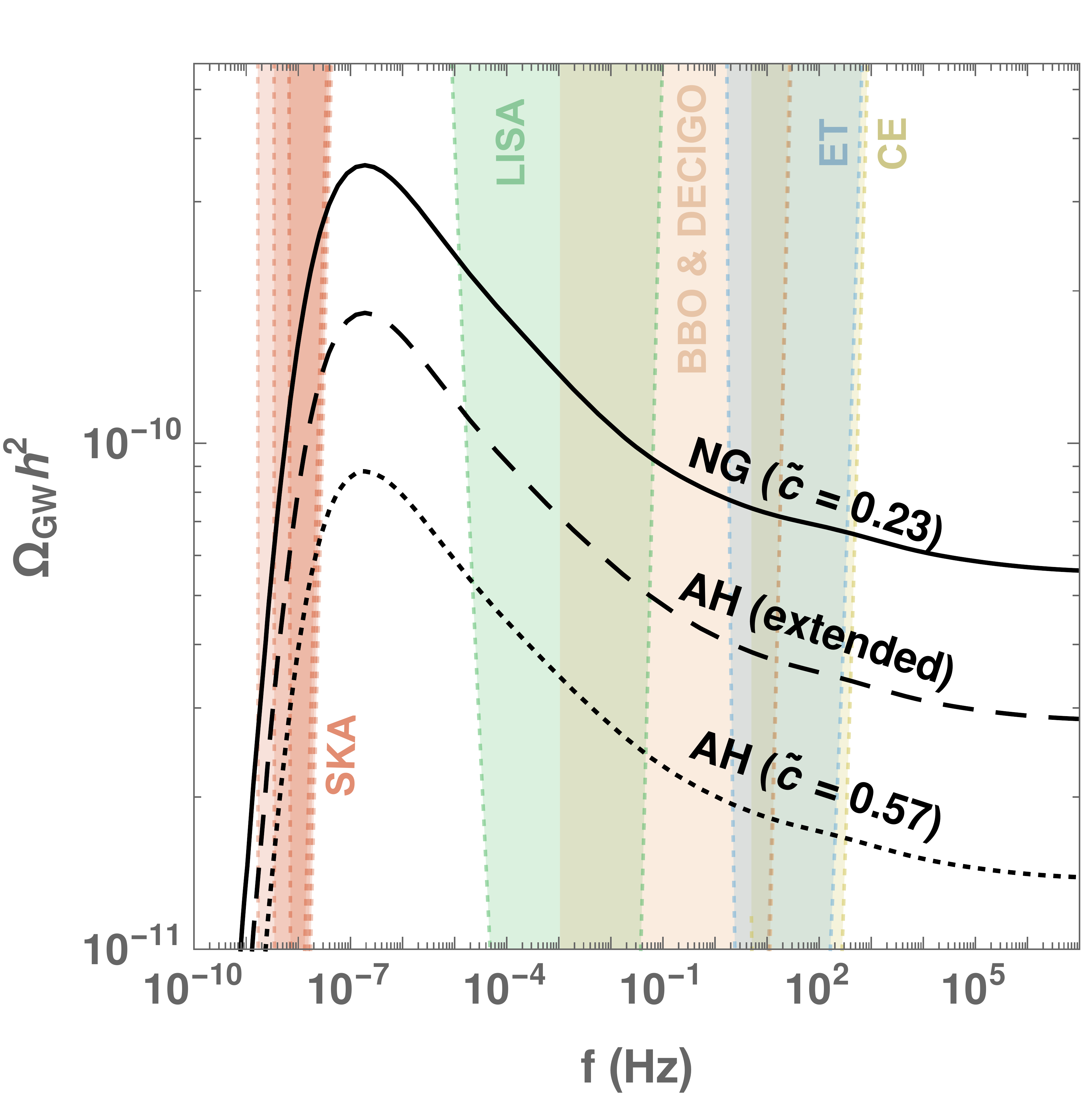}}}
			\raisebox{0cm}{\makebox{\includegraphics[height=0.4\textwidth, scale=1]{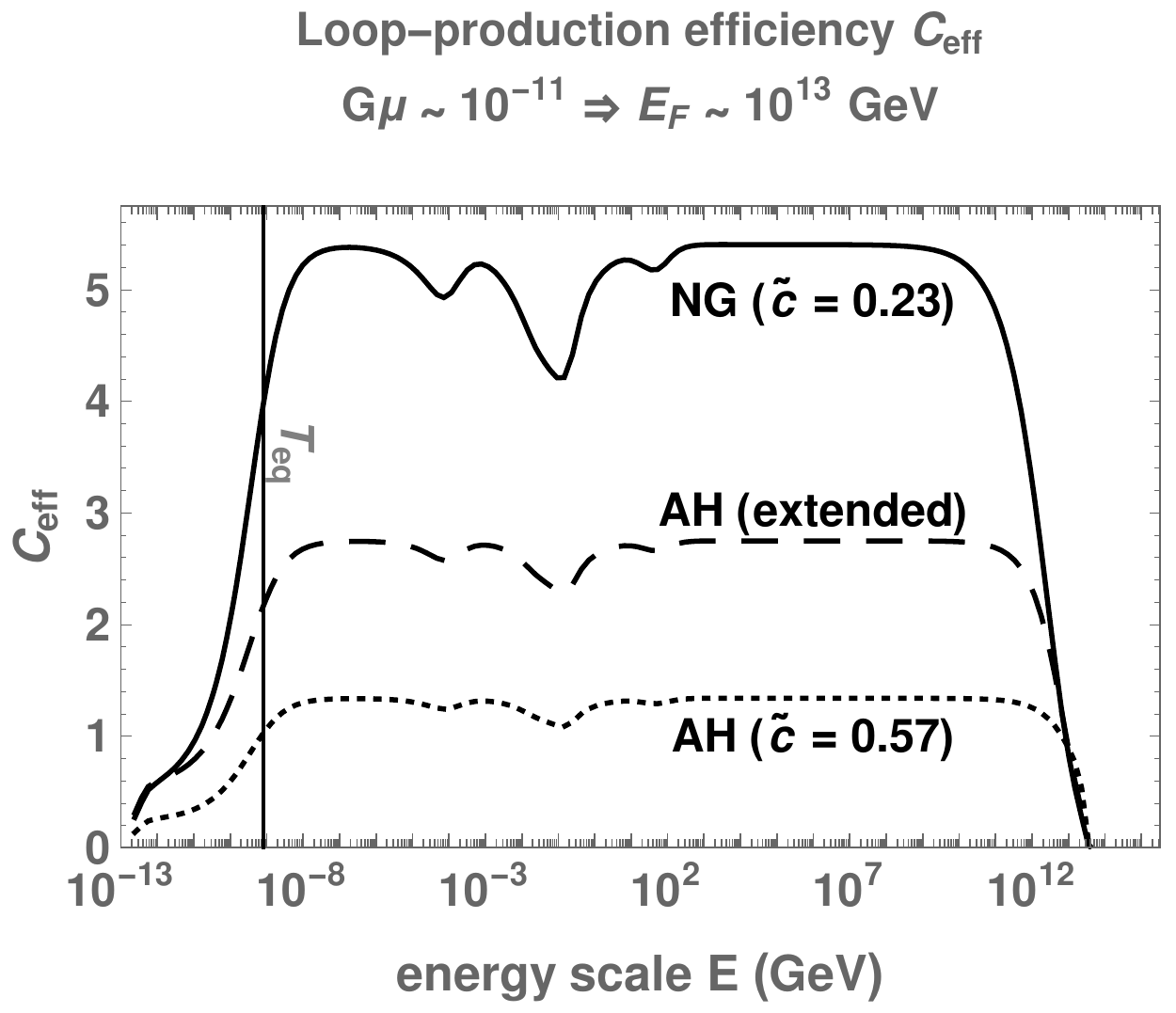}}}
			\hfill
			\caption{\it \small \textbf{Left:} GW spectra with different VOS modellings of the long-string network evolution. The VOS models are either based on Nambu-Goto simulations (solid line - $\tilde{c}=0.23$) \cite{Martins:2000cs} or abelian-Higgs (AH) field theory simulations (dashed line - $\tilde{c}=0.57$) \cite{Moore:2001px, Martins:2003vd}, possibly extended to include particle production \cite{Correia:2019bdl} (dotted line). \textbf{Right:} The corresponding loop-production efficiency for each VOS model.}
			\label{fig:VOSmodelS}
		\end{figure}
		\FloatBarrier

%
\subsection{VOS model from Abelian-Higgs  simulations with particle production}
\label{app:AH_extended_VOS}
In Abelian-Higgs simulations, the loops produced at the string core scale are non-linear lumps of field, called ``proto-loops", which decay fast into massive radiation. Therefore, a recent work \cite{Correia:2019bdl} extends the VOS model by including a term in Eq. \eqref{eq:VOS_L} to account for the emission of massive radiation at the string core scale. The energy-loss function $F(v)$ is modified as
\begin{align}
\label{eq:Fv_energy_loss_function}
\left.F(\bar{v})\right|_\textrm{original}=\frac{\tilde{c}\bar{v}}{2}\textrm{\hspace{1em}}\Rightarrow\textrm{\hspace{1em}}\left.F(\bar{v})\right|_\textrm{extended}=\frac{\tilde{c}\bar{v}+d[k_0-k(\bar{v})]^r}{2},
\end{align}
and the momentum operator $k(v)$, cf. Eq.~\eqref{eq:momentum_operator}, accounting for the amount of small-scale structures in the string, is modified to 
\begin{align}
\label{eq:kv_AH_extended}
 k(\bar{v})=\frac{2\sqrt{2}}{\pi}\frac{1-8\bar{v}^6}{1+8\bar{v}^6} \textrm{\hspace{1em}}\Rightarrow\textrm{\hspace{1em}} \left.k(\bar{v})\right|_\textrm{extended}=k_0\frac{1-(q \bar{v}^2)^\beta}{1+(q \bar{v}^2)^\beta},
\end{align}
where more free parameters have been introduced. With Abelian-Higgs simulations, one finds \cite{Correia:2019bdl}
\begin{equation}
\label{eq:AH_extended}
\textbf{AH extended:}\qquad \tilde{c}=0.31,
\end{equation}
as well as $d=0.26$, $k_0=1.27$, $r = 1.66$, $q = 2.27$, and $\beta = 1.54$. In Abelian-Higgs extended, the loop-chopping efficiency, cf. Eq.~\eqref{eq:AH_extended}, is smaller than the one in the original Abelian-Higgs model, cf. Eq.~\eqref{eq:AH_original}. Indeed, because of the additional energy loss through massive-radiation, less energy loss via loop-chopping is needed to maintain scaling.

In figure~\ref{fig:VOSmodelS}, we compare the GW spectra in the different VOS models.  The difference in amplitude comes from the difference in the number of loops, set by $C_{\rm eff}$. The larger the loop-chopping efficiency $\tilde{c}$, the smaller the loop-formation efficiency $C_{\rm eff}$. This counter-intuitive result can be better understood by looking at table.~\ref{tab:tableVOS}. A larger loop-chopping efficiency $\tilde{c}$ implies a larger loop formation rate only during the transient regime. In the scaling regime, a larger loop-chopping efficiency $\tilde{c}$ implies a more depleted long-string network and then a larger correlation length $\xi$. Hence, the long-string network is more sparse and so the rate of loop formation via string crossing is lower.
\begin{table}[]
\centering
\begin{tabular}{c|ccc}
 \hline \hline\\[-0.75em]
\begin{tabular}[c]{@{}c@{}}scaling in radiation\\ dominated universe\end{tabular}& \begin{tabular}[c]{@{}c@{}}NG\\ $\tilde{c}=0.23$\end{tabular} & \begin{tabular}[c]{@{}c@{}}AH \\ $\tilde{c}=0.57$\end{tabular} & \begin{tabular}[c]{@{}c@{}}AH extended\\ $\tilde{c}=0.31$\\ ($d,~k_0,~r,~q,~\beta$)\end{tabular} \\ \\[-0.75em]\hline\\[-0.75em]
$\bar{v}$                   & 0.66                                           & 0.62                                           & 0.59                                                    \\
$\xi$                  & 0.27                                           & 0.57                                           & 0.36                                                    \\
$C_\textrm{eff}$                & 5.4                                            & 1.3                                            & 2.8                                                     \\[0.25em] \hline \hline
\end{tabular}
\caption{\it \small Values of mean velocity $\bar{v}$, correlation length $\xi$, and loop-production efficiency $C_\textrm{eff}$ in radiation scaling regime with different VOS calibrations.}
\label{tab:tableVOS}
\end{table}

\section{GW spectrum from global strings}
\label{app:global_strings}

 The main distinction with loops from global strings is that they are short-lived whereas loops from local strings are long-lived. This results in different GW spectra in both frequency and amplitude as  we discuss in detail below.

\subsection{The presence of a massless mode}

For global string, the absence of gauge field implies the existence of a massless Goldstone mode, with logarithmically-divergent gradient energy, hence leading to the tension $\mu_{\rm g}$, cf. Eq.~\eqref{tension_string_exp}
\begin{equation}
\label{eq:string_tension_global_app}
\mu_{\rm g}  \equiv \mu_{\rm l} \, \ln\left( \frac{H^{-1}}{\delta} \right) \simeq \mu_{\rm l} \, \ln\left( \eta\,t \right)  , \quad \text{with} \quad \mu_{\rm l} \equiv 2 \pi \eta^2,
\end{equation}
where $\eta$ is the scalar field VEV, $\mu_{\rm l}$ is the tension of the would-be local string (when the gauge coupling is switched on) and $\delta \sim \eta^{-1}$ is the string thickness.
Goldstones are efficiently produced by loop dynamics with the power
\begin{equation}
P_{\rm Gold}=\Gamma_{\rm Gold}\, \eta^2,
\label{eq:power_goldstone_app}
\end{equation}
where $\Gamma_{\rm Gold} \approx 65$ \cite{Vilenkin:1986ku, Chang:2019mza}, causing loops to decay with a rate
\begin{equation}
\frac{d l_{\rm g}}{dt}= \frac{dE}{dt}\frac{dl}{dE} \equiv \kappa \equiv \frac{\Gamma_{\rm Gold}}{2 \pi \, \ln\left( \eta\,t \right) }.
\end{equation}
Therefore, the string length evolving upon both GW and Goldstone bosons emission reads
\begin{equation}
\label{eq:length_string_app}
l_{\rm g}(t) = \alpha t_i - \Gamma G \mu_{\rm g}  (t - t_i) - \kappa (t - t_i ).
\end{equation}

\subsection{Evolution of the global network}
\label{sec:VOS_global}
The Velocity-One-Scale equations, presented in App.~\ref{sec:VOS_proof},
\begin{align}
\label{eq:VOS_global_1}
&\frac{dL}{dt}=HL \,( 1+ \bar{v}^2)+\left.F(\bar{v})\right|_\textrm{global}, \\
\label{eq:VOS_global_2}
&\frac{d\bar{v}}{dt}=(1-\bar{v}^2)\left[\frac{k(\bar{v})}{L}-\frac{\bar{v}}{l_d}\right],
\end{align}
are modified to include the additional energy-loss due to Goldstone production. Namely, the energy-loss coefficient $F(\bar{v})$, cf. Eq.~\eqref{eq:Fv_energy_loss_function}, becomes
\begin{equation}
\label{eq:Fv_global}
\left.F(\bar{v})\right|_\textrm{local} =\frac{\tilde{c}\bar{v}+d[k_0-k(\bar{v})]^r}{2}  \textrm{\hspace{1em}} \Rightarrow\textrm{\hspace{1em}} \left.F(\bar{v})\right|_\textrm{global}=\left.F(\bar{v})\right|_\textrm{local} + \frac{s \, v^6}{2\ln\left( \eta\,t\right)},
\end{equation}
where the constant $s$ controlling the efficiency of the Goldstone production, is inferred from lattice simulations \cite{Klaer:2017qhr}, to be $s \simeq 70$ \cite{Martins:2018dqg}.
However, the momentum operator $k(v)$ in Eq.~\eqref{eq:VOS_global_2}, is unchanged with respect to the local case
\begin{align}
\label{eq:kv_AH_extended_2}
 k(\bar{v})=k_0\frac{1-(q \bar{v}^2)^\beta}{1+(q \bar{v}^2)^\beta}.
\end{align}
with $k_0=1.37$, $q=2.3$, $\beta = 1.5$, $\tilde{c} = 0.34$, $d=0.22$, $r=1.8$ \cite{Correia:2019bdl}.
Here through Eq.~\eqref{eq:Fv_global} and Eq.~\eqref{eq:kv_AH_extended_2}, we follow \cite{Klaer:2019fxc} and consider the extended VOS model based on Abelian-Higgs simulations, proposed in \cite{Correia:2019bdl} and already discussed in app.~\ref{app:AH_extended_VOS}, in which we have simply added the backreaction of Goldstone production on long strings in Eq.~\eqref{eq:Fv_global}.
We have checked that we can neglect the thermal friction due to the contact interaction of the particles in the plasma with the string, cf. Eq.~\eqref{VOS_frictionlength_intro} for which the interaction cross-section is given by the Everett formula in \cite{Vilenkin:2000jqa}. 

In order to later compute the GW spectrum, we defined the loop-formation efficiency, analog of the local case in Eq.~\eqref{eq:loopEfficiency}
\begin{equation}
C_{\rm eff}^\textrm{g} = \tilde{c}\, \bar{v}/\xi^3,
\end{equation}
with $\bar{v}$ and $\xi \equiv L/t$ obeying the VOS equations in Eq.~\eqref{eq:VOS_global_1} and Eq.~\eqref{eq:VOS_global_2}. Due to the logarithmic dependence of the string tension on the cosmic time, the scaling regime is slightly violated. Consequently, the loop-formation efficiency plotted in right panel of Fig.~\ref{fig:global_strings_nonst}, never reaches a constant value. Hence, in this study we model the network based on VOS evolution, rather than using the scaling solutions. Only for enormous value of $ \ln\left( \eta\,t \right) $ corresponding to cosmic times much larger than the age of the Universe today, we find that the solutions to the modified VOS equations in Eq.~\eqref{eq:VOS_global_1} and Eq.~\eqref{eq:VOS_global_2} reach a scaling regime $C_\textrm{eff}=$ 0.46, 2.24, 6.70 for matter-, radiation-, and kination-dominated universe, respectively. By comparing to the values found in \cite{Chang:2019mza}, our results agree only for the radiation case. Note that the dependence of the string network parameters $\bar{v}$ and $\xi \equiv L/t$ on the logarithmically-time-dependent string tension arises only through the term of Goldstone production energy loss in Eq.~\eqref{eq:Fv_global}.

\subsection{The GW spectrum}
Finally, the GW spectrum generated by loops of global strings is given by a similar expression as for the local case, cf. the mattress equation in Eq.~\eqref{eq:SGWB_CS_Formula}, 
	\begin{equation}
	\Omega^{\rm g}_{\rm{GW}}(f)\equiv\frac{f}{\rho_c}\left|\frac{d\rho_{\rm{GW}}^{\rm g}}{df}\right|=\sum_k{\Omega^{(k),\, \rm g}_{\rm{GW}}(f)},
	\label{eq:SGWB_CS_Formula_global}
	\end{equation}
	where
	\begin{equation}
\Omega^{(k), \, \rm g}_{\rm{GW}}(f)=\frac{1}{\rho_c}\cdot\frac{2k}{f}\cdot\frac{\mathcal{F}_{\alpha}\,\Gamma^{(k)}G\mu_{\rm g} ^2}{\alpha(\alpha+\Gamma G \mu_{\rm g} + \kappa)}\int^{t_0}_{t_F}d\tilde{t}~ \frac{C_{\rm{eff}}^{\rm g}(t_i^{\rm g})}{\,(t_i^{\rm g})^4}\left[\frac{a(\tilde{t})}{a(t_0)}\right]^5\left[\frac{a(t_i^{\rm g})}{a(\tilde{t})}\right]^3\Theta(t_i^{\rm g}-t_F).
	\label{kmode_omega}
	\end{equation}
 We have checked that we can safely neglect massive radiation. The loop formation time $t_i^{\rm g}$ is related to the emission time $\tilde{t}$ after using $l_{\rm g}(\tilde{t}) =0$ in Eq.~\eqref{eq:length_string_app}
\begin{equation}
t_i^{\rm g} = \frac{\Gamma G\mu_{\rm g} + \kappa} {\alpha + \Gamma G\mu_{\rm g} + \kappa}\tilde{t}.
\end{equation}

\begin{figure}[h!]
			\centering
			\raisebox{0cm}{\makebox{\includegraphics[width=0.65\textwidth, scale=1]{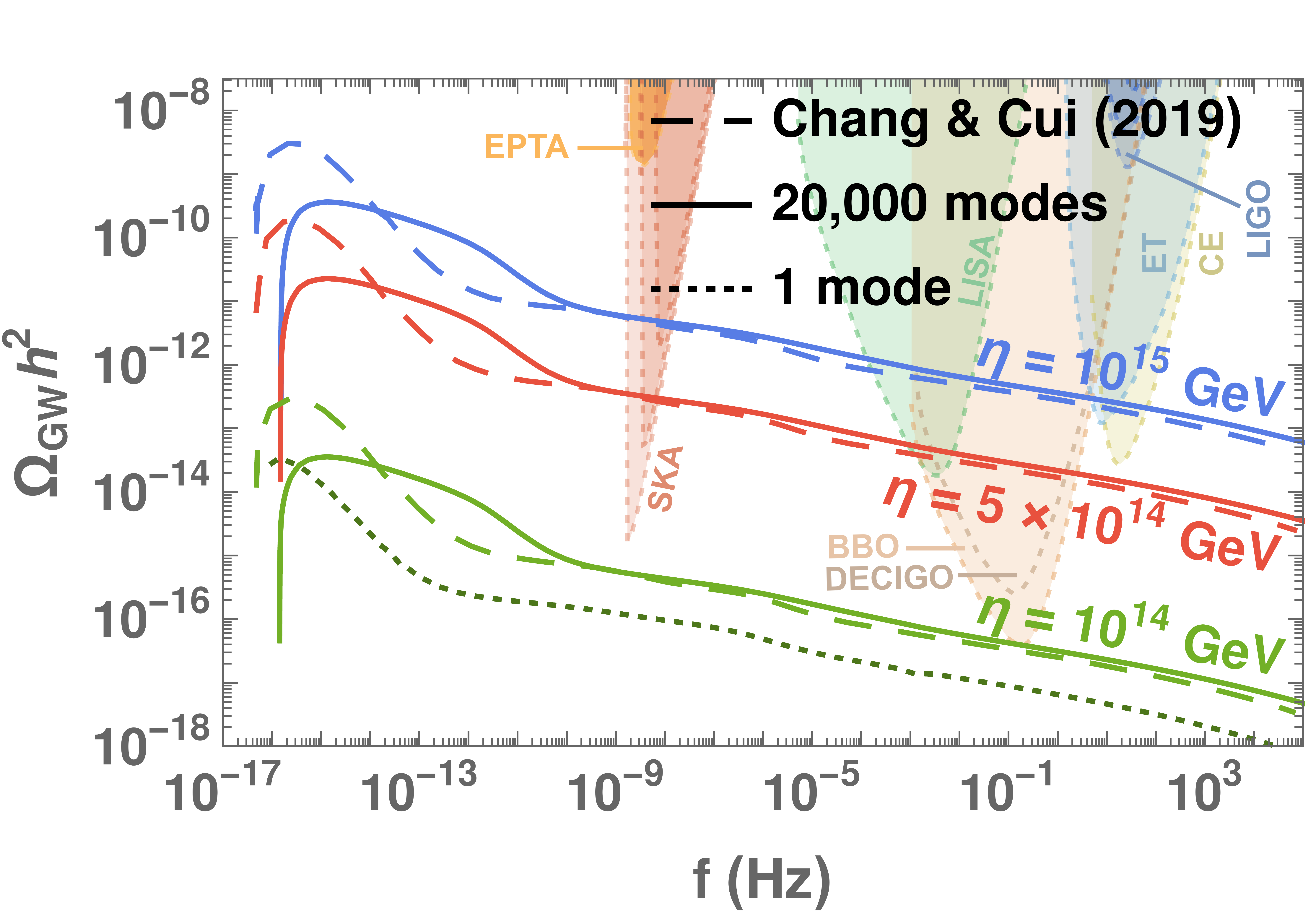}}}\\
			\hfill
		\caption{\it \small   GW spectrum from global cosmic strings assuming VOS network formed at energy scale $\eta$, evolved in the standard cosmology. The shape of the GW spectrum computed in \cite{Chang:2019mza} resembles the first mode $k=1$ of our spectrum.}
			\label{fig:global_strings_st}
		\end{figure}

We plot the GW spectrum from local strings in Fig.~\ref{fig:global_strings_st} and compare to the spectrum computed in \cite{Chang:2019mza}.
With respect to \cite{Chang:2019mza}, we find a lower value for $C_\textrm{eff}$ during the late matter-dominated universe ($0.46$ instead of $1.32$) which implies a smaller spectral bump, while the radiation contributions are considerably the same. Moreover, the shapes of the spectra are different. An explanation could be the summation over the high-frequency modes (up to $k=2\times 10^4$ in our case) which smoothens the spectrum. The spectrum from \cite{Chang:2019mza} resembles to the first-mode of our spectrum. 

The constraints on the inflation scale, $6 \times 10^{13}$~GeV, from the non-detection of the fundamental B-mode polarization patterns in the CMB \cite{Ade:2018gkx, Akrami:2018odb}, implies the upper bound $T_{\rm reh} \lesssim 5 \times 10^{16}$~GeV on the reheating temperature, assuming instantaneous reheating. Hence, assuming that the network is generated from a thermal phase transition, we are invited to impose $\eta \lesssim 5 \times 10^{16}$. 
However, a stronger restriction arises because of the CMB constraint on strings tensions
\begin{equation}
\left.G\mu_\textrm{g}\right|_{\textrm{CMB}}= \left.2\pi \left(\frac{\eta}{m_\textrm{pl}}\right)^2\log(\eta\, t_\textrm{CMB})\right.\lesssim 10^{-7} \quad \rightarrow \quad \eta \lesssim 1.4 \times 10^{15},
\end{equation}
where we use $t_\textrm{CMB}\simeq 374$ kyr. Hence, we restrict to $\eta \lesssim 10^{15}$~GeV as in \cite{Chang:2019mza}.

\subsection{Global versus local strings}
The local spectrum is recovered upon setting $ \ln\left( \eta\,t \right) =1$, $\kappa =0$ and $s=0$ in Eq.~\eqref{eq:string_tension_global_app}, Eq.~\eqref{eq:length_string_app} and Eq.~\eqref{eq:Fv_global}.
Because of the faster decay, in the global case, the main emission time $\tilde{t}_{\rm M}^{\rm g}$ is shorter than its local counterpart $\tilde{t}_{\rm M}^{\rm \,l}$, cf. Eq.~\eqref{eq:t_M},
\begin{align}
\label{eq:t_M_global}
\tilde{t}_{\rm M}^{\rm g} &\simeq \frac{\alpha + \Gamma G\mu_{\rm g} + \kappa }{ \Gamma G \mu_{\rm g} + \kappa}\frac{ t_i}{2}, \\ 
 & \sim \frac{\Gamma G \mu_{\rm l}}{ \alpha} ~\tilde{t}_{\rm M}^{\rm \,l},
\end{align}
where in the last line, we have assumed $\kappa \gg \alpha, \tilde{t}_{\rm M}^{\rm \,l}$.
Therefore, the frequency $f$ of a GW emitted by a loop produced at the temperature $T$, is lowered  compared to the local case computed in Eq.~\eqref{fdeltaApp}, by a factor
\begin{align}
\label{eq:global_vs_local_freq}
f\Big|_{\rm global} &\simeq (4.7\times10^{-6}\textrm{ Hz})\left(\frac{T}{\textrm{GeV}}\right)\left(\frac{0.1}{\alpha }\right)\left(\frac{g_*(T_i)}{g_*(T_0)}\right)^{1/4} ,\\
&\sim \left(\frac{\Gamma G \mu_{\rm l}}{\alpha}\right)^{1/2} f\Big|_{\rm local}.
\label{eq:global_vs_local_freq_2}
\end{align}
In contrast to the local case, the frequency is independent of the string scale and we explain its origin at the end of the section.
We can rewrite the GW spectrum in Eq.~\eqref{eq:SGWB_CS_Formula_global} as
\begin{equation}
\label{eq:Omega_global}
\Omega^{\rm g}_{\rm{GW}}(f) \simeq \frac{1}{\rho_c}\cdot\frac{2}{f} \cdot\frac{\mathcal{F}_\alpha\,\Gamma G\mu_{\rm g}^2}{\alpha(\alpha+\Gamma G \mu_{\rm g} + \kappa)}\cdot C_{\rm{eff}}^{\rm g}\cdot t_0^{-5/2}\cdot(\tilde{t}_{\rm M}^{\rm g})^{1+\frac{5}{2}-\frac{3}{2}}\cdot t_i^{-4 + \frac{3}{2}} .
\end{equation}
Then, from Eq.~\eqref{eq:t_M_global} and $\alpha \,t_{i}^{\rm g} \simeq 4\,a(\tilde{t}_{\rm M}^{\rm g})/a(t_0) f^{-1}$ in Eq.~\eqref{eq:tdelta_fdelta_eq_line1}, one obtains
\begin{align*}
&\tilde{t}_{M}^{\rm g} \simeq \frac{1}{t_0}\frac{4}{\alpha^2} \left( \frac{1}{f} \right)^2 \left( \frac{\alpha+\Gamma G \mu_{\rm g} + \kappa}{\Gamma G \mu_{\rm g} + \kappa}\right)^2, \\
&t_{i}^{\rm g} \simeq \frac{1}{t_0}\frac{8}{\alpha^2} \left( \frac{1}{f} \right)^2 \left( \frac{\alpha+\Gamma G \mu_{\rm g} + \kappa}{\Gamma G \mu_{\rm g} + \kappa}\right).
\end{align*}
From comparing the global GW spectrum $\Omega_{\rm GW}^{\rm g}$ to the local GW spectrum $\Omega_{\rm GW}^{\rm l} = \Omega_{\rm GW}^{\rm g}( \ln\left( \eta\,t \right) =1,\, \kappa=0, \, s=0)$, one obtains
\begin{equation}
\label{eq:ratio_global_to_local_GW}
\frac{\Omega_{\rm GW}^{\rm g}}{\Omega_{\rm GW}^{\rm l}} \sim \frac{\alpha}{\kappa}\left( \frac{\Gamma G \mu_{\rm l}}{\alpha} \right)^{3/2} \left( \frac{\mu_{\rm g}}{\mu_{\rm l}}\right)^2 \frac{C_{\rm eff}^{\rm g}}{C_{\rm eff}^{\rm l}},
\end{equation}
where we have assumed $\kappa \gg \alpha,\, \Gamma G \mu_{\rm g} $. 
Upon using Eq.~\eqref{eq:lewicki_formula_GW_spectrum_radiation} and Eq.~\eqref{eq:ratio_global_to_local_GW}, we get\footnote{Upon restoring the dependence on the different parameters appearing in Eq.~\eqref{eq:Omega_global}, we obtain
\begin{equation}
\label{eq:Omega_global_2}
\Omega_{\rm GW}^{\rm g} \simeq  25 \Delta_{R} \,\Omega_{r}h^2 \,C^{\rm g}_{\rm eff}(n=4) \, \mathcal{F}_\alpha \, \frac{\Gamma}{\Gamma_{\rm Gold}}  \log^3\left(\eta \,\tilde{t}_{M}^{\rm g}\right) \frac{\eta^4}{M_{\rm pl}^4},
\end{equation}
which can be compared with its local equivalent in Eq.~\eqref{eq:lewicki_formula_GW_spectrum_radiation}.}

\begin{equation}
\label{eq:ratio_global_to_local_GW_2}
\Omega_{\rm GW}^{\rm l} \simeq \Omega_{r}h^2\frac{\eta}{M_{\rm pl}}, \qquad \text{and}\qquad \Omega_{\rm GW}^{\rm g} \simeq \Omega_{r}h^2 \log^3\left(\eta \,\tilde{t}_{M}^{\rm g}\right) \frac{\eta^4}{M_{\rm pl}^4},
\end{equation}
where $\Omega_{r}h^2 \simeq 4.2\times 10^{-5}$ is the present radiation energy density of the universe \cite{Tanabashi:2018oca}.

As shown in Fig.~\ref{fig:global_strings_st}, in consequence of the strong dependence of the GW amplitude on the string scale $\eta$, only global networks above $\eta \gtrsim 5 \times 10^{14}~$GeV can be detected by LISA or CE whereas EPTA or BBO/DECIGO can probe $\eta \gtrsim  10^{14}~$GeV. Also note the logarithmic spectral tilt of the GW spectrum due to the logarithmic dependence of the global string tension on the cosmic time.

In summary, GW spectra from local and global strings manifest differences in frequency, Eq.~\eqref{eq:global_vs_local_freq_2}, and amplitude, Eq.~\eqref{eq:ratio_global_to_local_GW_2}, because local loops are long-lived whereas global loops are short-lived. In the local case, the dominant GW emission at $\tilde{t}_{\rm M}^{\rm l} \sim t_i^{\rm l}/G\mu_{\rm l}$ occurs much after the loop formation time $t_i^{\rm l}$, after one loop lifetime. However, the emitted frequency is fixed by $~(t_i^{\rm l})^{-1}$. Hence, as discussed in Sec.~\ref{sec:turning_point_scaling}, the observed frequency is exempted from a redshift factor given by $\sqrt{G\mu_{\rm l}}$. In the global case, the loops are short-lived and the time of dominant GW emission coincides with the time of loop formation. Hence, the emitted frequency redshifts more and the spectrum is shifted to the left. The GW spectrum is also reduced by the redshift factor $(G\mu)^{3/2}$.

\begin{figure}[h!]
			\centering
			\raisebox{0cm}{\makebox{\includegraphics[width=0.495\textwidth, scale=1]{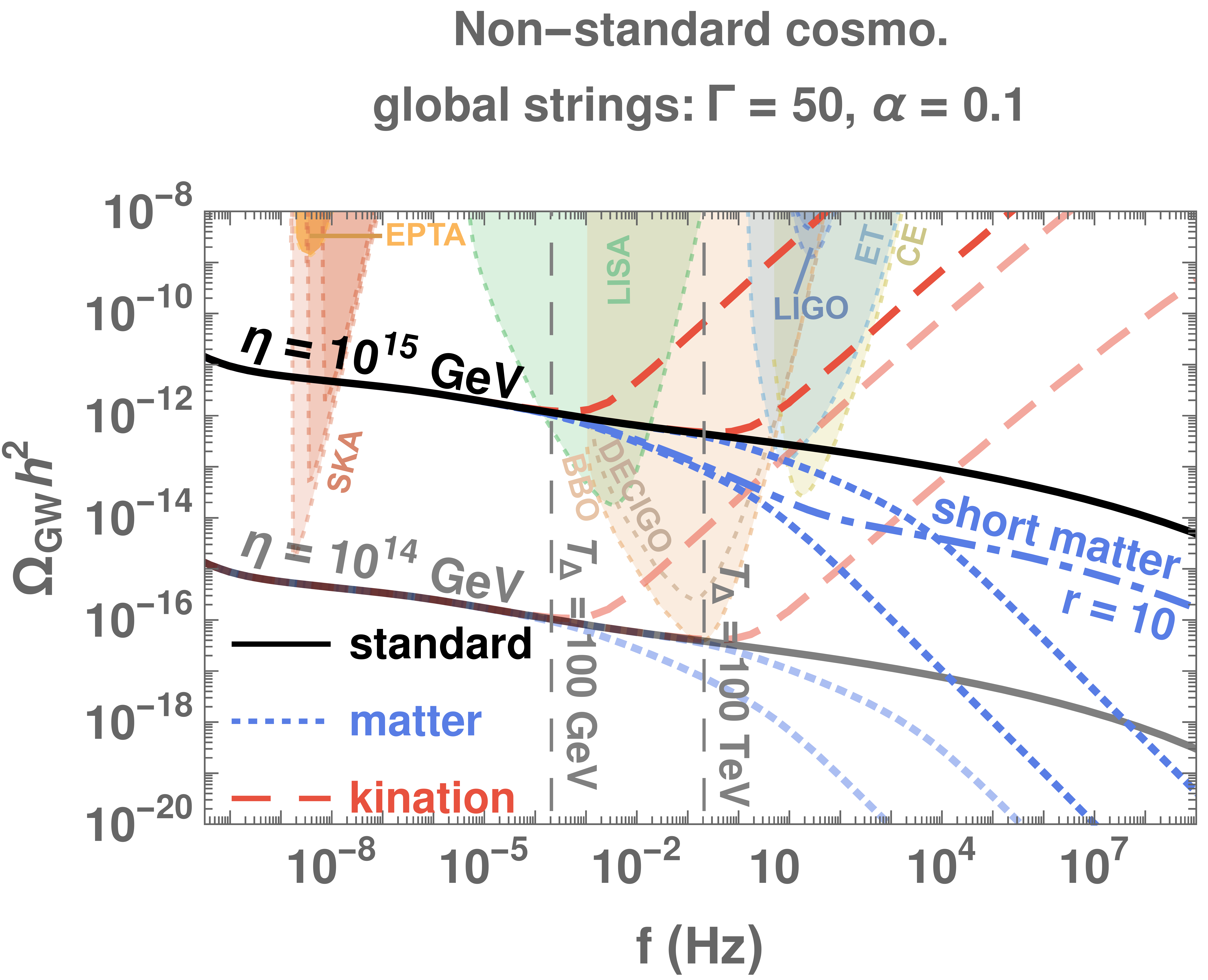}}}
			\raisebox{0cm}{\makebox{\includegraphics[width=0.475\textwidth, scale=1]{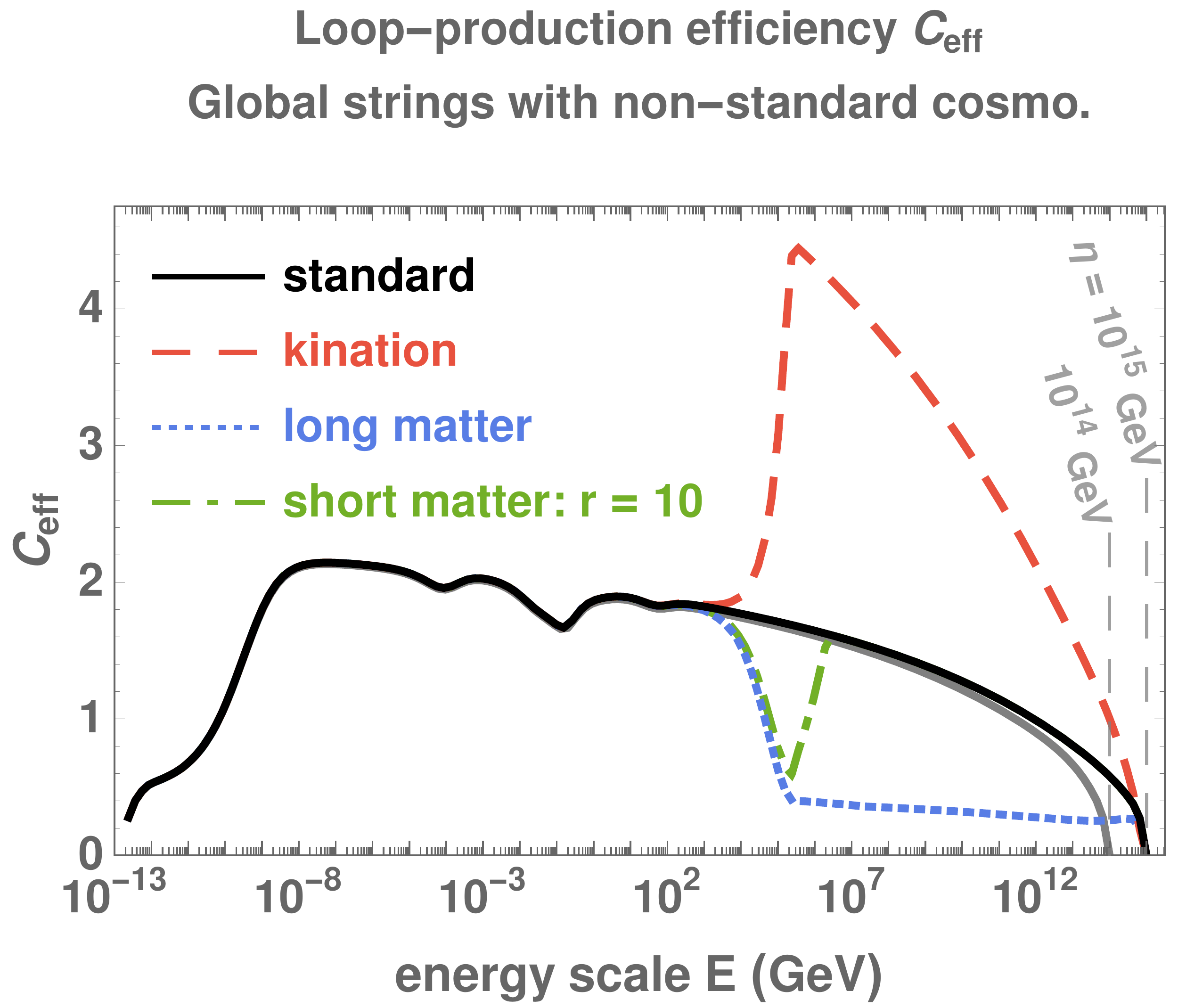}}}\\
			\hfill
		\caption{\it \small   \textbf{Left:} GW spectrum from the global cosmic strings assuming VOS network, evolving in the presence of a non-standard era, either long-lasting matter (dotted), intermediate matter (dot-dashed), or kination (dashed), ending at the temperature $T_\Delta=100$ GeV or $100$ TeV. The turning-point frequency is independent of the string tension. \textbf{Right:} The evolution of the loop-production efficiency for each cosmological background never reaches a plateau, in contrast to local strings in which case the scaling regime is an attactor solution, cf. right panel of Fig.~\ref{fig:VOSvsScaling_Ceff}. Indeed, for global strings the scaling behavior is logarithmically violated due to the inclusion of energy loss through Goldstone production in the VOS equations, cf. Sec.~\ref{sec:VOS_global}.}
			\label{fig:global_strings_nonst}
		\end{figure}
		
		\begin{figure}[h!]
			\centering
			\raisebox{0cm}{\makebox{\includegraphics[width=0.5\textwidth, scale=1]{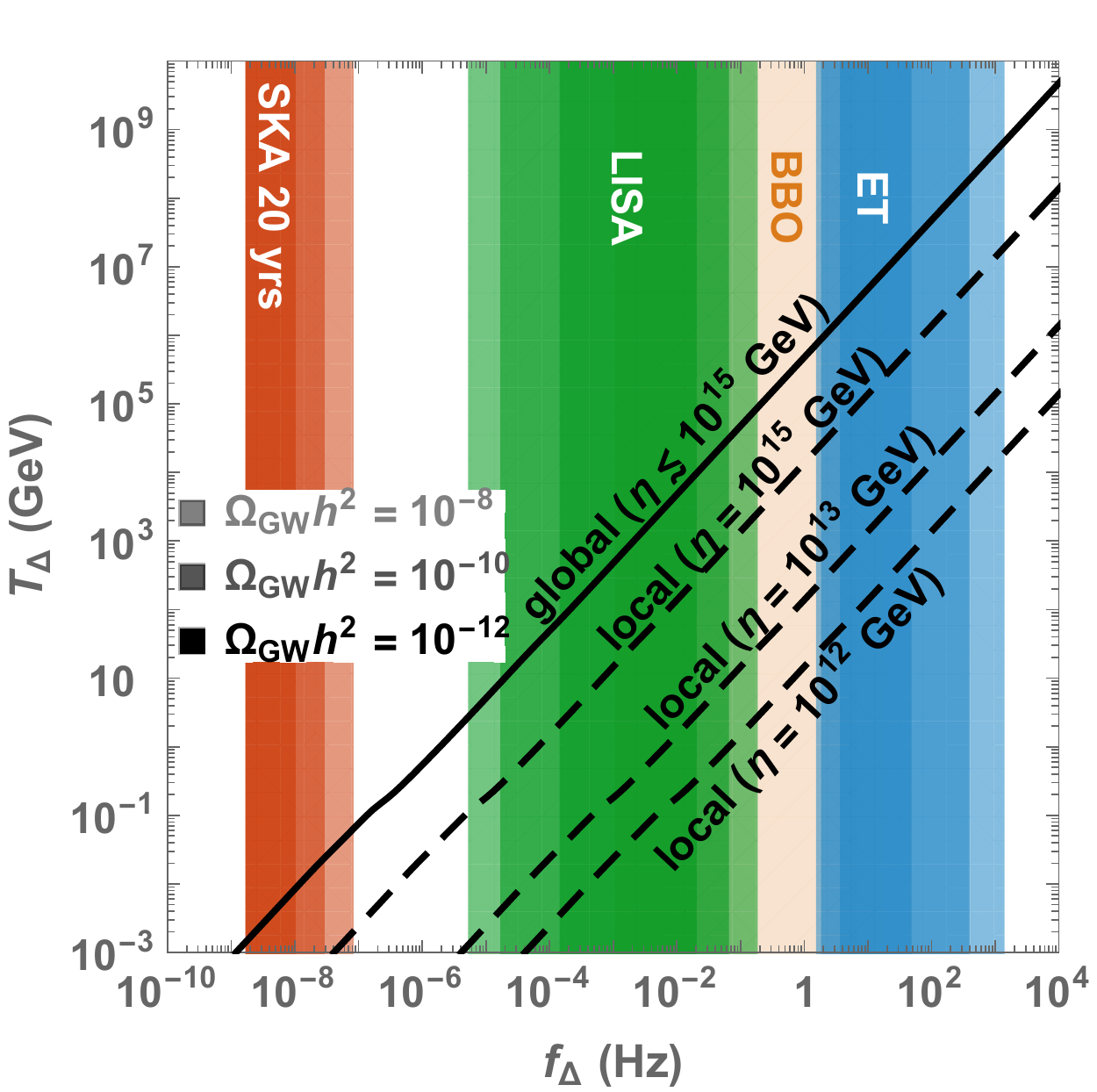}}}\\
			\hfill
		\caption{\it \small Detectability of the turning-points in the $T_\Delta - f_\Delta$ plane. The solid line represents turning-points for global strings formed at any energy scale $\eta$, while dashed lines are turning points from local strings. Shaded areas correspond to the frequencies probed by each observatory assuming SGWB of amplitudes $\Omega_{\textrm{GW}} h^2 = 10^{-8}$, $10^{-10}$ and $10^{-12}$. The plot is totally inspired from \cite{Chang:2019mza}.} 
			\label{fig:global_strings_contour}
		\end{figure}
		
		\subsection{As a probe of non-standard cosmology}

The impact of non-standard cosmology on the GW spectra of global strings is shown in Fig.~\ref{fig:global_strings_nonst}.
The frequency of the turning point corresponding to a change of cosmology at a temperature $T_\Delta$ is given by Eq.~\eqref{eq:global_vs_local_freq}. We report here a numerically-fitted version 
\begin{align}
f^{\textrm{glob}}_{\Delta}\simeq \textrm{ Hz} \left(\frac{T}{\textrm{GeV}}\right)\left(\frac{0.1}{\alpha}\right)\left(\frac{g_*(T)}{g_*(T_0)}\right)^{1/4} \times 
\begin{cases}
8.9 \times 10^{-7} &\textrm{for } 10 \% \\
7.0 \times 10^{-8} &\textrm{for } 1 \% \\
\end{cases},
\label{turning_point_globalstring}
\end{align}
where the detection criterion is defined as in Eq.~\eqref{10per_criterion}. In contrast to local strings, the turning-point is independent of the string tension. 

We now consider the reach of global strings for probing a non-standard cosmology. Fig.~\ref{fig:global_strings_contour} shows the detectability of the turning-points by future GW experiments. Due to the string-scale independence, the global-string detectability collapses onto a line. Because of the shift of the spectrum to lower frequencies by a factor $\sim \sqrt{G\mu}\sim \eta/M_{\rm pl}$, cf. Eq.~\eqref{eq:global_vs_local_freq_2}, GW from global string networks can probe earlier non-standard-cosmology and larger energy scales with respect to GW from local strings.

\end{subappendices}

%

\xintifboolexpr { \x = 2}
  {
  }
{
\medskip
\small
\bibliographystyle{JHEP}
\bibliography{thesis.bib}
}

%% file: chap9.tex
\chapterimage{spiderweb} 

\chapter{Probe heavy DM with GW from CS}
\label{chap:DM_GW_CS}

\begin{tikzpicture}[remember picture,overlay]
\node[text width=18cm,text=black,minimum width=\paperwidth,
minimum height=7em,anchor=north]%
 at (7,4) {This chapter is based on \cite{Gouttenoire:2019rtn}.};
 \draw (-2.0,2) -- (5,2);
\end{tikzpicture}

In the previous chapter, we have computed the GW spectrum from CS in non-standard cosmology and we have shown the possibility to use next generation of GW experiments to probe the existence of a early matter era or an early second period of inflation,
In this chapter, we use those findings to constrain particle physics models leading to such a change of cosmology with a particular focus on models of heavy Dark Matter.

In Sec.~\ref{sec:NSmatter}, we assume the existence of heavy unstable particles that can temporarily dominate the energy density of the universe, and therefore induce a period of matter domination within the radiation era after post-inflationary reheating. This leads to a modified expansion of the universe compared to the usually assumed single radiation era. Such modified cosmological history can be probed if during this period, there is an active source of gravitational waves, in which case the resulting GW spectrum would imprint any modification of the equation of state of the universe. 
Particularly well-motivated are the long-lasting GW production from cosmic string networks. We derive the improvement by many orders of magnitude of the current model-independent BBN constraints  on the abundance and lifetime of a particle, cf. Fig.~\ref{fig:mY_tauX_GWI_VS_BBN}, that can be inferred from the detection of GW produced by CS. In Sec.~\ref{sec:U(1)_model}, we provide unprecedented exclusion bounds on the model of $U(1)_D$ DM, introduced in Chap.~\ref{chap:secluded_DM}. We also study the scenario where the dark photon mass and the cosmic string network are generated by the spontaneous breaking of the same $U(1)$ symmetry. 

In Sec.~\ref{sec:model_indpt_inflation_GWCS} of the previous chapter, we have derived model-independent constraints on the energy scale and number of e-folds of a second period of inflation due to its imprint on the GW spectrum from CS. In Sec.~\ref{sec:SC_DM_CS} of the present chapter, we are able to exclude regions of parameter space in the model of Supercooled Composite DM, introduced in Chap.~\ref{chap:SC_conf_PT}, which predicts a short period of inflation.

\section{The imprints of an early era of matter domination}
\label{sec:NSmatter}

\begin{figure}[h!]
\centering
\raisebox{0cm}{\makebox{\includegraphics[height=0.55\textwidth, scale=1.]{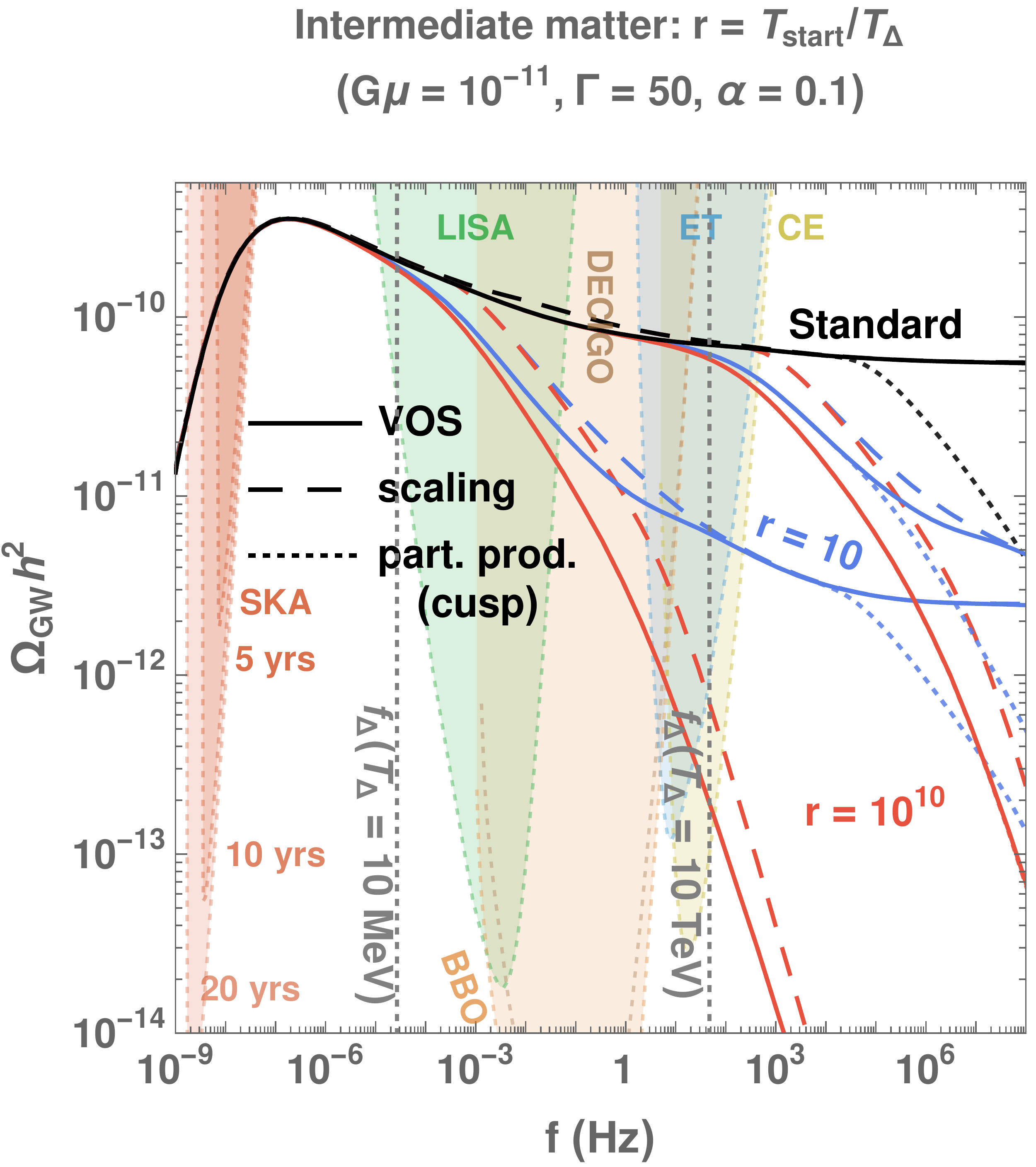}}}
\raisebox{0cm}{\makebox{\includegraphics[height=0.55\textwidth, scale=1.]{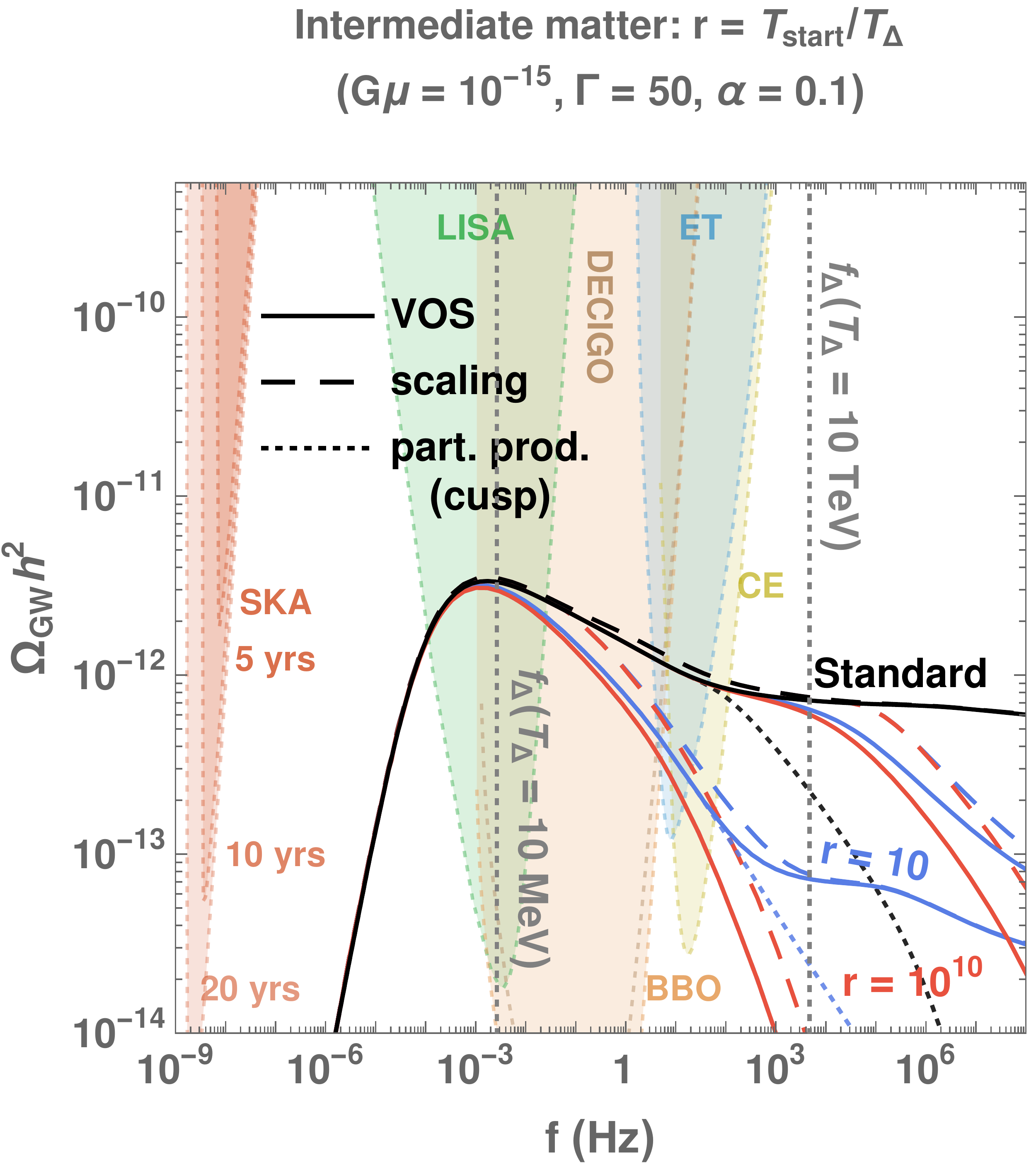}}}
\caption{\it \small \label{fig:OmegaGWCS}  SGWB generated by the gravitational decay of cosmic strings compared to the reach of different GW interferometers. We show the impact of a long (red) or a short (blue) intermediate matter era, starting at the temperature $r \, T_{\Delta}$ and ending at $T_{\Delta} = 10$~MeV or $T_{\Delta} = 10$~TeV. Black lines show the results obtained assuming  standard cosmological evolution. The dashed-lines assume that the scaling regime switches on instantaneously during the change of cosmology whereas the solid lines incorporate the transient behavior, solution of the VOS equations, as discussed in \cite{CS_VOS}. 
Limitations due to particle production assuming that the small-scale structures are dominated by cusps are shown with dotted lines \cite{CS_VOS}.
The dotted vertical lines indicate the relation in Eq.~\eqref{eq:f_delta_RD} between the temperature $T_{\Delta}$ and the frequency $f_{\Delta}$ of the turning point, where the matter-era-tilted spectrum meets the radiation-era-flat spectrum.  }
\end{figure}

\subsection{Modified spectral index}

The part of the spectrum coming from loops produced and emitting during radiation is flat since there is an exact cancellation between the red-tilted red-shift factor and the blue-tilted loop number density. However, in the case of a matter era, a mismatch induces a slope $f^{-1}$. The impact of a non-standard matter era is shown in Fig.~\ref{fig:OmegaGWCS}. The frequency detected today $f_{\Delta}$ of the turning point between the end of the matter domination and the beginning of the radiation-domination can be related to the temperature of the universe $T_\Delta$ when the change of cosmology occurs 
\begin{equation}
\label{eq:f_delta_RD}
f_{\Delta}=                  (2 \times 10^{-3} ~\textrm{Hz})\left( \frac{T_{\Delta}}{\textrm{GeV}} \right) \left( \frac{0.1 \times 50 \times 10^{-11}}{\alpha \Gamma G_{\mu}} \right)^{1/2} \left( \frac{g_{*}(T_{\Delta})}{g_{*}(T_{0})}\right)^{1/4}. 
\end{equation}
The GW  measured with frequency $f_{\Delta}$ have been emitted by loops produced during the change of cosmology at $T_{\Delta}$.
An extensive discussion of this frequency-temperature relation as provided  in \cite{CS_VOS}. 
The above formula  entirely relies on the assumptions that the back-reaction scale is $\Gamma G \mu$ as claimed by Blanco-Pillado et al. \cite{Lorenz:2010sm, Ringeval:2017eww, Auclair:2019zoz} and not much lower as claimed by Ringeval et al. \cite{Blanco-Pillado:2013qja, Blanco-Pillado:2017oxo} . 

\begin{figure}[h!]
\centering
\raisebox{0cm}{\makebox{\includegraphics[height=0.52\textwidth, scale=1]{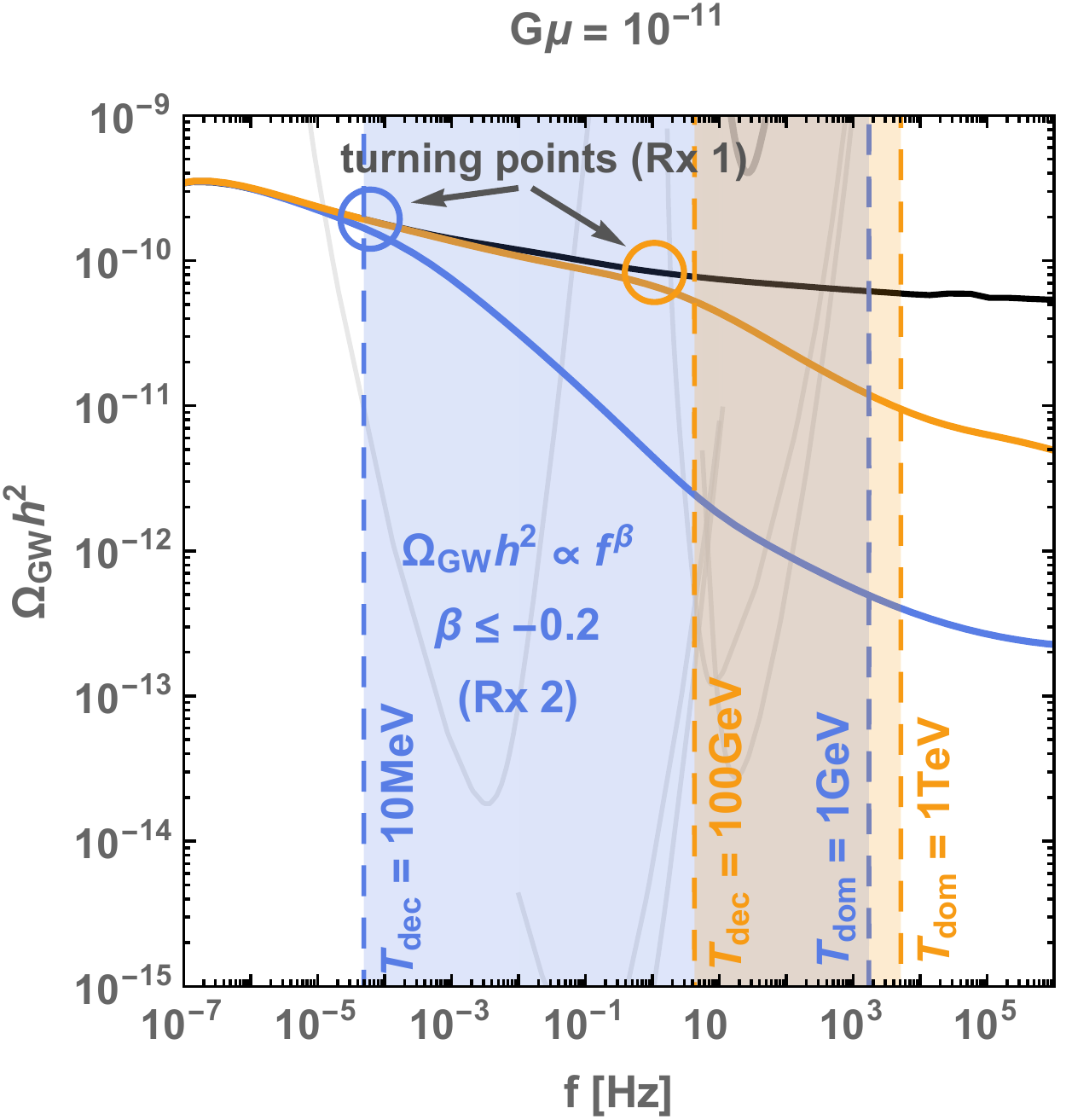}}}
\raisebox{0cm}{\makebox{\includegraphics[height=0.52\textwidth, scale=1]{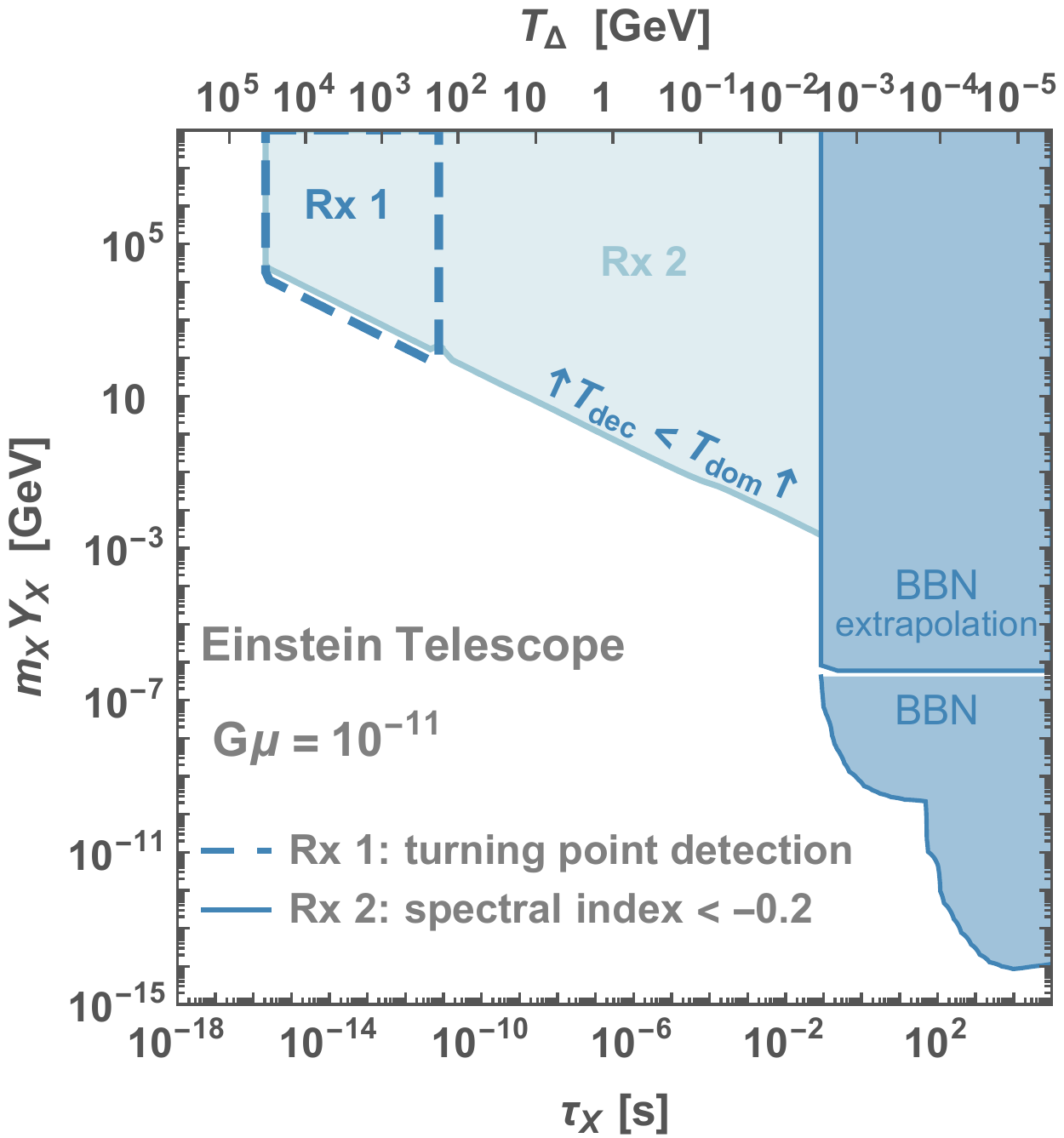}}}
\caption{\it \small \label{fig:FullMD}  \textbf{Left}: SGWB for $G\mu =10^{-11}$ assuming that a heavy cold particle dominates the energy density of the universe at the temperature $T_{\rm dom}$ and decays at the temperature $T_{\Delta} = T_{\rm dec}$. \textbf{Right}:  Considering the particular case of the Einstein Telescope, we illustrate how the constraints on the abundance and lifetime of a heavy relic depend on the choice of the prescription, Rx 1 or Rx 2 defined in Sec.~\ref{sec:triggerMatterGW}.}
\end{figure}

\subsection{How to detect a matter era with a GW interferometer}
\label{sec:triggerMatterGW}

For a first qualitative analysis, we start with two simple prescriptions for detecting a matter era from the measurement of a SGWB from CS by a GW interferometer.
\begin{itemize}
\item
\textbf{Rx 1} \textit{(turning-point prescription)}: The turning point, namely the frequency at which the spectral index of the GW spectrum changes, corresponding to the transition from the matter to the radiation era, defined in Eq.~\eqref{eq:f_delta_RD}, must be inside the interferometer window, as shown for instance  in Fig.~\ref{fig:OmegaGWCS}.
\item
\textbf{Rx 2} \textit{(spectra-index prescription)}: The measured spectral index must be smaller than $-0.2$, namely $\beta < -0.2$ where $\Omega_{\rm GW}h^2 \propto f^{\beta}$.
\end{itemize}
The same prescriptions can be applied to detect the presence of an inflationary era and more generally to detect any effect which induces a slope $\beta < -0.2$ in the GW spectrum from CS.
In Fig.~\ref{fig:FullMD}, we compare the above two prescriptions. The prescription Rx 1 is more conservative but enough to measure the lifetime of the particle. In our study, we use the prescription Rx 1 and, in right panel of Fig.~\ref{sec:triggerMatterGW}, we show how to extend the constraints with Rx 2.

We note here that the presence of the turning point and the changed spectral index at high frequencies would be similar in the case of a long intermediate inflation era instead of an intermediate matter era. Disentangling the two effects  deserves further studies. Interestingly, high-frequency burst signals due to cusp formation could be a way-out  \cite{CS_VOS}. In the analysis of this work, we interpret the suppression of the GW spectrum as due to an intermediate matter era.

\begin{figure}[p!tb]
\thispagestyle{empty}
\centering\vspace{-0cm}
\raisebox{0cm}{\makebox{\includegraphics[height=0.45\textwidth, scale=1]{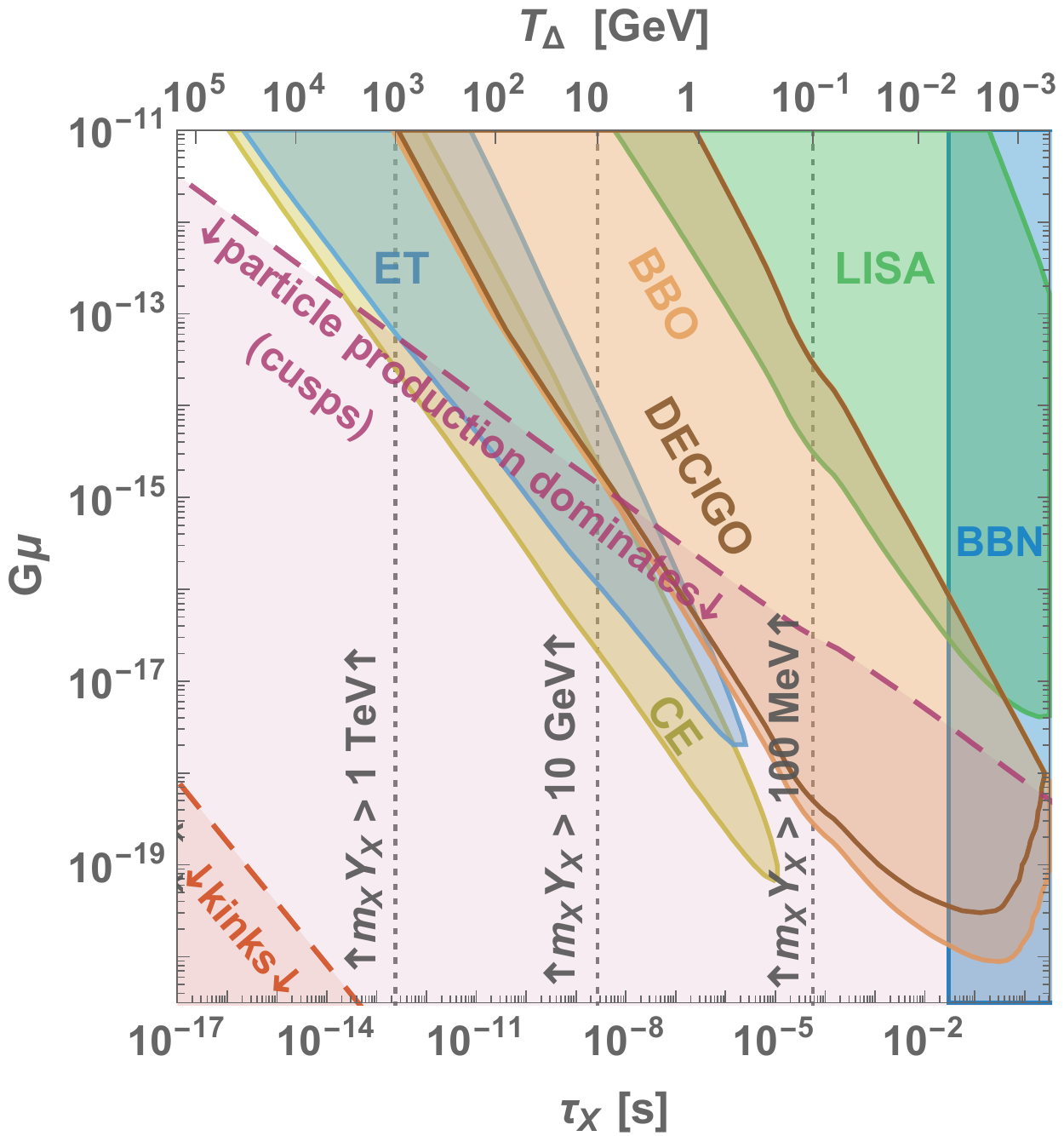}}}
\raisebox{0cm}{\makebox{\includegraphics[height=0.45\textwidth, scale=1]{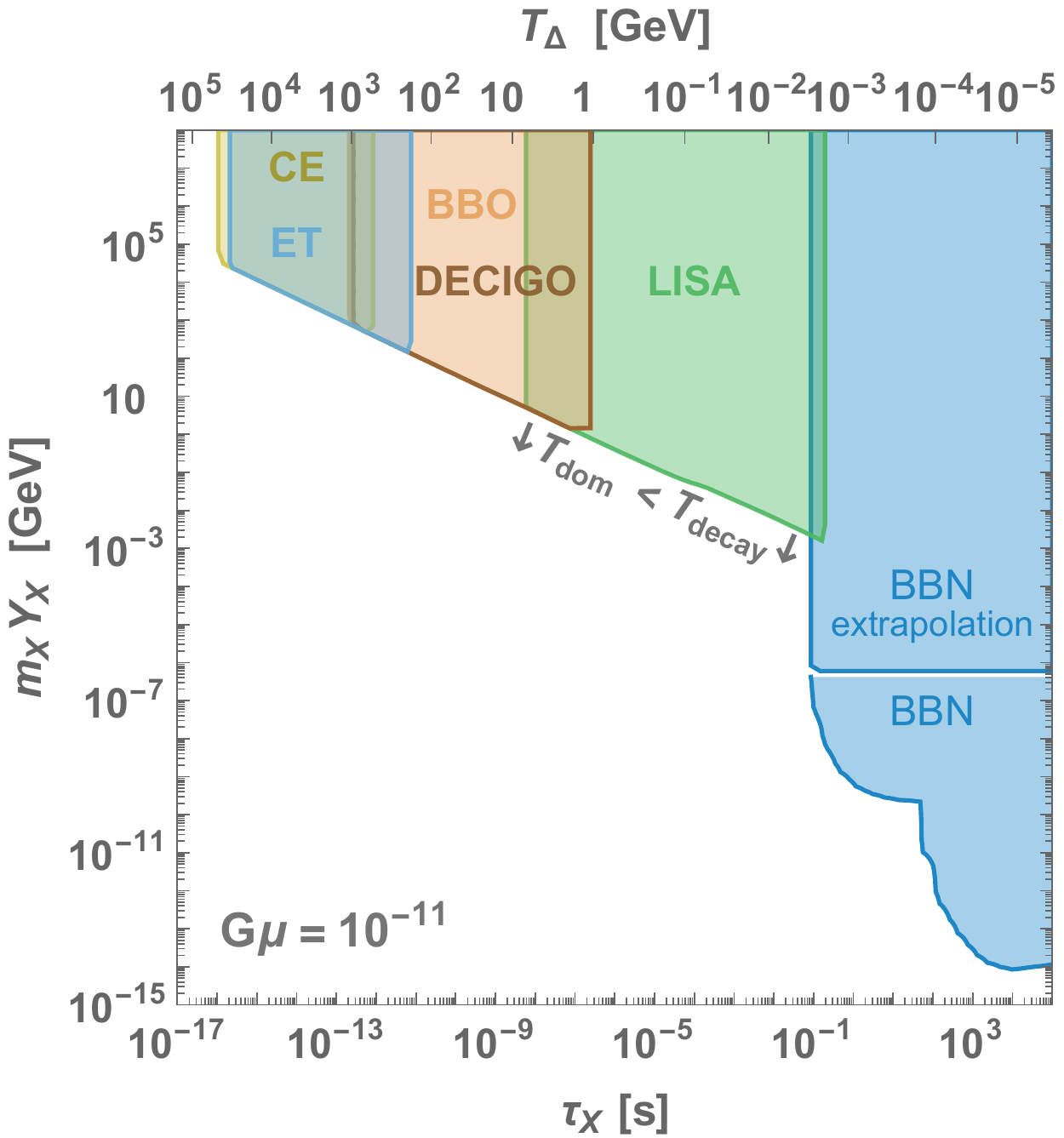}}}\vspace{0.1cm}
\raisebox{0cm}{\hspace{-0.0cm}\makebox{\includegraphics[height=0.45\textwidth, scale=1]{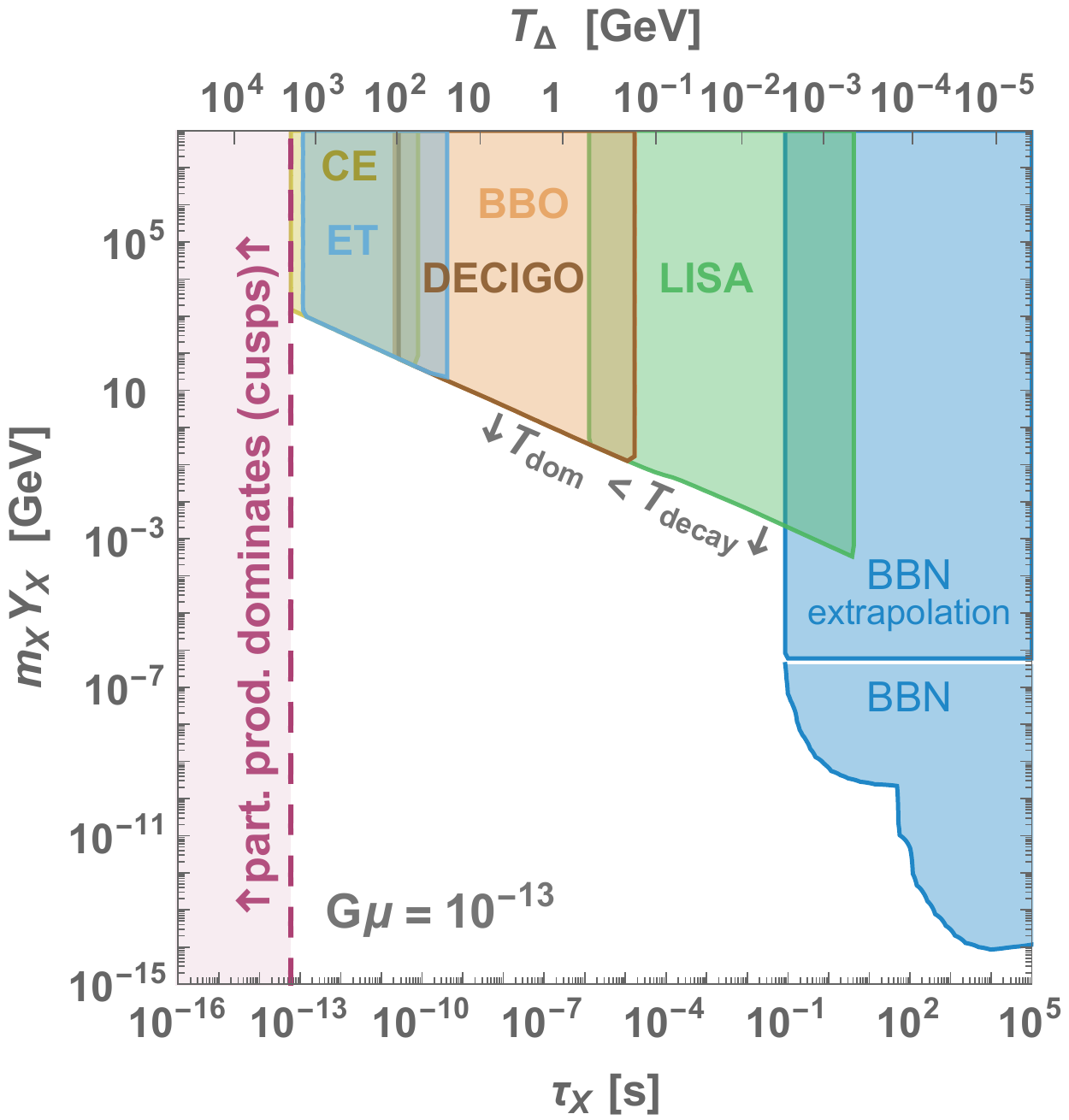}}}
\raisebox{0cm}{\hspace{0.0cm}\makebox{\includegraphics[height=0.45\textwidth, scale=1]{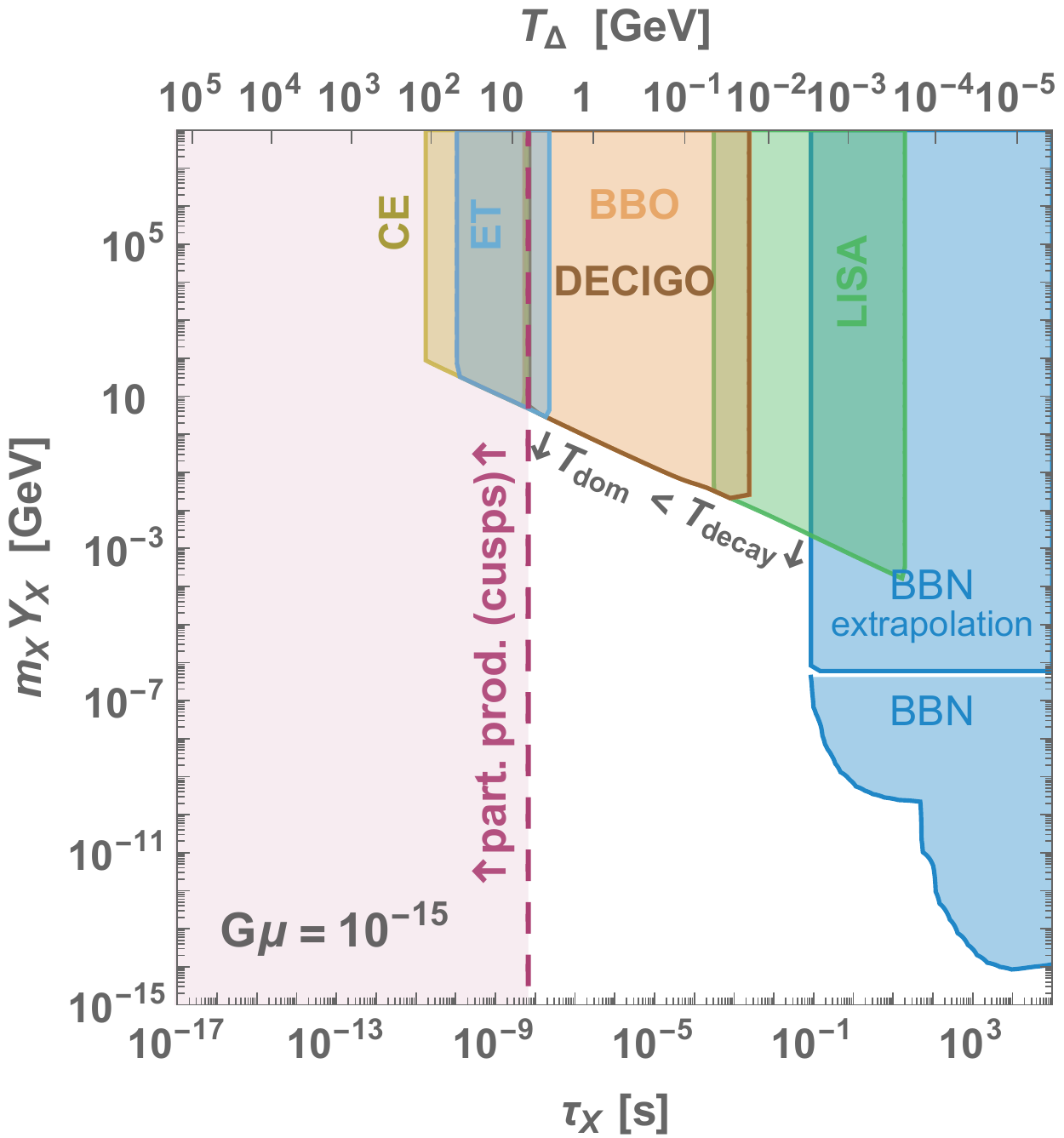}}}\vspace{0.1cm}
\raisebox{0cm}{\hspace{-0.0cm}\makebox{\includegraphics[height=0.45\textwidth, scale=1]{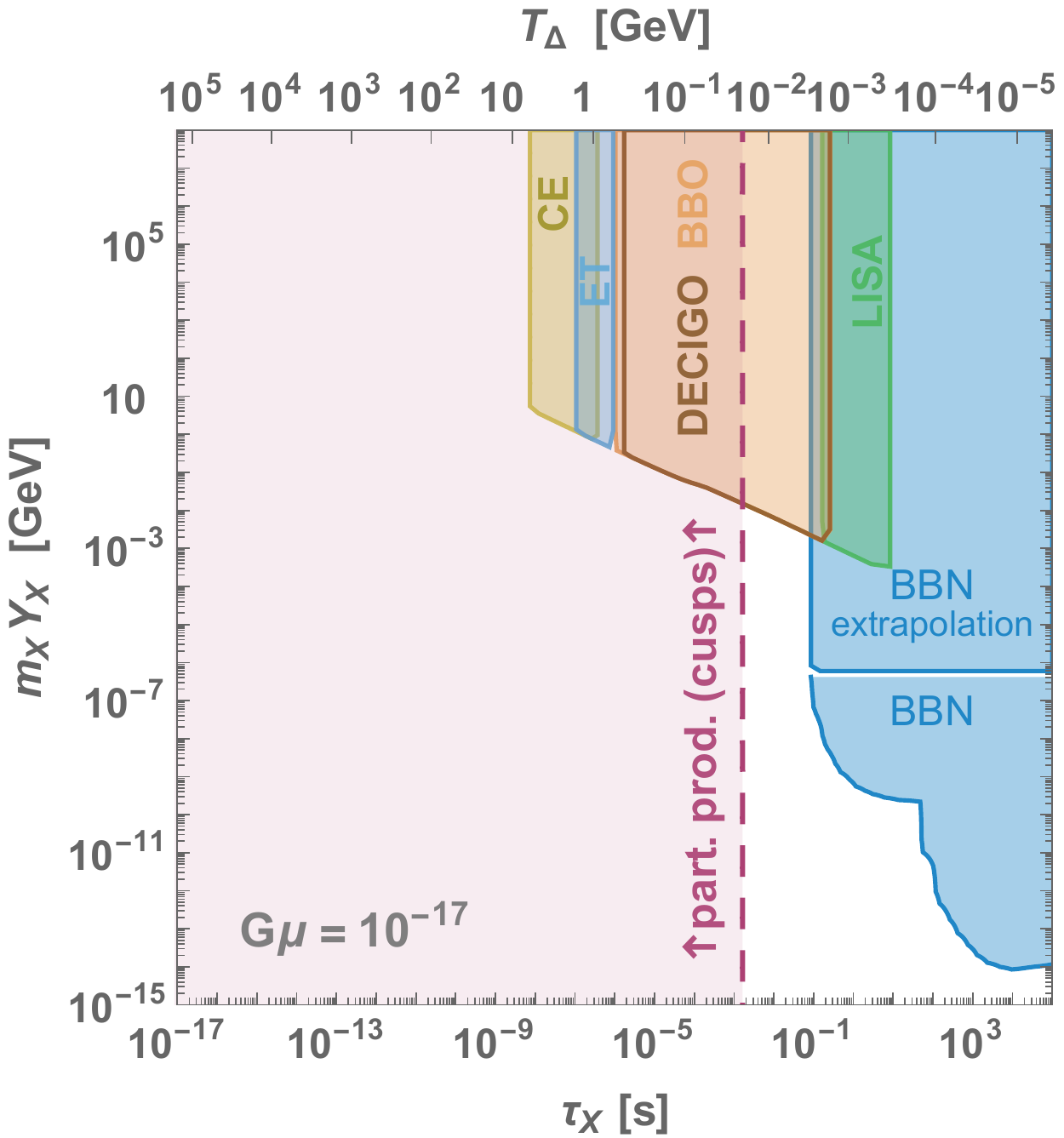}}}
\raisebox{0cm}{\hspace{0.0cm}\makebox{\includegraphics[height=0.45\textwidth, scale=1]{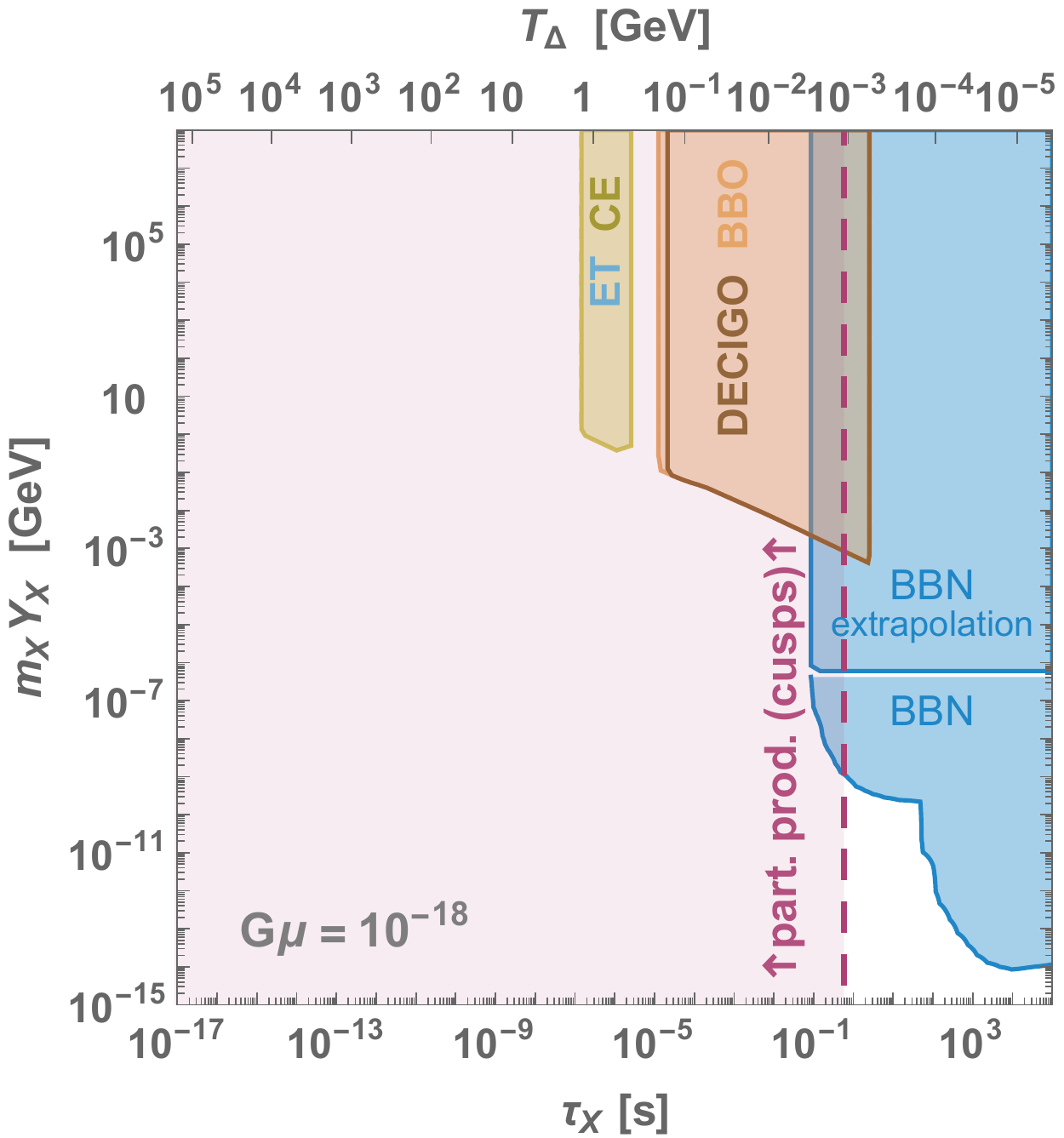}}}
\vspace{-0.3cm}
\caption{\it \small \label{fig:mY_tauX_GWI_VS_BBN} Constraints on the lifetime $\tau_X$ and would-be abundance $m_{X}Y_{X}$ of a heavy unstable particle inducing an early-matter era, assuming the observation of a SGWB from CS by a GW interferometer, cf. Sec.~\ref{eq:model_indpt}. We compare the new prospects with the current limits inferred from BBN \cite{Jedamzik:2006xz, Jedamzik:2009uy, Kawasaki:2017bqm}. We assume the detectability of the turning point in the GW spectrum at the frequency $f_{\Delta}$, induced by the decay of the particle at $T_{\rm dec}=T_{\Delta}$, cf. \textit{turning-point description Rx 1} in Sec.~\ref{sec:triggerMatterGW}. Limitation due to particle production in the cusp-domination case \cite{CS_VOS} are shown in purple.}
\end{figure}

\subsection{Model-independent constraints on particle physics parameters}
\label{eq:model_indpt}

A matter-dominated era may result from an oscillating scalar field  \cite{Turner:1983he}, such as a moduli field, or a relativistic plasma with a non-vanishing tensor bulk viscosity \cite{ Boyle:2007zx}, or simply a massive particle dominating the energy density of the universe. A matter-dominated era may be motivated by the possibillity to enhance structure growth at small scales, since density perturbations start to grow linearly earlier \cite{Erickcek:2011us, Fan:2014zua}, hence boosting the dark matter indirect detection signals  \cite{Delos:2019dyh,Blanco:2019eij}, or the possibility to enhance the primordial black holes production \cite{Polnarev:1986bi, Green:1997pr, Georg:2016yxa}.

We suppose an early-matter era is caused by the energy density of a cold particle $X$, meaning that $X$ is non-relativistic and is decoupled chemically and kinetically from the visible sector. The energy density $m_X n_X$ of $X$ dominates over the energy density of the SM radiation, with entropy $s_{\rm SM}$, at the temperature $T_{\rm dom}$
\begin{equation}
\label{eq:Tdom_def}
T_{\rm dom}=\frac{4}{3} m_{X} Y_{X}, \qquad Y_{X} \equiv n_{X}/s_{\rm SM}.
\end{equation}
Then, the cold relic decays when  its lifetime $\tau_X$ is equal to the age of the universe, corresponding to the temperature
\begin{equation}
\label{eq:T_dec_tauX}
T_{\rm dec} =1~\text{GeV} \left( \frac{80}{g_{\rm SM}}\right)^{1/4}\left( \frac{2.7\times 10^{-7}~\text{s}}{\tau_X} \right)^{1/2}.
\end{equation}
Note that the above relation between $T_{\rm dec}$ and $\tau_X$ only assumes that the decay is followed by a radiation dominated era and is independent of the previous thermal history of the universe.
$T_{\rm dec}$ is sometimes referred, mistakenly though \cite{Scherrer:1984fd}, as the reheating temperature following the decay.
We propose to use the third generation of GW interferometers to constrain cold relics responsible for early-matter domination. 
The constraints we will derive rely on the following assumptions:
\begin{itemize}
\item[1)] A SGWB from CS with tension $G\mu$ is measured by a GW interferometer $i$. 

\item[2)] The cold particle is abundant enough to lead to a matter-dominated era before it decays
\begin{equation}
T_{\rm dom} > T_{\rm dec} \ ,
\end{equation} 
where $T_{\rm dom}$ and $T_{\rm dec}$ satisfy Eq.~\eqref{eq:Tdom_def} and Eq.~\eqref{eq:T_dec_tauX}.

\item[3)]
The prescription Rx 1 of Sec.~\ref{sec:triggerMatterGW} is used, i.e.  the turning point in the GW spectrum  is in the observation window of the detector and 
\begin{equation}
\Omega_{\rm GW}(f_{\Delta}(T_{\rm dec}, \, G\mu), \, G\mu) h^2 \, > \,  \Omega_{\rm sens}^{(i)}h^2,  
\end{equation}
where $\Omega_{\rm GW}(f,\, G\mu)h^2$ is the predicted scale-invariant GW spectrum  from  Eq.~\eqref{eq:SGWB_CS_Formula}, and $ \Omega_{\rm sens}^{(i)}h^2$ is the power-law sensitivity curve of the detector $i$. 

\end{itemize}
Fig.~\ref{fig:mY_tauX_GWI_VS_BBN} shows  these new constraints  in comparison with the  current complementary constraints from BBN, usually represented in the plane $(\tau_{X}, \, m_{X}Y_{X})$ \cite{Jedamzik:2006xz, Jedamzik:2009uy, Kawasaki:2017bqm}.
We can translate the sensitivity of each interferometer to probe the particle lifetime into typical mass windows, assuming some decay width.

\subsection{Heavy dark photons}
\label{sec:U(1)_model}

\paragraph{The $U(1)_D$ dark photon:}

We consider a $U(1)_{\mathsmaller{\rm D}}$ gauge boson, $V_{\mu}$, the dark photon, of mass $\mV$, kinematically coupled to the $U(1)_{\mathsmaller{\rm Y}}$ gauge boson of the SM \cite{Holdom:1985ag, Foot:1991kb}
\begin{equation}
\mathcal{L} \supset - \frac{\epsilon}{2c_{w}} F_{\mathsmaller{\rm Y}\mu\nu} F_{\mathsmaller{\rm D}}^{\mu\nu},
\label{lagrangian_U(1)_D_model_DP}
\end{equation}
where $c_w$ is the cosine of the weak angle and $\epsilon$ is the dark-SM coupling constant. The decay width into SM, $\Gamma_{V}$, is computed in \cite{Cirelli:2018iax}. We here report the expression for $\mV \gtrsim 2 m_{Z}$
\begin{equation}
\Gamma_{V} \simeq \left(3\times 10^{-8}~\textrm{s}\right)^{-1} \, \left(\frac{\epsilon}{10^{-9}}\right)^2 \left( \frac{\mV}{1~\text{TeV}}\right).
\end{equation}
The dark photon leads to an early-matter-dominated era if it has a large energy density $\mV\, Y_V \gtrsim 10$ GeV and a long lifetime $\tauV \sim 10^{-8}$~s, cf. Fig.~\ref{fig:mY_tauX_GWI_VS_BBN} at $G\mu = 10^{-11}$.
Supposing that the dark photon abundance is close to thermal, $Y_{\mathsmaller{\rm V}} \sim 0.02$, cf. Eq.~\eqref{eq:DP_comoving_number}, this implies $\epsilon \lesssim 10^{-9}$. At such a low $\epsilon$, the dark sector and the SM sector may have never been at thermal equilibrium (cf. \cite{Hambye:2019dwd} or footnote 8 in \cite{Cirelli:2016rnw}) and may have their own distinct temperature. We assume that the dark sector and the SM have a different temperature by introducing the dark-to-SM temperature ratio \cite{Cirelli:2016rnw}
\begin{equation}
\label{eq:tempratio}
\rtilde \equiv \frac{\tilde{T}_D}{\tilde{T_{SM}}},
\end{equation} 
where quantities with a $\sim$ on top are evaluated at some high temperature $\tilde{T}$.
Thus, the dark photon abundance before its decay is given by 
\begin{equation}
\label{eq:DP_comoving_number}
Y_{\mathsmaller{\rm V}} = \frac{n_{\mathsmaller{\rm V}}}{\sSM} = \frac{45 \zeta(3)}{2\pi^4} \frac{\gtildeD}{\gtildeSM} \rtilde^3 \simeq 0.0169 \left(\frac{\gtildeD}{6.5}\right) \rtilde^3 ,
\end{equation}
where $\gtildeD$ and $\gtildeSM$ are the relativistic number of degrees of freedom in the dark sector and the SM at temperature $\tilde{T}$. Plugging Eq.~\eqref{eq:DP_comoving_number} into Eq.~\eqref{eq:Tdom_def} implies a simple relation between the temperature at which the dark photon dominates the universe $T_{\rm dom}$ and its mass $\mV$. 
We choose to be agnostic about the mechanism setting the abundances in the dark sector and we enclose all possibilities by introducing a dark-to-SM temperature ratio $\tilde{r}$\footnote{
Production of the dark photon in the early universe has been studied in the literature. For
a small kinetic mixing $\epsilon$, the abundance of the dark sector can be set non-thermally
either by freeze-in \cite{Hall:2009bx, Chu:2011be, Berger:2016vxi, Hambye:2019dwd}, or by a separate reheating mechanism. In the latter case, the temperature asymmetry in Eq.~\eqref{eq:tempratio} results from an asymmetric reheating \cite{ Hodges:1993yb,Berezhiani:1995am,Feng:2008mu,Adshead:2016xxj}. For moderate kinetic mixing $\epsilon \gtrsim 10^{-6} \sqrt{\MDM/\text{TeV}}$ \cite{Hambye:2019dwd}, the dark sector may have been at thermal equilibrium with the SM, but asymmetric temperatures can result from asymmetric changes in relativisitic degrees of freedom \cite{Cirelli:2016rnw}. On the other hand, a possibility for thermally equilibrating the $U(1)_D$ sector and the SM in the case  of a small kinetic mixing $\epsilon$ would be to introduce a dark Higgs $\phi$, mixing with the SM Higgs, which once at thermal equilibrium with SM, decays into dark photons.}

\begin{figure}[h!]
\centering
\raisebox{0cm}{\makebox{\includegraphics[height=0.5\textwidth, scale=1]{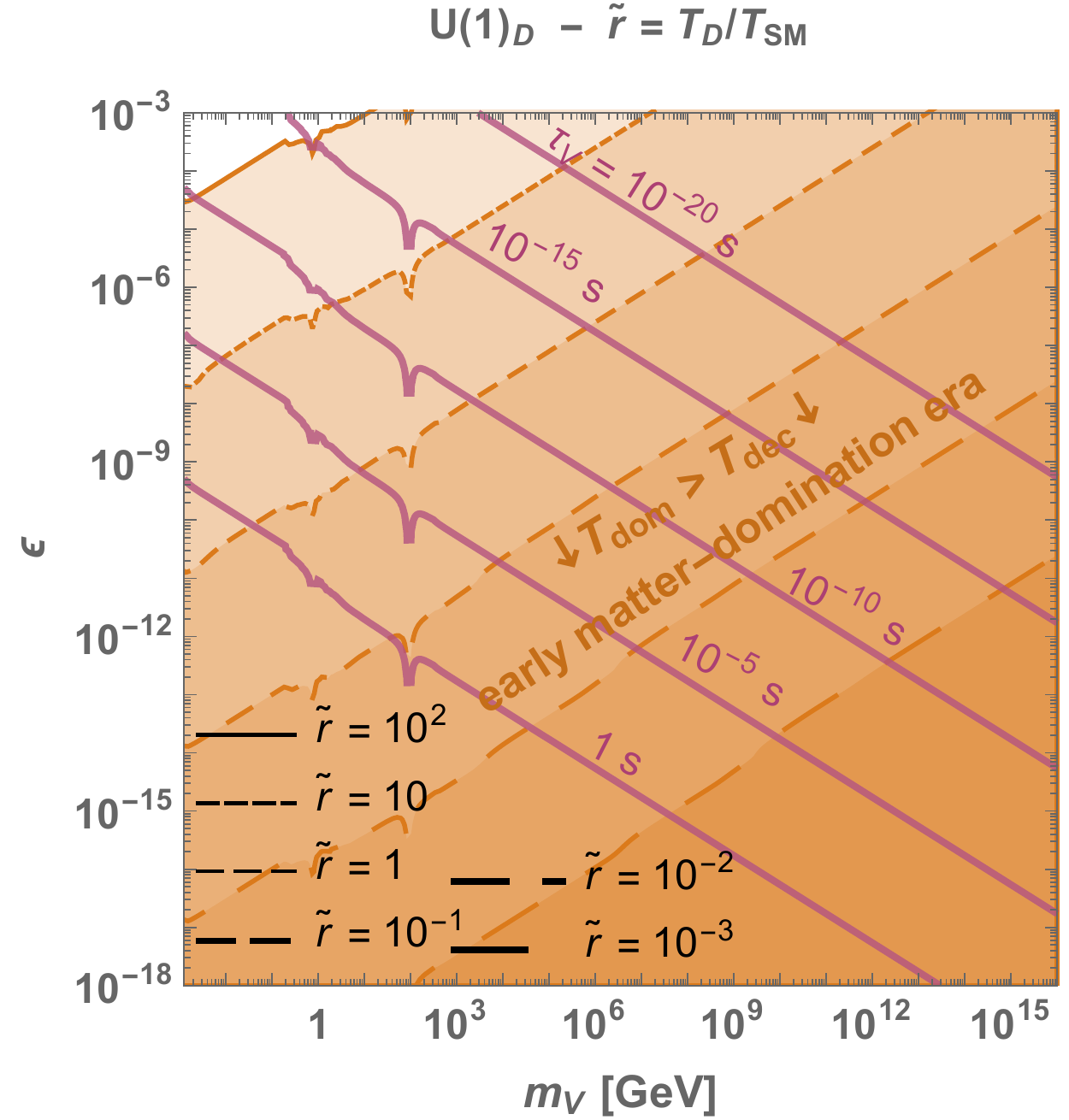}}}
\raisebox{0cm}{\makebox{\includegraphics[height=0.5\textwidth, scale=1]{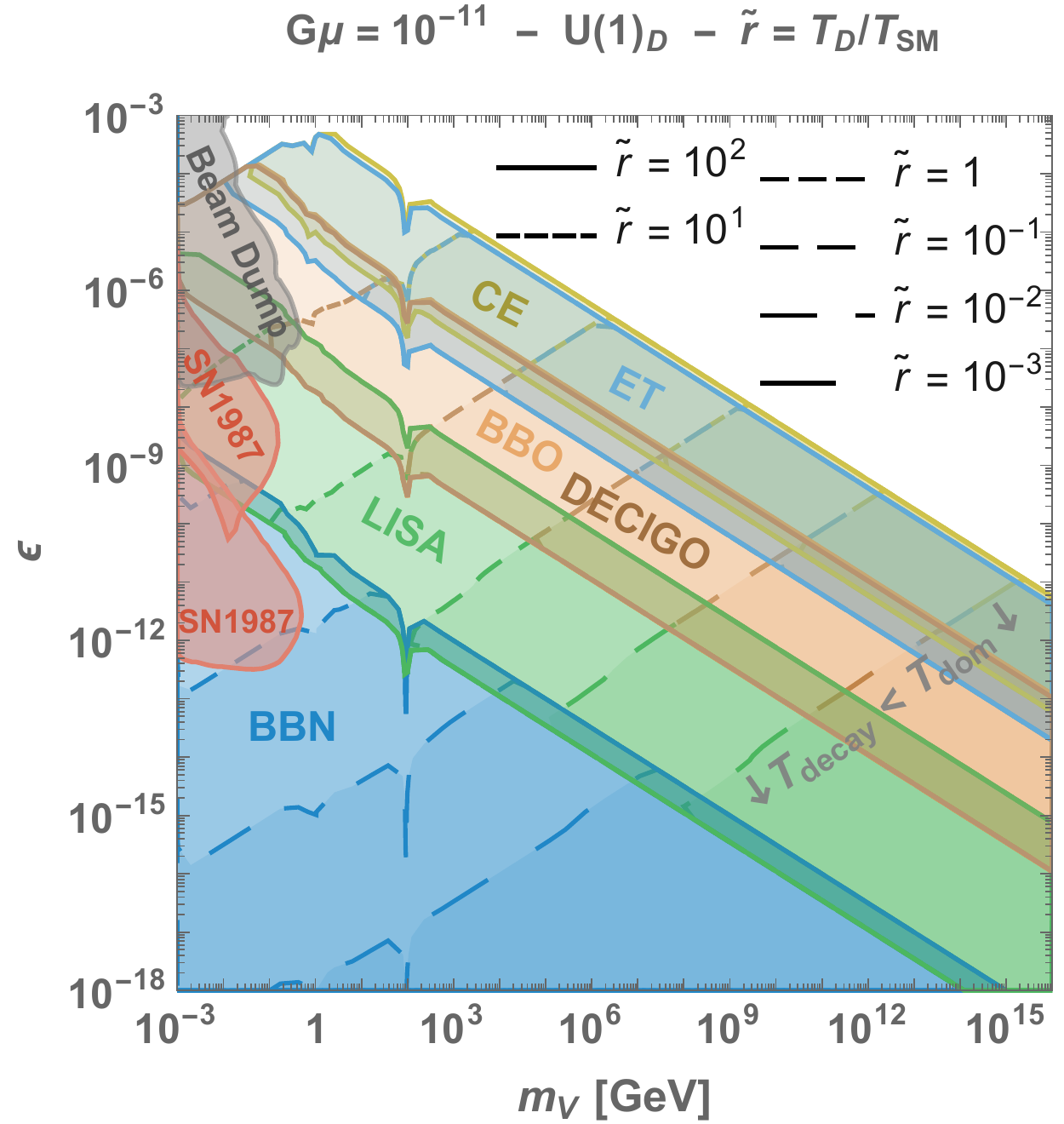}}}
\caption{\it \small \label{fig:GWconst_DarkPhoton1} \textbf{Left}: Constant dark photon lifetime $\tau_{V}$ contours. For a given dark-to-SM temperature ratio $\tilde{r}\equiv \TD/\TSM$, a non-standard early matter domination is induced below the corresponding orange line where the dark photon dominates the universe before it decays.
\textbf{Right}: Expected constraints on the dark photon mass $\mV$ and kinetic mixing $\epsilon$, assuming the measurement of a GW spectrum from CS with tension $G\mu = 10^{-11}$ by future GW interferometers. We use the \textit{turning-point prescription (Rx 1)} discussed in Sec.~\ref{sec:triggerMatterGW}.}
\end{figure}

As shown in the left panel of Fig.~\ref{fig:GWconst_DarkPhoton1}, low kinetic mixing $\epsilon$, large mass $\mV$ or large dark-to-SM temperature ratio $\tilde{r}$ lead to an early-matter-dominated era, triggered when $T_{\rm dom} \gtrsim T_{\rm dec}$. The non-detection with a future GW interferometer, of the imprint left by such a matter era, in the GW spectrum from CS, would exclude the existence of the dark photon for given values of the kinetic mixing, the dark photon mass and the dark-to-SM temperature ratio $(\epsilon, \, \mV, \, \tilde{r})$.  We show the GW-from-CS constraints on the dark photon in the right panel of Fig.~\ref{fig:GWconst_DarkPhoton1}, together with existing constraints coming from supernova SN1987 \cite{Kazanas:2014mca, Chang:2016ntp} and beam-dump experiments \cite{Cirelli:2016rnw}. Other constraints on lighter dark photons do not appear on the plot and are summarized in the reviews \cite{Jaeckel:2010ni,Essig:2013lka,Alexander:2016aln}.
We also include the BBN constraint which imposes the dark photon to decay before $\tauV \lesssim 0.1$~s \cite{Jedamzik:2006xz, Jedamzik:2009uy, Kawasaki:2017bqm} or later if the energy density fraction carried by the dark photon is smaller than $\sim 10\%$ \cite{Cirelli:2016rnw}.
Note, that only the BBN and the GW-from-CS constraints depend on the dark-to-SM temperature ratio $\tilde{r}$ which fixes the abundance of the dark photon in the early universe.

\begin{figure}[h!]
\centering
\raisebox{0cm}{\makebox{\includegraphics[height=0.48\textwidth, scale=1]{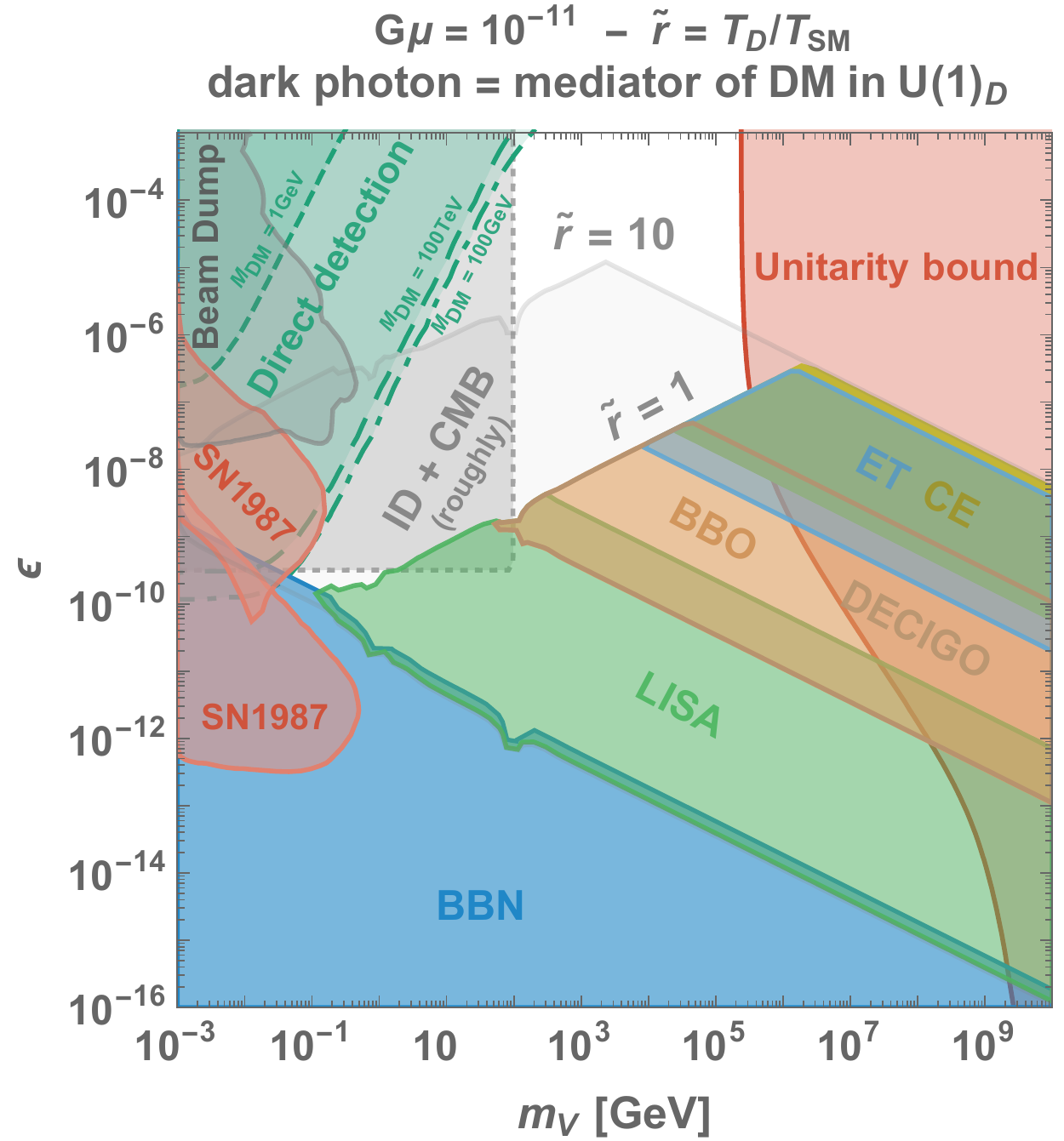}}}
\caption{\it \small \label{fig:GWconst_DarkPhoton2} 
Additional constraints when the dark photon is embedded in a DM model as the mediator of $U(1)_{D}$-charged DM (see text).
We compare the expected GW constraints from cosmic strings with the existing constraints on the $U(1)_{\rm D}$ DM model: Supernovae bounds from \cite{Kazanas:2014mca} and \cite{Chang:2016ntp}, direct detection bounds from \cite{Cirelli:2016rnw} and the indirect detection + CMB constraints are a rough estimate from \cite{Cirelli:2018iax}. Beam dump constraints are also taken from~\cite{Cirelli:2016rnw}. The unitarity bound on the DM mass $\MDM$ \cite{Griest:1989wd} can also be applied on the mediator mass because of the kinematic condition $\mV < \MDM$. The unitarity bound gets relaxed at small $\epsilon$ because of the larger entropy injection following the dark photon decay \cite{Cirelli:2018iax}. }
\end{figure}
We can appreciate the complementarity between the well-established supernova, beam dump, BBN constraints, and the expected constraints assuming the detection of a SGWB from CS by the GW interferometers. Indeed, whereas supernova and beam dump do not really constrain above $\mV \gtrsim 0.1$~GeV, the detection of a SGWB from CS with a string tension $G\mu \simeq 10^{-11}$ would exclude dark photon masses up to the maximal reheating temperature $\mV \sim 10^{16}$~GeV allowed by the maximal inflation scale $H_{\rm inf}\lesssim 6 \times 10^{13}$~GeV \cite{Ade:2018gkx, Akrami:2018odb}, and kinetic mixing as low as $\epsilon \sim 10^{-18}$.

\paragraph{The dark photon as a dark matter mediator:}
An interesting motivation for the dark photon is that it can play the role of a dark matter mediator. We can suppose that the dark sector also contains a Dirac fermion $\chi_{\mathsmaller{\rm D}}$ charged under $U(1)_{\mathsmaller{\rm D}}$, playing the role of DM \cite{Kors:2004dx, Feldman:2006wd, Fayet:2007ua, Ackerman:mha, Goodsell:2009xc, Morrissey:2009ur, Andreas:2011in, Goodsell:2011wn, Fayet:2016nyc, Cirelli:2016rnw, Cirelli:2018iax}
\begin{equation}
L \supset  \quad \bar{\chi}_{\mathsmaller{\rm D}}i\slashed{D}\chi_{\mathsmaller{\rm D}} - \MDM\bar{\chi}_{\mathsmaller{\rm D}}\chi_{\mathsmaller{\rm D}},
\label{lagrangian_U(1)_D_model}
\end{equation}
where $D_{\mu} = \partial_{\mu} + i g_{\mathsmaller{\rm D}} V_{\mathsmaller{\rm D}\mu}$ is the covariant derivative with $g_{\mathsmaller{\rm D}}$ the $U(1)_{\mathsmaller{\rm D}}$ gauge coupling constant. We suppose that the DM freezes-out by annihilating into pairs of dark photons,  we impose $\mV < \MDM$. We assume the dark photon to be non-relativistic when it decays but relativistic when it is produced, therefore, we set $\gtildeD = 3 + \frac{7}{8}\cdot 4 = 6.5$ in Eq.~\eqref{eq:DP_comoving_number}.

The unitarity bound on the DM mass $\MDM$ can be applied to the dark photon mass $\mV$ upon assuming $\mV < \MDM$. In the standard paradigm, the unitarity bound on s-wave annihilating dirac fermion DM is $\MDM \lesssim 140$~TeV \cite{Griest:1989wd, vonHarling:2014kha}. However, if long-lived and heavy, the decay of the mediator can, by injecting entropy, dilute the DM abundance and relax the unitarity bound to \cite{Cirelli:2018iax}
\begin{equation}
\MDM \lesssim 140~\text{TeV}~\sqrt{D},
\end{equation}
where $D$ is the dilution factor $D \simeq {T_{\rm dom}}/{T_{\rm dec}}$,
$T_{\rm dom}$ and $T_{\rm dec}$ are as defined in Eq.~\eqref{eq:Tdom_def} and Eq.~\eqref{eq:T_dec_tauX}.

In Fig.~\ref{fig:GWconst_DarkPhoton2}, we add the contraints on the dark photon when the later plays the role of the mediator of DM. They come  from direct detection \cite{Cirelli:2016rnw},  CMB \cite{Cirelli:2018iax},  indirect detection, using neutrino, gamma-rays, positrons-electrons and anti-protons \cite{Cirelli:2018iax}, as well as from unitarity \cite{Cirelli:2018iax}. 
They are complemented by the GW-from-CS constraints. For $\epsilon \lesssim 10^{-10}$, all the traditional indirect detection constraints evaporate and the unitarity bound is pushed to larger masses due to the entropy dilution following the dark photon decay such that the model is then currently only constrained by BBN. It is remarkable that  GW interferometers could probe this unconstrained region where $\epsilon<10^{-10}$ and $m_{V} > 1 $~GeV. In Fig.~\ref{fig:GWconst_DarkPhoton1} and ~\ref{fig:GWconst_DarkPhoton2}, we use the \textit{turning-point prescription (Rx 1)} discussed in Sec.~\ref{sec:triggerMatterGW}. Constraints are stronger when considering the \textit{spectral-index prescription (Rx 2)}, as shown in Fig.~\ref{fig:FullMD}.

\begin{figure}[h!]
\centering
\raisebox{0cm}{\makebox{\includegraphics[height=0.5\textwidth, scale=1]{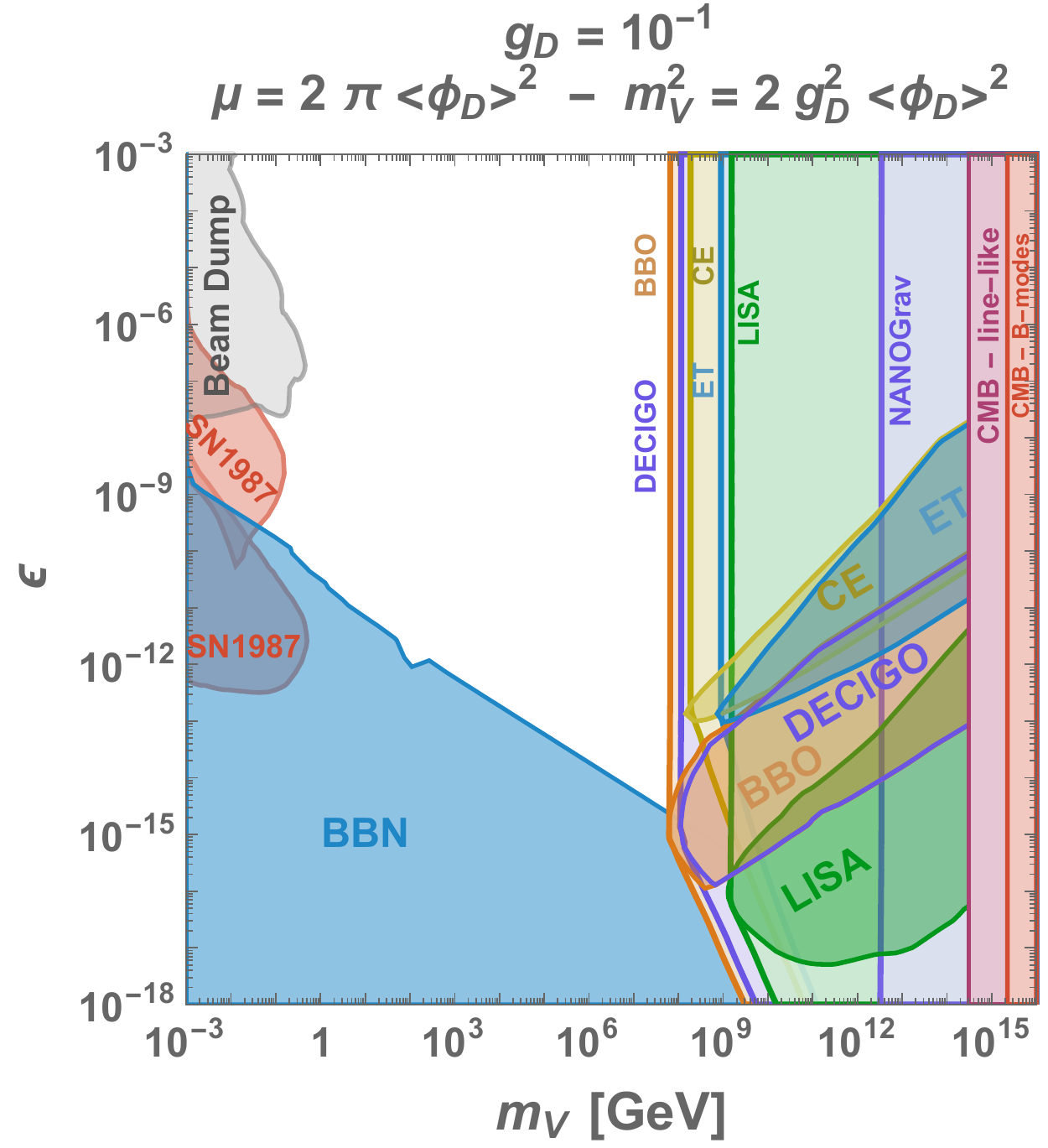}}}
\raisebox{0cm}{\makebox{\includegraphics[height=0.5\textwidth, scale=1]{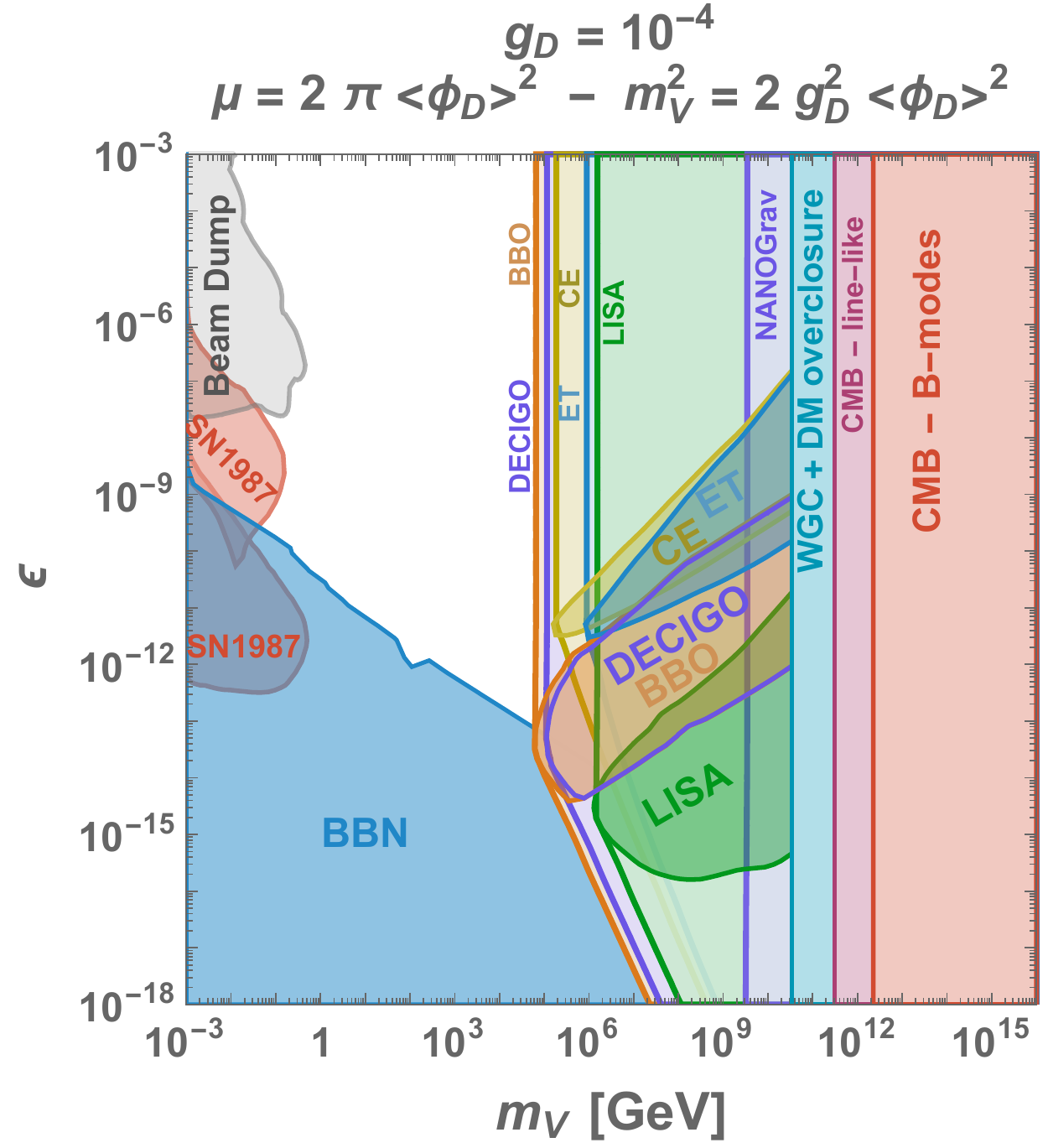}}}
\caption{\it \small \label{fig:U1_MD_gD} Scenario where the dark photon mass $\mV$ and the cosmic string network are generated by the spontaneous breaking of the same $U(1)$ symmetry, such that $\mV$ is related to the string tension $\mu$.  \textbf{Pale colors}: Constraints on the dark photon parameter space assuming the mere detection of the GW spectrum from CS by NANOGrav, LISA, ET, CE, DECIGO and BBO. \textbf{Opaque colors}: Constraints assuming the detection of the turning point in the GW spectrum induced by the transition from matter to radiation when the heavy dark photon decays. When combining Eq.~\eqref{eq:f_delta_RD} and Eq.~\eqref{eq:T_dec_tauX}, this last detection allows to measure the dark photon lifetime. The constraints described in the following part of this caption are independent of the GW emission. \textbf{Pale red}: The non-observation of the fundamental tensor B-modes in the CMB imposes the stringest upper bound on the energy density scale of inflation \cite{Akrami:2018odb}, $V_{\rm inf}  \lesssim 1.6 \times 10^{16}~\rm GeV$. This provides an upper-bound on the reheating temperature, which also must satisfy $\left< \phi_{\mathsmaller{\rm D}} \right> \lesssim T_{\rm reh}$ in order for the string network to be formed. Thus, we impose the CS formation to occur after the end of inflation with the following criteria: $\left< \phi_{\mathsmaller{\rm D}} \right> \lesssim V_{\rm inf}$.
\textbf{Pale purple}: Constraints from the non-observation of line-like temperature anisotropies in the CMB, e.g. \cite{Lizarraga:2016onn}, $G\mu \lesssim 2\times 10^{-7}$. \textbf{Pale sky blue}: In order to prevent DM overclosure, we must assume the $U(1)_D$ charged states to be heavier than the reheating temperature such that they are never produced. A possibility which is constrained by the Weak Gravity Conjecture, cf. main text, thus we impose $\left< \phi_{\mathsmaller{\rm D}} \right> \lesssim g_{\mathsmaller{\rm D}}M_{\rm pl}$. The last inequality implicitly assumes $\left< \phi_{\mathsmaller{\rm D}} \right> \lesssim T_{\rm reh} $. Note however that such a charged state could be unstable, e.g. if it is a dark Higgs, in which case the WGC constraint is relaxed. }
\end{figure}

\paragraph{Scenario where the cosmic string network and the dark photon mass have the same origin:}

As a last remark, we comment on the case where the spontaneous breaking of the $U(1)_D$ symmetry would be responsible for the formation of the cosmic string network, so that the dark photon mass is no longer a free parameter but is related to the string tension $\mu$, through the Abelian-Higgs relations \cite{Vilenkin:2000jqa} 
\begin{align}
\label{eq:mu_phi}
\mu = 2\pi <\phi_{\mathsmaller{\rm D}}>, \\
\label{eq:mV_phi}
\mV^2 = 2 g_{\mathsmaller{\rm D}}^2 <\phi_{\mathsmaller{\rm D}}>^2,
\end{align}
where $\phi_{\mathsmaller{\rm D}}$ is the scalar field whose vacuum expectation value $<\phi_{\mathsmaller{\rm D}}>$ breaks the $U(1)_D$ symmetry spontaneously. 

In this case, we find that most of the relevant parameter space is ruled out due to overabundance of dark matter.\footnote{The cross-section of a pair of $U(1)_D$ fermions annihilating into dark photons is given by $\sigma v \simeq \pi \alpha_{\mathsmaller{\rm D}}^2/m_{\psi}^2$ with $\alpha_{\mathsmaller{\rm D}}=g_{\mathsmaller{\rm D}}^2/4\pi$. It is way too weak to prevent universe overclosure, except if we tune the Yukawa coupling of the fermion, $\lambda$, defined by $m_{\psi}=\lambda \left< \phi \right>/\sqrt{2}$, to very small values.} The only viable solution would be to assume that the states which are charged under $U(1)_D$ and stable under decay, are heavier than the reheating temperature such that they are never produced. 
The Weak Gravity Conjecture (WGC) requires the existence of a charged state with mass smaller than \cite{ArkaniHamed:2006dz}
\begin{equation}
m_{\rm X} \lesssim g_{\mathsmaller{\rm D}} M_{\rm pl}.
\end{equation}
Hence, $g_{\mathsmaller{\rm D}} M_{\rm pl}$ sets the maximal reheating temperature, above which charged states responsible for universe overclosure might be produced.
Therefore, we should exclude the parameter space where the temperature of the $U(1)_D$ spontaneous breaking, taken as $\sim\left< \phi_{\mathsmaller{\rm D}} \right>$, is heavier than $g_{\mathsmaller{\rm D}} M_{\rm pl}$, cf. pale sky blue region in right plot of Fig.~\ref{fig:U1_MD_gD}.
Note that the WGC does not specify if the suggested charged state is stable under decay or not. For instance, it would be stable and overclose the universe if it is a $U(1)_D$ fermion but not if it is a $U(1)_D$ Higgs, which can still decay into a dark photon pair when $m_{\phi_{\mathsmaller{\rm D}}}\gtrsim 2\mV$. Hence, the WGC constraint in our parameter space has to be taken with a grain of salt.

Assuming a natural gauge coupling value, $g_{\mathsmaller{\rm D}}=10^{-1}$, we find that dark photons heavier than $\gtrsim 100~\rm PeV$ would be accompanied with a $U(1)_D$ cosmic string network producing an observable GW spectrum, see left plot of Fig.~\ref{fig:U1_MD_gD}.
In the case where $g_{\mathsmaller{\rm D}}=10^{-4}$, we could probe dark photon masses down to $\gtrsim 100$~TeV, see right plot of Fig.~\ref{fig:U1_MD_gD}.

On the same plot, we superpose the constraints, shown with pale colors, coming from the simple observation of the GW spectrum with future experiments (except NANOGrav which is already operating), and the constraints, shown with opaque colors, coming from the detection of the turning point where the spectral index of the GW spectrum changes due to the decay of the dark photon which was dominating the energy density of the universe.  The former detection would allow to measure the dark photon mass whereas the latter detection would allow to access its lifetime.

\section{Supercooled Composite Dark Matter}
\label{sec:SC_DM_CS}

\begin{figure}[h!]
			\centering
			\raisebox{0cm}{\makebox{\includegraphics[height=0.45\textwidth, scale=1]{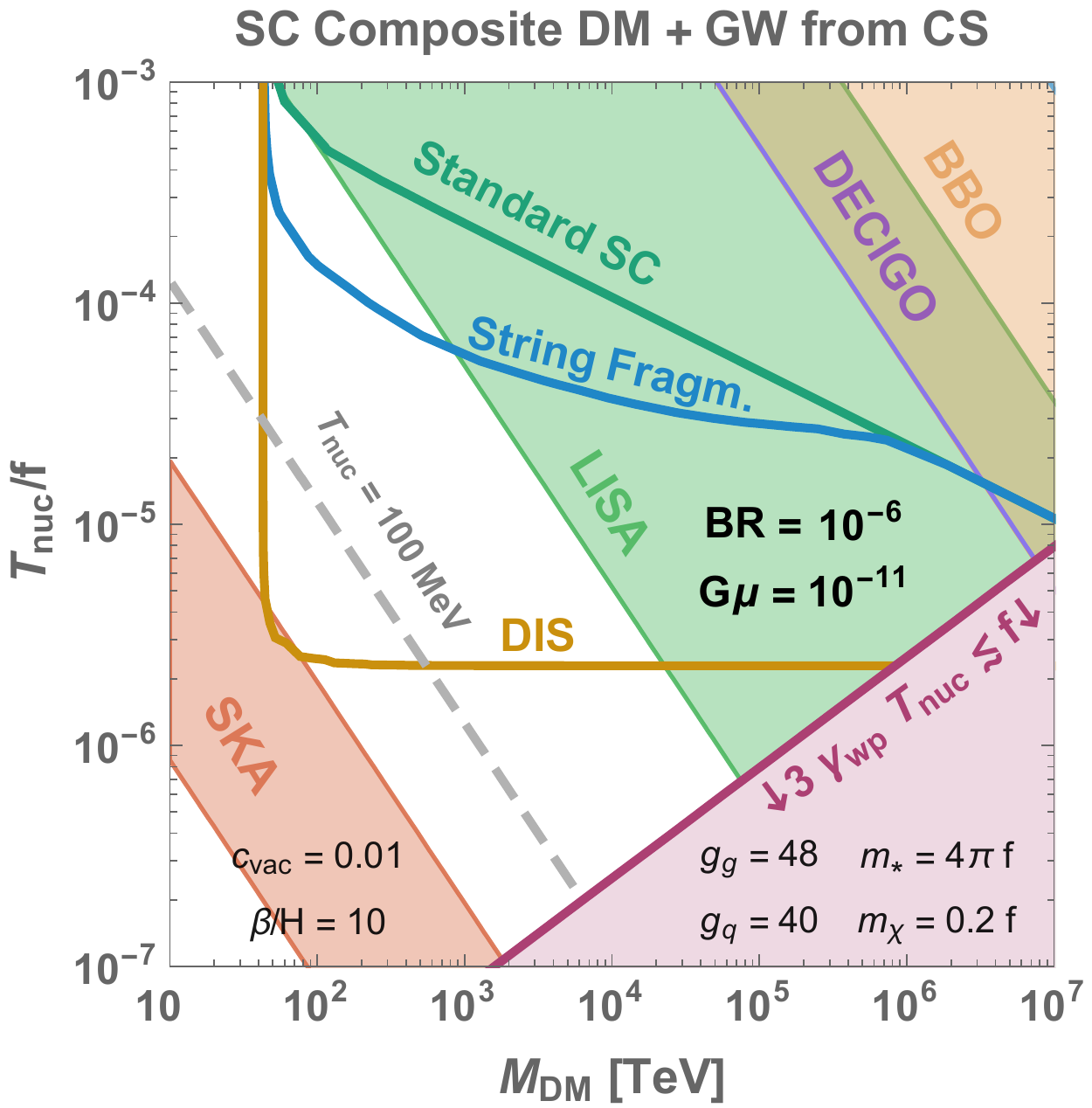}}}
			\raisebox{0cm}{\makebox{\includegraphics[height=0.45\textwidth, scale=1]{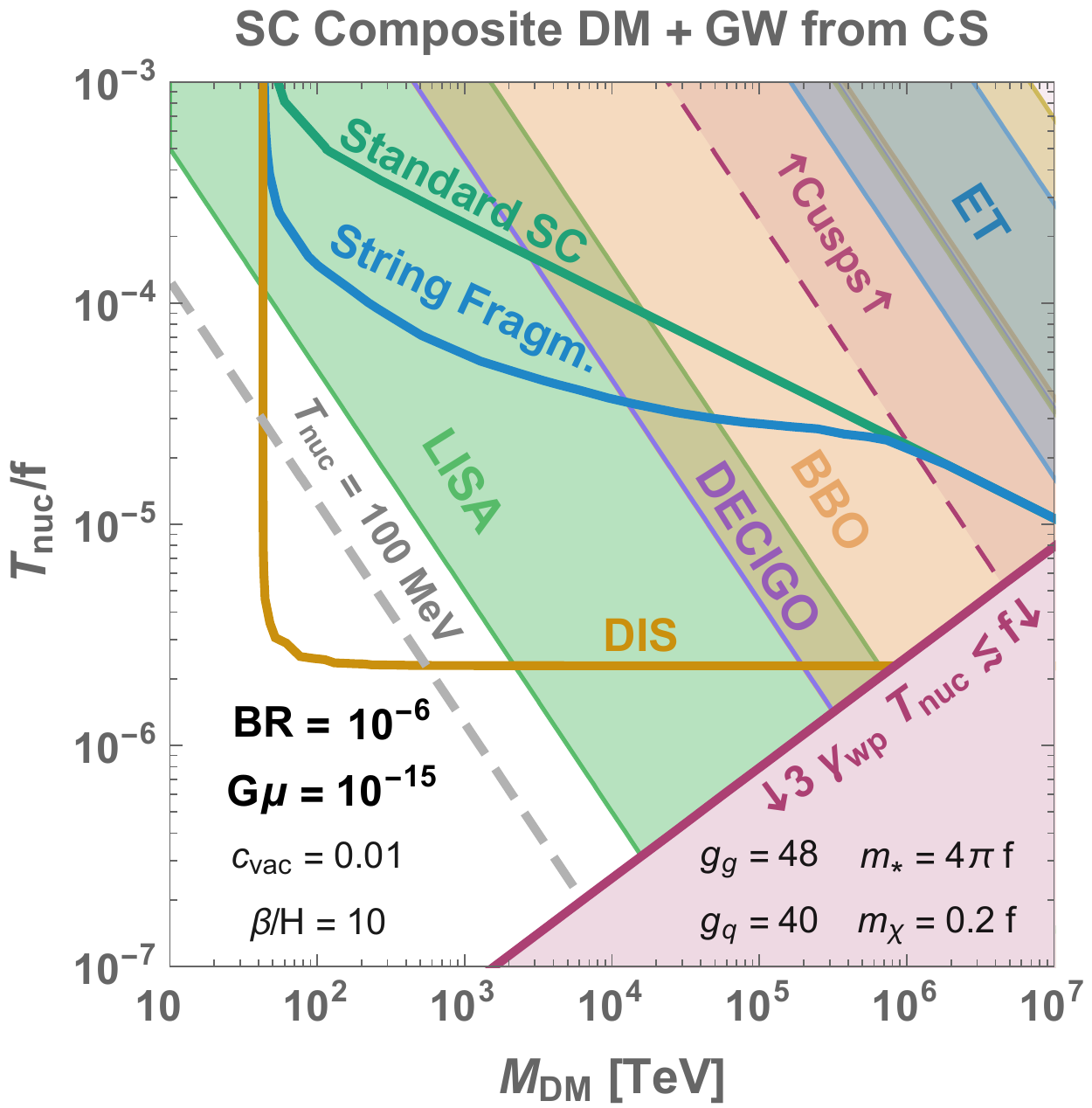}}}
		\caption{\it \small Regions of the parameter space of Supercooled Composite DM, cf. Chap.~\ref{fig:SCDM_GWCS}, which can be probed with the detection of a GW spectrum from CS with tension $G\mu = 10^{-11}$ (left) and $G\mu = 10^{-15}$ (right) by future GW observatories. We have considered the detectability of the turning point due to the change of cosmology, cf. turning-point prescription (Rx 1) in Sec.~\ref{sec:triggerMatterGW}. If instead, we use the detectability of the slope of the SGWB, cf. spectral-index description (Rx 2), the reach would be considerably enlarged to lower DM masses $\MDM$.  This figure is not contained in \cite{Gouttenoire:2019rtn} but instead it has been realized during the preparation of the thesis manuscript.}
			\label{fig:SCDM_GWCS}
		\end{figure}
				
In Fig.~\ref{fig:contour_power_inflation1}, \ref{fig:contour_power_inflation2} and \ref{fig:contour_power_inflation3} of the previous chapter, we were able to provide the energy scale and number of e-fold of a short period of inflation which can be probed by detecting the presence of a turning point in the GW spectrum from CS.
In Fig.~\ref{fig:SCDM_GWCS}, we give the corresponding parameter space of the model of Supercooled Composite DM, studied in Chap.~\ref{sec:supercool_potential} and published in \cite{Baldes:2020kam}. Indeed, in this model, DM of mass $\MDM \simeq 4\pi f$ is formed during a supercooled confining first-order phase transition with confining scale $f$ and nucleation temperature $\Tnuc$. The corresponding inflationary scale and number of e-folds are
\begin{equation}
E_{\rm inf} \simeq f,
\end{equation}
 and 
 \begin{equation}
 N_e = \textrm{log}~ \frac{f}{\Tnuc}.
 \end{equation}
We can see that a part of the parameter space leading to the correct DM abundance could be probed with GW interferometers if a GW spectrum from CS is detected. We stress that those constraints assume the turning-point prescription (Rx 1) defined in Sec.~\ref{sec:triggerMatterGW} and would be considerably enlarged to smaller DM masses if instead one assumes the spectral-index prescription (Rx 2). The advantage of the first prescription Rx 1 is that it allows to measure the temperature when the strings re-enter the Hubble horizon, cf. Eq.~\eqref{eq:Tre}
\begin{equation}
	T_\textrm{re}\simeq \frac{E_\textrm{inf}}{(0.1)\,g_*^{1/4}(T_\textrm{re})\,\exp(N_e)} \simeq \frac{\Tnuc}{(0.1)\,g_*^{1/4}(T_\textrm{re})}, 
\end{equation}
and then the nucleation temperature $\Tnuc$.\footnote{In the presence of multi-detection by different GW observatories of a SGWB from CS, the spectral-index prescription (Rx 2), defined in in Sec.~\ref{sec:triggerMatterGW}, could be enough to measure $T_\textrm{re}$.}
Some of the reachable parameter space falls in the region where the techni-quark can not trivially enter inside the bubbles of confined phase and where our modeling does not apply (purple). We leave the study of this regime for further studies. The detectability of SGWB generated by bubble collision are planned for a future study \cite{Baldes:2021aph}. The present section Sec.~\ref{sec:SC_DM_CS} was not included in any of the publications but instead it has been added during the preparation of the manuscript.

\section{Summary }

If future GW observatories have the sensitivity to detect stochastic GW backgrounds of primordial origin and to measure precise features in this spectrum, they can reveal very unique information about very high scale physics.
Particularly relevant sources of GW are cosmic strings. 
Cosmic strings are almost ubiquitous in many Grand-Unified Theories.  As they keep emitting throughout the whole cosmological history of the universe,  the resulting GW spectrum covers a wide range of frequencies and can be detected either by space-based or ground-based observatories.
An early era of matter domination due to new heavy particles or an intermediate period of inflation due to a supercool first-order phase transition generate clear signatures in the GW spectrum of cosmic strings. 

First, in this study we assume the existence of an early matter era due to the presence of a cold particle $X$ temporarily dominating the energy density of the universe and decaying before the onset of BBN. We compute its impact on the GW spectrum of CS beyond the scaling regime. 
We show that detecting such a feature and interpreting it in terms of a new heavy relic can lead to unparalleled constraints in the  $(\tau_X,\,m_XY_X)$ (lifetime, yield) plane.
In Fig.~\ref{fig:mY_tauX_GWI_VS_BBN}, we provide model-independent constraints which extend the usual BBN constraints on the lifetime $\tau_X$ by $15$ orders of magnitude for $G\mu=10^{-11}$, as we are able to constrain early matter dominated era ending when the temperature of the universe is between $50$~TeV and $1$~MeV.

We show that this new search strategy is likely to provide unprecedented constraints on a model of dark photon kinematically coupled to the Standard Model hypercharge, possibly embedded in the $U(1)_D$ dark-photon-mediated dark matter model. The constraints we obtain from GW on $U(1)_D$ dark matter falls in the large mass/small kinetic mixing ballpark which is otherwise unreachable  by any current and probably future direct/indirect detection and CMB constraints, cf. Fig.~\ref{fig:GWconst_DarkPhoton2}.  
{At last, we consider the possibility that  the dark photon mass and the cosmic string network are generated by the spontaneous breaking of the same $U(1)$ symmetry and show that we can use future GW interferometers to probe dark photon masses above $100 ~\rm PeV$, or even down to the $\rm TeV$ scale if we tune the gauge coupling to small values, see Fig.~\ref{fig:U1_MD_gD}.}
These are only a few minimal examples of particle physics models generating early matter eras. There are many other  well-motivated models which would deserve consideration in this respect.
We will present  the corresponding  constraints on axion-like-particles and primordial black holes in a separate study.

Finally, we show that the period of vacuum domination preceding the formation of supercooled composite DM in Chap.~\ref{chap:SC_conf_PT}, can leave a detectable imprint on the GW spectrum of the would-be CS network present at the same time.


%

\xintifboolexpr { \x = 2}
  {
  }
{
\medskip
\small
\bibliographystyle{JHEP}
\bibliography{thesis.bib}
}

%% file: chap10.tex
\chapterimage{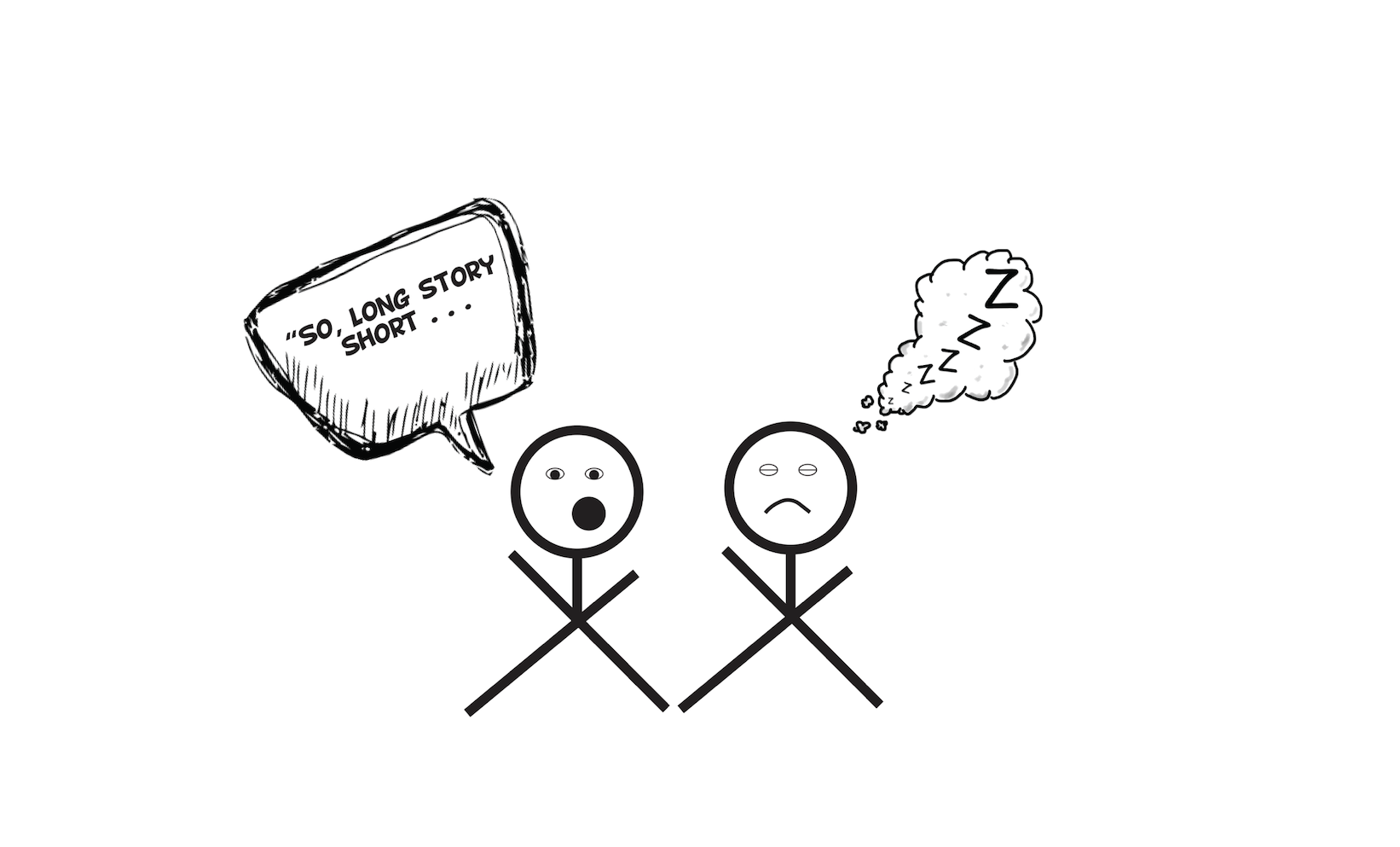} 
\chapter{Conclusion}



In the last two decades,  the LEP, Tevatron and LHC on one side, and WMAP and the Planck satellites on the other side, have confirmed the validity of the Standard Models of Elementary Particle and Cosmology with an unprecedented level of precision. However, the big questions, among which the nature of Dark Matter (DM) and the solution to the electroweak hierarchy problem, are left unanswered. 

The absence of experimental evidences in colliders, telescopes and laboratories for Weakly-Interacting Massive Particles (WIMPs) in the naturalness window $[100~\textrm{GeV},~\rm few~TeV]$ encourages searches beyond the TeV scale. A quest for such heavy particles is happening with currect telescopes Hess~II, Hawc, Veritas, Magic, Taiga, Antares, IceCube, Baikal, Ams, Calet, Dampe and will intensify with the future development of numerous multi-TeV telescopes, e.g.  Cta, Lhaaso, Km3net, Herd, Iss-Cream, and the high-sensitivity direct detection experiment Lux-Zeplin. 

The detection of Gravitational Waves (GW) by LIGO in 2015 has opened new directions for investigating compact massive objects in our universe such as black holes and neutron stars and for testing the laws of gravity in a new regime. At the same time, this new GW era is enabling new ways to explore high-energy particle physics beyond the Standard Model. LISA with peak sensitivity around $3~\rm mHz$ will offer a window on first-order phase transitions occurring around the electroweak scale while Einstein Telescope and Cosmic Explorer with peak sensitivity around $10~\rm Hz$ will offer views on the PeV scale. Additionally, the future detection of GW generated by a network of Cosmic Strings (CS) would allow to infer the equation of state of the universe up to $10^{-17}~$seconds after the Big-Bang.
 
On the theoretical side, new models motivating indirect-detection experiments beyond the WIMP windows but also GW experiments beyond the millihertz window are welcome.

\paragraph{Homeopathic Dark Matter.} 

We explored the possibility to have thermal Dark Matter with mass beyond the unitarity bound $\gtrsim 100$~TeV. We introduced a heavy unstable DM mediator which induces an early matter era and injects entropy when it decays into Standard Model. The dilution of the DM abundance allows to evade the unitarity bound. As a case of study, we considered the $U(1)_D$ model and studied the indirect-detection constraints, see Fig.~\ref{fig:ID_summary}. The use of a mediator avoids the necessity to resum electroweak Sudakov logarithms for computing cosmic-ray predictions from DM annhihilation.  We showed that the indirect-detection constraints are limited by the data range used by collaborations, and we hope that our study will motivate data analysis beyond the $100$~TeV threshold. We found that DM masses up to EeV are realizable, larger mass scales being constrained by big-bang nucleosynthesis. For such large mass scales, DM is far from being reached by standard methods: colliders, telescopes and laboratories. This motivates the use of GW from CS to detect the presence of an early matter era in the early universe.
 

\paragraph{Supercooled Composite Dark Matter.} 
We have explored an other possibility for having  thermal (or would-be thermal) DM beyond the unitarity bound. Instead of using the entropy injection due to reheating after a matter-dominated era, we have used the entropy injection due to reheating after a vacuum-dominated era. A well-known possibility to generate a short period of vacuum-domination relies on supercooled first-order phase transitions.  
Initially motivated by the electroweak hierarchy problem, we have considered a confining phase transition arising from a strongly-interacting sector.
We pointed out for the first time that when considering a confining phase transition with $\Tnuc/f \lesssim 0.1$ where $f$ and $\Tnuc$ are the confining scale and nucleation temperature respectively, cf. Fig.~\ref{fig:StdPictureRecovered}, the final abundance of composite states is enhanced by three new effects. First, the fragmentation of the string binding the incoming quark and the wall. Second, the ejection of a quark from the bubble in order to satisfy charge conservation. And third, deep-inelastic-scattering between the string fragments and the preheated soup arising from the decay of the scalar field after bubble wall collision. Supposing that DM is one of the composite states of the strong sector, we found that the needed amount of supercooling in order to fix the correct abundance is larger than in the non-confining scenario, see Fig.~\ref{fig:compositeDM_generic}. We obtained that DM masses up to EeV are realizable. The scenario with supra-EeV mass scales, where quarks are not energetic enough to enter the bubbles is left for further studies. In the follow-up \cite{Baldes:2021aph}, we have considered the case where the strong sector is successfully described by a light-dilaton effective-field-theory and we have studied the phenomenology of dilaton-mediated DM, including indirect detection, direct detection and GW. Far from the TeV scale, the indirect-detection and direct-detection constraints evaporate and GW is the only probe. It can either come from bubble collision or from the imprint of the vacuum-dominated era in the GW spectrum from CS.

\paragraph{Beyond the Standard Model with Cosmic Strings.} 
The presence of cosmic strings in our universe is motivated by many Grand Unification Theories, but also superstring theories. They are well-known to generate a GW spectrum over many decades in frequency which is flat during radiation-domination. We hope the paper \cite{Gouttenoire:2019kij} to become of standard reference for computing the GW spectrum from CS in standard cosmology or in the presence of an early matter-, kination- and inflation-era. As an input for the spectrum we computed the loop formation rate by solving numerically the evolution equation of the string network. We accounted for the deviation from the Nambu-Goto approximation due to the presence of small-scale structures which allow for the decay of string loops via particle production. We showed that the observation of a GW spectrum from CS could give information on the equation of state of the universe up to the PeV scale and even up to $10^{14}~$GeV in the special case of an inflationary era, see Fig.~\ref{fig:summary1}.

\paragraph{Probe Heavy Dark Matter with Gravitational Waves from Cosmic Strings.} 
The first two directions of research, homeopathic DM and supercooled composite DM, used the entropy injection following a matter- and vacuum-dominated era in order to have DM heavier than the unitarity bound at $100$~TeV. We showed that we can use the imprint of such non-standard cosmologies on the would-be GW spectrum from CS detected by future GW interferometers in order to probe models of heavy DM, see Fig.~\ref{fig:mY_tauX_GWI_VS_BBN} and \ref{fig:GWconst_DarkPhoton2} and \ref{fig:SCDM_GWCS}. 
This new method of detection can probe DM mass regions far beyond the reach of standard techniques based on colliders, indirect detection and direct detection.


%

{
\medskip
\small
\bibliographystyle{JHEP}
\bibliography{thesis.bib}
}